\documentclass[aps, prd,twocolumn,superscriptaddress,nofootinbib,preprintnumbers,floatfix]{revtex4-1}
\usepackage{adjustbox}
\usepackage{multirow}
\usepackage{amsmath}
\usepackage{array}
\usepackage{hyperref}
\usepackage{xfrac}
\usepackage[usenames,dvipsnames]{xcolor}
\usepackage{tablefootnote}

\newcommand{\aperp}{\alpha_\perp}	
\newcommand{\apara}{\alpha_\parallel}

\newcommand{\lya}{\mbox{Ly$\alpha$}}
\newcommand{\lyaf} {\lya\ forest}
\def\apjl{ApJL} 
\def\aj{AJ} 
\def\apj{ApJ} 
\def\pasp{PASP} 
\def\apjs{ApJS} 
\def\aap{A\&A} 
\def\nat{Nature} 
\def\mnras{MNRAS} 
\def\jcap{JCAP} 
\def\prd{{Phys.~Rev.~D}}        
\def\physrep{Physics Reports}   

\newcommand\LCDM{$\Lambda$CDM}
\newcommand\nuLCDM{$\nu\Lambda$CDM}
\newcommand\wCDM{$w$CDM}
\newcommand\wowaCDM{$w_0w_a$CDM}
\newcommand\oLCDM{o$\Lambda$CDM}
\newcommand\owCDM{o$w$CDM}
\newcommand\owowaCDM{o$w_0w_a$CDM}
\newcommand\nuwCDM{$\nu w$CDM}
\newcommand\nuowCDM{$\nu$o$w$CDM}
\newcommand\CMBTP{CMB T$\&$P}
\newcommand\Planck{{\it Planck}}
\newcommand\SN{SNe~Ia}

\usepackage{etoolbox}

\newtoggle{COMMENTS}
\toggletrue{COMMENTS}
\iftoggle{COMMENTS}{
\def\AS#1{{\bf \textcolor{blue}{[{AS: #1}]}}}
\def\AFR#1{{\bf \textcolor{red}{[{AFR: #1}]}}}
\def\ZZ#1{{\bf \textcolor{PineGreen}{[{ZZ: #1}]}}}
\def\EM#1{{\bf \textcolor{cyan}{[{EM: #1}]}}}
\def\KD#1{{\bf \textcolor{Periwinkle}{[{KD: #1}]}}}
}{
\def\AS#1{}
\def\AFR#1{}
\def\ZZ#1{}
\def\EM#1{}
\def\KD#1{}
}

\def\jcap{Journal of Cosmology and Astroparticle Physics}

\def\pr{^{\prime}}
\def\2pr{^{\prime \prime}}

\def\kmsmpc{{\rm km\, s^{-1} Mpc^{-1}}}
\newcommand{\fnl}{f_{\rm NL}}
\newcommand{\rhocrit}{\rho_{\rm crit}}

\newcommand{\neff}{N_{\rm eff}}

\newcommand{\hubunits}{\,{\rm km}\,{\rm s}^{-1}\,{\rm Mpc}^{-1}}

\def\greatsim{\mathrel{\raise.3ex\hbox{$>$\kern-.75em\lower1ex\hbox{$\sim$}}}}
\def\lesssim{\mathrel{\raise.3ex\hbox{$<$\kern-.75em\lower1ex\hbox{$\sim$}}}}
\def\gs{\mathrel{\raise0.27ex\hbox{$>$}\kern-0.70em 
\lower0.71ex\hbox{{$\scriptstyle \sim$}}}}
\def\ls{\mathrel{\raise0.27ex\hbox{$<$}\kern-0.70em 
\lower0.71ex\hbox{{$\scriptstyle \sim$}}}}

\newcolumntype{L}{>{$}l<{$}}
\newcolumntype{C}{>{$}c<{$}}

\newcommand{\APONMSU}{Apache Point Observatory and New Mexico State University, P.O. Box 59, Sunspot, NM 88349}

\newcommand{\BNL}{Brookhaven National Laboratory, Upton, NY~11973, USA}

\newcommand{\Caltech}{California Institute of Technology, Pasadena, CA 91125}

\newcommand{\CCAPP}{Center for Cosmology and AstroParticle Physics, The Ohio State University, Columbus, OH 43212}
\newcommand{\CCPP}{Center for Cosmology and Particle Physics, Department of Physics, New York University, 726 Broadway, Room 1005, New York, NY 10003, USA}
\newcommand{\CPPM}{Aix Marseille Univ, CNRS/IN2P3, CPPM, Marseille, France}

\newcommand{\CNCT}{Consejo Nacional de Ciencia y Tecnolog\'ia, Av. Insurgentes Sur 1582. Colonia Cr\'edito Constructor, Del. Benito Ju\'arez, C.P. 03940, M\'exico D.F. M\'exico}

\newcommand{\daa}{Department of Astronomy and Astrophysics, University of Toronto, ON, M5S3H4}

\newcommand{\dunlap}{Dunlap Institute for Astronomy and Astrophysics, University of Toronto, ON~M5S~3H4, Canada}

\newcommand{\ED}{University of Edinburgh, Edinburgh EH8~9YL, UK}
\newcommand{\EPFL}{Institute of Physics, Laboratory of Astrophysics, Ecole Polytechnique Fédérale de Lausanne (EPFL), Observatoire de Sauverny, 1290 Versoix, Switzerland}

\newcommand{\FNAL}{Fermi National Accelerator Laboratory, Batavia, IL~60510, USA}

\newcommand{\Haverford}{Haverford College, 370 Lancaster Ave, Haverford PA, 19041, USA}

\newcommand{\IAP}{Institut d'Astrophysique de Paris, CNRS \& Sorbonne University, UMR 7095, 98bis bd Arago, 75014 Paris, France}

\newcommand{\ICC}{ICC, University of Barcelona, IEEC-UB, Mart\' i i Franqu\` es, 1, E08028 Barcelona, Spain}

\newcommand{\ICCUK}{Institute for Computational Cosmology, Dept. of Physics, Univ. of Durham, South Road, Durham DH1 3LE, UK}

\newcommand{\ICFUNAM}{ICFUNAM - Instituto de Ciencias F\'{i}sicas, Universidad Nacional Aut\'onoma de M\'exico,  62210 Cuernavaca, Mor., M\'exico}

\newcommand{\IFAE}{Institut de F\'{i}sica d’Altes Energies, The Barcelona Institute of Science and Technology, Campus UAB, 08193 Bellaterra (Barcelona), Spain}

\newcommand{\IFT}{Instituto de Fisica Teorica UAM/CSIC, Universidad Autonoma de Madrid, 28049 Madrid, Spain}
\newcommand{\IFUNAM}{IFUNAM - Instituto de F\'{i}sica, Universidad Nacional Aut\'onoma de M\'exico, 04510 CDMX, M\'exico}

\newcommand{\IRFU}{IRFU, CEA, Universit\'e Paris-Saclay, F-91191 Gif-sur-Yvette, France}

\newcommand{\KASSI}{Korea Astronomy and Space Science Institute, Daejeon 34055, Korea}

\newcommand{\KIPAC}{Kavli Institute for Particle Astrophysics and Cosmology, Stanford 94305}

\newcommand{\KST}{University of Science and Technology, Daejeon 34113, Korea}

\newcommand{\LBL}{Lawrence Berkeley National Laboratory, Berkeley, CA~94720, USA}

\newcommand{\LJMU}{Liverpool John Moores University,  L3 5RF Liverpool , United Kingdom}

\newcommand{\LPNHE}{Sorbonne Universit\'e, Universit\'e Paris Diderot, CNRS/IN2P3, Laboratoire de Physique Nucl\'eaire et de Hautes Energies, LPNHE, 4 Place Jussieu, F-75252 Paris, France}

\newcommand{\MPE}{Max-Planck-Institut f\"{u}r extraterrestrische Physik (MPE), Giessenbachstrasse 1, D-85748 Garching bei M\"unchen, Germany}

\newcommand{\LAM}{Aix Marseille Univ, CNRS, CNES, LAM, Marseille, France}

\newcommand{\NAOC}{National Astronomical Observatories of China, Chinese Academy of Sciences, 20A Datun Road, Chaoyang District, Beijing 100012, China}

\newcommand{\NYU}{New York University, New York, NY 10003}
\newcommand{\OIRLab}{NSF's National Optical-Infrared Astronomy Research Laboratory, 950 N. Cherry Ave, Tucson, AZ 85719 USA}

\newcommand{\OSU}{The Ohio State University, Columbus, OH 43212}
\newcommand{\OU}{Department of Physics and Astronomy, Ohio University, Clippinger Labs, Athens, OH 45701, USA}

\newcommand{\Oxfordastro}{Sub-department of Astrophysics, Department of Physics, University of Oxford, Denys Wilkinson Building, Keble Road, Oxford OX1 3RH}

\newcommand{\PI}{Perimeter Institute, Waterloo, ON~N2L~2Y5, Canada}
\newcommand{\Pitt}{University of Pittsburgh and PITT PACC, Pittsburgh, PA 15260}

\newcommand{\Port}{Institute of Cosmology \& Gravitation, University of Portsmouth, Dennis Sciama Building, Burnaby Road, Portsmouth PO1 3FX, UK}

\newcommand{\PSU}{The Pennsylvania State University, University Park, PA 16802}

\newcommand{\PSUIGC}{Institute for Gravitation and the Cosmos, The Pennsylvania State University, University Park, PA 16802}

\newcommand{\SAAO}{South African Astronomical Observatory, PO Box 9, Observatory 7935, Cape Town, South Africa}
\newcommand{\SAIMSU}{Sternberg Astronomical Institute, Moscow State University, Moscow, 119992, Russia}

\newcommand{\Sejong}{Department of Physics and Astronomy, Sejong University, Seoul, 143-747, Korea}

\newcommand{\StAndrews}{School of Physics and Astronomy, University of St Andrews, North Haugh, St Andrews, KY16 9SS, UK}

\newcommand{\SUOT}{Swinburne University of Technology, Centre for Astrophysics and Supercomputing, Melbourne, VIC 3122, Australia}

\newcommand{\TCU}{Department of Physics \& Astronomy, Texas Christian University, Fort Worth, TX 76129, USA}

\newcommand{\UAM}{Universidad Aut\'onoma de Madrid, 28049, Madrid, Spain}

\newcommand{\UCL}{University College London, WC1E 6BT London, United Kingdom}

\newcommand{\UCT}{Department of Astronomy, University of Cape Town, Private Bag X3, Rondebosch 7701, South Africa}

\newcommand{\UGTO}{Divisi\'on de Ciencias e Ingenier\'ias, Universidad de Guanajuato, Le\'on 37150, M\'exico}

\newcommand{\Utah}{University of Utah, Department of Physics and Astronomy, 115 S 1400 E, Salt Lake City, UT 84112, USA}

\newcommand{\UWaterloo}{Department of Physics and Astronomy, University of Waterloo, 200 University Ave W, Waterloo, ON N2L 3G1, Canada}
\newcommand{\UWisc}{Department of Astronomy, University of Wisconsin-Madison, Madison, WI~53706, USA}

\newcommand{\UW}{University of Washington, Seattle 98195}
\newcommand{\UWastro}{Department of Astronomy, University of Washington, Box 351580, Seattle, WA 98195, USA}

\newcommand{\WCA}{Waterloo Centre for Astrophysics, University of Waterloo, Waterloo, ON~N2L~3G1, Canada}

\newcommand{\Wyoming}{Department of Physics and Astronomy, University of Wyoming, Laramie, WY 82071, USA}

\begin{document}

\title{The Completed SDSS-IV extended Baryon Oscillation Spectroscopic Survey:
Cosmological Implications from two Decades of Spectroscopic Surveys at the Apache Point observatory}

\author{Shadab Alam}
\affiliation{\ED}
\author{Marie Aubert}
\affiliation{\CPPM}
\author{Santiago Avila}
\affiliation{\UAM}
\affiliation{\IFT}
\author{Christophe Balland}
\affiliation{\LPNHE}
\author{Julian E. Bautista}
\affiliation{\Port}
\author{Matthew A. Bershady}
\affiliation{\UWisc}
\affiliation{\SAAO}
\affiliation{\UCT}
\author{Dmitry Bizyaev}
\affiliation{\APONMSU}
\affiliation{\SAIMSU}
\author{Michael R.~Blanton}
\affiliation{\CCPP}
\author{Adam S. Bolton}
\affiliation{\OIRLab}
\affiliation{\Utah}
\author{Jo Bovy}
\affiliation{\daa}
\affiliation{\dunlap}
\author{Jonathan Brinkmann}
\affiliation{\APONMSU}
\author{Joel R.~Brownstein}
\affiliation{\Utah}
\author{Etienne Burtin}
\affiliation{\IRFU}
\author{Sol\`ene Chabanier}
\affiliation{\IRFU}
\author{Michael J. Chapman}
\affiliation{\WCA}
\affiliation{\UWaterloo}
\author{Peter Doohyun Choi}
\affiliation{\Sejong}
\author{Chia-Hsun Chuang}
\affiliation{\KIPAC}
\author{Johan Comparat}
\affiliation{\MPE}
\author{Marie-Claude Cousinou}
\affiliation{\CPPM}
\author{Andrei Cuceu}
\affiliation{\UCL}
\author{Kyle S. Dawson}
\affiliation{\Utah}
\author{Sylvain de la Torre}
\affiliation{\LAM}
\author{Arnaud de~Mattia}
\affiliation{\IRFU}
\author{Victoria de Sainte Agathe}
\affiliation{\LPNHE}
\author{H\'elion~du~Mas~des~Bourboux}
\affiliation{\Utah}
\author{Stephanie Escoffier}
\affiliation{\CPPM}
\author{Thomas Etourneau}
\affiliation{\IRFU}
\author{James Farr}
\affiliation{\UCL}
\author{Andreu Font-Ribera}
\affiliation{\IFAE}
\affiliation{\UCL}
\author{Peter M. Frinchaboy}
\affiliation{\TCU}
\author{Sebastien Fromenteau}
\affiliation{\ICFUNAM}
\author{H\'ector Gil-Mar\'in}
\affiliation{\ICC}
\author{Jean-Marc Le Goff}
\affiliation{\IRFU}
\author{Alma X. Gonzalez-Morales}
\affiliation{\UGTO}
\affiliation{\CNCT}
\author{Violeta Gonzalez-Perez}
\affiliation{\LJMU}
\affiliation{\Port}
\author{Kathleen Grabowski}
\affiliation{\APONMSU}
\author{Julien Guy}
\affiliation{\LBL}
\author{Adam J. Hawken}
\affiliation{\CPPM}
\author{Jiamin Hou}
\affiliation{\MPE}
\author{Hui Kong}
\affiliation{\CCAPP}
\author{James Parker III}
\affiliation{\APONMSU}
\author{Mark Klaene}
\affiliation{\APONMSU}
\author{Jean-Paul Kneib}
\affiliation{\EPFL}
\author{Sicheng Lin}
\affiliation{\CCPP}
\author{Daniel Long}
\affiliation{\APONMSU}
\author{Brad W. Lyke}
\affiliation{\Wyoming}
\author{Axel de la Macorra}
\affiliation{\IFUNAM}
\author{Paul Martini}
\affiliation{\OSU}
\affiliation{\CCAPP}
\author{Karen Masters}
\affiliation{\Haverford}
\author{Faizan G. Mohammad}
\affiliation{\WCA}
\affiliation{\UWaterloo}
\author{Jeongin Moon}
\affiliation{\Sejong}
\author{Eva-Maria Mueller}
\affiliation{\Oxfordastro}
\author{Andrea Mu\~noz-Guti\'errez}
\affiliation{\IFUNAM}
\author{Adam D.~Myers}
\affiliation{\Wyoming}
\author{Seshadri Nadathur}
\affiliation{\Port}
\author{Richard Neveux}
\affiliation{\IRFU}
\author{Jeffrey A.~Newman}
\affiliation{\Pitt}
\author{Pasquier Noterdaeme}
\affiliation{\IAP}
\author{Audrey Oravetz}
\affiliation{\APONMSU}
\author{Daniel Oravetz}
\affiliation{\APONMSU}
\author{Nathalie Palanque-Delabrouille}
\affiliation{\IRFU}
\author{Kaike Pan}
\affiliation{\APONMSU}
\author{Romain Paviot}
\affiliation{\LAM}
\author{Will J. Percival}
\affiliation{\WCA}
\affiliation{\UWaterloo}
\affiliation{\PI}
\author{Ignasi P\'erez-R\`afols}
\affiliation{\LPNHE}
\author{Patrick Petitjean}
\affiliation{\IAP}
\author{Matthew M. Pieri}
\affiliation{\LAM}
\author{Abhishek Prakash}
\affiliation{\Caltech}
\author{Anand Raichoor}
\affiliation{\EPFL}
\author{Corentin Ravoux}
\affiliation{\IRFU}
\author{Mehdi Rezaie}
\affiliation{\OU}
\author{James Rich}
\affiliation{\IRFU}
\author{Ashley J. Ross}
\affiliation{\CCAPP}
\author{Graziano Rossi}
\affiliation{\Sejong}
\author{Rossana Ruggeri}
\affiliation{\SUOT}
\affiliation{\Port}
\author{Vanina Ruhlmann-Kleider}
\affiliation{\IRFU}
\author{Ariel G. S\'anchez}
\affiliation{\MPE}
\author{F.~Javier S\'{a}nchez}
\affiliation{\FNAL}
\author{Jos\'e R. S\'anchez-Gallego}
\affiliation{\UWastro}
\author{Conor Sayres}
\affiliation{\UW}
\author{Donald P. Schneider}
\affiliation{\PSU}
\affiliation{\PSUIGC}
\author{Hee-Jong Seo}
\affiliation{\OU}
\author{Arman Shafieloo}
\affiliation{\KASSI}
\affiliation{\KST}
\author{An\v{z}e Slosar}
\affiliation{\BNL}
\author{Alex Smith}
\affiliation{\IRFU}
\author{Julianna Stermer}
\affiliation{\LPNHE}
\author{Amelie Tamone}
\affiliation{\EPFL}
\author{Jeremy L. Tinker}
\affiliation{\NYU}
\author{Rita Tojeiro}
\affiliation{\StAndrews}
\author{Mariana Vargas-Maga\~na}
\affiliation{\IFUNAM}
\author{Andrei Variu}
\affiliation{\EPFL}
\author{Yuting Wang}
\affiliation{\NAOC}
\author{Benjamin A. Weaver}
\affiliation{\OIRLab}
\author{Anne-Marie Weijmans}
\affiliation{\StAndrews}
\author{Christophe Y\`eche}
\affiliation{\IRFU}
\author{Pauline Zarrouk}
\affiliation{\ICCUK}
\affiliation{\IRFU}
\author{Cheng Zhao}
\affiliation{\EPFL}
\author{Gong-Bo Zhao}
\affiliation{\NAOC}
\author{Zheng Zheng}
\affiliation{\Utah}

\begin{abstract}
We present the cosmological implications from final measurements of clustering using galaxies, quasars, and \lya\
forests from the completed Sloan Digital Sky Survey (SDSS) lineage of experiments in large-scale structure.
These experiments, composed of data from SDSS, SDSS-II, BOSS, and eBOSS, offer independent measurements of
baryon acoustic oscillation (BAO) measurements of angular-diameter distances and Hubble distances
relative to the sound horizon, $r_d$, from eight different
samples and six measurements of the growth rate parameter, $f\sigma_8$, from redshift-space distortions (RSD).  
This composite sample is the most constraining of its kind and allows us to perform a comprehensive assessment
of the cosmological model after two decades of dedicated spectroscopic observation.  We show that the BAO data
alone are able to rule out dark-energy-free models at more than eight standard deviations in an
extension to the flat, \LCDM\ model that allows for curvature. 
When combined with \Planck\ Cosmic Microwave Background (CMB) measurements
of temperature and polarization, under the same model, the BAO data provide nearly an order of magnitude improvement
on curvature constraints relative to primary CMB constraints alone.  Independent of distance measurements,
the SDSS RSD data complement weak lensing measurements from the Dark Energy Survey (DES) in
demonstrating a preference for a flat \LCDM\ cosmological model when combined with \Planck\ measurements.
The combined BAO and RSD measurements indicate $\sigma_8 = 0.85 \pm 0.03$,
implying a growth rate that is consistent with predictions from \Planck\
temperature and polarization data and with General Relativity.  When combining the results of SDSS BAO and RSD,
\Planck, Pantheon Type Ia supernovae (SNe~Ia), and DES weak lensing and clustering measurements, all multiple-parameter extensions
remain consistent with a \LCDM\ model.  Regardless of cosmological model, the precision on each of the three
parameters, $\Omega_\Lambda$, $H_0$, and $\sigma_8$, remains at roughly 1\%, showing changes of less than 0.6\% in the
central values between models.  In a model that allows for free curvature and a time-evolving
equation of state for dark energy, the combined samples produce a constraint $\Omega_k = -0.0022 \pm 0.0022$.
The dark energy constraints lead to $w_0 = -0.909 \pm 0.081$ and $w_a = -0.49^{+0.35}_{-0.30}$, corresponding to an equation of state of
$w_p = -1.018 \pm 0.032$ at a pivot redshift $z_p=0.29$ and a Dark Energy Figure of Merit of 94.  
The inverse distance ladder measurement under this model yields $H_0= 68.18 \pm 0.79 \, \kmsmpc$, remaining in tension with 
several direct determination methods; the BAO data allow Hubble constant estimates that are robust against the
assumption of the cosmological model.  In addition, the BAO data allow estimates of $H_0$ that 
are independent of the CMB data, with similar central values and precision under a \LCDM\ model.
Our most constraining combination 
of data gives the upper limit on the sum of neutrino masses at $\sum m_\nu <0.115$ eV (95\% confidence).  
Finally, we consider the improvements in cosmology constraints over the last decade by comparing our results 
to a sample representative of the period 2000--2010. We compute the relative gain across the five dimensions 
spanned by $w$, $\Omega_k$, $\sum m_\nu$, $H_0$, and $\sigma_8$ and find that the SDSS BAO and 
RSD data reduce the total posterior volume by a factor of 40 relative to the previous generation.  Adding again the \Planck, 
DES, and Pantheon SN~Ia samples leads to an overall contraction in the five-dimensional posterior volume
of three orders of magnitude.
\end{abstract}

\maketitle


\section{Introduction}\label{sec:intro}
\setcounter{footnote}{0} Understanding the energy content of the
Universe, the physical mechanisms behind cosmic expansion, and the
growth of structure are the primary challenges of
cosmology. Developmental milestones for the current standard model of
these properties, the spatially-flat \LCDM\ model, include
measurements of the expansion history using Type Ia supernovae
(SNe~Ia) in the 1990's, which provided the first evidence for cosmic
acceleration \citep{riess98a,perlmutter99a}, and studies of
perturbations in the cosmic microwave background (CMB), which provided
the first convincing evidence for a nearly flat geometry
\citep{miller99a,debernardis00a,balbi00a,jaffe01a} when assuming weak
priors and fitting results from the BOOMERanG \citep{netterfield02}
and MAXIMA \citep{hanany00a} CMB experiments. These observations
  showed that prior suggestions that the \LCDM\ model would solve some
  of the emerging problems in cosmology were particularly adroit
  \citep{Efstathiou90,Ostriker95}.
At around the same time as these observations, the first measurements
of the baryon and matter densities from the shape of the power
spectrum from the 2dFGRS \citep{colless01a} were published
\citep{percival01a}. The combination of the galaxy survey data and CMB
data is particularly strong for breaking degeneracies inherent to
either method individually: combining early 2dFGRS and CMB data meant
that, at around the turn of the century, the physical baryon and cold
dark matter densities were known to 10\% and 8\% respectively, and the
Hubble parameter to 7\% within the flat \LCDM\ model
\citep{percival02a}.

The first decade of the 21st century witnessed a strong advancement in
the precision with which the parameters of this standard model were
known, without demonstrating significant tension with this model. This
came through dedicated CMB experiments including ACT
\citep{fowler07a}, SPT \citep{carlstrom11a} and the
Wilkinson Microwave Anisotropy Probe \citep[WMAP;][]{bennett13a}. 
SN~Ia observations continued to improve in sample
size and analysis methodology
\citep{jha06a,riess07a,frieman08a,dawson09a,hicken09a,contreras10a,guy10a,conley11a,sullivan11a},
and direct measurements of the local expansion rate using Cepheid
variables and SNe~Ia led to estimates of $H_0$ with better than 4\%
precision \citep{riess09a,freedman12a}. During this same approximate
period, the 2dFGRS and Sloan Digital Sky Survey \citep[SDSS;][]{york00a} galaxy surveys became sufficiently large to clearly
measure the Baryon Acoustic Oscillation (BAO) feature in the
clustering of galaxies \citep{eisenstein05a,cole05a} and use this as a
robust cosmological probe \citep{percival07a}. Combined, these
experiments offered strong evidence supporting the simple six
parameter \LCDM\ cosmological model consisting of the baryon density
($\Omega_b$), dark matter density ($\Omega_c$), Hubble Constant
($H_0$), amplitude of primordial perturbations ($A_s$), power-law
spectral index of primordial density perturbations ($n_s$), and
reionization optical depth ($\tau$). The 5-year WMAP data
\citep{hinshaw08a}, combined with the SDSS-II BAO data
\citep{percival07a} and the union SN sample \citep{kowalski08b}, led
to measurements of the physical baryon and cold dark matter densities to 3\%,
and the Hubble parameter to 2\% \citep{komatsu09a} within the
framework of the \LCDM\ model.

The last ten years have seen significant advances in cosmology through
CMB observations, improved calibration of systematic errors in SNe~Ia
studies, and large area spectroscopic surveys.  
Gravitational lensing from the
CMB has provided important, high signal-to-noise measurements of
structure growth in the low redshift universe
\citep{2014A&A...571A..17P,plancklensing}.  CMB lensing has been
supplemented by increasingly robust and statistically sensitive
estimates of weak lensing based on galaxy shapes, including CFHTLenS
\citep{2012MNRAS.427..146H}, KiDS \citep{hildebrandt20a,joachimi20a}, Dark
Energy Survey \citep[DES;][]{2018MNRAS.481.1149Z,2018PhRvD..98d3526A},
and Hyper Suprime-Cam survey
\citep[HSC;][]{2018PASJ...70S..25M,2019PASJ..tmp...22H}.  Improvements
in the statistical precision from weak lensing programs have begun through
joint analysis of multiple data sets, using consistent priors, modeling of 
the power spectrum, and redshift distributions \citep[e.g.,][]{joudaki20a}.

While broad support for the flat \LCDM\ model remains amongst
measurements, a number of increasingly worrying discrepancies are
appearing. For example, measurements of the Hubble constant 
from the distance ladder technique \citep[SH0ES;][]{riess19a} and
those from the \Planck\ CMB measurements \citep{2018arXiv180706209P},
are in tension, as discussed further in
Section~\ref{subsec:H0}. Additionally, albeit at lower levels of
discordance, weak lensing measurements
\citep[e.g.,][]{hildebrandt20a,joachimi20a} tend to find lower levels
of cosmological structure than found from redshift-space distortions
or predicted from the amplitude of the \Planck\ CMB power spectrum
(see Section~\ref{sec:RSD-meas}).

The continuing development of massive spectroscopic surveys over the
last decade is of particular interest to this study.  Spectroscopy of
galaxies and quasars over wide areas allows precise measurements of
cosmic expansion history with BAO and measurements of the rate of
structure growth with redshift space distortions (RSD).  The largest
spectroscopic survey to date is the Baryon Oscillation Spectroscopic
Survey \citep[BOSS;][]{dawson13a}, which was the primary driver for
SDSS-III \citep{eisenstein11a}.  In operations spanning 2009--2014,
BOSS completed spectroscopy on more than 1.5 million galaxies as faint
as $i=19.9$ and more than 160,000 $z>2.1$ quasars as faint as
$g=22$. In 2012, the first BAO measurements from BOSS were published
\citep{anderson12a}, just before the final results from the WMAP CMB
experiment. At this point, the data were sufficient to set interesting
constraints on models that go beyond \LCDM. For example, an analysis
under a flat \LCDM\ model with neutrinos using the final WMAP data, 
an estimate of $H_0 =
73.8 \pm 2.4$ km s$^{-1}$ Mpc$^{-1}$ \citep{riess11a}, and the BOSS
BAO measurements \citep{anderson12a}, together with those from the
6dFGS, SDSS-II and WiggleZ surveys
\citep{beutler11a,padmanabhan12a,blake12a} led to a 95\% upper limit
of 0.44 eV on the sum of the neutrino masses \citep{hinshaw13a}.
Adding measurements of luminosity-distance ratios from a large sample of SNe~Ia
\citep{guy10a,conley11a,sullivan11a} led to constraints consistent
with a cosmological constant when allowing a
Chevallier-Polarski-Linder (CPL) parameterization
\citep{chevallier01a,linder03b} of dark energy, indicating $w_0 =
-1.17^{+0.13}_{-0.12}$ and $w_a = 0.35^{+0.50}_{-0.49}$ in a model
with a flat universe \citep{hinshaw13a}. Within the \LCDM\ model, the
errors on the physical baryon density, cold dark matter density were
now at the level of 2\%, and the Hubble Constant 1.3\%.

Final measurements of the CMB-calibrated BAO scale from the BOSS experiment led to 1\%
precision measurements of the cosmological distance scale for
redshifts $z < 0.75$ \citep{alam17a} and 2\% precision measurements at
$z = 2.33$ \citep{bautista17a,masdesbourboux17a}. By the time that the
final results from BOSS were ready, the \Planck\ satellite had released
its 2015 CMB measurements \citep{planck15a}, surpassing the precision
afforded by WMAP. In combination, the 2015 CMB power spectrum
measurements from the \Planck\ satellite together with BOSS constrain
the rate of structure growth at the level of 6\% and constrain the sum
of the neutrino masses to be less than 160 meV at 95\% confidence
\citep{alam17a}. With these data, the constraints on the physical
matter density and Hubble Constant within the \LCDM\ model were both
at the level of 0.6\%.

So far, there have been four generations of SDSS
conducted from the 2.5-meter Sloan
Telescope \citep{gunn06a} at the Apache Point Observatory. The
extended Baryon Oscillation Spectroscopic Survey
\citep[eBOSS;][]{dawson16a}\footnote{\url{https://www.sdss.org/surveys/eboss/}} 
is the cosmological survey within SDSS-IV \citep{blanton17a}.  Using the same
spectrographs used for BOSS \citep{smee13a}, eBOSS concluded 4.5 years of spectroscopic observations of
large-scale structure on March 1, 2019. eBOSS extends the
BOSS analysis using galaxies as direct tracers of the density field to
measure BAO and RSD to higher redshifts, and increases the number of
quasars used for \lya\ forest studies. It also marks the last use of the Sloan
Telescope for galaxy redshift surveys designed to
measure cosmological parameters using BAO and RSD techniques, with SDSS
now focusing on other exciting astronomical questions \citep{kollmeier17a}.

In this paper, we characterize the advances made in constraining the
cosmological model over the last decade, focusing specifically on the
impact of the BOSS and eBOSS spectroscopic programs. A summary of the key
results from this work, as well as a few additional figures, can be found
in the SDSS webpages\footnote{\url{https://www.sdss.org/science/cosmology-results-from-eboss/}}.
The study presented in this work is part of a coordinated release of the final
eBOSS measurements of BAO and RSD in the clustering of luminous red
galaxies \citep[LRG;][]{LRG_corr,gil-marin19a}, emission line galaxies
\citep[ELG;][]{raichoor19a,tamone19a,demattia19a}, and quasars
\citep{hou19a,neveux19a}.  At the highest redshifts ($z>2.1$), the
coordinated release of final eBOSS measurements includes measurements
of BAO in the \lya\ forest \citep{2019duMasdesBourbouxH}.  An
essential component of these studies is the construction of data
catalogs \citep{ross20a,lyke20a}, mock catalogs
\citep{lin20a,zhao20a,carr20a}, and galaxy mocks based on N-body simulations
for assessing theoretical systematic errors \citep{alam20a,avila20a,rossi20a,smith20}.
A summary table of the BAO and RSD measurements, with links to supporting studies
and legacy figures describing the measurements can be found in the SDSS
webpages\footnote{\url{https://www.sdss.org/science/final-bao-and-rsd-measurements/}}.

In all, the SDSS, BOSS, and eBOSS surveys provide galaxy and quasar samples from
which BAO can be measured covering all redshifts $z<2.2$, and \lya\
forest observations over $2<z<3.5$.  The aggregate precision of the
expansion history measurements is 0.70\% at redshifts $z<1$ and 1.19\%
at redshifts $z>1$, while the aggregate precision of the growth
measurements is 4.77\% over the redshift interval $0<z<1.5$.  With
this coverage and sensitivity, the SDSS experiment is unparalleled in
its ability to explore models of dark energy.

In Section~\ref{sec:cosmo}, we present the cosmological background and
the signatures in the key observational probes.  This section is
intended to provide a high level background that will put the SDSS
spectroscopic surveys into the broader context for relatively new
readers.  In Section~\ref{sec:data}, we present the data samples for
the cosmological analyses performed in this work.  In
Section~\ref{sec:expansion}, we discuss the impact of SDSS BAO
distance measurements on single parameter extensions to \LCDM,
relative to SNe~Ia and CMB probes.  We also demonstrate the key
contributions from BAO measurements in the well-known tension between
local measurements of $H_0$ and estimates extrapolated from high-redshift 
observations.  In Section~\ref{sec:growth}, we discuss RSD
and weak lensing measurements both in constraining the relative
abundance of dark energy and in testing predictions of growth under an
assumption of General Relativity (GR).  In
Section~\ref{sec:global_fit}, we present the cosmological model that
best describes all of the observational data used in this work.  We
conclude in Section~\ref{sec:conclusion} by presenting the substantial advances in our understanding
of the cosmological model that have been made in the last decade and
the role that the BOSS and eBOSS programs play in those advances.
\\
\\

\section{Cosmological Model and Observable Signatures}\label{sec:cosmo}
The BOSS and eBOSS surveys have fostered the development of the BAO technique to percent-level precision over a
larger redshift range than any other probe of the distance-redshift relation.
RSD measurements from BOSS and eBOSS
offer constraints on structure growth over nearly as large a redshift range.
Meanwhile, in the last ten years, the CMB maps produced by the \Planck\ satellite have allowed
precise constraints on the conditions of the Universe at the time of last scattering and on
the angular diameter distance to that epoch.
With probes of the late-time expansion history, the evolution of cosmic expansion
can be extrapolated from the CMB to today's epoch under models with freedom for curvature, dark energy density, dark
energy equation of state, and neutrino mass.

SNe~Ia measurements remain the most effective way to constrain expansion history at redshifts below $z<0.5$,
while the BOSS and eBOSS BAO measurements cover redshifts $0<z<2.5$ and rely on a model that is well-understood
because it is based in simple physics.  Large weak lensing surveys have measured cosmic shear to constrain
the local matter density and amplitude of fluctuations,
while RSD measure the change in the fluctuation amplitude with time through measurements of the gravitational infall of matter.
As will be presented in the analysis of this paper, the full suite of these complementary measurements allows a comprehensive
assessment of the cosmological model.

In this section, we provide an overview of the cosmological model and a pedagogical summary of the observational
signatures in BAO, RSD, CMB, SNe~Ia, and weak lensing surveys that we use to provide new constraints on that model.
This section is intended to provide the key details of the cosmological models and data sets that are explored in the
remainder of the paper.  The discussion will be familiar to the reader experienced in multi-probe cosmology
constraints and will offer the highlights for additional study for the less experienced reader.

\subsection{Background Models and Notation}

Throughout this paper we employ the standard cosmological model based on the Friedmann-Robertson-Walker metric
where the scale factor, $a$, is unity today and is related to redshift by $a(t) = (1+z)^{-1}$. 
The evolution of the scale factor with time describes the background expansion history of the Universe,
governed by the Friedmann equation, normally written as 
\begin{equation}
  H^2(a) = \frac{8\pi G}{3}\rho(a) - \frac{kc^2}{a^2}~.
\end{equation}
$H \equiv \dot{a}/a$ is the Hubble parameter and $\rho(a)$ is
the total energy density (radiation + matter + dark energy).
The curvature constant $k$ parameterizes the global curvature of space.
An open universe is represented by $k<0$ and a closed universe by $k>0$.
The curvature term can be expressed in terms of an effective energy density through $-kc^2/a^2=(8\pi G/3) \rho_k(a)$.
However, we note that a Universe that is globally flat ($k=0$) will appear to have a non-zero mean curvature
due to horizon-scale fluctuations in the matter density field.
These large-scale fluctuations place a fundamental limit on
constraints on the curvature term under inflationary models that best describe CMB fluctuations,
and the detectable limit is roughly one part in 10,000 \citep{vardanyan09}.

We define the dimensionless density parameter of each energy component ($x$) by the ratio
\begin{equation}
  \Omega_x = {\rho_x \over \rhocrit} = {8\pi G \over 3H^2} \rho_x ~
\end{equation}
so that $\sum \Omega_x  = 1$, where the sum is over all energy components
including the curvature. 
Density parameters and $\rhocrit$ always refer to values at $z=0$
unless a dependence on $a$ or $z$ is stated explicitly, 
e.g., $\Omega_x(z)$.  We will frequently refer to the present-day ($t_0$)
Hubble parameter $H_0$ through the dimensionless ratio
$h \equiv H_0/100\hubunits$.
The dimensionless quantity $\omega_x \equiv \Omega_x h^2$ is
proportional to the physical density of component $x$ at the
present day.

The energy components considered in our models are pressureless (cold)
dark matter (CDM), baryons, photons, neutrinos, and dark energy.
The densities of CDM and baryons scale as $a^{-3}$; we refer to the
density parameter of these two components together as $\Omega_{cb}$.
The energy density in radiation ($\Omega_{r}$) scales as $a^{-4}$; in the standard
cosmological model, $\Omega_{r}$ is dynamically negligible in the low redshift
universe probed by spectroscopic surveys.
However, the radiation density is dominant at very high redshift, where it receives contributions
from the electromagnetic CMB radiation ($\omega_\gamma$, known exquisitely well)
and from neutrinos (at temperature $T$ higher than the rest energy $m_\nu$):
\begin{equation}
 \omega_r(T > m_{\nu}) = \omega_\gamma + \omega_\nu 
    = \left[1 + \frac{7}{8} \left(\frac{4}{11} \right)^{4/3} \neff \right] \omega_\gamma ~,
\end{equation}
with $\neff = 3.044$ in the standard case with three neutrino species \citep{akita20a}
(note:  following precedent, we use $\neff = 3.046$ throughout, as in \citet{mangano05a}).
Other potential contributions to radiation energy density are traditionally
parameterized in terms of their change to the effective number of neutrino
species, $\Delta \neff$, regardless of whether they represent extra neutrino
species or other light degrees of freedom.

While the effect of neutrinos in cosmology has been detected through their
contribution to the radiation energy density in the CMB
\citep{2018arXiv180706209P}, we have not yet reached the sensitivity to
detect their mass. 
However, the detection of neutrino oscillations in terrestrial experiments
strongly implies that at least two species are massive and that
at least one species is now non-relativistic \citep[see][for a recent review]{DESALAS2018633}
The energy density of neutrinos with non-zero mass scales like
radiation at early times when the particles are ultra-relativistic.
Once cosmic expansion reduces their kinetic energy below the rest
mass, the neutrinos transition towards dark matter behavior.
For neutrinos with non-degenerate mass eigenstates, the transition to
non-relativistic energies will happen at different epochs
for the three eigenstates. 
In practice, for realistic neutrino masses, the transition occurs after the
epoch of the last scattering in the CMB, but before the epochs observed by spectroscopic
surveys.
Therefore, we can safely assume that the most massive neutrino species are
ultra-relativistic at epochs relevant for the CMB and act as dark matter
at epochs probed by galaxy surveys \citep{2006PhR...429..307L}.

At the current level of precision, cosmological measurements are sensitive
only to the sum of neutrino mass eigenvalues
\citep{astro-ph/0602133,2006PhR...429..307L,2014JCAP...05..023F,2009PhRvD..80l3509D,2010JCAP...05..035J},
thus allowing a simple modeling of neutrino masses with a single parameter, 
$\sum m_\nu$.
We use \nuLCDM\ to refer to the flat \LCDM\ model with this
extra free parameter.
Following standard convention, our total matter density at redshift $z=0$
therefore includes neutrinos, $\Omega_m=\Omega_{cb}+\Omega_\nu$. 

Finally, the dark energy component is approximately constant in time and
thus dominates the late-time evolution of the Universe
(all the other components scale at least with $a^{-2}$). 
Conventionally, the dark energy component is parameterized in terms of its
pressure-to-density ratio, $w=p_{\rm DE}/\rho_{\rm DE}$ ($c=1$ units).
We consider three basic phenomenological possibilities for $w$:
\begin{equation}
  w (a) =\begin{cases}
    -1 \\
    w  \\
    w_0 + w_a (1-a),\\
    \end{cases}
 \label{eq:darkenergy}
\end{equation}
corresponding to cosmological constant, constant equation of state, and equation of state in the form of 
the CPL parameterization, respectively.

For the three cases in equation~(\ref{eq:darkenergy}), the energy density of dark energy can be analytically integrated into
\begin{equation}
  \frac{\rho_{\rm DE}(a)}{\rho_{\rm DE,0}} = \begin{cases}
    1 \\
    a^{-3(1+w)} \\
    a^{-3(1+w_0+w_a)}\exp[{-3w_a(1-a)}]. \\
\end{cases}
\end{equation}
We describe these models as \LCDM, \wCDM\ and \wowaCDM, respectively.
By default, these models assume a flat geometry, but we also consider versions of these with free curvature.
Dark energy models where $\Omega_k$ is allowed as a free parameter are
referred to as \oLCDM, \owCDM\ and \owowaCDM.
All of these models are nested in the sense that they contain \LCDM\ as a special limit, with $w=w_0=-1$,
$w_a=0$, and $\Omega_k=0$. 

\subsection{Growth of Structure in the Cosmological Model}\label{subsec:growth}

The cosmic expansion history is determined by the mean energy densities of the components in the Universe
and their evolution with time.  The structure growth history reflects the evolution of density perturbations
against the background of cosmic expansion. Density perturbations in the matter are described by their relative perturbations
\begin{equation}
\delta(\mathbf{x},t) ~\equiv~ \frac{\rho_m(\mathbf{x},t)-\bar{\rho}_m(t)}{\bar{\rho}_m(t)},
\end{equation}
where $\bar{\rho}_m(t)$ is the mean matter density of the Universe and $\mathbf{x}$ is the comoving coordinate.
In this paper we ignore theoretical subtleties related to choice of gauge, because 
on the sub-horizon scales of interest, the Newtonian description is fully adequate.

To the first order in perturbation theory, the growth of fluctuations with time is specified by a single,
scale-independent growth factor, $D(t)$: 
\begin{equation}
\label{eqn:deltaxt}
\delta(\mathbf{x},t) ~=~ D(t) \delta(\mathbf{x},t_0),
\end{equation}
where $D(t_0)=1$ and $D(t)$ satisfies
\begin{equation}
\label{eqn:lingrowth}
\ddot D + 2H(z) \dot D - {3\over 2}\Omega_m H_0^2 (1+z)^3 D =0 ~.
\end{equation}
Strictly speaking, this equation only holds for a single fluid. However, it describes the low-redshift
universe very well, since gravitational evolution drives the multiple fluids towards a common over-density field.
Therefore, in cosmological models consistent with GR, the growth of density fluctuations can be
predicted uniquely for a given expansion history.
In this work, we use growth measurements to probe dark energy,
to measure the amplitude of the current matter density perturbations to test for tension in the cosmological model, 
as well as to test GR as the model for gravity on cosmological scales.

The linear growth rate is often expressed as a differential in the linear growth function with respect to the scale factor
\begin{equation}
\label{eqn:dlng.dlna}
f(z) \equiv {d\ln D \over d\ln a}
 ~.
\end{equation}
In standard cosmological models under GR, the growth rate can be approximated
as $f(z) \propto \Omega_m(z)^{0.55}$
\citep{wang98a,linder05a,linder07a}.  However, with the same expansion
history, theories of modified gravity may predict different rates
of structure growth, which motivates a simple parameterization to
modifications to $f(z) \propto \Omega_m(z)^{\gamma}$, where departures from $\gamma=0.55$ correspond to departures from GR.
Another strong prediction from GR is that the two metric potentials $\Psi$
and $\Phi$ (corresponding to time and space perturbations of the
metric) are the same ($\Psi=\Phi$).  This is not necessarily so in
theories of modified gravity, and the difference in the two potentials
(known as gravitational slip) can affect the difference between the trajectories 
of relativistic and non-relativist particles.

In this work, we follow the analysis of \citet{2019PhRvD..99l3505A} to test for more general deviations from GR.
Starting from scalar metric perturbations in the conformal Newtonian Gauge,
represented as $ds^2 = a^2(\tau)[(1 + 2\Psi)d\tau^2 - (1 - 2\Phi)\delta_{ij}dx_idx_j$]
with conformal time $\tau$ defined through $d\tau=dt/a(t)$,
this phenomenological model allows modification to the Poisson equations, or equivalently,
to the time-variation of the gravitational constant and gravitational slip \citep[e.g.,][]{zhao09a}.
A time-dependent parameter $\mu(a)$ plays a similar role to the $\gamma$ parameter in modifying the growth rate.
The model also allows a perturbation of the potential for massless particles relative to matter
particles through the time-dependent parameter $\Sigma(a)$.
These two parameters provide linear perturbations to the GR form of gravity according to the
relations
\begin{eqnarray}
\label{eq:constraint1}
k^2 \Psi = & -  4 \pi G a^2 (1+\mu(a)) \rho \delta \,, \\
\label{eq:constraint2}
k^2 (\Psi + \Phi) =  & - 8 \pi G a^2 (1+\Sigma(a))  \rho \delta \,,
\end{eqnarray}
where $k$ is the wavenumber and $\delta$ is the comoving-gauge density perturbation.
Both $\mu(a)$ and $\Sigma(a)$ are equal to zero at all redshifts in GR.
This parameterization has the advantage that the $\Sigma$ term can be constrained independently
by weak lensing with only mild degeneracy with $\mu$.
The RSD measurements probe the response of matter to a gravitational potential
and therefore provide independent constraints on the $\mu$ term.
Again, following \citet{2019PhRvD..99l3505A}, we describe the redshift evolution of $\mu$ and $\Sigma$ as
\begin{equation}
\mu(z) = \mu_0 \frac{\Omega_{\Lambda}(z)}{\Omega_\Lambda}\, , \, \, \, \, \Sigma(z) = \Sigma_0 \frac{\Omega_{\Lambda}(z)}{\Omega_\Lambda}\, .
\label{eq:musigform}
\end{equation}

Finally, neutrinos can affect the measured growth of fluctuations. While ultra-relativistic,
they free-stream out of over-densities and thus suppress growth on scales smaller than their
free-streaming length \citep[e.g.,][]{2006PhR...429..307L}. The dominant effect is a decrease in the amplitude of fluctuations at
low redshifts compared to extrapolations from the CMB under a model with zero neutrino mass. 

\subsection{Observable Signatures}\label{subsec:signatures}

\subsubsection{The CMB}

The temperature of the CMB is uniform across the sky to one part in 100,000;
beyond this level, anisotropies appear at all observable scales.
The angular power spectrum of the CMB can be predicted to high precision
based on an inflationary model and an expansion model.
The fluctuation modes corresponding to scales greater than one degree were larger
than the Hubble distance at the time of
the last scattering and capture the initial conditions imprinted at the end of inflation ($n_s$ and its derivative).
At smaller scales, the sound waves that propagate in the ionized universe due to photon-baryon coupling imprint the
characteristic acoustic oscillations into the CMB power spectrum.
The relative amplitudes of the peaks of the oscillations provide information on the energy
contents of the Universe while the spacing of the peaks provides a BAO
`standard ruler' whose length can be computed using straightforward physics.

Using the latest CMB results under an assumption of a \LCDM\ cosmology, the BAO feature
has a {\it comoving} scale of $147.18 \pm 0.29$ Mpc \citep{2018arXiv180706209P}, 
set by the distance $r_d$
traveled by sound waves between the end of inflation and the
decoupling of baryons from photons after recombination,
\begin{equation}
\label{eq:rd}
r_d = \int_{z_d}^\infty \frac{c_s(z)}{H(z)} dz,
\end{equation}
where $z_d$ is the redshift of the drag epoch and $c_s$ is the sound speed. 
Not to be confused with the redshift at the time of last scattering, the drag epoch corresponds to the time
when the baryons decouple from the photons, around a redshift $z=1020$. 
In the standard cosmological models explored here, $r_d$ can be computed given the physical densities of
dark matter ($\omega_c$), baryonic matter ($\omega_b$), and the radiation content of the Universe.
The radiation content can be determined from the
temperature of the CMB and the effective number of neutrino species ($N_{\rm eff}$).
Combined, these abundances determine the shape and position of the BAO peak in 
comoving space that can then be used as a standard ruler.
Because the CMB provides an image of the oscillations at the epoch of last photon scattering,
the BAO scale has not reached its maximum size, but it can still be measured at very high precision
to provide a constraint on the angular diameter distance to a redshift of $z \sim 1100$.

Because the proton-electron plasma does not recombine instantaneously,
the last scattering surface has a finite thickness.  Photon diffusion also results in damping at
the smallest scales, leading to a diffusion scale that depends on the expansion rate and
energy densities.  
The effect of damping on the power spectrum therefore allows constraints on the energy densities
of relativistic particles, primordial helium abundance, dark matter, and baryon matter at the time of last scattering.  
Finally, the signal from the CMB records the integrated ionization history of hydrogen and the integrated formation of structure
in the form of polarization and lensing signals.  Characterization of polarization and lensing in the CMB
thus provides information about the integrated optical depth ($\tau$) to
the surface of last scattering and the effects of neutrinos on the growth rate of structure.

One of the main challenges in interpreting the high signal-to-noise CMB data is the modeling of
foreground contamination, whether from dust emission, induced polarization, or radio sources.
For a review of experimental and analysis methods to extract cosmological information from the CMB,
see \citet{staggs18a} and \citet{2018arXiv180706205P}.

\subsubsection{BAO measurements from spectroscopic surveys}  \label{sec:BAO-meas}

The same sound waves that appear as acoustic oscillations in the CMB
appear in the clustering of matter at later times, although with a weaker amplitude
due to the coupling of baryonic matter with dark matter \citep[e.g.,][]{pardo20a}.  For this reason, survey volumes of several
${\rm Gpc}^3$ are required to reach percent-level precision constraints on the BAO feature.
The dark matter distribution that records the BAO feature cannot be probed directly, and is instead
traced by galaxies, quasars, or absorption line systems corresponding to neutral hydrogen
or other material in the intergalactic medium.
 
The cosmological parameters used to calibrate the
characteristic BAO scale $r_d$ are typically derived from CMB
observations.  The $r_d$ scale can also be derived from Big Bang Nucleosynthesis
(BBN) measurements (giving constraints on $\omega_b$) in combination with
measurements of expansion history (giving constraints on $\Omega_m$),
if the early universe is assumed to be a mixture of radiation,
baryonic matter, and cold dark matter with three neutrino species.
With a calibrated $r_d$, the BAO scale can be 
used to make absolute distance measurements as a function of redshift. Or $r_d$
can be treated as a nuisance parameter, allowing multiple BAO measurements
over a range of redshifts to be used for relative measures of the cosmic expansion history.

In a spectroscopic survey, the BAO feature appears in both the line-of-sight
direction and the transverse direction. Along the line-of-sight direction,
a measurement of the redshift
interval, $\Delta z$, over which the BAO feature extends, provides a means to
directly measure the Hubble parameter, $H(z) =c \Delta z/r_d$.
Equivalently, it measures the Hubble distance at redshift $z$,
\begin{equation}
 D_H(z) = \frac{c}{H(z)}.
\end{equation}
Along the transverse direction, the BAO scale corresponds to an angle,
$r_d=D_M(z)\Delta\theta$. Measuring the angle $\Delta\theta$ subtended by the BAO feature 
at a given redshift provides a means to estimate the (comoving) angular diameter
distance, $D_M(z)$, which depends on the expansion history and curvature as
\begin{equation}
  D_M(z) = {c \over H_0} S_k \left( {D_C(z) \over c/H_0} \right).
\label{eqn:dm}
\end{equation}
Here the line-of-sight comoving distance is
\begin{equation}
  D_C(z) = {c \over H_0}\int_0^z dz' {H_0 \over H(z')}
\label{eqn:dcomove}
\end{equation}
and
\begin{equation}
S_k (x) =
\begin{cases}
  \sin(\sqrt{-\Omega_k} x)/\sqrt{-\Omega_k} & \Omega_k<0, \\
  x & \Omega_k=0, \\
  \sinh(\sqrt{\Omega_k}x)/\sqrt{\Omega_k} & \Omega_k>0. \\
\end{cases}
\end{equation}
When considering the dependence of $r_d$ on cosmology, 
the quantities that the BAO measurements directly
constrain are $D_M(z)/r_d$ and $D_H(z)/r_d$.  The BAO measurements
were also historically summarized by a single quantity representing the
spherically-averaged distance,
\begin{equation}
 D_V(z) \equiv \left[z D^2_M(z) D_H(z)\right]^{1/3},
\end{equation}
or more directly $D_V(z)/r_d$. The powers of \sfrac{2}{3} and \sfrac{1}{3} approximately 
account for two transverse and one radial dimension and the extra factor of $z$ is a 
conventional normalization. Today we almost always specify the transverse and radial BAO as two
independent measurements with correlated error bars instead, unless the signal-to-noise ratio is low.

For measurements using discrete tracers with sufficiently high number density, the BAO feature in
clustering measurements can be sharpened through a process known as
`reconstruction' \citep{eisenstein07a}.  Reconstruction uses the
observed three-dimensional map of galaxy positions to infer their 
peculiar velocities.  Each galaxy tracer is then moved to a position that is
approximately where the galaxy would reside if there were no bulk
flows.  The process removes the dominant non-linear effect from the
BAO feature, which is smearing caused by the large-scale bulk flows.
Reconstruction recovers almost all theoretically available information
in the BAO.  In the SDSS analyses, the fitting to reconstructed data
is performed with minimal information from the broadband clustering signal,
in an attempt to isolate the BAO signal.

A review of BAO as a probe for cosmology is presented in Section~4 of \citet{weinberg13a}, and a 
discussion on the BAO measurement in practice can be found in Appendix~\ref{sec:appendix_a}.

\subsubsection{RSD measurements from galaxy surveys}  \label{sec:RSD-meas}

The galaxy redshifts used in spectroscopic BAO measurements
can also be used to study anisotropic clustering.
There are two primary ways in which anisotropy is introduced into the large-scale
clustering of matter:
the Alcock-Paczynski (AP) effect \citep{alcock79a} and the RSD effect from the growth of structure \citep{kaiser87a}.
The AP effect arises in clustering statistics as a deviation from physically isotropic
signal due to an incorrect translation of angular and radial (redshift) separations to physical ones
(e.g. from lack of knowledge of the true cosmology; see Appendix~\ref{sec:appendix_a}).
The AP effect thus serves as a way to measure the
product of $H(z)$ and $D_M(z)$, offering additional constraints on dark energy and curvature \citep[e.g.,][]{nadathur20a}.

The RSD effect arises from the growth of structure \citep{kaiser87a} and is
observed due to the bulk flow of matter in response to the
gravitational potential of matter overdensities. The peculiar velocities
introduce additional redshifts on top of those caused by cosmic expansion,
leading to an increase in the measured amplitude of radial
clustering relative to transverse clustering on large scales.
The resulting
anisotropy is correlated with the rate at which structure grows.  
The
growth rate $f(z)$ from equation~(\ref{eqn:dlng.dlna}) can also be
expressed as
\begin{equation}
f = \frac{\partial \ln\sigma_8}{\partial\ln a},
\end{equation}
where $\sigma_8(z)$ describes the amplitude of linear matter
fluctuations on a comoving scale of 8$h^{-1}{\rm Mpc}$.
The RSD measurements provide constraints on $f\sigma_8$, which
characterizes the amplitude of the velocity power spectrum.  

The AP and RSD signals are partially degenerate, which limits the AP
signal that can be extracted from measurements of clustering \citep[e.g.,][]{ballinger96a}.
A review of RSD and AP as a probe for cosmology is presented in Section~4 of \citet{weinberg13a},
while a discussion of the RSD measurement in practice can be found in Appendix~\ref{sec:appendix_a}.

\subsubsection{Weak lensing} 
\begin{table*}
  \centering
\caption{\rm Symbols and Definitions of Cosmological Parameters}
\begin{tabular}{|ll|}
\hline \hline
Parameter & Definition \\
\hline
 $\Omega_m$                  & density parameter of matter\\
 $\Omega_c$                  & density parameter of cold dark matter\\
 $\Omega_b$                  & density parameter of baryons\\
 $\Omega_\Lambda$            & density parameter of cosmological constant\\
 $\Omega_{\rm DE}$           & density parameter of dark energy\\
 $\Omega_k$                  & curvature parameter\\
 $\omega_c=\Omega_c h^2$     & physical density parameter of cold dark matter\\
 $\omega_b=\Omega_b h^2$     & physical density parameter of baryons\\
 $H_0$                       & current expansion rate (Hubble constant)\\
 $h  $                       & $H_0/100\, {\rm km\, s^{-1}Mpc^{-1}}$\\
 $\theta_{\rm MC}$           & approximate angular scale of sound horizon (CosmoMC)\\
 $A_s$                       & power of the primordial curvature perturbations at $k = 0.05\, {\rm Mpc}^{-1}$ \\
 $\sigma_8$                  & amplitude of matter fluctuation on 8$h^{-1}{\rm Mpc}$ comoving scale\\
 $n_s$                       & power-law index of the scalar spectrum\\
 $\tau$                      & Thomson scattering optical depth due to reionization\\
 $N_{\rm eff}$               & effective number of neutrino-like relativistic degrees of freedom\\
 $w$ ($w_0$)                 & dark energy equation of state, $w=p_{\rm DE}/\rho_{\rm DE}$ ($c=1$ units)\\
 $w_a$                       & time derivative of dark energy equation of state parameter (eq.~\ref{eq:darkenergy})\\
 $\sum m_\nu$                & sum of neutrino masses\\
\hline
\end{tabular}
\label{tab:parameters}
\end{table*}

As RSD probe the response of matter to a gravitational potential,
gravitational lensing probes the response of photons to a
gravitational potential.  Gravitational lensing can be
observed in several forms in cosmic surveys, we focus on the weak
lensing regime in this work.  More specifically, we use cosmic shear measurements of
weak lensing and galaxy-galaxy lensing measurements in performing cosmological constraints.

Cosmic shear shows up as distortions on the order of 1\% that appear
in the images of background galaxies due to lensing by the integrated foreground
mass distribution.  By introducing correlations of neighboring galaxy
shapes due to shared foregrounds, cosmic shear allows direct inference
of the gravitational potential gradients integrated along the line of sight.  If
these correlations are computed over discrete intervals 
over a range of redshifts, a smooth, three-dimensional mapping of the matter
distribution can be deduced.  The direct observable in lensing surveys
is the cosmic shear power spectrum, with an amplitude that scales
approximately as $\Omega_m^2 \sigma_8^2$ in the linear
regime. However, weak-lensing measurements are often in the non-linear
regime, and also depend on relative distances through the lens
equation. The relative balance between $\Omega_m$ and $\sigma_8$ in
the measurement depends on a number of factors within CDM models, as
described in \citet{Jain1997}. For the redshifts probed by current
surveys, around the benchmark \LCDM\ model, the redshift
evolution of the amplitude of the cosmic shear power spectrum is best
described by the approximate combination
\begin{equation}
S_8 \equiv \sigma_8(\Omega_m/0.3)^{0.5}.
\end{equation}
A review of cosmic shear methodology and its challenges as a probe for
cosmology can be found in Section~5 of \citet{weinberg13a}.

In addition to shear measurements, we also use galaxy-galaxy lensing
results in Section~\ref{sec:global_fit} to provide additional
information on the galaxy clustering measurements obtained
in photometric surveys.  Galaxy-galaxy lensing measurements probe
the local gravitational potential around specific classes of galaxies.
For the cosmology studies presented here, these measurements give insight into mass density
profiles, thus providing important information on the bias of the galaxies used as tracers
in the photometric clustering measurements.

\subsubsection{Type~Ia supernovae} 

Type~Ia supernovae are generally believed to occur when a white dwarf
approaches the Chandrasekhar mass limit due to mass accretion or merger.  This class
of SN is easily characterized with spectroscopy due to the strong calcium and silicon lines and lack of hydrogen and helium lines.  
While SNe~Ia are not perfect standard candles, their diversity can be described
by the SN light curve width (hereafter $X_1$) and SN color at maximum brightness (hereafter $C$).
The distance modulus, $\mu=5{\rm log_{10}}[D_L(z)/10\rm{pc}]$,
is then given by
\begin{equation}
 \label{eq:mu}
 \mu = m_B^{*} - (M_B - \alpha X_1 + \beta C) ,
 \end{equation}
where $m_B^*$ is the observed SN peak magnitude in rest-frame $B$ band \citep{astier06a}. 
Here $D_L$ is the luminosity distance, which follows the relation $D_L=D_M(1+z)$.
The quantity $M_B$ characterizes the SN~Ia absolute magnitude, while $\alpha$ and $\beta$ describe the
change in magnitude with diversity in width and color, respectively. 
The linear dependence between SN property and peak magnitude follows from the empirical observation
that brighter SNe~Ia are also slower to rise and/or bluer in color \citep[see][]{hamuy96a,phillips99a}.
Beyond those two dominant effects, a residual diversity related to host galaxy properties
was also found \citep[e.g.,][]{sullivan11a}, with brighter SNe occurring in more massive galaxies.
This effect is usually accounted for by considering that the SN~Ia absolute magnitude is different
depending on the host stellar mass, such as in \cite{betoule14a}:
\begin{equation}
 \label{eq:mb}
\begin{cases}
M_B = M_B ^1 \quad {\rm if} \quad M_{\rm stellar}<10^{10}M_{\odot}; \\
M_B = M_B^1 + \Delta_{M} \quad {\rm otherwise}.
\end{cases}
 \end{equation}
The model assumes that SNe~Ia with identical color, light curve shape and galactic environment
have on average the same intrinsic luminosity for all redshifts.
Note that the hypothesis of redshift independence can be checked with data for $\Delta_{M}$, $\alpha$, and $\beta$
and so far has been found to be consistent with observations \citep[e.g.,][]{scolnic18a}.

If the above model is sufficiently accurate, the measured SN distance modulus traces the
redshift dependence of luminosity distance.
The absolute magnitude can be calibrated using nearby SNe~Ia and Cepheid variables, giving a distance
ladder from which $H_0$ can be computed.  A review of supernova
astrophysics and their use in cosmology to constrain the dark energy
equation-of-state can be found in \citet{goobar11a}.

\subsection{Combining measurements} 

The measurements of the redshift-distance relation through BAO, AP, and SNe~Ia
provide tests of extended models for dark energy and cosmic expansion that are only weakly constrained with CMB data alone.
Generally speaking, the SNe~Ia data provide a high precision constraint of the luminosity distance--redshift relation
in the dark-energy dominated regime while the BAO and AP measurements sample the matter-dominated
regime and the epoch of matter-dark energy equality.
Likewise, the measurements of growth of structure through RSD and weak lensing allow additional tests
on the background expansion and on whether GR describes the rate of structure growth.
Measurements of the redshift-distance relation and growth of structure allow tests of the neutrino mass by constraining
the effects on both the cosmic expansion after the CMB formation and  
the amplitude of matter fluctuations relative to amplitude of CMB fluctuations.
The sensitivity of the latter approach is limited by our knowledge of
optical depth $\tau$ to the last scattering surface.
Alternative approaches to constrain the neutrino mass rely on measuring the
redshift-dependence of growth directly with clustering data or scale-dependence of the matter power spectrum
\citep{2006PhR...429..307L,2018arXiv180902120Y,2018PhRvD..97l3526C} but are not explored here.

\medskip
For fitting the measurements, model calculations throughout this paper are made with {\tt CosmoMC} \citep{lewis02a}.
Figures are produced with the {\tt GetDist} Python package \citep{lewis19a}.
The model parameters are summarized in 
Table~\ref{tab:parameters}, while parameterizations and priors are described in Appendix~\ref{sec:appendix_model}.
We stress that choice of parameterization is sometimes important -- the shape and visual overlap of
marginalized contours can be significantly impacted, especially in a prior-dominated regime.
In all cases that use information from the shape of the power spectrum, we hold $N_{\rm eff}$
fixed to its baseline value.
In the majority of the studies presented in this paper, the priors we assume on
free parameters do not impact the posterior distributions when CMB data are included in the likelihoods.
We refer to this series of priors as those with the `CMB' parameterization.
In the cases where we study the expansion history without the CMB (Section~\ref{sec:expansion}), 
we use the
`background' parameterization.
In all studies, the same priors are used for curvature, the dark energy equation
of state, or neutrino masses in the cases that those parameters are fit to the data.
Those priors are reported in the `extended' portion of the table in Appendix~\ref{sec:appendix_model}.

\section{Data and Methodology}\label{sec:data}
\begin{table*}
  \centering
 \caption{\rm Data sets for cosmology analyses.}
  \small
\begin{tabular}{|l|l|c|}
\hline \hline
Name & Data Combination & Cosmology Analysis \\
\hline
BAO & $D_M(z)/r_d$ and $D_H(z)/r_d$ from BAO measurements of all SDSS tracers & Section~\ref{sec:expansion} \\
RSD & $f\sigma_8(z)$ from all SDSS tracers, marginalizing over $D_M(z)/r_d$ and $D_H(z)/r_d$ & Section~\ref{sec:growth} \\
SDSS & $D_M(z)/r_d$, $D_H(z)/r_d$, and $f\sigma_8(z)$ of all SDSS tracers & Sections~\ref{sec:global_fit},\ref{sec:conclusion} \\
\CMBTP\ & \Planck\ TT, TE, EE, and lowE power spectra & Sections~\ref{sec:expansion},\ref{sec:growth} \\
CMB lens & \Planck\ lensing measurements & Section~\ref{sec:growth} \\ 
Planck & \Planck\ temperature, polarization, and lensing measurements & Sections~\ref{sec:global_fit},\ref{sec:conclusion} \\
SN & Pantheon SNe~Ia measurements & Sections~\ref{sec:expansion},\ref{sec:global_fit},\ref{sec:conclusion} \\
WL & DES cosmic shear correlation functions & Section~\ref{sec:growth} \\
DES & DES 3$\times$2 measurements (cosmic shear, galaxy clustering, and galaxy-galaxy lensing) & Sections~\ref{sec:global_fit},\ref{sec:conclusion} \\
\hline
\end{tabular}
  \label{tab:names}
\end{table*}

In this section, we provide an overview of the different measurements used
in our primary cosmological analysis, including: Baryon Acoustic Oscillations
(BAO), Redshift Space Distortions (RSD), Cosmic Microwave Background (CMB),
Supernovae (SN) and Weak Lensing (WL).
The samples we use in this work and the naming conventions we choose are summarized in Table~\ref{tab:names}.
We present the state-of-the-art results, and discuss how the different
probes have evolved during the last decade.

\subsection{SDSS BAO and RSD Measurements}

\begin{figure*}[htb!]
\centering
\includegraphics[width=0.75\textwidth, angle=0]{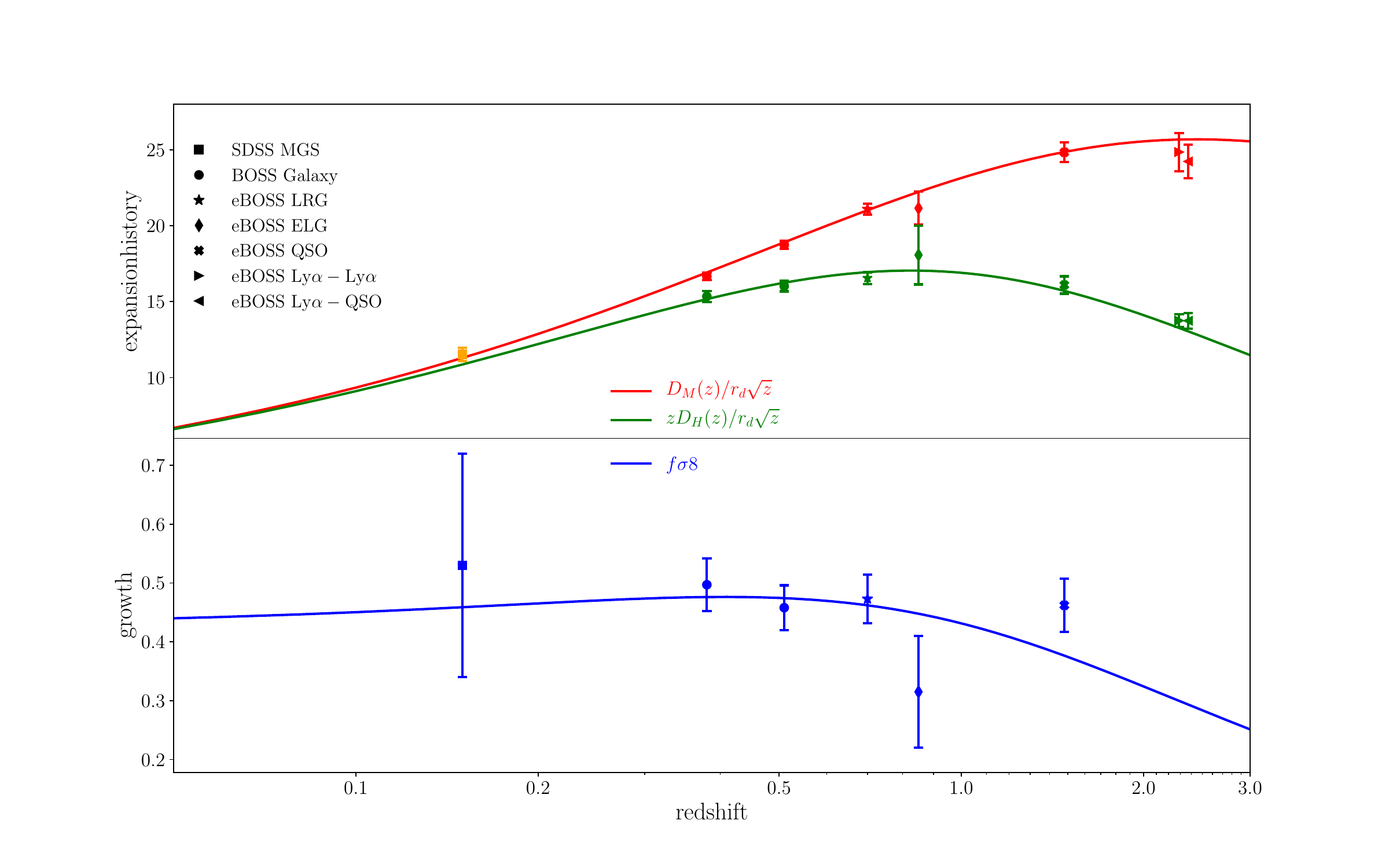}
\caption{{\bf Top: } Distance measurements from the SDSS lineage of BAO measurements presented as a
  function of redshift.  Measurements include those from SDSS
  MGS \citep{ross15a,howlett15a}, BOSS galaxies \citep{alam17a}, eBOSS LRGs \citep{LRG_corr,gil-marin19a},
  eBOSS ELGs \citep{tamone19a,demattia19a}, eBOSS quasars \citep{hou19a,neveux19a}, the BOSS+eBOSS $\lya$ auto-correlation,
  and the BOSS+eBOSS $\lya$-quasar cross-correlation measurements \citep{2019duMasdesBourbouxH}. 
Red points correspond to transverse BAO, while green points to radial BAO. The MGS $D_V$ measurement is plotted 
in orange with a translation to $D_M$ assuming a \LCDM\ model for illustrative purposes.
The red and green theory curves are not fit to the BAO data;
they are the \Planck\ bestfit predictions for a flat $\Lambda$CDM model.
  {\bf Bottom: } Growth rate measurements from the SDSS lineage of $f\sigma_8$ measurements as
  a function of redshift.  The measurements match the BAO samples
  except for $z>2$, where we do not report a measurement of the growth
  rate.  As for the upper panel, theory curve is not a fit, but a bestfit \Planck\ model.  }
    
\label{fig:sdssmeasurements}
\end{figure*}

The study presented in this work characterizes the impact of BAO and RSD measurements from spectroscopic
galaxy and quasar samples obtained over four generations of SDSS.
A summary of the BAO-only measurements is found in Table~\ref{tab:BAORSD_measurements}
and in the top panel of Figure~\ref{fig:sdssmeasurements}.
In these measurements, the broadband clustering signal that carries information on the AP effect or RSD
is effectively deweighted to capture only the BAO signature.  These measurements are used
to explore the impact of BAO measurements on models for dark energy in Section~\ref{sec:expansion}.
Results from the full-shape fits, without information from reconstructed BAO measurements,
are found in the central region of Table~\ref{tab:BAORSD_measurements}.
These measurements include information from the AP effect and are used to explore the impact of
growth measurements in Section~\ref{sec:growth}
A summary of the BAO and RSD measurements, including information from the AP effect and reconstruction,
is also found in Table~\ref{tab:BAORSD_measurements} and Figure~\ref{fig:sdssmeasurements}.
These measurements are used to 
perform the global cosmology fitting in Sections~\ref{sec:global_fit} and \ref{sec:conclusion}.
The background to each of these measurements is summarized below and described in detail in the relevant references.
All results in Table~\ref{tab:BAORSD_measurements} reflect the consensus values in the
cases where multiple measurements are made.

In this paper we only include large-scale structure based
  measurements from SDSS experiments. These are consistent with those
  from other experiments, including 6dFGS \citep{beutler11a} and
  WiggleZ \citep{blake12a}. However, the non-SDSS experiments do not
  add significantly to the measurements from SDSS: for example,
  \citet{Carter18} showed that 6dFGS only adds enough information to
  provide an improvement of $\sim16$\% on the SDSS MGS results at low
  redshift, while the WiggleZ sample has less than 10\% of the
  effective volume of the BOSS CMASS sample (see Table~1 of
  \citealt{Beutler16}). Given issues with overlap between samples and
  the resulting complicated covariance, we simply do not include these
  data.

\begin{table*}
\caption{\rm Clustering measurements\footnote{Uncertainties are Gaussian approximations to the likelihoods for each tracer ignoring
the correlations between measurements} for each of the BAO and RSD samples used in this paper.}
\begin{center}
\resizebox{\textwidth}{!}{%
\begin{tabular}{|l|c|c|c|c|c|c|c|c|}
\hline \hline
Parameter & MGS & BOSS Galaxy & BOSS Galaxy & eBOSS LRG & eBOSS ELG & eBOSS Quasar & \lya-\lya & \lya-Quasar \\
\hline
\multicolumn{9}{|c|}{}\\
\multicolumn{9}{|c|}{{\bf Sample Properties}} \\
\hline
redshift range & $0.07<z<0.2$ & $0.2<z<0.5$ & $0.4<z<0.6$ & $0.6<z<1.0$ & $0.6<z<1.1$ & $0.8<z<2.2$ & $z>2.1$ & $z>1.77$  \\
$N_{\rm tracers}$ & 63,163 & 604,001 & 686,370 & 377,458 & 173,736 & 343,708 & 210,005\footnote{The number of tracers reported for the \lya-\lya\
measurement corresponds to the number of sightlines, or forests.} & 341,468\footnote{Number of tracer quasars is used for the \lya-quasar study.} \\
$z_{\rm eff}$ & 0.15 & 0.38 & 0.51 & 0.70 & 0.85 & 1.48 & 2.33 & 2.33 \\
$V_{\rm eff}$ (Gpc$^3$)\footnote{The effective volume, $V_{\rm eff}$, is quoted here in Gpc$^3$ using a flat \LCDM\ model with
$\Omega_m=0.31$ and $h=0.676$.} & 0.24 & 3.7 & 4.2 & 2.7 & 0.6 & 0.6 & &  \\
\hline
\multicolumn{9}{|c|}{}\\
\multicolumn{9}{|c|}{{\bf BAO-Only Measurements\footnote{The measurements for MGS, the two BOSS galaxy samples, eBOSS LRG, and eBOSS ELG
are performed after reconstruction.}} (Section~\ref{sec:expansion})} \\
\hline
$D_V(z)/r_{\rm d}$ & $4.47 \pm 0.17$ &  & &  & $18.33_{-0.62}^{+0.57}$ & & &  \\
$D_M(z)/r_{\rm d}$ & & $10.23 \pm 0.17$ & $13.36 \pm 0.21$ & $17.86 \pm 0.33$ & & $30.69 \pm 0.80$ & $37.6 \pm 1.9$ & $37.3 \pm 1.7$ \\
$D_H(z)/r_{\rm d}$ & & $25.00 \pm 0.76$ & $22.33 \pm 0.58$ & $19.33 \pm 0.53$ & &  $13.26 \pm 0.55$ & $8.93 \pm 0.28$ & $9.08 \pm 0.34$ \\
\hline
\multicolumn{9}{|c|}{}\\
\multicolumn{9}{|c|}{{\bf RSD-Only Measurements} (Section~\ref{sec:growth})} \\
\hline
$f\sigma_8(z)$ & $0.53 \pm 0.16$ & $0.500 \pm 0.047$ & $0.455 \pm 0.039$ & $0.448 \pm 0.043$ & $0.315 \pm 0.095$ & $0.462 \pm 0.045$ & & \\
\hline
\multicolumn{9}{|c|}{}\\
\multicolumn{9}{|c|}{{\bf BAO+RSD Measurements} (Sections~\ref{sec:global_fit} and \ref{sec:conclusion})} \\
\hline
$D_V(z)/r_{\rm d}$ & $4.51  \pm 0.14$ & & & & & & &  \\
$D_M(z)/r_{\rm d}$ & & $10.27 \pm 0.15$ & $13.38 \pm 0.18$ & $17.65 \pm 0.30$ & $19.5 \pm 1.0$ & $30.21 \pm 0.79$ & $37.6 \pm 1.9$ & $37.3 \pm 1.7$  \\
$D_H(z)/r_{\rm d}$ & & $24.89 \pm 0.58$  & $22.43 \pm 0.48$ & $19.78 \pm 0.46$ & $19.6 \pm 2.1$ & $ 13.23 \pm 0.47 $ & $8.93 \pm 0.28$ & $9.08 \pm 0.34$  \\
$f\sigma_8(z)$ & $0.53 \pm 0.16$ & $0.497 \pm 0.045$ & $0.459 \pm 0.038$ & $0.473 \pm 0.041$ & $0.315 \pm 0.095$ & $0.462 \pm 0.045$ & & \\
\hline 
\end{tabular}
}
\end{center}
\label{tab:BAORSD_measurements}
\end{table*}

{\bf Main Galaxy Sample (MGS) ($0.07<z<0.2$):}  The first two generations of SDSS (SDSS-I and -II)
provided redshifts of nearly one million galaxies \citep{abazajian09a}.
SDSS galaxies were selected with $14.5 < r < 17.6$\footnote{\url{http://sdss.physics.nyu.edu/vagc/lss.html}}
over a contiguous footprint of 6,813 deg$^2$ to perform clustering measurements.
The sample was further refined to cover the redshift range $0.07<z<0.2$,
include the bright objects with $M_r < -21.2$, and include red objects with $g - r > 0.8$.
The resulting sample contains 63,163 galaxies intended to occupy the highest mass halos
while providing a roughly uniform number density over the full redshift interval.
The sample was used to perform a BAO measurement from the reconstructed
correlation function \citep{ross15a} and an RSD measurement from the anisotropic correlation function \citep{howlett15a},
both at an effective redshift $z_{\rm eff}=0.15$.
The BAO measurement was characterized with $D_V(z)/r_d$
and the RSD fit was performed using the post-reconstruction BAO fit as a prior.
The likelihoods from this work are found in the Supplementary Data associated with \citet{howlett15a}.
We refer to this sample as the `Main Galaxy Sample' (MGS) in the table and throughout the paper.

{\bf BOSS DR12 Galaxies ($0.2<z<0.6$):}  Over the period 2009--2014, 
BOSS performed spectroscopy to measure large-scale structure with galaxies over the redshift interval $0.2<z<0.75$.
BOSS obtained redshifts for 1,372,737 galaxies over 9,376 deg$^2$ from which the final
galaxy catalog was produced for clustering measurements \citep{reid16a}.  
The sample was divided into three redshift bins covering $0.2<z<0.5$, $0.4<z<0.6$, and $0.5<z<0.75$
for studies of BAO and RSD.  
For each redshift bin, seven different measurements of BAO, AP, and RSD were performed
\citep{ross17a,vargas18a,beutler17a,beutler17b,satpathy17a,sanchez17a,grieb17a}
based on the galaxy correlation function or power spectrum.
Following the methodology of \citet{sanchez17b}, these measurements were combined into a single consensus likelihood
spanning $D_M(z)/r_d$ and $D_H(z)/r_d$ for the BAO-only measurements
and $D_M(z)/r_d$, $D_H(z)/r_d$, and $f\sigma_8(z)$ for the combined BAO and RSD measurements.
These results were computed over all three redshift intervals after fully accounting
for systematic errors and covariances between parameters and between redshift bins \citep{alam17a}.
We refer to the $0.2<z<0.5$ and $0.4<z<0.6$ samples as the `BOSS Galaxies'.

{\bf eBOSS Galaxies and Quasars ($0.6<z<2.2$):}  eBOSS began full operations in July 2014
to perform spectroscopy on luminous red galaxies (LRGs), emission line galaxies (ELGs), and quasars
and concluded on March 1, 2019.  eBOSS obtained reliable redshifts for 174,816 LRGs over the interval
$0.6 < z < 1$ in an area of 4,103 deg$^2$.  The targets for spectroscopy were
selected from SDSS $riz$ imaging data and infrared sky maps from the Wide-field Infrared Survey Explorer \citep[WISE;][]{wright10a}.
The LRG selection \citep{prakash16a} was optimized to cover $0.6<z<1$ with a median redshift $z=0.72$.
The sample was supplemented with the galaxies in the $z>0.6$ tail of the BOSS DR12 redshift distribution,
but over the full 9,376 deg$^2$ of the BOSS footprint. The addition of BOSS galaxies more than doubles
the total sample size to 377,458 redshifts while slightly lowering the effective redshift.
This `eBOSS LRG' sample was used to measure $D_M(z)/r_d$ and $D_H(z)/r_d$
using a catalog of reconstructed galaxy positions.  In addition, the sample was used to perform a joint
$D_M(z)/r_d$, $D_H(z)/r_d$, and $f\sigma_8(z)$
measurement in both the correlation function \citep{LRG_corr} and the power spectrum \citep{gil-marin19a}.

Covering an area of 1,170 deg$^2$, eBOSS also obtained reliable redshifts for 173,736 ELGs over
the redshift range $0.6<z<1.1$.  These targets were identified in
$grz$ photometry from the Dark Energy Camera \citep[DECam;][]{flaugher15a} following
the selection algorithms presented in \citet{raichoor17a}.
These star-forming galaxies were spectroscopically confirmed with high efficiency
due to their strong emission lines that are easily detectable with the BOSS spectrographs \citep{smee13a} to $z=1.1$.
The `eBOSS ELG' sample reaches an effective redshift $z_{\rm eff}=0.85$.
We performed an isotropic BAO fit to measure $D_V(z)/r_d$ \citep{raichoor19a,demattia19a} and a combined RSD and BAO analysis 
to constrain $f\sigma_8(z)$, $D_H(z)/r_d$ and $D_M(z)/r_d$
from both the correlation function \citep{tamone19a} and the power spectrum \citep{demattia19a}.
Because the likelihoods are not well-described by a Gaussian distribution, we use the full likelihoods in the cosmology fitting.

Finally, the `eBOSS quasar' sample includes 343,708 reliable redshifts with $0.8 < z < 2.2$ over 4,699 deg$^2$.
The sample selection \citep{myers15a} was derived from WISE infrared and SDSS optical imaging data;
18\% of these quasars identified by the algorithm had been observed in SDSS-I, -II, or -III.
The sample was used to measure $D_M(z)/r_d$, $D_H(z)/r_d$, and $f\sigma_8(z)$
from both the correlation function \citep{hou19a} and the power spectrum \citep{neveux19a}.
The consensus BAO-only results were determined without reconstruction.
The full-shape $D_M(z)/r_d$, $D_H(z)/r_d$, and
$f\sigma_8(z)$ measurements were therefore not combined with the BAO-only measurements.

{\bf Lyman-$\alpha$ Forest Samples ($1.8<z<3.5$):}
The complete BOSS sample contains the spectra of 157,845 quasars at 
$2.0<z<3.5$ that are free of significant broad absorption lines.
These quasar targets were selected using a variety of techniques \citep{ross12a}
to measure fluctuations in the transmission of the Lyman-$\alpha$ (\lya)
forest due to fluctuations in the density of neutral hydrogen.
The auto-correlation of the \lyaf\ and its cross-correlation with 217,780
quasars at $1.8<z<3.5$ led to 2\% precision measurements of the BAO distance
scale at $z_{\rm eff} = 2.33$ \citep{bautista17a,masdesbourboux17a}.

Several techniques, such as those using photometric variability 
\citep{palanque-delabrouille16a}, were used to select new $z>2.1$ quasars to observe in eBOSS.
In addition, 42,859 quasars with low signal-to-noise BOSS spectra were
re-observed in eBOSS to better measure the fluctuations in the \lyaf.
Finally, improvements to the analysis methods enabled the use of a larger
wavelength range for determining the forest.
The final sample used to trace the \lya\ forest has 210,005 quasars at $z>2.1$,
consisting of the original sample from BOSS and the sample from eBOSS.
A total of 341,468 quasars with $z>1.77$ were used for cross-correlation studies with the \lyaf.

The final eBOSS results are presented in \citet{2019duMasdesBourbouxH}.
The auto- and cross-correlation measurements can also be combined into a single
estimate of $D_M(z)/r_d$ and $D_H(z)/r_d$ with associated covariances
\citep{2019duMasdesBourbouxH}, resulting in a 25\% reduction in
the area of the contours relative to the BOSS DR12 studies.
The uncertainties quoted in Table \ref{tab:BAORSD_measurements} correspond to
a Gaussian approximation of the real likelihood, but in our analysis we use
the full (non-Gaussian) likelihood.

In \citet{2019duMasdesBourbouxH}, we also presented a 4\% measurement of the redshift-space
distortion parameter of the \lya\ forest, $\beta_F$. However, $\beta_F$ can not be readily translated into
a measurement of $f \sigma_8$, since the response of \lya\ forest fluctuations to a velocity gradient
is unknown \citep{mcdonald99a, seljak12a}.

{\bf Summary of SDSS Measurements and Systematic Errors:}
From the BOSS and eBOSS clustering analyses with galaxies and quasars, the main
systematic errors in BAO and RSD estimates arise from modeling of the two-point statistics,
the choice of fiducial cosmology taken as a reference for coordinate transformation and power spectrum
template, and from the observational effects. The systematic errors also have larger effect on the RSD
analyses than the BAO analyses.  The estimation of the systematic errors was done in a similar fashion
for all tracers although some differences in the treatment remain and are outlined in the following. 

The modeling systematic errors are studied using accurate mocks based on N-body simulations for
which the cosmology is known \citep{rossi20a,smith20,alam20a}.
Special care is taken to estimate the effect of having a fixed fiducial cosmology for calculating distances
and shape of the template for the two-point statistics. In detail, we measure the range of the
differences between true and recovered values obtained by fitting to mocks where the true and fiducial
cosmologies do not match.  The distribution of cosmologies spanned by the mocks acts as a prior on `allowed cosmologies'. 
All galaxy and quasar tracers used both blind and non-blind mocks to assess their modeling systematic errors.
Variations of the Halo Occupation Distribution parameterizations are also taken into account.
For the BOSS Galaxy, ELG, and LRG samples, the modeling systematic error is further reduced by scaling the
$\sigma_8$ value according to the isotropic dilation factor measured independently in the
data and in each set of mocks (see Appendix~\ref{sec:appendix_a}). For the quasar sample, the redshift determination is an order
of magnitude less precise than for the galaxies and requires special modeling. The systematic effect of redshift errors
on the two-point statistics is estimated using the N-body mocks \citep{smith20} and is comparable in size
to the systematic errors in modeling. 

Observational effects are studied using approximate mocks that are modified to account for the observational
conditions \citep{lin20a,zhao20a}. This includes the dependence of the spectroscopic success rate on the signal-to-noise
ratio of the spectra, the treatment of fiber collisions, and the variations of the density of
targets for different photometric conditions in the imaging data. For the ELGs and quasars,
fiber collisions are taken into account at the model level and their effect is reduced.

More details about the sets of mocks used to estimate these errors are presented in the papers describing the
mocks and the papers describing the
individual measurements \citep{LRG_corr,gil-marin19a,raichoor19a,tamone19a,demattia19a,hou19a,neveux19a}.
In summary, for the LRG full-shape analysis, the overall systematic errors amount to about 40 to 60\% of the
statistical error depending on the parameters. The systematic errors for the ELG measurement reaches the
same level although with different sources of systematic effects. For the quasars, the systematic errors are at the level of 30\%
of the statistical error for all parameters.

Several tests for systematic errors were performed for the \lya\ BAO
studies, such as tests on mock spectra, modeling of the broadband signal
in the correlation function, and assessment of metal and sky
contributions to the \lya\ transmission estimates.  The central values
of the $D_M/r_d$ and $D_H/r_d$ estimates did not change significantly during these tests and no
additional systematic errors were included in the reported BAO results.
To account for the somewhat non-Gaussian errors on $D_M/r_d$ and $D_H/r_d$, we
generated 1000 realizations to estimate the translation of the $\Delta
\chi^2$ from each measurement in the parameter space to confidence
intervals on the BAO parameters.  The BAO measurements reported in
Table~\ref{tab:BAORSD_measurements} include this correction.

{\bf Summary of SDSS Likelihoods:}
The final $D_M(z)/r_d$, $D_H(z)/r_d$, and $f\sigma_8(z)$ measurements
cover eight distinct redshift intervals.
The systematic errors and consensus estimates are assessed in the studies that report the final measurements and incorporated
directly into the covariance matrices used in this study.  
Covariances between the two BOSS galaxy
measurements are propagated to this study through the same covariance matrix reported in \citet{alam17a}.

We find that the expected statistical correlation between clustering measurements derived from the
eBOSS samples is negligibly small and we thus include no covariance between them in our cosmological analyses.
This decision for the covariance between the quasar clustering measurements,
the Ly$\alpha$ auto-correlation measurements, and the Ly$\alpha$-quasar cross-correlation measurements was justified
using mock catalogs that demonstrated negligible correlation.
For the galaxy and quasar samples, the correlation within the overlapping volume can be estimated as
\begin{equation}
C_{\rm o} = \frac{P_1P_2}{(P_1+1/n_1)(P_2+1/n_2)},
\end{equation}
where $P$ represents the power-spectrum amplitude and $n$ is the number density.
We use the effective $P$ value in \cite{ross20a} and determine an effective
$1/n$ value based on the effective volume. For both the correlation between the quasars
and the ELGs and between the quasars and the LRGs, we find $C_{\rm o}$ is less than 0.1,
implying any correlation with the quasar sample is negligible.
Within their overlapped volume, the expected correlation between the ELGs and LRGs is higher, as each sample has a peak $nP > 1$.
However, over the full $0.6 < z < 1.0$ overlap range, we find $C_{\rm o} = 0.24$.
Accounting for the fact that the ELG footprint is significantly smaller than the LRG footprint
again reduces the expected correlation to less than 0.1.

Upon final acceptance for publication, the final likelihoods for the MGS, BOSS galaxy, and eBOSS
measurements will all be found on the public SDSS svn
repository\footnote{\url{https://svn.sdss.org/public/data/eboss/mcmc/trunk/likelihoods}} and in the Github
repository\footnote{\url{https://github.com/evamariam/CosmoMC_SDSS2020}}.
The full likelihood is reported for BAO-only studies in the MGS, ELG, and \lya\ forest samples.
The BAO-only results for the BOSS galaxy, eBOSS LRG, and eBOSS quasar samples are recorded
as a covariance matrix.  We refer to the combination of these measurements as the `BAO' measurements throughout
the paper.
The combined fits for BAO, AP, and RSD results are recorded as a full likelihood for the MGS and
ELG samples, while the results for the BOSS galaxy, eBOSS LRG, and eBOSS quasar samples are
recorded in a single covariance matrix.  We refer to these data samples as the `RSD' samples
when no information from reconstruction is used and the likelihoods are collapsed to a single dimension on $f\sigma_8$.
We refer to the full analyses of reconstructed BAO and full-shape AP+RSD fitting as the
`SDSS' sample.
In all cases, the likelihoods include both statistical and systematic errors.

\subsection{CMB, SNe, and WL Measurements}

The BAO measurements from the four generations of SDSS are complemented by relative distance measurements from SNe~Ia.
The SDSS RSD measurements are complemented by WL measurements from CMB and recent imaging programs.
CMB anisotropies from all-sky, space-based surveys are used throughout to provide a baseline of high redshift, cosmological measurements.
Finally, we compare the local value of the Hubble expansion parameter derived from various combinations of CMB,
BAO, SNe~Ia, and BBN to the most recent results using local measurements.
Neither the BBN nor the $H_0$ estimates are directly used in any other cosmological fitting
and are not discussed any further in this section,
although the BBN constraints on $\omega_b$ are used to inform priors in several growth measurements.
In the remainder of this section, we discuss the results from the CMB, SNe~Ia, and WL studies that
we use to assess progress in building the cosmological model.

The WMAP satellite launched on June 30, 2001 and ceased scientific operations on August 19, 2010.
The cosmological measurements based only on the final WMAP sample provide constraints of $\Omega_c h^2 = 0.1138 \pm 0.0045$
and $\Omega_b h^2 = 0.02264 \pm 0.00050$ in a flat \LCDM\ model \citep{bennett13a}.
The \Planck\ satellite \citep{planck11a} operated from 2009--2013 to measure CMB temperature and polarization
anisotropies to scales as small as $5\pr$.  These measurements allow very precise constraints on the
matter content and early expansion history of the
Universe, especially in the limit of a \LCDM\ cosmology.
An analysis under the assumption of a flat \LCDM\ model using only \Planck\ temperature and polarization data
leads to constraints $\Omega_c h^2 = 0.120\pm 0.001$, $\Omega_b h^2 = 0.0224\pm 0.0001$,
$n_s = 0.965\pm 0.004$, and $\tau = 0.054\pm 0.007$ \citep{2018arXiv180706209P}.
As the latest generation of CMB experiment, \Planck\ therefore provides a factor of 4.5 improvement over WMAP on the precision
of the dark matter density and a factor of 5 improvement on the precision of the baryonic matter density.
When computing constraints using the baseline \Planck\ measurements,
denoted \CMBTP\ throughout,
we use the {\tt Plik} likelihoods for the TT, TE, EE, and lowE power spectra \citep{2019arXiv190712875P}.
The data cover multipoles in the range $30 \le \ell \le 2508$ for the TT power spectrum
and $30 \le \ell \le 1996$ for the power spectra that include polarization.
When including additional lensing data from \Planck\, denoted `CMB lens',
we use the likelihoods from \citet{plancklensing} computed over lensing multipoles $8 \le \ell \le 400$.
When using temperature, polarization, and lensing data together, we refer to the sample simply as `Planck'.
The full likelihoods for \Planck\ and WMAP measurements are found in the \Planck\ public release of
2018 Cosmological parameters and MC
chains\footnote{\url{https://wiki.cosmos.esa.int/planck-legacy-archive/index.php/Cosmological_Parameters}
with a description of the CosmoMC implementation at \url{https://cosmologist.info/cosmomc/readme_planck.html} }
and the WMAP 2013 public
release\footnote{\url{https://lambda.gsfc.nasa.gov/product/map/current/likelihood_get.cfm}}, respectively.

At the time that eBOSS began observations, the leading SNe~Ia cosmology studies stemmed from the
`joint light-curve analysis' (JLA) sample.
These 740 SNe~Ia lightcurves were taken from low redshift surveys \citep{hicken09a,contreras10a},
the SDSS-II Supernova Survey \citep[2005--2007;][]{frieman08a,sako18a}, the Supernova Legacy Survey
\cite[SNLS, 2003--2008;][]{guy10a,conley11a,sullivan11a}, and 
high redshift space-based observations with the Hubble Space Telescope \citep{riess07a}.
A major effort in the analysis focused on reducing systematic uncertainties in the photometric calibration
of the SNLS and SDSS surveys.
For a flat \LCDM\ cosmology using only the SNe from this sample, the constraints on
the matter content of the local universe were found to be $\Omega_m =0.295 \pm 0.034$, including systematic errors \citep{betoule14a}.
More recently, the `Pantheon sample' of 1,048 SNe~Ia was used in a comprehensive cosmology analysis.
This sample includes the full set of spectroscopically confirmed SNe~Ia from PanStarrs \citep{kaiser10a}
supplemented by SNe~Ia observed at low redshift \citep{riess99a,jha06a,hicken09a,contreras10a,folatelli10a,stritzinger11a},
the SDSS and SNLS samples, and with HST \citep{suzuki12a,riess07a,rodney14a,graur14a,riess18a}.
While the increase in sample size since the JLA analysis is significant, the largest improvement in precision 
results from new cross-calibration of all ground-based measurements to the PanStarrs photometric system.
Using only this SN sample with the systematic uncertainties leads to a constraint
$\Omega_m = 0.298 \pm 0.022$ in a flat \LCDM\ model \citep{scolnic18a}.
Within the basis of a flat \LCDM, the Pantheon sample therefore offers a factor of 1.5 improvement in precision
over the JLA sample.  Systematic errors are still significant and dominated largely by photometric uncertainties of each sample,
the calibration uncertainties of the lightcurve model, and the assumption of no redshift dependence of $M_B$.
The statistical and systematic uncertainties are captured in a covariance matrix with an element for each supernova
following the methodology of \citet{conley11a}.
The statistical component of the uncertainties contributes only to the diagonal elements while the
off-diagonal elements are dominated by systematic errors arising from common uncertainties in bandpass
and zeropoint calibration.
We primarily use measurements of individual SNe~Ia from the Pantheon sample in making cosmological constraints
and refer to this as the `SN' sample.
The covariance matrix for both the JLA sample and the Pantheon sample can be found at the
Barbara A. Mikulski Archive for Space Telescopes (MAST)\footnote{\url{https://archive.stsci.edu/prepds/ps1cosmo/}}
and are included with the CosmoMC installation.

Several recent programs
\citep[e.g.,][]{2012MNRAS.427..146H,2018PASJ...70S..25M,2019PASJ..tmp...22H,hildebrandt20a,joudaki20a}
have reported cosmology constraints from measurements of cosmic shear.  Because we are not
able to account for covariances between these results due to shared systematic errors, we do not
attempt an analysis on the combined weak lensing results.  Instead, as an example of how weak
lensing data impact cosmological constraints, we focus here on the
results from the Dark Energy Survey (DES) conducted with the Dark Energy Camera \citep{flaugher15a}.
DES released an analysis of cosmic shear using the first year of data covering an area 
exceeding 1000 deg$^2$ with more than 20 million galaxy shape measurements.
Tomographic cosmic shear measurements were performed after assigning source galaxies to redshift bins
spanning the intervals $0.20<z<0.43$, $0.43<z<0.63$, $0.63<z<0.90$, and $0.90<z<1.30$.
The data are used under an assumption of a \LCDM\ model to constrain the combination
of $\Omega_m$ and $\sigma_8$ represented by $S_8 = 0.782 \pm 0.027$ at 68\% confidence \citep{troxel18a}.
As in the DES analysis, we only use scales in the cosmic shear correlation functions
that are expected to have contributions from baryonic effects of less than 2\%.
These studies are denoted `WL' in Section~\ref{sec:growth}.
In addition to cosmic shear measurements, we use the 3$\times$2pt DES Year 1 results
in the analysis presented in Sections~\ref{sec:global_fit} and \ref{sec:conclusion}, where
the additional correlation functions are computed from galaxy clustering
and galaxy-galaxy lensing. Following \citet{krause17a}, we
only use information from the correlation function on comoving scales larger than
$8 h^{-1}$Mpc for the galaxy clustering measurements and $12 h^{-1}$Mpc
for the galaxy-galaxy lensing.

We use the CosmoMC implementation of the DES
likelihood\footnote{\url{https://github.com/cmbant/CosmoMC/blob/master/batch3/DES_lensing.ini}}\footnote{\url{https://github.com/cmbant/CosmoMC/blob/master/batch3/DES.ini}}
with covariance matrix, power spectra measurements and nuisance parameters in agreement with \citet{troxel18a},
\citet{krause17a}, and \citet{2018PhRvD..98d3526A}.
The combined 3$\times$2pt sample is referenced simply as `DES'.

\section{Implications of Expansion History Measurements}\label{sec:expansion}

\begin{table*}[t!]
 \centering
 \small
 \caption{\rm Marginalized values and 68\% confidence limits in \LCDM\ and one-parameter
extensions using only expansion history and CMB temperature and polarization measurements.
 }
\begin{tabular}{|L|L|C|C|C|C|C|}\hline \hline 
& &  \Omega_\mathrm{DE} & H_0 \mathrm{[km/s/Mpc]} & \Omega_k & w &\Sigma m_\nu\,[\mathrm{eV}]\footnote{The reported $\sum m_\nu$ values correspond to the 95\% upper limits.}\\
\hline
 & \mathrm{BAO} & 0.701^{+0.017}_{-0.015} & -\footnote{BAO measure the dimensionless quantity $r_d H_0 /c$, and therefore can only
provide constraints on $H_0$ when combined with other probes.} & - & - & -\\
\multirow{5}{*}{$\Lambda \mathrm{CDM}$} & \mathrm{CMB \ T\&P} & 0.6836\pm 0.0084 & 67.29\pm 0.61 & - & - & -\\
 & \mathrm{CMB \ T\&P + BAO} & 0.6881\pm 0.0058 & 67.60\pm 0.43 & - & - & -\\
 & \mathrm{CMB \ T\&P + SN} & 0.6856\pm 0.0078 & 67.43\pm 0.57 & - & - & -\\
 & \mathrm{CMB \ T\&P + BAO + SN} & 0.6890\pm 0.0057 & 67.67\pm 0.43 & - & - & -\\
 \hline
 & \mathrm{BAO} & 0.636^{+0.085}_{-0.070} & - & 0.079^{+0.083}_{-0.10} & - & -\\
\multirow{5}{*}{$\mathrm{o}\Lambda\mathrm{CDM}$} & \mathrm{CMB \ T\&P} & 0.561^{+0.050}_{-0.041} & 54.5^{+3.3}_{-3.9} & -0.044^{+0.019}_{-0.014} & - & -\\
 & \mathrm{CMB \ T\&P + BAO} & 0.6884\pm 0.0059 & 67.62\pm 0.62 & 0.0000\pm 0.0018 & - & -\\
 & \mathrm{CMB \ T\&P + SN} & 0.670\pm 0.017 & 65.2\pm 2.2 & -0.0061^{+0.0062}_{-0.0054} & - & -\\
 & \mathrm{CMB \ T\&P + BAO + SN} & 0.6891\pm 0.0057 & 67.67\pm 0.60 & 0.0000\pm 0.0018 & - & -\\
 \hline
 & \mathrm{BAO} & 0.728^{+0.017}_{-0.038} & - & - & -0.69\pm 0.15 & -\\
\multirow{5}{*}{$\mathrm{wCDM}$} & \mathrm{CMB \ T\&P} & 0.801^{+0.057}_{-0.022} &  -\footnote{The constraints of \CMBTP\ in the \wCDM\ model are affected by the $H_0$ prior of $H_0<100$km/s/Mpc,
so no entry is provided.}  & - & -1.58^{+0.16}_{-0.35} & -\\
 & \mathrm{CMB \ T\&P + BAO} & 0.695\pm 0.013 & 68.5^{+1.3}_{-1.6} & - & -1.035^{+0.062}_{-0.052} & -\\
 & \mathrm{CMB \ T\&P + SN} & 0.692\pm 0.010 & 68.3\pm 1.1 & - & -1.035\pm 0.037 & -\\
 & \mathrm{CMB \ T\&P + BAO + SN} & 0.6929\pm 0.0076 & 68.23\pm 0.83 & - & -1.027\pm 0.033 & -\\
 \hline
\multirow{5}{*}{$\nu \Lambda \mathrm{CDM}$} & \mathrm{CMB \ T\&P} & 0.680^{+0.016}_{-0.0087} & 67.0^{+1.2}_{-0.67} & - & - & <0.268\ (95\%)\\
 & \mathrm{CMB \ T\&P + BAO} & 0.6892\pm 0.0065 & 67.72\pm 0.50 & - & - & <0.129\ (95\%)\\
 & \mathrm{CMB \ T\&P + SN} & 0.686^{+0.011}_{-0.0083} & 67.47^{+0.83}_{-0.65} & - & - & <0.174\ (95\%)\\
 & \mathrm{CMB \ T\&P + BAO + SN} & 0.6898^{+0.0065}_{-0.0056} & 67.76^{+0.49}_{-0.44} & - & - & <0.124\ (95\%)\\
 \hline
\end{tabular}
 \label{tab:main}
\end{table*}

In this section we discuss measurements of the background expansion
history, with an emphasis on the BAO measurements from SDSS.
We use the \Planck\ temperature and polarization data (\CMBTP),
the SN data from Pantheon, and the BAO data from SDSS.
The BAO data, summarized in the BAO-only section of
Table~\ref{tab:BAORSD_measurements}, include measurements from galaxy,
quasar and \lya\ forest samples.
It is this wide redshift range that enables the tight constraints on
cosmological parameters presented in this section.

We start in Section~\ref{subsec:extensions} with a discussion on the role
of BAO and SN measurements on single parameter extensions to the \LCDM.
By adding measurements of the expansion history, we show that we can break
parameter degeneracies present in the CMB results, leaving combined
fits that are always consistent with a flat \LCDM\ model.
The combined probes also offer some of the most competitive constraints on
neutrino mass without adding any information from growth of structure.
In Section~\ref{subsec:H0}, we show that the BAO data enable estimates
of $H_0$ that are robust against the assumption of cosmological model
and estimates that are independent of CMB anisotropies altogether.

\subsection{Impact of BAO Measurements on Models for Single Parameter Extensions to \LCDM} \label{subsec:extensions}

We first report the results in the simplest cosmology, that of a
spatially flat universe where dark energy can be explained by a 
cosmological constant (\LCDM).
As shown in Table~\ref{tab:main}, CMB data alone are sufficient
to constrain the dark energy density parameter to roughly 1\% precision.
Adding the BAO and SN data improves this constraint by a factor of 1.5 
for this simplest model of the expansion history.

\begin{figure*}
  \centering
  \includegraphics[width=0.9\textwidth, angle=0]{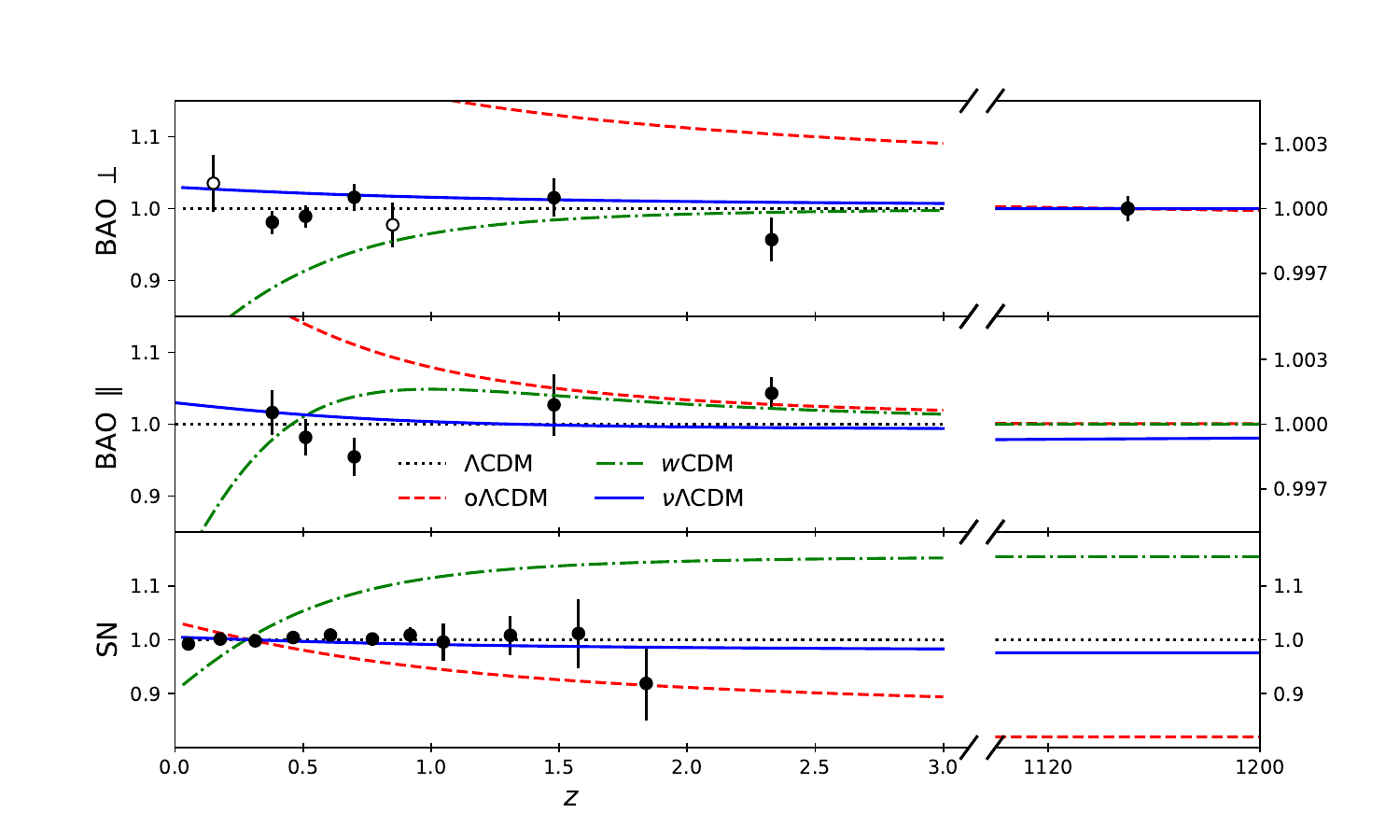}
  \caption{
Demonstration of BAO, SN, and CMB constraining power as a function of redshift.
To construct alternative models, we have fixed to their best-fit
\LCDM\ values the quantities that are best measured by the CMB:
$\Omega_b h^2$, $\Omega_c h^2$ and the angular acoustic scale $D_M(z=1150)/r_d$.
Because the sound horizon at decoupling is a function of $\Omega_b h^2$, $\Omega_c h^2$ and
$N_{\rm eff}$ only, the models have the same value of $r_d=147.16$\,Mpc.
{\bf Top: } The Hubble diagram residuals of BAO $D_M(z)$ measurements,
presented as the ratio of the measured value of $D_M(z)/r_d$ relative to the
prediction for that value based on the best-fit \LCDM\ model from CMB alone.
$D_V(z)$ measurements are shown as open circles.  We display the CMB determination of the angular position of the acoustic peak
as a measurement of transverse BAO, and we split the redshift scale
to include this data point.
{\bf Center: } The Hubble diagram residuals of BAO $D_H(z)=c/H(z)$ measurements,
normalized in the same manner as the $D_M(z)$ measurements.
{\bf Bottom: } The Hubble diagram residuals of \SN\ measurements, with relative
normalization of the luminosity distance estimates.
In order to increase the signal-to-noise, the supernovae data were binned into 11 bins
between redshifts 0.1 and 2.5.  Spacing was chosen to maintain a relatively constant signal-to-noise ratio.
Since the distance modulus varies significantly across the bin at low redshift,
we have averaged the signal by averaging the inverse covariance weighted deviations from the \LCDM\
model after the absolute normalization has been fitted. The covariance matrix was taken from the Pantheon dataset.
In each case, the residuals are computed relative to the best-fit \LCDM\ model
from CMB alone.
The curves represent the difference between the \LCDM\ model and
single-parameter extensions allowed by the CMB data.
The \oLCDM\ model favored by \Planck\ ($\Omega_k=-0.044$) is presented in
dashed red lines, the \wCDM\ model favored by \Planck\ ($w = -1.585$) is
presented in dot-dashed green lines, and a \LCDM\ model with non-zero
neutrino mass is presented in solid blue lines.
The model with massive neutrinos assumes a summed mass equal to $0.268$ eV,
corresponding to the \Planck\ 95\% upper limit.
  }
  \label{fig:hubble2}
\end{figure*}

\begin{figure*}
  \centering
  \includegraphics[width=0.45\textwidth, angle=0]{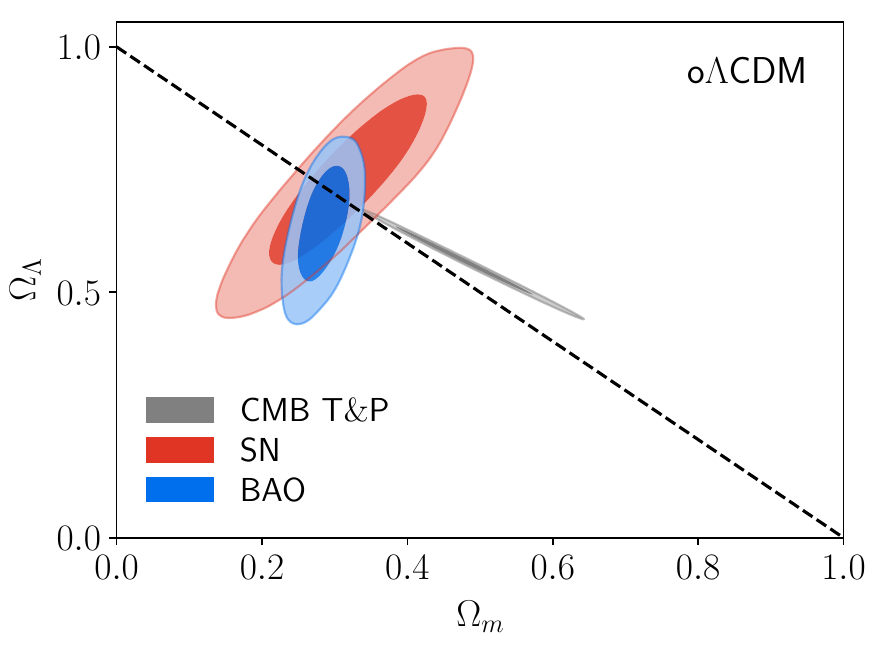}
  \includegraphics[width=0.45\textwidth, angle=0]{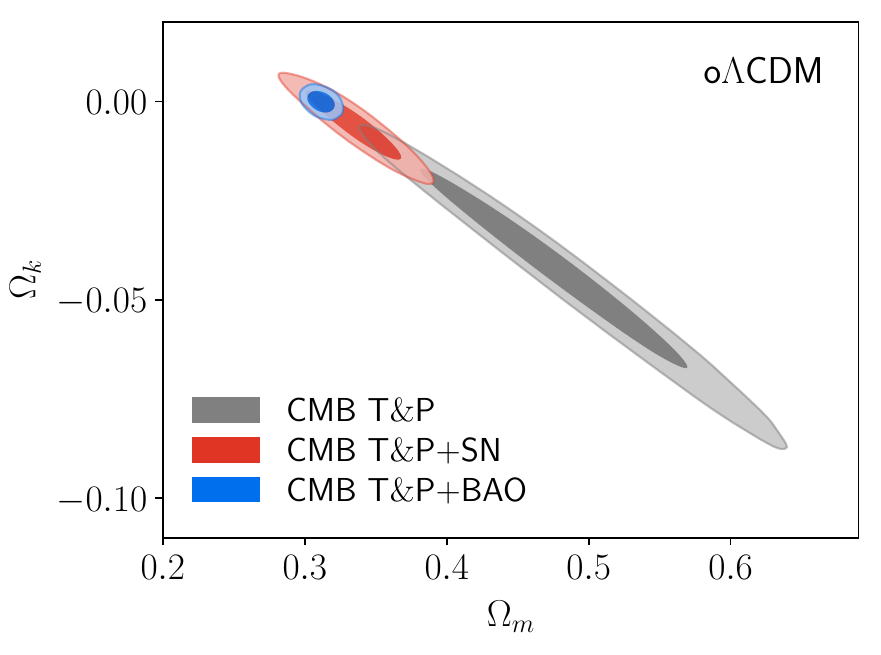}
  \caption{Cosmological constraints under the assumption of a model with a $w=-1$
cosmological constant with free curvature (\oLCDM, as in Table~\ref{tab:main}).
{\bf Left: } 68\% and 95\% constraints on $\Omega_m$--$\Omega_\Lambda$
from  the \Planck\ CMB temperature and polarization
data (gray), Pantheon \SN\ sample (red), and SDSS BAO-only measurements (blue).
The dashed line represents a model with zero curvature.
{\bf Right: } The $\Omega_m$--$\Omega_k$ constraints for the combination of
CMB (gray), CMB + SN (red), and CMB + BAO (blue). 
}
  \label{fig:SNvsBAO1}
\end{figure*}

\begin{figure*}[b!]
  \centering
  \includegraphics[width=0.45\textwidth, angle=0]{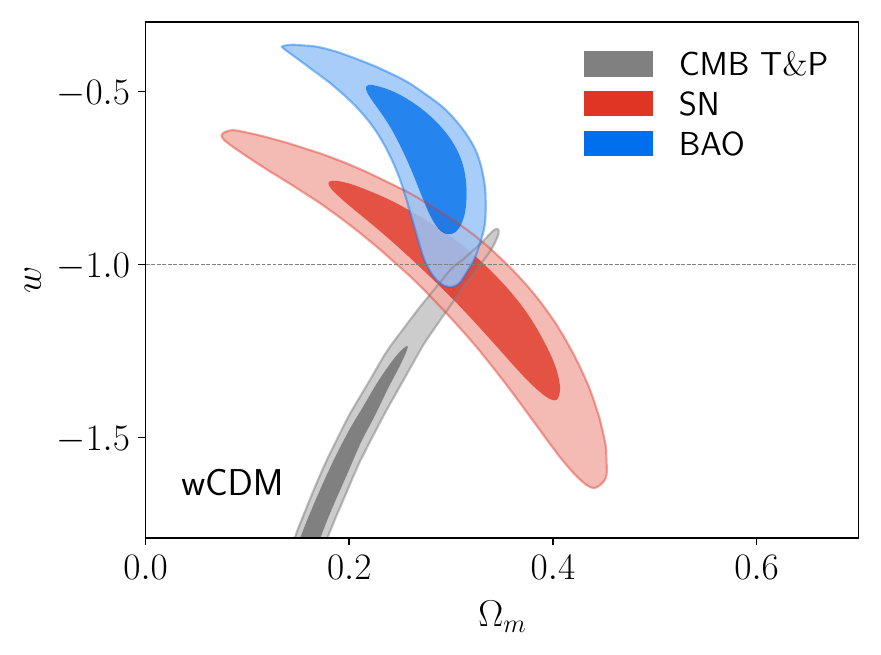}
  \includegraphics[width=0.45\textwidth, angle=0]{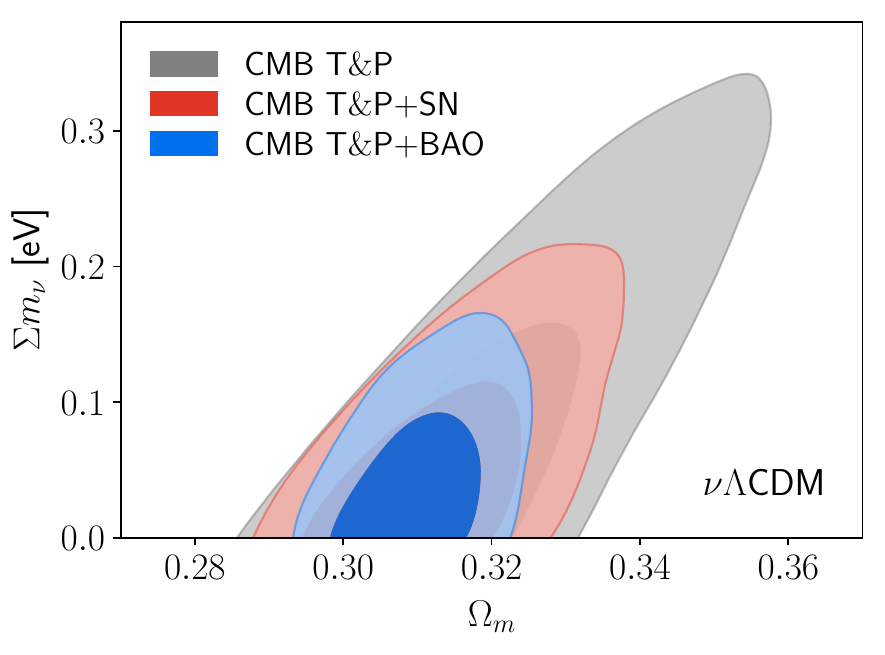}
  \caption{
Constraints on the \wCDM\ and \nuLCDM\ models, as in Table~\ref{tab:main}.
{\bf Left: } $w$--$\Omega_m$ constraints under the assumption of a flat
\wCDM\ cosmology from the \Planck\ CMB temperature
and polarization data (gray),  Pantheon \SN\ sample (red), and SDSS BAO-only measurements (blue).
{\bf Right: } $\sum m_\nu$--$\Omega_m$ constraints under the assumption of a
flat \LCDM\ cosmology where the summed neutrino mass is allowed as a free
parameter, for the combination of CMB (grey), CMB + SN (red), and CMB + BAO (blue).
  }
  \label{fig:SNvsBAO2}
\end{figure*}

Figure~\ref{fig:hubble2} shows the residuals of the BAO and \SN\
distances with respect to the \LCDM\ model favored by the \Planck\ temperature
and polarization data.
The BAO and \SN\ data have a combined $\chi^2/\mathrm{DoF}<1$
with respect to this model, indicating very good agreement.
In order to highlight how BAO and \SN\ data complement the CMB results in
models with a single-parameter extension to \LCDM, in Figure~\ref{fig:hubble2}
we also show the prediction for three models that are allowed by \Planck,
but are ruled out by measurements of the low redshift expansion history:
an \oLCDM\ model with the \Planck-favored value of $\Omega_k=-0.044$ 
(dashed red); a \wCDM\ model with the \Planck-favored value of $w=-1.585$ 
(dot-dashed green); and a \LCDM\ model with the \Planck\ 95\% upper limits on
the sum of the neutrino masses of $\sum m_\nu = 0.268$ eV
(solid blue)\footnote{The slight change in neutrino mass compared to the Planck
analysis is due to our use of an updated version of CosmoMC.}.

In the next sub-sections, we present in detail how BAO and \SN\ can break
strong degeneracies present in the CMB data when studying single-parameter
extensions to the \LCDM\ model.

\subsubsection{Expansion history and curvature}

The \Planck\ temperature and polarization data alone offer strong constraints within
the \oLCDM\ model, but with degenerate posteriors as shown in 
both panels of Figure~\ref{fig:SNvsBAO1}.
The consequences of these degeneracies are quantified in Table~\ref{tab:main}, where the uncertainty
on $\Omega_\Lambda$ in this model is five times larger than in a flat
\LCDM\ model.
The preference for a closed universe, with a significance
slightly above 95\% confidence, 
is discussed in detail in \cite{ade16a} and \cite{2018arXiv180706209P}.
As shown in Figure~\ref{fig:hubble2}, the predictions from the closed
universe favored by the CMB (dashed red lines) are disallowed at
high confidence by both the BAO and the SN data.

In an \oLCDM\ model, BAO measurements at different redshifts constrain
different combinations of ($\Omega_m$, $\Omega_k$, $r_d H_0 /c$).
When we combine BAO results over a wide redshift range, we are able to
break internal degeneracies and provide independent constraints on these
parameters \citep[e.g.,][]{nadathur20a}.
Table~\ref{tab:main} and the left panel of Figure~\ref{fig:SNvsBAO1} show that BAO measurements alone
lead to $\Omega_\Lambda = 0.636_{-0.070}^{+0.085}$, an $\sim 8\sigma$ confidence detection
of a cosmological constant without any information from the CMB or \SN\ data.
The \SN\ data alone also favor a flat geometry, but are not as constraining as BAO.
Using only \SN\ leads to a detection of $\Omega_\Lambda = 0.73 \pm 0.11$.

The right panel of Figure~\ref{fig:SNvsBAO1} demonstrates that including
either BAO or SN data reduces the parameter degeneracies in the CMB data.
The $\Omega_{\rm DE}$ results in the \CMBTP\ + BAO entries in
Table~\ref{tab:main} are almost the same in \LCDM\ and \oLCDM\ models.
The combination of BAO and CMB data favors a flat universe with
$\Omega_k=-0.0000 \pm 0.0018$.

\subsubsection{Expansion history and dark energy}

We next consider a flat \wCDM\ model, with an extra free parameter $w$ to
describe the equation of state of dark energy.
As with the \oLCDM\ model, the left panel of Figure~\ref{fig:SNvsBAO2} shows that the
CMB temperature and polarization data leave strong degeneracies between the $w$ and energy
density parameters that determine the expansion history.
Table~\ref{tab:main} shows
that the constraints on $\Omega_{\rm DE}$ are again degraded by a factor of about 
five with respect to the constraints in a \LCDM\ model,
with a shift in the central value that is opposite in direction to the shift
in the \oLCDM\ model.
The models with very negative values of $w$ favored by CMB 
(dot-dashed green lines in Figure~\ref{fig:hubble2}) are inconsistent with
both the BAO and the SN data.

As shown in the left panel of Figure~\ref{fig:SNvsBAO2}, the
\Planck\ \LCDM\ values ($w=-1$ and $\Omega_m=0.3164$) lie within the 95\% confidence intervals of both the
BAO data alone and the SN data alone.
The BAO data alone
are able to constrain the matter density without a strong degeneracy with $w$.
Even though the \SN\ contours have a strong degeneracy in $w$--$\Omega_m$,
the contours are perpendicular to the degeneracy direction of the CMB contours,
so the \CMBTP+SN combination results in very tight constraints on the
\wCDM\ model.
Each of the three combinations, \CMBTP+BAO, \CMBTP+SN, and even BAO+SN favor a model with a cosmological constant.
As shown in Table~\ref{tab:main}, the combination of all three datasets
results in a measurement of the equation of state of dark energy of
$w=-1.027 \pm 0.033$, consistent with a cosmological constant.

\subsubsection{Expansion history and neutrino masses}

We now turn our attention to a \nuLCDM\ model where the sum of the
neutrino masses is considered a free parameter.
As shown in the right panel of Figure~\ref{fig:SNvsBAO2} and in
Table~\ref{tab:main}, the \Planck\ temperature and polarization data offer
a 95\% upper bound on the summed neutrino mass of 268 meV.
Neutrinos lighter than $\sim 500$ meV are still relativistic at the time of
recombination, but they impact CMB observables by modifying the
late-time expansion, in particular the angular diameter distance to the 
epoch of recombination $D_M(z_{\rm rec})$.
Neutrino mass constraints from the CMB are therefore degenerate with other
cosmological parameters that modify $D_M(z_{\rm rec})$, like $\Omega_m$ or $H_0$.
Late-time measurements of the expansion can break this degeneracy,
as shown in blue lines of Figure~\ref{fig:hubble2} and in the right panel
of Figure~\ref{fig:SNvsBAO2}. 
Adding BAO or SN data reduces the upper bound
on the sum of neutrino masses by a factor of 2 and 1.5, respectively.
Combining the three datasets, we obtain a 95\% upper limit of 124 meV.

\bigskip

In this subsection we have shown that measurements of the
expansion history are very complementary to measurements of CMB temperature and polarization
anisotropies.
As shown in Figure~\ref{fig:hubble2}, both BAO and \SN\ are able to constrain
single-parameter extensions to \LCDM\ that can not be constrained by CMB alone.
As shown in Table~\ref{tab:main}, adding BAO to the CMB data reduces the
uncertainty on $\Omega_{\rm DE}$ in \oLCDM\ models by a factor of eight,
and it excludes models with curvature that would otherwise be favored by the CMB.
Similarly, adding SN and BAO to the CMB reduces the uncertainty on
$\Omega_{\rm DE}$ in \wCDM\ models by more than a factor of five,
and it excludes models with $w<-1$ favored by the CMB. 
For all models discussed, the combination of all three probes results in
a percent measurement of $\Omega_{\rm DE}$, consistent with
$\Omega_{\rm DE}=0.69$.

\subsection{BAO and the $H_0$ Tension}\label{subsec:H0}

\begin{figure*}[t!]
  \centering
   \includegraphics[width=0.45\textwidth, angle=0]{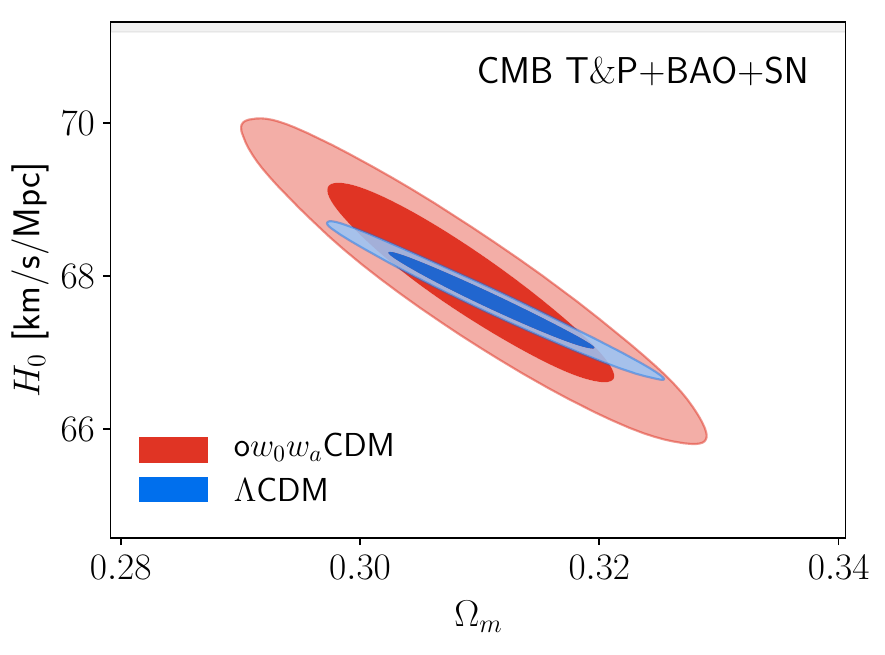}
   \includegraphics[width=0.45\textwidth, angle=0]{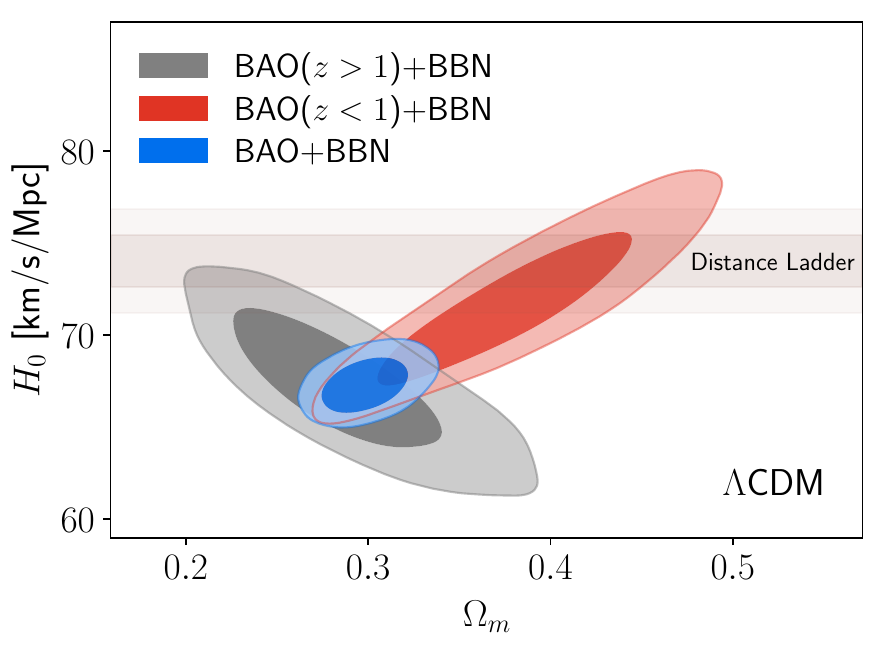}
  \caption{
    {\bf Left:} $H_0$ versus $\Omega_m$ from the inverse distance ladder
    (CMB+BAO+SN) under two different cosmological models.
    {\bf Right:} $H_0$ versus $\Omega_m$ from the combination of BAO and BBN,
    in a \LCDM\ model (blue).
    The red (gray) contours show the results when using only BAO measurements
    below (above) $z=1$.
    The horizontal shaded area shows the (68\%, 95\%) measurement of $H_0$
    from the distance ladder technique (\citep[SHOES;][]{riess19a}).
  }
\label{fig:Hocontours}
\end{figure*}

\begin{table*}[t!]
  \centering
 \caption{\rm Effect of BAO measurements on Hubble constant constraints.}
  \small
  \begin{tabular}{|p{3.5cm}ccp{5.5cm}|}
\hline \hline
    Dataset  & Cosmological model & $H_0$ ($\kmsmpc$) & Comments \\
    \hline
    \CMBTP+BAO+SN & \owowaCDM\ & $67.91\pm0.87$ & Inverse distance ladder\\
    BBN+BAO & \LCDM\ & $67.33\pm0.98$ & No CMB anisotropies\\
    \hline
    \CMBTP\ & \LCDM\ & $67.28\pm0.61$ & \Planck\ 2018 \citep{2018arXiv180706209P} \\
    \CMBTP\ & \oLCDM\ & $54.5^{+3.3}_{-3.9}$ & \Planck\ 2018 \citep{2018arXiv180706209P} \\
    Lensing time delays & \LCDM\ & $73.3 \pm 1.8$ & H0LiCOW \citep{1907.04869} \\
    Distance ladder & -  & $74.0 \pm 1.4$ & SH0ES \citep{riess19a} \\
    GW sirens & - &  $70 \pm 10 $ & LIGO \citep{1710.05835} \\
    TRGB & - & $69.6 \pm 1.9$ & LMC anchor \citep{freedman20a}  \\
    TFR & - & $76.2 \pm 4.3$ & Cosmicflows \citep{tully16a} \\
    Maser galaxies & - & $73.9 \pm 3.0$ & Megamaser Cosmology Project \citep{pesce20a} \\
    \hline
  \end{tabular}
  \label{tab:H0}
\end{table*}

The present-day expansion rate, $H_0$, is one of the basic parameters in
the cosmological model because it allows absolute estimates of the
age and the current energy content of the Universe. 
It is one of the three fundamental cosmological parameters that are not
dimensionless (the two other being the temperature of the CMB and the
neutrino masses).
Moreover, as discussed in \citet{hu2005} and in \citet{weinberg13a},
an accurate measurement of $H_0$ would allow a powerful test of dark energy
models and tightened constraints on cosmological parameters.
However, a statistically-significant tension has been demonstrated between
direct measurements of $H_0$ from the local distance ladder and those
estimates of $H_0$ inferred from the CMB \citep{riess16}.
This tension has persisted and even increased in significance,
despite significant effort to identify possible sources of systematic errors.

Measurements of the Hubble constant come in different flavors, as shown in the
compilation of studies presented in the bottom part of Table~\ref{tab:H0}.
An example of direct measurement, referred here as the {\it distance ladder},
uses parallaxes from local stars and other techniques to calibrate distances to Cepheid
variables, which are in turn used for absolute luminosity calibration of \SN\ hosted
by nearby galaxies \citep[e.g.,][]{riess19a}.
The calibrated luminosity is used to estimate the absolute luminosity
distance to a sample of \SN\ that covers a redshift range sufficient to
minimize the effect of peculiar velocities relative to the Hubble flow.
Similar efforts include the use of other distance indicators such as the tip
of the red giant branch \citep[TRGB; e.g.,][]{freedman20a}, Tully-Fisher
relation in galaxies \citep[TFR; e.g.,][]{neill14a}, or gravitational waves
from neutron star-neutron star mergers \citep[e.g.,][]{1710.05835}.
These measurements typically measure higher values of the Hubble constant.
For example, \citet{riess19a} perform a study using \SN\ distances calibrated
from 70 long-period Cepheids in the Large Magellanic Cloud. 
They find $H_0 = 74.03 \pm 1.42 \, \kmsmpc$, including systematic errors.

Other measurements of $H_0$ involve data at higher redshift, and need to
assume a cosmological model to extrapolate the constraints to redshift zero.
One example of this indirect measurement is that obtained using time delays in
strongly-lensed quasars \citep[e.g.,][]{birrer19a}.
Other indirect measurements of $H_0$ use CMB data under strong assumptions
about the model governing the expansion history from the last scattering
surface to today.
The CMB estimates typically give considerably lower values of the Hubble
constant.
The final \Planck\ data release, for example, finds
$H_0= 67.36 \pm 0.54 \, \kmsmpc$ \citep{2018arXiv180706209P} when assuming
the \LCDM\ model.

Explanations for the tension between direct measurements and CMB estimates
range from underestimated systematic errors or modeling of the primordial power spectrum
\citep[e.g.,][]{1907.12639,2001.09260,1909.10847,hazra19a}, to models for dark energy
\citep[e.g.,][]{li19a,alestas20a,divalentino20a},
to unmodeled pre-recombination physics that lead to a decreased sound horizon scale
\citep[e.g.,][]{1811.04083,1811.03624,2001.05874,1907.07569,1905.12618,1909.07986}.
See \cite{1908.03663} for a review of possible solutions to the tension.

We provide here two alternative analyses to show how BAO measurements
allow estimates of $H_0$ that are robust against the strict assumptions of
the CMB-only estimates.
First, we combine \Planck\ temperature and polarization, SN, and BAO data
and allow a very flexible expansion history to demonstrate that the tension
in $H_0$ estimates is not due to the assumptions of a \LCDM\ model.
Second, we present a measurement of $H_0$ that uses BAO and a BBN prior
that is independent of CMB anisotropies to demonstrate that the tension
is not due to systematic errors in the CMB data.
We finish this section presenting the combination of the BAO data with the
local distance ladder measurement, and we discuss the low value of $r_d$
inferred from this analysis.

\subsubsection{$H_0$ and the inverse distance ladder}
\label{subsub:H0_inverse}

In this subsection we present a cosmological measurement of $H_0$
using a model for expansion history that allows three additional free
parameters beyond \LCDM.
This approach is often referred as the {\it inverse distance ladder}, as it
relies on a calibrated distance measure at high redshift that is then
extrapolated to $z=0$.
Schematically, we use information from the CMB to calibrate the BAO distances.
Those in turn are used to calibrate the absolute luminosity of \SN.

Since the BAO feature follows $D_H(z)/r_d=c/H(z)/r_d$ and $D_M(z)/r_d$,
rather than $H(z)$ directly, this measurement relies on a calibration of the
sound horizon ($r_d$) at the drag epoch to extract the Hubble parameter.
Under the implicit assumption of a smooth expansion history, standard
pre-recombination physics, and a well-measured mean temperature of the CMB,
$r_d$ only depends on the cold dark matter density ($\Omega_c h^2$)
and the baryon density ($\Omega_b h^2$).
Thus, $r_d$ can be calibrated through constraints arising from the full CMB temperature and polarization
likelihoods, with little dependence on the late-time history of the universe,
as demonstrated in Figure~2 of \citet{Cuesta:2014asa}.

The extrapolation of $H(z)$ measurements from BAO to $z=0$ can be done using
a very flexible cosmology because both BAO and \SN\ relative
distance measurements constrain the evolving expansion rate.
The inclusion of BAO makes the technique robust to the assumed properties of
dark energy as was demonstrated in earlier BOSS analyses \citep{aubourg15a}.

We choose an \owowaCDM\ model to allow for a flexible expansion history of
the Universe.
Note that CMB alone can not constrain this model; as shown in
Table~\ref{tab:H0}, the uncertainties on $H_0$ from CMB constraints already increase
by a factor of about six when we consider only one parameter extensions,
such as models with curvature.
The combination of CMB, BAO and SN data, however, is able to provide a
very precise measurement of $H_0$ even in this flexible model.
Our results, presented in Table~\ref{tab:H0} and in the left panel of
Figure~\ref{fig:Hocontours}, have an uncertainty better than $1 \, \kmsmpc$
and are consistent with the low value of $H_0$ measured by the CMB
under the strict assumption of \LCDM.

\subsubsection{$H_0$ independent of CMB anisotropies}
\label{subsub:H0_BBN}

In the previous subsection, we showed that the value of $H_0$ measured
by the combination of CMB, BAO and SN data is robust under different
models for curvature and dark energy equation of state. 
In this section we return to the \LCDM\ model, and present a measurement
of $H_0$ that is independent of CMB anisotropies.

The combination of BAO measurements at different redshifts can provide a
precise measurement of the dimensionless quantity $r_d H_0 /c$.
To translate constraints on this dimensionless quantity to a measurement of
$H_0$, we use information on $\omega_b$ by including BBN constraints; $\omega_c$ and
$H_0$ are also left free as they can be determined in the fitting by the BAO
data\footnote{To estimate the radiation density we also use the absolute CMB
temperature measured by FIRAS, $T_0 = 2.7255$ K \citep{fixsen09}.}. 
We use the results of recent high resolution spectroscopic measurements of
seven quasar absorption systems that indicate a primordial deuterium abundance
${\rm D/H} = (2.527 \pm 0.030) \times 10^{-5}$ \citep{cooke18a}.
Using the empirically-derived reaction cross-section \citep{adelberger11},
the deuterium abundances imply $\omega_b = 0.02235 \pm 0.00037$ under an assumption
that $N_{\rm eff} = 3.046$.
The 68\% confidence interval reflects the combined deuterium abundance
and reaction rate uncertainties.

As can be seen in the right panel of Figure~\ref{fig:Hocontours}, we obtain
a tight constraint on $H_0$ only when we combine BAO measurements from
a wide redshift range. 
In particular, the line-of-sight BAO measurements above $z=1$ (from quasars
and the \lya\ forest) provide measurements of the expansion in the 
matter-dominated area, and their contours have different degeneracies in the
($\Omega_m$, $H_0$) plane.

As shown in Table~\ref{tab:H0}, the precision on $H_0$ when combining
BAO measurements with a BBN prior is $0.98 \, \kmsmpc$.
This result is consistent with the findings of \cite{1707.06547} and
\cite{1906.11628}, who used BAO data from SDSS DR12 and DR14, respectively.
The central value remains relatively unchanged from the results using
CMB, BAO and SN data in the \owowaCDM\ model, providing further evidence
that the tension is not due to peculiarities in the CMB anisotropy data.

\subsubsection{Sound horizon at drag epoch from low redshifts}
\label{subsub:H0_rd}

\begin{figure}
  \centering
  \includegraphics[width=0.45\textwidth, angle=0]{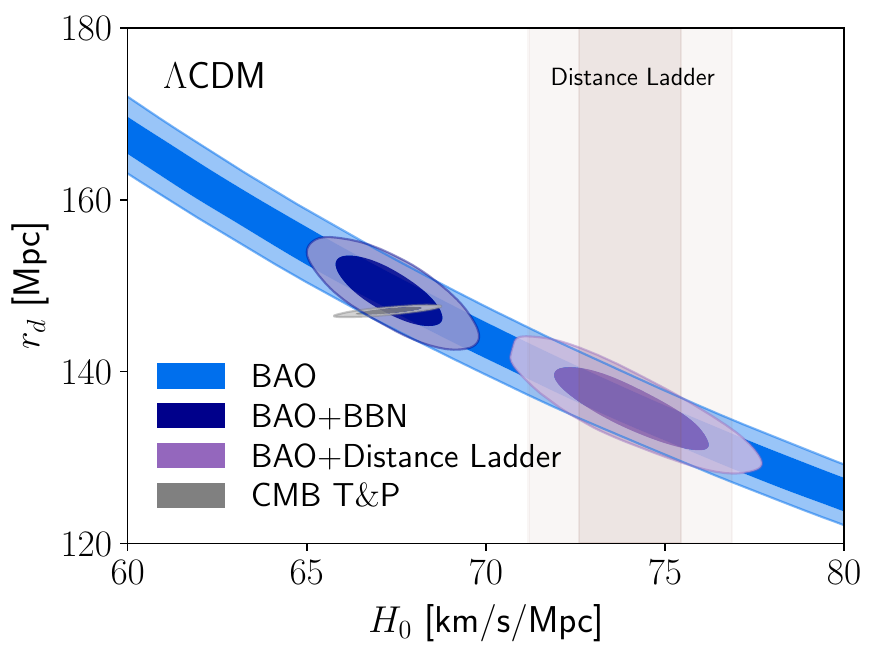}

  \caption{Cosmological constraints on $H_0$ and $r_d$ under the assumption of the \LCDM\ model
using BAO data (blue) in combination with $H_0$ distance ladder measurements (purple) and BBN data
(dark blue) in contrast to CMB measurements (grey). The shaded band refers to the $H_0$ distance ladder measurement. 
}
  \label{fig:H0_rd}
\end{figure}

As shown above, the BAO data in combination with information on the baryon density from the early Universe
can be used to extrapolate late universe expansion history to constrain the Hubble constant.
The BAO data can also be used to constrain the sound horizon at the drag epoch when combined with
local $H_0$ measurements \citep[e.g.,][]{Cuesta:2014asa}. 

Figure~\ref{fig:H0_rd} shows the 2D-contours of $H_0$ and $r_d$ for BAO data in combination with
different datasets under the assumption of a \LCDM\ model. The BAO data alone are completely
degenerate in the $H_0$--$r_d$ plane, however this degeneracy can be broken by either local $H_0$
measurements, by BBN, or by CMB data. The local $H_0$ measurements are clearly in tension with early
Universe measurements of the sound horizon. BAO and BBN data prefer a value
$r_d=149.3\pm2.7$ Mpc, in good agreement with the value $r_d=147.06\pm0.29$ Mpc
preferred by the CMB temperature and polarization data alone.
These estimates are much larger than the BAO and distance ladder constraint of $r_d=135.8\pm3.2$ Mpc.
These constraints on $r_d$ can also be translated into limits on the baryon density,
yielding $\omega_b=0.0310\pm0.0023$ for BAO and distance ladder data.
In comparison, the CMB best fit of $\omega_b=0.02236\pm0.00015$ or the BBN best-fit
of $\omega_b = 0.02235 \pm 0.00037$ are much lower.

Finally, dropping the assumption of a \LCDM\ model and including SN in our analysis of the distance
ladder, we find $r_d=135.2\pm3.1$ Mpc and $\omega_b=0.0375^{+0.0054}_{-0.0085}$ in a \owowaCDM\ model.
This extended distance ladder measurement shows that the discrepancy between low and high redshift
measurements of the sound horizon is independent of the assumption of the cosmological model.
However, we caution that we did not take the correlation between the SN data and the local
$H_0$ measurement into account for our analysis.

\bigskip

To summarize, the BAO data allow robust, consistent measurements of
$H_0$ that include freedom in expansion history beyond the strict
\LCDM\ assumptions in CMB-only estimates.
The BAO data allow robust, consistent measurements of
$H_0$ that are insensitive to the use of CMB anisotropies
altogether if using the \LCDM\ model.
In all cases, the central values remain below $H_0=68 \, \kmsmpc$, and the
uncertainties remain at $1 \, \kmsmpc$ or better.

On the other hand, the Cepheid distance ladder or strong lensing time delays
of quasars provide precise estimates of $H_0$ that favor larger values of $H_0$,
or smaller values of $r_d$ if being used to calibrate the BAO scale.
Combining their results as independent measurements produces an estimate
of $H_0=73.7 \pm 1.1 \, \kmsmpc$. This central value differs from those presented
in this work by more than four standard deviations whether we use a multiple parameter
model for expansion or the BBN measurements of $\omega_b$.
The consistency of the results highlights that the `$H_0$ tension' can
not be restricted to systematic errors in \Planck, or to the strict assumptions of the
\LCDM\ model.

Both the CMB analysis and those presented here are sensitive to the assumption
of standard pre-recombination physics that sets the scale of $r_d$.
As summarized in \citet{1908.03663}, there have been many attempts
to reconcile the $H_0$ tension by modifying the value of $r_d$,
with limited success.

\section{Implications of Growth Measurements}\label{sec:growth}
\begin{figure*}
  \centering
  \includegraphics[width=\textwidth,angle=0]{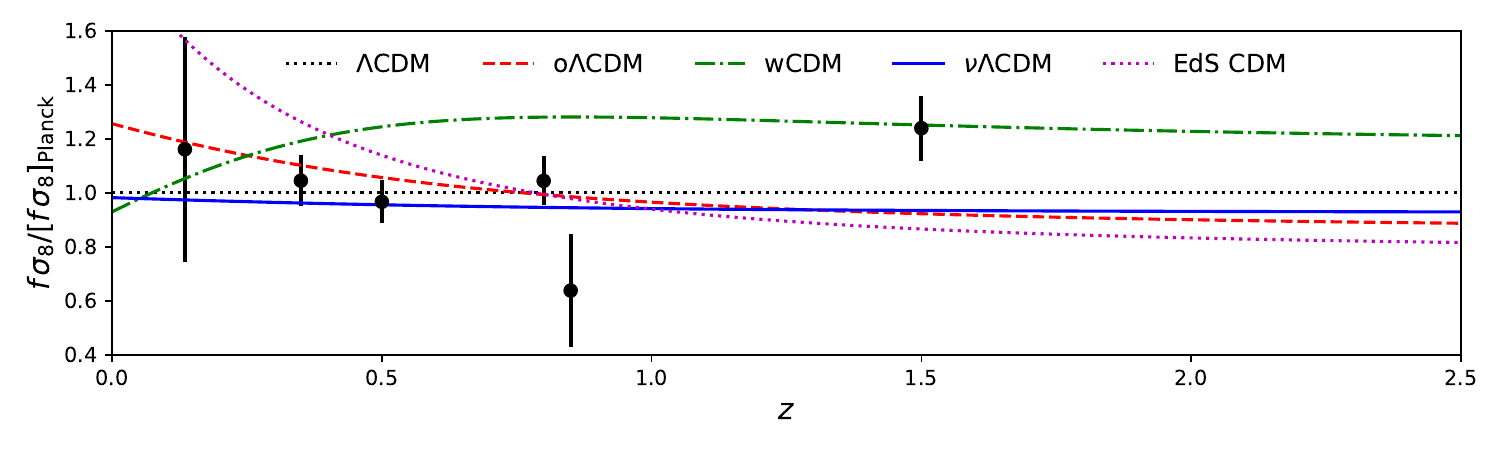}
  \caption{
The SDSS $f\sigma_8$ measurements as a function of redshift, normalized by the \Planck\ 2018 bestfit \LCDM\ model
(shown in dotted black).  The three colored curves represent the fractional deviations from
\LCDM\ for an \oLCDM\ model with $\Omega_k=-0.044$ (red), a \wCDM\ model with $w=-1.58$ (green), and
a \nuLCDM\ model with $\sum m_\nu = 0.268$ eV (blue). These are the same models as those in Figure \ref{fig:hubble2}.
An Einstein de Sitter model (magenta; $\Omega_m=1$, $\Omega_\Lambda=0$ and $\sigma_8(z=0)$ matching that of fiducial
model) is ruled out at high confidence, further demonstrating the long-standing preference for growth measurements
for models with lower matter densities.
}
        \label{fig:growth_residuals}
\end{figure*}

A key development of the BOSS and eBOSS surveys is the advancement of
RSD as a tool to make high precision measurements of structure growth over a wide redshift range.
In this section, we assess the impact of those growth measurements on the cosmological model.
We first compare the RSD measurements to DES weak lensing and \Planck\ lensing results 
to complement the CMB temperature and polarization data in dark energy models.
To achieve `RSD-only' constraints, we marginalize the dependency of $D_H$ and $D_M$
out of our `Full Shape' measurement and use only the results on the growth, $f\sigma_8(z)$.
In the second part of this section, we explore the use of growth measurements to constrain matter fluctuations and
to test the assumptions of GR in the cosmological model. 

\begin{table*}
 \centering
 \small
 \caption{\rm Marginalized values and 68\% confidence limits on curvature, dark energy parameters,
and the amplitude of density fluctuations using only growth
and CMB temperature and polarization measurements.}
\resizebox{\textwidth}{!}{%
\begin{tabular}{|C|C|C|C|C|C|C|}
 \hline \hline& & \Omega_m & \Omega_\mathrm{DE} & \sigma_8 & \Omega_k & w\\
\hline
\multirow{4}{*}{$\mathrm{o}\Lambda\mathrm{CDM}$} & \mathrm{CMB \ T}$\&$\mathrm{P} & 0.483^{+0.055}_{-0.069} & 0.561^{+0.050}_{-0.041} & 0.774^{+0.016}_{-0.014} & -0.044^{+0.019}_{-0.014} & -\\
 & \mathrm{CMB \ T}$\&$\mathrm{P+RSD} & 0.456^{+0.051}_{-0.066} & 0.580^{+0.048}_{-0.038} & 0.780^{+0.015}_{-0.013} & -0.037^{+0.018}_{-0.013} & -\\
 & \mathrm{CMB \ T}$\&$\mathrm{P+WL} & 0.310\pm 0.017 & 0.690\pm 0.013 & 0.806\pm 0.010 & -0.0004\pm 0.0048 & -\\
 & \mathrm{CMB \ T}$\&$\mathrm{P(+lens)+RSD+WL} & 0.313\pm 0.014 & 0.688\pm 0.011 & 0.8069\pm 0.0096 & -0.0010^{+0.0044}_{-0.0039} & -\\
 \hline
\multirow{4}{*}{$w\mathrm{CDM}$} & \mathrm{CMB \ T}$\&$\mathrm{P} & 0.199^{+0.022}_{-0.057} & 0.801^{+0.057}_{-0.022} & 0.970^{+0.096}_{-0.045} & - & -1.58^{+0.16}_{-0.35} \  \footnote{The lower bound on $w$ is affected by the $H_0$ prior.}\\\
 & \mathrm{CMB \ T}$\&$\mathrm{P+RSD} & 0.295^{+0.028}_{-0.034} & 0.705^{+0.034}_{-0.028} & 0.834\pm 0.030 & - & -1.08\pm 0.11\\
 & \mathrm{CMB \ T}$\&$\mathrm{P+WL} & 0.188^{+0.012}_{-0.046} & 0.812^{+0.046}_{-0.012} & 0.977^{+0.083}_{-0.037} & - & -1.61^{+0.13}_{-0.30}\\
 & \mathrm{CMB \ T}$\&$\mathrm{P(+lens)+RSD+WL} & 0.275^{+0.024}_{-0.028} & 0.725^{+0.028}_{-0.024} & 0.846\pm 0.028 & - & -1.14\pm 0.10\\
 \hline
\end{tabular}
}
  \label{tab:rsd_darkenergy}
\end{table*}

\begin{figure*}
  \centering
  \includegraphics[width=0.45\textwidth, angle=0]{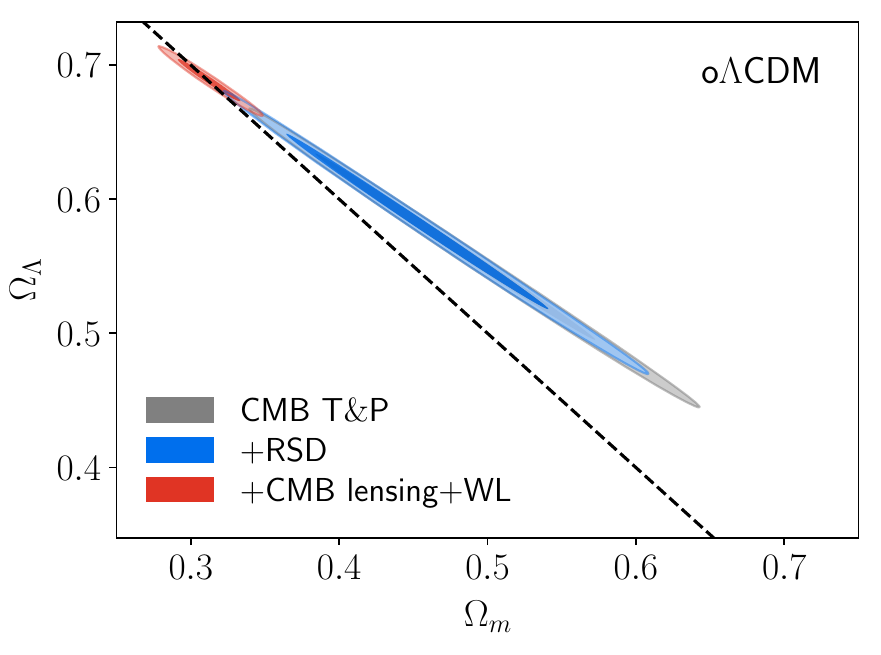}
  \includegraphics[width=0.45\textwidth, angle=0]{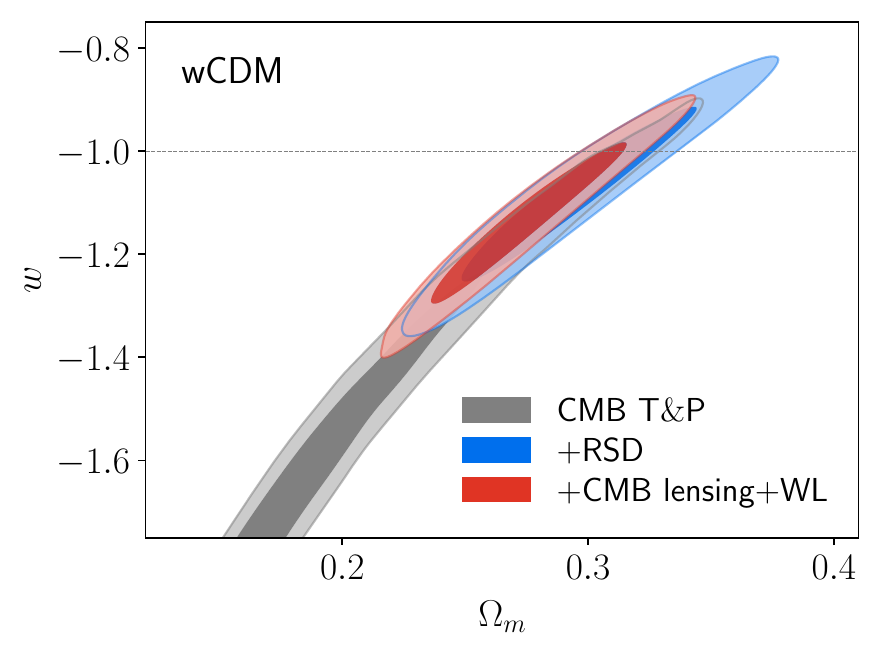}
  \caption{ 
Constraints from CMB temperature and polarization and growth measurements in one-parameter extensions to \LCDM,
as in Table~\ref{tab:rsd_darkenergy}.
{\bf Left:}  The $\Omega_m$--$\Omega_\Lambda$ constraints for a cosmological model under the assumption
of a $w=-1$ cosmological constant with free curvature (\oLCDM).
{\bf Right:}  The $w$--$\Omega_m$ constraints for a flat cosmological model where the
equation of state is allowed as a constant, free parameter.
In both cases, the gray contours represent the 68\% and 95\% confidence intervals
using only the \Planck\ temperature and polarization data, while the blue contours show
the results including RSD data. The combination of RSD, DES WL, \Planck\ lensing, and CMB is shown in red.
}
        \label{fig:growth_darkenergy}
\end{figure*}

\subsection{Impact of RSD Measurements on Models for Single Parameter Extensions to \LCDM}

The constraining power of RSD is illustrated in Figure~\ref{fig:growth_residuals}.
The low-redshift RSD measurements alone have
sensitivity to rule out Einstein-de Sitter ($\Omega_m=1$) models while the higher redshift
RSD measurements are sensitive to variations in the dark energy equation of state.
We first quantify how these RSD data offer complementary views to the WL data on
single parameter extensions to a \LCDM\ cosmology. 

\subsubsection{Expansion history and curvature}

We begin by exploring the constraints on a model with free curvature (\oLCDM)
using growth measurements combined with the \Planck\ CMB temperature and polarization data.
The marginalized 68\% constraints on key cosmological parameters are shown in the
top half of Table~\ref{tab:rsd_darkenergy}.  The two-dimensional contours on $\Omega_m$
and $\Omega_\Lambda$ are shown in the left panel of Figure~\ref{fig:growth_darkenergy}.
While the \Planck\ CMB data alone favor a model with negative curvature, the combination with all growth 
measurements (RSD, WL, and CMB lensing) reduces the $\Omega_k$ uncertainty by a factor of
four and leads to a model consistent with zero curvature ($\Omega_k=-0.0010^{+0.0044}_{-0.0039}$).

As shown in the residual diagram of growth measurements (Figure~\ref{fig:growth_residuals}),
the predictions for growth in a free-curvature o\LCDM\ universe have the largest deviations from a \LCDM\ prediction 
as redshift approaches $z=0$.  The RSD measurements in this regime are governed largely by the MGS sample,
with a precision of 36\% on $f\sigma_8$. As a consequence, relative to the CMB-only constraints on curvature,
those from the CMB+RSD measurements only result in a mild shift with a slightly
reduced uncertainty (Table~\ref{tab:rsd_darkenergy}). 
The DES WL data, on the other hand, offer an
independent measurement of the mass distribution, in particular using source galaxies over the redshift range
$0.20 < z < 0.43$.  While difficult to visualize in a manner similar to the RSD, the WL
measurements offer significantly higher precision estimates in the low redshift regime. 
The WL measurements, when combined with the CMB data, substantially shift the constraints on curvature 
to be consistent with flatness ($\Omega_k=-0.0004\pm0.0048$), with a factor of 3.4 reduction in uncertainty.
The constraining power of CMB lensing lies in between the RSD and low redshift WL; combining the \Planck\
lensing with the temperature and polarization data leads to a bestfit model consistent with the \LCDM\ model 
($\Omega_k=-0.011\pm 0.006$; \citep{2018arXiv180706209P}).

\subsubsection{Expansion history and dark energy}

We next explore the constraints on a flat \wCDM\ model,
where the equation of state $w$ for dark energy is constant but allowed to vary.
The \Planck\ temperature and polarization data prefer a value of $w$ much more negative than
$-1$, and adding CMB lensing  causes virtually no change \citep{2018arXiv180706209P}.
As shown in the right panel of Figure~\ref{fig:growth_darkenergy}, the
combination of growth measurements with CMB data provides constraints on $w$
that enclose the cosmological constant model within the 95\% contours.
Contrary to the case of \oLCDM,
it is the RSD data that have the largest impact in shifting the CMB contours.
As shown in Table~\ref{tab:rsd_darkenergy},
combining WL measurements with CMB does not significantly improve the precision on $w$, but RSD
measurements are able to improve the precision by more than a factor of two.

The constraining power of RSD on $w$ can be understood from Figure~\ref{fig:growth_residuals}.
A more negative value of $w$ causes increasingly slower structure growth toward lower redshifts.
The $f\sigma_8$ measurements with BOSS and eBOSS galaxies sample the growth rate in the redshift range $0.2<z<1.1$,
providing good constraints on the shape of $f\sigma_8(z)$ around its peak and thus constraints on $w$.
The CMB+RSD data are therefore able to rule out (at 4.5 standard deviations) the formal central value of $w=-1.585$ preferred by CMB
alone. The combination of all the growth measurements
with CMB prefers a model consistent with $w=-1$, at the level of 1.4$\sigma$, when using the one-dimensional, marginalized
likelihoods.

We see that CMB and growth measurements provide factor of 2.5--4 improvements on
the precision in extended \LCDM\ models when compared to CMB temperature and polarization data alone.
The growth data have the net effect of pulling the CMB data closer to a \LCDM\ model.

\begin{figure*}
  \centering
  \includegraphics[width=0.45\textwidth, angle=0]{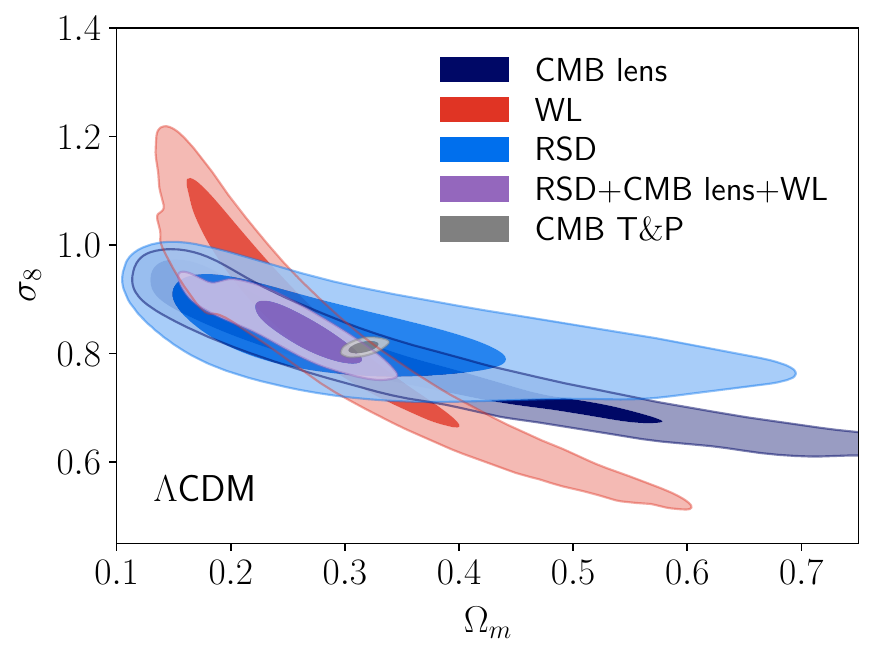}
  \includegraphics[width=0.45\textwidth, angle=0]{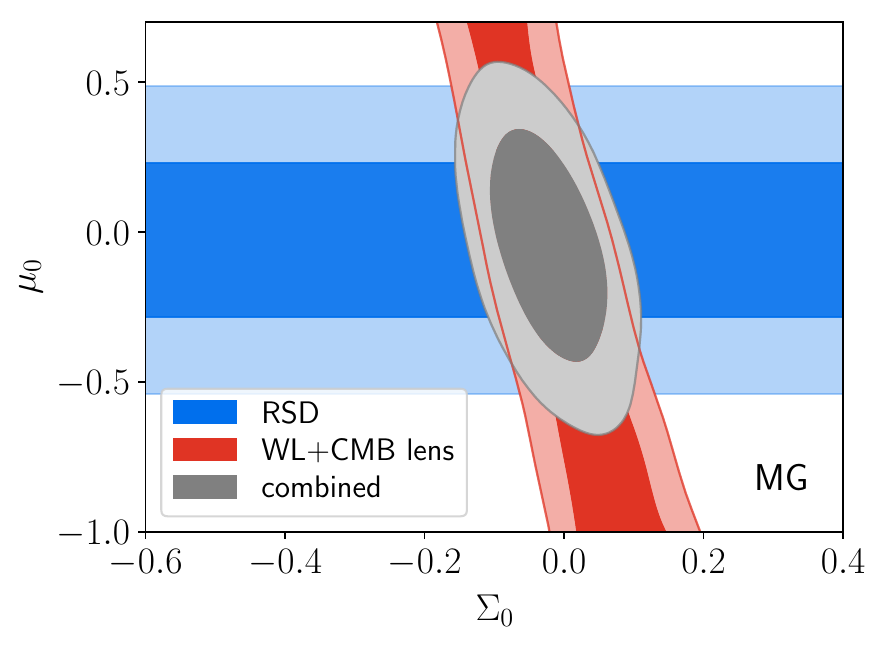}
  \caption{
{\bf Left:}  The $\Omega_m$--$\sigma_8$ constraints for a \LCDM\ cosmology.
The blue contours represent the 68\% and 95\%
constraints when using only the RSD sample.  The red contours represent the same when using
only the DES WL results.
The dark blue contours represent the constraints from \Planck\ lensing, while the light purple
represent the combination of RSD, DES WL, and \Planck\ lensing.
The gray contours represent the predictions from the \Planck\ temperature and polarization data under an assumption
of a \LCDM\ cosmological model. Note that the contours for low values of $\Omega_m$ are affected by the $H_0$ prior.
{\bf Right:}  The $\Sigma_0$--$\mu_0$ constraints for a cosmology with a fixed cosmological constant and
perturbations to GR as described in Equations~\ref{eq:constraint2} and \ref{eq:musigform}.
The blue contours represent the 68\% and 95\%
constraints when using only the RSD measurements.  The red contours represent the same when using
the DES WL and \Planck\ lensing data.
The gray contours represent the constraints when combining RSD, DES WL, and
\Planck\ lensing measurements.
}
        \label{fig:growth_amplitude}
\end{figure*}

\subsection{RSD Constraints on the Amplitude of Matter Fluctuations and Tests of Gravity}

Within the \LCDM\ model, RSD and lensing provide a means to integrate
the rate of structure growth to redshift zero and estimate the current amplitude
of matter fluctuations, $\sigma_8$.
This estimate of $\sigma_8$ can be compared to the predictions
when extrapolating the amplitude of the measured CMB power spectrum,
thus serving as a \LCDM\ consistency test
similar to that of the $H_0$ inverse distance ladder tests in Section~\ref{subsec:H0}.
Structure growth can also be used to test the basic assumptions of the \LCDM\ model
through modifications to GR.
In this case, the redshift evolution of matter density fluctuations
and the interaction of matter and photons with the resultant gravitational potential 
can be directly compared to predictions of GR.

Here, we use SDSS RSD measurements, DES WL measurements, and \Planck\ CMB lensing results to assess 
the amplitude of local matter fluctuations and perform a consistency test for GR.

\subsubsection{RSD constraints on the amplitude of matter fluctuations}

First we explore the constraints on $\Omega_m$--$\sigma_8$ from growth measurements assuming a \LCDM\ 
cosmology. As shown in the left panel of Figure~\ref{fig:growth_amplitude}, the constraints from each of 
the growth measurements (RSD, DES WL, and CMB lensing) are consistent with the predictions of models
informed only by the CMB, albeit with much larger contours.
Note that we have applied conservative priors on $n_s$ and $\omega_b$ for all
contours (see Appendix~\ref{sec:appendix_model}).

The degeneracy from growth measurements follows the direction of lower $\Omega_m$ (thus slower structure 
growth) and higher fluctuation amplitude $\sigma_8$. The differences in the degeneracy directions with RSD, WL, 
and CMB lensing measurements result from their different dependences on cosmology and different redshift
sensitivities. Given the differences, we do not seek to present constraints on
optimal combinations of the two parameters.
Among the three growth measurements, RSD appears to have the largest contour area but provide 
the tightest constraints on $\sigma_8$, with $\sigma_8=0.836_{-0.062}^{+0.053}$. 
The WL measurements lead to overall better 
constraints in the $\Omega_m$--$\sigma_8$ plane, while the marginalized constraints on $\Omega_m$ are 
comparable to RSD and those on $\sigma_8$ ($\sigma_8=0.857 _{-0.136}^{+0.163}$) are not as tight as RSD. CMB lensing 
results in constraints in a direction similar to that of RSD,
but with a stronger $\Omega_m$--$\sigma_8$ degeneracy and thus narrower contours than RSD.

The combination of RSD, WL, and CMB lensing is shown in the light purple contours in the
left panel of Figure~\ref{fig:growth_amplitude}.  The resulting constraints are greatly improved,
offering $\sigma_8=0.842_{+0.043}^{+0.034}$ and $\Omega_m =0.261_{-0.027}^{+0.036}$.  In addition, the 68\% confidence
intervals overlap the 68\% confidence intervals from the prediction based on 
CMB temperature and polarization data, indicating general consistency.

\subsubsection{RSD constraints on modified gravity}

The difference between the speed of gravity and the speed of light has been shown
to be negligible \citep{ligo17a}, as predicted by GR.
$f\sigma_8(z)$ measurements from RSD can be used to further test theories of gravity in the
context of structure formation.

Here we consider a phenomenological parameterization of gravity, as described in 
Section~\ref{subsec:growth}, allowing for the two metric potentials $\Psi$ and $\Phi$ to 
deviate from their GR prediction, independent of the speed of gravitational waves.
The parameter $\mu_0$ characterizes the deviation of $\Psi$, which 
determines the response of matter to the gravitational potential and thus can be probed by RSD.
The parameter $\Sigma_0$ characterizes the deviation of $\Psi+\Phi$, which determines the propagation of
light and thus can be probed by lensing. Therefore, the combination of varying $\mu_0$ and $\Sigma_0$ 
provides us with a null test of gravity along the degeneracy direction of our most potent probes of 
modified gravity, RSD and lensing (WL and CMB lensing). 

Assuming the fiducial cosmological model to be \LCDM\ with the background parameters fixed to
the baseline values (see Appendix~\ref{sec:appendix_model}),
we compute the constraints on $\mu_0$ and $\Sigma_0$ (right panel of Figure~\ref{fig:growth_amplitude}).
We use MGCAMB \citep{zucca19a}\footnote{\url{https://github.com/sfu-cosmo/MGCAMB}}
to allow for these phenomenological tests of general relativity.
As expected, WL+CMB lensing mainly constrain $\Sigma_0$, while RSD is only sensitive to $\mu_0$.
A combination of both probes is necessary to break degeneracies between
the two parameters. 
With the combined RSD and lensing, we find $\mu_0=-0.05 \pm  0.24$ and $\Sigma_0=-0.024\pm0.054$,
consistent with the GR prediction of $\mu_0=\Sigma_0=0$.

\section{Global Fits}\label{sec:global_fit}
After examining the impacts of expansion history and growth measurements alone, we now proceed
to combine the \Planck\ (including lensing), Pantheon SNe~Ia, SDSS, and DES 
data to determine the best fitting cosmological model.
The SDSS data consist of the combined BAO and RSD measurements found in the bottom
section of Table~\ref{tab:BAORSD_measurements}, while the DES data consist
of the cosmic shear, galaxy clustering, and galaxy-galaxy lensing data (i.e., 3$\times$2pt).
We refer to the results from the previous sections where there
is guidance on which datasets are providing the critical information.   

We start by establishing the parameters for the simplest cosmology, that
of a \LCDM\ universe with a fixed neutrino mass.  We examine the 
distribution of BAO and RSD measurements about this model to assess
potential tension with any of the individual measurements.

We then expand the cosmological model to include free parameters for
curvature, the dark energy equation of state, and the neutrino mass.
In all cases, the best fitting values of $H_0$ are determined at a precision
of better than 0.8 $\kmsmpc$.
We then show that the
addition of these free parameters does not lead to significant changes in
any of the \LCDM\ parameters and that the results remain consistent with a flat \LCDM\ universe.
Finally, we provide a physical interpretation of the cosmology constraints on summed
neutrino mass in the context of neutrino oscillation experiments.

\begin{table*}[t!]
 \caption{\rm Marginalized values and 68\% confidence limits for models using \Planck,
Pantheon SNe, SDSS BAO+RSD, and DES 3$\times$2pt data.}
\begin{center}
\resizebox{\textwidth}{!}{%
\begin{tabular}{|C|C|C|C|C|C|C|C|C|C|}
 \hline \hline 
 & \Omega_\Lambda & H_0 & \sigma_8 & \Omega_K & w_0 & w_a & \Sigma m_\nu\,[\mathrm{eV}]
 \footnote{The reported $\sum m_\nu$ values correspond to the 95\% upper limits.}\\
\hline
\Lambda \mathrm{CDM} & 0.6960\pm 0.0047 & 68.19\pm 0.37 & 0.8072\pm 0.0056 & - & - & - & -\\
 \hline
\mathrm{o}\Lambda\mathrm{CDM} & 0.6959\pm 0.0046 & 68.24\pm 0.54 & 0.8076\pm 0.0065 & 0.0002\pm 0.0017 & - & - & -\\
 \hline
w\mathrm{CDM} & 0.6996\pm 0.0066 & 68.68\pm 0.73 & 0.8130\pm 0.0093 & - & -1.021\pm 0.027 & - & -\\
 \hline
\mathrm{o}w\mathrm{CDM} & 0.6998\pm 0.0071 & 68.62\pm 0.75 & 0.8126\pm 0.0093 & -0.0003\pm 0.0019 & -1.023\pm 0.031 & - & -\\
 \hline
w_0 w_a\mathrm{CDM} & 0.6973\pm 0.0069 & 68.49\pm 0.75 & 0.8141\pm 0.0094 & - & -0.938\pm 0.073 & -0.32^{+0.28}_{-0.24} & -\\
 \hline
\mathrm{o}w_0 w_a\mathrm{CDM} & 0.6984\pm 0.0070 & 68.18\pm 0.79 & 0.8136\pm 0.0092 & -0.0022\pm 0.0022 & -0.909\pm 0.081 & -0.49^{+0.35}_{-0.30} & -\\
 \hline
\nu\Lambda \mathrm{CDM} & 0.6974^{+0.0054}_{-0.0049} & 68.33^{+0.45}_{-0.39} & 0.8113^{+0.0093}_{-0.0068} & - & - & - & <0.115\\
 \hline
\nu w \mathrm{CDM} & 0.6993\pm 0.0066 & 68.63\pm 0.72 & 0.813^{+0.011}_{-0.0099} & - & -1.018^{+0.033}_{-0.028} & - & <0.162\\
 \hline
\end{tabular}
}
  \label{tab:full}
\end{center}
\end{table*}

\subsection{\LCDM\ Model}

We start by finding the \LCDM\ model that best describes the full suite of data.
As shown in the first row of Table~\ref{tab:full}, the dark energy density is constrained
at the level of 0.7\%.
This precision is improved by a factor of 1.28 over the value found in Table~\ref{tab:main}
for a combination of CMB, BAO and SN and a factor of 1.79 over the CMB data alone,
indicating that the dark energy density constraints are dominated by the expansion history measurements. 

The precision of \LCDM\ parameter constraints allows us to evaluate the distribution
of SDSS measurements about the model. For the purpose here, we use Gaussian approximations to the measurements for the evaluation.
In comparison with the bestfit model, the 14 BAO-only measurements ($D_V/r_d$, $D_M/r_d$, and $D_H/r_d$; 
Table~\ref{tab:BAORSD_measurements}) give a value of $\chi^2$=11.0, with covariance among measurements taken into account. 
Similarly, the six RSD measurements (Table~\ref{tab:BAORSD_measurements}) give $\chi^2$=6.6. 
Finally, we consider the full set by combining the 17 BAO+RSD measurements with the four BAO measurements
from Ly$\alpha$--Ly$\alpha$ and Ly$\alpha$--Quasar correlations and obtain $\chi^2$=23.7. 
Based on the $\chi^2$ distribution with 14, 6, and 21 degrees of freedom, respectively, all sets
of measurements are fully consistent with the preferred model. 

\begin{figure}
  \centering
    \includegraphics[width=0.45\textwidth, angle=0]{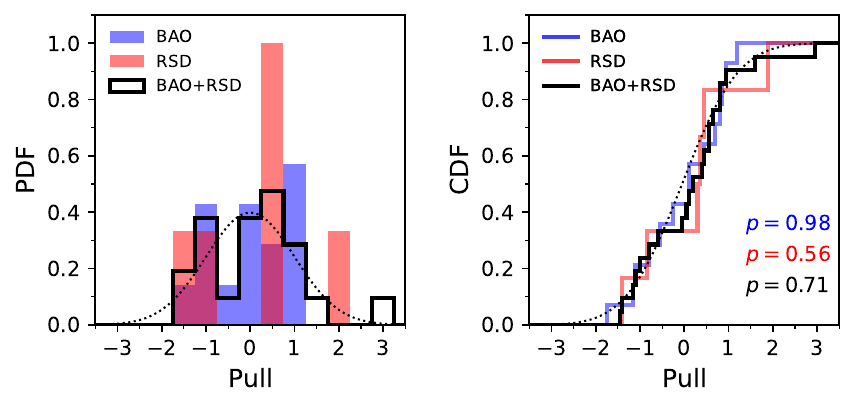}
  \caption{Distribution of residuals of the SDSS BAO (blue), RSD (red), and BAO+RSD (black) measurements with
respect to the bestfit \LCDM\ model. In all cases, the residuals are represented in the form of pulls, i.e.,
normalized by the measurement uncertainty. The left panel shows the probability distribution, and the right panel
the cumulative distribution. The $p$ values are from KS tests in comparison with the normal distribution
(dotted curves).
}
        \label{fig:residuals}
\end{figure}

To evaluate whether there is any statistically significant outlier in the measurements,
we compute the residual between each SDSS measurement and the value predicted by the preferred model.
In this pull distribution, the residuals are normalized by the measurement uncertainty, so one
would expect a Gaussian distribution with unity width if the measurements were distributed according
to the measurement uncertainties. We account for the correlations among measurements by diagonalizing 
each covariance matrix to produce statistically independent pull values. 
The resulting distribution of the pull values is shown in the left panel of 
Figure~\ref{fig:residuals}. For the BAO-only measurements, the pull with the largest deviation, 
$-1.75$, 
comes from the $z\sim 0.7$ eBOSS LRG sample. 
For the RSD measurements, the largest deviation, 1.91, is from the $z\sim 1.48$ eBOSS Quasar sample.
For the full set (labeled as `BAO+RSD'), the largest deviation is again from the eBOSS Quasar sample.
After accounting for the covariance between BAO and RSD measurements, the measurement differs from
the \LCDM\ prediction by 3.0 standard deviations.
Based on the Kolmogorov--Smirnov (KS) test with the cumulative distribution (right panel 
of Figure~\ref{fig:residuals}), the pull distributions for BAO, RSD, and BAO+RSD measurements are 
found to be consistent with the normal distribution, revealing no unexpected feature in the measurements
for a universe best described by the \LCDM\ model.
Additional diagnostics using the Hubble parameter can serve as
consistency check on dark energy constraints \citep[e.g.,][]{shafieloo12a}
as has been done to assess the BOSS BAO results \citep{sahni14a}.
Preliminary results also indicate that the eBOSS data are consistent
with a \LCDM\ model.

\begin{figure*}
  \centering
    \includegraphics[width=0.45\textwidth, angle=0]{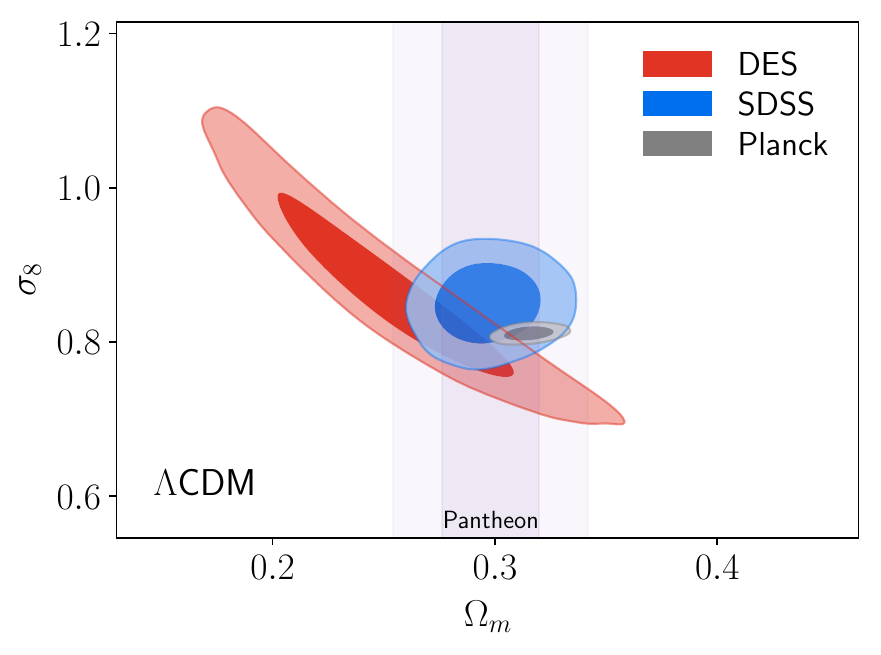}
    \includegraphics[width=0.45\textwidth, angle=0]{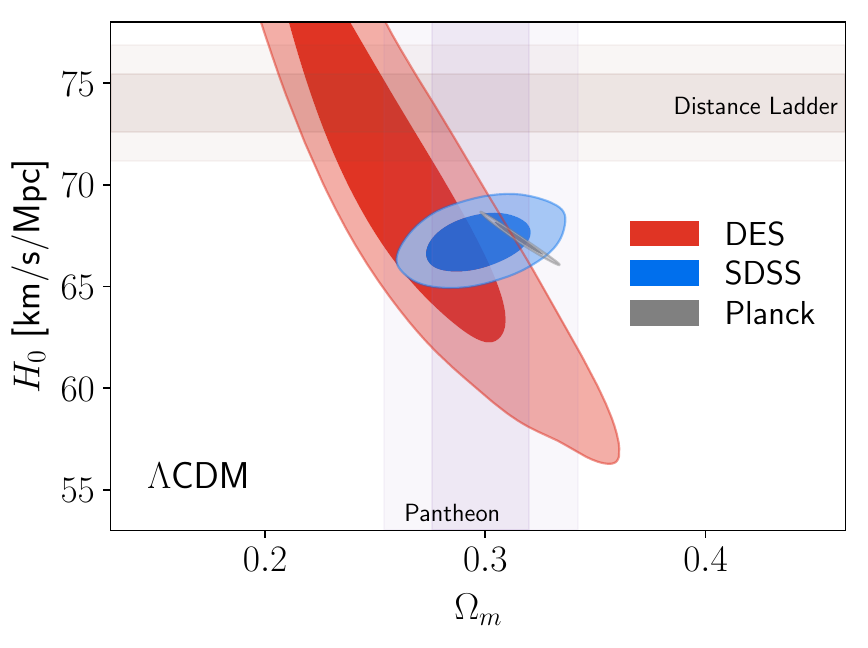}
  \caption{
    {\bf Left:} The $\Omega_m$--$\sigma_8$ constraints for a \LCDM\ model. 
    A BBN-inspired prior on $\omega_b$ and a
    prior of $n_s=0.96 \pm 0.02$ was assumed for the SDSS and DES contours.
    {\bf Right:} $H_0$ versus $\Omega_m$ under a \LCDM\ model. 
    In both panels, the 68\% and 95\% confidence intervals for the
    BAO+RSD data are shown in blue,
    the DES 3$\times$2pt data in red, and the \Planck\ CMB and lensing data in gray.
    The faint, vertical purple bands represent the Pantheon constraints of $\Omega_m=0.298\pm0.022$ \citep{scolnic18a}.
    In the right panel, the faint, brown horizontal bands represent the Cepheid/SNe~Ia
measurements from \citet{riess19a}, $H_0=74.03\pm1.42\, \kmsmpc$.
  }
\label{fig:dist+growth}
\end{figure*}

Within the \LCDM\ model, 
the SDSS, DES, and \Planck\ data offer tests of GR predictions on growth rates and model-dependent predictions for $H_0$.
The left panel of Figure~\ref{fig:dist+growth} shows the constraints from these three programs
on the amplitude of the (linear) power spectrum while the right panel shows the constraints on $H_0$.
In a \LCDM\ cosmology, the model describing the DES 3$\times$2pt data has a strong degeneracy between
$\sigma_8$ and $\Omega_m$ and a strong degeneracy between $H_0$ and $\Omega_m$.
In both cases, the DES data are described by somewhat lower values of $\Omega_m$
than the model describing the CMB data.
The mild tension that has been reported between lensing and CMB estimates of $\sigma_8$
can therefore be equally explained by the preference in WL estimates for a $\Omega_m$ value lower than 0.3.
The $\sigma_8$ and $\Omega_m$ degeneracy is similar to that found in the KiDS weak lensing analysis
\citep{heymans20a} and in the HSC weak lensing analysis \citep{2019PASJ..tmp...22H}, both of which
prefer lower values of $\sigma_8$ if evaluated with a CMB prior on $\Omega_m$.

Other studies have recently used the same BOSS galaxy data we use in order to obtain results on $\sigma_8$
and $\Omega_m$ when restricting to a \LCDM\ cosmology \citep{troster20a,damico20a,ivanov20a}.
These studies have found lower central values of $\sigma_8$ than what we find in our measurements
from SDSS clustering or what would be expected from \Planck\ measurements.
A recent example is that found in the KiDS weak lensing analysis \citep{heymans20a}.
In this study, BOSS galaxy clustering measurements over redshift intervals
$0.2 < z< 0.5$ and $0.5 < z < 0.75$ were used to further constrain $\sigma_8$ and $\Omega_m$
and reduce the degeneracy between the two parameters. 
They find $S_8 = 0.766^{+0.020}_{-0.014}$, lower by $8.3 \pm 2.6$\% relative to \Planck\ predictions.
Their result using only the BOSS DR12 clustering data leads to an estimate $\sigma_8= 0.75^{+0.045}_{-0.047}$,
about two standard deviations lower than the central value in result from all SDSS clustering ($\sigma_8 = 0.85 \pm 0.03$).

In addition to the difference in methodologies, the SDSS data set includes a larger number
of tracers over a larger redshift range.
We include results from six redshift regimes and include post-reconstruction BAO measurements where possible.
Even restricting to BOSS, the data we include is substantially different, as we use the
results from the BOSS $0.4 < z< 0.6$ data set and we do not directly use the $0.5 < z < 0.75$ BOSS galaxy clustering
data.  Instead, the BOSS $z > 0.6$ data were combined with the eBOSS LRG sample \cite{LRG_corr}.
The additional data generally favors $f \sigma_8$ values that are greater than those
preferred by \Planck\ (Figure~\ref{fig:growth_residuals}).
Our SDSS result thus represents the most complete representation of structure growth constraints from galaxy
clustering data and indicates no tension with \Planck\ CMB measurements, albeit with an implied prior that the
shape of the linear power spectrum matches that observed by the CMB.

\subsection{Constraints on Dark Energy and Curvature}

As was demonstrated in Section~\ref{sec:expansion}, the main strength of the BAO and SN
distance measurements is to constrain cosmological models with free curvature and varying dark energy equation of state, respectively.
As was shown in Section~\ref{sec:growth}, the main strength of the growth measurements is in
constraining possible deviations from GR.
We now explore the complementarity of distance and growth measurements by 
testing the same single parameter extensions to \LCDM\ that were presented in
Section~\ref{sec:expansion}, followed by models with increasing degrees of freedom.
The results are found in Table~\ref{tab:full}.

\subsubsection{$\Omega_{\rm DE}$, $H_0$, and $\sigma_8$ Parameters}

First, the central values of the three parameters, $\Omega_{\rm DE}$, $H_0$, and $\sigma_8$,
are all consistent with the prediction from the bestfit \LCDM\ model (Table~\ref{tab:full})
at 68\% confidence, regardless of the cosmological model that is assumed.
The largest fractional deviation from the \LCDM\ prediction is only 0.8\%, in the 
case of $\sigma_8$ in the \owowaCDM\ model. 
That measurement is fully consistent with the \LCDM\ prediction of $\sigma_8 = 0.8120 \pm 0.0073$
from CMB data alone \citep{2018arXiv180706209P}.
The robustness of $\sigma_8$ measurements to cosmological model provide further evidence that
the growth of structure can be described using GR in a \LCDM\ parameterization.

In addition, the precision on the three parameters does not degrade significantly between the
\LCDM\ model and expanded models.  When expanding to the \owowaCDM\ model,
the precision on the $\Omega_{\rm DE}$ and $H_0$ parameters degrades
by factors of 1.5 and 2.1, respectively.
The largest degradation for $\sigma_8$ precision occurs with the \nuwCDM\ model, leading
to a factor of 1.8 increase in the uncertainties.
The tight constraints offered in all models are a result of the complementarity between
observational probes.

As discussed in Section~\ref{sec:expansion}, interesting tensions appear between
the estimates of the current Hubble expansion rate from local measurements and
from extrapolations of the calibrated drag scale to $z=0$ using the CMB.
Those estimates of a low $H_0$ extrapolated
from early times are not changed with the addition of the growth data.
For even the most flexible \owowaCDM\ cosmology, we find $H_0 = 68.18 \pm 0.79 \, \kmsmpc$,
consistent with the findings in Section~\ref{sec:expansion}. The addition of the
growth data leads to a 9\% improvement in the precision on $H_0$ compared to the results
using \CMBTP+BAO+SN.

None of the extended models improve the best-fit $\chi^2$ by more than 4 compared to the  $\chi^2$ of the \LCDM\ model of $\chi^2=4366$, even with the introduction of additional degrees of freedom. We find a mild change in the $\chi^2$ for models with free dark energy equation of state, $w$, while models with curvature and varying time evolution of dark energy, $w_a$,  only show a marginal change. Hence, we do not detect strong evidence for the need to extend the \LCDM\ model.

\subsubsection{Curvature and Dark Energy}

When comparing the global results in an \oLCDM\ model to those from \CMBTP+BAO+SN in Table~\ref{tab:main},
we find that the addition of the RSD, Planck lensing, and DES data only provides improvements of 6\% on
the precision of curvature constraints.
The impact of growth measurements is larger in the \wCDM\ model; the additional data
provide improvements of about 22\% on the precision for a constant equation of state.
As discussed in Section~\ref{sec:growth} and in Figure~\ref{fig:growth_residuals},
the improvement is likely primarily from the RSD measurements.

When expanding to an evolving dark energy model with zero curvature,
we find that the bestfit models are still consistent with \LCDM. 
The \wowaCDM\ model does not improve the fit relative to \LCDM,
indicating that the additional free parameter is not providing critical new information.
Overall we find consistent constraints on $w_0$ with those in a \wCDM\ model.

\begin{figure*}
  \centering
    \includegraphics[width=0.75\textwidth, angle=0]{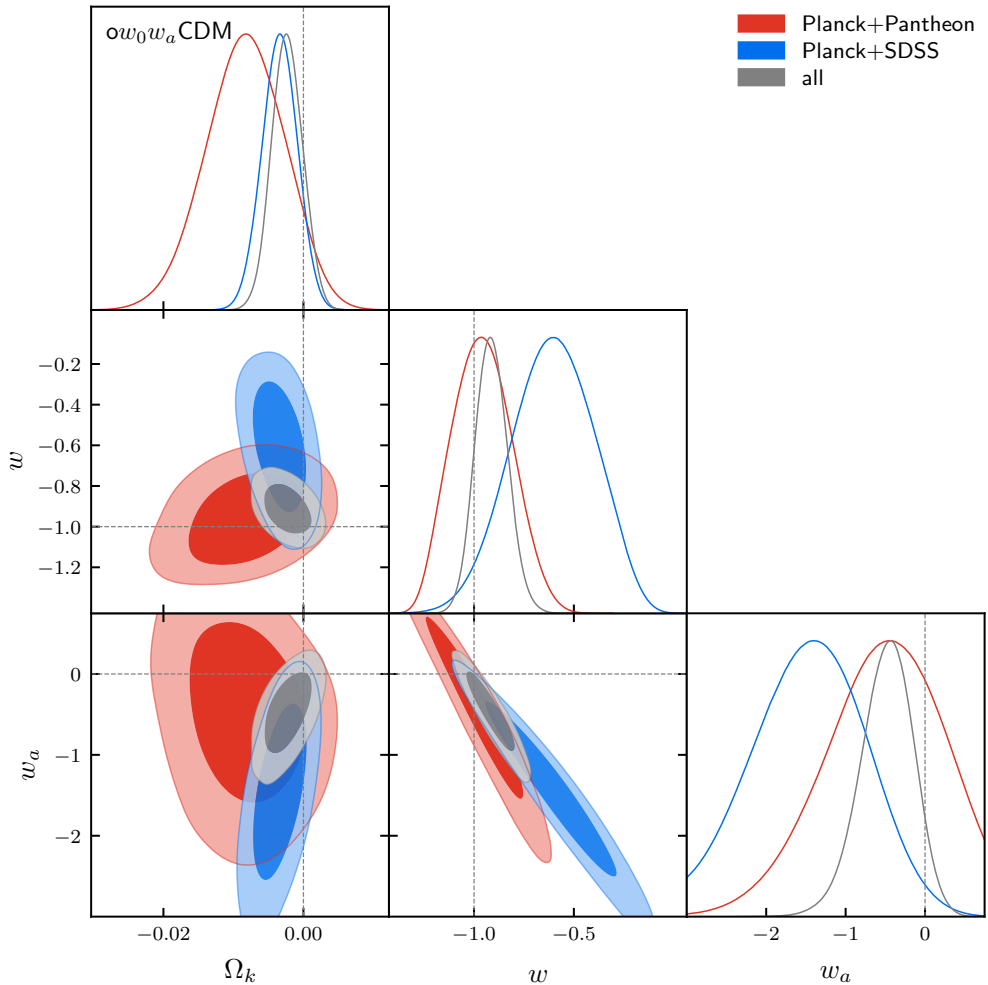}
  \caption{Two dimensional contours on $w_0$, $w_a$, and $\Omega_k$ under the assumption of
an \owowaCDM\ cosmological model.  The one-dimensional constraints on each independent parameter
are presented in the top panels.
The red contours represent the 68\% and 95\% constraints when using the full \Planck\ data (T$\&$P and lensing)
and the Pantheon SNe~Ia measurements.
The blue contours represent the constraints from \Planck\ and SDSS BAO+RSD, while the gray contours represent the
combination of all measurements presented in this work.
}
        \label{fig:w0waomegak}
\end{figure*}

The complementarity of BAO/RSD and SNe~Ia measurements is best demonstrated in expanded dark energy models
that also allow for free curvature.
We only find meaningful prior-independent constraints on the general \owowaCDM\ model
for the combination of all datasets, as shown in Figure~\ref{fig:w0waomegak}.
As shown in Table~\ref{tab:full},
the uncertainties on the two dark energy equation of state parameters in an \owowaCDM\ model are
relatively unchanged when compared to results under the assumption of a spatially flat universe (\wowaCDM).
The uncertainties on curvature are only increased by 29\% when compared to the single parameter
\oLCDM\ extension.

In Figure~\ref{fig:w0waomegak}, it can be seen that the $w_0$--$\Omega_k$ 
confidence intervals from the Planck+SDSS data are orthogonal to the
Planck+SNe contours.  As was demonstrated in Section~\ref{subsec:extensions}, the BAO data best constrain
the curvature while the combination of CMB and SNe~Ia best constrains the dark energy equation of state.
As shown in the one-dimensional likelihood of $\Omega_k$, the constraints on $\Omega_k$ are
roughly three times better using Planck+SDSS than those using Planck+SN.
The Planck+SN data perform slightly better than do the Planck+SDSS data in the $w_0$--$w_a$ plane,
but the net precision on both $w_0$ and $w_a$ increases by roughly a factor of two when
combining all measurements.
This statistical increase in constraining power is much larger than one
would expect due to the contribution of the Planck+BAO data to provide tight constraints on curvature.
Most importantly, the combination of all cosmological probes reveals again a preference for the \LCDM\ model.
From the one-dimensional marginalized distributions, $w_0=-1$ is at 1.1 standard deviations, $w_a=0$ at 1.3 
standard deviations, and $\Omega_k=0$ almost within the 68\% confidence interval.

In a related CPL parameterization for dark energy, we can define a pivot scale factor $a_p$,
or equivalently a pivot redshift $z_p$.
Instead of evaluating the equation of state at $z=0$, as is done throughout this paper, we can represent the
time-evolving equation of state as $w(a) = w_p + w_a (a_p-a)$.
Note that change of the pivot redshift does not change the model physically,
because the same linear relation can be described by the value and slope at any one point.
However, by choosing the pivot scale appropriate to the redshifts covered by the data,
constraints on $w_p$ and $w_a$ can be made to be nearly uncorrelated.
In doing so, we find constraints in the \wowaCDM\ model $w_p = -1.020 \pm 0.029$ and $w_a = -0.32^{+0.28}_{-0.24}$
when using a pivot redshift $z_p=0.35$.  The result demonstrates that we can constrain the
dark energy equation of state to 3\% precision at an earlier epoch in cosmic history.
This precision is only degraded by a factor of 1.07 relative to the constraint on $w$ in an \wCDM\ model,
indicating that the overall effect of adding the additional parameter for a time-varying
equation of state is minimal.

The results from joint fits can be used to compute a total Dark Energy
Figure of Merit \citep[FoM;][]{albrecht06a} for various sample combinations in a model with time-varying equation of state.
Computed as the inverse product of $w_p$ and $w_a$, the FoM associated with the full SDSS and \Planck\
data is 37.5 in the \wowaCDM\ model.  The FoM increases by 
a factor of 3.5 when adding the Pantheon SNe~Ia and the DES 3$\times$2pt data.
Demonstrating the complementarity of the BAO and SNe~Ia data in
constraining curvature and the dark energy equation of state, the Dark Energy FoM for all datasets only decreases
from 132 in the \wowaCDM\ model to 94 in the \owowaCDM\ model.

\subsection{Neutrino Mass}
\label{sec:mnu}

The existence of neutrino oscillations has been confirmed by numerous terrestrial experiments
\citep{1311.4750,1207.6632,1204.0626,1203.1669,0806.2237,hep-ex/0406035,nucl-ex/0204008,hep-ex/9807003}.
These experiments measure the difference between the squares of neutrino mass eigenstates, 
leading to two sets of possible solutions for individual masses, which are referred to as the normal
and inverted hierarchies. 
Both of these two solutions lead to degenerate neutrino masses if $\sum m_\nu \gtrsim 0.15$ eV,
but lead to different predictions at lower masses. For the normal hierarchy,
the minimum neutrino mass is given by two essentially massless neutrinos and one massive neutrino.
For the inverted hierarchy, the minimum mass is composed of one massless and two degenerate neutrinos.
The constraints for these two scenarios are
\begin{eqnarray}
  \sum m_\nu &>& 0.0588\,{\rm eV} \qquad \mbox{normal hierarchy},\\
  \sum m_\nu &>& 0.0995\,{\rm eV} \qquad \mbox{inverted hierarchy}
\end{eqnarray}
\citep{1811.05487,1907.12598}.

Throughout this paper we assume the neutrino masses to be at the minimum mass
$\sum m_\nu = 0.06\,$ eV with one massive and two massless neutrinos.
When allowing a free parameter to describe the neutrino mass, we continue to assume two massless and
one massive neutrinos, which is a good approximation for the masses of interest \citep{2006PhR...429..307L,fogli12a}.

Resolving the hierarchy problem remains a key goal of ground-based neutrino experiments
\citep[e.g.,][]{li13a,akhmedov13a,abe15a}.
Likewise, the goal of constraining the absolute mass of neutrinos has motivated
a series of terrestrial experiments. The tightest constraints of direct measurements
arise from the Katrin experiment \citep{1909.06048}, resulting in a 90\% upper
limit on the effective electron neutrino mass of $m(\nu_e) < 1.1$ eV.

\begin{table*}
  \centering
  \caption{\rm Constraints on neutrino masses and relative probabilities of neutrino models with
$\nu$\LCDM\ and $\nu w$CDM cosmological models. The 95\% upper limits are derived assuming a $\sum m_\nu>0$ prior.}
  \label{tab:neutrinos}
  \begin{tabular}{|p{7cm}|c|c|c|c|c}
\hline \hline   
    Data & 95\% upper limit [eV]& $P_{\rm inv}/P_{\rm norm}$ & $P_{\rm unphy}$ & Gaussian fit [eV]\\
\hline
 \Planck & 0.252 & 0.64 & 0.43 & \\  
 \Planck\ + BAO & 0.126 & 0.36 & 0.64 & $-0.026\pm0.074$ \\ 
 \Planck\ + BAO + RSD & 0.101 & 0.24 & 0.76 & $-0.026\pm0.060$ \\ 
 \Planck\ + SN & 0.170 & 0.49 & 0.56 & $-0.076\pm0.106$ \\ 
 \Planck\ + BAO + RSD + SN & 0.097 & 0.22 & 0.78 & $-0.024\pm0.057$ \\ 
 \Planck\ + BAO + RSD + SN + DES & 0.115 & 0.27 & 0.71 & $-0.014\pm0.061$ \\ 
 \Planck\  +  BAO + RSD + SN ($\nu w$CDM) & 0.138 & 0.40 & 0.61 & $-0.033\pm0.082$ \\ 
 \Planck\   +  BAO + RSD + SN + DES ($\nu w$CDM) & 0.162 & 0.48 & 0.56 & $-0.048\pm0.097$ \\ 
\hline
  \end{tabular}
\end{table*}

\begin{figure}
  \centering
   \includegraphics[width=0.45\textwidth, angle=0]{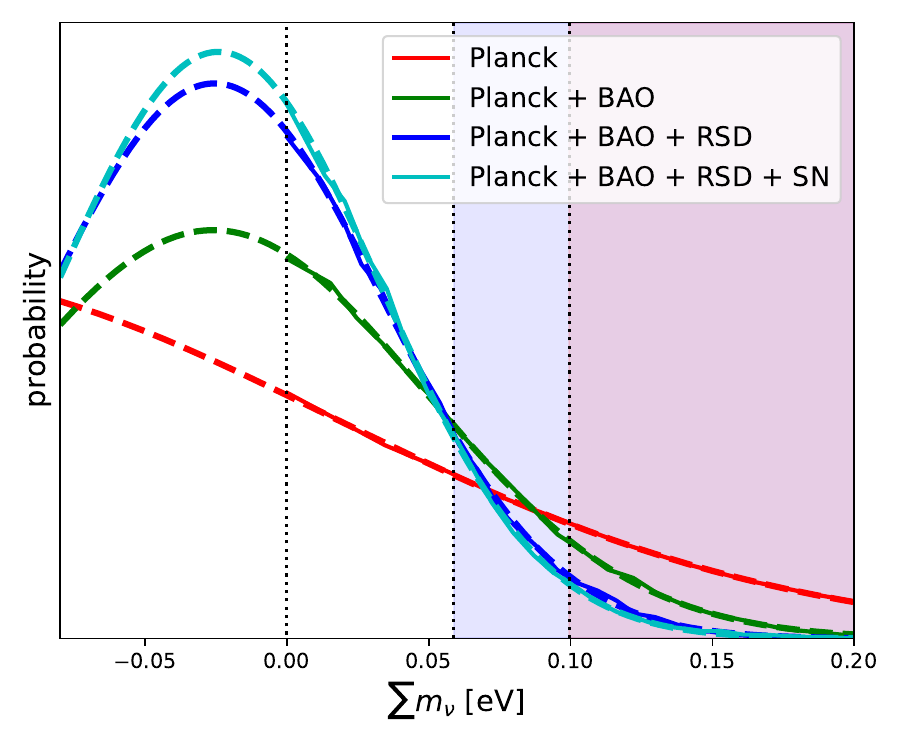}
   \caption{Posterior for sum of neutrino masses for selected combinations of
     data with a $\nu$\LCDM\ cosmology. Dashed curves show the implied Gaussian fits.  Shaded
     regions correspond to lower limits on normal and inverted
     hiearchies. Likelihood curves are normalized to have the same area under the curve for $\sum m_\nu>0$.
   }
  
\label{fig:neutrinos}
\end{figure}

Therefore, it is timely to address the status of neutrino mass
constraints before the advent of Stage IV dark energy experiments. We
show our results for the $\nu$\LCDM\ and $\nu w$CDM cosmological models 
in Table~\ref{tab:neutrinos} with several quantities. 
The 95\% upper limits are derived from Markov chains,
containing a prior $\sum m_\nu>0$.
In requiring only a mass that is positive, the cosmology constraints assume
no prior information from the neutrino oscillation experiments and offer
a fully independent measurement of neutrino mass.
Four selected data combinations are plotted in Figure~\ref{fig:neutrinos}.

It is useful to make
a Gaussian approximation to better characterize the central values
without influence from the prior and to provide a simple compression
of the information for other analyses.  These Gaussian fits are determined over the range
$0<\sum m_\nu<0.15\,{\rm eV}$ and are given in the last column of
Table~\ref{tab:neutrinos}.  The upper 95\% limits
from the fitted Gaussian posteriors are within 2\% of the chains for
$\Lambda$CDM and within 4\% when $w$ is free to vary. 
While we see that the preferred neutrino
mass implied by these fits is negative, a solution with $\sum m_\nu=0$
is usually within one standard deviation of the central value
and the minimal mass solution with the normal hierarchy is always within the 95\% contours.
Finally, the Gaussian variances on these fits are
essentially second derivatives of the log-likelihood and are akin
to Fisher matrix predictions. They can therefore be used to give an
insight into the constraining power of various probes that is free
from the vagaries of the most likely position.

Because of the ability to break degeneracies with $\Omega_m$, the strongest
improvement in neutrino mass precision over CMB-only constraints is caused
by the addition of BAO. This is due to reasons discussed at length in
\cite{aubourg15a} and demonstrated in Section~\ref{subsec:extensions}.
The RSD, which is the canonical neutrino probe for its ability to
measure the suppression of growth due to free-streaming,
improves the precision by another 26\%.  
Adding the RSD data is equivalent to an independent measurement with
an error of about $0.1$ eV in $\Lambda$CDM. 
The ability of BAO to improve upon CMB limits by breaking degeneracies
with matter densities is essentially exhausted with the
current generation of BAO experiments, as indicated
in the right panel of Figure~\ref{fig:SNvsBAO2}.
While currently not the dominant source of information on neutrinos, RSD 
should become the main probe with arrival of the new data in the next decade.

In Table \ref{tab:neutrinos} we also show several integrated
probabilities defined as
\begin{eqnarray}
  P_{\rm norm} &=& \int_{0.0588\,{\rm eV}}^\infty p(m_\nu) dm_\nu , \\
  P_{\rm inv} &=& \int_{0.0995\,{\rm eV}}^\infty p(m_\nu) dm_\nu , \\
  P_{\rm unphy} &=& \int_{0}^{0.0588\,{\rm eV}} p(m_\nu) dm_\nu . 
\end{eqnarray}
Note that these are not Bayesian evidences, because we do not account
for the prior volume. Nevertheless, the ratio of $P_{\rm inv}/P_{\rm norm}$
is the relative probability of the true mass lying in the range allowed by the
inverted/normal hierarchy and is equivalent to an evidence ratio when the
priors are very wide. The quantity $P_{\rm unphy}$ is the probability of the summed
neutrino mass lying in the unphysical region, with a mass lower than allowed
by the normal hierarchy. We see that these probabilities are always
inconclusive; there is no strong evidence from cosmology
on a preference for a normal hierarchy, an inverted hierarchy, or a model where the
neutrino mass is anomalously low (with or without allowing extrapolation
into the negative $\sum m_\nu$). We also note that a 95\% upper limit
of less than $0.0995$ eV would not constitute a 2$\sigma$ detection of
normal hierarchy, because much of that posterior volume belongs to the
unphysically low neutrino mass.

Evaluating the 95\% upper limits, the strongest constraint excluding lensing data is $\sum
m_\nu<0.097$\,eV, which degrades to $\sum m_\nu<0.115$\,eV upon
addition of lensing data. This reflects the shift toward a relatively low amplitude
of $\sigma_8$ in the lensing data with the larger values of $\Omega_m$ preferred by the
other probes.

Finally, we see that allowing the dark energy equation of state parameter ($w$) to be free
degrades the neutrino mass constraint by a factor of 1.4 to 1.6.
This effect is due to
a known degeneracy direction in the neutrino mass \citep{astro-ph/0505551}.
Nevertheless, the effect is not as dramatic as it used to be and with further data
it will become negligible.

\section{Conclusion}\label{sec:conclusion}
The eight distinct samples of SDSS BAO measurements fill a unique niche in their ability to
independently characterize dark energy and curvature in one-parameter extensions to \LCDM.
When combined with \Planck\ temperature and polarization data, the BAO measurements
allow an order of magnitude improvement on curvature constraints when compared
to \Planck\ data alone.  The BAO data provide strong evidence for a nearly flat geometry
and allow constraints on curvature that are now roughly one order of magnitude within the detectable
limit of $\sigma(\Omega_k) \sim 0.0001$ \citep{vardanyan09}. 
The SDSS BAO measurements demonstrate that the observed cosmic acceleration
is best described by a dark energy equation of state that is consistent with a cosmological constant
to better than 6\% precision when combined with the \Planck\ temperature and polarization data.
Finally, the SDSS BAO measurements allow robust estimates
of the current expansion rate, demonstrating $H_0 < 70 \, \kmsmpc$ at 95\% confidence under standard assumptions
of pre-recombination physics, regardless of cosmological model.  These $H_0$ results remain consistent, even
without the \Planck\ CMB data, as long as the \LCDM\ model is assumed.

Beyond the distance-redshift relation, we have also demonstrated the 
complementary role of the six independent SDSS RSD measurements to DES and \Planck\ lensing measurements.
The SDSS RSD measurements tighten \Planck\ temperature and polarization constraints on the dark energy equation
of state
by more than a factor of two; the DES WL measurements tighten constraints on curvature
by more than a factor of three.
Independent of any BAO or SNe~Ia information on the expansion history, the CMB, RSD, and WL
measurements present a history of structure growth that is best described by a
standard \LCDM\ cosmology and a GR model for gravity.

The tightest constraints on the cosmological model are found when combining current measurements
of the expansion history, CMB, and growth of structure.  This combination
reveals a dark energy density measured to 0.7\% precision under an assumed \LCDM\ model.
We find $\sim$1\% precision estimates on $\Omega_{\rm DE}$, $H_0$, and $\sigma_8$ with central values
that barely change under any extension explored in Section~\ref{sec:global_fit}.
The best fitting parameters in extended models remain consistent with a \LCDM\ cosmology;
the most flexible \owowaCDM\ model indicates constraints $\Omega_k = -0.0022 \pm 0.0022$,
$w_0 = -0.909 \pm 0.081$, and $w_a = -0.49^{+0.35}_{-0.30}$.
The Dark Energy FoM for the full combination, in a model that allows for curvature, is
94, about 38\% lower than what was predicted 14 years ago by the Dark Energy Task Force \citep{albrecht06a}.
However, the assumptions of the Dark Energy Task Force included the final DES cosmology results,
whereas we only included the results of the first year WL and clustering studies.
If the final DES studies of SNe~Ia, galaxy clusters, and WL can provide an additional
60\% increase in the FoM, then the Dark Energy Task Force predictions will be proven accurate.
The combination of measurements also provides an independent constraint on the summed neutrino mass,
leading to $\sum m_\nu<0.115$ eV at 95\% confidence ($\nu\Lambda$CDM), with a slight preference for
a normal hierarchy of mass eigenstates over an inverted hierarchy.
The dominant factors in this neutrino mass measurement are the constraints from CMB and BAO,
thus making the result robust against challenges in modeling the full-shape of the power
spectrum in clustering and lensing measurements.

At the high precision found here, cosmic acceleration remains most
consistent with predictions from a cosmological constant. A deviation from consistency
with a pure cosmological constant perhaps would have pointed
toward specific dark energy and modified gravity models.
However, since many of these models have
parameter choices that make them indistinguishable from \LCDM,
those models all can be made consistent with our observations.
Nevertheless, the observed consistency with flat \LCDM\ at the
higher precision of this work points increasingly towards a pure
cosmological constant solution, for example, as would be produced by a
vacuum energy fine-tuned to have a small value. This fine-tuning
represents a theoretical difficulty without any agreed-upon resolution
and one that may not be resolvable through fundamental physics
considerations alone \citep{weinberg89a, brax19a}. This difficulty
has been substantially sharpened by the observations presented
here.

\subsection{A Decade of Dark Energy}

\begin{figure*}
  \centering
  \includegraphics[width=0.9\textwidth,angle=0]{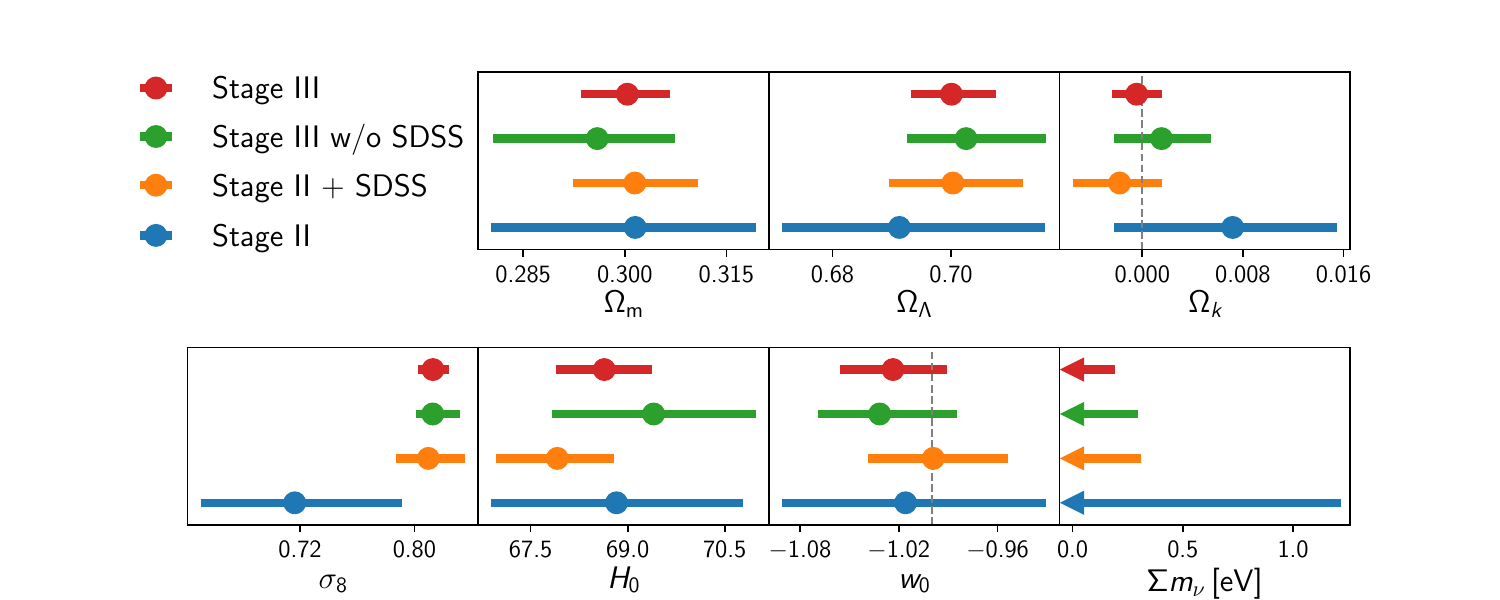}
  \caption{Central values and 68\% contours for each of the parameters describing
expansion history and growth of structure in a \nuowCDM\ model.
Results are shown for each data set combination presented in the text, where
Stage-II corresponds to a combination of the WMAP, JLA, and SDSS DR7 data
and Stage-III corresponds to a combination of the SDSS BAO+RSD, \Planck,
Pantheon SN~Ia, and DES 3$\times$2pt data.
}
  \label{fig:decade}
\end{figure*}

The profound insight offered into the cosmological model is only possible
after several generations of experimental effort.
Experiments designed to study the nature of dark energy have steadily improved
in technique, redshift coverage, and sample size.
In particular, the \Planck\ CMB experiment offered a significant boost
in spatial coverage and precision over WMAP while the BOSS and eBOSS programs
offered vast improvements in redshift range and statistical precision over the
preceding spectroscopic surveys.

As a baseline to assess the impact of the current generation of dark energy
experiments, we first characterize the dark energy constraints with the 
analogous programs that were concluding as BOSS was
achieving first light.
Representing the approximate period 2000--2010, we choose
the final WMAP sample \citep{bennett13a,hinshaw13a},
the JLA sample of SNe~Ia \citep{betoule14a},
and the 2.7\% precision measurement of isotropic BAO at $z=0.275$
\citep{percival10a} from SDSS DR7 \citep{abazajian09a} and
the 2-degree Field Galaxy Redshift Survey \citep{colless01a}.
Following the convention presented in the report from the Dark Energy Task Force \citep{albrecht06a},
we refer to this dataset as `Stage-II'.
Although some of these results were released as BOSS was nearing its conclusion,
the data are representative of the previous generation of dark energy study.

The most recent cosmology results are reflected in
the \Planck\ temperature, polarization, and lensing data, the Pantheon SNe~Ia sample,
the SDSS BAO+RSD measurements, and the DES 3$\times$2pt samples.
This dataset, referred to as `Stage-III', provides the main constraints
presented in this paper and in Table~\ref{tab:full}.

Finally we isolate the improvements over the Stage-II constraints from the SDSS BAO+RSD program.
We do so by replacing the SDSS DR7 BAO measurements with the SDSS BAO+RSD measurements,
while keeping the WMAP CMB and JLA SNe~Ia samples intact.
This combination is then referred to as `Stage-II+SDSS'.
In the same vein, we isolate the improvements over the Stage-II constraints from recent programs
other than SDSS.  Denoted `Stage-III w/o SDSS', the constraints are derived from the Stage-II
programs, \Planck\ temperature, polarization, and lensing data, the Pantheon SNe~Ia sample,
and the DES 3$\times$2pt samples.

Although the \owowaCDM\ model is the most flexible of all models explored in this work,
with regards to dark energy parameterization,
only the full Stage-III dataset is able to converge without strong priors that exclude unphysical
values of $w_a$ (e.g., see Figure~\ref{fig:w0waomegak}).
On the other hand, the three one-parameter extensions presented in Section~\ref{sec:expansion}
demonstrate the complementarity between the probes in constraining a constant dark energy
equation of state, curvature, and the neutrino mass.
We therefore quantify advances of the last decade by computing cosmological constraints in a \nuowCDM\ model.
The marginalized 68\% confidence intervals for each of the key cosmological parameters in this model,
for each of the relevant Stage-II and Stage-III sample combinations, is presented in Figure~\ref{fig:decade}.

The general effect of the Stage-III measurements is to push the Stage-II results closer to a \LCDM\
model in both curvature and the dark energy equation of state.
The Stage-III results also significantly reduce the upper bounds on the neutrino mass without any indication
for a central value that is larger than 0 eV.
With the exception of $\sigma_8$, the central values of all parameters in the Stage-III results overlap
the 68\% confidence intervals of the Stage-II results.  The precision on all parameters has increased
by at least a factor of 2.5.  The largest gains from Stage-II to Stage-III are found in the 
constraints on $\Omega_k$, $\sigma_8$, and $\sum m_\nu$, with improvements in precision by factors
of 4.5, 7.0, and 7.1, respectively.  

We compute the relative gain across the full volume of the 68\% confidence intervals on
$w$, $\Omega_k$, $\sum m_\nu$, $H_0$, and $\sigma_8$.
We use a figure of merit related to the inverse of the determinant of the covariance matrix for these five parameters. 
We define our figure of merit as ${\rm FoM} = |{\rm Cov}(\mathbf{p},\mathbf{p})|^{-1/(2N)}$, 
where $N=5$ is the number of free parameters (represented by $\mathbf{p}$).
This form properly tracks the typical gain in the 68\% confidence interval for each free parameter. 
We find ${\rm FoM} =$ 11, 24, and 44 for the Stage-II, Stage-II+SDSS, and Stage-III results, respectively.
The gain by a factor of 2 when adding the SDSS data to the Stage-II experiments demonstrates the
significant contribution of BAO and RSD measurements in advancing the cosmological model.
The SDSS BAO and RSD data reduce the total volume (within 68\% confidence) of 
the five dimensional likelihood surface by a factor of 40.

The SDSS BAO+RSD measurements have the most significant impact on the precision of $\Omega_{k}$, $H_0$, 
and $\sum m_\nu$.  In particular, the combination of Stage-II+SDSS leads to estimates of
$H_0= 67.91 \pm 0.91 \, \kmsmpc$, comparable to the tightest constraints on the local expansion rate
presented in Section~\ref{subsec:H0}.
This result is in disagreement with the combined Cepheid distance ladder and strong lensing time delay
results by more than four standard deviations, further reinforcing one of the biggest surprises
of the last decade of cosmology results.

\subsection{Beyond Dark Energy:  Cosmology from eBOSS} 

The spectroscopic samples from BOSS and eBOSS allow for a diverse array of cosmology studies beyond
the cosmic expansion history and growth of structure presented in this work.
These data have already been used to advance models for the summed neutrino mass and inflation.
In addition, new techniques have been developed to use combinations of tracers or new
tracers for direct measurements of BAO and RSD.

These data have been used to place constraints on neutrino masses and inflation parameters
through measurements of the one-dimensional flux power spectrum of the \lya\ forest
\citep{palanque-delabrouille13b,chabanier19a}.
When combining the recent eBOSS measurement with CMB and BAO measurements,
the sum of the neutrino masses is constrained with a 95\% upper
limit $\sum m_{\nu} <0.09$ eV \citep{palanque-delabrouille20a}.
These same \lya\ forest power spectrum measurements 
present evidence for a departure from a constant value for the power-law index of the
primordial power spectrum.  The model that best describes the \lya\ and Planck data
has a running that is non-zero at more than 95\% confidence,
$\alpha_s \equiv d n_s/d \ln k = -0.010 \pm 0.004$.

The eBOSS data have been used to further explore inflationary models through
tests for primordial non-Gaussianities of the local form, $\fnl$.
Recent measurements of the power spectrum in eBOSS quasars offer measurements
of $\fnl$ that are independent of the current Planck bispectrum limits.
The measurements find $-51<\fnl<21$ at 95\% confidence \citep{castorina19a}
and indicate that the full eBOSS dataset could reach $\sigma(\fnl) < 10$ using
a full range of scales and a larger redshift range.

The five-year eBOSS sample also provides an area that is sampled simultaneously
with LRGs, ELGs, and quasars.
The overlap in redshifts between samples enables techniques to combine multiple tracers and
reduce the effects of sample variance \citep{mcdonald09a,seljak09a}.
Projections for $f_{\rm NL}$ and RSD from eBOSS following the multi-tracer technique are
found in the work by \citet{zhao16a}.  
In an effort to understand the joint clustering across multiple tracers,
\citet{alam19a} detect one-halo conformity between the eBOSS LRG and ELG samples at a significance
of more than three standard deviations.  The result presents the challenges of
predicting multi-tracer clustering at high precision beyond what is possible with the basic halo model.
The first eBOSS multi-tracer cosmology study
is associated with this final eBOSS release \citep{wang20a,zhao20a}.  In the
configuration-space study, they find
an improvement in the RSD measurement precision of approximately $12\%$ over using the LRG samples presented in this work.

The eBOSS data have inspired several other advanced techniques in cosmology.
Tentative BAO measurements have been made at $z<1$ using the
cross-correlation between the MgII forest and galaxy and quasar tracers \citep{masdesbourboux19a},
between the CIV forest and quasars at $z>2$ \citep{blomqvist18a},
and finally, in the cross-correlation between spectroscopic quasars and high redshift
galaxies \citep{zarrouk20a} selected from the DESI Legacy Imaging Surveys \citep{dey19a}.
Voids in the clustering of galaxies and quasars have long been explored as a means
to constrain growth of structure and the distance-redshift relation through the AP effect.
The first void detections in eBOSS are presented using DR14 LRGs and quasars \citep{hawken19a}.
This analysis has been extended to the eBOSS DR16, including the ELG sample for the first time
in eBOSS and finding a linear redshift-space distortion parameter \citep{aubert20}
$\beta^{\rm LRG}(z=0.74)=0.415\pm0.087$,  $\beta^{\rm ELG}(z=0.85)=0.665\pm0.125$ and $\beta^{\rm QSO}(z=1.48)=0.313\pm0.134$,
consistent with other measurements from eBOSS DR16 using conventional clustering techniques presented in this paper.
\citet{ravoux20a} recently developed a parallel technique for void finding at higher redshifts.
They derived a three-dimensional map of large-scale matter fluctuations from
a region that was densely sampled with \lya\ forest quasars.  Covering a volume of $0.94 h^{-3}$Gpc$^3$
with a resolution of $13 h^{-1}{\rm Mpc}$, they identify voids and protocluster candidates in the cosmic web. 

Finally, the eBOSS data have enabled new techniques for controlling and assessing the selection
function for tracers of large-scale structure.
Among those new techniques are those results associated with the release of this paper,
such as forward modeling of the selection function from imaging surveys \citep{kong20a},
new models for fiber collisions \citep{mohammad20a},
and N-body mock catalogs for characterization of the ELG sample \citep{avila20a},

Having recently completed installation and commissioning, the Dark Energy Spectroscopic
Instrument \citep[DESI;][]{desi16a,desi16b} will obtain a sample of nearly 50 million galaxies
and quasars spanning the redshift range $0<z<3.5$.
The techniques developed in eBOSS to use the one-dimensional \lya\ forest flux power spectrum,
large-scale clustering, multi-tracer clustering, void detection, and new models for the selection
function and halo occupation statistics will be incorporated into the future DESI studies.
This next generation of the analyses developed within eBOSS will be an integral part of the
final cosmology results at the completion of DESI.

\subsection{Beyond Cosmology:  Astrophysics Results and Potential Studies with the eBOSS Spectra}

The final eBOSS data sample found in the SDSS Sixteenth Data Release \citep{dr16} is the result of
nearly two decades of development in the spectral data reduction and redshift classification software pipelines.
These catalogs of more than four million spectra and their classifications
have been extremely well-vetted and are ripe for further study.
These data have already been used to explore a range of astrophysical processes beyond the
cosmology that inspired the program, with continued potential for studies in galaxy evolution,
lensing and absorption systems, and quasar physics.

The high-redshift ELG sample is unique within the four generations of SDSS and allows
systematic studies of the internal dynamics, composition, and environment
of these star-forming galaxies.
An example of potential for spectroscopic studies in this large data sample is found in one of the 
earliest results from eBOSS.
\citet{zhu15a} constructed a composite spectrum based on 8,620 galaxies over the redshift
interval $0.6 < z < 1.2$.  This composite spectrum reveals blueshifted
lines, indicating outflows driven by star formation.  This high
signal-to-noise spectrum, along with smaller aperture emission line
measurements from Hubble Space Telescope and quasar absorption line
observations, can all be explained by a self-consistent outflow model.
The ELG spectra of roughly 180,000 galaxies were further investigated
to constrain the mass-metallicity relation at high
redshift \citep{huang19a}.  The results indicate that the $0.6<z<1.05$
ELGs follow the fundamental metallicity relation that is observed in
the local universe.  The local environment of the eBOSS galaxies can
also be studied through the illumination of the circumgalactic medium.
In a study of SDSS quasar spectra, the absorption due to MgII
and FeII in intervening LRGs and ELGs was explored over impact
parameters ranging from 10 kpc to 1 Mpc \citep{lan18a}.  The metal
absorption strengths were stronger along the minor-axis of the
galaxies due to outflowing gas and followed a steeper profile for ELGs
than for LRGs, indicating more recent enrichment of the circumgalactic
medium due to star formation.

The eBOSS spectra have also been used to identify previously unknown superpositions of multiple galaxy spectra.
In a search for serendipitous emission lines in the spectra of BOSS and eBOSS galaxy targets,
\citet{talbot20a} were able to identify 1,551 strong galaxy-galaxy gravitational lens candidates.
The full catalog of these lens candidates is being released as a value-added catalog to enable future 
studies\footnote{Spectroscopic Identification of Lensing Objects (SILO) VAC: \url{https://data.sdss.org/sas/dr16/eboss/spectro/lensing/silo}}.
Such a large sample can be used to study the demographics of background source galaxies,
for advanced modeling of the dark matter structure of lens galaxies with a diverse sample,
and for calibrating searches for lens galaxies with ground-based imaging programs.

Finally, eBOSS has unique spectroscopic programs in quasar astrophysics.
Three dedicated programs were coordinated with eBOSS to take advantage of the potential
for studies in quasar astrophysics:
\begin{itemize}
\item The Time Domain Spectroscopic Survey \citep[TDSS;][]{morganson15a,macleod18a}
characterized the spectra of variable stars and quasars identified in photometric imaging;
\item The Spectroscopic Identification of eROSITA Sources \citep[SPIDERS;][]{clerc16a,dwelly17a}
observed cluster galaxies and active galactic nuclei detected in
the ROSAT All-Sky Survey \citep{voges99a,voges00a} and with XMM-Newton \citep{jansen01a} observations; and
\item The SDSS-RM program monitored a sample of 849 quasars at more than 70 epochs
over five years.  The data enable the measurement of more black hole masses, over a larger range of redshift,
than any previous reverberation mapping program \citep{shen15a}.
\end{itemize}
Between the clustering quasar sample \citep{myers15a} and the three
quasar programs, quasar targets comprised the majority of all eBOSS
spectra.  Reverberation mapping studies have measured lags of broad
lines relative to the continuum for
H$\alpha$ \citep[17 quasars;][]{grier17a},
H$\beta$ \citep[42 quasars;][]{grier17a},
MgII \citep[57 quasars;][]{homayouni20a}, and 
CIV \citep[52 quasars, in the redshift range $1.4<z<2.8$;][]{grier19a}.
Arguably the biggest surprise in quasar astrophysics from SDSS was changing-look quasars that change
from broad line quasars with strong continua to narrow line systems
with weak continua over the course of a few years \citep{lamassa15a,macleod16a,dexter19a}.
This phenomenon had not previously been seen for luminous AGN.

Quasar astrophysics is just one of the topics that motivates the next generation
of the Sloan Digital Sky Survey, SDSS-V \citep{kollmeier17a}.
SDSS-V will provide single-object spectra of more than six million
sources across the whole sky with the
BOSS spectrographs and the infrared APOGEE spectrographs \citep{majewski17a,wilson19a}.
In addition, SDSS-V will perform spatially-resolved spectroscopy in the Milky Way and nearby galaxies
using new optical spectrographs on several small telescopes.
The SDSS-V program will produce the world's premier sample of spectra for studies of
Milky Way assembly history, origin of the chemical elements, mapping the local interstellar
medium, and time-domain spectroscopy.
Scheduled for observations over 2020--2025, the five year program will 
multiply the science returns from space-based missions such as {\it Gaia}
and eROSITA \citep{merloni12a} while setting the stage for 
spectroscopy coordinated with imaging from the Vera Rubin Observatory \citep{stubbs04a}.

\acknowledgments

This paper represents an effort by both the SDSS-III and SDSS-IV collaborations.
Funding for SDSS-III was provided by the Alfred
P. Sloan Foundation, the Participating Institutions, the
National Science Foundation, and the U.S. Department
of Energy Office of Science.
Funding for the Sloan Digital Sky Survey IV has been provided by
the Alfred P. Sloan Foundation, the U.S. Department of Energy Office of
Science, and the Participating Institutions. SDSS-IV acknowledges
support and resources from the Center for High-Performance Computing at
the University of Utah. The SDSS web site is www.sdss.org.

SDSS-IV is managed by the Astrophysical Research Consortium for the
Participating Institutions of the SDSS Collaboration including the
Brazilian Participation Group, the Carnegie Institution for Science,
Carnegie Mellon University, the Chilean Participation Group,
the French Participation Group, Harvard-Smithsonian Center for Astrophysics,
Instituto de Astrof\'isica de Canarias, The Johns Hopkins University,
Kavli Institute for the Physics and Mathematics of the Universe (IPMU) /
University of Tokyo, Lawrence Berkeley National Laboratory,
Leibniz Institut f\"ur Astrophysik Potsdam (AIP),
Max-Planck-Institut f\"ur Astronomie (MPIA Heidelberg),
Max-Planck-Institut f\"ur Astrophysik (MPA Garching),
Max-Planck-Institut f\"ur Extraterrestrische Physik (MPE),
National Astronomical Observatory of China, New Mexico State University,
New York University, University of Notre Dame,
Observat\'ario Nacional / MCTI, The Ohio State University,
Pennsylvania State University, Shanghai Astronomical Observatory,
United Kingdom Participation Group,
Universidad Nacional Aut\'onoma de M\'exico, University of Arizona,
University of Colorado Boulder, University of Portsmouth,
University of Utah, University of Virginia, University of Washington, University of Wisconsin,
Vanderbilt University, and Yale University.

\begin{appendix}
\section{BAO and RSD Measurements and Systematic Errors} \label{sec:appendix_a}
In galaxy redshift surveys, BAO and RSD are usually measured through two-point clustering 
statistics.  
To calculate the two-point clustering statistics, we
convert galaxy redshifts and angular positions into comoving coordinates using a
fiducial cosmological model, denoted by a superscript `fid'. For a
pair of galaxies at effective redshift $z_{\rm eff}$, with a small separation, the comoving transverse
and line-of-sight separations depend on the comoving angular diameter distance $D_M(z_{\rm eff})$ and 
the Hubble distance $D_H(z_{\rm eff}) = c/H(z_{\rm eff})$, respectively.
Conversion from radial comoving distance, $D_C$, to $D_M$ depends on the cosmological model
(see Section~\ref{sec:cosmo}).
Limiting ourselves to the \LCDM\ model, and working in units of
$h^{-1}$Mpc, this conversion depends solely on the value of
$\Omega_m^{\rm fid}$. Counts of galaxy pairs, in the form of the
correlation function or power spectrum, are then fitted with a fixed
model in which the BAO feature is located at $r_d^{\rm fid}$. Although
not necessary for the methodology we adopt, we use the same model
for computing $r_d^{\rm fid}$ as we do for the conversion of 
measured coordinates to comoving coordinates.

For the SDSS BAO measurements, we
parameterize the position of the BAO feature using a dimensionless dilation parameter in the transverse 
direction ($\aperp$) and in the radial ($\apara$) direction. 
The best-fit values and covariance between these
parameters are calculated by fitting the template spectrum to the
observed BAO positions in the monopole, quadrupole and hexadecapole
moments of the two-point statistics. Information other than the BAO peak
position is removed from the fit by marginalizing over a set of
simultaneously fitted parameters that model the shape of the multipole moments.

If the true BAO peak is located at $r_d$, which can be different from
$r_d^{\rm fid}$, then both $\aperp$ and $\apara$ will be shifted by
$r_d^{\rm fid}/r_d$. The location of the radial BAO peak in the data
depends on $D_H(z_{\rm eff})/D_H^{\rm fid}(z_{\rm eff})$, while the
location of the angular BAO feature depends on
$D_M(z_{\rm eff})/D_M^{\rm fid}(z_{\rm eff})$. Combining these, we have
\begin{eqnarray}
  \apara &=&  
    \frac{D_H(z_{\rm eff})/r_d}{D_H^{\rm fid}(z_{\rm eff})/ r_d^{\rm fid}},\\
  \aperp &=& 
    \frac{D_M(z_{\rm eff})/r_d}{D_M^{\rm fid}(z_{\rm eff})/ r_d^{\rm fid}}.
\end{eqnarray}
We see that $\aperp$ and $\apara$ combine information about the model
and distance-redshift relationship in a way that is perfectly
degenerate. To demonstrate this, we now consider the dependence of the
fit on $h/h^{\rm fid}$.

Working in units of $h^{-1}$Mpc, and assuming that the fiducial and
true physical densities match so $r_d=r_d^{\rm fid}$ in units of Mpc,
the ratio $h/h^{\rm fid}$ enters in the shift of the model to
$h^{-1}$Mpc, and hence via $r_d/r_d^{\rm fid}$. The ratios
$D_M^{\rm fid}(z_{\rm eff})/D_M(z_{\rm eff})$ and
$D_H^{\rm fid}(z_{\rm eff})/D_H(z_{\rm eff})$ are independent of $h$
and $h^{\rm fid}$ for fixed $\Omega_m$. Suppose instead that we had worked
in units of Mpc and measured the two-point functions and
model in these units. Then we would have to specify $h^{\rm fid}$ before
calculating the two-point measurements, and so the ratio $h/h^{\rm fid}$
enters into the calculation of
$D_M(z_{\rm eff})/D_M^{\rm fid}(z_{\rm eff})$ and
$D_H(z_{\rm eff})/D_H^{\rm fid}(z_{\rm eff})$. For models where
$r_d=r_d^{\rm fid}$, there is no $h$ dependence in the theoretical BAO
positions.

Note that the above thought experiment shows that we should always
work in the same basis when calculating the components of both
$\aperp$ and $\apara$. That is, we should not calculate
$r_d/r_d^{\rm fid}$ in Mpc and
$D_M^{\rm fid}(z_{\rm eff})/D_M(z_{\rm eff})$ in $h^{-1}{\rm Mpc}$, for
example, which would then ignore the $h/h^{\rm fid}$ dependence of the
fit.

Another way to see this is that the dimensionless quantities
$r_d/D_M(z_{\rm eff})$ and $r_d/D_H(z_{\rm eff})$ correspond to the size
of the BAO feature in observed quantities, namely angular separations
measured in radians and radial separations measured in redshift
differences. As long as we operate in units that make these ratios
truly dimensionless (i.e., without residual dependence on $h$), we are
performing a correct compression of the available information.

We present our baseline results as $D_M(z_{\rm eff})/r_d$ and $D_H(z_{\rm eff})/r_d$
to reflect what is measured in the data.  There is no dependence on fiducial
values, thus removing potential for ambiguity and the exact values
assumed in our fiducial model.

To measure RSD, we fit a template power spectrum or correlation
function decomposed into multipoles.  We allow the template to be shifted in
scale by the dimensionless parameters $\aperp$ and $\apara$, and
normalized using the parameters $b\sigma_8$ and $f\sigma_8$. Use of a
template spectrum means that our measurements necessarily depend on
the shape of this template and on the fiducial cosmology used to
create it. As $\sigma_8$ is defined as the r.m.s. fluctuations on
comoving scales of $8h^{-1}$Mpc (i.e., not the gauge independent
angular separations and redshift differences), we also need to
consider the model dependency of the scale on which the normalization
parameters are measured.

In the analysis procedure in SDSS clustering studies, we normalize the template to a predicted $\sigma_8$
and find the shifts in scale and normalization required to fit the data.
One complication is whether the template is
shifted in scale by the dilation parameters before or after the
normalization of the model is fixed. Shifting the template before
measuring the normalization is equivalent to fixing the scale on
which we measure $f\sigma_8$ and $b\sigma_8$ in the observed two-point
clustering, as determined by the fiducial cosmology in units of
$h^{-1}$Mpc. For data at $z=0$, this would result in no cosmological
dependence in the scale chosen. However for measurements at higher
redshifts such as those from eBOSS, we have a dependence on the
fiducial value of $\Omega_m$ used to calculate the distance-redshift
relationship.

If, instead, we do not shift the template before fixing the
normalization of the model, then we fix the scale 
in the units of the template.  The scale from the template can be different from that
preferred by the data, potentially bringing in a further dependence on
$h/h^{\rm fid}$ and changing the degeneracies with other cosmological
parameters. In general, we find a larger systematic
error contribution for our measurements in this case due to an increased
dependence on the fiducial cosmology.

We interpret the dilation parameters as measured from template fits in
the same way as those with the above BAO measurements,
assuming that the model
dependence arises only through $r_d$ and not through other scales in
the theoretical model. While the BAO scale provides most of the
dilation constraint, it is possible that some component arises from
other features, and therefore this should be considered an
approximation.

\begin{figure}
\centering
\includegraphics[scale=0.28, trim={80 0 0 0}, clip=false]{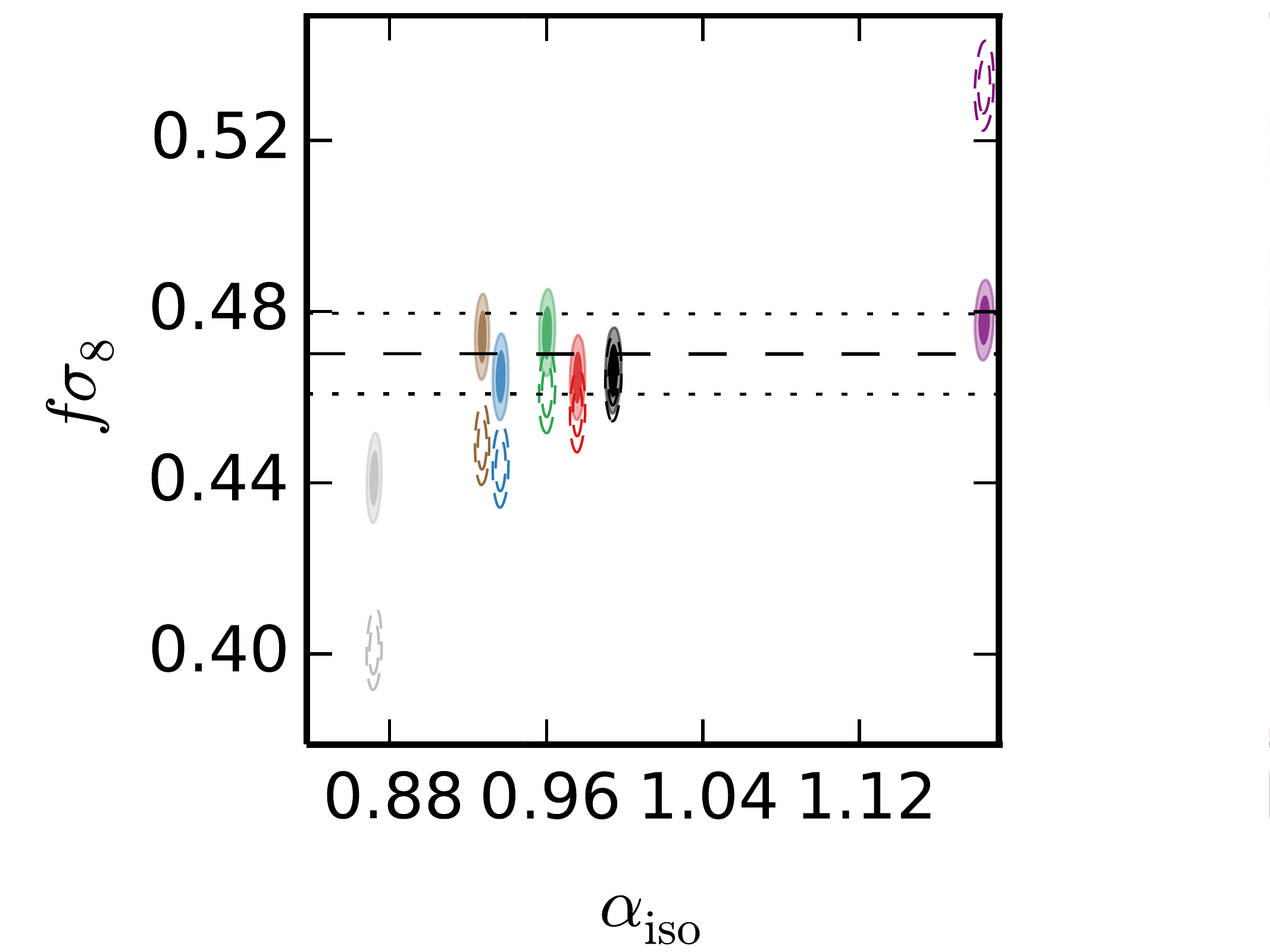}
\caption{Dependence of the measured values of $f\sigma_8$ on
  $\alpha_{\rm iso}=(\aperp^2\apara)^{1/3}$, which gives the offset
  between the template and the true cosmology, calculated from the set
  of \textsc{Nseries} mocks matching the BOSS CMASS NGC LRG sample at
  an effective redshift $z_{\rm eff}=0.56$. Filled ellipses give
  results fixing the scale at which $f\sigma_8$ is measured after
  re-scaling the template allowing for the shift in the best-fit
  $\alpha_{\rm iso}$; the empty ellipses show results where the
  template is not rescaled. The colors of the ellipses separate fits
  where different template models were used when analyzing the
  mocks. Only a weak dependence is seen when $f\sigma_8$ is
  constructed from the re-scaled $\sigma_8$, significantly smaller than
  the statistical errors on the measured values. Dotted lines mark a $2\%$
  deviation with respect to the expected $f\sigma_8$ value. Further
  details of these fits are in \citet{gil-marin19a}.
  }
\label{fig:fs8-stability}
\end{figure}

For our RSD measurements made using the BOSS and eBOSS galaxy samples, we have found that
rescaling the template before fixing the normalization of the model
significantly reduces the dependence on fiducial cosmology, and hence
the required systematic error.

The $\sigma_8$--based normalization
measurements we present and analyze retain a dependence on the
fiducial $\Omega_m$ that sets the scale on which they are measured, and
on the shape of the template, which links the scales on which
$f\sigma_8$ and $b\sigma_8$ are defined to those constrained by the
data. These dependencies are illustrated in Figure~7 of \cite{LRG_corr}
by comparing recovered measurements (from the LRG correlation function) 
with mock catalogues where the 
fiducial cosmology assumed is different from the true cosmology.
For
the power spectrum as measured from the LRGs, we show how measurements
of $f\sigma_8$ depend on $\alpha_{\rm iso}=(\aperp^2\apara)^{1/3}$ in
Figure~\ref{fig:fs8-stability}. From these tests, we see only a weak
dependence on the fiducial assumptions, given our baseline
procedure. The scatter in the measurements is included in our assumed
systematic error.

For our MGS and eBOSS quasar measurements, the contribution to the systematic
error from the fiducial cosmology is significantly reduced compared to
other contributions to the systematic error, and we instead adopt the
slightly simpler procedure where we do not rescale the template before
normalization. This results in slightly larger estimates of the
systematic error induced by the analysis method. More details about
the systematic errors and the dependencies on the fiducial assumptions
in the analysis method can be found in the papers describing the
individual measurements.

We have shown that the approximations we made to compress BOSS and eBOSS
data into combinations of parameters ($\aperp$, $\apara$, $f\sigma_8$,
and $b\sigma_8$) do not significantly impact the interpretation of the growth measurements. 
The compression therefore has minimal impact on our conclusions in testing cosmological models
after allowing for appropriate systematic errors.
This analysis of the clustering might not be the best approach for future
surveys including DESI and Euclid. In order not to compromise the
precision available with these future data, it may be best to directly
fit models to the two-point multipoles without an intermediate data
compression step.

\section{Model Parameterization and Parameter Priors} \label{sec:appendix_model}
For cosmological models considered in this study, it is possible to adopt a single parameterization. However, 
it is convenient to choose a parameterization depending on the investigation,
thus allowing different priors depending on the constraints provided by the probes being used.
We employ two sets of parameterization. The definitions of
parameters can be found in Table~\ref{tab:parameters}.

The first parameterization is labeled as the `CMB' parameterization, which is used whenever CMB
likelihoods are included in the analysis. It follows the natural degeneracy direction 
of the CMB constraints. The basic parameters in this parameterization include $\omega_c$, $\omega_b$, 
$\theta_{\rm MC}$, $A_s$, $n_s$, $\tau$, and $N_{\rm eff}$.

The second parameterization is labeled as the `background' parameterization.  This parameterization is used for chains 
without a CMB likelihood, such as in studying SN-only or BAO-only constraints. The basic parameters include
$\Omega_m$, $H_0$, $\Omega_b$ with the latter two used for BAO constraints.

In addition to the basic parameters, we also introduce extended parameters for testing extensions to the \LCDM\
model, which include $\Omega_k$, $w$ or $w_0$, $w_a$, and $\Sigma m_\nu$. 

In Table~\ref{tab:priors}, we list the parameters for each parameterization, along with the priors assigned
in the analyses and the sections/subsections where the priors are adopted. The baseline value of a parameter
refers to the value used whenever the parameter is fixed in an analysis.
In addition to the flat priors on parameters that are being varied in the analysis, the `CMB'
parameterization includes a prior on $H_0$ of $20\,\kmsmpc<H_0<100\,\kmsmpc$. 
We also applied a Gaussian prior on $n_s$ of $n_s=0.96\pm0.02$ and a BBN-inspired prior of
$\omega_b=0.0222\pm0.0005$ in all runs that include growth information but without CMB data,
i.e., in Figure \ref{fig:growth_amplitude} (left) and Figure \ref{fig:dist+growth}.  
In Section~4.2.2, for $H_0$ constraints without CMB data, a BBN-motivated Gaussian prior is
used for $\omega_b$ ($0.02235\pm 0.00037$).

\begin{table*}
  \centering
\caption{\rm Cosmological parameters and priors}
\begin{tabular}{|llccl|}
\hline \hline
Parameterization & Parameter\footnote{For each parameter, in parentheses is the range of the flat prior.} & Baseline\footnote{The baseline value of a parameter is the one adopted whenever the parameter is fixed in an analysis.} & Prior & Analysis \\
\hline
CMB        & $\omega_c$            & 0.12   & (0.001, 0.99) & whenever CMB likelihood is included,\\
           & $\omega_b$            & 0.0221 & (0.005, 0.1)  & as in Sections 4, 5, 6.\\
           & $100\theta_{\rm MC}$  & 1.0411 & (0.5, 10)     & ...\\
           & $\ln (10^{10}A_s)$    & 3.05    & (1.61, 3.91)  & ...\\
           & $n_s$                 & 0.96   & (0.8, 1.2)    & ...\\
           & $\tau$                & 0.06   & (0.01, 0.8)   & ...\\
           & $N_{\rm eff}$         & 3.046  & --            & ...\\
           &                       &        &               &    \\
\hline
           &                       &        &               & \\
Background & $\Omega_m$            & --     & (0.1, 0.9)    & Section~4.1 (SN-only, BAO-only)\\
           & $\Omega_b$            & --     & (0.001, 0.3)  & Section~4.1 (BAO-only)\\
           & $H_0$ (${\rm km\, s^{-1}Mpc^{-1}}$)              & --      & (20, 100)      & Section~4.2.2\\
           &                       &        &               & \\
\hline
           &                       &        &               & \\
Extended   & $\Omega_k$            & 0.0    & (-0.8, 0.8)   & Section~4.1.1, Section~6.2\\
           & $w$ ($w_0$)             &-1      & (-3, 1)       & Section~4.1.2, Section~6.2\\
           & $w_a$                 & 0      & (-3, 0.75)    & Section~6.2 \\
           & $\sum m_\nu$ (eV)     & 0.06   & (0, 5)         & Section~4.1.3, Section~6.3\\
\hline
\end{tabular}
\label{tab:priors}
\end{table*}

\end{appendix}

\bibliography{archive}

\begin{thebibliography}{254}%
\makeatletter
\providecommand \@ifxundefined [1]{%
 \@ifx{#1\undefined}
}%
\providecommand \@ifnum [1]{%
 \ifnum #1\expandafter \@firstoftwo
 \else \expandafter \@secondoftwo
 \fi
}%
\providecommand \@ifx [1]{%
 \ifx #1\expandafter \@firstoftwo
 \else \expandafter \@secondoftwo
 \fi
}%
\providecommand \natexlab [1]{#1}%
\providecommand \enquote  [1]{``#1''}%
\providecommand \bibnamefont  [1]{#1}%
\providecommand \bibfnamefont [1]{#1}%
\providecommand \citenamefont [1]{#1}%
\providecommand \href@noop [0]{\@secondoftwo}%
\providecommand \href [0]{\begingroup \@sanitize@url \@href}%
\providecommand \@href[1]{\@@startlink{#1}\@@href}%
\providecommand \@@href[1]{\endgroup#1\@@endlink}%
\providecommand \@sanitize@url [0]{\catcode `\\12\catcode `\$12\catcode
  `\&12\catcode `\#12\catcode `\^12\catcode `\_12\catcode `\%12\relax}%
\providecommand \@@startlink[1]{}%
\providecommand \@@endlink[0]{}%
\providecommand \url  [0]{\begingroup\@sanitize@url \@url }%
\providecommand \@url [1]{\endgroup\@href {#1}{\urlprefix }}%
\providecommand \urlprefix  [0]{URL }%
\providecommand \Eprint [0]{\href }%
\providecommand \doibase [0]{http://dx.doi.org/}%
\providecommand \selectlanguage [0]{\@gobble}%
\providecommand \bibinfo  [0]{\@secondoftwo}%
\providecommand \bibfield  [0]{\@secondoftwo}%
\providecommand \translation [1]{[#1]}%
\providecommand \BibitemOpen [0]{}%
\providecommand \bibitemStop [0]{}%
\providecommand \bibitemNoStop [0]{.\EOS\space}%
\providecommand \EOS [0]{\spacefactor3000\relax}%
\providecommand \BibitemShut  [1]{\csname bibitem#1\endcsname}%
\let\auto@bib@innerbib\@empty
\bibitem [{\citenamefont {{Riess}}\ \emph {et~al.}(1998)\citenamefont
  {{Riess}}, \citenamefont {{Filippenko}}, \citenamefont {{Challis}},
  \citenamefont {{Clocchiatti}}, \citenamefont {{Diercks}}, \citenamefont
  {{Garnavich}}, \citenamefont {{Gilliland}}, \citenamefont {{Hogan}},
  \citenamefont {{Jha}}, \citenamefont {{Kirshner}}, \citenamefont
  {{Leibundgut}}, \citenamefont {{Phillips}}, \citenamefont {{Reiss}},
  \citenamefont {{Schmidt}}, \citenamefont {{Schommer}}, \citenamefont
  {{Smith}}, \citenamefont {{Spyromilio}}, \citenamefont {{Stubbs}},
  \citenamefont {{Suntzeff}},\ and\ \citenamefont {{Tonry}}}]{riess98a}%
  \BibitemOpen
  \bibfield  {author} {\bibinfo {author} {\bibfnamefont {A.~G.}\ \bibnamefont
  {{Riess}}}, \bibinfo {author} {\bibfnamefont {A.~V.}\ \bibnamefont
  {{Filippenko}}}, \bibinfo {author} {\bibfnamefont {P.}~\bibnamefont
  {{Challis}}}, \bibinfo {author} {\bibfnamefont {A.}~\bibnamefont
  {{Clocchiatti}}}, \bibinfo {author} {\bibfnamefont {A.}~\bibnamefont
  {{Diercks}}}, \bibinfo {author} {\bibfnamefont {P.~M.}\ \bibnamefont
  {{Garnavich}}}, \bibinfo {author} {\bibfnamefont {R.~L.}\ \bibnamefont
  {{Gilliland}}}, \bibinfo {author} {\bibfnamefont {C.~J.}\ \bibnamefont
  {{Hogan}}}, \bibinfo {author} {\bibfnamefont {S.}~\bibnamefont {{Jha}}},
  \bibinfo {author} {\bibfnamefont {R.~P.}\ \bibnamefont {{Kirshner}}},
  \bibinfo {author} {\bibfnamefont {B.}~\bibnamefont {{Leibundgut}}}, \bibinfo
  {author} {\bibfnamefont {M.~M.}\ \bibnamefont {{Phillips}}}, \bibinfo
  {author} {\bibfnamefont {D.}~\bibnamefont {{Reiss}}}, \bibinfo {author}
  {\bibfnamefont {B.~P.}\ \bibnamefont {{Schmidt}}}, \bibinfo {author}
  {\bibfnamefont {R.~A.}\ \bibnamefont {{Schommer}}}, \bibinfo {author}
  {\bibfnamefont {R.~C.}\ \bibnamefont {{Smith}}}, \bibinfo {author}
  {\bibfnamefont {J.}~\bibnamefont {{Spyromilio}}}, \bibinfo {author}
  {\bibfnamefont {C.}~\bibnamefont {{Stubbs}}}, \bibinfo {author}
  {\bibfnamefont {N.~B.}\ \bibnamefont {{Suntzeff}}}, \ and\ \bibinfo {author}
  {\bibfnamefont {J.}~\bibnamefont {{Tonry}}},\ }\href@noop {} {\bibfield
  {journal} {\bibinfo  {journal} {\aj}\ }\textbf {\bibinfo {volume} {116}},\
  \bibinfo {pages} {1009} (\bibinfo {year} {1998})}\BibitemShut {NoStop}%
\bibitem [{\citenamefont {{Perlmutter}}\ \emph {et~al.}(1999)\citenamefont
  {{Perlmutter}}, \citenamefont {{Aldering}}, \citenamefont {{Goldhaber}},
  \citenamefont {{Knop}}, \citenamefont {{Nugent}}, \citenamefont {{Castro}},
  \citenamefont {{Deustua}}, \citenamefont {{Fabbro}}, \citenamefont
  {{Goobar}}, \citenamefont {{Groom}}, \citenamefont {{Hook}}, \citenamefont
  {{Kim}}, \citenamefont {{Kim}}, \citenamefont {{Lee}}, \citenamefont
  {{Nunes}}, \citenamefont {{Pain}}, \citenamefont {{Pennypacker}},
  \citenamefont {{Quimby}}, \citenamefont {{Lidman}}, \citenamefont {{Ellis}},
  \citenamefont {{Irwin}}, \citenamefont {{McMahon}}, \citenamefont
  {{Ruiz-Lapuente}}, \citenamefont {{Walton}}, \citenamefont {{Schaefer}},
  \citenamefont {{Boyle}}, \citenamefont {{Filippenko}}, \citenamefont
  {{Matheson}}, \citenamefont {{Fruchter}}, \citenamefont {{Panagia}},
  \citenamefont {{Newberg}}, \citenamefont {{Couch}},\ and\ \citenamefont {{The
  Supernova Cosmology Project}}}]{perlmutter99a}%
  \BibitemOpen
  \bibfield  {author} {\bibinfo {author} {\bibfnamefont {S.}~\bibnamefont
  {{Perlmutter}}}, \bibinfo {author} {\bibfnamefont {G.}~\bibnamefont
  {{Aldering}}}, \bibinfo {author} {\bibfnamefont {G.}~\bibnamefont
  {{Goldhaber}}}, \bibinfo {author} {\bibfnamefont {R.~A.}\ \bibnamefont
  {{Knop}}}, \bibinfo {author} {\bibfnamefont {P.}~\bibnamefont {{Nugent}}},
  \bibinfo {author} {\bibfnamefont {P.~G.}\ \bibnamefont {{Castro}}}, \bibinfo
  {author} {\bibfnamefont {S.}~\bibnamefont {{Deustua}}}, \bibinfo {author}
  {\bibfnamefont {S.}~\bibnamefont {{Fabbro}}}, \bibinfo {author}
  {\bibfnamefont {A.}~\bibnamefont {{Goobar}}}, \bibinfo {author}
  {\bibfnamefont {D.~E.}\ \bibnamefont {{Groom}}}, \bibinfo {author}
  {\bibfnamefont {I.~M.}\ \bibnamefont {{Hook}}}, \bibinfo {author}
  {\bibfnamefont {A.~G.}\ \bibnamefont {{Kim}}}, \bibinfo {author}
  {\bibfnamefont {M.~Y.}\ \bibnamefont {{Kim}}}, \bibinfo {author}
  {\bibfnamefont {J.~C.}\ \bibnamefont {{Lee}}}, \bibinfo {author}
  {\bibfnamefont {N.~J.}\ \bibnamefont {{Nunes}}}, \bibinfo {author}
  {\bibfnamefont {R.}~\bibnamefont {{Pain}}}, \bibinfo {author} {\bibfnamefont
  {C.~R.}\ \bibnamefont {{Pennypacker}}}, \bibinfo {author} {\bibfnamefont
  {R.}~\bibnamefont {{Quimby}}}, \bibinfo {author} {\bibfnamefont
  {C.}~\bibnamefont {{Lidman}}}, \bibinfo {author} {\bibfnamefont {R.~S.}\
  \bibnamefont {{Ellis}}}, \bibinfo {author} {\bibfnamefont {M.}~\bibnamefont
  {{Irwin}}}, \bibinfo {author} {\bibfnamefont {R.~G.}\ \bibnamefont
  {{McMahon}}}, \bibinfo {author} {\bibfnamefont {P.}~\bibnamefont
  {{Ruiz-Lapuente}}}, \bibinfo {author} {\bibfnamefont {N.}~\bibnamefont
  {{Walton}}}, \bibinfo {author} {\bibfnamefont {B.}~\bibnamefont
  {{Schaefer}}}, \bibinfo {author} {\bibfnamefont {B.~J.}\ \bibnamefont
  {{Boyle}}}, \bibinfo {author} {\bibfnamefont {A.~V.}\ \bibnamefont
  {{Filippenko}}}, \bibinfo {author} {\bibfnamefont {T.}~\bibnamefont
  {{Matheson}}}, \bibinfo {author} {\bibfnamefont {A.~S.}\ \bibnamefont
  {{Fruchter}}}, \bibinfo {author} {\bibfnamefont {N.}~\bibnamefont
  {{Panagia}}}, \bibinfo {author} {\bibfnamefont {H.~J.~M.}\ \bibnamefont
  {{Newberg}}}, \bibinfo {author} {\bibfnamefont {W.~J.}\ \bibnamefont
  {{Couch}}}, \ and\ \bibinfo {author} {\bibnamefont {{The Supernova Cosmology
  Project}}},\ }\href@noop {} {\bibfield  {journal} {\bibinfo  {journal}
  {\apj}\ }\textbf {\bibinfo {volume} {517}},\ \bibinfo {pages} {565} (\bibinfo
  {year} {1999})}\BibitemShut {NoStop}%
\bibitem [{\citenamefont {{Miller}}\ \emph {et~al.}(1999)\citenamefont
  {{Miller}}, \citenamefont {{Caldwell}}, \citenamefont {{Devlin}},
  \citenamefont {{Dorwart}}, \citenamefont {{Herbig}}, \citenamefont {{Nolta}},
  \citenamefont {{Page}}, \citenamefont {{Puchalla}}, \citenamefont
  {{Torbet}},\ and\ \citenamefont {{Tran}}}]{miller99a}%
  \BibitemOpen
  \bibfield  {author} {\bibinfo {author} {\bibfnamefont {A.~D.}\ \bibnamefont
  {{Miller}}}, \bibinfo {author} {\bibfnamefont {R.}~\bibnamefont
  {{Caldwell}}}, \bibinfo {author} {\bibfnamefont {M.~J.}\ \bibnamefont
  {{Devlin}}}, \bibinfo {author} {\bibfnamefont {W.~B.}\ \bibnamefont
  {{Dorwart}}}, \bibinfo {author} {\bibfnamefont {T.}~\bibnamefont {{Herbig}}},
  \bibinfo {author} {\bibfnamefont {M.~R.}\ \bibnamefont {{Nolta}}}, \bibinfo
  {author} {\bibfnamefont {L.~A.}\ \bibnamefont {{Page}}}, \bibinfo {author}
  {\bibfnamefont {J.}~\bibnamefont {{Puchalla}}}, \bibinfo {author}
  {\bibfnamefont {E.}~\bibnamefont {{Torbet}}}, \ and\ \bibinfo {author}
  {\bibfnamefont {H.~T.}\ \bibnamefont {{Tran}}},\ }\href {\doibase
  10.1086/312293} {\bibfield  {journal} {\bibinfo  {journal} {\apjl}\ }\textbf
  {\bibinfo {volume} {524}},\ \bibinfo {pages} {L1} (\bibinfo {year}
  {1999})}\BibitemShut {NoStop}%
\bibitem [{\citenamefont {{de Bernardis}}\ \emph {et~al.}(2000)\citenamefont
  {{de Bernardis}}, \citenamefont {{Ade}}, \citenamefont {{Bock}},
  \citenamefont {{Bond}}, \citenamefont {{Borrill}}, \citenamefont
  {{Boscaleri}}, \citenamefont {{Coble}}, \citenamefont {{Crill}},
  \citenamefont {{De Gasperis}}, \citenamefont {{Farese}}, \citenamefont
  {{Ferreira}}, \citenamefont {{Ganga}}, \citenamefont {{Giacometti}},
  \citenamefont {{Hivon}}, \citenamefont {{Hristov}}, \citenamefont
  {{Iacoangeli}}, \citenamefont {{Jaffe}}, \citenamefont {{Lange}},
  \citenamefont {{Martinis}}, \citenamefont {{Masi}}, \citenamefont {{Mason}},
  \citenamefont {{Mauskopf}}, \citenamefont {{Melchiorri}}, \citenamefont
  {{Miglio}}, \citenamefont {{Montroy}}, \citenamefont {{Netterfield}},
  \citenamefont {{Pascale}}, \citenamefont {{Piacentini}}, \citenamefont
  {{Pogosyan}}, \citenamefont {{Prunet}}, \citenamefont {{Rao}}, \citenamefont
  {{Romeo}}, \citenamefont {{Ruhl}}, \citenamefont {{Scaramuzzi}},
  \citenamefont {{Sforna}},\ and\ \citenamefont {{Vittorio}}}]{debernardis00a}%
  \BibitemOpen
  \bibfield  {author} {\bibinfo {author} {\bibfnamefont {P.}~\bibnamefont {{de
  Bernardis}}}, \bibinfo {author} {\bibfnamefont {P.~A.~R.}\ \bibnamefont
  {{Ade}}}, \bibinfo {author} {\bibfnamefont {J.~J.}\ \bibnamefont {{Bock}}},
  \bibinfo {author} {\bibfnamefont {J.~R.}\ \bibnamefont {{Bond}}}, \bibinfo
  {author} {\bibfnamefont {J.}~\bibnamefont {{Borrill}}}, \bibinfo {author}
  {\bibfnamefont {A.}~\bibnamefont {{Boscaleri}}}, \bibinfo {author}
  {\bibfnamefont {K.}~\bibnamefont {{Coble}}}, \bibinfo {author} {\bibfnamefont
  {B.~P.}\ \bibnamefont {{Crill}}}, \bibinfo {author} {\bibfnamefont
  {G.}~\bibnamefont {{De Gasperis}}}, \bibinfo {author} {\bibfnamefont {P.~C.}\
  \bibnamefont {{Farese}}}, \bibinfo {author} {\bibfnamefont {P.~G.}\
  \bibnamefont {{Ferreira}}}, \bibinfo {author} {\bibfnamefont
  {K.}~\bibnamefont {{Ganga}}}, \bibinfo {author} {\bibfnamefont
  {M.}~\bibnamefont {{Giacometti}}}, \bibinfo {author} {\bibfnamefont
  {E.}~\bibnamefont {{Hivon}}}, \bibinfo {author} {\bibfnamefont {V.~V.}\
  \bibnamefont {{Hristov}}}, \bibinfo {author} {\bibfnamefont {A.}~\bibnamefont
  {{Iacoangeli}}}, \bibinfo {author} {\bibfnamefont {A.~H.}\ \bibnamefont
  {{Jaffe}}}, \bibinfo {author} {\bibfnamefont {A.~E.}\ \bibnamefont
  {{Lange}}}, \bibinfo {author} {\bibfnamefont {L.}~\bibnamefont {{Martinis}}},
  \bibinfo {author} {\bibfnamefont {S.}~\bibnamefont {{Masi}}}, \bibinfo
  {author} {\bibfnamefont {P.~V.}\ \bibnamefont {{Mason}}}, \bibinfo {author}
  {\bibfnamefont {P.~D.}\ \bibnamefont {{Mauskopf}}}, \bibinfo {author}
  {\bibfnamefont {A.}~\bibnamefont {{Melchiorri}}}, \bibinfo {author}
  {\bibfnamefont {L.}~\bibnamefont {{Miglio}}}, \bibinfo {author}
  {\bibfnamefont {T.}~\bibnamefont {{Montroy}}}, \bibinfo {author}
  {\bibfnamefont {C.~B.}\ \bibnamefont {{Netterfield}}}, \bibinfo {author}
  {\bibfnamefont {E.}~\bibnamefont {{Pascale}}}, \bibinfo {author}
  {\bibfnamefont {F.}~\bibnamefont {{Piacentini}}}, \bibinfo {author}
  {\bibfnamefont {D.}~\bibnamefont {{Pogosyan}}}, \bibinfo {author}
  {\bibfnamefont {S.}~\bibnamefont {{Prunet}}}, \bibinfo {author}
  {\bibfnamefont {S.}~\bibnamefont {{Rao}}}, \bibinfo {author} {\bibfnamefont
  {G.}~\bibnamefont {{Romeo}}}, \bibinfo {author} {\bibfnamefont {J.~E.}\
  \bibnamefont {{Ruhl}}}, \bibinfo {author} {\bibfnamefont {F.}~\bibnamefont
  {{Scaramuzzi}}}, \bibinfo {author} {\bibfnamefont {D.}~\bibnamefont
  {{Sforna}}}, \ and\ \bibinfo {author} {\bibfnamefont {N.}~\bibnamefont
  {{Vittorio}}},\ }\href
  {http://adsabs.harvard.edu/cgi-bin/nph-bib_query?bibcode=2000Natur.404..955D&db_key=AST}
  {\bibfield  {journal} {\bibinfo  {journal} {\nat}\ }\textbf {\bibinfo
  {volume} {404}},\ \bibinfo {pages} {955} (\bibinfo {year}
  {2000})}\BibitemShut {NoStop}%
\bibitem [{\citenamefont {{Balbi}}\ \emph {et~al.}(2000)\citenamefont
  {{Balbi}}, \citenamefont {{Ade}}, \citenamefont {{Bock}}, \citenamefont
  {{Borrill}}, \citenamefont {{Boscaleri}}, \citenamefont {{De Bernardis}},
  \citenamefont {{Ferreira}}, \citenamefont {{Hanany}}, \citenamefont
  {{Hristov}}, \citenamefont {{Jaffe}}, \citenamefont {{Lee}}, \citenamefont
  {{Oh}}, \citenamefont {{Pascale}}, \citenamefont {{Rabii}}, \citenamefont
  {{Richards}}, \citenamefont {{Smoot}}, \citenamefont {{Stompor}},
  \citenamefont {{Winant}},\ and\ \citenamefont {{Wu}}}]{balbi00a}%
  \BibitemOpen
  \bibfield  {author} {\bibinfo {author} {\bibfnamefont {A.}~\bibnamefont
  {{Balbi}}}, \bibinfo {author} {\bibfnamefont {P.}~\bibnamefont {{Ade}}},
  \bibinfo {author} {\bibfnamefont {J.}~\bibnamefont {{Bock}}}, \bibinfo
  {author} {\bibfnamefont {J.}~\bibnamefont {{Borrill}}}, \bibinfo {author}
  {\bibfnamefont {A.}~\bibnamefont {{Boscaleri}}}, \bibinfo {author}
  {\bibfnamefont {P.}~\bibnamefont {{De Bernardis}}}, \bibinfo {author}
  {\bibfnamefont {P.~G.}\ \bibnamefont {{Ferreira}}}, \bibinfo {author}
  {\bibfnamefont {S.}~\bibnamefont {{Hanany}}}, \bibinfo {author}
  {\bibfnamefont {V.}~\bibnamefont {{Hristov}}}, \bibinfo {author}
  {\bibfnamefont {A.~H.}\ \bibnamefont {{Jaffe}}}, \bibinfo {author}
  {\bibfnamefont {A.~T.}\ \bibnamefont {{Lee}}}, \bibinfo {author}
  {\bibfnamefont {S.}~\bibnamefont {{Oh}}}, \bibinfo {author} {\bibfnamefont
  {E.}~\bibnamefont {{Pascale}}}, \bibinfo {author} {\bibfnamefont
  {B.}~\bibnamefont {{Rabii}}}, \bibinfo {author} {\bibfnamefont {P.~L.}\
  \bibnamefont {{Richards}}}, \bibinfo {author} {\bibfnamefont {G.~F.}\
  \bibnamefont {{Smoot}}}, \bibinfo {author} {\bibfnamefont {R.}~\bibnamefont
  {{Stompor}}}, \bibinfo {author} {\bibfnamefont {C.~D.}\ \bibnamefont
  {{Winant}}}, \ and\ \bibinfo {author} {\bibfnamefont {J.~H.~P.}\ \bibnamefont
  {{Wu}}},\ }\href {\doibase 10.1086/317323} {\bibfield  {journal} {\bibinfo
  {journal} {\apjl}\ }\textbf {\bibinfo {volume} {545}},\ \bibinfo {pages} {L1}
  (\bibinfo {year} {2000})}\BibitemShut {NoStop}%
\bibitem [{\citenamefont {{Jaffe}}\ \emph {et~al.}(2001)\citenamefont
  {{Jaffe}}, \citenamefont {{Ade}}, \citenamefont {{Balbi}}, \citenamefont
  {{Bock}}, \citenamefont {{Bond}}, \citenamefont {{Borrill}}, \citenamefont
  {{Boscaleri}}, \citenamefont {{Coble}}, \citenamefont {{Crill}},
  \citenamefont {{de Bernardis}}, \citenamefont {{Farese}}, \citenamefont
  {{Ferreira}}, \citenamefont {{Ganga}}, \citenamefont {{Giacometti}},
  \citenamefont {{Hanany}}, \citenamefont {{Hivon}}, \citenamefont {{Hristov}},
  \citenamefont {{Iacoangeli}}, \citenamefont {{Lange}}, \citenamefont {{Lee}},
  \citenamefont {{Martinis}}, \citenamefont {{Masi}}, \citenamefont
  {{Mauskopf}}, \citenamefont {{Melchiorri}}, \citenamefont {{Montroy}},
  \citenamefont {{Netterfield}}, \citenamefont {{Oh}}, \citenamefont
  {{Pascale}}, \citenamefont {{Piacentini}}, \citenamefont {{Pogosyan}},
  \citenamefont {{Prunet}}, \citenamefont {{Rabii}}, \citenamefont {{Rao}},
  \citenamefont {{Richards}}, \citenamefont {{Romeo}}, \citenamefont {{Ruhl}},
  \citenamefont {{Scaramuzzi}}, \citenamefont {{Sforna}}, \citenamefont
  {{Smoot}}, \citenamefont {{Stompor}}, \citenamefont {{Winant}},\ and\
  \citenamefont {{Wu}}}]{jaffe01a}%
  \BibitemOpen
  \bibfield  {author} {\bibinfo {author} {\bibfnamefont {A.~H.}\ \bibnamefont
  {{Jaffe}}}, \bibinfo {author} {\bibfnamefont {P.~A.}\ \bibnamefont {{Ade}}},
  \bibinfo {author} {\bibfnamefont {A.}~\bibnamefont {{Balbi}}}, \bibinfo
  {author} {\bibfnamefont {J.~J.}\ \bibnamefont {{Bock}}}, \bibinfo {author}
  {\bibfnamefont {J.~R.}\ \bibnamefont {{Bond}}}, \bibinfo {author}
  {\bibfnamefont {J.}~\bibnamefont {{Borrill}}}, \bibinfo {author}
  {\bibfnamefont {A.}~\bibnamefont {{Boscaleri}}}, \bibinfo {author}
  {\bibfnamefont {K.}~\bibnamefont {{Coble}}}, \bibinfo {author} {\bibfnamefont
  {B.~P.}\ \bibnamefont {{Crill}}}, \bibinfo {author} {\bibfnamefont
  {P.}~\bibnamefont {{de Bernardis}}}, \bibinfo {author} {\bibfnamefont
  {P.}~\bibnamefont {{Farese}}}, \bibinfo {author} {\bibfnamefont {P.~G.}\
  \bibnamefont {{Ferreira}}}, \bibinfo {author} {\bibfnamefont
  {K.}~\bibnamefont {{Ganga}}}, \bibinfo {author} {\bibfnamefont
  {M.}~\bibnamefont {{Giacometti}}}, \bibinfo {author} {\bibfnamefont
  {S.}~\bibnamefont {{Hanany}}}, \bibinfo {author} {\bibfnamefont
  {E.}~\bibnamefont {{Hivon}}}, \bibinfo {author} {\bibfnamefont {V.~V.}\
  \bibnamefont {{Hristov}}}, \bibinfo {author} {\bibfnamefont {A.}~\bibnamefont
  {{Iacoangeli}}}, \bibinfo {author} {\bibfnamefont {A.~E.}\ \bibnamefont
  {{Lange}}}, \bibinfo {author} {\bibfnamefont {A.~T.}\ \bibnamefont {{Lee}}},
  \bibinfo {author} {\bibfnamefont {L.}~\bibnamefont {{Martinis}}}, \bibinfo
  {author} {\bibfnamefont {S.}~\bibnamefont {{Masi}}}, \bibinfo {author}
  {\bibfnamefont {P.~D.}\ \bibnamefont {{Mauskopf}}}, \bibinfo {author}
  {\bibfnamefont {A.}~\bibnamefont {{Melchiorri}}}, \bibinfo {author}
  {\bibfnamefont {T.}~\bibnamefont {{Montroy}}}, \bibinfo {author}
  {\bibfnamefont {C.~B.}\ \bibnamefont {{Netterfield}}}, \bibinfo {author}
  {\bibfnamefont {S.}~\bibnamefont {{Oh}}}, \bibinfo {author} {\bibfnamefont
  {E.}~\bibnamefont {{Pascale}}}, \bibinfo {author} {\bibfnamefont
  {F.}~\bibnamefont {{Piacentini}}}, \bibinfo {author} {\bibfnamefont
  {D.}~\bibnamefont {{Pogosyan}}}, \bibinfo {author} {\bibfnamefont
  {S.}~\bibnamefont {{Prunet}}}, \bibinfo {author} {\bibfnamefont
  {B.}~\bibnamefont {{Rabii}}}, \bibinfo {author} {\bibfnamefont
  {S.}~\bibnamefont {{Rao}}}, \bibinfo {author} {\bibfnamefont {P.~L.}\
  \bibnamefont {{Richards}}}, \bibinfo {author} {\bibfnamefont
  {G.}~\bibnamefont {{Romeo}}}, \bibinfo {author} {\bibfnamefont {J.~E.}\
  \bibnamefont {{Ruhl}}}, \bibinfo {author} {\bibfnamefont {F.}~\bibnamefont
  {{Scaramuzzi}}}, \bibinfo {author} {\bibfnamefont {D.}~\bibnamefont
  {{Sforna}}}, \bibinfo {author} {\bibfnamefont {G.~F.}\ \bibnamefont
  {{Smoot}}}, \bibinfo {author} {\bibfnamefont {R.}~\bibnamefont {{Stompor}}},
  \bibinfo {author} {\bibfnamefont {C.~D.}\ \bibnamefont {{Winant}}}, \ and\
  \bibinfo {author} {\bibfnamefont {J.~H.}\ \bibnamefont {{Wu}}},\ }\href
  {\doibase 10.1103/PhysRevLett.86.3475} {\bibfield  {journal} {\bibinfo
  {journal} {Physical Review Letters}\ }\textbf {\bibinfo {volume} {86}},\
  \bibinfo {pages} {3475} (\bibinfo {year} {2001})}\BibitemShut {NoStop}%
\bibitem [{\citenamefont {{Netterfield}}\ \emph {et~al.}(2002)\citenamefont
  {{Netterfield}}, \citenamefont {{Ade}}, \citenamefont {{Bock}}, \citenamefont
  {{Bond}}, \citenamefont {{Borrill}}, \citenamefont {{Boscaleri}},
  \citenamefont {{Coble}}, \citenamefont {{Contaldi}}, \citenamefont {{Crill}},
  \citenamefont {{de Bernardis}}, \citenamefont {{Farese}}, \citenamefont
  {{Ganga}}, \citenamefont {{Giacometti}}, \citenamefont {{Hivon}},
  \citenamefont {{Hristov}}, \citenamefont {{Iacoangeli}}, \citenamefont
  {{Jaffe}}, \citenamefont {{Jones}}, \citenamefont {{Lange}}, \citenamefont
  {{Martinis}}, \citenamefont {{Masi}}, \citenamefont {{Mason}}, \citenamefont
  {{Mauskopf}}, \citenamefont {{Melchiorri}}, \citenamefont {{Montroy}},
  \citenamefont {{Pascale}}, \citenamefont {{Piacentini}}, \citenamefont
  {{Pogosyan}}, \citenamefont {{Pongetti}}, \citenamefont {{Prunet}},
  \citenamefont {{Romeo}}, \citenamefont {{Ruhl}},\ and\ \citenamefont
  {{Scaramuzzi}}}]{netterfield02}%
  \BibitemOpen
  \bibfield  {author} {\bibinfo {author} {\bibfnamefont {C.~B.}\ \bibnamefont
  {{Netterfield}}}, \bibinfo {author} {\bibfnamefont {P.~A.~R.}\ \bibnamefont
  {{Ade}}}, \bibinfo {author} {\bibfnamefont {J.~J.}\ \bibnamefont {{Bock}}},
  \bibinfo {author} {\bibfnamefont {J.~R.}\ \bibnamefont {{Bond}}}, \bibinfo
  {author} {\bibfnamefont {J.}~\bibnamefont {{Borrill}}}, \bibinfo {author}
  {\bibfnamefont {A.}~\bibnamefont {{Boscaleri}}}, \bibinfo {author}
  {\bibfnamefont {K.}~\bibnamefont {{Coble}}}, \bibinfo {author} {\bibfnamefont
  {C.~R.}\ \bibnamefont {{Contaldi}}}, \bibinfo {author} {\bibfnamefont
  {B.~P.}\ \bibnamefont {{Crill}}}, \bibinfo {author} {\bibfnamefont
  {P.}~\bibnamefont {{de Bernardis}}}, \bibinfo {author} {\bibfnamefont
  {P.}~\bibnamefont {{Farese}}}, \bibinfo {author} {\bibfnamefont
  {K.}~\bibnamefont {{Ganga}}}, \bibinfo {author} {\bibfnamefont
  {M.}~\bibnamefont {{Giacometti}}}, \bibinfo {author} {\bibfnamefont
  {E.}~\bibnamefont {{Hivon}}}, \bibinfo {author} {\bibfnamefont {V.~V.}\
  \bibnamefont {{Hristov}}}, \bibinfo {author} {\bibfnamefont {A.}~\bibnamefont
  {{Iacoangeli}}}, \bibinfo {author} {\bibfnamefont {A.~H.}\ \bibnamefont
  {{Jaffe}}}, \bibinfo {author} {\bibfnamefont {W.~C.}\ \bibnamefont
  {{Jones}}}, \bibinfo {author} {\bibfnamefont {A.~E.}\ \bibnamefont
  {{Lange}}}, \bibinfo {author} {\bibfnamefont {L.}~\bibnamefont {{Martinis}}},
  \bibinfo {author} {\bibfnamefont {S.}~\bibnamefont {{Masi}}}, \bibinfo
  {author} {\bibfnamefont {P.}~\bibnamefont {{Mason}}}, \bibinfo {author}
  {\bibfnamefont {P.~D.}\ \bibnamefont {{Mauskopf}}}, \bibinfo {author}
  {\bibfnamefont {A.}~\bibnamefont {{Melchiorri}}}, \bibinfo {author}
  {\bibfnamefont {T.}~\bibnamefont {{Montroy}}}, \bibinfo {author}
  {\bibfnamefont {E.}~\bibnamefont {{Pascale}}}, \bibinfo {author}
  {\bibfnamefont {F.}~\bibnamefont {{Piacentini}}}, \bibinfo {author}
  {\bibfnamefont {D.}~\bibnamefont {{Pogosyan}}}, \bibinfo {author}
  {\bibfnamefont {F.}~\bibnamefont {{Pongetti}}}, \bibinfo {author}
  {\bibfnamefont {S.}~\bibnamefont {{Prunet}}}, \bibinfo {author}
  {\bibfnamefont {G.}~\bibnamefont {{Romeo}}}, \bibinfo {author} {\bibfnamefont
  {J.~E.}\ \bibnamefont {{Ruhl}}}, \ and\ \bibinfo {author} {\bibfnamefont
  {F.}~\bibnamefont {{Scaramuzzi}}},\ }\href
  {http://adsabs.harvard.edu/cgi-bin/nph-bib_query?bibcode=2002ApJ...571..604N&db_key=AST}
  {\bibfield  {journal} {\bibinfo  {journal} {\apj}\ }\textbf {\bibinfo
  {volume} {571}},\ \bibinfo {pages} {604} (\bibinfo {year}
  {2002})}\BibitemShut {NoStop}%
\bibitem [{\citenamefont {{Hanany}}\ \emph {et~al.}(2000)\citenamefont
  {{Hanany}}, \citenamefont {{Ade}}, \citenamefont {{Balbi}}, \citenamefont
  {{Bock}}, \citenamefont {{Borrill}}, \citenamefont {{Boscaleri}},
  \citenamefont {{de Bernardis}}, \citenamefont {{Ferreira}}, \citenamefont
  {{Hristov}}, \citenamefont {{Jaffe}}, \citenamefont {{Lange}}, \citenamefont
  {{Lee}}, \citenamefont {{Mauskopf}}, \citenamefont {{Netterfield}},
  \citenamefont {{Oh}}, \citenamefont {{Pascale}}, \citenamefont {{Rabii}},
  \citenamefont {{Richards}}, \citenamefont {{Smoot}}, \citenamefont
  {{Stompor}}, \citenamefont {{Winant}},\ and\ \citenamefont
  {{Wu}}}]{hanany00a}%
  \BibitemOpen
  \bibfield  {author} {\bibinfo {author} {\bibfnamefont {S.}~\bibnamefont
  {{Hanany}}}, \bibinfo {author} {\bibfnamefont {P.}~\bibnamefont {{Ade}}},
  \bibinfo {author} {\bibfnamefont {A.}~\bibnamefont {{Balbi}}}, \bibinfo
  {author} {\bibfnamefont {J.}~\bibnamefont {{Bock}}}, \bibinfo {author}
  {\bibfnamefont {J.}~\bibnamefont {{Borrill}}}, \bibinfo {author}
  {\bibfnamefont {A.}~\bibnamefont {{Boscaleri}}}, \bibinfo {author}
  {\bibfnamefont {P.}~\bibnamefont {{de Bernardis}}}, \bibinfo {author}
  {\bibfnamefont {P.~G.}\ \bibnamefont {{Ferreira}}}, \bibinfo {author}
  {\bibfnamefont {V.~V.}\ \bibnamefont {{Hristov}}}, \bibinfo {author}
  {\bibfnamefont {A.~H.}\ \bibnamefont {{Jaffe}}}, \bibinfo {author}
  {\bibfnamefont {A.~E.}\ \bibnamefont {{Lange}}}, \bibinfo {author}
  {\bibfnamefont {A.~T.}\ \bibnamefont {{Lee}}}, \bibinfo {author}
  {\bibfnamefont {P.~D.}\ \bibnamefont {{Mauskopf}}}, \bibinfo {author}
  {\bibfnamefont {C.~B.}\ \bibnamefont {{Netterfield}}}, \bibinfo {author}
  {\bibfnamefont {S.}~\bibnamefont {{Oh}}}, \bibinfo {author} {\bibfnamefont
  {E.}~\bibnamefont {{Pascale}}}, \bibinfo {author} {\bibfnamefont
  {B.}~\bibnamefont {{Rabii}}}, \bibinfo {author} {\bibfnamefont {P.~L.}\
  \bibnamefont {{Richards}}}, \bibinfo {author} {\bibfnamefont {G.~F.}\
  \bibnamefont {{Smoot}}}, \bibinfo {author} {\bibfnamefont {R.}~\bibnamefont
  {{Stompor}}}, \bibinfo {author} {\bibfnamefont {C.~D.}\ \bibnamefont
  {{Winant}}}, \ and\ \bibinfo {author} {\bibfnamefont {J.~H.~P.}\ \bibnamefont
  {{Wu}}},\ }\href {\doibase 10.1086/317322} {\bibfield  {journal} {\bibinfo
  {journal} {\apjl}\ }\textbf {\bibinfo {volume} {545}},\ \bibinfo {pages} {L5}
  (\bibinfo {year} {2000})},\ \Eprint {http://arxiv.org/abs/astro-ph/0005123}
  {astro-ph/0005123} \BibitemShut {NoStop}%
\bibitem [{\citenamefont {{Efstathiou}}\ \emph {et~al.}(1990)\citenamefont
  {{Efstathiou}}, \citenamefont {{Sutherland}},\ and\ \citenamefont
  {{Maddox}}}]{Efstathiou90}%
  \BibitemOpen
  \bibfield  {author} {\bibinfo {author} {\bibfnamefont {G.}~\bibnamefont
  {{Efstathiou}}}, \bibinfo {author} {\bibfnamefont {W.~J.}\ \bibnamefont
  {{Sutherland}}}, \ and\ \bibinfo {author} {\bibfnamefont {S.~J.}\
  \bibnamefont {{Maddox}}},\ }\href {\doibase 10.1038/348705a0} {\bibfield
  {journal} {\bibinfo  {journal} {\nat}\ }\textbf {\bibinfo {volume} {348}},\
  \bibinfo {pages} {705} (\bibinfo {year} {1990})}\BibitemShut {NoStop}%
\bibitem [{\citenamefont {{Ostriker}}\ and\ \citenamefont
  {{Steinhardt}}(1995)}]{Ostriker95}%
  \BibitemOpen
  \bibfield  {author} {\bibinfo {author} {\bibfnamefont {J.~P.}\ \bibnamefont
  {{Ostriker}}}\ and\ \bibinfo {author} {\bibfnamefont {P.~J.}\ \bibnamefont
  {{Steinhardt}}},\ }\href {\doibase 10.1038/377600a0} {\bibfield  {journal}
  {\bibinfo  {journal} {\nat}\ }\textbf {\bibinfo {volume} {377}},\ \bibinfo
  {pages} {600} (\bibinfo {year} {1995})}\BibitemShut {NoStop}%
\bibitem [{\citenamefont {{Colless}}\ \emph {et~al.}(2001)\citenamefont
  {{Colless}}, \citenamefont {{Dalton}}, \citenamefont {{Maddox}},
  \citenamefont {{Sutherland}}, \citenamefont {{Norberg}}, \citenamefont
  {{Cole}}, \citenamefont {{Bland-Hawthorn}}, \citenamefont {{Bridges}},
  \citenamefont {{Cannon}}, \citenamefont {{Collins}}, \citenamefont {{Couch}},
  \citenamefont {{Cross}}, \citenamefont {{Deeley}}, \citenamefont {{De
  Propris}}, \citenamefont {{Driver}}, \citenamefont {{Efstathiou}},
  \citenamefont {{Ellis}}, \citenamefont {{Frenk}}, \citenamefont
  {{Glazebrook}}, \citenamefont {{Jackson}}, \citenamefont {{Lahav}},
  \citenamefont {{Lewis}}, \citenamefont {{Lumsden}}, \citenamefont
  {{Madgwick}}, \citenamefont {{Peacock}}, \citenamefont {{Peterson}},
  \citenamefont {{Price}}, \citenamefont {{Seaborne}},\ and\ \citenamefont
  {{Taylor}}}]{colless01a}%
  \BibitemOpen
  \bibfield  {author} {\bibinfo {author} {\bibfnamefont {M.}~\bibnamefont
  {{Colless}}}, \bibinfo {author} {\bibfnamefont {G.}~\bibnamefont {{Dalton}}},
  \bibinfo {author} {\bibfnamefont {S.}~\bibnamefont {{Maddox}}}, \bibinfo
  {author} {\bibfnamefont {W.}~\bibnamefont {{Sutherland}}}, \bibinfo {author}
  {\bibfnamefont {P.}~\bibnamefont {{Norberg}}}, \bibinfo {author}
  {\bibfnamefont {S.}~\bibnamefont {{Cole}}}, \bibinfo {author} {\bibfnamefont
  {J.}~\bibnamefont {{Bland-Hawthorn}}}, \bibinfo {author} {\bibfnamefont
  {T.}~\bibnamefont {{Bridges}}}, \bibinfo {author} {\bibfnamefont
  {R.}~\bibnamefont {{Cannon}}}, \bibinfo {author} {\bibfnamefont
  {C.}~\bibnamefont {{Collins}}}, \bibinfo {author} {\bibfnamefont
  {W.}~\bibnamefont {{Couch}}}, \bibinfo {author} {\bibfnamefont
  {N.}~\bibnamefont {{Cross}}}, \bibinfo {author} {\bibfnamefont
  {K.}~\bibnamefont {{Deeley}}}, \bibinfo {author} {\bibfnamefont
  {R.}~\bibnamefont {{De Propris}}}, \bibinfo {author} {\bibfnamefont {S.~P.}\
  \bibnamefont {{Driver}}}, \bibinfo {author} {\bibfnamefont {G.}~\bibnamefont
  {{Efstathiou}}}, \bibinfo {author} {\bibfnamefont {R.~S.}\ \bibnamefont
  {{Ellis}}}, \bibinfo {author} {\bibfnamefont {C.~S.}\ \bibnamefont
  {{Frenk}}}, \bibinfo {author} {\bibfnamefont {K.}~\bibnamefont
  {{Glazebrook}}}, \bibinfo {author} {\bibfnamefont {C.}~\bibnamefont
  {{Jackson}}}, \bibinfo {author} {\bibfnamefont {O.}~\bibnamefont {{Lahav}}},
  \bibinfo {author} {\bibfnamefont {I.}~\bibnamefont {{Lewis}}}, \bibinfo
  {author} {\bibfnamefont {S.}~\bibnamefont {{Lumsden}}}, \bibinfo {author}
  {\bibfnamefont {D.}~\bibnamefont {{Madgwick}}}, \bibinfo {author}
  {\bibfnamefont {J.~A.}\ \bibnamefont {{Peacock}}}, \bibinfo {author}
  {\bibfnamefont {B.~A.}\ \bibnamefont {{Peterson}}}, \bibinfo {author}
  {\bibfnamefont {I.}~\bibnamefont {{Price}}}, \bibinfo {author} {\bibfnamefont
  {M.}~\bibnamefont {{Seaborne}}}, \ and\ \bibinfo {author} {\bibfnamefont
  {K.}~\bibnamefont {{Taylor}}},\ }\href {\doibase
  10.1046/j.1365-8711.2001.04902.x} {\bibfield  {journal} {\bibinfo  {journal}
  {\mnras}\ }\textbf {\bibinfo {volume} {328}},\ \bibinfo {pages} {1039}
  (\bibinfo {year} {2001})},\ \Eprint
  {http://arxiv.org/abs/arXiv:astro-ph/0106498} {arXiv:astro-ph/0106498}
  \BibitemShut {NoStop}%
\bibitem [{\citenamefont {{Percival}}\ \emph {et~al.}(2001)\citenamefont
  {{Percival}}, \citenamefont {{Baugh}}, \citenamefont {{Bland-Hawthorn}},
  \citenamefont {{Bridges}}, \citenamefont {{Cannon}}, \citenamefont {{Cole}},
  \citenamefont {{Colless}}, \citenamefont {{Collins}}, \citenamefont
  {{Couch}}, \citenamefont {{Dalton}}, \citenamefont {{De Propris}},
  \citenamefont {{Driver}}, \citenamefont {{Efstathiou}}, \citenamefont
  {{Ellis}}, \citenamefont {{Frenk}}, \citenamefont {{Glazebrook}},
  \citenamefont {{Jackson}}, \citenamefont {{Lahav}}, \citenamefont {{Lewis}},
  \citenamefont {{Lumsden}}, \citenamefont {{Maddox}}, \citenamefont {{Moody}},
  \citenamefont {{Norberg}}, \citenamefont {{Peacock}}, \citenamefont
  {{Peterson}}, \citenamefont {{Sutherland}},\ and\ \citenamefont
  {{Taylor}}}]{percival01a}%
  \BibitemOpen
  \bibfield  {author} {\bibinfo {author} {\bibfnamefont {W.~J.}\ \bibnamefont
  {{Percival}}}, \bibinfo {author} {\bibfnamefont {C.~M.}\ \bibnamefont
  {{Baugh}}}, \bibinfo {author} {\bibfnamefont {J.}~\bibnamefont
  {{Bland-Hawthorn}}}, \bibinfo {author} {\bibfnamefont {T.}~\bibnamefont
  {{Bridges}}}, \bibinfo {author} {\bibfnamefont {R.}~\bibnamefont {{Cannon}}},
  \bibinfo {author} {\bibfnamefont {S.}~\bibnamefont {{Cole}}}, \bibinfo
  {author} {\bibfnamefont {M.}~\bibnamefont {{Colless}}}, \bibinfo {author}
  {\bibfnamefont {C.}~\bibnamefont {{Collins}}}, \bibinfo {author}
  {\bibfnamefont {W.}~\bibnamefont {{Couch}}}, \bibinfo {author} {\bibfnamefont
  {G.}~\bibnamefont {{Dalton}}}, \bibinfo {author} {\bibfnamefont
  {R.}~\bibnamefont {{De Propris}}}, \bibinfo {author} {\bibfnamefont {S.~P.}\
  \bibnamefont {{Driver}}}, \bibinfo {author} {\bibfnamefont {G.}~\bibnamefont
  {{Efstathiou}}}, \bibinfo {author} {\bibfnamefont {R.~S.}\ \bibnamefont
  {{Ellis}}}, \bibinfo {author} {\bibfnamefont {C.~S.}\ \bibnamefont
  {{Frenk}}}, \bibinfo {author} {\bibfnamefont {K.}~\bibnamefont
  {{Glazebrook}}}, \bibinfo {author} {\bibfnamefont {C.}~\bibnamefont
  {{Jackson}}}, \bibinfo {author} {\bibfnamefont {O.}~\bibnamefont {{Lahav}}},
  \bibinfo {author} {\bibfnamefont {I.}~\bibnamefont {{Lewis}}}, \bibinfo
  {author} {\bibfnamefont {S.}~\bibnamefont {{Lumsden}}}, \bibinfo {author}
  {\bibfnamefont {S.}~\bibnamefont {{Maddox}}}, \bibinfo {author}
  {\bibfnamefont {S.}~\bibnamefont {{Moody}}}, \bibinfo {author} {\bibfnamefont
  {P.}~\bibnamefont {{Norberg}}}, \bibinfo {author} {\bibfnamefont {J.~A.}\
  \bibnamefont {{Peacock}}}, \bibinfo {author} {\bibfnamefont {B.~A.}\
  \bibnamefont {{Peterson}}}, \bibinfo {author} {\bibfnamefont
  {W.}~\bibnamefont {{Sutherland}}}, \ and\ \bibinfo {author} {\bibfnamefont
  {K.}~\bibnamefont {{Taylor}}},\ }\href {\doibase
  10.1046/j.1365-8711.2001.04827.x} {\bibfield  {journal} {\bibinfo  {journal}
  {\mnras}\ }\textbf {\bibinfo {volume} {327}},\ \bibinfo {pages} {1297}
  (\bibinfo {year} {2001})},\ \Eprint
  {http://arxiv.org/abs/arXiv:astro-ph/0105252} {arXiv:astro-ph/0105252}
  \BibitemShut {NoStop}%
\bibitem [{\citenamefont {{Percival}}\ \emph {et~al.}(2002)\citenamefont
  {{Percival}}, \citenamefont {{Sutherland}}, \citenamefont {{Peacock}},
  \citenamefont {{Baugh}}, \citenamefont {{Bland-Hawthorn}}, \citenamefont
  {{Bridges}}, \citenamefont {{Cannon}}, \citenamefont {{Cole}}, \citenamefont
  {{Colless}}, \citenamefont {{Collins}}, \citenamefont {{Couch}},
  \citenamefont {{Dalton}}, \citenamefont {{De Propris}}, \citenamefont
  {{Driver}}, \citenamefont {{Efstathiou}}, \citenamefont {{Ellis}},
  \citenamefont {{Frenk}}, \citenamefont {{Glazebrook}}, \citenamefont
  {{Jackson}}, \citenamefont {{Lahav}}, \citenamefont {{Lewis}}, \citenamefont
  {{Lumsden}}, \citenamefont {{Maddox}}, \citenamefont {{Moody}}, \citenamefont
  {{Norberg}}, \citenamefont {{Peterson}},\ and\ \citenamefont
  {{Taylor}}}]{percival02a}%
  \BibitemOpen
  \bibfield  {author} {\bibinfo {author} {\bibfnamefont {W.~J.}\ \bibnamefont
  {{Percival}}}, \bibinfo {author} {\bibfnamefont {W.}~\bibnamefont
  {{Sutherland}}}, \bibinfo {author} {\bibfnamefont {J.~A.}\ \bibnamefont
  {{Peacock}}}, \bibinfo {author} {\bibfnamefont {C.~M.}\ \bibnamefont
  {{Baugh}}}, \bibinfo {author} {\bibfnamefont {J.}~\bibnamefont
  {{Bland-Hawthorn}}}, \bibinfo {author} {\bibfnamefont {T.}~\bibnamefont
  {{Bridges}}}, \bibinfo {author} {\bibfnamefont {R.}~\bibnamefont {{Cannon}}},
  \bibinfo {author} {\bibfnamefont {S.}~\bibnamefont {{Cole}}}, \bibinfo
  {author} {\bibfnamefont {M.}~\bibnamefont {{Colless}}}, \bibinfo {author}
  {\bibfnamefont {C.}~\bibnamefont {{Collins}}}, \bibinfo {author}
  {\bibfnamefont {W.}~\bibnamefont {{Couch}}}, \bibinfo {author} {\bibfnamefont
  {G.}~\bibnamefont {{Dalton}}}, \bibinfo {author} {\bibfnamefont
  {R.}~\bibnamefont {{De Propris}}}, \bibinfo {author} {\bibfnamefont {S.~P.}\
  \bibnamefont {{Driver}}}, \bibinfo {author} {\bibfnamefont {G.}~\bibnamefont
  {{Efstathiou}}}, \bibinfo {author} {\bibfnamefont {R.~S.}\ \bibnamefont
  {{Ellis}}}, \bibinfo {author} {\bibfnamefont {C.~S.}\ \bibnamefont
  {{Frenk}}}, \bibinfo {author} {\bibfnamefont {K.}~\bibnamefont
  {{Glazebrook}}}, \bibinfo {author} {\bibfnamefont {C.}~\bibnamefont
  {{Jackson}}}, \bibinfo {author} {\bibfnamefont {O.}~\bibnamefont {{Lahav}}},
  \bibinfo {author} {\bibfnamefont {I.}~\bibnamefont {{Lewis}}}, \bibinfo
  {author} {\bibfnamefont {S.}~\bibnamefont {{Lumsden}}}, \bibinfo {author}
  {\bibfnamefont {S.}~\bibnamefont {{Maddox}}}, \bibinfo {author}
  {\bibfnamefont {S.}~\bibnamefont {{Moody}}}, \bibinfo {author} {\bibfnamefont
  {P.}~\bibnamefont {{Norberg}}}, \bibinfo {author} {\bibfnamefont {B.~A.}\
  \bibnamefont {{Peterson}}}, \ and\ \bibinfo {author} {\bibfnamefont
  {K.}~\bibnamefont {{Taylor}}},\ }\href {\doibase
  10.1046/j.1365-8711.2002.06001.x} {\bibfield  {journal} {\bibinfo  {journal}
  {\mnras}\ }\textbf {\bibinfo {volume} {337}},\ \bibinfo {pages} {1068}
  (\bibinfo {year} {2002})},\ \Eprint
  {http://arxiv.org/abs/arXiv:astro-ph/0206256} {arXiv:astro-ph/0206256}
  \BibitemShut {NoStop}%
\bibitem [{\citenamefont {{Fowler}}\ \emph {et~al.}(2007)\citenamefont
  {{Fowler}}, \citenamefont {{Niemack}}, \citenamefont {{Dicker}},
  \citenamefont {{Aboobaker}}, \citenamefont {{Ade}}, \citenamefont
  {{Battistelli}}, \citenamefont {{Devlin}}, \citenamefont {{Fisher}},
  \citenamefont {{Halpern}}, \citenamefont {{Hargrave}}, \citenamefont
  {{Hincks}}, \citenamefont {{Kaul}}, \citenamefont {{Klein}}, \citenamefont
  {{Lau}}, \citenamefont {{Limon}}, \citenamefont {{Marriage}}, \citenamefont
  {{Mauskopf}}, \citenamefont {{Page}}, \citenamefont {{Staggs}}, \citenamefont
  {{Swetz}}, \citenamefont {{Switzer}}, \citenamefont {{Thornton}},\ and\
  \citenamefont {{Tucker}}}]{fowler07a}%
  \BibitemOpen
  \bibfield  {author} {\bibinfo {author} {\bibfnamefont {J.~W.}\ \bibnamefont
  {{Fowler}}}, \bibinfo {author} {\bibfnamefont {M.~D.}\ \bibnamefont
  {{Niemack}}}, \bibinfo {author} {\bibfnamefont {S.~R.}\ \bibnamefont
  {{Dicker}}}, \bibinfo {author} {\bibfnamefont {A.~M.}\ \bibnamefont
  {{Aboobaker}}}, \bibinfo {author} {\bibfnamefont {P.~A.~R.}\ \bibnamefont
  {{Ade}}}, \bibinfo {author} {\bibfnamefont {E.~S.}\ \bibnamefont
  {{Battistelli}}}, \bibinfo {author} {\bibfnamefont {M.~J.}\ \bibnamefont
  {{Devlin}}}, \bibinfo {author} {\bibfnamefont {R.~P.}\ \bibnamefont
  {{Fisher}}}, \bibinfo {author} {\bibfnamefont {M.}~\bibnamefont {{Halpern}}},
  \bibinfo {author} {\bibfnamefont {P.~C.}\ \bibnamefont {{Hargrave}}},
  \bibinfo {author} {\bibfnamefont {A.~D.}\ \bibnamefont {{Hincks}}}, \bibinfo
  {author} {\bibfnamefont {M.}~\bibnamefont {{Kaul}}}, \bibinfo {author}
  {\bibfnamefont {J.}~\bibnamefont {{Klein}}}, \bibinfo {author} {\bibfnamefont
  {J.~M.}\ \bibnamefont {{Lau}}}, \bibinfo {author} {\bibfnamefont
  {M.}~\bibnamefont {{Limon}}}, \bibinfo {author} {\bibfnamefont {T.~A.}\
  \bibnamefont {{Marriage}}}, \bibinfo {author} {\bibfnamefont {P.~D.}\
  \bibnamefont {{Mauskopf}}}, \bibinfo {author} {\bibfnamefont
  {L.}~\bibnamefont {{Page}}}, \bibinfo {author} {\bibfnamefont {S.~T.}\
  \bibnamefont {{Staggs}}}, \bibinfo {author} {\bibfnamefont {D.~S.}\
  \bibnamefont {{Swetz}}}, \bibinfo {author} {\bibfnamefont {E.~R.}\
  \bibnamefont {{Switzer}}}, \bibinfo {author} {\bibfnamefont {R.~J.}\
  \bibnamefont {{Thornton}}}, \ and\ \bibinfo {author} {\bibfnamefont {C.~E.}\
  \bibnamefont {{Tucker}}},\ }\href {\doibase 10.1364/AO.46.003444} {\bibfield
  {journal} {\bibinfo  {journal} {\ao}\ }\textbf {\bibinfo {volume} {46}},\
  \bibinfo {pages} {3444} (\bibinfo {year} {2007})},\ \Eprint
  {http://arxiv.org/abs/astro-ph/0701020} {astro-ph/0701020} \BibitemShut
  {NoStop}%
\bibitem [{\citenamefont {{Carlstrom}}\ \emph {et~al.}(2011)\citenamefont
  {{Carlstrom}}, \citenamefont {{Ade}}, \citenamefont {{Aird}}, \citenamefont
  {{Benson}}, \citenamefont {{Bleem}}, \citenamefont {{Busetti}}, \citenamefont
  {{Chang}}, \citenamefont {{Chauvin}}, \citenamefont {{Cho}}, \citenamefont
  {{Crawford}}, \citenamefont {{Crites}}, \citenamefont {{Dobbs}},
  \citenamefont {{Halverson}}, \citenamefont {{Heimsath}}, \citenamefont
  {{Holzapfel}}, \citenamefont {{Hrubes}}, \citenamefont {{Joy}}, \citenamefont
  {{Keisler}}, \citenamefont {{Lanting}}, \citenamefont {{Lee}}, \citenamefont
  {{Leitch}}, \citenamefont {{Leong}}, \citenamefont {{Lu}}, \citenamefont
  {{Lueker}}, \citenamefont {{Luong-Van}}, \citenamefont {{McMahon}},
  \citenamefont {{Mehl}}, \citenamefont {{Meyer}}, \citenamefont {{Mohr}},
  \citenamefont {{Montroy}}, \citenamefont {{Padin}}, \citenamefont {{Plagge}},
  \citenamefont {{Pryke}}, \citenamefont {{Ruhl}}, \citenamefont {{Schaffer}},
  \citenamefont {{Schwan}}, \citenamefont {{Shirokoff}}, \citenamefont
  {{Spieler}}, \citenamefont {{Staniszewski}}, \citenamefont {{Stark}},
  \citenamefont {{Tucker}}, \citenamefont {{Vanderlinde}}, \citenamefont
  {{Vieira}},\ and\ \citenamefont {{Williamson}}}]{carlstrom11a}%
  \BibitemOpen
  \bibfield  {author} {\bibinfo {author} {\bibfnamefont {J.~E.}\ \bibnamefont
  {{Carlstrom}}}, \bibinfo {author} {\bibfnamefont {P.~A.~R.}\ \bibnamefont
  {{Ade}}}, \bibinfo {author} {\bibfnamefont {K.~A.}\ \bibnamefont {{Aird}}},
  \bibinfo {author} {\bibfnamefont {B.~A.}\ \bibnamefont {{Benson}}}, \bibinfo
  {author} {\bibfnamefont {L.~E.}\ \bibnamefont {{Bleem}}}, \bibinfo {author}
  {\bibfnamefont {S.}~\bibnamefont {{Busetti}}}, \bibinfo {author}
  {\bibfnamefont {C.~L.}\ \bibnamefont {{Chang}}}, \bibinfo {author}
  {\bibfnamefont {E.}~\bibnamefont {{Chauvin}}}, \bibinfo {author}
  {\bibfnamefont {H.-M.}\ \bibnamefont {{Cho}}}, \bibinfo {author}
  {\bibfnamefont {T.~M.}\ \bibnamefont {{Crawford}}}, \bibinfo {author}
  {\bibfnamefont {A.~T.}\ \bibnamefont {{Crites}}}, \bibinfo {author}
  {\bibfnamefont {M.~A.}\ \bibnamefont {{Dobbs}}}, \bibinfo {author}
  {\bibfnamefont {N.~W.}\ \bibnamefont {{Halverson}}}, \bibinfo {author}
  {\bibfnamefont {S.}~\bibnamefont {{Heimsath}}}, \bibinfo {author}
  {\bibfnamefont {W.~L.}\ \bibnamefont {{Holzapfel}}}, \bibinfo {author}
  {\bibfnamefont {J.~D.}\ \bibnamefont {{Hrubes}}}, \bibinfo {author}
  {\bibfnamefont {M.}~\bibnamefont {{Joy}}}, \bibinfo {author} {\bibfnamefont
  {R.}~\bibnamefont {{Keisler}}}, \bibinfo {author} {\bibfnamefont {T.~M.}\
  \bibnamefont {{Lanting}}}, \bibinfo {author} {\bibfnamefont {A.~T.}\
  \bibnamefont {{Lee}}}, \bibinfo {author} {\bibfnamefont {E.~M.}\ \bibnamefont
  {{Leitch}}}, \bibinfo {author} {\bibfnamefont {J.}~\bibnamefont {{Leong}}},
  \bibinfo {author} {\bibfnamefont {W.}~\bibnamefont {{Lu}}}, \bibinfo {author}
  {\bibfnamefont {M.}~\bibnamefont {{Lueker}}}, \bibinfo {author}
  {\bibfnamefont {D.}~\bibnamefont {{Luong-Van}}}, \bibinfo {author}
  {\bibfnamefont {J.~J.}\ \bibnamefont {{McMahon}}}, \bibinfo {author}
  {\bibfnamefont {J.}~\bibnamefont {{Mehl}}}, \bibinfo {author} {\bibfnamefont
  {S.~S.}\ \bibnamefont {{Meyer}}}, \bibinfo {author} {\bibfnamefont {J.~J.}\
  \bibnamefont {{Mohr}}}, \bibinfo {author} {\bibfnamefont {T.~E.}\
  \bibnamefont {{Montroy}}}, \bibinfo {author} {\bibfnamefont {S.}~\bibnamefont
  {{Padin}}}, \bibinfo {author} {\bibfnamefont {T.}~\bibnamefont {{Plagge}}},
  \bibinfo {author} {\bibfnamefont {C.}~\bibnamefont {{Pryke}}}, \bibinfo
  {author} {\bibfnamefont {J.~E.}\ \bibnamefont {{Ruhl}}}, \bibinfo {author}
  {\bibfnamefont {K.~K.}\ \bibnamefont {{Schaffer}}}, \bibinfo {author}
  {\bibfnamefont {D.}~\bibnamefont {{Schwan}}}, \bibinfo {author}
  {\bibfnamefont {E.}~\bibnamefont {{Shirokoff}}}, \bibinfo {author}
  {\bibfnamefont {H.~G.}\ \bibnamefont {{Spieler}}}, \bibinfo {author}
  {\bibfnamefont {Z.}~\bibnamefont {{Staniszewski}}}, \bibinfo {author}
  {\bibfnamefont {A.~A.}\ \bibnamefont {{Stark}}}, \bibinfo {author}
  {\bibfnamefont {C.}~\bibnamefont {{Tucker}}}, \bibinfo {author}
  {\bibfnamefont {K.}~\bibnamefont {{Vanderlinde}}}, \bibinfo {author}
  {\bibfnamefont {J.~D.}\ \bibnamefont {{Vieira}}}, \ and\ \bibinfo {author}
  {\bibfnamefont {R.}~\bibnamefont {{Williamson}}},\ }\href {\doibase
  10.1086/659879} {\bibfield  {journal} {\bibinfo  {journal} {\pasp}\ }\textbf
  {\bibinfo {volume} {123}},\ \bibinfo {pages} {568} (\bibinfo {year}
  {2011})},\ \Eprint {http://arxiv.org/abs/0907.4445} {arXiv:0907.4445
  [astro-ph.IM]} \BibitemShut {NoStop}%
\bibitem [{\citenamefont {{Bennett}}\ \emph {et~al.}(2013)\citenamefont
  {{Bennett}}, \citenamefont {{Larson}}, \citenamefont {{Weiland}},
  \citenamefont {{Jarosik}}, \citenamefont {{Hinshaw}}, \citenamefont
  {{Odegard}}, \citenamefont {{Smith}}, \citenamefont {{Hill}}, \citenamefont
  {{Gold}}, \citenamefont {{Halpern}}, \citenamefont {{Komatsu}}, \citenamefont
  {{Nolta}}, \citenamefont {{Page}}, \citenamefont {{Spergel}}, \citenamefont
  {{Wollack}}, \citenamefont {{Dunkley}}, \citenamefont {{Kogut}},
  \citenamefont {{Limon}}, \citenamefont {{Meyer}}, \citenamefont {{Tucker}},\
  and\ \citenamefont {{Wright}}}]{bennett13a}%
  \BibitemOpen
  \bibfield  {author} {\bibinfo {author} {\bibfnamefont {C.~L.}\ \bibnamefont
  {{Bennett}}}, \bibinfo {author} {\bibfnamefont {D.}~\bibnamefont {{Larson}}},
  \bibinfo {author} {\bibfnamefont {J.~L.}\ \bibnamefont {{Weiland}}}, \bibinfo
  {author} {\bibfnamefont {N.}~\bibnamefont {{Jarosik}}}, \bibinfo {author}
  {\bibfnamefont {G.}~\bibnamefont {{Hinshaw}}}, \bibinfo {author}
  {\bibfnamefont {N.}~\bibnamefont {{Odegard}}}, \bibinfo {author}
  {\bibfnamefont {K.~M.}\ \bibnamefont {{Smith}}}, \bibinfo {author}
  {\bibfnamefont {R.~S.}\ \bibnamefont {{Hill}}}, \bibinfo {author}
  {\bibfnamefont {B.}~\bibnamefont {{Gold}}}, \bibinfo {author} {\bibfnamefont
  {M.}~\bibnamefont {{Halpern}}}, \bibinfo {author} {\bibfnamefont
  {E.}~\bibnamefont {{Komatsu}}}, \bibinfo {author} {\bibfnamefont {M.~R.}\
  \bibnamefont {{Nolta}}}, \bibinfo {author} {\bibfnamefont {L.}~\bibnamefont
  {{Page}}}, \bibinfo {author} {\bibfnamefont {D.~N.}\ \bibnamefont
  {{Spergel}}}, \bibinfo {author} {\bibfnamefont {E.}~\bibnamefont
  {{Wollack}}}, \bibinfo {author} {\bibfnamefont {J.}~\bibnamefont
  {{Dunkley}}}, \bibinfo {author} {\bibfnamefont {A.}~\bibnamefont {{Kogut}}},
  \bibinfo {author} {\bibfnamefont {M.}~\bibnamefont {{Limon}}}, \bibinfo
  {author} {\bibfnamefont {S.~S.}\ \bibnamefont {{Meyer}}}, \bibinfo {author}
  {\bibfnamefont {G.~S.}\ \bibnamefont {{Tucker}}}, \ and\ \bibinfo {author}
  {\bibfnamefont {E.~L.}\ \bibnamefont {{Wright}}},\ }\href {\doibase
  10.1088/0067-0049/208/2/20} {\bibfield  {journal} {\bibinfo  {journal}
  {\apjs}\ }\textbf {\bibinfo {volume} {208}},\ \bibinfo {eid} {20} (\bibinfo
  {year} {2013})},\ \Eprint {http://arxiv.org/abs/1212.5225} {arXiv:1212.5225}
  \BibitemShut {NoStop}%
\bibitem [{\citenamefont {{Jha}}\ \emph {et~al.}(2006)\citenamefont {{Jha}},
  \citenamefont {{Kirshner}}, \citenamefont {{Challis}}, \citenamefont
  {{Garnavich}}, \citenamefont {{Matheson}}, \citenamefont {{Soderberg}},
  \citenamefont {{Graves}}, \citenamefont {{Hicken}}, \citenamefont {{Alves}},
  \citenamefont {{Arce}}, \citenamefont {{Balog}}, \citenamefont {{Barmby}},
  \citenamefont {{Barton}}, \citenamefont {{Berlind}}, \citenamefont {{Bragg}},
  \citenamefont {{Brice{\~n}o}}, \citenamefont {{Brown}}, \citenamefont
  {{Buckley}}, \citenamefont {{Caldwell}}, \citenamefont {{Calkins}},
  \citenamefont {{Carter}}, \citenamefont {{Concannon}}, \citenamefont
  {{Donnelly}}, \citenamefont {{Eriksen}}, \citenamefont {{Fabricant}},
  \citenamefont {{Falco}}, \citenamefont {{Fiore}}, \citenamefont {{Garcia}},
  \citenamefont {{G{\'o}mez}}, \citenamefont {{Grogin}}, \citenamefont
  {{Groner}}, \citenamefont {{Groot}}, \citenamefont {{Haisch}}, \citenamefont
  {{Hartmann}}, \citenamefont {{Hergenrother}}, \citenamefont {{Holman}},
  \citenamefont {{Huchra}}, \citenamefont {{Jayawardhana}}, \citenamefont
  {{Jerius}}, \citenamefont {{Kannappan}}, \citenamefont {{Kim}}, \citenamefont
  {{Kleyna}}, \citenamefont {{Kochanek}}, \citenamefont {{Koranyi}},
  \citenamefont {{Krockenberger}}, \citenamefont {{Lada}}, \citenamefont
  {{Luhman}}, \citenamefont {{Luu}}, \citenamefont {{Macri}}, \citenamefont
  {{Mader}}, \citenamefont {{Mahdavi}}, \citenamefont {{Marengo}},
  \citenamefont {{Marsden}}, \citenamefont {{McLeod}}, \citenamefont
  {{McNamara}}, \citenamefont {{Megeath}}, \citenamefont {{Moraru}},
  \citenamefont {{Mossman}}, \citenamefont {{Muench}}, \citenamefont
  {{Mu{\~n}oz}}, \citenamefont {{Muzerolle}}, \citenamefont {{Naranjo}},
  \citenamefont {{Nelson-Patel}}, \citenamefont {{Pahre}}, \citenamefont
  {{Patten}}, \citenamefont {{Peters}}, \citenamefont {{Peters}}, \citenamefont
  {{Raymond}}, \citenamefont {{Rines}}, \citenamefont {{Schild}}, \citenamefont
  {{Sobczak}}, \citenamefont {{Spahr}}, \citenamefont {{Stauffer}},
  \citenamefont {{Stefanik}}, \citenamefont {{Szentgyorgyi}}, \citenamefont
  {{Tollestrup}}, \citenamefont {{V{\"a}is{\"a}nen}}, \citenamefont
  {{Vikhlinin}}, \citenamefont {{Wang}}, \citenamefont {{Willner}},
  \citenamefont {{Wolk}}, \citenamefont {{Zajac}}, \citenamefont {{Zhao}},\
  and\ \citenamefont {{Stanek}}}]{jha06a}%
  \BibitemOpen
  \bibfield  {author} {\bibinfo {author} {\bibfnamefont {S.}~\bibnamefont
  {{Jha}}}, \bibinfo {author} {\bibfnamefont {R.~P.}\ \bibnamefont
  {{Kirshner}}}, \bibinfo {author} {\bibfnamefont {P.}~\bibnamefont
  {{Challis}}}, \bibinfo {author} {\bibfnamefont {P.~M.}\ \bibnamefont
  {{Garnavich}}}, \bibinfo {author} {\bibfnamefont {T.}~\bibnamefont
  {{Matheson}}}, \bibinfo {author} {\bibfnamefont {A.~M.}\ \bibnamefont
  {{Soderberg}}}, \bibinfo {author} {\bibfnamefont {G.~J.~M.}\ \bibnamefont
  {{Graves}}}, \bibinfo {author} {\bibfnamefont {M.}~\bibnamefont {{Hicken}}},
  \bibinfo {author} {\bibfnamefont {J.~F.}\ \bibnamefont {{Alves}}}, \bibinfo
  {author} {\bibfnamefont {H.~G.}\ \bibnamefont {{Arce}}}, \bibinfo {author}
  {\bibfnamefont {Z.}~\bibnamefont {{Balog}}}, \bibinfo {author} {\bibfnamefont
  {P.}~\bibnamefont {{Barmby}}}, \bibinfo {author} {\bibfnamefont {E.~J.}\
  \bibnamefont {{Barton}}}, \bibinfo {author} {\bibfnamefont {P.}~\bibnamefont
  {{Berlind}}}, \bibinfo {author} {\bibfnamefont {A.~E.}\ \bibnamefont
  {{Bragg}}}, \bibinfo {author} {\bibfnamefont {C.}~\bibnamefont
  {{Brice{\~n}o}}}, \bibinfo {author} {\bibfnamefont {W.~R.}\ \bibnamefont
  {{Brown}}}, \bibinfo {author} {\bibfnamefont {J.~H.}\ \bibnamefont
  {{Buckley}}}, \bibinfo {author} {\bibfnamefont {N.}~\bibnamefont
  {{Caldwell}}}, \bibinfo {author} {\bibfnamefont {M.~L.}\ \bibnamefont
  {{Calkins}}}, \bibinfo {author} {\bibfnamefont {B.~J.}\ \bibnamefont
  {{Carter}}}, \bibinfo {author} {\bibfnamefont {K.~D.}\ \bibnamefont
  {{Concannon}}}, \bibinfo {author} {\bibfnamefont {R.~H.}\ \bibnamefont
  {{Donnelly}}}, \bibinfo {author} {\bibfnamefont {K.~A.}\ \bibnamefont
  {{Eriksen}}}, \bibinfo {author} {\bibfnamefont {D.~G.}\ \bibnamefont
  {{Fabricant}}}, \bibinfo {author} {\bibfnamefont {E.~E.}\ \bibnamefont
  {{Falco}}}, \bibinfo {author} {\bibfnamefont {F.}~\bibnamefont {{Fiore}}},
  \bibinfo {author} {\bibfnamefont {M.~R.}\ \bibnamefont {{Garcia}}}, \bibinfo
  {author} {\bibfnamefont {M.}~\bibnamefont {{G{\'o}mez}}}, \bibinfo {author}
  {\bibfnamefont {N.~A.}\ \bibnamefont {{Grogin}}}, \bibinfo {author}
  {\bibfnamefont {T.}~\bibnamefont {{Groner}}}, \bibinfo {author}
  {\bibfnamefont {P.~J.}\ \bibnamefont {{Groot}}}, \bibinfo {author}
  {\bibfnamefont {K.~E.}\ \bibnamefont {{Haisch}}, \bibfnamefont {Jr.}},
  \bibinfo {author} {\bibfnamefont {L.}~\bibnamefont {{Hartmann}}}, \bibinfo
  {author} {\bibfnamefont {C.~W.}\ \bibnamefont {{Hergenrother}}}, \bibinfo
  {author} {\bibfnamefont {M.~J.}\ \bibnamefont {{Holman}}}, \bibinfo {author}
  {\bibfnamefont {J.~P.}\ \bibnamefont {{Huchra}}}, \bibinfo {author}
  {\bibfnamefont {R.}~\bibnamefont {{Jayawardhana}}}, \bibinfo {author}
  {\bibfnamefont {D.}~\bibnamefont {{Jerius}}}, \bibinfo {author}
  {\bibfnamefont {S.~J.}\ \bibnamefont {{Kannappan}}}, \bibinfo {author}
  {\bibfnamefont {D.-W.}\ \bibnamefont {{Kim}}}, \bibinfo {author}
  {\bibfnamefont {J.~T.}\ \bibnamefont {{Kleyna}}}, \bibinfo {author}
  {\bibfnamefont {C.~S.}\ \bibnamefont {{Kochanek}}}, \bibinfo {author}
  {\bibfnamefont {D.~M.}\ \bibnamefont {{Koranyi}}}, \bibinfo {author}
  {\bibfnamefont {M.}~\bibnamefont {{Krockenberger}}}, \bibinfo {author}
  {\bibfnamefont {C.~J.}\ \bibnamefont {{Lada}}}, \bibinfo {author}
  {\bibfnamefont {K.~L.}\ \bibnamefont {{Luhman}}}, \bibinfo {author}
  {\bibfnamefont {J.~X.}\ \bibnamefont {{Luu}}}, \bibinfo {author}
  {\bibfnamefont {L.~M.}\ \bibnamefont {{Macri}}}, \bibinfo {author}
  {\bibfnamefont {J.~A.}\ \bibnamefont {{Mader}}}, \bibinfo {author}
  {\bibfnamefont {A.}~\bibnamefont {{Mahdavi}}}, \bibinfo {author}
  {\bibfnamefont {M.}~\bibnamefont {{Marengo}}}, \bibinfo {author}
  {\bibfnamefont {B.~G.}\ \bibnamefont {{Marsden}}}, \bibinfo {author}
  {\bibfnamefont {B.~A.}\ \bibnamefont {{McLeod}}}, \bibinfo {author}
  {\bibfnamefont {B.~R.}\ \bibnamefont {{McNamara}}}, \bibinfo {author}
  {\bibfnamefont {S.~T.}\ \bibnamefont {{Megeath}}}, \bibinfo {author}
  {\bibfnamefont {D.}~\bibnamefont {{Moraru}}}, \bibinfo {author}
  {\bibfnamefont {A.~E.}\ \bibnamefont {{Mossman}}}, \bibinfo {author}
  {\bibfnamefont {A.~A.}\ \bibnamefont {{Muench}}}, \bibinfo {author}
  {\bibfnamefont {J.~A.}\ \bibnamefont {{Mu{\~n}oz}}}, \bibinfo {author}
  {\bibfnamefont {J.}~\bibnamefont {{Muzerolle}}}, \bibinfo {author}
  {\bibfnamefont {O.}~\bibnamefont {{Naranjo}}}, \bibinfo {author}
  {\bibfnamefont {K.}~\bibnamefont {{Nelson-Patel}}}, \bibinfo {author}
  {\bibfnamefont {M.~A.}\ \bibnamefont {{Pahre}}}, \bibinfo {author}
  {\bibfnamefont {B.~M.}\ \bibnamefont {{Patten}}}, \bibinfo {author}
  {\bibfnamefont {J.}~\bibnamefont {{Peters}}}, \bibinfo {author}
  {\bibfnamefont {W.}~\bibnamefont {{Peters}}}, \bibinfo {author}
  {\bibfnamefont {J.~C.}\ \bibnamefont {{Raymond}}}, \bibinfo {author}
  {\bibfnamefont {K.}~\bibnamefont {{Rines}}}, \bibinfo {author} {\bibfnamefont
  {R.~E.}\ \bibnamefont {{Schild}}}, \bibinfo {author} {\bibfnamefont {G.~J.}\
  \bibnamefont {{Sobczak}}}, \bibinfo {author} {\bibfnamefont {T.~B.}\
  \bibnamefont {{Spahr}}}, \bibinfo {author} {\bibfnamefont {J.~R.}\
  \bibnamefont {{Stauffer}}}, \bibinfo {author} {\bibfnamefont {R.~P.}\
  \bibnamefont {{Stefanik}}}, \bibinfo {author} {\bibfnamefont {A.~H.}\
  \bibnamefont {{Szentgyorgyi}}}, \bibinfo {author} {\bibfnamefont {E.~V.}\
  \bibnamefont {{Tollestrup}}}, \bibinfo {author} {\bibfnamefont
  {P.}~\bibnamefont {{V{\"a}is{\"a}nen}}}, \bibinfo {author} {\bibfnamefont
  {A.}~\bibnamefont {{Vikhlinin}}}, \bibinfo {author} {\bibfnamefont
  {Z.}~\bibnamefont {{Wang}}}, \bibinfo {author} {\bibfnamefont {S.~P.}\
  \bibnamefont {{Willner}}}, \bibinfo {author} {\bibfnamefont {S.~J.}\
  \bibnamefont {{Wolk}}}, \bibinfo {author} {\bibfnamefont {J.~M.}\
  \bibnamefont {{Zajac}}}, \bibinfo {author} {\bibfnamefont {P.}~\bibnamefont
  {{Zhao}}}, \ and\ \bibinfo {author} {\bibfnamefont {K.~Z.}\ \bibnamefont
  {{Stanek}}},\ }\href {\doibase 10.1086/497989} {\bibfield  {journal}
  {\bibinfo  {journal} {\aj}\ }\textbf {\bibinfo {volume} {131}},\ \bibinfo
  {pages} {527} (\bibinfo {year} {2006})},\ \Eprint
  {http://arxiv.org/abs/arXiv:astro-ph/0509234} {arXiv:astro-ph/0509234}
  \BibitemShut {NoStop}%
\bibitem [{\citenamefont {{Riess}}\ \emph {et~al.}(2007)\citenamefont
  {{Riess}}, \citenamefont {{Strolger}}, \citenamefont {{Casertano}},
  \citenamefont {{Ferguson}}, \citenamefont {{Mobasher}}, \citenamefont
  {{Gold}}, \citenamefont {{Challis}}, \citenamefont {{Filippenko}},
  \citenamefont {{Jha}}, \citenamefont {{Li}}, \citenamefont {{Tonry}},
  \citenamefont {{Foley}}, \citenamefont {{Kirshner}}, \citenamefont
  {{Dickinson}}, \citenamefont {{MacDonald}}, \citenamefont {{Eisenstein}},
  \citenamefont {{Livio}}, \citenamefont {{Younger}}, \citenamefont {{Xu}},
  \citenamefont {{Dahl{\'e}n}},\ and\ \citenamefont {{Stern}}}]{riess07a}%
  \BibitemOpen
  \bibfield  {author} {\bibinfo {author} {\bibfnamefont {A.~G.}\ \bibnamefont
  {{Riess}}}, \bibinfo {author} {\bibfnamefont {L.-G.}\ \bibnamefont
  {{Strolger}}}, \bibinfo {author} {\bibfnamefont {S.}~\bibnamefont
  {{Casertano}}}, \bibinfo {author} {\bibfnamefont {H.~C.}\ \bibnamefont
  {{Ferguson}}}, \bibinfo {author} {\bibfnamefont {B.}~\bibnamefont
  {{Mobasher}}}, \bibinfo {author} {\bibfnamefont {B.}~\bibnamefont {{Gold}}},
  \bibinfo {author} {\bibfnamefont {P.~J.}\ \bibnamefont {{Challis}}}, \bibinfo
  {author} {\bibfnamefont {A.~V.}\ \bibnamefont {{Filippenko}}}, \bibinfo
  {author} {\bibfnamefont {S.}~\bibnamefont {{Jha}}}, \bibinfo {author}
  {\bibfnamefont {W.}~\bibnamefont {{Li}}}, \bibinfo {author} {\bibfnamefont
  {J.}~\bibnamefont {{Tonry}}}, \bibinfo {author} {\bibfnamefont
  {R.}~\bibnamefont {{Foley}}}, \bibinfo {author} {\bibfnamefont {R.~P.}\
  \bibnamefont {{Kirshner}}}, \bibinfo {author} {\bibfnamefont
  {M.}~\bibnamefont {{Dickinson}}}, \bibinfo {author} {\bibfnamefont
  {E.}~\bibnamefont {{MacDonald}}}, \bibinfo {author} {\bibfnamefont
  {D.}~\bibnamefont {{Eisenstein}}}, \bibinfo {author} {\bibfnamefont
  {M.}~\bibnamefont {{Livio}}}, \bibinfo {author} {\bibfnamefont
  {J.}~\bibnamefont {{Younger}}}, \bibinfo {author} {\bibfnamefont
  {C.}~\bibnamefont {{Xu}}}, \bibinfo {author} {\bibfnamefont {T.}~\bibnamefont
  {{Dahl{\'e}n}}}, \ and\ \bibinfo {author} {\bibfnamefont {D.}~\bibnamefont
  {{Stern}}},\ }\href {\doibase 10.1086/510378} {\bibfield  {journal} {\bibinfo
   {journal} {\apj}\ }\textbf {\bibinfo {volume} {659}},\ \bibinfo {pages} {98}
  (\bibinfo {year} {2007})},\ \Eprint
  {http://arxiv.org/abs/arXiv:astro-ph/0611572} {arXiv:astro-ph/0611572}
  \BibitemShut {NoStop}%
\bibitem [{\citenamefont {{Frieman}}\ \emph {et~al.}(2008)\citenamefont
  {{Frieman}}, \citenamefont {{Bassett}}, \citenamefont {{Becker}},
  \citenamefont {{Choi}}, \citenamefont {{Cinabro}}, \citenamefont {{DeJongh}},
  \citenamefont {{Depoy}}, \citenamefont {{Dilday}}, \citenamefont {{Doi}},
  \citenamefont {{Garnavich}}, \citenamefont {{Hogan}}, \citenamefont
  {{Holtzman}}, \citenamefont {{Im}}, \citenamefont {{Jha}}, \citenamefont
  {{Kessler}}, \citenamefont {{Konishi}}, \citenamefont {{Lampeitl}},
  \citenamefont {{Marriner}}, \citenamefont {{Marshall}}, \citenamefont
  {{McGinnis}}, \citenamefont {{Miknaitis}}, \citenamefont {{Nichol}},
  \citenamefont {{Prieto}}, \citenamefont {{Riess}}, \citenamefont
  {{Richmond}}, \citenamefont {{Romani}}, \citenamefont {{Sako}}, \citenamefont
  {{Schneider}}, \citenamefont {{Smith}}, \citenamefont {{Takanashi}},
  \citenamefont {{Tokita}}, \citenamefont {{van der Heyden}}, \citenamefont
  {{Yasuda}}, \citenamefont {{Zheng}}, \citenamefont {{Adelman-McCarthy}},
  \citenamefont {{Annis}}, \citenamefont {{Assef}}, \citenamefont
  {{Barentine}}, \citenamefont {{Bender}}, \citenamefont {{Blandford}},
  \citenamefont {{Boroski}}, \citenamefont {{Bremer}}, \citenamefont
  {{Brewington}}, \citenamefont {{Collins}}, \citenamefont {{Crotts}},
  \citenamefont {{Dembicky}}, \citenamefont {{Eastman}}, \citenamefont
  {{Edge}}, \citenamefont {{Edmondson}}, \citenamefont {{Elson}}, \citenamefont
  {{Eyler}}, \citenamefont {{Filippenko}}, \citenamefont {{Foley}},
  \citenamefont {{Frank}}, \citenamefont {{Goobar}}, \citenamefont {{Gueth}},
  \citenamefont {{Gunn}}, \citenamefont {{Harvanek}}, \citenamefont {{Hopp}},
  \citenamefont {{Ihara}}, \citenamefont {{Ivezi{\'c}}}, \citenamefont
  {{Kahn}}, \citenamefont {{Kaplan}}, \citenamefont {{Kent}}, \citenamefont
  {{Ketzeback}}, \citenamefont {{Kleinman}}, \citenamefont {{Kollatschny}},
  \citenamefont {{Kron}}, \citenamefont {{Krzesi{\'n}ski}}, \citenamefont
  {{Lamenti}}, \citenamefont {{Leloudas}}, \citenamefont {{Lin}}, \citenamefont
  {{Long}}, \citenamefont {{Lucey}}, \citenamefont {{Lupton}}, \citenamefont
  {{Malanushenko}}, \citenamefont {{Malanushenko}}, \citenamefont {{McMillan}},
  \citenamefont {{Mendez}}, \citenamefont {{Morgan}}, \citenamefont
  {{Morokuma}}, \citenamefont {{Nitta}}, \citenamefont {{Ostman}},
  \citenamefont {{Pan}}, \citenamefont {{Rockosi}}, \citenamefont {{Romer}},
  \citenamefont {{Ruiz-Lapuente}}, \citenamefont {{Saurage}}, \citenamefont
  {{Schlesinger}}, \citenamefont {{Snedden}}, \citenamefont {{Sollerman}},
  \citenamefont {{Stoughton}}, \citenamefont {{Stritzinger}}, \citenamefont
  {{Subba Rao}}, \citenamefont {{Tucker}}, \citenamefont {{Vaisanen}},
  \citenamefont {{Watson}}, \citenamefont {{Watters}}, \citenamefont
  {{Wheeler}}, \citenamefont {{Yanny}},\ and\ \citenamefont
  {{York}}}]{frieman08a}%
  \BibitemOpen
  \bibfield  {author} {\bibinfo {author} {\bibfnamefont {J.~A.}\ \bibnamefont
  {{Frieman}}}, \bibinfo {author} {\bibfnamefont {B.}~\bibnamefont
  {{Bassett}}}, \bibinfo {author} {\bibfnamefont {A.}~\bibnamefont {{Becker}}},
  \bibinfo {author} {\bibfnamefont {C.}~\bibnamefont {{Choi}}}, \bibinfo
  {author} {\bibfnamefont {D.}~\bibnamefont {{Cinabro}}}, \bibinfo {author}
  {\bibfnamefont {F.}~\bibnamefont {{DeJongh}}}, \bibinfo {author}
  {\bibfnamefont {D.~L.}\ \bibnamefont {{Depoy}}}, \bibinfo {author}
  {\bibfnamefont {B.}~\bibnamefont {{Dilday}}}, \bibinfo {author}
  {\bibfnamefont {M.}~\bibnamefont {{Doi}}}, \bibinfo {author} {\bibfnamefont
  {P.~M.}\ \bibnamefont {{Garnavich}}}, \bibinfo {author} {\bibfnamefont
  {C.~J.}\ \bibnamefont {{Hogan}}}, \bibinfo {author} {\bibfnamefont
  {J.}~\bibnamefont {{Holtzman}}}, \bibinfo {author} {\bibfnamefont
  {M.}~\bibnamefont {{Im}}}, \bibinfo {author} {\bibfnamefont {S.}~\bibnamefont
  {{Jha}}}, \bibinfo {author} {\bibfnamefont {R.}~\bibnamefont {{Kessler}}},
  \bibinfo {author} {\bibfnamefont {K.}~\bibnamefont {{Konishi}}}, \bibinfo
  {author} {\bibfnamefont {H.}~\bibnamefont {{Lampeitl}}}, \bibinfo {author}
  {\bibfnamefont {J.}~\bibnamefont {{Marriner}}}, \bibinfo {author}
  {\bibfnamefont {J.~L.}\ \bibnamefont {{Marshall}}}, \bibinfo {author}
  {\bibfnamefont {D.}~\bibnamefont {{McGinnis}}}, \bibinfo {author}
  {\bibfnamefont {G.}~\bibnamefont {{Miknaitis}}}, \bibinfo {author}
  {\bibfnamefont {R.~C.}\ \bibnamefont {{Nichol}}}, \bibinfo {author}
  {\bibfnamefont {J.~L.}\ \bibnamefont {{Prieto}}}, \bibinfo {author}
  {\bibfnamefont {A.~G.}\ \bibnamefont {{Riess}}}, \bibinfo {author}
  {\bibfnamefont {M.~W.}\ \bibnamefont {{Richmond}}}, \bibinfo {author}
  {\bibfnamefont {R.}~\bibnamefont {{Romani}}}, \bibinfo {author}
  {\bibfnamefont {M.}~\bibnamefont {{Sako}}}, \bibinfo {author} {\bibfnamefont
  {D.~P.}\ \bibnamefont {{Schneider}}}, \bibinfo {author} {\bibfnamefont
  {M.}~\bibnamefont {{Smith}}}, \bibinfo {author} {\bibfnamefont
  {N.}~\bibnamefont {{Takanashi}}}, \bibinfo {author} {\bibfnamefont
  {K.}~\bibnamefont {{Tokita}}}, \bibinfo {author} {\bibfnamefont
  {K.}~\bibnamefont {{van der Heyden}}}, \bibinfo {author} {\bibfnamefont
  {N.}~\bibnamefont {{Yasuda}}}, \bibinfo {author} {\bibfnamefont
  {C.}~\bibnamefont {{Zheng}}}, \bibinfo {author} {\bibfnamefont
  {J.}~\bibnamefont {{Adelman-McCarthy}}}, \bibinfo {author} {\bibfnamefont
  {J.}~\bibnamefont {{Annis}}}, \bibinfo {author} {\bibfnamefont {R.~J.}\
  \bibnamefont {{Assef}}}, \bibinfo {author} {\bibfnamefont {J.}~\bibnamefont
  {{Barentine}}}, \bibinfo {author} {\bibfnamefont {R.}~\bibnamefont
  {{Bender}}}, \bibinfo {author} {\bibfnamefont {R.~D.}\ \bibnamefont
  {{Blandford}}}, \bibinfo {author} {\bibfnamefont {W.~N.}\ \bibnamefont
  {{Boroski}}}, \bibinfo {author} {\bibfnamefont {M.}~\bibnamefont {{Bremer}}},
  \bibinfo {author} {\bibfnamefont {H.}~\bibnamefont {{Brewington}}}, \bibinfo
  {author} {\bibfnamefont {C.~A.}\ \bibnamefont {{Collins}}}, \bibinfo {author}
  {\bibfnamefont {A.}~\bibnamefont {{Crotts}}}, \bibinfo {author}
  {\bibfnamefont {J.}~\bibnamefont {{Dembicky}}}, \bibinfo {author}
  {\bibfnamefont {J.}~\bibnamefont {{Eastman}}}, \bibinfo {author}
  {\bibfnamefont {A.}~\bibnamefont {{Edge}}}, \bibinfo {author} {\bibfnamefont
  {E.}~\bibnamefont {{Edmondson}}}, \bibinfo {author} {\bibfnamefont
  {E.}~\bibnamefont {{Elson}}}, \bibinfo {author} {\bibfnamefont {M.~E.}\
  \bibnamefont {{Eyler}}}, \bibinfo {author} {\bibfnamefont {A.~V.}\
  \bibnamefont {{Filippenko}}}, \bibinfo {author} {\bibfnamefont {R.~J.}\
  \bibnamefont {{Foley}}}, \bibinfo {author} {\bibfnamefont {S.}~\bibnamefont
  {{Frank}}}, \bibinfo {author} {\bibfnamefont {A.}~\bibnamefont {{Goobar}}},
  \bibinfo {author} {\bibfnamefont {T.}~\bibnamefont {{Gueth}}}, \bibinfo
  {author} {\bibfnamefont {J.~E.}\ \bibnamefont {{Gunn}}}, \bibinfo {author}
  {\bibfnamefont {M.}~\bibnamefont {{Harvanek}}}, \bibinfo {author}
  {\bibfnamefont {U.}~\bibnamefont {{Hopp}}}, \bibinfo {author} {\bibfnamefont
  {Y.}~\bibnamefont {{Ihara}}}, \bibinfo {author} {\bibfnamefont {{\v
  Z}.}~\bibnamefont {{Ivezi{\'c}}}}, \bibinfo {author} {\bibfnamefont
  {S.}~\bibnamefont {{Kahn}}}, \bibinfo {author} {\bibfnamefont
  {J.}~\bibnamefont {{Kaplan}}}, \bibinfo {author} {\bibfnamefont
  {S.}~\bibnamefont {{Kent}}}, \bibinfo {author} {\bibfnamefont
  {W.}~\bibnamefont {{Ketzeback}}}, \bibinfo {author} {\bibfnamefont {S.~J.}\
  \bibnamefont {{Kleinman}}}, \bibinfo {author} {\bibfnamefont
  {W.}~\bibnamefont {{Kollatschny}}}, \bibinfo {author} {\bibfnamefont {R.~G.}\
  \bibnamefont {{Kron}}}, \bibinfo {author} {\bibfnamefont {J.}~\bibnamefont
  {{Krzesi{\'n}ski}}}, \bibinfo {author} {\bibfnamefont {D.}~\bibnamefont
  {{Lamenti}}}, \bibinfo {author} {\bibfnamefont {G.}~\bibnamefont
  {{Leloudas}}}, \bibinfo {author} {\bibfnamefont {H.}~\bibnamefont {{Lin}}},
  \bibinfo {author} {\bibfnamefont {D.~C.}\ \bibnamefont {{Long}}}, \bibinfo
  {author} {\bibfnamefont {J.}~\bibnamefont {{Lucey}}}, \bibinfo {author}
  {\bibfnamefont {R.~H.}\ \bibnamefont {{Lupton}}}, \bibinfo {author}
  {\bibfnamefont {E.}~\bibnamefont {{Malanushenko}}}, \bibinfo {author}
  {\bibfnamefont {V.}~\bibnamefont {{Malanushenko}}}, \bibinfo {author}
  {\bibfnamefont {R.~J.}\ \bibnamefont {{McMillan}}}, \bibinfo {author}
  {\bibfnamefont {J.}~\bibnamefont {{Mendez}}}, \bibinfo {author}
  {\bibfnamefont {C.~W.}\ \bibnamefont {{Morgan}}}, \bibinfo {author}
  {\bibfnamefont {T.}~\bibnamefont {{Morokuma}}}, \bibinfo {author}
  {\bibfnamefont {A.}~\bibnamefont {{Nitta}}}, \bibinfo {author} {\bibfnamefont
  {L.}~\bibnamefont {{Ostman}}}, \bibinfo {author} {\bibfnamefont
  {K.}~\bibnamefont {{Pan}}}, \bibinfo {author} {\bibfnamefont {C.~M.}\
  \bibnamefont {{Rockosi}}}, \bibinfo {author} {\bibfnamefont {A.~K.}\
  \bibnamefont {{Romer}}}, \bibinfo {author} {\bibfnamefont {P.}~\bibnamefont
  {{Ruiz-Lapuente}}}, \bibinfo {author} {\bibfnamefont {G.}~\bibnamefont
  {{Saurage}}}, \bibinfo {author} {\bibfnamefont {K.}~\bibnamefont
  {{Schlesinger}}}, \bibinfo {author} {\bibfnamefont {S.~A.}\ \bibnamefont
  {{Snedden}}}, \bibinfo {author} {\bibfnamefont {J.}~\bibnamefont
  {{Sollerman}}}, \bibinfo {author} {\bibfnamefont {C.}~\bibnamefont
  {{Stoughton}}}, \bibinfo {author} {\bibfnamefont {M.}~\bibnamefont
  {{Stritzinger}}}, \bibinfo {author} {\bibfnamefont {M.}~\bibnamefont {{Subba
  Rao}}}, \bibinfo {author} {\bibfnamefont {D.}~\bibnamefont {{Tucker}}},
  \bibinfo {author} {\bibfnamefont {P.}~\bibnamefont {{Vaisanen}}}, \bibinfo
  {author} {\bibfnamefont {L.~C.}\ \bibnamefont {{Watson}}}, \bibinfo {author}
  {\bibfnamefont {S.}~\bibnamefont {{Watters}}}, \bibinfo {author}
  {\bibfnamefont {J.~C.}\ \bibnamefont {{Wheeler}}}, \bibinfo {author}
  {\bibfnamefont {B.}~\bibnamefont {{Yanny}}}, \ and\ \bibinfo {author}
  {\bibfnamefont {D.}~\bibnamefont {{York}}},\ }\href {\doibase
  10.1088/0004-6256/135/1/338} {\bibfield  {journal} {\bibinfo  {journal}
  {\aj}\ }\textbf {\bibinfo {volume} {135}},\ \bibinfo {pages} {338} (\bibinfo
  {year} {2008})},\ \Eprint {http://arxiv.org/abs/0708.2749} {arXiv:0708.2749}
  \BibitemShut {NoStop}%
\bibitem [{\citenamefont {{Dawson}}\ \emph {et~al.}(2009)\citenamefont
  {{Dawson}}, \citenamefont {{Aldering}}, \citenamefont {{Amanullah}},
  \citenamefont {{Barbary}}, \citenamefont {{Barrientos}}, \citenamefont
  {{Brodwin}}, \citenamefont {{Connolly}}, \citenamefont {{Dey}}, \citenamefont
  {{Doi}}, \citenamefont {{Donahue}}, \citenamefont {{Eisenhardt}},
  \citenamefont {{Ellingson}}, \citenamefont {{Faccioli}}, \citenamefont
  {{Fadeyev}}, \citenamefont {{Fakhouri}}, \citenamefont {{Fruchter}},
  \citenamefont {{Gilbank}}, \citenamefont {{Gladders}}, \citenamefont
  {{Goldhaber}}, \citenamefont {{Gonzalez}}, \citenamefont {{Goobar}},
  \citenamefont {{Gude}}, \citenamefont {{Hattori}}, \citenamefont
  {{Hoekstra}}, \citenamefont {{Huang}}, \citenamefont {{Ihara}}, \citenamefont
  {{Jannuzi}}, \citenamefont {{Johnston}}, \citenamefont {{Kashikawa}},
  \citenamefont {{Koester}}, \citenamefont {{Konishi}}, \citenamefont
  {{Kowalski}}, \citenamefont {{Lidman}}, \citenamefont {{Linder}},
  \citenamefont {{Lubin}}, \citenamefont {{Meyers}}, \citenamefont
  {{Morokuma}}, \citenamefont {{Munshi}}, \citenamefont {{Mullis}},
  \citenamefont {{Oda}}, \citenamefont {{Panagia}}, \citenamefont
  {{Perlmutter}}, \citenamefont {{Postman}}, \citenamefont {{Pritchard}},
  \citenamefont {{Rhodes}}, \citenamefont {{Rosati}}, \citenamefont {{Rubin}},
  \citenamefont {{Schlegel}}, \citenamefont {{Spadafora}}, \citenamefont
  {{Stanford}}, \citenamefont {{Stanishev}}, \citenamefont {{Stern}},
  \citenamefont {{Strovink}}, \citenamefont {{Suzuki}}, \citenamefont
  {{Takanashi}}, \citenamefont {{Tokita}}, \citenamefont {{Wagner}},
  \citenamefont {{Wang}}, \citenamefont {{Yasuda}}, \citenamefont {{Yee}},\
  and\ \citenamefont {{Supernova Cosmology Project}}}]{dawson09a}%
  \BibitemOpen
  \bibfield  {author} {\bibinfo {author} {\bibfnamefont {K.~S.}\ \bibnamefont
  {{Dawson}}}, \bibinfo {author} {\bibfnamefont {G.}~\bibnamefont
  {{Aldering}}}, \bibinfo {author} {\bibfnamefont {R.}~\bibnamefont
  {{Amanullah}}}, \bibinfo {author} {\bibfnamefont {K.}~\bibnamefont
  {{Barbary}}}, \bibinfo {author} {\bibfnamefont {L.~F.}\ \bibnamefont
  {{Barrientos}}}, \bibinfo {author} {\bibfnamefont {M.}~\bibnamefont
  {{Brodwin}}}, \bibinfo {author} {\bibfnamefont {N.}~\bibnamefont
  {{Connolly}}}, \bibinfo {author} {\bibfnamefont {A.}~\bibnamefont {{Dey}}},
  \bibinfo {author} {\bibfnamefont {M.}~\bibnamefont {{Doi}}}, \bibinfo
  {author} {\bibfnamefont {M.}~\bibnamefont {{Donahue}}}, \bibinfo {author}
  {\bibfnamefont {P.}~\bibnamefont {{Eisenhardt}}}, \bibinfo {author}
  {\bibfnamefont {E.}~\bibnamefont {{Ellingson}}}, \bibinfo {author}
  {\bibfnamefont {L.}~\bibnamefont {{Faccioli}}}, \bibinfo {author}
  {\bibfnamefont {V.}~\bibnamefont {{Fadeyev}}}, \bibinfo {author}
  {\bibfnamefont {H.~K.}\ \bibnamefont {{Fakhouri}}}, \bibinfo {author}
  {\bibfnamefont {A.~S.}\ \bibnamefont {{Fruchter}}}, \bibinfo {author}
  {\bibfnamefont {D.~G.}\ \bibnamefont {{Gilbank}}}, \bibinfo {author}
  {\bibfnamefont {M.~D.}\ \bibnamefont {{Gladders}}}, \bibinfo {author}
  {\bibfnamefont {G.}~\bibnamefont {{Goldhaber}}}, \bibinfo {author}
  {\bibfnamefont {A.~H.}\ \bibnamefont {{Gonzalez}}}, \bibinfo {author}
  {\bibfnamefont {A.}~\bibnamefont {{Goobar}}}, \bibinfo {author}
  {\bibfnamefont {A.}~\bibnamefont {{Gude}}}, \bibinfo {author} {\bibfnamefont
  {T.}~\bibnamefont {{Hattori}}}, \bibinfo {author} {\bibfnamefont
  {H.}~\bibnamefont {{Hoekstra}}}, \bibinfo {author} {\bibfnamefont
  {X.}~\bibnamefont {{Huang}}}, \bibinfo {author} {\bibfnamefont
  {Y.}~\bibnamefont {{Ihara}}}, \bibinfo {author} {\bibfnamefont {B.~T.}\
  \bibnamefont {{Jannuzi}}}, \bibinfo {author} {\bibfnamefont {D.}~\bibnamefont
  {{Johnston}}}, \bibinfo {author} {\bibfnamefont {K.}~\bibnamefont
  {{Kashikawa}}}, \bibinfo {author} {\bibfnamefont {B.}~\bibnamefont
  {{Koester}}}, \bibinfo {author} {\bibfnamefont {K.}~\bibnamefont
  {{Konishi}}}, \bibinfo {author} {\bibfnamefont {M.}~\bibnamefont
  {{Kowalski}}}, \bibinfo {author} {\bibfnamefont {C.}~\bibnamefont
  {{Lidman}}}, \bibinfo {author} {\bibfnamefont {E.~V.}\ \bibnamefont
  {{Linder}}}, \bibinfo {author} {\bibfnamefont {L.}~\bibnamefont {{Lubin}}},
  \bibinfo {author} {\bibfnamefont {J.}~\bibnamefont {{Meyers}}}, \bibinfo
  {author} {\bibfnamefont {T.}~\bibnamefont {{Morokuma}}}, \bibinfo {author}
  {\bibfnamefont {F.}~\bibnamefont {{Munshi}}}, \bibinfo {author}
  {\bibfnamefont {C.}~\bibnamefont {{Mullis}}}, \bibinfo {author}
  {\bibfnamefont {T.}~\bibnamefont {{Oda}}}, \bibinfo {author} {\bibfnamefont
  {N.}~\bibnamefont {{Panagia}}}, \bibinfo {author} {\bibfnamefont
  {S.}~\bibnamefont {{Perlmutter}}}, \bibinfo {author} {\bibfnamefont
  {M.}~\bibnamefont {{Postman}}}, \bibinfo {author} {\bibfnamefont
  {T.}~\bibnamefont {{Pritchard}}}, \bibinfo {author} {\bibfnamefont
  {J.}~\bibnamefont {{Rhodes}}}, \bibinfo {author} {\bibfnamefont
  {P.}~\bibnamefont {{Rosati}}}, \bibinfo {author} {\bibfnamefont
  {D.}~\bibnamefont {{Rubin}}}, \bibinfo {author} {\bibfnamefont {D.~J.}\
  \bibnamefont {{Schlegel}}}, \bibinfo {author} {\bibfnamefont
  {A.}~\bibnamefont {{Spadafora}}}, \bibinfo {author} {\bibfnamefont {S.~A.}\
  \bibnamefont {{Stanford}}}, \bibinfo {author} {\bibfnamefont
  {V.}~\bibnamefont {{Stanishev}}}, \bibinfo {author} {\bibfnamefont
  {D.}~\bibnamefont {{Stern}}}, \bibinfo {author} {\bibfnamefont
  {M.}~\bibnamefont {{Strovink}}}, \bibinfo {author} {\bibfnamefont
  {N.}~\bibnamefont {{Suzuki}}}, \bibinfo {author} {\bibfnamefont
  {N.}~\bibnamefont {{Takanashi}}}, \bibinfo {author} {\bibfnamefont
  {K.}~\bibnamefont {{Tokita}}}, \bibinfo {author} {\bibfnamefont
  {M.}~\bibnamefont {{Wagner}}}, \bibinfo {author} {\bibfnamefont
  {L.}~\bibnamefont {{Wang}}}, \bibinfo {author} {\bibfnamefont
  {N.}~\bibnamefont {{Yasuda}}}, \bibinfo {author} {\bibfnamefont {H.~K.~C.}\
  \bibnamefont {{Yee}}}, \ and\ \bibinfo {author} {\bibfnamefont
  {T.}~\bibnamefont {{Supernova Cosmology Project}}},\ }\href {\doibase
  10.1088/0004-6256/138/5/1271} {\bibfield  {journal} {\bibinfo  {journal}
  {\aj}\ }\textbf {\bibinfo {volume} {138}},\ \bibinfo {pages} {1271} (\bibinfo
  {year} {2009})},\ \Eprint {http://arxiv.org/abs/0908.3928} {arXiv:0908.3928}
  \BibitemShut {NoStop}%
\bibitem [{\citenamefont {{Hicken}}\ \emph {et~al.}(2009)\citenamefont
  {{Hicken}}, \citenamefont {{Wood-Vasey}}, \citenamefont {{Blondin}},
  \citenamefont {{Challis}}, \citenamefont {{Jha}}, \citenamefont {{Kelly}},
  \citenamefont {{Rest}},\ and\ \citenamefont {{Kirshner}}}]{hicken09a}%
  \BibitemOpen
  \bibfield  {author} {\bibinfo {author} {\bibfnamefont {M.}~\bibnamefont
  {{Hicken}}}, \bibinfo {author} {\bibfnamefont {W.~M.}\ \bibnamefont
  {{Wood-Vasey}}}, \bibinfo {author} {\bibfnamefont {S.}~\bibnamefont
  {{Blondin}}}, \bibinfo {author} {\bibfnamefont {P.}~\bibnamefont
  {{Challis}}}, \bibinfo {author} {\bibfnamefont {S.}~\bibnamefont {{Jha}}},
  \bibinfo {author} {\bibfnamefont {P.~L.}\ \bibnamefont {{Kelly}}}, \bibinfo
  {author} {\bibfnamefont {A.}~\bibnamefont {{Rest}}}, \ and\ \bibinfo {author}
  {\bibfnamefont {R.~P.}\ \bibnamefont {{Kirshner}}},\ }\href {\doibase
  10.1088/0004-637X/700/2/1097} {\bibfield  {journal} {\bibinfo  {journal}
  {\apj}\ }\textbf {\bibinfo {volume} {700}},\ \bibinfo {pages} {1097}
  (\bibinfo {year} {2009})},\ \Eprint {http://arxiv.org/abs/0901.4804}
  {arXiv:0901.4804} \BibitemShut {NoStop}%
\bibitem [{\citenamefont {{Contreras}}\ \emph {et~al.}(2010)\citenamefont
  {{Contreras}}, \citenamefont {{Hamuy}}, \citenamefont {{Phillips}},
  \citenamefont {{Folatelli}}, \citenamefont {{Suntzeff}}, \citenamefont
  {{Persson}}, \citenamefont {{Stritzinger}}, \citenamefont {{Boldt}},
  \citenamefont {{Gonz{\'a}lez}}, \citenamefont {{Krzeminski}}, \citenamefont
  {{Morrell}}, \citenamefont {{Roth}}, \citenamefont {{Salgado}}, \citenamefont
  {{Jos{\'e} Maureira}}, \citenamefont {{Burns}}, \citenamefont {{Freedman}},
  \citenamefont {{Madore}}, \citenamefont {{Murphy}}, \citenamefont {{Wyatt}},
  \citenamefont {{Li}},\ and\ \citenamefont {{Filippenko}}}]{contreras10a}%
  \BibitemOpen
  \bibfield  {author} {\bibinfo {author} {\bibfnamefont {C.}~\bibnamefont
  {{Contreras}}}, \bibinfo {author} {\bibfnamefont {M.}~\bibnamefont
  {{Hamuy}}}, \bibinfo {author} {\bibfnamefont {M.~M.}\ \bibnamefont
  {{Phillips}}}, \bibinfo {author} {\bibfnamefont {G.}~\bibnamefont
  {{Folatelli}}}, \bibinfo {author} {\bibfnamefont {N.~B.}\ \bibnamefont
  {{Suntzeff}}}, \bibinfo {author} {\bibfnamefont {S.~E.}\ \bibnamefont
  {{Persson}}}, \bibinfo {author} {\bibfnamefont {M.}~\bibnamefont
  {{Stritzinger}}}, \bibinfo {author} {\bibfnamefont {L.}~\bibnamefont
  {{Boldt}}}, \bibinfo {author} {\bibfnamefont {S.}~\bibnamefont
  {{Gonz{\'a}lez}}}, \bibinfo {author} {\bibfnamefont {W.}~\bibnamefont
  {{Krzeminski}}}, \bibinfo {author} {\bibfnamefont {N.}~\bibnamefont
  {{Morrell}}}, \bibinfo {author} {\bibfnamefont {M.}~\bibnamefont {{Roth}}},
  \bibinfo {author} {\bibfnamefont {F.}~\bibnamefont {{Salgado}}}, \bibinfo
  {author} {\bibfnamefont {M.}~\bibnamefont {{Jos{\'e} Maureira}}}, \bibinfo
  {author} {\bibfnamefont {C.~R.}\ \bibnamefont {{Burns}}}, \bibinfo {author}
  {\bibfnamefont {W.~L.}\ \bibnamefont {{Freedman}}}, \bibinfo {author}
  {\bibfnamefont {B.~F.}\ \bibnamefont {{Madore}}}, \bibinfo {author}
  {\bibfnamefont {D.}~\bibnamefont {{Murphy}}}, \bibinfo {author}
  {\bibfnamefont {P.}~\bibnamefont {{Wyatt}}}, \bibinfo {author} {\bibfnamefont
  {W.}~\bibnamefont {{Li}}}, \ and\ \bibinfo {author} {\bibfnamefont {A.~V.}\
  \bibnamefont {{Filippenko}}},\ }\href {\doibase 10.1088/0004-6256/139/2/519}
  {\bibfield  {journal} {\bibinfo  {journal} {\aj}\ }\textbf {\bibinfo {volume}
  {139}},\ \bibinfo {pages} {519} (\bibinfo {year} {2010})},\ \Eprint
  {http://arxiv.org/abs/0910.3330} {arXiv:0910.3330 [astro-ph.CO]} \BibitemShut
  {NoStop}%
\bibitem [{\citenamefont {{Guy}}\ \emph {et~al.}(2010)\citenamefont {{Guy}},
  \citenamefont {{Sullivan}}, \citenamefont {{Conley}}, \citenamefont
  {{Regnault}}, \citenamefont {{Astier}}, \citenamefont {{Balland}},
  \citenamefont {{Basa}}, \citenamefont {{Carlberg}}, \citenamefont
  {{Fouchez}}, \citenamefont {{Hardin}}, \citenamefont {{Hook}}, \citenamefont
  {{Howell}}, \citenamefont {{Pain}}, \citenamefont {{Palanque-Delabrouille}},
  \citenamefont {{Perrett}}, \citenamefont {{Pritchet}}, \citenamefont
  {{Rich}}, \citenamefont {{Ruhlmann-Kleider}}, \citenamefont {{Balam}},
  \citenamefont {{Baumont}}, \citenamefont {{Ellis}}, \citenamefont {{Fabbro}},
  \citenamefont {{Fakhouri}}, \citenamefont {{Fourmanoit}}, \citenamefont
  {{Gonz{\'a}lez-Gait{\'a}n}}, \citenamefont {{Graham}}, \citenamefont
  {{Hsiao}}, \citenamefont {{Kronborg}}, \citenamefont {{Lidman}},
  \citenamefont {{Mourao}}, \citenamefont {{Perlmutter}}, \citenamefont
  {{Ripoche}}, \citenamefont {{Suzuki}},\ and\ \citenamefont
  {{Walker}}}]{guy10a}%
  \BibitemOpen
  \bibfield  {author} {\bibinfo {author} {\bibfnamefont {J.}~\bibnamefont
  {{Guy}}}, \bibinfo {author} {\bibfnamefont {M.}~\bibnamefont {{Sullivan}}},
  \bibinfo {author} {\bibfnamefont {A.}~\bibnamefont {{Conley}}}, \bibinfo
  {author} {\bibfnamefont {N.}~\bibnamefont {{Regnault}}}, \bibinfo {author}
  {\bibfnamefont {P.}~\bibnamefont {{Astier}}}, \bibinfo {author}
  {\bibfnamefont {C.}~\bibnamefont {{Balland}}}, \bibinfo {author}
  {\bibfnamefont {S.}~\bibnamefont {{Basa}}}, \bibinfo {author} {\bibfnamefont
  {R.~G.}\ \bibnamefont {{Carlberg}}}, \bibinfo {author} {\bibfnamefont
  {D.}~\bibnamefont {{Fouchez}}}, \bibinfo {author} {\bibfnamefont
  {D.}~\bibnamefont {{Hardin}}}, \bibinfo {author} {\bibfnamefont {I.~M.}\
  \bibnamefont {{Hook}}}, \bibinfo {author} {\bibfnamefont {D.~A.}\
  \bibnamefont {{Howell}}}, \bibinfo {author} {\bibfnamefont {R.}~\bibnamefont
  {{Pain}}}, \bibinfo {author} {\bibfnamefont {N.}~\bibnamefont
  {{Palanque-Delabrouille}}}, \bibinfo {author} {\bibfnamefont {K.~M.}\
  \bibnamefont {{Perrett}}}, \bibinfo {author} {\bibfnamefont {C.~J.}\
  \bibnamefont {{Pritchet}}}, \bibinfo {author} {\bibfnamefont
  {J.}~\bibnamefont {{Rich}}}, \bibinfo {author} {\bibfnamefont
  {V.}~\bibnamefont {{Ruhlmann-Kleider}}}, \bibinfo {author} {\bibfnamefont
  {D.}~\bibnamefont {{Balam}}}, \bibinfo {author} {\bibfnamefont
  {S.}~\bibnamefont {{Baumont}}}, \bibinfo {author} {\bibfnamefont {R.~S.}\
  \bibnamefont {{Ellis}}}, \bibinfo {author} {\bibfnamefont {S.}~\bibnamefont
  {{Fabbro}}}, \bibinfo {author} {\bibfnamefont {H.~K.}\ \bibnamefont
  {{Fakhouri}}}, \bibinfo {author} {\bibfnamefont {N.}~\bibnamefont
  {{Fourmanoit}}}, \bibinfo {author} {\bibfnamefont {S.}~\bibnamefont
  {{Gonz{\'a}lez-Gait{\'a}n}}}, \bibinfo {author} {\bibfnamefont {M.~L.}\
  \bibnamefont {{Graham}}}, \bibinfo {author} {\bibfnamefont {E.}~\bibnamefont
  {{Hsiao}}}, \bibinfo {author} {\bibfnamefont {T.}~\bibnamefont {{Kronborg}}},
  \bibinfo {author} {\bibfnamefont {C.}~\bibnamefont {{Lidman}}}, \bibinfo
  {author} {\bibfnamefont {A.~M.}\ \bibnamefont {{Mourao}}}, \bibinfo {author}
  {\bibfnamefont {S.}~\bibnamefont {{Perlmutter}}}, \bibinfo {author}
  {\bibfnamefont {P.}~\bibnamefont {{Ripoche}}}, \bibinfo {author}
  {\bibfnamefont {N.}~\bibnamefont {{Suzuki}}}, \ and\ \bibinfo {author}
  {\bibfnamefont {E.~S.}\ \bibnamefont {{Walker}}},\ }\href {\doibase
  10.1051/0004-6361/201014468} {\bibfield  {journal} {\bibinfo  {journal}
  {\aap}\ }\textbf {\bibinfo {volume} {523}},\ \bibinfo {pages} {A7} (\bibinfo
  {year} {2010})},\ \Eprint {http://arxiv.org/abs/1010.4743} {arXiv:1010.4743
  [astro-ph.CO]} \BibitemShut {NoStop}%
\bibitem [{\citenamefont {{Conley}}\ \emph {et~al.}(2011)\citenamefont
  {{Conley}}, \citenamefont {{Guy}}, \citenamefont {{Sullivan}}, \citenamefont
  {{Regnault}}, \citenamefont {{Astier}}, \citenamefont {{Balland}},
  \citenamefont {{Basa}}, \citenamefont {{Carlberg}}, \citenamefont
  {{Fouchez}}, \citenamefont {{Hardin}}, \citenamefont {{Hook}}, \citenamefont
  {{Howell}}, \citenamefont {{Pain}}, \citenamefont {{Palanque-Delabrouille}},
  \citenamefont {{Perrett}}, \citenamefont {{Pritchet}}, \citenamefont
  {{Rich}}, \citenamefont {{Ruhlmann-Kleider}}, \citenamefont {{Balam}},
  \citenamefont {{Baumont}}, \citenamefont {{Ellis}}, \citenamefont {{Fabbro}},
  \citenamefont {{Fakhouri}}, \citenamefont {{Fourmanoit}}, \citenamefont
  {{Gonz{\'a}lez-Gait{\'a}n}}, \citenamefont {{Graham}}, \citenamefont
  {{Hudson}}, \citenamefont {{Hsiao}}, \citenamefont {{Kronborg}},
  \citenamefont {{Lidman}}, \citenamefont {{Mourao}}, \citenamefont {{Neill}},
  \citenamefont {{Perlmutter}}, \citenamefont {{Ripoche}}, \citenamefont
  {{Suzuki}},\ and\ \citenamefont {{Walker}}}]{conley11a}%
  \BibitemOpen
  \bibfield  {author} {\bibinfo {author} {\bibfnamefont {A.}~\bibnamefont
  {{Conley}}}, \bibinfo {author} {\bibfnamefont {J.}~\bibnamefont {{Guy}}},
  \bibinfo {author} {\bibfnamefont {M.}~\bibnamefont {{Sullivan}}}, \bibinfo
  {author} {\bibfnamefont {N.}~\bibnamefont {{Regnault}}}, \bibinfo {author}
  {\bibfnamefont {P.}~\bibnamefont {{Astier}}}, \bibinfo {author}
  {\bibfnamefont {C.}~\bibnamefont {{Balland}}}, \bibinfo {author}
  {\bibfnamefont {S.}~\bibnamefont {{Basa}}}, \bibinfo {author} {\bibfnamefont
  {R.~G.}\ \bibnamefont {{Carlberg}}}, \bibinfo {author} {\bibfnamefont
  {D.}~\bibnamefont {{Fouchez}}}, \bibinfo {author} {\bibfnamefont
  {D.}~\bibnamefont {{Hardin}}}, \bibinfo {author} {\bibfnamefont {I.~M.}\
  \bibnamefont {{Hook}}}, \bibinfo {author} {\bibfnamefont {D.~A.}\
  \bibnamefont {{Howell}}}, \bibinfo {author} {\bibfnamefont {R.}~\bibnamefont
  {{Pain}}}, \bibinfo {author} {\bibfnamefont {N.}~\bibnamefont
  {{Palanque-Delabrouille}}}, \bibinfo {author} {\bibfnamefont {K.~M.}\
  \bibnamefont {{Perrett}}}, \bibinfo {author} {\bibfnamefont {C.~J.}\
  \bibnamefont {{Pritchet}}}, \bibinfo {author} {\bibfnamefont
  {J.}~\bibnamefont {{Rich}}}, \bibinfo {author} {\bibfnamefont
  {V.}~\bibnamefont {{Ruhlmann-Kleider}}}, \bibinfo {author} {\bibfnamefont
  {D.}~\bibnamefont {{Balam}}}, \bibinfo {author} {\bibfnamefont
  {S.}~\bibnamefont {{Baumont}}}, \bibinfo {author} {\bibfnamefont {R.~S.}\
  \bibnamefont {{Ellis}}}, \bibinfo {author} {\bibfnamefont {S.}~\bibnamefont
  {{Fabbro}}}, \bibinfo {author} {\bibfnamefont {H.~K.}\ \bibnamefont
  {{Fakhouri}}}, \bibinfo {author} {\bibfnamefont {N.}~\bibnamefont
  {{Fourmanoit}}}, \bibinfo {author} {\bibfnamefont {S.}~\bibnamefont
  {{Gonz{\'a}lez-Gait{\'a}n}}}, \bibinfo {author} {\bibfnamefont {M.~L.}\
  \bibnamefont {{Graham}}}, \bibinfo {author} {\bibfnamefont {M.~J.}\
  \bibnamefont {{Hudson}}}, \bibinfo {author} {\bibfnamefont {E.}~\bibnamefont
  {{Hsiao}}}, \bibinfo {author} {\bibfnamefont {T.}~\bibnamefont {{Kronborg}}},
  \bibinfo {author} {\bibfnamefont {C.}~\bibnamefont {{Lidman}}}, \bibinfo
  {author} {\bibfnamefont {A.~M.}\ \bibnamefont {{Mourao}}}, \bibinfo {author}
  {\bibfnamefont {J.~D.}\ \bibnamefont {{Neill}}}, \bibinfo {author}
  {\bibfnamefont {S.}~\bibnamefont {{Perlmutter}}}, \bibinfo {author}
  {\bibfnamefont {P.}~\bibnamefont {{Ripoche}}}, \bibinfo {author}
  {\bibfnamefont {N.}~\bibnamefont {{Suzuki}}}, \ and\ \bibinfo {author}
  {\bibfnamefont {E.~S.}\ \bibnamefont {{Walker}}},\ }\href {\doibase
  10.1088/0067-0049/192/1/1} {\bibfield  {journal} {\bibinfo  {journal}
  {\apjs}\ }\textbf {\bibinfo {volume} {192}},\ \bibinfo {pages} {1} (\bibinfo
  {year} {2011})}\BibitemShut {NoStop}%
\bibitem [{\citenamefont {{Sullivan}}\ \emph {et~al.}(2011)\citenamefont
  {{Sullivan}}, \citenamefont {{Guy}}, \citenamefont {{Conley}}, \citenamefont
  {{Regnault}}, \citenamefont {{Astier}}, \citenamefont {{Balland}},
  \citenamefont {{Basa}}, \citenamefont {{Carlberg}}, \citenamefont
  {{Fouchez}}, \citenamefont {{Hardin}}, \citenamefont {{Hook}}, \citenamefont
  {{Howell}}, \citenamefont {{Pain}}, \citenamefont {{Palanque-Delabrouille}},
  \citenamefont {{Perrett}}, \citenamefont {{Pritchet}}, \citenamefont
  {{Rich}}, \citenamefont {{Ruhlmann-Kleider}}, \citenamefont {{Balam}},
  \citenamefont {{Baumont}}, \citenamefont {{Ellis}}, \citenamefont {{Fabbro}},
  \citenamefont {{Fakhouri}}, \citenamefont {{Fourmanoit}}, \citenamefont
  {{Gonz{\'a}lez-Gait{\'a}n}}, \citenamefont {{Graham}}, \citenamefont
  {{Hudson}}, \citenamefont {{Hsiao}}, \citenamefont {{Kronborg}},
  \citenamefont {{Lidman}}, \citenamefont {{Mourao}}, \citenamefont {{Neill}},
  \citenamefont {{Perlmutter}}, \citenamefont {{Ripoche}}, \citenamefont
  {{Suzuki}},\ and\ \citenamefont {{Walker}}}]{sullivan11a}%
  \BibitemOpen
  \bibfield  {author} {\bibinfo {author} {\bibfnamefont {M.}~\bibnamefont
  {{Sullivan}}}, \bibinfo {author} {\bibfnamefont {J.}~\bibnamefont {{Guy}}},
  \bibinfo {author} {\bibfnamefont {A.}~\bibnamefont {{Conley}}}, \bibinfo
  {author} {\bibfnamefont {N.}~\bibnamefont {{Regnault}}}, \bibinfo {author}
  {\bibfnamefont {P.}~\bibnamefont {{Astier}}}, \bibinfo {author}
  {\bibfnamefont {C.}~\bibnamefont {{Balland}}}, \bibinfo {author}
  {\bibfnamefont {S.}~\bibnamefont {{Basa}}}, \bibinfo {author} {\bibfnamefont
  {R.~G.}\ \bibnamefont {{Carlberg}}}, \bibinfo {author} {\bibfnamefont
  {D.}~\bibnamefont {{Fouchez}}}, \bibinfo {author} {\bibfnamefont
  {D.}~\bibnamefont {{Hardin}}}, \bibinfo {author} {\bibfnamefont {I.~M.}\
  \bibnamefont {{Hook}}}, \bibinfo {author} {\bibfnamefont {D.~A.}\
  \bibnamefont {{Howell}}}, \bibinfo {author} {\bibfnamefont {R.}~\bibnamefont
  {{Pain}}}, \bibinfo {author} {\bibfnamefont {N.}~\bibnamefont
  {{Palanque-Delabrouille}}}, \bibinfo {author} {\bibfnamefont {K.~M.}\
  \bibnamefont {{Perrett}}}, \bibinfo {author} {\bibfnamefont {C.~J.}\
  \bibnamefont {{Pritchet}}}, \bibinfo {author} {\bibfnamefont
  {J.}~\bibnamefont {{Rich}}}, \bibinfo {author} {\bibfnamefont
  {V.}~\bibnamefont {{Ruhlmann-Kleider}}}, \bibinfo {author} {\bibfnamefont
  {D.}~\bibnamefont {{Balam}}}, \bibinfo {author} {\bibfnamefont
  {S.}~\bibnamefont {{Baumont}}}, \bibinfo {author} {\bibfnamefont {R.~S.}\
  \bibnamefont {{Ellis}}}, \bibinfo {author} {\bibfnamefont {S.}~\bibnamefont
  {{Fabbro}}}, \bibinfo {author} {\bibfnamefont {H.~K.}\ \bibnamefont
  {{Fakhouri}}}, \bibinfo {author} {\bibfnamefont {N.}~\bibnamefont
  {{Fourmanoit}}}, \bibinfo {author} {\bibfnamefont {S.}~\bibnamefont
  {{Gonz{\'a}lez-Gait{\'a}n}}}, \bibinfo {author} {\bibfnamefont {M.~L.}\
  \bibnamefont {{Graham}}}, \bibinfo {author} {\bibfnamefont {M.~J.}\
  \bibnamefont {{Hudson}}}, \bibinfo {author} {\bibfnamefont {E.}~\bibnamefont
  {{Hsiao}}}, \bibinfo {author} {\bibfnamefont {T.}~\bibnamefont {{Kronborg}}},
  \bibinfo {author} {\bibfnamefont {C.}~\bibnamefont {{Lidman}}}, \bibinfo
  {author} {\bibfnamefont {A.~M.}\ \bibnamefont {{Mourao}}}, \bibinfo {author}
  {\bibfnamefont {J.~D.}\ \bibnamefont {{Neill}}}, \bibinfo {author}
  {\bibfnamefont {S.}~\bibnamefont {{Perlmutter}}}, \bibinfo {author}
  {\bibfnamefont {P.}~\bibnamefont {{Ripoche}}}, \bibinfo {author}
  {\bibfnamefont {N.}~\bibnamefont {{Suzuki}}}, \ and\ \bibinfo {author}
  {\bibfnamefont {E.~S.}\ \bibnamefont {{Walker}}},\ }\href {\doibase
  10.1088/0004-637X/737/2/102} {\bibfield  {journal} {\bibinfo  {journal}
  {\apj}\ }\textbf {\bibinfo {volume} {737}},\ \bibinfo {eid} {102} (\bibinfo
  {year} {2011})},\ \Eprint {http://arxiv.org/abs/1104.1444} {arXiv:1104.1444
  [astro-ph.CO]} \BibitemShut {NoStop}%
\bibitem [{\citenamefont {{Riess}}\ \emph {et~al.}(2009)\citenamefont
  {{Riess}}, \citenamefont {{Macri}}, \citenamefont {{Casertano}},
  \citenamefont {{Sosey}}, \citenamefont {{Lampeitl}}, \citenamefont
  {{Ferguson}}, \citenamefont {{Filippenko}}, \citenamefont {{Jha}},
  \citenamefont {{Li}}, \citenamefont {{Chornock}},\ and\ \citenamefont
  {{Sarkar}}}]{riess09a}%
  \BibitemOpen
  \bibfield  {author} {\bibinfo {author} {\bibfnamefont {A.~G.}\ \bibnamefont
  {{Riess}}}, \bibinfo {author} {\bibfnamefont {L.}~\bibnamefont {{Macri}}},
  \bibinfo {author} {\bibfnamefont {S.}~\bibnamefont {{Casertano}}}, \bibinfo
  {author} {\bibfnamefont {M.}~\bibnamefont {{Sosey}}}, \bibinfo {author}
  {\bibfnamefont {H.}~\bibnamefont {{Lampeitl}}}, \bibinfo {author}
  {\bibfnamefont {H.~C.}\ \bibnamefont {{Ferguson}}}, \bibinfo {author}
  {\bibfnamefont {A.~V.}\ \bibnamefont {{Filippenko}}}, \bibinfo {author}
  {\bibfnamefont {S.~W.}\ \bibnamefont {{Jha}}}, \bibinfo {author}
  {\bibfnamefont {W.}~\bibnamefont {{Li}}}, \bibinfo {author} {\bibfnamefont
  {R.}~\bibnamefont {{Chornock}}}, \ and\ \bibinfo {author} {\bibfnamefont
  {D.}~\bibnamefont {{Sarkar}}},\ }\href {\doibase 10.1088/0004-637X/699/1/539}
  {\bibfield  {journal} {\bibinfo  {journal} {\apj}\ }\textbf {\bibinfo
  {volume} {699}},\ \bibinfo {pages} {539} (\bibinfo {year} {2009})},\ \Eprint
  {http://arxiv.org/abs/0905.0695} {arXiv:0905.0695} \BibitemShut {NoStop}%
\bibitem [{\citenamefont {{Freedman}}\ \emph {et~al.}(2012)\citenamefont
  {{Freedman}}, \citenamefont {{Madore}}, \citenamefont {{Scowcroft}},
  \citenamefont {{Burns}}, \citenamefont {{Monson}}, \citenamefont {{Persson}},
  \citenamefont {{Seibert}},\ and\ \citenamefont {{Rigby}}}]{freedman12a}%
  \BibitemOpen
  \bibfield  {author} {\bibinfo {author} {\bibfnamefont {W.~L.}\ \bibnamefont
  {{Freedman}}}, \bibinfo {author} {\bibfnamefont {B.~F.}\ \bibnamefont
  {{Madore}}}, \bibinfo {author} {\bibfnamefont {V.}~\bibnamefont
  {{Scowcroft}}}, \bibinfo {author} {\bibfnamefont {C.}~\bibnamefont
  {{Burns}}}, \bibinfo {author} {\bibfnamefont {A.}~\bibnamefont {{Monson}}},
  \bibinfo {author} {\bibfnamefont {S.~E.}\ \bibnamefont {{Persson}}}, \bibinfo
  {author} {\bibfnamefont {M.}~\bibnamefont {{Seibert}}}, \ and\ \bibinfo
  {author} {\bibfnamefont {J.}~\bibnamefont {{Rigby}}},\ }\href {\doibase
  10.1088/0004-637X/758/1/24} {\bibfield  {journal} {\bibinfo  {journal}
  {\apj}\ }\textbf {\bibinfo {volume} {758}},\ \bibinfo {eid} {24} (\bibinfo
  {year} {2012})},\ \Eprint {http://arxiv.org/abs/1208.3281} {arXiv:1208.3281}
  \BibitemShut {NoStop}%
\bibitem [{\citenamefont {{York}}\ \emph {et~al.}(2000)\citenamefont {{York}},
  \citenamefont {{Adelman}}, \citenamefont {{Anderson}}, \citenamefont
  {{Anderson}}, \citenamefont {{Annis}}, \citenamefont {{Bahcall}},
  \citenamefont {{Bakken}}, \citenamefont {{Barkhouser}}, \citenamefont
  {{Bastian}}, \citenamefont {{Berman}}, \citenamefont {{Boroski}},
  \citenamefont {{Bracker}}, \citenamefont {{Briegel}}, \citenamefont
  {{Briggs}}, \citenamefont {{Brinkmann}}, \citenamefont {{Brunner}},
  \citenamefont {{Burles}}, \citenamefont {{Carey}}, \citenamefont {{Carr}},
  \citenamefont {{Castander}}, \citenamefont {{Chen}}, \citenamefont
  {{Colestock}}, \citenamefont {{Connolly}}, \citenamefont {{Crocker}},
  \citenamefont {{Csabai}}, \citenamefont {{Czarapata}}, \citenamefont
  {{Davis}}, \citenamefont {{Doi}}, \citenamefont {{Dombeck}}, \citenamefont
  {{Eisenstein}}, \citenamefont {{Ellman}}, \citenamefont {{Elms}},
  \citenamefont {{Evans}}, \citenamefont {{Fan}}, \citenamefont {{Federwitz}},
  \citenamefont {{Fiscelli}}, \citenamefont {{Friedman}}, \citenamefont
  {{Frieman}}, \citenamefont {{Fukugita}}, \citenamefont {{Gillespie}},
  \citenamefont {{Gunn}}, \citenamefont {{Gurbani}}, \citenamefont {{de Haas}},
  \citenamefont {{Haldeman}}, \citenamefont {{Harris}}, \citenamefont
  {{Hayes}}, \citenamefont {{Heckman}}, \citenamefont {{Hennessy}},
  \citenamefont {{Hindsley}}, \citenamefont {{Holm}}, \citenamefont
  {{Holmgren}}, \citenamefont {{Huang}}, \citenamefont {{Hull}}, \citenamefont
  {{Husby}}, \citenamefont {{Ichikawa}}, \citenamefont {{Ichikawa}},
  \citenamefont {{Ivezi{\' c}}}, \citenamefont {{Kent}}, \citenamefont {{Kim}},
  \citenamefont {{Kinney}}, \citenamefont {{Klaene}}, \citenamefont
  {{Kleinman}}, \citenamefont {{Kleinman}}, \citenamefont {{Knapp}},
  \citenamefont {{Korienek}}, \citenamefont {{Kron}}, \citenamefont {{Kunszt}},
  \citenamefont {{Lamb}}, \citenamefont {{Lee}}, \citenamefont {{Leger}},
  \citenamefont {{Limmongkol}}, \citenamefont {{Lindenmeyer}}, \citenamefont
  {{Long}}, \citenamefont {{Loomis}}, \citenamefont {{Loveday}}, \citenamefont
  {{Lucinio}}, \citenamefont {{Lupton}}, \citenamefont {{MacKinnon}},
  \citenamefont {{Mannery}}, \citenamefont {{Mantsch}}, \citenamefont
  {{Margon}}, \citenamefont {{McGehee}}, \citenamefont {{McKay}}, \citenamefont
  {{Meiksin}}, \citenamefont {{Merelli}}, \citenamefont {{Monet}},
  \citenamefont {{Munn}}, \citenamefont {{Narayanan}}, \citenamefont {{Nash}},
  \citenamefont {{Neilsen}}, \citenamefont {{Neswold}}, \citenamefont
  {{Newberg}}, \citenamefont {{Nichol}}, \citenamefont {{Nicinski}},
  \citenamefont {{Nonino}}, \citenamefont {{Okada}}, \citenamefont {{Okamura}},
  \citenamefont {{Ostriker}}, \citenamefont {{Owen}}, \citenamefont {{Pauls}},
  \citenamefont {{Peoples}}, \citenamefont {{Peterson}}, \citenamefont
  {{Petravick}}, \citenamefont {{Pier}}, \citenamefont {{Pope}}, \citenamefont
  {{Pordes}}, \citenamefont {{Prosapio}}, \citenamefont {{Rechenmacher}},
  \citenamefont {{Quinn}}, \citenamefont {{Richards}}, \citenamefont
  {{Richmond}}, \citenamefont {{Rivetta}}, \citenamefont {{Rockosi}},
  \citenamefont {{Ruthmansdorfer}}, \citenamefont {{Sandford}}, \citenamefont
  {{Schlegel}}, \citenamefont {{Schneider}}, \citenamefont {{Sekiguchi}},
  \citenamefont {{Sergey}}, \citenamefont {{Shimasaku}}, \citenamefont
  {{Siegmund}}, \citenamefont {{Smee}}, \citenamefont {{Smith}}, \citenamefont
  {{Snedden}}, \citenamefont {{Stone}}, \citenamefont {{Stoughton}},
  \citenamefont {{Strauss}}, \citenamefont {{Stubbs}}, \citenamefont
  {{SubbaRao}}, \citenamefont {{Szalay}}, \citenamefont {{Szapudi}},
  \citenamefont {{Szokoly}}, \citenamefont {{Thakar}}, \citenamefont
  {{Tremonti}}, \citenamefont {{Tucker}}, \citenamefont {{Uomoto}},
  \citenamefont {{Vanden Berk}}, \citenamefont {{Vogeley}}, \citenamefont
  {{Waddell}}, \citenamefont {{Wang}}, \citenamefont {{Watanabe}},
  \citenamefont {{Weinberg}}, \citenamefont {{Yanny}},\ and\ \citenamefont
  {{Yasuda}}}]{york00a}%
  \BibitemOpen
  \bibfield  {author} {\bibinfo {author} {\bibfnamefont {D.~G.}\ \bibnamefont
  {{York}}}, \bibinfo {author} {\bibfnamefont {J.}~\bibnamefont {{Adelman}}},
  \bibinfo {author} {\bibfnamefont {J.~E.}\ \bibnamefont {{Anderson}}},
  \bibinfo {author} {\bibfnamefont {S.~F.}\ \bibnamefont {{Anderson}}},
  \bibinfo {author} {\bibfnamefont {J.}~\bibnamefont {{Annis}}}, \bibinfo
  {author} {\bibfnamefont {N.~A.}\ \bibnamefont {{Bahcall}}}, \bibinfo {author}
  {\bibfnamefont {J.~A.}\ \bibnamefont {{Bakken}}}, \bibinfo {author}
  {\bibfnamefont {R.}~\bibnamefont {{Barkhouser}}}, \bibinfo {author}
  {\bibfnamefont {S.}~\bibnamefont {{Bastian}}}, \bibinfo {author}
  {\bibfnamefont {E.}~\bibnamefont {{Berman}}}, \bibinfo {author}
  {\bibfnamefont {W.~N.}\ \bibnamefont {{Boroski}}}, \bibinfo {author}
  {\bibfnamefont {S.}~\bibnamefont {{Bracker}}}, \bibinfo {author}
  {\bibfnamefont {C.}~\bibnamefont {{Briegel}}}, \bibinfo {author}
  {\bibfnamefont {J.~W.}\ \bibnamefont {{Briggs}}}, \bibinfo {author}
  {\bibfnamefont {J.}~\bibnamefont {{Brinkmann}}}, \bibinfo {author}
  {\bibfnamefont {R.}~\bibnamefont {{Brunner}}}, \bibinfo {author}
  {\bibfnamefont {S.}~\bibnamefont {{Burles}}}, \bibinfo {author}
  {\bibfnamefont {L.}~\bibnamefont {{Carey}}}, \bibinfo {author} {\bibfnamefont
  {M.~A.}\ \bibnamefont {{Carr}}}, \bibinfo {author} {\bibfnamefont {F.~J.}\
  \bibnamefont {{Castander}}}, \bibinfo {author} {\bibfnamefont
  {B.}~\bibnamefont {{Chen}}}, \bibinfo {author} {\bibfnamefont {P.~L.}\
  \bibnamefont {{Colestock}}}, \bibinfo {author} {\bibfnamefont {A.~J.}\
  \bibnamefont {{Connolly}}}, \bibinfo {author} {\bibfnamefont {J.~H.}\
  \bibnamefont {{Crocker}}}, \bibinfo {author} {\bibfnamefont {I.}~\bibnamefont
  {{Csabai}}}, \bibinfo {author} {\bibfnamefont {P.~C.}\ \bibnamefont
  {{Czarapata}}}, \bibinfo {author} {\bibfnamefont {J.~E.}\ \bibnamefont
  {{Davis}}}, \bibinfo {author} {\bibfnamefont {M.}~\bibnamefont {{Doi}}},
  \bibinfo {author} {\bibfnamefont {T.}~\bibnamefont {{Dombeck}}}, \bibinfo
  {author} {\bibfnamefont {D.}~\bibnamefont {{Eisenstein}}}, \bibinfo {author}
  {\bibfnamefont {N.}~\bibnamefont {{Ellman}}}, \bibinfo {author}
  {\bibfnamefont {B.~R.}\ \bibnamefont {{Elms}}}, \bibinfo {author}
  {\bibfnamefont {M.~L.}\ \bibnamefont {{Evans}}}, \bibinfo {author}
  {\bibfnamefont {X.}~\bibnamefont {{Fan}}}, \bibinfo {author} {\bibfnamefont
  {G.~R.}\ \bibnamefont {{Federwitz}}}, \bibinfo {author} {\bibfnamefont
  {L.}~\bibnamefont {{Fiscelli}}}, \bibinfo {author} {\bibfnamefont
  {S.}~\bibnamefont {{Friedman}}}, \bibinfo {author} {\bibfnamefont {J.~A.}\
  \bibnamefont {{Frieman}}}, \bibinfo {author} {\bibfnamefont {M.}~\bibnamefont
  {{Fukugita}}}, \bibinfo {author} {\bibfnamefont {B.}~\bibnamefont
  {{Gillespie}}}, \bibinfo {author} {\bibfnamefont {J.~E.}\ \bibnamefont
  {{Gunn}}}, \bibinfo {author} {\bibfnamefont {V.~K.}\ \bibnamefont
  {{Gurbani}}}, \bibinfo {author} {\bibfnamefont {E.}~\bibnamefont {{de
  Haas}}}, \bibinfo {author} {\bibfnamefont {M.}~\bibnamefont {{Haldeman}}},
  \bibinfo {author} {\bibfnamefont {F.~H.}\ \bibnamefont {{Harris}}}, \bibinfo
  {author} {\bibfnamefont {J.}~\bibnamefont {{Hayes}}}, \bibinfo {author}
  {\bibfnamefont {T.~M.}\ \bibnamefont {{Heckman}}}, \bibinfo {author}
  {\bibfnamefont {G.~S.}\ \bibnamefont {{Hennessy}}}, \bibinfo {author}
  {\bibfnamefont {R.~B.}\ \bibnamefont {{Hindsley}}}, \bibinfo {author}
  {\bibfnamefont {S.}~\bibnamefont {{Holm}}}, \bibinfo {author} {\bibfnamefont
  {D.~J.}\ \bibnamefont {{Holmgren}}}, \bibinfo {author} {\bibfnamefont
  {C.}~\bibnamefont {{Huang}}}, \bibinfo {author} {\bibfnamefont
  {C.}~\bibnamefont {{Hull}}}, \bibinfo {author} {\bibfnamefont
  {D.}~\bibnamefont {{Husby}}}, \bibinfo {author} {\bibfnamefont
  {S.}~\bibnamefont {{Ichikawa}}}, \bibinfo {author} {\bibfnamefont
  {T.}~\bibnamefont {{Ichikawa}}}, \bibinfo {author} {\bibfnamefont {{\v
  Z}.}~\bibnamefont {{Ivezi{\' c}}}}, \bibinfo {author} {\bibfnamefont
  {S.}~\bibnamefont {{Kent}}}, \bibinfo {author} {\bibfnamefont {R.~S.~J.}\
  \bibnamefont {{Kim}}}, \bibinfo {author} {\bibfnamefont {E.}~\bibnamefont
  {{Kinney}}}, \bibinfo {author} {\bibfnamefont {M.}~\bibnamefont {{Klaene}}},
  \bibinfo {author} {\bibfnamefont {A.~N.}\ \bibnamefont {{Kleinman}}},
  \bibinfo {author} {\bibfnamefont {S.}~\bibnamefont {{Kleinman}}}, \bibinfo
  {author} {\bibfnamefont {G.~R.}\ \bibnamefont {{Knapp}}}, \bibinfo {author}
  {\bibfnamefont {J.}~\bibnamefont {{Korienek}}}, \bibinfo {author}
  {\bibfnamefont {R.~G.}\ \bibnamefont {{Kron}}}, \bibinfo {author}
  {\bibfnamefont {P.~Z.}\ \bibnamefont {{Kunszt}}}, \bibinfo {author}
  {\bibfnamefont {D.~Q.}\ \bibnamefont {{Lamb}}}, \bibinfo {author}
  {\bibfnamefont {B.}~\bibnamefont {{Lee}}}, \bibinfo {author} {\bibfnamefont
  {R.~F.}\ \bibnamefont {{Leger}}}, \bibinfo {author} {\bibfnamefont
  {S.}~\bibnamefont {{Limmongkol}}}, \bibinfo {author} {\bibfnamefont
  {C.}~\bibnamefont {{Lindenmeyer}}}, \bibinfo {author} {\bibfnamefont {D.~C.}\
  \bibnamefont {{Long}}}, \bibinfo {author} {\bibfnamefont {C.}~\bibnamefont
  {{Loomis}}}, \bibinfo {author} {\bibfnamefont {J.}~\bibnamefont {{Loveday}}},
  \bibinfo {author} {\bibfnamefont {R.}~\bibnamefont {{Lucinio}}}, \bibinfo
  {author} {\bibfnamefont {R.~H.}\ \bibnamefont {{Lupton}}}, \bibinfo {author}
  {\bibfnamefont {B.}~\bibnamefont {{MacKinnon}}}, \bibinfo {author}
  {\bibfnamefont {E.~J.}\ \bibnamefont {{Mannery}}}, \bibinfo {author}
  {\bibfnamefont {P.~M.}\ \bibnamefont {{Mantsch}}}, \bibinfo {author}
  {\bibfnamefont {B.}~\bibnamefont {{Margon}}}, \bibinfo {author}
  {\bibfnamefont {P.}~\bibnamefont {{McGehee}}}, \bibinfo {author}
  {\bibfnamefont {T.~A.}\ \bibnamefont {{McKay}}}, \bibinfo {author}
  {\bibfnamefont {A.}~\bibnamefont {{Meiksin}}}, \bibinfo {author}
  {\bibfnamefont {A.}~\bibnamefont {{Merelli}}}, \bibinfo {author}
  {\bibfnamefont {D.~G.}\ \bibnamefont {{Monet}}}, \bibinfo {author}
  {\bibfnamefont {J.~A.}\ \bibnamefont {{Munn}}}, \bibinfo {author}
  {\bibfnamefont {V.~K.}\ \bibnamefont {{Narayanan}}}, \bibinfo {author}
  {\bibfnamefont {T.}~\bibnamefont {{Nash}}}, \bibinfo {author} {\bibfnamefont
  {E.}~\bibnamefont {{Neilsen}}}, \bibinfo {author} {\bibfnamefont
  {R.}~\bibnamefont {{Neswold}}}, \bibinfo {author} {\bibfnamefont {H.~J.}\
  \bibnamefont {{Newberg}}}, \bibinfo {author} {\bibfnamefont {R.~C.}\
  \bibnamefont {{Nichol}}}, \bibinfo {author} {\bibfnamefont {T.}~\bibnamefont
  {{Nicinski}}}, \bibinfo {author} {\bibfnamefont {M.}~\bibnamefont
  {{Nonino}}}, \bibinfo {author} {\bibfnamefont {N.}~\bibnamefont {{Okada}}},
  \bibinfo {author} {\bibfnamefont {S.}~\bibnamefont {{Okamura}}}, \bibinfo
  {author} {\bibfnamefont {J.~P.}\ \bibnamefont {{Ostriker}}}, \bibinfo
  {author} {\bibfnamefont {R.}~\bibnamefont {{Owen}}}, \bibinfo {author}
  {\bibfnamefont {A.~G.}\ \bibnamefont {{Pauls}}}, \bibinfo {author}
  {\bibfnamefont {J.}~\bibnamefont {{Peoples}}}, \bibinfo {author}
  {\bibfnamefont {R.~L.}\ \bibnamefont {{Peterson}}}, \bibinfo {author}
  {\bibfnamefont {D.}~\bibnamefont {{Petravick}}}, \bibinfo {author}
  {\bibfnamefont {J.~R.}\ \bibnamefont {{Pier}}}, \bibinfo {author}
  {\bibfnamefont {A.}~\bibnamefont {{Pope}}}, \bibinfo {author} {\bibfnamefont
  {R.}~\bibnamefont {{Pordes}}}, \bibinfo {author} {\bibfnamefont
  {A.}~\bibnamefont {{Prosapio}}}, \bibinfo {author} {\bibfnamefont
  {R.}~\bibnamefont {{Rechenmacher}}}, \bibinfo {author} {\bibfnamefont
  {T.~R.}\ \bibnamefont {{Quinn}}}, \bibinfo {author} {\bibfnamefont {G.~T.}\
  \bibnamefont {{Richards}}}, \bibinfo {author} {\bibfnamefont {M.~W.}\
  \bibnamefont {{Richmond}}}, \bibinfo {author} {\bibfnamefont {C.~H.}\
  \bibnamefont {{Rivetta}}}, \bibinfo {author} {\bibfnamefont {C.~M.}\
  \bibnamefont {{Rockosi}}}, \bibinfo {author} {\bibfnamefont {K.}~\bibnamefont
  {{Ruthmansdorfer}}}, \bibinfo {author} {\bibfnamefont {D.}~\bibnamefont
  {{Sandford}}}, \bibinfo {author} {\bibfnamefont {D.~J.}\ \bibnamefont
  {{Schlegel}}}, \bibinfo {author} {\bibfnamefont {D.~P.}\ \bibnamefont
  {{Schneider}}}, \bibinfo {author} {\bibfnamefont {M.}~\bibnamefont
  {{Sekiguchi}}}, \bibinfo {author} {\bibfnamefont {G.}~\bibnamefont
  {{Sergey}}}, \bibinfo {author} {\bibfnamefont {K.}~\bibnamefont
  {{Shimasaku}}}, \bibinfo {author} {\bibfnamefont {W.~A.}\ \bibnamefont
  {{Siegmund}}}, \bibinfo {author} {\bibfnamefont {S.}~\bibnamefont {{Smee}}},
  \bibinfo {author} {\bibfnamefont {J.~A.}\ \bibnamefont {{Smith}}}, \bibinfo
  {author} {\bibfnamefont {S.}~\bibnamefont {{Snedden}}}, \bibinfo {author}
  {\bibfnamefont {R.}~\bibnamefont {{Stone}}}, \bibinfo {author} {\bibfnamefont
  {C.}~\bibnamefont {{Stoughton}}}, \bibinfo {author} {\bibfnamefont {M.~A.}\
  \bibnamefont {{Strauss}}}, \bibinfo {author} {\bibfnamefont {C.}~\bibnamefont
  {{Stubbs}}}, \bibinfo {author} {\bibfnamefont {M.}~\bibnamefont
  {{SubbaRao}}}, \bibinfo {author} {\bibfnamefont {A.~S.}\ \bibnamefont
  {{Szalay}}}, \bibinfo {author} {\bibfnamefont {I.}~\bibnamefont {{Szapudi}}},
  \bibinfo {author} {\bibfnamefont {G.~P.}\ \bibnamefont {{Szokoly}}}, \bibinfo
  {author} {\bibfnamefont {A.~R.}\ \bibnamefont {{Thakar}}}, \bibinfo {author}
  {\bibfnamefont {C.}~\bibnamefont {{Tremonti}}}, \bibinfo {author}
  {\bibfnamefont {D.~L.}\ \bibnamefont {{Tucker}}}, \bibinfo {author}
  {\bibfnamefont {A.}~\bibnamefont {{Uomoto}}}, \bibinfo {author}
  {\bibfnamefont {D.}~\bibnamefont {{Vanden Berk}}}, \bibinfo {author}
  {\bibfnamefont {M.~S.}\ \bibnamefont {{Vogeley}}}, \bibinfo {author}
  {\bibfnamefont {P.}~\bibnamefont {{Waddell}}}, \bibinfo {author}
  {\bibfnamefont {S.}~\bibnamefont {{Wang}}}, \bibinfo {author} {\bibfnamefont
  {M.}~\bibnamefont {{Watanabe}}}, \bibinfo {author} {\bibfnamefont {D.~H.}\
  \bibnamefont {{Weinberg}}}, \bibinfo {author} {\bibfnamefont
  {B.}~\bibnamefont {{Yanny}}}, \ and\ \bibinfo {author} {\bibfnamefont
  {N.}~\bibnamefont {{Yasuda}}},\ }\href@noop {} {\bibfield  {journal}
  {\bibinfo  {journal} {\aj}\ }\textbf {\bibinfo {volume} {120}},\ \bibinfo
  {pages} {1579} (\bibinfo {year} {2000})}\BibitemShut {NoStop}%
\bibitem [{\citenamefont {{Eisenstein}}\ \emph {et~al.}(2005)\citenamefont
  {{Eisenstein}}, \citenamefont {{Zehavi}}, \citenamefont {{Hogg}},
  \citenamefont {{Scoccimarro}}, \citenamefont {{Blanton}}, \citenamefont
  {{Nichol}}, \citenamefont {{Scranton}}, \citenamefont {{Seo}}, \citenamefont
  {{Tegmark}}, \citenamefont {{Zheng}}, \citenamefont {{Anderson}},
  \citenamefont {{Annis}}, \citenamefont {{Bahcall}}, \citenamefont
  {{Brinkmann}}, \citenamefont {{Burles}}, \citenamefont {{Castander}},
  \citenamefont {{Connolly}}, \citenamefont {{Csabai}}, \citenamefont {{Doi}},
  \citenamefont {{Fukugita}}, \citenamefont {{Frieman}}, \citenamefont
  {{Glazebrook}}, \citenamefont {{Gunn}}, \citenamefont {{Hendry}},
  \citenamefont {{Hennessy}}, \citenamefont {{Ivezi{\'c}}}, \citenamefont
  {{Kent}}, \citenamefont {{Knapp}}, \citenamefont {{Lin}}, \citenamefont
  {{Loh}}, \citenamefont {{Lupton}}, \citenamefont {{Margon}}, \citenamefont
  {{McKay}}, \citenamefont {{Meiksin}}, \citenamefont {{Munn}}, \citenamefont
  {{Pope}}, \citenamefont {{Richmond}}, \citenamefont {{Schlegel}},
  \citenamefont {{Schneider}}, \citenamefont {{Shimasaku}}, \citenamefont
  {{Stoughton}}, \citenamefont {{Strauss}}, \citenamefont {{SubbaRao}},
  \citenamefont {{Szalay}}, \citenamefont {{Szapudi}}, \citenamefont
  {{Tucker}}, \citenamefont {{Yanny}},\ and\ \citenamefont
  {{York}}}]{eisenstein05a}%
  \BibitemOpen
  \bibfield  {author} {\bibinfo {author} {\bibfnamefont {D.~J.}\ \bibnamefont
  {{Eisenstein}}}, \bibinfo {author} {\bibfnamefont {I.}~\bibnamefont
  {{Zehavi}}}, \bibinfo {author} {\bibfnamefont {D.~W.}\ \bibnamefont
  {{Hogg}}}, \bibinfo {author} {\bibfnamefont {R.}~\bibnamefont
  {{Scoccimarro}}}, \bibinfo {author} {\bibfnamefont {M.~R.}\ \bibnamefont
  {{Blanton}}}, \bibinfo {author} {\bibfnamefont {R.~C.}\ \bibnamefont
  {{Nichol}}}, \bibinfo {author} {\bibfnamefont {R.}~\bibnamefont
  {{Scranton}}}, \bibinfo {author} {\bibfnamefont {H.-J.}\ \bibnamefont
  {{Seo}}}, \bibinfo {author} {\bibfnamefont {M.}~\bibnamefont {{Tegmark}}},
  \bibinfo {author} {\bibfnamefont {Z.}~\bibnamefont {{Zheng}}}, \bibinfo
  {author} {\bibfnamefont {S.~F.}\ \bibnamefont {{Anderson}}}, \bibinfo
  {author} {\bibfnamefont {J.}~\bibnamefont {{Annis}}}, \bibinfo {author}
  {\bibfnamefont {N.}~\bibnamefont {{Bahcall}}}, \bibinfo {author}
  {\bibfnamefont {J.}~\bibnamefont {{Brinkmann}}}, \bibinfo {author}
  {\bibfnamefont {S.}~\bibnamefont {{Burles}}}, \bibinfo {author}
  {\bibfnamefont {F.~J.}\ \bibnamefont {{Castander}}}, \bibinfo {author}
  {\bibfnamefont {A.}~\bibnamefont {{Connolly}}}, \bibinfo {author}
  {\bibfnamefont {I.}~\bibnamefont {{Csabai}}}, \bibinfo {author}
  {\bibfnamefont {M.}~\bibnamefont {{Doi}}}, \bibinfo {author} {\bibfnamefont
  {M.}~\bibnamefont {{Fukugita}}}, \bibinfo {author} {\bibfnamefont {J.~A.}\
  \bibnamefont {{Frieman}}}, \bibinfo {author} {\bibfnamefont {K.}~\bibnamefont
  {{Glazebrook}}}, \bibinfo {author} {\bibfnamefont {J.~E.}\ \bibnamefont
  {{Gunn}}}, \bibinfo {author} {\bibfnamefont {J.~S.}\ \bibnamefont
  {{Hendry}}}, \bibinfo {author} {\bibfnamefont {G.}~\bibnamefont
  {{Hennessy}}}, \bibinfo {author} {\bibfnamefont {Z.}~\bibnamefont
  {{Ivezi{\'c}}}}, \bibinfo {author} {\bibfnamefont {S.}~\bibnamefont
  {{Kent}}}, \bibinfo {author} {\bibfnamefont {G.~R.}\ \bibnamefont {{Knapp}}},
  \bibinfo {author} {\bibfnamefont {H.}~\bibnamefont {{Lin}}}, \bibinfo
  {author} {\bibfnamefont {Y.-S.}\ \bibnamefont {{Loh}}}, \bibinfo {author}
  {\bibfnamefont {R.~H.}\ \bibnamefont {{Lupton}}}, \bibinfo {author}
  {\bibfnamefont {B.}~\bibnamefont {{Margon}}}, \bibinfo {author}
  {\bibfnamefont {T.~A.}\ \bibnamefont {{McKay}}}, \bibinfo {author}
  {\bibfnamefont {A.}~\bibnamefont {{Meiksin}}}, \bibinfo {author}
  {\bibfnamefont {J.~A.}\ \bibnamefont {{Munn}}}, \bibinfo {author}
  {\bibfnamefont {A.}~\bibnamefont {{Pope}}}, \bibinfo {author} {\bibfnamefont
  {M.~W.}\ \bibnamefont {{Richmond}}}, \bibinfo {author} {\bibfnamefont
  {D.}~\bibnamefont {{Schlegel}}}, \bibinfo {author} {\bibfnamefont {D.~P.}\
  \bibnamefont {{Schneider}}}, \bibinfo {author} {\bibfnamefont
  {K.}~\bibnamefont {{Shimasaku}}}, \bibinfo {author} {\bibfnamefont
  {C.}~\bibnamefont {{Stoughton}}}, \bibinfo {author} {\bibfnamefont {M.~A.}\
  \bibnamefont {{Strauss}}}, \bibinfo {author} {\bibfnamefont {M.}~\bibnamefont
  {{SubbaRao}}}, \bibinfo {author} {\bibfnamefont {A.~S.}\ \bibnamefont
  {{Szalay}}}, \bibinfo {author} {\bibfnamefont {I.}~\bibnamefont {{Szapudi}}},
  \bibinfo {author} {\bibfnamefont {D.~L.}\ \bibnamefont {{Tucker}}}, \bibinfo
  {author} {\bibfnamefont {B.}~\bibnamefont {{Yanny}}}, \ and\ \bibinfo
  {author} {\bibfnamefont {D.~G.}\ \bibnamefont {{York}}},\ }\href {\doibase
  10.1086/466512} {\bibfield  {journal} {\bibinfo  {journal} {\apj}\ }\textbf
  {\bibinfo {volume} {633}},\ \bibinfo {pages} {560} (\bibinfo {year}
  {2005})},\ \Eprint {http://arxiv.org/abs/arXiv:astro-ph/0501171}
  {arXiv:astro-ph/0501171} \BibitemShut {NoStop}%
\bibitem [{\citenamefont {{Cole}}\ \emph {et~al.}(2005)\citenamefont {{Cole}},
  \citenamefont {{Percival}}, \citenamefont {{Peacock}}, \citenamefont
  {{Norberg}}, \citenamefont {{Baugh}}, \citenamefont {{Frenk}}, \citenamefont
  {{Baldry}}, \citenamefont {{Bland-Hawthorn}}, \citenamefont {{Bridges}},
  \citenamefont {{Cannon}}, \citenamefont {{Colless}}, \citenamefont
  {{Collins}}, \citenamefont {{Couch}}, \citenamefont {{Cross}}, \citenamefont
  {{Dalton}}, \citenamefont {{Eke}}, \citenamefont {{De Propris}},
  \citenamefont {{Driver}}, \citenamefont {{Efstathiou}}, \citenamefont
  {{Ellis}}, \citenamefont {{Glazebrook}}, \citenamefont {{Jackson}},
  \citenamefont {{Jenkins}}, \citenamefont {{Lahav}}, \citenamefont {{Lewis}},
  \citenamefont {{Lumsden}}, \citenamefont {{Maddox}}, \citenamefont
  {{Madgwick}}, \citenamefont {{Peterson}}, \citenamefont {{Sutherland}},\ and\
  \citenamefont {{Taylor}}}]{cole05a}%
  \BibitemOpen
  \bibfield  {author} {\bibinfo {author} {\bibfnamefont {S.}~\bibnamefont
  {{Cole}}}, \bibinfo {author} {\bibfnamefont {W.~J.}\ \bibnamefont
  {{Percival}}}, \bibinfo {author} {\bibfnamefont {J.~A.}\ \bibnamefont
  {{Peacock}}}, \bibinfo {author} {\bibfnamefont {P.}~\bibnamefont
  {{Norberg}}}, \bibinfo {author} {\bibfnamefont {C.~M.}\ \bibnamefont
  {{Baugh}}}, \bibinfo {author} {\bibfnamefont {C.~S.}\ \bibnamefont
  {{Frenk}}}, \bibinfo {author} {\bibfnamefont {I.}~\bibnamefont {{Baldry}}},
  \bibinfo {author} {\bibfnamefont {J.}~\bibnamefont {{Bland-Hawthorn}}},
  \bibinfo {author} {\bibfnamefont {T.}~\bibnamefont {{Bridges}}}, \bibinfo
  {author} {\bibfnamefont {R.}~\bibnamefont {{Cannon}}}, \bibinfo {author}
  {\bibfnamefont {M.}~\bibnamefont {{Colless}}}, \bibinfo {author}
  {\bibfnamefont {C.}~\bibnamefont {{Collins}}}, \bibinfo {author}
  {\bibfnamefont {W.}~\bibnamefont {{Couch}}}, \bibinfo {author} {\bibfnamefont
  {N.~J.~G.}\ \bibnamefont {{Cross}}}, \bibinfo {author} {\bibfnamefont
  {G.}~\bibnamefont {{Dalton}}}, \bibinfo {author} {\bibfnamefont {V.~R.}\
  \bibnamefont {{Eke}}}, \bibinfo {author} {\bibfnamefont {R.}~\bibnamefont
  {{De Propris}}}, \bibinfo {author} {\bibfnamefont {S.~P.}\ \bibnamefont
  {{Driver}}}, \bibinfo {author} {\bibfnamefont {G.}~\bibnamefont
  {{Efstathiou}}}, \bibinfo {author} {\bibfnamefont {R.~S.}\ \bibnamefont
  {{Ellis}}}, \bibinfo {author} {\bibfnamefont {K.}~\bibnamefont
  {{Glazebrook}}}, \bibinfo {author} {\bibfnamefont {C.}~\bibnamefont
  {{Jackson}}}, \bibinfo {author} {\bibfnamefont {A.}~\bibnamefont
  {{Jenkins}}}, \bibinfo {author} {\bibfnamefont {O.}~\bibnamefont {{Lahav}}},
  \bibinfo {author} {\bibfnamefont {I.}~\bibnamefont {{Lewis}}}, \bibinfo
  {author} {\bibfnamefont {S.}~\bibnamefont {{Lumsden}}}, \bibinfo {author}
  {\bibfnamefont {S.}~\bibnamefont {{Maddox}}}, \bibinfo {author}
  {\bibfnamefont {D.}~\bibnamefont {{Madgwick}}}, \bibinfo {author}
  {\bibfnamefont {B.~A.}\ \bibnamefont {{Peterson}}}, \bibinfo {author}
  {\bibfnamefont {W.}~\bibnamefont {{Sutherland}}}, \ and\ \bibinfo {author}
  {\bibfnamefont {K.}~\bibnamefont {{Taylor}}},\ }\href {\doibase
  10.1111/j.1365-2966.2005.09318.x} {\bibfield  {journal} {\bibinfo  {journal}
  {\mnras}\ }\textbf {\bibinfo {volume} {362}},\ \bibinfo {pages} {505}
  (\bibinfo {year} {2005})},\ \Eprint
  {http://arxiv.org/abs/arXiv:astro-ph/0501174} {arXiv:astro-ph/0501174}
  \BibitemShut {NoStop}%
\bibitem [{\citenamefont {{Percival}}\ \emph {et~al.}(2007)\citenamefont
  {{Percival}}, \citenamefont {{Cole}}, \citenamefont {{Eisenstein}},
  \citenamefont {{Nichol}}, \citenamefont {{Peacock}}, \citenamefont {{Pope}},\
  and\ \citenamefont {{Szalay}}}]{percival07a}%
  \BibitemOpen
  \bibfield  {author} {\bibinfo {author} {\bibfnamefont {W.~J.}\ \bibnamefont
  {{Percival}}}, \bibinfo {author} {\bibfnamefont {S.}~\bibnamefont {{Cole}}},
  \bibinfo {author} {\bibfnamefont {D.~J.}\ \bibnamefont {{Eisenstein}}},
  \bibinfo {author} {\bibfnamefont {R.~C.}\ \bibnamefont {{Nichol}}}, \bibinfo
  {author} {\bibfnamefont {J.~A.}\ \bibnamefont {{Peacock}}}, \bibinfo {author}
  {\bibfnamefont {A.~C.}\ \bibnamefont {{Pope}}}, \ and\ \bibinfo {author}
  {\bibfnamefont {A.~S.}\ \bibnamefont {{Szalay}}},\ }\href {\doibase
  10.1111/j.1365-2966.2007.12268.x} {\bibfield  {journal} {\bibinfo  {journal}
  {\mnras}\ }\textbf {\bibinfo {volume} {381}},\ \bibinfo {pages} {1053}
  (\bibinfo {year} {2007})},\ \Eprint {http://arxiv.org/abs/0705.3323}
  {arXiv:0705.3323} \BibitemShut {NoStop}%
\bibitem [{\citenamefont {{Hinshaw}}\ \emph {et~al.}(2009)\citenamefont
  {{Hinshaw}}, \citenamefont {{Weiland}}, \citenamefont {{Hill}}, \citenamefont
  {{Odegard}}, \citenamefont {{Larson}}, \citenamefont {{Bennett}},
  \citenamefont {{Dunkley}}, \citenamefont {{Gold}}, \citenamefont {{Greason}},
  \citenamefont {{Jarosik}}, \citenamefont {{Komatsu}}, \citenamefont
  {{Nolta}}, \citenamefont {{Page}}, \citenamefont {{Spergel}}, \citenamefont
  {{Wollack}}, \citenamefont {{Halpern}}, \citenamefont {{Kogut}},
  \citenamefont {{Limon}}, \citenamefont {{Meyer}}, \citenamefont {{Tucker}},\
  and\ \citenamefont {{Wright}}}]{hinshaw08a}%
  \BibitemOpen
  \bibfield  {author} {\bibinfo {author} {\bibfnamefont {G.}~\bibnamefont
  {{Hinshaw}}}, \bibinfo {author} {\bibfnamefont {J.~L.}\ \bibnamefont
  {{Weiland}}}, \bibinfo {author} {\bibfnamefont {R.~S.}\ \bibnamefont
  {{Hill}}}, \bibinfo {author} {\bibfnamefont {N.}~\bibnamefont {{Odegard}}},
  \bibinfo {author} {\bibfnamefont {D.}~\bibnamefont {{Larson}}}, \bibinfo
  {author} {\bibfnamefont {C.~L.}\ \bibnamefont {{Bennett}}}, \bibinfo {author}
  {\bibfnamefont {J.}~\bibnamefont {{Dunkley}}}, \bibinfo {author}
  {\bibfnamefont {B.}~\bibnamefont {{Gold}}}, \bibinfo {author} {\bibfnamefont
  {M.~R.}\ \bibnamefont {{Greason}}}, \bibinfo {author} {\bibfnamefont
  {N.}~\bibnamefont {{Jarosik}}}, \bibinfo {author} {\bibfnamefont
  {E.}~\bibnamefont {{Komatsu}}}, \bibinfo {author} {\bibfnamefont {M.~R.}\
  \bibnamefont {{Nolta}}}, \bibinfo {author} {\bibfnamefont {L.}~\bibnamefont
  {{Page}}}, \bibinfo {author} {\bibfnamefont {D.~N.}\ \bibnamefont
  {{Spergel}}}, \bibinfo {author} {\bibfnamefont {E.}~\bibnamefont
  {{Wollack}}}, \bibinfo {author} {\bibfnamefont {M.}~\bibnamefont
  {{Halpern}}}, \bibinfo {author} {\bibfnamefont {A.}~\bibnamefont {{Kogut}}},
  \bibinfo {author} {\bibfnamefont {M.}~\bibnamefont {{Limon}}}, \bibinfo
  {author} {\bibfnamefont {S.~S.}\ \bibnamefont {{Meyer}}}, \bibinfo {author}
  {\bibfnamefont {G.~S.}\ \bibnamefont {{Tucker}}}, \ and\ \bibinfo {author}
  {\bibfnamefont {E.~L.}\ \bibnamefont {{Wright}}},\ }\href {\doibase
  10.1088/0067-0049/180/2/225} {\bibfield  {journal} {\bibinfo  {journal}
  {\apjs}\ }\textbf {\bibinfo {volume} {180}},\ \bibinfo {pages} {225}
  (\bibinfo {year} {2009})},\ \Eprint {http://arxiv.org/abs/0803.0732}
  {arXiv:0803.0732 [astro-ph]} \BibitemShut {NoStop}%
\bibitem [{\citenamefont {{Kowalski}}\ \emph {et~al.}(2008)\citenamefont
  {{Kowalski}}, \citenamefont {{Rubin}}, \citenamefont {{Aldering}},
  \citenamefont {{Agostinho}}, \citenamefont {{Amadon}}, \citenamefont
  {{Amanullah}}, \citenamefont {{Balland}}, \citenamefont {{Barbary}},
  \citenamefont {{Blanc}}, \citenamefont {{Challis}}, \citenamefont {{Conley}},
  \citenamefont {{Connolly}}, \citenamefont {{Covarrubias}}, \citenamefont
  {{Dawson}}, \citenamefont {{Deustua}}, \citenamefont {{Ellis}}, \citenamefont
  {{Fabbro}}, \citenamefont {{Fadeyev}}, \citenamefont {{Fan}}, \citenamefont
  {{Farris}}, \citenamefont {{Folatelli}}, \citenamefont {{Frye}},
  \citenamefont {{Garavini}}, \citenamefont {{Gates}}, \citenamefont
  {{Germany}}, \citenamefont {{Goldhaber}}, \citenamefont {{Goldman}},
  \citenamefont {{Goobar}}, \citenamefont {{Groom}}, \citenamefont
  {{Haissinski}}, \citenamefont {{Hardin}}, \citenamefont {{Hook}},
  \citenamefont {{Kent}}, \citenamefont {{Kim}}, \citenamefont {{Knop}},
  \citenamefont {{Lidman}}, \citenamefont {{Linder}}, \citenamefont {{Mendez}},
  \citenamefont {{Meyers}}, \citenamefont {{Miller}}, \citenamefont {{Moniez}},
  \citenamefont {{Mour{\~a}o}}, \citenamefont {{Newberg}}, \citenamefont
  {{Nobili}}, \citenamefont {{Nugent}}, \citenamefont {{Pain}}, \citenamefont
  {{Perdereau}}, \citenamefont {{Perlmutter}}, \citenamefont {{Phillips}},
  \citenamefont {{Prasad}}, \citenamefont {{Quimby}}, \citenamefont
  {{Regnault}}, \citenamefont {{Rich}}, \citenamefont {{Rubenstein}},
  \citenamefont {{Ruiz-Lapuente}}, \citenamefont {{Santos}}, \citenamefont
  {{Schaefer}}, \citenamefont {{Schommer}}, \citenamefont {{Smith}},
  \citenamefont {{Soderberg}}, \citenamefont {{Spadafora}}, \citenamefont
  {{Strolger}}, \citenamefont {{Strovink}}, \citenamefont {{Suntzeff}},
  \citenamefont {{Suzuki}}, \citenamefont {{Thomas}}, \citenamefont {{Walton}},
  \citenamefont {{Wang}}, \citenamefont {{Wood-Vasey}},\ and\ \citenamefont
  {{Yun}}}]{kowalski08b}%
  \BibitemOpen
  \bibfield  {author} {\bibinfo {author} {\bibfnamefont {M.}~\bibnamefont
  {{Kowalski}}}, \bibinfo {author} {\bibfnamefont {D.}~\bibnamefont {{Rubin}}},
  \bibinfo {author} {\bibfnamefont {G.}~\bibnamefont {{Aldering}}}, \bibinfo
  {author} {\bibfnamefont {R.~J.}\ \bibnamefont {{Agostinho}}}, \bibinfo
  {author} {\bibfnamefont {A.}~\bibnamefont {{Amadon}}}, \bibinfo {author}
  {\bibfnamefont {R.}~\bibnamefont {{Amanullah}}}, \bibinfo {author}
  {\bibfnamefont {C.}~\bibnamefont {{Balland}}}, \bibinfo {author}
  {\bibfnamefont {K.}~\bibnamefont {{Barbary}}}, \bibinfo {author}
  {\bibfnamefont {G.}~\bibnamefont {{Blanc}}}, \bibinfo {author} {\bibfnamefont
  {P.~J.}\ \bibnamefont {{Challis}}}, \bibinfo {author} {\bibfnamefont
  {A.}~\bibnamefont {{Conley}}}, \bibinfo {author} {\bibfnamefont {N.~V.}\
  \bibnamefont {{Connolly}}}, \bibinfo {author} {\bibfnamefont
  {R.}~\bibnamefont {{Covarrubias}}}, \bibinfo {author} {\bibfnamefont {K.~S.}\
  \bibnamefont {{Dawson}}}, \bibinfo {author} {\bibfnamefont {S.~E.}\
  \bibnamefont {{Deustua}}}, \bibinfo {author} {\bibfnamefont {R.}~\bibnamefont
  {{Ellis}}}, \bibinfo {author} {\bibfnamefont {S.}~\bibnamefont {{Fabbro}}},
  \bibinfo {author} {\bibfnamefont {V.}~\bibnamefont {{Fadeyev}}}, \bibinfo
  {author} {\bibfnamefont {X.}~\bibnamefont {{Fan}}}, \bibinfo {author}
  {\bibfnamefont {B.}~\bibnamefont {{Farris}}}, \bibinfo {author}
  {\bibfnamefont {G.}~\bibnamefont {{Folatelli}}}, \bibinfo {author}
  {\bibfnamefont {B.~L.}\ \bibnamefont {{Frye}}}, \bibinfo {author}
  {\bibfnamefont {G.}~\bibnamefont {{Garavini}}}, \bibinfo {author}
  {\bibfnamefont {E.~L.}\ \bibnamefont {{Gates}}}, \bibinfo {author}
  {\bibfnamefont {L.}~\bibnamefont {{Germany}}}, \bibinfo {author}
  {\bibfnamefont {G.}~\bibnamefont {{Goldhaber}}}, \bibinfo {author}
  {\bibfnamefont {B.}~\bibnamefont {{Goldman}}}, \bibinfo {author}
  {\bibfnamefont {A.}~\bibnamefont {{Goobar}}}, \bibinfo {author}
  {\bibfnamefont {D.~E.}\ \bibnamefont {{Groom}}}, \bibinfo {author}
  {\bibfnamefont {J.}~\bibnamefont {{Haissinski}}}, \bibinfo {author}
  {\bibfnamefont {D.}~\bibnamefont {{Hardin}}}, \bibinfo {author}
  {\bibfnamefont {I.}~\bibnamefont {{Hook}}}, \bibinfo {author} {\bibfnamefont
  {S.}~\bibnamefont {{Kent}}}, \bibinfo {author} {\bibfnamefont {A.~G.}\
  \bibnamefont {{Kim}}}, \bibinfo {author} {\bibfnamefont {R.~A.}\ \bibnamefont
  {{Knop}}}, \bibinfo {author} {\bibfnamefont {C.}~\bibnamefont {{Lidman}}},
  \bibinfo {author} {\bibfnamefont {E.~V.}\ \bibnamefont {{Linder}}}, \bibinfo
  {author} {\bibfnamefont {J.}~\bibnamefont {{Mendez}}}, \bibinfo {author}
  {\bibfnamefont {J.}~\bibnamefont {{Meyers}}}, \bibinfo {author}
  {\bibfnamefont {G.~J.}\ \bibnamefont {{Miller}}}, \bibinfo {author}
  {\bibfnamefont {M.}~\bibnamefont {{Moniez}}}, \bibinfo {author}
  {\bibfnamefont {A.~M.}\ \bibnamefont {{Mour{\~a}o}}}, \bibinfo {author}
  {\bibfnamefont {H.}~\bibnamefont {{Newberg}}}, \bibinfo {author}
  {\bibfnamefont {S.}~\bibnamefont {{Nobili}}}, \bibinfo {author}
  {\bibfnamefont {P.~E.}\ \bibnamefont {{Nugent}}}, \bibinfo {author}
  {\bibfnamefont {R.}~\bibnamefont {{Pain}}}, \bibinfo {author} {\bibfnamefont
  {O.}~\bibnamefont {{Perdereau}}}, \bibinfo {author} {\bibfnamefont
  {S.}~\bibnamefont {{Perlmutter}}}, \bibinfo {author} {\bibfnamefont {M.~M.}\
  \bibnamefont {{Phillips}}}, \bibinfo {author} {\bibfnamefont
  {V.}~\bibnamefont {{Prasad}}}, \bibinfo {author} {\bibfnamefont
  {R.}~\bibnamefont {{Quimby}}}, \bibinfo {author} {\bibfnamefont
  {N.}~\bibnamefont {{Regnault}}}, \bibinfo {author} {\bibfnamefont
  {J.}~\bibnamefont {{Rich}}}, \bibinfo {author} {\bibfnamefont {E.~P.}\
  \bibnamefont {{Rubenstein}}}, \bibinfo {author} {\bibfnamefont
  {P.}~\bibnamefont {{Ruiz-Lapuente}}}, \bibinfo {author} {\bibfnamefont
  {F.~D.}\ \bibnamefont {{Santos}}}, \bibinfo {author} {\bibfnamefont {B.~E.}\
  \bibnamefont {{Schaefer}}}, \bibinfo {author} {\bibfnamefont {R.~A.}\
  \bibnamefont {{Schommer}}}, \bibinfo {author} {\bibfnamefont {R.~C.}\
  \bibnamefont {{Smith}}}, \bibinfo {author} {\bibfnamefont {A.~M.}\
  \bibnamefont {{Soderberg}}}, \bibinfo {author} {\bibfnamefont {A.~L.}\
  \bibnamefont {{Spadafora}}}, \bibinfo {author} {\bibfnamefont {L.-G.}\
  \bibnamefont {{Strolger}}}, \bibinfo {author} {\bibfnamefont
  {M.}~\bibnamefont {{Strovink}}}, \bibinfo {author} {\bibfnamefont {N.~B.}\
  \bibnamefont {{Suntzeff}}}, \bibinfo {author} {\bibfnamefont
  {N.}~\bibnamefont {{Suzuki}}}, \bibinfo {author} {\bibfnamefont {R.~C.}\
  \bibnamefont {{Thomas}}}, \bibinfo {author} {\bibfnamefont {N.~A.}\
  \bibnamefont {{Walton}}}, \bibinfo {author} {\bibfnamefont {L.}~\bibnamefont
  {{Wang}}}, \bibinfo {author} {\bibfnamefont {W.~M.}\ \bibnamefont
  {{Wood-Vasey}}}, \ and\ \bibinfo {author} {\bibfnamefont {J.~L.}\
  \bibnamefont {{Yun}}},\ }\href {\doibase 10.1086/589937} {\bibfield
  {journal} {\bibinfo  {journal} {\apj}\ }\textbf {\bibinfo {volume} {686}},\
  \bibinfo {pages} {749} (\bibinfo {year} {2008})},\ \Eprint
  {http://arxiv.org/abs/0804.4142} {arXiv:0804.4142} \BibitemShut {NoStop}%
\bibitem [{\citenamefont {{Komatsu}}\ \emph {et~al.}(2009)\citenamefont
  {{Komatsu}}, \citenamefont {{Dunkley}}, \citenamefont {{Nolta}},
  \citenamefont {{Bennett}}, \citenamefont {{Gold}}, \citenamefont {{Hinshaw}},
  \citenamefont {{Jarosik}}, \citenamefont {{Larson}}, \citenamefont {{Limon}},
  \citenamefont {{Page}}, \citenamefont {{Spergel}}, \citenamefont {{Halpern}},
  \citenamefont {{Hill}}, \citenamefont {{Kogut}}, \citenamefont {{Meyer}},
  \citenamefont {{Tucker}}, \citenamefont {{Weiland}}, \citenamefont
  {{Wollack}},\ and\ \citenamefont {{Wright}}}]{komatsu09a}%
  \BibitemOpen
  \bibfield  {author} {\bibinfo {author} {\bibfnamefont {E.}~\bibnamefont
  {{Komatsu}}}, \bibinfo {author} {\bibfnamefont {J.}~\bibnamefont
  {{Dunkley}}}, \bibinfo {author} {\bibfnamefont {M.~R.}\ \bibnamefont
  {{Nolta}}}, \bibinfo {author} {\bibfnamefont {C.~L.}\ \bibnamefont
  {{Bennett}}}, \bibinfo {author} {\bibfnamefont {B.}~\bibnamefont {{Gold}}},
  \bibinfo {author} {\bibfnamefont {G.}~\bibnamefont {{Hinshaw}}}, \bibinfo
  {author} {\bibfnamefont {N.}~\bibnamefont {{Jarosik}}}, \bibinfo {author}
  {\bibfnamefont {D.}~\bibnamefont {{Larson}}}, \bibinfo {author}
  {\bibfnamefont {M.}~\bibnamefont {{Limon}}}, \bibinfo {author} {\bibfnamefont
  {L.}~\bibnamefont {{Page}}}, \bibinfo {author} {\bibfnamefont {D.~N.}\
  \bibnamefont {{Spergel}}}, \bibinfo {author} {\bibfnamefont {M.}~\bibnamefont
  {{Halpern}}}, \bibinfo {author} {\bibfnamefont {R.~S.}\ \bibnamefont
  {{Hill}}}, \bibinfo {author} {\bibfnamefont {A.}~\bibnamefont {{Kogut}}},
  \bibinfo {author} {\bibfnamefont {S.~S.}\ \bibnamefont {{Meyer}}}, \bibinfo
  {author} {\bibfnamefont {G.~S.}\ \bibnamefont {{Tucker}}}, \bibinfo {author}
  {\bibfnamefont {J.~L.}\ \bibnamefont {{Weiland}}}, \bibinfo {author}
  {\bibfnamefont {E.}~\bibnamefont {{Wollack}}}, \ and\ \bibinfo {author}
  {\bibfnamefont {E.~L.}\ \bibnamefont {{Wright}}},\ }\href {\doibase
  10.1088/0067-0049/180/2/330} {\bibfield  {journal} {\bibinfo  {journal}
  {\apjs}\ }\textbf {\bibinfo {volume} {180}},\ \bibinfo {pages} {330}
  (\bibinfo {year} {2009})},\ \Eprint {http://arxiv.org/abs/0803.0547}
  {arXiv:0803.0547} \BibitemShut {NoStop}%
\bibitem [{\citenamefont {{Planck Collaboration}}\ \emph
  {et~al.}(2014)\citenamefont {{Planck Collaboration}}, \citenamefont {{Ade}},
  \citenamefont {{Aghanim}}, \citenamefont {{Armitage-Caplan}}, \citenamefont
  {{Arnaud}}, \citenamefont {{Ashdown}}, \citenamefont {{Atrio-Barand ela}},
  \citenamefont {{Aumont}}, \citenamefont {{Baccigalupi}}, \citenamefont
  {{Banday}}, \citenamefont {{Barreiro}}, \citenamefont {{Bartlett}},
  \citenamefont {{Basak}}, \citenamefont {{Battaner}}, \citenamefont
  {{Benabed}}, \citenamefont {{Beno{\^\i}t}}, \citenamefont
  {{Benoit-L{\'e}vy}}, \citenamefont {{Bernard}}, \citenamefont {{Bersanelli}},
  \citenamefont {{Bielewicz}}, \citenamefont {{Bobin}}, \citenamefont {{Bock}},
  \citenamefont {{Bonaldi}}, \citenamefont {{Bonavera}}, \citenamefont
  {{Bond}}, \citenamefont {{Borrill}}, \citenamefont {{Bouchet}}, \citenamefont
  {{Bridges}}, \citenamefont {{Bucher}}, \citenamefont {{Burigana}},
  \citenamefont {{Butler}}, \citenamefont {{Cardoso}}, \citenamefont
  {{Catalano}}, \citenamefont {{Challinor}}, \citenamefont {{Chamballu}},
  \citenamefont {{Chiang}}, \citenamefont {{Chiang}}, \citenamefont
  {{Christensen}}, \citenamefont {{Church}}, \citenamefont {{Clements}},
  \citenamefont {{Colombi}}, \citenamefont {{Colombo}}, \citenamefont
  {{Couchot}}, \citenamefont {{Coulais}}, \citenamefont {{Crill}},
  \citenamefont {{Curto}}, \citenamefont {{Cuttaia}}, \citenamefont {{Danese}},
  \citenamefont {{Davies}}, \citenamefont {{Davis}}, \citenamefont {{de
  Bernardis}}, \citenamefont {{de Rosa}}, \citenamefont {{de Zotti}},
  \citenamefont {{D{\'e}chelette}}, \citenamefont {{Delabrouille}},
  \citenamefont {{Delouis}}, \citenamefont {{D{\'e}sert}}, \citenamefont
  {{Dickinson}}, \citenamefont {{Diego}}, \citenamefont {{Dole}}, \citenamefont
  {{Donzelli}}, \citenamefont {{Dor{\'e}}}, \citenamefont {{Douspis}},
  \citenamefont {{Dunkley}}, \citenamefont {{Dupac}}, \citenamefont
  {{Efstathiou}}, \citenamefont {{En{\ss}lin}}, \citenamefont {{Eriksen}},
  \citenamefont {{Finelli}}, \citenamefont {{Forni}}, \citenamefont
  {{Frailis}}, \citenamefont {{Franceschi}}, \citenamefont {{Galeotta}},
  \citenamefont {{Ganga}}, \citenamefont {{Giard}}, \citenamefont {{Giardino}},
  \citenamefont {{Giraud-H{\'e}raud}}, \citenamefont {{Gonz{\'a}lez-Nuevo}},
  \citenamefont {{G{\'o}rski}}, \citenamefont {{Gratton}}, \citenamefont
  {{Gregorio}}, \citenamefont {{Gruppuso}}, \citenamefont {{Gudmundsson}},
  \citenamefont {{Hansen}}, \citenamefont {{Hanson}}, \citenamefont
  {{Harrison}}, \citenamefont {{Henrot-Versill{\'e}}}, \citenamefont
  {{Hern{\'a}ndez-Monteagudo}}, \citenamefont {{Herranz}}, \citenamefont
  {{Hildebrand t}}, \citenamefont {{Hivon}}, \citenamefont {{Ho}},
  \citenamefont {{Hobson}}, \citenamefont {{Holmes}}, \citenamefont
  {{Hornstrup}}, \citenamefont {{Hovest}}, \citenamefont {{Huffenberger}},
  \citenamefont {{Jaffe}}, \citenamefont {{Jaffe}}, \citenamefont {{Jones}},
  \citenamefont {{Juvela}}, \citenamefont {{Keih{\"a}nen}}, \citenamefont
  {{Keskitalo}}, \citenamefont {{Kisner}}, \citenamefont {{Kneissl}},
  \citenamefont {{Knoche}}, \citenamefont {{Knox}}, \citenamefont {{Kunz}},
  \citenamefont {{Kurki-Suonio}}, \citenamefont {{Lagache}}, \citenamefont
  {{L{\"a}hteenm{\"a}ki}}, \citenamefont {{Lamarre}}, \citenamefont
  {{Lasenby}}, \citenamefont {{Laureijs}}, \citenamefont {{Lavabre}},
  \citenamefont {{Lawrence}}, \citenamefont {{Leahy}}, \citenamefont
  {{Leonardi}}, \citenamefont {{Le{\'o}n-Tavares}}, \citenamefont
  {{Lesgourgues}}, \citenamefont {{Lewis}}, \citenamefont {{Liguori}},
  \citenamefont {{Lilje}}, \citenamefont {{Linden-V{\o}rnle}}, \citenamefont
  {{L{\'o}pez-Caniego}}, \citenamefont {{Lubin}}, \citenamefont
  {{Mac{\'\i}as-P{\'e}rez}}, \citenamefont {{Maffei}}, \citenamefont {{Maino}},
  \citenamefont {{Mand olesi}}, \citenamefont {{Mangilli}}, \citenamefont
  {{Maris}}, \citenamefont {{Marshall}}, \citenamefont {{Martin}},
  \citenamefont {{Mart{\'\i}nez-Gonz{\'a}lez}}, \citenamefont {{Masi}},
  \citenamefont {{Massardi}}, \citenamefont {{Matarrese}}, \citenamefont
  {{Matthai}}, \citenamefont {{Mazzotta}}, \citenamefont {{Melchiorri}},
  \citenamefont {{Mendes}}, \citenamefont {{Mennella}}, \citenamefont
  {{Migliaccio}}, \citenamefont {{Mitra}}, \citenamefont
  {{Miville-Desch{\^e}nes}}, \citenamefont {{Moneti}}, \citenamefont
  {{Montier}}, \citenamefont {{Morgante}}, \citenamefont {{Mortlock}},
  \citenamefont {{Moss}}, \citenamefont {{Munshi}}, \citenamefont {{Murphy}},
  \citenamefont {{Naselsky}}, \citenamefont {{Nati}}, \citenamefont {{Natoli}},
  \citenamefont {{Netterfield}}, \citenamefont {{N{\o}rgaard-Nielsen}},
  \citenamefont {{Noviello}}, \citenamefont {{Novikov}}, \citenamefont
  {{Novikov}}, \citenamefont {{Osborne}}, \citenamefont {{Oxborrow}},
  \citenamefont {{Paci}}, \citenamefont {{Pagano}}, \citenamefont {{Pajot}},
  \citenamefont {{Paoletti}}, \citenamefont {{Partridge}}, \citenamefont
  {{Pasian}}, \citenamefont {{Patanchon}}, \citenamefont {{Perdereau}},
  \citenamefont {{Perotto}}, \citenamefont {{Perrotta}}, \citenamefont
  {{Piacentini}}, \citenamefont {{Piat}}, \citenamefont {{Pierpaoli}},
  \citenamefont {{Pietrobon}}, \citenamefont {{Plaszczynski}}, \citenamefont
  {{Pointecouteau}}, \citenamefont {{Polenta}}, \citenamefont {{Ponthieu}},
  \citenamefont {{Popa}}, \citenamefont {{Poutanen}}, \citenamefont {{Pratt}},
  \citenamefont {{Pr{\'e}zeau}}, \citenamefont {{Prunet}}, \citenamefont
  {{Puget}}, \citenamefont {{Pullen}}, \citenamefont {{Rachen}}, \citenamefont
  {{Rebolo}}, \citenamefont {{Reinecke}}, \citenamefont {{Remazeilles}},
  \citenamefont {{Renault}}, \citenamefont {{Ricciardi}}, \citenamefont
  {{Riller}}, \citenamefont {{Ristorcelli}}, \citenamefont {{Rocha}},
  \citenamefont {{Rosset}}, \citenamefont {{Roudier}}, \citenamefont
  {{Rowan-Robinson}}, \citenamefont {{Rubi{\~n}o-Mart{\'\i}n}}, \citenamefont
  {{Rusholme}}, \citenamefont {{Sandri}}, \citenamefont {{Santos}},
  \citenamefont {{Savini}}, \citenamefont {{Scott}}, \citenamefont
  {{Seiffert}}, \citenamefont {{Shellard}}, \citenamefont {{Smith}},
  \citenamefont {{Spencer}}, \citenamefont {{Starck}}, \citenamefont
  {{Stolyarov}}, \citenamefont {{Stompor}}, \citenamefont {{Sudiwala}},
  \citenamefont {{Sunyaev}}, \citenamefont {{Sureau}}, \citenamefont
  {{Sutton}}, \citenamefont {{Suur-Uski}}, \citenamefont {{Sygnet}},
  \citenamefont {{Tauber}}, \citenamefont {{Tavagnacco}}, \citenamefont
  {{Terenzi}}, \citenamefont {{Toffolatti}}, \citenamefont {{Tomasi}},
  \citenamefont {{Tristram}}, \citenamefont {{Tucci}}, \citenamefont
  {{Tuovinen}}, \citenamefont {{Umana}}, \citenamefont {{Valenziano}},
  \citenamefont {{Valiviita}}, \citenamefont {{Van Tent}}, \citenamefont
  {{Vielva}}, \citenamefont {{Villa}}, \citenamefont {{Vittorio}},
  \citenamefont {{Wade}}, \citenamefont {{Wandelt}}, \citenamefont {{White}},
  \citenamefont {{White}}, \citenamefont {{Yvon}}, \citenamefont {{Zacchei}},\
  and\ \citenamefont {{Zonca}}}]{2014A&A...571A..17P}%
  \BibitemOpen
  \bibfield  {author} {\bibinfo {author} {\bibnamefont {{Planck
  Collaboration}}}, \bibinfo {author} {\bibfnamefont {P.~A.~R.}\ \bibnamefont
  {{Ade}}}, \bibinfo {author} {\bibfnamefont {N.}~\bibnamefont {{Aghanim}}},
  \bibinfo {author} {\bibfnamefont {C.}~\bibnamefont {{Armitage-Caplan}}},
  \bibinfo {author} {\bibfnamefont {M.}~\bibnamefont {{Arnaud}}}, \bibinfo
  {author} {\bibfnamefont {M.}~\bibnamefont {{Ashdown}}}, \bibinfo {author}
  {\bibfnamefont {F.}~\bibnamefont {{Atrio-Barand ela}}}, \bibinfo {author}
  {\bibfnamefont {J.}~\bibnamefont {{Aumont}}}, \bibinfo {author}
  {\bibfnamefont {C.}~\bibnamefont {{Baccigalupi}}}, \bibinfo {author}
  {\bibfnamefont {A.~J.}\ \bibnamefont {{Banday}}}, \bibinfo {author}
  {\bibfnamefont {R.~B.}\ \bibnamefont {{Barreiro}}}, \bibinfo {author}
  {\bibfnamefont {J.~G.}\ \bibnamefont {{Bartlett}}}, \bibinfo {author}
  {\bibfnamefont {S.}~\bibnamefont {{Basak}}}, \bibinfo {author} {\bibfnamefont
  {E.}~\bibnamefont {{Battaner}}}, \bibinfo {author} {\bibfnamefont
  {K.}~\bibnamefont {{Benabed}}}, \bibinfo {author} {\bibfnamefont
  {A.}~\bibnamefont {{Beno{\^\i}t}}}, \bibinfo {author} {\bibfnamefont
  {A.}~\bibnamefont {{Benoit-L{\'e}vy}}}, \bibinfo {author} {\bibfnamefont
  {J.~P.}\ \bibnamefont {{Bernard}}}, \bibinfo {author} {\bibfnamefont
  {M.}~\bibnamefont {{Bersanelli}}}, \bibinfo {author} {\bibfnamefont
  {P.}~\bibnamefont {{Bielewicz}}}, \bibinfo {author} {\bibfnamefont
  {J.}~\bibnamefont {{Bobin}}}, \bibinfo {author} {\bibfnamefont {J.~J.}\
  \bibnamefont {{Bock}}}, \bibinfo {author} {\bibfnamefont {A.}~\bibnamefont
  {{Bonaldi}}}, \bibinfo {author} {\bibfnamefont {L.}~\bibnamefont
  {{Bonavera}}}, \bibinfo {author} {\bibfnamefont {J.~R.}\ \bibnamefont
  {{Bond}}}, \bibinfo {author} {\bibfnamefont {J.}~\bibnamefont {{Borrill}}},
  \bibinfo {author} {\bibfnamefont {F.~R.}\ \bibnamefont {{Bouchet}}}, \bibinfo
  {author} {\bibfnamefont {M.}~\bibnamefont {{Bridges}}}, \bibinfo {author}
  {\bibfnamefont {M.}~\bibnamefont {{Bucher}}}, \bibinfo {author}
  {\bibfnamefont {C.}~\bibnamefont {{Burigana}}}, \bibinfo {author}
  {\bibfnamefont {R.~C.}\ \bibnamefont {{Butler}}}, \bibinfo {author}
  {\bibfnamefont {J.~F.}\ \bibnamefont {{Cardoso}}}, \bibinfo {author}
  {\bibfnamefont {A.}~\bibnamefont {{Catalano}}}, \bibinfo {author}
  {\bibfnamefont {A.}~\bibnamefont {{Challinor}}}, \bibinfo {author}
  {\bibfnamefont {A.}~\bibnamefont {{Chamballu}}}, \bibinfo {author}
  {\bibfnamefont {H.~C.}\ \bibnamefont {{Chiang}}}, \bibinfo {author}
  {\bibfnamefont {L.~Y.}\ \bibnamefont {{Chiang}}}, \bibinfo {author}
  {\bibfnamefont {P.~R.}\ \bibnamefont {{Christensen}}}, \bibinfo {author}
  {\bibfnamefont {S.}~\bibnamefont {{Church}}}, \bibinfo {author}
  {\bibfnamefont {D.~L.}\ \bibnamefont {{Clements}}}, \bibinfo {author}
  {\bibfnamefont {S.}~\bibnamefont {{Colombi}}}, \bibinfo {author}
  {\bibfnamefont {L.~P.~L.}\ \bibnamefont {{Colombo}}}, \bibinfo {author}
  {\bibfnamefont {F.}~\bibnamefont {{Couchot}}}, \bibinfo {author}
  {\bibfnamefont {A.}~\bibnamefont {{Coulais}}}, \bibinfo {author}
  {\bibfnamefont {B.~P.}\ \bibnamefont {{Crill}}}, \bibinfo {author}
  {\bibfnamefont {A.}~\bibnamefont {{Curto}}}, \bibinfo {author} {\bibfnamefont
  {F.}~\bibnamefont {{Cuttaia}}}, \bibinfo {author} {\bibfnamefont
  {L.}~\bibnamefont {{Danese}}}, \bibinfo {author} {\bibfnamefont {R.~D.}\
  \bibnamefont {{Davies}}}, \bibinfo {author} {\bibfnamefont {R.~J.}\
  \bibnamefont {{Davis}}}, \bibinfo {author} {\bibfnamefont {P.}~\bibnamefont
  {{de Bernardis}}}, \bibinfo {author} {\bibfnamefont {A.}~\bibnamefont {{de
  Rosa}}}, \bibinfo {author} {\bibfnamefont {G.}~\bibnamefont {{de Zotti}}},
  \bibinfo {author} {\bibfnamefont {T.}~\bibnamefont {{D{\'e}chelette}}},
  \bibinfo {author} {\bibfnamefont {J.}~\bibnamefont {{Delabrouille}}},
  \bibinfo {author} {\bibfnamefont {J.~M.}\ \bibnamefont {{Delouis}}}, \bibinfo
  {author} {\bibfnamefont {F.~X.}\ \bibnamefont {{D{\'e}sert}}}, \bibinfo
  {author} {\bibfnamefont {C.}~\bibnamefont {{Dickinson}}}, \bibinfo {author}
  {\bibfnamefont {J.~M.}\ \bibnamefont {{Diego}}}, \bibinfo {author}
  {\bibfnamefont {H.}~\bibnamefont {{Dole}}}, \bibinfo {author} {\bibfnamefont
  {S.}~\bibnamefont {{Donzelli}}}, \bibinfo {author} {\bibfnamefont
  {O.}~\bibnamefont {{Dor{\'e}}}}, \bibinfo {author} {\bibfnamefont
  {M.}~\bibnamefont {{Douspis}}}, \bibinfo {author} {\bibfnamefont
  {J.}~\bibnamefont {{Dunkley}}}, \bibinfo {author} {\bibfnamefont
  {X.}~\bibnamefont {{Dupac}}}, \bibinfo {author} {\bibfnamefont
  {G.}~\bibnamefont {{Efstathiou}}}, \bibinfo {author} {\bibfnamefont {T.~A.}\
  \bibnamefont {{En{\ss}lin}}}, \bibinfo {author} {\bibfnamefont {H.~K.}\
  \bibnamefont {{Eriksen}}}, \bibinfo {author} {\bibfnamefont {F.}~\bibnamefont
  {{Finelli}}}, \bibinfo {author} {\bibfnamefont {O.}~\bibnamefont {{Forni}}},
  \bibinfo {author} {\bibfnamefont {M.}~\bibnamefont {{Frailis}}}, \bibinfo
  {author} {\bibfnamefont {E.}~\bibnamefont {{Franceschi}}}, \bibinfo {author}
  {\bibfnamefont {S.}~\bibnamefont {{Galeotta}}}, \bibinfo {author}
  {\bibfnamefont {K.}~\bibnamefont {{Ganga}}}, \bibinfo {author} {\bibfnamefont
  {M.}~\bibnamefont {{Giard}}}, \bibinfo {author} {\bibfnamefont
  {G.}~\bibnamefont {{Giardino}}}, \bibinfo {author} {\bibfnamefont
  {Y.}~\bibnamefont {{Giraud-H{\'e}raud}}}, \bibinfo {author} {\bibfnamefont
  {J.}~\bibnamefont {{Gonz{\'a}lez-Nuevo}}}, \bibinfo {author} {\bibfnamefont
  {K.~M.}\ \bibnamefont {{G{\'o}rski}}}, \bibinfo {author} {\bibfnamefont
  {S.}~\bibnamefont {{Gratton}}}, \bibinfo {author} {\bibfnamefont
  {A.}~\bibnamefont {{Gregorio}}}, \bibinfo {author} {\bibfnamefont
  {A.}~\bibnamefont {{Gruppuso}}}, \bibinfo {author} {\bibfnamefont {J.~E.}\
  \bibnamefont {{Gudmundsson}}}, \bibinfo {author} {\bibfnamefont {F.~K.}\
  \bibnamefont {{Hansen}}}, \bibinfo {author} {\bibfnamefont {D.}~\bibnamefont
  {{Hanson}}}, \bibinfo {author} {\bibfnamefont {D.}~\bibnamefont
  {{Harrison}}}, \bibinfo {author} {\bibfnamefont {S.}~\bibnamefont
  {{Henrot-Versill{\'e}}}}, \bibinfo {author} {\bibfnamefont {C.}~\bibnamefont
  {{Hern{\'a}ndez-Monteagudo}}}, \bibinfo {author} {\bibfnamefont
  {D.}~\bibnamefont {{Herranz}}}, \bibinfo {author} {\bibfnamefont {S.~R.}\
  \bibnamefont {{Hildebrand t}}}, \bibinfo {author} {\bibfnamefont
  {E.}~\bibnamefont {{Hivon}}}, \bibinfo {author} {\bibfnamefont
  {S.}~\bibnamefont {{Ho}}}, \bibinfo {author} {\bibfnamefont {M.}~\bibnamefont
  {{Hobson}}}, \bibinfo {author} {\bibfnamefont {W.~A.}\ \bibnamefont
  {{Holmes}}}, \bibinfo {author} {\bibfnamefont {A.}~\bibnamefont
  {{Hornstrup}}}, \bibinfo {author} {\bibfnamefont {W.}~\bibnamefont
  {{Hovest}}}, \bibinfo {author} {\bibfnamefont {K.~M.}\ \bibnamefont
  {{Huffenberger}}}, \bibinfo {author} {\bibfnamefont {A.~H.}\ \bibnamefont
  {{Jaffe}}}, \bibinfo {author} {\bibfnamefont {T.~R.}\ \bibnamefont
  {{Jaffe}}}, \bibinfo {author} {\bibfnamefont {W.~C.}\ \bibnamefont
  {{Jones}}}, \bibinfo {author} {\bibfnamefont {M.}~\bibnamefont {{Juvela}}},
  \bibinfo {author} {\bibfnamefont {E.}~\bibnamefont {{Keih{\"a}nen}}},
  \bibinfo {author} {\bibfnamefont {R.}~\bibnamefont {{Keskitalo}}}, \bibinfo
  {author} {\bibfnamefont {T.~S.}\ \bibnamefont {{Kisner}}}, \bibinfo {author}
  {\bibfnamefont {R.}~\bibnamefont {{Kneissl}}}, \bibinfo {author}
  {\bibfnamefont {J.}~\bibnamefont {{Knoche}}}, \bibinfo {author}
  {\bibfnamefont {L.}~\bibnamefont {{Knox}}}, \bibinfo {author} {\bibfnamefont
  {M.}~\bibnamefont {{Kunz}}}, \bibinfo {author} {\bibfnamefont
  {H.}~\bibnamefont {{Kurki-Suonio}}}, \bibinfo {author} {\bibfnamefont
  {G.}~\bibnamefont {{Lagache}}}, \bibinfo {author} {\bibfnamefont
  {A.}~\bibnamefont {{L{\"a}hteenm{\"a}ki}}}, \bibinfo {author} {\bibfnamefont
  {J.~M.}\ \bibnamefont {{Lamarre}}}, \bibinfo {author} {\bibfnamefont
  {A.}~\bibnamefont {{Lasenby}}}, \bibinfo {author} {\bibfnamefont {R.~J.}\
  \bibnamefont {{Laureijs}}}, \bibinfo {author} {\bibfnamefont
  {A.}~\bibnamefont {{Lavabre}}}, \bibinfo {author} {\bibfnamefont {C.~R.}\
  \bibnamefont {{Lawrence}}}, \bibinfo {author} {\bibfnamefont {J.~P.}\
  \bibnamefont {{Leahy}}}, \bibinfo {author} {\bibfnamefont {R.}~\bibnamefont
  {{Leonardi}}}, \bibinfo {author} {\bibfnamefont {J.}~\bibnamefont
  {{Le{\'o}n-Tavares}}}, \bibinfo {author} {\bibfnamefont {J.}~\bibnamefont
  {{Lesgourgues}}}, \bibinfo {author} {\bibfnamefont {A.}~\bibnamefont
  {{Lewis}}}, \bibinfo {author} {\bibfnamefont {M.}~\bibnamefont {{Liguori}}},
  \bibinfo {author} {\bibfnamefont {P.~B.}\ \bibnamefont {{Lilje}}}, \bibinfo
  {author} {\bibfnamefont {M.}~\bibnamefont {{Linden-V{\o}rnle}}}, \bibinfo
  {author} {\bibfnamefont {M.}~\bibnamefont {{L{\'o}pez-Caniego}}}, \bibinfo
  {author} {\bibfnamefont {P.~M.}\ \bibnamefont {{Lubin}}}, \bibinfo {author}
  {\bibfnamefont {J.~F.}\ \bibnamefont {{Mac{\'\i}as-P{\'e}rez}}}, \bibinfo
  {author} {\bibfnamefont {B.}~\bibnamefont {{Maffei}}}, \bibinfo {author}
  {\bibfnamefont {D.}~\bibnamefont {{Maino}}}, \bibinfo {author} {\bibfnamefont
  {N.}~\bibnamefont {{Mand olesi}}}, \bibinfo {author} {\bibfnamefont
  {A.}~\bibnamefont {{Mangilli}}}, \bibinfo {author} {\bibfnamefont
  {M.}~\bibnamefont {{Maris}}}, \bibinfo {author} {\bibfnamefont {D.~J.}\
  \bibnamefont {{Marshall}}}, \bibinfo {author} {\bibfnamefont {P.~G.}\
  \bibnamefont {{Martin}}}, \bibinfo {author} {\bibfnamefont {E.}~\bibnamefont
  {{Mart{\'\i}nez-Gonz{\'a}lez}}}, \bibinfo {author} {\bibfnamefont
  {S.}~\bibnamefont {{Masi}}}, \bibinfo {author} {\bibfnamefont
  {M.}~\bibnamefont {{Massardi}}}, \bibinfo {author} {\bibfnamefont
  {S.}~\bibnamefont {{Matarrese}}}, \bibinfo {author} {\bibfnamefont
  {F.}~\bibnamefont {{Matthai}}}, \bibinfo {author} {\bibfnamefont
  {P.}~\bibnamefont {{Mazzotta}}}, \bibinfo {author} {\bibfnamefont
  {A.}~\bibnamefont {{Melchiorri}}}, \bibinfo {author} {\bibfnamefont
  {L.}~\bibnamefont {{Mendes}}}, \bibinfo {author} {\bibfnamefont
  {A.}~\bibnamefont {{Mennella}}}, \bibinfo {author} {\bibfnamefont
  {M.}~\bibnamefont {{Migliaccio}}}, \bibinfo {author} {\bibfnamefont
  {S.}~\bibnamefont {{Mitra}}}, \bibinfo {author} {\bibfnamefont {M.~A.}\
  \bibnamefont {{Miville-Desch{\^e}nes}}}, \bibinfo {author} {\bibfnamefont
  {A.}~\bibnamefont {{Moneti}}}, \bibinfo {author} {\bibfnamefont
  {L.}~\bibnamefont {{Montier}}}, \bibinfo {author} {\bibfnamefont
  {G.}~\bibnamefont {{Morgante}}}, \bibinfo {author} {\bibfnamefont
  {D.}~\bibnamefont {{Mortlock}}}, \bibinfo {author} {\bibfnamefont
  {A.}~\bibnamefont {{Moss}}}, \bibinfo {author} {\bibfnamefont
  {D.}~\bibnamefont {{Munshi}}}, \bibinfo {author} {\bibfnamefont {J.~A.}\
  \bibnamefont {{Murphy}}}, \bibinfo {author} {\bibfnamefont {P.}~\bibnamefont
  {{Naselsky}}}, \bibinfo {author} {\bibfnamefont {F.}~\bibnamefont {{Nati}}},
  \bibinfo {author} {\bibfnamefont {P.}~\bibnamefont {{Natoli}}}, \bibinfo
  {author} {\bibfnamefont {C.~B.}\ \bibnamefont {{Netterfield}}}, \bibinfo
  {author} {\bibfnamefont {H.~U.}\ \bibnamefont {{N{\o}rgaard-Nielsen}}},
  \bibinfo {author} {\bibfnamefont {F.}~\bibnamefont {{Noviello}}}, \bibinfo
  {author} {\bibfnamefont {D.}~\bibnamefont {{Novikov}}}, \bibinfo {author}
  {\bibfnamefont {I.}~\bibnamefont {{Novikov}}}, \bibinfo {author}
  {\bibfnamefont {S.}~\bibnamefont {{Osborne}}}, \bibinfo {author}
  {\bibfnamefont {C.~A.}\ \bibnamefont {{Oxborrow}}}, \bibinfo {author}
  {\bibfnamefont {F.}~\bibnamefont {{Paci}}}, \bibinfo {author} {\bibfnamefont
  {L.}~\bibnamefont {{Pagano}}}, \bibinfo {author} {\bibfnamefont
  {F.}~\bibnamefont {{Pajot}}}, \bibinfo {author} {\bibfnamefont
  {D.}~\bibnamefont {{Paoletti}}}, \bibinfo {author} {\bibfnamefont
  {B.}~\bibnamefont {{Partridge}}}, \bibinfo {author} {\bibfnamefont
  {F.}~\bibnamefont {{Pasian}}}, \bibinfo {author} {\bibfnamefont
  {G.}~\bibnamefont {{Patanchon}}}, \bibinfo {author} {\bibfnamefont
  {O.}~\bibnamefont {{Perdereau}}}, \bibinfo {author} {\bibfnamefont
  {L.}~\bibnamefont {{Perotto}}}, \bibinfo {author} {\bibfnamefont
  {F.}~\bibnamefont {{Perrotta}}}, \bibinfo {author} {\bibfnamefont
  {F.}~\bibnamefont {{Piacentini}}}, \bibinfo {author} {\bibfnamefont
  {M.}~\bibnamefont {{Piat}}}, \bibinfo {author} {\bibfnamefont
  {E.}~\bibnamefont {{Pierpaoli}}}, \bibinfo {author} {\bibfnamefont
  {D.}~\bibnamefont {{Pietrobon}}}, \bibinfo {author} {\bibfnamefont
  {S.}~\bibnamefont {{Plaszczynski}}}, \bibinfo {author} {\bibfnamefont
  {E.}~\bibnamefont {{Pointecouteau}}}, \bibinfo {author} {\bibfnamefont
  {G.}~\bibnamefont {{Polenta}}}, \bibinfo {author} {\bibfnamefont
  {N.}~\bibnamefont {{Ponthieu}}}, \bibinfo {author} {\bibfnamefont
  {L.}~\bibnamefont {{Popa}}}, \bibinfo {author} {\bibfnamefont
  {T.}~\bibnamefont {{Poutanen}}}, \bibinfo {author} {\bibfnamefont {G.~W.}\
  \bibnamefont {{Pratt}}}, \bibinfo {author} {\bibfnamefont {G.}~\bibnamefont
  {{Pr{\'e}zeau}}}, \bibinfo {author} {\bibfnamefont {S.}~\bibnamefont
  {{Prunet}}}, \bibinfo {author} {\bibfnamefont {J.~L.}\ \bibnamefont
  {{Puget}}}, \bibinfo {author} {\bibfnamefont {A.~R.}\ \bibnamefont
  {{Pullen}}}, \bibinfo {author} {\bibfnamefont {J.~P.}\ \bibnamefont
  {{Rachen}}}, \bibinfo {author} {\bibfnamefont {R.}~\bibnamefont {{Rebolo}}},
  \bibinfo {author} {\bibfnamefont {M.}~\bibnamefont {{Reinecke}}}, \bibinfo
  {author} {\bibfnamefont {M.}~\bibnamefont {{Remazeilles}}}, \bibinfo {author}
  {\bibfnamefont {C.}~\bibnamefont {{Renault}}}, \bibinfo {author}
  {\bibfnamefont {S.}~\bibnamefont {{Ricciardi}}}, \bibinfo {author}
  {\bibfnamefont {T.}~\bibnamefont {{Riller}}}, \bibinfo {author}
  {\bibfnamefont {I.}~\bibnamefont {{Ristorcelli}}}, \bibinfo {author}
  {\bibfnamefont {G.}~\bibnamefont {{Rocha}}}, \bibinfo {author} {\bibfnamefont
  {C.}~\bibnamefont {{Rosset}}}, \bibinfo {author} {\bibfnamefont
  {G.}~\bibnamefont {{Roudier}}}, \bibinfo {author} {\bibfnamefont
  {M.}~\bibnamefont {{Rowan-Robinson}}}, \bibinfo {author} {\bibfnamefont
  {J.~A.}\ \bibnamefont {{Rubi{\~n}o-Mart{\'\i}n}}}, \bibinfo {author}
  {\bibfnamefont {B.}~\bibnamefont {{Rusholme}}}, \bibinfo {author}
  {\bibfnamefont {M.}~\bibnamefont {{Sandri}}}, \bibinfo {author}
  {\bibfnamefont {D.}~\bibnamefont {{Santos}}}, \bibinfo {author}
  {\bibfnamefont {G.}~\bibnamefont {{Savini}}}, \bibinfo {author}
  {\bibfnamefont {D.}~\bibnamefont {{Scott}}}, \bibinfo {author} {\bibfnamefont
  {M.~D.}\ \bibnamefont {{Seiffert}}}, \bibinfo {author} {\bibfnamefont
  {E.~P.~S.}\ \bibnamefont {{Shellard}}}, \bibinfo {author} {\bibfnamefont
  {K.}~\bibnamefont {{Smith}}}, \bibinfo {author} {\bibfnamefont {L.~D.}\
  \bibnamefont {{Spencer}}}, \bibinfo {author} {\bibfnamefont {J.~L.}\
  \bibnamefont {{Starck}}}, \bibinfo {author} {\bibfnamefont {V.}~\bibnamefont
  {{Stolyarov}}}, \bibinfo {author} {\bibfnamefont {R.}~\bibnamefont
  {{Stompor}}}, \bibinfo {author} {\bibfnamefont {R.}~\bibnamefont
  {{Sudiwala}}}, \bibinfo {author} {\bibfnamefont {R.}~\bibnamefont
  {{Sunyaev}}}, \bibinfo {author} {\bibfnamefont {F.}~\bibnamefont {{Sureau}}},
  \bibinfo {author} {\bibfnamefont {D.}~\bibnamefont {{Sutton}}}, \bibinfo
  {author} {\bibfnamefont {A.~S.}\ \bibnamefont {{Suur-Uski}}}, \bibinfo
  {author} {\bibfnamefont {J.~F.}\ \bibnamefont {{Sygnet}}}, \bibinfo {author}
  {\bibfnamefont {J.~A.}\ \bibnamefont {{Tauber}}}, \bibinfo {author}
  {\bibfnamefont {D.}~\bibnamefont {{Tavagnacco}}}, \bibinfo {author}
  {\bibfnamefont {L.}~\bibnamefont {{Terenzi}}}, \bibinfo {author}
  {\bibfnamefont {L.}~\bibnamefont {{Toffolatti}}}, \bibinfo {author}
  {\bibfnamefont {M.}~\bibnamefont {{Tomasi}}}, \bibinfo {author}
  {\bibfnamefont {M.}~\bibnamefont {{Tristram}}}, \bibinfo {author}
  {\bibfnamefont {M.}~\bibnamefont {{Tucci}}}, \bibinfo {author} {\bibfnamefont
  {J.}~\bibnamefont {{Tuovinen}}}, \bibinfo {author} {\bibfnamefont
  {G.}~\bibnamefont {{Umana}}}, \bibinfo {author} {\bibfnamefont
  {L.}~\bibnamefont {{Valenziano}}}, \bibinfo {author} {\bibfnamefont
  {J.}~\bibnamefont {{Valiviita}}}, \bibinfo {author} {\bibfnamefont
  {B.}~\bibnamefont {{Van Tent}}}, \bibinfo {author} {\bibfnamefont
  {P.}~\bibnamefont {{Vielva}}}, \bibinfo {author} {\bibfnamefont
  {F.}~\bibnamefont {{Villa}}}, \bibinfo {author} {\bibfnamefont
  {N.}~\bibnamefont {{Vittorio}}}, \bibinfo {author} {\bibfnamefont {L.~A.}\
  \bibnamefont {{Wade}}}, \bibinfo {author} {\bibfnamefont {B.~D.}\
  \bibnamefont {{Wandelt}}}, \bibinfo {author} {\bibfnamefont {M.}~\bibnamefont
  {{White}}}, \bibinfo {author} {\bibfnamefont {S.~D.~M.}\ \bibnamefont
  {{White}}}, \bibinfo {author} {\bibfnamefont {D.}~\bibnamefont {{Yvon}}},
  \bibinfo {author} {\bibfnamefont {A.}~\bibnamefont {{Zacchei}}}, \ and\
  \bibinfo {author} {\bibfnamefont {A.}~\bibnamefont {{Zonca}}},\ }\href
  {\doibase 10.1051/0004-6361/201321543} {\bibfield  {journal} {\bibinfo
  {journal} {\aap}\ }\textbf {\bibinfo {volume} {571}},\ \bibinfo {eid} {A17}
  (\bibinfo {year} {2014})},\ \Eprint {http://arxiv.org/abs/1303.5077}
  {arXiv:1303.5077 [astro-ph.CO]} \BibitemShut {NoStop}%
\bibitem [{\citenamefont {{Planck Collaboration}}\ \emph
  {et~al.}(2018{\natexlab{a}})\citenamefont {{Planck Collaboration}},
  \citenamefont {{Aghanim}}, \citenamefont {{Akrami}}, \citenamefont
  {{Ashdown}}, \citenamefont {{Aumont}}, \citenamefont {{Baccigalupi}},
  \citenamefont {{Ballardini}}, \citenamefont {{Banday}}, \citenamefont
  {{Barreiro}}, \citenamefont {{Bartolo}}, \citenamefont {{Basak}},
  \citenamefont {{Benabed}}, \citenamefont {{Bernard}}, \citenamefont
  {{Bersanelli}}, \citenamefont {{Bielewicz}}, \citenamefont {{Bock}},
  \citenamefont {{Bond}}, \citenamefont {{Borrill}}, \citenamefont {{Bouchet}},
  \citenamefont {{Boulanger}}, \citenamefont {{Bucher}}, \citenamefont
  {{Burigana}}, \citenamefont {{Calabrese}}, \citenamefont {{Cardoso}},
  \citenamefont {{Carron}}, \citenamefont {{Challinor}}, \citenamefont
  {{Chiang}}, \citenamefont {{Colombo}}, \citenamefont {{Combet}},
  \citenamefont {{Crill}}, \citenamefont {{Cuttaia}}, \citenamefont {{de
  Bernardis}}, \citenamefont {{de Zotti}}, \citenamefont {{Delabrouille}},
  \citenamefont {{Di Valentino}}, \citenamefont {{Diego}}, \citenamefont
  {{Dor{\'e}}}, \citenamefont {{Douspis}}, \citenamefont {{Ducout}},
  \citenamefont {{Dupac}}, \citenamefont {{Efstathiou}}, \citenamefont
  {{Elsner}}, \citenamefont {{En{\ss}lin}}, \citenamefont {{Eriksen}},
  \citenamefont {{Fantaye}}, \citenamefont {{Fernandez-Cobos}}, \citenamefont
  {{Forastieri}}, \citenamefont {{Frailis}}, \citenamefont {{Fraisse}},
  \citenamefont {{Franceschi}}, \citenamefont {{Frolov}}, \citenamefont
  {{Galeotta}}, \citenamefont {{Galli}}, \citenamefont {{Ganga}}, \citenamefont
  {{G{\'e}nova-Santos}}, \citenamefont {{Gerbino}}, \citenamefont {{Ghosh}},
  \citenamefont {{Gonz{\'a}lez-Nuevo}}, \citenamefont {{G{\'o}rski}},
  \citenamefont {{Gratton}}, \citenamefont {{Gruppuso}}, \citenamefont
  {{Gudmundsson}}, \citenamefont {{Hamann}}, \citenamefont {{Hand ley}},
  \citenamefont {{Hansen}}, \citenamefont {{Herranz}}, \citenamefont {{Hivon}},
  \citenamefont {{Huang}}, \citenamefont {{Jaffe}}, \citenamefont {{Jones}},
  \citenamefont {{Karakci}}, \citenamefont {{Keih{\"a}nen}}, \citenamefont
  {{Keskitalo}}, \citenamefont {{Kiiveri}}, \citenamefont {{Kim}},
  \citenamefont {{Knox}}, \citenamefont {{Krachmalnicoff}}, \citenamefont
  {{Kunz}}, \citenamefont {{Kurki-Suonio}}, \citenamefont {{Lagache}},
  \citenamefont {{Lamarre}}, \citenamefont {{Lasenby}}, \citenamefont
  {{Lattanzi}}, \citenamefont {{Lawrence}}, \citenamefont {{Le Jeune}},
  \citenamefont {{Levrier}}, \citenamefont {{Lewis}}, \citenamefont
  {{Liguori}}, \citenamefont {{Lilje}}, \citenamefont {{Lindholm}},
  \citenamefont {{L{\'o}pez-Caniego}}, \citenamefont {{Lubin}}, \citenamefont
  {{Ma}}, \citenamefont {{Mac{\'\i}as-P{\'e}rez}}, \citenamefont {{Maggio}},
  \citenamefont {{Maino}}, \citenamefont {{Mandolesi}}, \citenamefont
  {{Mangilli}}, \citenamefont {{Marcos-Caballero}}, \citenamefont {{Maris}},
  \citenamefont {{Martin}}, \citenamefont {{Mart{\'\i}nez-Gonz{\'a}lez}},
  \citenamefont {{Matarrese}}, \citenamefont {{Mauri}}, \citenamefont
  {{McEwen}}, \citenamefont {{Melchiorri}}, \citenamefont {{Mennella}},
  \citenamefont {{Migliaccio}}, \citenamefont {{Miville-Desch{\^e}nes}},
  \citenamefont {{Molinari}}, \citenamefont {{Moneti}}, \citenamefont
  {{Montier}}, \citenamefont {{Morgante}}, \citenamefont {{Moss}},
  \citenamefont {{Natoli}}, \citenamefont {{Pagano}}, \citenamefont
  {{Paoletti}}, \citenamefont {{Partridge}}, \citenamefont {{Patanchon}},
  \citenamefont {{Perrotta}}, \citenamefont {{Pettorino}}, \citenamefont
  {{Piacentini}}, \citenamefont {{Polastri}}, \citenamefont {{Polenta}},
  \citenamefont {{Puget}}, \citenamefont {{Rachen}}, \citenamefont
  {{Reinecke}}, \citenamefont {{Remazeilles}}, \citenamefont {{Renzi}},
  \citenamefont {{Rocha}}, \citenamefont {{Rosset}}, \citenamefont {{Roudier}},
  \citenamefont {{Rubi{\~n}o-Mart{\'\i}n}}, \citenamefont {{Ruiz-Granados}},
  \citenamefont {{Salvati}}, \citenamefont {{Sandri}}, \citenamefont
  {{Savelainen}}, \citenamefont {{Scott}}, \citenamefont {{Sirignano}},
  \citenamefont {{Sunyaev}}, \citenamefont {{Suur-Uski}}, \citenamefont
  {{Tauber}}, \citenamefont {{Tavagnacco}}, \citenamefont {{Tenti}},
  \citenamefont {{Toffolatti}}, \citenamefont {{Tomasi}}, \citenamefont
  {{Trombetti}}, \citenamefont {{Valiviita}}, \citenamefont {{Van Tent}},
  \citenamefont {{Vielva}}, \citenamefont {{Villa}}, \citenamefont
  {{Vittorio}}, \citenamefont {{Wandelt}}, \citenamefont {{Wehus}},
  \citenamefont {{White}}, \citenamefont {{White}}, \citenamefont {{Zacchei}},\
  and\ \citenamefont {{Zonca}}}]{plancklensing}%
  \BibitemOpen
  \bibfield  {author} {\bibinfo {author} {\bibnamefont {{Planck
  Collaboration}}}, \bibinfo {author} {\bibfnamefont {N.}~\bibnamefont
  {{Aghanim}}}, \bibinfo {author} {\bibfnamefont {Y.}~\bibnamefont {{Akrami}}},
  \bibinfo {author} {\bibfnamefont {M.}~\bibnamefont {{Ashdown}}}, \bibinfo
  {author} {\bibfnamefont {J.}~\bibnamefont {{Aumont}}}, \bibinfo {author}
  {\bibfnamefont {C.}~\bibnamefont {{Baccigalupi}}}, \bibinfo {author}
  {\bibfnamefont {M.}~\bibnamefont {{Ballardini}}}, \bibinfo {author}
  {\bibfnamefont {A.~J.}\ \bibnamefont {{Banday}}}, \bibinfo {author}
  {\bibfnamefont {R.~B.}\ \bibnamefont {{Barreiro}}}, \bibinfo {author}
  {\bibfnamefont {N.}~\bibnamefont {{Bartolo}}}, \bibinfo {author}
  {\bibfnamefont {S.}~\bibnamefont {{Basak}}}, \bibinfo {author} {\bibfnamefont
  {K.}~\bibnamefont {{Benabed}}}, \bibinfo {author} {\bibfnamefont {J.~P.}\
  \bibnamefont {{Bernard}}}, \bibinfo {author} {\bibfnamefont {M.}~\bibnamefont
  {{Bersanelli}}}, \bibinfo {author} {\bibfnamefont {P.}~\bibnamefont
  {{Bielewicz}}}, \bibinfo {author} {\bibfnamefont {J.~J.}\ \bibnamefont
  {{Bock}}}, \bibinfo {author} {\bibfnamefont {J.~R.}\ \bibnamefont {{Bond}}},
  \bibinfo {author} {\bibfnamefont {J.}~\bibnamefont {{Borrill}}}, \bibinfo
  {author} {\bibfnamefont {F.~R.}\ \bibnamefont {{Bouchet}}}, \bibinfo {author}
  {\bibfnamefont {F.}~\bibnamefont {{Boulanger}}}, \bibinfo {author}
  {\bibfnamefont {M.}~\bibnamefont {{Bucher}}}, \bibinfo {author}
  {\bibfnamefont {C.}~\bibnamefont {{Burigana}}}, \bibinfo {author}
  {\bibfnamefont {E.}~\bibnamefont {{Calabrese}}}, \bibinfo {author}
  {\bibfnamefont {J.~F.}\ \bibnamefont {{Cardoso}}}, \bibinfo {author}
  {\bibfnamefont {J.}~\bibnamefont {{Carron}}}, \bibinfo {author}
  {\bibfnamefont {A.}~\bibnamefont {{Challinor}}}, \bibinfo {author}
  {\bibfnamefont {H.~C.}\ \bibnamefont {{Chiang}}}, \bibinfo {author}
  {\bibfnamefont {L.~P.~L.}\ \bibnamefont {{Colombo}}}, \bibinfo {author}
  {\bibfnamefont {C.}~\bibnamefont {{Combet}}}, \bibinfo {author}
  {\bibfnamefont {B.~P.}\ \bibnamefont {{Crill}}}, \bibinfo {author}
  {\bibfnamefont {F.}~\bibnamefont {{Cuttaia}}}, \bibinfo {author}
  {\bibfnamefont {P.}~\bibnamefont {{de Bernardis}}}, \bibinfo {author}
  {\bibfnamefont {G.}~\bibnamefont {{de Zotti}}}, \bibinfo {author}
  {\bibfnamefont {J.}~\bibnamefont {{Delabrouille}}}, \bibinfo {author}
  {\bibfnamefont {E.}~\bibnamefont {{Di Valentino}}}, \bibinfo {author}
  {\bibfnamefont {J.~M.}\ \bibnamefont {{Diego}}}, \bibinfo {author}
  {\bibfnamefont {O.}~\bibnamefont {{Dor{\'e}}}}, \bibinfo {author}
  {\bibfnamefont {M.}~\bibnamefont {{Douspis}}}, \bibinfo {author}
  {\bibfnamefont {A.}~\bibnamefont {{Ducout}}}, \bibinfo {author}
  {\bibfnamefont {X.}~\bibnamefont {{Dupac}}}, \bibinfo {author} {\bibfnamefont
  {G.}~\bibnamefont {{Efstathiou}}}, \bibinfo {author} {\bibfnamefont
  {F.}~\bibnamefont {{Elsner}}}, \bibinfo {author} {\bibfnamefont {T.~A.}\
  \bibnamefont {{En{\ss}lin}}}, \bibinfo {author} {\bibfnamefont {H.~K.}\
  \bibnamefont {{Eriksen}}}, \bibinfo {author} {\bibfnamefont {Y.}~\bibnamefont
  {{Fantaye}}}, \bibinfo {author} {\bibfnamefont {R.}~\bibnamefont
  {{Fernandez-Cobos}}}, \bibinfo {author} {\bibfnamefont {F.}~\bibnamefont
  {{Forastieri}}}, \bibinfo {author} {\bibfnamefont {M.}~\bibnamefont
  {{Frailis}}}, \bibinfo {author} {\bibfnamefont {A.~A.}\ \bibnamefont
  {{Fraisse}}}, \bibinfo {author} {\bibfnamefont {E.}~\bibnamefont
  {{Franceschi}}}, \bibinfo {author} {\bibfnamefont {A.}~\bibnamefont
  {{Frolov}}}, \bibinfo {author} {\bibfnamefont {S.}~\bibnamefont
  {{Galeotta}}}, \bibinfo {author} {\bibfnamefont {S.}~\bibnamefont {{Galli}}},
  \bibinfo {author} {\bibfnamefont {K.}~\bibnamefont {{Ganga}}}, \bibinfo
  {author} {\bibfnamefont {R.~T.}\ \bibnamefont {{G{\'e}nova-Santos}}},
  \bibinfo {author} {\bibfnamefont {M.}~\bibnamefont {{Gerbino}}}, \bibinfo
  {author} {\bibfnamefont {T.}~\bibnamefont {{Ghosh}}}, \bibinfo {author}
  {\bibfnamefont {J.}~\bibnamefont {{Gonz{\'a}lez-Nuevo}}}, \bibinfo {author}
  {\bibfnamefont {K.~M.}\ \bibnamefont {{G{\'o}rski}}}, \bibinfo {author}
  {\bibfnamefont {S.}~\bibnamefont {{Gratton}}}, \bibinfo {author}
  {\bibfnamefont {A.}~\bibnamefont {{Gruppuso}}}, \bibinfo {author}
  {\bibfnamefont {J.~E.}\ \bibnamefont {{Gudmundsson}}}, \bibinfo {author}
  {\bibfnamefont {J.}~\bibnamefont {{Hamann}}}, \bibinfo {author}
  {\bibfnamefont {W.}~\bibnamefont {{Hand ley}}}, \bibinfo {author}
  {\bibfnamefont {F.~K.}\ \bibnamefont {{Hansen}}}, \bibinfo {author}
  {\bibfnamefont {D.}~\bibnamefont {{Herranz}}}, \bibinfo {author}
  {\bibfnamefont {E.}~\bibnamefont {{Hivon}}}, \bibinfo {author} {\bibfnamefont
  {Z.}~\bibnamefont {{Huang}}}, \bibinfo {author} {\bibfnamefont {A.~H.}\
  \bibnamefont {{Jaffe}}}, \bibinfo {author} {\bibfnamefont {W.~C.}\
  \bibnamefont {{Jones}}}, \bibinfo {author} {\bibfnamefont {A.}~\bibnamefont
  {{Karakci}}}, \bibinfo {author} {\bibfnamefont {E.}~\bibnamefont
  {{Keih{\"a}nen}}}, \bibinfo {author} {\bibfnamefont {R.}~\bibnamefont
  {{Keskitalo}}}, \bibinfo {author} {\bibfnamefont {K.}~\bibnamefont
  {{Kiiveri}}}, \bibinfo {author} {\bibfnamefont {J.}~\bibnamefont {{Kim}}},
  \bibinfo {author} {\bibfnamefont {L.}~\bibnamefont {{Knox}}}, \bibinfo
  {author} {\bibfnamefont {N.}~\bibnamefont {{Krachmalnicoff}}}, \bibinfo
  {author} {\bibfnamefont {M.}~\bibnamefont {{Kunz}}}, \bibinfo {author}
  {\bibfnamefont {H.}~\bibnamefont {{Kurki-Suonio}}}, \bibinfo {author}
  {\bibfnamefont {G.}~\bibnamefont {{Lagache}}}, \bibinfo {author}
  {\bibfnamefont {J.~M.}\ \bibnamefont {{Lamarre}}}, \bibinfo {author}
  {\bibfnamefont {A.}~\bibnamefont {{Lasenby}}}, \bibinfo {author}
  {\bibfnamefont {M.}~\bibnamefont {{Lattanzi}}}, \bibinfo {author}
  {\bibfnamefont {C.~R.}\ \bibnamefont {{Lawrence}}}, \bibinfo {author}
  {\bibfnamefont {M.}~\bibnamefont {{Le Jeune}}}, \bibinfo {author}
  {\bibfnamefont {F.}~\bibnamefont {{Levrier}}}, \bibinfo {author}
  {\bibfnamefont {A.}~\bibnamefont {{Lewis}}}, \bibinfo {author} {\bibfnamefont
  {M.}~\bibnamefont {{Liguori}}}, \bibinfo {author} {\bibfnamefont {P.~B.}\
  \bibnamefont {{Lilje}}}, \bibinfo {author} {\bibfnamefont {V.}~\bibnamefont
  {{Lindholm}}}, \bibinfo {author} {\bibfnamefont {M.}~\bibnamefont
  {{L{\'o}pez-Caniego}}}, \bibinfo {author} {\bibfnamefont {P.~M.}\
  \bibnamefont {{Lubin}}}, \bibinfo {author} {\bibfnamefont {Y.~Z.}\
  \bibnamefont {{Ma}}}, \bibinfo {author} {\bibfnamefont {J.~F.}\ \bibnamefont
  {{Mac{\'\i}as-P{\'e}rez}}}, \bibinfo {author} {\bibfnamefont
  {G.}~\bibnamefont {{Maggio}}}, \bibinfo {author} {\bibfnamefont
  {D.}~\bibnamefont {{Maino}}}, \bibinfo {author} {\bibfnamefont
  {N.}~\bibnamefont {{Mandolesi}}}, \bibinfo {author} {\bibfnamefont
  {A.}~\bibnamefont {{Mangilli}}}, \bibinfo {author} {\bibfnamefont
  {A.}~\bibnamefont {{Marcos-Caballero}}}, \bibinfo {author} {\bibfnamefont
  {M.}~\bibnamefont {{Maris}}}, \bibinfo {author} {\bibfnamefont {P.~G.}\
  \bibnamefont {{Martin}}}, \bibinfo {author} {\bibfnamefont {E.}~\bibnamefont
  {{Mart{\'\i}nez-Gonz{\'a}lez}}}, \bibinfo {author} {\bibfnamefont
  {S.}~\bibnamefont {{Matarrese}}}, \bibinfo {author} {\bibfnamefont
  {N.}~\bibnamefont {{Mauri}}}, \bibinfo {author} {\bibfnamefont {J.~D.}\
  \bibnamefont {{McEwen}}}, \bibinfo {author} {\bibfnamefont {A.}~\bibnamefont
  {{Melchiorri}}}, \bibinfo {author} {\bibfnamefont {A.}~\bibnamefont
  {{Mennella}}}, \bibinfo {author} {\bibfnamefont {M.}~\bibnamefont
  {{Migliaccio}}}, \bibinfo {author} {\bibfnamefont {M.~A.}\ \bibnamefont
  {{Miville-Desch{\^e}nes}}}, \bibinfo {author} {\bibfnamefont
  {D.}~\bibnamefont {{Molinari}}}, \bibinfo {author} {\bibfnamefont
  {A.}~\bibnamefont {{Moneti}}}, \bibinfo {author} {\bibfnamefont
  {L.}~\bibnamefont {{Montier}}}, \bibinfo {author} {\bibfnamefont
  {G.}~\bibnamefont {{Morgante}}}, \bibinfo {author} {\bibfnamefont
  {A.}~\bibnamefont {{Moss}}}, \bibinfo {author} {\bibfnamefont
  {P.}~\bibnamefont {{Natoli}}}, \bibinfo {author} {\bibfnamefont
  {L.}~\bibnamefont {{Pagano}}}, \bibinfo {author} {\bibfnamefont
  {D.}~\bibnamefont {{Paoletti}}}, \bibinfo {author} {\bibfnamefont
  {B.}~\bibnamefont {{Partridge}}}, \bibinfo {author} {\bibfnamefont
  {G.}~\bibnamefont {{Patanchon}}}, \bibinfo {author} {\bibfnamefont
  {F.}~\bibnamefont {{Perrotta}}}, \bibinfo {author} {\bibfnamefont
  {V.}~\bibnamefont {{Pettorino}}}, \bibinfo {author} {\bibfnamefont
  {F.}~\bibnamefont {{Piacentini}}}, \bibinfo {author} {\bibfnamefont
  {L.}~\bibnamefont {{Polastri}}}, \bibinfo {author} {\bibfnamefont
  {G.}~\bibnamefont {{Polenta}}}, \bibinfo {author} {\bibfnamefont {J.~L.}\
  \bibnamefont {{Puget}}}, \bibinfo {author} {\bibfnamefont {J.~P.}\
  \bibnamefont {{Rachen}}}, \bibinfo {author} {\bibfnamefont {M.}~\bibnamefont
  {{Reinecke}}}, \bibinfo {author} {\bibfnamefont {M.}~\bibnamefont
  {{Remazeilles}}}, \bibinfo {author} {\bibfnamefont {A.}~\bibnamefont
  {{Renzi}}}, \bibinfo {author} {\bibfnamefont {G.}~\bibnamefont {{Rocha}}},
  \bibinfo {author} {\bibfnamefont {C.}~\bibnamefont {{Rosset}}}, \bibinfo
  {author} {\bibfnamefont {G.}~\bibnamefont {{Roudier}}}, \bibinfo {author}
  {\bibfnamefont {J.~A.}\ \bibnamefont {{Rubi{\~n}o-Mart{\'\i}n}}}, \bibinfo
  {author} {\bibfnamefont {B.}~\bibnamefont {{Ruiz-Granados}}}, \bibinfo
  {author} {\bibfnamefont {L.}~\bibnamefont {{Salvati}}}, \bibinfo {author}
  {\bibfnamefont {M.}~\bibnamefont {{Sandri}}}, \bibinfo {author}
  {\bibfnamefont {M.}~\bibnamefont {{Savelainen}}}, \bibinfo {author}
  {\bibfnamefont {D.}~\bibnamefont {{Scott}}}, \bibinfo {author} {\bibfnamefont
  {C.}~\bibnamefont {{Sirignano}}}, \bibinfo {author} {\bibfnamefont
  {R.}~\bibnamefont {{Sunyaev}}}, \bibinfo {author} {\bibfnamefont {A.~S.}\
  \bibnamefont {{Suur-Uski}}}, \bibinfo {author} {\bibfnamefont {J.~A.}\
  \bibnamefont {{Tauber}}}, \bibinfo {author} {\bibfnamefont {D.}~\bibnamefont
  {{Tavagnacco}}}, \bibinfo {author} {\bibfnamefont {M.}~\bibnamefont
  {{Tenti}}}, \bibinfo {author} {\bibfnamefont {L.}~\bibnamefont
  {{Toffolatti}}}, \bibinfo {author} {\bibfnamefont {M.}~\bibnamefont
  {{Tomasi}}}, \bibinfo {author} {\bibfnamefont {T.}~\bibnamefont
  {{Trombetti}}}, \bibinfo {author} {\bibfnamefont {J.}~\bibnamefont
  {{Valiviita}}}, \bibinfo {author} {\bibfnamefont {B.}~\bibnamefont {{Van
  Tent}}}, \bibinfo {author} {\bibfnamefont {P.}~\bibnamefont {{Vielva}}},
  \bibinfo {author} {\bibfnamefont {F.}~\bibnamefont {{Villa}}}, \bibinfo
  {author} {\bibfnamefont {N.}~\bibnamefont {{Vittorio}}}, \bibinfo {author}
  {\bibfnamefont {B.~D.}\ \bibnamefont {{Wandelt}}}, \bibinfo {author}
  {\bibfnamefont {I.~K.}\ \bibnamefont {{Wehus}}}, \bibinfo {author}
  {\bibfnamefont {M.}~\bibnamefont {{White}}}, \bibinfo {author} {\bibfnamefont
  {S.~D.~M.}\ \bibnamefont {{White}}}, \bibinfo {author} {\bibfnamefont
  {A.}~\bibnamefont {{Zacchei}}}, \ and\ \bibinfo {author} {\bibfnamefont
  {A.}~\bibnamefont {{Zonca}}},\ }\href@noop {} {\bibfield  {journal} {\bibinfo
   {journal} {ArXiv e-prints}\ ,\ \bibinfo {eid} {arXiv:1807.06210}} (\bibinfo
  {year} {2018}{\natexlab{a}})},\ \Eprint {http://arxiv.org/abs/1807.06210}
  {arXiv:1807.06210 [astro-ph.CO]} \BibitemShut {NoStop}%
\bibitem [{\citenamefont {{Heymans}}\ \emph {et~al.}(2012)\citenamefont
  {{Heymans}}, \citenamefont {{Van Waerbeke}}, \citenamefont {{Miller}},
  \citenamefont {{Erben}}, \citenamefont {{Hildebrandt}}, \citenamefont
  {{Hoekstra}}, \citenamefont {{Kitching}}, \citenamefont {{Mellier}},
  \citenamefont {{Simon}}, \citenamefont {{Bonnett}}, \citenamefont {{Coupon}},
  \citenamefont {{Fu}}, \citenamefont {{Harnois D{\'e}raps}}, \citenamefont
  {{Hudson}}, \citenamefont {{Kilbinger}}, \citenamefont {{Kuijken}},
  \citenamefont {{Rowe}}, \citenamefont {{Schrabback}}, \citenamefont
  {{Semboloni}}, \citenamefont {{van Uitert}}, \citenamefont {{Vafaei}},\ and\
  \citenamefont {{Velander}}}]{2012MNRAS.427..146H}%
  \BibitemOpen
  \bibfield  {author} {\bibinfo {author} {\bibfnamefont {C.}~\bibnamefont
  {{Heymans}}}, \bibinfo {author} {\bibfnamefont {L.}~\bibnamefont {{Van
  Waerbeke}}}, \bibinfo {author} {\bibfnamefont {L.}~\bibnamefont {{Miller}}},
  \bibinfo {author} {\bibfnamefont {T.}~\bibnamefont {{Erben}}}, \bibinfo
  {author} {\bibfnamefont {H.}~\bibnamefont {{Hildebrandt}}}, \bibinfo {author}
  {\bibfnamefont {H.}~\bibnamefont {{Hoekstra}}}, \bibinfo {author}
  {\bibfnamefont {T.~D.}\ \bibnamefont {{Kitching}}}, \bibinfo {author}
  {\bibfnamefont {Y.}~\bibnamefont {{Mellier}}}, \bibinfo {author}
  {\bibfnamefont {P.}~\bibnamefont {{Simon}}}, \bibinfo {author} {\bibfnamefont
  {C.}~\bibnamefont {{Bonnett}}}, \bibinfo {author} {\bibfnamefont
  {J.}~\bibnamefont {{Coupon}}}, \bibinfo {author} {\bibfnamefont
  {L.}~\bibnamefont {{Fu}}}, \bibinfo {author} {\bibfnamefont {J.}~\bibnamefont
  {{Harnois D{\'e}raps}}}, \bibinfo {author} {\bibfnamefont {M.~J.}\
  \bibnamefont {{Hudson}}}, \bibinfo {author} {\bibfnamefont {M.}~\bibnamefont
  {{Kilbinger}}}, \bibinfo {author} {\bibfnamefont {K.}~\bibnamefont
  {{Kuijken}}}, \bibinfo {author} {\bibfnamefont {B.}~\bibnamefont {{Rowe}}},
  \bibinfo {author} {\bibfnamefont {T.}~\bibnamefont {{Schrabback}}}, \bibinfo
  {author} {\bibfnamefont {E.}~\bibnamefont {{Semboloni}}}, \bibinfo {author}
  {\bibfnamefont {E.}~\bibnamefont {{van Uitert}}}, \bibinfo {author}
  {\bibfnamefont {S.}~\bibnamefont {{Vafaei}}}, \ and\ \bibinfo {author}
  {\bibfnamefont {M.}~\bibnamefont {{Velander}}},\ }\href {\doibase
  10.1111/j.1365-2966.2012.21952.x} {\bibfield  {journal} {\bibinfo  {journal}
  {\mnras}\ }\textbf {\bibinfo {volume} {427}},\ \bibinfo {pages} {146}
  (\bibinfo {year} {2012})},\ \Eprint {http://arxiv.org/abs/1210.0032}
  {arXiv:1210.0032 [astro-ph.CO]} \BibitemShut {NoStop}%
\bibitem [{\citenamefont {{Hildebrandt}}\ \emph {et~al.}(2020)\citenamefont
  {{Hildebrandt}}, \citenamefont {{K{\"o}hlinger}}, \citenamefont {{van den
  Busch}}, \citenamefont {{Joachimi}}, \citenamefont {{Heymans}}, \citenamefont
  {{Kannawadi}}, \citenamefont {{Wright}}, \citenamefont {{Asgari}},
  \citenamefont {{Blake}}, \citenamefont {{Hoekstra}}, \citenamefont
  {{Joudaki}}, \citenamefont {{Kuijken}}, \citenamefont {{Miller}},
  \citenamefont {{Morrison}}, \citenamefont {{Tr{\"o}ster}}, \citenamefont
  {{Amon}}, \citenamefont {{Archidiacono}}, \citenamefont {{Brieden}},
  \citenamefont {{Choi}}, \citenamefont {{de Jong}}, \citenamefont {{Erben}},
  \citenamefont {{Giblin}}, \citenamefont {{Mead}}, \citenamefont {{Peacock}},
  \citenamefont {{Radovich}}, \citenamefont {{Schneider}}, \citenamefont
  {{Sif{\'o}n}},\ and\ \citenamefont {{Tewes}}}]{hildebrandt20a}%
  \BibitemOpen
  \bibfield  {author} {\bibinfo {author} {\bibfnamefont {H.}~\bibnamefont
  {{Hildebrandt}}}, \bibinfo {author} {\bibfnamefont {F.}~\bibnamefont
  {{K{\"o}hlinger}}}, \bibinfo {author} {\bibfnamefont {J.~L.}\ \bibnamefont
  {{van den Busch}}}, \bibinfo {author} {\bibfnamefont {B.}~\bibnamefont
  {{Joachimi}}}, \bibinfo {author} {\bibfnamefont {C.}~\bibnamefont
  {{Heymans}}}, \bibinfo {author} {\bibfnamefont {A.}~\bibnamefont
  {{Kannawadi}}}, \bibinfo {author} {\bibfnamefont {A.~H.}\ \bibnamefont
  {{Wright}}}, \bibinfo {author} {\bibfnamefont {M.}~\bibnamefont {{Asgari}}},
  \bibinfo {author} {\bibfnamefont {C.}~\bibnamefont {{Blake}}}, \bibinfo
  {author} {\bibfnamefont {H.}~\bibnamefont {{Hoekstra}}}, \bibinfo {author}
  {\bibfnamefont {S.}~\bibnamefont {{Joudaki}}}, \bibinfo {author}
  {\bibfnamefont {K.}~\bibnamefont {{Kuijken}}}, \bibinfo {author}
  {\bibfnamefont {L.}~\bibnamefont {{Miller}}}, \bibinfo {author}
  {\bibfnamefont {C.~B.}\ \bibnamefont {{Morrison}}}, \bibinfo {author}
  {\bibfnamefont {T.}~\bibnamefont {{Tr{\"o}ster}}}, \bibinfo {author}
  {\bibfnamefont {A.}~\bibnamefont {{Amon}}}, \bibinfo {author} {\bibfnamefont
  {M.}~\bibnamefont {{Archidiacono}}}, \bibinfo {author} {\bibfnamefont
  {S.}~\bibnamefont {{Brieden}}}, \bibinfo {author} {\bibfnamefont
  {A.}~\bibnamefont {{Choi}}}, \bibinfo {author} {\bibfnamefont {J.~T.~A.}\
  \bibnamefont {{de Jong}}}, \bibinfo {author} {\bibfnamefont {T.}~\bibnamefont
  {{Erben}}}, \bibinfo {author} {\bibfnamefont {B.}~\bibnamefont {{Giblin}}},
  \bibinfo {author} {\bibfnamefont {A.}~\bibnamefont {{Mead}}}, \bibinfo
  {author} {\bibfnamefont {J.~A.}\ \bibnamefont {{Peacock}}}, \bibinfo {author}
  {\bibfnamefont {M.}~\bibnamefont {{Radovich}}}, \bibinfo {author}
  {\bibfnamefont {P.}~\bibnamefont {{Schneider}}}, \bibinfo {author}
  {\bibfnamefont {C.}~\bibnamefont {{Sif{\'o}n}}}, \ and\ \bibinfo {author}
  {\bibfnamefont {M.}~\bibnamefont {{Tewes}}},\ }\href {\doibase
  10.1051/0004-6361/201834878} {\bibfield  {journal} {\bibinfo  {journal}
  {\aap}\ }\textbf {\bibinfo {volume} {633}},\ \bibinfo {eid} {A69} (\bibinfo
  {year} {2020})},\ \Eprint {http://arxiv.org/abs/1812.06076} {arXiv:1812.06076
  [astro-ph.CO]} \BibitemShut {NoStop}%
\bibitem [{\citenamefont {{Joachimi}}\ \emph {et~al.}(2020)\citenamefont
  {{Joachimi}}, \citenamefont {{Lin}}, \citenamefont {{Asgari}}, \citenamefont
  {{Tr{\"o}ster}}, \citenamefont {{Heymans}}, \citenamefont {{Hildebrandt}},
  \citenamefont {{K{\"o}hlinger}}, \citenamefont {{S{\'a}nchez}}, \citenamefont
  {{Wright}}, \citenamefont {{Bilicki}}, \citenamefont {{Blake}}, \citenamefont
  {{van den Busch}}, \citenamefont {{Crocce}}, \citenamefont {{Dvornik}},
  \citenamefont {{Erben}}, \citenamefont {{Getman}}, \citenamefont {{Giblin}},
  \citenamefont {{Hoekstra}}, \citenamefont {{Kannawadi}}, \citenamefont
  {{Kuijken}}, \citenamefont {{Napolitano}}, \citenamefont {{Schneider}},
  \citenamefont {{Scoccimarro}}, \citenamefont {{Sellentin}}, \citenamefont
  {{Shan}}, \citenamefont {{von Wietersheim-Kramsta}},\ and\ \citenamefont
  {{Zuntz}}}]{joachimi20a}%
  \BibitemOpen
  \bibfield  {author} {\bibinfo {author} {\bibfnamefont {B.}~\bibnamefont
  {{Joachimi}}}, \bibinfo {author} {\bibfnamefont {C.~A.}\ \bibnamefont
  {{Lin}}}, \bibinfo {author} {\bibfnamefont {M.}~\bibnamefont {{Asgari}}},
  \bibinfo {author} {\bibfnamefont {T.}~\bibnamefont {{Tr{\"o}ster}}}, \bibinfo
  {author} {\bibfnamefont {C.}~\bibnamefont {{Heymans}}}, \bibinfo {author}
  {\bibfnamefont {H.}~\bibnamefont {{Hildebrandt}}}, \bibinfo {author}
  {\bibfnamefont {F.}~\bibnamefont {{K{\"o}hlinger}}}, \bibinfo {author}
  {\bibfnamefont {A.~G.}\ \bibnamefont {{S{\'a}nchez}}}, \bibinfo {author}
  {\bibfnamefont {A.~H.}\ \bibnamefont {{Wright}}}, \bibinfo {author}
  {\bibfnamefont {M.}~\bibnamefont {{Bilicki}}}, \bibinfo {author}
  {\bibfnamefont {C.}~\bibnamefont {{Blake}}}, \bibinfo {author} {\bibfnamefont
  {J.~L.}\ \bibnamefont {{van den Busch}}}, \bibinfo {author} {\bibfnamefont
  {M.}~\bibnamefont {{Crocce}}}, \bibinfo {author} {\bibfnamefont
  {A.}~\bibnamefont {{Dvornik}}}, \bibinfo {author} {\bibfnamefont
  {T.}~\bibnamefont {{Erben}}}, \bibinfo {author} {\bibfnamefont
  {F.}~\bibnamefont {{Getman}}}, \bibinfo {author} {\bibfnamefont
  {B.}~\bibnamefont {{Giblin}}}, \bibinfo {author} {\bibfnamefont
  {H.}~\bibnamefont {{Hoekstra}}}, \bibinfo {author} {\bibfnamefont
  {A.}~\bibnamefont {{Kannawadi}}}, \bibinfo {author} {\bibfnamefont
  {K.}~\bibnamefont {{Kuijken}}}, \bibinfo {author} {\bibfnamefont {N.~R.}\
  \bibnamefont {{Napolitano}}}, \bibinfo {author} {\bibfnamefont
  {P.}~\bibnamefont {{Schneider}}}, \bibinfo {author} {\bibfnamefont
  {R.}~\bibnamefont {{Scoccimarro}}}, \bibinfo {author} {\bibfnamefont
  {E.}~\bibnamefont {{Sellentin}}}, \bibinfo {author} {\bibfnamefont {H.~Y.}\
  \bibnamefont {{Shan}}}, \bibinfo {author} {\bibfnamefont {M.}~\bibnamefont
  {{von Wietersheim-Kramsta}}}, \ and\ \bibinfo {author} {\bibfnamefont
  {J.}~\bibnamefont {{Zuntz}}},\ }\href@noop {} {\bibfield  {journal} {\bibinfo
   {journal} {arXiv e-prints}\ ,\ \bibinfo {eid} {arXiv:2007.01844}} (\bibinfo
  {year} {2020})},\ \Eprint {http://arxiv.org/abs/2007.01844} {arXiv:2007.01844
  [astro-ph.CO]} \BibitemShut {NoStop}%
\bibitem [{\citenamefont {{Zuntz}}\ \emph {et~al.}(2018)\citenamefont
  {{Zuntz}}, \citenamefont {{Sheldon}}, \citenamefont {{Samuroff}},
  \citenamefont {{Troxel}}, \citenamefont {{Jarvis}}, \citenamefont
  {{MacCrann}}, \citenamefont {{Gruen}}, \citenamefont {{Prat}}, \citenamefont
  {{S{\'a}nchez}}, \citenamefont {{Choi}}, \citenamefont {{Bridle}},
  \citenamefont {{Bernstein}}, \citenamefont {{Dodelson}}, \citenamefont
  {{Drlica-Wagner}}, \citenamefont {{Fang}}, \citenamefont {{Gruendl}},
  \citenamefont {{Hoyle}}, \citenamefont {{Huff}}, \citenamefont {{Jain}},
  \citenamefont {{Kirk}}, \citenamefont {{Kacprzak}}, \citenamefont
  {{Krawiec}}, \citenamefont {{Plazas}}, \citenamefont {{Rollins}},
  \citenamefont {{Rykoff}}, \citenamefont {{Sevilla-Noarbe}}, \citenamefont
  {{Soergel}}, \citenamefont {{Varga}}, \citenamefont {{Abbott}}, \citenamefont
  {{Abdalla}}, \citenamefont {{Allam}}, \citenamefont {{Annis}}, \citenamefont
  {{Bechtol}}, \citenamefont {{Benoit-L{\'e}vy}}, \citenamefont {{Bertin}},
  \citenamefont {{Buckley-Geer}}, \citenamefont {{Burke}}, \citenamefont
  {{Carnero Rosell}}, \citenamefont {{Carrasco Kind}}, \citenamefont
  {{Carretero}}, \citenamefont {{Castander}}, \citenamefont {{Crocce}},
  \citenamefont {{Cunha}}, \citenamefont {{D'Andrea}}, \citenamefont {{da
  Costa}}, \citenamefont {{Davis}}, \citenamefont {{Desai}}, \citenamefont
  {{Diehl}}, \citenamefont {{Dietrich}}, \citenamefont {{Doel}}, \citenamefont
  {{Eifler}}, \citenamefont {{Estrada}}, \citenamefont {{Evrard}},
  \citenamefont {{Neto}}, \citenamefont {{Fernandez}}, \citenamefont
  {{Flaugher}}, \citenamefont {{Fosalba}}, \citenamefont {{Frieman}},
  \citenamefont {{Garc{\'\i}a-Bellido}}, \citenamefont {{Gaztanaga}},
  \citenamefont {{Gerdes}}, \citenamefont {{Giannantonio}}, \citenamefont
  {{Gschwend}}, \citenamefont {{Gutierrez}}, \citenamefont {{Hartley}},
  \citenamefont {{Honscheid}}, \citenamefont {{James}}, \citenamefont
  {{Jeltema}}, \citenamefont {{Johnson}}, \citenamefont {{Johnson}},
  \citenamefont {{Kuehn}}, \citenamefont {{Kuhlmann}}, \citenamefont
  {{Kuropatkin}}, \citenamefont {{Lahav}}, \citenamefont {{Li}}, \citenamefont
  {{Lima}}, \citenamefont {{Maia}}, \citenamefont {{March}}, \citenamefont
  {{Martini}}, \citenamefont {{Melchior}}, \citenamefont {{Menanteau}},
  \citenamefont {{Miller}}, \citenamefont {{Miquel}}, \citenamefont {{Mohr}},
  \citenamefont {{Neilsen}}, \citenamefont {{Nichol}}, \citenamefont
  {{Ogando}}, \citenamefont {{Roe}}, \citenamefont {{Romer}}, \citenamefont
  {{Roodman}}, \citenamefont {{Sanchez}}, \citenamefont {{Scarpine}},
  \citenamefont {{Schindler}}, \citenamefont {{Schubnell}}, \citenamefont
  {{Smith}}, \citenamefont {{Smith}}, \citenamefont {{Soares-Santos}},
  \citenamefont {{Sobreira}}, \citenamefont {{Suchyta}}, \citenamefont
  {{Swanson}}, \citenamefont {{Tarle}}, \citenamefont {{Thomas}}, \citenamefont
  {{Tucker}}, \citenamefont {{Vikram}}, \citenamefont {{Walker}}, \citenamefont
  {{Wechsler}},\ and\ \citenamefont {{Zhang}}}]{2018MNRAS.481.1149Z}%
  \BibitemOpen
  \bibfield  {author} {\bibinfo {author} {\bibfnamefont {J.}~\bibnamefont
  {{Zuntz}}}, \bibinfo {author} {\bibfnamefont {E.}~\bibnamefont {{Sheldon}}},
  \bibinfo {author} {\bibfnamefont {S.}~\bibnamefont {{Samuroff}}}, \bibinfo
  {author} {\bibfnamefont {M.~A.}\ \bibnamefont {{Troxel}}}, \bibinfo {author}
  {\bibfnamefont {M.}~\bibnamefont {{Jarvis}}}, \bibinfo {author}
  {\bibfnamefont {N.}~\bibnamefont {{MacCrann}}}, \bibinfo {author}
  {\bibfnamefont {D.}~\bibnamefont {{Gruen}}}, \bibinfo {author} {\bibfnamefont
  {J.}~\bibnamefont {{Prat}}}, \bibinfo {author} {\bibfnamefont
  {C.}~\bibnamefont {{S{\'a}nchez}}}, \bibinfo {author} {\bibfnamefont
  {A.}~\bibnamefont {{Choi}}}, \bibinfo {author} {\bibfnamefont {S.~L.}\
  \bibnamefont {{Bridle}}}, \bibinfo {author} {\bibfnamefont {G.~M.}\
  \bibnamefont {{Bernstein}}}, \bibinfo {author} {\bibfnamefont
  {S.}~\bibnamefont {{Dodelson}}}, \bibinfo {author} {\bibfnamefont
  {A.}~\bibnamefont {{Drlica-Wagner}}}, \bibinfo {author} {\bibfnamefont
  {Y.}~\bibnamefont {{Fang}}}, \bibinfo {author} {\bibfnamefont {R.~A.}\
  \bibnamefont {{Gruendl}}}, \bibinfo {author} {\bibfnamefont {B.}~\bibnamefont
  {{Hoyle}}}, \bibinfo {author} {\bibfnamefont {E.~M.}\ \bibnamefont {{Huff}}},
  \bibinfo {author} {\bibfnamefont {B.}~\bibnamefont {{Jain}}}, \bibinfo
  {author} {\bibfnamefont {D.}~\bibnamefont {{Kirk}}}, \bibinfo {author}
  {\bibfnamefont {T.}~\bibnamefont {{Kacprzak}}}, \bibinfo {author}
  {\bibfnamefont {C.}~\bibnamefont {{Krawiec}}}, \bibinfo {author}
  {\bibfnamefont {A.~A.}\ \bibnamefont {{Plazas}}}, \bibinfo {author}
  {\bibfnamefont {R.~P.}\ \bibnamefont {{Rollins}}}, \bibinfo {author}
  {\bibfnamefont {E.~S.}\ \bibnamefont {{Rykoff}}}, \bibinfo {author}
  {\bibfnamefont {I.}~\bibnamefont {{Sevilla-Noarbe}}}, \bibinfo {author}
  {\bibfnamefont {B.}~\bibnamefont {{Soergel}}}, \bibinfo {author}
  {\bibfnamefont {T.~N.}\ \bibnamefont {{Varga}}}, \bibinfo {author}
  {\bibfnamefont {T.~M.~C.}\ \bibnamefont {{Abbott}}}, \bibinfo {author}
  {\bibfnamefont {F.~B.}\ \bibnamefont {{Abdalla}}}, \bibinfo {author}
  {\bibfnamefont {S.}~\bibnamefont {{Allam}}}, \bibinfo {author} {\bibfnamefont
  {J.}~\bibnamefont {{Annis}}}, \bibinfo {author} {\bibfnamefont
  {K.}~\bibnamefont {{Bechtol}}}, \bibinfo {author} {\bibfnamefont
  {A.}~\bibnamefont {{Benoit-L{\'e}vy}}}, \bibinfo {author} {\bibfnamefont
  {E.}~\bibnamefont {{Bertin}}}, \bibinfo {author} {\bibfnamefont
  {E.}~\bibnamefont {{Buckley-Geer}}}, \bibinfo {author} {\bibfnamefont
  {D.~L.}\ \bibnamefont {{Burke}}}, \bibinfo {author} {\bibfnamefont
  {A.}~\bibnamefont {{Carnero Rosell}}}, \bibinfo {author} {\bibfnamefont
  {M.}~\bibnamefont {{Carrasco Kind}}}, \bibinfo {author} {\bibfnamefont
  {J.}~\bibnamefont {{Carretero}}}, \bibinfo {author} {\bibfnamefont {F.~J.}\
  \bibnamefont {{Castander}}}, \bibinfo {author} {\bibfnamefont
  {M.}~\bibnamefont {{Crocce}}}, \bibinfo {author} {\bibfnamefont {C.~E.}\
  \bibnamefont {{Cunha}}}, \bibinfo {author} {\bibfnamefont {C.~B.}\
  \bibnamefont {{D'Andrea}}}, \bibinfo {author} {\bibfnamefont {L.~N.}\
  \bibnamefont {{da Costa}}}, \bibinfo {author} {\bibfnamefont
  {C.}~\bibnamefont {{Davis}}}, \bibinfo {author} {\bibfnamefont
  {S.}~\bibnamefont {{Desai}}}, \bibinfo {author} {\bibfnamefont {H.~T.}\
  \bibnamefont {{Diehl}}}, \bibinfo {author} {\bibfnamefont {J.~P.}\
  \bibnamefont {{Dietrich}}}, \bibinfo {author} {\bibfnamefont
  {P.}~\bibnamefont {{Doel}}}, \bibinfo {author} {\bibfnamefont {T.~F.}\
  \bibnamefont {{Eifler}}}, \bibinfo {author} {\bibfnamefont {J.}~\bibnamefont
  {{Estrada}}}, \bibinfo {author} {\bibfnamefont {A.~E.}\ \bibnamefont
  {{Evrard}}}, \bibinfo {author} {\bibfnamefont {A.~F.}\ \bibnamefont
  {{Neto}}}, \bibinfo {author} {\bibfnamefont {E.}~\bibnamefont {{Fernandez}}},
  \bibinfo {author} {\bibfnamefont {B.}~\bibnamefont {{Flaugher}}}, \bibinfo
  {author} {\bibfnamefont {P.}~\bibnamefont {{Fosalba}}}, \bibinfo {author}
  {\bibfnamefont {J.}~\bibnamefont {{Frieman}}}, \bibinfo {author}
  {\bibfnamefont {J.}~\bibnamefont {{Garc{\'\i}a-Bellido}}}, \bibinfo {author}
  {\bibfnamefont {E.}~\bibnamefont {{Gaztanaga}}}, \bibinfo {author}
  {\bibfnamefont {D.~W.}\ \bibnamefont {{Gerdes}}}, \bibinfo {author}
  {\bibfnamefont {T.}~\bibnamefont {{Giannantonio}}}, \bibinfo {author}
  {\bibfnamefont {J.}~\bibnamefont {{Gschwend}}}, \bibinfo {author}
  {\bibfnamefont {G.}~\bibnamefont {{Gutierrez}}}, \bibinfo {author}
  {\bibfnamefont {W.~G.}\ \bibnamefont {{Hartley}}}, \bibinfo {author}
  {\bibfnamefont {K.}~\bibnamefont {{Honscheid}}}, \bibinfo {author}
  {\bibfnamefont {D.~J.}\ \bibnamefont {{James}}}, \bibinfo {author}
  {\bibfnamefont {T.}~\bibnamefont {{Jeltema}}}, \bibinfo {author}
  {\bibfnamefont {M.~W.~G.}\ \bibnamefont {{Johnson}}}, \bibinfo {author}
  {\bibfnamefont {M.~D.}\ \bibnamefont {{Johnson}}}, \bibinfo {author}
  {\bibfnamefont {K.}~\bibnamefont {{Kuehn}}}, \bibinfo {author} {\bibfnamefont
  {S.}~\bibnamefont {{Kuhlmann}}}, \bibinfo {author} {\bibfnamefont
  {N.}~\bibnamefont {{Kuropatkin}}}, \bibinfo {author} {\bibfnamefont
  {O.}~\bibnamefont {{Lahav}}}, \bibinfo {author} {\bibfnamefont {T.~S.}\
  \bibnamefont {{Li}}}, \bibinfo {author} {\bibfnamefont {M.}~\bibnamefont
  {{Lima}}}, \bibinfo {author} {\bibfnamefont {M.~A.~G.}\ \bibnamefont
  {{Maia}}}, \bibinfo {author} {\bibfnamefont {M.}~\bibnamefont {{March}}},
  \bibinfo {author} {\bibfnamefont {P.}~\bibnamefont {{Martini}}}, \bibinfo
  {author} {\bibfnamefont {P.}~\bibnamefont {{Melchior}}}, \bibinfo {author}
  {\bibfnamefont {F.}~\bibnamefont {{Menanteau}}}, \bibinfo {author}
  {\bibfnamefont {C.~J.}\ \bibnamefont {{Miller}}}, \bibinfo {author}
  {\bibfnamefont {R.}~\bibnamefont {{Miquel}}}, \bibinfo {author}
  {\bibfnamefont {J.~J.}\ \bibnamefont {{Mohr}}}, \bibinfo {author}
  {\bibfnamefont {E.}~\bibnamefont {{Neilsen}}}, \bibinfo {author}
  {\bibfnamefont {R.~C.}\ \bibnamefont {{Nichol}}}, \bibinfo {author}
  {\bibfnamefont {R.~L.~C.}\ \bibnamefont {{Ogando}}}, \bibinfo {author}
  {\bibfnamefont {N.}~\bibnamefont {{Roe}}}, \bibinfo {author} {\bibfnamefont
  {A.~K.}\ \bibnamefont {{Romer}}}, \bibinfo {author} {\bibfnamefont
  {A.}~\bibnamefont {{Roodman}}}, \bibinfo {author} {\bibfnamefont
  {E.}~\bibnamefont {{Sanchez}}}, \bibinfo {author} {\bibfnamefont
  {V.}~\bibnamefont {{Scarpine}}}, \bibinfo {author} {\bibfnamefont
  {R.}~\bibnamefont {{Schindler}}}, \bibinfo {author} {\bibfnamefont
  {M.}~\bibnamefont {{Schubnell}}}, \bibinfo {author} {\bibfnamefont
  {M.}~\bibnamefont {{Smith}}}, \bibinfo {author} {\bibfnamefont {R.~C.}\
  \bibnamefont {{Smith}}}, \bibinfo {author} {\bibfnamefont {M.}~\bibnamefont
  {{Soares-Santos}}}, \bibinfo {author} {\bibfnamefont {F.}~\bibnamefont
  {{Sobreira}}}, \bibinfo {author} {\bibfnamefont {E.}~\bibnamefont
  {{Suchyta}}}, \bibinfo {author} {\bibfnamefont {M.~E.~C.}\ \bibnamefont
  {{Swanson}}}, \bibinfo {author} {\bibfnamefont {G.}~\bibnamefont {{Tarle}}},
  \bibinfo {author} {\bibfnamefont {D.}~\bibnamefont {{Thomas}}}, \bibinfo
  {author} {\bibfnamefont {D.~L.}\ \bibnamefont {{Tucker}}}, \bibinfo {author}
  {\bibfnamefont {V.}~\bibnamefont {{Vikram}}}, \bibinfo {author}
  {\bibfnamefont {A.~R.}\ \bibnamefont {{Walker}}}, \bibinfo {author}
  {\bibfnamefont {R.~H.}\ \bibnamefont {{Wechsler}}}, \ and\ \bibinfo {author}
  {\bibfnamefont {Y.}~\bibnamefont {{Zhang}}},\ }\href {\doibase
  10.1093/mnras/sty2219} {\bibfield  {journal} {\bibinfo  {journal} {\mnras}\
  }\textbf {\bibinfo {volume} {481}},\ \bibinfo {pages} {1149} (\bibinfo {year}
  {2018})},\ \Eprint {http://arxiv.org/abs/1708.01533} {arXiv:1708.01533
  [astro-ph.CO]} \BibitemShut {NoStop}%
\bibitem [{\citenamefont {{Abbott}}\ \emph {et~al.}(2018)\citenamefont
  {{Abbott}}, \citenamefont {{Abdalla}}, \citenamefont {{Alarcon}},
  \citenamefont {{Aleksi{\'c}}}, \citenamefont {{Allam}}, \citenamefont
  {{Allen}}, \citenamefont {{Amara}}, \citenamefont {{Annis}}, \citenamefont
  {{Asorey}}, \citenamefont {{Avila}}, \citenamefont {{Bacon}}, \citenamefont
  {{Balbinot}}, \citenamefont {{Banerji}}, \citenamefont {{Banik}},
  \citenamefont {{Barkhouse}}, \citenamefont {{Baumer}}, \citenamefont
  {{Baxter}}, \citenamefont {{Bechtol}}, \citenamefont {{Becker}},
  \citenamefont {{Benoit-L{\'e}vy}}, \citenamefont {{Benson}}, \citenamefont
  {{Bernstein}}, \citenamefont {{Bertin}}, \citenamefont {{Blazek}},
  \citenamefont {{Bridle}}, \citenamefont {{Brooks}}, \citenamefont {{Brout}},
  \citenamefont {{Buckley-Geer}}, \citenamefont {{Burke}}, \citenamefont
  {{Busha}}, \citenamefont {{Campos}}, \citenamefont {{Capozzi}}, \citenamefont
  {{Carnero Rosell}}, \citenamefont {{Carrasco Kind}}, \citenamefont
  {{Carretero}}, \citenamefont {{Castander}}, \citenamefont {{Cawthon}},
  \citenamefont {{Chang}}, \citenamefont {{Chen}}, \citenamefont {{Childress}},
  \citenamefont {{Choi}}, \citenamefont {{Conselice}}, \citenamefont
  {{Crittenden}}, \citenamefont {{Crocce}}, \citenamefont {{Cunha}},
  \citenamefont {{D'Andrea}}, \citenamefont {{da Costa}}, \citenamefont
  {{Das}}, \citenamefont {{Davis}}, \citenamefont {{Davis}}, \citenamefont {{De
  Vicente}}, \citenamefont {{DePoy}}, \citenamefont {{DeRose}}, \citenamefont
  {{Desai}}, \citenamefont {{Diehl}}, \citenamefont {{Dietrich}}, \citenamefont
  {{Dodelson}}, \citenamefont {{Doel}}, \citenamefont {{Drlica-Wagner}},
  \citenamefont {{Eifler}}, \citenamefont {{Elliott}}, \citenamefont
  {{Elsner}}, \citenamefont {{Elvin-Poole}}, \citenamefont {{Estrada}},
  \citenamefont {{Evrard}}, \citenamefont {{Fang}}, \citenamefont
  {{Fernandez}}, \citenamefont {{Fert{\'e}}}, \citenamefont {{Finley}},
  \citenamefont {{Flaugher}}, \citenamefont {{Fosalba}}, \citenamefont
  {{Friedrich}}, \citenamefont {{Frieman}}, \citenamefont
  {{Garc{\'\i}a-Bellido}}, \citenamefont {{Garcia-Fernandez}}, \citenamefont
  {{Gatti}}, \citenamefont {{Gaztanaga}}, \citenamefont {{Gerdes}},
  \citenamefont {{Giannantonio}}, \citenamefont {{Gill}}, \citenamefont
  {{Glazebrook}}, \citenamefont {{Goldstein}}, \citenamefont {{Gruen}},
  \citenamefont {{Gruendl}}, \citenamefont {{Gschwend}}, \citenamefont
  {{Gutierrez}}, \citenamefont {{Hamilton}}, \citenamefont {{Hartley}},
  \citenamefont {{Hinton}}, \citenamefont {{Honscheid}}, \citenamefont
  {{Hoyle}}, \citenamefont {{Huterer}}, \citenamefont {{Jain}}, \citenamefont
  {{James}}, \citenamefont {{Jarvis}}, \citenamefont {{Jeltema}}, \citenamefont
  {{Johnson}}, \citenamefont {{Johnson}}, \citenamefont {{Kacprzak}},
  \citenamefont {{Kent}}, \citenamefont {{Kim}}, \citenamefont {{King}},
  \citenamefont {{Kirk}}, \citenamefont {{Kokron}}, \citenamefont {{Kovacs}},
  \citenamefont {{Krause}}, \citenamefont {{Krawiec}}, \citenamefont
  {{Kremin}}, \citenamefont {{Kuehn}}, \citenamefont {{Kuhlmann}},
  \citenamefont {{Kuropatkin}}, \citenamefont {{Lacasa}}, \citenamefont
  {{Lahav}}, \citenamefont {{Li}}, \citenamefont {{Liddle}}, \citenamefont
  {{Lidman}}, \citenamefont {{Lima}}, \citenamefont {{Lin}}, \citenamefont
  {{MacCrann}}, \citenamefont {{Maia}}, \citenamefont {{Makler}}, \citenamefont
  {{Manera}}, \citenamefont {{March}}, \citenamefont {{Marshall}},
  \citenamefont {{Martini}}, \citenamefont {{McMahon}}, \citenamefont
  {{Melchior}}, \citenamefont {{Menanteau}}, \citenamefont {{Miquel}},
  \citenamefont {{Miranda}}, \citenamefont {{Mudd}}, \citenamefont {{Muir}},
  \citenamefont {{M{\"o}ller}}, \citenamefont {{Neilsen}}, \citenamefont
  {{Nichol}}, \citenamefont {{Nord}}, \citenamefont {{Nugent}}, \citenamefont
  {{Ogando}}, \citenamefont {{Palmese}}, \citenamefont {{Peacock}},
  \citenamefont {{Peiris}}, \citenamefont {{Peoples}}, \citenamefont
  {{Percival}}, \citenamefont {{Petravick}}, \citenamefont {{Plazas}},
  \citenamefont {{Porredon}}, \citenamefont {{Prat}}, \citenamefont {{Pujol}},
  \citenamefont {{Rau}}, \citenamefont {{Refregier}}, \citenamefont {{Ricker}},
  \citenamefont {{Roe}}, \citenamefont {{Rollins}}, \citenamefont {{Romer}},
  \citenamefont {{Roodman}}, \citenamefont {{Rosenfeld}}, \citenamefont
  {{Ross}}, \citenamefont {{Rozo}}, \citenamefont {{Rykoff}}, \citenamefont
  {{Sako}}, \citenamefont {{Salvador}}, \citenamefont {{Samuroff}},
  \citenamefont {{S{\'a}nchez}}, \citenamefont {{Sanchez}}, \citenamefont
  {{Santiago}}, \citenamefont {{Scarpine}}, \citenamefont {{Schindler}},
  \citenamefont {{Scolnic}}, \citenamefont {{Secco}}, \citenamefont
  {{Serrano}}, \citenamefont {{Sevilla-Noarbe}}, \citenamefont {{Sheldon}},
  \citenamefont {{Smith}}, \citenamefont {{Smith}}, \citenamefont {{Smith}},
  \citenamefont {{Soares-Santos}}, \citenamefont {{Sobreira}}, \citenamefont
  {{Suchyta}}, \citenamefont {{Tarle}}, \citenamefont {{Thomas}}, \citenamefont
  {{Troxel}}, \citenamefont {{Tucker}}, \citenamefont {{Tucker}}, \citenamefont
  {{Uddin}}, \citenamefont {{Varga}}, \citenamefont {{Vielzeuf}}, \citenamefont
  {{Vikram}}, \citenamefont {{Vivas}}, \citenamefont {{Walker}}, \citenamefont
  {{Wang}}, \citenamefont {{Wechsler}}, \citenamefont {{Weller}}, \citenamefont
  {{Wester}}, \citenamefont {{Wolf}}, \citenamefont {{Yanny}}, \citenamefont
  {{Yuan}}, \citenamefont {{Zenteno}}, \citenamefont {{Zhang}}, \citenamefont
  {{Zhang}}, \citenamefont {{Zuntz}},\ and\ \citenamefont {{Dark Energy Survey
  Collaboration}}}]{2018PhRvD..98d3526A}%
  \BibitemOpen
  \bibfield  {author} {\bibinfo {author} {\bibfnamefont {T.~M.~C.}\
  \bibnamefont {{Abbott}}}, \bibinfo {author} {\bibfnamefont {F.~B.}\
  \bibnamefont {{Abdalla}}}, \bibinfo {author} {\bibfnamefont {A.}~\bibnamefont
  {{Alarcon}}}, \bibinfo {author} {\bibfnamefont {J.}~\bibnamefont
  {{Aleksi{\'c}}}}, \bibinfo {author} {\bibfnamefont {S.}~\bibnamefont
  {{Allam}}}, \bibinfo {author} {\bibfnamefont {S.}~\bibnamefont {{Allen}}},
  \bibinfo {author} {\bibfnamefont {A.}~\bibnamefont {{Amara}}}, \bibinfo
  {author} {\bibfnamefont {J.}~\bibnamefont {{Annis}}}, \bibinfo {author}
  {\bibfnamefont {J.}~\bibnamefont {{Asorey}}}, \bibinfo {author}
  {\bibfnamefont {S.}~\bibnamefont {{Avila}}}, \bibinfo {author} {\bibfnamefont
  {D.}~\bibnamefont {{Bacon}}}, \bibinfo {author} {\bibfnamefont
  {E.}~\bibnamefont {{Balbinot}}}, \bibinfo {author} {\bibfnamefont
  {M.}~\bibnamefont {{Banerji}}}, \bibinfo {author} {\bibfnamefont
  {N.}~\bibnamefont {{Banik}}}, \bibinfo {author} {\bibfnamefont
  {W.}~\bibnamefont {{Barkhouse}}}, \bibinfo {author} {\bibfnamefont
  {M.}~\bibnamefont {{Baumer}}}, \bibinfo {author} {\bibfnamefont
  {E.}~\bibnamefont {{Baxter}}}, \bibinfo {author} {\bibfnamefont
  {K.}~\bibnamefont {{Bechtol}}}, \bibinfo {author} {\bibfnamefont {M.~R.}\
  \bibnamefont {{Becker}}}, \bibinfo {author} {\bibfnamefont {A.}~\bibnamefont
  {{Benoit-L{\'e}vy}}}, \bibinfo {author} {\bibfnamefont {B.~A.}\ \bibnamefont
  {{Benson}}}, \bibinfo {author} {\bibfnamefont {G.~M.}\ \bibnamefont
  {{Bernstein}}}, \bibinfo {author} {\bibfnamefont {E.}~\bibnamefont
  {{Bertin}}}, \bibinfo {author} {\bibfnamefont {J.}~\bibnamefont {{Blazek}}},
  \bibinfo {author} {\bibfnamefont {S.~L.}\ \bibnamefont {{Bridle}}}, \bibinfo
  {author} {\bibfnamefont {D.}~\bibnamefont {{Brooks}}}, \bibinfo {author}
  {\bibfnamefont {D.}~\bibnamefont {{Brout}}}, \bibinfo {author} {\bibfnamefont
  {E.}~\bibnamefont {{Buckley-Geer}}}, \bibinfo {author} {\bibfnamefont
  {D.~L.}\ \bibnamefont {{Burke}}}, \bibinfo {author} {\bibfnamefont {M.~T.}\
  \bibnamefont {{Busha}}}, \bibinfo {author} {\bibfnamefont {A.}~\bibnamefont
  {{Campos}}}, \bibinfo {author} {\bibfnamefont {D.}~\bibnamefont {{Capozzi}}},
  \bibinfo {author} {\bibfnamefont {A.}~\bibnamefont {{Carnero Rosell}}},
  \bibinfo {author} {\bibfnamefont {M.}~\bibnamefont {{Carrasco Kind}}},
  \bibinfo {author} {\bibfnamefont {J.}~\bibnamefont {{Carretero}}}, \bibinfo
  {author} {\bibfnamefont {F.~J.}\ \bibnamefont {{Castander}}}, \bibinfo
  {author} {\bibfnamefont {R.}~\bibnamefont {{Cawthon}}}, \bibinfo {author}
  {\bibfnamefont {C.}~\bibnamefont {{Chang}}}, \bibinfo {author} {\bibfnamefont
  {N.}~\bibnamefont {{Chen}}}, \bibinfo {author} {\bibfnamefont
  {M.}~\bibnamefont {{Childress}}}, \bibinfo {author} {\bibfnamefont
  {A.}~\bibnamefont {{Choi}}}, \bibinfo {author} {\bibfnamefont
  {C.}~\bibnamefont {{Conselice}}}, \bibinfo {author} {\bibfnamefont
  {R.}~\bibnamefont {{Crittenden}}}, \bibinfo {author} {\bibfnamefont
  {M.}~\bibnamefont {{Crocce}}}, \bibinfo {author} {\bibfnamefont {C.~E.}\
  \bibnamefont {{Cunha}}}, \bibinfo {author} {\bibfnamefont {C.~B.}\
  \bibnamefont {{D'Andrea}}}, \bibinfo {author} {\bibfnamefont {L.~N.}\
  \bibnamefont {{da Costa}}}, \bibinfo {author} {\bibfnamefont
  {R.}~\bibnamefont {{Das}}}, \bibinfo {author} {\bibfnamefont {T.~M.}\
  \bibnamefont {{Davis}}}, \bibinfo {author} {\bibfnamefont {C.}~\bibnamefont
  {{Davis}}}, \bibinfo {author} {\bibfnamefont {J.}~\bibnamefont {{De
  Vicente}}}, \bibinfo {author} {\bibfnamefont {D.~L.}\ \bibnamefont
  {{DePoy}}}, \bibinfo {author} {\bibfnamefont {J.}~\bibnamefont {{DeRose}}},
  \bibinfo {author} {\bibfnamefont {S.}~\bibnamefont {{Desai}}}, \bibinfo
  {author} {\bibfnamefont {H.~T.}\ \bibnamefont {{Diehl}}}, \bibinfo {author}
  {\bibfnamefont {J.~P.}\ \bibnamefont {{Dietrich}}}, \bibinfo {author}
  {\bibfnamefont {S.}~\bibnamefont {{Dodelson}}}, \bibinfo {author}
  {\bibfnamefont {P.}~\bibnamefont {{Doel}}}, \bibinfo {author} {\bibfnamefont
  {A.}~\bibnamefont {{Drlica-Wagner}}}, \bibinfo {author} {\bibfnamefont
  {T.~F.}\ \bibnamefont {{Eifler}}}, \bibinfo {author} {\bibfnamefont {A.~E.}\
  \bibnamefont {{Elliott}}}, \bibinfo {author} {\bibfnamefont {F.}~\bibnamefont
  {{Elsner}}}, \bibinfo {author} {\bibfnamefont {J.}~\bibnamefont
  {{Elvin-Poole}}}, \bibinfo {author} {\bibfnamefont {J.}~\bibnamefont
  {{Estrada}}}, \bibinfo {author} {\bibfnamefont {A.~E.}\ \bibnamefont
  {{Evrard}}}, \bibinfo {author} {\bibfnamefont {Y.}~\bibnamefont {{Fang}}},
  \bibinfo {author} {\bibfnamefont {E.}~\bibnamefont {{Fernandez}}}, \bibinfo
  {author} {\bibfnamefont {A.}~\bibnamefont {{Fert{\'e}}}}, \bibinfo {author}
  {\bibfnamefont {D.~A.}\ \bibnamefont {{Finley}}}, \bibinfo {author}
  {\bibfnamefont {B.}~\bibnamefont {{Flaugher}}}, \bibinfo {author}
  {\bibfnamefont {P.}~\bibnamefont {{Fosalba}}}, \bibinfo {author}
  {\bibfnamefont {O.}~\bibnamefont {{Friedrich}}}, \bibinfo {author}
  {\bibfnamefont {J.}~\bibnamefont {{Frieman}}}, \bibinfo {author}
  {\bibfnamefont {J.}~\bibnamefont {{Garc{\'\i}a-Bellido}}}, \bibinfo {author}
  {\bibfnamefont {M.}~\bibnamefont {{Garcia-Fernandez}}}, \bibinfo {author}
  {\bibfnamefont {M.}~\bibnamefont {{Gatti}}}, \bibinfo {author} {\bibfnamefont
  {E.}~\bibnamefont {{Gaztanaga}}}, \bibinfo {author} {\bibfnamefont {D.~W.}\
  \bibnamefont {{Gerdes}}}, \bibinfo {author} {\bibfnamefont {T.}~\bibnamefont
  {{Giannantonio}}}, \bibinfo {author} {\bibfnamefont {M.~S.~S.}\ \bibnamefont
  {{Gill}}}, \bibinfo {author} {\bibfnamefont {K.}~\bibnamefont
  {{Glazebrook}}}, \bibinfo {author} {\bibfnamefont {D.~A.}\ \bibnamefont
  {{Goldstein}}}, \bibinfo {author} {\bibfnamefont {D.}~\bibnamefont
  {{Gruen}}}, \bibinfo {author} {\bibfnamefont {R.~A.}\ \bibnamefont
  {{Gruendl}}}, \bibinfo {author} {\bibfnamefont {J.}~\bibnamefont
  {{Gschwend}}}, \bibinfo {author} {\bibfnamefont {G.}~\bibnamefont
  {{Gutierrez}}}, \bibinfo {author} {\bibfnamefont {S.}~\bibnamefont
  {{Hamilton}}}, \bibinfo {author} {\bibfnamefont {W.~G.}\ \bibnamefont
  {{Hartley}}}, \bibinfo {author} {\bibfnamefont {S.~R.}\ \bibnamefont
  {{Hinton}}}, \bibinfo {author} {\bibfnamefont {K.}~\bibnamefont
  {{Honscheid}}}, \bibinfo {author} {\bibfnamefont {B.}~\bibnamefont
  {{Hoyle}}}, \bibinfo {author} {\bibfnamefont {D.}~\bibnamefont {{Huterer}}},
  \bibinfo {author} {\bibfnamefont {B.}~\bibnamefont {{Jain}}}, \bibinfo
  {author} {\bibfnamefont {D.~J.}\ \bibnamefont {{James}}}, \bibinfo {author}
  {\bibfnamefont {M.}~\bibnamefont {{Jarvis}}}, \bibinfo {author}
  {\bibfnamefont {T.}~\bibnamefont {{Jeltema}}}, \bibinfo {author}
  {\bibfnamefont {M.~D.}\ \bibnamefont {{Johnson}}}, \bibinfo {author}
  {\bibfnamefont {M.~W.~G.}\ \bibnamefont {{Johnson}}}, \bibinfo {author}
  {\bibfnamefont {T.}~\bibnamefont {{Kacprzak}}}, \bibinfo {author}
  {\bibfnamefont {S.}~\bibnamefont {{Kent}}}, \bibinfo {author} {\bibfnamefont
  {A.~G.}\ \bibnamefont {{Kim}}}, \bibinfo {author} {\bibfnamefont
  {A.}~\bibnamefont {{King}}}, \bibinfo {author} {\bibfnamefont
  {D.}~\bibnamefont {{Kirk}}}, \bibinfo {author} {\bibfnamefont
  {N.}~\bibnamefont {{Kokron}}}, \bibinfo {author} {\bibfnamefont
  {A.}~\bibnamefont {{Kovacs}}}, \bibinfo {author} {\bibfnamefont
  {E.}~\bibnamefont {{Krause}}}, \bibinfo {author} {\bibfnamefont
  {C.}~\bibnamefont {{Krawiec}}}, \bibinfo {author} {\bibfnamefont
  {A.}~\bibnamefont {{Kremin}}}, \bibinfo {author} {\bibfnamefont
  {K.}~\bibnamefont {{Kuehn}}}, \bibinfo {author} {\bibfnamefont
  {S.}~\bibnamefont {{Kuhlmann}}}, \bibinfo {author} {\bibfnamefont
  {N.}~\bibnamefont {{Kuropatkin}}}, \bibinfo {author} {\bibfnamefont
  {F.}~\bibnamefont {{Lacasa}}}, \bibinfo {author} {\bibfnamefont
  {O.}~\bibnamefont {{Lahav}}}, \bibinfo {author} {\bibfnamefont {T.~S.}\
  \bibnamefont {{Li}}}, \bibinfo {author} {\bibfnamefont {A.~R.}\ \bibnamefont
  {{Liddle}}}, \bibinfo {author} {\bibfnamefont {C.}~\bibnamefont {{Lidman}}},
  \bibinfo {author} {\bibfnamefont {M.}~\bibnamefont {{Lima}}}, \bibinfo
  {author} {\bibfnamefont {H.}~\bibnamefont {{Lin}}}, \bibinfo {author}
  {\bibfnamefont {N.}~\bibnamefont {{MacCrann}}}, \bibinfo {author}
  {\bibfnamefont {M.~A.~G.}\ \bibnamefont {{Maia}}}, \bibinfo {author}
  {\bibfnamefont {M.}~\bibnamefont {{Makler}}}, \bibinfo {author}
  {\bibfnamefont {M.}~\bibnamefont {{Manera}}}, \bibinfo {author}
  {\bibfnamefont {M.}~\bibnamefont {{March}}}, \bibinfo {author} {\bibfnamefont
  {J.~L.}\ \bibnamefont {{Marshall}}}, \bibinfo {author} {\bibfnamefont
  {P.}~\bibnamefont {{Martini}}}, \bibinfo {author} {\bibfnamefont {R.~G.}\
  \bibnamefont {{McMahon}}}, \bibinfo {author} {\bibfnamefont {P.}~\bibnamefont
  {{Melchior}}}, \bibinfo {author} {\bibfnamefont {F.}~\bibnamefont
  {{Menanteau}}}, \bibinfo {author} {\bibfnamefont {R.}~\bibnamefont
  {{Miquel}}}, \bibinfo {author} {\bibfnamefont {V.}~\bibnamefont {{Miranda}}},
  \bibinfo {author} {\bibfnamefont {D.}~\bibnamefont {{Mudd}}}, \bibinfo
  {author} {\bibfnamefont {J.}~\bibnamefont {{Muir}}}, \bibinfo {author}
  {\bibfnamefont {A.}~\bibnamefont {{M{\"o}ller}}}, \bibinfo {author}
  {\bibfnamefont {E.}~\bibnamefont {{Neilsen}}}, \bibinfo {author}
  {\bibfnamefont {R.~C.}\ \bibnamefont {{Nichol}}}, \bibinfo {author}
  {\bibfnamefont {B.}~\bibnamefont {{Nord}}}, \bibinfo {author} {\bibfnamefont
  {P.}~\bibnamefont {{Nugent}}}, \bibinfo {author} {\bibfnamefont {R.~L.~C.}\
  \bibnamefont {{Ogando}}}, \bibinfo {author} {\bibfnamefont {A.}~\bibnamefont
  {{Palmese}}}, \bibinfo {author} {\bibfnamefont {J.}~\bibnamefont
  {{Peacock}}}, \bibinfo {author} {\bibfnamefont {H.~V.}\ \bibnamefont
  {{Peiris}}}, \bibinfo {author} {\bibfnamefont {J.}~\bibnamefont {{Peoples}}},
  \bibinfo {author} {\bibfnamefont {W.~J.}\ \bibnamefont {{Percival}}},
  \bibinfo {author} {\bibfnamefont {D.}~\bibnamefont {{Petravick}}}, \bibinfo
  {author} {\bibfnamefont {A.~A.}\ \bibnamefont {{Plazas}}}, \bibinfo {author}
  {\bibfnamefont {A.}~\bibnamefont {{Porredon}}}, \bibinfo {author}
  {\bibfnamefont {J.}~\bibnamefont {{Prat}}}, \bibinfo {author} {\bibfnamefont
  {A.}~\bibnamefont {{Pujol}}}, \bibinfo {author} {\bibfnamefont {M.~M.}\
  \bibnamefont {{Rau}}}, \bibinfo {author} {\bibfnamefont {A.}~\bibnamefont
  {{Refregier}}}, \bibinfo {author} {\bibfnamefont {P.~M.}\ \bibnamefont
  {{Ricker}}}, \bibinfo {author} {\bibfnamefont {N.}~\bibnamefont {{Roe}}},
  \bibinfo {author} {\bibfnamefont {R.~P.}\ \bibnamefont {{Rollins}}}, \bibinfo
  {author} {\bibfnamefont {A.~K.}\ \bibnamefont {{Romer}}}, \bibinfo {author}
  {\bibfnamefont {A.}~\bibnamefont {{Roodman}}}, \bibinfo {author}
  {\bibfnamefont {R.}~\bibnamefont {{Rosenfeld}}}, \bibinfo {author}
  {\bibfnamefont {A.~J.}\ \bibnamefont {{Ross}}}, \bibinfo {author}
  {\bibfnamefont {E.}~\bibnamefont {{Rozo}}}, \bibinfo {author} {\bibfnamefont
  {E.~S.}\ \bibnamefont {{Rykoff}}}, \bibinfo {author} {\bibfnamefont
  {M.}~\bibnamefont {{Sako}}}, \bibinfo {author} {\bibfnamefont {A.~I.}\
  \bibnamefont {{Salvador}}}, \bibinfo {author} {\bibfnamefont
  {S.}~\bibnamefont {{Samuroff}}}, \bibinfo {author} {\bibfnamefont
  {C.}~\bibnamefont {{S{\'a}nchez}}}, \bibinfo {author} {\bibfnamefont
  {E.}~\bibnamefont {{Sanchez}}}, \bibinfo {author} {\bibfnamefont
  {B.}~\bibnamefont {{Santiago}}}, \bibinfo {author} {\bibfnamefont
  {V.}~\bibnamefont {{Scarpine}}}, \bibinfo {author} {\bibfnamefont
  {R.}~\bibnamefont {{Schindler}}}, \bibinfo {author} {\bibfnamefont
  {D.}~\bibnamefont {{Scolnic}}}, \bibinfo {author} {\bibfnamefont {L.~F.}\
  \bibnamefont {{Secco}}}, \bibinfo {author} {\bibfnamefont {S.}~\bibnamefont
  {{Serrano}}}, \bibinfo {author} {\bibfnamefont {I.}~\bibnamefont
  {{Sevilla-Noarbe}}}, \bibinfo {author} {\bibfnamefont {E.}~\bibnamefont
  {{Sheldon}}}, \bibinfo {author} {\bibfnamefont {R.~C.}\ \bibnamefont
  {{Smith}}}, \bibinfo {author} {\bibfnamefont {M.}~\bibnamefont {{Smith}}},
  \bibinfo {author} {\bibfnamefont {J.}~\bibnamefont {{Smith}}}, \bibinfo
  {author} {\bibfnamefont {M.}~\bibnamefont {{Soares-Santos}}}, \bibinfo
  {author} {\bibfnamefont {F.}~\bibnamefont {{Sobreira}}}, \bibinfo {author}
  {\bibfnamefont {E.}~\bibnamefont {{Suchyta}}}, \bibinfo {author}
  {\bibfnamefont {G.}~\bibnamefont {{Tarle}}}, \bibinfo {author} {\bibfnamefont
  {D.}~\bibnamefont {{Thomas}}}, \bibinfo {author} {\bibfnamefont {M.~A.}\
  \bibnamefont {{Troxel}}}, \bibinfo {author} {\bibfnamefont {D.~L.}\
  \bibnamefont {{Tucker}}}, \bibinfo {author} {\bibfnamefont {B.~E.}\
  \bibnamefont {{Tucker}}}, \bibinfo {author} {\bibfnamefont {S.~A.}\
  \bibnamefont {{Uddin}}}, \bibinfo {author} {\bibfnamefont {T.~N.}\
  \bibnamefont {{Varga}}}, \bibinfo {author} {\bibfnamefont {P.}~\bibnamefont
  {{Vielzeuf}}}, \bibinfo {author} {\bibfnamefont {V.}~\bibnamefont
  {{Vikram}}}, \bibinfo {author} {\bibfnamefont {A.~K.}\ \bibnamefont
  {{Vivas}}}, \bibinfo {author} {\bibfnamefont {A.~R.}\ \bibnamefont
  {{Walker}}}, \bibinfo {author} {\bibfnamefont {M.}~\bibnamefont {{Wang}}},
  \bibinfo {author} {\bibfnamefont {R.~H.}\ \bibnamefont {{Wechsler}}},
  \bibinfo {author} {\bibfnamefont {J.}~\bibnamefont {{Weller}}}, \bibinfo
  {author} {\bibfnamefont {W.}~\bibnamefont {{Wester}}}, \bibinfo {author}
  {\bibfnamefont {R.~C.}\ \bibnamefont {{Wolf}}}, \bibinfo {author}
  {\bibfnamefont {B.}~\bibnamefont {{Yanny}}}, \bibinfo {author} {\bibfnamefont
  {F.}~\bibnamefont {{Yuan}}}, \bibinfo {author} {\bibfnamefont
  {A.}~\bibnamefont {{Zenteno}}}, \bibinfo {author} {\bibfnamefont
  {B.}~\bibnamefont {{Zhang}}}, \bibinfo {author} {\bibfnamefont
  {Y.}~\bibnamefont {{Zhang}}}, \bibinfo {author} {\bibfnamefont
  {J.}~\bibnamefont {{Zuntz}}}, \ and\ \bibinfo {author} {\bibnamefont {{Dark
  Energy Survey Collaboration}}},\ }\href {\doibase 10.1103/PhysRevD.98.043526}
  {\bibfield  {journal} {\bibinfo  {journal} {\prd}\ }\textbf {\bibinfo
  {volume} {98}},\ \bibinfo {eid} {043526} (\bibinfo {year} {2018})},\ \Eprint
  {http://arxiv.org/abs/1708.01530} {arXiv:1708.01530 [astro-ph.CO]}
  \BibitemShut {NoStop}%
\bibitem [{\citenamefont {{Mandelbaum}}\ \emph {et~al.}(2018)\citenamefont
  {{Mandelbaum}}, \citenamefont {{Miyatake}}, \citenamefont {{Hamana}},
  \citenamefont {{Oguri}}, \citenamefont {{Simet}}, \citenamefont
  {{Armstrong}}, \citenamefont {{Bosch}}, \citenamefont {{Murata}},
  \citenamefont {{Lanusse}}, \citenamefont {{Leauthaud}}, \citenamefont
  {{Coupon}}, \citenamefont {{More}}, \citenamefont {{Takada}}, \citenamefont
  {{Miyazaki}}, \citenamefont {{Speagle}}, \citenamefont {{Shirasaki}},
  \citenamefont {{Sif{\'o}n}}, \citenamefont {{Huang}}, \citenamefont
  {{Nishizawa}}, \citenamefont {{Medezinski}}, \citenamefont {{Okura}},
  \citenamefont {{Okabe}}, \citenamefont {{Czakon}}, \citenamefont
  {{Takahashi}}, \citenamefont {{Coulton}}, \citenamefont {{Hikage}},
  \citenamefont {{Komiyama}}, \citenamefont {{Lupton}}, \citenamefont
  {{Strauss}}, \citenamefont {{Tanaka}},\ and\ \citenamefont
  {{Utsumi}}}]{2018PASJ...70S..25M}%
  \BibitemOpen
  \bibfield  {author} {\bibinfo {author} {\bibfnamefont {R.}~\bibnamefont
  {{Mandelbaum}}}, \bibinfo {author} {\bibfnamefont {H.}~\bibnamefont
  {{Miyatake}}}, \bibinfo {author} {\bibfnamefont {T.}~\bibnamefont
  {{Hamana}}}, \bibinfo {author} {\bibfnamefont {M.}~\bibnamefont {{Oguri}}},
  \bibinfo {author} {\bibfnamefont {M.}~\bibnamefont {{Simet}}}, \bibinfo
  {author} {\bibfnamefont {R.}~\bibnamefont {{Armstrong}}}, \bibinfo {author}
  {\bibfnamefont {J.}~\bibnamefont {{Bosch}}}, \bibinfo {author} {\bibfnamefont
  {R.}~\bibnamefont {{Murata}}}, \bibinfo {author} {\bibfnamefont
  {F.}~\bibnamefont {{Lanusse}}}, \bibinfo {author} {\bibfnamefont
  {A.}~\bibnamefont {{Leauthaud}}}, \bibinfo {author} {\bibfnamefont
  {J.}~\bibnamefont {{Coupon}}}, \bibinfo {author} {\bibfnamefont
  {S.}~\bibnamefont {{More}}}, \bibinfo {author} {\bibfnamefont
  {M.}~\bibnamefont {{Takada}}}, \bibinfo {author} {\bibfnamefont
  {S.}~\bibnamefont {{Miyazaki}}}, \bibinfo {author} {\bibfnamefont {J.~S.}\
  \bibnamefont {{Speagle}}}, \bibinfo {author} {\bibfnamefont {M.}~\bibnamefont
  {{Shirasaki}}}, \bibinfo {author} {\bibfnamefont {C.}~\bibnamefont
  {{Sif{\'o}n}}}, \bibinfo {author} {\bibfnamefont {S.}~\bibnamefont
  {{Huang}}}, \bibinfo {author} {\bibfnamefont {A.~J.}\ \bibnamefont
  {{Nishizawa}}}, \bibinfo {author} {\bibfnamefont {E.}~\bibnamefont
  {{Medezinski}}}, \bibinfo {author} {\bibfnamefont {Y.}~\bibnamefont
  {{Okura}}}, \bibinfo {author} {\bibfnamefont {N.}~\bibnamefont {{Okabe}}},
  \bibinfo {author} {\bibfnamefont {N.}~\bibnamefont {{Czakon}}}, \bibinfo
  {author} {\bibfnamefont {R.}~\bibnamefont {{Takahashi}}}, \bibinfo {author}
  {\bibfnamefont {W.~R.}\ \bibnamefont {{Coulton}}}, \bibinfo {author}
  {\bibfnamefont {C.}~\bibnamefont {{Hikage}}}, \bibinfo {author}
  {\bibfnamefont {Y.}~\bibnamefont {{Komiyama}}}, \bibinfo {author}
  {\bibfnamefont {R.~H.}\ \bibnamefont {{Lupton}}}, \bibinfo {author}
  {\bibfnamefont {M.~A.}\ \bibnamefont {{Strauss}}}, \bibinfo {author}
  {\bibfnamefont {M.}~\bibnamefont {{Tanaka}}}, \ and\ \bibinfo {author}
  {\bibfnamefont {Y.}~\bibnamefont {{Utsumi}}},\ }\href {\doibase
  10.1093/pasj/psx130} {\bibfield  {journal} {\bibinfo  {journal} {Publications
  of the Astronomical Society of Japan}\ }\textbf {\bibinfo {volume} {70}},\
  \bibinfo {eid} {S25} (\bibinfo {year} {2018})},\ \Eprint
  {http://arxiv.org/abs/1705.06745} {arXiv:1705.06745 [astro-ph.CO]}
  \BibitemShut {NoStop}%
\bibitem [{\citenamefont {{Hikage}}\ \emph {et~al.}(2019)\citenamefont
  {{Hikage}}, \citenamefont {{Oguri}}, \citenamefont {{Hamana}}, \citenamefont
  {{More}}, \citenamefont {{Mandelbaum}}, \citenamefont {{Takada}},
  \citenamefont {{K{\"o}hlinger}}, \citenamefont {{Miyatake}}, \citenamefont
  {{Nishizawa}}, \citenamefont {{Aihara}}, \citenamefont {{Armstrong}},
  \citenamefont {{Bosch}}, \citenamefont {{Coupon}}, \citenamefont {{Ducout}},
  \citenamefont {{Ho}}, \citenamefont {{Hsieh}}, \citenamefont {{Komiyama}},
  \citenamefont {{Lanusse}}, \citenamefont {{Leauthaud}}, \citenamefont
  {{Lupton}}, \citenamefont {{Medezinski}}, \citenamefont {{Mineo}},
  \citenamefont {{Miyama}}, \citenamefont {{Miyazaki}}, \citenamefont
  {{Murata}}, \citenamefont {{Murayama}}, \citenamefont {{Shirasaki}},
  \citenamefont {{Sif{\'o}n}}, \citenamefont {{Simet}}, \citenamefont
  {{Speagle}}, \citenamefont {{Spergel}}, \citenamefont {{Strauss}},
  \citenamefont {{Sugiyama}}, \citenamefont {{Tanaka}}, \citenamefont
  {{Utsumi}}, \citenamefont {{Wang}},\ and\ \citenamefont
  {{Yamada}}}]{2019PASJ..tmp...22H}%
  \BibitemOpen
  \bibfield  {author} {\bibinfo {author} {\bibfnamefont {C.}~\bibnamefont
  {{Hikage}}}, \bibinfo {author} {\bibfnamefont {M.}~\bibnamefont {{Oguri}}},
  \bibinfo {author} {\bibfnamefont {T.}~\bibnamefont {{Hamana}}}, \bibinfo
  {author} {\bibfnamefont {S.}~\bibnamefont {{More}}}, \bibinfo {author}
  {\bibfnamefont {R.}~\bibnamefont {{Mandelbaum}}}, \bibinfo {author}
  {\bibfnamefont {M.}~\bibnamefont {{Takada}}}, \bibinfo {author}
  {\bibfnamefont {F.}~\bibnamefont {{K{\"o}hlinger}}}, \bibinfo {author}
  {\bibfnamefont {H.}~\bibnamefont {{Miyatake}}}, \bibinfo {author}
  {\bibfnamefont {A.~J.}\ \bibnamefont {{Nishizawa}}}, \bibinfo {author}
  {\bibfnamefont {H.}~\bibnamefont {{Aihara}}}, \bibinfo {author}
  {\bibfnamefont {R.}~\bibnamefont {{Armstrong}}}, \bibinfo {author}
  {\bibfnamefont {J.}~\bibnamefont {{Bosch}}}, \bibinfo {author} {\bibfnamefont
  {J.}~\bibnamefont {{Coupon}}}, \bibinfo {author} {\bibfnamefont
  {A.}~\bibnamefont {{Ducout}}}, \bibinfo {author} {\bibfnamefont
  {P.}~\bibnamefont {{Ho}}}, \bibinfo {author} {\bibfnamefont {B.-C.}\
  \bibnamefont {{Hsieh}}}, \bibinfo {author} {\bibfnamefont {Y.}~\bibnamefont
  {{Komiyama}}}, \bibinfo {author} {\bibfnamefont {F.}~\bibnamefont
  {{Lanusse}}}, \bibinfo {author} {\bibfnamefont {A.}~\bibnamefont
  {{Leauthaud}}}, \bibinfo {author} {\bibfnamefont {R.~H.}\ \bibnamefont
  {{Lupton}}}, \bibinfo {author} {\bibfnamefont {E.}~\bibnamefont
  {{Medezinski}}}, \bibinfo {author} {\bibfnamefont {S.}~\bibnamefont
  {{Mineo}}}, \bibinfo {author} {\bibfnamefont {S.}~\bibnamefont {{Miyama}}},
  \bibinfo {author} {\bibfnamefont {S.}~\bibnamefont {{Miyazaki}}}, \bibinfo
  {author} {\bibfnamefont {R.}~\bibnamefont {{Murata}}}, \bibinfo {author}
  {\bibfnamefont {H.}~\bibnamefont {{Murayama}}}, \bibinfo {author}
  {\bibfnamefont {M.}~\bibnamefont {{Shirasaki}}}, \bibinfo {author}
  {\bibfnamefont {C.}~\bibnamefont {{Sif{\'o}n}}}, \bibinfo {author}
  {\bibfnamefont {M.}~\bibnamefont {{Simet}}}, \bibinfo {author} {\bibfnamefont
  {J.}~\bibnamefont {{Speagle}}}, \bibinfo {author} {\bibfnamefont {D.~N.}\
  \bibnamefont {{Spergel}}}, \bibinfo {author} {\bibfnamefont {M.~A.}\
  \bibnamefont {{Strauss}}}, \bibinfo {author} {\bibfnamefont {N.}~\bibnamefont
  {{Sugiyama}}}, \bibinfo {author} {\bibfnamefont {M.}~\bibnamefont
  {{Tanaka}}}, \bibinfo {author} {\bibfnamefont {Y.}~\bibnamefont {{Utsumi}}},
  \bibinfo {author} {\bibfnamefont {S.-Y.}\ \bibnamefont {{Wang}}}, \ and\
  \bibinfo {author} {\bibfnamefont {Y.}~\bibnamefont {{Yamada}}},\ }\href
  {\doibase 10.1093/pasj/psz010} {\bibfield  {journal} {\bibinfo  {journal}
  {Publications of the Astronomical Society of Japan}\ ,\ \bibinfo {pages}
  {22}} (\bibinfo {year} {2019})},\ \Eprint {http://arxiv.org/abs/1809.09148}
  {arXiv:1809.09148 [astro-ph.CO]} \BibitemShut {NoStop}%
\bibitem [{\citenamefont {{Joudaki}}\ \emph {et~al.}(2020)\citenamefont
  {{Joudaki}}, \citenamefont {{Hildebrandt}}, \citenamefont {{Traykova}},
  \citenamefont {{Chisari}}, \citenamefont {{Heymans}}, \citenamefont
  {{Kannawadi}}, \citenamefont {{Kuijken}}, \citenamefont {{Wright}},
  \citenamefont {{Asgari}}, \citenamefont {{Erben}}, \citenamefont
  {{Hoekstra}}, \citenamefont {{Joachimi}}, \citenamefont {{Miller}},
  \citenamefont {{Tr{\"o}ster}},\ and\ \citenamefont {{van den
  Busch}}}]{joudaki20a}%
  \BibitemOpen
  \bibfield  {author} {\bibinfo {author} {\bibfnamefont {S.}~\bibnamefont
  {{Joudaki}}}, \bibinfo {author} {\bibfnamefont {H.}~\bibnamefont
  {{Hildebrandt}}}, \bibinfo {author} {\bibfnamefont {D.}~\bibnamefont
  {{Traykova}}}, \bibinfo {author} {\bibfnamefont {N.~E.}\ \bibnamefont
  {{Chisari}}}, \bibinfo {author} {\bibfnamefont {C.}~\bibnamefont
  {{Heymans}}}, \bibinfo {author} {\bibfnamefont {A.}~\bibnamefont
  {{Kannawadi}}}, \bibinfo {author} {\bibfnamefont {K.}~\bibnamefont
  {{Kuijken}}}, \bibinfo {author} {\bibfnamefont {A.~H.}\ \bibnamefont
  {{Wright}}}, \bibinfo {author} {\bibfnamefont {M.}~\bibnamefont {{Asgari}}},
  \bibinfo {author} {\bibfnamefont {T.}~\bibnamefont {{Erben}}}, \bibinfo
  {author} {\bibfnamefont {H.}~\bibnamefont {{Hoekstra}}}, \bibinfo {author}
  {\bibfnamefont {B.}~\bibnamefont {{Joachimi}}}, \bibinfo {author}
  {\bibfnamefont {L.}~\bibnamefont {{Miller}}}, \bibinfo {author}
  {\bibfnamefont {T.}~\bibnamefont {{Tr{\"o}ster}}}, \ and\ \bibinfo {author}
  {\bibfnamefont {J.~L.}\ \bibnamefont {{van den Busch}}},\ }\href {\doibase
  10.1051/0004-6361/201936154} {\bibfield  {journal} {\bibinfo  {journal}
  {\aap}\ }\textbf {\bibinfo {volume} {638}},\ \bibinfo {eid} {L1} (\bibinfo
  {year} {2020})},\ \Eprint {http://arxiv.org/abs/1906.09262} {arXiv:1906.09262
  [astro-ph.CO]} \BibitemShut {NoStop}%
\bibitem [{\citenamefont {{Riess}}\ \emph {et~al.}(2019)\citenamefont
  {{Riess}}, \citenamefont {{Casertano}}, \citenamefont {{Yuan}}, \citenamefont
  {{Macri}},\ and\ \citenamefont {{Scolnic}}}]{riess19a}%
  \BibitemOpen
  \bibfield  {author} {\bibinfo {author} {\bibfnamefont {A.~G.}\ \bibnamefont
  {{Riess}}}, \bibinfo {author} {\bibfnamefont {S.}~\bibnamefont
  {{Casertano}}}, \bibinfo {author} {\bibfnamefont {W.}~\bibnamefont {{Yuan}}},
  \bibinfo {author} {\bibfnamefont {L.~M.}\ \bibnamefont {{Macri}}}, \ and\
  \bibinfo {author} {\bibfnamefont {D.}~\bibnamefont {{Scolnic}}},\ }\href
  {\doibase 10.3847/1538-4357/ab1422} {\bibfield  {journal} {\bibinfo
  {journal} {\apj}\ }\textbf {\bibinfo {volume} {876}},\ \bibinfo {eid} {85}
  (\bibinfo {year} {2019})},\ \Eprint {http://arxiv.org/abs/1903.07603}
  {arXiv:1903.07603 [astro-ph.CO]} \BibitemShut {NoStop}%
\bibitem [{\citenamefont {{Planck Collaboration}}\ \emph
  {et~al.}(2018{\natexlab{b}})\citenamefont {{Planck Collaboration}},
  \citenamefont {{Aghanim}}, \citenamefont {{Akrami}}, \citenamefont
  {{Ashdown}}, \citenamefont {{Aumont}}, \citenamefont {{Baccigalupi}},
  \citenamefont {{Ballardini}}, \citenamefont {{Banday}}, \citenamefont
  {{Barreiro}}, \citenamefont {{Bartolo}}, \citenamefont {{Basak}},
  \citenamefont {{Battye}}, \citenamefont {{Benabed}}, \citenamefont
  {{Bernard}}, \citenamefont {{Bersanelli}}, \citenamefont {{Bielewicz}},
  \citenamefont {{Bock}}, \citenamefont {{Bond}}, \citenamefont {{Borrill}},
  \citenamefont {{Bouchet}}, \citenamefont {{Boulanger}}, \citenamefont
  {{Bucher}}, \citenamefont {{Burigana}}, \citenamefont {{Butler}},
  \citenamefont {{Calabrese}}, \citenamefont {{Cardoso}}, \citenamefont
  {{Carron}}, \citenamefont {{Challinor}}, \citenamefont {{Chiang}},
  \citenamefont {{Chluba}}, \citenamefont {{Colombo}}, \citenamefont
  {{Combet}}, \citenamefont {{Contreras}}, \citenamefont {{Crill}},
  \citenamefont {{Cuttaia}}, \citenamefont {{de Bernardis}}, \citenamefont {{de
  Zotti}}, \citenamefont {{Delabrouille}}, \citenamefont {{Delouis}},
  \citenamefont {{Di Valentino}}, \citenamefont {{Diego}}, \citenamefont
  {{Dor{\'e}}}, \citenamefont {{Douspis}}, \citenamefont {{Ducout}},
  \citenamefont {{Dupac}}, \citenamefont {{Dusini}}, \citenamefont
  {{Efstathiou}}, \citenamefont {{Elsner}}, \citenamefont {{En{\ss}lin}},
  \citenamefont {{Eriksen}}, \citenamefont {{Fantaye}}, \citenamefont
  {{Farhang}}, \citenamefont {{Fergusson}}, \citenamefont {{Fernandez-Cobos}},
  \citenamefont {{Finelli}}, \citenamefont {{Forastieri}}, \citenamefont
  {{Frailis}}, \citenamefont {{Franceschi}}, \citenamefont {{Frolov}},
  \citenamefont {{Galeotta}}, \citenamefont {{Galli}}, \citenamefont {{Ganga}},
  \citenamefont {{G{\'e}nova-Santos}}, \citenamefont {{Gerbino}}, \citenamefont
  {{Ghosh}}, \citenamefont {{Gonz{\'a}lez-Nuevo}}, \citenamefont
  {{G{\'o}rski}}, \citenamefont {{Gratton}}, \citenamefont {{Gruppuso}},
  \citenamefont {{Gudmundsson}}, \citenamefont {{Hamann}}, \citenamefont {{Hand
  ley}}, \citenamefont {{Herranz}}, \citenamefont {{Hivon}}, \citenamefont
  {{Huang}}, \citenamefont {{Jaffe}}, \citenamefont {{Jones}}, \citenamefont
  {{Karakci}}, \citenamefont {{Keih{\"a}nen}}, \citenamefont {{Keskitalo}},
  \citenamefont {{Kiiveri}}, \citenamefont {{Kim}}, \citenamefont {{Kisner}},
  \citenamefont {{Knox}}, \citenamefont {{Krachmalnicoff}}, \citenamefont
  {{Kunz}}, \citenamefont {{Kurki-Suonio}}, \citenamefont {{Lagache}},
  \citenamefont {{Lamarre}}, \citenamefont {{Lasenby}}, \citenamefont
  {{Lattanzi}}, \citenamefont {{Lawrence}}, \citenamefont {{Le Jeune}},
  \citenamefont {{Lemos}}, \citenamefont {{Lesgourgues}}, \citenamefont
  {{Levrier}}, \citenamefont {{Lewis}}, \citenamefont {{Liguori}},
  \citenamefont {{Lilje}}, \citenamefont {{Lilley}}, \citenamefont
  {{Lindholm}}, \citenamefont {{L{\'o}pez-Caniego}}, \citenamefont {{Lubin}},
  \citenamefont {{Ma}}, \citenamefont {{Mac{\'\i}as-P{\'e}rez}}, \citenamefont
  {{Maggio}}, \citenamefont {{Maino}}, \citenamefont {{Mandolesi}},
  \citenamefont {{Mangilli}}, \citenamefont {{Marcos-Caballero}}, \citenamefont
  {{Maris}}, \citenamefont {{Martin}}, \citenamefont {{Martinelli}},
  \citenamefont {{Mart{\'\i}nez-Gonz{\'a}lez}}, \citenamefont {{Matarrese}},
  \citenamefont {{Mauri}}, \citenamefont {{McEwen}}, \citenamefont
  {{Meinhold}}, \citenamefont {{Melchiorri}}, \citenamefont {{Mennella}},
  \citenamefont {{Migliaccio}}, \citenamefont {{Millea}}, \citenamefont
  {{Mitra}}, \citenamefont {{Miville-Desch{\^e}nes}}, \citenamefont
  {{Molinari}}, \citenamefont {{Montier}}, \citenamefont {{Morgante}},
  \citenamefont {{Moss}}, \citenamefont {{Natoli}}, \citenamefont
  {{N{\o}rgaard-Nielsen}}, \citenamefont {{Pagano}}, \citenamefont
  {{Paoletti}}, \citenamefont {{Partridge}}, \citenamefont {{Patanchon}},
  \citenamefont {{Peiris}}, \citenamefont {{Perrotta}}, \citenamefont
  {{Pettorino}}, \citenamefont {{Piacentini}}, \citenamefont {{Polastri}},
  \citenamefont {{Polenta}}, \citenamefont {{Puget}}, \citenamefont {{Rachen}},
  \citenamefont {{Reinecke}}, \citenamefont {{Remazeilles}}, \citenamefont
  {{Renzi}}, \citenamefont {{Rocha}}, \citenamefont {{Rosset}}, \citenamefont
  {{Roudier}}, \citenamefont {{Rubi{\~n}o-Mart{\'\i}n}}, \citenamefont
  {{Ruiz-Granados}}, \citenamefont {{Salvati}}, \citenamefont {{Sandri}},
  \citenamefont {{Savelainen}}, \citenamefont {{Scott}}, \citenamefont
  {{Shellard}}, \citenamefont {{Sirignano}}, \citenamefont {{Sirri}},
  \citenamefont {{Spencer}}, \citenamefont {{Sunyaev}}, \citenamefont
  {{Suur-Uski}}, \citenamefont {{Tauber}}, \citenamefont {{Tavagnacco}},
  \citenamefont {{Tenti}}, \citenamefont {{Toffolatti}}, \citenamefont
  {{Tomasi}}, \citenamefont {{Trombetti}}, \citenamefont {{Valenziano}},
  \citenamefont {{Valiviita}}, \citenamefont {{Van Tent}}, \citenamefont
  {{Vibert}}, \citenamefont {{Vielva}}, \citenamefont {{Villa}}, \citenamefont
  {{Vittorio}}, \citenamefont {{Wand elt}}, \citenamefont {{Wehus}},
  \citenamefont {{White}}, \citenamefont {{White}}, \citenamefont {{Zacchei}},\
  and\ \citenamefont {{Zonca}}}]{2018arXiv180706209P}%
  \BibitemOpen
  \bibfield  {author} {\bibinfo {author} {\bibnamefont {{Planck
  Collaboration}}}, \bibinfo {author} {\bibfnamefont {N.}~\bibnamefont
  {{Aghanim}}}, \bibinfo {author} {\bibfnamefont {Y.}~\bibnamefont {{Akrami}}},
  \bibinfo {author} {\bibfnamefont {M.}~\bibnamefont {{Ashdown}}}, \bibinfo
  {author} {\bibfnamefont {J.}~\bibnamefont {{Aumont}}}, \bibinfo {author}
  {\bibfnamefont {C.}~\bibnamefont {{Baccigalupi}}}, \bibinfo {author}
  {\bibfnamefont {M.}~\bibnamefont {{Ballardini}}}, \bibinfo {author}
  {\bibfnamefont {A.~J.}\ \bibnamefont {{Banday}}}, \bibinfo {author}
  {\bibfnamefont {R.~B.}\ \bibnamefont {{Barreiro}}}, \bibinfo {author}
  {\bibfnamefont {N.}~\bibnamefont {{Bartolo}}}, \bibinfo {author}
  {\bibfnamefont {S.}~\bibnamefont {{Basak}}}, \bibinfo {author} {\bibfnamefont
  {R.}~\bibnamefont {{Battye}}}, \bibinfo {author} {\bibfnamefont
  {K.}~\bibnamefont {{Benabed}}}, \bibinfo {author} {\bibfnamefont {J.~P.}\
  \bibnamefont {{Bernard}}}, \bibinfo {author} {\bibfnamefont {M.}~\bibnamefont
  {{Bersanelli}}}, \bibinfo {author} {\bibfnamefont {P.}~\bibnamefont
  {{Bielewicz}}}, \bibinfo {author} {\bibfnamefont {J.~J.}\ \bibnamefont
  {{Bock}}}, \bibinfo {author} {\bibfnamefont {J.~R.}\ \bibnamefont {{Bond}}},
  \bibinfo {author} {\bibfnamefont {J.}~\bibnamefont {{Borrill}}}, \bibinfo
  {author} {\bibfnamefont {F.~R.}\ \bibnamefont {{Bouchet}}}, \bibinfo {author}
  {\bibfnamefont {F.}~\bibnamefont {{Boulanger}}}, \bibinfo {author}
  {\bibfnamefont {M.}~\bibnamefont {{Bucher}}}, \bibinfo {author}
  {\bibfnamefont {C.}~\bibnamefont {{Burigana}}}, \bibinfo {author}
  {\bibfnamefont {R.~C.}\ \bibnamefont {{Butler}}}, \bibinfo {author}
  {\bibfnamefont {E.}~\bibnamefont {{Calabrese}}}, \bibinfo {author}
  {\bibfnamefont {J.~F.}\ \bibnamefont {{Cardoso}}}, \bibinfo {author}
  {\bibfnamefont {J.}~\bibnamefont {{Carron}}}, \bibinfo {author}
  {\bibfnamefont {A.}~\bibnamefont {{Challinor}}}, \bibinfo {author}
  {\bibfnamefont {H.~C.}\ \bibnamefont {{Chiang}}}, \bibinfo {author}
  {\bibfnamefont {J.}~\bibnamefont {{Chluba}}}, \bibinfo {author}
  {\bibfnamefont {L.~P.~L.}\ \bibnamefont {{Colombo}}}, \bibinfo {author}
  {\bibfnamefont {C.}~\bibnamefont {{Combet}}}, \bibinfo {author}
  {\bibfnamefont {D.}~\bibnamefont {{Contreras}}}, \bibinfo {author}
  {\bibfnamefont {B.~P.}\ \bibnamefont {{Crill}}}, \bibinfo {author}
  {\bibfnamefont {F.}~\bibnamefont {{Cuttaia}}}, \bibinfo {author}
  {\bibfnamefont {P.}~\bibnamefont {{de Bernardis}}}, \bibinfo {author}
  {\bibfnamefont {G.}~\bibnamefont {{de Zotti}}}, \bibinfo {author}
  {\bibfnamefont {J.}~\bibnamefont {{Delabrouille}}}, \bibinfo {author}
  {\bibfnamefont {J.~M.}\ \bibnamefont {{Delouis}}}, \bibinfo {author}
  {\bibfnamefont {E.}~\bibnamefont {{Di Valentino}}}, \bibinfo {author}
  {\bibfnamefont {J.~M.}\ \bibnamefont {{Diego}}}, \bibinfo {author}
  {\bibfnamefont {O.}~\bibnamefont {{Dor{\'e}}}}, \bibinfo {author}
  {\bibfnamefont {M.}~\bibnamefont {{Douspis}}}, \bibinfo {author}
  {\bibfnamefont {A.}~\bibnamefont {{Ducout}}}, \bibinfo {author}
  {\bibfnamefont {X.}~\bibnamefont {{Dupac}}}, \bibinfo {author} {\bibfnamefont
  {S.}~\bibnamefont {{Dusini}}}, \bibinfo {author} {\bibfnamefont
  {G.}~\bibnamefont {{Efstathiou}}}, \bibinfo {author} {\bibfnamefont
  {F.}~\bibnamefont {{Elsner}}}, \bibinfo {author} {\bibfnamefont {T.~A.}\
  \bibnamefont {{En{\ss}lin}}}, \bibinfo {author} {\bibfnamefont {H.~K.}\
  \bibnamefont {{Eriksen}}}, \bibinfo {author} {\bibfnamefont {Y.}~\bibnamefont
  {{Fantaye}}}, \bibinfo {author} {\bibfnamefont {M.}~\bibnamefont
  {{Farhang}}}, \bibinfo {author} {\bibfnamefont {J.}~\bibnamefont
  {{Fergusson}}}, \bibinfo {author} {\bibfnamefont {R.}~\bibnamefont
  {{Fernandez-Cobos}}}, \bibinfo {author} {\bibfnamefont {F.}~\bibnamefont
  {{Finelli}}}, \bibinfo {author} {\bibfnamefont {F.}~\bibnamefont
  {{Forastieri}}}, \bibinfo {author} {\bibfnamefont {M.}~\bibnamefont
  {{Frailis}}}, \bibinfo {author} {\bibfnamefont {E.}~\bibnamefont
  {{Franceschi}}}, \bibinfo {author} {\bibfnamefont {A.}~\bibnamefont
  {{Frolov}}}, \bibinfo {author} {\bibfnamefont {S.}~\bibnamefont
  {{Galeotta}}}, \bibinfo {author} {\bibfnamefont {S.}~\bibnamefont {{Galli}}},
  \bibinfo {author} {\bibfnamefont {K.}~\bibnamefont {{Ganga}}}, \bibinfo
  {author} {\bibfnamefont {R.~T.}\ \bibnamefont {{G{\'e}nova-Santos}}},
  \bibinfo {author} {\bibfnamefont {M.}~\bibnamefont {{Gerbino}}}, \bibinfo
  {author} {\bibfnamefont {T.}~\bibnamefont {{Ghosh}}}, \bibinfo {author}
  {\bibfnamefont {J.}~\bibnamefont {{Gonz{\'a}lez-Nuevo}}}, \bibinfo {author}
  {\bibfnamefont {K.~M.}\ \bibnamefont {{G{\'o}rski}}}, \bibinfo {author}
  {\bibfnamefont {S.}~\bibnamefont {{Gratton}}}, \bibinfo {author}
  {\bibfnamefont {A.}~\bibnamefont {{Gruppuso}}}, \bibinfo {author}
  {\bibfnamefont {J.~E.}\ \bibnamefont {{Gudmundsson}}}, \bibinfo {author}
  {\bibfnamefont {J.}~\bibnamefont {{Hamann}}}, \bibinfo {author}
  {\bibfnamefont {W.}~\bibnamefont {{Hand ley}}}, \bibinfo {author}
  {\bibfnamefont {D.}~\bibnamefont {{Herranz}}}, \bibinfo {author}
  {\bibfnamefont {E.}~\bibnamefont {{Hivon}}}, \bibinfo {author} {\bibfnamefont
  {Z.}~\bibnamefont {{Huang}}}, \bibinfo {author} {\bibfnamefont {A.~H.}\
  \bibnamefont {{Jaffe}}}, \bibinfo {author} {\bibfnamefont {W.~C.}\
  \bibnamefont {{Jones}}}, \bibinfo {author} {\bibfnamefont {A.}~\bibnamefont
  {{Karakci}}}, \bibinfo {author} {\bibfnamefont {E.}~\bibnamefont
  {{Keih{\"a}nen}}}, \bibinfo {author} {\bibfnamefont {R.}~\bibnamefont
  {{Keskitalo}}}, \bibinfo {author} {\bibfnamefont {K.}~\bibnamefont
  {{Kiiveri}}}, \bibinfo {author} {\bibfnamefont {J.}~\bibnamefont {{Kim}}},
  \bibinfo {author} {\bibfnamefont {T.~S.}\ \bibnamefont {{Kisner}}}, \bibinfo
  {author} {\bibfnamefont {L.}~\bibnamefont {{Knox}}}, \bibinfo {author}
  {\bibfnamefont {N.}~\bibnamefont {{Krachmalnicoff}}}, \bibinfo {author}
  {\bibfnamefont {M.}~\bibnamefont {{Kunz}}}, \bibinfo {author} {\bibfnamefont
  {H.}~\bibnamefont {{Kurki-Suonio}}}, \bibinfo {author} {\bibfnamefont
  {G.}~\bibnamefont {{Lagache}}}, \bibinfo {author} {\bibfnamefont {J.~M.}\
  \bibnamefont {{Lamarre}}}, \bibinfo {author} {\bibfnamefont {A.}~\bibnamefont
  {{Lasenby}}}, \bibinfo {author} {\bibfnamefont {M.}~\bibnamefont
  {{Lattanzi}}}, \bibinfo {author} {\bibfnamefont {C.~R.}\ \bibnamefont
  {{Lawrence}}}, \bibinfo {author} {\bibfnamefont {M.}~\bibnamefont {{Le
  Jeune}}}, \bibinfo {author} {\bibfnamefont {P.}~\bibnamefont {{Lemos}}},
  \bibinfo {author} {\bibfnamefont {J.}~\bibnamefont {{Lesgourgues}}}, \bibinfo
  {author} {\bibfnamefont {F.}~\bibnamefont {{Levrier}}}, \bibinfo {author}
  {\bibfnamefont {A.}~\bibnamefont {{Lewis}}}, \bibinfo {author} {\bibfnamefont
  {M.}~\bibnamefont {{Liguori}}}, \bibinfo {author} {\bibfnamefont {P.~B.}\
  \bibnamefont {{Lilje}}}, \bibinfo {author} {\bibfnamefont {M.}~\bibnamefont
  {{Lilley}}}, \bibinfo {author} {\bibfnamefont {V.}~\bibnamefont
  {{Lindholm}}}, \bibinfo {author} {\bibfnamefont {M.}~\bibnamefont
  {{L{\'o}pez-Caniego}}}, \bibinfo {author} {\bibfnamefont {P.~M.}\
  \bibnamefont {{Lubin}}}, \bibinfo {author} {\bibfnamefont {Y.~Z.}\
  \bibnamefont {{Ma}}}, \bibinfo {author} {\bibfnamefont {J.~F.}\ \bibnamefont
  {{Mac{\'\i}as-P{\'e}rez}}}, \bibinfo {author} {\bibfnamefont
  {G.}~\bibnamefont {{Maggio}}}, \bibinfo {author} {\bibfnamefont
  {D.}~\bibnamefont {{Maino}}}, \bibinfo {author} {\bibfnamefont
  {N.}~\bibnamefont {{Mandolesi}}}, \bibinfo {author} {\bibfnamefont
  {A.}~\bibnamefont {{Mangilli}}}, \bibinfo {author} {\bibfnamefont
  {A.}~\bibnamefont {{Marcos-Caballero}}}, \bibinfo {author} {\bibfnamefont
  {M.}~\bibnamefont {{Maris}}}, \bibinfo {author} {\bibfnamefont {P.~G.}\
  \bibnamefont {{Martin}}}, \bibinfo {author} {\bibfnamefont {M.}~\bibnamefont
  {{Martinelli}}}, \bibinfo {author} {\bibfnamefont {E.}~\bibnamefont
  {{Mart{\'\i}nez-Gonz{\'a}lez}}}, \bibinfo {author} {\bibfnamefont
  {S.}~\bibnamefont {{Matarrese}}}, \bibinfo {author} {\bibfnamefont
  {N.}~\bibnamefont {{Mauri}}}, \bibinfo {author} {\bibfnamefont {J.~D.}\
  \bibnamefont {{McEwen}}}, \bibinfo {author} {\bibfnamefont {P.~R.}\
  \bibnamefont {{Meinhold}}}, \bibinfo {author} {\bibfnamefont
  {A.}~\bibnamefont {{Melchiorri}}}, \bibinfo {author} {\bibfnamefont
  {A.}~\bibnamefont {{Mennella}}}, \bibinfo {author} {\bibfnamefont
  {M.}~\bibnamefont {{Migliaccio}}}, \bibinfo {author} {\bibfnamefont
  {M.}~\bibnamefont {{Millea}}}, \bibinfo {author} {\bibfnamefont
  {S.}~\bibnamefont {{Mitra}}}, \bibinfo {author} {\bibfnamefont {M.~A.}\
  \bibnamefont {{Miville-Desch{\^e}nes}}}, \bibinfo {author} {\bibfnamefont
  {D.}~\bibnamefont {{Molinari}}}, \bibinfo {author} {\bibfnamefont
  {L.}~\bibnamefont {{Montier}}}, \bibinfo {author} {\bibfnamefont
  {G.}~\bibnamefont {{Morgante}}}, \bibinfo {author} {\bibfnamefont
  {A.}~\bibnamefont {{Moss}}}, \bibinfo {author} {\bibfnamefont
  {P.}~\bibnamefont {{Natoli}}}, \bibinfo {author} {\bibfnamefont {H.~U.}\
  \bibnamefont {{N{\o}rgaard-Nielsen}}}, \bibinfo {author} {\bibfnamefont
  {L.}~\bibnamefont {{Pagano}}}, \bibinfo {author} {\bibfnamefont
  {D.}~\bibnamefont {{Paoletti}}}, \bibinfo {author} {\bibfnamefont
  {B.}~\bibnamefont {{Partridge}}}, \bibinfo {author} {\bibfnamefont
  {G.}~\bibnamefont {{Patanchon}}}, \bibinfo {author} {\bibfnamefont {H.~V.}\
  \bibnamefont {{Peiris}}}, \bibinfo {author} {\bibfnamefont {F.}~\bibnamefont
  {{Perrotta}}}, \bibinfo {author} {\bibfnamefont {V.}~\bibnamefont
  {{Pettorino}}}, \bibinfo {author} {\bibfnamefont {F.}~\bibnamefont
  {{Piacentini}}}, \bibinfo {author} {\bibfnamefont {L.}~\bibnamefont
  {{Polastri}}}, \bibinfo {author} {\bibfnamefont {G.}~\bibnamefont
  {{Polenta}}}, \bibinfo {author} {\bibfnamefont {J.~L.}\ \bibnamefont
  {{Puget}}}, \bibinfo {author} {\bibfnamefont {J.~P.}\ \bibnamefont
  {{Rachen}}}, \bibinfo {author} {\bibfnamefont {M.}~\bibnamefont
  {{Reinecke}}}, \bibinfo {author} {\bibfnamefont {M.}~\bibnamefont
  {{Remazeilles}}}, \bibinfo {author} {\bibfnamefont {A.}~\bibnamefont
  {{Renzi}}}, \bibinfo {author} {\bibfnamefont {G.}~\bibnamefont {{Rocha}}},
  \bibinfo {author} {\bibfnamefont {C.}~\bibnamefont {{Rosset}}}, \bibinfo
  {author} {\bibfnamefont {G.}~\bibnamefont {{Roudier}}}, \bibinfo {author}
  {\bibfnamefont {J.~A.}\ \bibnamefont {{Rubi{\~n}o-Mart{\'\i}n}}}, \bibinfo
  {author} {\bibfnamefont {B.}~\bibnamefont {{Ruiz-Granados}}}, \bibinfo
  {author} {\bibfnamefont {L.}~\bibnamefont {{Salvati}}}, \bibinfo {author}
  {\bibfnamefont {M.}~\bibnamefont {{Sandri}}}, \bibinfo {author}
  {\bibfnamefont {M.}~\bibnamefont {{Savelainen}}}, \bibinfo {author}
  {\bibfnamefont {D.}~\bibnamefont {{Scott}}}, \bibinfo {author} {\bibfnamefont
  {E.~P.~S.}\ \bibnamefont {{Shellard}}}, \bibinfo {author} {\bibfnamefont
  {C.}~\bibnamefont {{Sirignano}}}, \bibinfo {author} {\bibfnamefont
  {G.}~\bibnamefont {{Sirri}}}, \bibinfo {author} {\bibfnamefont {L.~D.}\
  \bibnamefont {{Spencer}}}, \bibinfo {author} {\bibfnamefont {R.}~\bibnamefont
  {{Sunyaev}}}, \bibinfo {author} {\bibfnamefont {A.~S.}\ \bibnamefont
  {{Suur-Uski}}}, \bibinfo {author} {\bibfnamefont {J.~A.}\ \bibnamefont
  {{Tauber}}}, \bibinfo {author} {\bibfnamefont {D.}~\bibnamefont
  {{Tavagnacco}}}, \bibinfo {author} {\bibfnamefont {M.}~\bibnamefont
  {{Tenti}}}, \bibinfo {author} {\bibfnamefont {L.}~\bibnamefont
  {{Toffolatti}}}, \bibinfo {author} {\bibfnamefont {M.}~\bibnamefont
  {{Tomasi}}}, \bibinfo {author} {\bibfnamefont {T.}~\bibnamefont
  {{Trombetti}}}, \bibinfo {author} {\bibfnamefont {L.}~\bibnamefont
  {{Valenziano}}}, \bibinfo {author} {\bibfnamefont {J.}~\bibnamefont
  {{Valiviita}}}, \bibinfo {author} {\bibfnamefont {B.}~\bibnamefont {{Van
  Tent}}}, \bibinfo {author} {\bibfnamefont {L.}~\bibnamefont {{Vibert}}},
  \bibinfo {author} {\bibfnamefont {P.}~\bibnamefont {{Vielva}}}, \bibinfo
  {author} {\bibfnamefont {F.}~\bibnamefont {{Villa}}}, \bibinfo {author}
  {\bibfnamefont {N.}~\bibnamefont {{Vittorio}}}, \bibinfo {author}
  {\bibfnamefont {B.~D.}\ \bibnamefont {{Wand elt}}}, \bibinfo {author}
  {\bibfnamefont {I.~K.}\ \bibnamefont {{Wehus}}}, \bibinfo {author}
  {\bibfnamefont {M.}~\bibnamefont {{White}}}, \bibinfo {author} {\bibfnamefont
  {S.~D.~M.}\ \bibnamefont {{White}}}, \bibinfo {author} {\bibfnamefont
  {A.}~\bibnamefont {{Zacchei}}}, \ and\ \bibinfo {author} {\bibfnamefont
  {A.}~\bibnamefont {{Zonca}}},\ }\href@noop {} {\bibfield  {journal} {\bibinfo
   {journal} {ArXiv e-prints}\ ,\ \bibinfo {eid} {arXiv:1807.06209}} (\bibinfo
  {year} {2018}{\natexlab{b}})},\ \Eprint {http://arxiv.org/abs/1807.06209}
  {arXiv:1807.06209 [astro-ph.CO]} \BibitemShut {NoStop}%
\bibitem [{\citenamefont {{Dawson}}\ \emph {et~al.}(2013)\citenamefont
  {{Dawson}}, \citenamefont {{Schlegel}}, \citenamefont {{Ahn}}, \citenamefont
  {{Anderson}}, \citenamefont {{Aubourg}}, \citenamefont {{Bailey}},
  \citenamefont {{Barkhouser}}, \citenamefont {{Bautista}}, \citenamefont
  {{Beifiori}}, \citenamefont {{Berlind}}, \citenamefont {{Bhardwaj}},
  \citenamefont {{Bizyaev}}, \citenamefont {{Blake}}, \citenamefont
  {{Blanton}}, \citenamefont {{Blomqvist}}, \citenamefont {{Bolton}},
  \citenamefont {{Borde}}, \citenamefont {{Bovy}}, \citenamefont {{Brandt}},
  \citenamefont {{Brewington}}, \citenamefont {{Brinkmann}}, \citenamefont
  {{Brown}}, \citenamefont {{Brownstein}}, \citenamefont {{Bundy}},
  \citenamefont {{Busca}}, \citenamefont {{Carithers}}, \citenamefont
  {{Carnero}}, \citenamefont {{Carr}}, \citenamefont {{Chen}}, \citenamefont
  {{Comparat}}, \citenamefont {{Connolly}}, \citenamefont {{Cope}},
  \citenamefont {{Croft}}, \citenamefont {{Cuesta}}, \citenamefont {{da
  Costa}}, \citenamefont {{Davenport}}, \citenamefont {{Delubac}},
  \citenamefont {{de Putter}}, \citenamefont {{Dhital}}, \citenamefont
  {{Ealet}}, \citenamefont {{Ebelke}}, \citenamefont {{Eisenstein}},
  \citenamefont {{Escoffier}}, \citenamefont {{Fan}}, \citenamefont {{Filiz
  Ak}}, \citenamefont {{Finley}}, \citenamefont {{Font-Ribera}}, \citenamefont
  {{G{\'e}nova-Santos}}, \citenamefont {{Gunn}}, \citenamefont {{Guo}},
  \citenamefont {{Haggard}}, \citenamefont {{Hall}}, \citenamefont
  {{Hamilton}}, \citenamefont {{Harris}}, \citenamefont {{Harris}},
  \citenamefont {{Ho}}, \citenamefont {{Hogg}}, \citenamefont {{Holder}},
  \citenamefont {{Honscheid}}, \citenamefont {{Huehnerhoff}}, \citenamefont
  {{Jordan}}, \citenamefont {{Jordan}}, \citenamefont {{Kauffmann}},
  \citenamefont {{Kazin}}, \citenamefont {{Kirkby}}, \citenamefont {{Klaene}},
  \citenamefont {{Kneib}}, \citenamefont {{Le Goff}}, \citenamefont {{Lee}},
  \citenamefont {{Long}}, \citenamefont {{Loomis}}, \citenamefont {{Lundgren}},
  \citenamefont {{Lupton}}, \citenamefont {{Maia}}, \citenamefont {{Makler}},
  \citenamefont {{Malanushenko}}, \citenamefont {{Malanushenko}}, \citenamefont
  {{Mandelbaum}}, \citenamefont {{Manera}}, \citenamefont {{Maraston}},
  \citenamefont {{Margala}}, \citenamefont {{Masters}}, \citenamefont
  {{McBride}}, \citenamefont {{McDonald}}, \citenamefont {{McGreer}},
  \citenamefont {{McMahon}}, \citenamefont {{Mena}}, \citenamefont
  {{Miralda-Escud{\'e}}}, \citenamefont {{Montero-Dorta}}, \citenamefont
  {{Montesano}}, \citenamefont {{Muna}}, \citenamefont {{Myers}}, \citenamefont
  {{Naugle}}, \citenamefont {{Nichol}}, \citenamefont {{Noterdaeme}},
  \citenamefont {{Nuza}}, \citenamefont {{Olmstead}}, \citenamefont
  {{Oravetz}}, \citenamefont {{Oravetz}}, \citenamefont {{Owen}}, \citenamefont
  {{Padmanabhan}}, \citenamefont {{Palanque-Delabrouille}}, \citenamefont
  {{Pan}}, \citenamefont {{Parejko}}, \citenamefont {{P{\^a}ris}},
  \citenamefont {{Percival}}, \citenamefont {{P{\'e}rez-Fournon}},
  \citenamefont {{P{\'e}rez-R{\`a}fols}}, \citenamefont {{Petitjean}},
  \citenamefont {{Pfaffenberger}}, \citenamefont {{Pforr}}, \citenamefont
  {{Pieri}}, \citenamefont {{Prada}}, \citenamefont {{Price-Whelan}},
  \citenamefont {{Raddick}}, \citenamefont {{Rebolo}}, \citenamefont {{Rich}},
  \citenamefont {{Richards}}, \citenamefont {{Rockosi}}, \citenamefont {{Roe}},
  \citenamefont {{Ross}}, \citenamefont {{Ross}}, \citenamefont {{Rossi}},
  \citenamefont {{Rubi{\~n}o-Martin}}, \citenamefont {{Samushia}},
  \citenamefont {{S{\'a}nchez}}, \citenamefont {{Sayres}}, \citenamefont
  {{Schmidt}}, \citenamefont {{Schneider}}, \citenamefont {{Sc{\'o}ccola}},
  \citenamefont {{Seo}}, \citenamefont {{Shelden}}, \citenamefont {{Sheldon}},
  \citenamefont {{Shen}}, \citenamefont {{Shu}}, \citenamefont {{Slosar}},
  \citenamefont {{Smee}}, \citenamefont {{Snedden}}, \citenamefont
  {{Stauffer}}, \citenamefont {{Steele}}, \citenamefont {{Strauss}},
  \citenamefont {{Streblyanska}}, \citenamefont {{Suzuki}}, \citenamefont
  {{Swanson}}, \citenamefont {{Tal}}, \citenamefont {{Tanaka}}, \citenamefont
  {{Thomas}}, \citenamefont {{Tinker}}, \citenamefont {{Tojeiro}},
  \citenamefont {{Tremonti}}, \citenamefont {{Vargas Maga{\~n}a}},
  \citenamefont {{Verde}}, \citenamefont {{Viel}}, \citenamefont {{Wake}},
  \citenamefont {{Watson}}, \citenamefont {{Weaver}}, \citenamefont
  {{Weinberg}}, \citenamefont {{Weiner}}, \citenamefont {{West}}, \citenamefont
  {{White}}, \citenamefont {{Wood-Vasey}}, \citenamefont {{Yeche}},
  \citenamefont {{Zehavi}}, \citenamefont {{Zhao}},\ and\ \citenamefont
  {{Zheng}}}]{dawson13a}%
  \BibitemOpen
  \bibfield  {author} {\bibinfo {author} {\bibfnamefont {K.~S.}\ \bibnamefont
  {{Dawson}}}, \bibinfo {author} {\bibfnamefont {D.~J.}\ \bibnamefont
  {{Schlegel}}}, \bibinfo {author} {\bibfnamefont {C.~P.}\ \bibnamefont
  {{Ahn}}}, \bibinfo {author} {\bibfnamefont {S.~F.}\ \bibnamefont
  {{Anderson}}}, \bibinfo {author} {\bibfnamefont {{\'E}.}~\bibnamefont
  {{Aubourg}}}, \bibinfo {author} {\bibfnamefont {S.}~\bibnamefont {{Bailey}}},
  \bibinfo {author} {\bibfnamefont {R.~H.}\ \bibnamefont {{Barkhouser}}},
  \bibinfo {author} {\bibfnamefont {J.~E.}\ \bibnamefont {{Bautista}}},
  \bibinfo {author} {\bibfnamefont {A.}~\bibnamefont {{Beifiori}}}, \bibinfo
  {author} {\bibfnamefont {A.~A.}\ \bibnamefont {{Berlind}}}, \bibinfo {author}
  {\bibfnamefont {V.}~\bibnamefont {{Bhardwaj}}}, \bibinfo {author}
  {\bibfnamefont {D.}~\bibnamefont {{Bizyaev}}}, \bibinfo {author}
  {\bibfnamefont {C.~H.}\ \bibnamefont {{Blake}}}, \bibinfo {author}
  {\bibfnamefont {M.~R.}\ \bibnamefont {{Blanton}}}, \bibinfo {author}
  {\bibfnamefont {M.}~\bibnamefont {{Blomqvist}}}, \bibinfo {author}
  {\bibfnamefont {A.~S.}\ \bibnamefont {{Bolton}}}, \bibinfo {author}
  {\bibfnamefont {A.}~\bibnamefont {{Borde}}}, \bibinfo {author} {\bibfnamefont
  {J.}~\bibnamefont {{Bovy}}}, \bibinfo {author} {\bibfnamefont {W.~N.}\
  \bibnamefont {{Brandt}}}, \bibinfo {author} {\bibfnamefont {H.}~\bibnamefont
  {{Brewington}}}, \bibinfo {author} {\bibfnamefont {J.}~\bibnamefont
  {{Brinkmann}}}, \bibinfo {author} {\bibfnamefont {P.~J.}\ \bibnamefont
  {{Brown}}}, \bibinfo {author} {\bibfnamefont {J.~R.}\ \bibnamefont
  {{Brownstein}}}, \bibinfo {author} {\bibfnamefont {K.}~\bibnamefont
  {{Bundy}}}, \bibinfo {author} {\bibfnamefont {N.~G.}\ \bibnamefont
  {{Busca}}}, \bibinfo {author} {\bibfnamefont {W.}~\bibnamefont
  {{Carithers}}}, \bibinfo {author} {\bibfnamefont {A.~R.}\ \bibnamefont
  {{Carnero}}}, \bibinfo {author} {\bibfnamefont {M.~A.}\ \bibnamefont
  {{Carr}}}, \bibinfo {author} {\bibfnamefont {Y.}~\bibnamefont {{Chen}}},
  \bibinfo {author} {\bibfnamefont {J.}~\bibnamefont {{Comparat}}}, \bibinfo
  {author} {\bibfnamefont {N.}~\bibnamefont {{Connolly}}}, \bibinfo {author}
  {\bibfnamefont {F.}~\bibnamefont {{Cope}}}, \bibinfo {author} {\bibfnamefont
  {R.~A.~C.}\ \bibnamefont {{Croft}}}, \bibinfo {author} {\bibfnamefont
  {A.~J.}\ \bibnamefont {{Cuesta}}}, \bibinfo {author} {\bibfnamefont {L.~N.}\
  \bibnamefont {{da Costa}}}, \bibinfo {author} {\bibfnamefont {J.~R.~A.}\
  \bibnamefont {{Davenport}}}, \bibinfo {author} {\bibfnamefont
  {T.}~\bibnamefont {{Delubac}}}, \bibinfo {author} {\bibfnamefont
  {R.}~\bibnamefont {{de Putter}}}, \bibinfo {author} {\bibfnamefont
  {S.}~\bibnamefont {{Dhital}}}, \bibinfo {author} {\bibfnamefont
  {A.}~\bibnamefont {{Ealet}}}, \bibinfo {author} {\bibfnamefont {G.~L.}\
  \bibnamefont {{Ebelke}}}, \bibinfo {author} {\bibfnamefont {D.~J.}\
  \bibnamefont {{Eisenstein}}}, \bibinfo {author} {\bibfnamefont
  {S.}~\bibnamefont {{Escoffier}}}, \bibinfo {author} {\bibfnamefont
  {X.}~\bibnamefont {{Fan}}}, \bibinfo {author} {\bibfnamefont
  {N.}~\bibnamefont {{Filiz Ak}}}, \bibinfo {author} {\bibfnamefont
  {H.}~\bibnamefont {{Finley}}}, \bibinfo {author} {\bibfnamefont
  {A.}~\bibnamefont {{Font-Ribera}}}, \bibinfo {author} {\bibfnamefont
  {R.}~\bibnamefont {{G{\'e}nova-Santos}}}, \bibinfo {author} {\bibfnamefont
  {J.~E.}\ \bibnamefont {{Gunn}}}, \bibinfo {author} {\bibfnamefont
  {H.}~\bibnamefont {{Guo}}}, \bibinfo {author} {\bibfnamefont
  {D.}~\bibnamefont {{Haggard}}}, \bibinfo {author} {\bibfnamefont {P.~B.}\
  \bibnamefont {{Hall}}}, \bibinfo {author} {\bibfnamefont {J.-C.}\
  \bibnamefont {{Hamilton}}}, \bibinfo {author} {\bibfnamefont
  {B.}~\bibnamefont {{Harris}}}, \bibinfo {author} {\bibfnamefont {D.~W.}\
  \bibnamefont {{Harris}}}, \bibinfo {author} {\bibfnamefont {S.}~\bibnamefont
  {{Ho}}}, \bibinfo {author} {\bibfnamefont {D.~W.}\ \bibnamefont {{Hogg}}},
  \bibinfo {author} {\bibfnamefont {D.}~\bibnamefont {{Holder}}}, \bibinfo
  {author} {\bibfnamefont {K.}~\bibnamefont {{Honscheid}}}, \bibinfo {author}
  {\bibfnamefont {J.}~\bibnamefont {{Huehnerhoff}}}, \bibinfo {author}
  {\bibfnamefont {B.}~\bibnamefont {{Jordan}}}, \bibinfo {author}
  {\bibfnamefont {W.~P.}\ \bibnamefont {{Jordan}}}, \bibinfo {author}
  {\bibfnamefont {G.}~\bibnamefont {{Kauffmann}}}, \bibinfo {author}
  {\bibfnamefont {E.~A.}\ \bibnamefont {{Kazin}}}, \bibinfo {author}
  {\bibfnamefont {D.}~\bibnamefont {{Kirkby}}}, \bibinfo {author}
  {\bibfnamefont {M.~A.}\ \bibnamefont {{Klaene}}}, \bibinfo {author}
  {\bibfnamefont {J.-P.}\ \bibnamefont {{Kneib}}}, \bibinfo {author}
  {\bibfnamefont {J.-M.}\ \bibnamefont {{Le Goff}}}, \bibinfo {author}
  {\bibfnamefont {K.-G.}\ \bibnamefont {{Lee}}}, \bibinfo {author}
  {\bibfnamefont {D.~C.}\ \bibnamefont {{Long}}}, \bibinfo {author}
  {\bibfnamefont {C.~P.}\ \bibnamefont {{Loomis}}}, \bibinfo {author}
  {\bibfnamefont {B.}~\bibnamefont {{Lundgren}}}, \bibinfo {author}
  {\bibfnamefont {R.~H.}\ \bibnamefont {{Lupton}}}, \bibinfo {author}
  {\bibfnamefont {M.~A.~G.}\ \bibnamefont {{Maia}}}, \bibinfo {author}
  {\bibfnamefont {M.}~\bibnamefont {{Makler}}}, \bibinfo {author}
  {\bibfnamefont {E.}~\bibnamefont {{Malanushenko}}}, \bibinfo {author}
  {\bibfnamefont {V.}~\bibnamefont {{Malanushenko}}}, \bibinfo {author}
  {\bibfnamefont {R.}~\bibnamefont {{Mandelbaum}}}, \bibinfo {author}
  {\bibfnamefont {M.}~\bibnamefont {{Manera}}}, \bibinfo {author}
  {\bibfnamefont {C.}~\bibnamefont {{Maraston}}}, \bibinfo {author}
  {\bibfnamefont {D.}~\bibnamefont {{Margala}}}, \bibinfo {author}
  {\bibfnamefont {K.~L.}\ \bibnamefont {{Masters}}}, \bibinfo {author}
  {\bibfnamefont {C.~K.}\ \bibnamefont {{McBride}}}, \bibinfo {author}
  {\bibfnamefont {P.}~\bibnamefont {{McDonald}}}, \bibinfo {author}
  {\bibfnamefont {I.~D.}\ \bibnamefont {{McGreer}}}, \bibinfo {author}
  {\bibfnamefont {R.~G.}\ \bibnamefont {{McMahon}}}, \bibinfo {author}
  {\bibfnamefont {O.}~\bibnamefont {{Mena}}}, \bibinfo {author} {\bibfnamefont
  {J.}~\bibnamefont {{Miralda-Escud{\'e}}}}, \bibinfo {author} {\bibfnamefont
  {A.~D.}\ \bibnamefont {{Montero-Dorta}}}, \bibinfo {author} {\bibfnamefont
  {F.}~\bibnamefont {{Montesano}}}, \bibinfo {author} {\bibfnamefont
  {D.}~\bibnamefont {{Muna}}}, \bibinfo {author} {\bibfnamefont {A.~D.}\
  \bibnamefont {{Myers}}}, \bibinfo {author} {\bibfnamefont {T.}~\bibnamefont
  {{Naugle}}}, \bibinfo {author} {\bibfnamefont {R.~C.}\ \bibnamefont
  {{Nichol}}}, \bibinfo {author} {\bibfnamefont {P.}~\bibnamefont
  {{Noterdaeme}}}, \bibinfo {author} {\bibfnamefont {S.~E.}\ \bibnamefont
  {{Nuza}}}, \bibinfo {author} {\bibfnamefont {M.~D.}\ \bibnamefont
  {{Olmstead}}}, \bibinfo {author} {\bibfnamefont {A.}~\bibnamefont
  {{Oravetz}}}, \bibinfo {author} {\bibfnamefont {D.~J.}\ \bibnamefont
  {{Oravetz}}}, \bibinfo {author} {\bibfnamefont {R.}~\bibnamefont {{Owen}}},
  \bibinfo {author} {\bibfnamefont {N.}~\bibnamefont {{Padmanabhan}}}, \bibinfo
  {author} {\bibfnamefont {N.}~\bibnamefont {{Palanque-Delabrouille}}},
  \bibinfo {author} {\bibfnamefont {K.}~\bibnamefont {{Pan}}}, \bibinfo
  {author} {\bibfnamefont {J.~K.}\ \bibnamefont {{Parejko}}}, \bibinfo {author}
  {\bibfnamefont {I.}~\bibnamefont {{P{\^a}ris}}}, \bibinfo {author}
  {\bibfnamefont {W.~J.}\ \bibnamefont {{Percival}}}, \bibinfo {author}
  {\bibfnamefont {I.}~\bibnamefont {{P{\'e}rez-Fournon}}}, \bibinfo {author}
  {\bibfnamefont {I.}~\bibnamefont {{P{\'e}rez-R{\`a}fols}}}, \bibinfo {author}
  {\bibfnamefont {P.}~\bibnamefont {{Petitjean}}}, \bibinfo {author}
  {\bibfnamefont {R.}~\bibnamefont {{Pfaffenberger}}}, \bibinfo {author}
  {\bibfnamefont {J.}~\bibnamefont {{Pforr}}}, \bibinfo {author} {\bibfnamefont
  {M.~M.}\ \bibnamefont {{Pieri}}}, \bibinfo {author} {\bibfnamefont
  {F.}~\bibnamefont {{Prada}}}, \bibinfo {author} {\bibfnamefont {A.~M.}\
  \bibnamefont {{Price-Whelan}}}, \bibinfo {author} {\bibfnamefont {M.~J.}\
  \bibnamefont {{Raddick}}}, \bibinfo {author} {\bibfnamefont {R.}~\bibnamefont
  {{Rebolo}}}, \bibinfo {author} {\bibfnamefont {J.}~\bibnamefont {{Rich}}},
  \bibinfo {author} {\bibfnamefont {G.~T.}\ \bibnamefont {{Richards}}},
  \bibinfo {author} {\bibfnamefont {C.~M.}\ \bibnamefont {{Rockosi}}}, \bibinfo
  {author} {\bibfnamefont {N.~A.}\ \bibnamefont {{Roe}}}, \bibinfo {author}
  {\bibfnamefont {A.~J.}\ \bibnamefont {{Ross}}}, \bibinfo {author}
  {\bibfnamefont {N.~P.}\ \bibnamefont {{Ross}}}, \bibinfo {author}
  {\bibfnamefont {G.}~\bibnamefont {{Rossi}}}, \bibinfo {author} {\bibfnamefont
  {J.~A.}\ \bibnamefont {{Rubi{\~n}o-Martin}}}, \bibinfo {author}
  {\bibfnamefont {L.}~\bibnamefont {{Samushia}}}, \bibinfo {author}
  {\bibfnamefont {A.~G.}\ \bibnamefont {{S{\'a}nchez}}}, \bibinfo {author}
  {\bibfnamefont {C.}~\bibnamefont {{Sayres}}}, \bibinfo {author}
  {\bibfnamefont {S.~J.}\ \bibnamefont {{Schmidt}}}, \bibinfo {author}
  {\bibfnamefont {D.~P.}\ \bibnamefont {{Schneider}}}, \bibinfo {author}
  {\bibfnamefont {C.~G.}\ \bibnamefont {{Sc{\'o}ccola}}}, \bibinfo {author}
  {\bibfnamefont {H.-J.}\ \bibnamefont {{Seo}}}, \bibinfo {author}
  {\bibfnamefont {A.}~\bibnamefont {{Shelden}}}, \bibinfo {author}
  {\bibfnamefont {E.}~\bibnamefont {{Sheldon}}}, \bibinfo {author}
  {\bibfnamefont {Y.}~\bibnamefont {{Shen}}}, \bibinfo {author} {\bibfnamefont
  {Y.}~\bibnamefont {{Shu}}}, \bibinfo {author} {\bibfnamefont
  {A.}~\bibnamefont {{Slosar}}}, \bibinfo {author} {\bibfnamefont {S.~A.}\
  \bibnamefont {{Smee}}}, \bibinfo {author} {\bibfnamefont {S.~A.}\
  \bibnamefont {{Snedden}}}, \bibinfo {author} {\bibfnamefont {F.}~\bibnamefont
  {{Stauffer}}}, \bibinfo {author} {\bibfnamefont {O.}~\bibnamefont
  {{Steele}}}, \bibinfo {author} {\bibfnamefont {M.~A.}\ \bibnamefont
  {{Strauss}}}, \bibinfo {author} {\bibfnamefont {A.}~\bibnamefont
  {{Streblyanska}}}, \bibinfo {author} {\bibfnamefont {N.}~\bibnamefont
  {{Suzuki}}}, \bibinfo {author} {\bibfnamefont {M.~E.~C.}\ \bibnamefont
  {{Swanson}}}, \bibinfo {author} {\bibfnamefont {T.}~\bibnamefont {{Tal}}},
  \bibinfo {author} {\bibfnamefont {M.}~\bibnamefont {{Tanaka}}}, \bibinfo
  {author} {\bibfnamefont {D.}~\bibnamefont {{Thomas}}}, \bibinfo {author}
  {\bibfnamefont {J.~L.}\ \bibnamefont {{Tinker}}}, \bibinfo {author}
  {\bibfnamefont {R.}~\bibnamefont {{Tojeiro}}}, \bibinfo {author}
  {\bibfnamefont {C.~A.}\ \bibnamefont {{Tremonti}}}, \bibinfo {author}
  {\bibfnamefont {M.}~\bibnamefont {{Vargas Maga{\~n}a}}}, \bibinfo {author}
  {\bibfnamefont {L.}~\bibnamefont {{Verde}}}, \bibinfo {author} {\bibfnamefont
  {M.}~\bibnamefont {{Viel}}}, \bibinfo {author} {\bibfnamefont {D.~A.}\
  \bibnamefont {{Wake}}}, \bibinfo {author} {\bibfnamefont {M.}~\bibnamefont
  {{Watson}}}, \bibinfo {author} {\bibfnamefont {B.~A.}\ \bibnamefont
  {{Weaver}}}, \bibinfo {author} {\bibfnamefont {D.~H.}\ \bibnamefont
  {{Weinberg}}}, \bibinfo {author} {\bibfnamefont {B.~J.}\ \bibnamefont
  {{Weiner}}}, \bibinfo {author} {\bibfnamefont {A.~A.}\ \bibnamefont
  {{West}}}, \bibinfo {author} {\bibfnamefont {M.}~\bibnamefont {{White}}},
  \bibinfo {author} {\bibfnamefont {W.~M.}\ \bibnamefont {{Wood-Vasey}}},
  \bibinfo {author} {\bibfnamefont {C.}~\bibnamefont {{Yeche}}}, \bibinfo
  {author} {\bibfnamefont {I.}~\bibnamefont {{Zehavi}}}, \bibinfo {author}
  {\bibfnamefont {G.-B.}\ \bibnamefont {{Zhao}}}, \ and\ \bibinfo {author}
  {\bibfnamefont {Z.}~\bibnamefont {{Zheng}}},\ }\href {\doibase
  10.1088/0004-6256/145/1/10} {\bibfield  {journal} {\bibinfo  {journal} {\aj}\
  }\textbf {\bibinfo {volume} {145}},\ \bibinfo {eid} {10} (\bibinfo {year}
  {2013})},\ \Eprint {http://arxiv.org/abs/1208.0022} {arXiv:1208.0022
  [astro-ph.CO]} \BibitemShut {NoStop}%
\bibitem [{\citenamefont {{Eisenstein}}\ \emph {et~al.}(2011)\citenamefont
  {{Eisenstein}}, \citenamefont {{Weinberg}}, \citenamefont {{Agol}},
  \citenamefont {{Aihara}}, \citenamefont {{Allende Prieto}}, \citenamefont
  {{Anderson}}, \citenamefont {{Arns}}, \citenamefont {{Aubourg}},
  \citenamefont {{Bailey}}, \citenamefont {{Balbinot}},\ and\ \citenamefont
  {et~al.}}]{eisenstein11a}%
  \BibitemOpen
  \bibfield  {author} {\bibinfo {author} {\bibfnamefont {D.~J.}\ \bibnamefont
  {{Eisenstein}}}, \bibinfo {author} {\bibfnamefont {D.~H.}\ \bibnamefont
  {{Weinberg}}}, \bibinfo {author} {\bibfnamefont {E.}~\bibnamefont {{Agol}}},
  \bibinfo {author} {\bibfnamefont {H.}~\bibnamefont {{Aihara}}}, \bibinfo
  {author} {\bibfnamefont {C.}~\bibnamefont {{Allende Prieto}}}, \bibinfo
  {author} {\bibfnamefont {S.~F.}\ \bibnamefont {{Anderson}}}, \bibinfo
  {author} {\bibfnamefont {J.~A.}\ \bibnamefont {{Arns}}}, \bibinfo {author}
  {\bibfnamefont {{\'E}.}~\bibnamefont {{Aubourg}}}, \bibinfo {author}
  {\bibfnamefont {S.}~\bibnamefont {{Bailey}}}, \bibinfo {author}
  {\bibfnamefont {E.}~\bibnamefont {{Balbinot}}}, \ and\ \bibinfo {author}
  {\bibnamefont {et~al.}},\ }\href {\doibase 10.1088/0004-6256/142/3/72}
  {\bibfield  {journal} {\bibinfo  {journal} {\aj}\ }\textbf {\bibinfo {volume}
  {142}},\ \bibinfo {eid} {72} (\bibinfo {year} {2011})},\ \Eprint
  {http://arxiv.org/abs/1101.1529} {arXiv:1101.1529 [astro-ph.IM]} \BibitemShut
  {NoStop}%
\bibitem [{\citenamefont {{Anderson}}\ \emph {et~al.}(2012)\citenamefont
  {{Anderson}}, \citenamefont {{Aubourg}}, \citenamefont {{Bailey}},
  \citenamefont {{Bizyaev}}, \citenamefont {{Blanton}}, \citenamefont
  {{Bolton}}, \citenamefont {{Brinkmann}}, \citenamefont {{Brownstein}},
  \citenamefont {{Burden}}, \citenamefont {{Cuesta}}, \citenamefont {{da
  Costa}}, \citenamefont {{Dawson}}, \citenamefont {{de Putter}}, \citenamefont
  {{Eisenstein}}, \citenamefont {{Gunn}}, \citenamefont {{Guo}}, \citenamefont
  {{Hamilton}}, \citenamefont {{Harding}}, \citenamefont {{Ho}}, \citenamefont
  {{Honscheid}}, \citenamefont {{Kazin}}, \citenamefont {{Kirkby}},
  \citenamefont {{Kneib}}, \citenamefont {{Labatie}}, \citenamefont {{Loomis}},
  \citenamefont {{Lupton}}, \citenamefont {{Malanushenko}}, \citenamefont
  {{Malanushenko}}, \citenamefont {{Mandelbaum}}, \citenamefont {{Manera}},
  \citenamefont {{Maraston}}, \citenamefont {{McBride}}, \citenamefont
  {{Mehta}}, \citenamefont {{Mena}}, \citenamefont {{Montesano}}, \citenamefont
  {{Muna}}, \citenamefont {{Nichol}}, \citenamefont {{Nuza}}, \citenamefont
  {{Olmstead}}, \citenamefont {{Oravetz}}, \citenamefont {{Padmanabhan}},
  \citenamefont {{Palanque-Delabrouille}}, \citenamefont {{Pan}}, \citenamefont
  {{Parejko}}, \citenamefont {{Paris}}, \citenamefont {{Percival}},
  \citenamefont {{Petitjean}}, \citenamefont {{Prada}}, \citenamefont {{Reid}},
  \citenamefont {{Roe}}, \citenamefont {{Ross}}, \citenamefont {{Ross}},
  \citenamefont {{Samushia}}, \citenamefont {{Sanchez}}, \citenamefont
  {{Schneider}}, \citenamefont {{Scoccola}}, \citenamefont {{Seo}},
  \citenamefont {{Sheldon}}, \citenamefont {{Simmons}}, \citenamefont
  {{Skibba}}, \citenamefont {{Strauss}}, \citenamefont {{Swanson}},
  \citenamefont {{Thomas}}, \citenamefont {{Tinker}}, \citenamefont
  {{Tojeiro}}, \citenamefont {{Vargas Magana}}, \citenamefont {{Verde}},
  \citenamefont {{Wagner}}, \citenamefont {{Wake}}, \citenamefont {{Weaver}},
  \citenamefont {{Weinberg}}, \citenamefont {{White}}, \citenamefont {{Xu}},
  \citenamefont {{Yeche}}, \citenamefont {{Zehavi}},\ and\ \citenamefont
  {{Zhao}}}]{anderson12a}%
  \BibitemOpen
  \bibfield  {author} {\bibinfo {author} {\bibfnamefont {L.}~\bibnamefont
  {{Anderson}}}, \bibinfo {author} {\bibfnamefont {E.}~\bibnamefont
  {{Aubourg}}}, \bibinfo {author} {\bibfnamefont {S.}~\bibnamefont {{Bailey}}},
  \bibinfo {author} {\bibfnamefont {D.}~\bibnamefont {{Bizyaev}}}, \bibinfo
  {author} {\bibfnamefont {M.}~\bibnamefont {{Blanton}}}, \bibinfo {author}
  {\bibfnamefont {A.~S.}\ \bibnamefont {{Bolton}}}, \bibinfo {author}
  {\bibfnamefont {J.}~\bibnamefont {{Brinkmann}}}, \bibinfo {author}
  {\bibfnamefont {J.~R.}\ \bibnamefont {{Brownstein}}}, \bibinfo {author}
  {\bibfnamefont {A.}~\bibnamefont {{Burden}}}, \bibinfo {author}
  {\bibfnamefont {A.~J.}\ \bibnamefont {{Cuesta}}}, \bibinfo {author}
  {\bibfnamefont {L.~N.~A.}\ \bibnamefont {{da Costa}}}, \bibinfo {author}
  {\bibfnamefont {K.~S.}\ \bibnamefont {{Dawson}}}, \bibinfo {author}
  {\bibfnamefont {R.}~\bibnamefont {{de Putter}}}, \bibinfo {author}
  {\bibfnamefont {D.~J.}\ \bibnamefont {{Eisenstein}}}, \bibinfo {author}
  {\bibfnamefont {J.~E.}\ \bibnamefont {{Gunn}}}, \bibinfo {author}
  {\bibfnamefont {H.}~\bibnamefont {{Guo}}}, \bibinfo {author} {\bibfnamefont
  {J.-C.}\ \bibnamefont {{Hamilton}}}, \bibinfo {author} {\bibfnamefont
  {P.}~\bibnamefont {{Harding}}}, \bibinfo {author} {\bibfnamefont
  {S.}~\bibnamefont {{Ho}}}, \bibinfo {author} {\bibfnamefont {K.}~\bibnamefont
  {{Honscheid}}}, \bibinfo {author} {\bibfnamefont {E.}~\bibnamefont
  {{Kazin}}}, \bibinfo {author} {\bibfnamefont {D.}~\bibnamefont {{Kirkby}}},
  \bibinfo {author} {\bibfnamefont {J.-P.}\ \bibnamefont {{Kneib}}}, \bibinfo
  {author} {\bibfnamefont {A.}~\bibnamefont {{Labatie}}}, \bibinfo {author}
  {\bibfnamefont {C.}~\bibnamefont {{Loomis}}}, \bibinfo {author}
  {\bibfnamefont {R.~H.}\ \bibnamefont {{Lupton}}}, \bibinfo {author}
  {\bibfnamefont {E.}~\bibnamefont {{Malanushenko}}}, \bibinfo {author}
  {\bibfnamefont {V.}~\bibnamefont {{Malanushenko}}}, \bibinfo {author}
  {\bibfnamefont {R.}~\bibnamefont {{Mandelbaum}}}, \bibinfo {author}
  {\bibfnamefont {M.}~\bibnamefont {{Manera}}}, \bibinfo {author}
  {\bibfnamefont {C.}~\bibnamefont {{Maraston}}}, \bibinfo {author}
  {\bibfnamefont {C.~K.}\ \bibnamefont {{McBride}}}, \bibinfo {author}
  {\bibfnamefont {K.~T.}\ \bibnamefont {{Mehta}}}, \bibinfo {author}
  {\bibfnamefont {O.}~\bibnamefont {{Mena}}}, \bibinfo {author} {\bibfnamefont
  {F.}~\bibnamefont {{Montesano}}}, \bibinfo {author} {\bibfnamefont
  {D.}~\bibnamefont {{Muna}}}, \bibinfo {author} {\bibfnamefont {R.~C.}\
  \bibnamefont {{Nichol}}}, \bibinfo {author} {\bibfnamefont {S.~E.}\
  \bibnamefont {{Nuza}}}, \bibinfo {author} {\bibfnamefont {M.~D.}\
  \bibnamefont {{Olmstead}}}, \bibinfo {author} {\bibfnamefont
  {D.}~\bibnamefont {{Oravetz}}}, \bibinfo {author} {\bibfnamefont
  {N.}~\bibnamefont {{Padmanabhan}}}, \bibinfo {author} {\bibfnamefont
  {N.}~\bibnamefont {{Palanque-Delabrouille}}}, \bibinfo {author}
  {\bibfnamefont {K.}~\bibnamefont {{Pan}}}, \bibinfo {author} {\bibfnamefont
  {J.}~\bibnamefont {{Parejko}}}, \bibinfo {author} {\bibfnamefont
  {I.}~\bibnamefont {{Paris}}}, \bibinfo {author} {\bibfnamefont {W.~J.}\
  \bibnamefont {{Percival}}}, \bibinfo {author} {\bibfnamefont
  {P.}~\bibnamefont {{Petitjean}}}, \bibinfo {author} {\bibfnamefont
  {F.}~\bibnamefont {{Prada}}}, \bibinfo {author} {\bibfnamefont
  {B.}~\bibnamefont {{Reid}}}, \bibinfo {author} {\bibfnamefont {N.~A.}\
  \bibnamefont {{Roe}}}, \bibinfo {author} {\bibfnamefont {A.~J.}\ \bibnamefont
  {{Ross}}}, \bibinfo {author} {\bibfnamefont {N.~P.}\ \bibnamefont {{Ross}}},
  \bibinfo {author} {\bibfnamefont {L.}~\bibnamefont {{Samushia}}}, \bibinfo
  {author} {\bibfnamefont {A.~G.}\ \bibnamefont {{Sanchez}}}, \bibinfo {author}
  {\bibfnamefont {D.~J.~S.~D.~P.}\ \bibnamefont {{Schneider}}}, \bibinfo
  {author} {\bibfnamefont {C.~G.}\ \bibnamefont {{Scoccola}}}, \bibinfo
  {author} {\bibfnamefont {H.-J.}\ \bibnamefont {{Seo}}}, \bibinfo {author}
  {\bibfnamefont {E.~S.}\ \bibnamefont {{Sheldon}}}, \bibinfo {author}
  {\bibfnamefont {A.}~\bibnamefont {{Simmons}}}, \bibinfo {author}
  {\bibfnamefont {R.~A.}\ \bibnamefont {{Skibba}}}, \bibinfo {author}
  {\bibfnamefont {M.~A.}\ \bibnamefont {{Strauss}}}, \bibinfo {author}
  {\bibfnamefont {M.~E.~C.}\ \bibnamefont {{Swanson}}}, \bibinfo {author}
  {\bibfnamefont {D.}~\bibnamefont {{Thomas}}}, \bibinfo {author}
  {\bibfnamefont {J.~L.}\ \bibnamefont {{Tinker}}}, \bibinfo {author}
  {\bibfnamefont {R.}~\bibnamefont {{Tojeiro}}}, \bibinfo {author}
  {\bibfnamefont {M.}~\bibnamefont {{Vargas Magana}}}, \bibinfo {author}
  {\bibfnamefont {L.}~\bibnamefont {{Verde}}}, \bibinfo {author} {\bibfnamefont
  {C.}~\bibnamefont {{Wagner}}}, \bibinfo {author} {\bibfnamefont {D.~A.}\
  \bibnamefont {{Wake}}}, \bibinfo {author} {\bibfnamefont {B.~A.}\
  \bibnamefont {{Weaver}}}, \bibinfo {author} {\bibfnamefont {D.~H.}\
  \bibnamefont {{Weinberg}}}, \bibinfo {author} {\bibfnamefont
  {M.}~\bibnamefont {{White}}}, \bibinfo {author} {\bibfnamefont
  {X.}~\bibnamefont {{Xu}}}, \bibinfo {author} {\bibfnamefont {C.}~\bibnamefont
  {{Yeche}}}, \bibinfo {author} {\bibfnamefont {I.}~\bibnamefont {{Zehavi}}}, \
  and\ \bibinfo {author} {\bibfnamefont {G.-B.}\ \bibnamefont {{Zhao}}},\
  }\href {\doibase 10.1111/j.1365-2966.2012.22066.x} {\bibfield  {journal}
  {\bibinfo  {journal} {\mnras}\ }\textbf {\bibinfo {volume} {427}},\ \bibinfo
  {pages} {3435} (\bibinfo {year} {2012})},\ \Eprint
  {http://arxiv.org/abs/1203.6594} {arXiv:1203.6594 [astro-ph.CO]} \BibitemShut
  {NoStop}%
\bibitem [{\citenamefont {{Riess}}\ \emph {et~al.}(2011)\citenamefont
  {{Riess}}, \citenamefont {{Macri}}, \citenamefont {{Casertano}},
  \citenamefont {{Lampeitl}}, \citenamefont {{Ferguson}}, \citenamefont
  {{Filippenko}}, \citenamefont {{Jha}}, \citenamefont {{Li}},\ and\
  \citenamefont {{Chornock}}}]{riess11a}%
  \BibitemOpen
  \bibfield  {author} {\bibinfo {author} {\bibfnamefont {A.~G.}\ \bibnamefont
  {{Riess}}}, \bibinfo {author} {\bibfnamefont {L.}~\bibnamefont {{Macri}}},
  \bibinfo {author} {\bibfnamefont {S.}~\bibnamefont {{Casertano}}}, \bibinfo
  {author} {\bibfnamefont {H.}~\bibnamefont {{Lampeitl}}}, \bibinfo {author}
  {\bibfnamefont {H.~C.}\ \bibnamefont {{Ferguson}}}, \bibinfo {author}
  {\bibfnamefont {A.~V.}\ \bibnamefont {{Filippenko}}}, \bibinfo {author}
  {\bibfnamefont {S.~W.}\ \bibnamefont {{Jha}}}, \bibinfo {author}
  {\bibfnamefont {W.}~\bibnamefont {{Li}}}, \ and\ \bibinfo {author}
  {\bibfnamefont {R.}~\bibnamefont {{Chornock}}},\ }\href {\doibase
  10.1088/0004-637X/730/2/119} {\bibfield  {journal} {\bibinfo  {journal}
  {\apj}\ }\textbf {\bibinfo {volume} {730}},\ \bibinfo {pages} {119} (\bibinfo
  {year} {2011})},\ \Eprint {http://arxiv.org/abs/1103.2976} {arXiv:1103.2976
  [astro-ph.CO]} \BibitemShut {NoStop}%
\bibitem [{\citenamefont {{Beutler}}\ \emph {et~al.}(2011)\citenamefont
  {{Beutler}}, \citenamefont {{Blake}}, \citenamefont {{Colless}},
  \citenamefont {{Jones}}, \citenamefont {{Staveley-Smith}}, \citenamefont
  {{Campbell}}, \citenamefont {{Parker}}, \citenamefont {{Saunders}},\ and\
  \citenamefont {{Watson}}}]{beutler11a}%
  \BibitemOpen
  \bibfield  {author} {\bibinfo {author} {\bibfnamefont {F.}~\bibnamefont
  {{Beutler}}}, \bibinfo {author} {\bibfnamefont {C.}~\bibnamefont {{Blake}}},
  \bibinfo {author} {\bibfnamefont {M.}~\bibnamefont {{Colless}}}, \bibinfo
  {author} {\bibfnamefont {D.~H.}\ \bibnamefont {{Jones}}}, \bibinfo {author}
  {\bibfnamefont {L.}~\bibnamefont {{Staveley-Smith}}}, \bibinfo {author}
  {\bibfnamefont {L.}~\bibnamefont {{Campbell}}}, \bibinfo {author}
  {\bibfnamefont {Q.}~\bibnamefont {{Parker}}}, \bibinfo {author}
  {\bibfnamefont {W.}~\bibnamefont {{Saunders}}}, \ and\ \bibinfo {author}
  {\bibfnamefont {F.}~\bibnamefont {{Watson}}},\ }\href {\doibase
  10.1111/j.1365-2966.2011.19250.x} {\bibfield  {journal} {\bibinfo  {journal}
  {\mnras}\ }\textbf {\bibinfo {volume} {416}},\ \bibinfo {pages} {3017}
  (\bibinfo {year} {2011})},\ \Eprint {http://arxiv.org/abs/1106.3366}
  {arXiv:1106.3366 [astro-ph.CO]} \BibitemShut {NoStop}%
\bibitem [{\citenamefont {{Padmanabhan}}\ \emph {et~al.}(2012)\citenamefont
  {{Padmanabhan}}, \citenamefont {{Xu}}, \citenamefont {{Eisenstein}},
  \citenamefont {{Scalzo}}, \citenamefont {{Cuesta}}, \citenamefont {{Mehta}},\
  and\ \citenamefont {{Kazin}}}]{padmanabhan12a}%
  \BibitemOpen
  \bibfield  {author} {\bibinfo {author} {\bibfnamefont {N.}~\bibnamefont
  {{Padmanabhan}}}, \bibinfo {author} {\bibfnamefont {X.}~\bibnamefont {{Xu}}},
  \bibinfo {author} {\bibfnamefont {D.~J.}\ \bibnamefont {{Eisenstein}}},
  \bibinfo {author} {\bibfnamefont {R.}~\bibnamefont {{Scalzo}}}, \bibinfo
  {author} {\bibfnamefont {A.~J.}\ \bibnamefont {{Cuesta}}}, \bibinfo {author}
  {\bibfnamefont {K.~T.}\ \bibnamefont {{Mehta}}}, \ and\ \bibinfo {author}
  {\bibfnamefont {E.}~\bibnamefont {{Kazin}}},\ }\href {\doibase
  10.1111/j.1365-2966.2012.21888.x} {\bibfield  {journal} {\bibinfo  {journal}
  {\mnras}\ }\textbf {\bibinfo {volume} {427}},\ \bibinfo {pages} {2132}
  (\bibinfo {year} {2012})},\ \Eprint {http://arxiv.org/abs/1202.0090}
  {arXiv:1202.0090} \BibitemShut {NoStop}%
\bibitem [{\citenamefont {{Blake}}\ \emph {et~al.}(2012)\citenamefont
  {{Blake}}, \citenamefont {{Brough}}, \citenamefont {{Colless}}, \citenamefont
  {{Contreras}}, \citenamefont {{Couch}}, \citenamefont {{Croom}},
  \citenamefont {{Croton}}, \citenamefont {{Davis}}, \citenamefont
  {{Drinkwater}}, \citenamefont {{Forster}}, \citenamefont {{Gilbank}},
  \citenamefont {{Gladders}}, \citenamefont {{Glazebrook}}, \citenamefont
  {{Jelliffe}}, \citenamefont {{Jurek}}, \citenamefont {{Li}}, \citenamefont
  {{Madore}}, \citenamefont {{Martin}}, \citenamefont {{Pimbblet}},
  \citenamefont {{Poole}}, \citenamefont {{Pracy}}, \citenamefont {{Sharp}},
  \citenamefont {{Wisnioski}}, \citenamefont {{Woods}}, \citenamefont
  {{Wyder}},\ and\ \citenamefont {{Yee}}}]{blake12a}%
  \BibitemOpen
  \bibfield  {author} {\bibinfo {author} {\bibfnamefont {C.}~\bibnamefont
  {{Blake}}}, \bibinfo {author} {\bibfnamefont {S.}~\bibnamefont {{Brough}}},
  \bibinfo {author} {\bibfnamefont {M.}~\bibnamefont {{Colless}}}, \bibinfo
  {author} {\bibfnamefont {C.}~\bibnamefont {{Contreras}}}, \bibinfo {author}
  {\bibfnamefont {W.}~\bibnamefont {{Couch}}}, \bibinfo {author} {\bibfnamefont
  {S.}~\bibnamefont {{Croom}}}, \bibinfo {author} {\bibfnamefont
  {D.}~\bibnamefont {{Croton}}}, \bibinfo {author} {\bibfnamefont {T.~M.}\
  \bibnamefont {{Davis}}}, \bibinfo {author} {\bibfnamefont {M.~J.}\
  \bibnamefont {{Drinkwater}}}, \bibinfo {author} {\bibfnamefont
  {K.}~\bibnamefont {{Forster}}}, \bibinfo {author} {\bibfnamefont
  {D.}~\bibnamefont {{Gilbank}}}, \bibinfo {author} {\bibfnamefont
  {M.}~\bibnamefont {{Gladders}}}, \bibinfo {author} {\bibfnamefont
  {K.}~\bibnamefont {{Glazebrook}}}, \bibinfo {author} {\bibfnamefont
  {B.}~\bibnamefont {{Jelliffe}}}, \bibinfo {author} {\bibfnamefont {R.~J.}\
  \bibnamefont {{Jurek}}}, \bibinfo {author} {\bibfnamefont {I.-h.}\
  \bibnamefont {{Li}}}, \bibinfo {author} {\bibfnamefont {B.}~\bibnamefont
  {{Madore}}}, \bibinfo {author} {\bibfnamefont {D.~C.}\ \bibnamefont
  {{Martin}}}, \bibinfo {author} {\bibfnamefont {K.}~\bibnamefont
  {{Pimbblet}}}, \bibinfo {author} {\bibfnamefont {G.~B.}\ \bibnamefont
  {{Poole}}}, \bibinfo {author} {\bibfnamefont {M.}~\bibnamefont {{Pracy}}},
  \bibinfo {author} {\bibfnamefont {R.}~\bibnamefont {{Sharp}}}, \bibinfo
  {author} {\bibfnamefont {E.}~\bibnamefont {{Wisnioski}}}, \bibinfo {author}
  {\bibfnamefont {D.}~\bibnamefont {{Woods}}}, \bibinfo {author} {\bibfnamefont
  {T.~K.}\ \bibnamefont {{Wyder}}}, \ and\ \bibinfo {author} {\bibfnamefont
  {H.~K.~C.}\ \bibnamefont {{Yee}}},\ }\href {\doibase
  10.1111/j.1365-2966.2012.21473.x} {\bibfield  {journal} {\bibinfo  {journal}
  {\mnras}\ }\textbf {\bibinfo {volume} {425}},\ \bibinfo {pages} {405}
  (\bibinfo {year} {2012})},\ \Eprint {http://arxiv.org/abs/1204.3674}
  {arXiv:1204.3674} \BibitemShut {NoStop}%
\bibitem [{\citenamefont {{Hinshaw}}\ \emph {et~al.}(2013)\citenamefont
  {{Hinshaw}}, \citenamefont {{Larson}}, \citenamefont {{Komatsu}},
  \citenamefont {{Spergel}}, \citenamefont {{Bennett}}, \citenamefont
  {{Dunkley}}, \citenamefont {{Nolta}}, \citenamefont {{Halpern}},
  \citenamefont {{Hill}}, \citenamefont {{Odegard}}, \citenamefont {{Page}},
  \citenamefont {{Smith}}, \citenamefont {{Weiland}}, \citenamefont {{Gold}},
  \citenamefont {{Jarosik}}, \citenamefont {{Kogut}}, \citenamefont {{Limon}},
  \citenamefont {{Meyer}}, \citenamefont {{Tucker}}, \citenamefont
  {{Wollack}},\ and\ \citenamefont {{Wright}}}]{hinshaw13a}%
  \BibitemOpen
  \bibfield  {author} {\bibinfo {author} {\bibfnamefont {G.}~\bibnamefont
  {{Hinshaw}}}, \bibinfo {author} {\bibfnamefont {D.}~\bibnamefont {{Larson}}},
  \bibinfo {author} {\bibfnamefont {E.}~\bibnamefont {{Komatsu}}}, \bibinfo
  {author} {\bibfnamefont {D.~N.}\ \bibnamefont {{Spergel}}}, \bibinfo {author}
  {\bibfnamefont {C.~L.}\ \bibnamefont {{Bennett}}}, \bibinfo {author}
  {\bibfnamefont {J.}~\bibnamefont {{Dunkley}}}, \bibinfo {author}
  {\bibfnamefont {M.~R.}\ \bibnamefont {{Nolta}}}, \bibinfo {author}
  {\bibfnamefont {M.}~\bibnamefont {{Halpern}}}, \bibinfo {author}
  {\bibfnamefont {R.~S.}\ \bibnamefont {{Hill}}}, \bibinfo {author}
  {\bibfnamefont {N.}~\bibnamefont {{Odegard}}}, \bibinfo {author}
  {\bibfnamefont {L.}~\bibnamefont {{Page}}}, \bibinfo {author} {\bibfnamefont
  {K.~M.}\ \bibnamefont {{Smith}}}, \bibinfo {author} {\bibfnamefont {J.~L.}\
  \bibnamefont {{Weiland}}}, \bibinfo {author} {\bibfnamefont {B.}~\bibnamefont
  {{Gold}}}, \bibinfo {author} {\bibfnamefont {N.}~\bibnamefont {{Jarosik}}},
  \bibinfo {author} {\bibfnamefont {A.}~\bibnamefont {{Kogut}}}, \bibinfo
  {author} {\bibfnamefont {M.}~\bibnamefont {{Limon}}}, \bibinfo {author}
  {\bibfnamefont {S.~S.}\ \bibnamefont {{Meyer}}}, \bibinfo {author}
  {\bibfnamefont {G.~S.}\ \bibnamefont {{Tucker}}}, \bibinfo {author}
  {\bibfnamefont {E.}~\bibnamefont {{Wollack}}}, \ and\ \bibinfo {author}
  {\bibfnamefont {E.~L.}\ \bibnamefont {{Wright}}},\ }\href {\doibase
  10.1088/0067-0049/208/2/19} {\bibfield  {journal} {\bibinfo  {journal}
  {\apjs}\ }\textbf {\bibinfo {volume} {208}},\ \bibinfo {eid} {19} (\bibinfo
  {year} {2013})},\ \Eprint {http://arxiv.org/abs/1212.5226} {arXiv:1212.5226}
  \BibitemShut {NoStop}%
\bibitem [{\citenamefont {{Chevallier}}\ and\ \citenamefont
  {{Polarski}}(2001)}]{chevallier01a}%
  \BibitemOpen
  \bibfield  {author} {\bibinfo {author} {\bibfnamefont {M.}~\bibnamefont
  {{Chevallier}}}\ and\ \bibinfo {author} {\bibfnamefont {D.}~\bibnamefont
  {{Polarski}}},\ }\href {\doibase 10.1142/S0218271801000822} {\bibfield
  {journal} {\bibinfo  {journal} {International Journal of Modern Physics D}\
  }\textbf {\bibinfo {volume} {10}},\ \bibinfo {pages} {213} (\bibinfo {year}
  {2001})},\ \Eprint {http://arxiv.org/abs/gr-qc/0009008} {gr-qc/0009008}
  \BibitemShut {NoStop}%
\bibitem [{\citenamefont {{Linder}}(2003)}]{linder03b}%
  \BibitemOpen
  \bibfield  {author} {\bibinfo {author} {\bibfnamefont {E.~V.}\ \bibnamefont
  {{Linder}}},\ }\href {\doibase 10.1103/PhysRevLett.90.091301} {\bibfield
  {journal} {\bibinfo  {journal} {Physical Review Letters}\ }\textbf {\bibinfo
  {volume} {90}},\ \bibinfo {pages} {091301} (\bibinfo {year} {2003})},\
  \Eprint {http://arxiv.org/abs/arXiv:astro-ph/0208512}
  {arXiv:astro-ph/0208512} \BibitemShut {NoStop}%
\bibitem [{\citenamefont {{Alam}}\ \emph {et~al.}(2017)\citenamefont {{Alam}},
  \citenamefont {{Ata}}, \citenamefont {{Bailey}}, \citenamefont {{Beutler}},
  \citenamefont {{Bizyaev}}, \citenamefont {{Blazek}}, \citenamefont
  {{Bolton}}, \citenamefont {{Brownstein}}, \citenamefont {{Burden}},
  \citenamefont {{Chuang}}, \citenamefont {{Comparat}}, \citenamefont
  {{Cuesta}}, \citenamefont {{Dawson}}, \citenamefont {{Eisenstein}},
  \citenamefont {{Escoffier}}, \citenamefont {{Gil-Mar{\'{\i}}n}},
  \citenamefont {{Grieb}}, \citenamefont {{Hand}}, \citenamefont {{Ho}},
  \citenamefont {{Kinemuchi}}, \citenamefont {{Kirkby}}, \citenamefont
  {{Kitaura}}, \citenamefont {{Malanushenko}}, \citenamefont {{Malanushenko}},
  \citenamefont {{Maraston}}, \citenamefont {{McBride}}, \citenamefont
  {{Nichol}}, \citenamefont {{Olmstead}}, \citenamefont {{Oravetz}},
  \citenamefont {{Padmanabhan}}, \citenamefont {{Palanque-Delabrouille}},
  \citenamefont {{Pan}}, \citenamefont {{Pellejero-Ibanez}}, \citenamefont
  {{Percival}}, \citenamefont {{Petitjean}}, \citenamefont {{Prada}},
  \citenamefont {{Price-Whelan}}, \citenamefont {{Reid}}, \citenamefont
  {{Rodr{\'{\i}}guez-Torres}}, \citenamefont {{Roe}}, \citenamefont {{Ross}},
  \citenamefont {{Ross}}, \citenamefont {{Rossi}}, \citenamefont
  {{Rubi{\~n}o-Mart{\'{\i}}n}}, \citenamefont {{Saito}}, \citenamefont
  {{Salazar-Albornoz}}, \citenamefont {{Samushia}}, \citenamefont
  {{S{\'a}nchez}}, \citenamefont {{Satpathy}}, \citenamefont {{Schlegel}},
  \citenamefont {{Schneider}}, \citenamefont {{Sc{\'o}ccola}}, \citenamefont
  {{Seo}}, \citenamefont {{Sheldon}}, \citenamefont {{Simmons}}, \citenamefont
  {{Slosar}}, \citenamefont {{Strauss}}, \citenamefont {{Swanson}},
  \citenamefont {{Thomas}}, \citenamefont {{Tinker}}, \citenamefont
  {{Tojeiro}}, \citenamefont {{Maga{\~n}a}}, \citenamefont {{Vazquez}},
  \citenamefont {{Verde}}, \citenamefont {{Wake}}, \citenamefont {{Wang}},
  \citenamefont {{Weinberg}}, \citenamefont {{White}}, \citenamefont
  {{Wood-Vasey}}, \citenamefont {{Y{\`e}che}}, \citenamefont {{Zehavi}},
  \citenamefont {{Zhai}},\ and\ \citenamefont {{Zhao}}}]{alam17a}%
  \BibitemOpen
  \bibfield  {author} {\bibinfo {author} {\bibfnamefont {S.}~\bibnamefont
  {{Alam}}}, \bibinfo {author} {\bibfnamefont {M.}~\bibnamefont {{Ata}}},
  \bibinfo {author} {\bibfnamefont {S.}~\bibnamefont {{Bailey}}}, \bibinfo
  {author} {\bibfnamefont {F.}~\bibnamefont {{Beutler}}}, \bibinfo {author}
  {\bibfnamefont {D.}~\bibnamefont {{Bizyaev}}}, \bibinfo {author}
  {\bibfnamefont {J.~A.}\ \bibnamefont {{Blazek}}}, \bibinfo {author}
  {\bibfnamefont {A.~S.}\ \bibnamefont {{Bolton}}}, \bibinfo {author}
  {\bibfnamefont {J.~R.}\ \bibnamefont {{Brownstein}}}, \bibinfo {author}
  {\bibfnamefont {A.}~\bibnamefont {{Burden}}}, \bibinfo {author}
  {\bibfnamefont {C.-H.}\ \bibnamefont {{Chuang}}}, \bibinfo {author}
  {\bibfnamefont {J.}~\bibnamefont {{Comparat}}}, \bibinfo {author}
  {\bibfnamefont {A.~J.}\ \bibnamefont {{Cuesta}}}, \bibinfo {author}
  {\bibfnamefont {K.~S.}\ \bibnamefont {{Dawson}}}, \bibinfo {author}
  {\bibfnamefont {D.~J.}\ \bibnamefont {{Eisenstein}}}, \bibinfo {author}
  {\bibfnamefont {S.}~\bibnamefont {{Escoffier}}}, \bibinfo {author}
  {\bibfnamefont {H.}~\bibnamefont {{Gil-Mar{\'{\i}}n}}}, \bibinfo {author}
  {\bibfnamefont {J.~N.}\ \bibnamefont {{Grieb}}}, \bibinfo {author}
  {\bibfnamefont {N.}~\bibnamefont {{Hand}}}, \bibinfo {author} {\bibfnamefont
  {S.}~\bibnamefont {{Ho}}}, \bibinfo {author} {\bibfnamefont {K.}~\bibnamefont
  {{Kinemuchi}}}, \bibinfo {author} {\bibfnamefont {D.}~\bibnamefont
  {{Kirkby}}}, \bibinfo {author} {\bibfnamefont {F.}~\bibnamefont {{Kitaura}}},
  \bibinfo {author} {\bibfnamefont {E.}~\bibnamefont {{Malanushenko}}},
  \bibinfo {author} {\bibfnamefont {V.}~\bibnamefont {{Malanushenko}}},
  \bibinfo {author} {\bibfnamefont {C.}~\bibnamefont {{Maraston}}}, \bibinfo
  {author} {\bibfnamefont {C.~K.}\ \bibnamefont {{McBride}}}, \bibinfo {author}
  {\bibfnamefont {R.~C.}\ \bibnamefont {{Nichol}}}, \bibinfo {author}
  {\bibfnamefont {M.~D.}\ \bibnamefont {{Olmstead}}}, \bibinfo {author}
  {\bibfnamefont {D.}~\bibnamefont {{Oravetz}}}, \bibinfo {author}
  {\bibfnamefont {N.}~\bibnamefont {{Padmanabhan}}}, \bibinfo {author}
  {\bibfnamefont {N.}~\bibnamefont {{Palanque-Delabrouille}}}, \bibinfo
  {author} {\bibfnamefont {K.}~\bibnamefont {{Pan}}}, \bibinfo {author}
  {\bibfnamefont {M.}~\bibnamefont {{Pellejero-Ibanez}}}, \bibinfo {author}
  {\bibfnamefont {W.~J.}\ \bibnamefont {{Percival}}}, \bibinfo {author}
  {\bibfnamefont {P.}~\bibnamefont {{Petitjean}}}, \bibinfo {author}
  {\bibfnamefont {F.}~\bibnamefont {{Prada}}}, \bibinfo {author} {\bibfnamefont
  {A.~M.}\ \bibnamefont {{Price-Whelan}}}, \bibinfo {author} {\bibfnamefont
  {B.~A.}\ \bibnamefont {{Reid}}}, \bibinfo {author} {\bibfnamefont {S.~A.}\
  \bibnamefont {{Rodr{\'{\i}}guez-Torres}}}, \bibinfo {author} {\bibfnamefont
  {N.~A.}\ \bibnamefont {{Roe}}}, \bibinfo {author} {\bibfnamefont {A.~J.}\
  \bibnamefont {{Ross}}}, \bibinfo {author} {\bibfnamefont {N.~P.}\
  \bibnamefont {{Ross}}}, \bibinfo {author} {\bibfnamefont {G.}~\bibnamefont
  {{Rossi}}}, \bibinfo {author} {\bibfnamefont {J.~A.}\ \bibnamefont
  {{Rubi{\~n}o-Mart{\'{\i}}n}}}, \bibinfo {author} {\bibfnamefont
  {S.}~\bibnamefont {{Saito}}}, \bibinfo {author} {\bibfnamefont
  {S.}~\bibnamefont {{Salazar-Albornoz}}}, \bibinfo {author} {\bibfnamefont
  {L.}~\bibnamefont {{Samushia}}}, \bibinfo {author} {\bibfnamefont {A.~G.}\
  \bibnamefont {{S{\'a}nchez}}}, \bibinfo {author} {\bibfnamefont
  {S.}~\bibnamefont {{Satpathy}}}, \bibinfo {author} {\bibfnamefont {D.~J.}\
  \bibnamefont {{Schlegel}}}, \bibinfo {author} {\bibfnamefont {D.~P.}\
  \bibnamefont {{Schneider}}}, \bibinfo {author} {\bibfnamefont {C.~G.}\
  \bibnamefont {{Sc{\'o}ccola}}}, \bibinfo {author} {\bibfnamefont {H.-J.}\
  \bibnamefont {{Seo}}}, \bibinfo {author} {\bibfnamefont {E.~S.}\ \bibnamefont
  {{Sheldon}}}, \bibinfo {author} {\bibfnamefont {A.}~\bibnamefont
  {{Simmons}}}, \bibinfo {author} {\bibfnamefont {A.}~\bibnamefont {{Slosar}}},
  \bibinfo {author} {\bibfnamefont {M.~A.}\ \bibnamefont {{Strauss}}}, \bibinfo
  {author} {\bibfnamefont {M.~E.~C.}\ \bibnamefont {{Swanson}}}, \bibinfo
  {author} {\bibfnamefont {D.}~\bibnamefont {{Thomas}}}, \bibinfo {author}
  {\bibfnamefont {J.~L.}\ \bibnamefont {{Tinker}}}, \bibinfo {author}
  {\bibfnamefont {R.}~\bibnamefont {{Tojeiro}}}, \bibinfo {author}
  {\bibfnamefont {M.~V.}\ \bibnamefont {{Maga{\~n}a}}}, \bibinfo {author}
  {\bibfnamefont {J.~A.}\ \bibnamefont {{Vazquez}}}, \bibinfo {author}
  {\bibfnamefont {L.}~\bibnamefont {{Verde}}}, \bibinfo {author} {\bibfnamefont
  {D.~A.}\ \bibnamefont {{Wake}}}, \bibinfo {author} {\bibfnamefont
  {Y.}~\bibnamefont {{Wang}}}, \bibinfo {author} {\bibfnamefont {D.~H.}\
  \bibnamefont {{Weinberg}}}, \bibinfo {author} {\bibfnamefont
  {M.}~\bibnamefont {{White}}}, \bibinfo {author} {\bibfnamefont {W.~M.}\
  \bibnamefont {{Wood-Vasey}}}, \bibinfo {author} {\bibfnamefont
  {C.}~\bibnamefont {{Y{\`e}che}}}, \bibinfo {author} {\bibfnamefont
  {I.}~\bibnamefont {{Zehavi}}}, \bibinfo {author} {\bibfnamefont
  {Z.}~\bibnamefont {{Zhai}}}, \ and\ \bibinfo {author} {\bibfnamefont {G.-B.}\
  \bibnamefont {{Zhao}}},\ }\href {\doibase 10.1093/mnras/stx721} {\bibfield
  {journal} {\bibinfo  {journal} {\mnras}\ }\textbf {\bibinfo {volume} {470}},\
  \bibinfo {pages} {2617} (\bibinfo {year} {2017})},\ \Eprint
  {http://arxiv.org/abs/1607.03155} {arXiv:1607.03155} \BibitemShut {NoStop}%
\bibitem [{\citenamefont {{Bautista}}\ \emph {et~al.}(2017)\citenamefont
  {{Bautista}}, \citenamefont {{Busca}}, \citenamefont {{Guy}}, \citenamefont
  {{Rich}}, \citenamefont {{Blomqvist}}, \citenamefont {{du Mas des Bourboux}},
  \citenamefont {{Pieri}}, \citenamefont {{Font-Ribera}}, \citenamefont
  {{Bailey}}, \citenamefont {{Delubac}}, \citenamefont {{Kirkby}},
  \citenamefont {{Le Goff}}, \citenamefont {{Margala}}, \citenamefont
  {{Slosar}}, \citenamefont {{Vazquez}}, \citenamefont {{Brownstein}},
  \citenamefont {{Dawson}}, \citenamefont {{Eisenstein}}, \citenamefont
  {{Miralda-Escud{\'e}}}, \citenamefont {{Noterdaeme}}, \citenamefont
  {{Palanque-Delabrouille}}, \citenamefont {{P{\^a}ris}}, \citenamefont
  {{Petitjean}}, \citenamefont {{Ross}}, \citenamefont {{Schneider}},
  \citenamefont {{Weinberg}},\ and\ \citenamefont {{Y{\`e}che}}}]{bautista17a}%
  \BibitemOpen
  \bibfield  {author} {\bibinfo {author} {\bibfnamefont {J.~E.}\ \bibnamefont
  {{Bautista}}}, \bibinfo {author} {\bibfnamefont {N.~G.}\ \bibnamefont
  {{Busca}}}, \bibinfo {author} {\bibfnamefont {J.}~\bibnamefont {{Guy}}},
  \bibinfo {author} {\bibfnamefont {J.}~\bibnamefont {{Rich}}}, \bibinfo
  {author} {\bibfnamefont {M.}~\bibnamefont {{Blomqvist}}}, \bibinfo {author}
  {\bibfnamefont {H.}~\bibnamefont {{du Mas des Bourboux}}}, \bibinfo {author}
  {\bibfnamefont {M.~M.}\ \bibnamefont {{Pieri}}}, \bibinfo {author}
  {\bibfnamefont {A.}~\bibnamefont {{Font-Ribera}}}, \bibinfo {author}
  {\bibfnamefont {S.}~\bibnamefont {{Bailey}}}, \bibinfo {author}
  {\bibfnamefont {T.}~\bibnamefont {{Delubac}}}, \bibinfo {author}
  {\bibfnamefont {D.}~\bibnamefont {{Kirkby}}}, \bibinfo {author}
  {\bibfnamefont {J.-M.}\ \bibnamefont {{Le Goff}}}, \bibinfo {author}
  {\bibfnamefont {D.}~\bibnamefont {{Margala}}}, \bibinfo {author}
  {\bibfnamefont {A.}~\bibnamefont {{Slosar}}}, \bibinfo {author}
  {\bibfnamefont {J.~A.}\ \bibnamefont {{Vazquez}}}, \bibinfo {author}
  {\bibfnamefont {J.~R.}\ \bibnamefont {{Brownstein}}}, \bibinfo {author}
  {\bibfnamefont {K.~S.}\ \bibnamefont {{Dawson}}}, \bibinfo {author}
  {\bibfnamefont {D.~J.}\ \bibnamefont {{Eisenstein}}}, \bibinfo {author}
  {\bibfnamefont {J.}~\bibnamefont {{Miralda-Escud{\'e}}}}, \bibinfo {author}
  {\bibfnamefont {P.}~\bibnamefont {{Noterdaeme}}}, \bibinfo {author}
  {\bibfnamefont {N.}~\bibnamefont {{Palanque-Delabrouille}}}, \bibinfo
  {author} {\bibfnamefont {I.}~\bibnamefont {{P{\^a}ris}}}, \bibinfo {author}
  {\bibfnamefont {P.}~\bibnamefont {{Petitjean}}}, \bibinfo {author}
  {\bibfnamefont {N.~P.}\ \bibnamefont {{Ross}}}, \bibinfo {author}
  {\bibfnamefont {D.~P.}\ \bibnamefont {{Schneider}}}, \bibinfo {author}
  {\bibfnamefont {D.~H.}\ \bibnamefont {{Weinberg}}}, \ and\ \bibinfo {author}
  {\bibfnamefont {C.}~\bibnamefont {{Y{\`e}che}}},\ }\href {\doibase
  10.1051/0004-6361/201730533} {\bibfield  {journal} {\bibinfo  {journal}
  {\aap}\ }\textbf {\bibinfo {volume} {603}},\ \bibinfo {eid} {A12} (\bibinfo
  {year} {2017})},\ \Eprint {http://arxiv.org/abs/1702.00176}
  {arXiv:1702.00176} \BibitemShut {NoStop}%
\bibitem [{\citenamefont {{du Mas des Bourboux}}\ \emph
  {et~al.}(2017)\citenamefont {{du Mas des Bourboux}}, \citenamefont {{Le
  Goff}}, \citenamefont {{Blomqvist}}, \citenamefont {{Busca}}, \citenamefont
  {{Guy}}, \citenamefont {{Rich}}, \citenamefont {{Y{\`e}che}}, \citenamefont
  {{Bautista}}, \citenamefont {{Burtin}}, \citenamefont {{Dawson}},
  \citenamefont {{Eisenstein}}, \citenamefont {{Font-Ribera}}, \citenamefont
  {{Kirkby}}, \citenamefont {{Miralda-Escud{\'e}}}, \citenamefont
  {{Noterdaeme}}, \citenamefont {{Palanque-Delabrouille}}, \citenamefont
  {{P{\^a}ris}}, \citenamefont {{Petitjean}}, \citenamefont
  {{P{\'e}rez-R{\`a}fols}}, \citenamefont {{Pieri}}, \citenamefont {{Ross}},
  \citenamefont {{Schlegel}}, \citenamefont {{Schneider}}, \citenamefont
  {{Slosar}}, \citenamefont {{Weinberg}},\ and\ \citenamefont
  {{Zarrouk}}}]{masdesbourboux17a}%
  \BibitemOpen
  \bibfield  {author} {\bibinfo {author} {\bibfnamefont {H.}~\bibnamefont {{du
  Mas des Bourboux}}}, \bibinfo {author} {\bibfnamefont {J.-M.}\ \bibnamefont
  {{Le Goff}}}, \bibinfo {author} {\bibfnamefont {M.}~\bibnamefont
  {{Blomqvist}}}, \bibinfo {author} {\bibfnamefont {N.~G.}\ \bibnamefont
  {{Busca}}}, \bibinfo {author} {\bibfnamefont {J.}~\bibnamefont {{Guy}}},
  \bibinfo {author} {\bibfnamefont {J.}~\bibnamefont {{Rich}}}, \bibinfo
  {author} {\bibfnamefont {C.}~\bibnamefont {{Y{\`e}che}}}, \bibinfo {author}
  {\bibfnamefont {J.~E.}\ \bibnamefont {{Bautista}}}, \bibinfo {author}
  {\bibfnamefont {{\'E}.}~\bibnamefont {{Burtin}}}, \bibinfo {author}
  {\bibfnamefont {K.~S.}\ \bibnamefont {{Dawson}}}, \bibinfo {author}
  {\bibfnamefont {D.~J.}\ \bibnamefont {{Eisenstein}}}, \bibinfo {author}
  {\bibfnamefont {A.}~\bibnamefont {{Font-Ribera}}}, \bibinfo {author}
  {\bibfnamefont {D.}~\bibnamefont {{Kirkby}}}, \bibinfo {author}
  {\bibfnamefont {J.}~\bibnamefont {{Miralda-Escud{\'e}}}}, \bibinfo {author}
  {\bibfnamefont {P.}~\bibnamefont {{Noterdaeme}}}, \bibinfo {author}
  {\bibfnamefont {N.}~\bibnamefont {{Palanque-Delabrouille}}}, \bibinfo
  {author} {\bibfnamefont {I.}~\bibnamefont {{P{\^a}ris}}}, \bibinfo {author}
  {\bibfnamefont {P.}~\bibnamefont {{Petitjean}}}, \bibinfo {author}
  {\bibfnamefont {I.}~\bibnamefont {{P{\'e}rez-R{\`a}fols}}}, \bibinfo {author}
  {\bibfnamefont {M.~M.}\ \bibnamefont {{Pieri}}}, \bibinfo {author}
  {\bibfnamefont {N.~P.}\ \bibnamefont {{Ross}}}, \bibinfo {author}
  {\bibfnamefont {D.~J.}\ \bibnamefont {{Schlegel}}}, \bibinfo {author}
  {\bibfnamefont {D.~P.}\ \bibnamefont {{Schneider}}}, \bibinfo {author}
  {\bibfnamefont {A.}~\bibnamefont {{Slosar}}}, \bibinfo {author}
  {\bibfnamefont {D.~H.}\ \bibnamefont {{Weinberg}}}, \ and\ \bibinfo {author}
  {\bibfnamefont {P.}~\bibnamefont {{Zarrouk}}},\ }\href {\doibase
  10.1051/0004-6361/201731731} {\bibfield  {journal} {\bibinfo  {journal}
  {\aap}\ }\textbf {\bibinfo {volume} {608}},\ \bibinfo {eid} {A130} (\bibinfo
  {year} {2017})},\ \Eprint {http://arxiv.org/abs/1708.02225} {arXiv:1708.02225
  [astro-ph.CO]} \BibitemShut {NoStop}%
\bibitem [{\citenamefont {{Planck Collaboration}}\ \emph
  {et~al.}(2016{\natexlab{a}})\citenamefont {{Planck Collaboration}},
  \citenamefont {{Adam}}, \citenamefont {{Ade}}, \citenamefont {{Aghanim}},
  \citenamefont {{Akrami}}, \citenamefont {{Alves}}, \citenamefont
  {{Arg{\"u}eso}}, \citenamefont {{Arnaud}}, \citenamefont {{Arroja}},
  \citenamefont {{Ashdown}},\ and\ \citenamefont {et~al.}}]{planck15a}%
  \BibitemOpen
  \bibfield  {author} {\bibinfo {author} {\bibnamefont {{Planck
  Collaboration}}}, \bibinfo {author} {\bibfnamefont {R.}~\bibnamefont
  {{Adam}}}, \bibinfo {author} {\bibfnamefont {P.~A.~R.}\ \bibnamefont
  {{Ade}}}, \bibinfo {author} {\bibfnamefont {N.}~\bibnamefont {{Aghanim}}},
  \bibinfo {author} {\bibfnamefont {Y.}~\bibnamefont {{Akrami}}}, \bibinfo
  {author} {\bibfnamefont {M.~I.~R.}\ \bibnamefont {{Alves}}}, \bibinfo
  {author} {\bibfnamefont {F.}~\bibnamefont {{Arg{\"u}eso}}}, \bibinfo {author}
  {\bibfnamefont {M.}~\bibnamefont {{Arnaud}}}, \bibinfo {author}
  {\bibfnamefont {F.}~\bibnamefont {{Arroja}}}, \bibinfo {author}
  {\bibfnamefont {M.}~\bibnamefont {{Ashdown}}}, \ and\ \bibinfo {author}
  {\bibnamefont {et~al.}},\ }\href {\doibase 10.1051/0004-6361/201527101}
  {\bibfield  {journal} {\bibinfo  {journal} {\aap}\ }\textbf {\bibinfo
  {volume} {594}},\ \bibinfo {eid} {A1} (\bibinfo {year}
  {2016}{\natexlab{a}})},\ \Eprint {http://arxiv.org/abs/1502.01582}
  {arXiv:1502.01582 [astro-ph.CO]} \BibitemShut {NoStop}%
\bibitem [{\citenamefont {{Gunn}}\ \emph {et~al.}(2006)\citenamefont {{Gunn}},
  \citenamefont {{Siegmund}}, \citenamefont {{Mannery}}, \citenamefont
  {{Owen}}, \citenamefont {{Hull}}, \citenamefont {{Leger}}, \citenamefont
  {{Carey}}, \citenamefont {{Knapp}}, \citenamefont {{York}}, \citenamefont
  {{Boroski}}, \citenamefont {{Kent}}, \citenamefont {{Lupton}}, \citenamefont
  {{Rockosi}}, \citenamefont {{Evans}}, \citenamefont {{Waddell}},
  \citenamefont {{Anderson}}, \citenamefont {{Annis}}, \citenamefont
  {{Barentine}}, \citenamefont {{Bartoszek}}, \citenamefont {{Bastian}},
  \citenamefont {{Bracker}}, \citenamefont {{Brewington}}, \citenamefont
  {{Briegel}}, \citenamefont {{Brinkmann}}, \citenamefont {{Brown}},
  \citenamefont {{Carr}}, \citenamefont {{Czarapata}}, \citenamefont
  {{Drennan}}, \citenamefont {{Dombeck}}, \citenamefont {{Federwitz}},
  \citenamefont {{Gillespie}}, \citenamefont {{Gonzales}}, \citenamefont
  {{Hansen}}, \citenamefont {{Harvanek}}, \citenamefont {{Hayes}},
  \citenamefont {{Jordan}}, \citenamefont {{Kinney}}, \citenamefont {{Klaene}},
  \citenamefont {{Kleinman}}, \citenamefont {{Kron}}, \citenamefont
  {{Kresinski}}, \citenamefont {{Lee}}, \citenamefont {{Limmongkol}},
  \citenamefont {{Lindenmeyer}}, \citenamefont {{Long}}, \citenamefont
  {{Loomis}}, \citenamefont {{McGehee}}, \citenamefont {{Mantsch}},
  \citenamefont {{Neilsen}}, \citenamefont {{Neswold}}, \citenamefont
  {{Newman}}, \citenamefont {{Nitta}}, \citenamefont {{Peoples}}, \citenamefont
  {{Pier}}, \citenamefont {{Prieto}}, \citenamefont {{Prosapio}}, \citenamefont
  {{Rivetta}}, \citenamefont {{Schneider}}, \citenamefont {{Snedden}},\ and\
  \citenamefont {{Wang}}}]{gunn06a}%
  \BibitemOpen
  \bibfield  {author} {\bibinfo {author} {\bibfnamefont {J.~E.}\ \bibnamefont
  {{Gunn}}}, \bibinfo {author} {\bibfnamefont {W.~A.}\ \bibnamefont
  {{Siegmund}}}, \bibinfo {author} {\bibfnamefont {E.~J.}\ \bibnamefont
  {{Mannery}}}, \bibinfo {author} {\bibfnamefont {R.~E.}\ \bibnamefont
  {{Owen}}}, \bibinfo {author} {\bibfnamefont {C.~L.}\ \bibnamefont {{Hull}}},
  \bibinfo {author} {\bibfnamefont {R.~F.}\ \bibnamefont {{Leger}}}, \bibinfo
  {author} {\bibfnamefont {L.~N.}\ \bibnamefont {{Carey}}}, \bibinfo {author}
  {\bibfnamefont {G.~R.}\ \bibnamefont {{Knapp}}}, \bibinfo {author}
  {\bibfnamefont {D.~G.}\ \bibnamefont {{York}}}, \bibinfo {author}
  {\bibfnamefont {W.~N.}\ \bibnamefont {{Boroski}}}, \bibinfo {author}
  {\bibfnamefont {S.~M.}\ \bibnamefont {{Kent}}}, \bibinfo {author}
  {\bibfnamefont {R.~H.}\ \bibnamefont {{Lupton}}}, \bibinfo {author}
  {\bibfnamefont {C.~M.}\ \bibnamefont {{Rockosi}}}, \bibinfo {author}
  {\bibfnamefont {M.~L.}\ \bibnamefont {{Evans}}}, \bibinfo {author}
  {\bibfnamefont {P.}~\bibnamefont {{Waddell}}}, \bibinfo {author}
  {\bibfnamefont {J.~E.}\ \bibnamefont {{Anderson}}}, \bibinfo {author}
  {\bibfnamefont {J.}~\bibnamefont {{Annis}}}, \bibinfo {author} {\bibfnamefont
  {J.~C.}\ \bibnamefont {{Barentine}}}, \bibinfo {author} {\bibfnamefont
  {L.~M.}\ \bibnamefont {{Bartoszek}}}, \bibinfo {author} {\bibfnamefont
  {S.}~\bibnamefont {{Bastian}}}, \bibinfo {author} {\bibfnamefont {S.~B.}\
  \bibnamefont {{Bracker}}}, \bibinfo {author} {\bibfnamefont {H.~J.}\
  \bibnamefont {{Brewington}}}, \bibinfo {author} {\bibfnamefont {C.~I.}\
  \bibnamefont {{Briegel}}}, \bibinfo {author} {\bibfnamefont {J.}~\bibnamefont
  {{Brinkmann}}}, \bibinfo {author} {\bibfnamefont {Y.~J.}\ \bibnamefont
  {{Brown}}}, \bibinfo {author} {\bibfnamefont {M.~A.}\ \bibnamefont {{Carr}}},
  \bibinfo {author} {\bibfnamefont {P.~C.}\ \bibnamefont {{Czarapata}}},
  \bibinfo {author} {\bibfnamefont {C.~C.}\ \bibnamefont {{Drennan}}}, \bibinfo
  {author} {\bibfnamefont {T.}~\bibnamefont {{Dombeck}}}, \bibinfo {author}
  {\bibfnamefont {G.~R.}\ \bibnamefont {{Federwitz}}}, \bibinfo {author}
  {\bibfnamefont {B.~A.}\ \bibnamefont {{Gillespie}}}, \bibinfo {author}
  {\bibfnamefont {C.}~\bibnamefont {{Gonzales}}}, \bibinfo {author}
  {\bibfnamefont {S.~U.}\ \bibnamefont {{Hansen}}}, \bibinfo {author}
  {\bibfnamefont {M.}~\bibnamefont {{Harvanek}}}, \bibinfo {author}
  {\bibfnamefont {J.}~\bibnamefont {{Hayes}}}, \bibinfo {author} {\bibfnamefont
  {W.}~\bibnamefont {{Jordan}}}, \bibinfo {author} {\bibfnamefont
  {E.}~\bibnamefont {{Kinney}}}, \bibinfo {author} {\bibfnamefont
  {M.}~\bibnamefont {{Klaene}}}, \bibinfo {author} {\bibfnamefont {S.~J.}\
  \bibnamefont {{Kleinman}}}, \bibinfo {author} {\bibfnamefont {R.~G.}\
  \bibnamefont {{Kron}}}, \bibinfo {author} {\bibfnamefont {J.}~\bibnamefont
  {{Kresinski}}}, \bibinfo {author} {\bibfnamefont {G.}~\bibnamefont {{Lee}}},
  \bibinfo {author} {\bibfnamefont {S.}~\bibnamefont {{Limmongkol}}}, \bibinfo
  {author} {\bibfnamefont {C.~W.}\ \bibnamefont {{Lindenmeyer}}}, \bibinfo
  {author} {\bibfnamefont {D.~C.}\ \bibnamefont {{Long}}}, \bibinfo {author}
  {\bibfnamefont {C.~L.}\ \bibnamefont {{Loomis}}}, \bibinfo {author}
  {\bibfnamefont {P.~M.}\ \bibnamefont {{McGehee}}}, \bibinfo {author}
  {\bibfnamefont {P.~M.}\ \bibnamefont {{Mantsch}}}, \bibinfo {author}
  {\bibfnamefont {E.~H.}\ \bibnamefont {{Neilsen}}, \bibfnamefont {Jr.}},
  \bibinfo {author} {\bibfnamefont {R.~M.}\ \bibnamefont {{Neswold}}}, \bibinfo
  {author} {\bibfnamefont {P.~R.}\ \bibnamefont {{Newman}}}, \bibinfo {author}
  {\bibfnamefont {A.}~\bibnamefont {{Nitta}}}, \bibinfo {author} {\bibfnamefont
  {J.}~\bibnamefont {{Peoples}}, \bibfnamefont {Jr.}}, \bibinfo {author}
  {\bibfnamefont {J.~R.}\ \bibnamefont {{Pier}}}, \bibinfo {author}
  {\bibfnamefont {P.~S.}\ \bibnamefont {{Prieto}}}, \bibinfo {author}
  {\bibfnamefont {A.}~\bibnamefont {{Prosapio}}}, \bibinfo {author}
  {\bibfnamefont {C.}~\bibnamefont {{Rivetta}}}, \bibinfo {author}
  {\bibfnamefont {D.~P.}\ \bibnamefont {{Schneider}}}, \bibinfo {author}
  {\bibfnamefont {S.}~\bibnamefont {{Snedden}}}, \ and\ \bibinfo {author}
  {\bibfnamefont {S.-i.}\ \bibnamefont {{Wang}}},\ }\href {\doibase
  10.1086/500975} {\bibfield  {journal} {\bibinfo  {journal} {\aj}\ }\textbf
  {\bibinfo {volume} {131}},\ \bibinfo {pages} {2332} (\bibinfo {year}
  {2006})},\ \Eprint {http://arxiv.org/abs/arXiv:astro-ph/0602326}
  {arXiv:astro-ph/0602326} \BibitemShut {NoStop}%
\bibitem [{\citenamefont {{Dawson}}\ \emph {et~al.}(2016)\citenamefont
  {{Dawson}}, \citenamefont {{Kneib}}, \citenamefont {{Percival}},
  \citenamefont {{Alam}}, \citenamefont {{Albareti}}, \citenamefont
  {{Anderson}}, \citenamefont {{Armengaud}}, \citenamefont {{Aubourg}},
  \citenamefont {{Bailey}}, \citenamefont {{Bautista}}, \citenamefont
  {{Berlind}}, \citenamefont {{Bershady}}, \citenamefont {{Beutler}},
  \citenamefont {{Bizyaev}}, \citenamefont {{Blanton}}, \citenamefont
  {{Blomqvist}}, \citenamefont {{Bolton}}, \citenamefont {{Bovy}},
  \citenamefont {{Brandt}}, \citenamefont {{Brinkmann}}, \citenamefont
  {{Brownstein}}, \citenamefont {{Burtin}}, \citenamefont {{Busca}},
  \citenamefont {{Cai}}, \citenamefont {{Chuang}}, \citenamefont {{Clerc}},
  \citenamefont {{Comparat}}, \citenamefont {{Cope}}, \citenamefont {{Croft}},
  \citenamefont {{Cruz-Gonzalez}}, \citenamefont {{da Costa}}, \citenamefont
  {{Cousinou}}, \citenamefont {{Darling}}, \citenamefont {{de la Macorra}},
  \citenamefont {{de la Torre}}, \citenamefont {{Delubac}}, \citenamefont {{du
  Mas des Bourboux}}, \citenamefont {{Dwelly}}, \citenamefont {{Ealet}},
  \citenamefont {{Eisenstein}}, \citenamefont {{Eracleous}}, \citenamefont
  {{Escoffier}}, \citenamefont {{Fan}}, \citenamefont {{Finoguenov}},
  \citenamefont {{Font-Ribera}}, \citenamefont {{Frinchaboy}}, \citenamefont
  {{Gaulme}}, \citenamefont {{Georgakakis}}, \citenamefont {{Green}},
  \citenamefont {{Guo}}, \citenamefont {{Guy}}, \citenamefont {{Ho}},
  \citenamefont {{Holder}}, \citenamefont {{Huehnerhoff}}, \citenamefont
  {{Hutchinson}}, \citenamefont {{Jing}}, \citenamefont {{Jullo}},
  \citenamefont {{Kamble}}, \citenamefont {{Kinemuchi}}, \citenamefont
  {{Kirkby}}, \citenamefont {{Kitaura}}, \citenamefont {{Klaene}},
  \citenamefont {{Laher}}, \citenamefont {{Lang}}, \citenamefont {{Laurent}},
  \citenamefont {{Le Goff}}, \citenamefont {{Li}}, \citenamefont {{Liang}},
  \citenamefont {{Lima}}, \citenamefont {{Lin}}, \citenamefont {{Lin}},
  \citenamefont {{Lin}}, \citenamefont {{Long}}, \citenamefont {{Lundgren}},
  \citenamefont {{MacDonald}}, \citenamefont {{Geimba Maia}}, \citenamefont
  {{Malanushenko}}, \citenamefont {{Malanushenko}}, \citenamefont
  {{Mariappan}}, \citenamefont {{McBride}}, \citenamefont {{McGreer}},
  \citenamefont {{M{\'e}nard}}, \citenamefont {{Merloni}}, \citenamefont
  {{Meza}}, \citenamefont {{Montero-Dorta}}, \citenamefont {{Muna}},
  \citenamefont {{Myers}}, \citenamefont {{Nandra}}, \citenamefont {{Naugle}},
  \citenamefont {{Newman}}, \citenamefont {{Noterdaeme}}, \citenamefont
  {{Nugent}}, \citenamefont {{Ogando}}, \citenamefont {{Olmstead}},
  \citenamefont {{Oravetz}}, \citenamefont {{Oravetz}}, \citenamefont
  {{Padmanabhan}}, \citenamefont {{Palanque-Delabrouille}}, \citenamefont
  {{Pan}}, \citenamefont {{Parejko}}, \citenamefont {{P{\^a}ris}},
  \citenamefont {{Peacock}}, \citenamefont {{Petitjean}}, \citenamefont
  {{Pieri}}, \citenamefont {{Pisani}}, \citenamefont {{Prada}}, \citenamefont
  {{Prakash}}, \citenamefont {{Raichoor}}, \citenamefont {{Reid}},
  \citenamefont {{Rich}}, \citenamefont {{Ridl}}, \citenamefont
  {{Rodriguez-Torres}}, \citenamefont {{Carnero Rosell}}, \citenamefont
  {{Ross}}, \citenamefont {{Rossi}}, \citenamefont {{Ruan}}, \citenamefont
  {{Salvato}}, \citenamefont {{Sayres}}, \citenamefont {{Schneider}},
  \citenamefont {{Schlegel}}, \citenamefont {{Seljak}}, \citenamefont {{Seo}},
  \citenamefont {{Sesar}}, \citenamefont {{Shandera}}, \citenamefont {{Shu}},
  \citenamefont {{Slosar}}, \citenamefont {{Sobreira}}, \citenamefont
  {{Streblyanska}}, \citenamefont {{Suzuki}}, \citenamefont {{Taylor}},
  \citenamefont {{Tao}}, \citenamefont {{Tinker}}, \citenamefont {{Tojeiro}},
  \citenamefont {{Vargas-Maga{\~n}a}}, \citenamefont {{Wang}}, \citenamefont
  {{Weaver}}, \citenamefont {{Weinberg}}, \citenamefont {{White}},
  \citenamefont {{Wood-Vasey}}, \citenamefont {{Yeche}}, \citenamefont
  {{Zhai}}, \citenamefont {{Zhao}}, \citenamefont {{Zhao}}, \citenamefont
  {{Zheng}}, \citenamefont {{Ben Zhu}},\ and\ \citenamefont
  {{Zou}}}]{dawson16a}%
  \BibitemOpen
  \bibfield  {author} {\bibinfo {author} {\bibfnamefont {K.~S.}\ \bibnamefont
  {{Dawson}}}, \bibinfo {author} {\bibfnamefont {J.-P.}\ \bibnamefont
  {{Kneib}}}, \bibinfo {author} {\bibfnamefont {W.~J.}\ \bibnamefont
  {{Percival}}}, \bibinfo {author} {\bibfnamefont {S.}~\bibnamefont {{Alam}}},
  \bibinfo {author} {\bibfnamefont {F.~D.}\ \bibnamefont {{Albareti}}},
  \bibinfo {author} {\bibfnamefont {S.~F.}\ \bibnamefont {{Anderson}}},
  \bibinfo {author} {\bibfnamefont {E.}~\bibnamefont {{Armengaud}}}, \bibinfo
  {author} {\bibfnamefont {{\'E}.}~\bibnamefont {{Aubourg}}}, \bibinfo {author}
  {\bibfnamefont {S.}~\bibnamefont {{Bailey}}}, \bibinfo {author}
  {\bibfnamefont {J.~E.}\ \bibnamefont {{Bautista}}}, \bibinfo {author}
  {\bibfnamefont {A.~A.}\ \bibnamefont {{Berlind}}}, \bibinfo {author}
  {\bibfnamefont {M.~A.}\ \bibnamefont {{Bershady}}}, \bibinfo {author}
  {\bibfnamefont {F.}~\bibnamefont {{Beutler}}}, \bibinfo {author}
  {\bibfnamefont {D.}~\bibnamefont {{Bizyaev}}}, \bibinfo {author}
  {\bibfnamefont {M.~R.}\ \bibnamefont {{Blanton}}}, \bibinfo {author}
  {\bibfnamefont {M.}~\bibnamefont {{Blomqvist}}}, \bibinfo {author}
  {\bibfnamefont {A.~S.}\ \bibnamefont {{Bolton}}}, \bibinfo {author}
  {\bibfnamefont {J.}~\bibnamefont {{Bovy}}}, \bibinfo {author} {\bibfnamefont
  {W.~N.}\ \bibnamefont {{Brandt}}}, \bibinfo {author} {\bibfnamefont
  {J.}~\bibnamefont {{Brinkmann}}}, \bibinfo {author} {\bibfnamefont {J.~R.}\
  \bibnamefont {{Brownstein}}}, \bibinfo {author} {\bibfnamefont
  {E.}~\bibnamefont {{Burtin}}}, \bibinfo {author} {\bibfnamefont {N.~G.}\
  \bibnamefont {{Busca}}}, \bibinfo {author} {\bibfnamefont {Z.}~\bibnamefont
  {{Cai}}}, \bibinfo {author} {\bibfnamefont {C.-H.}\ \bibnamefont {{Chuang}}},
  \bibinfo {author} {\bibfnamefont {N.}~\bibnamefont {{Clerc}}}, \bibinfo
  {author} {\bibfnamefont {J.}~\bibnamefont {{Comparat}}}, \bibinfo {author}
  {\bibfnamefont {F.}~\bibnamefont {{Cope}}}, \bibinfo {author} {\bibfnamefont
  {R.~A.~C.}\ \bibnamefont {{Croft}}}, \bibinfo {author} {\bibfnamefont
  {I.}~\bibnamefont {{Cruz-Gonzalez}}}, \bibinfo {author} {\bibfnamefont
  {L.~N.}\ \bibnamefont {{da Costa}}}, \bibinfo {author} {\bibfnamefont
  {M.-C.}\ \bibnamefont {{Cousinou}}}, \bibinfo {author} {\bibfnamefont
  {J.}~\bibnamefont {{Darling}}}, \bibinfo {author} {\bibfnamefont
  {A.}~\bibnamefont {{de la Macorra}}}, \bibinfo {author} {\bibfnamefont
  {S.}~\bibnamefont {{de la Torre}}}, \bibinfo {author} {\bibfnamefont
  {T.}~\bibnamefont {{Delubac}}}, \bibinfo {author} {\bibfnamefont
  {H.}~\bibnamefont {{du Mas des Bourboux}}}, \bibinfo {author} {\bibfnamefont
  {T.}~\bibnamefont {{Dwelly}}}, \bibinfo {author} {\bibfnamefont
  {A.}~\bibnamefont {{Ealet}}}, \bibinfo {author} {\bibfnamefont {D.~J.}\
  \bibnamefont {{Eisenstein}}}, \bibinfo {author} {\bibfnamefont
  {M.}~\bibnamefont {{Eracleous}}}, \bibinfo {author} {\bibfnamefont
  {S.}~\bibnamefont {{Escoffier}}}, \bibinfo {author} {\bibfnamefont
  {X.}~\bibnamefont {{Fan}}}, \bibinfo {author} {\bibfnamefont
  {A.}~\bibnamefont {{Finoguenov}}}, \bibinfo {author} {\bibfnamefont
  {A.}~\bibnamefont {{Font-Ribera}}}, \bibinfo {author} {\bibfnamefont
  {P.}~\bibnamefont {{Frinchaboy}}}, \bibinfo {author} {\bibfnamefont
  {P.}~\bibnamefont {{Gaulme}}}, \bibinfo {author} {\bibfnamefont
  {A.}~\bibnamefont {{Georgakakis}}}, \bibinfo {author} {\bibfnamefont
  {P.}~\bibnamefont {{Green}}}, \bibinfo {author} {\bibfnamefont
  {H.}~\bibnamefont {{Guo}}}, \bibinfo {author} {\bibfnamefont
  {J.}~\bibnamefont {{Guy}}}, \bibinfo {author} {\bibfnamefont
  {S.}~\bibnamefont {{Ho}}}, \bibinfo {author} {\bibfnamefont {D.}~\bibnamefont
  {{Holder}}}, \bibinfo {author} {\bibfnamefont {J.}~\bibnamefont
  {{Huehnerhoff}}}, \bibinfo {author} {\bibfnamefont {T.}~\bibnamefont
  {{Hutchinson}}}, \bibinfo {author} {\bibfnamefont {Y.}~\bibnamefont
  {{Jing}}}, \bibinfo {author} {\bibfnamefont {E.}~\bibnamefont {{Jullo}}},
  \bibinfo {author} {\bibfnamefont {V.}~\bibnamefont {{Kamble}}}, \bibinfo
  {author} {\bibfnamefont {K.}~\bibnamefont {{Kinemuchi}}}, \bibinfo {author}
  {\bibfnamefont {D.}~\bibnamefont {{Kirkby}}}, \bibinfo {author}
  {\bibfnamefont {F.-S.}\ \bibnamefont {{Kitaura}}}, \bibinfo {author}
  {\bibfnamefont {M.~A.}\ \bibnamefont {{Klaene}}}, \bibinfo {author}
  {\bibfnamefont {R.~R.}\ \bibnamefont {{Laher}}}, \bibinfo {author}
  {\bibfnamefont {D.}~\bibnamefont {{Lang}}}, \bibinfo {author} {\bibfnamefont
  {P.}~\bibnamefont {{Laurent}}}, \bibinfo {author} {\bibfnamefont {J.-M.}\
  \bibnamefont {{Le Goff}}}, \bibinfo {author} {\bibfnamefont {C.}~\bibnamefont
  {{Li}}}, \bibinfo {author} {\bibfnamefont {Y.}~\bibnamefont {{Liang}}},
  \bibinfo {author} {\bibfnamefont {M.}~\bibnamefont {{Lima}}}, \bibinfo
  {author} {\bibfnamefont {Q.}~\bibnamefont {{Lin}}}, \bibinfo {author}
  {\bibfnamefont {W.}~\bibnamefont {{Lin}}}, \bibinfo {author} {\bibfnamefont
  {Y.-T.}\ \bibnamefont {{Lin}}}, \bibinfo {author} {\bibfnamefont {D.~C.}\
  \bibnamefont {{Long}}}, \bibinfo {author} {\bibfnamefont {B.}~\bibnamefont
  {{Lundgren}}}, \bibinfo {author} {\bibfnamefont {N.}~\bibnamefont
  {{MacDonald}}}, \bibinfo {author} {\bibfnamefont {M.~A.}\ \bibnamefont
  {{Geimba Maia}}}, \bibinfo {author} {\bibfnamefont {E.}~\bibnamefont
  {{Malanushenko}}}, \bibinfo {author} {\bibfnamefont {V.}~\bibnamefont
  {{Malanushenko}}}, \bibinfo {author} {\bibfnamefont {V.}~\bibnamefont
  {{Mariappan}}}, \bibinfo {author} {\bibfnamefont {C.~K.}\ \bibnamefont
  {{McBride}}}, \bibinfo {author} {\bibfnamefont {I.~D.}\ \bibnamefont
  {{McGreer}}}, \bibinfo {author} {\bibfnamefont {B.}~\bibnamefont
  {{M{\'e}nard}}}, \bibinfo {author} {\bibfnamefont {A.}~\bibnamefont
  {{Merloni}}}, \bibinfo {author} {\bibfnamefont {A.}~\bibnamefont {{Meza}}},
  \bibinfo {author} {\bibfnamefont {A.~D.}\ \bibnamefont {{Montero-Dorta}}},
  \bibinfo {author} {\bibfnamefont {D.}~\bibnamefont {{Muna}}}, \bibinfo
  {author} {\bibfnamefont {A.~D.}\ \bibnamefont {{Myers}}}, \bibinfo {author}
  {\bibfnamefont {K.}~\bibnamefont {{Nandra}}}, \bibinfo {author}
  {\bibfnamefont {T.}~\bibnamefont {{Naugle}}}, \bibinfo {author}
  {\bibfnamefont {J.~A.}\ \bibnamefont {{Newman}}}, \bibinfo {author}
  {\bibfnamefont {P.}~\bibnamefont {{Noterdaeme}}}, \bibinfo {author}
  {\bibfnamefont {P.}~\bibnamefont {{Nugent}}}, \bibinfo {author}
  {\bibfnamefont {R.}~\bibnamefont {{Ogando}}}, \bibinfo {author}
  {\bibfnamefont {M.~D.}\ \bibnamefont {{Olmstead}}}, \bibinfo {author}
  {\bibfnamefont {A.}~\bibnamefont {{Oravetz}}}, \bibinfo {author}
  {\bibfnamefont {D.~J.}\ \bibnamefont {{Oravetz}}}, \bibinfo {author}
  {\bibfnamefont {N.}~\bibnamefont {{Padmanabhan}}}, \bibinfo {author}
  {\bibfnamefont {N.}~\bibnamefont {{Palanque-Delabrouille}}}, \bibinfo
  {author} {\bibfnamefont {K.}~\bibnamefont {{Pan}}}, \bibinfo {author}
  {\bibfnamefont {J.~K.}\ \bibnamefont {{Parejko}}}, \bibinfo {author}
  {\bibfnamefont {I.}~\bibnamefont {{P{\^a}ris}}}, \bibinfo {author}
  {\bibfnamefont {J.~A.}\ \bibnamefont {{Peacock}}}, \bibinfo {author}
  {\bibfnamefont {P.}~\bibnamefont {{Petitjean}}}, \bibinfo {author}
  {\bibfnamefont {M.~M.}\ \bibnamefont {{Pieri}}}, \bibinfo {author}
  {\bibfnamefont {A.}~\bibnamefont {{Pisani}}}, \bibinfo {author}
  {\bibfnamefont {F.}~\bibnamefont {{Prada}}}, \bibinfo {author} {\bibfnamefont
  {A.}~\bibnamefont {{Prakash}}}, \bibinfo {author} {\bibfnamefont
  {A.}~\bibnamefont {{Raichoor}}}, \bibinfo {author} {\bibfnamefont
  {B.}~\bibnamefont {{Reid}}}, \bibinfo {author} {\bibfnamefont
  {J.}~\bibnamefont {{Rich}}}, \bibinfo {author} {\bibfnamefont
  {J.}~\bibnamefont {{Ridl}}}, \bibinfo {author} {\bibfnamefont
  {S.}~\bibnamefont {{Rodriguez-Torres}}}, \bibinfo {author} {\bibfnamefont
  {A.}~\bibnamefont {{Carnero Rosell}}}, \bibinfo {author} {\bibfnamefont
  {A.~J.}\ \bibnamefont {{Ross}}}, \bibinfo {author} {\bibfnamefont
  {G.}~\bibnamefont {{Rossi}}}, \bibinfo {author} {\bibfnamefont
  {J.}~\bibnamefont {{Ruan}}}, \bibinfo {author} {\bibfnamefont
  {M.}~\bibnamefont {{Salvato}}}, \bibinfo {author} {\bibfnamefont
  {C.}~\bibnamefont {{Sayres}}}, \bibinfo {author} {\bibfnamefont {D.~P.}\
  \bibnamefont {{Schneider}}}, \bibinfo {author} {\bibfnamefont {D.~J.}\
  \bibnamefont {{Schlegel}}}, \bibinfo {author} {\bibfnamefont
  {U.}~\bibnamefont {{Seljak}}}, \bibinfo {author} {\bibfnamefont {H.-J.}\
  \bibnamefont {{Seo}}}, \bibinfo {author} {\bibfnamefont {B.}~\bibnamefont
  {{Sesar}}}, \bibinfo {author} {\bibfnamefont {S.}~\bibnamefont {{Shandera}}},
  \bibinfo {author} {\bibfnamefont {Y.}~\bibnamefont {{Shu}}}, \bibinfo
  {author} {\bibfnamefont {A.}~\bibnamefont {{Slosar}}}, \bibinfo {author}
  {\bibfnamefont {F.}~\bibnamefont {{Sobreira}}}, \bibinfo {author}
  {\bibfnamefont {A.}~\bibnamefont {{Streblyanska}}}, \bibinfo {author}
  {\bibfnamefont {N.}~\bibnamefont {{Suzuki}}}, \bibinfo {author}
  {\bibfnamefont {D.}~\bibnamefont {{Taylor}}}, \bibinfo {author}
  {\bibfnamefont {C.}~\bibnamefont {{Tao}}}, \bibinfo {author} {\bibfnamefont
  {J.~L.}\ \bibnamefont {{Tinker}}}, \bibinfo {author} {\bibfnamefont
  {R.}~\bibnamefont {{Tojeiro}}}, \bibinfo {author} {\bibfnamefont
  {M.}~\bibnamefont {{Vargas-Maga{\~n}a}}}, \bibinfo {author} {\bibfnamefont
  {Y.}~\bibnamefont {{Wang}}}, \bibinfo {author} {\bibfnamefont {B.~A.}\
  \bibnamefont {{Weaver}}}, \bibinfo {author} {\bibfnamefont {D.~H.}\
  \bibnamefont {{Weinberg}}}, \bibinfo {author} {\bibfnamefont
  {M.}~\bibnamefont {{White}}}, \bibinfo {author} {\bibfnamefont {W.~M.}\
  \bibnamefont {{Wood-Vasey}}}, \bibinfo {author} {\bibfnamefont
  {C.}~\bibnamefont {{Yeche}}}, \bibinfo {author} {\bibfnamefont
  {Z.}~\bibnamefont {{Zhai}}}, \bibinfo {author} {\bibfnamefont
  {C.}~\bibnamefont {{Zhao}}}, \bibinfo {author} {\bibfnamefont {G.-b.}\
  \bibnamefont {{Zhao}}}, \bibinfo {author} {\bibfnamefont {Z.}~\bibnamefont
  {{Zheng}}}, \bibinfo {author} {\bibfnamefont {G.}~\bibnamefont {{Ben Zhu}}},
  \ and\ \bibinfo {author} {\bibfnamefont {H.}~\bibnamefont {{Zou}}},\ }\href
  {\doibase 10.3847/0004-6256/151/2/44} {\bibfield  {journal} {\bibinfo
  {journal} {\aj}\ }\textbf {\bibinfo {volume} {151}},\ \bibinfo {eid} {44}
  (\bibinfo {year} {2016})},\ \Eprint {http://arxiv.org/abs/1508.04473}
  {arXiv:1508.04473} \BibitemShut {NoStop}%
\bibitem [{\citenamefont {{Blanton}}\ \emph {et~al.}(2017)\citenamefont
  {{Blanton}}, \citenamefont {{Bershady}}, \citenamefont {{Abolfathi}},
  \citenamefont {{Albareti}}, \citenamefont {{Allende Prieto}}, \citenamefont
  {{Almeida}}, \citenamefont {{Alonso-Garc{\'{\i}}a}}, \citenamefont
  {{Anders}}, \citenamefont {{Anderson}}, \citenamefont {{Andrews}},\ and\
  \citenamefont {et~al.}}]{blanton17a}%
  \BibitemOpen
  \bibfield  {author} {\bibinfo {author} {\bibfnamefont {M.~R.}\ \bibnamefont
  {{Blanton}}}, \bibinfo {author} {\bibfnamefont {M.~A.}\ \bibnamefont
  {{Bershady}}}, \bibinfo {author} {\bibfnamefont {B.}~\bibnamefont
  {{Abolfathi}}}, \bibinfo {author} {\bibfnamefont {F.~D.}\ \bibnamefont
  {{Albareti}}}, \bibinfo {author} {\bibfnamefont {C.}~\bibnamefont {{Allende
  Prieto}}}, \bibinfo {author} {\bibfnamefont {A.}~\bibnamefont {{Almeida}}},
  \bibinfo {author} {\bibfnamefont {J.}~\bibnamefont {{Alonso-Garc{\'{\i}}a}}},
  \bibinfo {author} {\bibfnamefont {F.}~\bibnamefont {{Anders}}}, \bibinfo
  {author} {\bibfnamefont {S.~F.}\ \bibnamefont {{Anderson}}}, \bibinfo
  {author} {\bibfnamefont {B.}~\bibnamefont {{Andrews}}}, \ and\ \bibinfo
  {author} {\bibnamefont {et~al.}},\ }\href {\doibase 10.3847/1538-3881/aa7567}
  {\bibfield  {journal} {\bibinfo  {journal} {\aj}\ }\textbf {\bibinfo {volume}
  {154}},\ \bibinfo {eid} {28} (\bibinfo {year} {2017})},\ \Eprint
  {http://arxiv.org/abs/1703.00052} {arXiv:1703.00052} \BibitemShut {NoStop}%
\bibitem [{\citenamefont {{Smee}}\ \emph {et~al.}(2013)\citenamefont {{Smee}},
  \citenamefont {{Gunn}}, \citenamefont {{Uomoto}}, \citenamefont {{Roe}},
  \citenamefont {{Schlegel}}, \citenamefont {{Rockosi}}, \citenamefont
  {{Carr}}, \citenamefont {{Leger}}, \citenamefont {{Dawson}}, \citenamefont
  {{Olmstead}}, \citenamefont {{Brinkmann}}, \citenamefont {{Owen}},
  \citenamefont {{Barkhouser}}, \citenamefont {{Honscheid}}, \citenamefont
  {{Harding}}, \citenamefont {{Long}}, \citenamefont {{Lupton}}, \citenamefont
  {{Loomis}}, \citenamefont {{Anderson}}, \citenamefont {{Annis}},
  \citenamefont {{Bernardi}}, \citenamefont {{Bhardwaj}}, \citenamefont
  {{Bizyaev}}, \citenamefont {{Bolton}}, \citenamefont {{Brewington}},
  \citenamefont {{Briggs}}, \citenamefont {{Burles}}, \citenamefont {{Burns}},
  \citenamefont {{Castander}}, \citenamefont {{Connolly}}, \citenamefont
  {{Davenport}}, \citenamefont {{Ebelke}}, \citenamefont {{Epps}},
  \citenamefont {{Feldman}}, \citenamefont {{Friedman}}, \citenamefont
  {{Frieman}}, \citenamefont {{Heckman}}, \citenamefont {{Hull}}, \citenamefont
  {{Knapp}}, \citenamefont {{Lawrence}}, \citenamefont {{Loveday}},
  \citenamefont {{Mannery}}, \citenamefont {{Malanushenko}}, \citenamefont
  {{Malanushenko}}, \citenamefont {{Merrelli}}, \citenamefont {{Muna}},
  \citenamefont {{Newman}}, \citenamefont {{Nichol}}, \citenamefont
  {{Oravetz}}, \citenamefont {{Pan}}, \citenamefont {{Pope}}, \citenamefont
  {{Ricketts}}, \citenamefont {{Shelden}}, \citenamefont {{Sandford}},
  \citenamefont {{Siegmund}}, \citenamefont {{Simmons}}, \citenamefont
  {{Smith}}, \citenamefont {{Snedden}}, \citenamefont {{Schneider}},
  \citenamefont {{SubbaRao}}, \citenamefont {{Tremonti}}, \citenamefont
  {{Waddell}},\ and\ \citenamefont {{York}}}]{smee13a}%
  \BibitemOpen
  \bibfield  {author} {\bibinfo {author} {\bibfnamefont {S.~A.}\ \bibnamefont
  {{Smee}}}, \bibinfo {author} {\bibfnamefont {J.~E.}\ \bibnamefont {{Gunn}}},
  \bibinfo {author} {\bibfnamefont {A.}~\bibnamefont {{Uomoto}}}, \bibinfo
  {author} {\bibfnamefont {N.}~\bibnamefont {{Roe}}}, \bibinfo {author}
  {\bibfnamefont {D.}~\bibnamefont {{Schlegel}}}, \bibinfo {author}
  {\bibfnamefont {C.~M.}\ \bibnamefont {{Rockosi}}}, \bibinfo {author}
  {\bibfnamefont {M.~A.}\ \bibnamefont {{Carr}}}, \bibinfo {author}
  {\bibfnamefont {F.}~\bibnamefont {{Leger}}}, \bibinfo {author} {\bibfnamefont
  {K.~S.}\ \bibnamefont {{Dawson}}}, \bibinfo {author} {\bibfnamefont {M.~D.}\
  \bibnamefont {{Olmstead}}}, \bibinfo {author} {\bibfnamefont
  {J.}~\bibnamefont {{Brinkmann}}}, \bibinfo {author} {\bibfnamefont
  {R.}~\bibnamefont {{Owen}}}, \bibinfo {author} {\bibfnamefont {R.~H.}\
  \bibnamefont {{Barkhouser}}}, \bibinfo {author} {\bibfnamefont
  {K.}~\bibnamefont {{Honscheid}}}, \bibinfo {author} {\bibfnamefont
  {P.}~\bibnamefont {{Harding}}}, \bibinfo {author} {\bibfnamefont
  {D.}~\bibnamefont {{Long}}}, \bibinfo {author} {\bibfnamefont {R.~H.}\
  \bibnamefont {{Lupton}}}, \bibinfo {author} {\bibfnamefont {C.}~\bibnamefont
  {{Loomis}}}, \bibinfo {author} {\bibfnamefont {L.}~\bibnamefont
  {{Anderson}}}, \bibinfo {author} {\bibfnamefont {J.}~\bibnamefont {{Annis}}},
  \bibinfo {author} {\bibfnamefont {M.}~\bibnamefont {{Bernardi}}}, \bibinfo
  {author} {\bibfnamefont {V.}~\bibnamefont {{Bhardwaj}}}, \bibinfo {author}
  {\bibfnamefont {D.}~\bibnamefont {{Bizyaev}}}, \bibinfo {author}
  {\bibfnamefont {A.~S.}\ \bibnamefont {{Bolton}}}, \bibinfo {author}
  {\bibfnamefont {H.}~\bibnamefont {{Brewington}}}, \bibinfo {author}
  {\bibfnamefont {J.~W.}\ \bibnamefont {{Briggs}}}, \bibinfo {author}
  {\bibfnamefont {S.}~\bibnamefont {{Burles}}}, \bibinfo {author}
  {\bibfnamefont {J.~G.}\ \bibnamefont {{Burns}}}, \bibinfo {author}
  {\bibfnamefont {F.~J.}\ \bibnamefont {{Castander}}}, \bibinfo {author}
  {\bibfnamefont {A.}~\bibnamefont {{Connolly}}}, \bibinfo {author}
  {\bibfnamefont {J.~R.~A.}\ \bibnamefont {{Davenport}}}, \bibinfo {author}
  {\bibfnamefont {G.}~\bibnamefont {{Ebelke}}}, \bibinfo {author}
  {\bibfnamefont {H.}~\bibnamefont {{Epps}}}, \bibinfo {author} {\bibfnamefont
  {P.~D.}\ \bibnamefont {{Feldman}}}, \bibinfo {author} {\bibfnamefont {S.~D.}\
  \bibnamefont {{Friedman}}}, \bibinfo {author} {\bibfnamefont
  {J.}~\bibnamefont {{Frieman}}}, \bibinfo {author} {\bibfnamefont
  {T.}~\bibnamefont {{Heckman}}}, \bibinfo {author} {\bibfnamefont {C.~L.}\
  \bibnamefont {{Hull}}}, \bibinfo {author} {\bibfnamefont {G.~R.}\
  \bibnamefont {{Knapp}}}, \bibinfo {author} {\bibfnamefont {D.~M.}\
  \bibnamefont {{Lawrence}}}, \bibinfo {author} {\bibfnamefont
  {J.}~\bibnamefont {{Loveday}}}, \bibinfo {author} {\bibfnamefont {E.~J.}\
  \bibnamefont {{Mannery}}}, \bibinfo {author} {\bibfnamefont {E.}~\bibnamefont
  {{Malanushenko}}}, \bibinfo {author} {\bibfnamefont {V.}~\bibnamefont
  {{Malanushenko}}}, \bibinfo {author} {\bibfnamefont {A.~J.}\ \bibnamefont
  {{Merrelli}}}, \bibinfo {author} {\bibfnamefont {D.}~\bibnamefont {{Muna}}},
  \bibinfo {author} {\bibfnamefont {P.~R.}\ \bibnamefont {{Newman}}}, \bibinfo
  {author} {\bibfnamefont {R.~C.}\ \bibnamefont {{Nichol}}}, \bibinfo {author}
  {\bibfnamefont {D.}~\bibnamefont {{Oravetz}}}, \bibinfo {author}
  {\bibfnamefont {K.}~\bibnamefont {{Pan}}}, \bibinfo {author} {\bibfnamefont
  {A.~C.}\ \bibnamefont {{Pope}}}, \bibinfo {author} {\bibfnamefont {P.~G.}\
  \bibnamefont {{Ricketts}}}, \bibinfo {author} {\bibfnamefont
  {A.}~\bibnamefont {{Shelden}}}, \bibinfo {author} {\bibfnamefont
  {D.}~\bibnamefont {{Sandford}}}, \bibinfo {author} {\bibfnamefont
  {W.}~\bibnamefont {{Siegmund}}}, \bibinfo {author} {\bibfnamefont
  {A.}~\bibnamefont {{Simmons}}}, \bibinfo {author} {\bibfnamefont {D.~S.}\
  \bibnamefont {{Smith}}}, \bibinfo {author} {\bibfnamefont {S.}~\bibnamefont
  {{Snedden}}}, \bibinfo {author} {\bibfnamefont {D.~P.}\ \bibnamefont
  {{Schneider}}}, \bibinfo {author} {\bibfnamefont {M.}~\bibnamefont
  {{SubbaRao}}}, \bibinfo {author} {\bibfnamefont {C.}~\bibnamefont
  {{Tremonti}}}, \bibinfo {author} {\bibfnamefont {P.}~\bibnamefont
  {{Waddell}}}, \ and\ \bibinfo {author} {\bibfnamefont {D.~G.}\ \bibnamefont
  {{York}}},\ }\href {\doibase 10.1088/0004-6256/146/2/32} {\bibfield
  {journal} {\bibinfo  {journal} {\aj}\ }\textbf {\bibinfo {volume} {146}},\
  \bibinfo {eid} {32} (\bibinfo {year} {2013})},\ \Eprint
  {http://arxiv.org/abs/1208.2233} {arXiv:1208.2233 [astro-ph.IM]} \BibitemShut
  {NoStop}%
\bibitem [{\citenamefont {{Kollmeier}}\ \emph {et~al.}(2017)\citenamefont
  {{Kollmeier}}, \citenamefont {{Zasowski}}, \citenamefont {{Rix}},
  \citenamefont {{Johns}}, \citenamefont {{Anderson}}, \citenamefont {{Drory}},
  \citenamefont {{Johnson}}, \citenamefont {{Pogge}}, \citenamefont {{Bird}},
  \citenamefont {{Blanc}}, \citenamefont {{Brownstein}}, \citenamefont
  {{Crane}}, \citenamefont {{De Lee}}, \citenamefont {{Klaene}}, \citenamefont
  {{Kreckel}}, \citenamefont {{MacDonald}}, \citenamefont {{Merloni}},
  \citenamefont {{Ness}}, \citenamefont {{O'Brien}}, \citenamefont
  {{Sanchez-Gallego}}, \citenamefont {{Sayres}}, \citenamefont {{Shen}},
  \citenamefont {{Thakar}}, \citenamefont {{Tkachenko}}, \citenamefont
  {{Aerts}}, \citenamefont {{Blanton}}, \citenamefont {{Eisenstein}},
  \citenamefont {{Holtzman}}, \citenamefont {{Maoz}}, \citenamefont {{Nandra}},
  \citenamefont {{Rockosi}}, \citenamefont {{Weinberg}}, \citenamefont
  {{Bovy}}, \citenamefont {{Casey}}, \citenamefont {{Chaname}}, \citenamefont
  {{Clerc}}, \citenamefont {{Conroy}}, \citenamefont {{Eracleous}},
  \citenamefont {{G{\"a}nsicke}}, \citenamefont {{Hekker}}, \citenamefont
  {{Horne}}, \citenamefont {{Kauffmann}}, \citenamefont {{McQuinn}},
  \citenamefont {{Pellegrini}}, \citenamefont {{Schinnerer}}, \citenamefont
  {{Schlafly}}, \citenamefont {{Schwope}}, \citenamefont {{Seibert}},
  \citenamefont {{Teske}},\ and\ \citenamefont {{van Saders}}}]{kollmeier17a}%
  \BibitemOpen
  \bibfield  {author} {\bibinfo {author} {\bibfnamefont {J.~A.}\ \bibnamefont
  {{Kollmeier}}}, \bibinfo {author} {\bibfnamefont {G.}~\bibnamefont
  {{Zasowski}}}, \bibinfo {author} {\bibfnamefont {H.-W.}\ \bibnamefont
  {{Rix}}}, \bibinfo {author} {\bibfnamefont {M.}~\bibnamefont {{Johns}}},
  \bibinfo {author} {\bibfnamefont {S.~F.}\ \bibnamefont {{Anderson}}},
  \bibinfo {author} {\bibfnamefont {N.}~\bibnamefont {{Drory}}}, \bibinfo
  {author} {\bibfnamefont {J.~A.}\ \bibnamefont {{Johnson}}}, \bibinfo {author}
  {\bibfnamefont {R.~W.}\ \bibnamefont {{Pogge}}}, \bibinfo {author}
  {\bibfnamefont {J.~C.}\ \bibnamefont {{Bird}}}, \bibinfo {author}
  {\bibfnamefont {G.~A.}\ \bibnamefont {{Blanc}}}, \bibinfo {author}
  {\bibfnamefont {J.~R.}\ \bibnamefont {{Brownstein}}}, \bibinfo {author}
  {\bibfnamefont {J.~D.}\ \bibnamefont {{Crane}}}, \bibinfo {author}
  {\bibfnamefont {N.~M.}\ \bibnamefont {{De Lee}}}, \bibinfo {author}
  {\bibfnamefont {M.~A.}\ \bibnamefont {{Klaene}}}, \bibinfo {author}
  {\bibfnamefont {K.}~\bibnamefont {{Kreckel}}}, \bibinfo {author}
  {\bibfnamefont {N.}~\bibnamefont {{MacDonald}}}, \bibinfo {author}
  {\bibfnamefont {A.}~\bibnamefont {{Merloni}}}, \bibinfo {author}
  {\bibfnamefont {M.~K.}\ \bibnamefont {{Ness}}}, \bibinfo {author}
  {\bibfnamefont {T.}~\bibnamefont {{O'Brien}}}, \bibinfo {author}
  {\bibfnamefont {J.~R.}\ \bibnamefont {{Sanchez-Gallego}}}, \bibinfo {author}
  {\bibfnamefont {C.~C.}\ \bibnamefont {{Sayres}}}, \bibinfo {author}
  {\bibfnamefont {Y.}~\bibnamefont {{Shen}}}, \bibinfo {author} {\bibfnamefont
  {A.~R.}\ \bibnamefont {{Thakar}}}, \bibinfo {author} {\bibfnamefont
  {A.}~\bibnamefont {{Tkachenko}}}, \bibinfo {author} {\bibfnamefont
  {C.}~\bibnamefont {{Aerts}}}, \bibinfo {author} {\bibfnamefont {M.~R.}\
  \bibnamefont {{Blanton}}}, \bibinfo {author} {\bibfnamefont {D.~J.}\
  \bibnamefont {{Eisenstein}}}, \bibinfo {author} {\bibfnamefont {J.~A.}\
  \bibnamefont {{Holtzman}}}, \bibinfo {author} {\bibfnamefont
  {D.}~\bibnamefont {{Maoz}}}, \bibinfo {author} {\bibfnamefont
  {K.}~\bibnamefont {{Nandra}}}, \bibinfo {author} {\bibfnamefont
  {C.}~\bibnamefont {{Rockosi}}}, \bibinfo {author} {\bibfnamefont {D.~H.}\
  \bibnamefont {{Weinberg}}}, \bibinfo {author} {\bibfnamefont
  {J.}~\bibnamefont {{Bovy}}}, \bibinfo {author} {\bibfnamefont {A.~R.}\
  \bibnamefont {{Casey}}}, \bibinfo {author} {\bibfnamefont {J.}~\bibnamefont
  {{Chaname}}}, \bibinfo {author} {\bibfnamefont {N.}~\bibnamefont {{Clerc}}},
  \bibinfo {author} {\bibfnamefont {C.}~\bibnamefont {{Conroy}}}, \bibinfo
  {author} {\bibfnamefont {M.}~\bibnamefont {{Eracleous}}}, \bibinfo {author}
  {\bibfnamefont {B.~T.}\ \bibnamefont {{G{\"a}nsicke}}}, \bibinfo {author}
  {\bibfnamefont {S.}~\bibnamefont {{Hekker}}}, \bibinfo {author}
  {\bibfnamefont {K.}~\bibnamefont {{Horne}}}, \bibinfo {author} {\bibfnamefont
  {J.}~\bibnamefont {{Kauffmann}}}, \bibinfo {author} {\bibfnamefont
  {K.~B.~W.}\ \bibnamefont {{McQuinn}}}, \bibinfo {author} {\bibfnamefont
  {E.~W.}\ \bibnamefont {{Pellegrini}}}, \bibinfo {author} {\bibfnamefont
  {E.}~\bibnamefont {{Schinnerer}}}, \bibinfo {author} {\bibfnamefont {E.~F.}\
  \bibnamefont {{Schlafly}}}, \bibinfo {author} {\bibfnamefont {A.~D.}\
  \bibnamefont {{Schwope}}}, \bibinfo {author} {\bibfnamefont {M.}~\bibnamefont
  {{Seibert}}}, \bibinfo {author} {\bibfnamefont {J.~K.}\ \bibnamefont
  {{Teske}}}, \ and\ \bibinfo {author} {\bibfnamefont {J.~L.}\ \bibnamefont
  {{van Saders}}},\ }\href@noop {} {\bibfield  {journal} {\bibinfo  {journal}
  {arXiv e-prints}\ ,\ \bibinfo {eid} {arXiv:1711.03234}} (\bibinfo {year}
  {2017})},\ \Eprint {http://arxiv.org/abs/1711.03234} {arXiv:1711.03234
  [astro-ph.GA]} \BibitemShut {NoStop}%
\bibitem [{\citenamefont {{Bautista}}\ \emph {et~al.}(2021)\citenamefont
  {{Bautista}}, \citenamefont {{Paviot}}, \citenamefont {{Vargas Maga{\~n}a}},
  \citenamefont {{de la Torre}}, \citenamefont {{Fromenteau}}, \citenamefont
  {{Gil-Mar{\'\i}n}}, \citenamefont {{Ross}}, \citenamefont {{Burtin}},
  \citenamefont {{Dawson}}, \citenamefont {{Hou}}, \citenamefont {{Kneib}},
  \citenamefont {{de Mattia}}, \citenamefont {{Percival}}, \citenamefont
  {{Rossi}}, \citenamefont {{Tojeiro}}, \citenamefont {{Zhao}}, \citenamefont
  {{Zhao}}, \citenamefont {{Alam}}, \citenamefont {{Brownstein}}, \citenamefont
  {{Chapman}}, \citenamefont {{Choi}}, \citenamefont {{Chuang}}, \citenamefont
  {{Escoffier}}, \citenamefont {{de la Macorra}}, \citenamefont {{du Mas des
  Bourboux}}, \citenamefont {{Mohammad}}, \citenamefont {{Moon}}, \citenamefont
  {{M{\"u}ller}}, \citenamefont {{Nadathur}}, \citenamefont {{Newman}},
  \citenamefont {{Schneider}}, \citenamefont {{Seo}},\ and\ \citenamefont
  {{Wang}}}]{LRG_corr}%
  \BibitemOpen
  \bibfield  {author} {\bibinfo {author} {\bibfnamefont {J.~E.}\ \bibnamefont
  {{Bautista}}}, \bibinfo {author} {\bibfnamefont {R.}~\bibnamefont
  {{Paviot}}}, \bibinfo {author} {\bibfnamefont {M.}~\bibnamefont {{Vargas
  Maga{\~n}a}}}, \bibinfo {author} {\bibfnamefont {S.}~\bibnamefont {{de la
  Torre}}}, \bibinfo {author} {\bibfnamefont {S.}~\bibnamefont {{Fromenteau}}},
  \bibinfo {author} {\bibfnamefont {H.}~\bibnamefont {{Gil-Mar{\'\i}n}}},
  \bibinfo {author} {\bibfnamefont {A.~J.}\ \bibnamefont {{Ross}}}, \bibinfo
  {author} {\bibfnamefont {E.}~\bibnamefont {{Burtin}}}, \bibinfo {author}
  {\bibfnamefont {K.~S.}\ \bibnamefont {{Dawson}}}, \bibinfo {author}
  {\bibfnamefont {J.}~\bibnamefont {{Hou}}}, \bibinfo {author} {\bibfnamefont
  {J.-P.}\ \bibnamefont {{Kneib}}}, \bibinfo {author} {\bibfnamefont
  {A.}~\bibnamefont {{de Mattia}}}, \bibinfo {author} {\bibfnamefont {W.~J.}\
  \bibnamefont {{Percival}}}, \bibinfo {author} {\bibfnamefont
  {G.}~\bibnamefont {{Rossi}}}, \bibinfo {author} {\bibfnamefont
  {R.}~\bibnamefont {{Tojeiro}}}, \bibinfo {author} {\bibfnamefont
  {C.}~\bibnamefont {{Zhao}}}, \bibinfo {author} {\bibfnamefont {G.-B.}\
  \bibnamefont {{Zhao}}}, \bibinfo {author} {\bibfnamefont {S.}~\bibnamefont
  {{Alam}}}, \bibinfo {author} {\bibfnamefont {J.}~\bibnamefont
  {{Brownstein}}}, \bibinfo {author} {\bibfnamefont {M.~J.}\ \bibnamefont
  {{Chapman}}}, \bibinfo {author} {\bibfnamefont {P.~D.}\ \bibnamefont
  {{Choi}}}, \bibinfo {author} {\bibfnamefont {C.-H.}\ \bibnamefont
  {{Chuang}}}, \bibinfo {author} {\bibfnamefont {S.}~\bibnamefont
  {{Escoffier}}}, \bibinfo {author} {\bibfnamefont {A.}~\bibnamefont {{de la
  Macorra}}}, \bibinfo {author} {\bibfnamefont {H.}~\bibnamefont {{du Mas des
  Bourboux}}}, \bibinfo {author} {\bibfnamefont {F.~G.}\ \bibnamefont
  {{Mohammad}}}, \bibinfo {author} {\bibfnamefont {J.}~\bibnamefont {{Moon}}},
  \bibinfo {author} {\bibfnamefont {E.-M.}\ \bibnamefont {{M{\"u}ller}}},
  \bibinfo {author} {\bibfnamefont {S.}~\bibnamefont {{Nadathur}}}, \bibinfo
  {author} {\bibfnamefont {J.~A.}\ \bibnamefont {{Newman}}}, \bibinfo {author}
  {\bibfnamefont {D.}~\bibnamefont {{Schneider}}}, \bibinfo {author}
  {\bibfnamefont {H.-J.}\ \bibnamefont {{Seo}}}, \ and\ \bibinfo {author}
  {\bibfnamefont {Y.}~\bibnamefont {{Wang}}},\ }\href {\doibase
  10.1093/mnras/staa2800} {\bibfield  {journal} {\bibinfo  {journal} {\mnras}\
  }\textbf {\bibinfo {volume} {500}},\ \bibinfo {pages} {736} (\bibinfo {year}
  {2021})},\ \Eprint {http://arxiv.org/abs/2007.08993} {arXiv:2007.08993
  [astro-ph.CO]} \BibitemShut {NoStop}%
\bibitem [{\citenamefont {{Gil-Mar{\'\i}n}}\ \emph {et~al.}(2020)\citenamefont
  {{Gil-Mar{\'\i}n}}, \citenamefont {{Bautista}}, \citenamefont {{Paviot}},
  \citenamefont {{Vargas-Maga{\~n}a}}, \citenamefont {{de la Torre}},
  \citenamefont {{Fromenteau}}, \citenamefont {{Alam}}, \citenamefont
  {{{\'A}vila}}, \citenamefont {{Burtin}}, \citenamefont {{Chuang}},
  \citenamefont {{Dawson}}, \citenamefont {{Hou}}, \citenamefont {{de Mattia}},
  \citenamefont {{Mohammad}}, \citenamefont {{M{\"u}ller}}, \citenamefont
  {{Nadathur}}, \citenamefont {{Neveux}}, \citenamefont {{Percival}},
  \citenamefont {{Raichoor}}, \citenamefont {{Rezaie}}, \citenamefont {{Ross}},
  \citenamefont {{Rossi}}, \citenamefont {{Ruhlmann-Kleider}}, \citenamefont
  {{Smith}}, \citenamefont {{Tamone}}, \citenamefont {{Tinker}}, \citenamefont
  {{Tojeiro}}, \citenamefont {{Wang}}, \citenamefont {{Zhao}}, \citenamefont
  {{Zhao}}, \citenamefont {{Brinkmann}}, \citenamefont {{Brownstein}},
  \citenamefont {{Choi}}, \citenamefont {{Escoffier}}, \citenamefont {{de la
  Macorra}}, \citenamefont {{Moon}}, \citenamefont {{Newman}}, \citenamefont
  {{Schneider}}, \citenamefont {{Seo}},\ and\ \citenamefont
  {{Vivek}}}]{gil-marin19a}%
  \BibitemOpen
  \bibfield  {author} {\bibinfo {author} {\bibfnamefont {H.}~\bibnamefont
  {{Gil-Mar{\'\i}n}}}, \bibinfo {author} {\bibfnamefont {J.~E.}\ \bibnamefont
  {{Bautista}}}, \bibinfo {author} {\bibfnamefont {R.}~\bibnamefont
  {{Paviot}}}, \bibinfo {author} {\bibfnamefont {M.}~\bibnamefont
  {{Vargas-Maga{\~n}a}}}, \bibinfo {author} {\bibfnamefont {S.}~\bibnamefont
  {{de la Torre}}}, \bibinfo {author} {\bibfnamefont {S.}~\bibnamefont
  {{Fromenteau}}}, \bibinfo {author} {\bibfnamefont {S.}~\bibnamefont
  {{Alam}}}, \bibinfo {author} {\bibfnamefont {S.}~\bibnamefont {{{\'A}vila}}},
  \bibinfo {author} {\bibfnamefont {E.}~\bibnamefont {{Burtin}}}, \bibinfo
  {author} {\bibfnamefont {C.-H.}\ \bibnamefont {{Chuang}}}, \bibinfo {author}
  {\bibfnamefont {K.~S.}\ \bibnamefont {{Dawson}}}, \bibinfo {author}
  {\bibfnamefont {J.}~\bibnamefont {{Hou}}}, \bibinfo {author} {\bibfnamefont
  {A.}~\bibnamefont {{de Mattia}}}, \bibinfo {author} {\bibfnamefont {F.~G.}\
  \bibnamefont {{Mohammad}}}, \bibinfo {author} {\bibfnamefont {E.-M.}\
  \bibnamefont {{M{\"u}ller}}}, \bibinfo {author} {\bibfnamefont
  {S.}~\bibnamefont {{Nadathur}}}, \bibinfo {author} {\bibfnamefont
  {R.}~\bibnamefont {{Neveux}}}, \bibinfo {author} {\bibfnamefont {W.~J.}\
  \bibnamefont {{Percival}}}, \bibinfo {author} {\bibfnamefont
  {A.}~\bibnamefont {{Raichoor}}}, \bibinfo {author} {\bibfnamefont
  {M.}~\bibnamefont {{Rezaie}}}, \bibinfo {author} {\bibfnamefont {A.~J.}\
  \bibnamefont {{Ross}}}, \bibinfo {author} {\bibfnamefont {G.}~\bibnamefont
  {{Rossi}}}, \bibinfo {author} {\bibfnamefont {V.}~\bibnamefont
  {{Ruhlmann-Kleider}}}, \bibinfo {author} {\bibfnamefont {A.}~\bibnamefont
  {{Smith}}}, \bibinfo {author} {\bibfnamefont {A.}~\bibnamefont {{Tamone}}},
  \bibinfo {author} {\bibfnamefont {J.~L.}\ \bibnamefont {{Tinker}}}, \bibinfo
  {author} {\bibfnamefont {R.}~\bibnamefont {{Tojeiro}}}, \bibinfo {author}
  {\bibfnamefont {Y.}~\bibnamefont {{Wang}}}, \bibinfo {author} {\bibfnamefont
  {G.-B.}\ \bibnamefont {{Zhao}}}, \bibinfo {author} {\bibfnamefont
  {C.}~\bibnamefont {{Zhao}}}, \bibinfo {author} {\bibfnamefont
  {J.}~\bibnamefont {{Brinkmann}}}, \bibinfo {author} {\bibfnamefont {J.~R.}\
  \bibnamefont {{Brownstein}}}, \bibinfo {author} {\bibfnamefont {P.~D.}\
  \bibnamefont {{Choi}}}, \bibinfo {author} {\bibfnamefont {S.}~\bibnamefont
  {{Escoffier}}}, \bibinfo {author} {\bibfnamefont {A.}~\bibnamefont {{de la
  Macorra}}}, \bibinfo {author} {\bibfnamefont {J.}~\bibnamefont {{Moon}}},
  \bibinfo {author} {\bibfnamefont {J.~A.}\ \bibnamefont {{Newman}}}, \bibinfo
  {author} {\bibfnamefont {D.~P.}\ \bibnamefont {{Schneider}}}, \bibinfo
  {author} {\bibfnamefont {H.-J.}\ \bibnamefont {{Seo}}}, \ and\ \bibinfo
  {author} {\bibfnamefont {M.}~\bibnamefont {{Vivek}}},\ }\href {\doibase
  10.1093/mnras/staa2455} {\bibfield  {journal} {\bibinfo  {journal} {\mnras}\
  }\textbf {\bibinfo {volume} {498}},\ \bibinfo {pages} {2492} (\bibinfo {year}
  {2020})},\ \Eprint {http://arxiv.org/abs/2007.08994} {arXiv:2007.08994
  [astro-ph.CO]} \BibitemShut {NoStop}%
\bibitem [{\citenamefont {{Raichoor}}\ \emph {et~al.}(2021)\citenamefont
  {{Raichoor}}, \citenamefont {{de Mattia}}, \citenamefont {{Ross}},
  \citenamefont {{Zhao}}, \citenamefont {{Alam}}, \citenamefont {{Avila}},
  \citenamefont {{Bautista}}, \citenamefont {{Brinkmann}}, \citenamefont
  {{Brownstein}}, \citenamefont {{Burtin}}, \citenamefont {{Chapman}},
  \citenamefont {{Chuang}}, \citenamefont {{Comparat}}, \citenamefont
  {{Dawson}}, \citenamefont {{Dey}}, \citenamefont {{du Mas des Bourboux}},
  \citenamefont {{Elvin-Poole}}, \citenamefont {{Gonzalez-Perez}},
  \citenamefont {{Gorgoni}}, \citenamefont {{Kneib}}, \citenamefont {{Kong}},
  \citenamefont {{Lang}}, \citenamefont {{Moustakas}}, \citenamefont {{Myers}},
  \citenamefont {{M{\"u}ller}}, \citenamefont {{Nadathur}}, \citenamefont
  {{Newman}}, \citenamefont {{Percival}}, \citenamefont {{Rezaie}},
  \citenamefont {{Rossi}}, \citenamefont {{Ruhlmann-Kleider}}, \citenamefont
  {{Schlegel}}, \citenamefont {{Schneider}}, \citenamefont {{Seo}},
  \citenamefont {{Tamone}}, \citenamefont {{Tinker}}, \citenamefont
  {{Tojeiro}}, \citenamefont {{Vivek}}, \citenamefont {{Y{\`e}che}},\ and\
  \citenamefont {{Zhao}}}]{raichoor19a}%
  \BibitemOpen
  \bibfield  {author} {\bibinfo {author} {\bibfnamefont {A.}~\bibnamefont
  {{Raichoor}}}, \bibinfo {author} {\bibfnamefont {A.}~\bibnamefont {{de
  Mattia}}}, \bibinfo {author} {\bibfnamefont {A.~J.}\ \bibnamefont {{Ross}}},
  \bibinfo {author} {\bibfnamefont {C.}~\bibnamefont {{Zhao}}}, \bibinfo
  {author} {\bibfnamefont {S.}~\bibnamefont {{Alam}}}, \bibinfo {author}
  {\bibfnamefont {S.}~\bibnamefont {{Avila}}}, \bibinfo {author} {\bibfnamefont
  {J.}~\bibnamefont {{Bautista}}}, \bibinfo {author} {\bibfnamefont
  {J.}~\bibnamefont {{Brinkmann}}}, \bibinfo {author} {\bibfnamefont {J.~R.}\
  \bibnamefont {{Brownstein}}}, \bibinfo {author} {\bibfnamefont
  {E.}~\bibnamefont {{Burtin}}}, \bibinfo {author} {\bibfnamefont {M.~J.}\
  \bibnamefont {{Chapman}}}, \bibinfo {author} {\bibfnamefont {C.-H.}\
  \bibnamefont {{Chuang}}}, \bibinfo {author} {\bibfnamefont {J.}~\bibnamefont
  {{Comparat}}}, \bibinfo {author} {\bibfnamefont {K.~S.}\ \bibnamefont
  {{Dawson}}}, \bibinfo {author} {\bibfnamefont {A.}~\bibnamefont {{Dey}}},
  \bibinfo {author} {\bibfnamefont {H.}~\bibnamefont {{du Mas des Bourboux}}},
  \bibinfo {author} {\bibfnamefont {J.}~\bibnamefont {{Elvin-Poole}}}, \bibinfo
  {author} {\bibfnamefont {V.}~\bibnamefont {{Gonzalez-Perez}}}, \bibinfo
  {author} {\bibfnamefont {C.}~\bibnamefont {{Gorgoni}}}, \bibinfo {author}
  {\bibfnamefont {J.-P.}\ \bibnamefont {{Kneib}}}, \bibinfo {author}
  {\bibfnamefont {H.}~\bibnamefont {{Kong}}}, \bibinfo {author} {\bibfnamefont
  {D.}~\bibnamefont {{Lang}}}, \bibinfo {author} {\bibfnamefont
  {J.}~\bibnamefont {{Moustakas}}}, \bibinfo {author} {\bibfnamefont {A.~D.}\
  \bibnamefont {{Myers}}}, \bibinfo {author} {\bibfnamefont {E.-M.}\
  \bibnamefont {{M{\"u}ller}}}, \bibinfo {author} {\bibfnamefont
  {S.}~\bibnamefont {{Nadathur}}}, \bibinfo {author} {\bibfnamefont {J.~A.}\
  \bibnamefont {{Newman}}}, \bibinfo {author} {\bibfnamefont {W.~J.}\
  \bibnamefont {{Percival}}}, \bibinfo {author} {\bibfnamefont
  {M.}~\bibnamefont {{Rezaie}}}, \bibinfo {author} {\bibfnamefont
  {G.}~\bibnamefont {{Rossi}}}, \bibinfo {author} {\bibfnamefont
  {V.}~\bibnamefont {{Ruhlmann-Kleider}}}, \bibinfo {author} {\bibfnamefont
  {D.~J.}\ \bibnamefont {{Schlegel}}}, \bibinfo {author} {\bibfnamefont
  {D.~P.}\ \bibnamefont {{Schneider}}}, \bibinfo {author} {\bibfnamefont
  {H.-J.}\ \bibnamefont {{Seo}}}, \bibinfo {author} {\bibfnamefont
  {A.}~\bibnamefont {{Tamone}}}, \bibinfo {author} {\bibfnamefont {J.~L.}\
  \bibnamefont {{Tinker}}}, \bibinfo {author} {\bibfnamefont {R.}~\bibnamefont
  {{Tojeiro}}}, \bibinfo {author} {\bibfnamefont {M.}~\bibnamefont {{Vivek}}},
  \bibinfo {author} {\bibfnamefont {C.}~\bibnamefont {{Y{\`e}che}}}, \ and\
  \bibinfo {author} {\bibfnamefont {G.-B.}\ \bibnamefont {{Zhao}}},\ }\href
  {\doibase 10.1093/mnras/staa3336} {\bibfield  {journal} {\bibinfo  {journal}
  {\mnras}\ }\textbf {\bibinfo {volume} {500}},\ \bibinfo {pages} {3254}
  (\bibinfo {year} {2021})},\ \Eprint {http://arxiv.org/abs/2007.09007}
  {arXiv:2007.09007 [astro-ph.CO]} \BibitemShut {NoStop}%
\bibitem [{\citenamefont {{Tamone}}\ \emph {et~al.}(2020)\citenamefont
  {{Tamone}}, \citenamefont {{Raichoor}}, \citenamefont {{Zhao}}, \citenamefont
  {{de Mattia}}, \citenamefont {{Gorgoni}}, \citenamefont {{Burtin}},
  \citenamefont {{Ruhlmann-Kleider}}, \citenamefont {{Ross}}, \citenamefont
  {{Alam}}, \citenamefont {{Percival}}, \citenamefont {{Avila}}, \citenamefont
  {{Chapman}}, \citenamefont {{Chuang}}, \citenamefont {{Comparat}},
  \citenamefont {{Dawson}}, \citenamefont {{de la Torre}}, \citenamefont {{du
  Mas des Bourboux}}, \citenamefont {{Escoffier}}, \citenamefont
  {{Gonzalez-Perez}}, \citenamefont {{Hou}}, \citenamefont {{Kneib}},
  \citenamefont {{Mohammad}}, \citenamefont {{Mueller}}, \citenamefont
  {{Paviot}}, \citenamefont {{Rossi}}, \citenamefont {{Schneider}},
  \citenamefont {{Wang}},\ and\ \citenamefont {{Zhao}}}]{tamone19a}%
  \BibitemOpen
  \bibfield  {author} {\bibinfo {author} {\bibfnamefont {A.}~\bibnamefont
  {{Tamone}}}, \bibinfo {author} {\bibfnamefont {A.}~\bibnamefont
  {{Raichoor}}}, \bibinfo {author} {\bibfnamefont {C.}~\bibnamefont {{Zhao}}},
  \bibinfo {author} {\bibfnamefont {A.}~\bibnamefont {{de Mattia}}}, \bibinfo
  {author} {\bibfnamefont {C.}~\bibnamefont {{Gorgoni}}}, \bibinfo {author}
  {\bibfnamefont {E.}~\bibnamefont {{Burtin}}}, \bibinfo {author}
  {\bibfnamefont {V.}~\bibnamefont {{Ruhlmann-Kleider}}}, \bibinfo {author}
  {\bibfnamefont {A.~J.}\ \bibnamefont {{Ross}}}, \bibinfo {author}
  {\bibfnamefont {S.}~\bibnamefont {{Alam}}}, \bibinfo {author} {\bibfnamefont
  {W.~J.}\ \bibnamefont {{Percival}}}, \bibinfo {author} {\bibfnamefont
  {S.}~\bibnamefont {{Avila}}}, \bibinfo {author} {\bibfnamefont {M.~J.}\
  \bibnamefont {{Chapman}}}, \bibinfo {author} {\bibfnamefont {C.-H.}\
  \bibnamefont {{Chuang}}}, \bibinfo {author} {\bibfnamefont {J.}~\bibnamefont
  {{Comparat}}}, \bibinfo {author} {\bibfnamefont {K.~S.}\ \bibnamefont
  {{Dawson}}}, \bibinfo {author} {\bibfnamefont {S.}~\bibnamefont {{de la
  Torre}}}, \bibinfo {author} {\bibfnamefont {H.}~\bibnamefont {{du Mas des
  Bourboux}}}, \bibinfo {author} {\bibfnamefont {S.}~\bibnamefont
  {{Escoffier}}}, \bibinfo {author} {\bibfnamefont {V.}~\bibnamefont
  {{Gonzalez-Perez}}}, \bibinfo {author} {\bibfnamefont {J.}~\bibnamefont
  {{Hou}}}, \bibinfo {author} {\bibfnamefont {J.-P.}\ \bibnamefont {{Kneib}}},
  \bibinfo {author} {\bibfnamefont {F.~G.}\ \bibnamefont {{Mohammad}}},
  \bibinfo {author} {\bibfnamefont {E.-M.}\ \bibnamefont {{Mueller}}}, \bibinfo
  {author} {\bibfnamefont {R.}~\bibnamefont {{Paviot}}}, \bibinfo {author}
  {\bibfnamefont {G.}~\bibnamefont {{Rossi}}}, \bibinfo {author} {\bibfnamefont
  {D.~P.}\ \bibnamefont {{Schneider}}}, \bibinfo {author} {\bibfnamefont
  {Y.}~\bibnamefont {{Wang}}}, \ and\ \bibinfo {author} {\bibfnamefont {G.-B.}\
  \bibnamefont {{Zhao}}},\ }\href {\doibase 10.1093/mnras/staa3050} {\bibfield
  {journal} {\bibinfo  {journal} {\mnras}\ }\textbf {\bibinfo {volume} {499}},\
  \bibinfo {pages} {5527} (\bibinfo {year} {2020})},\ \Eprint
  {http://arxiv.org/abs/2007.09009} {arXiv:2007.09009 [astro-ph.CO]}
  \BibitemShut {NoStop}%
\bibitem [{\citenamefont {{de Mattia}}\ \emph {et~al.}(2020)\citenamefont {{de
  Mattia}}, \citenamefont {{Ruhlmann-Kleider}}, \citenamefont {{Raichoor}},
  \citenamefont {{Ross}}, \citenamefont {{Tamone}}, \citenamefont {{Zhao}},
  \citenamefont {{Alam}}, \citenamefont {{Avila}}, \citenamefont {{Burtin}},
  \citenamefont {{Bautista}}, \citenamefont {{Beutler}}, \citenamefont
  {{Brinkmann}}, \citenamefont {{Brownstein}}, \citenamefont {{Chapman}},
  \citenamefont {{Chuang}}, \citenamefont {{Comparat}}, \citenamefont {{du Mas
  des Bourboux}}, \citenamefont {{Dawson}}, \citenamefont {{de la Macorra}},
  \citenamefont {{Gil-Mar{\'\i}n}}, \citenamefont {{Gonzalez-Perez}},
  \citenamefont {{Gorgoni}}, \citenamefont {{Hou}}, \citenamefont {{Kong}},
  \citenamefont {{Lin}}, \citenamefont {{Nadathur}}, \citenamefont {{Newman}},
  \citenamefont {{Mueller}}, \citenamefont {{Percival}}, \citenamefont
  {{Rezaie}}, \citenamefont {{Rossi}}, \citenamefont {{Schneider}},
  \citenamefont {{Tiwari}}, \citenamefont {{Vivek}}, \citenamefont {{Wang}},\
  and\ \citenamefont {{Zhao}}}]{demattia19a}%
  \BibitemOpen
  \bibfield  {author} {\bibinfo {author} {\bibfnamefont {A.}~\bibnamefont {{de
  Mattia}}}, \bibinfo {author} {\bibfnamefont {V.}~\bibnamefont
  {{Ruhlmann-Kleider}}}, \bibinfo {author} {\bibfnamefont {A.}~\bibnamefont
  {{Raichoor}}}, \bibinfo {author} {\bibfnamefont {A.~J.}\ \bibnamefont
  {{Ross}}}, \bibinfo {author} {\bibfnamefont {A.}~\bibnamefont {{Tamone}}},
  \bibinfo {author} {\bibfnamefont {C.}~\bibnamefont {{Zhao}}}, \bibinfo
  {author} {\bibfnamefont {S.}~\bibnamefont {{Alam}}}, \bibinfo {author}
  {\bibfnamefont {S.}~\bibnamefont {{Avila}}}, \bibinfo {author} {\bibfnamefont
  {E.}~\bibnamefont {{Burtin}}}, \bibinfo {author} {\bibfnamefont
  {J.}~\bibnamefont {{Bautista}}}, \bibinfo {author} {\bibfnamefont
  {F.}~\bibnamefont {{Beutler}}}, \bibinfo {author} {\bibfnamefont
  {J.}~\bibnamefont {{Brinkmann}}}, \bibinfo {author} {\bibfnamefont {J.~R.}\
  \bibnamefont {{Brownstein}}}, \bibinfo {author} {\bibfnamefont {M.~J.}\
  \bibnamefont {{Chapman}}}, \bibinfo {author} {\bibfnamefont {C.-H.}\
  \bibnamefont {{Chuang}}}, \bibinfo {author} {\bibfnamefont {J.}~\bibnamefont
  {{Comparat}}}, \bibinfo {author} {\bibfnamefont {H.}~\bibnamefont {{du Mas
  des Bourboux}}}, \bibinfo {author} {\bibfnamefont {K.~S.}\ \bibnamefont
  {{Dawson}}}, \bibinfo {author} {\bibfnamefont {A.}~\bibnamefont {{de la
  Macorra}}}, \bibinfo {author} {\bibfnamefont {H.}~\bibnamefont
  {{Gil-Mar{\'\i}n}}}, \bibinfo {author} {\bibfnamefont {V.}~\bibnamefont
  {{Gonzalez-Perez}}}, \bibinfo {author} {\bibfnamefont {C.}~\bibnamefont
  {{Gorgoni}}}, \bibinfo {author} {\bibfnamefont {J.}~\bibnamefont {{Hou}}},
  \bibinfo {author} {\bibfnamefont {H.}~\bibnamefont {{Kong}}}, \bibinfo
  {author} {\bibfnamefont {S.}~\bibnamefont {{Lin}}}, \bibinfo {author}
  {\bibfnamefont {S.}~\bibnamefont {{Nadathur}}}, \bibinfo {author}
  {\bibfnamefont {J.~A.}\ \bibnamefont {{Newman}}}, \bibinfo {author}
  {\bibfnamefont {E.-M.}\ \bibnamefont {{Mueller}}}, \bibinfo {author}
  {\bibfnamefont {W.~J.}\ \bibnamefont {{Percival}}}, \bibinfo {author}
  {\bibfnamefont {M.}~\bibnamefont {{Rezaie}}}, \bibinfo {author}
  {\bibfnamefont {G.}~\bibnamefont {{Rossi}}}, \bibinfo {author} {\bibfnamefont
  {D.~P.}\ \bibnamefont {{Schneider}}}, \bibinfo {author} {\bibfnamefont
  {P.}~\bibnamefont {{Tiwari}}}, \bibinfo {author} {\bibfnamefont
  {M.}~\bibnamefont {{Vivek}}}, \bibinfo {author} {\bibfnamefont
  {Y.}~\bibnamefont {{Wang}}}, \ and\ \bibinfo {author} {\bibfnamefont {G.-B.}\
  \bibnamefont {{Zhao}}},\ }\href {\doibase 10.1093/mnras/staa3891} {\bibfield
  {journal} {\bibinfo  {journal} {\mnras}\ } (\bibinfo {year} {2020}),\
  10.1093/mnras/staa3891},\ \Eprint {http://arxiv.org/abs/2007.09008}
  {arXiv:2007.09008 [astro-ph.CO]} \BibitemShut {NoStop}%
\bibitem [{\citenamefont {{Hou}}\ \emph {et~al.}(2021)\citenamefont {{Hou}},
  \citenamefont {{S{\'a}nchez}}, \citenamefont {{Ross}}, \citenamefont
  {{Smith}}, \citenamefont {{Neveux}}, \citenamefont {{Bautista}},
  \citenamefont {{Burtin}}, \citenamefont {{Zhao}}, \citenamefont
  {{Scoccimarro}}, \citenamefont {{Dawson}}, \citenamefont {{de Mattia}},
  \citenamefont {{de la Macorra}}, \citenamefont {{du Mas des Bourboux}},
  \citenamefont {{Eisenstein}}, \citenamefont {{Gil-Mar{\'\i}n}}, \citenamefont
  {{Lyke}}, \citenamefont {{Mohammad}}, \citenamefont {{Mueller}},
  \citenamefont {{Percival}}, \citenamefont {{Rossi}}, \citenamefont {{Vargas
  Maga{\~n}a}}, \citenamefont {{Zarrouk}}, \citenamefont {{Zhao}},
  \citenamefont {{Brinkmann}}, \citenamefont {{Brownstein}}, \citenamefont
  {{Chuang}}, \citenamefont {{Myers}}, \citenamefont {{Newman}}, \citenamefont
  {{Schneider}},\ and\ \citenamefont {{Vivek}}}]{hou19a}%
  \BibitemOpen
  \bibfield  {author} {\bibinfo {author} {\bibfnamefont {J.}~\bibnamefont
  {{Hou}}}, \bibinfo {author} {\bibfnamefont {A.~G.}\ \bibnamefont
  {{S{\'a}nchez}}}, \bibinfo {author} {\bibfnamefont {A.~J.}\ \bibnamefont
  {{Ross}}}, \bibinfo {author} {\bibfnamefont {A.}~\bibnamefont {{Smith}}},
  \bibinfo {author} {\bibfnamefont {R.}~\bibnamefont {{Neveux}}}, \bibinfo
  {author} {\bibfnamefont {J.}~\bibnamefont {{Bautista}}}, \bibinfo {author}
  {\bibfnamefont {E.}~\bibnamefont {{Burtin}}}, \bibinfo {author}
  {\bibfnamefont {C.}~\bibnamefont {{Zhao}}}, \bibinfo {author} {\bibfnamefont
  {R.}~\bibnamefont {{Scoccimarro}}}, \bibinfo {author} {\bibfnamefont {K.~S.}\
  \bibnamefont {{Dawson}}}, \bibinfo {author} {\bibfnamefont {A.}~\bibnamefont
  {{de Mattia}}}, \bibinfo {author} {\bibfnamefont {A.}~\bibnamefont {{de la
  Macorra}}}, \bibinfo {author} {\bibfnamefont {H.}~\bibnamefont {{du Mas des
  Bourboux}}}, \bibinfo {author} {\bibfnamefont {D.~J.}\ \bibnamefont
  {{Eisenstein}}}, \bibinfo {author} {\bibfnamefont {H.}~\bibnamefont
  {{Gil-Mar{\'\i}n}}}, \bibinfo {author} {\bibfnamefont {B.~W.}\ \bibnamefont
  {{Lyke}}}, \bibinfo {author} {\bibfnamefont {F.~G.}\ \bibnamefont
  {{Mohammad}}}, \bibinfo {author} {\bibfnamefont {E.-M.}\ \bibnamefont
  {{Mueller}}}, \bibinfo {author} {\bibfnamefont {W.~J.}\ \bibnamefont
  {{Percival}}}, \bibinfo {author} {\bibfnamefont {G.}~\bibnamefont {{Rossi}}},
  \bibinfo {author} {\bibfnamefont {M.}~\bibnamefont {{Vargas Maga{\~n}a}}},
  \bibinfo {author} {\bibfnamefont {P.}~\bibnamefont {{Zarrouk}}}, \bibinfo
  {author} {\bibfnamefont {G.-B.}\ \bibnamefont {{Zhao}}}, \bibinfo {author}
  {\bibfnamefont {J.}~\bibnamefont {{Brinkmann}}}, \bibinfo {author}
  {\bibfnamefont {J.~R.}\ \bibnamefont {{Brownstein}}}, \bibinfo {author}
  {\bibfnamefont {C.-H.}\ \bibnamefont {{Chuang}}}, \bibinfo {author}
  {\bibfnamefont {A.~D.}\ \bibnamefont {{Myers}}}, \bibinfo {author}
  {\bibfnamefont {J.~A.}\ \bibnamefont {{Newman}}}, \bibinfo {author}
  {\bibfnamefont {D.~P.}\ \bibnamefont {{Schneider}}}, \ and\ \bibinfo {author}
  {\bibfnamefont {M.}~\bibnamefont {{Vivek}}},\ }\href {\doibase
  10.1093/mnras/staa3234} {\bibfield  {journal} {\bibinfo  {journal} {\mnras}\
  }\textbf {\bibinfo {volume} {500}},\ \bibinfo {pages} {1201} (\bibinfo {year}
  {2021})},\ \Eprint {http://arxiv.org/abs/2007.08998} {arXiv:2007.08998
  [astro-ph.CO]} \BibitemShut {NoStop}%
\bibitem [{\citenamefont {{Neveux}}\ \emph {et~al.}(2020)\citenamefont
  {{Neveux}}, \citenamefont {{Burtin}}, \citenamefont {{de Mattia}},
  \citenamefont {{Smith}}, \citenamefont {{Ross}}, \citenamefont {{Hou}},
  \citenamefont {{Bautista}}, \citenamefont {{Brinkmann}}, \citenamefont
  {{Chuang}}, \citenamefont {{Dawson}}, \citenamefont {{Gil-Mar{\'\i}n}},
  \citenamefont {{Lyke}}, \citenamefont {{de la Macorra}}, \citenamefont {{du
  Mas des Bourboux}}, \citenamefont {{M{\"u}ller}}, \citenamefont {{Myers}},
  \citenamefont {{Newman}}, \citenamefont {{Percival}}, \citenamefont
  {{Rossi}}, \citenamefont {{Schneider}}, \citenamefont {{Vivek}},
  \citenamefont {{Zarrouk}}, \citenamefont {{Zhao}},\ and\ \citenamefont
  {{Zhao}}}]{neveux19a}%
  \BibitemOpen
  \bibfield  {author} {\bibinfo {author} {\bibfnamefont {R.}~\bibnamefont
  {{Neveux}}}, \bibinfo {author} {\bibfnamefont {E.}~\bibnamefont {{Burtin}}},
  \bibinfo {author} {\bibfnamefont {A.}~\bibnamefont {{de Mattia}}}, \bibinfo
  {author} {\bibfnamefont {A.}~\bibnamefont {{Smith}}}, \bibinfo {author}
  {\bibfnamefont {A.~J.}\ \bibnamefont {{Ross}}}, \bibinfo {author}
  {\bibfnamefont {J.}~\bibnamefont {{Hou}}}, \bibinfo {author} {\bibfnamefont
  {J.}~\bibnamefont {{Bautista}}}, \bibinfo {author} {\bibfnamefont
  {J.}~\bibnamefont {{Brinkmann}}}, \bibinfo {author} {\bibfnamefont {C.-H.}\
  \bibnamefont {{Chuang}}}, \bibinfo {author} {\bibfnamefont {K.~S.}\
  \bibnamefont {{Dawson}}}, \bibinfo {author} {\bibfnamefont {H.}~\bibnamefont
  {{Gil-Mar{\'\i}n}}}, \bibinfo {author} {\bibfnamefont {B.~W.}\ \bibnamefont
  {{Lyke}}}, \bibinfo {author} {\bibfnamefont {A.}~\bibnamefont {{de la
  Macorra}}}, \bibinfo {author} {\bibfnamefont {H.}~\bibnamefont {{du Mas des
  Bourboux}}}, \bibinfo {author} {\bibfnamefont {E.-M.}\ \bibnamefont
  {{M{\"u}ller}}}, \bibinfo {author} {\bibfnamefont {A.~D.}\ \bibnamefont
  {{Myers}}}, \bibinfo {author} {\bibfnamefont {J.~A.}\ \bibnamefont
  {{Newman}}}, \bibinfo {author} {\bibfnamefont {W.~J.}\ \bibnamefont
  {{Percival}}}, \bibinfo {author} {\bibfnamefont {G.}~\bibnamefont {{Rossi}}},
  \bibinfo {author} {\bibfnamefont {D.}~\bibnamefont {{Schneider}}}, \bibinfo
  {author} {\bibfnamefont {M.}~\bibnamefont {{Vivek}}}, \bibinfo {author}
  {\bibfnamefont {P.}~\bibnamefont {{Zarrouk}}}, \bibinfo {author}
  {\bibfnamefont {C.}~\bibnamefont {{Zhao}}}, \ and\ \bibinfo {author}
  {\bibfnamefont {G.-B.}\ \bibnamefont {{Zhao}}},\ }\href@noop {} {\bibfield
  {journal} {\bibinfo  {journal} {arXiv e-prints}\ ,\ \bibinfo {eid}
  {arXiv:2007.08999}} (\bibinfo {year} {2020})},\ \Eprint
  {http://arxiv.org/abs/2007.08999} {arXiv:2007.08999 [astro-ph.CO]}
  \BibitemShut {NoStop}%
\bibitem [{\citenamefont {{du Mas des Bourboux}}\ \emph
  {et~al.}(2020)\citenamefont {{du Mas des Bourboux}}, \citenamefont {{Rich}},
  \citenamefont {{Font-Ribera}}, \citenamefont {{de Sainte Agathe}},
  \citenamefont {{Farr}}, \citenamefont {{Etourneau}}, \citenamefont {{Le
  Goff}}, \citenamefont {{Cuceu}}, \citenamefont {{Balland}}, \citenamefont
  {{Bautista}}, \citenamefont {{Blomqvist}}, \citenamefont {{Brinkmann}},
  \citenamefont {{Brownstein}}, \citenamefont {{Chabanier}}, \citenamefont
  {{Chaussidon}}, \citenamefont {{Dawson}}, \citenamefont
  {{Gonz{\'a}lez-Morales}}, \citenamefont {{Guy}}, \citenamefont {{Lyke}},
  \citenamefont {{de la Macorra}}, \citenamefont {{Mueller}}, \citenamefont
  {{Myers}}, \citenamefont {{Nitschelm}}, \citenamefont {{Mu{\~n}oz
  Guti{\'e}rrez}}, \citenamefont {{Palanque-Delabrouille}}, \citenamefont
  {{Parker}}, \citenamefont {{Percival}}, \citenamefont
  {{P{\'e}rez-R{\`a}fols}}, \citenamefont {{Petitjean}}, \citenamefont
  {{Pieri}}, \citenamefont {{Ravoux}}, \citenamefont {{Rossi}}, \citenamefont
  {{Schneider}}, \citenamefont {{Seo}}, \citenamefont {{Slosar}}, \citenamefont
  {{Stermer}}, \citenamefont {{Vivek}}, \citenamefont {{Y{\`e}che}},\ and\
  \citenamefont {{Youles}}}]{2019duMasdesBourbouxH}%
  \BibitemOpen
  \bibfield  {author} {\bibinfo {author} {\bibfnamefont {H.}~\bibnamefont {{du
  Mas des Bourboux}}}, \bibinfo {author} {\bibfnamefont {J.}~\bibnamefont
  {{Rich}}}, \bibinfo {author} {\bibfnamefont {A.}~\bibnamefont
  {{Font-Ribera}}}, \bibinfo {author} {\bibfnamefont {V.}~\bibnamefont {{de
  Sainte Agathe}}}, \bibinfo {author} {\bibfnamefont {J.}~\bibnamefont
  {{Farr}}}, \bibinfo {author} {\bibfnamefont {T.}~\bibnamefont {{Etourneau}}},
  \bibinfo {author} {\bibfnamefont {J.-M.}\ \bibnamefont {{Le Goff}}}, \bibinfo
  {author} {\bibfnamefont {A.}~\bibnamefont {{Cuceu}}}, \bibinfo {author}
  {\bibfnamefont {C.}~\bibnamefont {{Balland}}}, \bibinfo {author}
  {\bibfnamefont {J.~E.}\ \bibnamefont {{Bautista}}}, \bibinfo {author}
  {\bibfnamefont {M.}~\bibnamefont {{Blomqvist}}}, \bibinfo {author}
  {\bibfnamefont {J.}~\bibnamefont {{Brinkmann}}}, \bibinfo {author}
  {\bibfnamefont {J.~R.}\ \bibnamefont {{Brownstein}}}, \bibinfo {author}
  {\bibfnamefont {S.}~\bibnamefont {{Chabanier}}}, \bibinfo {author}
  {\bibfnamefont {E.}~\bibnamefont {{Chaussidon}}}, \bibinfo {author}
  {\bibfnamefont {K.}~\bibnamefont {{Dawson}}}, \bibinfo {author}
  {\bibfnamefont {A.~X.}\ \bibnamefont {{Gonz{\'a}lez-Morales}}}, \bibinfo
  {author} {\bibfnamefont {J.}~\bibnamefont {{Guy}}}, \bibinfo {author}
  {\bibfnamefont {B.~W.}\ \bibnamefont {{Lyke}}}, \bibinfo {author}
  {\bibfnamefont {A.}~\bibnamefont {{de la Macorra}}}, \bibinfo {author}
  {\bibfnamefont {E.-M.}\ \bibnamefont {{Mueller}}}, \bibinfo {author}
  {\bibfnamefont {A.~D.}\ \bibnamefont {{Myers}}}, \bibinfo {author}
  {\bibfnamefont {C.}~\bibnamefont {{Nitschelm}}}, \bibinfo {author}
  {\bibfnamefont {A.}~\bibnamefont {{Mu{\~n}oz Guti{\'e}rrez}}}, \bibinfo
  {author} {\bibfnamefont {N.}~\bibnamefont {{Palanque-Delabrouille}}},
  \bibinfo {author} {\bibfnamefont {J.}~\bibnamefont {{Parker}}}, \bibinfo
  {author} {\bibfnamefont {W.~J.}\ \bibnamefont {{Percival}}}, \bibinfo
  {author} {\bibfnamefont {I.}~\bibnamefont {{P{\'e}rez-R{\`a}fols}}}, \bibinfo
  {author} {\bibfnamefont {P.}~\bibnamefont {{Petitjean}}}, \bibinfo {author}
  {\bibfnamefont {M.~M.}\ \bibnamefont {{Pieri}}}, \bibinfo {author}
  {\bibfnamefont {C.}~\bibnamefont {{Ravoux}}}, \bibinfo {author}
  {\bibfnamefont {G.}~\bibnamefont {{Rossi}}}, \bibinfo {author} {\bibfnamefont
  {D.~P.}\ \bibnamefont {{Schneider}}}, \bibinfo {author} {\bibfnamefont
  {H.-J.}\ \bibnamefont {{Seo}}}, \bibinfo {author} {\bibfnamefont
  {A.}~\bibnamefont {{Slosar}}}, \bibinfo {author} {\bibfnamefont
  {J.}~\bibnamefont {{Stermer}}}, \bibinfo {author} {\bibfnamefont
  {M.}~\bibnamefont {{Vivek}}}, \bibinfo {author} {\bibfnamefont
  {C.}~\bibnamefont {{Y{\`e}che}}}, \ and\ \bibinfo {author} {\bibfnamefont
  {S.}~\bibnamefont {{Youles}}},\ }\href {\doibase 10.3847/1538-4357/abb085}
  {\bibfield  {journal} {\bibinfo  {journal} {\apj}\ }\textbf {\bibinfo
  {volume} {901}},\ \bibinfo {eid} {153} (\bibinfo {year} {2020})},\ \Eprint
  {http://arxiv.org/abs/2007.08995} {arXiv:2007.08995 [astro-ph.CO]}
  \BibitemShut {NoStop}%
\bibitem [{\citenamefont {{Ross}}\ \emph {et~al.}(2020)\citenamefont {{Ross}},
  \citenamefont {{Bautista}}, \citenamefont {{Tojeiro}}, \citenamefont
  {{Alam}}, \citenamefont {{Bailey}}, \citenamefont {{Burtin}}, \citenamefont
  {{Comparat}}, \citenamefont {{Dawson}}, \citenamefont {{de Mattia}},
  \citenamefont {{du Mas des Bourboux}}, \citenamefont {{Gil-Mar{\'\i}n}},
  \citenamefont {{Hou}}, \citenamefont {{Kong}}, \citenamefont {{Lyke}},
  \citenamefont {{Mohammad}}, \citenamefont {{Moustakas}}, \citenamefont
  {{Mueller}}, \citenamefont {{Myers}}, \citenamefont {{Percival}},
  \citenamefont {{Raichoor}}, \citenamefont {{Rezaie}}, \citenamefont {{Seo}},
  \citenamefont {{Smith}}, \citenamefont {{Tinker}}, \citenamefont {{Zarrouk}},
  \citenamefont {{Zhao}}, \citenamefont {{Zhao}}, \citenamefont {{Bizyaev}},
  \citenamefont {{Brinkmann}}, \citenamefont {{Brownstein}}, \citenamefont
  {{Rosell}}, \citenamefont {{Chabanier}}, \citenamefont {{Choi}},
  \citenamefont {{Chuang}}, \citenamefont {{Cruz-Gonzalez}}, \citenamefont {{de
  la Macorra}}, \citenamefont {{de la Torre}}, \citenamefont {{Escoffier}},
  \citenamefont {{Fromenteau}}, \citenamefont {{Higley}}, \citenamefont
  {{Jullo}}, \citenamefont {{Kneib}}, \citenamefont {{McLane}}, \citenamefont
  {{Mu{\~n}oz-Guti{\'e}rrez}}, \citenamefont {{Neveux}}, \citenamefont
  {{Newman}}, \citenamefont {{Nitschelm}}, \citenamefont
  {{Palanque-Delabrouille}}, \citenamefont {{Paviot}}, \citenamefont
  {{Pullen}}, \citenamefont {{Rossi}}, \citenamefont {{Ruhlmann-Kleider}},
  \citenamefont {{Schneider}}, \citenamefont {{Maga{\~n}a}}, \citenamefont
  {{Vivek}},\ and\ \citenamefont {{Zhang}}}]{ross20a}%
  \BibitemOpen
  \bibfield  {author} {\bibinfo {author} {\bibfnamefont {A.~J.}\ \bibnamefont
  {{Ross}}}, \bibinfo {author} {\bibfnamefont {J.}~\bibnamefont {{Bautista}}},
  \bibinfo {author} {\bibfnamefont {R.}~\bibnamefont {{Tojeiro}}}, \bibinfo
  {author} {\bibfnamefont {S.}~\bibnamefont {{Alam}}}, \bibinfo {author}
  {\bibfnamefont {S.}~\bibnamefont {{Bailey}}}, \bibinfo {author}
  {\bibfnamefont {E.}~\bibnamefont {{Burtin}}}, \bibinfo {author}
  {\bibfnamefont {J.}~\bibnamefont {{Comparat}}}, \bibinfo {author}
  {\bibfnamefont {K.~S.}\ \bibnamefont {{Dawson}}}, \bibinfo {author}
  {\bibfnamefont {A.}~\bibnamefont {{de Mattia}}}, \bibinfo {author}
  {\bibfnamefont {H.}~\bibnamefont {{du Mas des Bourboux}}}, \bibinfo {author}
  {\bibfnamefont {H.}~\bibnamefont {{Gil-Mar{\'\i}n}}}, \bibinfo {author}
  {\bibfnamefont {J.}~\bibnamefont {{Hou}}}, \bibinfo {author} {\bibfnamefont
  {H.}~\bibnamefont {{Kong}}}, \bibinfo {author} {\bibfnamefont {B.~W.}\
  \bibnamefont {{Lyke}}}, \bibinfo {author} {\bibfnamefont {F.~G.}\
  \bibnamefont {{Mohammad}}}, \bibinfo {author} {\bibfnamefont
  {J.}~\bibnamefont {{Moustakas}}}, \bibinfo {author} {\bibfnamefont {E.-M.}\
  \bibnamefont {{Mueller}}}, \bibinfo {author} {\bibfnamefont {A.~D.}\
  \bibnamefont {{Myers}}}, \bibinfo {author} {\bibfnamefont {W.~J.}\
  \bibnamefont {{Percival}}}, \bibinfo {author} {\bibfnamefont
  {A.}~\bibnamefont {{Raichoor}}}, \bibinfo {author} {\bibfnamefont
  {M.}~\bibnamefont {{Rezaie}}}, \bibinfo {author} {\bibfnamefont {H.-J.}\
  \bibnamefont {{Seo}}}, \bibinfo {author} {\bibfnamefont {A.}~\bibnamefont
  {{Smith}}}, \bibinfo {author} {\bibfnamefont {J.~L.}\ \bibnamefont
  {{Tinker}}}, \bibinfo {author} {\bibfnamefont {P.}~\bibnamefont {{Zarrouk}}},
  \bibinfo {author} {\bibfnamefont {C.}~\bibnamefont {{Zhao}}}, \bibinfo
  {author} {\bibfnamefont {G.-B.}\ \bibnamefont {{Zhao}}}, \bibinfo {author}
  {\bibfnamefont {D.}~\bibnamefont {{Bizyaev}}}, \bibinfo {author}
  {\bibfnamefont {J.}~\bibnamefont {{Brinkmann}}}, \bibinfo {author}
  {\bibfnamefont {J.~R.}\ \bibnamefont {{Brownstein}}}, \bibinfo {author}
  {\bibfnamefont {A.~C.}\ \bibnamefont {{Rosell}}}, \bibinfo {author}
  {\bibfnamefont {S.}~\bibnamefont {{Chabanier}}}, \bibinfo {author}
  {\bibfnamefont {P.~D.}\ \bibnamefont {{Choi}}}, \bibinfo {author}
  {\bibfnamefont {C.-H.}\ \bibnamefont {{Chuang}}}, \bibinfo {author}
  {\bibfnamefont {I.}~\bibnamefont {{Cruz-Gonzalez}}}, \bibinfo {author}
  {\bibfnamefont {A.}~\bibnamefont {{de la Macorra}}}, \bibinfo {author}
  {\bibfnamefont {S.}~\bibnamefont {{de la Torre}}}, \bibinfo {author}
  {\bibfnamefont {S.}~\bibnamefont {{Escoffier}}}, \bibinfo {author}
  {\bibfnamefont {S.}~\bibnamefont {{Fromenteau}}}, \bibinfo {author}
  {\bibfnamefont {A.}~\bibnamefont {{Higley}}}, \bibinfo {author}
  {\bibfnamefont {E.}~\bibnamefont {{Jullo}}}, \bibinfo {author} {\bibfnamefont
  {J.-P.}\ \bibnamefont {{Kneib}}}, \bibinfo {author} {\bibfnamefont {J.~N.}\
  \bibnamefont {{McLane}}}, \bibinfo {author} {\bibfnamefont {A.}~\bibnamefont
  {{Mu{\~n}oz-Guti{\'e}rrez}}}, \bibinfo {author} {\bibfnamefont
  {R.}~\bibnamefont {{Neveux}}}, \bibinfo {author} {\bibfnamefont {J.~A.}\
  \bibnamefont {{Newman}}}, \bibinfo {author} {\bibfnamefont {C.}~\bibnamefont
  {{Nitschelm}}}, \bibinfo {author} {\bibfnamefont {N.}~\bibnamefont
  {{Palanque-Delabrouille}}}, \bibinfo {author} {\bibfnamefont
  {R.}~\bibnamefont {{Paviot}}}, \bibinfo {author} {\bibfnamefont {A.~R.}\
  \bibnamefont {{Pullen}}}, \bibinfo {author} {\bibfnamefont {G.}~\bibnamefont
  {{Rossi}}}, \bibinfo {author} {\bibfnamefont {V.}~\bibnamefont
  {{Ruhlmann-Kleider}}}, \bibinfo {author} {\bibfnamefont {D.~P.}\ \bibnamefont
  {{Schneider}}}, \bibinfo {author} {\bibfnamefont {M.~V.}\ \bibnamefont
  {{Maga{\~n}a}}}, \bibinfo {author} {\bibfnamefont {M.}~\bibnamefont
  {{Vivek}}}, \ and\ \bibinfo {author} {\bibfnamefont {Y.}~\bibnamefont
  {{Zhang}}},\ }\href {\doibase 10.1093/mnras/staa2416} {\bibfield  {journal}
  {\bibinfo  {journal} {\mnras}\ }\textbf {\bibinfo {volume} {498}},\ \bibinfo
  {pages} {2354} (\bibinfo {year} {2020})},\ \Eprint
  {http://arxiv.org/abs/2007.09000} {arXiv:2007.09000 [astro-ph.CO]}
  \BibitemShut {NoStop}%
\bibitem [{\citenamefont {{Lyke}}\ \emph {et~al.}(2020)\citenamefont {{Lyke}},
  \citenamefont {{Higley}}, \citenamefont {{McLane}}, \citenamefont
  {{Schurhammer}}, \citenamefont {{Myers}}, \citenamefont {{Ross}},
  \citenamefont {{Dawson}}, \citenamefont {{Chabanier}}, \citenamefont
  {{Martini}}, \citenamefont {{Busca}}, \citenamefont {{Mas des Bourboux}},
  \citenamefont {{Salvato}}, \citenamefont {{Streblyanska}}, \citenamefont
  {{Zarrouk}}, \citenamefont {{Burtin}}, \citenamefont {{Anderson}},
  \citenamefont {{Bautista}}, \citenamefont {{Bizyaev}}, \citenamefont
  {{Brandt}}, \citenamefont {{Brinkmann}}, \citenamefont {{Brownstein}},
  \citenamefont {{Comparat}}, \citenamefont {{Green}}, \citenamefont {{de la
  Macorra}}, \citenamefont {{Mu{\~n}oz Guti{\'e}rrez}}, \citenamefont {{Hou}},
  \citenamefont {{Newman}}, \citenamefont {{Palanque-Delabrouille}},
  \citenamefont {{P{\^a}ris}}, \citenamefont {{Percival}}, \citenamefont
  {{Petitjean}}, \citenamefont {{Rich}}, \citenamefont {{Rossi}}, \citenamefont
  {{Schneider}}, \citenamefont {{Smith}}, \citenamefont {{Vivek}},\ and\
  \citenamefont {{Weaver}}}]{lyke20a}%
  \BibitemOpen
  \bibfield  {author} {\bibinfo {author} {\bibfnamefont {B.~W.}\ \bibnamefont
  {{Lyke}}}, \bibinfo {author} {\bibfnamefont {A.~N.}\ \bibnamefont
  {{Higley}}}, \bibinfo {author} {\bibfnamefont {J.~N.}\ \bibnamefont
  {{McLane}}}, \bibinfo {author} {\bibfnamefont {D.~P.}\ \bibnamefont
  {{Schurhammer}}}, \bibinfo {author} {\bibfnamefont {A.~D.}\ \bibnamefont
  {{Myers}}}, \bibinfo {author} {\bibfnamefont {A.~J.}\ \bibnamefont {{Ross}}},
  \bibinfo {author} {\bibfnamefont {K.}~\bibnamefont {{Dawson}}}, \bibinfo
  {author} {\bibfnamefont {S.}~\bibnamefont {{Chabanier}}}, \bibinfo {author}
  {\bibfnamefont {P.}~\bibnamefont {{Martini}}}, \bibinfo {author}
  {\bibfnamefont {N.~G.}\ \bibnamefont {{Busca}}}, \bibinfo {author}
  {\bibfnamefont {H.~d.}\ \bibnamefont {{Mas des Bourboux}}}, \bibinfo {author}
  {\bibfnamefont {M.}~\bibnamefont {{Salvato}}}, \bibinfo {author}
  {\bibfnamefont {A.}~\bibnamefont {{Streblyanska}}}, \bibinfo {author}
  {\bibfnamefont {P.}~\bibnamefont {{Zarrouk}}}, \bibinfo {author}
  {\bibfnamefont {E.}~\bibnamefont {{Burtin}}}, \bibinfo {author}
  {\bibfnamefont {S.~F.}\ \bibnamefont {{Anderson}}}, \bibinfo {author}
  {\bibfnamefont {J.}~\bibnamefont {{Bautista}}}, \bibinfo {author}
  {\bibfnamefont {D.}~\bibnamefont {{Bizyaev}}}, \bibinfo {author}
  {\bibfnamefont {W.~N.}\ \bibnamefont {{Brandt}}}, \bibinfo {author}
  {\bibfnamefont {J.}~\bibnamefont {{Brinkmann}}}, \bibinfo {author}
  {\bibfnamefont {J.~R.}\ \bibnamefont {{Brownstein}}}, \bibinfo {author}
  {\bibfnamefont {J.}~\bibnamefont {{Comparat}}}, \bibinfo {author}
  {\bibfnamefont {P.}~\bibnamefont {{Green}}}, \bibinfo {author} {\bibfnamefont
  {A.}~\bibnamefont {{de la Macorra}}}, \bibinfo {author} {\bibfnamefont
  {A.}~\bibnamefont {{Mu{\~n}oz Guti{\'e}rrez}}}, \bibinfo {author}
  {\bibfnamefont {J.}~\bibnamefont {{Hou}}}, \bibinfo {author} {\bibfnamefont
  {J.~A.}\ \bibnamefont {{Newman}}}, \bibinfo {author} {\bibfnamefont
  {N.}~\bibnamefont {{Palanque-Delabrouille}}}, \bibinfo {author}
  {\bibfnamefont {I.}~\bibnamefont {{P{\^a}ris}}}, \bibinfo {author}
  {\bibfnamefont {W.~J.}\ \bibnamefont {{Percival}}}, \bibinfo {author}
  {\bibfnamefont {P.}~\bibnamefont {{Petitjean}}}, \bibinfo {author}
  {\bibfnamefont {J.}~\bibnamefont {{Rich}}}, \bibinfo {author} {\bibfnamefont
  {G.}~\bibnamefont {{Rossi}}}, \bibinfo {author} {\bibfnamefont {D.~P.}\
  \bibnamefont {{Schneider}}}, \bibinfo {author} {\bibfnamefont
  {A.}~\bibnamefont {{Smith}}}, \bibinfo {author} {\bibfnamefont
  {M.}~\bibnamefont {{Vivek}}}, \ and\ \bibinfo {author} {\bibfnamefont
  {B.~A.}\ \bibnamefont {{Weaver}}},\ }\href {\doibase
  10.3847/1538-4365/aba623} {\bibfield  {journal} {\bibinfo  {journal} {\apjs}\
  }\textbf {\bibinfo {volume} {250}},\ \bibinfo {eid} {8} (\bibinfo {year}
  {2020})},\ \Eprint {http://arxiv.org/abs/2007.09001} {arXiv:2007.09001
  [astro-ph.GA]} \BibitemShut {NoStop}%
\bibitem [{\citenamefont {{Lin}}\ \emph {et~al.}(2020)\citenamefont {{Lin}},
  \citenamefont {{Tinker}}, \citenamefont {{Klypin}}, \citenamefont {{Prada}},
  \citenamefont {{Blanton}}, \citenamefont {{Comparat}}, \citenamefont
  {{Dawson}}, \citenamefont {{de Mattia}}, \citenamefont {{du Mas des
  Bourboux}}, \citenamefont {{Percival}}, \citenamefont {{Raichoor}},
  \citenamefont {{Rossi}}, \citenamefont {{Smith}},\ and\ \citenamefont
  {{Zhao}}}]{lin20a}%
  \BibitemOpen
  \bibfield  {author} {\bibinfo {author} {\bibfnamefont {S.}~\bibnamefont
  {{Lin}}}, \bibinfo {author} {\bibfnamefont {J.~L.}\ \bibnamefont {{Tinker}}},
  \bibinfo {author} {\bibfnamefont {A.}~\bibnamefont {{Klypin}}}, \bibinfo
  {author} {\bibfnamefont {F.}~\bibnamefont {{Prada}}}, \bibinfo {author}
  {\bibfnamefont {M.~R.}\ \bibnamefont {{Blanton}}}, \bibinfo {author}
  {\bibfnamefont {J.}~\bibnamefont {{Comparat}}}, \bibinfo {author}
  {\bibfnamefont {K.~S.}\ \bibnamefont {{Dawson}}}, \bibinfo {author}
  {\bibfnamefont {A.}~\bibnamefont {{de Mattia}}}, \bibinfo {author}
  {\bibfnamefont {H.}~\bibnamefont {{du Mas des Bourboux}}}, \bibinfo {author}
  {\bibfnamefont {W.~J.}\ \bibnamefont {{Percival}}}, \bibinfo {author}
  {\bibfnamefont {A.}~\bibnamefont {{Raichoor}}}, \bibinfo {author}
  {\bibfnamefont {G.}~\bibnamefont {{Rossi}}}, \bibinfo {author} {\bibfnamefont
  {A.}~\bibnamefont {{Smith}}}, \ and\ \bibinfo {author} {\bibfnamefont
  {C.}~\bibnamefont {{Zhao}}},\ }\href {\doibase 10.1093/mnras/staa2571}
  {\bibfield  {journal} {\bibinfo  {journal} {\mnras}\ }\textbf {\bibinfo
  {volume} {498}},\ \bibinfo {pages} {5251} (\bibinfo {year} {2020})},\ \Eprint
  {http://arxiv.org/abs/2007.08996} {arXiv:2007.08996 [astro-ph.CO]}
  \BibitemShut {NoStop}%
\bibitem [{\citenamefont {{Zhao}}\ \emph {et~al.}(2020)\citenamefont {{Zhao}},
  \citenamefont {{Chuang}}, \citenamefont {{Bautista}}, \citenamefont {{de
  Mattia}}, \citenamefont {{Raichoor}}, \citenamefont {{Ross}}, \citenamefont
  {{Hou}}, \citenamefont {{Neveux}}, \citenamefont {{Tao}}, \citenamefont
  {{Burtin}}, \citenamefont {{Dawson}}, \citenamefont {{de la Torre}},
  \citenamefont {{Gil-Mar{\'\i}n}}, \citenamefont {{Kneib}}, \citenamefont
  {{Percival}}, \citenamefont {{Rossi}}, \citenamefont {{Tamone}},
  \citenamefont {{Tinker}}, \citenamefont {{Zhao}}, \citenamefont {{Alam}},\
  and\ \citenamefont {{Mueller}}}]{zhao20a}%
  \BibitemOpen
  \bibfield  {author} {\bibinfo {author} {\bibfnamefont {C.}~\bibnamefont
  {{Zhao}}}, \bibinfo {author} {\bibfnamefont {C.-H.}\ \bibnamefont
  {{Chuang}}}, \bibinfo {author} {\bibfnamefont {J.}~\bibnamefont
  {{Bautista}}}, \bibinfo {author} {\bibfnamefont {A.}~\bibnamefont {{de
  Mattia}}}, \bibinfo {author} {\bibfnamefont {A.}~\bibnamefont {{Raichoor}}},
  \bibinfo {author} {\bibfnamefont {A.~J.}\ \bibnamefont {{Ross}}}, \bibinfo
  {author} {\bibfnamefont {J.}~\bibnamefont {{Hou}}}, \bibinfo {author}
  {\bibfnamefont {R.}~\bibnamefont {{Neveux}}}, \bibinfo {author}
  {\bibfnamefont {C.}~\bibnamefont {{Tao}}}, \bibinfo {author} {\bibfnamefont
  {E.}~\bibnamefont {{Burtin}}}, \bibinfo {author} {\bibfnamefont {K.~S.}\
  \bibnamefont {{Dawson}}}, \bibinfo {author} {\bibfnamefont {S.}~\bibnamefont
  {{de la Torre}}}, \bibinfo {author} {\bibfnamefont {H.}~\bibnamefont
  {{Gil-Mar{\'\i}n}}}, \bibinfo {author} {\bibfnamefont {J.-P.}\ \bibnamefont
  {{Kneib}}}, \bibinfo {author} {\bibfnamefont {W.~J.}\ \bibnamefont
  {{Percival}}}, \bibinfo {author} {\bibfnamefont {G.}~\bibnamefont {{Rossi}}},
  \bibinfo {author} {\bibfnamefont {A.}~\bibnamefont {{Tamone}}}, \bibinfo
  {author} {\bibfnamefont {J.~L.}\ \bibnamefont {{Tinker}}}, \bibinfo {author}
  {\bibfnamefont {G.-B.}\ \bibnamefont {{Zhao}}}, \bibinfo {author}
  {\bibfnamefont {S.}~\bibnamefont {{Alam}}}, \ and\ \bibinfo {author}
  {\bibfnamefont {E.-M.}\ \bibnamefont {{Mueller}}},\ }\href@noop {} {\bibfield
   {journal} {\bibinfo  {journal} {arXiv e-prints}\ ,\ \bibinfo {eid}
  {arXiv:2007.08997}} (\bibinfo {year} {2020})},\ \Eprint
  {http://arxiv.org/abs/2007.08997} {arXiv:2007.08997 [astro-ph.CO]}
  \BibitemShut {NoStop}%
\bibitem [{\citenamefont {{Farr}}\ \emph {et~al.}(2020)\citenamefont {{Farr}},
  \citenamefont {{Font-Ribera}}, \citenamefont {{du Mas des Bourboux}},
  \citenamefont {{Mu{\~n}oz-Guti{\'e}rrez}}, \citenamefont {{S{\'a}nchez}},
  \citenamefont {{Pontzen}}, \citenamefont {{Xochitl Gonz{\'a}lez-Morales}},
  \citenamefont {{Alonso}}, \citenamefont {{Brooks}}, \citenamefont {{Doel}},
  \citenamefont {{Etourneau}}, \citenamefont {{Guy}}, \citenamefont {{Le
  Goff}}, \citenamefont {{de la Macorra}}, \citenamefont
  {{Palanque-Delabrouille}}, \citenamefont {{P{\'e}rez-R{\`a}fols}},
  \citenamefont {{Rich}}, \citenamefont {{Slosar}}, \citenamefont {{Tarle}},
  \citenamefont {{Yutong}},\ and\ \citenamefont {{Zhang}}}]{carr20a}%
  \BibitemOpen
  \bibfield  {author} {\bibinfo {author} {\bibfnamefont {J.}~\bibnamefont
  {{Farr}}}, \bibinfo {author} {\bibfnamefont {A.}~\bibnamefont
  {{Font-Ribera}}}, \bibinfo {author} {\bibfnamefont {H.}~\bibnamefont {{du Mas
  des Bourboux}}}, \bibinfo {author} {\bibfnamefont {A.}~\bibnamefont
  {{Mu{\~n}oz-Guti{\'e}rrez}}}, \bibinfo {author} {\bibfnamefont {F.~J.}\
  \bibnamefont {{S{\'a}nchez}}}, \bibinfo {author} {\bibfnamefont
  {A.}~\bibnamefont {{Pontzen}}}, \bibinfo {author} {\bibfnamefont
  {A.}~\bibnamefont {{Xochitl Gonz{\'a}lez-Morales}}}, \bibinfo {author}
  {\bibfnamefont {D.}~\bibnamefont {{Alonso}}}, \bibinfo {author}
  {\bibfnamefont {D.}~\bibnamefont {{Brooks}}}, \bibinfo {author}
  {\bibfnamefont {P.}~\bibnamefont {{Doel}}}, \bibinfo {author} {\bibfnamefont
  {T.}~\bibnamefont {{Etourneau}}}, \bibinfo {author} {\bibfnamefont
  {J.}~\bibnamefont {{Guy}}}, \bibinfo {author} {\bibfnamefont {J.-M.}\
  \bibnamefont {{Le Goff}}}, \bibinfo {author} {\bibfnamefont {A.}~\bibnamefont
  {{de la Macorra}}}, \bibinfo {author} {\bibfnamefont {N.}~\bibnamefont
  {{Palanque-Delabrouille}}}, \bibinfo {author} {\bibfnamefont
  {I.}~\bibnamefont {{P{\'e}rez-R{\`a}fols}}}, \bibinfo {author} {\bibfnamefont
  {J.}~\bibnamefont {{Rich}}}, \bibinfo {author} {\bibfnamefont
  {A.}~\bibnamefont {{Slosar}}}, \bibinfo {author} {\bibfnamefont
  {G.}~\bibnamefont {{Tarle}}}, \bibinfo {author} {\bibfnamefont
  {D.}~\bibnamefont {{Yutong}}}, \ and\ \bibinfo {author} {\bibfnamefont
  {K.}~\bibnamefont {{Zhang}}},\ }\href {\doibase
  10.1088/1475-7516/2020/03/068} {\bibfield  {journal} {\bibinfo  {journal}
  {\jcap}\ }\textbf {\bibinfo {volume} {2020}},\ \bibinfo {eid} {068} (\bibinfo
  {year} {2020})},\ \Eprint {http://arxiv.org/abs/1912.02763} {arXiv:1912.02763
  [astro-ph.CO]} \BibitemShut {NoStop}%
\bibitem [{\citenamefont {{Alam}}\ \emph
  {et~al.}(2020{\natexlab{a}})\citenamefont {{Alam}}, \citenamefont {{de
  Mattia}}, \citenamefont {{Tamone}}, \citenamefont {{{\'A}vila}},
  \citenamefont {{Peacock}}, \citenamefont {{Gonzalez-Perez}}, \citenamefont
  {{Smith}}, \citenamefont {{Raichoor}}, \citenamefont {{Ross}}, \citenamefont
  {{Bautista}}, \citenamefont {{Burtin}}, \citenamefont {{Comparat}},
  \citenamefont {{Dawson}}, \citenamefont {{du Mas des Bourboux}},
  \citenamefont {{Escoffier}}, \citenamefont {{Gil-Mar{\'\i}n}}, \citenamefont
  {{Habib}}, \citenamefont {{Heitmann}}, \citenamefont {{Hou}}, \citenamefont
  {{Mohammad}}, \citenamefont {{Mueller}}, \citenamefont {{Neveux}},
  \citenamefont {{Paviot}}, \citenamefont {{Percival}}, \citenamefont
  {{Rossi}}, \citenamefont {{Ruhlmann-Kleider}}, \citenamefont {{Tojeiro}},
  \citenamefont {{Vargas Maga{\~n}a}}, \citenamefont {{Zhao}},\ and\
  \citenamefont {{Zhao}}}]{alam20a}%
  \BibitemOpen
  \bibfield  {author} {\bibinfo {author} {\bibfnamefont {S.}~\bibnamefont
  {{Alam}}}, \bibinfo {author} {\bibfnamefont {A.}~\bibnamefont {{de Mattia}}},
  \bibinfo {author} {\bibfnamefont {A.}~\bibnamefont {{Tamone}}}, \bibinfo
  {author} {\bibfnamefont {S.}~\bibnamefont {{{\'A}vila}}}, \bibinfo {author}
  {\bibfnamefont {J.~A.}\ \bibnamefont {{Peacock}}}, \bibinfo {author}
  {\bibfnamefont {V.}~\bibnamefont {{Gonzalez-Perez}}}, \bibinfo {author}
  {\bibfnamefont {A.}~\bibnamefont {{Smith}}}, \bibinfo {author} {\bibfnamefont
  {A.}~\bibnamefont {{Raichoor}}}, \bibinfo {author} {\bibfnamefont {A.~J.}\
  \bibnamefont {{Ross}}}, \bibinfo {author} {\bibfnamefont {J.~E.}\
  \bibnamefont {{Bautista}}}, \bibinfo {author} {\bibfnamefont
  {E.}~\bibnamefont {{Burtin}}}, \bibinfo {author} {\bibfnamefont
  {J.}~\bibnamefont {{Comparat}}}, \bibinfo {author} {\bibfnamefont {K.~S.}\
  \bibnamefont {{Dawson}}}, \bibinfo {author} {\bibfnamefont {H.}~\bibnamefont
  {{du Mas des Bourboux}}}, \bibinfo {author} {\bibfnamefont {S.}~\bibnamefont
  {{Escoffier}}}, \bibinfo {author} {\bibfnamefont {H.}~\bibnamefont
  {{Gil-Mar{\'\i}n}}}, \bibinfo {author} {\bibfnamefont {S.}~\bibnamefont
  {{Habib}}}, \bibinfo {author} {\bibfnamefont {K.}~\bibnamefont {{Heitmann}}},
  \bibinfo {author} {\bibfnamefont {J.}~\bibnamefont {{Hou}}}, \bibinfo
  {author} {\bibfnamefont {F.~G.}\ \bibnamefont {{Mohammad}}}, \bibinfo
  {author} {\bibfnamefont {E.-M.}\ \bibnamefont {{Mueller}}}, \bibinfo {author}
  {\bibfnamefont {R.}~\bibnamefont {{Neveux}}}, \bibinfo {author}
  {\bibfnamefont {R.}~\bibnamefont {{Paviot}}}, \bibinfo {author}
  {\bibfnamefont {W.~J.}\ \bibnamefont {{Percival}}}, \bibinfo {author}
  {\bibfnamefont {G.}~\bibnamefont {{Rossi}}}, \bibinfo {author} {\bibfnamefont
  {V.}~\bibnamefont {{Ruhlmann-Kleider}}}, \bibinfo {author} {\bibfnamefont
  {R.}~\bibnamefont {{Tojeiro}}}, \bibinfo {author} {\bibfnamefont
  {M.}~\bibnamefont {{Vargas Maga{\~n}a}}}, \bibinfo {author} {\bibfnamefont
  {C.}~\bibnamefont {{Zhao}}}, \ and\ \bibinfo {author} {\bibfnamefont {G.-B.}\
  \bibnamefont {{Zhao}}},\ }\href@noop {} {\bibfield  {journal} {\bibinfo
  {journal} {arXiv e-prints}\ ,\ \bibinfo {eid} {arXiv:2007.09004}} (\bibinfo
  {year} {2020}{\natexlab{a}})},\ \Eprint {http://arxiv.org/abs/2007.09004}
  {arXiv:2007.09004 [astro-ph.CO]} \BibitemShut {NoStop}%
\bibitem [{\citenamefont {{Avila}}\ \emph {et~al.}(2020)\citenamefont
  {{Avila}}, \citenamefont {{Gonzalez-Perez}}, \citenamefont {{Mohammad}},
  \citenamefont {{de Mattia}}, \citenamefont {{Zhao}}, \citenamefont
  {{Raichoor}}, \citenamefont {{Tamone}}, \citenamefont {{Alam}}, \citenamefont
  {{Bautista}}, \citenamefont {{Bianchi}}, \citenamefont {{Burtin}},
  \citenamefont {{Chapman}}, \citenamefont {{Chuang}}, \citenamefont
  {{Comparat}}, \citenamefont {{Dawson}}, \citenamefont {{Divers}},
  \citenamefont {{du Mas des Bourboux}}, \citenamefont {{Gil-Marin}},
  \citenamefont {{Mueller}}, \citenamefont {{Habib}}, \citenamefont
  {{Heitmann}}, \citenamefont {{Ruhlmann-Kleider}}, \citenamefont {{Padilla}},
  \citenamefont {{Percival}}, \citenamefont {{Ross}}, \citenamefont {{Seo}},
  \citenamefont {{Schneider}},\ and\ \citenamefont {{Zhao}}}]{avila20a}%
  \BibitemOpen
  \bibfield  {author} {\bibinfo {author} {\bibfnamefont {S.}~\bibnamefont
  {{Avila}}}, \bibinfo {author} {\bibfnamefont {V.}~\bibnamefont
  {{Gonzalez-Perez}}}, \bibinfo {author} {\bibfnamefont {F.~G.}\ \bibnamefont
  {{Mohammad}}}, \bibinfo {author} {\bibfnamefont {A.}~\bibnamefont {{de
  Mattia}}}, \bibinfo {author} {\bibfnamefont {C.}~\bibnamefont {{Zhao}}},
  \bibinfo {author} {\bibfnamefont {A.}~\bibnamefont {{Raichoor}}}, \bibinfo
  {author} {\bibfnamefont {A.}~\bibnamefont {{Tamone}}}, \bibinfo {author}
  {\bibfnamefont {S.}~\bibnamefont {{Alam}}}, \bibinfo {author} {\bibfnamefont
  {J.}~\bibnamefont {{Bautista}}}, \bibinfo {author} {\bibfnamefont
  {D.}~\bibnamefont {{Bianchi}}}, \bibinfo {author} {\bibfnamefont
  {E.}~\bibnamefont {{Burtin}}}, \bibinfo {author} {\bibfnamefont {M.~J.}\
  \bibnamefont {{Chapman}}}, \bibinfo {author} {\bibfnamefont {C.~H.}\
  \bibnamefont {{Chuang}}}, \bibinfo {author} {\bibfnamefont {J.}~\bibnamefont
  {{Comparat}}}, \bibinfo {author} {\bibfnamefont {K.}~\bibnamefont
  {{Dawson}}}, \bibinfo {author} {\bibfnamefont {T.}~\bibnamefont {{Divers}}},
  \bibinfo {author} {\bibfnamefont {H.}~\bibnamefont {{du Mas des Bourboux}}},
  \bibinfo {author} {\bibfnamefont {H.}~\bibnamefont {{Gil-Marin}}}, \bibinfo
  {author} {\bibfnamefont {E.~M.}\ \bibnamefont {{Mueller}}}, \bibinfo {author}
  {\bibfnamefont {S.}~\bibnamefont {{Habib}}}, \bibinfo {author} {\bibfnamefont
  {K.}~\bibnamefont {{Heitmann}}}, \bibinfo {author} {\bibfnamefont
  {V.}~\bibnamefont {{Ruhlmann-Kleider}}}, \bibinfo {author} {\bibfnamefont
  {N.}~\bibnamefont {{Padilla}}}, \bibinfo {author} {\bibfnamefont {W.~J.}\
  \bibnamefont {{Percival}}}, \bibinfo {author} {\bibfnamefont {A.~J.}\
  \bibnamefont {{Ross}}}, \bibinfo {author} {\bibfnamefont {H.~J.}\
  \bibnamefont {{Seo}}}, \bibinfo {author} {\bibfnamefont {D.~P.}\ \bibnamefont
  {{Schneider}}}, \ and\ \bibinfo {author} {\bibfnamefont {G.}~\bibnamefont
  {{Zhao}}},\ }\href {\doibase 10.1093/mnras/staa2951} {\bibfield  {journal}
  {\bibinfo  {journal} {\mnras}\ }\textbf {\bibinfo {volume} {499}},\ \bibinfo
  {pages} {5486} (\bibinfo {year} {2020})},\ \Eprint
  {http://arxiv.org/abs/2007.09012} {arXiv:2007.09012 [astro-ph.CO]}
  \BibitemShut {NoStop}%
\bibitem [{\citenamefont {{Rossi}}\ \emph {et~al.}(2020)\citenamefont
  {{Rossi}}, \citenamefont {{Choi}}, \citenamefont {{Moon}}, \citenamefont
  {{Bautista}}, \citenamefont {{Gil-Mar{\'\i}n}}, \citenamefont {{Paviot}},
  \citenamefont {{Vargas-Maga{\~n}a}}, \citenamefont {{de la Torre}},
  \citenamefont {{Fromenteau}}, \citenamefont {{Ross}}, \citenamefont
  {{{\'A}vila}}, \citenamefont {{Burtin}}, \citenamefont {{Dawson}},
  \citenamefont {{Escoffier}}, \citenamefont {{Habib}}, \citenamefont
  {{Heitmann}}, \citenamefont {{Hou}}, \citenamefont {{Mueller}}, \citenamefont
  {{Percival}}, \citenamefont {{Smith}}, \citenamefont {{Zhao}},\ and\
  \citenamefont {{Zhao}}}]{rossi20a}%
  \BibitemOpen
  \bibfield  {author} {\bibinfo {author} {\bibfnamefont {G.}~\bibnamefont
  {{Rossi}}}, \bibinfo {author} {\bibfnamefont {P.~D.}\ \bibnamefont {{Choi}}},
  \bibinfo {author} {\bibfnamefont {J.}~\bibnamefont {{Moon}}}, \bibinfo
  {author} {\bibfnamefont {J.~E.}\ \bibnamefont {{Bautista}}}, \bibinfo
  {author} {\bibfnamefont {H.}~\bibnamefont {{Gil-Mar{\'\i}n}}}, \bibinfo
  {author} {\bibfnamefont {R.}~\bibnamefont {{Paviot}}}, \bibinfo {author}
  {\bibfnamefont {M.}~\bibnamefont {{Vargas-Maga{\~n}a}}}, \bibinfo {author}
  {\bibfnamefont {S.}~\bibnamefont {{de la Torre}}}, \bibinfo {author}
  {\bibfnamefont {S.}~\bibnamefont {{Fromenteau}}}, \bibinfo {author}
  {\bibfnamefont {A.~J.}\ \bibnamefont {{Ross}}}, \bibinfo {author}
  {\bibfnamefont {S.}~\bibnamefont {{{\'A}vila}}}, \bibinfo {author}
  {\bibfnamefont {E.}~\bibnamefont {{Burtin}}}, \bibinfo {author}
  {\bibfnamefont {K.~S.}\ \bibnamefont {{Dawson}}}, \bibinfo {author}
  {\bibfnamefont {S.}~\bibnamefont {{Escoffier}}}, \bibinfo {author}
  {\bibfnamefont {S.}~\bibnamefont {{Habib}}}, \bibinfo {author} {\bibfnamefont
  {K.}~\bibnamefont {{Heitmann}}}, \bibinfo {author} {\bibfnamefont
  {J.}~\bibnamefont {{Hou}}}, \bibinfo {author} {\bibfnamefont {E.-M.}\
  \bibnamefont {{Mueller}}}, \bibinfo {author} {\bibfnamefont {W.~J.}\
  \bibnamefont {{Percival}}}, \bibinfo {author} {\bibfnamefont
  {A.}~\bibnamefont {{Smith}}}, \bibinfo {author} {\bibfnamefont
  {C.}~\bibnamefont {{Zhao}}}, \ and\ \bibinfo {author} {\bibfnamefont {G.-B.}\
  \bibnamefont {{Zhao}}},\ }\href {\doibase 10.1093/mnras/staa3955} {\bibfield
  {journal} {\bibinfo  {journal} {\mnras}\ } (\bibinfo {year} {2020}),\
  10.1093/mnras/staa3955},\ \Eprint {http://arxiv.org/abs/2007.09002}
  {arXiv:2007.09002 [astro-ph.CO]} \BibitemShut {NoStop}%
\bibitem [{\citenamefont {{Smith}}\ \emph {et~al.}(2020)\citenamefont
  {{Smith}}, \citenamefont {{Burtin}}, \citenamefont {{Hou}}, \citenamefont
  {{Neveux}}, \citenamefont {{Ross}}, \citenamefont {{Alam}}, \citenamefont
  {{Brinkmann}}, \citenamefont {{Dawson}}, \citenamefont {{Habib}},
  \citenamefont {{Heitmann}}, \citenamefont {{Kneib}}, \citenamefont {{Lyke}},
  \citenamefont {{du Mas des Bourboux}}, \citenamefont {{Mueller}},
  \citenamefont {{Myers}}, \citenamefont {{Percival}}, \citenamefont {{Rossi}},
  \citenamefont {{Schneider}}, \citenamefont {{Zarrouk}},\ and\ \citenamefont
  {{Zhao}}}]{smith20}%
  \BibitemOpen
  \bibfield  {author} {\bibinfo {author} {\bibfnamefont {A.}~\bibnamefont
  {{Smith}}}, \bibinfo {author} {\bibfnamefont {E.}~\bibnamefont {{Burtin}}},
  \bibinfo {author} {\bibfnamefont {J.}~\bibnamefont {{Hou}}}, \bibinfo
  {author} {\bibfnamefont {R.}~\bibnamefont {{Neveux}}}, \bibinfo {author}
  {\bibfnamefont {A.~J.}\ \bibnamefont {{Ross}}}, \bibinfo {author}
  {\bibfnamefont {S.}~\bibnamefont {{Alam}}}, \bibinfo {author} {\bibfnamefont
  {J.}~\bibnamefont {{Brinkmann}}}, \bibinfo {author} {\bibfnamefont {K.~S.}\
  \bibnamefont {{Dawson}}}, \bibinfo {author} {\bibfnamefont {S.}~\bibnamefont
  {{Habib}}}, \bibinfo {author} {\bibfnamefont {K.}~\bibnamefont {{Heitmann}}},
  \bibinfo {author} {\bibfnamefont {J.-P.}\ \bibnamefont {{Kneib}}}, \bibinfo
  {author} {\bibfnamefont {B.~W.}\ \bibnamefont {{Lyke}}}, \bibinfo {author}
  {\bibfnamefont {H.}~\bibnamefont {{du Mas des Bourboux}}}, \bibinfo {author}
  {\bibfnamefont {E.-M.}\ \bibnamefont {{Mueller}}}, \bibinfo {author}
  {\bibfnamefont {A.~D.}\ \bibnamefont {{Myers}}}, \bibinfo {author}
  {\bibfnamefont {W.~J.}\ \bibnamefont {{Percival}}}, \bibinfo {author}
  {\bibfnamefont {G.}~\bibnamefont {{Rossi}}}, \bibinfo {author} {\bibfnamefont
  {D.~P.}\ \bibnamefont {{Schneider}}}, \bibinfo {author} {\bibfnamefont
  {P.}~\bibnamefont {{Zarrouk}}}, \ and\ \bibinfo {author} {\bibfnamefont
  {G.-B.}\ \bibnamefont {{Zhao}}},\ }\href {\doibase 10.1093/mnras/staa2825}
  {\bibfield  {journal} {\bibinfo  {journal} {\mnras}\ }\textbf {\bibinfo
  {volume} {499}},\ \bibinfo {pages} {269} (\bibinfo {year} {2020})},\ \Eprint
  {http://arxiv.org/abs/2007.09003} {arXiv:2007.09003 [astro-ph.CO]}
  \BibitemShut {NoStop}%
\bibitem [{\citenamefont {{Vardanyan}}\ \emph {et~al.}(2009)\citenamefont
  {{Vardanyan}}, \citenamefont {{Trotta}},\ and\ \citenamefont
  {{Silk}}}]{vardanyan09}%
  \BibitemOpen
  \bibfield  {author} {\bibinfo {author} {\bibfnamefont {M.}~\bibnamefont
  {{Vardanyan}}}, \bibinfo {author} {\bibfnamefont {R.}~\bibnamefont
  {{Trotta}}}, \ and\ \bibinfo {author} {\bibfnamefont {J.}~\bibnamefont
  {{Silk}}},\ }\href {\doibase 10.1111/j.1365-2966.2009.14938.x} {\bibfield
  {journal} {\bibinfo  {journal} {\mnras}\ }\textbf {\bibinfo {volume} {397}},\
  \bibinfo {pages} {431} (\bibinfo {year} {2009})},\ \Eprint
  {http://arxiv.org/abs/0901.3354} {arXiv:0901.3354 [astro-ph.CO]} \BibitemShut
  {NoStop}%
\bibitem [{\citenamefont {{Akita}}\ and\ \citenamefont
  {{Yamaguchi}}(2020)}]{akita20a}%
  \BibitemOpen
  \bibfield  {author} {\bibinfo {author} {\bibfnamefont {K.}~\bibnamefont
  {{Akita}}}\ and\ \bibinfo {author} {\bibfnamefont {M.}~\bibnamefont
  {{Yamaguchi}}},\ }\href {\doibase 10.1088/1475-7516/2020/08/012} {\bibfield
  {journal} {\bibinfo  {journal} {\jcap}\ }\textbf {\bibinfo {volume} {2020}},\
  \bibinfo {eid} {012} (\bibinfo {year} {2020})},\ \Eprint
  {http://arxiv.org/abs/2005.07047} {arXiv:2005.07047 [hep-ph]} \BibitemShut
  {NoStop}%
\bibitem [{\citenamefont {{Mangano}}\ \emph {et~al.}(2005)\citenamefont
  {{Mangano}}, \citenamefont {{Miele}}, \citenamefont {{Pastor}}, \citenamefont
  {{Pinto}}, \citenamefont {{Pisanti}},\ and\ \citenamefont
  {{Serpico}}}]{mangano05a}%
  \BibitemOpen
  \bibfield  {author} {\bibinfo {author} {\bibfnamefont {G.}~\bibnamefont
  {{Mangano}}}, \bibinfo {author} {\bibfnamefont {G.}~\bibnamefont {{Miele}}},
  \bibinfo {author} {\bibfnamefont {S.}~\bibnamefont {{Pastor}}}, \bibinfo
  {author} {\bibfnamefont {T.}~\bibnamefont {{Pinto}}}, \bibinfo {author}
  {\bibfnamefont {O.}~\bibnamefont {{Pisanti}}}, \ and\ \bibinfo {author}
  {\bibfnamefont {P.~D.}\ \bibnamefont {{Serpico}}},\ }\href {\doibase
  10.1016/j.nuclphysb.2005.09.041} {\bibfield  {journal} {\bibinfo  {journal}
  {Nuclear Physics B}\ }\textbf {\bibinfo {volume} {729}},\ \bibinfo {pages}
  {221} (\bibinfo {year} {2005})},\ \Eprint
  {http://arxiv.org/abs/hep-ph/0506164} {arXiv:hep-ph/0506164 [hep-ph]}
  \BibitemShut {NoStop}%
\bibitem [{\citenamefont {de~Salas}\ \emph {et~al.}(2018)\citenamefont
  {de~Salas}, \citenamefont {Forero}, \citenamefont {Ternes}, \citenamefont
  {Tórtola},\ and\ \citenamefont {Valle}}]{DESALAS2018633}%
  \BibitemOpen
  \bibfield  {author} {\bibinfo {author} {\bibfnamefont {P.}~\bibnamefont
  {de~Salas}}, \bibinfo {author} {\bibfnamefont {D.}~\bibnamefont {Forero}},
  \bibinfo {author} {\bibfnamefont {C.}~\bibnamefont {Ternes}}, \bibinfo
  {author} {\bibfnamefont {M.}~\bibnamefont {Tórtola}}, \ and\ \bibinfo
  {author} {\bibfnamefont {J.}~\bibnamefont {Valle}},\ }\href {\doibase
  https://doi.org/10.1016/j.physletb.2018.06.019} {\bibfield  {journal}
  {\bibinfo  {journal} {Physics Letters B}\ }\textbf {\bibinfo {volume}
  {782}},\ \bibinfo {pages} {633 } (\bibinfo {year} {2018})}\BibitemShut
  {NoStop}%
\bibitem [{\citenamefont {{Lesgourgues}}\ and\ \citenamefont
  {{Pastor}}(2006)}]{2006PhR...429..307L}%
  \BibitemOpen
  \bibfield  {author} {\bibinfo {author} {\bibfnamefont {J.}~\bibnamefont
  {{Lesgourgues}}}\ and\ \bibinfo {author} {\bibfnamefont {S.}~\bibnamefont
  {{Pastor}}},\ }\href {\doibase 10.1016/j.physrep.2006.04.001} {\bibfield
  {journal} {\bibinfo  {journal} {\physrep}\ }\textbf {\bibinfo {volume}
  {429}},\ \bibinfo {pages} {307} (\bibinfo {year} {2006})},\ \Eprint
  {http://arxiv.org/abs/astro-ph/0603494} {arXiv:astro-ph/0603494 [astro-ph]}
  \BibitemShut {NoStop}%
\bibitem [{\citenamefont {{Slosar}}(2006)}]{astro-ph/0602133}%
  \BibitemOpen
  \bibfield  {author} {\bibinfo {author} {\bibfnamefont {A.}~\bibnamefont
  {{Slosar}}},\ }\href {\doibase 10.1103/PhysRevD.73.123501} {\bibfield
  {journal} {\bibinfo  {journal} {\prd}\ }\textbf {\bibinfo {volume} {73}},\
  \bibinfo {eid} {123501} (\bibinfo {year} {2006})},\ \Eprint
  {http://arxiv.org/abs/astro-ph/0602133} {arXiv:astro-ph/0602133 [astro-ph]}
  \BibitemShut {NoStop}%
\bibitem [{\citenamefont {{Font-Ribera}}\ \emph {et~al.}(2014)\citenamefont
  {{Font-Ribera}}, \citenamefont {{McDonald}}, \citenamefont {{Mostek}},
  \citenamefont {{Reid}}, \citenamefont {{Seo}},\ and\ \citenamefont
  {{Slosar}}}]{2014JCAP...05..023F}%
  \BibitemOpen
  \bibfield  {author} {\bibinfo {author} {\bibfnamefont {A.}~\bibnamefont
  {{Font-Ribera}}}, \bibinfo {author} {\bibfnamefont {P.}~\bibnamefont
  {{McDonald}}}, \bibinfo {author} {\bibfnamefont {N.}~\bibnamefont
  {{Mostek}}}, \bibinfo {author} {\bibfnamefont {B.~A.}\ \bibnamefont
  {{Reid}}}, \bibinfo {author} {\bibfnamefont {H.-J.}\ \bibnamefont {{Seo}}}, \
  and\ \bibinfo {author} {\bibfnamefont {A.}~\bibnamefont {{Slosar}}},\ }\href
  {\doibase 10.1088/1475-7516/2014/05/023} {\bibfield  {journal} {\bibinfo
  {journal} {\jcap}\ }\textbf {\bibinfo {volume} {2014}},\ \bibinfo {eid} {023}
  (\bibinfo {year} {2014})},\ \Eprint {http://arxiv.org/abs/1308.4164}
  {arXiv:1308.4164 [astro-ph.CO]} \BibitemShut {NoStop}%
\bibitem [{\citenamefont {{de Bernardis}}\ \emph {et~al.}(2009)\citenamefont
  {{de Bernardis}}, \citenamefont {{Kitching}}, \citenamefont {{Heavens}},\
  and\ \citenamefont {{Melchiorri}}}]{2009PhRvD..80l3509D}%
  \BibitemOpen
  \bibfield  {author} {\bibinfo {author} {\bibfnamefont {F.}~\bibnamefont {{de
  Bernardis}}}, \bibinfo {author} {\bibfnamefont {T.~D.}\ \bibnamefont
  {{Kitching}}}, \bibinfo {author} {\bibfnamefont {A.}~\bibnamefont
  {{Heavens}}}, \ and\ \bibinfo {author} {\bibfnamefont {A.}~\bibnamefont
  {{Melchiorri}}},\ }\href {\doibase 10.1103/PhysRevD.80.123509} {\bibfield
  {journal} {\bibinfo  {journal} {\prd}\ }\textbf {\bibinfo {volume} {80}},\
  \bibinfo {eid} {123509} (\bibinfo {year} {2009})},\ \Eprint
  {http://arxiv.org/abs/0907.1917} {arXiv:0907.1917 [astro-ph.CO]} \BibitemShut
  {NoStop}%
\bibitem [{\citenamefont {{Jimenez}}\ \emph {et~al.}(2010)\citenamefont
  {{Jimenez}}, \citenamefont {{Kitching}}, \citenamefont {{Pe{\~n}a-Garay}},\
  and\ \citenamefont {{Verde}}}]{2010JCAP...05..035J}%
  \BibitemOpen
  \bibfield  {author} {\bibinfo {author} {\bibfnamefont {R.}~\bibnamefont
  {{Jimenez}}}, \bibinfo {author} {\bibfnamefont {T.}~\bibnamefont
  {{Kitching}}}, \bibinfo {author} {\bibfnamefont {C.}~\bibnamefont
  {{Pe{\~n}a-Garay}}}, \ and\ \bibinfo {author} {\bibfnamefont
  {L.}~\bibnamefont {{Verde}}},\ }\href {\doibase
  10.1088/1475-7516/2010/05/035} {\bibfield  {journal} {\bibinfo  {journal}
  {\jcap}\ }\textbf {\bibinfo {volume} {2010}},\ \bibinfo {eid} {035} (\bibinfo
  {year} {2010})},\ \Eprint {http://arxiv.org/abs/1003.5918} {arXiv:1003.5918
  [astro-ph.CO]} \BibitemShut {NoStop}%
\bibitem [{\citenamefont {{Wang}}\ and\ \citenamefont
  {{Steinhardt}}(1998)}]{wang98a}%
  \BibitemOpen
  \bibfield  {author} {\bibinfo {author} {\bibfnamefont {L.}~\bibnamefont
  {{Wang}}}\ and\ \bibinfo {author} {\bibfnamefont {P.~J.}\ \bibnamefont
  {{Steinhardt}}},\ }\href {\doibase 10.1086/306436} {\bibfield  {journal}
  {\bibinfo  {journal} {\apj}\ }\textbf {\bibinfo {volume} {508}},\ \bibinfo
  {pages} {483} (\bibinfo {year} {1998})},\ \Eprint
  {http://arxiv.org/abs/astro-ph/9804015} {arXiv:astro-ph/9804015 [astro-ph]}
  \BibitemShut {NoStop}%
\bibitem [{\citenamefont {{Linder}}(2005)}]{linder05a}%
  \BibitemOpen
  \bibfield  {author} {\bibinfo {author} {\bibfnamefont {E.~V.}\ \bibnamefont
  {{Linder}}},\ }\href {\doibase 10.1103/PhysRevD.72.043529} {\bibfield
  {journal} {\bibinfo  {journal} {\prd}\ }\textbf {\bibinfo {volume} {72}},\
  \bibinfo {eid} {043529} (\bibinfo {year} {2005})},\ \Eprint
  {http://arxiv.org/abs/arXiv:astro-ph/0507263} {arXiv:astro-ph/0507263}
  \BibitemShut {NoStop}%
\bibitem [{\citenamefont {{Linder}}\ and\ \citenamefont
  {{Cahn}}(2007)}]{linder07a}%
  \BibitemOpen
  \bibfield  {author} {\bibinfo {author} {\bibfnamefont {E.~V.}\ \bibnamefont
  {{Linder}}}\ and\ \bibinfo {author} {\bibfnamefont {R.~N.}\ \bibnamefont
  {{Cahn}}},\ }\href {\doibase 10.1016/j.astropartphys.2007.09.003} {\bibfield
  {journal} {\bibinfo  {journal} {Astroparticle Physics}\ }\textbf {\bibinfo
  {volume} {28}},\ \bibinfo {pages} {481} (\bibinfo {year} {2007})},\ \Eprint
  {http://arxiv.org/abs/astro-ph/0701317} {arXiv:astro-ph/0701317 [astro-ph]}
  \BibitemShut {NoStop}%
\bibitem [{\citenamefont {{Abbott}}\ \emph {et~al.}(2019)\citenamefont
  {{Abbott}}, \citenamefont {{Abdalla}}, \citenamefont {{Avila}}, \citenamefont
  {{Banerji}}, \citenamefont {{Baxter}}, \citenamefont {{Bechtol}},
  \citenamefont {{Becker}}, \citenamefont {{Bertin}}, \citenamefont {{Blazek}},
  \citenamefont {{Bridle}}, \citenamefont {{Brooks}}, \citenamefont {{Brout}},
  \citenamefont {{Burke}}, \citenamefont {{Campos}}, \citenamefont {{Carnero
  Rosell}}, \citenamefont {{Carrasco Kind}}, \citenamefont {{Carretero}},
  \citenamefont {{Castander}}, \citenamefont {{Cawthon}}, \citenamefont
  {{Chang}}, \citenamefont {{Chen}}, \citenamefont {{Crocce}}, \citenamefont
  {{Cunha}}, \citenamefont {{da Costa}}, \citenamefont {{Davis}}, \citenamefont
  {{De Vicente}}, \citenamefont {{DeRose}}, \citenamefont {{Desai}},
  \citenamefont {{Di Valentino}}, \citenamefont {{Diehl}}, \citenamefont
  {{Dietrich}}, \citenamefont {{Dodelson}}, \citenamefont {{Doel}},
  \citenamefont {{Drlica-Wagner}}, \citenamefont {{Eifler}}, \citenamefont
  {{Elvin-Poole}}, \citenamefont {{Evrard}}, \citenamefont {{Fernand ez}},
  \citenamefont {{Fert{\'e}}}, \citenamefont {{Flaugher}}, \citenamefont
  {{Fosalba}}, \citenamefont {{Frieman}}, \citenamefont
  {{Garc{\'\i}a-Bellido}}, \citenamefont {{Gaztanaga}}, \citenamefont
  {{Gerdes}}, \citenamefont {{Giannantonio}}, \citenamefont {{Gruen}},
  \citenamefont {{Gruendl}}, \citenamefont {{Gschwend}}, \citenamefont
  {{Gutierrez}}, \citenamefont {{Hartley}}, \citenamefont {{Hollowood}},
  \citenamefont {{Honscheid}}, \citenamefont {{Hoyle}}, \citenamefont
  {{Huterer}}, \citenamefont {{Jain}}, \citenamefont {{Jeltema}}, \citenamefont
  {{Johnson}}, \citenamefont {{Johnson}}, \citenamefont {{Kim}}, \citenamefont
  {{Krause}}, \citenamefont {{Kuehn}}, \citenamefont {{Kuropatkin}},
  \citenamefont {{Lahav}}, \citenamefont {{Lee}}, \citenamefont {{Lemos}},
  \citenamefont {{Leonard}}, \citenamefont {{Li}}, \citenamefont {{Liddle}},
  \citenamefont {{Lima}}, \citenamefont {{Lin}}, \citenamefont {{Maia}},
  \citenamefont {{Marshall}}, \citenamefont {{Martini}}, \citenamefont
  {{Menanteau}}, \citenamefont {{Miller}}, \citenamefont {{Miquel}},
  \citenamefont {{Miranda}}, \citenamefont {{Mohr}}, \citenamefont {{Muir}},
  \citenamefont {{Nichol}}, \citenamefont {{Nord}}, \citenamefont {{Ogando}},
  \citenamefont {{Plazas}}, \citenamefont {{Raveri}}, \citenamefont
  {{Rollins}}, \citenamefont {{Romer}}, \citenamefont {{Roodman}},
  \citenamefont {{Rosenfeld}}, \citenamefont {{Samuroff}}, \citenamefont
  {{Sanchez}}, \citenamefont {{Scarpine}}, \citenamefont {{Schindler}},
  \citenamefont {{Schubnell}}, \citenamefont {{Scolnic}}, \citenamefont
  {{Secco}}, \citenamefont {{Serrano}}, \citenamefont {{Sevilla-Noarbe}},
  \citenamefont {{Smith}}, \citenamefont {{Soares-Santos}}, \citenamefont
  {{Sobreira}}, \citenamefont {{Suchyta}}, \citenamefont {{Swanson}},
  \citenamefont {{Tarle}}, \citenamefont {{Thomas}}, \citenamefont {{Troxel}},
  \citenamefont {{Vikram}}, \citenamefont {{Walker}}, \citenamefont
  {{Weaverdyck}}, \citenamefont {{Wechsler}}, \citenamefont {{Weller}},
  \citenamefont {{Yanny}}, \citenamefont {{Zhang}}, \citenamefont {{Zuntz}},\
  and\ \citenamefont {{DES Collaboration}}}]{2019PhRvD..99l3505A}%
  \BibitemOpen
  \bibfield  {author} {\bibinfo {author} {\bibfnamefont {T.~M.~C.}\
  \bibnamefont {{Abbott}}}, \bibinfo {author} {\bibfnamefont {F.~B.}\
  \bibnamefont {{Abdalla}}}, \bibinfo {author} {\bibfnamefont {S.}~\bibnamefont
  {{Avila}}}, \bibinfo {author} {\bibfnamefont {M.}~\bibnamefont {{Banerji}}},
  \bibinfo {author} {\bibfnamefont {E.}~\bibnamefont {{Baxter}}}, \bibinfo
  {author} {\bibfnamefont {K.}~\bibnamefont {{Bechtol}}}, \bibinfo {author}
  {\bibfnamefont {M.~R.}\ \bibnamefont {{Becker}}}, \bibinfo {author}
  {\bibfnamefont {E.}~\bibnamefont {{Bertin}}}, \bibinfo {author}
  {\bibfnamefont {J.}~\bibnamefont {{Blazek}}}, \bibinfo {author}
  {\bibfnamefont {S.~L.}\ \bibnamefont {{Bridle}}}, \bibinfo {author}
  {\bibfnamefont {D.}~\bibnamefont {{Brooks}}}, \bibinfo {author}
  {\bibfnamefont {D.}~\bibnamefont {{Brout}}}, \bibinfo {author} {\bibfnamefont
  {D.~L.}\ \bibnamefont {{Burke}}}, \bibinfo {author} {\bibfnamefont
  {A.}~\bibnamefont {{Campos}}}, \bibinfo {author} {\bibfnamefont
  {A.}~\bibnamefont {{Carnero Rosell}}}, \bibinfo {author} {\bibfnamefont
  {M.}~\bibnamefont {{Carrasco Kind}}}, \bibinfo {author} {\bibfnamefont
  {J.}~\bibnamefont {{Carretero}}}, \bibinfo {author} {\bibfnamefont {F.~J.}\
  \bibnamefont {{Castander}}}, \bibinfo {author} {\bibfnamefont
  {R.}~\bibnamefont {{Cawthon}}}, \bibinfo {author} {\bibfnamefont
  {C.}~\bibnamefont {{Chang}}}, \bibinfo {author} {\bibfnamefont
  {A.}~\bibnamefont {{Chen}}}, \bibinfo {author} {\bibfnamefont
  {M.}~\bibnamefont {{Crocce}}}, \bibinfo {author} {\bibfnamefont {C.~E.}\
  \bibnamefont {{Cunha}}}, \bibinfo {author} {\bibfnamefont {L.~N.}\
  \bibnamefont {{da Costa}}}, \bibinfo {author} {\bibfnamefont
  {C.}~\bibnamefont {{Davis}}}, \bibinfo {author} {\bibfnamefont
  {J.}~\bibnamefont {{De Vicente}}}, \bibinfo {author} {\bibfnamefont
  {J.}~\bibnamefont {{DeRose}}}, \bibinfo {author} {\bibfnamefont
  {S.}~\bibnamefont {{Desai}}}, \bibinfo {author} {\bibfnamefont
  {E.}~\bibnamefont {{Di Valentino}}}, \bibinfo {author} {\bibfnamefont
  {H.~T.}\ \bibnamefont {{Diehl}}}, \bibinfo {author} {\bibfnamefont {J.~P.}\
  \bibnamefont {{Dietrich}}}, \bibinfo {author} {\bibfnamefont
  {S.}~\bibnamefont {{Dodelson}}}, \bibinfo {author} {\bibfnamefont
  {P.}~\bibnamefont {{Doel}}}, \bibinfo {author} {\bibfnamefont
  {A.}~\bibnamefont {{Drlica-Wagner}}}, \bibinfo {author} {\bibfnamefont
  {T.~F.}\ \bibnamefont {{Eifler}}}, \bibinfo {author} {\bibfnamefont
  {J.}~\bibnamefont {{Elvin-Poole}}}, \bibinfo {author} {\bibfnamefont {A.~E.}\
  \bibnamefont {{Evrard}}}, \bibinfo {author} {\bibfnamefont {E.}~\bibnamefont
  {{Fernand ez}}}, \bibinfo {author} {\bibfnamefont {A.}~\bibnamefont
  {{Fert{\'e}}}}, \bibinfo {author} {\bibfnamefont {B.}~\bibnamefont
  {{Flaugher}}}, \bibinfo {author} {\bibfnamefont {P.}~\bibnamefont
  {{Fosalba}}}, \bibinfo {author} {\bibfnamefont {J.}~\bibnamefont
  {{Frieman}}}, \bibinfo {author} {\bibfnamefont {J.}~\bibnamefont
  {{Garc{\'\i}a-Bellido}}}, \bibinfo {author} {\bibfnamefont {E.}~\bibnamefont
  {{Gaztanaga}}}, \bibinfo {author} {\bibfnamefont {D.~W.}\ \bibnamefont
  {{Gerdes}}}, \bibinfo {author} {\bibfnamefont {T.}~\bibnamefont
  {{Giannantonio}}}, \bibinfo {author} {\bibfnamefont {D.}~\bibnamefont
  {{Gruen}}}, \bibinfo {author} {\bibfnamefont {R.~A.}\ \bibnamefont
  {{Gruendl}}}, \bibinfo {author} {\bibfnamefont {J.}~\bibnamefont
  {{Gschwend}}}, \bibinfo {author} {\bibfnamefont {G.}~\bibnamefont
  {{Gutierrez}}}, \bibinfo {author} {\bibfnamefont {W.~G.}\ \bibnamefont
  {{Hartley}}}, \bibinfo {author} {\bibfnamefont {D.~L.}\ \bibnamefont
  {{Hollowood}}}, \bibinfo {author} {\bibfnamefont {K.}~\bibnamefont
  {{Honscheid}}}, \bibinfo {author} {\bibfnamefont {B.}~\bibnamefont
  {{Hoyle}}}, \bibinfo {author} {\bibfnamefont {D.}~\bibnamefont {{Huterer}}},
  \bibinfo {author} {\bibfnamefont {B.}~\bibnamefont {{Jain}}}, \bibinfo
  {author} {\bibfnamefont {T.}~\bibnamefont {{Jeltema}}}, \bibinfo {author}
  {\bibfnamefont {M.~W.~G.}\ \bibnamefont {{Johnson}}}, \bibinfo {author}
  {\bibfnamefont {M.~D.}\ \bibnamefont {{Johnson}}}, \bibinfo {author}
  {\bibfnamefont {A.~G.}\ \bibnamefont {{Kim}}}, \bibinfo {author}
  {\bibfnamefont {E.}~\bibnamefont {{Krause}}}, \bibinfo {author}
  {\bibfnamefont {K.}~\bibnamefont {{Kuehn}}}, \bibinfo {author} {\bibfnamefont
  {N.}~\bibnamefont {{Kuropatkin}}}, \bibinfo {author} {\bibfnamefont
  {O.}~\bibnamefont {{Lahav}}}, \bibinfo {author} {\bibfnamefont
  {S.}~\bibnamefont {{Lee}}}, \bibinfo {author} {\bibfnamefont
  {P.}~\bibnamefont {{Lemos}}}, \bibinfo {author} {\bibfnamefont {C.~D.}\
  \bibnamefont {{Leonard}}}, \bibinfo {author} {\bibfnamefont {T.~S.}\
  \bibnamefont {{Li}}}, \bibinfo {author} {\bibfnamefont {A.~R.}\ \bibnamefont
  {{Liddle}}}, \bibinfo {author} {\bibfnamefont {M.}~\bibnamefont {{Lima}}},
  \bibinfo {author} {\bibfnamefont {H.}~\bibnamefont {{Lin}}}, \bibinfo
  {author} {\bibfnamefont {M.~A.~G.}\ \bibnamefont {{Maia}}}, \bibinfo {author}
  {\bibfnamefont {J.~L.}\ \bibnamefont {{Marshall}}}, \bibinfo {author}
  {\bibfnamefont {P.}~\bibnamefont {{Martini}}}, \bibinfo {author}
  {\bibfnamefont {F.}~\bibnamefont {{Menanteau}}}, \bibinfo {author}
  {\bibfnamefont {C.~J.}\ \bibnamefont {{Miller}}}, \bibinfo {author}
  {\bibfnamefont {R.}~\bibnamefont {{Miquel}}}, \bibinfo {author}
  {\bibfnamefont {V.}~\bibnamefont {{Miranda}}}, \bibinfo {author}
  {\bibfnamefont {J.~J.}\ \bibnamefont {{Mohr}}}, \bibinfo {author}
  {\bibfnamefont {J.}~\bibnamefont {{Muir}}}, \bibinfo {author} {\bibfnamefont
  {R.~C.}\ \bibnamefont {{Nichol}}}, \bibinfo {author} {\bibfnamefont
  {B.}~\bibnamefont {{Nord}}}, \bibinfo {author} {\bibfnamefont {R.~L.~C.}\
  \bibnamefont {{Ogando}}}, \bibinfo {author} {\bibfnamefont {A.~A.}\
  \bibnamefont {{Plazas}}}, \bibinfo {author} {\bibfnamefont {M.}~\bibnamefont
  {{Raveri}}}, \bibinfo {author} {\bibfnamefont {R.~P.}\ \bibnamefont
  {{Rollins}}}, \bibinfo {author} {\bibfnamefont {A.~K.}\ \bibnamefont
  {{Romer}}}, \bibinfo {author} {\bibfnamefont {A.}~\bibnamefont {{Roodman}}},
  \bibinfo {author} {\bibfnamefont {R.}~\bibnamefont {{Rosenfeld}}}, \bibinfo
  {author} {\bibfnamefont {S.}~\bibnamefont {{Samuroff}}}, \bibinfo {author}
  {\bibfnamefont {E.}~\bibnamefont {{Sanchez}}}, \bibinfo {author}
  {\bibfnamefont {V.}~\bibnamefont {{Scarpine}}}, \bibinfo {author}
  {\bibfnamefont {R.}~\bibnamefont {{Schindler}}}, \bibinfo {author}
  {\bibfnamefont {M.}~\bibnamefont {{Schubnell}}}, \bibinfo {author}
  {\bibfnamefont {D.}~\bibnamefont {{Scolnic}}}, \bibinfo {author}
  {\bibfnamefont {L.~F.}\ \bibnamefont {{Secco}}}, \bibinfo {author}
  {\bibfnamefont {S.}~\bibnamefont {{Serrano}}}, \bibinfo {author}
  {\bibfnamefont {I.}~\bibnamefont {{Sevilla-Noarbe}}}, \bibinfo {author}
  {\bibfnamefont {M.}~\bibnamefont {{Smith}}}, \bibinfo {author} {\bibfnamefont
  {M.}~\bibnamefont {{Soares-Santos}}}, \bibinfo {author} {\bibfnamefont
  {F.}~\bibnamefont {{Sobreira}}}, \bibinfo {author} {\bibfnamefont
  {E.}~\bibnamefont {{Suchyta}}}, \bibinfo {author} {\bibfnamefont {M.~E.~C.}\
  \bibnamefont {{Swanson}}}, \bibinfo {author} {\bibfnamefont {G.}~\bibnamefont
  {{Tarle}}}, \bibinfo {author} {\bibfnamefont {D.}~\bibnamefont {{Thomas}}},
  \bibinfo {author} {\bibfnamefont {M.~A.}\ \bibnamefont {{Troxel}}}, \bibinfo
  {author} {\bibfnamefont {V.}~\bibnamefont {{Vikram}}}, \bibinfo {author}
  {\bibfnamefont {A.~R.}\ \bibnamefont {{Walker}}}, \bibinfo {author}
  {\bibfnamefont {N.}~\bibnamefont {{Weaverdyck}}}, \bibinfo {author}
  {\bibfnamefont {R.~H.}\ \bibnamefont {{Wechsler}}}, \bibinfo {author}
  {\bibfnamefont {J.}~\bibnamefont {{Weller}}}, \bibinfo {author}
  {\bibfnamefont {B.}~\bibnamefont {{Yanny}}}, \bibinfo {author} {\bibfnamefont
  {Y.}~\bibnamefont {{Zhang}}}, \bibinfo {author} {\bibfnamefont
  {J.}~\bibnamefont {{Zuntz}}}, \ and\ \bibinfo {author} {\bibnamefont {{DES
  Collaboration}}},\ }\href {\doibase 10.1103/PhysRevD.99.123505} {\bibfield
  {journal} {\bibinfo  {journal} {\prd}\ }\textbf {\bibinfo {volume} {99}},\
  \bibinfo {eid} {123505} (\bibinfo {year} {2019})},\ \Eprint
  {http://arxiv.org/abs/1810.02499} {arXiv:1810.02499 [astro-ph.CO]}
  \BibitemShut {NoStop}%
\bibitem [{\citenamefont {{Zhao}}\ \emph {et~al.}(2009)\citenamefont {{Zhao}},
  \citenamefont {{Pogosian}}, \citenamefont {{Silvestri}},\ and\ \citenamefont
  {{Zylberberg}}}]{zhao09a}%
  \BibitemOpen
  \bibfield  {author} {\bibinfo {author} {\bibfnamefont {G.-B.}\ \bibnamefont
  {{Zhao}}}, \bibinfo {author} {\bibfnamefont {L.}~\bibnamefont {{Pogosian}}},
  \bibinfo {author} {\bibfnamefont {A.}~\bibnamefont {{Silvestri}}}, \ and\
  \bibinfo {author} {\bibfnamefont {J.}~\bibnamefont {{Zylberberg}}},\ }\href
  {\doibase 10.1103/PhysRevD.79.083513} {\bibfield  {journal} {\bibinfo
  {journal} {\prd}\ }\textbf {\bibinfo {volume} {79}},\ \bibinfo {eid} {083513}
  (\bibinfo {year} {2009})},\ \Eprint {http://arxiv.org/abs/0809.3791}
  {arXiv:0809.3791 [astro-ph]} \BibitemShut {NoStop}%
\bibitem [{\citenamefont {{Staggs}}\ \emph {et~al.}(2018)\citenamefont
  {{Staggs}}, \citenamefont {{Dunkley}},\ and\ \citenamefont
  {{Page}}}]{staggs18a}%
  \BibitemOpen
  \bibfield  {author} {\bibinfo {author} {\bibfnamefont {S.}~\bibnamefont
  {{Staggs}}}, \bibinfo {author} {\bibfnamefont {J.}~\bibnamefont {{Dunkley}}},
  \ and\ \bibinfo {author} {\bibfnamefont {L.}~\bibnamefont {{Page}}},\ }\href
  {\doibase 10.1088/1361-6633/aa94d5} {\bibfield  {journal} {\bibinfo
  {journal} {Reports on Progress in Physics}\ }\textbf {\bibinfo {volume}
  {81}},\ \bibinfo {eid} {044901} (\bibinfo {year} {2018})}\BibitemShut
  {NoStop}%
\bibitem [{\citenamefont {{Planck Collaboration}}\ \emph
  {et~al.}(2018{\natexlab{c}})\citenamefont {{Planck Collaboration}},
  \citenamefont {{Akrami}}, \citenamefont {{Arroja}}, \citenamefont
  {{Ashdown}}, \citenamefont {{Aumont}}, \citenamefont {{Baccigalupi}},
  \citenamefont {{Ballardini}}, \citenamefont {{Banday}}, \citenamefont
  {{Barreiro}}, \citenamefont {{Bartolo}}, \citenamefont {{Basak}},
  \citenamefont {{Battye}}, \citenamefont {{Benabed}}, \citenamefont
  {{Bernard}}, \citenamefont {{Bersanelli}}, \citenamefont {{Bielewicz}},
  \citenamefont {{Bock}}, \citenamefont {{Bond}}, \citenamefont {{Borrill}},
  \citenamefont {{Bouchet}}, \citenamefont {{Boulanger}}, \citenamefont
  {{Bucher}}, \citenamefont {{Burigana}}, \citenamefont {{Butler}},
  \citenamefont {{Calabrese}}, \citenamefont {{Cardoso}}, \citenamefont
  {{Carron}}, \citenamefont {{Casaponsa}}, \citenamefont {{Challinor}},
  \citenamefont {{Chiang}}, \citenamefont {{Colombo}}, \citenamefont
  {{Combet}}, \citenamefont {{Contreras}}, \citenamefont {{Crill}},
  \citenamefont {{Cuttaia}}, \citenamefont {{de Bernardis}}, \citenamefont {{de
  Zotti}}, \citenamefont {{Delabrouille}}, \citenamefont {{Delouis}},
  \citenamefont {{D{\'e}sert}}, \citenamefont {{Di Valentino}}, \citenamefont
  {{Dickinson}}, \citenamefont {{Diego}}, \citenamefont {{Donzelli}},
  \citenamefont {{Dor{\'e}}}, \citenamefont {{Douspis}}, \citenamefont
  {{Ducout}}, \citenamefont {{Dupac}}, \citenamefont {{Efstathiou}},
  \citenamefont {{Elsner}}, \citenamefont {{En{\ss}lin}}, \citenamefont
  {{Eriksen}}, \citenamefont {{Falgarone}}, \citenamefont {{Fantaye}},
  \citenamefont {{Fergusson}}, \citenamefont {{Fernandez-Cobos}}, \citenamefont
  {{Finelli}}, \citenamefont {{Forastieri}}, \citenamefont {{Frailis}},
  \citenamefont {{Franceschi}}, \citenamefont {{Frolov}}, \citenamefont
  {{Galeotta}}, \citenamefont {{Galli}}, \citenamefont {{Ganga}}, \citenamefont
  {{G{\'e}nova-Santos}}, \citenamefont {{Gerbino}}, \citenamefont {{Ghosh}},
  \citenamefont {{Gonz{\'a}lez-Nuevo}}, \citenamefont {{G{\'o}rski}},
  \citenamefont {{Gratton}}, \citenamefont {{Gruppuso}}, \citenamefont
  {{Gudmundsson}}, \citenamefont {{Hamann}}, \citenamefont {{Hand ley}},
  \citenamefont {{Hansen}}, \citenamefont {{Helou}}, \citenamefont {{Herranz}},
  \citenamefont {{Hivon}}, \citenamefont {{Huang}}, \citenamefont {{Jaffe}},
  \citenamefont {{Jones}}, \citenamefont {{Karakci}}, \citenamefont
  {{Keih{\"a}nen}}, \citenamefont {{Keskitalo}}, \citenamefont {{Kiiveri}},
  \citenamefont {{Kim}}, \citenamefont {{Kisner}}, \citenamefont {{Knox}},
  \citenamefont {{Krachmalnicoff}}, \citenamefont {{Kunz}}, \citenamefont
  {{Kurki-Suonio}}, \citenamefont {{Lagache}}, \citenamefont {{Lamarre}},
  \citenamefont {{Langer}}, \citenamefont {{Lasenby}}, \citenamefont
  {{Lattanzi}}, \citenamefont {{Lawrence}}, \citenamefont {{Le Jeune}},
  \citenamefont {{Leahy}}, \citenamefont {{Lesgourgues}}, \citenamefont
  {{Levrier}}, \citenamefont {{Lewis}}, \citenamefont {{Liguori}},
  \citenamefont {{Lilje}}, \citenamefont {{Lilley}}, \citenamefont
  {{Lindholm}}, \citenamefont {{L{\'o}pez-Caniego}}, \citenamefont {{Lubin}},
  \citenamefont {{Ma}}, \citenamefont {{Mac{\'\i}as-P{\'e}rez}}, \citenamefont
  {{Maggio}}, \citenamefont {{Maino}}, \citenamefont {{Mand olesi}},
  \citenamefont {{Mangilli}}, \citenamefont {{Marcos-Caballero}}, \citenamefont
  {{Maris}}, \citenamefont {{Martin}}, \citenamefont
  {{Mart{\'\i}nez-Gonz{\'a}lez}}, \citenamefont {{Matarrese}}, \citenamefont
  {{Mauri}}, \citenamefont {{McEwen}}, \citenamefont {{Meerburg}},
  \citenamefont {{Meinhold}}, \citenamefont {{Melchiorri}}, \citenamefont
  {{Mennella}}, \citenamefont {{Migliaccio}}, \citenamefont {{Millea}},
  \citenamefont {{Mitra}}, \citenamefont {{Miville-Desch{\^e}nes}},
  \citenamefont {{Molinari}}, \citenamefont {{Moneti}}, \citenamefont
  {{Montier}}, \citenamefont {{Morgante}}, \citenamefont {{Moss}},
  \citenamefont {{Mottet}}, \citenamefont {{M{\"u}nchmeyer}}, \citenamefont
  {{Natoli}}, \citenamefont {{N{\o}rgaard-Nielsen}}, \citenamefont
  {{Oxborrow}}, \citenamefont {{Pagano}}, \citenamefont {{Paoletti}},
  \citenamefont {{Partridge}}, \citenamefont {{Patanchon}}, \citenamefont
  {{Pearson}}, \citenamefont {{Peel}}, \citenamefont {{Peiris}}, \citenamefont
  {{Perrotta}}, \citenamefont {{Pettorino}}, \citenamefont {{Piacentini}},
  \citenamefont {{Polastri}}, \citenamefont {{Polenta}}, \citenamefont
  {{Puget}}, \citenamefont {{Rachen}}, \citenamefont {{Reinecke}},
  \citenamefont {{Remazeilles}}, \citenamefont {{Renzi}}, \citenamefont
  {{Rocha}}, \citenamefont {{Rosset}}, \citenamefont {{Roudier}}, \citenamefont
  {{Rubi{\~n}o-Mart{\'\i}n}}, \citenamefont {{Ruiz-Granados}}, \citenamefont
  {{Salvati}}, \citenamefont {{Sandri}}, \citenamefont {{Savelainen}},
  \citenamefont {{Scott}}, \citenamefont {{Shellard}}, \citenamefont
  {{Shiraishi}}, \citenamefont {{Sirignano}}, \citenamefont {{Sirri}},
  \citenamefont {{Spencer}}, \citenamefont {{Sunyaev}}, \citenamefont
  {{Suur-Uski}}, \citenamefont {{Tauber}}, \citenamefont {{Tavagnacco}},
  \citenamefont {{Tenti}}, \citenamefont {{Terenzi}}, \citenamefont
  {{Toffolatti}}, \citenamefont {{Tomasi}}, \citenamefont {{Trombetti}},
  \citenamefont {{Valiviita}}, \citenamefont {{Van Tent}}, \citenamefont
  {{Vibert}}, \citenamefont {{Vielva}}, \citenamefont {{Villa}}, \citenamefont
  {{Vittorio}}, \citenamefont {{Wandelt}}, \citenamefont {{Wehus}},
  \citenamefont {{White}}, \citenamefont {{White}}, \citenamefont {{Zacchei}},\
  and\ \citenamefont {{Zonca}}}]{2018arXiv180706205P}%
  \BibitemOpen
  \bibfield  {author} {\bibinfo {author} {\bibnamefont {{Planck
  Collaboration}}}, \bibinfo {author} {\bibfnamefont {Y.}~\bibnamefont
  {{Akrami}}}, \bibinfo {author} {\bibfnamefont {F.}~\bibnamefont {{Arroja}}},
  \bibinfo {author} {\bibfnamefont {M.}~\bibnamefont {{Ashdown}}}, \bibinfo
  {author} {\bibfnamefont {J.}~\bibnamefont {{Aumont}}}, \bibinfo {author}
  {\bibfnamefont {C.}~\bibnamefont {{Baccigalupi}}}, \bibinfo {author}
  {\bibfnamefont {M.}~\bibnamefont {{Ballardini}}}, \bibinfo {author}
  {\bibfnamefont {A.~J.}\ \bibnamefont {{Banday}}}, \bibinfo {author}
  {\bibfnamefont {R.~B.}\ \bibnamefont {{Barreiro}}}, \bibinfo {author}
  {\bibfnamefont {N.}~\bibnamefont {{Bartolo}}}, \bibinfo {author}
  {\bibfnamefont {S.}~\bibnamefont {{Basak}}}, \bibinfo {author} {\bibfnamefont
  {R.}~\bibnamefont {{Battye}}}, \bibinfo {author} {\bibfnamefont
  {K.}~\bibnamefont {{Benabed}}}, \bibinfo {author} {\bibfnamefont {J.~P.}\
  \bibnamefont {{Bernard}}}, \bibinfo {author} {\bibfnamefont {M.}~\bibnamefont
  {{Bersanelli}}}, \bibinfo {author} {\bibfnamefont {P.}~\bibnamefont
  {{Bielewicz}}}, \bibinfo {author} {\bibfnamefont {J.~J.}\ \bibnamefont
  {{Bock}}}, \bibinfo {author} {\bibfnamefont {J.~R.}\ \bibnamefont {{Bond}}},
  \bibinfo {author} {\bibfnamefont {J.}~\bibnamefont {{Borrill}}}, \bibinfo
  {author} {\bibfnamefont {F.~R.}\ \bibnamefont {{Bouchet}}}, \bibinfo {author}
  {\bibfnamefont {F.}~\bibnamefont {{Boulanger}}}, \bibinfo {author}
  {\bibfnamefont {M.}~\bibnamefont {{Bucher}}}, \bibinfo {author}
  {\bibfnamefont {C.}~\bibnamefont {{Burigana}}}, \bibinfo {author}
  {\bibfnamefont {R.~C.}\ \bibnamefont {{Butler}}}, \bibinfo {author}
  {\bibfnamefont {E.}~\bibnamefont {{Calabrese}}}, \bibinfo {author}
  {\bibfnamefont {J.~F.}\ \bibnamefont {{Cardoso}}}, \bibinfo {author}
  {\bibfnamefont {J.}~\bibnamefont {{Carron}}}, \bibinfo {author}
  {\bibfnamefont {B.}~\bibnamefont {{Casaponsa}}}, \bibinfo {author}
  {\bibfnamefont {A.}~\bibnamefont {{Challinor}}}, \bibinfo {author}
  {\bibfnamefont {H.~C.}\ \bibnamefont {{Chiang}}}, \bibinfo {author}
  {\bibfnamefont {L.~P.~L.}\ \bibnamefont {{Colombo}}}, \bibinfo {author}
  {\bibfnamefont {C.}~\bibnamefont {{Combet}}}, \bibinfo {author}
  {\bibfnamefont {D.}~\bibnamefont {{Contreras}}}, \bibinfo {author}
  {\bibfnamefont {B.~P.}\ \bibnamefont {{Crill}}}, \bibinfo {author}
  {\bibfnamefont {F.}~\bibnamefont {{Cuttaia}}}, \bibinfo {author}
  {\bibfnamefont {P.}~\bibnamefont {{de Bernardis}}}, \bibinfo {author}
  {\bibfnamefont {G.}~\bibnamefont {{de Zotti}}}, \bibinfo {author}
  {\bibfnamefont {J.}~\bibnamefont {{Delabrouille}}}, \bibinfo {author}
  {\bibfnamefont {J.~M.}\ \bibnamefont {{Delouis}}}, \bibinfo {author}
  {\bibfnamefont {F.~X.}\ \bibnamefont {{D{\'e}sert}}}, \bibinfo {author}
  {\bibfnamefont {E.}~\bibnamefont {{Di Valentino}}}, \bibinfo {author}
  {\bibfnamefont {C.}~\bibnamefont {{Dickinson}}}, \bibinfo {author}
  {\bibfnamefont {J.~M.}\ \bibnamefont {{Diego}}}, \bibinfo {author}
  {\bibfnamefont {S.}~\bibnamefont {{Donzelli}}}, \bibinfo {author}
  {\bibfnamefont {O.}~\bibnamefont {{Dor{\'e}}}}, \bibinfo {author}
  {\bibfnamefont {M.}~\bibnamefont {{Douspis}}}, \bibinfo {author}
  {\bibfnamefont {A.}~\bibnamefont {{Ducout}}}, \bibinfo {author}
  {\bibfnamefont {X.}~\bibnamefont {{Dupac}}}, \bibinfo {author} {\bibfnamefont
  {G.}~\bibnamefont {{Efstathiou}}}, \bibinfo {author} {\bibfnamefont
  {F.}~\bibnamefont {{Elsner}}}, \bibinfo {author} {\bibfnamefont {T.~A.}\
  \bibnamefont {{En{\ss}lin}}}, \bibinfo {author} {\bibfnamefont {H.~K.}\
  \bibnamefont {{Eriksen}}}, \bibinfo {author} {\bibfnamefont {E.}~\bibnamefont
  {{Falgarone}}}, \bibinfo {author} {\bibfnamefont {Y.}~\bibnamefont
  {{Fantaye}}}, \bibinfo {author} {\bibfnamefont {J.}~\bibnamefont
  {{Fergusson}}}, \bibinfo {author} {\bibfnamefont {R.}~\bibnamefont
  {{Fernandez-Cobos}}}, \bibinfo {author} {\bibfnamefont {F.}~\bibnamefont
  {{Finelli}}}, \bibinfo {author} {\bibfnamefont {F.}~\bibnamefont
  {{Forastieri}}}, \bibinfo {author} {\bibfnamefont {M.}~\bibnamefont
  {{Frailis}}}, \bibinfo {author} {\bibfnamefont {E.}~\bibnamefont
  {{Franceschi}}}, \bibinfo {author} {\bibfnamefont {A.}~\bibnamefont
  {{Frolov}}}, \bibinfo {author} {\bibfnamefont {S.}~\bibnamefont
  {{Galeotta}}}, \bibinfo {author} {\bibfnamefont {S.}~\bibnamefont {{Galli}}},
  \bibinfo {author} {\bibfnamefont {K.}~\bibnamefont {{Ganga}}}, \bibinfo
  {author} {\bibfnamefont {R.~T.}\ \bibnamefont {{G{\'e}nova-Santos}}},
  \bibinfo {author} {\bibfnamefont {M.}~\bibnamefont {{Gerbino}}}, \bibinfo
  {author} {\bibfnamefont {T.}~\bibnamefont {{Ghosh}}}, \bibinfo {author}
  {\bibfnamefont {J.}~\bibnamefont {{Gonz{\'a}lez-Nuevo}}}, \bibinfo {author}
  {\bibfnamefont {K.~M.}\ \bibnamefont {{G{\'o}rski}}}, \bibinfo {author}
  {\bibfnamefont {S.}~\bibnamefont {{Gratton}}}, \bibinfo {author}
  {\bibfnamefont {A.}~\bibnamefont {{Gruppuso}}}, \bibinfo {author}
  {\bibfnamefont {J.~E.}\ \bibnamefont {{Gudmundsson}}}, \bibinfo {author}
  {\bibfnamefont {J.}~\bibnamefont {{Hamann}}}, \bibinfo {author}
  {\bibfnamefont {W.}~\bibnamefont {{Hand ley}}}, \bibinfo {author}
  {\bibfnamefont {F.~K.}\ \bibnamefont {{Hansen}}}, \bibinfo {author}
  {\bibfnamefont {G.}~\bibnamefont {{Helou}}}, \bibinfo {author} {\bibfnamefont
  {D.}~\bibnamefont {{Herranz}}}, \bibinfo {author} {\bibfnamefont
  {E.}~\bibnamefont {{Hivon}}}, \bibinfo {author} {\bibfnamefont
  {Z.}~\bibnamefont {{Huang}}}, \bibinfo {author} {\bibfnamefont {A.~H.}\
  \bibnamefont {{Jaffe}}}, \bibinfo {author} {\bibfnamefont {W.~C.}\
  \bibnamefont {{Jones}}}, \bibinfo {author} {\bibfnamefont {A.}~\bibnamefont
  {{Karakci}}}, \bibinfo {author} {\bibfnamefont {E.}~\bibnamefont
  {{Keih{\"a}nen}}}, \bibinfo {author} {\bibfnamefont {R.}~\bibnamefont
  {{Keskitalo}}}, \bibinfo {author} {\bibfnamefont {K.}~\bibnamefont
  {{Kiiveri}}}, \bibinfo {author} {\bibfnamefont {J.}~\bibnamefont {{Kim}}},
  \bibinfo {author} {\bibfnamefont {T.~S.}\ \bibnamefont {{Kisner}}}, \bibinfo
  {author} {\bibfnamefont {L.}~\bibnamefont {{Knox}}}, \bibinfo {author}
  {\bibfnamefont {N.}~\bibnamefont {{Krachmalnicoff}}}, \bibinfo {author}
  {\bibfnamefont {M.}~\bibnamefont {{Kunz}}}, \bibinfo {author} {\bibfnamefont
  {H.}~\bibnamefont {{Kurki-Suonio}}}, \bibinfo {author} {\bibfnamefont
  {G.}~\bibnamefont {{Lagache}}}, \bibinfo {author} {\bibfnamefont {J.~M.}\
  \bibnamefont {{Lamarre}}}, \bibinfo {author} {\bibfnamefont {M.}~\bibnamefont
  {{Langer}}}, \bibinfo {author} {\bibfnamefont {A.}~\bibnamefont {{Lasenby}}},
  \bibinfo {author} {\bibfnamefont {M.}~\bibnamefont {{Lattanzi}}}, \bibinfo
  {author} {\bibfnamefont {C.~R.}\ \bibnamefont {{Lawrence}}}, \bibinfo
  {author} {\bibfnamefont {M.}~\bibnamefont {{Le Jeune}}}, \bibinfo {author}
  {\bibfnamefont {J.~P.}\ \bibnamefont {{Leahy}}}, \bibinfo {author}
  {\bibfnamefont {J.}~\bibnamefont {{Lesgourgues}}}, \bibinfo {author}
  {\bibfnamefont {F.}~\bibnamefont {{Levrier}}}, \bibinfo {author}
  {\bibfnamefont {A.}~\bibnamefont {{Lewis}}}, \bibinfo {author} {\bibfnamefont
  {M.}~\bibnamefont {{Liguori}}}, \bibinfo {author} {\bibfnamefont {P.~B.}\
  \bibnamefont {{Lilje}}}, \bibinfo {author} {\bibfnamefont {M.}~\bibnamefont
  {{Lilley}}}, \bibinfo {author} {\bibfnamefont {V.}~\bibnamefont
  {{Lindholm}}}, \bibinfo {author} {\bibfnamefont {M.}~\bibnamefont
  {{L{\'o}pez-Caniego}}}, \bibinfo {author} {\bibfnamefont {P.~M.}\
  \bibnamefont {{Lubin}}}, \bibinfo {author} {\bibfnamefont {Y.~Z.}\
  \bibnamefont {{Ma}}}, \bibinfo {author} {\bibfnamefont {J.~F.}\ \bibnamefont
  {{Mac{\'\i}as-P{\'e}rez}}}, \bibinfo {author} {\bibfnamefont
  {G.}~\bibnamefont {{Maggio}}}, \bibinfo {author} {\bibfnamefont
  {D.}~\bibnamefont {{Maino}}}, \bibinfo {author} {\bibfnamefont
  {N.}~\bibnamefont {{Mand olesi}}}, \bibinfo {author} {\bibfnamefont
  {A.}~\bibnamefont {{Mangilli}}}, \bibinfo {author} {\bibfnamefont
  {A.}~\bibnamefont {{Marcos-Caballero}}}, \bibinfo {author} {\bibfnamefont
  {M.}~\bibnamefont {{Maris}}}, \bibinfo {author} {\bibfnamefont {P.~G.}\
  \bibnamefont {{Martin}}}, \bibinfo {author} {\bibfnamefont {E.}~\bibnamefont
  {{Mart{\'\i}nez-Gonz{\'a}lez}}}, \bibinfo {author} {\bibfnamefont
  {S.}~\bibnamefont {{Matarrese}}}, \bibinfo {author} {\bibfnamefont
  {N.}~\bibnamefont {{Mauri}}}, \bibinfo {author} {\bibfnamefont {J.~D.}\
  \bibnamefont {{McEwen}}}, \bibinfo {author} {\bibfnamefont {P.~D.}\
  \bibnamefont {{Meerburg}}}, \bibinfo {author} {\bibfnamefont {P.~R.}\
  \bibnamefont {{Meinhold}}}, \bibinfo {author} {\bibfnamefont
  {A.}~\bibnamefont {{Melchiorri}}}, \bibinfo {author} {\bibfnamefont
  {A.}~\bibnamefont {{Mennella}}}, \bibinfo {author} {\bibfnamefont
  {M.}~\bibnamefont {{Migliaccio}}}, \bibinfo {author} {\bibfnamefont
  {M.}~\bibnamefont {{Millea}}}, \bibinfo {author} {\bibfnamefont
  {S.}~\bibnamefont {{Mitra}}}, \bibinfo {author} {\bibfnamefont {M.~A.}\
  \bibnamefont {{Miville-Desch{\^e}nes}}}, \bibinfo {author} {\bibfnamefont
  {D.}~\bibnamefont {{Molinari}}}, \bibinfo {author} {\bibfnamefont
  {A.}~\bibnamefont {{Moneti}}}, \bibinfo {author} {\bibfnamefont
  {L.}~\bibnamefont {{Montier}}}, \bibinfo {author} {\bibfnamefont
  {G.}~\bibnamefont {{Morgante}}}, \bibinfo {author} {\bibfnamefont
  {A.}~\bibnamefont {{Moss}}}, \bibinfo {author} {\bibfnamefont
  {S.}~\bibnamefont {{Mottet}}}, \bibinfo {author} {\bibfnamefont
  {M.}~\bibnamefont {{M{\"u}nchmeyer}}}, \bibinfo {author} {\bibfnamefont
  {P.}~\bibnamefont {{Natoli}}}, \bibinfo {author} {\bibfnamefont {H.~U.}\
  \bibnamefont {{N{\o}rgaard-Nielsen}}}, \bibinfo {author} {\bibfnamefont
  {C.~A.}\ \bibnamefont {{Oxborrow}}}, \bibinfo {author} {\bibfnamefont
  {L.}~\bibnamefont {{Pagano}}}, \bibinfo {author} {\bibfnamefont
  {D.}~\bibnamefont {{Paoletti}}}, \bibinfo {author} {\bibfnamefont
  {B.}~\bibnamefont {{Partridge}}}, \bibinfo {author} {\bibfnamefont
  {G.}~\bibnamefont {{Patanchon}}}, \bibinfo {author} {\bibfnamefont {T.~J.}\
  \bibnamefont {{Pearson}}}, \bibinfo {author} {\bibfnamefont {M.}~\bibnamefont
  {{Peel}}}, \bibinfo {author} {\bibfnamefont {H.~V.}\ \bibnamefont
  {{Peiris}}}, \bibinfo {author} {\bibfnamefont {F.}~\bibnamefont
  {{Perrotta}}}, \bibinfo {author} {\bibfnamefont {V.}~\bibnamefont
  {{Pettorino}}}, \bibinfo {author} {\bibfnamefont {F.}~\bibnamefont
  {{Piacentini}}}, \bibinfo {author} {\bibfnamefont {L.}~\bibnamefont
  {{Polastri}}}, \bibinfo {author} {\bibfnamefont {G.}~\bibnamefont
  {{Polenta}}}, \bibinfo {author} {\bibfnamefont {J.~L.}\ \bibnamefont
  {{Puget}}}, \bibinfo {author} {\bibfnamefont {J.~P.}\ \bibnamefont
  {{Rachen}}}, \bibinfo {author} {\bibfnamefont {M.}~\bibnamefont
  {{Reinecke}}}, \bibinfo {author} {\bibfnamefont {M.}~\bibnamefont
  {{Remazeilles}}}, \bibinfo {author} {\bibfnamefont {A.}~\bibnamefont
  {{Renzi}}}, \bibinfo {author} {\bibfnamefont {G.}~\bibnamefont {{Rocha}}},
  \bibinfo {author} {\bibfnamefont {C.}~\bibnamefont {{Rosset}}}, \bibinfo
  {author} {\bibfnamefont {G.}~\bibnamefont {{Roudier}}}, \bibinfo {author}
  {\bibfnamefont {J.~A.}\ \bibnamefont {{Rubi{\~n}o-Mart{\'\i}n}}}, \bibinfo
  {author} {\bibfnamefont {B.}~\bibnamefont {{Ruiz-Granados}}}, \bibinfo
  {author} {\bibfnamefont {L.}~\bibnamefont {{Salvati}}}, \bibinfo {author}
  {\bibfnamefont {M.}~\bibnamefont {{Sandri}}}, \bibinfo {author}
  {\bibfnamefont {M.}~\bibnamefont {{Savelainen}}}, \bibinfo {author}
  {\bibfnamefont {D.}~\bibnamefont {{Scott}}}, \bibinfo {author} {\bibfnamefont
  {E.~P.~S.}\ \bibnamefont {{Shellard}}}, \bibinfo {author} {\bibfnamefont
  {M.}~\bibnamefont {{Shiraishi}}}, \bibinfo {author} {\bibfnamefont
  {C.}~\bibnamefont {{Sirignano}}}, \bibinfo {author} {\bibfnamefont
  {G.}~\bibnamefont {{Sirri}}}, \bibinfo {author} {\bibfnamefont {L.~D.}\
  \bibnamefont {{Spencer}}}, \bibinfo {author} {\bibfnamefont {R.}~\bibnamefont
  {{Sunyaev}}}, \bibinfo {author} {\bibfnamefont {A.~S.}\ \bibnamefont
  {{Suur-Uski}}}, \bibinfo {author} {\bibfnamefont {J.~A.}\ \bibnamefont
  {{Tauber}}}, \bibinfo {author} {\bibfnamefont {D.}~\bibnamefont
  {{Tavagnacco}}}, \bibinfo {author} {\bibfnamefont {M.}~\bibnamefont
  {{Tenti}}}, \bibinfo {author} {\bibfnamefont {L.}~\bibnamefont {{Terenzi}}},
  \bibinfo {author} {\bibfnamefont {L.}~\bibnamefont {{Toffolatti}}}, \bibinfo
  {author} {\bibfnamefont {M.}~\bibnamefont {{Tomasi}}}, \bibinfo {author}
  {\bibfnamefont {T.}~\bibnamefont {{Trombetti}}}, \bibinfo {author}
  {\bibfnamefont {J.}~\bibnamefont {{Valiviita}}}, \bibinfo {author}
  {\bibfnamefont {B.}~\bibnamefont {{Van Tent}}}, \bibinfo {author}
  {\bibfnamefont {L.}~\bibnamefont {{Vibert}}}, \bibinfo {author}
  {\bibfnamefont {P.}~\bibnamefont {{Vielva}}}, \bibinfo {author}
  {\bibfnamefont {F.}~\bibnamefont {{Villa}}}, \bibinfo {author} {\bibfnamefont
  {N.}~\bibnamefont {{Vittorio}}}, \bibinfo {author} {\bibfnamefont {B.~D.}\
  \bibnamefont {{Wandelt}}}, \bibinfo {author} {\bibfnamefont {I.~K.}\
  \bibnamefont {{Wehus}}}, \bibinfo {author} {\bibfnamefont {M.}~\bibnamefont
  {{White}}}, \bibinfo {author} {\bibfnamefont {S.~D.~M.}\ \bibnamefont
  {{White}}}, \bibinfo {author} {\bibfnamefont {A.}~\bibnamefont {{Zacchei}}},
  \ and\ \bibinfo {author} {\bibfnamefont {A.}~\bibnamefont {{Zonca}}},\
  }\href@noop {} {\bibfield  {journal} {\bibinfo  {journal} {ArXiv e-prints}\
  ,\ \bibinfo {eid} {arXiv:1807.06205}} (\bibinfo {year}
  {2018}{\natexlab{c}})},\ \Eprint {http://arxiv.org/abs/1807.06205}
  {arXiv:1807.06205 [astro-ph.CO]} \BibitemShut {NoStop}%
\bibitem [{\citenamefont {{Pardo}}\ and\ \citenamefont
  {{Spergel}}(2020)}]{pardo20a}%
  \BibitemOpen
  \bibfield  {author} {\bibinfo {author} {\bibfnamefont {K.}~\bibnamefont
  {{Pardo}}}\ and\ \bibinfo {author} {\bibfnamefont {D.~N.}\ \bibnamefont
  {{Spergel}}},\ }\href {\doibase 10.1103/PhysRevLett.125.211101} {\bibfield
  {journal} {\bibinfo  {journal} {\prl}\ }\textbf {\bibinfo {volume} {125}},\
  \bibinfo {eid} {211101} (\bibinfo {year} {2020})},\ \Eprint
  {http://arxiv.org/abs/2007.00555} {arXiv:2007.00555 [astro-ph.CO]}
  \BibitemShut {NoStop}%
\bibitem [{\citenamefont {{Eisenstein}}\ \emph {et~al.}(2007)\citenamefont
  {{Eisenstein}}, \citenamefont {{Seo}}, \citenamefont {{Sirko}},\ and\
  \citenamefont {{Spergel}}}]{eisenstein07a}%
  \BibitemOpen
  \bibfield  {author} {\bibinfo {author} {\bibfnamefont {D.~J.}\ \bibnamefont
  {{Eisenstein}}}, \bibinfo {author} {\bibfnamefont {H.-J.}\ \bibnamefont
  {{Seo}}}, \bibinfo {author} {\bibfnamefont {E.}~\bibnamefont {{Sirko}}}, \
  and\ \bibinfo {author} {\bibfnamefont {D.~N.}\ \bibnamefont {{Spergel}}},\
  }\href {\doibase 10.1086/518712} {\bibfield  {journal} {\bibinfo  {journal}
  {\apj}\ }\textbf {\bibinfo {volume} {664}},\ \bibinfo {pages} {675} (\bibinfo
  {year} {2007})},\ \Eprint {http://arxiv.org/abs/astro-ph/0604362}
  {astro-ph/0604362} \BibitemShut {NoStop}%
\bibitem [{\citenamefont {{Weinberg}}\ \emph {et~al.}(2013)\citenamefont
  {{Weinberg}}, \citenamefont {{Mortonson}}, \citenamefont {{Eisenstein}},
  \citenamefont {{Hirata}}, \citenamefont {{Riess}},\ and\ \citenamefont
  {{Rozo}}}]{weinberg13a}%
  \BibitemOpen
  \bibfield  {author} {\bibinfo {author} {\bibfnamefont {D.~H.}\ \bibnamefont
  {{Weinberg}}}, \bibinfo {author} {\bibfnamefont {M.~J.}\ \bibnamefont
  {{Mortonson}}}, \bibinfo {author} {\bibfnamefont {D.~J.}\ \bibnamefont
  {{Eisenstein}}}, \bibinfo {author} {\bibfnamefont {C.}~\bibnamefont
  {{Hirata}}}, \bibinfo {author} {\bibfnamefont {A.~G.}\ \bibnamefont
  {{Riess}}}, \ and\ \bibinfo {author} {\bibfnamefont {E.}~\bibnamefont
  {{Rozo}}},\ }\href {\doibase 10.1016/j.physrep.2013.05.001} {\bibfield
  {journal} {\bibinfo  {journal} {\physrep}\ }\textbf {\bibinfo {volume}
  {530}},\ \bibinfo {pages} {87} (\bibinfo {year} {2013})},\ \Eprint
  {http://arxiv.org/abs/1201.2434} {arXiv:1201.2434} \BibitemShut {NoStop}%
\bibitem [{\citenamefont {{Alcock}}\ and\ \citenamefont
  {{Paczynski}}(1979)}]{alcock79a}%
  \BibitemOpen
  \bibfield  {author} {\bibinfo {author} {\bibfnamefont {C.}~\bibnamefont
  {{Alcock}}}\ and\ \bibinfo {author} {\bibfnamefont {B.}~\bibnamefont
  {{Paczynski}}},\ }\href@noop {} {\bibfield  {journal} {\bibinfo  {journal}
  {\nat}\ }\textbf {\bibinfo {volume} {281}},\ \bibinfo {pages} {358} (\bibinfo
  {year} {1979})}\BibitemShut {NoStop}%
\bibitem [{\citenamefont {{Kaiser}}(1987)}]{kaiser87a}%
  \BibitemOpen
  \bibfield  {author} {\bibinfo {author} {\bibfnamefont {N.}~\bibnamefont
  {{Kaiser}}},\ }\href@noop {} {\bibfield  {journal} {\bibinfo  {journal}
  {\mnras}\ }\textbf {\bibinfo {volume} {227}},\ \bibinfo {pages} {1} (\bibinfo
  {year} {1987})}\BibitemShut {NoStop}%
\bibitem [{\citenamefont {{Nadathur}}\ \emph {et~al.}(2020)\citenamefont
  {{Nadathur}}, \citenamefont {{Percival}}, \citenamefont {{Beutler}},\ and\
  \citenamefont {{Winther}}}]{nadathur20a}%
  \BibitemOpen
  \bibfield  {author} {\bibinfo {author} {\bibfnamefont {S.}~\bibnamefont
  {{Nadathur}}}, \bibinfo {author} {\bibfnamefont {W.~J.}\ \bibnamefont
  {{Percival}}}, \bibinfo {author} {\bibfnamefont {F.}~\bibnamefont
  {{Beutler}}}, \ and\ \bibinfo {author} {\bibfnamefont {H.~A.}\ \bibnamefont
  {{Winther}}},\ }\href {\doibase 10.1103/PhysRevLett.124.221301} {\bibfield
  {journal} {\bibinfo  {journal} {\prl}\ }\textbf {\bibinfo {volume} {124}},\
  \bibinfo {eid} {221301} (\bibinfo {year} {2020})},\ \Eprint
  {http://arxiv.org/abs/2001.11044} {arXiv:2001.11044 [astro-ph.CO]}
  \BibitemShut {NoStop}%
\bibitem [{\citenamefont {{Ballinger}}\ \emph {et~al.}(1996)\citenamefont
  {{Ballinger}}, \citenamefont {{Peacock}},\ and\ \citenamefont
  {{Heavens}}}]{ballinger96a}%
  \BibitemOpen
  \bibfield  {author} {\bibinfo {author} {\bibfnamefont {W.~E.}\ \bibnamefont
  {{Ballinger}}}, \bibinfo {author} {\bibfnamefont {J.~A.}\ \bibnamefont
  {{Peacock}}}, \ and\ \bibinfo {author} {\bibfnamefont {A.~F.}\ \bibnamefont
  {{Heavens}}},\ }\href {\doibase 10.1093/mnras/282.3.877} {\bibfield
  {journal} {\bibinfo  {journal} {\mnras}\ }\textbf {\bibinfo {volume} {282}},\
  \bibinfo {pages} {877} (\bibinfo {year} {1996})},\ \Eprint
  {http://arxiv.org/abs/astro-ph/9605017} {arXiv:astro-ph/9605017 [astro-ph]}
  \BibitemShut {NoStop}%
\bibitem [{\citenamefont {{Jain}}\ and\ \citenamefont
  {{Seljak}}(1997)}]{Jain1997}%
  \BibitemOpen
  \bibfield  {author} {\bibinfo {author} {\bibfnamefont {B.}~\bibnamefont
  {{Jain}}}\ and\ \bibinfo {author} {\bibfnamefont {U.}~\bibnamefont
  {{Seljak}}},\ }\href {\doibase 10.1086/304372} {\bibfield  {journal}
  {\bibinfo  {journal} {\apj}\ }\textbf {\bibinfo {volume} {484}},\ \bibinfo
  {pages} {560} (\bibinfo {year} {1997})},\ \Eprint
  {http://arxiv.org/abs/astro-ph/9611077} {arXiv:astro-ph/9611077 [astro-ph]}
  \BibitemShut {NoStop}%
\bibitem [{\citenamefont {{Astier}}\ \emph {et~al.}(2006)\citenamefont
  {{Astier}}, \citenamefont {{Guy}}, \citenamefont {{Regnault}}, \citenamefont
  {{Pain}}, \citenamefont {{Aubourg}}, \citenamefont {{Balam}}, \citenamefont
  {{Basa}}, \citenamefont {{Carlberg}}, \citenamefont {{Fabbro}}, \citenamefont
  {{Fouchez}}, \citenamefont {{Hook}}, \citenamefont {{Howell}}, \citenamefont
  {{Lafoux}}, \citenamefont {{Neill}}, \citenamefont {{Palanque-Delabrouille}},
  \citenamefont {{Perrett}}, \citenamefont {{Pritchet}}, \citenamefont
  {{Rich}}, \citenamefont {{Sullivan}}, \citenamefont {{Taillet}},
  \citenamefont {{Aldering}}, \citenamefont {{Antilogus}}, \citenamefont
  {{Arsenijevic}}, \citenamefont {{Balland}}, \citenamefont {{Baumont}},
  \citenamefont {{Bronder}}, \citenamefont {{Courtois}}, \citenamefont
  {{Ellis}}, \citenamefont {{Filiol}}, \citenamefont {{Gon{\c c}alves}},
  \citenamefont {{Goobar}}, \citenamefont {{Guide}}, \citenamefont {{Hardin}},
  \citenamefont {{Lusset}}, \citenamefont {{Lidman}}, \citenamefont
  {{McMahon}}, \citenamefont {{Mouchet}}, \citenamefont {{Mourao}},
  \citenamefont {{Perlmutter}}, \citenamefont {{Ripoche}}, \citenamefont
  {{Tao}},\ and\ \citenamefont {{Walton}}}]{astier06a}%
  \BibitemOpen
  \bibfield  {author} {\bibinfo {author} {\bibfnamefont {P.}~\bibnamefont
  {{Astier}}}, \bibinfo {author} {\bibfnamefont {J.}~\bibnamefont {{Guy}}},
  \bibinfo {author} {\bibfnamefont {N.}~\bibnamefont {{Regnault}}}, \bibinfo
  {author} {\bibfnamefont {R.}~\bibnamefont {{Pain}}}, \bibinfo {author}
  {\bibfnamefont {E.}~\bibnamefont {{Aubourg}}}, \bibinfo {author}
  {\bibfnamefont {D.}~\bibnamefont {{Balam}}}, \bibinfo {author} {\bibfnamefont
  {S.}~\bibnamefont {{Basa}}}, \bibinfo {author} {\bibfnamefont {R.~G.}\
  \bibnamefont {{Carlberg}}}, \bibinfo {author} {\bibfnamefont
  {S.}~\bibnamefont {{Fabbro}}}, \bibinfo {author} {\bibfnamefont
  {D.}~\bibnamefont {{Fouchez}}}, \bibinfo {author} {\bibfnamefont {I.~M.}\
  \bibnamefont {{Hook}}}, \bibinfo {author} {\bibfnamefont {D.~A.}\
  \bibnamefont {{Howell}}}, \bibinfo {author} {\bibfnamefont {H.}~\bibnamefont
  {{Lafoux}}}, \bibinfo {author} {\bibfnamefont {J.~D.}\ \bibnamefont
  {{Neill}}}, \bibinfo {author} {\bibfnamefont {N.}~\bibnamefont
  {{Palanque-Delabrouille}}}, \bibinfo {author} {\bibfnamefont
  {K.}~\bibnamefont {{Perrett}}}, \bibinfo {author} {\bibfnamefont {C.~J.}\
  \bibnamefont {{Pritchet}}}, \bibinfo {author} {\bibfnamefont
  {J.}~\bibnamefont {{Rich}}}, \bibinfo {author} {\bibfnamefont
  {M.}~\bibnamefont {{Sullivan}}}, \bibinfo {author} {\bibfnamefont
  {R.}~\bibnamefont {{Taillet}}}, \bibinfo {author} {\bibfnamefont
  {G.}~\bibnamefont {{Aldering}}}, \bibinfo {author} {\bibfnamefont
  {P.}~\bibnamefont {{Antilogus}}}, \bibinfo {author} {\bibfnamefont
  {V.}~\bibnamefont {{Arsenijevic}}}, \bibinfo {author} {\bibfnamefont
  {C.}~\bibnamefont {{Balland}}}, \bibinfo {author} {\bibfnamefont
  {S.}~\bibnamefont {{Baumont}}}, \bibinfo {author} {\bibfnamefont
  {J.}~\bibnamefont {{Bronder}}}, \bibinfo {author} {\bibfnamefont
  {H.}~\bibnamefont {{Courtois}}}, \bibinfo {author} {\bibfnamefont {R.~S.}\
  \bibnamefont {{Ellis}}}, \bibinfo {author} {\bibfnamefont {M.}~\bibnamefont
  {{Filiol}}}, \bibinfo {author} {\bibfnamefont {A.~C.}\ \bibnamefont {{Gon{\c
  c}alves}}}, \bibinfo {author} {\bibfnamefont {A.}~\bibnamefont {{Goobar}}},
  \bibinfo {author} {\bibfnamefont {D.}~\bibnamefont {{Guide}}}, \bibinfo
  {author} {\bibfnamefont {D.}~\bibnamefont {{Hardin}}}, \bibinfo {author}
  {\bibfnamefont {V.}~\bibnamefont {{Lusset}}}, \bibinfo {author}
  {\bibfnamefont {C.}~\bibnamefont {{Lidman}}}, \bibinfo {author}
  {\bibfnamefont {R.}~\bibnamefont {{McMahon}}}, \bibinfo {author}
  {\bibfnamefont {M.}~\bibnamefont {{Mouchet}}}, \bibinfo {author}
  {\bibfnamefont {A.}~\bibnamefont {{Mourao}}}, \bibinfo {author}
  {\bibfnamefont {S.}~\bibnamefont {{Perlmutter}}}, \bibinfo {author}
  {\bibfnamefont {P.}~\bibnamefont {{Ripoche}}}, \bibinfo {author}
  {\bibfnamefont {C.}~\bibnamefont {{Tao}}}, \ and\ \bibinfo {author}
  {\bibfnamefont {N.}~\bibnamefont {{Walton}}},\ }\href {\doibase
  10.1051/0004-6361:20054185} {\bibfield  {journal} {\bibinfo  {journal}
  {\aap}\ }\textbf {\bibinfo {volume} {447}},\ \bibinfo {pages} {31} (\bibinfo
  {year} {2006})},\ \Eprint {http://arxiv.org/abs/arXiv:astro-ph/0510447}
  {arXiv:astro-ph/0510447} \BibitemShut {NoStop}%
\bibitem [{\citenamefont {{Hamuy}}\ \emph {et~al.}(1996)\citenamefont
  {{Hamuy}}, \citenamefont {{Phillips}}, \citenamefont {{Suntzeff}},
  \citenamefont {{Schommer}}, \citenamefont {{Maza}},\ and\ \citenamefont
  {{Aviles}}}]{hamuy96a}%
  \BibitemOpen
  \bibfield  {author} {\bibinfo {author} {\bibfnamefont {M.}~\bibnamefont
  {{Hamuy}}}, \bibinfo {author} {\bibfnamefont {M.~M.}\ \bibnamefont
  {{Phillips}}}, \bibinfo {author} {\bibfnamefont {N.~B.}\ \bibnamefont
  {{Suntzeff}}}, \bibinfo {author} {\bibfnamefont {R.~A.}\ \bibnamefont
  {{Schommer}}}, \bibinfo {author} {\bibfnamefont {J.}~\bibnamefont {{Maza}}},
  \ and\ \bibinfo {author} {\bibfnamefont {R.}~\bibnamefont {{Aviles}}},\
  }\href@noop {} {\bibfield  {journal} {\bibinfo  {journal} {\aj}\ }\textbf
  {\bibinfo {volume} {112}},\ \bibinfo {pages} {2391} (\bibinfo {year}
  {1996})}\BibitemShut {NoStop}%
\bibitem [{\citenamefont {{Phillips}}\ \emph {et~al.}(1999)\citenamefont
  {{Phillips}}, \citenamefont {{Lira}}, \citenamefont {{Suntzeff}},
  \citenamefont {{Schommer}}, \citenamefont {{Hamuy}},\ and\ \citenamefont
  {{Maza}}}]{phillips99a}%
  \BibitemOpen
  \bibfield  {author} {\bibinfo {author} {\bibfnamefont {M.~M.}\ \bibnamefont
  {{Phillips}}}, \bibinfo {author} {\bibfnamefont {P.}~\bibnamefont {{Lira}}},
  \bibinfo {author} {\bibfnamefont {N.~B.}\ \bibnamefont {{Suntzeff}}},
  \bibinfo {author} {\bibfnamefont {R.~A.}\ \bibnamefont {{Schommer}}},
  \bibinfo {author} {\bibfnamefont {M.}~\bibnamefont {{Hamuy}}}, \ and\
  \bibinfo {author} {\bibfnamefont {J.}~\bibnamefont {{Maza}}},\ }\href
  {\doibase 10.1086/301032} {\bibfield  {journal} {\bibinfo  {journal} {\aj}\
  }\textbf {\bibinfo {volume} {118}},\ \bibinfo {pages} {1766} (\bibinfo {year}
  {1999})},\ \Eprint {http://arxiv.org/abs/arXiv:astro-ph/9907052}
  {arXiv:astro-ph/9907052} \BibitemShut {NoStop}%
\bibitem [{\citenamefont {{Betoule}}\ \emph {et~al.}(2014)\citenamefont
  {{Betoule}}, \citenamefont {{Kessler}}, \citenamefont {{Guy}}, \citenamefont
  {{Mosher}}, \citenamefont {{Hardin}}, \citenamefont {{Biswas}}, \citenamefont
  {{Astier}}, \citenamefont {{El-Hage}}, \citenamefont {{Konig}}, \citenamefont
  {{Kuhlmann}}, \citenamefont {{Marriner}}, \citenamefont {{Pain}},
  \citenamefont {{Regnault}}, \citenamefont {{Balland}}, \citenamefont
  {{Bassett}}, \citenamefont {{Brown}}, \citenamefont {{Campbell}},
  \citenamefont {{Carlberg}}, \citenamefont {{Cellier-Holzem}}, \citenamefont
  {{Cinabro}}, \citenamefont {{Conley}}, \citenamefont {{D'Andrea}},
  \citenamefont {{DePoy}}, \citenamefont {{Doi}}, \citenamefont {{Ellis}},
  \citenamefont {{Fabbro}}, \citenamefont {{Filippenko}}, \citenamefont
  {{Foley}}, \citenamefont {{Frieman}}, \citenamefont {{Fouchez}},
  \citenamefont {{Galbany}}, \citenamefont {{Goobar}}, \citenamefont {{Gupta}},
  \citenamefont {{Hill}}, \citenamefont {{Hlozek}}, \citenamefont {{Hogan}},
  \citenamefont {{Hook}}, \citenamefont {{Howell}}, \citenamefont {{Jha}},
  \citenamefont {{Le Guillou}}, \citenamefont {{Leloudas}}, \citenamefont
  {{Lidman}}, \citenamefont {{Marshall}}, \citenamefont {{M{\"o}ller}},
  \citenamefont {{Mour{\~a}o}}, \citenamefont {{Neveu}}, \citenamefont
  {{Nichol}}, \citenamefont {{Olmstead}}, \citenamefont
  {{Palanque-Delabrouille}}, \citenamefont {{Perlmutter}}, \citenamefont
  {{Prieto}}, \citenamefont {{Pritchet}}, \citenamefont {{Richmond}},
  \citenamefont {{Riess}}, \citenamefont {{Ruhlmann-Kleider}}, \citenamefont
  {{Sako}}, \citenamefont {{Schahmaneche}}, \citenamefont {{Schneider}},
  \citenamefont {{Smith}}, \citenamefont {{Sollerman}}, \citenamefont
  {{Sullivan}}, \citenamefont {{Walton}},\ and\ \citenamefont
  {{Wheeler}}}]{betoule14a}%
  \BibitemOpen
  \bibfield  {author} {\bibinfo {author} {\bibfnamefont {M.}~\bibnamefont
  {{Betoule}}}, \bibinfo {author} {\bibfnamefont {R.}~\bibnamefont
  {{Kessler}}}, \bibinfo {author} {\bibfnamefont {J.}~\bibnamefont {{Guy}}},
  \bibinfo {author} {\bibfnamefont {J.}~\bibnamefont {{Mosher}}}, \bibinfo
  {author} {\bibfnamefont {D.}~\bibnamefont {{Hardin}}}, \bibinfo {author}
  {\bibfnamefont {R.}~\bibnamefont {{Biswas}}}, \bibinfo {author}
  {\bibfnamefont {P.}~\bibnamefont {{Astier}}}, \bibinfo {author}
  {\bibfnamefont {P.}~\bibnamefont {{El-Hage}}}, \bibinfo {author}
  {\bibfnamefont {M.}~\bibnamefont {{Konig}}}, \bibinfo {author} {\bibfnamefont
  {S.}~\bibnamefont {{Kuhlmann}}}, \bibinfo {author} {\bibfnamefont
  {J.}~\bibnamefont {{Marriner}}}, \bibinfo {author} {\bibfnamefont
  {R.}~\bibnamefont {{Pain}}}, \bibinfo {author} {\bibfnamefont
  {N.}~\bibnamefont {{Regnault}}}, \bibinfo {author} {\bibfnamefont
  {C.}~\bibnamefont {{Balland}}}, \bibinfo {author} {\bibfnamefont {B.~A.}\
  \bibnamefont {{Bassett}}}, \bibinfo {author} {\bibfnamefont {P.~J.}\
  \bibnamefont {{Brown}}}, \bibinfo {author} {\bibfnamefont {H.}~\bibnamefont
  {{Campbell}}}, \bibinfo {author} {\bibfnamefont {R.~G.}\ \bibnamefont
  {{Carlberg}}}, \bibinfo {author} {\bibfnamefont {F.}~\bibnamefont
  {{Cellier-Holzem}}}, \bibinfo {author} {\bibfnamefont {D.}~\bibnamefont
  {{Cinabro}}}, \bibinfo {author} {\bibfnamefont {A.}~\bibnamefont {{Conley}}},
  \bibinfo {author} {\bibfnamefont {C.~B.}\ \bibnamefont {{D'Andrea}}},
  \bibinfo {author} {\bibfnamefont {D.~L.}\ \bibnamefont {{DePoy}}}, \bibinfo
  {author} {\bibfnamefont {M.}~\bibnamefont {{Doi}}}, \bibinfo {author}
  {\bibfnamefont {R.~S.}\ \bibnamefont {{Ellis}}}, \bibinfo {author}
  {\bibfnamefont {S.}~\bibnamefont {{Fabbro}}}, \bibinfo {author}
  {\bibfnamefont {A.~V.}\ \bibnamefont {{Filippenko}}}, \bibinfo {author}
  {\bibfnamefont {R.~J.}\ \bibnamefont {{Foley}}}, \bibinfo {author}
  {\bibfnamefont {J.~A.}\ \bibnamefont {{Frieman}}}, \bibinfo {author}
  {\bibfnamefont {D.}~\bibnamefont {{Fouchez}}}, \bibinfo {author}
  {\bibfnamefont {L.}~\bibnamefont {{Galbany}}}, \bibinfo {author}
  {\bibfnamefont {A.}~\bibnamefont {{Goobar}}}, \bibinfo {author}
  {\bibfnamefont {R.~R.}\ \bibnamefont {{Gupta}}}, \bibinfo {author}
  {\bibfnamefont {G.~J.}\ \bibnamefont {{Hill}}}, \bibinfo {author}
  {\bibfnamefont {R.}~\bibnamefont {{Hlozek}}}, \bibinfo {author}
  {\bibfnamefont {C.~J.}\ \bibnamefont {{Hogan}}}, \bibinfo {author}
  {\bibfnamefont {I.~M.}\ \bibnamefont {{Hook}}}, \bibinfo {author}
  {\bibfnamefont {D.~A.}\ \bibnamefont {{Howell}}}, \bibinfo {author}
  {\bibfnamefont {S.~W.}\ \bibnamefont {{Jha}}}, \bibinfo {author}
  {\bibfnamefont {L.}~\bibnamefont {{Le Guillou}}}, \bibinfo {author}
  {\bibfnamefont {G.}~\bibnamefont {{Leloudas}}}, \bibinfo {author}
  {\bibfnamefont {C.}~\bibnamefont {{Lidman}}}, \bibinfo {author}
  {\bibfnamefont {J.~L.}\ \bibnamefont {{Marshall}}}, \bibinfo {author}
  {\bibfnamefont {A.}~\bibnamefont {{M{\"o}ller}}}, \bibinfo {author}
  {\bibfnamefont {A.~M.}\ \bibnamefont {{Mour{\~a}o}}}, \bibinfo {author}
  {\bibfnamefont {J.}~\bibnamefont {{Neveu}}}, \bibinfo {author} {\bibfnamefont
  {R.}~\bibnamefont {{Nichol}}}, \bibinfo {author} {\bibfnamefont {M.~D.}\
  \bibnamefont {{Olmstead}}}, \bibinfo {author} {\bibfnamefont
  {N.}~\bibnamefont {{Palanque-Delabrouille}}}, \bibinfo {author}
  {\bibfnamefont {S.}~\bibnamefont {{Perlmutter}}}, \bibinfo {author}
  {\bibfnamefont {J.~L.}\ \bibnamefont {{Prieto}}}, \bibinfo {author}
  {\bibfnamefont {C.~J.}\ \bibnamefont {{Pritchet}}}, \bibinfo {author}
  {\bibfnamefont {M.}~\bibnamefont {{Richmond}}}, \bibinfo {author}
  {\bibfnamefont {A.~G.}\ \bibnamefont {{Riess}}}, \bibinfo {author}
  {\bibfnamefont {V.}~\bibnamefont {{Ruhlmann-Kleider}}}, \bibinfo {author}
  {\bibfnamefont {M.}~\bibnamefont {{Sako}}}, \bibinfo {author} {\bibfnamefont
  {K.}~\bibnamefont {{Schahmaneche}}}, \bibinfo {author} {\bibfnamefont
  {D.~P.}\ \bibnamefont {{Schneider}}}, \bibinfo {author} {\bibfnamefont
  {M.}~\bibnamefont {{Smith}}}, \bibinfo {author} {\bibfnamefont
  {J.}~\bibnamefont {{Sollerman}}}, \bibinfo {author} {\bibfnamefont
  {M.}~\bibnamefont {{Sullivan}}}, \bibinfo {author} {\bibfnamefont {N.~A.}\
  \bibnamefont {{Walton}}}, \ and\ \bibinfo {author} {\bibfnamefont {C.~J.}\
  \bibnamefont {{Wheeler}}},\ }\href {\doibase 10.1051/0004-6361/201423413}
  {\bibfield  {journal} {\bibinfo  {journal} {\aap}\ }\textbf {\bibinfo
  {volume} {568}},\ \bibinfo {eid} {A22} (\bibinfo {year} {2014})},\ \Eprint
  {http://arxiv.org/abs/1401.4064} {arXiv:1401.4064} \BibitemShut {NoStop}%
\bibitem [{\citenamefont {{Scolnic}}\ \emph {et~al.}(2018)\citenamefont
  {{Scolnic}}, \citenamefont {{Jones}}, \citenamefont {{Rest}}, \citenamefont
  {{Pan}}, \citenamefont {{Chornock}}, \citenamefont {{Foley}}, \citenamefont
  {{Huber}}, \citenamefont {{Kessler}}, \citenamefont {{Narayan}},
  \citenamefont {{Riess}}, \citenamefont {{Rodney}}, \citenamefont {{Berger}},
  \citenamefont {{Brout}}, \citenamefont {{Challis}}, \citenamefont {{Drout}},
  \citenamefont {{Finkbeiner}}, \citenamefont {{Lunnan}}, \citenamefont
  {{Kirshner}}, \citenamefont {{Sanders}}, \citenamefont {{Schlafly}},
  \citenamefont {{Smartt}}, \citenamefont {{Stubbs}}, \citenamefont {{Tonry}},
  \citenamefont {{Wood-Vasey}}, \citenamefont {{Foley}}, \citenamefont
  {{Hand}}, \citenamefont {{Johnson}}, \citenamefont {{Burgett}}, \citenamefont
  {{Chambers}}, \citenamefont {{Draper}}, \citenamefont {{Hodapp}},
  \citenamefont {{Kaiser}}, \citenamefont {{Kudritzki}}, \citenamefont
  {{Magnier}}, \citenamefont {{Metcalfe}}, \citenamefont {{Bresolin}},
  \citenamefont {{Gall}}, \citenamefont {{Kotak}}, \citenamefont {{McCrum}},\
  and\ \citenamefont {{Smith}}}]{scolnic18a}%
  \BibitemOpen
  \bibfield  {author} {\bibinfo {author} {\bibfnamefont {D.~M.}\ \bibnamefont
  {{Scolnic}}}, \bibinfo {author} {\bibfnamefont {D.~O.}\ \bibnamefont
  {{Jones}}}, \bibinfo {author} {\bibfnamefont {A.}~\bibnamefont {{Rest}}},
  \bibinfo {author} {\bibfnamefont {Y.~C.}\ \bibnamefont {{Pan}}}, \bibinfo
  {author} {\bibfnamefont {R.}~\bibnamefont {{Chornock}}}, \bibinfo {author}
  {\bibfnamefont {R.~J.}\ \bibnamefont {{Foley}}}, \bibinfo {author}
  {\bibfnamefont {M.~E.}\ \bibnamefont {{Huber}}}, \bibinfo {author}
  {\bibfnamefont {R.}~\bibnamefont {{Kessler}}}, \bibinfo {author}
  {\bibfnamefont {G.}~\bibnamefont {{Narayan}}}, \bibinfo {author}
  {\bibfnamefont {A.~G.}\ \bibnamefont {{Riess}}}, \bibinfo {author}
  {\bibfnamefont {S.}~\bibnamefont {{Rodney}}}, \bibinfo {author}
  {\bibfnamefont {E.}~\bibnamefont {{Berger}}}, \bibinfo {author}
  {\bibfnamefont {D.~J.}\ \bibnamefont {{Brout}}}, \bibinfo {author}
  {\bibfnamefont {P.~J.}\ \bibnamefont {{Challis}}}, \bibinfo {author}
  {\bibfnamefont {M.}~\bibnamefont {{Drout}}}, \bibinfo {author} {\bibfnamefont
  {D.}~\bibnamefont {{Finkbeiner}}}, \bibinfo {author} {\bibfnamefont
  {R.}~\bibnamefont {{Lunnan}}}, \bibinfo {author} {\bibfnamefont {R.~P.}\
  \bibnamefont {{Kirshner}}}, \bibinfo {author} {\bibfnamefont {N.~E.}\
  \bibnamefont {{Sanders}}}, \bibinfo {author} {\bibfnamefont {E.}~\bibnamefont
  {{Schlafly}}}, \bibinfo {author} {\bibfnamefont {S.}~\bibnamefont
  {{Smartt}}}, \bibinfo {author} {\bibfnamefont {C.~W.}\ \bibnamefont
  {{Stubbs}}}, \bibinfo {author} {\bibfnamefont {J.}~\bibnamefont {{Tonry}}},
  \bibinfo {author} {\bibfnamefont {W.~M.}\ \bibnamefont {{Wood-Vasey}}},
  \bibinfo {author} {\bibfnamefont {M.}~\bibnamefont {{Foley}}}, \bibinfo
  {author} {\bibfnamefont {J.}~\bibnamefont {{Hand}}}, \bibinfo {author}
  {\bibfnamefont {E.}~\bibnamefont {{Johnson}}}, \bibinfo {author}
  {\bibfnamefont {W.~S.}\ \bibnamefont {{Burgett}}}, \bibinfo {author}
  {\bibfnamefont {K.~C.}\ \bibnamefont {{Chambers}}}, \bibinfo {author}
  {\bibfnamefont {P.~W.}\ \bibnamefont {{Draper}}}, \bibinfo {author}
  {\bibfnamefont {K.~W.}\ \bibnamefont {{Hodapp}}}, \bibinfo {author}
  {\bibfnamefont {N.}~\bibnamefont {{Kaiser}}}, \bibinfo {author}
  {\bibfnamefont {R.~P.}\ \bibnamefont {{Kudritzki}}}, \bibinfo {author}
  {\bibfnamefont {E.~A.}\ \bibnamefont {{Magnier}}}, \bibinfo {author}
  {\bibfnamefont {N.}~\bibnamefont {{Metcalfe}}}, \bibinfo {author}
  {\bibfnamefont {F.}~\bibnamefont {{Bresolin}}}, \bibinfo {author}
  {\bibfnamefont {E.}~\bibnamefont {{Gall}}}, \bibinfo {author} {\bibfnamefont
  {R.}~\bibnamefont {{Kotak}}}, \bibinfo {author} {\bibfnamefont
  {M.}~\bibnamefont {{McCrum}}}, \ and\ \bibinfo {author} {\bibfnamefont
  {K.~W.}\ \bibnamefont {{Smith}}},\ }\href {\doibase 10.3847/1538-4357/aab9bb}
  {\bibfield  {journal} {\bibinfo  {journal} {\apj}\ }\textbf {\bibinfo
  {volume} {859}},\ \bibinfo {eid} {101} (\bibinfo {year} {2018})},\ \Eprint
  {http://arxiv.org/abs/1710.00845} {arXiv:1710.00845} \BibitemShut {NoStop}%
\bibitem [{\citenamefont {{Goobar}}\ and\ \citenamefont
  {{Leibundgut}}(2011)}]{goobar11a}%
  \BibitemOpen
  \bibfield  {author} {\bibinfo {author} {\bibfnamefont {A.}~\bibnamefont
  {{Goobar}}}\ and\ \bibinfo {author} {\bibfnamefont {B.}~\bibnamefont
  {{Leibundgut}}},\ }\href {\doibase 10.1146/annurev-nucl-102010-130434}
  {\bibfield  {journal} {\bibinfo  {journal} {Annual Review of Nuclear and
  Particle Science}\ }\textbf {\bibinfo {volume} {61}},\ \bibinfo {pages} {251}
  (\bibinfo {year} {2011})},\ \Eprint {http://arxiv.org/abs/1102.1431}
  {arXiv:1102.1431} \BibitemShut {NoStop}%
\bibitem [{\citenamefont {{Yu}}\ \emph {et~al.}(2018)\citenamefont {{Yu}},
  \citenamefont {{Knight}}, \citenamefont {{Sherwin}}, \citenamefont
  {{Ferraro}}, \citenamefont {{Knox}},\ and\ \citenamefont
  {{Schmittfull}}}]{2018arXiv180902120Y}%
  \BibitemOpen
  \bibfield  {author} {\bibinfo {author} {\bibfnamefont {B.}~\bibnamefont
  {{Yu}}}, \bibinfo {author} {\bibfnamefont {R.~Z.}\ \bibnamefont {{Knight}}},
  \bibinfo {author} {\bibfnamefont {B.~D.}\ \bibnamefont {{Sherwin}}}, \bibinfo
  {author} {\bibfnamefont {S.}~\bibnamefont {{Ferraro}}}, \bibinfo {author}
  {\bibfnamefont {L.}~\bibnamefont {{Knox}}}, \ and\ \bibinfo {author}
  {\bibfnamefont {M.}~\bibnamefont {{Schmittfull}}},\ }\href@noop {} {\bibfield
   {journal} {\bibinfo  {journal} {ArXiv e-prints}\ ,\ \bibinfo {eid}
  {arXiv:1809.02120}} (\bibinfo {year} {2018})},\ \Eprint
  {http://arxiv.org/abs/1809.02120} {arXiv:1809.02120 [astro-ph.CO]}
  \BibitemShut {NoStop}%
\bibitem [{\citenamefont {{Chiang}}\ \emph {et~al.}(2018)\citenamefont
  {{Chiang}}, \citenamefont {{Hu}}, \citenamefont {{Li}},\ and\ \citenamefont
  {{LoVerde}}}]{2018PhRvD..97l3526C}%
  \BibitemOpen
  \bibfield  {author} {\bibinfo {author} {\bibfnamefont {C.-T.}\ \bibnamefont
  {{Chiang}}}, \bibinfo {author} {\bibfnamefont {W.}~\bibnamefont {{Hu}}},
  \bibinfo {author} {\bibfnamefont {Y.}~\bibnamefont {{Li}}}, \ and\ \bibinfo
  {author} {\bibfnamefont {M.}~\bibnamefont {{LoVerde}}},\ }\href {\doibase
  10.1103/PhysRevD.97.123526} {\bibfield  {journal} {\bibinfo  {journal}
  {\prd}\ }\textbf {\bibinfo {volume} {97}},\ \bibinfo {eid} {123526} (\bibinfo
  {year} {2018})},\ \Eprint {http://arxiv.org/abs/1710.01310} {arXiv:1710.01310
  [astro-ph.CO]} \BibitemShut {NoStop}%
\bibitem [{\citenamefont {{Lewis}}\ and\ \citenamefont
  {{Bridle}}(2002)}]{lewis02a}%
  \BibitemOpen
  \bibfield  {author} {\bibinfo {author} {\bibfnamefont {A.}~\bibnamefont
  {{Lewis}}}\ and\ \bibinfo {author} {\bibfnamefont {S.}~\bibnamefont
  {{Bridle}}},\ }\href {\doibase 10.1103/PhysRevD.66.103511} {\bibfield
  {journal} {\bibinfo  {journal} {\prd}\ }\textbf {\bibinfo {volume} {66}},\
  \bibinfo {eid} {103511} (\bibinfo {year} {2002})},\ \Eprint
  {http://arxiv.org/abs/astro-ph/0205436} {astro-ph/0205436} \BibitemShut
  {NoStop}%
\bibitem [{\citenamefont {{Lewis}}(2019)}]{lewis19a}%
  \BibitemOpen
  \bibfield  {author} {\bibinfo {author} {\bibfnamefont {A.}~\bibnamefont
  {{Lewis}}},\ }\href@noop {} {\bibfield  {journal} {\bibinfo  {journal} {ArXiv
  e-prints}\ ,\ \bibinfo {eid} {arXiv:1910.13970}} (\bibinfo {year} {2019})},\
  \Eprint {http://arxiv.org/abs/1910.13970} {arXiv:1910.13970 [astro-ph.IM]}
  \BibitemShut {NoStop}%
\bibitem [{\citenamefont {{Carter}}\ \emph {et~al.}(2018)\citenamefont
  {{Carter}}, \citenamefont {{Beutler}}, \citenamefont {{Percival}},
  \citenamefont {{Blake}}, \citenamefont {{Koda}},\ and\ \citenamefont
  {{Ross}}}]{Carter18}%
  \BibitemOpen
  \bibfield  {author} {\bibinfo {author} {\bibfnamefont {P.}~\bibnamefont
  {{Carter}}}, \bibinfo {author} {\bibfnamefont {F.}~\bibnamefont {{Beutler}}},
  \bibinfo {author} {\bibfnamefont {W.~J.}\ \bibnamefont {{Percival}}},
  \bibinfo {author} {\bibfnamefont {C.}~\bibnamefont {{Blake}}}, \bibinfo
  {author} {\bibfnamefont {J.}~\bibnamefont {{Koda}}}, \ and\ \bibinfo {author}
  {\bibfnamefont {A.~J.}\ \bibnamefont {{Ross}}},\ }\href {\doibase
  10.1093/mnras/sty2405} {\bibfield  {journal} {\bibinfo  {journal} {\mnras}\
  }\textbf {\bibinfo {volume} {481}},\ \bibinfo {pages} {2371} (\bibinfo {year}
  {2018})},\ \Eprint {http://arxiv.org/abs/1803.01746} {arXiv:1803.01746
  [astro-ph.CO]} \BibitemShut {NoStop}%
\bibitem [{\citenamefont {{Beutler}}\ \emph {et~al.}(2016)\citenamefont
  {{Beutler}}, \citenamefont {{Blake}}, \citenamefont {{Koda}}, \citenamefont
  {{Mar{\'\i}n}}, \citenamefont {{Seo}}, \citenamefont {{Cuesta}},\ and\
  \citenamefont {{Schneider}}}]{Beutler16}%
  \BibitemOpen
  \bibfield  {author} {\bibinfo {author} {\bibfnamefont {F.}~\bibnamefont
  {{Beutler}}}, \bibinfo {author} {\bibfnamefont {C.}~\bibnamefont {{Blake}}},
  \bibinfo {author} {\bibfnamefont {J.}~\bibnamefont {{Koda}}}, \bibinfo
  {author} {\bibfnamefont {F.~A.}\ \bibnamefont {{Mar{\'\i}n}}}, \bibinfo
  {author} {\bibfnamefont {H.-J.}\ \bibnamefont {{Seo}}}, \bibinfo {author}
  {\bibfnamefont {A.~J.}\ \bibnamefont {{Cuesta}}}, \ and\ \bibinfo {author}
  {\bibfnamefont {D.~P.}\ \bibnamefont {{Schneider}}},\ }\href {\doibase
  10.1093/mnras/stv1943} {\bibfield  {journal} {\bibinfo  {journal} {\mnras}\
  }\textbf {\bibinfo {volume} {455}},\ \bibinfo {pages} {3230} (\bibinfo {year}
  {2016})},\ \Eprint {http://arxiv.org/abs/1506.03900} {arXiv:1506.03900
  [astro-ph.CO]} \BibitemShut {NoStop}%
\bibitem [{\citenamefont {{Ross}}\ \emph {et~al.}(2015)\citenamefont {{Ross}},
  \citenamefont {{Samushia}}, \citenamefont {{Howlett}}, \citenamefont
  {{Percival}}, \citenamefont {{Burden}},\ and\ \citenamefont
  {{Manera}}}]{ross15a}%
  \BibitemOpen
  \bibfield  {author} {\bibinfo {author} {\bibfnamefont {A.~J.}\ \bibnamefont
  {{Ross}}}, \bibinfo {author} {\bibfnamefont {L.}~\bibnamefont {{Samushia}}},
  \bibinfo {author} {\bibfnamefont {C.}~\bibnamefont {{Howlett}}}, \bibinfo
  {author} {\bibfnamefont {W.~J.}\ \bibnamefont {{Percival}}}, \bibinfo
  {author} {\bibfnamefont {A.}~\bibnamefont {{Burden}}}, \ and\ \bibinfo
  {author} {\bibfnamefont {M.}~\bibnamefont {{Manera}}},\ }\href {\doibase
  10.1093/mnras/stv154} {\bibfield  {journal} {\bibinfo  {journal} {\mnras}\
  }\textbf {\bibinfo {volume} {449}},\ \bibinfo {pages} {835} (\bibinfo {year}
  {2015})},\ \Eprint {http://arxiv.org/abs/1409.3242} {arXiv:1409.3242}
  \BibitemShut {NoStop}%
\bibitem [{\citenamefont {{Howlett}}\ \emph {et~al.}(2015)\citenamefont
  {{Howlett}}, \citenamefont {{Ross}}, \citenamefont {{Samushia}},
  \citenamefont {{Percival}},\ and\ \citenamefont {{Manera}}}]{howlett15a}%
  \BibitemOpen
  \bibfield  {author} {\bibinfo {author} {\bibfnamefont {C.}~\bibnamefont
  {{Howlett}}}, \bibinfo {author} {\bibfnamefont {A.~J.}\ \bibnamefont
  {{Ross}}}, \bibinfo {author} {\bibfnamefont {L.}~\bibnamefont {{Samushia}}},
  \bibinfo {author} {\bibfnamefont {W.~J.}\ \bibnamefont {{Percival}}}, \ and\
  \bibinfo {author} {\bibfnamefont {M.}~\bibnamefont {{Manera}}},\ }\href
  {\doibase 10.1093/mnras/stu2693} {\bibfield  {journal} {\bibinfo  {journal}
  {\mnras}\ }\textbf {\bibinfo {volume} {449}},\ \bibinfo {pages} {848}
  (\bibinfo {year} {2015})},\ \Eprint {http://arxiv.org/abs/1409.3238}
  {arXiv:1409.3238} \BibitemShut {NoStop}%
\bibitem [{\citenamefont {{Abazajian}}\ \emph {et~al.}(2009)\citenamefont
  {{Abazajian}}, \citenamefont {{Adelman-McCarthy}}, \citenamefont
  {{Ag{\"u}eros}}, \citenamefont {{Allam}}, \citenamefont {{Allende Prieto}},
  \citenamefont {{An}}, \citenamefont {{Anderson}}, \citenamefont {{Anderson}},
  \citenamefont {{Annis}}, \citenamefont {{Bahcall}},\ and\ \citenamefont
  {et~al.}}]{abazajian09a}%
  \BibitemOpen
  \bibfield  {author} {\bibinfo {author} {\bibfnamefont {K.~N.}\ \bibnamefont
  {{Abazajian}}}, \bibinfo {author} {\bibfnamefont {J.~K.}\ \bibnamefont
  {{Adelman-McCarthy}}}, \bibinfo {author} {\bibfnamefont {M.~A.}\ \bibnamefont
  {{Ag{\"u}eros}}}, \bibinfo {author} {\bibfnamefont {S.~S.}\ \bibnamefont
  {{Allam}}}, \bibinfo {author} {\bibfnamefont {C.}~\bibnamefont {{Allende
  Prieto}}}, \bibinfo {author} {\bibfnamefont {D.}~\bibnamefont {{An}}},
  \bibinfo {author} {\bibfnamefont {K.~S.~J.}\ \bibnamefont {{Anderson}}},
  \bibinfo {author} {\bibfnamefont {S.~F.}\ \bibnamefont {{Anderson}}},
  \bibinfo {author} {\bibfnamefont {J.}~\bibnamefont {{Annis}}}, \bibinfo
  {author} {\bibfnamefont {N.~A.}\ \bibnamefont {{Bahcall}}}, \ and\ \bibinfo
  {author} {\bibnamefont {et~al.}},\ }\href {\doibase
  10.1088/0067-0049/182/2/543} {\bibfield  {journal} {\bibinfo  {journal}
  {\apjs}\ }\textbf {\bibinfo {volume} {182}},\ \bibinfo {pages} {543}
  (\bibinfo {year} {2009})},\ \Eprint {http://arxiv.org/abs/0812.0649}
  {arXiv:0812.0649} \BibitemShut {NoStop}%
\bibitem [{\citenamefont {{Reid}}\ \emph {et~al.}(2016)\citenamefont {{Reid}},
  \citenamefont {{Ho}}, \citenamefont {{Padmanabhan}}, \citenamefont
  {{Percival}}, \citenamefont {{Tinker}}, \citenamefont {{Tojeiro}},
  \citenamefont {{White}}, \citenamefont {{Eisenstein}}, \citenamefont
  {{Maraston}}, \citenamefont {{Ross}}, \citenamefont {{S{\'a}nchez}},
  \citenamefont {{Schlegel}}, \citenamefont {{Sheldon}}, \citenamefont
  {{Strauss}}, \citenamefont {{Thomas}}, \citenamefont {{Wake}}, \citenamefont
  {{Beutler}}, \citenamefont {{Bizyaev}}, \citenamefont {{Bolton}},
  \citenamefont {{Brownstein}}, \citenamefont {{Chuang}}, \citenamefont
  {{Dawson}}, \citenamefont {{Harding}}, \citenamefont {{Kitaura}},
  \citenamefont {{Leauthaud}}, \citenamefont {{Masters}}, \citenamefont
  {{McBride}}, \citenamefont {{More}}, \citenamefont {{Olmstead}},
  \citenamefont {{Oravetz}}, \citenamefont {{Nuza}}, \citenamefont {{Pan}},
  \citenamefont {{Parejko}}, \citenamefont {{Pforr}}, \citenamefont {{Prada}},
  \citenamefont {{Rodr{\'{\i}}guez-Torres}}, \citenamefont
  {{Salazar-Albornoz}}, \citenamefont {{Samushia}}, \citenamefont
  {{Schneider}}, \citenamefont {{Sc{\'o}ccola}}, \citenamefont {{Simmons}},\
  and\ \citenamefont {{Vargas-Magana}}}]{reid16a}%
  \BibitemOpen
  \bibfield  {author} {\bibinfo {author} {\bibfnamefont {B.}~\bibnamefont
  {{Reid}}}, \bibinfo {author} {\bibfnamefont {S.}~\bibnamefont {{Ho}}},
  \bibinfo {author} {\bibfnamefont {N.}~\bibnamefont {{Padmanabhan}}}, \bibinfo
  {author} {\bibfnamefont {W.~J.}\ \bibnamefont {{Percival}}}, \bibinfo
  {author} {\bibfnamefont {J.}~\bibnamefont {{Tinker}}}, \bibinfo {author}
  {\bibfnamefont {R.}~\bibnamefont {{Tojeiro}}}, \bibinfo {author}
  {\bibfnamefont {M.}~\bibnamefont {{White}}}, \bibinfo {author} {\bibfnamefont
  {D.~J.}\ \bibnamefont {{Eisenstein}}}, \bibinfo {author} {\bibfnamefont
  {C.}~\bibnamefont {{Maraston}}}, \bibinfo {author} {\bibfnamefont {A.~J.}\
  \bibnamefont {{Ross}}}, \bibinfo {author} {\bibfnamefont {A.~G.}\
  \bibnamefont {{S{\'a}nchez}}}, \bibinfo {author} {\bibfnamefont
  {D.}~\bibnamefont {{Schlegel}}}, \bibinfo {author} {\bibfnamefont
  {E.}~\bibnamefont {{Sheldon}}}, \bibinfo {author} {\bibfnamefont {M.~A.}\
  \bibnamefont {{Strauss}}}, \bibinfo {author} {\bibfnamefont {D.}~\bibnamefont
  {{Thomas}}}, \bibinfo {author} {\bibfnamefont {D.}~\bibnamefont {{Wake}}},
  \bibinfo {author} {\bibfnamefont {F.}~\bibnamefont {{Beutler}}}, \bibinfo
  {author} {\bibfnamefont {D.}~\bibnamefont {{Bizyaev}}}, \bibinfo {author}
  {\bibfnamefont {A.~S.}\ \bibnamefont {{Bolton}}}, \bibinfo {author}
  {\bibfnamefont {J.~R.}\ \bibnamefont {{Brownstein}}}, \bibinfo {author}
  {\bibfnamefont {C.-H.}\ \bibnamefont {{Chuang}}}, \bibinfo {author}
  {\bibfnamefont {K.}~\bibnamefont {{Dawson}}}, \bibinfo {author}
  {\bibfnamefont {P.}~\bibnamefont {{Harding}}}, \bibinfo {author}
  {\bibfnamefont {F.-S.}\ \bibnamefont {{Kitaura}}}, \bibinfo {author}
  {\bibfnamefont {A.}~\bibnamefont {{Leauthaud}}}, \bibinfo {author}
  {\bibfnamefont {K.}~\bibnamefont {{Masters}}}, \bibinfo {author}
  {\bibfnamefont {C.~K.}\ \bibnamefont {{McBride}}}, \bibinfo {author}
  {\bibfnamefont {S.}~\bibnamefont {{More}}}, \bibinfo {author} {\bibfnamefont
  {M.~D.}\ \bibnamefont {{Olmstead}}}, \bibinfo {author} {\bibfnamefont
  {D.}~\bibnamefont {{Oravetz}}}, \bibinfo {author} {\bibfnamefont {S.~E.}\
  \bibnamefont {{Nuza}}}, \bibinfo {author} {\bibfnamefont {K.}~\bibnamefont
  {{Pan}}}, \bibinfo {author} {\bibfnamefont {J.}~\bibnamefont {{Parejko}}},
  \bibinfo {author} {\bibfnamefont {J.}~\bibnamefont {{Pforr}}}, \bibinfo
  {author} {\bibfnamefont {F.}~\bibnamefont {{Prada}}}, \bibinfo {author}
  {\bibfnamefont {S.}~\bibnamefont {{Rodr{\'{\i}}guez-Torres}}}, \bibinfo
  {author} {\bibfnamefont {S.}~\bibnamefont {{Salazar-Albornoz}}}, \bibinfo
  {author} {\bibfnamefont {L.}~\bibnamefont {{Samushia}}}, \bibinfo {author}
  {\bibfnamefont {D.~P.}\ \bibnamefont {{Schneider}}}, \bibinfo {author}
  {\bibfnamefont {C.~G.}\ \bibnamefont {{Sc{\'o}ccola}}}, \bibinfo {author}
  {\bibfnamefont {A.}~\bibnamefont {{Simmons}}}, \ and\ \bibinfo {author}
  {\bibfnamefont {M.}~\bibnamefont {{Vargas-Magana}}},\ }\href {\doibase
  10.1093/mnras/stv2382} {\bibfield  {journal} {\bibinfo  {journal} {\mnras}\
  }\textbf {\bibinfo {volume} {455}},\ \bibinfo {pages} {1553} (\bibinfo {year}
  {2016})},\ \Eprint {http://arxiv.org/abs/1509.06529} {arXiv:1509.06529}
  \BibitemShut {NoStop}%
\bibitem [{\citenamefont {{Ross}}\ \emph {et~al.}(2017)\citenamefont {{Ross}},
  \citenamefont {{Beutler}}, \citenamefont {{Chuang}}, \citenamefont
  {{Pellejero-Ibanez}}, \citenamefont {{Seo}}, \citenamefont
  {{Vargas-Maga{\~n}a}}, \citenamefont {{Cuesta}}, \citenamefont {{Percival}},
  \citenamefont {{Burden}}, \citenamefont {{S{\'a}nchez}}, \citenamefont
  {{Grieb}}, \citenamefont {{Reid}}, \citenamefont {{Brownstein}},
  \citenamefont {{Dawson}}, \citenamefont {{Eisenstein}}, \citenamefont {{Ho}},
  \citenamefont {{Kitaura}}, \citenamefont {{Nichol}}, \citenamefont
  {{Olmstead}}, \citenamefont {{Prada}}, \citenamefont
  {{Rodr{\'{\i}}guez-Torres}}, \citenamefont {{Saito}}, \citenamefont
  {{Salazar-Albornoz}}, \citenamefont {{Schneider}}, \citenamefont {{Thomas}},
  \citenamefont {{Tinker}}, \citenamefont {{Tojeiro}}, \citenamefont {{Wang}},
  \citenamefont {{White}},\ and\ \citenamefont {{Zhao}}}]{ross17a}%
  \BibitemOpen
  \bibfield  {author} {\bibinfo {author} {\bibfnamefont {A.~J.}\ \bibnamefont
  {{Ross}}}, \bibinfo {author} {\bibfnamefont {F.}~\bibnamefont {{Beutler}}},
  \bibinfo {author} {\bibfnamefont {C.-H.}\ \bibnamefont {{Chuang}}}, \bibinfo
  {author} {\bibfnamefont {M.}~\bibnamefont {{Pellejero-Ibanez}}}, \bibinfo
  {author} {\bibfnamefont {H.-J.}\ \bibnamefont {{Seo}}}, \bibinfo {author}
  {\bibfnamefont {M.}~\bibnamefont {{Vargas-Maga{\~n}a}}}, \bibinfo {author}
  {\bibfnamefont {A.~J.}\ \bibnamefont {{Cuesta}}}, \bibinfo {author}
  {\bibfnamefont {W.~J.}\ \bibnamefont {{Percival}}}, \bibinfo {author}
  {\bibfnamefont {A.}~\bibnamefont {{Burden}}}, \bibinfo {author}
  {\bibfnamefont {A.~G.}\ \bibnamefont {{S{\'a}nchez}}}, \bibinfo {author}
  {\bibfnamefont {J.~N.}\ \bibnamefont {{Grieb}}}, \bibinfo {author}
  {\bibfnamefont {B.}~\bibnamefont {{Reid}}}, \bibinfo {author} {\bibfnamefont
  {J.~R.}\ \bibnamefont {{Brownstein}}}, \bibinfo {author} {\bibfnamefont
  {K.~S.}\ \bibnamefont {{Dawson}}}, \bibinfo {author} {\bibfnamefont {D.~J.}\
  \bibnamefont {{Eisenstein}}}, \bibinfo {author} {\bibfnamefont
  {S.}~\bibnamefont {{Ho}}}, \bibinfo {author} {\bibfnamefont {F.-S.}\
  \bibnamefont {{Kitaura}}}, \bibinfo {author} {\bibfnamefont {R.~C.}\
  \bibnamefont {{Nichol}}}, \bibinfo {author} {\bibfnamefont {M.~D.}\
  \bibnamefont {{Olmstead}}}, \bibinfo {author} {\bibfnamefont
  {F.}~\bibnamefont {{Prada}}}, \bibinfo {author} {\bibfnamefont {S.~A.}\
  \bibnamefont {{Rodr{\'{\i}}guez-Torres}}}, \bibinfo {author} {\bibfnamefont
  {S.}~\bibnamefont {{Saito}}}, \bibinfo {author} {\bibfnamefont
  {S.}~\bibnamefont {{Salazar-Albornoz}}}, \bibinfo {author} {\bibfnamefont
  {D.~P.}\ \bibnamefont {{Schneider}}}, \bibinfo {author} {\bibfnamefont
  {D.}~\bibnamefont {{Thomas}}}, \bibinfo {author} {\bibfnamefont
  {J.}~\bibnamefont {{Tinker}}}, \bibinfo {author} {\bibfnamefont
  {R.}~\bibnamefont {{Tojeiro}}}, \bibinfo {author} {\bibfnamefont
  {Y.}~\bibnamefont {{Wang}}}, \bibinfo {author} {\bibfnamefont
  {M.}~\bibnamefont {{White}}}, \ and\ \bibinfo {author} {\bibfnamefont
  {G.-b.}\ \bibnamefont {{Zhao}}},\ }\href {\doibase 10.1093/mnras/stw2372}
  {\bibfield  {journal} {\bibinfo  {journal} {\mnras}\ }\textbf {\bibinfo
  {volume} {464}},\ \bibinfo {pages} {1168} (\bibinfo {year} {2017})},\ \Eprint
  {http://arxiv.org/abs/1607.03145} {arXiv:1607.03145} \BibitemShut {NoStop}%
\bibitem [{\citenamefont {{Vargas-Maga{\~n}a}}\ \emph
  {et~al.}(2018)\citenamefont {{Vargas-Maga{\~n}a}}, \citenamefont {{Ho}},
  \citenamefont {{Cuesta}}, \citenamefont {{O'Connell}}, \citenamefont
  {{Ross}}, \citenamefont {{Eisenstein}}, \citenamefont {{Percival}},
  \citenamefont {{Grieb}}, \citenamefont {{S{\'a}nchez}}, \citenamefont
  {{Tinker}}, \citenamefont {{Tojeiro}}, \citenamefont {{Beutler}},
  \citenamefont {{Chuang}}, \citenamefont {{Kitaura}}, \citenamefont {{Prada}},
  \citenamefont {{Rodr{\'{\i}}guez-Torres}}, \citenamefont {{Rossi}},
  \citenamefont {{Seo}}, \citenamefont {{Brownstein}}, \citenamefont
  {{Olmstead}},\ and\ \citenamefont {{Thomas}}}]{vargas18a}%
  \BibitemOpen
  \bibfield  {author} {\bibinfo {author} {\bibfnamefont {M.}~\bibnamefont
  {{Vargas-Maga{\~n}a}}}, \bibinfo {author} {\bibfnamefont {S.}~\bibnamefont
  {{Ho}}}, \bibinfo {author} {\bibfnamefont {A.~J.}\ \bibnamefont {{Cuesta}}},
  \bibinfo {author} {\bibfnamefont {R.}~\bibnamefont {{O'Connell}}}, \bibinfo
  {author} {\bibfnamefont {A.~J.}\ \bibnamefont {{Ross}}}, \bibinfo {author}
  {\bibfnamefont {D.~J.}\ \bibnamefont {{Eisenstein}}}, \bibinfo {author}
  {\bibfnamefont {W.~J.}\ \bibnamefont {{Percival}}}, \bibinfo {author}
  {\bibfnamefont {J.~N.}\ \bibnamefont {{Grieb}}}, \bibinfo {author}
  {\bibfnamefont {A.~G.}\ \bibnamefont {{S{\'a}nchez}}}, \bibinfo {author}
  {\bibfnamefont {J.~L.}\ \bibnamefont {{Tinker}}}, \bibinfo {author}
  {\bibfnamefont {R.}~\bibnamefont {{Tojeiro}}}, \bibinfo {author}
  {\bibfnamefont {F.}~\bibnamefont {{Beutler}}}, \bibinfo {author}
  {\bibfnamefont {C.-H.}\ \bibnamefont {{Chuang}}}, \bibinfo {author}
  {\bibfnamefont {F.-S.}\ \bibnamefont {{Kitaura}}}, \bibinfo {author}
  {\bibfnamefont {F.}~\bibnamefont {{Prada}}}, \bibinfo {author} {\bibfnamefont
  {S.~A.}\ \bibnamefont {{Rodr{\'{\i}}guez-Torres}}}, \bibinfo {author}
  {\bibfnamefont {G.}~\bibnamefont {{Rossi}}}, \bibinfo {author} {\bibfnamefont
  {H.-J.}\ \bibnamefont {{Seo}}}, \bibinfo {author} {\bibfnamefont {J.~R.}\
  \bibnamefont {{Brownstein}}}, \bibinfo {author} {\bibfnamefont
  {M.}~\bibnamefont {{Olmstead}}}, \ and\ \bibinfo {author} {\bibfnamefont
  {D.}~\bibnamefont {{Thomas}}},\ }\href {\doibase 10.1093/mnras/sty571}
  {\bibfield  {journal} {\bibinfo  {journal} {\mnras}\ }\textbf {\bibinfo
  {volume} {477}},\ \bibinfo {pages} {1153} (\bibinfo {year} {2018})},\ \Eprint
  {http://arxiv.org/abs/1610.03506} {arXiv:1610.03506} \BibitemShut {NoStop}%
\bibitem [{\citenamefont {{Beutler}}\ \emph
  {et~al.}(2017{\natexlab{a}})\citenamefont {{Beutler}}, \citenamefont {{Seo}},
  \citenamefont {{Ross}}, \citenamefont {{McDonald}}, \citenamefont {{Saito}},
  \citenamefont {{Bolton}}, \citenamefont {{Brownstein}}, \citenamefont
  {{Chuang}}, \citenamefont {{Cuesta}}, \citenamefont {{Eisenstein}},
  \citenamefont {{Font-Ribera}}, \citenamefont {{Grieb}}, \citenamefont
  {{Hand}}, \citenamefont {{Kitaura}}, \citenamefont {{Modi}}, \citenamefont
  {{Nichol}}, \citenamefont {{Percival}}, \citenamefont {{Prada}},
  \citenamefont {{Rodriguez-Torres}}, \citenamefont {{Roe}}, \citenamefont
  {{Ross}}, \citenamefont {{Salazar-Albornoz}}, \citenamefont {{S{\'a}nchez}},
  \citenamefont {{Schneider}}, \citenamefont {{Slosar}}, \citenamefont
  {{Tinker}}, \citenamefont {{Tojeiro}}, \citenamefont {{Vargas-Maga{\~n}a}},\
  and\ \citenamefont {{Vazquez}}}]{beutler17a}%
  \BibitemOpen
  \bibfield  {author} {\bibinfo {author} {\bibfnamefont {F.}~\bibnamefont
  {{Beutler}}}, \bibinfo {author} {\bibfnamefont {H.-J.}\ \bibnamefont
  {{Seo}}}, \bibinfo {author} {\bibfnamefont {A.~J.}\ \bibnamefont {{Ross}}},
  \bibinfo {author} {\bibfnamefont {P.}~\bibnamefont {{McDonald}}}, \bibinfo
  {author} {\bibfnamefont {S.}~\bibnamefont {{Saito}}}, \bibinfo {author}
  {\bibfnamefont {A.~S.}\ \bibnamefont {{Bolton}}}, \bibinfo {author}
  {\bibfnamefont {J.~R.}\ \bibnamefont {{Brownstein}}}, \bibinfo {author}
  {\bibfnamefont {C.-H.}\ \bibnamefont {{Chuang}}}, \bibinfo {author}
  {\bibfnamefont {A.~J.}\ \bibnamefont {{Cuesta}}}, \bibinfo {author}
  {\bibfnamefont {D.~J.}\ \bibnamefont {{Eisenstein}}}, \bibinfo {author}
  {\bibfnamefont {A.}~\bibnamefont {{Font-Ribera}}}, \bibinfo {author}
  {\bibfnamefont {J.~N.}\ \bibnamefont {{Grieb}}}, \bibinfo {author}
  {\bibfnamefont {N.}~\bibnamefont {{Hand}}}, \bibinfo {author} {\bibfnamefont
  {F.-S.}\ \bibnamefont {{Kitaura}}}, \bibinfo {author} {\bibfnamefont
  {C.}~\bibnamefont {{Modi}}}, \bibinfo {author} {\bibfnamefont {R.~C.}\
  \bibnamefont {{Nichol}}}, \bibinfo {author} {\bibfnamefont {W.~J.}\
  \bibnamefont {{Percival}}}, \bibinfo {author} {\bibfnamefont
  {F.}~\bibnamefont {{Prada}}}, \bibinfo {author} {\bibfnamefont
  {S.}~\bibnamefont {{Rodriguez-Torres}}}, \bibinfo {author} {\bibfnamefont
  {N.~A.}\ \bibnamefont {{Roe}}}, \bibinfo {author} {\bibfnamefont {N.~P.}\
  \bibnamefont {{Ross}}}, \bibinfo {author} {\bibfnamefont {S.}~\bibnamefont
  {{Salazar-Albornoz}}}, \bibinfo {author} {\bibfnamefont {A.~G.}\ \bibnamefont
  {{S{\'a}nchez}}}, \bibinfo {author} {\bibfnamefont {D.~P.}\ \bibnamefont
  {{Schneider}}}, \bibinfo {author} {\bibfnamefont {A.}~\bibnamefont
  {{Slosar}}}, \bibinfo {author} {\bibfnamefont {J.}~\bibnamefont {{Tinker}}},
  \bibinfo {author} {\bibfnamefont {R.}~\bibnamefont {{Tojeiro}}}, \bibinfo
  {author} {\bibfnamefont {M.}~\bibnamefont {{Vargas-Maga{\~n}a}}}, \ and\
  \bibinfo {author} {\bibfnamefont {J.~A.}\ \bibnamefont {{Vazquez}}},\ }\href
  {\doibase 10.1093/mnras/stw2373} {\bibfield  {journal} {\bibinfo  {journal}
  {\mnras}\ }\textbf {\bibinfo {volume} {464}},\ \bibinfo {pages} {3409}
  (\bibinfo {year} {2017}{\natexlab{a}})},\ \Eprint
  {http://arxiv.org/abs/1607.03149} {arXiv:1607.03149} \BibitemShut {NoStop}%
\bibitem [{\citenamefont {{Beutler}}\ \emph
  {et~al.}(2017{\natexlab{b}})\citenamefont {{Beutler}}, \citenamefont {{Seo}},
  \citenamefont {{Saito}}, \citenamefont {{Chuang}}, \citenamefont {{Cuesta}},
  \citenamefont {{Eisenstein}}, \citenamefont {{Gil-Mar{\'{\i}}n}},
  \citenamefont {{Grieb}}, \citenamefont {{Hand}}, \citenamefont {{Kitaura}},
  \citenamefont {{Modi}}, \citenamefont {{Nichol}}, \citenamefont {{Olmstead}},
  \citenamefont {{Percival}}, \citenamefont {{Prada}}, \citenamefont
  {{S{\'a}nchez}}, \citenamefont {{Rodriguez-Torres}}, \citenamefont {{Ross}},
  \citenamefont {{Ross}}, \citenamefont {{Schneider}}, \citenamefont
  {{Tinker}}, \citenamefont {{Tojeiro}},\ and\ \citenamefont
  {{Vargas-Maga{\~n}a}}}]{beutler17b}%
  \BibitemOpen
  \bibfield  {author} {\bibinfo {author} {\bibfnamefont {F.}~\bibnamefont
  {{Beutler}}}, \bibinfo {author} {\bibfnamefont {H.-J.}\ \bibnamefont
  {{Seo}}}, \bibinfo {author} {\bibfnamefont {S.}~\bibnamefont {{Saito}}},
  \bibinfo {author} {\bibfnamefont {C.-H.}\ \bibnamefont {{Chuang}}}, \bibinfo
  {author} {\bibfnamefont {A.~J.}\ \bibnamefont {{Cuesta}}}, \bibinfo {author}
  {\bibfnamefont {D.~J.}\ \bibnamefont {{Eisenstein}}}, \bibinfo {author}
  {\bibfnamefont {H.}~\bibnamefont {{Gil-Mar{\'{\i}}n}}}, \bibinfo {author}
  {\bibfnamefont {J.~N.}\ \bibnamefont {{Grieb}}}, \bibinfo {author}
  {\bibfnamefont {N.}~\bibnamefont {{Hand}}}, \bibinfo {author} {\bibfnamefont
  {F.-S.}\ \bibnamefont {{Kitaura}}}, \bibinfo {author} {\bibfnamefont
  {C.}~\bibnamefont {{Modi}}}, \bibinfo {author} {\bibfnamefont {R.~C.}\
  \bibnamefont {{Nichol}}}, \bibinfo {author} {\bibfnamefont {M.~D.}\
  \bibnamefont {{Olmstead}}}, \bibinfo {author} {\bibfnamefont {W.~J.}\
  \bibnamefont {{Percival}}}, \bibinfo {author} {\bibfnamefont
  {F.}~\bibnamefont {{Prada}}}, \bibinfo {author} {\bibfnamefont {A.~G.}\
  \bibnamefont {{S{\'a}nchez}}}, \bibinfo {author} {\bibfnamefont
  {S.}~\bibnamefont {{Rodriguez-Torres}}}, \bibinfo {author} {\bibfnamefont
  {A.~J.}\ \bibnamefont {{Ross}}}, \bibinfo {author} {\bibfnamefont {N.~P.}\
  \bibnamefont {{Ross}}}, \bibinfo {author} {\bibfnamefont {D.~P.}\
  \bibnamefont {{Schneider}}}, \bibinfo {author} {\bibfnamefont
  {J.}~\bibnamefont {{Tinker}}}, \bibinfo {author} {\bibfnamefont
  {R.}~\bibnamefont {{Tojeiro}}}, \ and\ \bibinfo {author} {\bibfnamefont
  {M.}~\bibnamefont {{Vargas-Maga{\~n}a}}},\ }\href {\doibase
  10.1093/mnras/stw3298} {\bibfield  {journal} {\bibinfo  {journal} {\mnras}\
  }\textbf {\bibinfo {volume} {466}},\ \bibinfo {pages} {2242} (\bibinfo {year}
  {2017}{\natexlab{b}})},\ \Eprint {http://arxiv.org/abs/1607.03150}
  {arXiv:1607.03150} \BibitemShut {NoStop}%
\bibitem [{\citenamefont {{Satpathy}}\ \emph {et~al.}(2017)\citenamefont
  {{Satpathy}}, \citenamefont {{Alam}}, \citenamefont {{Ho}}, \citenamefont
  {{White}}, \citenamefont {{Bahcall}}, \citenamefont {{Beutler}},
  \citenamefont {{Brownstein}}, \citenamefont {{Chuang}}, \citenamefont
  {{Eisenstein}}, \citenamefont {{Grieb}}, \citenamefont {{Kitaura}},
  \citenamefont {{Olmstead}}, \citenamefont {{Percival}}, \citenamefont
  {{Salazar-Albornoz}}, \citenamefont {{S{\'a}nchez}}, \citenamefont {{Seo}},
  \citenamefont {{Thomas}}, \citenamefont {{Tinker}},\ and\ \citenamefont
  {{Tojeiro}}}]{satpathy17a}%
  \BibitemOpen
  \bibfield  {author} {\bibinfo {author} {\bibfnamefont {S.}~\bibnamefont
  {{Satpathy}}}, \bibinfo {author} {\bibfnamefont {S.}~\bibnamefont {{Alam}}},
  \bibinfo {author} {\bibfnamefont {S.}~\bibnamefont {{Ho}}}, \bibinfo {author}
  {\bibfnamefont {M.}~\bibnamefont {{White}}}, \bibinfo {author} {\bibfnamefont
  {N.~A.}\ \bibnamefont {{Bahcall}}}, \bibinfo {author} {\bibfnamefont
  {F.}~\bibnamefont {{Beutler}}}, \bibinfo {author} {\bibfnamefont {J.~R.}\
  \bibnamefont {{Brownstein}}}, \bibinfo {author} {\bibfnamefont {C.-H.}\
  \bibnamefont {{Chuang}}}, \bibinfo {author} {\bibfnamefont {D.~J.}\
  \bibnamefont {{Eisenstein}}}, \bibinfo {author} {\bibfnamefont {J.~N.}\
  \bibnamefont {{Grieb}}}, \bibinfo {author} {\bibfnamefont {F.}~\bibnamefont
  {{Kitaura}}}, \bibinfo {author} {\bibfnamefont {M.~D.}\ \bibnamefont
  {{Olmstead}}}, \bibinfo {author} {\bibfnamefont {W.~J.}\ \bibnamefont
  {{Percival}}}, \bibinfo {author} {\bibfnamefont {S.}~\bibnamefont
  {{Salazar-Albornoz}}}, \bibinfo {author} {\bibfnamefont {A.~G.}\ \bibnamefont
  {{S{\'a}nchez}}}, \bibinfo {author} {\bibfnamefont {H.-J.}\ \bibnamefont
  {{Seo}}}, \bibinfo {author} {\bibfnamefont {D.}~\bibnamefont {{Thomas}}},
  \bibinfo {author} {\bibfnamefont {J.~L.}\ \bibnamefont {{Tinker}}}, \ and\
  \bibinfo {author} {\bibfnamefont {R.}~\bibnamefont {{Tojeiro}}},\ }\href
  {\doibase 10.1093/mnras/stx883} {\bibfield  {journal} {\bibinfo  {journal}
  {\mnras}\ }\textbf {\bibinfo {volume} {469}},\ \bibinfo {pages} {1369}
  (\bibinfo {year} {2017})},\ \Eprint {http://arxiv.org/abs/1607.03148}
  {arXiv:1607.03148} \BibitemShut {NoStop}%
\bibitem [{\citenamefont {{S{\'a}nchez}}\ \emph
  {et~al.}(2017{\natexlab{a}})\citenamefont {{S{\'a}nchez}}, \citenamefont
  {{Scoccimarro}}, \citenamefont {{Crocce}}, \citenamefont {{Grieb}},
  \citenamefont {{Salazar-Albornoz}}, \citenamefont {{Dalla Vecchia}},
  \citenamefont {{Lippich}}, \citenamefont {{Beutler}}, \citenamefont
  {{Brownstein}}, \citenamefont {{Chuang}}, \citenamefont {{Eisenstein}},
  \citenamefont {{Kitaura}}, \citenamefont {{Olmstead}}, \citenamefont
  {{Percival}}, \citenamefont {{Prada}}, \citenamefont
  {{Rodr{\'{\i}}guez-Torres}}, \citenamefont {{Ross}}, \citenamefont
  {{Samushia}}, \citenamefont {{Seo}}, \citenamefont {{Tinker}}, \citenamefont
  {{Tojeiro}}, \citenamefont {{Vargas-Maga{\~n}a}}, \citenamefont {{Wang}},\
  and\ \citenamefont {{Zhao}}}]{sanchez17a}%
  \BibitemOpen
  \bibfield  {author} {\bibinfo {author} {\bibfnamefont {A.~G.}\ \bibnamefont
  {{S{\'a}nchez}}}, \bibinfo {author} {\bibfnamefont {R.}~\bibnamefont
  {{Scoccimarro}}}, \bibinfo {author} {\bibfnamefont {M.}~\bibnamefont
  {{Crocce}}}, \bibinfo {author} {\bibfnamefont {J.~N.}\ \bibnamefont
  {{Grieb}}}, \bibinfo {author} {\bibfnamefont {S.}~\bibnamefont
  {{Salazar-Albornoz}}}, \bibinfo {author} {\bibfnamefont {C.}~\bibnamefont
  {{Dalla Vecchia}}}, \bibinfo {author} {\bibfnamefont {M.}~\bibnamefont
  {{Lippich}}}, \bibinfo {author} {\bibfnamefont {F.}~\bibnamefont
  {{Beutler}}}, \bibinfo {author} {\bibfnamefont {J.~R.}\ \bibnamefont
  {{Brownstein}}}, \bibinfo {author} {\bibfnamefont {C.-H.}\ \bibnamefont
  {{Chuang}}}, \bibinfo {author} {\bibfnamefont {D.~J.}\ \bibnamefont
  {{Eisenstein}}}, \bibinfo {author} {\bibfnamefont {F.-S.}\ \bibnamefont
  {{Kitaura}}}, \bibinfo {author} {\bibfnamefont {M.~D.}\ \bibnamefont
  {{Olmstead}}}, \bibinfo {author} {\bibfnamefont {W.~J.}\ \bibnamefont
  {{Percival}}}, \bibinfo {author} {\bibfnamefont {F.}~\bibnamefont {{Prada}}},
  \bibinfo {author} {\bibfnamefont {S.}~\bibnamefont
  {{Rodr{\'{\i}}guez-Torres}}}, \bibinfo {author} {\bibfnamefont {A.~J.}\
  \bibnamefont {{Ross}}}, \bibinfo {author} {\bibfnamefont {L.}~\bibnamefont
  {{Samushia}}}, \bibinfo {author} {\bibfnamefont {H.-J.}\ \bibnamefont
  {{Seo}}}, \bibinfo {author} {\bibfnamefont {J.}~\bibnamefont {{Tinker}}},
  \bibinfo {author} {\bibfnamefont {R.}~\bibnamefont {{Tojeiro}}}, \bibinfo
  {author} {\bibfnamefont {M.}~\bibnamefont {{Vargas-Maga{\~n}a}}}, \bibinfo
  {author} {\bibfnamefont {Y.}~\bibnamefont {{Wang}}}, \ and\ \bibinfo {author}
  {\bibfnamefont {G.-B.}\ \bibnamefont {{Zhao}}},\ }\href {\doibase
  10.1093/mnras/stw2443} {\bibfield  {journal} {\bibinfo  {journal} {\mnras}\
  }\textbf {\bibinfo {volume} {464}},\ \bibinfo {pages} {1640} (\bibinfo {year}
  {2017}{\natexlab{a}})},\ \Eprint {http://arxiv.org/abs/1607.03147}
  {arXiv:1607.03147} \BibitemShut {NoStop}%
\bibitem [{\citenamefont {{Grieb}}\ \emph {et~al.}(2017)\citenamefont
  {{Grieb}}, \citenamefont {{S{\'a}nchez}}, \citenamefont {{Salazar-Albornoz}},
  \citenamefont {{Scoccimarro}}, \citenamefont {{Crocce}}, \citenamefont
  {{Dalla Vecchia}}, \citenamefont {{Montesano}}, \citenamefont
  {{Gil-Mar{\'{\i}}n}}, \citenamefont {{Ross}}, \citenamefont {{Beutler}},
  \citenamefont {{Rodr{\'{\i}}guez-Torres}}, \citenamefont {{Chuang}},
  \citenamefont {{Prada}}, \citenamefont {{Kitaura}}, \citenamefont {{Cuesta}},
  \citenamefont {{Eisenstein}}, \citenamefont {{Percival}}, \citenamefont
  {{Vargas-Maga{\~n}a}}, \citenamefont {{Tinker}}, \citenamefont {{Tojeiro}},
  \citenamefont {{Brownstein}}, \citenamefont {{Maraston}}, \citenamefont
  {{Nichol}}, \citenamefont {{Olmstead}}, \citenamefont {{Samushia}},
  \citenamefont {{Seo}}, \citenamefont {{Streblyanska}},\ and\ \citenamefont
  {{Zhao}}}]{grieb17a}%
  \BibitemOpen
  \bibfield  {author} {\bibinfo {author} {\bibfnamefont {J.~N.}\ \bibnamefont
  {{Grieb}}}, \bibinfo {author} {\bibfnamefont {A.~G.}\ \bibnamefont
  {{S{\'a}nchez}}}, \bibinfo {author} {\bibfnamefont {S.}~\bibnamefont
  {{Salazar-Albornoz}}}, \bibinfo {author} {\bibfnamefont {R.}~\bibnamefont
  {{Scoccimarro}}}, \bibinfo {author} {\bibfnamefont {M.}~\bibnamefont
  {{Crocce}}}, \bibinfo {author} {\bibfnamefont {C.}~\bibnamefont {{Dalla
  Vecchia}}}, \bibinfo {author} {\bibfnamefont {F.}~\bibnamefont
  {{Montesano}}}, \bibinfo {author} {\bibfnamefont {H.}~\bibnamefont
  {{Gil-Mar{\'{\i}}n}}}, \bibinfo {author} {\bibfnamefont {A.~J.}\ \bibnamefont
  {{Ross}}}, \bibinfo {author} {\bibfnamefont {F.}~\bibnamefont {{Beutler}}},
  \bibinfo {author} {\bibfnamefont {S.}~\bibnamefont
  {{Rodr{\'{\i}}guez-Torres}}}, \bibinfo {author} {\bibfnamefont {C.-H.}\
  \bibnamefont {{Chuang}}}, \bibinfo {author} {\bibfnamefont {F.}~\bibnamefont
  {{Prada}}}, \bibinfo {author} {\bibfnamefont {F.-S.}\ \bibnamefont
  {{Kitaura}}}, \bibinfo {author} {\bibfnamefont {A.~J.}\ \bibnamefont
  {{Cuesta}}}, \bibinfo {author} {\bibfnamefont {D.~J.}\ \bibnamefont
  {{Eisenstein}}}, \bibinfo {author} {\bibfnamefont {W.~J.}\ \bibnamefont
  {{Percival}}}, \bibinfo {author} {\bibfnamefont {M.}~\bibnamefont
  {{Vargas-Maga{\~n}a}}}, \bibinfo {author} {\bibfnamefont {J.~L.}\
  \bibnamefont {{Tinker}}}, \bibinfo {author} {\bibfnamefont {R.}~\bibnamefont
  {{Tojeiro}}}, \bibinfo {author} {\bibfnamefont {J.~R.}\ \bibnamefont
  {{Brownstein}}}, \bibinfo {author} {\bibfnamefont {C.}~\bibnamefont
  {{Maraston}}}, \bibinfo {author} {\bibfnamefont {R.~C.}\ \bibnamefont
  {{Nichol}}}, \bibinfo {author} {\bibfnamefont {M.~D.}\ \bibnamefont
  {{Olmstead}}}, \bibinfo {author} {\bibfnamefont {L.}~\bibnamefont
  {{Samushia}}}, \bibinfo {author} {\bibfnamefont {H.-J.}\ \bibnamefont
  {{Seo}}}, \bibinfo {author} {\bibfnamefont {A.}~\bibnamefont
  {{Streblyanska}}}, \ and\ \bibinfo {author} {\bibfnamefont {G.-b.}\
  \bibnamefont {{Zhao}}},\ }\href {\doibase 10.1093/mnras/stw3384} {\bibfield
  {journal} {\bibinfo  {journal} {\mnras}\ }\textbf {\bibinfo {volume} {467}},\
  \bibinfo {pages} {2085} (\bibinfo {year} {2017})},\ \Eprint
  {http://arxiv.org/abs/1607.03143} {arXiv:1607.03143} \BibitemShut {NoStop}%
\bibitem [{\citenamefont {{S{\'a}nchez}}\ \emph
  {et~al.}(2017{\natexlab{b}})\citenamefont {{S{\'a}nchez}}, \citenamefont
  {{Grieb}}, \citenamefont {{Salazar-Albornoz}}, \citenamefont {{Alam}},
  \citenamefont {{Beutler}}, \citenamefont {{Ross}}, \citenamefont
  {{Brownstein}}, \citenamefont {{Chuang}}, \citenamefont {{Cuesta}},
  \citenamefont {{Eisenstein}}, \citenamefont {{Kitaura}}, \citenamefont
  {{Percival}}, \citenamefont {{Prada}}, \citenamefont
  {{Rodr{\'{\i}}guez-Torres}}, \citenamefont {{Seo}}, \citenamefont {{Tinker}},
  \citenamefont {{Tojeiro}}, \citenamefont {{Vargas-Maga{\~n}a}}, \citenamefont
  {{Vazquez}},\ and\ \citenamefont {{Zhao}}}]{sanchez17b}%
  \BibitemOpen
  \bibfield  {author} {\bibinfo {author} {\bibfnamefont {A.~G.}\ \bibnamefont
  {{S{\'a}nchez}}}, \bibinfo {author} {\bibfnamefont {J.~N.}\ \bibnamefont
  {{Grieb}}}, \bibinfo {author} {\bibfnamefont {S.}~\bibnamefont
  {{Salazar-Albornoz}}}, \bibinfo {author} {\bibfnamefont {S.}~\bibnamefont
  {{Alam}}}, \bibinfo {author} {\bibfnamefont {F.}~\bibnamefont {{Beutler}}},
  \bibinfo {author} {\bibfnamefont {A.~J.}\ \bibnamefont {{Ross}}}, \bibinfo
  {author} {\bibfnamefont {J.~R.}\ \bibnamefont {{Brownstein}}}, \bibinfo
  {author} {\bibfnamefont {C.-H.}\ \bibnamefont {{Chuang}}}, \bibinfo {author}
  {\bibfnamefont {A.~J.}\ \bibnamefont {{Cuesta}}}, \bibinfo {author}
  {\bibfnamefont {D.~J.}\ \bibnamefont {{Eisenstein}}}, \bibinfo {author}
  {\bibfnamefont {F.-S.}\ \bibnamefont {{Kitaura}}}, \bibinfo {author}
  {\bibfnamefont {W.~J.}\ \bibnamefont {{Percival}}}, \bibinfo {author}
  {\bibfnamefont {F.}~\bibnamefont {{Prada}}}, \bibinfo {author} {\bibfnamefont
  {S.}~\bibnamefont {{Rodr{\'{\i}}guez-Torres}}}, \bibinfo {author}
  {\bibfnamefont {H.-J.}\ \bibnamefont {{Seo}}}, \bibinfo {author}
  {\bibfnamefont {J.}~\bibnamefont {{Tinker}}}, \bibinfo {author}
  {\bibfnamefont {R.}~\bibnamefont {{Tojeiro}}}, \bibinfo {author}
  {\bibfnamefont {M.}~\bibnamefont {{Vargas-Maga{\~n}a}}}, \bibinfo {author}
  {\bibfnamefont {J.~A.}\ \bibnamefont {{Vazquez}}}, \ and\ \bibinfo {author}
  {\bibfnamefont {G.-B.}\ \bibnamefont {{Zhao}}},\ }\href {\doibase
  10.1093/mnras/stw2495} {\bibfield  {journal} {\bibinfo  {journal} {\mnras}\
  }\textbf {\bibinfo {volume} {464}},\ \bibinfo {pages} {1493} (\bibinfo {year}
  {2017}{\natexlab{b}})},\ \Eprint {http://arxiv.org/abs/1607.03146}
  {arXiv:1607.03146} \BibitemShut {NoStop}%
\bibitem [{\citenamefont {{Wright}}\ \emph {et~al.}(2010)\citenamefont
  {{Wright}}, \citenamefont {{Eisenhardt}}, \citenamefont {{Mainzer}},
  \citenamefont {{Ressler}}, \citenamefont {{Cutri}}, \citenamefont
  {{Jarrett}}, \citenamefont {{Kirkpatrick}}, \citenamefont {{Padgett}},
  \citenamefont {{McMillan}}, \citenamefont {{Skrutskie}}, \citenamefont
  {{Stanford}}, \citenamefont {{Cohen}}, \citenamefont {{Walker}},
  \citenamefont {{Mather}}, \citenamefont {{Leisawitz}}, \citenamefont
  {{Gautier}}, \citenamefont {{McLean}}, \citenamefont {{Benford}},
  \citenamefont {{Lonsdale}}, \citenamefont {{Blain}}, \citenamefont
  {{Mendez}}, \citenamefont {{Irace}}, \citenamefont {{Duval}}, \citenamefont
  {{Liu}}, \citenamefont {{Royer}}, \citenamefont {{Heinrichsen}},
  \citenamefont {{Howard}}, \citenamefont {{Shannon}}, \citenamefont
  {{Kendall}}, \citenamefont {{Walsh}}, \citenamefont {{Larsen}}, \citenamefont
  {{Cardon}}, \citenamefont {{Schick}}, \citenamefont {{Schwalm}},
  \citenamefont {{Abid}}, \citenamefont {{Fabinsky}}, \citenamefont {{Naes}},\
  and\ \citenamefont {{Tsai}}}]{wright10a}%
  \BibitemOpen
  \bibfield  {author} {\bibinfo {author} {\bibfnamefont {E.~L.}\ \bibnamefont
  {{Wright}}}, \bibinfo {author} {\bibfnamefont {P.~R.~M.}\ \bibnamefont
  {{Eisenhardt}}}, \bibinfo {author} {\bibfnamefont {A.~K.}\ \bibnamefont
  {{Mainzer}}}, \bibinfo {author} {\bibfnamefont {M.~E.}\ \bibnamefont
  {{Ressler}}}, \bibinfo {author} {\bibfnamefont {R.~M.}\ \bibnamefont
  {{Cutri}}}, \bibinfo {author} {\bibfnamefont {T.}~\bibnamefont {{Jarrett}}},
  \bibinfo {author} {\bibfnamefont {J.~D.}\ \bibnamefont {{Kirkpatrick}}},
  \bibinfo {author} {\bibfnamefont {D.}~\bibnamefont {{Padgett}}}, \bibinfo
  {author} {\bibfnamefont {R.~S.}\ \bibnamefont {{McMillan}}}, \bibinfo
  {author} {\bibfnamefont {M.}~\bibnamefont {{Skrutskie}}}, \bibinfo {author}
  {\bibfnamefont {S.~A.}\ \bibnamefont {{Stanford}}}, \bibinfo {author}
  {\bibfnamefont {M.}~\bibnamefont {{Cohen}}}, \bibinfo {author} {\bibfnamefont
  {R.~G.}\ \bibnamefont {{Walker}}}, \bibinfo {author} {\bibfnamefont {J.~C.}\
  \bibnamefont {{Mather}}}, \bibinfo {author} {\bibfnamefont {D.}~\bibnamefont
  {{Leisawitz}}}, \bibinfo {author} {\bibfnamefont {T.~N.}\ \bibnamefont
  {{Gautier}}, \bibfnamefont {III}}, \bibinfo {author} {\bibfnamefont
  {I.}~\bibnamefont {{McLean}}}, \bibinfo {author} {\bibfnamefont
  {D.}~\bibnamefont {{Benford}}}, \bibinfo {author} {\bibfnamefont {C.~J.}\
  \bibnamefont {{Lonsdale}}}, \bibinfo {author} {\bibfnamefont
  {A.}~\bibnamefont {{Blain}}}, \bibinfo {author} {\bibfnamefont
  {B.}~\bibnamefont {{Mendez}}}, \bibinfo {author} {\bibfnamefont {W.~R.}\
  \bibnamefont {{Irace}}}, \bibinfo {author} {\bibfnamefont {V.}~\bibnamefont
  {{Duval}}}, \bibinfo {author} {\bibfnamefont {F.}~\bibnamefont {{Liu}}},
  \bibinfo {author} {\bibfnamefont {D.}~\bibnamefont {{Royer}}}, \bibinfo
  {author} {\bibfnamefont {I.}~\bibnamefont {{Heinrichsen}}}, \bibinfo {author}
  {\bibfnamefont {J.}~\bibnamefont {{Howard}}}, \bibinfo {author}
  {\bibfnamefont {M.}~\bibnamefont {{Shannon}}}, \bibinfo {author}
  {\bibfnamefont {M.}~\bibnamefont {{Kendall}}}, \bibinfo {author}
  {\bibfnamefont {A.~L.}\ \bibnamefont {{Walsh}}}, \bibinfo {author}
  {\bibfnamefont {M.}~\bibnamefont {{Larsen}}}, \bibinfo {author}
  {\bibfnamefont {J.~G.}\ \bibnamefont {{Cardon}}}, \bibinfo {author}
  {\bibfnamefont {S.}~\bibnamefont {{Schick}}}, \bibinfo {author}
  {\bibfnamefont {M.}~\bibnamefont {{Schwalm}}}, \bibinfo {author}
  {\bibfnamefont {M.}~\bibnamefont {{Abid}}}, \bibinfo {author} {\bibfnamefont
  {B.}~\bibnamefont {{Fabinsky}}}, \bibinfo {author} {\bibfnamefont
  {L.}~\bibnamefont {{Naes}}}, \ and\ \bibinfo {author} {\bibfnamefont {C.-W.}\
  \bibnamefont {{Tsai}}},\ }\href {\doibase 10.1088/0004-6256/140/6/1868}
  {\bibfield  {journal} {\bibinfo  {journal} {\aj}\ }\textbf {\bibinfo {volume}
  {140}},\ \bibinfo {eid} {1868} (\bibinfo {year} {2010})},\ \Eprint
  {http://arxiv.org/abs/1008.0031} {arXiv:1008.0031 [astro-ph.IM]} \BibitemShut
  {NoStop}%
\bibitem [{\citenamefont {{Prakash}}\ \emph {et~al.}(2016)\citenamefont
  {{Prakash}}, \citenamefont {{Licquia}}, \citenamefont {{Newman}},
  \citenamefont {{Ross}}, \citenamefont {{Myers}}, \citenamefont {{Dawson}},
  \citenamefont {{Kneib}}, \citenamefont {{Percival}}, \citenamefont
  {{Bautista}}, \citenamefont {{Comparat}}, \citenamefont {{Tinker}},
  \citenamefont {{Schlegel}}, \citenamefont {{Tojeiro}}, \citenamefont {{Ho}},
  \citenamefont {{Lang}}, \citenamefont {{Rao}}, \citenamefont {{McBride}},
  \citenamefont {{Ben Zhu}}, \citenamefont {{Brownstein}}, \citenamefont
  {{Bailey}}, \citenamefont {{Bolton}}, \citenamefont {{Delubac}},
  \citenamefont {{Mariappan}}, \citenamefont {{Blanton}}, \citenamefont
  {{Reid}}, \citenamefont {{Schneider}}, \citenamefont {{Seo}}, \citenamefont
  {{Carnero Rosell}},\ and\ \citenamefont {{Prada}}}]{prakash16a}%
  \BibitemOpen
  \bibfield  {author} {\bibinfo {author} {\bibfnamefont {A.}~\bibnamefont
  {{Prakash}}}, \bibinfo {author} {\bibfnamefont {T.~C.}\ \bibnamefont
  {{Licquia}}}, \bibinfo {author} {\bibfnamefont {J.~A.}\ \bibnamefont
  {{Newman}}}, \bibinfo {author} {\bibfnamefont {A.~J.}\ \bibnamefont
  {{Ross}}}, \bibinfo {author} {\bibfnamefont {A.~D.}\ \bibnamefont {{Myers}}},
  \bibinfo {author} {\bibfnamefont {K.~S.}\ \bibnamefont {{Dawson}}}, \bibinfo
  {author} {\bibfnamefont {J.-P.}\ \bibnamefont {{Kneib}}}, \bibinfo {author}
  {\bibfnamefont {W.~J.}\ \bibnamefont {{Percival}}}, \bibinfo {author}
  {\bibfnamefont {J.~E.}\ \bibnamefont {{Bautista}}}, \bibinfo {author}
  {\bibfnamefont {J.}~\bibnamefont {{Comparat}}}, \bibinfo {author}
  {\bibfnamefont {J.~L.}\ \bibnamefont {{Tinker}}}, \bibinfo {author}
  {\bibfnamefont {D.~J.}\ \bibnamefont {{Schlegel}}}, \bibinfo {author}
  {\bibfnamefont {R.}~\bibnamefont {{Tojeiro}}}, \bibinfo {author}
  {\bibfnamefont {S.}~\bibnamefont {{Ho}}}, \bibinfo {author} {\bibfnamefont
  {D.}~\bibnamefont {{Lang}}}, \bibinfo {author} {\bibfnamefont {S.~M.}\
  \bibnamefont {{Rao}}}, \bibinfo {author} {\bibfnamefont {C.~K.}\ \bibnamefont
  {{McBride}}}, \bibinfo {author} {\bibfnamefont {G.}~\bibnamefont {{Ben
  Zhu}}}, \bibinfo {author} {\bibfnamefont {J.~R.}\ \bibnamefont
  {{Brownstein}}}, \bibinfo {author} {\bibfnamefont {S.}~\bibnamefont
  {{Bailey}}}, \bibinfo {author} {\bibfnamefont {A.~S.}\ \bibnamefont
  {{Bolton}}}, \bibinfo {author} {\bibfnamefont {T.}~\bibnamefont {{Delubac}}},
  \bibinfo {author} {\bibfnamefont {V.}~\bibnamefont {{Mariappan}}}, \bibinfo
  {author} {\bibfnamefont {M.~R.}\ \bibnamefont {{Blanton}}}, \bibinfo {author}
  {\bibfnamefont {B.}~\bibnamefont {{Reid}}}, \bibinfo {author} {\bibfnamefont
  {D.~P.}\ \bibnamefont {{Schneider}}}, \bibinfo {author} {\bibfnamefont
  {H.-J.}\ \bibnamefont {{Seo}}}, \bibinfo {author} {\bibfnamefont
  {A.}~\bibnamefont {{Carnero Rosell}}}, \ and\ \bibinfo {author}
  {\bibfnamefont {F.}~\bibnamefont {{Prada}}},\ }\href {\doibase
  10.3847/0067-0049/224/2/34} {\bibfield  {journal} {\bibinfo  {journal}
  {\apjs}\ }\textbf {\bibinfo {volume} {224}},\ \bibinfo {eid} {34} (\bibinfo
  {year} {2016})},\ \Eprint {http://arxiv.org/abs/1508.04478}
  {arXiv:1508.04478} \BibitemShut {NoStop}%
\bibitem [{\citenamefont {{Flaugher}}\ \emph {et~al.}(2015)\citenamefont
  {{Flaugher}}, \citenamefont {{Diehl}}, \citenamefont {{Honscheid}},
  \citenamefont {{Abbott}}, \citenamefont {{Alvarez}}, \citenamefont
  {{Angstadt}}, \citenamefont {{Annis}}, \citenamefont {{Antonik}},
  \citenamefont {{Ballester}}, \citenamefont {{Beaufore}}, \citenamefont
  {{Bernstein}}, \citenamefont {{Bernstein}}, \citenamefont {{Bigelow}},
  \citenamefont {{Bonati}}, \citenamefont {{Boprie}}, \citenamefont {{Brooks}},
  \citenamefont {{Buckley-Geer}}, \citenamefont {{Campa}}, \citenamefont
  {{Cardiel-Sas}}, \citenamefont {{Castander}}, \citenamefont {{Castilla}},
  \citenamefont {{Cease}}, \citenamefont {{Cela-Ruiz}}, \citenamefont
  {{Chappa}}, \citenamefont {{Chi}}, \citenamefont {{Cooper}}, \citenamefont
  {{da Costa}}, \citenamefont {{Dede}}, \citenamefont {{Derylo}}, \citenamefont
  {{DePoy}}, \citenamefont {{de Vicente}}, \citenamefont {{Doel}},
  \citenamefont {{Drlica-Wagner}}, \citenamefont {{Eiting}}, \citenamefont
  {{Elliott}}, \citenamefont {{Emes}}, \citenamefont {{Estrada}}, \citenamefont
  {{Fausti Neto}}, \citenamefont {{Finley}}, \citenamefont {{Flores}},
  \citenamefont {{Frieman}}, \citenamefont {{Gerdes}}, \citenamefont
  {{Gladders}}, \citenamefont {{Gregory}}, \citenamefont {{Gutierrez}},
  \citenamefont {{Hao}}, \citenamefont {{Holland}}, \citenamefont {{Holm}},
  \citenamefont {{Huffman}}, \citenamefont {{Jackson}}, \citenamefont
  {{James}}, \citenamefont {{Jonas}}, \citenamefont {{Karcher}}, \citenamefont
  {{Karliner}}, \citenamefont {{Kent}}, \citenamefont {{Kessler}},
  \citenamefont {{Kozlovsky}}, \citenamefont {{Kron}}, \citenamefont {{Kubik}},
  \citenamefont {{Kuehn}}, \citenamefont {{Kuhlmann}}, \citenamefont {{Kuk}},
  \citenamefont {{Lahav}}, \citenamefont {{Lathrop}}, \citenamefont {{Lee}},
  \citenamefont {{Levi}}, \citenamefont {{Lewis}}, \citenamefont {{Li}},
  \citenamefont {{Mandrichenko}}, \citenamefont {{Marshall}}, \citenamefont
  {{Martinez}}, \citenamefont {{Merritt}}, \citenamefont {{Miquel}},
  \citenamefont {{Mu{\~n}oz}}, \citenamefont {{Neilsen}}, \citenamefont
  {{Nichol}}, \citenamefont {{Nord}}, \citenamefont {{Ogando}}, \citenamefont
  {{Olsen}}, \citenamefont {{Palaio}}, \citenamefont {{Patton}}, \citenamefont
  {{Peoples}}, \citenamefont {{Plazas}}, \citenamefont {{Rauch}}, \citenamefont
  {{Reil}}, \citenamefont {{Rheault}}, \citenamefont {{Roe}}, \citenamefont
  {{Rogers}}, \citenamefont {{Roodman}}, \citenamefont {{Sanchez}},
  \citenamefont {{Scarpine}}, \citenamefont {{Schindler}}, \citenamefont
  {{Schmidt}}, \citenamefont {{Schmitt}}, \citenamefont {{Schubnell}},
  \citenamefont {{Schultz}}, \citenamefont {{Schurter}}, \citenamefont
  {{Scott}}, \citenamefont {{Serrano}}, \citenamefont {{Shaw}}, \citenamefont
  {{Smith}}, \citenamefont {{Soares-Santos}}, \citenamefont {{Stefanik}},
  \citenamefont {{Stuermer}}, \citenamefont {{Suchyta}}, \citenamefont
  {{Sypniewski}}, \citenamefont {{Tarle}}, \citenamefont {{Thaler}},
  \citenamefont {{Tighe}}, \citenamefont {{Tran}}, \citenamefont {{Tucker}},
  \citenamefont {{Walker}}, \citenamefont {{Wang}}, \citenamefont {{Watson}},
  \citenamefont {{Weaverdyck}}, \citenamefont {{Wester}}, \citenamefont
  {{Woods}}, \citenamefont {{Yanny}},\ and\ \citenamefont {{DES
  Collaboration}}}]{flaugher15a}%
  \BibitemOpen
  \bibfield  {author} {\bibinfo {author} {\bibfnamefont {B.}~\bibnamefont
  {{Flaugher}}}, \bibinfo {author} {\bibfnamefont {H.~T.}\ \bibnamefont
  {{Diehl}}}, \bibinfo {author} {\bibfnamefont {K.}~\bibnamefont
  {{Honscheid}}}, \bibinfo {author} {\bibfnamefont {T.~M.~C.}\ \bibnamefont
  {{Abbott}}}, \bibinfo {author} {\bibfnamefont {O.}~\bibnamefont {{Alvarez}}},
  \bibinfo {author} {\bibfnamefont {R.}~\bibnamefont {{Angstadt}}}, \bibinfo
  {author} {\bibfnamefont {J.~T.}\ \bibnamefont {{Annis}}}, \bibinfo {author}
  {\bibfnamefont {M.}~\bibnamefont {{Antonik}}}, \bibinfo {author}
  {\bibfnamefont {O.}~\bibnamefont {{Ballester}}}, \bibinfo {author}
  {\bibfnamefont {L.}~\bibnamefont {{Beaufore}}}, \bibinfo {author}
  {\bibfnamefont {G.~M.}\ \bibnamefont {{Bernstein}}}, \bibinfo {author}
  {\bibfnamefont {R.~A.}\ \bibnamefont {{Bernstein}}}, \bibinfo {author}
  {\bibfnamefont {B.}~\bibnamefont {{Bigelow}}}, \bibinfo {author}
  {\bibfnamefont {M.}~\bibnamefont {{Bonati}}}, \bibinfo {author}
  {\bibfnamefont {D.}~\bibnamefont {{Boprie}}}, \bibinfo {author}
  {\bibfnamefont {D.}~\bibnamefont {{Brooks}}}, \bibinfo {author}
  {\bibfnamefont {E.~J.}\ \bibnamefont {{Buckley-Geer}}}, \bibinfo {author}
  {\bibfnamefont {J.}~\bibnamefont {{Campa}}}, \bibinfo {author} {\bibfnamefont
  {L.}~\bibnamefont {{Cardiel-Sas}}}, \bibinfo {author} {\bibfnamefont {F.~J.}\
  \bibnamefont {{Castander}}}, \bibinfo {author} {\bibfnamefont
  {J.}~\bibnamefont {{Castilla}}}, \bibinfo {author} {\bibfnamefont
  {H.}~\bibnamefont {{Cease}}}, \bibinfo {author} {\bibfnamefont {J.~M.}\
  \bibnamefont {{Cela-Ruiz}}}, \bibinfo {author} {\bibfnamefont
  {S.}~\bibnamefont {{Chappa}}}, \bibinfo {author} {\bibfnamefont
  {E.}~\bibnamefont {{Chi}}}, \bibinfo {author} {\bibfnamefont
  {C.}~\bibnamefont {{Cooper}}}, \bibinfo {author} {\bibfnamefont {L.~N.}\
  \bibnamefont {{da Costa}}}, \bibinfo {author} {\bibfnamefont
  {E.}~\bibnamefont {{Dede}}}, \bibinfo {author} {\bibfnamefont
  {G.}~\bibnamefont {{Derylo}}}, \bibinfo {author} {\bibfnamefont {D.~L.}\
  \bibnamefont {{DePoy}}}, \bibinfo {author} {\bibfnamefont {J.}~\bibnamefont
  {{de Vicente}}}, \bibinfo {author} {\bibfnamefont {P.}~\bibnamefont
  {{Doel}}}, \bibinfo {author} {\bibfnamefont {A.}~\bibnamefont
  {{Drlica-Wagner}}}, \bibinfo {author} {\bibfnamefont {J.}~\bibnamefont
  {{Eiting}}}, \bibinfo {author} {\bibfnamefont {A.~E.}\ \bibnamefont
  {{Elliott}}}, \bibinfo {author} {\bibfnamefont {J.}~\bibnamefont {{Emes}}},
  \bibinfo {author} {\bibfnamefont {J.}~\bibnamefont {{Estrada}}}, \bibinfo
  {author} {\bibfnamefont {A.}~\bibnamefont {{Fausti Neto}}}, \bibinfo {author}
  {\bibfnamefont {D.~A.}\ \bibnamefont {{Finley}}}, \bibinfo {author}
  {\bibfnamefont {R.}~\bibnamefont {{Flores}}}, \bibinfo {author}
  {\bibfnamefont {J.}~\bibnamefont {{Frieman}}}, \bibinfo {author}
  {\bibfnamefont {D.}~\bibnamefont {{Gerdes}}}, \bibinfo {author}
  {\bibfnamefont {M.~D.}\ \bibnamefont {{Gladders}}}, \bibinfo {author}
  {\bibfnamefont {B.}~\bibnamefont {{Gregory}}}, \bibinfo {author}
  {\bibfnamefont {G.~R.}\ \bibnamefont {{Gutierrez}}}, \bibinfo {author}
  {\bibfnamefont {J.}~\bibnamefont {{Hao}}}, \bibinfo {author} {\bibfnamefont
  {S.~E.}\ \bibnamefont {{Holland}}}, \bibinfo {author} {\bibfnamefont
  {S.}~\bibnamefont {{Holm}}}, \bibinfo {author} {\bibfnamefont
  {D.}~\bibnamefont {{Huffman}}}, \bibinfo {author} {\bibfnamefont
  {C.}~\bibnamefont {{Jackson}}}, \bibinfo {author} {\bibfnamefont {D.~J.}\
  \bibnamefont {{James}}}, \bibinfo {author} {\bibfnamefont {M.}~\bibnamefont
  {{Jonas}}}, \bibinfo {author} {\bibfnamefont {A.}~\bibnamefont {{Karcher}}},
  \bibinfo {author} {\bibfnamefont {I.}~\bibnamefont {{Karliner}}}, \bibinfo
  {author} {\bibfnamefont {S.}~\bibnamefont {{Kent}}}, \bibinfo {author}
  {\bibfnamefont {R.}~\bibnamefont {{Kessler}}}, \bibinfo {author}
  {\bibfnamefont {M.}~\bibnamefont {{Kozlovsky}}}, \bibinfo {author}
  {\bibfnamefont {R.~G.}\ \bibnamefont {{Kron}}}, \bibinfo {author}
  {\bibfnamefont {D.}~\bibnamefont {{Kubik}}}, \bibinfo {author} {\bibfnamefont
  {K.}~\bibnamefont {{Kuehn}}}, \bibinfo {author} {\bibfnamefont
  {S.}~\bibnamefont {{Kuhlmann}}}, \bibinfo {author} {\bibfnamefont
  {K.}~\bibnamefont {{Kuk}}}, \bibinfo {author} {\bibfnamefont
  {O.}~\bibnamefont {{Lahav}}}, \bibinfo {author} {\bibfnamefont
  {A.}~\bibnamefont {{Lathrop}}}, \bibinfo {author} {\bibfnamefont
  {J.}~\bibnamefont {{Lee}}}, \bibinfo {author} {\bibfnamefont {M.~E.}\
  \bibnamefont {{Levi}}}, \bibinfo {author} {\bibfnamefont {P.}~\bibnamefont
  {{Lewis}}}, \bibinfo {author} {\bibfnamefont {T.~S.}\ \bibnamefont {{Li}}},
  \bibinfo {author} {\bibfnamefont {I.}~\bibnamefont {{Mandrichenko}}},
  \bibinfo {author} {\bibfnamefont {J.~L.}\ \bibnamefont {{Marshall}}},
  \bibinfo {author} {\bibfnamefont {G.}~\bibnamefont {{Martinez}}}, \bibinfo
  {author} {\bibfnamefont {K.~W.}\ \bibnamefont {{Merritt}}}, \bibinfo {author}
  {\bibfnamefont {R.}~\bibnamefont {{Miquel}}}, \bibinfo {author}
  {\bibfnamefont {F.}~\bibnamefont {{Mu{\~n}oz}}}, \bibinfo {author}
  {\bibfnamefont {E.~H.}\ \bibnamefont {{Neilsen}}}, \bibinfo {author}
  {\bibfnamefont {R.~C.}\ \bibnamefont {{Nichol}}}, \bibinfo {author}
  {\bibfnamefont {B.}~\bibnamefont {{Nord}}}, \bibinfo {author} {\bibfnamefont
  {R.}~\bibnamefont {{Ogando}}}, \bibinfo {author} {\bibfnamefont
  {J.}~\bibnamefont {{Olsen}}}, \bibinfo {author} {\bibfnamefont
  {N.}~\bibnamefont {{Palaio}}}, \bibinfo {author} {\bibfnamefont
  {K.}~\bibnamefont {{Patton}}}, \bibinfo {author} {\bibfnamefont
  {J.}~\bibnamefont {{Peoples}}}, \bibinfo {author} {\bibfnamefont {A.~A.}\
  \bibnamefont {{Plazas}}}, \bibinfo {author} {\bibfnamefont {J.}~\bibnamefont
  {{Rauch}}}, \bibinfo {author} {\bibfnamefont {K.}~\bibnamefont {{Reil}}},
  \bibinfo {author} {\bibfnamefont {J.-P.}\ \bibnamefont {{Rheault}}}, \bibinfo
  {author} {\bibfnamefont {N.~A.}\ \bibnamefont {{Roe}}}, \bibinfo {author}
  {\bibfnamefont {H.}~\bibnamefont {{Rogers}}}, \bibinfo {author}
  {\bibfnamefont {A.}~\bibnamefont {{Roodman}}}, \bibinfo {author}
  {\bibfnamefont {E.}~\bibnamefont {{Sanchez}}}, \bibinfo {author}
  {\bibfnamefont {V.}~\bibnamefont {{Scarpine}}}, \bibinfo {author}
  {\bibfnamefont {R.~H.}\ \bibnamefont {{Schindler}}}, \bibinfo {author}
  {\bibfnamefont {R.}~\bibnamefont {{Schmidt}}}, \bibinfo {author}
  {\bibfnamefont {R.}~\bibnamefont {{Schmitt}}}, \bibinfo {author}
  {\bibfnamefont {M.}~\bibnamefont {{Schubnell}}}, \bibinfo {author}
  {\bibfnamefont {K.}~\bibnamefont {{Schultz}}}, \bibinfo {author}
  {\bibfnamefont {P.}~\bibnamefont {{Schurter}}}, \bibinfo {author}
  {\bibfnamefont {L.}~\bibnamefont {{Scott}}}, \bibinfo {author} {\bibfnamefont
  {S.}~\bibnamefont {{Serrano}}}, \bibinfo {author} {\bibfnamefont {T.~M.}\
  \bibnamefont {{Shaw}}}, \bibinfo {author} {\bibfnamefont {R.~C.}\
  \bibnamefont {{Smith}}}, \bibinfo {author} {\bibfnamefont {M.}~\bibnamefont
  {{Soares-Santos}}}, \bibinfo {author} {\bibfnamefont {A.}~\bibnamefont
  {{Stefanik}}}, \bibinfo {author} {\bibfnamefont {W.}~\bibnamefont
  {{Stuermer}}}, \bibinfo {author} {\bibfnamefont {E.}~\bibnamefont
  {{Suchyta}}}, \bibinfo {author} {\bibfnamefont {A.}~\bibnamefont
  {{Sypniewski}}}, \bibinfo {author} {\bibfnamefont {G.}~\bibnamefont
  {{Tarle}}}, \bibinfo {author} {\bibfnamefont {J.}~\bibnamefont {{Thaler}}},
  \bibinfo {author} {\bibfnamefont {R.}~\bibnamefont {{Tighe}}}, \bibinfo
  {author} {\bibfnamefont {C.}~\bibnamefont {{Tran}}}, \bibinfo {author}
  {\bibfnamefont {D.}~\bibnamefont {{Tucker}}}, \bibinfo {author}
  {\bibfnamefont {A.~R.}\ \bibnamefont {{Walker}}}, \bibinfo {author}
  {\bibfnamefont {G.}~\bibnamefont {{Wang}}}, \bibinfo {author} {\bibfnamefont
  {M.}~\bibnamefont {{Watson}}}, \bibinfo {author} {\bibfnamefont
  {C.}~\bibnamefont {{Weaverdyck}}}, \bibinfo {author} {\bibfnamefont
  {W.}~\bibnamefont {{Wester}}}, \bibinfo {author} {\bibfnamefont
  {R.}~\bibnamefont {{Woods}}}, \bibinfo {author} {\bibfnamefont
  {B.}~\bibnamefont {{Yanny}}}, \ and\ \bibinfo {author} {\bibnamefont {{DES
  Collaboration}}},\ }\href {\doibase 10.1088/0004-6256/150/5/150} {\bibfield
  {journal} {\bibinfo  {journal} {\aj}\ }\textbf {\bibinfo {volume} {150}},\
  \bibinfo {eid} {150} (\bibinfo {year} {2015})},\ \Eprint
  {http://arxiv.org/abs/1504.02900} {arXiv:1504.02900 [astro-ph.IM]}
  \BibitemShut {NoStop}%
\bibitem [{\citenamefont {{Raichoor}}\ \emph {et~al.}(2017)\citenamefont
  {{Raichoor}}, \citenamefont {{Comparat}}, \citenamefont {{Delubac}},
  \citenamefont {{Kneib}}, \citenamefont {{Y{\`e}che}}, \citenamefont
  {{Dawson}}, \citenamefont {{Percival}}, \citenamefont {{Dey}}, \citenamefont
  {{Lang}}, \citenamefont {{Schlegel}}, \citenamefont {{Gorgoni}},
  \citenamefont {{Bautista}}, \citenamefont {{Brownstein}}, \citenamefont
  {{Mariappan}}, \citenamefont {{Seo}}, \citenamefont {{Tinker}}, \citenamefont
  {{Ross}}, \citenamefont {{Wang}}, \citenamefont {{Zhao}}, \citenamefont
  {{Moustakas}}, \citenamefont {{Palanque-Delabrouille}}, \citenamefont
  {{Jullo}}, \citenamefont {{Newmann}}, \citenamefont {{Prada}},\ and\
  \citenamefont {{Zhu}}}]{raichoor17a}%
  \BibitemOpen
  \bibfield  {author} {\bibinfo {author} {\bibfnamefont {A.}~\bibnamefont
  {{Raichoor}}}, \bibinfo {author} {\bibfnamefont {J.}~\bibnamefont
  {{Comparat}}}, \bibinfo {author} {\bibfnamefont {T.}~\bibnamefont
  {{Delubac}}}, \bibinfo {author} {\bibfnamefont {J.-P.}\ \bibnamefont
  {{Kneib}}}, \bibinfo {author} {\bibfnamefont {C.}~\bibnamefont
  {{Y{\`e}che}}}, \bibinfo {author} {\bibfnamefont {K.~S.}\ \bibnamefont
  {{Dawson}}}, \bibinfo {author} {\bibfnamefont {W.~J.}\ \bibnamefont
  {{Percival}}}, \bibinfo {author} {\bibfnamefont {A.}~\bibnamefont {{Dey}}},
  \bibinfo {author} {\bibfnamefont {D.}~\bibnamefont {{Lang}}}, \bibinfo
  {author} {\bibfnamefont {D.~J.}\ \bibnamefont {{Schlegel}}}, \bibinfo
  {author} {\bibfnamefont {C.}~\bibnamefont {{Gorgoni}}}, \bibinfo {author}
  {\bibfnamefont {J.}~\bibnamefont {{Bautista}}}, \bibinfo {author}
  {\bibfnamefont {J.~R.}\ \bibnamefont {{Brownstein}}}, \bibinfo {author}
  {\bibfnamefont {V.}~\bibnamefont {{Mariappan}}}, \bibinfo {author}
  {\bibfnamefont {H.-J.}\ \bibnamefont {{Seo}}}, \bibinfo {author}
  {\bibfnamefont {J.~L.}\ \bibnamefont {{Tinker}}}, \bibinfo {author}
  {\bibfnamefont {A.~J.}\ \bibnamefont {{Ross}}}, \bibinfo {author}
  {\bibfnamefont {Y.}~\bibnamefont {{Wang}}}, \bibinfo {author} {\bibfnamefont
  {G.-B.}\ \bibnamefont {{Zhao}}}, \bibinfo {author} {\bibfnamefont
  {J.}~\bibnamefont {{Moustakas}}}, \bibinfo {author} {\bibfnamefont
  {N.}~\bibnamefont {{Palanque-Delabrouille}}}, \bibinfo {author}
  {\bibfnamefont {E.}~\bibnamefont {{Jullo}}}, \bibinfo {author} {\bibfnamefont
  {J.~A.}\ \bibnamefont {{Newmann}}}, \bibinfo {author} {\bibfnamefont
  {F.}~\bibnamefont {{Prada}}}, \ and\ \bibinfo {author} {\bibfnamefont
  {G.~B.}\ \bibnamefont {{Zhu}}},\ }\href {\doibase 10.1093/mnras/stx1790}
  {\bibfield  {journal} {\bibinfo  {journal} {\mnras}\ }\textbf {\bibinfo
  {volume} {471}},\ \bibinfo {pages} {3955} (\bibinfo {year} {2017})},\ \Eprint
  {http://arxiv.org/abs/1704.00338} {arXiv:1704.00338} \BibitemShut {NoStop}%
\bibitem [{\citenamefont {{Myers}}\ \emph {et~al.}(2015)\citenamefont
  {{Myers}}, \citenamefont {{Palanque-Delabrouille}}, \citenamefont
  {{Prakash}}, \citenamefont {{P{\^a}ris}}, \citenamefont {{Yeche}},
  \citenamefont {{Dawson}}, \citenamefont {{Bovy}}, \citenamefont {{Lang}},
  \citenamefont {{Schlegel}}, \citenamefont {{Newman}}, \citenamefont
  {{Petitjean}}, \citenamefont {{Kneib}}, \citenamefont {{Laurent}},
  \citenamefont {{Percival}}, \citenamefont {{Ross}}, \citenamefont {{Seo}},
  \citenamefont {{Tinker}}, \citenamefont {{Armengaud}}, \citenamefont
  {{Brownstein}}, \citenamefont {{Burtin}}, \citenamefont {{Cai}},
  \citenamefont {{Comparat}}, \citenamefont {{Kasliwal}}, \citenamefont
  {{Kulkarni}}, \citenamefont {{Laher}}, \citenamefont {{Levitan}},
  \citenamefont {{McBride}}, \citenamefont {{McGreer}}, \citenamefont
  {{Miller}}, \citenamefont {{Nugent}}, \citenamefont {{Ofek}}, \citenamefont
  {{Rossi}}, \citenamefont {{Ruan}}, \citenamefont {{Schneider}}, \citenamefont
  {{Sesar}}, \citenamefont {{Streblyanska}},\ and\ \citenamefont
  {{Surace}}}]{myers15a}%
  \BibitemOpen
  \bibfield  {author} {\bibinfo {author} {\bibfnamefont {A.~D.}\ \bibnamefont
  {{Myers}}}, \bibinfo {author} {\bibfnamefont {N.}~\bibnamefont
  {{Palanque-Delabrouille}}}, \bibinfo {author} {\bibfnamefont
  {A.}~\bibnamefont {{Prakash}}}, \bibinfo {author} {\bibfnamefont
  {I.}~\bibnamefont {{P{\^a}ris}}}, \bibinfo {author} {\bibfnamefont
  {C.}~\bibnamefont {{Yeche}}}, \bibinfo {author} {\bibfnamefont {K.~S.}\
  \bibnamefont {{Dawson}}}, \bibinfo {author} {\bibfnamefont {J.}~\bibnamefont
  {{Bovy}}}, \bibinfo {author} {\bibfnamefont {D.}~\bibnamefont {{Lang}}},
  \bibinfo {author} {\bibfnamefont {D.~J.}\ \bibnamefont {{Schlegel}}},
  \bibinfo {author} {\bibfnamefont {J.~A.}\ \bibnamefont {{Newman}}}, \bibinfo
  {author} {\bibfnamefont {P.}~\bibnamefont {{Petitjean}}}, \bibinfo {author}
  {\bibfnamefont {J.-P.}\ \bibnamefont {{Kneib}}}, \bibinfo {author}
  {\bibfnamefont {P.}~\bibnamefont {{Laurent}}}, \bibinfo {author}
  {\bibfnamefont {W.~J.}\ \bibnamefont {{Percival}}}, \bibinfo {author}
  {\bibfnamefont {A.~J.}\ \bibnamefont {{Ross}}}, \bibinfo {author}
  {\bibfnamefont {H.-J.}\ \bibnamefont {{Seo}}}, \bibinfo {author}
  {\bibfnamefont {J.~L.}\ \bibnamefont {{Tinker}}}, \bibinfo {author}
  {\bibfnamefont {E.}~\bibnamefont {{Armengaud}}}, \bibinfo {author}
  {\bibfnamefont {J.}~\bibnamefont {{Brownstein}}}, \bibinfo {author}
  {\bibfnamefont {E.}~\bibnamefont {{Burtin}}}, \bibinfo {author}
  {\bibfnamefont {Z.}~\bibnamefont {{Cai}}}, \bibinfo {author} {\bibfnamefont
  {J.}~\bibnamefont {{Comparat}}}, \bibinfo {author} {\bibfnamefont
  {M.}~\bibnamefont {{Kasliwal}}}, \bibinfo {author} {\bibfnamefont {S.~R.}\
  \bibnamefont {{Kulkarni}}}, \bibinfo {author} {\bibfnamefont
  {R.}~\bibnamefont {{Laher}}}, \bibinfo {author} {\bibfnamefont
  {D.}~\bibnamefont {{Levitan}}}, \bibinfo {author} {\bibfnamefont {C.~K.}\
  \bibnamefont {{McBride}}}, \bibinfo {author} {\bibfnamefont {I.~D.}\
  \bibnamefont {{McGreer}}}, \bibinfo {author} {\bibfnamefont {A.~A.}\
  \bibnamefont {{Miller}}}, \bibinfo {author} {\bibfnamefont {P.}~\bibnamefont
  {{Nugent}}}, \bibinfo {author} {\bibfnamefont {E.}~\bibnamefont {{Ofek}}},
  \bibinfo {author} {\bibfnamefont {G.}~\bibnamefont {{Rossi}}}, \bibinfo
  {author} {\bibfnamefont {J.}~\bibnamefont {{Ruan}}}, \bibinfo {author}
  {\bibfnamefont {D.~P.}\ \bibnamefont {{Schneider}}}, \bibinfo {author}
  {\bibfnamefont {B.}~\bibnamefont {{Sesar}}}, \bibinfo {author} {\bibfnamefont
  {A.}~\bibnamefont {{Streblyanska}}}, \ and\ \bibinfo {author} {\bibfnamefont
  {J.}~\bibnamefont {{Surace}}},\ }\href {\doibase 10.1088/0067-0049/221/2/27}
  {\bibfield  {journal} {\bibinfo  {journal} {\apjs}\ }\textbf {\bibinfo
  {volume} {221}},\ \bibinfo {eid} {27} (\bibinfo {year} {2015})},\ \Eprint
  {http://arxiv.org/abs/1508.04472} {arXiv:1508.04472} \BibitemShut {NoStop}%
\bibitem [{\citenamefont {{Ross}}\ \emph {et~al.}(2012)\citenamefont {{Ross}},
  \citenamefont {{Myers}}, \citenamefont {{Sheldon}}, \citenamefont
  {{Y{\`e}che}}, \citenamefont {{Strauss}}, \citenamefont {{Bovy}},
  \citenamefont {{Kirkpatrick}}, \citenamefont {{Richards}}, \citenamefont
  {{Aubourg}}, \citenamefont {{Blanton}}, \citenamefont {{Brandt}},
  \citenamefont {{Carithers}}, \citenamefont {{Croft}}, \citenamefont {{da
  Silva}}, \citenamefont {{Dawson}}, \citenamefont {{Eisenstein}},
  \citenamefont {{Hennawi}}, \citenamefont {{Ho}}, \citenamefont {{Hogg}},
  \citenamefont {{Lee}}, \citenamefont {{Lundgren}}, \citenamefont {{McMahon}},
  \citenamefont {{Miralda-Escud{\'e}}}, \citenamefont
  {{Palanque-Delabrouille}}, \citenamefont {{P{\^a}ris}}, \citenamefont
  {{Petitjean}}, \citenamefont {{Pieri}}, \citenamefont {{Rich}}, \citenamefont
  {{Roe}}, \citenamefont {{Schiminovich}}, \citenamefont {{Schlegel}},
  \citenamefont {{Schneider}}, \citenamefont {{Slosar}}, \citenamefont
  {{Suzuki}}, \citenamefont {{Tinker}}, \citenamefont {{Weinberg}},
  \citenamefont {{Weyant}}, \citenamefont {{White}},\ and\ \citenamefont
  {{Wood-Vasey}}}]{ross12a}%
  \BibitemOpen
  \bibfield  {author} {\bibinfo {author} {\bibfnamefont {N.~P.}\ \bibnamefont
  {{Ross}}}, \bibinfo {author} {\bibfnamefont {A.~D.}\ \bibnamefont {{Myers}}},
  \bibinfo {author} {\bibfnamefont {E.~S.}\ \bibnamefont {{Sheldon}}}, \bibinfo
  {author} {\bibfnamefont {C.}~\bibnamefont {{Y{\`e}che}}}, \bibinfo {author}
  {\bibfnamefont {M.~A.}\ \bibnamefont {{Strauss}}}, \bibinfo {author}
  {\bibfnamefont {J.}~\bibnamefont {{Bovy}}}, \bibinfo {author} {\bibfnamefont
  {J.~A.}\ \bibnamefont {{Kirkpatrick}}}, \bibinfo {author} {\bibfnamefont
  {G.~T.}\ \bibnamefont {{Richards}}}, \bibinfo {author} {\bibfnamefont
  {{\'E}.}~\bibnamefont {{Aubourg}}}, \bibinfo {author} {\bibfnamefont {M.~R.}\
  \bibnamefont {{Blanton}}}, \bibinfo {author} {\bibfnamefont {W.~N.}\
  \bibnamefont {{Brandt}}}, \bibinfo {author} {\bibfnamefont {W.~C.}\
  \bibnamefont {{Carithers}}}, \bibinfo {author} {\bibfnamefont {R.~A.~C.}\
  \bibnamefont {{Croft}}}, \bibinfo {author} {\bibfnamefont {R.}~\bibnamefont
  {{da Silva}}}, \bibinfo {author} {\bibfnamefont {K.}~\bibnamefont
  {{Dawson}}}, \bibinfo {author} {\bibfnamefont {D.~J.}\ \bibnamefont
  {{Eisenstein}}}, \bibinfo {author} {\bibfnamefont {J.~F.}\ \bibnamefont
  {{Hennawi}}}, \bibinfo {author} {\bibfnamefont {S.}~\bibnamefont {{Ho}}},
  \bibinfo {author} {\bibfnamefont {D.~W.}\ \bibnamefont {{Hogg}}}, \bibinfo
  {author} {\bibfnamefont {K.-G.}\ \bibnamefont {{Lee}}}, \bibinfo {author}
  {\bibfnamefont {B.}~\bibnamefont {{Lundgren}}}, \bibinfo {author}
  {\bibfnamefont {R.~G.}\ \bibnamefont {{McMahon}}}, \bibinfo {author}
  {\bibfnamefont {J.}~\bibnamefont {{Miralda-Escud{\'e}}}}, \bibinfo {author}
  {\bibfnamefont {N.}~\bibnamefont {{Palanque-Delabrouille}}}, \bibinfo
  {author} {\bibfnamefont {I.}~\bibnamefont {{P{\^a}ris}}}, \bibinfo {author}
  {\bibfnamefont {P.}~\bibnamefont {{Petitjean}}}, \bibinfo {author}
  {\bibfnamefont {M.~M.}\ \bibnamefont {{Pieri}}}, \bibinfo {author}
  {\bibfnamefont {J.}~\bibnamefont {{Rich}}}, \bibinfo {author} {\bibfnamefont
  {N.~A.}\ \bibnamefont {{Roe}}}, \bibinfo {author} {\bibfnamefont
  {D.}~\bibnamefont {{Schiminovich}}}, \bibinfo {author} {\bibfnamefont
  {D.~J.}\ \bibnamefont {{Schlegel}}}, \bibinfo {author} {\bibfnamefont
  {D.~P.}\ \bibnamefont {{Schneider}}}, \bibinfo {author} {\bibfnamefont
  {A.}~\bibnamefont {{Slosar}}}, \bibinfo {author} {\bibfnamefont
  {N.}~\bibnamefont {{Suzuki}}}, \bibinfo {author} {\bibfnamefont {J.~L.}\
  \bibnamefont {{Tinker}}}, \bibinfo {author} {\bibfnamefont {D.~H.}\
  \bibnamefont {{Weinberg}}}, \bibinfo {author} {\bibfnamefont
  {A.}~\bibnamefont {{Weyant}}}, \bibinfo {author} {\bibfnamefont
  {M.}~\bibnamefont {{White}}}, \ and\ \bibinfo {author} {\bibfnamefont
  {W.~M.}\ \bibnamefont {{Wood-Vasey}}},\ }\href {\doibase
  10.1088/0067-0049/199/1/3} {\bibfield  {journal} {\bibinfo  {journal}
  {\apjs}\ }\textbf {\bibinfo {volume} {199}},\ \bibinfo {eid} {3} (\bibinfo
  {year} {2012})},\ \Eprint {http://arxiv.org/abs/1105.0606} {arXiv:1105.0606
  [astro-ph.CO]} \BibitemShut {NoStop}%
\bibitem [{\citenamefont {{Palanque-Delabrouille}}\ \emph
  {et~al.}(2016)\citenamefont {{Palanque-Delabrouille}}, \citenamefont
  {{Magneville}}, \citenamefont {{Y{\`e}che}}, \citenamefont {{P{\^a}ris}},
  \citenamefont {{Petitjean}}, \citenamefont {{Burtin}}, \citenamefont
  {{Dawson}}, \citenamefont {{McGreer}}, \citenamefont {{Myers}}, \citenamefont
  {{Rossi}}, \citenamefont {{Schlegel}}, \citenamefont {{Schneider}},
  \citenamefont {{Streblyanska}},\ and\ \citenamefont
  {{Tinker}}}]{palanque-delabrouille16a}%
  \BibitemOpen
  \bibfield  {author} {\bibinfo {author} {\bibfnamefont {N.}~\bibnamefont
  {{Palanque-Delabrouille}}}, \bibinfo {author} {\bibfnamefont
  {C.}~\bibnamefont {{Magneville}}}, \bibinfo {author} {\bibfnamefont
  {C.}~\bibnamefont {{Y{\`e}che}}}, \bibinfo {author} {\bibfnamefont
  {I.}~\bibnamefont {{P{\^a}ris}}}, \bibinfo {author} {\bibfnamefont
  {P.}~\bibnamefont {{Petitjean}}}, \bibinfo {author} {\bibfnamefont
  {E.}~\bibnamefont {{Burtin}}}, \bibinfo {author} {\bibfnamefont
  {K.}~\bibnamefont {{Dawson}}}, \bibinfo {author} {\bibfnamefont
  {I.}~\bibnamefont {{McGreer}}}, \bibinfo {author} {\bibfnamefont {A.~D.}\
  \bibnamefont {{Myers}}}, \bibinfo {author} {\bibfnamefont {G.}~\bibnamefont
  {{Rossi}}}, \bibinfo {author} {\bibfnamefont {D.}~\bibnamefont {{Schlegel}}},
  \bibinfo {author} {\bibfnamefont {D.}~\bibnamefont {{Schneider}}}, \bibinfo
  {author} {\bibfnamefont {A.}~\bibnamefont {{Streblyanska}}}, \ and\ \bibinfo
  {author} {\bibfnamefont {J.}~\bibnamefont {{Tinker}}},\ }\href {\doibase
  10.1051/0004-6361/201527392} {\bibfield  {journal} {\bibinfo  {journal}
  {\aap}\ }\textbf {\bibinfo {volume} {587}},\ \bibinfo {eid} {A41} (\bibinfo
  {year} {2016})},\ \Eprint {http://arxiv.org/abs/1509.05607}
  {arXiv:1509.05607} \BibitemShut {NoStop}%
\bibitem [{\citenamefont {{McDonald}}\ and\ \citenamefont
  {{Miralda-Escud{\'e}}}(1999)}]{mcdonald99a}%
  \BibitemOpen
  \bibfield  {author} {\bibinfo {author} {\bibfnamefont {P.}~\bibnamefont
  {{McDonald}}}\ and\ \bibinfo {author} {\bibfnamefont {J.}~\bibnamefont
  {{Miralda-Escud{\'e}}}},\ }\href {\doibase 10.1086/307264} {\bibfield
  {journal} {\bibinfo  {journal} {\apj}\ }\textbf {\bibinfo {volume} {518}},\
  \bibinfo {pages} {24} (\bibinfo {year} {1999})},\ \Eprint
  {http://arxiv.org/abs/astro-ph/9807137} {arXiv:astro-ph/9807137 [astro-ph]}
  \BibitemShut {NoStop}%
\bibitem [{\citenamefont {{Seljak}}(2012)}]{seljak12a}%
  \BibitemOpen
  \bibfield  {author} {\bibinfo {author} {\bibfnamefont {U.}~\bibnamefont
  {{Seljak}}},\ }\href {\doibase 10.1088/1475-7516/2012/03/004} {\bibfield
  {journal} {\bibinfo  {journal} {\jcap}\ }\textbf {\bibinfo {volume} {2012}},\
  \bibinfo {eid} {004} (\bibinfo {year} {2012})},\ \Eprint
  {http://arxiv.org/abs/1201.0594} {arXiv:1201.0594 [astro-ph.CO]} \BibitemShut
  {NoStop}%
\bibitem [{\citenamefont {{Planck Collaboration}}\ \emph
  {et~al.}(2011)\citenamefont {{Planck Collaboration}}, \citenamefont {{Ade}},
  \citenamefont {{Aghanim}}, \citenamefont {{Arnaud}}, \citenamefont
  {{Ashdown}}, \citenamefont {{Aumont}}, \citenamefont {{Baccigalupi}},
  \citenamefont {{Baker}}, \citenamefont {{Balbi}}, \citenamefont {{Banday}},\
  and\ \citenamefont {et~al.}}]{planck11a}%
  \BibitemOpen
  \bibfield  {author} {\bibinfo {author} {\bibnamefont {{Planck
  Collaboration}}}, \bibinfo {author} {\bibfnamefont {P.~A.~R.}\ \bibnamefont
  {{Ade}}}, \bibinfo {author} {\bibfnamefont {N.}~\bibnamefont {{Aghanim}}},
  \bibinfo {author} {\bibfnamefont {M.}~\bibnamefont {{Arnaud}}}, \bibinfo
  {author} {\bibfnamefont {M.}~\bibnamefont {{Ashdown}}}, \bibinfo {author}
  {\bibfnamefont {J.}~\bibnamefont {{Aumont}}}, \bibinfo {author}
  {\bibfnamefont {C.}~\bibnamefont {{Baccigalupi}}}, \bibinfo {author}
  {\bibfnamefont {M.}~\bibnamefont {{Baker}}}, \bibinfo {author} {\bibfnamefont
  {A.}~\bibnamefont {{Balbi}}}, \bibinfo {author} {\bibfnamefont {A.~J.}\
  \bibnamefont {{Banday}}}, \ and\ \bibinfo {author} {\bibnamefont {et~al.}},\
  }\href {\doibase 10.1051/0004-6361/201116464} {\bibfield  {journal} {\bibinfo
   {journal} {\aap}\ }\textbf {\bibinfo {volume} {536}},\ \bibinfo {eid} {A1}
  (\bibinfo {year} {2011})},\ \Eprint {http://arxiv.org/abs/1101.2022}
  {arXiv:1101.2022 [astro-ph.IM]} \BibitemShut {NoStop}%
\bibitem [{\citenamefont {{Planck Collaboration}}\ \emph
  {et~al.}(2019)\citenamefont {{Planck Collaboration}}, \citenamefont
  {{Aghanim}}, \citenamefont {{Akrami}}, \citenamefont {{Ashdown}},
  \citenamefont {{Aumont}}, \citenamefont {{Baccigalupi}}, \citenamefont
  {{Ballardini}}, \citenamefont {{Banday}}, \citenamefont {{Barreiro}},
  \citenamefont {{Bartolo}}, \citenamefont {{Basak}}, \citenamefont
  {{Benabed}}, \citenamefont {{Bernard}}, \citenamefont {{Bersanelli}},
  \citenamefont {{Bielewicz}}, \citenamefont {{Bock}}, \citenamefont {{Bond}},
  \citenamefont {{Borrill}}, \citenamefont {{Bouchet}}, \citenamefont
  {{Boulanger}}, \citenamefont {{Bucher}}, \citenamefont {{Burigana}},
  \citenamefont {{Butler}}, \citenamefont {{Calabrese}}, \citenamefont
  {{Cardoso}}, \citenamefont {{Carron}}, \citenamefont {{Casaponsa}},
  \citenamefont {{Challinor}}, \citenamefont {{Chiang}}, \citenamefont
  {{Colombo}}, \citenamefont {{Combet}}, \citenamefont {{Crill}}, \citenamefont
  {{Cuttaia}}, \citenamefont {{de Bernardis}}, \citenamefont {{de Rosa}},
  \citenamefont {{de Zotti}}, \citenamefont {{Delabrouille}}, \citenamefont
  {{Delouis}}, \citenamefont {{Di Valentino}}, \citenamefont {{Diego}},
  \citenamefont {{Dor{\'e}}}, \citenamefont {{Douspis}}, \citenamefont
  {{Ducout}}, \citenamefont {{Dupac}}, \citenamefont {{Dusini}}, \citenamefont
  {{Efstathiou}}, \citenamefont {{Elsner}}, \citenamefont {{En{\ss}lin}},
  \citenamefont {{Eriksen}}, \citenamefont {{Fantaye}}, \citenamefont {{Fernand
  ez-Cobos}}, \citenamefont {{Finelli}}, \citenamefont {{Frailis}},
  \citenamefont {{Fraisse}}, \citenamefont {{Franceschi}}, \citenamefont
  {{Frolov}}, \citenamefont {{Galeotta}}, \citenamefont {{Galli}},
  \citenamefont {{Ganga}}, \citenamefont {{G{\'e}nova-Santos}}, \citenamefont
  {{Gerbino}}, \citenamefont {{Ghosh}}, \citenamefont {{Giraud-H{\'e}raud}},
  \citenamefont {{Gonz{\'a}lez-Nuevo}}, \citenamefont {{G{\'o}rski}},
  \citenamefont {{Gratton}}, \citenamefont {{Gruppuso}}, \citenamefont
  {{Gudmundsson}}, \citenamefont {{Hamann}}, \citenamefont {{Handley}},
  \citenamefont {{Hansen}}, \citenamefont {{Herranz}}, \citenamefont {{Hivon}},
  \citenamefont {{Huang}}, \citenamefont {{Jaffe}}, \citenamefont {{Jones}},
  \citenamefont {{Keih{\"a}nen}}, \citenamefont {{Keskitalo}}, \citenamefont
  {{Kiiveri}}, \citenamefont {{Kim}}, \citenamefont {{Kisner}}, \citenamefont
  {{Krachmalnicoff}}, \citenamefont {{Kunz}}, \citenamefont {{Kurki-Suonio}},
  \citenamefont {{Lagache}}, \citenamefont {{Lamarre}}, \citenamefont
  {{Lasenby}}, \citenamefont {{Lattanzi}}, \citenamefont {{Lawrence}},
  \citenamefont {{Le Jeune}}, \citenamefont {{Levrier}}, \citenamefont
  {{Lewis}}, \citenamefont {{Liguori}}, \citenamefont {{Lilje}}, \citenamefont
  {{Lilley}}, \citenamefont {{Lindholm}}, \citenamefont {{L{\'o}pez-Caniego}},
  \citenamefont {{Lubin}}, \citenamefont {{Ma}}, \citenamefont
  {{Mac{\'\i}as-P{\'e}rez}}, \citenamefont {{Maggio}}, \citenamefont {{Maino}},
  \citenamefont {{Mandolesi}}, \citenamefont {{Mangilli}}, \citenamefont
  {{Marcos-Caballero}}, \citenamefont {{Maris}}, \citenamefont {{Martin}},
  \citenamefont {{Mart{\'\i}nez-Gonz{\'a}lez}}, \citenamefont {{Matarrese}},
  \citenamefont {{Mauri}}, \citenamefont {{McEwen}}, \citenamefont
  {{Meinhold}}, \citenamefont {{Melchiorri}}, \citenamefont {{Mennella}},
  \citenamefont {{Migliaccio}}, \citenamefont {{Millea}}, \citenamefont
  {{Miville-Desch{\^e}nes}}, \citenamefont {{Molinari}}, \citenamefont
  {{Moneti}}, \citenamefont {{Montier}}, \citenamefont {{Morgante}},
  \citenamefont {{Moss}}, \citenamefont {{Natoli}}, \citenamefont
  {{N{\o}rgaard-Nielsen}}, \citenamefont {{Pagano}}, \citenamefont
  {{Paoletti}}, \citenamefont {{Partridge}}, \citenamefont {{Patanchon}},
  \citenamefont {{Peiris}}, \citenamefont {{Perrotta}}, \citenamefont
  {{Pettorino}}, \citenamefont {{Piacentini}}, \citenamefont {{Polenta}},
  \citenamefont {{Puget}}, \citenamefont {{Rachen}}, \citenamefont
  {{Reinecke}}, \citenamefont {{Remazeilles}}, \citenamefont {{Renzi}},
  \citenamefont {{Rocha}}, \citenamefont {{Rosset}}, \citenamefont {{Roudier}},
  \citenamefont {{Rubi{\~n}o-Mart{\'\i}n}}, \citenamefont {{Ruiz-Granados}},
  \citenamefont {{Salvati}}, \citenamefont {{Sandri}}, \citenamefont
  {{Savelainen}}, \citenamefont {{Scott}}, \citenamefont {{Shellard}},
  \citenamefont {{Sirignano}}, \citenamefont {{Sirri}}, \citenamefont
  {{Spencer}}, \citenamefont {{Sunyaev}}, \citenamefont {{Suur-Uski}},
  \citenamefont {{Tauber}}, \citenamefont {{Tavagnacco}}, \citenamefont
  {{Tenti}}, \citenamefont {{Toffolatti}}, \citenamefont {{Tomasi}},
  \citenamefont {{Trombetti}}, \citenamefont {{Valiviita}}, \citenamefont {{Van
  Tent}}, \citenamefont {{Vielva}}, \citenamefont {{Villa}}, \citenamefont
  {{Vittorio}}, \citenamefont {{Wandelt}}, \citenamefont {{Wehus}},
  \citenamefont {{Zacchei}},\ and\ \citenamefont
  {{Zonca}}}]{2019arXiv190712875P}%
  \BibitemOpen
  \bibfield  {author} {\bibinfo {author} {\bibnamefont {{Planck
  Collaboration}}}, \bibinfo {author} {\bibfnamefont {N.}~\bibnamefont
  {{Aghanim}}}, \bibinfo {author} {\bibfnamefont {Y.}~\bibnamefont {{Akrami}}},
  \bibinfo {author} {\bibfnamefont {M.}~\bibnamefont {{Ashdown}}}, \bibinfo
  {author} {\bibfnamefont {J.}~\bibnamefont {{Aumont}}}, \bibinfo {author}
  {\bibfnamefont {C.}~\bibnamefont {{Baccigalupi}}}, \bibinfo {author}
  {\bibfnamefont {M.}~\bibnamefont {{Ballardini}}}, \bibinfo {author}
  {\bibfnamefont {A.~J.}\ \bibnamefont {{Banday}}}, \bibinfo {author}
  {\bibfnamefont {R.~B.}\ \bibnamefont {{Barreiro}}}, \bibinfo {author}
  {\bibfnamefont {N.}~\bibnamefont {{Bartolo}}}, \bibinfo {author}
  {\bibfnamefont {S.}~\bibnamefont {{Basak}}}, \bibinfo {author} {\bibfnamefont
  {K.}~\bibnamefont {{Benabed}}}, \bibinfo {author} {\bibfnamefont {J.~P.}\
  \bibnamefont {{Bernard}}}, \bibinfo {author} {\bibfnamefont {M.}~\bibnamefont
  {{Bersanelli}}}, \bibinfo {author} {\bibfnamefont {P.}~\bibnamefont
  {{Bielewicz}}}, \bibinfo {author} {\bibfnamefont {J.~J.}\ \bibnamefont
  {{Bock}}}, \bibinfo {author} {\bibfnamefont {J.~R.}\ \bibnamefont {{Bond}}},
  \bibinfo {author} {\bibfnamefont {J.}~\bibnamefont {{Borrill}}}, \bibinfo
  {author} {\bibfnamefont {F.~R.}\ \bibnamefont {{Bouchet}}}, \bibinfo {author}
  {\bibfnamefont {F.}~\bibnamefont {{Boulanger}}}, \bibinfo {author}
  {\bibfnamefont {M.}~\bibnamefont {{Bucher}}}, \bibinfo {author}
  {\bibfnamefont {C.}~\bibnamefont {{Burigana}}}, \bibinfo {author}
  {\bibfnamefont {R.~C.}\ \bibnamefont {{Butler}}}, \bibinfo {author}
  {\bibfnamefont {E.}~\bibnamefont {{Calabrese}}}, \bibinfo {author}
  {\bibfnamefont {J.~F.}\ \bibnamefont {{Cardoso}}}, \bibinfo {author}
  {\bibfnamefont {J.}~\bibnamefont {{Carron}}}, \bibinfo {author}
  {\bibfnamefont {B.}~\bibnamefont {{Casaponsa}}}, \bibinfo {author}
  {\bibfnamefont {A.}~\bibnamefont {{Challinor}}}, \bibinfo {author}
  {\bibfnamefont {H.~C.}\ \bibnamefont {{Chiang}}}, \bibinfo {author}
  {\bibfnamefont {L.~P.~L.}\ \bibnamefont {{Colombo}}}, \bibinfo {author}
  {\bibfnamefont {C.}~\bibnamefont {{Combet}}}, \bibinfo {author}
  {\bibfnamefont {B.~P.}\ \bibnamefont {{Crill}}}, \bibinfo {author}
  {\bibfnamefont {F.}~\bibnamefont {{Cuttaia}}}, \bibinfo {author}
  {\bibfnamefont {P.}~\bibnamefont {{de Bernardis}}}, \bibinfo {author}
  {\bibfnamefont {A.}~\bibnamefont {{de Rosa}}}, \bibinfo {author}
  {\bibfnamefont {G.}~\bibnamefont {{de Zotti}}}, \bibinfo {author}
  {\bibfnamefont {J.}~\bibnamefont {{Delabrouille}}}, \bibinfo {author}
  {\bibfnamefont {J.~M.}\ \bibnamefont {{Delouis}}}, \bibinfo {author}
  {\bibfnamefont {E.}~\bibnamefont {{Di Valentino}}}, \bibinfo {author}
  {\bibfnamefont {J.~M.}\ \bibnamefont {{Diego}}}, \bibinfo {author}
  {\bibfnamefont {O.}~\bibnamefont {{Dor{\'e}}}}, \bibinfo {author}
  {\bibfnamefont {M.}~\bibnamefont {{Douspis}}}, \bibinfo {author}
  {\bibfnamefont {A.}~\bibnamefont {{Ducout}}}, \bibinfo {author}
  {\bibfnamefont {X.}~\bibnamefont {{Dupac}}}, \bibinfo {author} {\bibfnamefont
  {S.}~\bibnamefont {{Dusini}}}, \bibinfo {author} {\bibfnamefont
  {G.}~\bibnamefont {{Efstathiou}}}, \bibinfo {author} {\bibfnamefont
  {F.}~\bibnamefont {{Elsner}}}, \bibinfo {author} {\bibfnamefont {T.~A.}\
  \bibnamefont {{En{\ss}lin}}}, \bibinfo {author} {\bibfnamefont {H.~K.}\
  \bibnamefont {{Eriksen}}}, \bibinfo {author} {\bibfnamefont {Y.}~\bibnamefont
  {{Fantaye}}}, \bibinfo {author} {\bibfnamefont {R.}~\bibnamefont {{Fernand
  ez-Cobos}}}, \bibinfo {author} {\bibfnamefont {F.}~\bibnamefont {{Finelli}}},
  \bibinfo {author} {\bibfnamefont {M.}~\bibnamefont {{Frailis}}}, \bibinfo
  {author} {\bibfnamefont {A.~A.}\ \bibnamefont {{Fraisse}}}, \bibinfo {author}
  {\bibfnamefont {E.}~\bibnamefont {{Franceschi}}}, \bibinfo {author}
  {\bibfnamefont {A.}~\bibnamefont {{Frolov}}}, \bibinfo {author}
  {\bibfnamefont {S.}~\bibnamefont {{Galeotta}}}, \bibinfo {author}
  {\bibfnamefont {S.}~\bibnamefont {{Galli}}}, \bibinfo {author} {\bibfnamefont
  {K.}~\bibnamefont {{Ganga}}}, \bibinfo {author} {\bibfnamefont {R.~T.}\
  \bibnamefont {{G{\'e}nova-Santos}}}, \bibinfo {author} {\bibfnamefont
  {M.}~\bibnamefont {{Gerbino}}}, \bibinfo {author} {\bibfnamefont
  {T.}~\bibnamefont {{Ghosh}}}, \bibinfo {author} {\bibfnamefont
  {Y.}~\bibnamefont {{Giraud-H{\'e}raud}}}, \bibinfo {author} {\bibfnamefont
  {J.}~\bibnamefont {{Gonz{\'a}lez-Nuevo}}}, \bibinfo {author} {\bibfnamefont
  {K.~M.}\ \bibnamefont {{G{\'o}rski}}}, \bibinfo {author} {\bibfnamefont
  {S.}~\bibnamefont {{Gratton}}}, \bibinfo {author} {\bibfnamefont
  {A.}~\bibnamefont {{Gruppuso}}}, \bibinfo {author} {\bibfnamefont {J.~E.}\
  \bibnamefont {{Gudmundsson}}}, \bibinfo {author} {\bibfnamefont
  {J.}~\bibnamefont {{Hamann}}}, \bibinfo {author} {\bibfnamefont
  {W.}~\bibnamefont {{Handley}}}, \bibinfo {author} {\bibfnamefont {F.~K.}\
  \bibnamefont {{Hansen}}}, \bibinfo {author} {\bibfnamefont {D.}~\bibnamefont
  {{Herranz}}}, \bibinfo {author} {\bibfnamefont {E.}~\bibnamefont {{Hivon}}},
  \bibinfo {author} {\bibfnamefont {Z.}~\bibnamefont {{Huang}}}, \bibinfo
  {author} {\bibfnamefont {A.~H.}\ \bibnamefont {{Jaffe}}}, \bibinfo {author}
  {\bibfnamefont {W.~C.}\ \bibnamefont {{Jones}}}, \bibinfo {author}
  {\bibfnamefont {E.}~\bibnamefont {{Keih{\"a}nen}}}, \bibinfo {author}
  {\bibfnamefont {R.}~\bibnamefont {{Keskitalo}}}, \bibinfo {author}
  {\bibfnamefont {K.}~\bibnamefont {{Kiiveri}}}, \bibinfo {author}
  {\bibfnamefont {J.}~\bibnamefont {{Kim}}}, \bibinfo {author} {\bibfnamefont
  {T.~S.}\ \bibnamefont {{Kisner}}}, \bibinfo {author} {\bibfnamefont
  {N.}~\bibnamefont {{Krachmalnicoff}}}, \bibinfo {author} {\bibfnamefont
  {M.}~\bibnamefont {{Kunz}}}, \bibinfo {author} {\bibfnamefont
  {H.}~\bibnamefont {{Kurki-Suonio}}}, \bibinfo {author} {\bibfnamefont
  {G.}~\bibnamefont {{Lagache}}}, \bibinfo {author} {\bibfnamefont {J.~M.}\
  \bibnamefont {{Lamarre}}}, \bibinfo {author} {\bibfnamefont {A.}~\bibnamefont
  {{Lasenby}}}, \bibinfo {author} {\bibfnamefont {M.}~\bibnamefont
  {{Lattanzi}}}, \bibinfo {author} {\bibfnamefont {C.~R.}\ \bibnamefont
  {{Lawrence}}}, \bibinfo {author} {\bibfnamefont {M.}~\bibnamefont {{Le
  Jeune}}}, \bibinfo {author} {\bibfnamefont {F.}~\bibnamefont {{Levrier}}},
  \bibinfo {author} {\bibfnamefont {A.}~\bibnamefont {{Lewis}}}, \bibinfo
  {author} {\bibfnamefont {M.}~\bibnamefont {{Liguori}}}, \bibinfo {author}
  {\bibfnamefont {P.~B.}\ \bibnamefont {{Lilje}}}, \bibinfo {author}
  {\bibfnamefont {M.}~\bibnamefont {{Lilley}}}, \bibinfo {author}
  {\bibfnamefont {V.}~\bibnamefont {{Lindholm}}}, \bibinfo {author}
  {\bibfnamefont {M.}~\bibnamefont {{L{\'o}pez-Caniego}}}, \bibinfo {author}
  {\bibfnamefont {P.~M.}\ \bibnamefont {{Lubin}}}, \bibinfo {author}
  {\bibfnamefont {Y.~Z.}\ \bibnamefont {{Ma}}}, \bibinfo {author}
  {\bibfnamefont {J.~F.}\ \bibnamefont {{Mac{\'\i}as-P{\'e}rez}}}, \bibinfo
  {author} {\bibfnamefont {G.}~\bibnamefont {{Maggio}}}, \bibinfo {author}
  {\bibfnamefont {D.}~\bibnamefont {{Maino}}}, \bibinfo {author} {\bibfnamefont
  {N.}~\bibnamefont {{Mandolesi}}}, \bibinfo {author} {\bibfnamefont
  {A.}~\bibnamefont {{Mangilli}}}, \bibinfo {author} {\bibfnamefont
  {A.}~\bibnamefont {{Marcos-Caballero}}}, \bibinfo {author} {\bibfnamefont
  {M.}~\bibnamefont {{Maris}}}, \bibinfo {author} {\bibfnamefont {P.~G.}\
  \bibnamefont {{Martin}}}, \bibinfo {author} {\bibfnamefont {E.}~\bibnamefont
  {{Mart{\'\i}nez-Gonz{\'a}lez}}}, \bibinfo {author} {\bibfnamefont
  {S.}~\bibnamefont {{Matarrese}}}, \bibinfo {author} {\bibfnamefont
  {N.}~\bibnamefont {{Mauri}}}, \bibinfo {author} {\bibfnamefont {J.~D.}\
  \bibnamefont {{McEwen}}}, \bibinfo {author} {\bibfnamefont {P.~R.}\
  \bibnamefont {{Meinhold}}}, \bibinfo {author} {\bibfnamefont
  {A.}~\bibnamefont {{Melchiorri}}}, \bibinfo {author} {\bibfnamefont
  {A.}~\bibnamefont {{Mennella}}}, \bibinfo {author} {\bibfnamefont
  {M.}~\bibnamefont {{Migliaccio}}}, \bibinfo {author} {\bibfnamefont
  {M.}~\bibnamefont {{Millea}}}, \bibinfo {author} {\bibfnamefont {M.~A.}\
  \bibnamefont {{Miville-Desch{\^e}nes}}}, \bibinfo {author} {\bibfnamefont
  {D.}~\bibnamefont {{Molinari}}}, \bibinfo {author} {\bibfnamefont
  {A.}~\bibnamefont {{Moneti}}}, \bibinfo {author} {\bibfnamefont
  {L.}~\bibnamefont {{Montier}}}, \bibinfo {author} {\bibfnamefont
  {G.}~\bibnamefont {{Morgante}}}, \bibinfo {author} {\bibfnamefont
  {A.}~\bibnamefont {{Moss}}}, \bibinfo {author} {\bibfnamefont
  {P.}~\bibnamefont {{Natoli}}}, \bibinfo {author} {\bibfnamefont {H.~U.}\
  \bibnamefont {{N{\o}rgaard-Nielsen}}}, \bibinfo {author} {\bibfnamefont
  {L.}~\bibnamefont {{Pagano}}}, \bibinfo {author} {\bibfnamefont
  {D.}~\bibnamefont {{Paoletti}}}, \bibinfo {author} {\bibfnamefont
  {B.}~\bibnamefont {{Partridge}}}, \bibinfo {author} {\bibfnamefont
  {G.}~\bibnamefont {{Patanchon}}}, \bibinfo {author} {\bibfnamefont {H.~V.}\
  \bibnamefont {{Peiris}}}, \bibinfo {author} {\bibfnamefont {F.}~\bibnamefont
  {{Perrotta}}}, \bibinfo {author} {\bibfnamefont {V.}~\bibnamefont
  {{Pettorino}}}, \bibinfo {author} {\bibfnamefont {F.}~\bibnamefont
  {{Piacentini}}}, \bibinfo {author} {\bibfnamefont {G.}~\bibnamefont
  {{Polenta}}}, \bibinfo {author} {\bibfnamefont {J.~L.}\ \bibnamefont
  {{Puget}}}, \bibinfo {author} {\bibfnamefont {J.~P.}\ \bibnamefont
  {{Rachen}}}, \bibinfo {author} {\bibfnamefont {M.}~\bibnamefont
  {{Reinecke}}}, \bibinfo {author} {\bibfnamefont {M.}~\bibnamefont
  {{Remazeilles}}}, \bibinfo {author} {\bibfnamefont {A.}~\bibnamefont
  {{Renzi}}}, \bibinfo {author} {\bibfnamefont {G.}~\bibnamefont {{Rocha}}},
  \bibinfo {author} {\bibfnamefont {C.}~\bibnamefont {{Rosset}}}, \bibinfo
  {author} {\bibfnamefont {G.}~\bibnamefont {{Roudier}}}, \bibinfo {author}
  {\bibfnamefont {J.~A.}\ \bibnamefont {{Rubi{\~n}o-Mart{\'\i}n}}}, \bibinfo
  {author} {\bibfnamefont {B.}~\bibnamefont {{Ruiz-Granados}}}, \bibinfo
  {author} {\bibfnamefont {L.}~\bibnamefont {{Salvati}}}, \bibinfo {author}
  {\bibfnamefont {M.}~\bibnamefont {{Sandri}}}, \bibinfo {author}
  {\bibfnamefont {M.}~\bibnamefont {{Savelainen}}}, \bibinfo {author}
  {\bibfnamefont {D.}~\bibnamefont {{Scott}}}, \bibinfo {author} {\bibfnamefont
  {E.~P.~S.}\ \bibnamefont {{Shellard}}}, \bibinfo {author} {\bibfnamefont
  {C.}~\bibnamefont {{Sirignano}}}, \bibinfo {author} {\bibfnamefont
  {G.}~\bibnamefont {{Sirri}}}, \bibinfo {author} {\bibfnamefont {L.~D.}\
  \bibnamefont {{Spencer}}}, \bibinfo {author} {\bibfnamefont {R.}~\bibnamefont
  {{Sunyaev}}}, \bibinfo {author} {\bibfnamefont {A.~S.}\ \bibnamefont
  {{Suur-Uski}}}, \bibinfo {author} {\bibfnamefont {J.~A.}\ \bibnamefont
  {{Tauber}}}, \bibinfo {author} {\bibfnamefont {D.}~\bibnamefont
  {{Tavagnacco}}}, \bibinfo {author} {\bibfnamefont {M.}~\bibnamefont
  {{Tenti}}}, \bibinfo {author} {\bibfnamefont {L.}~\bibnamefont
  {{Toffolatti}}}, \bibinfo {author} {\bibfnamefont {M.}~\bibnamefont
  {{Tomasi}}}, \bibinfo {author} {\bibfnamefont {T.}~\bibnamefont
  {{Trombetti}}}, \bibinfo {author} {\bibfnamefont {J.}~\bibnamefont
  {{Valiviita}}}, \bibinfo {author} {\bibfnamefont {B.}~\bibnamefont {{Van
  Tent}}}, \bibinfo {author} {\bibfnamefont {P.}~\bibnamefont {{Vielva}}},
  \bibinfo {author} {\bibfnamefont {F.}~\bibnamefont {{Villa}}}, \bibinfo
  {author} {\bibfnamefont {N.}~\bibnamefont {{Vittorio}}}, \bibinfo {author}
  {\bibfnamefont {B.~D.}\ \bibnamefont {{Wandelt}}}, \bibinfo {author}
  {\bibfnamefont {I.~K.}\ \bibnamefont {{Wehus}}}, \bibinfo {author}
  {\bibfnamefont {A.}~\bibnamefont {{Zacchei}}}, \ and\ \bibinfo {author}
  {\bibfnamefont {A.}~\bibnamefont {{Zonca}}},\ }\href@noop {} {\bibfield
  {journal} {\bibinfo  {journal} {ArXiv e-prints}\ ,\ \bibinfo {eid}
  {arXiv:1907.12875}} (\bibinfo {year} {2019})},\ \Eprint
  {http://arxiv.org/abs/1907.12875} {arXiv:1907.12875 [astro-ph.CO]}
  \BibitemShut {NoStop}%
\bibitem [{\citenamefont {{Sako}}\ \emph {et~al.}(2018)\citenamefont {{Sako}},
  \citenamefont {{Bassett}}, \citenamefont {{Becker}}, \citenamefont {{Brown}},
  \citenamefont {{Campbell}}, \citenamefont {{Wolf}}, \citenamefont
  {{Cinabro}}, \citenamefont {{D'Andrea}}, \citenamefont {{Dawson}},
  \citenamefont {{DeJongh}}, \citenamefont {{Depoy}}, \citenamefont {{Dilday}},
  \citenamefont {{Doi}}, \citenamefont {{Filippenko}}, \citenamefont
  {{Fischer}}, \citenamefont {{Foley}}, \citenamefont {{Frieman}},
  \citenamefont {{Galbany}}, \citenamefont {{Garnavich}}, \citenamefont
  {{Goobar}}, \citenamefont {{Gupta}}, \citenamefont {{Hill}}, \citenamefont
  {{Hayden}}, \citenamefont {{Hlozek}}, \citenamefont {{Holtzman}},
  \citenamefont {{Hopp}}, \citenamefont {{Jha}}, \citenamefont {{Kessler}},
  \citenamefont {{Kollatschny}}, \citenamefont {{Leloudas}}, \citenamefont
  {{Marriner}}, \citenamefont {{Marshall}}, \citenamefont {{Miquel}},
  \citenamefont {{Morokuma}}, \citenamefont {{Mosher}}, \citenamefont
  {{Nichol}}, \citenamefont {{Nordin}}, \citenamefont {{Olmstead}},
  \citenamefont {{{\"O}stman}}, \citenamefont {{Prieto}}, \citenamefont
  {{Richmond}}, \citenamefont {{Romani}}, \citenamefont {{Sollerman}},
  \citenamefont {{Stritzinger}}, \citenamefont {{Schneider}}, \citenamefont
  {{Smith}}, \citenamefont {{Wheeler}}, \citenamefont {{Yasuda}},\ and\
  \citenamefont {{Zheng}}}]{sako18a}%
  \BibitemOpen
  \bibfield  {author} {\bibinfo {author} {\bibfnamefont {M.}~\bibnamefont
  {{Sako}}}, \bibinfo {author} {\bibfnamefont {B.}~\bibnamefont {{Bassett}}},
  \bibinfo {author} {\bibfnamefont {A.~C.}\ \bibnamefont {{Becker}}}, \bibinfo
  {author} {\bibfnamefont {P.~J.}\ \bibnamefont {{Brown}}}, \bibinfo {author}
  {\bibfnamefont {H.}~\bibnamefont {{Campbell}}}, \bibinfo {author}
  {\bibfnamefont {R.}~\bibnamefont {{Wolf}}}, \bibinfo {author} {\bibfnamefont
  {D.}~\bibnamefont {{Cinabro}}}, \bibinfo {author} {\bibfnamefont {C.~B.}\
  \bibnamefont {{D'Andrea}}}, \bibinfo {author} {\bibfnamefont {K.~S.}\
  \bibnamefont {{Dawson}}}, \bibinfo {author} {\bibfnamefont {F.}~\bibnamefont
  {{DeJongh}}}, \bibinfo {author} {\bibfnamefont {D.~L.}\ \bibnamefont
  {{Depoy}}}, \bibinfo {author} {\bibfnamefont {B.}~\bibnamefont {{Dilday}}},
  \bibinfo {author} {\bibfnamefont {M.}~\bibnamefont {{Doi}}}, \bibinfo
  {author} {\bibfnamefont {A.~V.}\ \bibnamefont {{Filippenko}}}, \bibinfo
  {author} {\bibfnamefont {J.~A.}\ \bibnamefont {{Fischer}}}, \bibinfo {author}
  {\bibfnamefont {R.~J.}\ \bibnamefont {{Foley}}}, \bibinfo {author}
  {\bibfnamefont {J.~A.}\ \bibnamefont {{Frieman}}}, \bibinfo {author}
  {\bibfnamefont {L.}~\bibnamefont {{Galbany}}}, \bibinfo {author}
  {\bibfnamefont {P.~M.}\ \bibnamefont {{Garnavich}}}, \bibinfo {author}
  {\bibfnamefont {A.}~\bibnamefont {{Goobar}}}, \bibinfo {author}
  {\bibfnamefont {R.~R.}\ \bibnamefont {{Gupta}}}, \bibinfo {author}
  {\bibfnamefont {G.~J.}\ \bibnamefont {{Hill}}}, \bibinfo {author}
  {\bibfnamefont {B.~T.}\ \bibnamefont {{Hayden}}}, \bibinfo {author}
  {\bibfnamefont {R.}~\bibnamefont {{Hlozek}}}, \bibinfo {author}
  {\bibfnamefont {J.~A.}\ \bibnamefont {{Holtzman}}}, \bibinfo {author}
  {\bibfnamefont {U.}~\bibnamefont {{Hopp}}}, \bibinfo {author} {\bibfnamefont
  {S.~W.}\ \bibnamefont {{Jha}}}, \bibinfo {author} {\bibfnamefont
  {R.}~\bibnamefont {{Kessler}}}, \bibinfo {author} {\bibfnamefont
  {W.}~\bibnamefont {{Kollatschny}}}, \bibinfo {author} {\bibfnamefont
  {G.}~\bibnamefont {{Leloudas}}}, \bibinfo {author} {\bibfnamefont
  {J.}~\bibnamefont {{Marriner}}}, \bibinfo {author} {\bibfnamefont {J.~L.}\
  \bibnamefont {{Marshall}}}, \bibinfo {author} {\bibfnamefont
  {R.}~\bibnamefont {{Miquel}}}, \bibinfo {author} {\bibfnamefont
  {T.}~\bibnamefont {{Morokuma}}}, \bibinfo {author} {\bibfnamefont
  {J.}~\bibnamefont {{Mosher}}}, \bibinfo {author} {\bibfnamefont {R.~C.}\
  \bibnamefont {{Nichol}}}, \bibinfo {author} {\bibfnamefont {J.}~\bibnamefont
  {{Nordin}}}, \bibinfo {author} {\bibfnamefont {M.~D.}\ \bibnamefont
  {{Olmstead}}}, \bibinfo {author} {\bibfnamefont {L.}~\bibnamefont
  {{{\"O}stman}}}, \bibinfo {author} {\bibfnamefont {J.~L.}\ \bibnamefont
  {{Prieto}}}, \bibinfo {author} {\bibfnamefont {M.}~\bibnamefont
  {{Richmond}}}, \bibinfo {author} {\bibfnamefont {R.~W.}\ \bibnamefont
  {{Romani}}}, \bibinfo {author} {\bibfnamefont {J.}~\bibnamefont
  {{Sollerman}}}, \bibinfo {author} {\bibfnamefont {M.}~\bibnamefont
  {{Stritzinger}}}, \bibinfo {author} {\bibfnamefont {D.~P.}\ \bibnamefont
  {{Schneider}}}, \bibinfo {author} {\bibfnamefont {M.}~\bibnamefont
  {{Smith}}}, \bibinfo {author} {\bibfnamefont {J.~C.}\ \bibnamefont
  {{Wheeler}}}, \bibinfo {author} {\bibfnamefont {N.}~\bibnamefont {{Yasuda}}},
  \ and\ \bibinfo {author} {\bibfnamefont {C.}~\bibnamefont {{Zheng}}},\ }\href
  {\doibase 10.1088/1538-3873/aab4e0} {\bibfield  {journal} {\bibinfo
  {journal} {\pasp}\ }\textbf {\bibinfo {volume} {130}},\ \bibinfo {pages}
  {064002} (\bibinfo {year} {2018})},\ \Eprint {http://arxiv.org/abs/1401.3317}
  {arXiv:1401.3317} \BibitemShut {NoStop}%
\bibitem [{\citenamefont {{Kaiser}}\ \emph {et~al.}(2010)\citenamefont
  {{Kaiser}}, \citenamefont {{Burgett}}, \citenamefont {{Chambers}},
  \citenamefont {{Denneau}}, \citenamefont {{Heasley}}, \citenamefont
  {{Jedicke}}, \citenamefont {{Magnier}}, \citenamefont {{Morgan}},
  \citenamefont {{Onaka}},\ and\ \citenamefont {{Tonry}}}]{kaiser10a}%
  \BibitemOpen
  \bibfield  {author} {\bibinfo {author} {\bibfnamefont {N.}~\bibnamefont
  {{Kaiser}}}, \bibinfo {author} {\bibfnamefont {W.}~\bibnamefont {{Burgett}}},
  \bibinfo {author} {\bibfnamefont {K.}~\bibnamefont {{Chambers}}}, \bibinfo
  {author} {\bibfnamefont {L.}~\bibnamefont {{Denneau}}}, \bibinfo {author}
  {\bibfnamefont {J.}~\bibnamefont {{Heasley}}}, \bibinfo {author}
  {\bibfnamefont {R.}~\bibnamefont {{Jedicke}}}, \bibinfo {author}
  {\bibfnamefont {E.}~\bibnamefont {{Magnier}}}, \bibinfo {author}
  {\bibfnamefont {J.}~\bibnamefont {{Morgan}}}, \bibinfo {author}
  {\bibfnamefont {P.}~\bibnamefont {{Onaka}}}, \ and\ \bibinfo {author}
  {\bibfnamefont {J.}~\bibnamefont {{Tonry}}},\ }in\ \href {\doibase
  10.1117/12.859188} {\emph {\bibinfo {booktitle} {Society of Photo-Optical
  Instrumentation Engineers (SPIE) Conference Series}}},\ \bibinfo {series}
  {Society of Photo-Optical Instrumentation Engineers (SPIE) Conference
  Series}, Vol.\ \bibinfo {volume} {7733}\ (\bibinfo {year} {2010})\BibitemShut
  {NoStop}%
\bibitem [{\citenamefont {{Riess}}\ \emph {et~al.}(1999)\citenamefont
  {{Riess}}, \citenamefont {{Kirshner}}, \citenamefont {{Schmidt}},
  \citenamefont {{Jha}}, \citenamefont {{Challis}}, \citenamefont
  {{Garnavich}}, \citenamefont {{Esin}}, \citenamefont {{Carpenter}},
  \citenamefont {{Grashius}}, \citenamefont {{Schild}}, \citenamefont
  {{Berlind}}, \citenamefont {{Huchra}}, \citenamefont {{Prosser}},
  \citenamefont {{Falco}}, \citenamefont {{Benson}}, \citenamefont
  {{Brice{\~n}o}}, \citenamefont {{Brown}}, \citenamefont {{Caldwell}},
  \citenamefont {{dell'Antonio}}, \citenamefont {{Filippenko}}, \citenamefont
  {{Goodman}}, \citenamefont {{Grogin}}, \citenamefont {{Groner}},
  \citenamefont {{Hughes}}, \citenamefont {{Green}}, \citenamefont {{Jansen}},
  \citenamefont {{Kleyna}}, \citenamefont {{Luu}}, \citenamefont {{Macri}},
  \citenamefont {{McLeod}}, \citenamefont {{McLeod}}, \citenamefont
  {{McNamara}}, \citenamefont {{McLean}}, \citenamefont {{Milone}},
  \citenamefont {{Mohr}}, \citenamefont {{Moraru}}, \citenamefont {{Peng}},
  \citenamefont {{Peters}}, \citenamefont {{Prestwich}}, \citenamefont
  {{Stanek}}, \citenamefont {{Szentgyorgyi}},\ and\ \citenamefont
  {{Zhao}}}]{riess99a}%
  \BibitemOpen
  \bibfield  {author} {\bibinfo {author} {\bibfnamefont {A.~G.}\ \bibnamefont
  {{Riess}}}, \bibinfo {author} {\bibfnamefont {R.~P.}\ \bibnamefont
  {{Kirshner}}}, \bibinfo {author} {\bibfnamefont {B.~P.}\ \bibnamefont
  {{Schmidt}}}, \bibinfo {author} {\bibfnamefont {S.}~\bibnamefont {{Jha}}},
  \bibinfo {author} {\bibfnamefont {P.}~\bibnamefont {{Challis}}}, \bibinfo
  {author} {\bibfnamefont {P.~M.}\ \bibnamefont {{Garnavich}}}, \bibinfo
  {author} {\bibfnamefont {A.~A.}\ \bibnamefont {{Esin}}}, \bibinfo {author}
  {\bibfnamefont {C.}~\bibnamefont {{Carpenter}}}, \bibinfo {author}
  {\bibfnamefont {R.}~\bibnamefont {{Grashius}}}, \bibinfo {author}
  {\bibfnamefont {R.~E.}\ \bibnamefont {{Schild}}}, \bibinfo {author}
  {\bibfnamefont {P.~L.}\ \bibnamefont {{Berlind}}}, \bibinfo {author}
  {\bibfnamefont {J.~P.}\ \bibnamefont {{Huchra}}}, \bibinfo {author}
  {\bibfnamefont {C.~F.}\ \bibnamefont {{Prosser}}}, \bibinfo {author}
  {\bibfnamefont {E.~E.}\ \bibnamefont {{Falco}}}, \bibinfo {author}
  {\bibfnamefont {P.~J.}\ \bibnamefont {{Benson}}}, \bibinfo {author}
  {\bibfnamefont {C.}~\bibnamefont {{Brice{\~n}o}}}, \bibinfo {author}
  {\bibfnamefont {W.~R.}\ \bibnamefont {{Brown}}}, \bibinfo {author}
  {\bibfnamefont {N.}~\bibnamefont {{Caldwell}}}, \bibinfo {author}
  {\bibfnamefont {I.~P.}\ \bibnamefont {{dell'Antonio}}}, \bibinfo {author}
  {\bibfnamefont {A.~V.}\ \bibnamefont {{Filippenko}}}, \bibinfo {author}
  {\bibfnamefont {A.~A.}\ \bibnamefont {{Goodman}}}, \bibinfo {author}
  {\bibfnamefont {N.~A.}\ \bibnamefont {{Grogin}}}, \bibinfo {author}
  {\bibfnamefont {T.}~\bibnamefont {{Groner}}}, \bibinfo {author}
  {\bibfnamefont {J.~P.}\ \bibnamefont {{Hughes}}}, \bibinfo {author}
  {\bibfnamefont {P.~J.}\ \bibnamefont {{Green}}}, \bibinfo {author}
  {\bibfnamefont {R.~A.}\ \bibnamefont {{Jansen}}}, \bibinfo {author}
  {\bibfnamefont {J.~T.}\ \bibnamefont {{Kleyna}}}, \bibinfo {author}
  {\bibfnamefont {J.~X.}\ \bibnamefont {{Luu}}}, \bibinfo {author}
  {\bibfnamefont {L.~M.}\ \bibnamefont {{Macri}}}, \bibinfo {author}
  {\bibfnamefont {B.~A.}\ \bibnamefont {{McLeod}}}, \bibinfo {author}
  {\bibfnamefont {K.~K.}\ \bibnamefont {{McLeod}}}, \bibinfo {author}
  {\bibfnamefont {B.~R.}\ \bibnamefont {{McNamara}}}, \bibinfo {author}
  {\bibfnamefont {B.}~\bibnamefont {{McLean}}}, \bibinfo {author}
  {\bibfnamefont {A.~A.~E.}\ \bibnamefont {{Milone}}}, \bibinfo {author}
  {\bibfnamefont {J.~J.}\ \bibnamefont {{Mohr}}}, \bibinfo {author}
  {\bibfnamefont {D.}~\bibnamefont {{Moraru}}}, \bibinfo {author}
  {\bibfnamefont {C.}~\bibnamefont {{Peng}}}, \bibinfo {author} {\bibfnamefont
  {J.}~\bibnamefont {{Peters}}}, \bibinfo {author} {\bibfnamefont {A.~H.}\
  \bibnamefont {{Prestwich}}}, \bibinfo {author} {\bibfnamefont {K.~Z.}\
  \bibnamefont {{Stanek}}}, \bibinfo {author} {\bibfnamefont {A.}~\bibnamefont
  {{Szentgyorgyi}}}, \ and\ \bibinfo {author} {\bibfnamefont {P.}~\bibnamefont
  {{Zhao}}},\ }\href {\doibase 10.1086/300738} {\bibfield  {journal} {\bibinfo
  {journal} {\aj}\ }\textbf {\bibinfo {volume} {117}},\ \bibinfo {pages} {707}
  (\bibinfo {year} {1999})},\ \Eprint
  {http://arxiv.org/abs/arXiv:astro-ph/9810291} {arXiv:astro-ph/9810291}
  \BibitemShut {NoStop}%
\bibitem [{\citenamefont {{Folatelli}}\ \emph {et~al.}(2010)\citenamefont
  {{Folatelli}}, \citenamefont {{Phillips}}, \citenamefont {{Burns}},
  \citenamefont {{Contreras}}, \citenamefont {{Hamuy}}, \citenamefont
  {{Freedman}}, \citenamefont {{Persson}}, \citenamefont {{Stritzinger}},
  \citenamefont {{Suntzeff}}, \citenamefont {{Krisciunas}}, \citenamefont
  {{Boldt}}, \citenamefont {{Gonz{\'a}lez}}, \citenamefont {{Krzeminski}},
  \citenamefont {{Morrell}}, \citenamefont {{Roth}}, \citenamefont {{Salgado}},
  \citenamefont {{Madore}}, \citenamefont {{Murphy}}, \citenamefont {{Wyatt}},
  \citenamefont {{Li}}, \citenamefont {{Filippenko}},\ and\ \citenamefont
  {{Miller}}}]{folatelli10a}%
  \BibitemOpen
  \bibfield  {author} {\bibinfo {author} {\bibfnamefont {G.}~\bibnamefont
  {{Folatelli}}}, \bibinfo {author} {\bibfnamefont {M.~M.}\ \bibnamefont
  {{Phillips}}}, \bibinfo {author} {\bibfnamefont {C.~R.}\ \bibnamefont
  {{Burns}}}, \bibinfo {author} {\bibfnamefont {C.}~\bibnamefont
  {{Contreras}}}, \bibinfo {author} {\bibfnamefont {M.}~\bibnamefont
  {{Hamuy}}}, \bibinfo {author} {\bibfnamefont {W.~L.}\ \bibnamefont
  {{Freedman}}}, \bibinfo {author} {\bibfnamefont {S.~E.}\ \bibnamefont
  {{Persson}}}, \bibinfo {author} {\bibfnamefont {M.}~\bibnamefont
  {{Stritzinger}}}, \bibinfo {author} {\bibfnamefont {N.~B.}\ \bibnamefont
  {{Suntzeff}}}, \bibinfo {author} {\bibfnamefont {K.}~\bibnamefont
  {{Krisciunas}}}, \bibinfo {author} {\bibfnamefont {L.}~\bibnamefont
  {{Boldt}}}, \bibinfo {author} {\bibfnamefont {S.}~\bibnamefont
  {{Gonz{\'a}lez}}}, \bibinfo {author} {\bibfnamefont {W.}~\bibnamefont
  {{Krzeminski}}}, \bibinfo {author} {\bibfnamefont {N.}~\bibnamefont
  {{Morrell}}}, \bibinfo {author} {\bibfnamefont {M.}~\bibnamefont {{Roth}}},
  \bibinfo {author} {\bibfnamefont {F.}~\bibnamefont {{Salgado}}}, \bibinfo
  {author} {\bibfnamefont {B.~F.}\ \bibnamefont {{Madore}}}, \bibinfo {author}
  {\bibfnamefont {D.}~\bibnamefont {{Murphy}}}, \bibinfo {author}
  {\bibfnamefont {P.}~\bibnamefont {{Wyatt}}}, \bibinfo {author} {\bibfnamefont
  {W.}~\bibnamefont {{Li}}}, \bibinfo {author} {\bibfnamefont {A.~V.}\
  \bibnamefont {{Filippenko}}}, \ and\ \bibinfo {author} {\bibfnamefont
  {N.}~\bibnamefont {{Miller}}},\ }\href {\doibase 10.1088/0004-6256/139/1/120}
  {\bibfield  {journal} {\bibinfo  {journal} {\aj}\ }\textbf {\bibinfo {volume}
  {139}},\ \bibinfo {pages} {120} (\bibinfo {year} {2010})},\ \Eprint
  {http://arxiv.org/abs/0910.3317} {arXiv:0910.3317 [astro-ph.CO]} \BibitemShut
  {NoStop}%
\bibitem [{\citenamefont {{Stritzinger}}\ \emph {et~al.}(2011)\citenamefont
  {{Stritzinger}}, \citenamefont {{Phillips}}, \citenamefont {{Boldt}},
  \citenamefont {{Burns}}, \citenamefont {{Campillay}}, \citenamefont
  {{Contreras}}, \citenamefont {{Gonzalez}}, \citenamefont {{Folatelli}},
  \citenamefont {{Morrell}}, \citenamefont {{Krzeminski}}, \citenamefont
  {{Roth}}, \citenamefont {{Salgado}}, \citenamefont {{DePoy}}, \citenamefont
  {{Hamuy}}, \citenamefont {{Freedman}}, \citenamefont {{Madore}},
  \citenamefont {{Marshall}}, \citenamefont {{Persson}}, \citenamefont
  {{Rheault}}, \citenamefont {{Suntzeff}}, \citenamefont {{Villanueva}},
  \citenamefont {{Li}},\ and\ \citenamefont {{Filippenko}}}]{stritzinger11a}%
  \BibitemOpen
  \bibfield  {author} {\bibinfo {author} {\bibfnamefont {M.~D.}\ \bibnamefont
  {{Stritzinger}}}, \bibinfo {author} {\bibfnamefont {M.~M.}\ \bibnamefont
  {{Phillips}}}, \bibinfo {author} {\bibfnamefont {L.~N.}\ \bibnamefont
  {{Boldt}}}, \bibinfo {author} {\bibfnamefont {C.}~\bibnamefont {{Burns}}},
  \bibinfo {author} {\bibfnamefont {A.}~\bibnamefont {{Campillay}}}, \bibinfo
  {author} {\bibfnamefont {C.}~\bibnamefont {{Contreras}}}, \bibinfo {author}
  {\bibfnamefont {S.}~\bibnamefont {{Gonzalez}}}, \bibinfo {author}
  {\bibfnamefont {G.}~\bibnamefont {{Folatelli}}}, \bibinfo {author}
  {\bibfnamefont {N.}~\bibnamefont {{Morrell}}}, \bibinfo {author}
  {\bibfnamefont {W.}~\bibnamefont {{Krzeminski}}}, \bibinfo {author}
  {\bibfnamefont {M.}~\bibnamefont {{Roth}}}, \bibinfo {author} {\bibfnamefont
  {F.}~\bibnamefont {{Salgado}}}, \bibinfo {author} {\bibfnamefont {D.~L.}\
  \bibnamefont {{DePoy}}}, \bibinfo {author} {\bibfnamefont {M.}~\bibnamefont
  {{Hamuy}}}, \bibinfo {author} {\bibfnamefont {W.~L.}\ \bibnamefont
  {{Freedman}}}, \bibinfo {author} {\bibfnamefont {B.~F.}\ \bibnamefont
  {{Madore}}}, \bibinfo {author} {\bibfnamefont {J.~L.}\ \bibnamefont
  {{Marshall}}}, \bibinfo {author} {\bibfnamefont {S.~E.}\ \bibnamefont
  {{Persson}}}, \bibinfo {author} {\bibfnamefont {J.-P.}\ \bibnamefont
  {{Rheault}}}, \bibinfo {author} {\bibfnamefont {N.~B.}\ \bibnamefont
  {{Suntzeff}}}, \bibinfo {author} {\bibfnamefont {S.}~\bibnamefont
  {{Villanueva}}}, \bibinfo {author} {\bibfnamefont {W.}~\bibnamefont {{Li}}},
  \ and\ \bibinfo {author} {\bibfnamefont {A.~V.}\ \bibnamefont
  {{Filippenko}}},\ }\href {\doibase 10.1088/0004-6256/142/5/156} {\bibfield
  {journal} {\bibinfo  {journal} {\aj}\ }\textbf {\bibinfo {volume} {142}},\
  \bibinfo {eid} {156} (\bibinfo {year} {2011})},\ \Eprint
  {http://arxiv.org/abs/1108.3108} {arXiv:1108.3108} \BibitemShut {NoStop}%
\bibitem [{\citenamefont {{Suzuki}}\ \emph {et~al.}(2012)\citenamefont
  {{Suzuki}}, \citenamefont {{Rubin}}, \citenamefont {{Lidman}}, \citenamefont
  {{Aldering}}, \citenamefont {{Amanullah}}, \citenamefont {{Barbary}},
  \citenamefont {{Barrientos}}, \citenamefont {{Botyanszki}}, \citenamefont
  {{Brodwin}}, \citenamefont {{Connolly}}, \citenamefont {{Dawson}},
  \citenamefont {{Dey}}, \citenamefont {{Doi}}, \citenamefont {{Donahue}},
  \citenamefont {{Deustua}}, \citenamefont {{Eisenhardt}}, \citenamefont
  {{Ellingson}}, \citenamefont {{Faccioli}}, \citenamefont {{Fadeyev}},
  \citenamefont {{Fakhouri}}, \citenamefont {{Fruchter}}, \citenamefont
  {{Gilbank}}, \citenamefont {{Gladders}}, \citenamefont {{Goldhaber}},
  \citenamefont {{Gonzalez}}, \citenamefont {{Goobar}}, \citenamefont {{Gude}},
  \citenamefont {{Hattori}}, \citenamefont {{Hoekstra}}, \citenamefont
  {{Hsiao}}, \citenamefont {{Huang}}, \citenamefont {{Ihara}}, \citenamefont
  {{Jee}}, \citenamefont {{Johnston}}, \citenamefont {{Kashikawa}},
  \citenamefont {{Koester}}, \citenamefont {{Konishi}}, \citenamefont
  {{Kowalski}}, \citenamefont {{Linder}}, \citenamefont {{Lubin}},
  \citenamefont {{Melbourne}}, \citenamefont {{Meyers}}, \citenamefont
  {{Morokuma}}, \citenamefont {{Munshi}}, \citenamefont {{Mullis}},
  \citenamefont {{Oda}}, \citenamefont {{Panagia}}, \citenamefont
  {{Perlmutter}}, \citenamefont {{Postman}}, \citenamefont {{Pritchard}},
  \citenamefont {{Rhodes}}, \citenamefont {{Ripoche}}, \citenamefont
  {{Rosati}}, \citenamefont {{Schlegel}}, \citenamefont {{Spadafora}},
  \citenamefont {{Stanford}}, \citenamefont {{Stanishev}}, \citenamefont
  {{Stern}}, \citenamefont {{Strovink}}, \citenamefont {{Takanashi}},
  \citenamefont {{Tokita}}, \citenamefont {{Wagner}}, \citenamefont {{Wang}},
  \citenamefont {{Yasuda}}, \citenamefont {{Yee}},\ and\ \citenamefont
  {{Supernova Cosmology Project}}}]{suzuki12a}%
  \BibitemOpen
  \bibfield  {author} {\bibinfo {author} {\bibfnamefont {N.}~\bibnamefont
  {{Suzuki}}}, \bibinfo {author} {\bibfnamefont {D.}~\bibnamefont {{Rubin}}},
  \bibinfo {author} {\bibfnamefont {C.}~\bibnamefont {{Lidman}}}, \bibinfo
  {author} {\bibfnamefont {G.}~\bibnamefont {{Aldering}}}, \bibinfo {author}
  {\bibfnamefont {R.}~\bibnamefont {{Amanullah}}}, \bibinfo {author}
  {\bibfnamefont {K.}~\bibnamefont {{Barbary}}}, \bibinfo {author}
  {\bibfnamefont {L.~F.}\ \bibnamefont {{Barrientos}}}, \bibinfo {author}
  {\bibfnamefont {J.}~\bibnamefont {{Botyanszki}}}, \bibinfo {author}
  {\bibfnamefont {M.}~\bibnamefont {{Brodwin}}}, \bibinfo {author}
  {\bibfnamefont {N.}~\bibnamefont {{Connolly}}}, \bibinfo {author}
  {\bibfnamefont {K.~S.}\ \bibnamefont {{Dawson}}}, \bibinfo {author}
  {\bibfnamefont {A.}~\bibnamefont {{Dey}}}, \bibinfo {author} {\bibfnamefont
  {M.}~\bibnamefont {{Doi}}}, \bibinfo {author} {\bibfnamefont
  {M.}~\bibnamefont {{Donahue}}}, \bibinfo {author} {\bibfnamefont
  {S.}~\bibnamefont {{Deustua}}}, \bibinfo {author} {\bibfnamefont
  {P.}~\bibnamefont {{Eisenhardt}}}, \bibinfo {author} {\bibfnamefont
  {E.}~\bibnamefont {{Ellingson}}}, \bibinfo {author} {\bibfnamefont
  {L.}~\bibnamefont {{Faccioli}}}, \bibinfo {author} {\bibfnamefont
  {V.}~\bibnamefont {{Fadeyev}}}, \bibinfo {author} {\bibfnamefont {H.~K.}\
  \bibnamefont {{Fakhouri}}}, \bibinfo {author} {\bibfnamefont {A.~S.}\
  \bibnamefont {{Fruchter}}}, \bibinfo {author} {\bibfnamefont {D.~G.}\
  \bibnamefont {{Gilbank}}}, \bibinfo {author} {\bibfnamefont {M.~D.}\
  \bibnamefont {{Gladders}}}, \bibinfo {author} {\bibfnamefont
  {G.}~\bibnamefont {{Goldhaber}}}, \bibinfo {author} {\bibfnamefont {A.~H.}\
  \bibnamefont {{Gonzalez}}}, \bibinfo {author} {\bibfnamefont
  {A.}~\bibnamefont {{Goobar}}}, \bibinfo {author} {\bibfnamefont
  {A.}~\bibnamefont {{Gude}}}, \bibinfo {author} {\bibfnamefont
  {T.}~\bibnamefont {{Hattori}}}, \bibinfo {author} {\bibfnamefont
  {H.}~\bibnamefont {{Hoekstra}}}, \bibinfo {author} {\bibfnamefont
  {E.}~\bibnamefont {{Hsiao}}}, \bibinfo {author} {\bibfnamefont
  {X.}~\bibnamefont {{Huang}}}, \bibinfo {author} {\bibfnamefont
  {Y.}~\bibnamefont {{Ihara}}}, \bibinfo {author} {\bibfnamefont {M.~J.}\
  \bibnamefont {{Jee}}}, \bibinfo {author} {\bibfnamefont {D.}~\bibnamefont
  {{Johnston}}}, \bibinfo {author} {\bibfnamefont {N.}~\bibnamefont
  {{Kashikawa}}}, \bibinfo {author} {\bibfnamefont {B.}~\bibnamefont
  {{Koester}}}, \bibinfo {author} {\bibfnamefont {K.}~\bibnamefont
  {{Konishi}}}, \bibinfo {author} {\bibfnamefont {M.}~\bibnamefont
  {{Kowalski}}}, \bibinfo {author} {\bibfnamefont {E.~V.}\ \bibnamefont
  {{Linder}}}, \bibinfo {author} {\bibfnamefont {L.}~\bibnamefont {{Lubin}}},
  \bibinfo {author} {\bibfnamefont {J.}~\bibnamefont {{Melbourne}}}, \bibinfo
  {author} {\bibfnamefont {J.}~\bibnamefont {{Meyers}}}, \bibinfo {author}
  {\bibfnamefont {T.}~\bibnamefont {{Morokuma}}}, \bibinfo {author}
  {\bibfnamefont {F.}~\bibnamefont {{Munshi}}}, \bibinfo {author}
  {\bibfnamefont {C.}~\bibnamefont {{Mullis}}}, \bibinfo {author}
  {\bibfnamefont {T.}~\bibnamefont {{Oda}}}, \bibinfo {author} {\bibfnamefont
  {N.}~\bibnamefont {{Panagia}}}, \bibinfo {author} {\bibfnamefont
  {S.}~\bibnamefont {{Perlmutter}}}, \bibinfo {author} {\bibfnamefont
  {M.}~\bibnamefont {{Postman}}}, \bibinfo {author} {\bibfnamefont
  {T.}~\bibnamefont {{Pritchard}}}, \bibinfo {author} {\bibfnamefont
  {J.}~\bibnamefont {{Rhodes}}}, \bibinfo {author} {\bibfnamefont
  {P.}~\bibnamefont {{Ripoche}}}, \bibinfo {author} {\bibfnamefont
  {P.}~\bibnamefont {{Rosati}}}, \bibinfo {author} {\bibfnamefont {D.~J.}\
  \bibnamefont {{Schlegel}}}, \bibinfo {author} {\bibfnamefont
  {A.}~\bibnamefont {{Spadafora}}}, \bibinfo {author} {\bibfnamefont {S.~A.}\
  \bibnamefont {{Stanford}}}, \bibinfo {author} {\bibfnamefont
  {V.}~\bibnamefont {{Stanishev}}}, \bibinfo {author} {\bibfnamefont
  {D.}~\bibnamefont {{Stern}}}, \bibinfo {author} {\bibfnamefont
  {M.}~\bibnamefont {{Strovink}}}, \bibinfo {author} {\bibfnamefont
  {N.}~\bibnamefont {{Takanashi}}}, \bibinfo {author} {\bibfnamefont
  {K.}~\bibnamefont {{Tokita}}}, \bibinfo {author} {\bibfnamefont
  {M.}~\bibnamefont {{Wagner}}}, \bibinfo {author} {\bibfnamefont
  {L.}~\bibnamefont {{Wang}}}, \bibinfo {author} {\bibfnamefont
  {N.}~\bibnamefont {{Yasuda}}}, \bibinfo {author} {\bibfnamefont {H.~K.~C.}\
  \bibnamefont {{Yee}}}, \ and\ \bibinfo {author} {\bibfnamefont
  {T.}~\bibnamefont {{Supernova Cosmology Project}}},\ }\href {\doibase
  10.1088/0004-637X/746/1/85} {\bibfield  {journal} {\bibinfo  {journal}
  {\apj}\ }\textbf {\bibinfo {volume} {746}},\ \bibinfo {eid} {85} (\bibinfo
  {year} {2012})},\ \Eprint {http://arxiv.org/abs/1105.3470} {arXiv:1105.3470
  [astro-ph.CO]} \BibitemShut {NoStop}%
\bibitem [{\citenamefont {{Rodney}}\ \emph {et~al.}(2014)\citenamefont
  {{Rodney}}, \citenamefont {{Riess}}, \citenamefont {{Strolger}},
  \citenamefont {{Dahlen}}, \citenamefont {{Graur}}, \citenamefont
  {{Casertano}}, \citenamefont {{Dickinson}}, \citenamefont {{Ferguson}},
  \citenamefont {{Garnavich}}, \citenamefont {{Hayden}}, \citenamefont {{Jha}},
  \citenamefont {{Jones}}, \citenamefont {{Kirshner}}, \citenamefont
  {{Koekemoer}}, \citenamefont {{McCully}}, \citenamefont {{Mobasher}},
  \citenamefont {{Patel}}, \citenamefont {{Weiner}}, \citenamefont {{Cenko}},
  \citenamefont {{Clubb}}, \citenamefont {{Cooper}}, \citenamefont
  {{Filippenko}}, \citenamefont {{Frederiksen}}, \citenamefont {{Hjorth}},
  \citenamefont {{Leibundgut}}, \citenamefont {{Matheson}}, \citenamefont
  {{Nayyeri}}, \citenamefont {{Penner}}, \citenamefont {{Trump}}, \citenamefont
  {{Silverman}}, \citenamefont {{U}}, \citenamefont {{Azalee Bostroem}},
  \citenamefont {{Challis}}, \citenamefont {{Rajan}}, \citenamefont {{Wolff}},
  \citenamefont {{Faber}}, \citenamefont {{Grogin}},\ and\ \citenamefont
  {{Kocevski}}}]{rodney14a}%
  \BibitemOpen
  \bibfield  {author} {\bibinfo {author} {\bibfnamefont {S.~A.}\ \bibnamefont
  {{Rodney}}}, \bibinfo {author} {\bibfnamefont {A.~G.}\ \bibnamefont
  {{Riess}}}, \bibinfo {author} {\bibfnamefont {L.-G.}\ \bibnamefont
  {{Strolger}}}, \bibinfo {author} {\bibfnamefont {T.}~\bibnamefont
  {{Dahlen}}}, \bibinfo {author} {\bibfnamefont {O.}~\bibnamefont {{Graur}}},
  \bibinfo {author} {\bibfnamefont {S.}~\bibnamefont {{Casertano}}}, \bibinfo
  {author} {\bibfnamefont {M.~E.}\ \bibnamefont {{Dickinson}}}, \bibinfo
  {author} {\bibfnamefont {H.~C.}\ \bibnamefont {{Ferguson}}}, \bibinfo
  {author} {\bibfnamefont {P.}~\bibnamefont {{Garnavich}}}, \bibinfo {author}
  {\bibfnamefont {B.}~\bibnamefont {{Hayden}}}, \bibinfo {author}
  {\bibfnamefont {S.~W.}\ \bibnamefont {{Jha}}}, \bibinfo {author}
  {\bibfnamefont {D.~O.}\ \bibnamefont {{Jones}}}, \bibinfo {author}
  {\bibfnamefont {R.~P.}\ \bibnamefont {{Kirshner}}}, \bibinfo {author}
  {\bibfnamefont {A.~M.}\ \bibnamefont {{Koekemoer}}}, \bibinfo {author}
  {\bibfnamefont {C.}~\bibnamefont {{McCully}}}, \bibinfo {author}
  {\bibfnamefont {B.}~\bibnamefont {{Mobasher}}}, \bibinfo {author}
  {\bibfnamefont {B.}~\bibnamefont {{Patel}}}, \bibinfo {author} {\bibfnamefont
  {B.~J.}\ \bibnamefont {{Weiner}}}, \bibinfo {author} {\bibfnamefont {S.~B.}\
  \bibnamefont {{Cenko}}}, \bibinfo {author} {\bibfnamefont {K.~I.}\
  \bibnamefont {{Clubb}}}, \bibinfo {author} {\bibfnamefont {M.}~\bibnamefont
  {{Cooper}}}, \bibinfo {author} {\bibfnamefont {A.~V.}\ \bibnamefont
  {{Filippenko}}}, \bibinfo {author} {\bibfnamefont {T.~F.}\ \bibnamefont
  {{Frederiksen}}}, \bibinfo {author} {\bibfnamefont {J.}~\bibnamefont
  {{Hjorth}}}, \bibinfo {author} {\bibfnamefont {B.}~\bibnamefont
  {{Leibundgut}}}, \bibinfo {author} {\bibfnamefont {T.}~\bibnamefont
  {{Matheson}}}, \bibinfo {author} {\bibfnamefont {H.}~\bibnamefont
  {{Nayyeri}}}, \bibinfo {author} {\bibfnamefont {K.}~\bibnamefont {{Penner}}},
  \bibinfo {author} {\bibfnamefont {J.}~\bibnamefont {{Trump}}}, \bibinfo
  {author} {\bibfnamefont {J.~M.}\ \bibnamefont {{Silverman}}}, \bibinfo
  {author} {\bibfnamefont {V.}~\bibnamefont {{U}}}, \bibinfo {author}
  {\bibfnamefont {K.}~\bibnamefont {{Azalee Bostroem}}}, \bibinfo {author}
  {\bibfnamefont {P.}~\bibnamefont {{Challis}}}, \bibinfo {author}
  {\bibfnamefont {A.}~\bibnamefont {{Rajan}}}, \bibinfo {author} {\bibfnamefont
  {S.}~\bibnamefont {{Wolff}}}, \bibinfo {author} {\bibfnamefont {S.~M.}\
  \bibnamefont {{Faber}}}, \bibinfo {author} {\bibfnamefont {N.~A.}\
  \bibnamefont {{Grogin}}}, \ and\ \bibinfo {author} {\bibfnamefont
  {D.}~\bibnamefont {{Kocevski}}},\ }\href {\doibase
  10.1088/0004-6256/148/1/13} {\bibfield  {journal} {\bibinfo  {journal} {\aj}\
  }\textbf {\bibinfo {volume} {148}},\ \bibinfo {eid} {13} (\bibinfo {year}
  {2014})},\ \Eprint {http://arxiv.org/abs/1401.7978} {arXiv:1401.7978}
  \BibitemShut {NoStop}%
\bibitem [{\citenamefont {{Graur}}\ \emph {et~al.}(2014)\citenamefont
  {{Graur}}, \citenamefont {{Rodney}}, \citenamefont {{Maoz}}, \citenamefont
  {{Riess}}, \citenamefont {{Jha}}, \citenamefont {{Postman}}, \citenamefont
  {{Dahlen}}, \citenamefont {{Holoien}}, \citenamefont {{McCully}},
  \citenamefont {{Patel}}, \citenamefont {{Strolger}}, \citenamefont
  {{Ben{\'{\i}}tez}}, \citenamefont {{Coe}}, \citenamefont {{Jouvel}},
  \citenamefont {{Medezinski}}, \citenamefont {{Molino}}, \citenamefont
  {{Nonino}}, \citenamefont {{Bradley}}, \citenamefont {{Koekemoer}},
  \citenamefont {{Balestra}}, \citenamefont {{Cenko}}, \citenamefont {{Clubb}},
  \citenamefont {{Dickinson}}, \citenamefont {{Filippenko}}, \citenamefont
  {{Frederiksen}}, \citenamefont {{Garnavich}}, \citenamefont {{Hjorth}},
  \citenamefont {{Jones}}, \citenamefont {{Leibundgut}}, \citenamefont
  {{Matheson}}, \citenamefont {{Mobasher}}, \citenamefont {{Rosati}},
  \citenamefont {{Silverman}}, \citenamefont {{U}}, \citenamefont
  {{Jedruszczuk}}, \citenamefont {{Li}}, \citenamefont {{Lin}}, \citenamefont
  {{Mirmelstein}}, \citenamefont {{Neustadt}}, \citenamefont {{Ovadia}},\ and\
  \citenamefont {{Rogers}}}]{graur14a}%
  \BibitemOpen
  \bibfield  {author} {\bibinfo {author} {\bibfnamefont {O.}~\bibnamefont
  {{Graur}}}, \bibinfo {author} {\bibfnamefont {S.~A.}\ \bibnamefont
  {{Rodney}}}, \bibinfo {author} {\bibfnamefont {D.}~\bibnamefont {{Maoz}}},
  \bibinfo {author} {\bibfnamefont {A.~G.}\ \bibnamefont {{Riess}}}, \bibinfo
  {author} {\bibfnamefont {S.~W.}\ \bibnamefont {{Jha}}}, \bibinfo {author}
  {\bibfnamefont {M.}~\bibnamefont {{Postman}}}, \bibinfo {author}
  {\bibfnamefont {T.}~\bibnamefont {{Dahlen}}}, \bibinfo {author}
  {\bibfnamefont {T.~W.-S.}\ \bibnamefont {{Holoien}}}, \bibinfo {author}
  {\bibfnamefont {C.}~\bibnamefont {{McCully}}}, \bibinfo {author}
  {\bibfnamefont {B.}~\bibnamefont {{Patel}}}, \bibinfo {author} {\bibfnamefont
  {L.-G.}\ \bibnamefont {{Strolger}}}, \bibinfo {author} {\bibfnamefont
  {N.}~\bibnamefont {{Ben{\'{\i}}tez}}}, \bibinfo {author} {\bibfnamefont
  {D.}~\bibnamefont {{Coe}}}, \bibinfo {author} {\bibfnamefont
  {S.}~\bibnamefont {{Jouvel}}}, \bibinfo {author} {\bibfnamefont
  {E.}~\bibnamefont {{Medezinski}}}, \bibinfo {author} {\bibfnamefont
  {A.}~\bibnamefont {{Molino}}}, \bibinfo {author} {\bibfnamefont
  {M.}~\bibnamefont {{Nonino}}}, \bibinfo {author} {\bibfnamefont
  {L.}~\bibnamefont {{Bradley}}}, \bibinfo {author} {\bibfnamefont
  {A.}~\bibnamefont {{Koekemoer}}}, \bibinfo {author} {\bibfnamefont
  {I.}~\bibnamefont {{Balestra}}}, \bibinfo {author} {\bibfnamefont {S.~B.}\
  \bibnamefont {{Cenko}}}, \bibinfo {author} {\bibfnamefont {K.~I.}\
  \bibnamefont {{Clubb}}}, \bibinfo {author} {\bibfnamefont {M.~E.}\
  \bibnamefont {{Dickinson}}}, \bibinfo {author} {\bibfnamefont {A.~V.}\
  \bibnamefont {{Filippenko}}}, \bibinfo {author} {\bibfnamefont {T.~F.}\
  \bibnamefont {{Frederiksen}}}, \bibinfo {author} {\bibfnamefont
  {P.}~\bibnamefont {{Garnavich}}}, \bibinfo {author} {\bibfnamefont
  {J.}~\bibnamefont {{Hjorth}}}, \bibinfo {author} {\bibfnamefont {D.~O.}\
  \bibnamefont {{Jones}}}, \bibinfo {author} {\bibfnamefont {B.}~\bibnamefont
  {{Leibundgut}}}, \bibinfo {author} {\bibfnamefont {T.}~\bibnamefont
  {{Matheson}}}, \bibinfo {author} {\bibfnamefont {B.}~\bibnamefont
  {{Mobasher}}}, \bibinfo {author} {\bibfnamefont {P.}~\bibnamefont
  {{Rosati}}}, \bibinfo {author} {\bibfnamefont {J.~M.}\ \bibnamefont
  {{Silverman}}}, \bibinfo {author} {\bibfnamefont {V.}~\bibnamefont {{U}}},
  \bibinfo {author} {\bibfnamefont {K.}~\bibnamefont {{Jedruszczuk}}}, \bibinfo
  {author} {\bibfnamefont {C.}~\bibnamefont {{Li}}}, \bibinfo {author}
  {\bibfnamefont {K.}~\bibnamefont {{Lin}}}, \bibinfo {author} {\bibfnamefont
  {M.}~\bibnamefont {{Mirmelstein}}}, \bibinfo {author} {\bibfnamefont
  {J.}~\bibnamefont {{Neustadt}}}, \bibinfo {author} {\bibfnamefont
  {A.}~\bibnamefont {{Ovadia}}}, \ and\ \bibinfo {author} {\bibfnamefont
  {E.~H.}\ \bibnamefont {{Rogers}}},\ }\href {\doibase
  10.1088/0004-637X/783/1/28} {\bibfield  {journal} {\bibinfo  {journal}
  {\apj}\ }\textbf {\bibinfo {volume} {783}},\ \bibinfo {eid} {28} (\bibinfo
  {year} {2014})},\ \Eprint {http://arxiv.org/abs/1310.3495} {arXiv:1310.3495}
  \BibitemShut {NoStop}%
\bibitem [{\citenamefont {{Riess}}\ \emph {et~al.}(2018)\citenamefont
  {{Riess}}, \citenamefont {{Rodney}}, \citenamefont {{Scolnic}}, \citenamefont
  {{Shafer}}, \citenamefont {{Strolger}}, \citenamefont {{Ferguson}},
  \citenamefont {{Postman}}, \citenamefont {{Graur}}, \citenamefont {{Maoz}},
  \citenamefont {{Jha}}, \citenamefont {{Mobasher}}, \citenamefont
  {{Casertano}}, \citenamefont {{Hayden}}, \citenamefont {{Molino}},
  \citenamefont {{Hjorth}}, \citenamefont {{Garnavich}}, \citenamefont
  {{Jones}}, \citenamefont {{Kirshner}}, \citenamefont {{Koekemoer}},
  \citenamefont {{Grogin}}, \citenamefont {{Brammer}}, \citenamefont
  {{Hemmati}}, \citenamefont {{Dickinson}}, \citenamefont {{Challis}},
  \citenamefont {{Wolff}}, \citenamefont {{Clubb}}, \citenamefont
  {{Filippenko}}, \citenamefont {{Nayyeri}}, \citenamefont {{U}}, \citenamefont
  {{Koo}}, \citenamefont {{Faber}}, \citenamefont {{Kocevski}}, \citenamefont
  {{Bradley}},\ and\ \citenamefont {{Coe}}}]{riess18a}%
  \BibitemOpen
  \bibfield  {author} {\bibinfo {author} {\bibfnamefont {A.~G.}\ \bibnamefont
  {{Riess}}}, \bibinfo {author} {\bibfnamefont {S.~A.}\ \bibnamefont
  {{Rodney}}}, \bibinfo {author} {\bibfnamefont {D.~M.}\ \bibnamefont
  {{Scolnic}}}, \bibinfo {author} {\bibfnamefont {D.~L.}\ \bibnamefont
  {{Shafer}}}, \bibinfo {author} {\bibfnamefont {L.-G.}\ \bibnamefont
  {{Strolger}}}, \bibinfo {author} {\bibfnamefont {H.~C.}\ \bibnamefont
  {{Ferguson}}}, \bibinfo {author} {\bibfnamefont {M.}~\bibnamefont
  {{Postman}}}, \bibinfo {author} {\bibfnamefont {O.}~\bibnamefont {{Graur}}},
  \bibinfo {author} {\bibfnamefont {D.}~\bibnamefont {{Maoz}}}, \bibinfo
  {author} {\bibfnamefont {S.~W.}\ \bibnamefont {{Jha}}}, \bibinfo {author}
  {\bibfnamefont {B.}~\bibnamefont {{Mobasher}}}, \bibinfo {author}
  {\bibfnamefont {S.}~\bibnamefont {{Casertano}}}, \bibinfo {author}
  {\bibfnamefont {B.}~\bibnamefont {{Hayden}}}, \bibinfo {author}
  {\bibfnamefont {A.}~\bibnamefont {{Molino}}}, \bibinfo {author}
  {\bibfnamefont {J.}~\bibnamefont {{Hjorth}}}, \bibinfo {author}
  {\bibfnamefont {P.~M.}\ \bibnamefont {{Garnavich}}}, \bibinfo {author}
  {\bibfnamefont {D.~O.}\ \bibnamefont {{Jones}}}, \bibinfo {author}
  {\bibfnamefont {R.~P.}\ \bibnamefont {{Kirshner}}}, \bibinfo {author}
  {\bibfnamefont {A.~M.}\ \bibnamefont {{Koekemoer}}}, \bibinfo {author}
  {\bibfnamefont {N.~A.}\ \bibnamefont {{Grogin}}}, \bibinfo {author}
  {\bibfnamefont {G.}~\bibnamefont {{Brammer}}}, \bibinfo {author}
  {\bibfnamefont {S.}~\bibnamefont {{Hemmati}}}, \bibinfo {author}
  {\bibfnamefont {M.}~\bibnamefont {{Dickinson}}}, \bibinfo {author}
  {\bibfnamefont {P.~M.}\ \bibnamefont {{Challis}}}, \bibinfo {author}
  {\bibfnamefont {S.}~\bibnamefont {{Wolff}}}, \bibinfo {author} {\bibfnamefont
  {K.~I.}\ \bibnamefont {{Clubb}}}, \bibinfo {author} {\bibfnamefont {A.~V.}\
  \bibnamefont {{Filippenko}}}, \bibinfo {author} {\bibfnamefont
  {H.}~\bibnamefont {{Nayyeri}}}, \bibinfo {author} {\bibfnamefont
  {V.}~\bibnamefont {{U}}}, \bibinfo {author} {\bibfnamefont {D.~C.}\
  \bibnamefont {{Koo}}}, \bibinfo {author} {\bibfnamefont {S.~M.}\ \bibnamefont
  {{Faber}}}, \bibinfo {author} {\bibfnamefont {D.}~\bibnamefont {{Kocevski}}},
  \bibinfo {author} {\bibfnamefont {L.}~\bibnamefont {{Bradley}}}, \ and\
  \bibinfo {author} {\bibfnamefont {D.}~\bibnamefont {{Coe}}},\ }\href
  {\doibase 10.3847/1538-4357/aaa5a9} {\bibfield  {journal} {\bibinfo
  {journal} {\apj}\ }\textbf {\bibinfo {volume} {853}},\ \bibinfo {eid} {126}
  (\bibinfo {year} {2018})},\ \Eprint {http://arxiv.org/abs/1710.00844}
  {arXiv:1710.00844} \BibitemShut {NoStop}%
\bibitem [{\citenamefont {{Troxel}}\ \emph {et~al.}(2018)\citenamefont
  {{Troxel}}, \citenamefont {{MacCrann}}, \citenamefont {{Zuntz}},
  \citenamefont {{Eifler}}, \citenamefont {{Krause}}, \citenamefont
  {{Dodelson}}, \citenamefont {{Gruen}}, \citenamefont {{Blazek}},
  \citenamefont {{Friedrich}}, \citenamefont {{Samuroff}}, \citenamefont
  {{Prat}}, \citenamefont {{Secco}}, \citenamefont {{Davis}}, \citenamefont
  {{Fert{\'e}}}, \citenamefont {{DeRose}}, \citenamefont {{Alarcon}},
  \citenamefont {{Amara}}, \citenamefont {{Baxter}}, \citenamefont {{Becker}},
  \citenamefont {{Bernstein}}, \citenamefont {{Bridle}}, \citenamefont
  {{Cawthon}}, \citenamefont {{Chang}}, \citenamefont {{Choi}}, \citenamefont
  {{De Vicente}}, \citenamefont {{Drlica-Wagner}}, \citenamefont
  {{Elvin-Poole}}, \citenamefont {{Frieman}}, \citenamefont {{Gatti}},
  \citenamefont {{Hartley}}, \citenamefont {{Honscheid}}, \citenamefont
  {{Hoyle}}, \citenamefont {{Huff}}, \citenamefont {{Huterer}}, \citenamefont
  {{Jain}}, \citenamefont {{Jarvis}}, \citenamefont {{Kacprzak}}, \citenamefont
  {{Kirk}}, \citenamefont {{Kokron}}, \citenamefont {{Krawiec}}, \citenamefont
  {{Lahav}}, \citenamefont {{Liddle}}, \citenamefont {{Peacock}}, \citenamefont
  {{Rau}}, \citenamefont {{Refregier}}, \citenamefont {{Rollins}},
  \citenamefont {{Rozo}}, \citenamefont {{Rykoff}}, \citenamefont
  {{S{\'a}nchez}}, \citenamefont {{Sevilla-Noarbe}}, \citenamefont {{Sheldon}},
  \citenamefont {{Stebbins}}, \citenamefont {{Varga}}, \citenamefont
  {{Vielzeuf}}, \citenamefont {{Wang}}, \citenamefont {{Wechsler}},
  \citenamefont {{Yanny}}, \citenamefont {{Abbott}}, \citenamefont {{Abdalla}},
  \citenamefont {{Allam}}, \citenamefont {{Annis}}, \citenamefont {{Bechtol}},
  \citenamefont {{Benoit-L{\'e}vy}}, \citenamefont {{Bertin}}, \citenamefont
  {{Brooks}}, \citenamefont {{Buckley-Geer}}, \citenamefont {{Burke}},
  \citenamefont {{Carnero Rosell}}, \citenamefont {{Carrasco Kind}},
  \citenamefont {{Carretero}}, \citenamefont {{Castander}}, \citenamefont
  {{Crocce}}, \citenamefont {{Cunha}}, \citenamefont {{D'Andrea}},
  \citenamefont {{da Costa}}, \citenamefont {{DePoy}}, \citenamefont {{Desai}},
  \citenamefont {{Diehl}}, \citenamefont {{Dietrich}}, \citenamefont {{Doel}},
  \citenamefont {{Fernandez}}, \citenamefont {{Flaugher}}, \citenamefont
  {{Fosalba}}, \citenamefont {{Garc{\'{\i}}a-Bellido}}, \citenamefont
  {{Gaztanaga}}, \citenamefont {{Gerdes}}, \citenamefont {{Giannantonio}},
  \citenamefont {{Goldstein}}, \citenamefont {{Gruendl}}, \citenamefont
  {{Gschwend}}, \citenamefont {{Gutierrez}}, \citenamefont {{James}},
  \citenamefont {{Jeltema}}, \citenamefont {{Johnson}}, \citenamefont
  {{Johnson}}, \citenamefont {{Kent}}, \citenamefont {{Kuehn}}, \citenamefont
  {{Kuhlmann}}, \citenamefont {{Kuropatkin}}, \citenamefont {{Li}},
  \citenamefont {{Lima}}, \citenamefont {{Lin}}, \citenamefont {{Maia}},
  \citenamefont {{March}}, \citenamefont {{Marshall}}, \citenamefont
  {{Martini}}, \citenamefont {{Melchior}}, \citenamefont {{Menanteau}},
  \citenamefont {{Miquel}}, \citenamefont {{Mohr}}, \citenamefont {{Neilsen}},
  \citenamefont {{Nichol}}, \citenamefont {{Nord}}, \citenamefont
  {{Petravick}}, \citenamefont {{Plazas}}, \citenamefont {{Romer}},
  \citenamefont {{Roodman}}, \citenamefont {{Sako}}, \citenamefont {{Sanchez}},
  \citenamefont {{Scarpine}}, \citenamefont {{Schindler}}, \citenamefont
  {{Schubnell}}, \citenamefont {{Smith}}, \citenamefont {{Smith}},
  \citenamefont {{Soares-Santos}}, \citenamefont {{Sobreira}}, \citenamefont
  {{Suchyta}}, \citenamefont {{Swanson}}, \citenamefont {{Tarle}},
  \citenamefont {{Thomas}}, \citenamefont {{Tucker}}, \citenamefont {{Vikram}},
  \citenamefont {{Walker}}, \citenamefont {{Weller}}, \citenamefont {{Zhang}},\
  and\ \citenamefont {{DES Collaboration}}}]{troxel18a}%
  \BibitemOpen
  \bibfield  {author} {\bibinfo {author} {\bibfnamefont {M.~A.}\ \bibnamefont
  {{Troxel}}}, \bibinfo {author} {\bibfnamefont {N.}~\bibnamefont
  {{MacCrann}}}, \bibinfo {author} {\bibfnamefont {J.}~\bibnamefont {{Zuntz}}},
  \bibinfo {author} {\bibfnamefont {T.~F.}\ \bibnamefont {{Eifler}}}, \bibinfo
  {author} {\bibfnamefont {E.}~\bibnamefont {{Krause}}}, \bibinfo {author}
  {\bibfnamefont {S.}~\bibnamefont {{Dodelson}}}, \bibinfo {author}
  {\bibfnamefont {D.}~\bibnamefont {{Gruen}}}, \bibinfo {author} {\bibfnamefont
  {J.}~\bibnamefont {{Blazek}}}, \bibinfo {author} {\bibfnamefont
  {O.}~\bibnamefont {{Friedrich}}}, \bibinfo {author} {\bibfnamefont
  {S.}~\bibnamefont {{Samuroff}}}, \bibinfo {author} {\bibfnamefont
  {J.}~\bibnamefont {{Prat}}}, \bibinfo {author} {\bibfnamefont {L.~F.}\
  \bibnamefont {{Secco}}}, \bibinfo {author} {\bibfnamefont {C.}~\bibnamefont
  {{Davis}}}, \bibinfo {author} {\bibfnamefont {A.}~\bibnamefont
  {{Fert{\'e}}}}, \bibinfo {author} {\bibfnamefont {J.}~\bibnamefont
  {{DeRose}}}, \bibinfo {author} {\bibfnamefont {A.}~\bibnamefont {{Alarcon}}},
  \bibinfo {author} {\bibfnamefont {A.}~\bibnamefont {{Amara}}}, \bibinfo
  {author} {\bibfnamefont {E.}~\bibnamefont {{Baxter}}}, \bibinfo {author}
  {\bibfnamefont {M.~R.}\ \bibnamefont {{Becker}}}, \bibinfo {author}
  {\bibfnamefont {G.~M.}\ \bibnamefont {{Bernstein}}}, \bibinfo {author}
  {\bibfnamefont {S.~L.}\ \bibnamefont {{Bridle}}}, \bibinfo {author}
  {\bibfnamefont {R.}~\bibnamefont {{Cawthon}}}, \bibinfo {author}
  {\bibfnamefont {C.}~\bibnamefont {{Chang}}}, \bibinfo {author} {\bibfnamefont
  {A.}~\bibnamefont {{Choi}}}, \bibinfo {author} {\bibfnamefont
  {J.}~\bibnamefont {{De Vicente}}}, \bibinfo {author} {\bibfnamefont
  {A.}~\bibnamefont {{Drlica-Wagner}}}, \bibinfo {author} {\bibfnamefont
  {J.}~\bibnamefont {{Elvin-Poole}}}, \bibinfo {author} {\bibfnamefont
  {J.}~\bibnamefont {{Frieman}}}, \bibinfo {author} {\bibfnamefont
  {M.}~\bibnamefont {{Gatti}}}, \bibinfo {author} {\bibfnamefont {W.~G.}\
  \bibnamefont {{Hartley}}}, \bibinfo {author} {\bibfnamefont {K.}~\bibnamefont
  {{Honscheid}}}, \bibinfo {author} {\bibfnamefont {B.}~\bibnamefont
  {{Hoyle}}}, \bibinfo {author} {\bibfnamefont {E.~M.}\ \bibnamefont {{Huff}}},
  \bibinfo {author} {\bibfnamefont {D.}~\bibnamefont {{Huterer}}}, \bibinfo
  {author} {\bibfnamefont {B.}~\bibnamefont {{Jain}}}, \bibinfo {author}
  {\bibfnamefont {M.}~\bibnamefont {{Jarvis}}}, \bibinfo {author}
  {\bibfnamefont {T.}~\bibnamefont {{Kacprzak}}}, \bibinfo {author}
  {\bibfnamefont {D.}~\bibnamefont {{Kirk}}}, \bibinfo {author} {\bibfnamefont
  {N.}~\bibnamefont {{Kokron}}}, \bibinfo {author} {\bibfnamefont
  {C.}~\bibnamefont {{Krawiec}}}, \bibinfo {author} {\bibfnamefont
  {O.}~\bibnamefont {{Lahav}}}, \bibinfo {author} {\bibfnamefont {A.~R.}\
  \bibnamefont {{Liddle}}}, \bibinfo {author} {\bibfnamefont {J.}~\bibnamefont
  {{Peacock}}}, \bibinfo {author} {\bibfnamefont {M.~M.}\ \bibnamefont
  {{Rau}}}, \bibinfo {author} {\bibfnamefont {A.}~\bibnamefont {{Refregier}}},
  \bibinfo {author} {\bibfnamefont {R.~P.}\ \bibnamefont {{Rollins}}}, \bibinfo
  {author} {\bibfnamefont {E.}~\bibnamefont {{Rozo}}}, \bibinfo {author}
  {\bibfnamefont {E.~S.}\ \bibnamefont {{Rykoff}}}, \bibinfo {author}
  {\bibfnamefont {C.}~\bibnamefont {{S{\'a}nchez}}}, \bibinfo {author}
  {\bibfnamefont {I.}~\bibnamefont {{Sevilla-Noarbe}}}, \bibinfo {author}
  {\bibfnamefont {E.}~\bibnamefont {{Sheldon}}}, \bibinfo {author}
  {\bibfnamefont {A.}~\bibnamefont {{Stebbins}}}, \bibinfo {author}
  {\bibfnamefont {T.~N.}\ \bibnamefont {{Varga}}}, \bibinfo {author}
  {\bibfnamefont {P.}~\bibnamefont {{Vielzeuf}}}, \bibinfo {author}
  {\bibfnamefont {M.}~\bibnamefont {{Wang}}}, \bibinfo {author} {\bibfnamefont
  {R.~H.}\ \bibnamefont {{Wechsler}}}, \bibinfo {author} {\bibfnamefont
  {B.}~\bibnamefont {{Yanny}}}, \bibinfo {author} {\bibfnamefont {T.~M.~C.}\
  \bibnamefont {{Abbott}}}, \bibinfo {author} {\bibfnamefont {F.~B.}\
  \bibnamefont {{Abdalla}}}, \bibinfo {author} {\bibfnamefont {S.}~\bibnamefont
  {{Allam}}}, \bibinfo {author} {\bibfnamefont {J.}~\bibnamefont {{Annis}}},
  \bibinfo {author} {\bibfnamefont {K.}~\bibnamefont {{Bechtol}}}, \bibinfo
  {author} {\bibfnamefont {A.}~\bibnamefont {{Benoit-L{\'e}vy}}}, \bibinfo
  {author} {\bibfnamefont {E.}~\bibnamefont {{Bertin}}}, \bibinfo {author}
  {\bibfnamefont {D.}~\bibnamefont {{Brooks}}}, \bibinfo {author}
  {\bibfnamefont {E.}~\bibnamefont {{Buckley-Geer}}}, \bibinfo {author}
  {\bibfnamefont {D.~L.}\ \bibnamefont {{Burke}}}, \bibinfo {author}
  {\bibfnamefont {A.}~\bibnamefont {{Carnero Rosell}}}, \bibinfo {author}
  {\bibfnamefont {M.}~\bibnamefont {{Carrasco Kind}}}, \bibinfo {author}
  {\bibfnamefont {J.}~\bibnamefont {{Carretero}}}, \bibinfo {author}
  {\bibfnamefont {F.~J.}\ \bibnamefont {{Castander}}}, \bibinfo {author}
  {\bibfnamefont {M.}~\bibnamefont {{Crocce}}}, \bibinfo {author}
  {\bibfnamefont {C.~E.}\ \bibnamefont {{Cunha}}}, \bibinfo {author}
  {\bibfnamefont {C.~B.}\ \bibnamefont {{D'Andrea}}}, \bibinfo {author}
  {\bibfnamefont {L.~N.}\ \bibnamefont {{da Costa}}}, \bibinfo {author}
  {\bibfnamefont {D.~L.}\ \bibnamefont {{DePoy}}}, \bibinfo {author}
  {\bibfnamefont {S.}~\bibnamefont {{Desai}}}, \bibinfo {author} {\bibfnamefont
  {H.~T.}\ \bibnamefont {{Diehl}}}, \bibinfo {author} {\bibfnamefont {J.~P.}\
  \bibnamefont {{Dietrich}}}, \bibinfo {author} {\bibfnamefont
  {P.}~\bibnamefont {{Doel}}}, \bibinfo {author} {\bibfnamefont
  {E.}~\bibnamefont {{Fernandez}}}, \bibinfo {author} {\bibfnamefont
  {B.}~\bibnamefont {{Flaugher}}}, \bibinfo {author} {\bibfnamefont
  {P.}~\bibnamefont {{Fosalba}}}, \bibinfo {author} {\bibfnamefont
  {J.}~\bibnamefont {{Garc{\'{\i}}a-Bellido}}}, \bibinfo {author}
  {\bibfnamefont {E.}~\bibnamefont {{Gaztanaga}}}, \bibinfo {author}
  {\bibfnamefont {D.~W.}\ \bibnamefont {{Gerdes}}}, \bibinfo {author}
  {\bibfnamefont {T.}~\bibnamefont {{Giannantonio}}}, \bibinfo {author}
  {\bibfnamefont {D.~A.}\ \bibnamefont {{Goldstein}}}, \bibinfo {author}
  {\bibfnamefont {R.~A.}\ \bibnamefont {{Gruendl}}}, \bibinfo {author}
  {\bibfnamefont {J.}~\bibnamefont {{Gschwend}}}, \bibinfo {author}
  {\bibfnamefont {G.}~\bibnamefont {{Gutierrez}}}, \bibinfo {author}
  {\bibfnamefont {D.~J.}\ \bibnamefont {{James}}}, \bibinfo {author}
  {\bibfnamefont {T.}~\bibnamefont {{Jeltema}}}, \bibinfo {author}
  {\bibfnamefont {M.~W.~G.}\ \bibnamefont {{Johnson}}}, \bibinfo {author}
  {\bibfnamefont {M.~D.}\ \bibnamefont {{Johnson}}}, \bibinfo {author}
  {\bibfnamefont {S.}~\bibnamefont {{Kent}}}, \bibinfo {author} {\bibfnamefont
  {K.}~\bibnamefont {{Kuehn}}}, \bibinfo {author} {\bibfnamefont
  {S.}~\bibnamefont {{Kuhlmann}}}, \bibinfo {author} {\bibfnamefont
  {N.}~\bibnamefont {{Kuropatkin}}}, \bibinfo {author} {\bibfnamefont {T.~S.}\
  \bibnamefont {{Li}}}, \bibinfo {author} {\bibfnamefont {M.}~\bibnamefont
  {{Lima}}}, \bibinfo {author} {\bibfnamefont {H.}~\bibnamefont {{Lin}}},
  \bibinfo {author} {\bibfnamefont {M.~A.~G.}\ \bibnamefont {{Maia}}}, \bibinfo
  {author} {\bibfnamefont {M.}~\bibnamefont {{March}}}, \bibinfo {author}
  {\bibfnamefont {J.~L.}\ \bibnamefont {{Marshall}}}, \bibinfo {author}
  {\bibfnamefont {P.}~\bibnamefont {{Martini}}}, \bibinfo {author}
  {\bibfnamefont {P.}~\bibnamefont {{Melchior}}}, \bibinfo {author}
  {\bibfnamefont {F.}~\bibnamefont {{Menanteau}}}, \bibinfo {author}
  {\bibfnamefont {R.}~\bibnamefont {{Miquel}}}, \bibinfo {author}
  {\bibfnamefont {J.~J.}\ \bibnamefont {{Mohr}}}, \bibinfo {author}
  {\bibfnamefont {E.}~\bibnamefont {{Neilsen}}}, \bibinfo {author}
  {\bibfnamefont {R.~C.}\ \bibnamefont {{Nichol}}}, \bibinfo {author}
  {\bibfnamefont {B.}~\bibnamefont {{Nord}}}, \bibinfo {author} {\bibfnamefont
  {D.}~\bibnamefont {{Petravick}}}, \bibinfo {author} {\bibfnamefont {A.~A.}\
  \bibnamefont {{Plazas}}}, \bibinfo {author} {\bibfnamefont {A.~K.}\
  \bibnamefont {{Romer}}}, \bibinfo {author} {\bibfnamefont {A.}~\bibnamefont
  {{Roodman}}}, \bibinfo {author} {\bibfnamefont {M.}~\bibnamefont {{Sako}}},
  \bibinfo {author} {\bibfnamefont {E.}~\bibnamefont {{Sanchez}}}, \bibinfo
  {author} {\bibfnamefont {V.}~\bibnamefont {{Scarpine}}}, \bibinfo {author}
  {\bibfnamefont {R.}~\bibnamefont {{Schindler}}}, \bibinfo {author}
  {\bibfnamefont {M.}~\bibnamefont {{Schubnell}}}, \bibinfo {author}
  {\bibfnamefont {M.}~\bibnamefont {{Smith}}}, \bibinfo {author} {\bibfnamefont
  {R.~C.}\ \bibnamefont {{Smith}}}, \bibinfo {author} {\bibfnamefont
  {M.}~\bibnamefont {{Soares-Santos}}}, \bibinfo {author} {\bibfnamefont
  {F.}~\bibnamefont {{Sobreira}}}, \bibinfo {author} {\bibfnamefont
  {E.}~\bibnamefont {{Suchyta}}}, \bibinfo {author} {\bibfnamefont {M.~E.~C.}\
  \bibnamefont {{Swanson}}}, \bibinfo {author} {\bibfnamefont {G.}~\bibnamefont
  {{Tarle}}}, \bibinfo {author} {\bibfnamefont {D.}~\bibnamefont {{Thomas}}},
  \bibinfo {author} {\bibfnamefont {D.~L.}\ \bibnamefont {{Tucker}}}, \bibinfo
  {author} {\bibfnamefont {V.}~\bibnamefont {{Vikram}}}, \bibinfo {author}
  {\bibfnamefont {A.~R.}\ \bibnamefont {{Walker}}}, \bibinfo {author}
  {\bibfnamefont {J.}~\bibnamefont {{Weller}}}, \bibinfo {author}
  {\bibfnamefont {Y.}~\bibnamefont {{Zhang}}}, \ and\ \bibinfo {author}
  {\bibnamefont {{DES Collaboration}}},\ }\href {\doibase
  10.1103/PhysRevD.98.043528} {\bibfield  {journal} {\bibinfo  {journal}
  {\prd}\ }\textbf {\bibinfo {volume} {98}},\ \bibinfo {eid} {043528} (\bibinfo
  {year} {2018})},\ \Eprint {http://arxiv.org/abs/1708.01538}
  {arXiv:1708.01538} \BibitemShut {NoStop}%
\bibitem [{\citenamefont {{Krause}}\ \emph {et~al.}(2017)\citenamefont
  {{Krause}}, \citenamefont {{Eifler}}, \citenamefont {{Zuntz}}, \citenamefont
  {{Friedrich}}, \citenamefont {{Troxel}}, \citenamefont {{Dodelson}},
  \citenamefont {{Blazek}}, \citenamefont {{Secco}}, \citenamefont
  {{MacCrann}}, \citenamefont {{Baxter}}, \citenamefont {{Chang}},
  \citenamefont {{Chen}}, \citenamefont {{Crocce}}, \citenamefont {{DeRose}},
  \citenamefont {{Ferte}}, \citenamefont {{Kokron}}, \citenamefont {{Lacasa}},
  \citenamefont {{Miranda}}, \citenamefont {{Omori}}, \citenamefont
  {{Porredon}}, \citenamefont {{Rosenfeld}}, \citenamefont {{Samuroff}},
  \citenamefont {{Wang}}, \citenamefont {{Wechsler}}, \citenamefont {{Abbott}},
  \citenamefont {{Abdalla}}, \citenamefont {{Allam}}, \citenamefont {{Annis}},
  \citenamefont {{Bechtol}}, \citenamefont {{Benoit-Levy}}, \citenamefont
  {{Bernstein}}, \citenamefont {{Brooks}}, \citenamefont {{Burke}},
  \citenamefont {{Capozzi}}, \citenamefont {{Carrasco Kind}}, \citenamefont
  {{Carretero}}, \citenamefont {{D'Andrea}}, \citenamefont {{da Costa}},
  \citenamefont {{Davis}}, \citenamefont {{DePoy}}, \citenamefont {{Desai}},
  \citenamefont {{Diehl}}, \citenamefont {{Dietrich}}, \citenamefont
  {{Evrard}}, \citenamefont {{Flaugher}}, \citenamefont {{Fosalba}},
  \citenamefont {{Frieman}}, \citenamefont {{Garcia-Bellido}}, \citenamefont
  {{Gaztanaga}}, \citenamefont {{Giannantonio}}, \citenamefont {{Gruen}},
  \citenamefont {{Gruendl}}, \citenamefont {{Gschwend}}, \citenamefont
  {{Gutierrez}}, \citenamefont {{Honscheid}}, \citenamefont {{James}},
  \citenamefont {{Jeltema}}, \citenamefont {{Kuehn}}, \citenamefont
  {{Kuhlmann}}, \citenamefont {{Lahav}}, \citenamefont {{Lima}}, \citenamefont
  {{Maia}}, \citenamefont {{March}}, \citenamefont {{Marshall}}, \citenamefont
  {{Martini}}, \citenamefont {{Menanteau}}, \citenamefont {{Miquel}},
  \citenamefont {{Nichol}}, \citenamefont {{Plazas}}, \citenamefont {{Romer}},
  \citenamefont {{Rykoff}}, \citenamefont {{Sanchez}}, \citenamefont
  {{Scarpine}}, \citenamefont {{Schindler}}, \citenamefont {{Schubnell}},
  \citenamefont {{Sevilla-Noarbe}}, \citenamefont {{Smith}}, \citenamefont
  {{Soares-Santos}}, \citenamefont {{Sobreira}}, \citenamefont {{Suchyta}},
  \citenamefont {{Swanson}}, \citenamefont {{Tarle}}, \citenamefont {{Tucker}},
  \citenamefont {{Vikram}}, \citenamefont {{Walker}},\ and\ \citenamefont
  {{Weller}}}]{krause17a}%
  \BibitemOpen
  \bibfield  {author} {\bibinfo {author} {\bibfnamefont {E.}~\bibnamefont
  {{Krause}}}, \bibinfo {author} {\bibfnamefont {T.~F.}\ \bibnamefont
  {{Eifler}}}, \bibinfo {author} {\bibfnamefont {J.}~\bibnamefont {{Zuntz}}},
  \bibinfo {author} {\bibfnamefont {O.}~\bibnamefont {{Friedrich}}}, \bibinfo
  {author} {\bibfnamefont {M.~A.}\ \bibnamefont {{Troxel}}}, \bibinfo {author}
  {\bibfnamefont {S.}~\bibnamefont {{Dodelson}}}, \bibinfo {author}
  {\bibfnamefont {J.}~\bibnamefont {{Blazek}}}, \bibinfo {author}
  {\bibfnamefont {L.~F.}\ \bibnamefont {{Secco}}}, \bibinfo {author}
  {\bibfnamefont {N.}~\bibnamefont {{MacCrann}}}, \bibinfo {author}
  {\bibfnamefont {E.}~\bibnamefont {{Baxter}}}, \bibinfo {author}
  {\bibfnamefont {C.}~\bibnamefont {{Chang}}}, \bibinfo {author} {\bibfnamefont
  {N.}~\bibnamefont {{Chen}}}, \bibinfo {author} {\bibfnamefont
  {M.}~\bibnamefont {{Crocce}}}, \bibinfo {author} {\bibfnamefont
  {J.}~\bibnamefont {{DeRose}}}, \bibinfo {author} {\bibfnamefont
  {A.}~\bibnamefont {{Ferte}}}, \bibinfo {author} {\bibfnamefont
  {N.}~\bibnamefont {{Kokron}}}, \bibinfo {author} {\bibfnamefont
  {F.}~\bibnamefont {{Lacasa}}}, \bibinfo {author} {\bibfnamefont
  {V.}~\bibnamefont {{Miranda}}}, \bibinfo {author} {\bibfnamefont
  {Y.}~\bibnamefont {{Omori}}}, \bibinfo {author} {\bibfnamefont
  {A.}~\bibnamefont {{Porredon}}}, \bibinfo {author} {\bibfnamefont
  {R.}~\bibnamefont {{Rosenfeld}}}, \bibinfo {author} {\bibfnamefont
  {S.}~\bibnamefont {{Samuroff}}}, \bibinfo {author} {\bibfnamefont
  {M.}~\bibnamefont {{Wang}}}, \bibinfo {author} {\bibfnamefont {R.~H.}\
  \bibnamefont {{Wechsler}}}, \bibinfo {author} {\bibfnamefont {T.~M.~C.}\
  \bibnamefont {{Abbott}}}, \bibinfo {author} {\bibfnamefont {F.~B.}\
  \bibnamefont {{Abdalla}}}, \bibinfo {author} {\bibfnamefont {S.}~\bibnamefont
  {{Allam}}}, \bibinfo {author} {\bibfnamefont {J.}~\bibnamefont {{Annis}}},
  \bibinfo {author} {\bibfnamefont {K.}~\bibnamefont {{Bechtol}}}, \bibinfo
  {author} {\bibfnamefont {A.}~\bibnamefont {{Benoit-Levy}}}, \bibinfo {author}
  {\bibfnamefont {G.~M.}\ \bibnamefont {{Bernstein}}}, \bibinfo {author}
  {\bibfnamefont {D.}~\bibnamefont {{Brooks}}}, \bibinfo {author}
  {\bibfnamefont {D.~L.}\ \bibnamefont {{Burke}}}, \bibinfo {author}
  {\bibfnamefont {D.}~\bibnamefont {{Capozzi}}}, \bibinfo {author}
  {\bibfnamefont {M.}~\bibnamefont {{Carrasco Kind}}}, \bibinfo {author}
  {\bibfnamefont {J.}~\bibnamefont {{Carretero}}}, \bibinfo {author}
  {\bibfnamefont {C.~B.}\ \bibnamefont {{D'Andrea}}}, \bibinfo {author}
  {\bibfnamefont {L.~N.}\ \bibnamefont {{da Costa}}}, \bibinfo {author}
  {\bibfnamefont {C.}~\bibnamefont {{Davis}}}, \bibinfo {author} {\bibfnamefont
  {D.~L.}\ \bibnamefont {{DePoy}}}, \bibinfo {author} {\bibfnamefont
  {S.}~\bibnamefont {{Desai}}}, \bibinfo {author} {\bibfnamefont {H.~T.}\
  \bibnamefont {{Diehl}}}, \bibinfo {author} {\bibfnamefont {J.~P.}\
  \bibnamefont {{Dietrich}}}, \bibinfo {author} {\bibfnamefont {A.~E.}\
  \bibnamefont {{Evrard}}}, \bibinfo {author} {\bibfnamefont {B.}~\bibnamefont
  {{Flaugher}}}, \bibinfo {author} {\bibfnamefont {P.}~\bibnamefont
  {{Fosalba}}}, \bibinfo {author} {\bibfnamefont {J.}~\bibnamefont
  {{Frieman}}}, \bibinfo {author} {\bibfnamefont {J.}~\bibnamefont
  {{Garcia-Bellido}}}, \bibinfo {author} {\bibfnamefont {E.}~\bibnamefont
  {{Gaztanaga}}}, \bibinfo {author} {\bibfnamefont {T.}~\bibnamefont
  {{Giannantonio}}}, \bibinfo {author} {\bibfnamefont {D.}~\bibnamefont
  {{Gruen}}}, \bibinfo {author} {\bibfnamefont {R.~A.}\ \bibnamefont
  {{Gruendl}}}, \bibinfo {author} {\bibfnamefont {J.}~\bibnamefont
  {{Gschwend}}}, \bibinfo {author} {\bibfnamefont {G.}~\bibnamefont
  {{Gutierrez}}}, \bibinfo {author} {\bibfnamefont {K.}~\bibnamefont
  {{Honscheid}}}, \bibinfo {author} {\bibfnamefont {D.~J.}\ \bibnamefont
  {{James}}}, \bibinfo {author} {\bibfnamefont {T.}~\bibnamefont {{Jeltema}}},
  \bibinfo {author} {\bibfnamefont {K.}~\bibnamefont {{Kuehn}}}, \bibinfo
  {author} {\bibfnamefont {S.}~\bibnamefont {{Kuhlmann}}}, \bibinfo {author}
  {\bibfnamefont {O.}~\bibnamefont {{Lahav}}}, \bibinfo {author} {\bibfnamefont
  {M.}~\bibnamefont {{Lima}}}, \bibinfo {author} {\bibfnamefont {M.~A.~G.}\
  \bibnamefont {{Maia}}}, \bibinfo {author} {\bibfnamefont {M.}~\bibnamefont
  {{March}}}, \bibinfo {author} {\bibfnamefont {J.~L.}\ \bibnamefont
  {{Marshall}}}, \bibinfo {author} {\bibfnamefont {P.}~\bibnamefont
  {{Martini}}}, \bibinfo {author} {\bibfnamefont {F.}~\bibnamefont
  {{Menanteau}}}, \bibinfo {author} {\bibfnamefont {R.}~\bibnamefont
  {{Miquel}}}, \bibinfo {author} {\bibfnamefont {R.~C.}\ \bibnamefont
  {{Nichol}}}, \bibinfo {author} {\bibfnamefont {A.~A.}\ \bibnamefont
  {{Plazas}}}, \bibinfo {author} {\bibfnamefont {A.~K.}\ \bibnamefont
  {{Romer}}}, \bibinfo {author} {\bibfnamefont {E.~S.}\ \bibnamefont
  {{Rykoff}}}, \bibinfo {author} {\bibfnamefont {E.}~\bibnamefont {{Sanchez}}},
  \bibinfo {author} {\bibfnamefont {V.}~\bibnamefont {{Scarpine}}}, \bibinfo
  {author} {\bibfnamefont {R.}~\bibnamefont {{Schindler}}}, \bibinfo {author}
  {\bibfnamefont {M.}~\bibnamefont {{Schubnell}}}, \bibinfo {author}
  {\bibfnamefont {I.}~\bibnamefont {{Sevilla-Noarbe}}}, \bibinfo {author}
  {\bibfnamefont {M.}~\bibnamefont {{Smith}}}, \bibinfo {author} {\bibfnamefont
  {M.}~\bibnamefont {{Soares-Santos}}}, \bibinfo {author} {\bibfnamefont
  {F.}~\bibnamefont {{Sobreira}}}, \bibinfo {author} {\bibfnamefont
  {E.}~\bibnamefont {{Suchyta}}}, \bibinfo {author} {\bibfnamefont {M.~E.~C.}\
  \bibnamefont {{Swanson}}}, \bibinfo {author} {\bibfnamefont {G.}~\bibnamefont
  {{Tarle}}}, \bibinfo {author} {\bibfnamefont {D.~L.}\ \bibnamefont
  {{Tucker}}}, \bibinfo {author} {\bibfnamefont {V.}~\bibnamefont {{Vikram}}},
  \bibinfo {author} {\bibfnamefont {A.~R.}\ \bibnamefont {{Walker}}}, \ and\
  \bibinfo {author} {\bibfnamefont {J.}~\bibnamefont {{Weller}}},\ }\href@noop
  {} {\bibfield  {journal} {\bibinfo  {journal} {ArXiv e-prints}\ ,\ \bibinfo
  {eid} {arXiv:1706.09359}} (\bibinfo {year} {2017})},\ \Eprint
  {http://arxiv.org/abs/1706.09359} {arXiv:1706.09359 [astro-ph.CO]}
  \BibitemShut {NoStop}%
\bibitem [{\citenamefont {{Planck Collaboration}}\ \emph
  {et~al.}(2016{\natexlab{b}})\citenamefont {{Planck Collaboration}},
  \citenamefont {{Ade}}, \citenamefont {{Aghanim}}, \citenamefont {{Arnaud}},
  \citenamefont {{Ashdown}}, \citenamefont {{Aumont}}, \citenamefont
  {{Baccigalupi}}, \citenamefont {{Banday}}, \citenamefont {{Barreiro}},
  \citenamefont {{Bartlett}},\ and\ \citenamefont {et~al.}}]{ade16a}%
  \BibitemOpen
  \bibfield  {author} {\bibinfo {author} {\bibnamefont {{Planck
  Collaboration}}}, \bibinfo {author} {\bibfnamefont {P.~A.~R.}\ \bibnamefont
  {{Ade}}}, \bibinfo {author} {\bibfnamefont {N.}~\bibnamefont {{Aghanim}}},
  \bibinfo {author} {\bibfnamefont {M.}~\bibnamefont {{Arnaud}}}, \bibinfo
  {author} {\bibfnamefont {M.}~\bibnamefont {{Ashdown}}}, \bibinfo {author}
  {\bibfnamefont {J.}~\bibnamefont {{Aumont}}}, \bibinfo {author}
  {\bibfnamefont {C.}~\bibnamefont {{Baccigalupi}}}, \bibinfo {author}
  {\bibfnamefont {A.~J.}\ \bibnamefont {{Banday}}}, \bibinfo {author}
  {\bibfnamefont {R.~B.}\ \bibnamefont {{Barreiro}}}, \bibinfo {author}
  {\bibfnamefont {J.~G.}\ \bibnamefont {{Bartlett}}}, \ and\ \bibinfo {author}
  {\bibnamefont {et~al.}},\ }\href {\doibase 10.1051/0004-6361/201525830}
  {\bibfield  {journal} {\bibinfo  {journal} {\aap}\ }\textbf {\bibinfo
  {volume} {594}},\ \bibinfo {eid} {A13} (\bibinfo {year}
  {2016}{\natexlab{b}})},\ \Eprint {http://arxiv.org/abs/1502.01589}
  {arXiv:1502.01589} \BibitemShut {NoStop}%
\bibitem [{\citenamefont {{Wong}}\ \emph {et~al.}(2020)\citenamefont {{Wong}},
  \citenamefont {{Suyu}}, \citenamefont {{Chen}}, \citenamefont {{Rusu}},
  \citenamefont {{Millon}}, \citenamefont {{Sluse}}, \citenamefont {{Bonvin}},
  \citenamefont {{Fassnacht}}, \citenamefont {{Taubenberger}}, \citenamefont
  {{Auger}}, \citenamefont {{Birrer}}, \citenamefont {{Chan}}, \citenamefont
  {{Courbin}}, \citenamefont {{Hilbert}}, \citenamefont {{Tihhonova}},
  \citenamefont {{Treu}}, \citenamefont {{Agnello}}, \citenamefont {{Ding}},
  \citenamefont {{Jee}}, \citenamefont {{Komatsu}}, \citenamefont {{Shajib}},
  \citenamefont {{Sonnenfeld}}, \citenamefont {{Bland ford}}, \citenamefont
  {{Koopmans}}, \citenamefont {{Marshall}},\ and\ \citenamefont
  {{Meylan}}}]{1907.04869}%
  \BibitemOpen
  \bibfield  {author} {\bibinfo {author} {\bibfnamefont {K.~C.}\ \bibnamefont
  {{Wong}}}, \bibinfo {author} {\bibfnamefont {S.~H.}\ \bibnamefont {{Suyu}}},
  \bibinfo {author} {\bibfnamefont {G.~C.~F.}\ \bibnamefont {{Chen}}}, \bibinfo
  {author} {\bibfnamefont {C.~E.}\ \bibnamefont {{Rusu}}}, \bibinfo {author}
  {\bibfnamefont {M.}~\bibnamefont {{Millon}}}, \bibinfo {author}
  {\bibfnamefont {D.}~\bibnamefont {{Sluse}}}, \bibinfo {author} {\bibfnamefont
  {V.}~\bibnamefont {{Bonvin}}}, \bibinfo {author} {\bibfnamefont {C.~D.}\
  \bibnamefont {{Fassnacht}}}, \bibinfo {author} {\bibfnamefont
  {S.}~\bibnamefont {{Taubenberger}}}, \bibinfo {author} {\bibfnamefont
  {M.~W.}\ \bibnamefont {{Auger}}}, \bibinfo {author} {\bibfnamefont
  {S.}~\bibnamefont {{Birrer}}}, \bibinfo {author} {\bibfnamefont {J.~H.~H.}\
  \bibnamefont {{Chan}}}, \bibinfo {author} {\bibfnamefont {F.}~\bibnamefont
  {{Courbin}}}, \bibinfo {author} {\bibfnamefont {S.}~\bibnamefont
  {{Hilbert}}}, \bibinfo {author} {\bibfnamefont {O.}~\bibnamefont
  {{Tihhonova}}}, \bibinfo {author} {\bibfnamefont {T.}~\bibnamefont {{Treu}}},
  \bibinfo {author} {\bibfnamefont {A.}~\bibnamefont {{Agnello}}}, \bibinfo
  {author} {\bibfnamefont {X.}~\bibnamefont {{Ding}}}, \bibinfo {author}
  {\bibfnamefont {I.}~\bibnamefont {{Jee}}}, \bibinfo {author} {\bibfnamefont
  {E.}~\bibnamefont {{Komatsu}}}, \bibinfo {author} {\bibfnamefont {A.~J.}\
  \bibnamefont {{Shajib}}}, \bibinfo {author} {\bibfnamefont {A.}~\bibnamefont
  {{Sonnenfeld}}}, \bibinfo {author} {\bibfnamefont {R.~D.}\ \bibnamefont
  {{Bland ford}}}, \bibinfo {author} {\bibfnamefont {L.~V.~E.}\ \bibnamefont
  {{Koopmans}}}, \bibinfo {author} {\bibfnamefont {P.~J.}\ \bibnamefont
  {{Marshall}}}, \ and\ \bibinfo {author} {\bibfnamefont {G.}~\bibnamefont
  {{Meylan}}},\ }\href {\doibase 10.1093/mnras/stz3094} {\bibfield  {journal}
  {\bibinfo  {journal} {\mnras}\ } (\bibinfo {year} {2020}),\
  10.1093/mnras/stz3094},\ \Eprint {http://arxiv.org/abs/1907.04869}
  {arXiv:1907.04869 [astro-ph.CO]} \BibitemShut {NoStop}%
\bibitem [{\citenamefont {{Abbott}}\ \emph
  {et~al.}(2017{\natexlab{a}})\citenamefont {{Abbott}}, \citenamefont
  {{Abbott}}, \citenamefont {{Abbott}}, \citenamefont {{Acernese}},
  \citenamefont {{Ackley}}, \citenamefont {{Adams}}, \citenamefont {{Adams}},
  \citenamefont {{Addesso}}, \citenamefont {{Adhikari}}, \citenamefont
  {{Adya}},\ and\ \citenamefont {et~al.}}]{1710.05835}%
  \BibitemOpen
  \bibfield  {author} {\bibinfo {author} {\bibfnamefont {B.~P.}\ \bibnamefont
  {{Abbott}}}, \bibinfo {author} {\bibfnamefont {R.}~\bibnamefont {{Abbott}}},
  \bibinfo {author} {\bibfnamefont {T.~D.}\ \bibnamefont {{Abbott}}}, \bibinfo
  {author} {\bibfnamefont {F.}~\bibnamefont {{Acernese}}}, \bibinfo {author}
  {\bibfnamefont {K.}~\bibnamefont {{Ackley}}}, \bibinfo {author}
  {\bibfnamefont {C.}~\bibnamefont {{Adams}}}, \bibinfo {author} {\bibfnamefont
  {T.}~\bibnamefont {{Adams}}}, \bibinfo {author} {\bibfnamefont
  {P.}~\bibnamefont {{Addesso}}}, \bibinfo {author} {\bibfnamefont {R.~X.}\
  \bibnamefont {{Adhikari}}}, \bibinfo {author} {\bibfnamefont {V.~B.}\
  \bibnamefont {{Adya}}}, \ and\ \bibinfo {author} {\bibnamefont {et~al.}},\
  }\href {\doibase 10.1038/nature24471} {\bibfield  {journal} {\bibinfo
  {journal} {\nat}\ }\textbf {\bibinfo {volume} {551}},\ \bibinfo {pages} {85}
  (\bibinfo {year} {2017}{\natexlab{a}})},\ \Eprint
  {http://arxiv.org/abs/1710.05835} {arXiv:1710.05835 [astro-ph.CO]}
  \BibitemShut {NoStop}%
\bibitem [{\citenamefont {{Freedman}}\ \emph {et~al.}(2020)\citenamefont
  {{Freedman}}, \citenamefont {{Madore}}, \citenamefont {{Hoyt}}, \citenamefont
  {{Jang}}, \citenamefont {{Beaton}}, \citenamefont {{Lee}}, \citenamefont
  {{Monson}}, \citenamefont {{Neeley}},\ and\ \citenamefont
  {{Rich}}}]{freedman20a}%
  \BibitemOpen
  \bibfield  {author} {\bibinfo {author} {\bibfnamefont {W.~L.}\ \bibnamefont
  {{Freedman}}}, \bibinfo {author} {\bibfnamefont {B.~F.}\ \bibnamefont
  {{Madore}}}, \bibinfo {author} {\bibfnamefont {T.}~\bibnamefont {{Hoyt}}},
  \bibinfo {author} {\bibfnamefont {I.~S.}\ \bibnamefont {{Jang}}}, \bibinfo
  {author} {\bibfnamefont {R.}~\bibnamefont {{Beaton}}}, \bibinfo {author}
  {\bibfnamefont {M.~G.}\ \bibnamefont {{Lee}}}, \bibinfo {author}
  {\bibfnamefont {A.}~\bibnamefont {{Monson}}}, \bibinfo {author}
  {\bibfnamefont {J.}~\bibnamefont {{Neeley}}}, \ and\ \bibinfo {author}
  {\bibfnamefont {J.}~\bibnamefont {{Rich}}},\ }\href {\doibase
  10.3847/1538-4357/ab7339} {\bibfield  {journal} {\bibinfo  {journal} {\apj}\
  }\textbf {\bibinfo {volume} {891}},\ \bibinfo {eid} {57} (\bibinfo {year}
  {2020})},\ \Eprint {http://arxiv.org/abs/2002.01550} {arXiv:2002.01550
  [astro-ph.GA]} \BibitemShut {NoStop}%
\bibitem [{\citenamefont {{Tully}}\ \emph {et~al.}(2016)\citenamefont
  {{Tully}}, \citenamefont {{Courtois}},\ and\ \citenamefont
  {{Sorce}}}]{tully16a}%
  \BibitemOpen
  \bibfield  {author} {\bibinfo {author} {\bibfnamefont {R.~B.}\ \bibnamefont
  {{Tully}}}, \bibinfo {author} {\bibfnamefont {H.~M.}\ \bibnamefont
  {{Courtois}}}, \ and\ \bibinfo {author} {\bibfnamefont {J.~G.}\ \bibnamefont
  {{Sorce}}},\ }\href {\doibase 10.3847/0004-6256/152/2/50} {\bibfield
  {journal} {\bibinfo  {journal} {\aj}\ }\textbf {\bibinfo {volume} {152}},\
  \bibinfo {eid} {50} (\bibinfo {year} {2016})},\ \Eprint
  {http://arxiv.org/abs/1605.01765} {arXiv:1605.01765 [astro-ph.CO]}
  \BibitemShut {NoStop}%
\bibitem [{\citenamefont {{Pesce}}\ \emph {et~al.}(2020)\citenamefont
  {{Pesce}}, \citenamefont {{Braatz}}, \citenamefont {{Reid}}, \citenamefont
  {{Riess}}, \citenamefont {{Scolnic}}, \citenamefont {{Condon}}, \citenamefont
  {{Gao}}, \citenamefont {{Henkel}}, \citenamefont {{Impellizzeri}},
  \citenamefont {{Kuo}},\ and\ \citenamefont {{Lo}}}]{pesce20a}%
  \BibitemOpen
  \bibfield  {author} {\bibinfo {author} {\bibfnamefont {D.~W.}\ \bibnamefont
  {{Pesce}}}, \bibinfo {author} {\bibfnamefont {J.~A.}\ \bibnamefont
  {{Braatz}}}, \bibinfo {author} {\bibfnamefont {M.~J.}\ \bibnamefont
  {{Reid}}}, \bibinfo {author} {\bibfnamefont {A.~G.}\ \bibnamefont {{Riess}}},
  \bibinfo {author} {\bibfnamefont {D.}~\bibnamefont {{Scolnic}}}, \bibinfo
  {author} {\bibfnamefont {J.~J.}\ \bibnamefont {{Condon}}}, \bibinfo {author}
  {\bibfnamefont {F.}~\bibnamefont {{Gao}}}, \bibinfo {author} {\bibfnamefont
  {C.}~\bibnamefont {{Henkel}}}, \bibinfo {author} {\bibfnamefont {C.~M.~V.}\
  \bibnamefont {{Impellizzeri}}}, \bibinfo {author} {\bibfnamefont {C.~Y.}\
  \bibnamefont {{Kuo}}}, \ and\ \bibinfo {author} {\bibfnamefont {K.~Y.}\
  \bibnamefont {{Lo}}},\ }\href {\doibase 10.3847/2041-8213/ab75f0} {\bibfield
  {journal} {\bibinfo  {journal} {\apjl}\ }\textbf {\bibinfo {volume} {891}},\
  \bibinfo {eid} {L1} (\bibinfo {year} {2020})},\ \Eprint
  {http://arxiv.org/abs/2001.09213} {arXiv:2001.09213 [astro-ph.CO]}
  \BibitemShut {NoStop}%
\bibitem [{\citenamefont {{Hu}}(2005)}]{hu2005}%
  \BibitemOpen
  \bibfield  {author} {\bibinfo {author} {\bibfnamefont {W.}~\bibnamefont
  {{Hu}}},\ }in\ \href@noop {} {\emph {\bibinfo {booktitle} {Observing Dark
  Energy}}},\ \bibinfo {series} {Astronomical Society of the Pacific Conference
  Series}, Vol.\ \bibinfo {volume} {339},\ \bibinfo {editor} {edited by\
  \bibinfo {editor} {\bibfnamefont {S.~C.}\ \bibnamefont {{Wolff}}}\ and\
  \bibinfo {editor} {\bibfnamefont {T.~R.}\ \bibnamefont {{Lauer}}}}\ (\bibinfo
  {year} {2005})\ p.\ \bibinfo {pages} {215},\ \Eprint
  {http://arxiv.org/abs/astro-ph/0407158} {arXiv:astro-ph/0407158 [astro-ph]}
  \BibitemShut {NoStop}%
\bibitem [{\citenamefont {{Riess}}\ \emph {et~al.}(2016)\citenamefont
  {{Riess}}, \citenamefont {{Macri}}, \citenamefont {{Hoffmann}}, \citenamefont
  {{Scolnic}}, \citenamefont {{Casertano}}, \citenamefont {{Filippenko}},
  \citenamefont {{Tucker}}, \citenamefont {{Reid}}, \citenamefont {{Jones}},
  \citenamefont {{Silverman}}, \citenamefont {{Chornock}}, \citenamefont
  {{Challis}}, \citenamefont {{Yuan}}, \citenamefont {{Brown}},\ and\
  \citenamefont {{Foley}}}]{riess16}%
  \BibitemOpen
  \bibfield  {author} {\bibinfo {author} {\bibfnamefont {A.~G.}\ \bibnamefont
  {{Riess}}}, \bibinfo {author} {\bibfnamefont {L.~M.}\ \bibnamefont
  {{Macri}}}, \bibinfo {author} {\bibfnamefont {S.~L.}\ \bibnamefont
  {{Hoffmann}}}, \bibinfo {author} {\bibfnamefont {D.}~\bibnamefont
  {{Scolnic}}}, \bibinfo {author} {\bibfnamefont {S.}~\bibnamefont
  {{Casertano}}}, \bibinfo {author} {\bibfnamefont {A.~V.}\ \bibnamefont
  {{Filippenko}}}, \bibinfo {author} {\bibfnamefont {B.~E.}\ \bibnamefont
  {{Tucker}}}, \bibinfo {author} {\bibfnamefont {M.~J.}\ \bibnamefont
  {{Reid}}}, \bibinfo {author} {\bibfnamefont {D.~O.}\ \bibnamefont {{Jones}}},
  \bibinfo {author} {\bibfnamefont {J.~M.}\ \bibnamefont {{Silverman}}},
  \bibinfo {author} {\bibfnamefont {R.}~\bibnamefont {{Chornock}}}, \bibinfo
  {author} {\bibfnamefont {P.}~\bibnamefont {{Challis}}}, \bibinfo {author}
  {\bibfnamefont {W.}~\bibnamefont {{Yuan}}}, \bibinfo {author} {\bibfnamefont
  {P.~J.}\ \bibnamefont {{Brown}}}, \ and\ \bibinfo {author} {\bibfnamefont
  {R.~J.}\ \bibnamefont {{Foley}}},\ }\href {\doibase
  10.3847/0004-637X/826/1/56} {\bibfield  {journal} {\bibinfo  {journal}
  {\apj}\ }\textbf {\bibinfo {volume} {826}},\ \bibinfo {eid} {56} (\bibinfo
  {year} {2016})},\ \Eprint {http://arxiv.org/abs/1604.01424} {arXiv:1604.01424
  [astro-ph.CO]} \BibitemShut {NoStop}%
\bibitem [{\citenamefont {{Neill}}\ \emph {et~al.}(2014)\citenamefont
  {{Neill}}, \citenamefont {{Seibert}}, \citenamefont {{Tully}}, \citenamefont
  {{Courtois}}, \citenamefont {{Sorce}}, \citenamefont {{Jarrett}},
  \citenamefont {{Scowcroft}},\ and\ \citenamefont {{Masci}}}]{neill14a}%
  \BibitemOpen
  \bibfield  {author} {\bibinfo {author} {\bibfnamefont {J.~D.}\ \bibnamefont
  {{Neill}}}, \bibinfo {author} {\bibfnamefont {M.}~\bibnamefont {{Seibert}}},
  \bibinfo {author} {\bibfnamefont {R.~B.}\ \bibnamefont {{Tully}}}, \bibinfo
  {author} {\bibfnamefont {H.}~\bibnamefont {{Courtois}}}, \bibinfo {author}
  {\bibfnamefont {J.~G.}\ \bibnamefont {{Sorce}}}, \bibinfo {author}
  {\bibfnamefont {T.~H.}\ \bibnamefont {{Jarrett}}}, \bibinfo {author}
  {\bibfnamefont {V.}~\bibnamefont {{Scowcroft}}}, \ and\ \bibinfo {author}
  {\bibfnamefont {F.~J.}\ \bibnamefont {{Masci}}},\ }\href {\doibase
  10.1088/0004-637X/792/2/129} {\bibfield  {journal} {\bibinfo  {journal}
  {\apj}\ }\textbf {\bibinfo {volume} {792}},\ \bibinfo {eid} {129} (\bibinfo
  {year} {2014})},\ \Eprint {http://arxiv.org/abs/1407.7528} {arXiv:1407.7528
  [astro-ph.CO]} \BibitemShut {NoStop}%
\bibitem [{\citenamefont {{Birrer}}\ \emph {et~al.}(2019)\citenamefont
  {{Birrer}}, \citenamefont {{Treu}}, \citenamefont {{Rusu}}, \citenamefont
  {{Bonvin}}, \citenamefont {{Fassnacht}}, \citenamefont {{Chan}},
  \citenamefont {{Agnello}}, \citenamefont {{Shajib}}, \citenamefont {{Chen}},
  \citenamefont {{Auger}}, \citenamefont {{Courbin}}, \citenamefont
  {{Hilbert}}, \citenamefont {{Sluse}}, \citenamefont {{Suyu}}, \citenamefont
  {{Wong}}, \citenamefont {{Marshall}}, \citenamefont {{Lemaux}},\ and\
  \citenamefont {{Meylan}}}]{birrer19a}%
  \BibitemOpen
  \bibfield  {author} {\bibinfo {author} {\bibfnamefont {S.}~\bibnamefont
  {{Birrer}}}, \bibinfo {author} {\bibfnamefont {T.}~\bibnamefont {{Treu}}},
  \bibinfo {author} {\bibfnamefont {C.~E.}\ \bibnamefont {{Rusu}}}, \bibinfo
  {author} {\bibfnamefont {V.}~\bibnamefont {{Bonvin}}}, \bibinfo {author}
  {\bibfnamefont {C.~D.}\ \bibnamefont {{Fassnacht}}}, \bibinfo {author}
  {\bibfnamefont {J.~H.~H.}\ \bibnamefont {{Chan}}}, \bibinfo {author}
  {\bibfnamefont {A.}~\bibnamefont {{Agnello}}}, \bibinfo {author}
  {\bibfnamefont {A.~J.}\ \bibnamefont {{Shajib}}}, \bibinfo {author}
  {\bibfnamefont {G.~C.-F.}\ \bibnamefont {{Chen}}}, \bibinfo {author}
  {\bibfnamefont {M.}~\bibnamefont {{Auger}}}, \bibinfo {author} {\bibfnamefont
  {F.}~\bibnamefont {{Courbin}}}, \bibinfo {author} {\bibfnamefont
  {S.}~\bibnamefont {{Hilbert}}}, \bibinfo {author} {\bibfnamefont
  {D.}~\bibnamefont {{Sluse}}}, \bibinfo {author} {\bibfnamefont {S.~H.}\
  \bibnamefont {{Suyu}}}, \bibinfo {author} {\bibfnamefont {K.~C.}\
  \bibnamefont {{Wong}}}, \bibinfo {author} {\bibfnamefont {P.}~\bibnamefont
  {{Marshall}}}, \bibinfo {author} {\bibfnamefont {B.~C.}\ \bibnamefont
  {{Lemaux}}}, \ and\ \bibinfo {author} {\bibfnamefont {G.}~\bibnamefont
  {{Meylan}}},\ }\href {\doibase 10.1093/mnras/stz200} {\bibfield  {journal}
  {\bibinfo  {journal} {\mnras}\ }\textbf {\bibinfo {volume} {484}},\ \bibinfo
  {pages} {4726} (\bibinfo {year} {2019})},\ \Eprint
  {http://arxiv.org/abs/1809.01274} {arXiv:1809.01274} \BibitemShut {NoStop}%
\bibitem [{\citenamefont {{Davis}}\ \emph {et~al.}(2019)\citenamefont
  {{Davis}}, \citenamefont {{Hinton}}, \citenamefont {{Howlett}},\ and\
  \citenamefont {{Calcino}}}]{1907.12639}%
  \BibitemOpen
  \bibfield  {author} {\bibinfo {author} {\bibfnamefont {T.~M.}\ \bibnamefont
  {{Davis}}}, \bibinfo {author} {\bibfnamefont {S.~R.}\ \bibnamefont
  {{Hinton}}}, \bibinfo {author} {\bibfnamefont {C.}~\bibnamefont {{Howlett}}},
  \ and\ \bibinfo {author} {\bibfnamefont {J.}~\bibnamefont {{Calcino}}},\
  }\href {\doibase 10.1093/mnras/stz2652} {\bibfield  {journal} {\bibinfo
  {journal} {\mnras}\ }\textbf {\bibinfo {volume} {490}},\ \bibinfo {pages}
  {2948} (\bibinfo {year} {2019})},\ \Eprint {http://arxiv.org/abs/1907.12639}
  {arXiv:1907.12639 [astro-ph.CO]} \BibitemShut {NoStop}%
\bibitem [{\citenamefont {{Dhawan}}\ \emph {et~al.}(2020)\citenamefont
  {{Dhawan}}, \citenamefont {{Brout}}, \citenamefont {{Scolnic}}, \citenamefont
  {{Goobar}}, \citenamefont {{Riess}},\ and\ \citenamefont
  {{Miranda}}}]{2001.09260}%
  \BibitemOpen
  \bibfield  {author} {\bibinfo {author} {\bibfnamefont {S.}~\bibnamefont
  {{Dhawan}}}, \bibinfo {author} {\bibfnamefont {D.}~\bibnamefont {{Brout}}},
  \bibinfo {author} {\bibfnamefont {D.}~\bibnamefont {{Scolnic}}}, \bibinfo
  {author} {\bibfnamefont {A.}~\bibnamefont {{Goobar}}}, \bibinfo {author}
  {\bibfnamefont {A.~G.}\ \bibnamefont {{Riess}}}, \ and\ \bibinfo {author}
  {\bibfnamefont {V.}~\bibnamefont {{Miranda}}},\ }\href {\doibase
  10.3847/1538-4357/ab7fb0} {\bibfield  {journal} {\bibinfo  {journal} {\apj}\
  }\textbf {\bibinfo {volume} {894}},\ \bibinfo {eid} {54} (\bibinfo {year}
  {2020})},\ \Eprint {http://arxiv.org/abs/2001.09260} {arXiv:2001.09260
  [astro-ph.CO]} \BibitemShut {NoStop}%
\bibitem [{\citenamefont {{Anderson}}(2019)}]{1909.10847}%
  \BibitemOpen
  \bibfield  {author} {\bibinfo {author} {\bibfnamefont {R.~I.}\ \bibnamefont
  {{Anderson}}},\ }\href {\doibase 10.1051/0004-6361/201936585} {\bibfield
  {journal} {\bibinfo  {journal} {\aap}\ }\textbf {\bibinfo {volume} {631}},\
  \bibinfo {eid} {A165} (\bibinfo {year} {2019})},\ \Eprint
  {http://arxiv.org/abs/1909.10847} {arXiv:1909.10847 [astro-ph.CO]}
  \BibitemShut {NoStop}%
\bibitem [{\citenamefont {{Hazra}}\ \emph {et~al.}(2019)\citenamefont
  {{Hazra}}, \citenamefont {{Shafieloo}},\ and\ \citenamefont
  {{Souradeep}}}]{hazra19a}%
  \BibitemOpen
  \bibfield  {author} {\bibinfo {author} {\bibfnamefont {D.~K.}\ \bibnamefont
  {{Hazra}}}, \bibinfo {author} {\bibfnamefont {A.}~\bibnamefont
  {{Shafieloo}}}, \ and\ \bibinfo {author} {\bibfnamefont {T.}~\bibnamefont
  {{Souradeep}}},\ }\href {\doibase 10.1088/1475-7516/2019/04/036} {\bibfield
  {journal} {\bibinfo  {journal} {\jcap}\ }\textbf {\bibinfo {volume} {2019}},\
  \bibinfo {eid} {036} (\bibinfo {year} {2019})},\ \Eprint
  {http://arxiv.org/abs/1810.08101} {arXiv:1810.08101 [astro-ph.CO]}
  \BibitemShut {NoStop}%
\bibitem [{\citenamefont {{Li}}\ and\ \citenamefont
  {{Shafieloo}}(2019)}]{li19a}%
  \BibitemOpen
  \bibfield  {author} {\bibinfo {author} {\bibfnamefont {X.}~\bibnamefont
  {{Li}}}\ and\ \bibinfo {author} {\bibfnamefont {A.}~\bibnamefont
  {{Shafieloo}}},\ }\href {\doibase 10.3847/2041-8213/ab3e09} {\bibfield
  {journal} {\bibinfo  {journal} {\apjl}\ }\textbf {\bibinfo {volume} {883}},\
  \bibinfo {eid} {L3} (\bibinfo {year} {2019})},\ \Eprint
  {http://arxiv.org/abs/1906.08275} {arXiv:1906.08275 [astro-ph.CO]}
  \BibitemShut {NoStop}%
\bibitem [{\citenamefont {{Alestas}}\ \emph {et~al.}(2020)\citenamefont
  {{Alestas}}, \citenamefont {{Kazantzidis}},\ and\ \citenamefont
  {{Perivolaropoulos}}}]{alestas20a}%
  \BibitemOpen
  \bibfield  {author} {\bibinfo {author} {\bibfnamefont {G.}~\bibnamefont
  {{Alestas}}}, \bibinfo {author} {\bibfnamefont {L.}~\bibnamefont
  {{Kazantzidis}}}, \ and\ \bibinfo {author} {\bibfnamefont {L.}~\bibnamefont
  {{Perivolaropoulos}}},\ }\href {\doibase 10.1103/PhysRevD.101.123516}
  {\bibfield  {journal} {\bibinfo  {journal} {\prd}\ }\textbf {\bibinfo
  {volume} {101}},\ \bibinfo {eid} {123516} (\bibinfo {year} {2020})},\ \Eprint
  {http://arxiv.org/abs/2004.08363} {arXiv:2004.08363 [astro-ph.CO]}
  \BibitemShut {NoStop}%
\bibitem [{\citenamefont {{Di Valentino}}\ \emph {et~al.}(2020)\citenamefont
  {{Di Valentino}}, \citenamefont {{Melchiorri}}, \citenamefont {{Mena}},\ and\
  \citenamefont {{Vagnozzi}}}]{divalentino20a}%
  \BibitemOpen
  \bibfield  {author} {\bibinfo {author} {\bibfnamefont {E.}~\bibnamefont {{Di
  Valentino}}}, \bibinfo {author} {\bibfnamefont {A.}~\bibnamefont
  {{Melchiorri}}}, \bibinfo {author} {\bibfnamefont {O.}~\bibnamefont
  {{Mena}}}, \ and\ \bibinfo {author} {\bibfnamefont {S.}~\bibnamefont
  {{Vagnozzi}}},\ }\href {\doibase 10.1103/PhysRevD.101.063502} {\bibfield
  {journal} {\bibinfo  {journal} {\prd}\ }\textbf {\bibinfo {volume} {101}},\
  \bibinfo {eid} {063502} (\bibinfo {year} {2020})},\ \Eprint
  {http://arxiv.org/abs/1910.09853} {arXiv:1910.09853 [astro-ph.CO]}
  \BibitemShut {NoStop}%
\bibitem [{\citenamefont {{Poulin}}\ \emph {et~al.}(2019)\citenamefont
  {{Poulin}}, \citenamefont {{Smith}}, \citenamefont {{Karwal}},\ and\
  \citenamefont {{Kamionkowski}}}]{1811.04083}%
  \BibitemOpen
  \bibfield  {author} {\bibinfo {author} {\bibfnamefont {V.}~\bibnamefont
  {{Poulin}}}, \bibinfo {author} {\bibfnamefont {T.~L.}\ \bibnamefont
  {{Smith}}}, \bibinfo {author} {\bibfnamefont {T.}~\bibnamefont {{Karwal}}}, \
  and\ \bibinfo {author} {\bibfnamefont {M.}~\bibnamefont {{Kamionkowski}}},\
  }\href {\doibase 10.1103/PhysRevLett.122.221301} {\bibfield  {journal}
  {\bibinfo  {journal} {\prl}\ }\textbf {\bibinfo {volume} {122}},\ \bibinfo
  {eid} {221301} (\bibinfo {year} {2019})},\ \Eprint
  {http://arxiv.org/abs/1811.04083} {arXiv:1811.04083 [astro-ph.CO]}
  \BibitemShut {NoStop}%
\bibitem [{\citenamefont {{Chiang}}\ and\ \citenamefont
  {{Slosar}}(2018)}]{1811.03624}%
  \BibitemOpen
  \bibfield  {author} {\bibinfo {author} {\bibfnamefont {C.-T.}\ \bibnamefont
  {{Chiang}}}\ and\ \bibinfo {author} {\bibfnamefont {A.}~\bibnamefont
  {{Slosar}}},\ }\href@noop {} {\bibfield  {journal} {\bibinfo  {journal}
  {ArXiv e-prints}\ ,\ \bibinfo {eid} {arXiv:1811.03624}} (\bibinfo {year}
  {2018})},\ \Eprint {http://arxiv.org/abs/1811.03624} {arXiv:1811.03624
  [astro-ph.CO]} \BibitemShut {NoStop}%
\bibitem [{\citenamefont {{Beradze}}\ and\ \citenamefont
  {{Gogberashvili}}(2020)}]{2001.05874}%
  \BibitemOpen
  \bibfield  {author} {\bibinfo {author} {\bibfnamefont {R.}~\bibnamefont
  {{Beradze}}}\ and\ \bibinfo {author} {\bibfnamefont {M.}~\bibnamefont
  {{Gogberashvili}}},\ }\href@noop {} {\bibfield  {journal} {\bibinfo
  {journal} {ArXiv e-prints}\ ,\ \bibinfo {eid} {arXiv:2001.05874}} (\bibinfo
  {year} {2020})},\ \Eprint {http://arxiv.org/abs/2001.05874} {arXiv:2001.05874
  [astro-ph.CO]} \BibitemShut {NoStop}%
\bibitem [{\citenamefont {{Vagnozzi}}(2020)}]{1907.07569}%
  \BibitemOpen
  \bibfield  {author} {\bibinfo {author} {\bibfnamefont {S.}~\bibnamefont
  {{Vagnozzi}}},\ }\href {\doibase 10.1103/PhysRevD.102.023518} {\bibfield
  {journal} {\bibinfo  {journal} {\prd}\ }\textbf {\bibinfo {volume} {102}},\
  \bibinfo {eid} {023518} (\bibinfo {year} {2020})},\ \Eprint
  {http://arxiv.org/abs/1907.07569} {arXiv:1907.07569 [astro-ph.CO]}
  \BibitemShut {NoStop}%
\bibitem [{\citenamefont {{Lin}}\ \emph {et~al.}(2019)\citenamefont {{Lin}},
  \citenamefont {{Benevento}}, \citenamefont {{Hu}},\ and\ \citenamefont
  {{Raveri}}}]{1905.12618}%
  \BibitemOpen
  \bibfield  {author} {\bibinfo {author} {\bibfnamefont {M.-X.}\ \bibnamefont
  {{Lin}}}, \bibinfo {author} {\bibfnamefont {G.}~\bibnamefont {{Benevento}}},
  \bibinfo {author} {\bibfnamefont {W.}~\bibnamefont {{Hu}}}, \ and\ \bibinfo
  {author} {\bibfnamefont {M.}~\bibnamefont {{Raveri}}},\ }\href {\doibase
  10.1103/PhysRevD.100.063542} {\bibfield  {journal} {\bibinfo  {journal}
  {\prd}\ }\textbf {\bibinfo {volume} {100}},\ \bibinfo {eid} {063542}
  (\bibinfo {year} {2019})},\ \Eprint {http://arxiv.org/abs/1905.12618}
  {arXiv:1905.12618 [astro-ph.CO]} \BibitemShut {NoStop}%
\bibitem [{\citenamefont {{Arendse}}\ \emph {et~al.}(2020)\citenamefont
  {{Arendse}}, \citenamefont {{Wojtak}}, \citenamefont {{Agnello}},
  \citenamefont {{Chen}}, \citenamefont {{Fassnacht}}, \citenamefont {{Sluse}},
  \citenamefont {{Hilbert}}, \citenamefont {{Millon}}, \citenamefont
  {{Bonvin}}, \citenamefont {{Wong}}, \citenamefont {{Courbin}}, \citenamefont
  {{Suyu}}, \citenamefont {{Birrer}}, \citenamefont {{Treu}},\ and\
  \citenamefont {{Koopmans}}}]{1909.07986}%
  \BibitemOpen
  \bibfield  {author} {\bibinfo {author} {\bibfnamefont {N.}~\bibnamefont
  {{Arendse}}}, \bibinfo {author} {\bibfnamefont {R.~J.}\ \bibnamefont
  {{Wojtak}}}, \bibinfo {author} {\bibfnamefont {A.}~\bibnamefont {{Agnello}}},
  \bibinfo {author} {\bibfnamefont {G.~C.~F.}\ \bibnamefont {{Chen}}}, \bibinfo
  {author} {\bibfnamefont {C.~D.}\ \bibnamefont {{Fassnacht}}}, \bibinfo
  {author} {\bibfnamefont {D.}~\bibnamefont {{Sluse}}}, \bibinfo {author}
  {\bibfnamefont {S.}~\bibnamefont {{Hilbert}}}, \bibinfo {author}
  {\bibfnamefont {M.}~\bibnamefont {{Millon}}}, \bibinfo {author}
  {\bibfnamefont {V.}~\bibnamefont {{Bonvin}}}, \bibinfo {author}
  {\bibfnamefont {K.~C.}\ \bibnamefont {{Wong}}}, \bibinfo {author}
  {\bibfnamefont {F.}~\bibnamefont {{Courbin}}}, \bibinfo {author}
  {\bibfnamefont {S.~H.}\ \bibnamefont {{Suyu}}}, \bibinfo {author}
  {\bibfnamefont {S.}~\bibnamefont {{Birrer}}}, \bibinfo {author}
  {\bibfnamefont {T.}~\bibnamefont {{Treu}}}, \ and\ \bibinfo {author}
  {\bibfnamefont {L.~V.~E.}\ \bibnamefont {{Koopmans}}},\ }\href {\doibase
  10.1051/0004-6361/201936720} {\bibfield  {journal} {\bibinfo  {journal}
  {\aap}\ }\textbf {\bibinfo {volume} {639}},\ \bibinfo {eid} {A57} (\bibinfo
  {year} {2020})},\ \Eprint {http://arxiv.org/abs/1909.07986} {arXiv:1909.07986
  [astro-ph.CO]} \BibitemShut {NoStop}%
\bibitem [{\citenamefont {{Knox}}\ and\ \citenamefont
  {{Millea}}(2020)}]{1908.03663}%
  \BibitemOpen
  \bibfield  {author} {\bibinfo {author} {\bibfnamefont {L.}~\bibnamefont
  {{Knox}}}\ and\ \bibinfo {author} {\bibfnamefont {M.}~\bibnamefont
  {{Millea}}},\ }\href {\doibase 10.1103/PhysRevD.101.043533} {\bibfield
  {journal} {\bibinfo  {journal} {\prd}\ }\textbf {\bibinfo {volume} {101}},\
  \bibinfo {eid} {043533} (\bibinfo {year} {2020})},\ \Eprint
  {http://arxiv.org/abs/1908.03663} {arXiv:1908.03663 [astro-ph.CO]}
  \BibitemShut {NoStop}%
\bibitem [{\citenamefont {Cuesta}\ \emph {et~al.}(2015)\citenamefont {Cuesta},
  \citenamefont {Verde}, \citenamefont {Riess},\ and\ \citenamefont
  {Jimenez}}]{Cuesta:2014asa}%
  \BibitemOpen
  \bibfield  {author} {\bibinfo {author} {\bibfnamefont {A.~J.}\ \bibnamefont
  {Cuesta}}, \bibinfo {author} {\bibfnamefont {L.}~\bibnamefont {Verde}},
  \bibinfo {author} {\bibfnamefont {A.}~\bibnamefont {Riess}}, \ and\ \bibinfo
  {author} {\bibfnamefont {R.}~\bibnamefont {Jimenez}},\ }\href {\doibase
  10.1093/mnras/stv261} {\bibfield  {journal} {\bibinfo  {journal} {Mon. Not.
  Roy. Astron. Soc.}\ }\textbf {\bibinfo {volume} {448}},\ \bibinfo {pages}
  {3463} (\bibinfo {year} {2015})},\ \Eprint {http://arxiv.org/abs/1411.1094}
  {arXiv:1411.1094 [astro-ph.CO]} \BibitemShut {NoStop}%
\bibitem [{\citenamefont {{Aubourg}}\ \emph {et~al.}(2015)\citenamefont
  {{Aubourg}}, \citenamefont {{Bailey}}, \citenamefont {{Bautista}},
  \citenamefont {{Beutler}}, \citenamefont {{Bhardwaj}}, \citenamefont
  {{Bizyaev}}, \citenamefont {{Blanton}}, \citenamefont {{Blomqvist}},
  \citenamefont {{Bolton}}, \citenamefont {{Bovy}}, \citenamefont
  {{Brewington}}, \citenamefont {{Brinkmann}}, \citenamefont {{Brownstein}},
  \citenamefont {{Burden}}, \citenamefont {{Busca}}, \citenamefont
  {{Carithers}}, \citenamefont {{Chuang}}, \citenamefont {{Comparat}},
  \citenamefont {{Croft}}, \citenamefont {{Cuesta}}, \citenamefont {{Dawson}},
  \citenamefont {{Delubac}}, \citenamefont {{Eisenstein}}, \citenamefont
  {{Font-Ribera}}, \citenamefont {{Ge}}, \citenamefont {{Le Goff}},
  \citenamefont {{Gontcho}}, \citenamefont {{Gott}}, \citenamefont {{Gunn}},
  \citenamefont {{Guo}}, \citenamefont {{Guy}}, \citenamefont {{Hamilton}},
  \citenamefont {{Ho}}, \citenamefont {{Honscheid}}, \citenamefont {{Howlett}},
  \citenamefont {{Kirkby}}, \citenamefont {{Kitaura}}, \citenamefont {{Kneib}},
  \citenamefont {{Lee}}, \citenamefont {{Long}}, \citenamefont {{Lupton}},
  \citenamefont {{Maga{\~n}a}}, \citenamefont {{Malanushenko}}, \citenamefont
  {{Malanushenko}}, \citenamefont {{Manera}}, \citenamefont {{Maraston}},
  \citenamefont {{Margala}}, \citenamefont {{McBride}}, \citenamefont
  {{Miralda-Escud{\'e}}}, \citenamefont {{Myers}}, \citenamefont {{Nichol}},
  \citenamefont {{Noterdaeme}}, \citenamefont {{Nuza}}, \citenamefont
  {{Olmstead}}, \citenamefont {{Oravetz}}, \citenamefont {{P{\^a}ris}},
  \citenamefont {{Padmanabhan}}, \citenamefont {{Palanque-Delabrouille}},
  \citenamefont {{Pan}}, \citenamefont {{Pellejero-Ibanez}}, \citenamefont
  {{Percival}}, \citenamefont {{Petitjean}}, \citenamefont {{Pieri}},
  \citenamefont {{Prada}}, \citenamefont {{Reid}}, \citenamefont {{Rich}},
  \citenamefont {{Roe}}, \citenamefont {{Ross}}, \citenamefont {{Ross}},
  \citenamefont {{Rossi}}, \citenamefont {{Rubi{\~n}o-Mart{\'{\i}}n}},
  \citenamefont {{S{\'a}nchez}}, \citenamefont {{Samushia}}, \citenamefont
  {{G{\'e}nova-Santos}}, \citenamefont {{Sc{\'o}ccola}}, \citenamefont
  {{Schlegel}}, \citenamefont {{Schneider}}, \citenamefont {{Seo}},
  \citenamefont {{Sheldon}}, \citenamefont {{Simmons}}, \citenamefont
  {{Skibba}}, \citenamefont {{Slosar}}, \citenamefont {{Strauss}},
  \citenamefont {{Thomas}}, \citenamefont {{Tinker}}, \citenamefont
  {{Tojeiro}}, \citenamefont {{Vazquez}}, \citenamefont {{Viel}}, \citenamefont
  {{Wake}}, \citenamefont {{Weaver}}, \citenamefont {{Weinberg}}, \citenamefont
  {{Wood-Vasey}}, \citenamefont {{Y{\`e}che}}, \citenamefont {{Zehavi}},
  \citenamefont {{Zhao}},\ and\ \citenamefont {{BOSS
  Collaboration}}}]{aubourg15a}%
  \BibitemOpen
  \bibfield  {author} {\bibinfo {author} {\bibfnamefont {{\'E}.}~\bibnamefont
  {{Aubourg}}}, \bibinfo {author} {\bibfnamefont {S.}~\bibnamefont {{Bailey}}},
  \bibinfo {author} {\bibfnamefont {J.~E.}\ \bibnamefont {{Bautista}}},
  \bibinfo {author} {\bibfnamefont {F.}~\bibnamefont {{Beutler}}}, \bibinfo
  {author} {\bibfnamefont {V.}~\bibnamefont {{Bhardwaj}}}, \bibinfo {author}
  {\bibfnamefont {D.}~\bibnamefont {{Bizyaev}}}, \bibinfo {author}
  {\bibfnamefont {M.}~\bibnamefont {{Blanton}}}, \bibinfo {author}
  {\bibfnamefont {M.}~\bibnamefont {{Blomqvist}}}, \bibinfo {author}
  {\bibfnamefont {A.~S.}\ \bibnamefont {{Bolton}}}, \bibinfo {author}
  {\bibfnamefont {J.}~\bibnamefont {{Bovy}}}, \bibinfo {author} {\bibfnamefont
  {H.}~\bibnamefont {{Brewington}}}, \bibinfo {author} {\bibfnamefont
  {J.}~\bibnamefont {{Brinkmann}}}, \bibinfo {author} {\bibfnamefont {J.~R.}\
  \bibnamefont {{Brownstein}}}, \bibinfo {author} {\bibfnamefont
  {A.}~\bibnamefont {{Burden}}}, \bibinfo {author} {\bibfnamefont {N.~G.}\
  \bibnamefont {{Busca}}}, \bibinfo {author} {\bibfnamefont {W.}~\bibnamefont
  {{Carithers}}}, \bibinfo {author} {\bibfnamefont {C.-H.}\ \bibnamefont
  {{Chuang}}}, \bibinfo {author} {\bibfnamefont {J.}~\bibnamefont
  {{Comparat}}}, \bibinfo {author} {\bibfnamefont {R.~A.~C.}\ \bibnamefont
  {{Croft}}}, \bibinfo {author} {\bibfnamefont {A.~J.}\ \bibnamefont
  {{Cuesta}}}, \bibinfo {author} {\bibfnamefont {K.~S.}\ \bibnamefont
  {{Dawson}}}, \bibinfo {author} {\bibfnamefont {T.}~\bibnamefont {{Delubac}}},
  \bibinfo {author} {\bibfnamefont {D.~J.}\ \bibnamefont {{Eisenstein}}},
  \bibinfo {author} {\bibfnamefont {A.}~\bibnamefont {{Font-Ribera}}}, \bibinfo
  {author} {\bibfnamefont {J.}~\bibnamefont {{Ge}}}, \bibinfo {author}
  {\bibfnamefont {J.-M.}\ \bibnamefont {{Le Goff}}}, \bibinfo {author}
  {\bibfnamefont {S.~G.~A.}\ \bibnamefont {{Gontcho}}}, \bibinfo {author}
  {\bibfnamefont {J.~R.}\ \bibnamefont {{Gott}}}, \bibinfo {author}
  {\bibfnamefont {J.~E.}\ \bibnamefont {{Gunn}}}, \bibinfo {author}
  {\bibfnamefont {H.}~\bibnamefont {{Guo}}}, \bibinfo {author} {\bibfnamefont
  {J.}~\bibnamefont {{Guy}}}, \bibinfo {author} {\bibfnamefont {J.-C.}\
  \bibnamefont {{Hamilton}}}, \bibinfo {author} {\bibfnamefont
  {S.}~\bibnamefont {{Ho}}}, \bibinfo {author} {\bibfnamefont {K.}~\bibnamefont
  {{Honscheid}}}, \bibinfo {author} {\bibfnamefont {C.}~\bibnamefont
  {{Howlett}}}, \bibinfo {author} {\bibfnamefont {D.}~\bibnamefont {{Kirkby}}},
  \bibinfo {author} {\bibfnamefont {F.~S.}\ \bibnamefont {{Kitaura}}}, \bibinfo
  {author} {\bibfnamefont {J.-P.}\ \bibnamefont {{Kneib}}}, \bibinfo {author}
  {\bibfnamefont {K.-G.}\ \bibnamefont {{Lee}}}, \bibinfo {author}
  {\bibfnamefont {D.}~\bibnamefont {{Long}}}, \bibinfo {author} {\bibfnamefont
  {R.~H.}\ \bibnamefont {{Lupton}}}, \bibinfo {author} {\bibfnamefont {M.~V.}\
  \bibnamefont {{Maga{\~n}a}}}, \bibinfo {author} {\bibfnamefont
  {V.}~\bibnamefont {{Malanushenko}}}, \bibinfo {author} {\bibfnamefont
  {E.}~\bibnamefont {{Malanushenko}}}, \bibinfo {author} {\bibfnamefont
  {M.}~\bibnamefont {{Manera}}}, \bibinfo {author} {\bibfnamefont
  {C.}~\bibnamefont {{Maraston}}}, \bibinfo {author} {\bibfnamefont
  {D.}~\bibnamefont {{Margala}}}, \bibinfo {author} {\bibfnamefont {C.~K.}\
  \bibnamefont {{McBride}}}, \bibinfo {author} {\bibfnamefont {J.}~\bibnamefont
  {{Miralda-Escud{\'e}}}}, \bibinfo {author} {\bibfnamefont {A.~D.}\
  \bibnamefont {{Myers}}}, \bibinfo {author} {\bibfnamefont {R.~C.}\
  \bibnamefont {{Nichol}}}, \bibinfo {author} {\bibfnamefont {P.}~\bibnamefont
  {{Noterdaeme}}}, \bibinfo {author} {\bibfnamefont {S.~E.}\ \bibnamefont
  {{Nuza}}}, \bibinfo {author} {\bibfnamefont {M.~D.}\ \bibnamefont
  {{Olmstead}}}, \bibinfo {author} {\bibfnamefont {D.}~\bibnamefont
  {{Oravetz}}}, \bibinfo {author} {\bibfnamefont {I.}~\bibnamefont
  {{P{\^a}ris}}}, \bibinfo {author} {\bibfnamefont {N.}~\bibnamefont
  {{Padmanabhan}}}, \bibinfo {author} {\bibfnamefont {N.}~\bibnamefont
  {{Palanque-Delabrouille}}}, \bibinfo {author} {\bibfnamefont
  {K.}~\bibnamefont {{Pan}}}, \bibinfo {author} {\bibfnamefont
  {M.}~\bibnamefont {{Pellejero-Ibanez}}}, \bibinfo {author} {\bibfnamefont
  {W.~J.}\ \bibnamefont {{Percival}}}, \bibinfo {author} {\bibfnamefont
  {P.}~\bibnamefont {{Petitjean}}}, \bibinfo {author} {\bibfnamefont {M.~M.}\
  \bibnamefont {{Pieri}}}, \bibinfo {author} {\bibfnamefont {F.}~\bibnamefont
  {{Prada}}}, \bibinfo {author} {\bibfnamefont {B.}~\bibnamefont {{Reid}}},
  \bibinfo {author} {\bibfnamefont {J.}~\bibnamefont {{Rich}}}, \bibinfo
  {author} {\bibfnamefont {N.~A.}\ \bibnamefont {{Roe}}}, \bibinfo {author}
  {\bibfnamefont {A.~J.}\ \bibnamefont {{Ross}}}, \bibinfo {author}
  {\bibfnamefont {N.~P.}\ \bibnamefont {{Ross}}}, \bibinfo {author}
  {\bibfnamefont {G.}~\bibnamefont {{Rossi}}}, \bibinfo {author} {\bibfnamefont
  {J.~A.}\ \bibnamefont {{Rubi{\~n}o-Mart{\'{\i}}n}}}, \bibinfo {author}
  {\bibfnamefont {A.~G.}\ \bibnamefont {{S{\'a}nchez}}}, \bibinfo {author}
  {\bibfnamefont {L.}~\bibnamefont {{Samushia}}}, \bibinfo {author}
  {\bibfnamefont {R.~T.}\ \bibnamefont {{G{\'e}nova-Santos}}}, \bibinfo
  {author} {\bibfnamefont {C.~G.}\ \bibnamefont {{Sc{\'o}ccola}}}, \bibinfo
  {author} {\bibfnamefont {D.~J.}\ \bibnamefont {{Schlegel}}}, \bibinfo
  {author} {\bibfnamefont {D.~P.}\ \bibnamefont {{Schneider}}}, \bibinfo
  {author} {\bibfnamefont {H.-J.}\ \bibnamefont {{Seo}}}, \bibinfo {author}
  {\bibfnamefont {E.}~\bibnamefont {{Sheldon}}}, \bibinfo {author}
  {\bibfnamefont {A.}~\bibnamefont {{Simmons}}}, \bibinfo {author}
  {\bibfnamefont {R.~A.}\ \bibnamefont {{Skibba}}}, \bibinfo {author}
  {\bibfnamefont {A.}~\bibnamefont {{Slosar}}}, \bibinfo {author}
  {\bibfnamefont {M.~A.}\ \bibnamefont {{Strauss}}}, \bibinfo {author}
  {\bibfnamefont {D.}~\bibnamefont {{Thomas}}}, \bibinfo {author}
  {\bibfnamefont {J.~L.}\ \bibnamefont {{Tinker}}}, \bibinfo {author}
  {\bibfnamefont {R.}~\bibnamefont {{Tojeiro}}}, \bibinfo {author}
  {\bibfnamefont {J.~A.}\ \bibnamefont {{Vazquez}}}, \bibinfo {author}
  {\bibfnamefont {M.}~\bibnamefont {{Viel}}}, \bibinfo {author} {\bibfnamefont
  {D.~A.}\ \bibnamefont {{Wake}}}, \bibinfo {author} {\bibfnamefont {B.~A.}\
  \bibnamefont {{Weaver}}}, \bibinfo {author} {\bibfnamefont {D.~H.}\
  \bibnamefont {{Weinberg}}}, \bibinfo {author} {\bibfnamefont {W.~M.}\
  \bibnamefont {{Wood-Vasey}}}, \bibinfo {author} {\bibfnamefont
  {C.}~\bibnamefont {{Y{\`e}che}}}, \bibinfo {author} {\bibfnamefont
  {I.}~\bibnamefont {{Zehavi}}}, \bibinfo {author} {\bibfnamefont {G.-B.}\
  \bibnamefont {{Zhao}}}, \ and\ \bibinfo {author} {\bibnamefont {{BOSS
  Collaboration}}},\ }\href {\doibase 10.1103/PhysRevD.92.123516} {\bibfield
  {journal} {\bibinfo  {journal} {\prd}\ }\textbf {\bibinfo {volume} {92}},\
  \bibinfo {eid} {123516} (\bibinfo {year} {2015})},\ \Eprint
  {http://arxiv.org/abs/1411.1074} {arXiv:1411.1074} \BibitemShut {NoStop}%
\bibitem [{\citenamefont {{Fixsen}}(2009)}]{fixsen09}%
  \BibitemOpen
  \bibfield  {author} {\bibinfo {author} {\bibfnamefont {D.~J.}\ \bibnamefont
  {{Fixsen}}},\ }\href {\doibase 10.1088/0004-637X/707/2/916} {\bibfield
  {journal} {\bibinfo  {journal} {\apj}\ }\textbf {\bibinfo {volume} {707}},\
  \bibinfo {pages} {916} (\bibinfo {year} {2009})},\ \Eprint
  {http://arxiv.org/abs/0911.1955} {arXiv:0911.1955 [astro-ph.CO]} \BibitemShut
  {NoStop}%
\bibitem [{\citenamefont {{Cooke}}\ \emph {et~al.}(2018)\citenamefont
  {{Cooke}}, \citenamefont {{Pettini}},\ and\ \citenamefont
  {{Steidel}}}]{cooke18a}%
  \BibitemOpen
  \bibfield  {author} {\bibinfo {author} {\bibfnamefont {R.~J.}\ \bibnamefont
  {{Cooke}}}, \bibinfo {author} {\bibfnamefont {M.}~\bibnamefont {{Pettini}}},
  \ and\ \bibinfo {author} {\bibfnamefont {C.~C.}\ \bibnamefont {{Steidel}}},\
  }\href {\doibase 10.3847/1538-4357/aaab53} {\bibfield  {journal} {\bibinfo
  {journal} {\apj}\ }\textbf {\bibinfo {volume} {855}},\ \bibinfo {eid} {102}
  (\bibinfo {year} {2018})},\ \Eprint {http://arxiv.org/abs/1710.11129}
  {arXiv:1710.11129 [astro-ph.CO]} \BibitemShut {NoStop}%
\bibitem [{\citenamefont {{Adelberger}}\ \emph {et~al.}(2011)\citenamefont
  {{Adelberger}}, \citenamefont {{Garc{\'\i}a}}, \citenamefont {{Robertson}},
  \citenamefont {{Snover}}, \citenamefont {{Balantekin}}, \citenamefont
  {{Heeger}}, \citenamefont {{Ramsey-Musolf}}, \citenamefont {{Bemmerer}},
  \citenamefont {{Junghans}}, \citenamefont {{Bertulani}}, \citenamefont
  {{Chen}}, \citenamefont {{Costantini}}, \citenamefont {{Prati}},
  \citenamefont {{Couder}}, \citenamefont {{Uberseder}}, \citenamefont
  {{Wiescher}}, \citenamefont {{Cyburt}}, \citenamefont {{Davids}},
  \citenamefont {{Freedman}}, \citenamefont {{Gai}}, \citenamefont {{Gazit}},
  \citenamefont {{Gialanella}}, \citenamefont {{Imbriani}}, \citenamefont
  {{Greife}}, \citenamefont {{Hass}}, \citenamefont {{Haxton}}, \citenamefont
  {{Itahashi}}, \citenamefont {{Kubodera}}, \citenamefont {{Langanke}},
  \citenamefont {{Leitner}}, \citenamefont {{Leitner}}, \citenamefont
  {{Vetter}}, \citenamefont {{Winslow}}, \citenamefont {{Marcucci}},
  \citenamefont {{Motobayashi}}, \citenamefont {{Mukhamedzhanov}},
  \citenamefont {{Tribble}}, \citenamefont {{Nollett}}, \citenamefont
  {{Nunes}}, \citenamefont {{Park}}, \citenamefont {{Parker}}, \citenamefont
  {{Schiavilla}}, \citenamefont {{Simpson}}, \citenamefont {{Spitaleri}},
  \citenamefont {{Strieder}}, \citenamefont {{Trautvetter}}, \citenamefont
  {{Suemmerer}},\ and\ \citenamefont {{Typel}}}]{adelberger11}%
  \BibitemOpen
  \bibfield  {author} {\bibinfo {author} {\bibfnamefont {E.~G.}\ \bibnamefont
  {{Adelberger}}}, \bibinfo {author} {\bibfnamefont {A.}~\bibnamefont
  {{Garc{\'\i}a}}}, \bibinfo {author} {\bibfnamefont {R.~G.~H.}\ \bibnamefont
  {{Robertson}}}, \bibinfo {author} {\bibfnamefont {K.~A.}\ \bibnamefont
  {{Snover}}}, \bibinfo {author} {\bibfnamefont {A.~B.}\ \bibnamefont
  {{Balantekin}}}, \bibinfo {author} {\bibfnamefont {K.}~\bibnamefont
  {{Heeger}}}, \bibinfo {author} {\bibfnamefont {M.~J.}\ \bibnamefont
  {{Ramsey-Musolf}}}, \bibinfo {author} {\bibfnamefont {D.}~\bibnamefont
  {{Bemmerer}}}, \bibinfo {author} {\bibfnamefont {A.}~\bibnamefont
  {{Junghans}}}, \bibinfo {author} {\bibfnamefont {C.~A.}\ \bibnamefont
  {{Bertulani}}}, \bibinfo {author} {\bibfnamefont {J.~W.}\ \bibnamefont
  {{Chen}}}, \bibinfo {author} {\bibfnamefont {H.}~\bibnamefont
  {{Costantini}}}, \bibinfo {author} {\bibfnamefont {P.}~\bibnamefont
  {{Prati}}}, \bibinfo {author} {\bibfnamefont {M.}~\bibnamefont {{Couder}}},
  \bibinfo {author} {\bibfnamefont {E.}~\bibnamefont {{Uberseder}}}, \bibinfo
  {author} {\bibfnamefont {M.}~\bibnamefont {{Wiescher}}}, \bibinfo {author}
  {\bibfnamefont {R.}~\bibnamefont {{Cyburt}}}, \bibinfo {author}
  {\bibfnamefont {B.}~\bibnamefont {{Davids}}}, \bibinfo {author}
  {\bibfnamefont {S.~J.}\ \bibnamefont {{Freedman}}}, \bibinfo {author}
  {\bibfnamefont {M.}~\bibnamefont {{Gai}}}, \bibinfo {author} {\bibfnamefont
  {D.}~\bibnamefont {{Gazit}}}, \bibinfo {author} {\bibfnamefont
  {L.}~\bibnamefont {{Gialanella}}}, \bibinfo {author} {\bibfnamefont
  {G.}~\bibnamefont {{Imbriani}}}, \bibinfo {author} {\bibfnamefont
  {U.}~\bibnamefont {{Greife}}}, \bibinfo {author} {\bibfnamefont
  {M.}~\bibnamefont {{Hass}}}, \bibinfo {author} {\bibfnamefont {W.~C.}\
  \bibnamefont {{Haxton}}}, \bibinfo {author} {\bibfnamefont {T.}~\bibnamefont
  {{Itahashi}}}, \bibinfo {author} {\bibfnamefont {K.}~\bibnamefont
  {{Kubodera}}}, \bibinfo {author} {\bibfnamefont {K.}~\bibnamefont
  {{Langanke}}}, \bibinfo {author} {\bibfnamefont {D.}~\bibnamefont
  {{Leitner}}}, \bibinfo {author} {\bibfnamefont {M.}~\bibnamefont
  {{Leitner}}}, \bibinfo {author} {\bibfnamefont {P.}~\bibnamefont {{Vetter}}},
  \bibinfo {author} {\bibfnamefont {L.}~\bibnamefont {{Winslow}}}, \bibinfo
  {author} {\bibfnamefont {L.~E.}\ \bibnamefont {{Marcucci}}}, \bibinfo
  {author} {\bibfnamefont {T.}~\bibnamefont {{Motobayashi}}}, \bibinfo {author}
  {\bibfnamefont {A.}~\bibnamefont {{Mukhamedzhanov}}}, \bibinfo {author}
  {\bibfnamefont {R.~E.}\ \bibnamefont {{Tribble}}}, \bibinfo {author}
  {\bibfnamefont {K.~M.}\ \bibnamefont {{Nollett}}}, \bibinfo {author}
  {\bibfnamefont {F.~M.}\ \bibnamefont {{Nunes}}}, \bibinfo {author}
  {\bibfnamefont {T.~S.}\ \bibnamefont {{Park}}}, \bibinfo {author}
  {\bibfnamefont {P.~D.}\ \bibnamefont {{Parker}}}, \bibinfo {author}
  {\bibfnamefont {R.}~\bibnamefont {{Schiavilla}}}, \bibinfo {author}
  {\bibfnamefont {E.~C.}\ \bibnamefont {{Simpson}}}, \bibinfo {author}
  {\bibfnamefont {C.}~\bibnamefont {{Spitaleri}}}, \bibinfo {author}
  {\bibfnamefont {F.}~\bibnamefont {{Strieder}}}, \bibinfo {author}
  {\bibfnamefont {H.~P.}\ \bibnamefont {{Trautvetter}}}, \bibinfo {author}
  {\bibfnamefont {K.}~\bibnamefont {{Suemmerer}}}, \ and\ \bibinfo {author}
  {\bibfnamefont {S.}~\bibnamefont {{Typel}}},\ }\href {\doibase
  10.1103/RevModPhys.83.195} {\bibfield  {journal} {\bibinfo  {journal}
  {Reviews of Modern Physics}\ }\textbf {\bibinfo {volume} {83}},\ \bibinfo
  {pages} {195} (\bibinfo {year} {2011})},\ \Eprint
  {http://arxiv.org/abs/1004.2318} {arXiv:1004.2318 [nucl-ex]} \BibitemShut
  {NoStop}%
\bibitem [{\citenamefont {{Addison}}\ \emph {et~al.}(2018)\citenamefont
  {{Addison}}, \citenamefont {{Watts}}, \citenamefont {{Bennett}},
  \citenamefont {{Halpern}}, \citenamefont {{Hinshaw}},\ and\ \citenamefont
  {{Weiland}}}]{1707.06547}%
  \BibitemOpen
  \bibfield  {author} {\bibinfo {author} {\bibfnamefont {G.~E.}\ \bibnamefont
  {{Addison}}}, \bibinfo {author} {\bibfnamefont {D.~J.}\ \bibnamefont
  {{Watts}}}, \bibinfo {author} {\bibfnamefont {C.~L.}\ \bibnamefont
  {{Bennett}}}, \bibinfo {author} {\bibfnamefont {M.}~\bibnamefont
  {{Halpern}}}, \bibinfo {author} {\bibfnamefont {G.}~\bibnamefont
  {{Hinshaw}}}, \ and\ \bibinfo {author} {\bibfnamefont {J.~L.}\ \bibnamefont
  {{Weiland}}},\ }\href {\doibase 10.3847/1538-4357/aaa1ed} {\bibfield
  {journal} {\bibinfo  {journal} {\apj}\ }\textbf {\bibinfo {volume} {853}},\
  \bibinfo {eid} {119} (\bibinfo {year} {2018})},\ \Eprint
  {http://arxiv.org/abs/1707.06547} {arXiv:1707.06547 [astro-ph.CO]}
  \BibitemShut {NoStop}%
\bibitem [{\citenamefont {{Cuceu}}\ \emph {et~al.}(2019)\citenamefont
  {{Cuceu}}, \citenamefont {{Farr}}, \citenamefont {{Lemos}},\ and\
  \citenamefont {{Font-Ribera}}}]{1906.11628}%
  \BibitemOpen
  \bibfield  {author} {\bibinfo {author} {\bibfnamefont {A.}~\bibnamefont
  {{Cuceu}}}, \bibinfo {author} {\bibfnamefont {J.}~\bibnamefont {{Farr}}},
  \bibinfo {author} {\bibfnamefont {P.}~\bibnamefont {{Lemos}}}, \ and\
  \bibinfo {author} {\bibfnamefont {A.}~\bibnamefont {{Font-Ribera}}},\ }\href
  {\doibase 10.1088/1475-7516/2019/10/044} {\bibfield  {journal} {\bibinfo
  {journal} {\jcap}\ }\textbf {\bibinfo {volume} {2019}},\ \bibinfo {eid} {044}
  (\bibinfo {year} {2019})},\ \Eprint {http://arxiv.org/abs/1906.11628}
  {arXiv:1906.11628 [astro-ph.CO]} \BibitemShut {NoStop}%
\bibitem [{\citenamefont {{Abbott}}\ \emph
  {et~al.}(2017{\natexlab{b}})\citenamefont {{Abbott}}, \citenamefont
  {{Abbott}}, \citenamefont {{Abbott}}, \citenamefont {{Acernese}},
  \citenamefont {{Ackley}}, \citenamefont {{Adams}}, \citenamefont {{Adams}},
  \citenamefont {{Addesso}}, \citenamefont {{Adhikari}}, \citenamefont
  {{Adya}}, \citenamefont {{Affeldt}}, \citenamefont {{Afrough}}, \citenamefont
  {{Agarwal}}, \citenamefont {{Agathos}}, \citenamefont {{Agatsuma}},
  \citenamefont {{Aggarwal}}, \citenamefont {{Aguiar}}, \citenamefont
  {{Aiello}}, \citenamefont {{Ain}}, \citenamefont {{Ajith}}, \citenamefont
  {{Allen}}, \citenamefont {{Allen}}, \citenamefont {{Allocca}}, \citenamefont
  {{Aloy}}, \citenamefont {{Altin}}, \citenamefont {{Amato}}, \citenamefont
  {{Ananyeva}}, \citenamefont {{Anderson}}, \citenamefont {{Anderson}},
  \citenamefont {{Angelova}}, \citenamefont {{Antier}}, \citenamefont
  {{Appert}}, \citenamefont {{Arai}}, \citenamefont {{Araya}}, \citenamefont
  {{Areeda}}, \citenamefont {{Arnaud}}, \citenamefont {{Arun}}, \citenamefont
  {{Ascenzi}}, \citenamefont {{Ashton}}, \citenamefont {{Ast}}, \citenamefont
  {{Aston}}, \citenamefont {{Astone}}, \citenamefont {{Atallah}}, \citenamefont
  {{Aufmuth}}, \citenamefont {{Aulbert}}, \citenamefont {{AultONeal}},
  \citenamefont {{Austin}}, \citenamefont {{Avila-Alvarez}}, \citenamefont
  {{Babak}}, \citenamefont {{Bacon}}, \citenamefont {{Bader}}, \citenamefont
  {{Bae}}, \citenamefont {{Baker}}, \citenamefont {{Baldaccini}}, \citenamefont
  {{Ballardin}}, \citenamefont {{Ballmer}}, \citenamefont {{Banagiri}},
  \citenamefont {{Barayoga}}, \citenamefont {{Barclay}}, \citenamefont
  {{Barish}}, \citenamefont {{Barker}}, \citenamefont {{Barkett}},
  \citenamefont {{Barone}}, \citenamefont {{Barr}}, \citenamefont {{Barsotti}},
  \citenamefont {{Barsuglia}}, \citenamefont {{Barta}}, \citenamefont
  {{Bartlett}}, \citenamefont {{Bartos}}, \citenamefont {{Bassiri}},
  \citenamefont {{Basti}}, \citenamefont {{Batch}}, \citenamefont {{Bawaj}},
  \citenamefont {{Bayley}}, \citenamefont {{Bazzan}}, \citenamefont
  {{B{\'e}csy}}, \citenamefont {{Beer}}, \citenamefont {{Bejger}},
  \citenamefont {{Belahcene}}, \citenamefont {{Bell}}, \citenamefont
  {{Berger}}, \citenamefont {{Bergmann}}, \citenamefont {{Bero}}, \citenamefont
  {{Berry}}, \citenamefont {{Bersanetti}}, \citenamefont {{Bertolini}},
  \citenamefont {{Betzwieser}}, \citenamefont {{Bhagwat}}, \citenamefont
  {{Bhandare}}, \citenamefont {{Bilenko}}, \citenamefont {{Billingsley}},
  \citenamefont {{Billman}}, \citenamefont {{Birch}}, \citenamefont {{Birney}},
  \citenamefont {{Birnholtz}}, \citenamefont {{Biscans}}, \citenamefont
  {{Biscoveanu}}, \citenamefont {{Bisht}}, \citenamefont {{Bitossi}},
  \citenamefont {{Biwer}}, \citenamefont {{Bizouard}}, \citenamefont
  {{Blackburn}}, \citenamefont {{Blackman}}, \citenamefont {{Blair}},
  \citenamefont {{Blair}}, \citenamefont {{Blair}}, \citenamefont {{Bloemen}},
  \citenamefont {{Walet}}, \citenamefont {{Walker}}, \citenamefont {{Wallace}},
  \citenamefont {{Walsh}}, \citenamefont {{Wang}}, \citenamefont {{Wang}},
  \citenamefont {{Wang}}, \citenamefont {{Wang}}, \citenamefont {{Wang}},
  \citenamefont {{Ward}}, \citenamefont {{Warner}}, \citenamefont {{Was}},
  \citenamefont {{Watchi}}, \citenamefont {{Weaver}}, \citenamefont {{Wei}},
  \citenamefont {{Weinert}}, \citenamefont {{Weinstein}}, \citenamefont
  {{Weiss}}, \citenamefont {{Wen}}, \citenamefont {{Wessel}}, \citenamefont
  {{We{\ss}els}}, \citenamefont {{Westerweck}}, \citenamefont {{Westphal}},
  \citenamefont {{Wette}}, \citenamefont {{Whelan}}, \citenamefont
  {{Whitcomb}}, \citenamefont {{Whiting}}, \citenamefont {{Whittle}},
  \citenamefont {{Wilken}}, \citenamefont {{Williams}}, \citenamefont
  {{Williams}}, \citenamefont {{Williamson}}, \citenamefont {{Willis}},
  \citenamefont {{Willke}}, \citenamefont {{Wimmer}}, \citenamefont
  {{Winkler}}, \citenamefont {{Wipf}}, \citenamefont {{Wittel}}, \citenamefont
  {{Woan}}, \citenamefont {{Woehler}}, \citenamefont {{Wofford}}, \citenamefont
  {{Wong}}, \citenamefont {{Worden}}, \citenamefont {{Wright}}, \citenamefont
  {{Wu}}, \citenamefont {{Wysocki}}, \citenamefont {{Xiao}}, \citenamefont
  {{Yamamoto}}, \citenamefont {{Yancey}}, \citenamefont {{Yang}}, \citenamefont
  {{Yap}}, \citenamefont {{Yazback}}, \citenamefont {{Yu}}, \citenamefont
  {{Yu}}, \citenamefont {{Yvert}}, \citenamefont {{Zadro{\.z}ny}},
  \citenamefont {{Zanolin}}, \citenamefont {{Zelenova}}, \citenamefont
  {{Zendri}}, \citenamefont {{Zevin}}, \citenamefont {{Zhang}}, \citenamefont
  {{Zhang}}, \citenamefont {{Zhang}}, \citenamefont {{Zhang}}, \citenamefont
  {{Zhao}}, \citenamefont {{Zhou}}, \citenamefont {{Zhou}}, \citenamefont
  {{Zhu}}, \citenamefont {{Zhu}}, \citenamefont {{Zimmerman}}, \citenamefont
  {{Zucker}}, \citenamefont {{Zweizig}}, \citenamefont {{(LIGO Scientific
  Collaboration}}, \citenamefont {{Virgo Collaboration}}, \citenamefont
  {{Burns}}, \citenamefont {{Veres}}, \citenamefont {{Kocevski}}, \citenamefont
  {{Racusin}}, \citenamefont {{Goldstein}}, \citenamefont {{Connaughton}},
  \citenamefont {{Briggs}}, \citenamefont {{Blackburn}}, \citenamefont
  {{Hamburg}}, \citenamefont {{Hui}}, \citenamefont {{von Kienlin}},
  \citenamefont {{McEnery}}, \citenamefont {{Preece}}, \citenamefont
  {{Wilson-Hodge}}, \citenamefont {{Bissaldi}}, \citenamefont {{Cleveland}},
  \citenamefont {{Gibby}}, \citenamefont {{Giles}}, \citenamefont {{Kippen}},
  \citenamefont {{McBreen}}, \citenamefont {{Meegan}}, \citenamefont
  {{Paciesas}}, \citenamefont {{Poolakkil}}, \citenamefont {{Roberts}},
  \citenamefont {{Stanbro}}, \citenamefont {{Gamma-ray Burst Monitor}},
  \citenamefont {{Savchenko}}, \citenamefont {{Ferrigno}}, \citenamefont
  {{Kuulkers}}, \citenamefont {{Bazzano}}, \citenamefont {{Bozzo}},
  \citenamefont {{Brandt}}, \citenamefont {{Chenevez}}, \citenamefont
  {{Courvoisier}}, \citenamefont {{Diehl}}, \citenamefont {{Domingo}},
  \citenamefont {{Hanlon}}, \citenamefont {{Jourdain}}, \citenamefont
  {{Laurent}}, \citenamefont {{Lebrun}}, \citenamefont {{Lutovinov}},
  \citenamefont {{Mereghetti}}, \citenamefont {{Natalucci}}, \citenamefont
  {{Rodi}}, \citenamefont {{Roques}}, \citenamefont {{Sunyaev}}, \citenamefont
  {{Ubertini}},\ and\ \citenamefont {{(INTEGRAL}}}]{ligo17a}%
  \BibitemOpen
  \bibfield  {author} {\bibinfo {author} {\bibfnamefont {B.~P.}\ \bibnamefont
  {{Abbott}}}, \bibinfo {author} {\bibfnamefont {R.}~\bibnamefont {{Abbott}}},
  \bibinfo {author} {\bibfnamefont {T.~D.}\ \bibnamefont {{Abbott}}}, \bibinfo
  {author} {\bibfnamefont {F.}~\bibnamefont {{Acernese}}}, \bibinfo {author}
  {\bibfnamefont {K.}~\bibnamefont {{Ackley}}}, \bibinfo {author}
  {\bibfnamefont {C.}~\bibnamefont {{Adams}}}, \bibinfo {author} {\bibfnamefont
  {T.}~\bibnamefont {{Adams}}}, \bibinfo {author} {\bibfnamefont
  {P.}~\bibnamefont {{Addesso}}}, \bibinfo {author} {\bibfnamefont {R.~X.}\
  \bibnamefont {{Adhikari}}}, \bibinfo {author} {\bibfnamefont {V.~B.}\
  \bibnamefont {{Adya}}}, \bibinfo {author} {\bibfnamefont {C.}~\bibnamefont
  {{Affeldt}}}, \bibinfo {author} {\bibfnamefont {M.}~\bibnamefont
  {{Afrough}}}, \bibinfo {author} {\bibfnamefont {B.}~\bibnamefont
  {{Agarwal}}}, \bibinfo {author} {\bibfnamefont {M.}~\bibnamefont
  {{Agathos}}}, \bibinfo {author} {\bibfnamefont {K.}~\bibnamefont
  {{Agatsuma}}}, \bibinfo {author} {\bibfnamefont {N.}~\bibnamefont
  {{Aggarwal}}}, \bibinfo {author} {\bibfnamefont {O.~D.}\ \bibnamefont
  {{Aguiar}}}, \bibinfo {author} {\bibfnamefont {L.}~\bibnamefont {{Aiello}}},
  \bibinfo {author} {\bibfnamefont {A.}~\bibnamefont {{Ain}}}, \bibinfo
  {author} {\bibfnamefont {P.}~\bibnamefont {{Ajith}}}, \bibinfo {author}
  {\bibfnamefont {B.}~\bibnamefont {{Allen}}}, \bibinfo {author} {\bibfnamefont
  {G.}~\bibnamefont {{Allen}}}, \bibinfo {author} {\bibfnamefont
  {A.}~\bibnamefont {{Allocca}}}, \bibinfo {author} {\bibfnamefont {M.~A.}\
  \bibnamefont {{Aloy}}}, \bibinfo {author} {\bibfnamefont {P.~A.}\
  \bibnamefont {{Altin}}}, \bibinfo {author} {\bibfnamefont {A.}~\bibnamefont
  {{Amato}}}, \bibinfo {author} {\bibfnamefont {A.}~\bibnamefont {{Ananyeva}}},
  \bibinfo {author} {\bibfnamefont {S.~B.}\ \bibnamefont {{Anderson}}},
  \bibinfo {author} {\bibfnamefont {W.~G.}\ \bibnamefont {{Anderson}}},
  \bibinfo {author} {\bibfnamefont {S.~V.}\ \bibnamefont {{Angelova}}},
  \bibinfo {author} {\bibfnamefont {S.}~\bibnamefont {{Antier}}}, \bibinfo
  {author} {\bibfnamefont {S.}~\bibnamefont {{Appert}}}, \bibinfo {author}
  {\bibfnamefont {K.}~\bibnamefont {{Arai}}}, \bibinfo {author} {\bibfnamefont
  {M.~C.}\ \bibnamefont {{Araya}}}, \bibinfo {author} {\bibfnamefont {J.~S.}\
  \bibnamefont {{Areeda}}}, \bibinfo {author} {\bibfnamefont {N.}~\bibnamefont
  {{Arnaud}}}, \bibinfo {author} {\bibfnamefont {K.~G.}\ \bibnamefont
  {{Arun}}}, \bibinfo {author} {\bibfnamefont {S.}~\bibnamefont {{Ascenzi}}},
  \bibinfo {author} {\bibfnamefont {G.}~\bibnamefont {{Ashton}}}, \bibinfo
  {author} {\bibfnamefont {M.}~\bibnamefont {{Ast}}}, \bibinfo {author}
  {\bibfnamefont {S.~M.}\ \bibnamefont {{Aston}}}, \bibinfo {author}
  {\bibfnamefont {P.}~\bibnamefont {{Astone}}}, \bibinfo {author}
  {\bibfnamefont {D.~V.}\ \bibnamefont {{Atallah}}}, \bibinfo {author}
  {\bibfnamefont {P.}~\bibnamefont {{Aufmuth}}}, \bibinfo {author}
  {\bibfnamefont {C.}~\bibnamefont {{Aulbert}}}, \bibinfo {author}
  {\bibfnamefont {K.}~\bibnamefont {{AultONeal}}}, \bibinfo {author}
  {\bibfnamefont {C.}~\bibnamefont {{Austin}}}, \bibinfo {author}
  {\bibfnamefont {A.}~\bibnamefont {{Avila-Alvarez}}}, \bibinfo {author}
  {\bibfnamefont {S.}~\bibnamefont {{Babak}}}, \bibinfo {author} {\bibfnamefont
  {P.}~\bibnamefont {{Bacon}}}, \bibinfo {author} {\bibfnamefont {M.~K.~M.}\
  \bibnamefont {{Bader}}}, \bibinfo {author} {\bibfnamefont {S.}~\bibnamefont
  {{Bae}}}, \bibinfo {author} {\bibfnamefont {P.~T.}\ \bibnamefont {{Baker}}},
  \bibinfo {author} {\bibfnamefont {F.}~\bibnamefont {{Baldaccini}}}, \bibinfo
  {author} {\bibfnamefont {G.}~\bibnamefont {{Ballardin}}}, \bibinfo {author}
  {\bibfnamefont {S.~W.}\ \bibnamefont {{Ballmer}}}, \bibinfo {author}
  {\bibfnamefont {S.}~\bibnamefont {{Banagiri}}}, \bibinfo {author}
  {\bibfnamefont {J.~C.}\ \bibnamefont {{Barayoga}}}, \bibinfo {author}
  {\bibfnamefont {S.~E.}\ \bibnamefont {{Barclay}}}, \bibinfo {author}
  {\bibfnamefont {B.~C.}\ \bibnamefont {{Barish}}}, \bibinfo {author}
  {\bibfnamefont {D.}~\bibnamefont {{Barker}}}, \bibinfo {author}
  {\bibfnamefont {K.}~\bibnamefont {{Barkett}}}, \bibinfo {author}
  {\bibfnamefont {F.}~\bibnamefont {{Barone}}}, \bibinfo {author}
  {\bibfnamefont {B.}~\bibnamefont {{Barr}}}, \bibinfo {author} {\bibfnamefont
  {L.}~\bibnamefont {{Barsotti}}}, \bibinfo {author} {\bibfnamefont
  {M.}~\bibnamefont {{Barsuglia}}}, \bibinfo {author} {\bibfnamefont
  {D.}~\bibnamefont {{Barta}}}, \bibinfo {author} {\bibfnamefont
  {J.}~\bibnamefont {{Bartlett}}}, \bibinfo {author} {\bibfnamefont
  {I.}~\bibnamefont {{Bartos}}}, \bibinfo {author} {\bibfnamefont
  {R.}~\bibnamefont {{Bassiri}}}, \bibinfo {author} {\bibfnamefont
  {A.}~\bibnamefont {{Basti}}}, \bibinfo {author} {\bibfnamefont {J.~C.}\
  \bibnamefont {{Batch}}}, \bibinfo {author} {\bibfnamefont {M.}~\bibnamefont
  {{Bawaj}}}, \bibinfo {author} {\bibfnamefont {J.~C.}\ \bibnamefont
  {{Bayley}}}, \bibinfo {author} {\bibfnamefont {M.}~\bibnamefont {{Bazzan}}},
  \bibinfo {author} {\bibfnamefont {B.}~\bibnamefont {{B{\'e}csy}}}, \bibinfo
  {author} {\bibfnamefont {C.}~\bibnamefont {{Beer}}}, \bibinfo {author}
  {\bibfnamefont {M.}~\bibnamefont {{Bejger}}}, \bibinfo {author}
  {\bibfnamefont {I.}~\bibnamefont {{Belahcene}}}, \bibinfo {author}
  {\bibfnamefont {A.~S.}\ \bibnamefont {{Bell}}}, \bibinfo {author}
  {\bibfnamefont {B.~K.}\ \bibnamefont {{Berger}}}, \bibinfo {author}
  {\bibfnamefont {G.}~\bibnamefont {{Bergmann}}}, \bibinfo {author}
  {\bibfnamefont {J.~J.}\ \bibnamefont {{Bero}}}, \bibinfo {author}
  {\bibfnamefont {C.~P.~L.}\ \bibnamefont {{Berry}}}, \bibinfo {author}
  {\bibfnamefont {D.}~\bibnamefont {{Bersanetti}}}, \bibinfo {author}
  {\bibfnamefont {A.}~\bibnamefont {{Bertolini}}}, \bibinfo {author}
  {\bibfnamefont {J.}~\bibnamefont {{Betzwieser}}}, \bibinfo {author}
  {\bibfnamefont {S.}~\bibnamefont {{Bhagwat}}}, \bibinfo {author}
  {\bibfnamefont {R.}~\bibnamefont {{Bhandare}}}, \bibinfo {author}
  {\bibfnamefont {I.~A.}\ \bibnamefont {{Bilenko}}}, \bibinfo {author}
  {\bibfnamefont {G.}~\bibnamefont {{Billingsley}}}, \bibinfo {author}
  {\bibfnamefont {C.~R.}\ \bibnamefont {{Billman}}}, \bibinfo {author}
  {\bibfnamefont {J.}~\bibnamefont {{Birch}}}, \bibinfo {author} {\bibfnamefont
  {R.}~\bibnamefont {{Birney}}}, \bibinfo {author} {\bibfnamefont
  {O.}~\bibnamefont {{Birnholtz}}}, \bibinfo {author} {\bibfnamefont
  {S.}~\bibnamefont {{Biscans}}}, \bibinfo {author} {\bibfnamefont
  {S.}~\bibnamefont {{Biscoveanu}}}, \bibinfo {author} {\bibfnamefont
  {A.}~\bibnamefont {{Bisht}}}, \bibinfo {author} {\bibfnamefont
  {M.}~\bibnamefont {{Bitossi}}}, \bibinfo {author} {\bibfnamefont
  {C.}~\bibnamefont {{Biwer}}}, \bibinfo {author} {\bibfnamefont {M.~A.}\
  \bibnamefont {{Bizouard}}}, \bibinfo {author} {\bibfnamefont {J.~K.}\
  \bibnamefont {{Blackburn}}}, \bibinfo {author} {\bibfnamefont
  {J.}~\bibnamefont {{Blackman}}}, \bibinfo {author} {\bibfnamefont {C.~D.}\
  \bibnamefont {{Blair}}}, \bibinfo {author} {\bibfnamefont {D.~G.}\
  \bibnamefont {{Blair}}}, \bibinfo {author} {\bibfnamefont {R.~M.}\
  \bibnamefont {{Blair}}}, \bibinfo {author} {\bibfnamefont {S.}~\bibnamefont
  {{Bloemen}}}, \bibinfo {author} {\bibfnamefont {R.}~\bibnamefont {{Walet}}},
  \bibinfo {author} {\bibfnamefont {M.}~\bibnamefont {{Walker}}}, \bibinfo
  {author} {\bibfnamefont {L.}~\bibnamefont {{Wallace}}}, \bibinfo {author}
  {\bibfnamefont {S.}~\bibnamefont {{Walsh}}}, \bibinfo {author} {\bibfnamefont
  {G.}~\bibnamefont {{Wang}}}, \bibinfo {author} {\bibfnamefont
  {H.}~\bibnamefont {{Wang}}}, \bibinfo {author} {\bibfnamefont {J.~Z.}\
  \bibnamefont {{Wang}}}, \bibinfo {author} {\bibfnamefont {W.~H.}\
  \bibnamefont {{Wang}}}, \bibinfo {author} {\bibfnamefont {Y.~F.}\
  \bibnamefont {{Wang}}}, \bibinfo {author} {\bibfnamefont {R.~L.}\
  \bibnamefont {{Ward}}}, \bibinfo {author} {\bibfnamefont {J.}~\bibnamefont
  {{Warner}}}, \bibinfo {author} {\bibfnamefont {M.}~\bibnamefont {{Was}}},
  \bibinfo {author} {\bibfnamefont {J.}~\bibnamefont {{Watchi}}}, \bibinfo
  {author} {\bibfnamefont {B.}~\bibnamefont {{Weaver}}}, \bibinfo {author}
  {\bibfnamefont {L.~W.}\ \bibnamefont {{Wei}}}, \bibinfo {author}
  {\bibfnamefont {M.}~\bibnamefont {{Weinert}}}, \bibinfo {author}
  {\bibfnamefont {A.~J.}\ \bibnamefont {{Weinstein}}}, \bibinfo {author}
  {\bibfnamefont {R.}~\bibnamefont {{Weiss}}}, \bibinfo {author} {\bibfnamefont
  {L.}~\bibnamefont {{Wen}}}, \bibinfo {author} {\bibfnamefont {E.~K.}\
  \bibnamefont {{Wessel}}}, \bibinfo {author} {\bibfnamefont {P.}~\bibnamefont
  {{We{\ss}els}}}, \bibinfo {author} {\bibfnamefont {J.}~\bibnamefont
  {{Westerweck}}}, \bibinfo {author} {\bibfnamefont {T.}~\bibnamefont
  {{Westphal}}}, \bibinfo {author} {\bibfnamefont {K.}~\bibnamefont {{Wette}}},
  \bibinfo {author} {\bibfnamefont {J.~T.}\ \bibnamefont {{Whelan}}}, \bibinfo
  {author} {\bibfnamefont {S.~E.}\ \bibnamefont {{Whitcomb}}}, \bibinfo
  {author} {\bibfnamefont {B.~F.}\ \bibnamefont {{Whiting}}}, \bibinfo {author}
  {\bibfnamefont {C.}~\bibnamefont {{Whittle}}}, \bibinfo {author}
  {\bibfnamefont {D.}~\bibnamefont {{Wilken}}}, \bibinfo {author}
  {\bibfnamefont {D.}~\bibnamefont {{Williams}}}, \bibinfo {author}
  {\bibfnamefont {R.~D.}\ \bibnamefont {{Williams}}}, \bibinfo {author}
  {\bibfnamefont {A.~R.}\ \bibnamefont {{Williamson}}}, \bibinfo {author}
  {\bibfnamefont {J.~L.}\ \bibnamefont {{Willis}}}, \bibinfo {author}
  {\bibfnamefont {B.}~\bibnamefont {{Willke}}}, \bibinfo {author}
  {\bibfnamefont {M.~H.}\ \bibnamefont {{Wimmer}}}, \bibinfo {author}
  {\bibfnamefont {W.}~\bibnamefont {{Winkler}}}, \bibinfo {author}
  {\bibfnamefont {C.~C.}\ \bibnamefont {{Wipf}}}, \bibinfo {author}
  {\bibfnamefont {H.}~\bibnamefont {{Wittel}}}, \bibinfo {author}
  {\bibfnamefont {G.}~\bibnamefont {{Woan}}}, \bibinfo {author} {\bibfnamefont
  {J.}~\bibnamefont {{Woehler}}}, \bibinfo {author} {\bibfnamefont
  {J.}~\bibnamefont {{Wofford}}}, \bibinfo {author} {\bibfnamefont {K.~W.~K.}\
  \bibnamefont {{Wong}}}, \bibinfo {author} {\bibfnamefont {J.}~\bibnamefont
  {{Worden}}}, \bibinfo {author} {\bibfnamefont {J.~L.}\ \bibnamefont
  {{Wright}}}, \bibinfo {author} {\bibfnamefont {D.~S.}\ \bibnamefont {{Wu}}},
  \bibinfo {author} {\bibfnamefont {D.~M.}\ \bibnamefont {{Wysocki}}}, \bibinfo
  {author} {\bibfnamefont {S.}~\bibnamefont {{Xiao}}}, \bibinfo {author}
  {\bibfnamefont {H.}~\bibnamefont {{Yamamoto}}}, \bibinfo {author}
  {\bibfnamefont {C.~C.}\ \bibnamefont {{Yancey}}}, \bibinfo {author}
  {\bibfnamefont {L.}~\bibnamefont {{Yang}}}, \bibinfo {author} {\bibfnamefont
  {M.~J.}\ \bibnamefont {{Yap}}}, \bibinfo {author} {\bibfnamefont
  {M.}~\bibnamefont {{Yazback}}}, \bibinfo {author} {\bibfnamefont
  {H.}~\bibnamefont {{Yu}}}, \bibinfo {author} {\bibfnamefont {H.}~\bibnamefont
  {{Yu}}}, \bibinfo {author} {\bibfnamefont {M.}~\bibnamefont {{Yvert}}},
  \bibinfo {author} {\bibfnamefont {A.}~\bibnamefont {{Zadro{\.z}ny}}},
  \bibinfo {author} {\bibfnamefont {M.}~\bibnamefont {{Zanolin}}}, \bibinfo
  {author} {\bibfnamefont {T.}~\bibnamefont {{Zelenova}}}, \bibinfo {author}
  {\bibfnamefont {J.~P.}\ \bibnamefont {{Zendri}}}, \bibinfo {author}
  {\bibfnamefont {M.}~\bibnamefont {{Zevin}}}, \bibinfo {author} {\bibfnamefont
  {L.}~\bibnamefont {{Zhang}}}, \bibinfo {author} {\bibfnamefont
  {M.}~\bibnamefont {{Zhang}}}, \bibinfo {author} {\bibfnamefont
  {T.}~\bibnamefont {{Zhang}}}, \bibinfo {author} {\bibfnamefont {Y.~H.}\
  \bibnamefont {{Zhang}}}, \bibinfo {author} {\bibfnamefont {C.}~\bibnamefont
  {{Zhao}}}, \bibinfo {author} {\bibfnamefont {M.}~\bibnamefont {{Zhou}}},
  \bibinfo {author} {\bibfnamefont {Z.}~\bibnamefont {{Zhou}}}, \bibinfo
  {author} {\bibfnamefont {S.~J.}\ \bibnamefont {{Zhu}}}, \bibinfo {author}
  {\bibfnamefont {X.~J.}\ \bibnamefont {{Zhu}}}, \bibinfo {author}
  {\bibfnamefont {A.~B.}\ \bibnamefont {{Zimmerman}}}, \bibinfo {author}
  {\bibfnamefont {M.~E.}\ \bibnamefont {{Zucker}}}, \bibinfo {author}
  {\bibfnamefont {J.}~\bibnamefont {{Zweizig}}}, \bibinfo {author}
  {\bibnamefont {{(LIGO Scientific Collaboration}}}, \bibinfo {author}
  {\bibnamefont {{Virgo Collaboration}}}, \bibinfo {author} {\bibfnamefont
  {E.}~\bibnamefont {{Burns}}}, \bibinfo {author} {\bibfnamefont
  {P.}~\bibnamefont {{Veres}}}, \bibinfo {author} {\bibfnamefont
  {D.}~\bibnamefont {{Kocevski}}}, \bibinfo {author} {\bibfnamefont
  {J.}~\bibnamefont {{Racusin}}}, \bibinfo {author} {\bibfnamefont
  {A.}~\bibnamefont {{Goldstein}}}, \bibinfo {author} {\bibfnamefont
  {V.}~\bibnamefont {{Connaughton}}}, \bibinfo {author} {\bibfnamefont {M.~S.}\
  \bibnamefont {{Briggs}}}, \bibinfo {author} {\bibfnamefont {L.}~\bibnamefont
  {{Blackburn}}}, \bibinfo {author} {\bibfnamefont {R.}~\bibnamefont
  {{Hamburg}}}, \bibinfo {author} {\bibfnamefont {C.~M.}\ \bibnamefont
  {{Hui}}}, \bibinfo {author} {\bibfnamefont {A.}~\bibnamefont {{von
  Kienlin}}}, \bibinfo {author} {\bibfnamefont {J.}~\bibnamefont {{McEnery}}},
  \bibinfo {author} {\bibfnamefont {R.~D.}\ \bibnamefont {{Preece}}}, \bibinfo
  {author} {\bibfnamefont {C.~A.}\ \bibnamefont {{Wilson-Hodge}}}, \bibinfo
  {author} {\bibfnamefont {E.}~\bibnamefont {{Bissaldi}}}, \bibinfo {author}
  {\bibfnamefont {W.~H.}\ \bibnamefont {{Cleveland}}}, \bibinfo {author}
  {\bibfnamefont {M.~H.}\ \bibnamefont {{Gibby}}}, \bibinfo {author}
  {\bibfnamefont {M.~M.}\ \bibnamefont {{Giles}}}, \bibinfo {author}
  {\bibfnamefont {R.~M.}\ \bibnamefont {{Kippen}}}, \bibinfo {author}
  {\bibfnamefont {S.}~\bibnamefont {{McBreen}}}, \bibinfo {author}
  {\bibfnamefont {C.~A.}\ \bibnamefont {{Meegan}}}, \bibinfo {author}
  {\bibfnamefont {W.~S.}\ \bibnamefont {{Paciesas}}}, \bibinfo {author}
  {\bibfnamefont {S.}~\bibnamefont {{Poolakkil}}}, \bibinfo {author}
  {\bibfnamefont {O.~J.}\ \bibnamefont {{Roberts}}}, \bibinfo {author}
  {\bibfnamefont {M.}~\bibnamefont {{Stanbro}}}, \bibinfo {author}
  {\bibfnamefont {F.}~\bibnamefont {{Gamma-ray Burst Monitor}}}, \bibinfo
  {author} {\bibfnamefont {V.}~\bibnamefont {{Savchenko}}}, \bibinfo {author}
  {\bibfnamefont {C.}~\bibnamefont {{Ferrigno}}}, \bibinfo {author}
  {\bibfnamefont {E.}~\bibnamefont {{Kuulkers}}}, \bibinfo {author}
  {\bibfnamefont {A.}~\bibnamefont {{Bazzano}}}, \bibinfo {author}
  {\bibfnamefont {E.}~\bibnamefont {{Bozzo}}}, \bibinfo {author} {\bibfnamefont
  {S.}~\bibnamefont {{Brandt}}}, \bibinfo {author} {\bibfnamefont
  {J.}~\bibnamefont {{Chenevez}}}, \bibinfo {author} {\bibfnamefont {T.~J.~L.}\
  \bibnamefont {{Courvoisier}}}, \bibinfo {author} {\bibfnamefont
  {R.}~\bibnamefont {{Diehl}}}, \bibinfo {author} {\bibfnamefont
  {A.}~\bibnamefont {{Domingo}}}, \bibinfo {author} {\bibfnamefont
  {L.}~\bibnamefont {{Hanlon}}}, \bibinfo {author} {\bibfnamefont
  {E.}~\bibnamefont {{Jourdain}}}, \bibinfo {author} {\bibfnamefont
  {P.}~\bibnamefont {{Laurent}}}, \bibinfo {author} {\bibfnamefont
  {F.}~\bibnamefont {{Lebrun}}}, \bibinfo {author} {\bibfnamefont
  {A.}~\bibnamefont {{Lutovinov}}}, \bibinfo {author} {\bibfnamefont
  {S.}~\bibnamefont {{Mereghetti}}}, \bibinfo {author} {\bibfnamefont
  {L.}~\bibnamefont {{Natalucci}}}, \bibinfo {author} {\bibfnamefont
  {J.}~\bibnamefont {{Rodi}}}, \bibinfo {author} {\bibfnamefont {J.~P.}\
  \bibnamefont {{Roques}}}, \bibinfo {author} {\bibfnamefont {R.}~\bibnamefont
  {{Sunyaev}}}, \bibinfo {author} {\bibfnamefont {P.}~\bibnamefont
  {{Ubertini}}}, \ and\ \bibinfo {author} {\bibnamefont {{(INTEGRAL}}},\ }\href
  {\doibase 10.3847/2041-8213/aa920c} {\bibfield  {journal} {\bibinfo
  {journal} {\apjl}\ }\textbf {\bibinfo {volume} {848}},\ \bibinfo {eid} {L13}
  (\bibinfo {year} {2017}{\natexlab{b}})},\ \Eprint
  {http://arxiv.org/abs/1710.05834} {arXiv:1710.05834 [astro-ph.HE]}
  \BibitemShut {NoStop}%
\bibitem [{\citenamefont {{Zucca}}\ \emph {et~al.}(2019)\citenamefont
  {{Zucca}}, \citenamefont {{Pogosian}}, \citenamefont {{Silvestri}},\ and\
  \citenamefont {{Zhao}}}]{zucca19a}%
  \BibitemOpen
  \bibfield  {author} {\bibinfo {author} {\bibfnamefont {A.}~\bibnamefont
  {{Zucca}}}, \bibinfo {author} {\bibfnamefont {L.}~\bibnamefont {{Pogosian}}},
  \bibinfo {author} {\bibfnamefont {A.}~\bibnamefont {{Silvestri}}}, \ and\
  \bibinfo {author} {\bibfnamefont {G.~B.}\ \bibnamefont {{Zhao}}},\ }\href
  {\doibase 10.1088/1475-7516/2019/05/001} {\bibfield  {journal} {\bibinfo
  {journal} {\jcap}\ }\textbf {\bibinfo {volume} {2019}},\ \bibinfo {eid} {001}
  (\bibinfo {year} {2019})},\ \Eprint {http://arxiv.org/abs/1901.05956}
  {arXiv:1901.05956 [astro-ph.CO]} \BibitemShut {NoStop}%
\bibitem [{\citenamefont {{Shafieloo}}\ \emph {et~al.}(2012)\citenamefont
  {{Shafieloo}}, \citenamefont {{Sahni}},\ and\ \citenamefont
  {{Starobinsky}}}]{shafieloo12a}%
  \BibitemOpen
  \bibfield  {author} {\bibinfo {author} {\bibfnamefont {A.}~\bibnamefont
  {{Shafieloo}}}, \bibinfo {author} {\bibfnamefont {V.}~\bibnamefont
  {{Sahni}}}, \ and\ \bibinfo {author} {\bibfnamefont {A.~A.}\ \bibnamefont
  {{Starobinsky}}},\ }\href {\doibase 10.1103/PhysRevD.86.103527} {\bibfield
  {journal} {\bibinfo  {journal} {\prd}\ }\textbf {\bibinfo {volume} {86}},\
  \bibinfo {eid} {103527} (\bibinfo {year} {2012})},\ \Eprint
  {http://arxiv.org/abs/1205.2870} {arXiv:1205.2870 [astro-ph.CO]} \BibitemShut
  {NoStop}%
\bibitem [{\citenamefont {{Sahni}}\ \emph {et~al.}(2014)\citenamefont
  {{Sahni}}, \citenamefont {{Shafieloo}},\ and\ \citenamefont
  {{Starobinsky}}}]{sahni14a}%
  \BibitemOpen
  \bibfield  {author} {\bibinfo {author} {\bibfnamefont {V.}~\bibnamefont
  {{Sahni}}}, \bibinfo {author} {\bibfnamefont {A.}~\bibnamefont
  {{Shafieloo}}}, \ and\ \bibinfo {author} {\bibfnamefont {A.~A.}\ \bibnamefont
  {{Starobinsky}}},\ }\href {\doibase 10.1088/2041-8205/793/2/L40} {\bibfield
  {journal} {\bibinfo  {journal} {\apjl}\ }\textbf {\bibinfo {volume} {793}},\
  \bibinfo {eid} {L40} (\bibinfo {year} {2014})},\ \Eprint
  {http://arxiv.org/abs/1406.2209} {arXiv:1406.2209 [astro-ph.CO]} \BibitemShut
  {NoStop}%
\bibitem [{\citenamefont {{Heymans}}\ \emph {et~al.}(2020)\citenamefont
  {{Heymans}}, \citenamefont {{Tr{\"o}ster}}, \citenamefont {{Asgari}},
  \citenamefont {{Blake}}, \citenamefont {{Hildebrandt}}, \citenamefont
  {{Joachimi}}, \citenamefont {{Kuijken}}, \citenamefont {{Lin}}, \citenamefont
  {{S{\'a}nchez}}, \citenamefont {{van den Busch}}, \citenamefont {{Wright}},
  \citenamefont {{Amon}}, \citenamefont {{Bilicki}}, \citenamefont {{de Jong}},
  \citenamefont {{Crocce}}, \citenamefont {{Dvornik}}, \citenamefont {{Erben}},
  \citenamefont {{Getman}}, \citenamefont {{Giblin}}, \citenamefont
  {{Glazebrook}}, \citenamefont {{Hoekstra}}, \citenamefont {{Joudaki}},
  \citenamefont {{Kannawadi}}, \citenamefont {{Lidman}}, \citenamefont
  {{K{\"o}hlinger}}, \citenamefont {{Miller}}, \citenamefont {{Napolitano}},
  \citenamefont {{Parkinson}}, \citenamefont {{Schneider}}, \citenamefont
  {{Shan}},\ and\ \citenamefont {{Wolf}}}]{heymans20a}%
  \BibitemOpen
  \bibfield  {author} {\bibinfo {author} {\bibfnamefont {C.}~\bibnamefont
  {{Heymans}}}, \bibinfo {author} {\bibfnamefont {T.}~\bibnamefont
  {{Tr{\"o}ster}}}, \bibinfo {author} {\bibfnamefont {M.}~\bibnamefont
  {{Asgari}}}, \bibinfo {author} {\bibfnamefont {C.}~\bibnamefont {{Blake}}},
  \bibinfo {author} {\bibfnamefont {H.}~\bibnamefont {{Hildebrandt}}}, \bibinfo
  {author} {\bibfnamefont {B.}~\bibnamefont {{Joachimi}}}, \bibinfo {author}
  {\bibfnamefont {K.}~\bibnamefont {{Kuijken}}}, \bibinfo {author}
  {\bibfnamefont {C.-A.}\ \bibnamefont {{Lin}}}, \bibinfo {author}
  {\bibfnamefont {A.~G.}\ \bibnamefont {{S{\'a}nchez}}}, \bibinfo {author}
  {\bibfnamefont {J.~L.}\ \bibnamefont {{van den Busch}}}, \bibinfo {author}
  {\bibfnamefont {A.~H.}\ \bibnamefont {{Wright}}}, \bibinfo {author}
  {\bibfnamefont {A.}~\bibnamefont {{Amon}}}, \bibinfo {author} {\bibfnamefont
  {M.}~\bibnamefont {{Bilicki}}}, \bibinfo {author} {\bibfnamefont
  {J.}~\bibnamefont {{de Jong}}}, \bibinfo {author} {\bibfnamefont
  {M.}~\bibnamefont {{Crocce}}}, \bibinfo {author} {\bibfnamefont
  {A.}~\bibnamefont {{Dvornik}}}, \bibinfo {author} {\bibfnamefont
  {T.}~\bibnamefont {{Erben}}}, \bibinfo {author} {\bibfnamefont
  {F.}~\bibnamefont {{Getman}}}, \bibinfo {author} {\bibfnamefont
  {B.}~\bibnamefont {{Giblin}}}, \bibinfo {author} {\bibfnamefont
  {K.}~\bibnamefont {{Glazebrook}}}, \bibinfo {author} {\bibfnamefont
  {H.}~\bibnamefont {{Hoekstra}}}, \bibinfo {author} {\bibfnamefont
  {S.}~\bibnamefont {{Joudaki}}}, \bibinfo {author} {\bibfnamefont
  {A.}~\bibnamefont {{Kannawadi}}}, \bibinfo {author} {\bibfnamefont
  {C.}~\bibnamefont {{Lidman}}}, \bibinfo {author} {\bibfnamefont
  {F.}~\bibnamefont {{K{\"o}hlinger}}}, \bibinfo {author} {\bibfnamefont
  {L.}~\bibnamefont {{Miller}}}, \bibinfo {author} {\bibfnamefont {N.~R.}\
  \bibnamefont {{Napolitano}}}, \bibinfo {author} {\bibfnamefont
  {D.}~\bibnamefont {{Parkinson}}}, \bibinfo {author} {\bibfnamefont
  {P.}~\bibnamefont {{Schneider}}}, \bibinfo {author} {\bibfnamefont
  {H.}~\bibnamefont {{Shan}}}, \ and\ \bibinfo {author} {\bibfnamefont
  {C.}~\bibnamefont {{Wolf}}},\ }\href@noop {} {\bibfield  {journal} {\bibinfo
  {journal} {arXiv e-prints}\ ,\ \bibinfo {eid} {arXiv:2007.15632}} (\bibinfo
  {year} {2020})},\ \Eprint {http://arxiv.org/abs/2007.15632} {arXiv:2007.15632
  [astro-ph.CO]} \BibitemShut {NoStop}%
\bibitem [{\citenamefont {{Tr{\"o}ster}}\ \emph {et~al.}(2020)\citenamefont
  {{Tr{\"o}ster}}, \citenamefont {{S{\'a}nchez}}, \citenamefont {{Asgari}},
  \citenamefont {{Blake}}, \citenamefont {{Crocce}}, \citenamefont {{Heymans}},
  \citenamefont {{Hildebrandt}}, \citenamefont {{Joachimi}}, \citenamefont
  {{Joudaki}}, \citenamefont {{Kannawadi}}, \citenamefont {{Lin}},\ and\
  \citenamefont {{Wright}}}]{troster20a}%
  \BibitemOpen
  \bibfield  {author} {\bibinfo {author} {\bibfnamefont {T.}~\bibnamefont
  {{Tr{\"o}ster}}}, \bibinfo {author} {\bibfnamefont {A.~G.}\ \bibnamefont
  {{S{\'a}nchez}}}, \bibinfo {author} {\bibfnamefont {M.}~\bibnamefont
  {{Asgari}}}, \bibinfo {author} {\bibfnamefont {C.}~\bibnamefont {{Blake}}},
  \bibinfo {author} {\bibfnamefont {M.}~\bibnamefont {{Crocce}}}, \bibinfo
  {author} {\bibfnamefont {C.}~\bibnamefont {{Heymans}}}, \bibinfo {author}
  {\bibfnamefont {H.}~\bibnamefont {{Hildebrandt}}}, \bibinfo {author}
  {\bibfnamefont {B.}~\bibnamefont {{Joachimi}}}, \bibinfo {author}
  {\bibfnamefont {S.}~\bibnamefont {{Joudaki}}}, \bibinfo {author}
  {\bibfnamefont {A.}~\bibnamefont {{Kannawadi}}}, \bibinfo {author}
  {\bibfnamefont {C.-A.}\ \bibnamefont {{Lin}}}, \ and\ \bibinfo {author}
  {\bibfnamefont {A.}~\bibnamefont {{Wright}}},\ }\href {\doibase
  10.1051/0004-6361/201936772} {\bibfield  {journal} {\bibinfo  {journal}
  {\aap}\ }\textbf {\bibinfo {volume} {633}},\ \bibinfo {eid} {L10} (\bibinfo
  {year} {2020})},\ \Eprint {http://arxiv.org/abs/1909.11006} {arXiv:1909.11006
  [astro-ph.CO]} \BibitemShut {NoStop}%
\bibitem [{\citenamefont {{d'Amico}}\ \emph {et~al.}(2020)\citenamefont
  {{d'Amico}}, \citenamefont {{Gleyzes}}, \citenamefont {{Kokron}},
  \citenamefont {{Markovic}}, \citenamefont {{Senatore}}, \citenamefont
  {{Zhang}}, \citenamefont {{Beutler}},\ and\ \citenamefont
  {{Gil-Mar{\'\i}n}}}]{damico20a}%
  \BibitemOpen
  \bibfield  {author} {\bibinfo {author} {\bibfnamefont {G.}~\bibnamefont
  {{d'Amico}}}, \bibinfo {author} {\bibfnamefont {J.}~\bibnamefont
  {{Gleyzes}}}, \bibinfo {author} {\bibfnamefont {N.}~\bibnamefont {{Kokron}}},
  \bibinfo {author} {\bibfnamefont {K.}~\bibnamefont {{Markovic}}}, \bibinfo
  {author} {\bibfnamefont {L.}~\bibnamefont {{Senatore}}}, \bibinfo {author}
  {\bibfnamefont {P.}~\bibnamefont {{Zhang}}}, \bibinfo {author} {\bibfnamefont
  {F.}~\bibnamefont {{Beutler}}}, \ and\ \bibinfo {author} {\bibfnamefont
  {H.}~\bibnamefont {{Gil-Mar{\'\i}n}}},\ }\href {\doibase
  10.1088/1475-7516/2020/05/005} {\bibfield  {journal} {\bibinfo  {journal}
  {\jcap}\ }\textbf {\bibinfo {volume} {2020}},\ \bibinfo {eid} {005} (\bibinfo
  {year} {2020})},\ \Eprint {http://arxiv.org/abs/1909.05271} {arXiv:1909.05271
  [astro-ph.CO]} \BibitemShut {NoStop}%
\bibitem [{\citenamefont {{Ivanov}}\ \emph {et~al.}(2020)\citenamefont
  {{Ivanov}}, \citenamefont {{Simonovi{\'c}}},\ and\ \citenamefont
  {{Zaldarriaga}}}]{ivanov20a}%
  \BibitemOpen
  \bibfield  {author} {\bibinfo {author} {\bibfnamefont {M.~M.}\ \bibnamefont
  {{Ivanov}}}, \bibinfo {author} {\bibfnamefont {M.}~\bibnamefont
  {{Simonovi{\'c}}}}, \ and\ \bibinfo {author} {\bibfnamefont {M.}~\bibnamefont
  {{Zaldarriaga}}},\ }\href {\doibase 10.1088/1475-7516/2020/05/042} {\bibfield
   {journal} {\bibinfo  {journal} {\jcap}\ }\textbf {\bibinfo {volume}
  {2020}},\ \bibinfo {eid} {042} (\bibinfo {year} {2020})},\ \Eprint
  {http://arxiv.org/abs/1909.05277} {arXiv:1909.05277 [astro-ph.CO]}
  \BibitemShut {NoStop}%
\bibitem [{\citenamefont {{Albrecht}}\ \emph {et~al.}(2006)\citenamefont
  {{Albrecht}}, \citenamefont {{Bernstein}}, \citenamefont {{Cahn}},
  \citenamefont {{Freedman}}, \citenamefont {{Hewitt}}, \citenamefont {{Hu}},
  \citenamefont {{Huth}}, \citenamefont {{Kamionkowski}}, \citenamefont
  {{Kolb}}, \citenamefont {{Knox}}, \citenamefont {{Mather}}, \citenamefont
  {{Staggs}},\ and\ \citenamefont {{Suntzeff}}}]{albrecht06a}%
  \BibitemOpen
  \bibfield  {author} {\bibinfo {author} {\bibfnamefont {A.}~\bibnamefont
  {{Albrecht}}}, \bibinfo {author} {\bibfnamefont {G.}~\bibnamefont
  {{Bernstein}}}, \bibinfo {author} {\bibfnamefont {R.}~\bibnamefont {{Cahn}}},
  \bibinfo {author} {\bibfnamefont {W.~L.}\ \bibnamefont {{Freedman}}},
  \bibinfo {author} {\bibfnamefont {J.}~\bibnamefont {{Hewitt}}}, \bibinfo
  {author} {\bibfnamefont {W.}~\bibnamefont {{Hu}}}, \bibinfo {author}
  {\bibfnamefont {J.}~\bibnamefont {{Huth}}}, \bibinfo {author} {\bibfnamefont
  {M.}~\bibnamefont {{Kamionkowski}}}, \bibinfo {author} {\bibfnamefont
  {E.~W.}\ \bibnamefont {{Kolb}}}, \bibinfo {author} {\bibfnamefont
  {L.}~\bibnamefont {{Knox}}}, \bibinfo {author} {\bibfnamefont {J.~C.}\
  \bibnamefont {{Mather}}}, \bibinfo {author} {\bibfnamefont {S.}~\bibnamefont
  {{Staggs}}}, \ and\ \bibinfo {author} {\bibfnamefont {N.~B.}\ \bibnamefont
  {{Suntzeff}}},\ }\href@noop {} {\bibfield  {journal} {\bibinfo  {journal}
  {arXiv}\ } (\bibinfo {year} {2006})},\ \Eprint
  {http://arxiv.org/abs/astro-ph/0609591} {astro-ph/0609591} \BibitemShut
  {NoStop}%
\bibitem [{\citenamefont {{Abe}}\ \emph {et~al.}(2014)\citenamefont {{Abe}},
  \citenamefont {{Adam}}, \citenamefont {{Aihara}}, \citenamefont {{Akiri}},
  \citenamefont {{Andreopoulos}}, \citenamefont {{Aoki}}, \citenamefont
  {{Ariga}}, \citenamefont {{Ariga}}, \citenamefont {{Assylbekov}},
  \citenamefont {{Autiero}}, \citenamefont {{Barbi}}, \citenamefont {{Barker}},
  \citenamefont {{Barr}}, \citenamefont {{Bass}}, \citenamefont {{Batkiewicz}},
  \citenamefont {{Bay}}, \citenamefont {{Bentham}}, \citenamefont {{Berardi}},
  \citenamefont {{Berger}}, \citenamefont {{Berkman}}, \citenamefont
  {{Bertram}}, \citenamefont {{Bhadra}}, \citenamefont {{Blaszczyk}},
  \citenamefont {{Blondel}}, \citenamefont {{Bojechko}}, \citenamefont
  {{Bordoni}}, \citenamefont {{Boyd}}, \citenamefont {{Brailsford}},
  \citenamefont {{Bravar}}, \citenamefont {{Bronner}}, \citenamefont
  {{Buchanan}}, \citenamefont {{Calland}}, \citenamefont {{Caravaca
  Rodr{\'\i}guez}}, \citenamefont {{Cartwright}}, \citenamefont {{Castillo}},
  \citenamefont {{Catanesi}}, \citenamefont {{Cervera}}, \citenamefont
  {{Cherdack}}, \citenamefont {{Christodoulou}}, \citenamefont {{Clifton}},
  \citenamefont {{Coleman}}, \citenamefont {{Coleman}}, \citenamefont
  {{Collazuol}}, \citenamefont {{Connolly}}, \citenamefont {{Cremonesi}},
  \citenamefont {{Dabrowska}}, \citenamefont {{Danko}}, \citenamefont {{Das}},
  \citenamefont {{Davis}}, \citenamefont {{de Perio}}, \citenamefont {{De
  Rosa}}, \citenamefont {{Dealtry}}, \citenamefont {{Dennis}}, \citenamefont
  {{Densham}}, \citenamefont {{Di Lodovico}}, \citenamefont {{Di Luise}},
  \citenamefont {{Drapier}}, \citenamefont {{Duboyski}}, \citenamefont
  {{Duffy}}, \citenamefont {{Dufour}}, \citenamefont {{Dumarchez}},
  \citenamefont {{Dytman}}, \citenamefont {{Dziewiecki}}, \citenamefont
  {{Emery}}, \citenamefont {{Ereditato}}, \citenamefont {{Escudero}},
  \citenamefont {{Finch}}, \citenamefont {{Floetotto}}, \citenamefont
  {{Friend}}, \citenamefont {{Fujii}}, \citenamefont {{Fukuda}}, \citenamefont
  {{Furmanski}}, \citenamefont {{Galymov}}, \citenamefont {{Gaudin}},
  \citenamefont {{Giffin}}, \citenamefont {{Giganti}}, \citenamefont {{Gilje}},
  \citenamefont {{Goeldi}}, \citenamefont {{Golan}}, \citenamefont
  {{Gomez-Cadenas}}, \citenamefont {{Gonin}}, \citenamefont {{Grant}},
  \citenamefont {{Gudin}}, \citenamefont {{Hadley}}, \citenamefont {{Haesler}},
  \citenamefont {{Haigh}}, \citenamefont {{Hamilton}}, \citenamefont
  {{Hansen}}, \citenamefont {{Hara}}, \citenamefont {{Hartz}}, \citenamefont
  {{Hasegawa}}, \citenamefont {{Hastings}}, \citenamefont {{Hayato}},
  \citenamefont {{Hearty}}, \citenamefont {{Helmer}}, \citenamefont
  {{Hierholzer}}, \citenamefont {{Hignight}}, \citenamefont {{Hillairet}},
  \citenamefont {{Himmel}}, \citenamefont {{Hiraki}}, \citenamefont {{Hirota}},
  \citenamefont {{Holeczek}}, \citenamefont {{Horikawa}}, \citenamefont
  {{Huang}}, \citenamefont {{Ichikawa}}, \citenamefont {{Ieki}}, \citenamefont
  {{Ieva}}, \citenamefont {{Ikeda}}, \citenamefont {{Imber}}, \citenamefont
  {{Insler}}, \citenamefont {{Irvine}}, \citenamefont {{Ishida}}, \citenamefont
  {{Ishii}}, \citenamefont {{Ives}}, \citenamefont {{Iyogi}}, \citenamefont
  {{Izmaylov}}, \citenamefont {{Jacob}}, \citenamefont {{Jamieson}},
  \citenamefont {{Johnson}}, \citenamefont {{Jo}}, \citenamefont {{Jonsson}},
  \citenamefont {{Jung}}, \citenamefont {{Kaboth}}, \citenamefont {{Kajita}},
  \citenamefont {{Kakuno}}, \citenamefont {{Kameda}}, \citenamefont
  {{Kanazawa}}, \citenamefont {{Karlen}}, \citenamefont {{Karpikov}},
  \citenamefont {{Kearns}}, \citenamefont {{Khabibullin}}, \citenamefont
  {{Khotjantsev}}, \citenamefont {{Kielczewska}}, \citenamefont {{Kikawa}},
  \citenamefont {{Kilinski}}, \citenamefont {{Kim}}, \citenamefont {{Kisiel}},
  \citenamefont {{Kitching}}, \citenamefont {{Kobayashi}}, \citenamefont
  {{Koch}}, \citenamefont {{Kolaceke}}, \citenamefont {{Konaka}}, \citenamefont
  {{Kormos}}, \citenamefont {{Korzenev}}, \citenamefont {{Koseki}},
  \citenamefont {{Koshio}}, \citenamefont {{Kreslo}}, \citenamefont {{Kropp}},
  \citenamefont {{Kubo}}, \citenamefont {{Kudenko}}, \citenamefont
  {{Kumaratunga}}, \citenamefont {{Kurjata}}, \citenamefont {{Kutter}},
  \citenamefont {{Lagoda}}, \citenamefont {{Laihem}}, \citenamefont {{Lamont}},
  \citenamefont {{Laveder}}, \citenamefont {{Lawe}}, \citenamefont {{Lazos}},
  \citenamefont {{Lee}}, \citenamefont {{Licciardi}}, \citenamefont
  {{Lindner}}, \citenamefont {{Lister}}, \citenamefont {{Litchfield}},
  \citenamefont {{Longhin}}, \citenamefont {{Ludovici}}, \citenamefont
  {{Macaire}}, \citenamefont {{Magaletti}}, \citenamefont {{Mahn}},
  \citenamefont {{Malek}}, \citenamefont {{Manly}}, \citenamefont {{Marino}},
  \citenamefont {{Marteau}}, \citenamefont {{Martin}}, \citenamefont
  {{Maruyama}}, \citenamefont {{Marzec}}, \citenamefont {{Mathie}},
  \citenamefont {{Matveev}}, \citenamefont {{Mavrokoridis}}, \citenamefont
  {{Mazzucato}}, \citenamefont {{McCarthy}}, \citenamefont {{McCauley}},
  \citenamefont {{McFarland }}, \citenamefont {{McGrew}}, \citenamefont
  {{Metelko}}, \citenamefont {{Mezzetto}}, \citenamefont {{Mijakowski}},
  \citenamefont {{Miller}}, \citenamefont {{Minamino}}, \citenamefont
  {{Mineev}}, \citenamefont {{Mine}}, \citenamefont {{Missert}}, \citenamefont
  {{Miura}}, \citenamefont {{Monfregola}}, \citenamefont {{Moriyama}},
  \citenamefont {{Mueller}}, \citenamefont {{Murakami}}, \citenamefont
  {{Murdoch}}, \citenamefont {{Murphy}}, \citenamefont {{Myslik}},
  \citenamefont {{Nagasaki}}, \citenamefont {{Nakadaira}}, \citenamefont
  {{Nakahata}}, \citenamefont {{Nakai}}, \citenamefont {{Nakamura}},
  \citenamefont {{Nakayama}}, \citenamefont {{Nakaya}}, \citenamefont
  {{Nakayoshi}}, \citenamefont {{Naples}}, \citenamefont {{Nielsen}},
  \citenamefont {{Nirkko}}, \citenamefont {{Nishikawa}}, \citenamefont
  {{Nishimura}}, \citenamefont {{O'Keeffe}}, \citenamefont {{Ohta}},
  \citenamefont {{Okumura}}, \citenamefont {{Okusawa}}, \citenamefont
  {{Oryszczak}}, \citenamefont {{Oser}}, \citenamefont {{Owen}}, \citenamefont
  {{Oyama}}, \citenamefont {{Palladino}}, \citenamefont {{Paolone}},
  \citenamefont {{Payne}}, \citenamefont {{Pearce}}, \citenamefont
  {{Perevozchikov}}, \citenamefont {{Perkin}}, \citenamefont {{Petrov}},
  \citenamefont {{Pickard}}, \citenamefont {{Pinzon Guerra}}, \citenamefont
  {{Pistillo}}, \citenamefont {{Plonski}}, \citenamefont {{Poplawska}},
  \citenamefont {{Popov}}, \citenamefont {{Posiadala}}, \citenamefont
  {{Poutissou}}, \citenamefont {{Poutissou}}, \citenamefont {{Przewlocki}},
  \citenamefont {{Quilain}}, \citenamefont {{Radicioni}}, \citenamefont
  {{Ratoff}}, \citenamefont {{Ravonel}}, \citenamefont {{Rayner}},
  \citenamefont {{Redij}}, \citenamefont {{Reeves}}, \citenamefont
  {{Reinherz-Aronis}}, \citenamefont {{Retiere}}, \citenamefont {{Robert}},
  \citenamefont {{Rodrigues}}, \citenamefont {{Rojas}}, \citenamefont
  {{Rondio}}, \citenamefont {{Roth}}, \citenamefont {{Rubbia}}, \citenamefont
  {{Ruterbories}}, \citenamefont {{Sacco}}, \citenamefont {{Sakashita}},
  \citenamefont {{S{\'a}nchez}}, \citenamefont {{Sato}}, \citenamefont
  {{Scantamburlo}}, \citenamefont {{Scholberg}}, \citenamefont {{Schwehr}},
  \citenamefont {{Scott}}, \citenamefont {{Seiya}}, \citenamefont
  {{Sekiguchi}}, \citenamefont {{Sekiya}}, \citenamefont {{Sgalaberna}},
  \citenamefont {{Shiozawa}}, \citenamefont {{Short}}, \citenamefont
  {{Shustrov}}, \citenamefont {{Sinclair}}, \citenamefont {{Smith}},
  \citenamefont {{Smith}}, \citenamefont {{Smy}}, \citenamefont {{Sobczyk}},
  \citenamefont {{Sobel}}, \citenamefont {{Sorel}}, \citenamefont
  {{Southwell}}, \citenamefont {{Stamoulis}}, \citenamefont {{Steinmann}},
  \citenamefont {{Still}}, \citenamefont {{Suda}}, \citenamefont {{Suzuki}},
  \citenamefont {{Suzuki}}, \citenamefont {{Suzuki}}, \citenamefont {{Suzuki}},
  \citenamefont {{Szeglowski}}, \citenamefont {{Tacik}}, \citenamefont
  {{Tada}}, \citenamefont {{Takahashi}}, \citenamefont {{Takeda}},
  \citenamefont {{Takeuchi}}, \citenamefont {{Tanaka}}, \citenamefont
  {{Tanaka}}, \citenamefont {{Tanaka}}, \citenamefont {{Terhorst}},
  \citenamefont {{Terri}}, \citenamefont {{Thompson}}, \citenamefont
  {{Thorley}}, \citenamefont {{Tobayama}}, \citenamefont {{Toki}},
  \citenamefont {{Tomura}}, \citenamefont {{Totsuka}}, \citenamefont
  {{Touramanis}}, \citenamefont {{Tsukamoto}}, \citenamefont {{Tzanov}},
  \citenamefont {{Uchida}}, \citenamefont {{Ueno}}, \citenamefont {{Vacheret}},
  \citenamefont {{Vagins}}, \citenamefont {{Vasseur}}, \citenamefont
  {{Wachala}}, \citenamefont {{Waldron}}, \citenamefont {{Walter}},
  \citenamefont {{Wark}}, \citenamefont {{Wascko}}, \citenamefont {{Weber}},
  \citenamefont {{Wendell}}, \citenamefont {{Wilkes}}, \citenamefont
  {{Wilking}}, \citenamefont {{Wilkinson}}, \citenamefont {{Williamson}},
  \citenamefont {{Wilson}}, \citenamefont {{Wilson}}, \citenamefont
  {{Wongjirad}}, \citenamefont {{Yamada}}, \citenamefont {{Yamamoto}},
  \citenamefont {{Yanagisawa}}, \citenamefont {{Yen}}, \citenamefont
  {{Yershov}}, \citenamefont {{Yokoyama}}, \citenamefont {{Yuan}},
  \citenamefont {{Zalewska}}, \citenamefont {{Zalipska}}, \citenamefont
  {{Zambelli}}, \citenamefont {{Zaremba}}, \citenamefont {{Ziembicki}},
  \citenamefont {{Zimmerman}}, \citenamefont {{Zito}}, \citenamefont
  {{{\.Z}muda}},\ and\ \citenamefont {{T2K Collaboration}}}]{1311.4750}%
  \BibitemOpen
  \bibfield  {author} {\bibinfo {author} {\bibfnamefont {K.}~\bibnamefont
  {{Abe}}}, \bibinfo {author} {\bibfnamefont {J.}~\bibnamefont {{Adam}}},
  \bibinfo {author} {\bibfnamefont {H.}~\bibnamefont {{Aihara}}}, \bibinfo
  {author} {\bibfnamefont {T.}~\bibnamefont {{Akiri}}}, \bibinfo {author}
  {\bibfnamefont {C.}~\bibnamefont {{Andreopoulos}}}, \bibinfo {author}
  {\bibfnamefont {S.}~\bibnamefont {{Aoki}}}, \bibinfo {author} {\bibfnamefont
  {A.}~\bibnamefont {{Ariga}}}, \bibinfo {author} {\bibfnamefont
  {T.}~\bibnamefont {{Ariga}}}, \bibinfo {author} {\bibfnamefont
  {S.}~\bibnamefont {{Assylbekov}}}, \bibinfo {author} {\bibfnamefont
  {D.}~\bibnamefont {{Autiero}}}, \bibinfo {author} {\bibfnamefont
  {M.}~\bibnamefont {{Barbi}}}, \bibinfo {author} {\bibfnamefont {G.~J.}\
  \bibnamefont {{Barker}}}, \bibinfo {author} {\bibfnamefont {G.}~\bibnamefont
  {{Barr}}}, \bibinfo {author} {\bibfnamefont {M.}~\bibnamefont {{Bass}}},
  \bibinfo {author} {\bibfnamefont {M.}~\bibnamefont {{Batkiewicz}}}, \bibinfo
  {author} {\bibfnamefont {F.}~\bibnamefont {{Bay}}}, \bibinfo {author}
  {\bibfnamefont {S.~W.}\ \bibnamefont {{Bentham}}}, \bibinfo {author}
  {\bibfnamefont {V.}~\bibnamefont {{Berardi}}}, \bibinfo {author}
  {\bibfnamefont {B.~E.}\ \bibnamefont {{Berger}}}, \bibinfo {author}
  {\bibfnamefont {S.}~\bibnamefont {{Berkman}}}, \bibinfo {author}
  {\bibfnamefont {I.}~\bibnamefont {{Bertram}}}, \bibinfo {author}
  {\bibfnamefont {S.}~\bibnamefont {{Bhadra}}}, \bibinfo {author}
  {\bibfnamefont {F.~d.~M.}\ \bibnamefont {{Blaszczyk}}}, \bibinfo {author}
  {\bibfnamefont {A.}~\bibnamefont {{Blondel}}}, \bibinfo {author}
  {\bibfnamefont {C.}~\bibnamefont {{Bojechko}}}, \bibinfo {author}
  {\bibfnamefont {S.}~\bibnamefont {{Bordoni}}}, \bibinfo {author}
  {\bibfnamefont {S.~B.}\ \bibnamefont {{Boyd}}}, \bibinfo {author}
  {\bibfnamefont {D.}~\bibnamefont {{Brailsford}}}, \bibinfo {author}
  {\bibfnamefont {A.}~\bibnamefont {{Bravar}}}, \bibinfo {author}
  {\bibfnamefont {C.}~\bibnamefont {{Bronner}}}, \bibinfo {author}
  {\bibfnamefont {N.}~\bibnamefont {{Buchanan}}}, \bibinfo {author}
  {\bibfnamefont {R.~G.}\ \bibnamefont {{Calland}}}, \bibinfo {author}
  {\bibfnamefont {J.}~\bibnamefont {{Caravaca Rodr{\'\i}guez}}}, \bibinfo
  {author} {\bibfnamefont {S.~L.}\ \bibnamefont {{Cartwright}}}, \bibinfo
  {author} {\bibfnamefont {R.}~\bibnamefont {{Castillo}}}, \bibinfo {author}
  {\bibfnamefont {M.~G.}\ \bibnamefont {{Catanesi}}}, \bibinfo {author}
  {\bibfnamefont {A.}~\bibnamefont {{Cervera}}}, \bibinfo {author}
  {\bibfnamefont {D.}~\bibnamefont {{Cherdack}}}, \bibinfo {author}
  {\bibfnamefont {G.}~\bibnamefont {{Christodoulou}}}, \bibinfo {author}
  {\bibfnamefont {A.}~\bibnamefont {{Clifton}}}, \bibinfo {author}
  {\bibfnamefont {J.}~\bibnamefont {{Coleman}}}, \bibinfo {author}
  {\bibfnamefont {S.~J.}\ \bibnamefont {{Coleman}}}, \bibinfo {author}
  {\bibfnamefont {G.}~\bibnamefont {{Collazuol}}}, \bibinfo {author}
  {\bibfnamefont {K.}~\bibnamefont {{Connolly}}}, \bibinfo {author}
  {\bibfnamefont {L.}~\bibnamefont {{Cremonesi}}}, \bibinfo {author}
  {\bibfnamefont {A.}~\bibnamefont {{Dabrowska}}}, \bibinfo {author}
  {\bibfnamefont {I.}~\bibnamefont {{Danko}}}, \bibinfo {author} {\bibfnamefont
  {R.}~\bibnamefont {{Das}}}, \bibinfo {author} {\bibfnamefont
  {S.}~\bibnamefont {{Davis}}}, \bibinfo {author} {\bibfnamefont
  {P.}~\bibnamefont {{de Perio}}}, \bibinfo {author} {\bibfnamefont
  {G.}~\bibnamefont {{De Rosa}}}, \bibinfo {author} {\bibfnamefont
  {T.}~\bibnamefont {{Dealtry}}}, \bibinfo {author} {\bibfnamefont {S.~R.}\
  \bibnamefont {{Dennis}}}, \bibinfo {author} {\bibfnamefont {C.}~\bibnamefont
  {{Densham}}}, \bibinfo {author} {\bibfnamefont {F.}~\bibnamefont {{Di
  Lodovico}}}, \bibinfo {author} {\bibfnamefont {S.}~\bibnamefont {{Di
  Luise}}}, \bibinfo {author} {\bibfnamefont {O.}~\bibnamefont {{Drapier}}},
  \bibinfo {author} {\bibfnamefont {T.}~\bibnamefont {{Duboyski}}}, \bibinfo
  {author} {\bibfnamefont {K.}~\bibnamefont {{Duffy}}}, \bibinfo {author}
  {\bibfnamefont {F.}~\bibnamefont {{Dufour}}}, \bibinfo {author}
  {\bibfnamefont {J.}~\bibnamefont {{Dumarchez}}}, \bibinfo {author}
  {\bibfnamefont {S.}~\bibnamefont {{Dytman}}}, \bibinfo {author}
  {\bibfnamefont {M.}~\bibnamefont {{Dziewiecki}}}, \bibinfo {author}
  {\bibfnamefont {S.}~\bibnamefont {{Emery}}}, \bibinfo {author} {\bibfnamefont
  {A.}~\bibnamefont {{Ereditato}}}, \bibinfo {author} {\bibfnamefont
  {L.}~\bibnamefont {{Escudero}}}, \bibinfo {author} {\bibfnamefont {A.~J.}\
  \bibnamefont {{Finch}}}, \bibinfo {author} {\bibfnamefont {L.}~\bibnamefont
  {{Floetotto}}}, \bibinfo {author} {\bibfnamefont {M.}~\bibnamefont
  {{Friend}}}, \bibinfo {author} {\bibfnamefont {Y.}~\bibnamefont {{Fujii}}},
  \bibinfo {author} {\bibfnamefont {Y.}~\bibnamefont {{Fukuda}}}, \bibinfo
  {author} {\bibfnamefont {A.~P.}\ \bibnamefont {{Furmanski}}}, \bibinfo
  {author} {\bibfnamefont {V.}~\bibnamefont {{Galymov}}}, \bibinfo {author}
  {\bibfnamefont {A.}~\bibnamefont {{Gaudin}}}, \bibinfo {author}
  {\bibfnamefont {S.}~\bibnamefont {{Giffin}}}, \bibinfo {author}
  {\bibfnamefont {C.}~\bibnamefont {{Giganti}}}, \bibinfo {author}
  {\bibfnamefont {K.}~\bibnamefont {{Gilje}}}, \bibinfo {author} {\bibfnamefont
  {D.}~\bibnamefont {{Goeldi}}}, \bibinfo {author} {\bibfnamefont
  {T.}~\bibnamefont {{Golan}}}, \bibinfo {author} {\bibfnamefont {J.~J.}\
  \bibnamefont {{Gomez-Cadenas}}}, \bibinfo {author} {\bibfnamefont
  {M.}~\bibnamefont {{Gonin}}}, \bibinfo {author} {\bibfnamefont
  {N.}~\bibnamefont {{Grant}}}, \bibinfo {author} {\bibfnamefont
  {D.}~\bibnamefont {{Gudin}}}, \bibinfo {author} {\bibfnamefont {D.~R.}\
  \bibnamefont {{Hadley}}}, \bibinfo {author} {\bibfnamefont {A.}~\bibnamefont
  {{Haesler}}}, \bibinfo {author} {\bibfnamefont {M.~D.}\ \bibnamefont
  {{Haigh}}}, \bibinfo {author} {\bibfnamefont {P.}~\bibnamefont {{Hamilton}}},
  \bibinfo {author} {\bibfnamefont {D.}~\bibnamefont {{Hansen}}}, \bibinfo
  {author} {\bibfnamefont {T.}~\bibnamefont {{Hara}}}, \bibinfo {author}
  {\bibfnamefont {M.}~\bibnamefont {{Hartz}}}, \bibinfo {author} {\bibfnamefont
  {T.}~\bibnamefont {{Hasegawa}}}, \bibinfo {author} {\bibfnamefont {N.~C.}\
  \bibnamefont {{Hastings}}}, \bibinfo {author} {\bibfnamefont
  {Y.}~\bibnamefont {{Hayato}}}, \bibinfo {author} {\bibfnamefont
  {C.}~\bibnamefont {{Hearty}}}, \bibinfo {author} {\bibfnamefont {R.~L.}\
  \bibnamefont {{Helmer}}}, \bibinfo {author} {\bibfnamefont {M.}~\bibnamefont
  {{Hierholzer}}}, \bibinfo {author} {\bibfnamefont {J.}~\bibnamefont
  {{Hignight}}}, \bibinfo {author} {\bibfnamefont {A.}~\bibnamefont
  {{Hillairet}}}, \bibinfo {author} {\bibfnamefont {A.}~\bibnamefont
  {{Himmel}}}, \bibinfo {author} {\bibfnamefont {T.}~\bibnamefont {{Hiraki}}},
  \bibinfo {author} {\bibfnamefont {S.}~\bibnamefont {{Hirota}}}, \bibinfo
  {author} {\bibfnamefont {J.}~\bibnamefont {{Holeczek}}}, \bibinfo {author}
  {\bibfnamefont {S.}~\bibnamefont {{Horikawa}}}, \bibinfo {author}
  {\bibfnamefont {K.}~\bibnamefont {{Huang}}}, \bibinfo {author} {\bibfnamefont
  {A.~K.}\ \bibnamefont {{Ichikawa}}}, \bibinfo {author} {\bibfnamefont
  {K.}~\bibnamefont {{Ieki}}}, \bibinfo {author} {\bibfnamefont
  {M.}~\bibnamefont {{Ieva}}}, \bibinfo {author} {\bibfnamefont
  {M.}~\bibnamefont {{Ikeda}}}, \bibinfo {author} {\bibfnamefont
  {J.}~\bibnamefont {{Imber}}}, \bibinfo {author} {\bibfnamefont
  {J.}~\bibnamefont {{Insler}}}, \bibinfo {author} {\bibfnamefont {T.~J.}\
  \bibnamefont {{Irvine}}}, \bibinfo {author} {\bibfnamefont {T.}~\bibnamefont
  {{Ishida}}}, \bibinfo {author} {\bibfnamefont {T.}~\bibnamefont {{Ishii}}},
  \bibinfo {author} {\bibfnamefont {S.~J.}\ \bibnamefont {{Ives}}}, \bibinfo
  {author} {\bibfnamefont {K.}~\bibnamefont {{Iyogi}}}, \bibinfo {author}
  {\bibfnamefont {A.}~\bibnamefont {{Izmaylov}}}, \bibinfo {author}
  {\bibfnamefont {A.}~\bibnamefont {{Jacob}}}, \bibinfo {author} {\bibfnamefont
  {B.}~\bibnamefont {{Jamieson}}}, \bibinfo {author} {\bibfnamefont {R.~A.}\
  \bibnamefont {{Johnson}}}, \bibinfo {author} {\bibfnamefont {J.~H.}\
  \bibnamefont {{Jo}}}, \bibinfo {author} {\bibfnamefont {P.}~\bibnamefont
  {{Jonsson}}}, \bibinfo {author} {\bibfnamefont {C.~K.}\ \bibnamefont
  {{Jung}}}, \bibinfo {author} {\bibfnamefont {A.~C.}\ \bibnamefont
  {{Kaboth}}}, \bibinfo {author} {\bibfnamefont {T.}~\bibnamefont {{Kajita}}},
  \bibinfo {author} {\bibfnamefont {H.}~\bibnamefont {{Kakuno}}}, \bibinfo
  {author} {\bibfnamefont {J.}~\bibnamefont {{Kameda}}}, \bibinfo {author}
  {\bibfnamefont {Y.}~\bibnamefont {{Kanazawa}}}, \bibinfo {author}
  {\bibfnamefont {D.}~\bibnamefont {{Karlen}}}, \bibinfo {author}
  {\bibfnamefont {I.}~\bibnamefont {{Karpikov}}}, \bibinfo {author}
  {\bibfnamefont {E.}~\bibnamefont {{Kearns}}}, \bibinfo {author}
  {\bibfnamefont {M.}~\bibnamefont {{Khabibullin}}}, \bibinfo {author}
  {\bibfnamefont {A.}~\bibnamefont {{Khotjantsev}}}, \bibinfo {author}
  {\bibfnamefont {D.}~\bibnamefont {{Kielczewska}}}, \bibinfo {author}
  {\bibfnamefont {T.}~\bibnamefont {{Kikawa}}}, \bibinfo {author}
  {\bibfnamefont {A.}~\bibnamefont {{Kilinski}}}, \bibinfo {author}
  {\bibfnamefont {J.}~\bibnamefont {{Kim}}}, \bibinfo {author} {\bibfnamefont
  {J.}~\bibnamefont {{Kisiel}}}, \bibinfo {author} {\bibfnamefont
  {P.}~\bibnamefont {{Kitching}}}, \bibinfo {author} {\bibfnamefont
  {T.}~\bibnamefont {{Kobayashi}}}, \bibinfo {author} {\bibfnamefont
  {L.}~\bibnamefont {{Koch}}}, \bibinfo {author} {\bibfnamefont
  {A.}~\bibnamefont {{Kolaceke}}}, \bibinfo {author} {\bibfnamefont
  {A.}~\bibnamefont {{Konaka}}}, \bibinfo {author} {\bibfnamefont {L.~L.}\
  \bibnamefont {{Kormos}}}, \bibinfo {author} {\bibfnamefont {A.}~\bibnamefont
  {{Korzenev}}}, \bibinfo {author} {\bibfnamefont {K.}~\bibnamefont
  {{Koseki}}}, \bibinfo {author} {\bibfnamefont {Y.}~\bibnamefont {{Koshio}}},
  \bibinfo {author} {\bibfnamefont {I.}~\bibnamefont {{Kreslo}}}, \bibinfo
  {author} {\bibfnamefont {W.}~\bibnamefont {{Kropp}}}, \bibinfo {author}
  {\bibfnamefont {H.}~\bibnamefont {{Kubo}}}, \bibinfo {author} {\bibfnamefont
  {Y.}~\bibnamefont {{Kudenko}}}, \bibinfo {author} {\bibfnamefont
  {S.}~\bibnamefont {{Kumaratunga}}}, \bibinfo {author} {\bibfnamefont
  {R.}~\bibnamefont {{Kurjata}}}, \bibinfo {author} {\bibfnamefont
  {T.}~\bibnamefont {{Kutter}}}, \bibinfo {author} {\bibfnamefont
  {J.}~\bibnamefont {{Lagoda}}}, \bibinfo {author} {\bibfnamefont
  {K.}~\bibnamefont {{Laihem}}}, \bibinfo {author} {\bibfnamefont
  {I.}~\bibnamefont {{Lamont}}}, \bibinfo {author} {\bibfnamefont
  {M.}~\bibnamefont {{Laveder}}}, \bibinfo {author} {\bibfnamefont
  {M.}~\bibnamefont {{Lawe}}}, \bibinfo {author} {\bibfnamefont
  {M.}~\bibnamefont {{Lazos}}}, \bibinfo {author} {\bibfnamefont {K.~P.}\
  \bibnamefont {{Lee}}}, \bibinfo {author} {\bibfnamefont {C.}~\bibnamefont
  {{Licciardi}}}, \bibinfo {author} {\bibfnamefont {T.}~\bibnamefont
  {{Lindner}}}, \bibinfo {author} {\bibfnamefont {C.}~\bibnamefont {{Lister}}},
  \bibinfo {author} {\bibfnamefont {R.~P.}\ \bibnamefont {{Litchfield}}},
  \bibinfo {author} {\bibfnamefont {A.}~\bibnamefont {{Longhin}}}, \bibinfo
  {author} {\bibfnamefont {L.}~\bibnamefont {{Ludovici}}}, \bibinfo {author}
  {\bibfnamefont {M.}~\bibnamefont {{Macaire}}}, \bibinfo {author}
  {\bibfnamefont {L.}~\bibnamefont {{Magaletti}}}, \bibinfo {author}
  {\bibfnamefont {K.}~\bibnamefont {{Mahn}}}, \bibinfo {author} {\bibfnamefont
  {M.}~\bibnamefont {{Malek}}}, \bibinfo {author} {\bibfnamefont
  {S.}~\bibnamefont {{Manly}}}, \bibinfo {author} {\bibfnamefont {A.~D.}\
  \bibnamefont {{Marino}}}, \bibinfo {author} {\bibfnamefont {J.}~\bibnamefont
  {{Marteau}}}, \bibinfo {author} {\bibfnamefont {J.~F.}\ \bibnamefont
  {{Martin}}}, \bibinfo {author} {\bibfnamefont {T.}~\bibnamefont
  {{Maruyama}}}, \bibinfo {author} {\bibfnamefont {J.}~\bibnamefont
  {{Marzec}}}, \bibinfo {author} {\bibfnamefont {E.~L.}\ \bibnamefont
  {{Mathie}}}, \bibinfo {author} {\bibfnamefont {V.}~\bibnamefont {{Matveev}}},
  \bibinfo {author} {\bibfnamefont {K.}~\bibnamefont {{Mavrokoridis}}},
  \bibinfo {author} {\bibfnamefont {E.}~\bibnamefont {{Mazzucato}}}, \bibinfo
  {author} {\bibfnamefont {M.}~\bibnamefont {{McCarthy}}}, \bibinfo {author}
  {\bibfnamefont {N.}~\bibnamefont {{McCauley}}}, \bibinfo {author}
  {\bibfnamefont {K.~S.}\ \bibnamefont {{McFarland }}}, \bibinfo {author}
  {\bibfnamefont {C.}~\bibnamefont {{McGrew}}}, \bibinfo {author}
  {\bibfnamefont {C.}~\bibnamefont {{Metelko}}}, \bibinfo {author}
  {\bibfnamefont {M.}~\bibnamefont {{Mezzetto}}}, \bibinfo {author}
  {\bibfnamefont {P.}~\bibnamefont {{Mijakowski}}}, \bibinfo {author}
  {\bibfnamefont {C.~A.}\ \bibnamefont {{Miller}}}, \bibinfo {author}
  {\bibfnamefont {A.}~\bibnamefont {{Minamino}}}, \bibinfo {author}
  {\bibfnamefont {O.}~\bibnamefont {{Mineev}}}, \bibinfo {author}
  {\bibfnamefont {S.}~\bibnamefont {{Mine}}}, \bibinfo {author} {\bibfnamefont
  {A.}~\bibnamefont {{Missert}}}, \bibinfo {author} {\bibfnamefont
  {M.}~\bibnamefont {{Miura}}}, \bibinfo {author} {\bibfnamefont
  {L.}~\bibnamefont {{Monfregola}}}, \bibinfo {author} {\bibfnamefont
  {S.}~\bibnamefont {{Moriyama}}}, \bibinfo {author} {\bibfnamefont {T.~A.}\
  \bibnamefont {{Mueller}}}, \bibinfo {author} {\bibfnamefont {A.}~\bibnamefont
  {{Murakami}}}, \bibinfo {author} {\bibfnamefont {M.}~\bibnamefont
  {{Murdoch}}}, \bibinfo {author} {\bibfnamefont {S.}~\bibnamefont {{Murphy}}},
  \bibinfo {author} {\bibfnamefont {J.}~\bibnamefont {{Myslik}}}, \bibinfo
  {author} {\bibfnamefont {T.}~\bibnamefont {{Nagasaki}}}, \bibinfo {author}
  {\bibfnamefont {T.}~\bibnamefont {{Nakadaira}}}, \bibinfo {author}
  {\bibfnamefont {M.}~\bibnamefont {{Nakahata}}}, \bibinfo {author}
  {\bibfnamefont {T.}~\bibnamefont {{Nakai}}}, \bibinfo {author} {\bibfnamefont
  {K.}~\bibnamefont {{Nakamura}}}, \bibinfo {author} {\bibfnamefont
  {S.}~\bibnamefont {{Nakayama}}}, \bibinfo {author} {\bibfnamefont
  {T.}~\bibnamefont {{Nakaya}}}, \bibinfo {author} {\bibfnamefont
  {K.}~\bibnamefont {{Nakayoshi}}}, \bibinfo {author} {\bibfnamefont
  {D.}~\bibnamefont {{Naples}}}, \bibinfo {author} {\bibfnamefont
  {C.}~\bibnamefont {{Nielsen}}}, \bibinfo {author} {\bibfnamefont
  {M.}~\bibnamefont {{Nirkko}}}, \bibinfo {author} {\bibfnamefont
  {K.}~\bibnamefont {{Nishikawa}}}, \bibinfo {author} {\bibfnamefont
  {Y.}~\bibnamefont {{Nishimura}}}, \bibinfo {author} {\bibfnamefont {H.~M.}\
  \bibnamefont {{O'Keeffe}}}, \bibinfo {author} {\bibfnamefont
  {R.}~\bibnamefont {{Ohta}}}, \bibinfo {author} {\bibfnamefont
  {K.}~\bibnamefont {{Okumura}}}, \bibinfo {author} {\bibfnamefont
  {T.}~\bibnamefont {{Okusawa}}}, \bibinfo {author} {\bibfnamefont
  {W.}~\bibnamefont {{Oryszczak}}}, \bibinfo {author} {\bibfnamefont {S.~M.}\
  \bibnamefont {{Oser}}}, \bibinfo {author} {\bibfnamefont {R.~A.}\
  \bibnamefont {{Owen}}}, \bibinfo {author} {\bibfnamefont {Y.}~\bibnamefont
  {{Oyama}}}, \bibinfo {author} {\bibfnamefont {V.}~\bibnamefont
  {{Palladino}}}, \bibinfo {author} {\bibfnamefont {V.}~\bibnamefont
  {{Paolone}}}, \bibinfo {author} {\bibfnamefont {D.}~\bibnamefont {{Payne}}},
  \bibinfo {author} {\bibfnamefont {G.~F.}\ \bibnamefont {{Pearce}}}, \bibinfo
  {author} {\bibfnamefont {O.}~\bibnamefont {{Perevozchikov}}}, \bibinfo
  {author} {\bibfnamefont {J.~D.}\ \bibnamefont {{Perkin}}}, \bibinfo {author}
  {\bibfnamefont {Y.}~\bibnamefont {{Petrov}}}, \bibinfo {author}
  {\bibfnamefont {L.~J.}\ \bibnamefont {{Pickard}}}, \bibinfo {author}
  {\bibfnamefont {E.~S.}\ \bibnamefont {{Pinzon Guerra}}}, \bibinfo {author}
  {\bibfnamefont {C.}~\bibnamefont {{Pistillo}}}, \bibinfo {author}
  {\bibfnamefont {P.}~\bibnamefont {{Plonski}}}, \bibinfo {author}
  {\bibfnamefont {E.}~\bibnamefont {{Poplawska}}}, \bibinfo {author}
  {\bibfnamefont {B.}~\bibnamefont {{Popov}}}, \bibinfo {author} {\bibfnamefont
  {M.}~\bibnamefont {{Posiadala}}}, \bibinfo {author} {\bibfnamefont {J.~M.}\
  \bibnamefont {{Poutissou}}}, \bibinfo {author} {\bibfnamefont
  {R.}~\bibnamefont {{Poutissou}}}, \bibinfo {author} {\bibfnamefont
  {P.}~\bibnamefont {{Przewlocki}}}, \bibinfo {author} {\bibfnamefont
  {B.}~\bibnamefont {{Quilain}}}, \bibinfo {author} {\bibfnamefont
  {E.}~\bibnamefont {{Radicioni}}}, \bibinfo {author} {\bibfnamefont {P.~N.}\
  \bibnamefont {{Ratoff}}}, \bibinfo {author} {\bibfnamefont {M.}~\bibnamefont
  {{Ravonel}}}, \bibinfo {author} {\bibfnamefont {M.~A.~M.}\ \bibnamefont
  {{Rayner}}}, \bibinfo {author} {\bibfnamefont {A.}~\bibnamefont {{Redij}}},
  \bibinfo {author} {\bibfnamefont {M.}~\bibnamefont {{Reeves}}}, \bibinfo
  {author} {\bibfnamefont {E.}~\bibnamefont {{Reinherz-Aronis}}}, \bibinfo
  {author} {\bibfnamefont {F.}~\bibnamefont {{Retiere}}}, \bibinfo {author}
  {\bibfnamefont {A.}~\bibnamefont {{Robert}}}, \bibinfo {author}
  {\bibfnamefont {P.~A.}\ \bibnamefont {{Rodrigues}}}, \bibinfo {author}
  {\bibfnamefont {P.}~\bibnamefont {{Rojas}}}, \bibinfo {author} {\bibfnamefont
  {E.}~\bibnamefont {{Rondio}}}, \bibinfo {author} {\bibfnamefont
  {S.}~\bibnamefont {{Roth}}}, \bibinfo {author} {\bibfnamefont
  {A.}~\bibnamefont {{Rubbia}}}, \bibinfo {author} {\bibfnamefont
  {D.}~\bibnamefont {{Ruterbories}}}, \bibinfo {author} {\bibfnamefont
  {R.}~\bibnamefont {{Sacco}}}, \bibinfo {author} {\bibfnamefont
  {K.}~\bibnamefont {{Sakashita}}}, \bibinfo {author} {\bibfnamefont
  {F.}~\bibnamefont {{S{\'a}nchez}}}, \bibinfo {author} {\bibfnamefont
  {F.}~\bibnamefont {{Sato}}}, \bibinfo {author} {\bibfnamefont
  {E.}~\bibnamefont {{Scantamburlo}}}, \bibinfo {author} {\bibfnamefont
  {K.}~\bibnamefont {{Scholberg}}}, \bibinfo {author} {\bibfnamefont
  {J.}~\bibnamefont {{Schwehr}}}, \bibinfo {author} {\bibfnamefont
  {M.}~\bibnamefont {{Scott}}}, \bibinfo {author} {\bibfnamefont
  {Y.}~\bibnamefont {{Seiya}}}, \bibinfo {author} {\bibfnamefont
  {T.}~\bibnamefont {{Sekiguchi}}}, \bibinfo {author} {\bibfnamefont
  {H.}~\bibnamefont {{Sekiya}}}, \bibinfo {author} {\bibfnamefont
  {D.}~\bibnamefont {{Sgalaberna}}}, \bibinfo {author} {\bibfnamefont
  {M.}~\bibnamefont {{Shiozawa}}}, \bibinfo {author} {\bibfnamefont
  {S.}~\bibnamefont {{Short}}}, \bibinfo {author} {\bibfnamefont
  {Y.}~\bibnamefont {{Shustrov}}}, \bibinfo {author} {\bibfnamefont
  {P.}~\bibnamefont {{Sinclair}}}, \bibinfo {author} {\bibfnamefont
  {B.}~\bibnamefont {{Smith}}}, \bibinfo {author} {\bibfnamefont {R.~J.}\
  \bibnamefont {{Smith}}}, \bibinfo {author} {\bibfnamefont {M.}~\bibnamefont
  {{Smy}}}, \bibinfo {author} {\bibfnamefont {J.~T.}\ \bibnamefont
  {{Sobczyk}}}, \bibinfo {author} {\bibfnamefont {H.}~\bibnamefont {{Sobel}}},
  \bibinfo {author} {\bibfnamefont {M.}~\bibnamefont {{Sorel}}}, \bibinfo
  {author} {\bibfnamefont {L.}~\bibnamefont {{Southwell}}}, \bibinfo {author}
  {\bibfnamefont {P.}~\bibnamefont {{Stamoulis}}}, \bibinfo {author}
  {\bibfnamefont {J.}~\bibnamefont {{Steinmann}}}, \bibinfo {author}
  {\bibfnamefont {B.}~\bibnamefont {{Still}}}, \bibinfo {author} {\bibfnamefont
  {Y.}~\bibnamefont {{Suda}}}, \bibinfo {author} {\bibfnamefont
  {A.}~\bibnamefont {{Suzuki}}}, \bibinfo {author} {\bibfnamefont
  {K.}~\bibnamefont {{Suzuki}}}, \bibinfo {author} {\bibfnamefont {S.~Y.}\
  \bibnamefont {{Suzuki}}}, \bibinfo {author} {\bibfnamefont {Y.}~\bibnamefont
  {{Suzuki}}}, \bibinfo {author} {\bibfnamefont {T.}~\bibnamefont
  {{Szeglowski}}}, \bibinfo {author} {\bibfnamefont {R.}~\bibnamefont
  {{Tacik}}}, \bibinfo {author} {\bibfnamefont {M.}~\bibnamefont {{Tada}}},
  \bibinfo {author} {\bibfnamefont {S.}~\bibnamefont {{Takahashi}}}, \bibinfo
  {author} {\bibfnamefont {A.}~\bibnamefont {{Takeda}}}, \bibinfo {author}
  {\bibfnamefont {Y.}~\bibnamefont {{Takeuchi}}}, \bibinfo {author}
  {\bibfnamefont {H.~K.}\ \bibnamefont {{Tanaka}}}, \bibinfo {author}
  {\bibfnamefont {H.~A.}\ \bibnamefont {{Tanaka}}}, \bibinfo {author}
  {\bibfnamefont {M.~M.}\ \bibnamefont {{Tanaka}}}, \bibinfo {author}
  {\bibfnamefont {D.}~\bibnamefont {{Terhorst}}}, \bibinfo {author}
  {\bibfnamefont {R.}~\bibnamefont {{Terri}}}, \bibinfo {author} {\bibfnamefont
  {L.~F.}\ \bibnamefont {{Thompson}}}, \bibinfo {author} {\bibfnamefont
  {A.}~\bibnamefont {{Thorley}}}, \bibinfo {author} {\bibfnamefont
  {S.}~\bibnamefont {{Tobayama}}}, \bibinfo {author} {\bibfnamefont
  {W.}~\bibnamefont {{Toki}}}, \bibinfo {author} {\bibfnamefont
  {T.}~\bibnamefont {{Tomura}}}, \bibinfo {author} {\bibfnamefont
  {Y.}~\bibnamefont {{Totsuka}}}, \bibinfo {author} {\bibfnamefont
  {C.}~\bibnamefont {{Touramanis}}}, \bibinfo {author} {\bibfnamefont
  {T.}~\bibnamefont {{Tsukamoto}}}, \bibinfo {author} {\bibfnamefont
  {M.}~\bibnamefont {{Tzanov}}}, \bibinfo {author} {\bibfnamefont
  {Y.}~\bibnamefont {{Uchida}}}, \bibinfo {author} {\bibfnamefont
  {K.}~\bibnamefont {{Ueno}}}, \bibinfo {author} {\bibfnamefont
  {A.}~\bibnamefont {{Vacheret}}}, \bibinfo {author} {\bibfnamefont
  {M.}~\bibnamefont {{Vagins}}}, \bibinfo {author} {\bibfnamefont
  {G.}~\bibnamefont {{Vasseur}}}, \bibinfo {author} {\bibfnamefont
  {T.}~\bibnamefont {{Wachala}}}, \bibinfo {author} {\bibfnamefont {A.~V.}\
  \bibnamefont {{Waldron}}}, \bibinfo {author} {\bibfnamefont {C.~W.}\
  \bibnamefont {{Walter}}}, \bibinfo {author} {\bibfnamefont {D.}~\bibnamefont
  {{Wark}}}, \bibinfo {author} {\bibfnamefont {M.~O.}\ \bibnamefont
  {{Wascko}}}, \bibinfo {author} {\bibfnamefont {A.}~\bibnamefont {{Weber}}},
  \bibinfo {author} {\bibfnamefont {R.}~\bibnamefont {{Wendell}}}, \bibinfo
  {author} {\bibfnamefont {R.~J.}\ \bibnamefont {{Wilkes}}}, \bibinfo {author}
  {\bibfnamefont {M.~J.}\ \bibnamefont {{Wilking}}}, \bibinfo {author}
  {\bibfnamefont {C.}~\bibnamefont {{Wilkinson}}}, \bibinfo {author}
  {\bibfnamefont {Z.}~\bibnamefont {{Williamson}}}, \bibinfo {author}
  {\bibfnamefont {J.~R.}\ \bibnamefont {{Wilson}}}, \bibinfo {author}
  {\bibfnamefont {R.~J.}\ \bibnamefont {{Wilson}}}, \bibinfo {author}
  {\bibfnamefont {T.}~\bibnamefont {{Wongjirad}}}, \bibinfo {author}
  {\bibfnamefont {Y.}~\bibnamefont {{Yamada}}}, \bibinfo {author}
  {\bibfnamefont {K.}~\bibnamefont {{Yamamoto}}}, \bibinfo {author}
  {\bibfnamefont {C.}~\bibnamefont {{Yanagisawa}}}, \bibinfo {author}
  {\bibfnamefont {S.}~\bibnamefont {{Yen}}}, \bibinfo {author} {\bibfnamefont
  {N.}~\bibnamefont {{Yershov}}}, \bibinfo {author} {\bibfnamefont
  {M.}~\bibnamefont {{Yokoyama}}}, \bibinfo {author} {\bibfnamefont
  {T.}~\bibnamefont {{Yuan}}}, \bibinfo {author} {\bibfnamefont
  {A.}~\bibnamefont {{Zalewska}}}, \bibinfo {author} {\bibfnamefont
  {J.}~\bibnamefont {{Zalipska}}}, \bibinfo {author} {\bibfnamefont
  {L.}~\bibnamefont {{Zambelli}}}, \bibinfo {author} {\bibfnamefont
  {K.}~\bibnamefont {{Zaremba}}}, \bibinfo {author} {\bibfnamefont
  {M.}~\bibnamefont {{Ziembicki}}}, \bibinfo {author} {\bibfnamefont {E.~D.}\
  \bibnamefont {{Zimmerman}}}, \bibinfo {author} {\bibfnamefont
  {M.}~\bibnamefont {{Zito}}}, \bibinfo {author} {\bibfnamefont
  {J.}~\bibnamefont {{{\.Z}muda}}}, \ and\ \bibinfo {author} {\bibnamefont
  {{T2K Collaboration}}},\ }\href {\doibase 10.1103/PhysRevLett.112.061802}
  {\bibfield  {journal} {\bibinfo  {journal} {\prl}\ }\textbf {\bibinfo
  {volume} {112}},\ \bibinfo {eid} {061802} (\bibinfo {year} {2014})},\ \Eprint
  {http://arxiv.org/abs/1311.4750} {arXiv:1311.4750 [hep-ex]} \BibitemShut
  {NoStop}%
\bibitem [{\citenamefont {{Abe}}\ \emph {et~al.}(2012)\citenamefont {{Abe}},
  \citenamefont {{Aberle}}, \citenamefont {{dos Anjos}}, \citenamefont
  {{Barriere}}, \citenamefont {{Bergevin}}, \citenamefont {{Bernstein}},
  \citenamefont {{Bezerra}}, \citenamefont {{Bezrukhov}}, \citenamefont
  {{Blucher}}, \citenamefont {{Bowden}}, \citenamefont {{Buck}}, \citenamefont
  {{Busenitz}}, \citenamefont {{Cabrera}}, \citenamefont {{Caden}},
  \citenamefont {{Camilleri}}, \citenamefont {{Carr}}, \citenamefont
  {{Cerrada}}, \citenamefont {{Chang}}, \citenamefont {{Chimenti}},
  \citenamefont {{Classen}}, \citenamefont {{Collin}}, \citenamefont
  {{Conover}}, \citenamefont {{Conrad}}, \citenamefont {{Crespo-Anad{\'o}n}},
  \citenamefont {{Crum}}, \citenamefont {{Cucoanes}}, \citenamefont
  {{D'Agostino}}, \citenamefont {{Damon}}, \citenamefont {{Dawson}},
  \citenamefont {{Dazeley}}, \citenamefont {{Dietrich}}, \citenamefont
  {{Djurcic}}, \citenamefont {{Dracos}}, \citenamefont {{Durand}},
  \citenamefont {{Ebert}}, \citenamefont {{Efremenko}}, \citenamefont
  {{Elnimr}}, \citenamefont {{Etenko}}, \citenamefont {{Fallot}}, \citenamefont
  {{Fechner}}, \citenamefont {{von Feilitzsch}}, \citenamefont {{Felde}},
  \citenamefont {{Franco}}, \citenamefont {{Franke}}, \citenamefont {{Franke}},
  \citenamefont {{Furuta}}, \citenamefont {{Gama}}, \citenamefont
  {{Gil-Botella}}, \citenamefont {{Giot}}, \citenamefont {{G{\"o}ger-Neff}},
  \citenamefont {{Gonzalez}}, \citenamefont {{Goodman}}, \citenamefont
  {{Goon}}, \citenamefont {{Greiner}}, \citenamefont {{Haag}}, \citenamefont
  {{Hagner}}, \citenamefont {{Hara}}, \citenamefont {{Hartmann}}, \citenamefont
  {{Haser}}, \citenamefont {{Hatzikoutelis}}, \citenamefont {{Hayakawa}},
  \citenamefont {{Hofmann}}, \citenamefont {{Horton-Smith}}, \citenamefont
  {{Hourlier}}, \citenamefont {{Ishitsuka}}, \citenamefont {{Jochum}},
  \citenamefont {{Jollet}}, \citenamefont {{Jones}}, \citenamefont {{Kaether}},
  \citenamefont {{Kalousis}}, \citenamefont {{Kamyshkov}}, \citenamefont
  {{Kaplan}}, \citenamefont {{Kawasaki}}, \citenamefont {{Keefer}},
  \citenamefont {{Kemp}}, \citenamefont {{de Kerret}}, \citenamefont {{Kibe}},
  \citenamefont {{Konno}}, \citenamefont {{Kryn}}, \citenamefont {{Kuze}},
  \citenamefont {{Lachenmaier}}, \citenamefont {{Lane}}, \citenamefont
  {{Langbrandtner}}, \citenamefont {{Lasserre}}, \citenamefont {{Letourneau}},
  \citenamefont {{Lhuillier}}, \citenamefont {{Lima}}, \citenamefont
  {{Lindner}}, \citenamefont {{L{\'o}pez-Castan{\~o}}}, \citenamefont
  {{LoSecco}}, \citenamefont {{Lubsandorzhiev}}, \citenamefont {{Lucht}},
  \citenamefont {{McKee}}, \citenamefont {{Maeda}}, \citenamefont {{Maesano}},
  \citenamefont {{Mariani}}, \citenamefont {{Maricic}}, \citenamefont
  {{Martino}}, \citenamefont {{Matsubara}}, \citenamefont {{Mention}},
  \citenamefont {{Meregaglia}}, \citenamefont {{Miletic}}, \citenamefont
  {{Milincic}}, \citenamefont {{Miyata}}, \citenamefont {{Mueller}},
  \citenamefont {{Nagasaka}}, \citenamefont {{Nakajima}}, \citenamefont
  {{Novella}}, \citenamefont {{Obolensky}}, \citenamefont {{Oberauer}},
  \citenamefont {{Onillon}}, \citenamefont {{Osborn}}, \citenamefont
  {{Ostrovskiy}}, \citenamefont {{Palomares}}, \citenamefont {{Pepe}},
  \citenamefont {{Perasso}}, \citenamefont {{Perrin}}, \citenamefont
  {{Pfahler}}, \citenamefont {{Porta}}, \citenamefont {{Potzel}}, \citenamefont
  {{Reichenbacher}}, \citenamefont {{Reinhold}}, \citenamefont {{Remoto}},
  \citenamefont {{R{\"o}hling}}, \citenamefont {{Roncin}}, \citenamefont
  {{Roth}}, \citenamefont {{Sakamoto}}, \citenamefont {{Santorelli}},
  \citenamefont {{Sato}}, \citenamefont {{Sch{\"o}nert}}, \citenamefont
  {{Schoppmann}}, \citenamefont {{Schwetz}}, \citenamefont {{Shaevitz}},
  \citenamefont {{Shimojima}}, \citenamefont {{Shrestha}}, \citenamefont
  {{Sida}}, \citenamefont {{Sinev}}, \citenamefont {{Skorokhvatov}},
  \citenamefont {{Smith}}, \citenamefont {{Spitz}}, \citenamefont {{Stahl}},
  \citenamefont {{Stancu}}, \citenamefont {{Stokes}}, \citenamefont {{Strait}},
  \citenamefont {{St{\"u}ken}}, \citenamefont {{Suekane}}, \citenamefont
  {{Sukhotin}}, \citenamefont {{Sumiyoshi}}, \citenamefont {{Sun}},
  \citenamefont {{Svoboda}}, \citenamefont {{Terao}}, \citenamefont
  {{Tonazzo}}, \citenamefont {{Toups}}, \citenamefont {{Trinh Thi}},
  \citenamefont {{Valdiviesso}}, \citenamefont {{Veyssiere}}, \citenamefont
  {{Wagner}}, \citenamefont {{Watanabe}}, \citenamefont {{White}},
  \citenamefont {{Wiebusch}}, \citenamefont {{Winslow}}, \citenamefont
  {{Worcester}}, \citenamefont {{Wurm}}, \citenamefont {{Yermia}},\ and\
  \citenamefont {{Zimmer}}}]{1207.6632}%
  \BibitemOpen
  \bibfield  {author} {\bibinfo {author} {\bibfnamefont {Y.}~\bibnamefont
  {{Abe}}}, \bibinfo {author} {\bibfnamefont {C.}~\bibnamefont {{Aberle}}},
  \bibinfo {author} {\bibfnamefont {J.~C.}\ \bibnamefont {{dos Anjos}}},
  \bibinfo {author} {\bibfnamefont {J.~C.}\ \bibnamefont {{Barriere}}},
  \bibinfo {author} {\bibfnamefont {M.}~\bibnamefont {{Bergevin}}}, \bibinfo
  {author} {\bibfnamefont {A.}~\bibnamefont {{Bernstein}}}, \bibinfo {author}
  {\bibfnamefont {T.~J.~C.}\ \bibnamefont {{Bezerra}}}, \bibinfo {author}
  {\bibfnamefont {L.}~\bibnamefont {{Bezrukhov}}}, \bibinfo {author}
  {\bibfnamefont {E.}~\bibnamefont {{Blucher}}}, \bibinfo {author}
  {\bibfnamefont {N.~S.}\ \bibnamefont {{Bowden}}}, \bibinfo {author}
  {\bibfnamefont {C.}~\bibnamefont {{Buck}}}, \bibinfo {author} {\bibfnamefont
  {J.}~\bibnamefont {{Busenitz}}}, \bibinfo {author} {\bibfnamefont
  {A.}~\bibnamefont {{Cabrera}}}, \bibinfo {author} {\bibfnamefont
  {E.}~\bibnamefont {{Caden}}}, \bibinfo {author} {\bibfnamefont
  {L.}~\bibnamefont {{Camilleri}}}, \bibinfo {author} {\bibfnamefont
  {R.}~\bibnamefont {{Carr}}}, \bibinfo {author} {\bibfnamefont
  {M.}~\bibnamefont {{Cerrada}}}, \bibinfo {author} {\bibfnamefont {P.~J.}\
  \bibnamefont {{Chang}}}, \bibinfo {author} {\bibfnamefont {P.}~\bibnamefont
  {{Chimenti}}}, \bibinfo {author} {\bibfnamefont {T.}~\bibnamefont
  {{Classen}}}, \bibinfo {author} {\bibfnamefont {A.~P.}\ \bibnamefont
  {{Collin}}}, \bibinfo {author} {\bibfnamefont {E.}~\bibnamefont {{Conover}}},
  \bibinfo {author} {\bibfnamefont {J.~M.}\ \bibnamefont {{Conrad}}}, \bibinfo
  {author} {\bibfnamefont {J.~I.}\ \bibnamefont {{Crespo-Anad{\'o}n}}},
  \bibinfo {author} {\bibfnamefont {K.}~\bibnamefont {{Crum}}}, \bibinfo
  {author} {\bibfnamefont {A.}~\bibnamefont {{Cucoanes}}}, \bibinfo {author}
  {\bibfnamefont {M.~V.}\ \bibnamefont {{D'Agostino}}}, \bibinfo {author}
  {\bibfnamefont {E.}~\bibnamefont {{Damon}}}, \bibinfo {author} {\bibfnamefont
  {J.~V.}\ \bibnamefont {{Dawson}}}, \bibinfo {author} {\bibfnamefont
  {S.}~\bibnamefont {{Dazeley}}}, \bibinfo {author} {\bibfnamefont
  {D.}~\bibnamefont {{Dietrich}}}, \bibinfo {author} {\bibfnamefont
  {Z.}~\bibnamefont {{Djurcic}}}, \bibinfo {author} {\bibfnamefont
  {M.}~\bibnamefont {{Dracos}}}, \bibinfo {author} {\bibfnamefont
  {V.}~\bibnamefont {{Durand}}}, \bibinfo {author} {\bibfnamefont
  {J.}~\bibnamefont {{Ebert}}}, \bibinfo {author} {\bibfnamefont
  {Y.}~\bibnamefont {{Efremenko}}}, \bibinfo {author} {\bibfnamefont
  {M.}~\bibnamefont {{Elnimr}}}, \bibinfo {author} {\bibfnamefont
  {A.}~\bibnamefont {{Etenko}}}, \bibinfo {author} {\bibfnamefont
  {M.}~\bibnamefont {{Fallot}}}, \bibinfo {author} {\bibfnamefont
  {M.}~\bibnamefont {{Fechner}}}, \bibinfo {author} {\bibfnamefont
  {F.}~\bibnamefont {{von Feilitzsch}}}, \bibinfo {author} {\bibfnamefont
  {J.}~\bibnamefont {{Felde}}}, \bibinfo {author} {\bibfnamefont
  {D.}~\bibnamefont {{Franco}}}, \bibinfo {author} {\bibfnamefont {A.~J.}\
  \bibnamefont {{Franke}}}, \bibinfo {author} {\bibfnamefont {M.}~\bibnamefont
  {{Franke}}}, \bibinfo {author} {\bibfnamefont {H.}~\bibnamefont {{Furuta}}},
  \bibinfo {author} {\bibfnamefont {R.}~\bibnamefont {{Gama}}}, \bibinfo
  {author} {\bibfnamefont {I.}~\bibnamefont {{Gil-Botella}}}, \bibinfo {author}
  {\bibfnamefont {L.}~\bibnamefont {{Giot}}}, \bibinfo {author} {\bibfnamefont
  {M.}~\bibnamefont {{G{\"o}ger-Neff}}}, \bibinfo {author} {\bibfnamefont
  {L.~F.~G.}\ \bibnamefont {{Gonzalez}}}, \bibinfo {author} {\bibfnamefont
  {M.~C.}\ \bibnamefont {{Goodman}}}, \bibinfo {author} {\bibfnamefont {J.~T.}\
  \bibnamefont {{Goon}}}, \bibinfo {author} {\bibfnamefont {D.}~\bibnamefont
  {{Greiner}}}, \bibinfo {author} {\bibfnamefont {N.}~\bibnamefont {{Haag}}},
  \bibinfo {author} {\bibfnamefont {C.}~\bibnamefont {{Hagner}}}, \bibinfo
  {author} {\bibfnamefont {T.}~\bibnamefont {{Hara}}}, \bibinfo {author}
  {\bibfnamefont {F.~X.}\ \bibnamefont {{Hartmann}}}, \bibinfo {author}
  {\bibfnamefont {J.}~\bibnamefont {{Haser}}}, \bibinfo {author} {\bibfnamefont
  {A.}~\bibnamefont {{Hatzikoutelis}}}, \bibinfo {author} {\bibfnamefont
  {T.}~\bibnamefont {{Hayakawa}}}, \bibinfo {author} {\bibfnamefont
  {M.}~\bibnamefont {{Hofmann}}}, \bibinfo {author} {\bibfnamefont {G.~A.}\
  \bibnamefont {{Horton-Smith}}}, \bibinfo {author} {\bibfnamefont
  {A.}~\bibnamefont {{Hourlier}}}, \bibinfo {author} {\bibfnamefont
  {M.}~\bibnamefont {{Ishitsuka}}}, \bibinfo {author} {\bibfnamefont
  {J.}~\bibnamefont {{Jochum}}}, \bibinfo {author} {\bibfnamefont
  {C.}~\bibnamefont {{Jollet}}}, \bibinfo {author} {\bibfnamefont {C.~L.}\
  \bibnamefont {{Jones}}}, \bibinfo {author} {\bibfnamefont {F.}~\bibnamefont
  {{Kaether}}}, \bibinfo {author} {\bibfnamefont {L.~N.}\ \bibnamefont
  {{Kalousis}}}, \bibinfo {author} {\bibfnamefont {Y.}~\bibnamefont
  {{Kamyshkov}}}, \bibinfo {author} {\bibfnamefont {D.~M.}\ \bibnamefont
  {{Kaplan}}}, \bibinfo {author} {\bibfnamefont {T.}~\bibnamefont
  {{Kawasaki}}}, \bibinfo {author} {\bibfnamefont {G.}~\bibnamefont
  {{Keefer}}}, \bibinfo {author} {\bibfnamefont {E.}~\bibnamefont {{Kemp}}},
  \bibinfo {author} {\bibfnamefont {H.}~\bibnamefont {{de Kerret}}}, \bibinfo
  {author} {\bibfnamefont {Y.}~\bibnamefont {{Kibe}}}, \bibinfo {author}
  {\bibfnamefont {T.}~\bibnamefont {{Konno}}}, \bibinfo {author} {\bibfnamefont
  {D.}~\bibnamefont {{Kryn}}}, \bibinfo {author} {\bibfnamefont
  {M.}~\bibnamefont {{Kuze}}}, \bibinfo {author} {\bibfnamefont
  {T.}~\bibnamefont {{Lachenmaier}}}, \bibinfo {author} {\bibfnamefont {C.~E.}\
  \bibnamefont {{Lane}}}, \bibinfo {author} {\bibfnamefont {C.}~\bibnamefont
  {{Langbrandtner}}}, \bibinfo {author} {\bibfnamefont {T.}~\bibnamefont
  {{Lasserre}}}, \bibinfo {author} {\bibfnamefont {A.}~\bibnamefont
  {{Letourneau}}}, \bibinfo {author} {\bibfnamefont {D.}~\bibnamefont
  {{Lhuillier}}}, \bibinfo {author} {\bibfnamefont {J.}~\bibnamefont {{Lima}},
  \bibfnamefont {H.~P.}}, \bibinfo {author} {\bibfnamefont {M.}~\bibnamefont
  {{Lindner}}}, \bibinfo {author} {\bibfnamefont {J.~M.}\ \bibnamefont
  {{L{\'o}pez-Castan{\~o}}}}, \bibinfo {author} {\bibfnamefont {J.~M.}\
  \bibnamefont {{LoSecco}}}, \bibinfo {author} {\bibfnamefont {B.~K.}\
  \bibnamefont {{Lubsandorzhiev}}}, \bibinfo {author} {\bibfnamefont
  {S.}~\bibnamefont {{Lucht}}}, \bibinfo {author} {\bibfnamefont
  {D.}~\bibnamefont {{McKee}}}, \bibinfo {author} {\bibfnamefont
  {J.}~\bibnamefont {{Maeda}}}, \bibinfo {author} {\bibfnamefont {C.~N.}\
  \bibnamefont {{Maesano}}}, \bibinfo {author} {\bibfnamefont {C.}~\bibnamefont
  {{Mariani}}}, \bibinfo {author} {\bibfnamefont {J.}~\bibnamefont
  {{Maricic}}}, \bibinfo {author} {\bibfnamefont {J.}~\bibnamefont
  {{Martino}}}, \bibinfo {author} {\bibfnamefont {T.}~\bibnamefont
  {{Matsubara}}}, \bibinfo {author} {\bibfnamefont {G.}~\bibnamefont
  {{Mention}}}, \bibinfo {author} {\bibfnamefont {A.}~\bibnamefont
  {{Meregaglia}}}, \bibinfo {author} {\bibfnamefont {T.}~\bibnamefont
  {{Miletic}}}, \bibinfo {author} {\bibfnamefont {R.}~\bibnamefont
  {{Milincic}}}, \bibinfo {author} {\bibfnamefont {H.}~\bibnamefont
  {{Miyata}}}, \bibinfo {author} {\bibfnamefont {T.~A.}\ \bibnamefont
  {{Mueller}}}, \bibinfo {author} {\bibfnamefont {Y.}~\bibnamefont
  {{Nagasaka}}}, \bibinfo {author} {\bibfnamefont {K.}~\bibnamefont
  {{Nakajima}}}, \bibinfo {author} {\bibfnamefont {P.}~\bibnamefont
  {{Novella}}}, \bibinfo {author} {\bibfnamefont {M.}~\bibnamefont
  {{Obolensky}}}, \bibinfo {author} {\bibfnamefont {L.}~\bibnamefont
  {{Oberauer}}}, \bibinfo {author} {\bibfnamefont {A.}~\bibnamefont
  {{Onillon}}}, \bibinfo {author} {\bibfnamefont {A.}~\bibnamefont {{Osborn}}},
  \bibinfo {author} {\bibfnamefont {I.}~\bibnamefont {{Ostrovskiy}}}, \bibinfo
  {author} {\bibfnamefont {C.}~\bibnamefont {{Palomares}}}, \bibinfo {author}
  {\bibfnamefont {I.~M.}\ \bibnamefont {{Pepe}}}, \bibinfo {author}
  {\bibfnamefont {S.}~\bibnamefont {{Perasso}}}, \bibinfo {author}
  {\bibfnamefont {P.}~\bibnamefont {{Perrin}}}, \bibinfo {author}
  {\bibfnamefont {P.}~\bibnamefont {{Pfahler}}}, \bibinfo {author}
  {\bibfnamefont {A.}~\bibnamefont {{Porta}}}, \bibinfo {author} {\bibfnamefont
  {W.}~\bibnamefont {{Potzel}}}, \bibinfo {author} {\bibfnamefont
  {J.}~\bibnamefont {{Reichenbacher}}}, \bibinfo {author} {\bibfnamefont
  {B.}~\bibnamefont {{Reinhold}}}, \bibinfo {author} {\bibfnamefont
  {A.}~\bibnamefont {{Remoto}}}, \bibinfo {author} {\bibfnamefont
  {M.}~\bibnamefont {{R{\"o}hling}}}, \bibinfo {author} {\bibfnamefont
  {R.}~\bibnamefont {{Roncin}}}, \bibinfo {author} {\bibfnamefont
  {S.}~\bibnamefont {{Roth}}}, \bibinfo {author} {\bibfnamefont
  {Y.}~\bibnamefont {{Sakamoto}}}, \bibinfo {author} {\bibfnamefont
  {R.}~\bibnamefont {{Santorelli}}}, \bibinfo {author} {\bibfnamefont
  {F.}~\bibnamefont {{Sato}}}, \bibinfo {author} {\bibfnamefont
  {S.}~\bibnamefont {{Sch{\"o}nert}}}, \bibinfo {author} {\bibfnamefont
  {S.}~\bibnamefont {{Schoppmann}}}, \bibinfo {author} {\bibfnamefont
  {T.}~\bibnamefont {{Schwetz}}}, \bibinfo {author} {\bibfnamefont {M.~H.}\
  \bibnamefont {{Shaevitz}}}, \bibinfo {author} {\bibfnamefont
  {S.}~\bibnamefont {{Shimojima}}}, \bibinfo {author} {\bibfnamefont
  {D.}~\bibnamefont {{Shrestha}}}, \bibinfo {author} {\bibfnamefont {J.~L.}\
  \bibnamefont {{Sida}}}, \bibinfo {author} {\bibfnamefont {V.}~\bibnamefont
  {{Sinev}}}, \bibinfo {author} {\bibfnamefont {M.}~\bibnamefont
  {{Skorokhvatov}}}, \bibinfo {author} {\bibfnamefont {E.}~\bibnamefont
  {{Smith}}}, \bibinfo {author} {\bibfnamefont {J.}~\bibnamefont {{Spitz}}},
  \bibinfo {author} {\bibfnamefont {A.}~\bibnamefont {{Stahl}}}, \bibinfo
  {author} {\bibfnamefont {I.}~\bibnamefont {{Stancu}}}, \bibinfo {author}
  {\bibfnamefont {L.~F.~F.}\ \bibnamefont {{Stokes}}}, \bibinfo {author}
  {\bibfnamefont {M.}~\bibnamefont {{Strait}}}, \bibinfo {author}
  {\bibfnamefont {A.}~\bibnamefont {{St{\"u}ken}}}, \bibinfo {author}
  {\bibfnamefont {F.}~\bibnamefont {{Suekane}}}, \bibinfo {author}
  {\bibfnamefont {S.}~\bibnamefont {{Sukhotin}}}, \bibinfo {author}
  {\bibfnamefont {T.}~\bibnamefont {{Sumiyoshi}}}, \bibinfo {author}
  {\bibfnamefont {Y.}~\bibnamefont {{Sun}}}, \bibinfo {author} {\bibfnamefont
  {R.}~\bibnamefont {{Svoboda}}}, \bibinfo {author} {\bibfnamefont
  {K.}~\bibnamefont {{Terao}}}, \bibinfo {author} {\bibfnamefont
  {A.}~\bibnamefont {{Tonazzo}}}, \bibinfo {author} {\bibfnamefont
  {M.}~\bibnamefont {{Toups}}}, \bibinfo {author} {\bibfnamefont {H.~H.}\
  \bibnamefont {{Trinh Thi}}}, \bibinfo {author} {\bibfnamefont
  {G.}~\bibnamefont {{Valdiviesso}}}, \bibinfo {author} {\bibfnamefont
  {C.}~\bibnamefont {{Veyssiere}}}, \bibinfo {author} {\bibfnamefont
  {S.}~\bibnamefont {{Wagner}}}, \bibinfo {author} {\bibfnamefont
  {H.}~\bibnamefont {{Watanabe}}}, \bibinfo {author} {\bibfnamefont
  {B.}~\bibnamefont {{White}}}, \bibinfo {author} {\bibfnamefont
  {C.}~\bibnamefont {{Wiebusch}}}, \bibinfo {author} {\bibfnamefont
  {L.}~\bibnamefont {{Winslow}}}, \bibinfo {author} {\bibfnamefont
  {M.}~\bibnamefont {{Worcester}}}, \bibinfo {author} {\bibfnamefont
  {M.}~\bibnamefont {{Wurm}}}, \bibinfo {author} {\bibfnamefont
  {F.}~\bibnamefont {{Yermia}}}, \ and\ \bibinfo {author} {\bibfnamefont
  {V.}~\bibnamefont {{Zimmer}}},\ }\href {\doibase 10.1103/PhysRevD.86.052008}
  {\bibfield  {journal} {\bibinfo  {journal} {\prd}\ }\textbf {\bibinfo
  {volume} {86}},\ \bibinfo {eid} {052008} (\bibinfo {year} {2012})},\ \Eprint
  {http://arxiv.org/abs/1207.6632} {arXiv:1207.6632 [hep-ex]} \BibitemShut
  {NoStop}%
\bibitem [{\citenamefont {{Ahn}}\ \emph {et~al.}(2012)\citenamefont {{Ahn}},
  \citenamefont {{Chebotaryov}}, \citenamefont {{Choi}}, \citenamefont
  {{Choi}}, \citenamefont {{Choi}}, \citenamefont {{Choi}}, \citenamefont
  {{Jang}}, \citenamefont {{Jang}}, \citenamefont {{Jeon}}, \citenamefont
  {{Jeong}}, \citenamefont {{Joo}}, \citenamefont {{Kim}}, \citenamefont
  {{Kim}}, \citenamefont {{Kim}}, \citenamefont {{Kim}}, \citenamefont {{Kim}},
  \citenamefont {{Kim}}, \citenamefont {{Kim}}, \citenamefont {{Kim}},
  \citenamefont {{Kim}}, \citenamefont {{Lee}}, \citenamefont {{Lee}},
  \citenamefont {{Lim}}, \citenamefont {{Ma}}, \citenamefont {{Pac}},
  \citenamefont {{Park}}, \citenamefont {{Park}}, \citenamefont {{Park}},
  \citenamefont {{Shin}}, \citenamefont {{Siyeon}}, \citenamefont {{Yang}},
  \citenamefont {{Yeo}}, \citenamefont {{Yi}},\ and\ \citenamefont
  {{Yu}}}]{1204.0626}%
  \BibitemOpen
  \bibfield  {author} {\bibinfo {author} {\bibfnamefont {J.~K.}\ \bibnamefont
  {{Ahn}}}, \bibinfo {author} {\bibfnamefont {S.}~\bibnamefont
  {{Chebotaryov}}}, \bibinfo {author} {\bibfnamefont {J.~H.}\ \bibnamefont
  {{Choi}}}, \bibinfo {author} {\bibfnamefont {S.}~\bibnamefont {{Choi}}},
  \bibinfo {author} {\bibfnamefont {W.}~\bibnamefont {{Choi}}}, \bibinfo
  {author} {\bibfnamefont {Y.}~\bibnamefont {{Choi}}}, \bibinfo {author}
  {\bibfnamefont {H.~I.}\ \bibnamefont {{Jang}}}, \bibinfo {author}
  {\bibfnamefont {J.~S.}\ \bibnamefont {{Jang}}}, \bibinfo {author}
  {\bibfnamefont {E.~J.}\ \bibnamefont {{Jeon}}}, \bibinfo {author}
  {\bibfnamefont {I.~S.}\ \bibnamefont {{Jeong}}}, \bibinfo {author}
  {\bibfnamefont {K.~K.}\ \bibnamefont {{Joo}}}, \bibinfo {author}
  {\bibfnamefont {B.~R.}\ \bibnamefont {{Kim}}}, \bibinfo {author}
  {\bibfnamefont {B.~C.}\ \bibnamefont {{Kim}}}, \bibinfo {author}
  {\bibfnamefont {H.~S.}\ \bibnamefont {{Kim}}}, \bibinfo {author}
  {\bibfnamefont {J.~Y.}\ \bibnamefont {{Kim}}}, \bibinfo {author}
  {\bibfnamefont {S.~B.}\ \bibnamefont {{Kim}}}, \bibinfo {author}
  {\bibfnamefont {S.~H.}\ \bibnamefont {{Kim}}}, \bibinfo {author}
  {\bibfnamefont {S.~Y.}\ \bibnamefont {{Kim}}}, \bibinfo {author}
  {\bibfnamefont {W.}~\bibnamefont {{Kim}}}, \bibinfo {author} {\bibfnamefont
  {Y.~D.}\ \bibnamefont {{Kim}}}, \bibinfo {author} {\bibfnamefont
  {J.}~\bibnamefont {{Lee}}}, \bibinfo {author} {\bibfnamefont {J.~K.}\
  \bibnamefont {{Lee}}}, \bibinfo {author} {\bibfnamefont {I.~T.}\ \bibnamefont
  {{Lim}}}, \bibinfo {author} {\bibfnamefont {K.~J.}\ \bibnamefont {{Ma}}},
  \bibinfo {author} {\bibfnamefont {M.~Y.}\ \bibnamefont {{Pac}}}, \bibinfo
  {author} {\bibfnamefont {I.~G.}\ \bibnamefont {{Park}}}, \bibinfo {author}
  {\bibfnamefont {J.~S.}\ \bibnamefont {{Park}}}, \bibinfo {author}
  {\bibfnamefont {K.~S.}\ \bibnamefont {{Park}}}, \bibinfo {author}
  {\bibfnamefont {J.~W.}\ \bibnamefont {{Shin}}}, \bibinfo {author}
  {\bibfnamefont {K.}~\bibnamefont {{Siyeon}}}, \bibinfo {author}
  {\bibfnamefont {B.~S.}\ \bibnamefont {{Yang}}}, \bibinfo {author}
  {\bibfnamefont {I.~S.}\ \bibnamefont {{Yeo}}}, \bibinfo {author}
  {\bibfnamefont {S.~H.}\ \bibnamefont {{Yi}}}, \ and\ \bibinfo {author}
  {\bibfnamefont {I.}~\bibnamefont {{Yu}}},\ }\href {\doibase
  10.1103/PhysRevLett.108.191802} {\bibfield  {journal} {\bibinfo  {journal}
  {\prl}\ }\textbf {\bibinfo {volume} {108}},\ \bibinfo {eid} {191802}
  (\bibinfo {year} {2012})},\ \Eprint {http://arxiv.org/abs/1204.0626}
  {arXiv:1204.0626 [hep-ex]} \BibitemShut {NoStop}%
\bibitem [{\citenamefont {{An}}\ \emph {et~al.}(2012)\citenamefont {{An}},
  \citenamefont {{Bai}}, \citenamefont {{Balantekin}}, \citenamefont {{Band}},
  \citenamefont {{Beavis}}, \citenamefont {{Beriguete}}, \citenamefont
  {{Bishai}}, \citenamefont {{Blyth}}, \citenamefont {{Boddy}}, \citenamefont
  {{Brown}},\ and\ \citenamefont {et~al.}}]{1203.1669}%
  \BibitemOpen
  \bibfield  {author} {\bibinfo {author} {\bibfnamefont {F.~P.}\ \bibnamefont
  {{An}}}, \bibinfo {author} {\bibfnamefont {J.~Z.}\ \bibnamefont {{Bai}}},
  \bibinfo {author} {\bibfnamefont {A.~B.}\ \bibnamefont {{Balantekin}}},
  \bibinfo {author} {\bibfnamefont {H.~R.}\ \bibnamefont {{Band}}}, \bibinfo
  {author} {\bibfnamefont {D.}~\bibnamefont {{Beavis}}}, \bibinfo {author}
  {\bibfnamefont {W.}~\bibnamefont {{Beriguete}}}, \bibinfo {author}
  {\bibfnamefont {M.}~\bibnamefont {{Bishai}}}, \bibinfo {author}
  {\bibfnamefont {S.}~\bibnamefont {{Blyth}}}, \bibinfo {author} {\bibfnamefont
  {K.}~\bibnamefont {{Boddy}}}, \bibinfo {author} {\bibfnamefont {R.~L.}\
  \bibnamefont {{Brown}}}, \ and\ \bibinfo {author} {\bibnamefont {et~al.}},\
  }\href {\doibase 10.1103/PhysRevLett.108.171803} {\bibfield  {journal}
  {\bibinfo  {journal} {\prl}\ }\textbf {\bibinfo {volume} {108}},\ \bibinfo
  {eid} {171803} (\bibinfo {year} {2012})},\ \Eprint
  {http://arxiv.org/abs/1203.1669} {arXiv:1203.1669 [hep-ex]} \BibitemShut
  {NoStop}%
\bibitem [{\citenamefont {{Adamson}}\ \emph {et~al.}(2008)\citenamefont
  {{Adamson}}, \citenamefont {{Andreopoulos}}, \citenamefont {{Arms}},
  \citenamefont {{Armstrong}}, \citenamefont {{Auty}}, \citenamefont {{Ayres}},
  \citenamefont {{Baller}}, \citenamefont {{Barnes}}, \citenamefont {{Barr}},
  \citenamefont {{Barrett}}, \citenamefont {{Becker}}, \citenamefont
  {{Belias}}, \citenamefont {{Bernstein}}, \citenamefont {{Bhattacharya}},
  \citenamefont {{Bishai}}, \citenamefont {{Blake}}, \citenamefont {{Bock}},
  \citenamefont {{Boehm}}, \citenamefont {{Boehnlein}}, \citenamefont
  {{Bogert}}, \citenamefont {{Bower}}, \citenamefont {{Buckley-Geer}},
  \citenamefont {{Cavanaugh}}, \citenamefont {{Chapman}}, \citenamefont
  {{Cherdack}}, \citenamefont {{Childress}}, \citenamefont {{Choudhary}},
  \citenamefont {{Cobb}}, \citenamefont {{Coleman}}, \citenamefont {{Culling}},
  \citenamefont {{de Jong}}, \citenamefont {{Dierckxsens}}, \citenamefont
  {{Diwan}}, \citenamefont {{Dorman}}, \citenamefont {{Dytman}}, \citenamefont
  {{Escobar}}, \citenamefont {{Evans}}, \citenamefont {{Falk Harris}},
  \citenamefont {{Feldman}}, \citenamefont {{Frohne}}, \citenamefont
  {{Gallagher}}, \citenamefont {{Godley}}, \citenamefont {{Goodman}},
  \citenamefont {{Gouffon}}, \citenamefont {{Gran}}, \citenamefont
  {{Grashorn}}, \citenamefont {{Grossman}}, \citenamefont {{Grzelak}},
  \citenamefont {{Habig}}, \citenamefont {{Harris}}, \citenamefont {{Harris}},
  \citenamefont {{Hartnell}}, \citenamefont {{Hatcher}}, \citenamefont
  {{Heller}}, \citenamefont {{Himmel}}, \citenamefont {{Holin}}, \citenamefont
  {{Hylen}}, \citenamefont {{Irwin}}, \citenamefont {{Ishitsuka}},
  \citenamefont {{Jaffe}}, \citenamefont {{James}}, \citenamefont {{Jensen}},
  \citenamefont {{Kafka}}, \citenamefont {{Kasahara}}, \citenamefont {{Kim}},
  \citenamefont {{Kim}}, \citenamefont {{Koizumi}}, \citenamefont {{Kopp}},
  \citenamefont {{Kordosky}}, \citenamefont {{Koskinen}}, \citenamefont
  {{Kotelnikov}}, \citenamefont {{Kreymer}}, \citenamefont {{Kumaratunga}},
  \citenamefont {{Lang}}, \citenamefont {{Ling}}, \citenamefont {{Litchfield}},
  \citenamefont {{Litchfield}}, \citenamefont {{Loiacono}}, \citenamefont
  {{Lucas}}, \citenamefont {{Ma}}, \citenamefont {{Mann}}, \citenamefont
  {{Marchionni}}, \citenamefont {{Marshak}}, \citenamefont {{Marshall}},
  \citenamefont {{Mayer}}, \citenamefont {{McGowan}}, \citenamefont {{Meier}},
  \citenamefont {{Merzon}}, \citenamefont {{Messier}}, \citenamefont
  {{Metelko}}, \citenamefont {{Michael}}, \citenamefont {{Miller}},
  \citenamefont {{Miller}}, \citenamefont {{Mishra}}, \citenamefont {{Moore}},
  \citenamefont {{Morf{\'\i}n}}, \citenamefont {{Mualem}}, \citenamefont
  {{Mufson}}, \citenamefont {{Murgia}}, \citenamefont {{Musser}}, \citenamefont
  {{Naples}}, \citenamefont {{Nelson}}, \citenamefont {{Newman}}, \citenamefont
  {{Nichol}}, \citenamefont {{Nicholls}}, \citenamefont {{Ochoa-Ricoux}},
  \citenamefont {{Oliver}}, \citenamefont {{Ospanov}}, \citenamefont {{Paley}},
  \citenamefont {{Paolone}}, \citenamefont {{Para}}, \citenamefont {{Patzak}},
  \citenamefont {{Pavlovi{\'c}}}, \citenamefont {{Pawloski}}, \citenamefont
  {{Pearce}}, \citenamefont {{Peck}}, \citenamefont {{Peterson}}, \citenamefont
  {{Petyt}}, \citenamefont {{Pittam}}, \citenamefont {{Plunkett}},
  \citenamefont {{Rahaman}}, \citenamefont {{Rameika}}, \citenamefont
  {{Raufer}}, \citenamefont {{Rebel}}, \citenamefont {{Reichenbacher}},
  \citenamefont {{Rodrigues}}, \citenamefont {{Rosenfeld}}, \citenamefont
  {{Rubin}}, \citenamefont {{Ruddick}}, \citenamefont {{Ryabov}}, \citenamefont
  {{Sanchez}}, \citenamefont {{Saoulidou}}, \citenamefont {{Schneps}},
  \citenamefont {{Schreiner}}, \citenamefont {{Seun}}, \citenamefont
  {{Shanahan}}, \citenamefont {{Smart}}, \citenamefont {{Smith}}, \citenamefont
  {{Sousa}}, \citenamefont {{Speakman}}, \citenamefont {{Stamoulis}},
  \citenamefont {{Strait}}, \citenamefont {{Symes}}, \citenamefont {{Tagg}},
  \citenamefont {{Talaga}}, \citenamefont {{Tavera}}, \citenamefont {{Thomas}},
  \citenamefont {{Thompson}}, \citenamefont {{Thomson}}, \citenamefont
  {{Thron}}, \citenamefont {{Tinti}}, \citenamefont {{Trostin}}, \citenamefont
  {{Tsarev}}, \citenamefont {{Tzanakos}}, \citenamefont {{Urheim}},
  \citenamefont {{Vahle}}, \citenamefont {{Viren}}, \citenamefont {{Ward}},
  \citenamefont {{Ward}}, \citenamefont {{Watabe}}, \citenamefont {{Weber}},
  \citenamefont {{Webb}}, \citenamefont {{Wehmann}}, \citenamefont {{West}},
  \citenamefont {{White}}, \citenamefont {{Wojcicki}}, \citenamefont
  {{Wright}}, \citenamefont {{Yang}}, \citenamefont {{Zois}}, \citenamefont
  {{Zhang}},\ and\ \citenamefont {{Zwaska}}}]{0806.2237}%
  \BibitemOpen
  \bibfield  {author} {\bibinfo {author} {\bibfnamefont {P.}~\bibnamefont
  {{Adamson}}}, \bibinfo {author} {\bibfnamefont {C.}~\bibnamefont
  {{Andreopoulos}}}, \bibinfo {author} {\bibfnamefont {K.~E.}\ \bibnamefont
  {{Arms}}}, \bibinfo {author} {\bibfnamefont {R.}~\bibnamefont {{Armstrong}}},
  \bibinfo {author} {\bibfnamefont {D.~J.}\ \bibnamefont {{Auty}}}, \bibinfo
  {author} {\bibfnamefont {D.~S.}\ \bibnamefont {{Ayres}}}, \bibinfo {author}
  {\bibfnamefont {B.}~\bibnamefont {{Baller}}}, \bibinfo {author}
  {\bibfnamefont {J.}~\bibnamefont {{Barnes}}, \bibfnamefont {P.~D.}}, \bibinfo
  {author} {\bibfnamefont {G.}~\bibnamefont {{Barr}}}, \bibinfo {author}
  {\bibfnamefont {W.~L.}\ \bibnamefont {{Barrett}}}, \bibinfo {author}
  {\bibfnamefont {B.~R.}\ \bibnamefont {{Becker}}}, \bibinfo {author}
  {\bibfnamefont {A.}~\bibnamefont {{Belias}}}, \bibinfo {author}
  {\bibfnamefont {R.~H.}\ \bibnamefont {{Bernstein}}}, \bibinfo {author}
  {\bibfnamefont {D.}~\bibnamefont {{Bhattacharya}}}, \bibinfo {author}
  {\bibfnamefont {M.}~\bibnamefont {{Bishai}}}, \bibinfo {author}
  {\bibfnamefont {A.}~\bibnamefont {{Blake}}}, \bibinfo {author} {\bibfnamefont
  {G.~J.}\ \bibnamefont {{Bock}}}, \bibinfo {author} {\bibfnamefont
  {J.}~\bibnamefont {{Boehm}}}, \bibinfo {author} {\bibfnamefont {D.~J.}\
  \bibnamefont {{Boehnlein}}}, \bibinfo {author} {\bibfnamefont
  {D.}~\bibnamefont {{Bogert}}}, \bibinfo {author} {\bibfnamefont
  {C.}~\bibnamefont {{Bower}}}, \bibinfo {author} {\bibfnamefont
  {E.}~\bibnamefont {{Buckley-Geer}}}, \bibinfo {author} {\bibfnamefont
  {S.}~\bibnamefont {{Cavanaugh}}}, \bibinfo {author} {\bibfnamefont {J.~D.}\
  \bibnamefont {{Chapman}}}, \bibinfo {author} {\bibfnamefont {D.}~\bibnamefont
  {{Cherdack}}}, \bibinfo {author} {\bibfnamefont {S.}~\bibnamefont
  {{Childress}}}, \bibinfo {author} {\bibfnamefont {B.~C.}\ \bibnamefont
  {{Choudhary}}}, \bibinfo {author} {\bibfnamefont {J.~H.}\ \bibnamefont
  {{Cobb}}}, \bibinfo {author} {\bibfnamefont {S.~J.}\ \bibnamefont
  {{Coleman}}}, \bibinfo {author} {\bibfnamefont {A.~J.}\ \bibnamefont
  {{Culling}}}, \bibinfo {author} {\bibfnamefont {J.~K.}\ \bibnamefont {{de
  Jong}}}, \bibinfo {author} {\bibfnamefont {M.}~\bibnamefont {{Dierckxsens}}},
  \bibinfo {author} {\bibfnamefont {M.~V.}\ \bibnamefont {{Diwan}}}, \bibinfo
  {author} {\bibfnamefont {M.}~\bibnamefont {{Dorman}}}, \bibinfo {author}
  {\bibfnamefont {S.~A.}\ \bibnamefont {{Dytman}}}, \bibinfo {author}
  {\bibfnamefont {C.~O.}\ \bibnamefont {{Escobar}}}, \bibinfo {author}
  {\bibfnamefont {J.~J.}\ \bibnamefont {{Evans}}}, \bibinfo {author}
  {\bibfnamefont {E.}~\bibnamefont {{Falk Harris}}}, \bibinfo {author}
  {\bibfnamefont {G.~J.}\ \bibnamefont {{Feldman}}}, \bibinfo {author}
  {\bibfnamefont {M.~V.}\ \bibnamefont {{Frohne}}}, \bibinfo {author}
  {\bibfnamefont {H.~R.}\ \bibnamefont {{Gallagher}}}, \bibinfo {author}
  {\bibfnamefont {A.}~\bibnamefont {{Godley}}}, \bibinfo {author}
  {\bibfnamefont {M.~C.}\ \bibnamefont {{Goodman}}}, \bibinfo {author}
  {\bibfnamefont {P.}~\bibnamefont {{Gouffon}}}, \bibinfo {author}
  {\bibfnamefont {R.}~\bibnamefont {{Gran}}}, \bibinfo {author} {\bibfnamefont
  {E.~W.}\ \bibnamefont {{Grashorn}}}, \bibinfo {author} {\bibfnamefont
  {N.}~\bibnamefont {{Grossman}}}, \bibinfo {author} {\bibfnamefont
  {K.}~\bibnamefont {{Grzelak}}}, \bibinfo {author} {\bibfnamefont
  {A.}~\bibnamefont {{Habig}}}, \bibinfo {author} {\bibfnamefont
  {D.}~\bibnamefont {{Harris}}}, \bibinfo {author} {\bibfnamefont {P.~G.}\
  \bibnamefont {{Harris}}}, \bibinfo {author} {\bibfnamefont {J.}~\bibnamefont
  {{Hartnell}}}, \bibinfo {author} {\bibfnamefont {R.}~\bibnamefont
  {{Hatcher}}}, \bibinfo {author} {\bibfnamefont {K.}~\bibnamefont {{Heller}}},
  \bibinfo {author} {\bibfnamefont {A.}~\bibnamefont {{Himmel}}}, \bibinfo
  {author} {\bibfnamefont {A.}~\bibnamefont {{Holin}}}, \bibinfo {author}
  {\bibfnamefont {J.}~\bibnamefont {{Hylen}}}, \bibinfo {author} {\bibfnamefont
  {G.~M.}\ \bibnamefont {{Irwin}}}, \bibinfo {author} {\bibfnamefont
  {M.}~\bibnamefont {{Ishitsuka}}}, \bibinfo {author} {\bibfnamefont {D.~E.}\
  \bibnamefont {{Jaffe}}}, \bibinfo {author} {\bibfnamefont {C.}~\bibnamefont
  {{James}}}, \bibinfo {author} {\bibfnamefont {D.}~\bibnamefont {{Jensen}}},
  \bibinfo {author} {\bibfnamefont {T.}~\bibnamefont {{Kafka}}}, \bibinfo
  {author} {\bibfnamefont {S.~M.~S.}\ \bibnamefont {{Kasahara}}}, \bibinfo
  {author} {\bibfnamefont {J.~J.}\ \bibnamefont {{Kim}}}, \bibinfo {author}
  {\bibfnamefont {M.~S.}\ \bibnamefont {{Kim}}}, \bibinfo {author}
  {\bibfnamefont {G.}~\bibnamefont {{Koizumi}}}, \bibinfo {author}
  {\bibfnamefont {S.}~\bibnamefont {{Kopp}}}, \bibinfo {author} {\bibfnamefont
  {M.}~\bibnamefont {{Kordosky}}}, \bibinfo {author} {\bibfnamefont {D.~J.}\
  \bibnamefont {{Koskinen}}}, \bibinfo {author} {\bibfnamefont {S.~K.}\
  \bibnamefont {{Kotelnikov}}}, \bibinfo {author} {\bibfnamefont
  {A.}~\bibnamefont {{Kreymer}}}, \bibinfo {author} {\bibfnamefont
  {S.}~\bibnamefont {{Kumaratunga}}}, \bibinfo {author} {\bibfnamefont
  {K.}~\bibnamefont {{Lang}}}, \bibinfo {author} {\bibfnamefont
  {J.}~\bibnamefont {{Ling}}}, \bibinfo {author} {\bibfnamefont {P.~J.}\
  \bibnamefont {{Litchfield}}}, \bibinfo {author} {\bibfnamefont {R.~P.}\
  \bibnamefont {{Litchfield}}}, \bibinfo {author} {\bibfnamefont
  {L.}~\bibnamefont {{Loiacono}}}, \bibinfo {author} {\bibfnamefont
  {P.}~\bibnamefont {{Lucas}}}, \bibinfo {author} {\bibfnamefont
  {J.}~\bibnamefont {{Ma}}}, \bibinfo {author} {\bibfnamefont {W.~A.}\
  \bibnamefont {{Mann}}}, \bibinfo {author} {\bibfnamefont {A.}~\bibnamefont
  {{Marchionni}}}, \bibinfo {author} {\bibfnamefont {M.~L.}\ \bibnamefont
  {{Marshak}}}, \bibinfo {author} {\bibfnamefont {J.~S.}\ \bibnamefont
  {{Marshall}}}, \bibinfo {author} {\bibfnamefont {N.}~\bibnamefont {{Mayer}}},
  \bibinfo {author} {\bibfnamefont {A.~M.}\ \bibnamefont {{McGowan}}}, \bibinfo
  {author} {\bibfnamefont {J.~R.}\ \bibnamefont {{Meier}}}, \bibinfo {author}
  {\bibfnamefont {G.~I.}\ \bibnamefont {{Merzon}}}, \bibinfo {author}
  {\bibfnamefont {M.~D.}\ \bibnamefont {{Messier}}}, \bibinfo {author}
  {\bibfnamefont {C.~J.}\ \bibnamefont {{Metelko}}}, \bibinfo {author}
  {\bibfnamefont {D.~G.}\ \bibnamefont {{Michael}}}, \bibinfo {author}
  {\bibfnamefont {J.~L.}\ \bibnamefont {{Miller}}}, \bibinfo {author}
  {\bibfnamefont {W.~H.}\ \bibnamefont {{Miller}}}, \bibinfo {author}
  {\bibfnamefont {S.~R.}\ \bibnamefont {{Mishra}}}, \bibinfo {author}
  {\bibfnamefont {C.~D.}\ \bibnamefont {{Moore}}}, \bibinfo {author}
  {\bibfnamefont {J.}~\bibnamefont {{Morf{\'\i}n}}}, \bibinfo {author}
  {\bibfnamefont {L.}~\bibnamefont {{Mualem}}}, \bibinfo {author}
  {\bibfnamefont {S.}~\bibnamefont {{Mufson}}}, \bibinfo {author}
  {\bibfnamefont {S.}~\bibnamefont {{Murgia}}}, \bibinfo {author}
  {\bibfnamefont {J.}~\bibnamefont {{Musser}}}, \bibinfo {author}
  {\bibfnamefont {D.}~\bibnamefont {{Naples}}}, \bibinfo {author}
  {\bibfnamefont {J.~K.}\ \bibnamefont {{Nelson}}}, \bibinfo {author}
  {\bibfnamefont {H.~B.}\ \bibnamefont {{Newman}}}, \bibinfo {author}
  {\bibfnamefont {R.~J.}\ \bibnamefont {{Nichol}}}, \bibinfo {author}
  {\bibfnamefont {T.~C.}\ \bibnamefont {{Nicholls}}}, \bibinfo {author}
  {\bibfnamefont {J.~P.}\ \bibnamefont {{Ochoa-Ricoux}}}, \bibinfo {author}
  {\bibfnamefont {W.~P.}\ \bibnamefont {{Oliver}}}, \bibinfo {author}
  {\bibfnamefont {R.}~\bibnamefont {{Ospanov}}}, \bibinfo {author}
  {\bibfnamefont {J.}~\bibnamefont {{Paley}}}, \bibinfo {author} {\bibfnamefont
  {V.}~\bibnamefont {{Paolone}}}, \bibinfo {author} {\bibfnamefont
  {A.}~\bibnamefont {{Para}}}, \bibinfo {author} {\bibfnamefont
  {T.}~\bibnamefont {{Patzak}}}, \bibinfo {author} {\bibfnamefont
  {{\v{Z}}.}~\bibnamefont {{Pavlovi{\'c}}}}, \bibinfo {author} {\bibfnamefont
  {G.}~\bibnamefont {{Pawloski}}}, \bibinfo {author} {\bibfnamefont {G.~F.}\
  \bibnamefont {{Pearce}}}, \bibinfo {author} {\bibfnamefont {C.~W.}\
  \bibnamefont {{Peck}}}, \bibinfo {author} {\bibfnamefont {E.~A.}\
  \bibnamefont {{Peterson}}}, \bibinfo {author} {\bibfnamefont {D.~A.}\
  \bibnamefont {{Petyt}}}, \bibinfo {author} {\bibfnamefont {R.}~\bibnamefont
  {{Pittam}}}, \bibinfo {author} {\bibfnamefont {R.~K.}\ \bibnamefont
  {{Plunkett}}}, \bibinfo {author} {\bibfnamefont {A.}~\bibnamefont
  {{Rahaman}}}, \bibinfo {author} {\bibfnamefont {R.~A.}\ \bibnamefont
  {{Rameika}}}, \bibinfo {author} {\bibfnamefont {T.~M.}\ \bibnamefont
  {{Raufer}}}, \bibinfo {author} {\bibfnamefont {B.}~\bibnamefont {{Rebel}}},
  \bibinfo {author} {\bibfnamefont {J.}~\bibnamefont {{Reichenbacher}}},
  \bibinfo {author} {\bibfnamefont {P.~A.}\ \bibnamefont {{Rodrigues}}},
  \bibinfo {author} {\bibfnamefont {C.}~\bibnamefont {{Rosenfeld}}}, \bibinfo
  {author} {\bibfnamefont {H.~A.}\ \bibnamefont {{Rubin}}}, \bibinfo {author}
  {\bibfnamefont {K.}~\bibnamefont {{Ruddick}}}, \bibinfo {author}
  {\bibfnamefont {V.~A.}\ \bibnamefont {{Ryabov}}}, \bibinfo {author}
  {\bibfnamefont {M.~C.}\ \bibnamefont {{Sanchez}}}, \bibinfo {author}
  {\bibfnamefont {N.}~\bibnamefont {{Saoulidou}}}, \bibinfo {author}
  {\bibfnamefont {J.}~\bibnamefont {{Schneps}}}, \bibinfo {author}
  {\bibfnamefont {P.}~\bibnamefont {{Schreiner}}}, \bibinfo {author}
  {\bibfnamefont {S.~M.}\ \bibnamefont {{Seun}}}, \bibinfo {author}
  {\bibfnamefont {P.}~\bibnamefont {{Shanahan}}}, \bibinfo {author}
  {\bibfnamefont {W.}~\bibnamefont {{Smart}}}, \bibinfo {author} {\bibfnamefont
  {C.}~\bibnamefont {{Smith}}}, \bibinfo {author} {\bibfnamefont
  {A.}~\bibnamefont {{Sousa}}}, \bibinfo {author} {\bibfnamefont
  {B.}~\bibnamefont {{Speakman}}}, \bibinfo {author} {\bibfnamefont
  {P.}~\bibnamefont {{Stamoulis}}}, \bibinfo {author} {\bibfnamefont
  {M.}~\bibnamefont {{Strait}}}, \bibinfo {author} {\bibfnamefont
  {P.}~\bibnamefont {{Symes}}}, \bibinfo {author} {\bibfnamefont
  {N.}~\bibnamefont {{Tagg}}}, \bibinfo {author} {\bibfnamefont {R.~L.}\
  \bibnamefont {{Talaga}}}, \bibinfo {author} {\bibfnamefont {M.~A.}\
  \bibnamefont {{Tavera}}}, \bibinfo {author} {\bibfnamefont {J.}~\bibnamefont
  {{Thomas}}}, \bibinfo {author} {\bibfnamefont {J.}~\bibnamefont
  {{Thompson}}}, \bibinfo {author} {\bibfnamefont {M.~A.}\ \bibnamefont
  {{Thomson}}}, \bibinfo {author} {\bibfnamefont {J.~L.}\ \bibnamefont
  {{Thron}}}, \bibinfo {author} {\bibfnamefont {G.}~\bibnamefont {{Tinti}}},
  \bibinfo {author} {\bibfnamefont {I.}~\bibnamefont {{Trostin}}}, \bibinfo
  {author} {\bibfnamefont {V.~A.}\ \bibnamefont {{Tsarev}}}, \bibinfo {author}
  {\bibfnamefont {G.}~\bibnamefont {{Tzanakos}}}, \bibinfo {author}
  {\bibfnamefont {J.}~\bibnamefont {{Urheim}}}, \bibinfo {author}
  {\bibfnamefont {P.}~\bibnamefont {{Vahle}}}, \bibinfo {author} {\bibfnamefont
  {B.}~\bibnamefont {{Viren}}}, \bibinfo {author} {\bibfnamefont {C.~P.}\
  \bibnamefont {{Ward}}}, \bibinfo {author} {\bibfnamefont {D.~R.}\
  \bibnamefont {{Ward}}}, \bibinfo {author} {\bibfnamefont {M.}~\bibnamefont
  {{Watabe}}}, \bibinfo {author} {\bibfnamefont {A.}~\bibnamefont {{Weber}}},
  \bibinfo {author} {\bibfnamefont {R.~C.}\ \bibnamefont {{Webb}}}, \bibinfo
  {author} {\bibfnamefont {A.}~\bibnamefont {{Wehmann}}}, \bibinfo {author}
  {\bibfnamefont {N.}~\bibnamefont {{West}}}, \bibinfo {author} {\bibfnamefont
  {C.}~\bibnamefont {{White}}}, \bibinfo {author} {\bibfnamefont {S.~G.}\
  \bibnamefont {{Wojcicki}}}, \bibinfo {author} {\bibfnamefont {D.~M.}\
  \bibnamefont {{Wright}}}, \bibinfo {author} {\bibfnamefont {T.}~\bibnamefont
  {{Yang}}}, \bibinfo {author} {\bibfnamefont {M.}~\bibnamefont {{Zois}}},
  \bibinfo {author} {\bibfnamefont {K.}~\bibnamefont {{Zhang}}}, \ and\
  \bibinfo {author} {\bibfnamefont {R.}~\bibnamefont {{Zwaska}}},\ }\href
  {\doibase 10.1103/PhysRevLett.101.131802} {\bibfield  {journal} {\bibinfo
  {journal} {\prl}\ }\textbf {\bibinfo {volume} {101}},\ \bibinfo {eid}
  {131802} (\bibinfo {year} {2008})},\ \Eprint {http://arxiv.org/abs/0806.2237}
  {arXiv:0806.2237 [hep-ex]} \BibitemShut {NoStop}%
\bibitem [{\citenamefont {{Araki}}\ \emph {et~al.}(2005)\citenamefont
  {{Araki}}, \citenamefont {{Eguchi}}, \citenamefont {{Enomoto}}, \citenamefont
  {{Furuno}}, \citenamefont {{Ichimura}}, \citenamefont {{Ikeda}},
  \citenamefont {{Inoue}}, \citenamefont {{Ishihara}}, \citenamefont
  {{Iwamoto}}, \citenamefont {{Kawashima}}, \citenamefont {{Kishimoto}},
  \citenamefont {{Koga}}, \citenamefont {{Koseki}}, \citenamefont {{Maeda}},
  \citenamefont {{Mitsui}}, \citenamefont {{Motoki}}, \citenamefont
  {{Nakajima}}, \citenamefont {{Ogawa}}, \citenamefont {{Owada}}, \citenamefont
  {{Ricol}}, \citenamefont {{Shimizu}}, \citenamefont {{Shirai}}, \citenamefont
  {{Suekane}}, \citenamefont {{Suzuki}}, \citenamefont {{Tada}}, \citenamefont
  {{Tajima}}, \citenamefont {{Tamae}}, \citenamefont {{Tsuda}}, \citenamefont
  {{Watanabe}}, \citenamefont {{Busenitz}}, \citenamefont {{Classen}},
  \citenamefont {{Djurcic}}, \citenamefont {{Keefer}}, \citenamefont
  {{McKinny}}, \citenamefont {{Mei}}, \citenamefont {{Piepke}}, \citenamefont
  {{Yakushev}}, \citenamefont {{Berger}}, \citenamefont {{Chan}}, \citenamefont
  {{Decowski}}, \citenamefont {{Dwyer}}, \citenamefont {{Freedman}},
  \citenamefont {{Fu}}, \citenamefont {{Fujikawa}}, \citenamefont {{Goldman}},
  \citenamefont {{Gray}}, \citenamefont {{Heeger}}, \citenamefont {{Lesko}},
  \citenamefont {{Luk}}, \citenamefont {{Murayama}}, \citenamefont {{Poon}},
  \citenamefont {{Steiner}}, \citenamefont {{Winslow}}, \citenamefont
  {{Horton-Smith}}, \citenamefont {{Mauger}}, \citenamefont {{McKeown}},
  \citenamefont {{Vogel}}, \citenamefont {{Lane}}, \citenamefont {{Miletic}},
  \citenamefont {{Gorham}}, \citenamefont {{Guillian}}, \citenamefont
  {{Learned}}, \citenamefont {{Maricic}}, \citenamefont {{Matsuno}},
  \citenamefont {{Pakvasa}}, \citenamefont {{Dazeley}}, \citenamefont
  {{Hatakeyama}}, \citenamefont {{Rojas}}, \citenamefont {{Svoboda}},
  \citenamefont {{Dieterle}}, \citenamefont {{Detwiler}}, \citenamefont
  {{Gratta}}, \citenamefont {{Ishii}}, \citenamefont {{Tolich}}, \citenamefont
  {{Uchida}}, \citenamefont {{Batygov}}, \citenamefont {{Bugg}}, \citenamefont
  {{Efremenko}}, \citenamefont {{Kamyshkov}}, \citenamefont {{Kozlov}},
  \citenamefont {{Nakamura}}, \citenamefont {{Gould}}, \citenamefont
  {{Karwowski}}, \citenamefont {{Markoff}}, \citenamefont {{Messimore}},
  \citenamefont {{Nakamura}}, \citenamefont {{Rohm}}, \citenamefont {{Tornow}},
  \citenamefont {{Wendell}}, \citenamefont {{Young}}, \citenamefont {{Chen}},
  \citenamefont {{Wang}},\ and\ \citenamefont {{Piquemal}}}]{hep-ex/0406035}%
  \BibitemOpen
  \bibfield  {author} {\bibinfo {author} {\bibfnamefont {T.}~\bibnamefont
  {{Araki}}}, \bibinfo {author} {\bibfnamefont {K.}~\bibnamefont {{Eguchi}}},
  \bibinfo {author} {\bibfnamefont {S.}~\bibnamefont {{Enomoto}}}, \bibinfo
  {author} {\bibfnamefont {K.}~\bibnamefont {{Furuno}}}, \bibinfo {author}
  {\bibfnamefont {K.}~\bibnamefont {{Ichimura}}}, \bibinfo {author}
  {\bibfnamefont {H.}~\bibnamefont {{Ikeda}}}, \bibinfo {author} {\bibfnamefont
  {K.}~\bibnamefont {{Inoue}}}, \bibinfo {author} {\bibfnamefont
  {K.}~\bibnamefont {{Ishihara}}}, \bibinfo {author} {\bibfnamefont
  {T.}~\bibnamefont {{Iwamoto}}}, \bibinfo {author} {\bibfnamefont
  {T.}~\bibnamefont {{Kawashima}}}, \bibinfo {author} {\bibfnamefont
  {Y.}~\bibnamefont {{Kishimoto}}}, \bibinfo {author} {\bibfnamefont
  {M.}~\bibnamefont {{Koga}}}, \bibinfo {author} {\bibfnamefont
  {Y.}~\bibnamefont {{Koseki}}}, \bibinfo {author} {\bibfnamefont
  {T.}~\bibnamefont {{Maeda}}}, \bibinfo {author} {\bibfnamefont
  {T.}~\bibnamefont {{Mitsui}}}, \bibinfo {author} {\bibfnamefont
  {M.}~\bibnamefont {{Motoki}}}, \bibinfo {author} {\bibfnamefont
  {K.}~\bibnamefont {{Nakajima}}}, \bibinfo {author} {\bibfnamefont
  {H.}~\bibnamefont {{Ogawa}}}, \bibinfo {author} {\bibfnamefont
  {K.}~\bibnamefont {{Owada}}}, \bibinfo {author} {\bibfnamefont {J.~S.}\
  \bibnamefont {{Ricol}}}, \bibinfo {author} {\bibfnamefont {I.}~\bibnamefont
  {{Shimizu}}}, \bibinfo {author} {\bibfnamefont {J.}~\bibnamefont {{Shirai}}},
  \bibinfo {author} {\bibfnamefont {F.}~\bibnamefont {{Suekane}}}, \bibinfo
  {author} {\bibfnamefont {A.}~\bibnamefont {{Suzuki}}}, \bibinfo {author}
  {\bibfnamefont {K.}~\bibnamefont {{Tada}}}, \bibinfo {author} {\bibfnamefont
  {O.}~\bibnamefont {{Tajima}}}, \bibinfo {author} {\bibfnamefont
  {K.}~\bibnamefont {{Tamae}}}, \bibinfo {author} {\bibfnamefont
  {Y.}~\bibnamefont {{Tsuda}}}, \bibinfo {author} {\bibfnamefont
  {H.}~\bibnamefont {{Watanabe}}}, \bibinfo {author} {\bibfnamefont
  {J.}~\bibnamefont {{Busenitz}}}, \bibinfo {author} {\bibfnamefont
  {T.}~\bibnamefont {{Classen}}}, \bibinfo {author} {\bibfnamefont
  {Z.}~\bibnamefont {{Djurcic}}}, \bibinfo {author} {\bibfnamefont
  {G.}~\bibnamefont {{Keefer}}}, \bibinfo {author} {\bibfnamefont
  {K.}~\bibnamefont {{McKinny}}}, \bibinfo {author} {\bibfnamefont {D.~M.}\
  \bibnamefont {{Mei}}}, \bibinfo {author} {\bibfnamefont {A.}~\bibnamefont
  {{Piepke}}}, \bibinfo {author} {\bibfnamefont {E.}~\bibnamefont
  {{Yakushev}}}, \bibinfo {author} {\bibfnamefont {B.~E.}\ \bibnamefont
  {{Berger}}}, \bibinfo {author} {\bibfnamefont {Y.~D.}\ \bibnamefont
  {{Chan}}}, \bibinfo {author} {\bibfnamefont {M.~P.}\ \bibnamefont
  {{Decowski}}}, \bibinfo {author} {\bibfnamefont {D.~A.}\ \bibnamefont
  {{Dwyer}}}, \bibinfo {author} {\bibfnamefont {S.~J.}\ \bibnamefont
  {{Freedman}}}, \bibinfo {author} {\bibfnamefont {Y.}~\bibnamefont {{Fu}}},
  \bibinfo {author} {\bibfnamefont {B.~K.}\ \bibnamefont {{Fujikawa}}},
  \bibinfo {author} {\bibfnamefont {J.}~\bibnamefont {{Goldman}}}, \bibinfo
  {author} {\bibfnamefont {F.}~\bibnamefont {{Gray}}}, \bibinfo {author}
  {\bibfnamefont {K.~M.}\ \bibnamefont {{Heeger}}}, \bibinfo {author}
  {\bibfnamefont {K.~T.}\ \bibnamefont {{Lesko}}}, \bibinfo {author}
  {\bibfnamefont {K.~B.}\ \bibnamefont {{Luk}}}, \bibinfo {author}
  {\bibfnamefont {H.}~\bibnamefont {{Murayama}}}, \bibinfo {author}
  {\bibfnamefont {A.~W.}\ \bibnamefont {{Poon}}}, \bibinfo {author}
  {\bibfnamefont {H.~M.}\ \bibnamefont {{Steiner}}}, \bibinfo {author}
  {\bibfnamefont {L.~A.}\ \bibnamefont {{Winslow}}}, \bibinfo {author}
  {\bibfnamefont {G.~A.}\ \bibnamefont {{Horton-Smith}}}, \bibinfo {author}
  {\bibfnamefont {C.}~\bibnamefont {{Mauger}}}, \bibinfo {author}
  {\bibfnamefont {R.~D.}\ \bibnamefont {{McKeown}}}, \bibinfo {author}
  {\bibfnamefont {P.}~\bibnamefont {{Vogel}}}, \bibinfo {author} {\bibfnamefont
  {C.~E.}\ \bibnamefont {{Lane}}}, \bibinfo {author} {\bibfnamefont
  {T.}~\bibnamefont {{Miletic}}}, \bibinfo {author} {\bibfnamefont {P.~W.}\
  \bibnamefont {{Gorham}}}, \bibinfo {author} {\bibfnamefont {G.}~\bibnamefont
  {{Guillian}}}, \bibinfo {author} {\bibfnamefont {J.~G.}\ \bibnamefont
  {{Learned}}}, \bibinfo {author} {\bibfnamefont {J.}~\bibnamefont
  {{Maricic}}}, \bibinfo {author} {\bibfnamefont {S.}~\bibnamefont
  {{Matsuno}}}, \bibinfo {author} {\bibfnamefont {S.}~\bibnamefont
  {{Pakvasa}}}, \bibinfo {author} {\bibfnamefont {S.}~\bibnamefont
  {{Dazeley}}}, \bibinfo {author} {\bibfnamefont {S.}~\bibnamefont
  {{Hatakeyama}}}, \bibinfo {author} {\bibfnamefont {A.}~\bibnamefont
  {{Rojas}}}, \bibinfo {author} {\bibfnamefont {R.}~\bibnamefont {{Svoboda}}},
  \bibinfo {author} {\bibfnamefont {B.~D.}\ \bibnamefont {{Dieterle}}},
  \bibinfo {author} {\bibfnamefont {J.}~\bibnamefont {{Detwiler}}}, \bibinfo
  {author} {\bibfnamefont {G.}~\bibnamefont {{Gratta}}}, \bibinfo {author}
  {\bibfnamefont {K.}~\bibnamefont {{Ishii}}}, \bibinfo {author} {\bibfnamefont
  {N.}~\bibnamefont {{Tolich}}}, \bibinfo {author} {\bibfnamefont
  {Y.}~\bibnamefont {{Uchida}}}, \bibinfo {author} {\bibfnamefont
  {M.}~\bibnamefont {{Batygov}}}, \bibinfo {author} {\bibfnamefont
  {W.}~\bibnamefont {{Bugg}}}, \bibinfo {author} {\bibfnamefont
  {Y.}~\bibnamefont {{Efremenko}}}, \bibinfo {author} {\bibfnamefont
  {Y.}~\bibnamefont {{Kamyshkov}}}, \bibinfo {author} {\bibfnamefont
  {A.}~\bibnamefont {{Kozlov}}}, \bibinfo {author} {\bibfnamefont
  {Y.}~\bibnamefont {{Nakamura}}}, \bibinfo {author} {\bibfnamefont {C.~R.}\
  \bibnamefont {{Gould}}}, \bibinfo {author} {\bibfnamefont {H.~J.}\
  \bibnamefont {{Karwowski}}}, \bibinfo {author} {\bibfnamefont {D.~M.}\
  \bibnamefont {{Markoff}}}, \bibinfo {author} {\bibfnamefont {J.~A.}\
  \bibnamefont {{Messimore}}}, \bibinfo {author} {\bibfnamefont
  {K.}~\bibnamefont {{Nakamura}}}, \bibinfo {author} {\bibfnamefont {R.~M.}\
  \bibnamefont {{Rohm}}}, \bibinfo {author} {\bibfnamefont {W.}~\bibnamefont
  {{Tornow}}}, \bibinfo {author} {\bibfnamefont {R.}~\bibnamefont {{Wendell}}},
  \bibinfo {author} {\bibfnamefont {A.~R.}\ \bibnamefont {{Young}}}, \bibinfo
  {author} {\bibfnamefont {M.~J.}\ \bibnamefont {{Chen}}}, \bibinfo {author}
  {\bibfnamefont {Y.~F.}\ \bibnamefont {{Wang}}}, \ and\ \bibinfo {author}
  {\bibfnamefont {F.}~\bibnamefont {{Piquemal}}},\ }\href {\doibase
  10.1103/PhysRevLett.94.081801} {\bibfield  {journal} {\bibinfo  {journal}
  {\prl}\ }\textbf {\bibinfo {volume} {94}},\ \bibinfo {eid} {081801} (\bibinfo
  {year} {2005})},\ \Eprint {http://arxiv.org/abs/hep-ex/0406035}
  {arXiv:hep-ex/0406035 [hep-ex]} \BibitemShut {NoStop}%
\bibitem [{\citenamefont {{Ahmad}}\ \emph {et~al.}(2002)\citenamefont
  {{Ahmad}}, \citenamefont {{Allen}}, \citenamefont {{Andersen}}, \citenamefont
  {{Anglin}}, \citenamefont {{Barton}}, \citenamefont {{Beier}}, \citenamefont
  {{Bercovitch}}, \citenamefont {{Bigu}}, \citenamefont {{Biller}},
  \citenamefont {{Black}}, \citenamefont {{Blevis}}, \citenamefont
  {{Boardman}}, \citenamefont {{Boger}}, \citenamefont {{Bonvin}},
  \citenamefont {{Boulay}}, \citenamefont {{Bowler}}, \citenamefont {{Bowles}},
  \citenamefont {{Brice}}, \citenamefont {{Browne}}, \citenamefont {{Bullard}},
  \citenamefont {{B{\"u}hler}}, \citenamefont {{Cameron}}, \citenamefont
  {{Chan}}, \citenamefont {{Chen}}, \citenamefont {{Chen}}, \citenamefont
  {{Chen}}, \citenamefont {{Cleveland}}, \citenamefont {{Clifford}},
  \citenamefont {{Cowan}}, \citenamefont {{Cowen}}, \citenamefont {{Cox}},
  \citenamefont {{Dai}}, \citenamefont {{Dalnoki-Veress}}, \citenamefont
  {{Davidson}}, \citenamefont {{Doe}}, \citenamefont {{Doucas}}, \citenamefont
  {{Dragowsky}}, \citenamefont {{Duba}}, \citenamefont {{Duncan}},
  \citenamefont {{Dunford}}, \citenamefont {{Dunmore}}, \citenamefont
  {{Earle}}, \citenamefont {{Elliott}}, \citenamefont {{Evans}}, \citenamefont
  {{Ewan}}, \citenamefont {{Farine}}, \citenamefont {{Fergani}}, \citenamefont
  {{Ferraris}}, \citenamefont {{Ford}}, \citenamefont {{Formaggio}},
  \citenamefont {{Fowler}}, \citenamefont {{Frame}}, \citenamefont {{Frank}},
  \citenamefont {{Frati}}, \citenamefont {{Gagnon}}, \citenamefont {{Germani}},
  \citenamefont {{Gil}}, \citenamefont {{Graham}}, \citenamefont {{Grant}},
  \citenamefont {{Hahn}}, \citenamefont {{Hallin}}, \citenamefont {{Hallman}},
  \citenamefont {{Hamer}}, \citenamefont {{Hamian}}, \citenamefont {{Handler}},
  \citenamefont {{Haq}}, \citenamefont {{Hargrove}}, \citenamefont {{Harvey}},
  \citenamefont {{Hazama}}, \citenamefont {{Heeger}}, \citenamefont
  {{Heintzelman}}, \citenamefont {{Heise}}, \citenamefont {{Helmer}},
  \citenamefont {{Hepburn}}, \citenamefont {{Heron}}, \citenamefont {{Hewett}},
  \citenamefont {{Hime}}, \citenamefont {{Howe}}, \citenamefont {{Hykawy}},
  \citenamefont {{Isaac}}, \citenamefont {{Jagam}}, \citenamefont {{Jelley}},
  \citenamefont {{Jillings}}, \citenamefont {{Jonkmans}}, \citenamefont
  {{Kazkaz}}, \citenamefont {{Keener}}, \citenamefont {{Klein}}, \citenamefont
  {{Knox}}, \citenamefont {{Komar}}, \citenamefont {{Kouzes}}, \citenamefont
  {{Kutter}}, \citenamefont {{Kyba}}, \citenamefont {{Law}}, \citenamefont
  {{Lawson}}, \citenamefont {{Lay}}, \citenamefont {{Lee}}, \citenamefont
  {{Lesko}}, \citenamefont {{Leslie}}, \citenamefont {{Levine}}, \citenamefont
  {{Locke}}, \citenamefont {{Luoma}}, \citenamefont {{Lyon}}, \citenamefont
  {{Majerus}}, \citenamefont {{Mak}}, \citenamefont {{Maneira}}, \citenamefont
  {{Manor}}, \citenamefont {{Marino}}, \citenamefont {{McCauley}},
  \citenamefont {{McDonald}}, \citenamefont {{McDonald}}, \citenamefont
  {{McFarlane}}, \citenamefont {{McGregor}}, \citenamefont {{Meijer Drees}},
  \citenamefont {{Mifflin}}, \citenamefont {{Miller}}, \citenamefont
  {{Milton}}, \citenamefont {{Moffat}}, \citenamefont {{Moorhead}},
  \citenamefont {{Nally}}, \citenamefont {{Neubauer}}, \citenamefont
  {{Newcomer}}, \citenamefont {{Ng}}, \citenamefont {{Noble}}, \citenamefont
  {{Norman}}, \citenamefont {{Novikov}}, \citenamefont {{O'Neill}},
  \citenamefont {{Okada}}, \citenamefont {{Ollerhead}}, \citenamefont
  {{Omori}}, \citenamefont {{Orrell}}, \citenamefont {{Oser}}, \citenamefont
  {{Poon}}, \citenamefont {{Radcliffe}}, \citenamefont {{Roberge}},
  \citenamefont {{Robertson}}, \citenamefont {{Robertson}}, \citenamefont
  {{Rosendahl}}, \citenamefont {{Rowley}}, \citenamefont {{Rusu}},
  \citenamefont {{Saettler}}, \citenamefont {{Schaffer}}, \citenamefont
  {{Schwendener}}, \citenamefont {{Sch{\"u}lke}}, \citenamefont {{Seifert}},
  \citenamefont {{Shatkay}}, \citenamefont {{Simpson}}, \citenamefont {{Sims}},
  \citenamefont {{Sinclair}}, \citenamefont {{Skensved}}, \citenamefont
  {{Smith}}, \citenamefont {{Smith}}, \citenamefont {{Spreitzer}},
  \citenamefont {{Starinsky}}, \citenamefont {{Steiger}}, \citenamefont
  {{Stokstad}}, \citenamefont {{Stonehill}}, \citenamefont {{Storey}},
  \citenamefont {{Sur}}, \citenamefont {{Tafirout}}, \citenamefont {{Tagg}},
  \citenamefont {{Tanner}}, \citenamefont {{Taplin}}, \citenamefont
  {{Thorman}}, \citenamefont {{Thornewell}}, \citenamefont {{Trent}},
  \citenamefont {{Tserkovnyak}}, \citenamefont {{van Berg}}, \citenamefont
  {{van de Water}}, \citenamefont {{Virtue}}, \citenamefont {{Waltham}},
  \citenamefont {{Wang}}, \citenamefont {{Wark}}, \citenamefont {{West}},
  \citenamefont {{Wilhelmy}}, \citenamefont {{Wilkerson}}, \citenamefont
  {{Wilson}}, \citenamefont {{Wittich}}, \citenamefont {{Wouters}},\ and\
  \citenamefont {{Yeh}}}]{nucl-ex/0204008}%
  \BibitemOpen
  \bibfield  {author} {\bibinfo {author} {\bibfnamefont {Q.~R.}\ \bibnamefont
  {{Ahmad}}}, \bibinfo {author} {\bibfnamefont {R.~C.}\ \bibnamefont
  {{Allen}}}, \bibinfo {author} {\bibfnamefont {T.~C.}\ \bibnamefont
  {{Andersen}}}, \bibinfo {author} {\bibfnamefont {J.~D.}\ \bibnamefont
  {{Anglin}}}, \bibinfo {author} {\bibfnamefont {J.~C.}\ \bibnamefont
  {{Barton}}}, \bibinfo {author} {\bibfnamefont {E.~W.}\ \bibnamefont
  {{Beier}}}, \bibinfo {author} {\bibfnamefont {M.}~\bibnamefont
  {{Bercovitch}}}, \bibinfo {author} {\bibfnamefont {J.}~\bibnamefont
  {{Bigu}}}, \bibinfo {author} {\bibfnamefont {S.~D.}\ \bibnamefont
  {{Biller}}}, \bibinfo {author} {\bibfnamefont {R.~A.}\ \bibnamefont
  {{Black}}}, \bibinfo {author} {\bibfnamefont {I.}~\bibnamefont {{Blevis}}},
  \bibinfo {author} {\bibfnamefont {R.~J.}\ \bibnamefont {{Boardman}}},
  \bibinfo {author} {\bibfnamefont {J.}~\bibnamefont {{Boger}}}, \bibinfo
  {author} {\bibfnamefont {E.}~\bibnamefont {{Bonvin}}}, \bibinfo {author}
  {\bibfnamefont {M.~G.}\ \bibnamefont {{Boulay}}}, \bibinfo {author}
  {\bibfnamefont {M.~G.}\ \bibnamefont {{Bowler}}}, \bibinfo {author}
  {\bibfnamefont {T.~J.}\ \bibnamefont {{Bowles}}}, \bibinfo {author}
  {\bibfnamefont {S.~J.}\ \bibnamefont {{Brice}}}, \bibinfo {author}
  {\bibfnamefont {M.~C.}\ \bibnamefont {{Browne}}}, \bibinfo {author}
  {\bibfnamefont {T.~V.}\ \bibnamefont {{Bullard}}}, \bibinfo {author}
  {\bibfnamefont {G.}~\bibnamefont {{B{\"u}hler}}}, \bibinfo {author}
  {\bibfnamefont {J.}~\bibnamefont {{Cameron}}}, \bibinfo {author}
  {\bibfnamefont {Y.~D.}\ \bibnamefont {{Chan}}}, \bibinfo {author}
  {\bibfnamefont {H.~H.}\ \bibnamefont {{Chen}}}, \bibinfo {author}
  {\bibfnamefont {M.}~\bibnamefont {{Chen}}}, \bibinfo {author} {\bibfnamefont
  {X.}~\bibnamefont {{Chen}}}, \bibinfo {author} {\bibfnamefont {B.~T.}\
  \bibnamefont {{Cleveland}}}, \bibinfo {author} {\bibfnamefont {E.~T.}\
  \bibnamefont {{Clifford}}}, \bibinfo {author} {\bibfnamefont {J.~H.}\
  \bibnamefont {{Cowan}}}, \bibinfo {author} {\bibfnamefont {D.~F.}\
  \bibnamefont {{Cowen}}}, \bibinfo {author} {\bibfnamefont {G.~A.}\
  \bibnamefont {{Cox}}}, \bibinfo {author} {\bibfnamefont {X.}~\bibnamefont
  {{Dai}}}, \bibinfo {author} {\bibfnamefont {F.}~\bibnamefont
  {{Dalnoki-Veress}}}, \bibinfo {author} {\bibfnamefont {W.~F.}\ \bibnamefont
  {{Davidson}}}, \bibinfo {author} {\bibfnamefont {P.~J.}\ \bibnamefont
  {{Doe}}}, \bibinfo {author} {\bibfnamefont {G.}~\bibnamefont {{Doucas}}},
  \bibinfo {author} {\bibfnamefont {M.~R.}\ \bibnamefont {{Dragowsky}}},
  \bibinfo {author} {\bibfnamefont {C.~A.}\ \bibnamefont {{Duba}}}, \bibinfo
  {author} {\bibfnamefont {F.~A.}\ \bibnamefont {{Duncan}}}, \bibinfo {author}
  {\bibfnamefont {M.}~\bibnamefont {{Dunford}}}, \bibinfo {author}
  {\bibfnamefont {J.~A.}\ \bibnamefont {{Dunmore}}}, \bibinfo {author}
  {\bibfnamefont {E.~D.}\ \bibnamefont {{Earle}}}, \bibinfo {author}
  {\bibfnamefont {S.~R.}\ \bibnamefont {{Elliott}}}, \bibinfo {author}
  {\bibfnamefont {H.~C.}\ \bibnamefont {{Evans}}}, \bibinfo {author}
  {\bibfnamefont {G.~T.}\ \bibnamefont {{Ewan}}}, \bibinfo {author}
  {\bibfnamefont {J.}~\bibnamefont {{Farine}}}, \bibinfo {author}
  {\bibfnamefont {H.}~\bibnamefont {{Fergani}}}, \bibinfo {author}
  {\bibfnamefont {A.~P.}\ \bibnamefont {{Ferraris}}}, \bibinfo {author}
  {\bibfnamefont {R.~J.}\ \bibnamefont {{Ford}}}, \bibinfo {author}
  {\bibfnamefont {J.~A.}\ \bibnamefont {{Formaggio}}}, \bibinfo {author}
  {\bibfnamefont {M.~M.}\ \bibnamefont {{Fowler}}}, \bibinfo {author}
  {\bibfnamefont {K.}~\bibnamefont {{Frame}}}, \bibinfo {author} {\bibfnamefont
  {E.~D.}\ \bibnamefont {{Frank}}}, \bibinfo {author} {\bibfnamefont
  {W.}~\bibnamefont {{Frati}}}, \bibinfo {author} {\bibfnamefont
  {N.}~\bibnamefont {{Gagnon}}}, \bibinfo {author} {\bibfnamefont {J.~V.}\
  \bibnamefont {{Germani}}}, \bibinfo {author} {\bibfnamefont {S.}~\bibnamefont
  {{Gil}}}, \bibinfo {author} {\bibfnamefont {K.}~\bibnamefont {{Graham}}},
  \bibinfo {author} {\bibfnamefont {D.~R.}\ \bibnamefont {{Grant}}}, \bibinfo
  {author} {\bibfnamefont {R.~L.}\ \bibnamefont {{Hahn}}}, \bibinfo {author}
  {\bibfnamefont {A.~L.}\ \bibnamefont {{Hallin}}}, \bibinfo {author}
  {\bibfnamefont {E.~D.}\ \bibnamefont {{Hallman}}}, \bibinfo {author}
  {\bibfnamefont {A.~S.}\ \bibnamefont {{Hamer}}}, \bibinfo {author}
  {\bibfnamefont {A.~A.}\ \bibnamefont {{Hamian}}}, \bibinfo {author}
  {\bibfnamefont {W.~B.}\ \bibnamefont {{Handler}}}, \bibinfo {author}
  {\bibfnamefont {R.~U.}\ \bibnamefont {{Haq}}}, \bibinfo {author}
  {\bibfnamefont {C.~K.}\ \bibnamefont {{Hargrove}}}, \bibinfo {author}
  {\bibfnamefont {P.~J.}\ \bibnamefont {{Harvey}}}, \bibinfo {author}
  {\bibfnamefont {R.}~\bibnamefont {{Hazama}}}, \bibinfo {author}
  {\bibfnamefont {K.~M.}\ \bibnamefont {{Heeger}}}, \bibinfo {author}
  {\bibfnamefont {W.~J.}\ \bibnamefont {{Heintzelman}}}, \bibinfo {author}
  {\bibfnamefont {J.}~\bibnamefont {{Heise}}}, \bibinfo {author} {\bibfnamefont
  {R.~L.}\ \bibnamefont {{Helmer}}}, \bibinfo {author} {\bibfnamefont {J.~D.}\
  \bibnamefont {{Hepburn}}}, \bibinfo {author} {\bibfnamefont {H.}~\bibnamefont
  {{Heron}}}, \bibinfo {author} {\bibfnamefont {J.}~\bibnamefont {{Hewett}}},
  \bibinfo {author} {\bibfnamefont {A.}~\bibnamefont {{Hime}}}, \bibinfo
  {author} {\bibfnamefont {M.}~\bibnamefont {{Howe}}}, \bibinfo {author}
  {\bibfnamefont {J.~G.}\ \bibnamefont {{Hykawy}}}, \bibinfo {author}
  {\bibfnamefont {M.~C.}\ \bibnamefont {{Isaac}}}, \bibinfo {author}
  {\bibfnamefont {P.}~\bibnamefont {{Jagam}}}, \bibinfo {author} {\bibfnamefont
  {N.~A.}\ \bibnamefont {{Jelley}}}, \bibinfo {author} {\bibfnamefont
  {C.}~\bibnamefont {{Jillings}}}, \bibinfo {author} {\bibfnamefont
  {G.}~\bibnamefont {{Jonkmans}}}, \bibinfo {author} {\bibfnamefont
  {K.}~\bibnamefont {{Kazkaz}}}, \bibinfo {author} {\bibfnamefont {P.~T.}\
  \bibnamefont {{Keener}}}, \bibinfo {author} {\bibfnamefont {J.~R.}\
  \bibnamefont {{Klein}}}, \bibinfo {author} {\bibfnamefont {A.~B.}\
  \bibnamefont {{Knox}}}, \bibinfo {author} {\bibfnamefont {R.~J.}\
  \bibnamefont {{Komar}}}, \bibinfo {author} {\bibfnamefont {R.}~\bibnamefont
  {{Kouzes}}}, \bibinfo {author} {\bibfnamefont {T.}~\bibnamefont {{Kutter}}},
  \bibinfo {author} {\bibfnamefont {C.~C.}\ \bibnamefont {{Kyba}}}, \bibinfo
  {author} {\bibfnamefont {J.}~\bibnamefont {{Law}}}, \bibinfo {author}
  {\bibfnamefont {I.~T.}\ \bibnamefont {{Lawson}}}, \bibinfo {author}
  {\bibfnamefont {M.}~\bibnamefont {{Lay}}}, \bibinfo {author} {\bibfnamefont
  {H.~W.}\ \bibnamefont {{Lee}}}, \bibinfo {author} {\bibfnamefont {K.~T.}\
  \bibnamefont {{Lesko}}}, \bibinfo {author} {\bibfnamefont {J.~R.}\
  \bibnamefont {{Leslie}}}, \bibinfo {author} {\bibfnamefont {I.}~\bibnamefont
  {{Levine}}}, \bibinfo {author} {\bibfnamefont {W.}~\bibnamefont {{Locke}}},
  \bibinfo {author} {\bibfnamefont {S.}~\bibnamefont {{Luoma}}}, \bibinfo
  {author} {\bibfnamefont {J.}~\bibnamefont {{Lyon}}}, \bibinfo {author}
  {\bibfnamefont {S.}~\bibnamefont {{Majerus}}}, \bibinfo {author}
  {\bibfnamefont {H.~B.}\ \bibnamefont {{Mak}}}, \bibinfo {author}
  {\bibfnamefont {J.}~\bibnamefont {{Maneira}}}, \bibinfo {author}
  {\bibfnamefont {J.}~\bibnamefont {{Manor}}}, \bibinfo {author} {\bibfnamefont
  {A.~D.}\ \bibnamefont {{Marino}}}, \bibinfo {author} {\bibfnamefont
  {N.}~\bibnamefont {{McCauley}}}, \bibinfo {author} {\bibfnamefont {A.~B.}\
  \bibnamefont {{McDonald}}}, \bibinfo {author} {\bibfnamefont {D.~S.}\
  \bibnamefont {{McDonald}}}, \bibinfo {author} {\bibfnamefont
  {K.}~\bibnamefont {{McFarlane}}}, \bibinfo {author} {\bibfnamefont
  {G.}~\bibnamefont {{McGregor}}}, \bibinfo {author} {\bibfnamefont
  {R.}~\bibnamefont {{Meijer Drees}}}, \bibinfo {author} {\bibfnamefont
  {C.}~\bibnamefont {{Mifflin}}}, \bibinfo {author} {\bibfnamefont {G.~G.}\
  \bibnamefont {{Miller}}}, \bibinfo {author} {\bibfnamefont {G.}~\bibnamefont
  {{Milton}}}, \bibinfo {author} {\bibfnamefont {B.~A.}\ \bibnamefont
  {{Moffat}}}, \bibinfo {author} {\bibfnamefont {M.}~\bibnamefont
  {{Moorhead}}}, \bibinfo {author} {\bibfnamefont {C.~W.}\ \bibnamefont
  {{Nally}}}, \bibinfo {author} {\bibfnamefont {M.~S.}\ \bibnamefont
  {{Neubauer}}}, \bibinfo {author} {\bibfnamefont {F.~M.}\ \bibnamefont
  {{Newcomer}}}, \bibinfo {author} {\bibfnamefont {H.~S.}\ \bibnamefont
  {{Ng}}}, \bibinfo {author} {\bibfnamefont {A.~J.}\ \bibnamefont {{Noble}}},
  \bibinfo {author} {\bibfnamefont {E.~B.}\ \bibnamefont {{Norman}}}, \bibinfo
  {author} {\bibfnamefont {V.~M.}\ \bibnamefont {{Novikov}}}, \bibinfo {author}
  {\bibfnamefont {M.}~\bibnamefont {{O'Neill}}}, \bibinfo {author}
  {\bibfnamefont {C.~E.}\ \bibnamefont {{Okada}}}, \bibinfo {author}
  {\bibfnamefont {R.~W.}\ \bibnamefont {{Ollerhead}}}, \bibinfo {author}
  {\bibfnamefont {M.}~\bibnamefont {{Omori}}}, \bibinfo {author} {\bibfnamefont
  {J.~L.}\ \bibnamefont {{Orrell}}}, \bibinfo {author} {\bibfnamefont {S.~M.}\
  \bibnamefont {{Oser}}}, \bibinfo {author} {\bibfnamefont {A.~W.}\
  \bibnamefont {{Poon}}}, \bibinfo {author} {\bibfnamefont {T.~J.}\
  \bibnamefont {{Radcliffe}}}, \bibinfo {author} {\bibfnamefont
  {A.}~\bibnamefont {{Roberge}}}, \bibinfo {author} {\bibfnamefont {B.~C.}\
  \bibnamefont {{Robertson}}}, \bibinfo {author} {\bibfnamefont {R.~G.}\
  \bibnamefont {{Robertson}}}, \bibinfo {author} {\bibfnamefont {S.~S.}\
  \bibnamefont {{Rosendahl}}}, \bibinfo {author} {\bibfnamefont {J.~K.}\
  \bibnamefont {{Rowley}}}, \bibinfo {author} {\bibfnamefont {V.~L.}\
  \bibnamefont {{Rusu}}}, \bibinfo {author} {\bibfnamefont {E.}~\bibnamefont
  {{Saettler}}}, \bibinfo {author} {\bibfnamefont {K.~K.}\ \bibnamefont
  {{Schaffer}}}, \bibinfo {author} {\bibfnamefont {M.~H.}\ \bibnamefont
  {{Schwendener}}}, \bibinfo {author} {\bibfnamefont {A.}~\bibnamefont
  {{Sch{\"u}lke}}}, \bibinfo {author} {\bibfnamefont {H.}~\bibnamefont
  {{Seifert}}}, \bibinfo {author} {\bibfnamefont {M.}~\bibnamefont
  {{Shatkay}}}, \bibinfo {author} {\bibfnamefont {J.~J.}\ \bibnamefont
  {{Simpson}}}, \bibinfo {author} {\bibfnamefont {C.~J.}\ \bibnamefont
  {{Sims}}}, \bibinfo {author} {\bibfnamefont {D.}~\bibnamefont {{Sinclair}}},
  \bibinfo {author} {\bibfnamefont {P.}~\bibnamefont {{Skensved}}}, \bibinfo
  {author} {\bibfnamefont {A.~R.}\ \bibnamefont {{Smith}}}, \bibinfo {author}
  {\bibfnamefont {M.~W.}\ \bibnamefont {{Smith}}}, \bibinfo {author}
  {\bibfnamefont {T.}~\bibnamefont {{Spreitzer}}}, \bibinfo {author}
  {\bibfnamefont {N.}~\bibnamefont {{Starinsky}}}, \bibinfo {author}
  {\bibfnamefont {T.~D.}\ \bibnamefont {{Steiger}}}, \bibinfo {author}
  {\bibfnamefont {R.~G.}\ \bibnamefont {{Stokstad}}}, \bibinfo {author}
  {\bibfnamefont {L.~C.}\ \bibnamefont {{Stonehill}}}, \bibinfo {author}
  {\bibfnamefont {R.~S.}\ \bibnamefont {{Storey}}}, \bibinfo {author}
  {\bibfnamefont {B.}~\bibnamefont {{Sur}}}, \bibinfo {author} {\bibfnamefont
  {R.}~\bibnamefont {{Tafirout}}}, \bibinfo {author} {\bibfnamefont
  {N.}~\bibnamefont {{Tagg}}}, \bibinfo {author} {\bibfnamefont {N.~W.}\
  \bibnamefont {{Tanner}}}, \bibinfo {author} {\bibfnamefont {R.~K.}\
  \bibnamefont {{Taplin}}}, \bibinfo {author} {\bibfnamefont {M.}~\bibnamefont
  {{Thorman}}}, \bibinfo {author} {\bibfnamefont {P.~M.}\ \bibnamefont
  {{Thornewell}}}, \bibinfo {author} {\bibfnamefont {P.~T.}\ \bibnamefont
  {{Trent}}}, \bibinfo {author} {\bibfnamefont {Y.~I.}\ \bibnamefont
  {{Tserkovnyak}}}, \bibinfo {author} {\bibfnamefont {R.}~\bibnamefont {{van
  Berg}}}, \bibinfo {author} {\bibfnamefont {R.~G.}\ \bibnamefont {{van de
  Water}}}, \bibinfo {author} {\bibfnamefont {C.~J.}\ \bibnamefont {{Virtue}}},
  \bibinfo {author} {\bibfnamefont {C.~E.}\ \bibnamefont {{Waltham}}}, \bibinfo
  {author} {\bibfnamefont {J.~X.}\ \bibnamefont {{Wang}}}, \bibinfo {author}
  {\bibfnamefont {D.~L.}\ \bibnamefont {{Wark}}}, \bibinfo {author}
  {\bibfnamefont {N.}~\bibnamefont {{West}}}, \bibinfo {author} {\bibfnamefont
  {J.~B.}\ \bibnamefont {{Wilhelmy}}}, \bibinfo {author} {\bibfnamefont
  {J.~F.}\ \bibnamefont {{Wilkerson}}}, \bibinfo {author} {\bibfnamefont
  {J.~R.}\ \bibnamefont {{Wilson}}}, \bibinfo {author} {\bibfnamefont
  {P.}~\bibnamefont {{Wittich}}}, \bibinfo {author} {\bibfnamefont {J.~M.}\
  \bibnamefont {{Wouters}}}, \ and\ \bibinfo {author} {\bibfnamefont
  {M.}~\bibnamefont {{Yeh}}},\ }\href {\doibase 10.1103/PhysRevLett.89.011301}
  {\bibfield  {journal} {\bibinfo  {journal} {\prl}\ }\textbf {\bibinfo
  {volume} {89}},\ \bibinfo {eid} {011301} (\bibinfo {year} {2002})},\ \Eprint
  {http://arxiv.org/abs/nucl-ex/0204008} {arXiv:nucl-ex/0204008 [nucl-ex]}
  \BibitemShut {NoStop}%
\bibitem [{\citenamefont {{Fukuda}}\ \emph {et~al.}(1998)\citenamefont
  {{Fukuda}}, \citenamefont {{Hayakawa}}, \citenamefont {{Ichihara}},
  \citenamefont {{Inoue}}, \citenamefont {{Ishihara}}, \citenamefont
  {{Ishino}}, \citenamefont {{Itow}}, \citenamefont {{Kajita}}, \citenamefont
  {{Kameda}}, \citenamefont {{Kasuga}}, \citenamefont {{Kobayashi}},
  \citenamefont {{Kobayashi}}, \citenamefont {{Koshio}}, \citenamefont
  {{Miura}}, \citenamefont {{Nakahata}}, \citenamefont {{Nakayama}},
  \citenamefont {{Okada}}, \citenamefont {{Okumura}}, \citenamefont
  {{Sakurai}}, \citenamefont {{Shiozawa}}, \citenamefont {{Suzuki}},
  \citenamefont {{Takeuchi}}, \citenamefont {{Totsuka}}, \citenamefont
  {{Yamada}}, \citenamefont {{Earl}}, \citenamefont {{Habig}}, \citenamefont
  {{Kearns}}, \citenamefont {{Messier}}, \citenamefont {{Scholberg}},
  \citenamefont {{Stone}}, \citenamefont {{Sulak}}, \citenamefont {{Walter}},
  \citenamefont {{Goldhaber}}, \citenamefont {{Barszczxak}}, \citenamefont
  {{Casper}}, \citenamefont {{Gajewski}}, \citenamefont {{Halverson}},
  \citenamefont {{Hsu}}, \citenamefont {{Kropp}}, \citenamefont {{Price}},
  \citenamefont {{Reines}}, \citenamefont {{Smy}}, \citenamefont {{Sobel}},
  \citenamefont {{Vagins}}, \citenamefont {{Ganezer}}, \citenamefont {{Keig}},
  \citenamefont {{Ellsworth}}, \citenamefont {{Tasaka}}, \citenamefont
  {{Flanagan}}, \citenamefont {{Kibayashi}}, \citenamefont {{Learned}},
  \citenamefont {{Matsuno}}, \citenamefont {{Stenger}}, \citenamefont
  {{Takemori}}, \citenamefont {{Ishii}}, \citenamefont {{Kanzaki}},
  \citenamefont {{Kobayashi}}, \citenamefont {{Mine}}, \citenamefont
  {{Nakamura}}, \citenamefont {{Nishikawa}}, \citenamefont {{Oyama}},
  \citenamefont {{Sakai}}, \citenamefont {{Sakuda}}, \citenamefont {{Sasaki}},
  \citenamefont {{Echigo}}, \citenamefont {{Kohama}}, \citenamefont {{Suzuki}},
  \citenamefont {{Haines}}, \citenamefont {{Blaufuss}}, \citenamefont {{Kim}},
  \citenamefont {{Sanford}}, \citenamefont {{Svoboda}}, \citenamefont {{Chen}},
  \citenamefont {{Conner}}, \citenamefont {{Goodman}}, \citenamefont
  {{Sullivan}}, \citenamefont {{Hill}}, \citenamefont {{Jung}}, \citenamefont
  {{Martens}}, \citenamefont {{Mauger}}, \citenamefont {{McGrew}},
  \citenamefont {{Sharkey}}, \citenamefont {{Viren}}, \citenamefont
  {{Yanagisawa}}, \citenamefont {{Doki}}, \citenamefont {{Miyano}},
  \citenamefont {{Okazawa}}, \citenamefont {{Saji}}, \citenamefont
  {{Takahata}}, \citenamefont {{Nagashima}}, \citenamefont {{Takita}},
  \citenamefont {{Yamaguchi}}, \citenamefont {{Yoshida}}, \citenamefont
  {{Kim}}, \citenamefont {{Etoh}}, \citenamefont {{Fujita}}, \citenamefont
  {{Hasegawa}}, \citenamefont {{Hasegawa}}, \citenamefont {{Hatakeyama}},
  \citenamefont {{Iwamoto}}, \citenamefont {{Koga}}, \citenamefont
  {{Maruyama}}, \citenamefont {{Ogawa}}, \citenamefont {{Shirai}},
  \citenamefont {{Suzuki}}, \citenamefont {{Tsushima}}, \citenamefont
  {{Koshiba}}, \citenamefont {{Nemoto}}, \citenamefont {{Nishijima}},
  \citenamefont {{Futagami}}, \citenamefont {{Hayato}}, \citenamefont
  {{Kanaya}}, \citenamefont {{Kaneyuki}}, \citenamefont {{Watanabe}},
  \citenamefont {{Kielczewska}}, \citenamefont {{Doyle}}, \citenamefont
  {{George}}, \citenamefont {{Stachyra}}, \citenamefont {{Wai}}, \citenamefont
  {{Wilkes}},\ and\ \citenamefont {{Young}}}]{hep-ex/9807003}%
  \BibitemOpen
  \bibfield  {author} {\bibinfo {author} {\bibfnamefont {Y.}~\bibnamefont
  {{Fukuda}}}, \bibinfo {author} {\bibfnamefont {T.}~\bibnamefont
  {{Hayakawa}}}, \bibinfo {author} {\bibfnamefont {E.}~\bibnamefont
  {{Ichihara}}}, \bibinfo {author} {\bibfnamefont {K.}~\bibnamefont {{Inoue}}},
  \bibinfo {author} {\bibfnamefont {K.}~\bibnamefont {{Ishihara}}}, \bibinfo
  {author} {\bibfnamefont {H.}~\bibnamefont {{Ishino}}}, \bibinfo {author}
  {\bibfnamefont {Y.}~\bibnamefont {{Itow}}}, \bibinfo {author} {\bibfnamefont
  {T.}~\bibnamefont {{Kajita}}}, \bibinfo {author} {\bibfnamefont
  {J.}~\bibnamefont {{Kameda}}}, \bibinfo {author} {\bibfnamefont
  {S.}~\bibnamefont {{Kasuga}}}, \bibinfo {author} {\bibfnamefont
  {K.}~\bibnamefont {{Kobayashi}}}, \bibinfo {author} {\bibfnamefont
  {Y.}~\bibnamefont {{Kobayashi}}}, \bibinfo {author} {\bibfnamefont
  {Y.}~\bibnamefont {{Koshio}}}, \bibinfo {author} {\bibfnamefont
  {M.}~\bibnamefont {{Miura}}}, \bibinfo {author} {\bibfnamefont
  {M.}~\bibnamefont {{Nakahata}}}, \bibinfo {author} {\bibfnamefont
  {S.}~\bibnamefont {{Nakayama}}}, \bibinfo {author} {\bibfnamefont
  {A.}~\bibnamefont {{Okada}}}, \bibinfo {author} {\bibfnamefont
  {K.}~\bibnamefont {{Okumura}}}, \bibinfo {author} {\bibfnamefont
  {N.}~\bibnamefont {{Sakurai}}}, \bibinfo {author} {\bibfnamefont
  {M.}~\bibnamefont {{Shiozawa}}}, \bibinfo {author} {\bibfnamefont
  {Y.}~\bibnamefont {{Suzuki}}}, \bibinfo {author} {\bibfnamefont
  {Y.}~\bibnamefont {{Takeuchi}}}, \bibinfo {author} {\bibfnamefont
  {Y.}~\bibnamefont {{Totsuka}}}, \bibinfo {author} {\bibfnamefont
  {S.}~\bibnamefont {{Yamada}}}, \bibinfo {author} {\bibfnamefont
  {M.}~\bibnamefont {{Earl}}}, \bibinfo {author} {\bibfnamefont
  {A.}~\bibnamefont {{Habig}}}, \bibinfo {author} {\bibfnamefont
  {E.}~\bibnamefont {{Kearns}}}, \bibinfo {author} {\bibfnamefont {M.~D.}\
  \bibnamefont {{Messier}}}, \bibinfo {author} {\bibfnamefont {K.}~\bibnamefont
  {{Scholberg}}}, \bibinfo {author} {\bibfnamefont {J.~L.}\ \bibnamefont
  {{Stone}}}, \bibinfo {author} {\bibfnamefont {L.~R.}\ \bibnamefont
  {{Sulak}}}, \bibinfo {author} {\bibfnamefont {C.~W.}\ \bibnamefont
  {{Walter}}}, \bibinfo {author} {\bibfnamefont {M.}~\bibnamefont
  {{Goldhaber}}}, \bibinfo {author} {\bibfnamefont {T.}~\bibnamefont
  {{Barszczxak}}}, \bibinfo {author} {\bibfnamefont {D.}~\bibnamefont
  {{Casper}}}, \bibinfo {author} {\bibfnamefont {W.}~\bibnamefont
  {{Gajewski}}}, \bibinfo {author} {\bibfnamefont {P.~G.}\ \bibnamefont
  {{Halverson}}}, \bibinfo {author} {\bibfnamefont {J.}~\bibnamefont {{Hsu}}},
  \bibinfo {author} {\bibfnamefont {W.~R.}\ \bibnamefont {{Kropp}}}, \bibinfo
  {author} {\bibfnamefont {L.~R.}\ \bibnamefont {{Price}}}, \bibinfo {author}
  {\bibfnamefont {F.}~\bibnamefont {{Reines}}}, \bibinfo {author}
  {\bibfnamefont {M.}~\bibnamefont {{Smy}}}, \bibinfo {author} {\bibfnamefont
  {H.~W.}\ \bibnamefont {{Sobel}}}, \bibinfo {author} {\bibfnamefont {M.~R.}\
  \bibnamefont {{Vagins}}}, \bibinfo {author} {\bibfnamefont {K.~S.}\
  \bibnamefont {{Ganezer}}}, \bibinfo {author} {\bibfnamefont {W.~E.}\
  \bibnamefont {{Keig}}}, \bibinfo {author} {\bibfnamefont {R.~W.}\
  \bibnamefont {{Ellsworth}}}, \bibinfo {author} {\bibfnamefont
  {S.}~\bibnamefont {{Tasaka}}}, \bibinfo {author} {\bibfnamefont {J.~W.}\
  \bibnamefont {{Flanagan}}}, \bibinfo {author} {\bibfnamefont
  {A.}~\bibnamefont {{Kibayashi}}}, \bibinfo {author} {\bibfnamefont {J.~G.}\
  \bibnamefont {{Learned}}}, \bibinfo {author} {\bibfnamefont {S.}~\bibnamefont
  {{Matsuno}}}, \bibinfo {author} {\bibfnamefont {V.~J.}\ \bibnamefont
  {{Stenger}}}, \bibinfo {author} {\bibfnamefont {D.}~\bibnamefont
  {{Takemori}}}, \bibinfo {author} {\bibfnamefont {T.}~\bibnamefont {{Ishii}}},
  \bibinfo {author} {\bibfnamefont {J.}~\bibnamefont {{Kanzaki}}}, \bibinfo
  {author} {\bibfnamefont {T.}~\bibnamefont {{Kobayashi}}}, \bibinfo {author}
  {\bibfnamefont {S.}~\bibnamefont {{Mine}}}, \bibinfo {author} {\bibfnamefont
  {K.}~\bibnamefont {{Nakamura}}}, \bibinfo {author} {\bibfnamefont
  {K.}~\bibnamefont {{Nishikawa}}}, \bibinfo {author} {\bibfnamefont
  {Y.}~\bibnamefont {{Oyama}}}, \bibinfo {author} {\bibfnamefont
  {A.}~\bibnamefont {{Sakai}}}, \bibinfo {author} {\bibfnamefont
  {M.}~\bibnamefont {{Sakuda}}}, \bibinfo {author} {\bibfnamefont
  {O.}~\bibnamefont {{Sasaki}}}, \bibinfo {author} {\bibfnamefont
  {S.}~\bibnamefont {{Echigo}}}, \bibinfo {author} {\bibfnamefont
  {M.}~\bibnamefont {{Kohama}}}, \bibinfo {author} {\bibfnamefont {A.~T.}\
  \bibnamefont {{Suzuki}}}, \bibinfo {author} {\bibfnamefont {T.~J.}\
  \bibnamefont {{Haines}}}, \bibinfo {author} {\bibfnamefont {E.}~\bibnamefont
  {{Blaufuss}}}, \bibinfo {author} {\bibfnamefont {B.~K.}\ \bibnamefont
  {{Kim}}}, \bibinfo {author} {\bibfnamefont {R.}~\bibnamefont {{Sanford}}},
  \bibinfo {author} {\bibfnamefont {R.}~\bibnamefont {{Svoboda}}}, \bibinfo
  {author} {\bibfnamefont {M.~L.}\ \bibnamefont {{Chen}}}, \bibinfo {author}
  {\bibfnamefont {Z.}~\bibnamefont {{Conner}}}, \bibinfo {author}
  {\bibfnamefont {J.~A.}\ \bibnamefont {{Goodman}}}, \bibinfo {author}
  {\bibfnamefont {G.~W.}\ \bibnamefont {{Sullivan}}}, \bibinfo {author}
  {\bibfnamefont {J.}~\bibnamefont {{Hill}}}, \bibinfo {author} {\bibfnamefont
  {C.~K.}\ \bibnamefont {{Jung}}}, \bibinfo {author} {\bibfnamefont
  {K.}~\bibnamefont {{Martens}}}, \bibinfo {author} {\bibfnamefont
  {C.}~\bibnamefont {{Mauger}}}, \bibinfo {author} {\bibfnamefont
  {C.}~\bibnamefont {{McGrew}}}, \bibinfo {author} {\bibfnamefont
  {E.}~\bibnamefont {{Sharkey}}}, \bibinfo {author} {\bibfnamefont
  {B.}~\bibnamefont {{Viren}}}, \bibinfo {author} {\bibfnamefont
  {C.}~\bibnamefont {{Yanagisawa}}}, \bibinfo {author} {\bibfnamefont
  {W.}~\bibnamefont {{Doki}}}, \bibinfo {author} {\bibfnamefont
  {K.}~\bibnamefont {{Miyano}}}, \bibinfo {author} {\bibfnamefont
  {H.}~\bibnamefont {{Okazawa}}}, \bibinfo {author} {\bibfnamefont
  {C.}~\bibnamefont {{Saji}}}, \bibinfo {author} {\bibfnamefont
  {M.}~\bibnamefont {{Takahata}}}, \bibinfo {author} {\bibfnamefont
  {Y.}~\bibnamefont {{Nagashima}}}, \bibinfo {author} {\bibfnamefont
  {M.}~\bibnamefont {{Takita}}}, \bibinfo {author} {\bibfnamefont
  {T.}~\bibnamefont {{Yamaguchi}}}, \bibinfo {author} {\bibfnamefont
  {M.}~\bibnamefont {{Yoshida}}}, \bibinfo {author} {\bibfnamefont {S.~B.}\
  \bibnamefont {{Kim}}}, \bibinfo {author} {\bibfnamefont {M.}~\bibnamefont
  {{Etoh}}}, \bibinfo {author} {\bibfnamefont {K.}~\bibnamefont {{Fujita}}},
  \bibinfo {author} {\bibfnamefont {A.}~\bibnamefont {{Hasegawa}}}, \bibinfo
  {author} {\bibfnamefont {T.}~\bibnamefont {{Hasegawa}}}, \bibinfo {author}
  {\bibfnamefont {S.}~\bibnamefont {{Hatakeyama}}}, \bibinfo {author}
  {\bibfnamefont {T.}~\bibnamefont {{Iwamoto}}}, \bibinfo {author}
  {\bibfnamefont {M.}~\bibnamefont {{Koga}}}, \bibinfo {author} {\bibfnamefont
  {T.}~\bibnamefont {{Maruyama}}}, \bibinfo {author} {\bibfnamefont
  {H.}~\bibnamefont {{Ogawa}}}, \bibinfo {author} {\bibfnamefont
  {J.}~\bibnamefont {{Shirai}}}, \bibinfo {author} {\bibfnamefont
  {A.}~\bibnamefont {{Suzuki}}}, \bibinfo {author} {\bibfnamefont
  {F.}~\bibnamefont {{Tsushima}}}, \bibinfo {author} {\bibfnamefont
  {M.}~\bibnamefont {{Koshiba}}}, \bibinfo {author} {\bibfnamefont
  {M.}~\bibnamefont {{Nemoto}}}, \bibinfo {author} {\bibfnamefont
  {K.}~\bibnamefont {{Nishijima}}}, \bibinfo {author} {\bibfnamefont
  {T.}~\bibnamefont {{Futagami}}}, \bibinfo {author} {\bibfnamefont
  {Y.}~\bibnamefont {{Hayato}}}, \bibinfo {author} {\bibfnamefont
  {Y.}~\bibnamefont {{Kanaya}}}, \bibinfo {author} {\bibfnamefont
  {K.}~\bibnamefont {{Kaneyuki}}}, \bibinfo {author} {\bibfnamefont
  {Y.}~\bibnamefont {{Watanabe}}}, \bibinfo {author} {\bibfnamefont
  {D.}~\bibnamefont {{Kielczewska}}}, \bibinfo {author} {\bibfnamefont {R.~A.}\
  \bibnamefont {{Doyle}}}, \bibinfo {author} {\bibfnamefont {J.~S.}\
  \bibnamefont {{George}}}, \bibinfo {author} {\bibfnamefont {A.~L.}\
  \bibnamefont {{Stachyra}}}, \bibinfo {author} {\bibfnamefont {L.~L.}\
  \bibnamefont {{Wai}}}, \bibinfo {author} {\bibfnamefont {R.~J.}\ \bibnamefont
  {{Wilkes}}}, \ and\ \bibinfo {author} {\bibfnamefont {K.~K.}\ \bibnamefont
  {{Young}}},\ }\href {\doibase 10.1103/PhysRevLett.81.1562} {\bibfield
  {journal} {\bibinfo  {journal} {\prl}\ }\textbf {\bibinfo {volume} {81}},\
  \bibinfo {pages} {1562} (\bibinfo {year} {1998})},\ \Eprint
  {http://arxiv.org/abs/hep-ex/9807003} {arXiv:hep-ex/9807003 [hep-ex]}
  \BibitemShut {NoStop}%
\bibitem [{\citenamefont {{Esteban}}\ \emph {et~al.}(2019)\citenamefont
  {{Esteban}}, \citenamefont {{Gonzalez-Garcia}}, \citenamefont {{Hernand
  ez-Cabezudo}}, \citenamefont {{Maltoni}},\ and\ \citenamefont
  {{Schwetz}}}]{1811.05487}%
  \BibitemOpen
  \bibfield  {author} {\bibinfo {author} {\bibfnamefont {I.}~\bibnamefont
  {{Esteban}}}, \bibinfo {author} {\bibfnamefont {M.~C.}\ \bibnamefont
  {{Gonzalez-Garcia}}}, \bibinfo {author} {\bibfnamefont {A.}~\bibnamefont
  {{Hernand ez-Cabezudo}}}, \bibinfo {author} {\bibfnamefont {M.}~\bibnamefont
  {{Maltoni}}}, \ and\ \bibinfo {author} {\bibfnamefont {T.}~\bibnamefont
  {{Schwetz}}},\ }\href {\doibase 10.1007/JHEP01(2019)106} {\bibfield
  {journal} {\bibinfo  {journal} {Journal of High Energy Physics}\ }\textbf
  {\bibinfo {volume} {2019}},\ \bibinfo {eid} {106} (\bibinfo {year} {2019})},\
  \Eprint {http://arxiv.org/abs/1811.05487} {arXiv:1811.05487 [hep-ph]}
  \BibitemShut {NoStop}%
\bibitem [{\citenamefont {{Choudhury}}\ and\ \citenamefont
  {{Hannestad}}(2020)}]{1907.12598}%
  \BibitemOpen
  \bibfield  {author} {\bibinfo {author} {\bibfnamefont {S.~R.}\ \bibnamefont
  {{Choudhury}}}\ and\ \bibinfo {author} {\bibfnamefont {S.}~\bibnamefont
  {{Hannestad}}},\ }\href {\doibase 10.1088/1475-7516/2020/07/037} {\bibfield
  {journal} {\bibinfo  {journal} {\jcap}\ }\textbf {\bibinfo {volume} {2020}},\
  \bibinfo {eid} {037} (\bibinfo {year} {2020})},\ \Eprint
  {http://arxiv.org/abs/1907.12598} {arXiv:1907.12598 [astro-ph.CO]}
  \BibitemShut {NoStop}%
\bibitem [{\citenamefont {{Fogli}}\ \emph {et~al.}(2012)\citenamefont
  {{Fogli}}, \citenamefont {{Lisi}}, \citenamefont {{Marrone}}, \citenamefont
  {{Montanino}}, \citenamefont {{Palazzo}},\ and\ \citenamefont
  {{Rotunno}}}]{fogli12a}%
  \BibitemOpen
  \bibfield  {author} {\bibinfo {author} {\bibfnamefont {G.~L.}\ \bibnamefont
  {{Fogli}}}, \bibinfo {author} {\bibfnamefont {E.}~\bibnamefont {{Lisi}}},
  \bibinfo {author} {\bibfnamefont {A.}~\bibnamefont {{Marrone}}}, \bibinfo
  {author} {\bibfnamefont {D.}~\bibnamefont {{Montanino}}}, \bibinfo {author}
  {\bibfnamefont {A.}~\bibnamefont {{Palazzo}}}, \ and\ \bibinfo {author}
  {\bibfnamefont {A.~M.}\ \bibnamefont {{Rotunno}}},\ }\href {\doibase
  10.1103/PhysRevD.86.013012} {\bibfield  {journal} {\bibinfo  {journal}
  {\prd}\ }\textbf {\bibinfo {volume} {86}},\ \bibinfo {eid} {013012} (\bibinfo
  {year} {2012})},\ \Eprint {http://arxiv.org/abs/1205.5254} {arXiv:1205.5254
  [hep-ph]} \BibitemShut {NoStop}%
\bibitem [{\citenamefont {{Li}}\ \emph {et~al.}(2013)\citenamefont {{Li}},
  \citenamefont {{Cao}}, \citenamefont {{Wang}},\ and\ \citenamefont
  {{Zhan}}}]{li13a}%
  \BibitemOpen
  \bibfield  {author} {\bibinfo {author} {\bibfnamefont {Y.-F.}\ \bibnamefont
  {{Li}}}, \bibinfo {author} {\bibfnamefont {J.}~\bibnamefont {{Cao}}},
  \bibinfo {author} {\bibfnamefont {Y.}~\bibnamefont {{Wang}}}, \ and\ \bibinfo
  {author} {\bibfnamefont {L.}~\bibnamefont {{Zhan}}},\ }\href {\doibase
  10.1103/PhysRevD.88.013008} {\bibfield  {journal} {\bibinfo  {journal}
  {\prd}\ }\textbf {\bibinfo {volume} {88}},\ \bibinfo {eid} {013008} (\bibinfo
  {year} {2013})},\ \Eprint {http://arxiv.org/abs/1303.6733} {arXiv:1303.6733
  [hep-ex]} \BibitemShut {NoStop}%
\bibitem [{\citenamefont {{Akhmedov}}\ \emph {et~al.}(2013)\citenamefont
  {{Akhmedov}}, \citenamefont {{Razzaque}},\ and\ \citenamefont
  {{Smirnov}}}]{akhmedov13a}%
  \BibitemOpen
  \bibfield  {author} {\bibinfo {author} {\bibfnamefont {E.~K.}\ \bibnamefont
  {{Akhmedov}}}, \bibinfo {author} {\bibfnamefont {S.}~\bibnamefont
  {{Razzaque}}}, \ and\ \bibinfo {author} {\bibfnamefont {A.~Y.}\ \bibnamefont
  {{Smirnov}}},\ }\href {\doibase 10.1007/JHEP02(2013)082} {\bibfield
  {journal} {\bibinfo  {journal} {Journal of High Energy Physics}\ }\textbf
  {\bibinfo {volume} {2013}},\ \bibinfo {eid} {82} (\bibinfo {year} {2013})},\
  \Eprint {http://arxiv.org/abs/1205.7071} {arXiv:1205.7071 [hep-ph]}
  \BibitemShut {NoStop}%
\bibitem [{\citenamefont {{Abe}}\ \emph {et~al.}(2015)\citenamefont {{Abe}},
  \citenamefont {{Aihara}}, \citenamefont {{Andreopoulos}}, \citenamefont
  {{Anghel}}, \citenamefont {{Ariga}}, \citenamefont {{Ariga}}, \citenamefont
  {{Asfandiyarov}}, \citenamefont {{Askins}}, \citenamefont {{Back}},
  \citenamefont {{Ballett}}, \citenamefont {{Barbi}}, \citenamefont {{Barker}},
  \citenamefont {{Barr}}, \citenamefont {{Bay}}, \citenamefont {{Beltrame}},
  \citenamefont {{Berardi}}, \citenamefont {{Bergevin}}, \citenamefont
  {{Berkman}}, \citenamefont {{Berry}}, \citenamefont {{Bhadra}}, \citenamefont
  {{Blaszczyk}}, \citenamefont {{Blondel}}, \citenamefont {{Bolognesi}},
  \citenamefont {{Boyd}}, \citenamefont {{Bravar}}, \citenamefont {{Bronner}},
  \citenamefont {{Cafagna}}, \citenamefont {{Carminati}}, \citenamefont
  {{Cartwright}}, \citenamefont {{Catanesi}}, \citenamefont {{Choi}},
  \citenamefont {{Choi}}, \citenamefont {{Collazuol}}, \citenamefont {{Cowan}},
  \citenamefont {{Cremonesi}}, \citenamefont {{Davies}}, \citenamefont {{De
  Rosa}}, \citenamefont {{Densham}}, \citenamefont {{Detwiler}}, \citenamefont
  {{Dewhurst}}, \citenamefont {{Di Lodovico}}, \citenamefont {{Di Luise}},
  \citenamefont {{Drapier}}, \citenamefont {{Emery}}, \citenamefont
  {{Ereditato}}, \citenamefont {{Fern{\'a}ndez}}, \citenamefont {{Feusels}},
  \citenamefont {{Finch}}, \citenamefont {{Fitton}}, \citenamefont {{Friend}},
  \citenamefont {{Fujii}}, \citenamefont {{Fukuda}}, \citenamefont {{Fukuda}},
  \citenamefont {{Galymov}}, \citenamefont {{Ganezer}}, \citenamefont
  {{Gonin}}, \citenamefont {{Gumplinger}}, \citenamefont {{Hadley}},
  \citenamefont {{Haegel}}, \citenamefont {{Haesler}}, \citenamefont {{Haga}},
  \citenamefont {{Hartfiel}}, \citenamefont {{Hartz}}, \citenamefont
  {{Hayato}}, \citenamefont {{Hierholzer}}, \citenamefont {{Hill}},
  \citenamefont {{Himmel}}, \citenamefont {{Hirota}}, \citenamefont
  {{Horiuchi}}, \citenamefont {{Huang}}, \citenamefont {{Ichikawa}},
  \citenamefont {{Iijima}}, \citenamefont {{Ikeda}}, \citenamefont {{Imber}},
  \citenamefont {{Inoue}}, \citenamefont {{Insler}}, \citenamefont {{Intonti}},
  \citenamefont {{Irvine}}, \citenamefont {{Ishida}}, \citenamefont {{Ishino}},
  \citenamefont {{Ishitsuka}}, \citenamefont {{Itow}}, \citenamefont
  {{Izmaylov}}, \citenamefont {{Jamieson}}, \citenamefont {{Jang}},
  \citenamefont {{Jiang}}, \citenamefont {{Joo}}, \citenamefont {{Jung}},
  \citenamefont {{Kaboth}}, \citenamefont {{Kajita}}, \citenamefont {{Kameda}},
  \citenamefont {{Karadhzov}}, \citenamefont {{Katori}}, \citenamefont
  {{Kearns}}, \citenamefont {{Khabibullin}}, \citenamefont {{Khotjantsev}},
  \citenamefont {{Kim}}, \citenamefont {{Kim}}, \citenamefont {{Kishimoto}},
  \citenamefont {{Kobayashi}}, \citenamefont {{Koga}}, \citenamefont
  {{Konaka}}, \citenamefont {{Kormos}}, \citenamefont {{Korzenev}},
  \citenamefont {{Koshio}}, \citenamefont {{Kropp}}, \citenamefont {{Kudenko}},
  \citenamefont {{Kutter}}, \citenamefont {{Kuze}}, \citenamefont {{Labarga}},
  \citenamefont {{Lagoda}}, \citenamefont {{Laveder}}, \citenamefont {{Lawe}},
  \citenamefont {{Learned}}, \citenamefont {{Lim}}, \citenamefont {{Lindner}},
  \citenamefont {{Longhin}}, \citenamefont {{Ludovici}}, \citenamefont {{Ma}},
  \citenamefont {{Magaletti}}, \citenamefont {{Mahn}}, \citenamefont {{Malek}},
  \citenamefont {{Mariani}}, \citenamefont {{Marti}}, \citenamefont {{Martin}},
  \citenamefont {{Martin}}, \citenamefont {{Martins}}, \citenamefont
  {{Mazzucato}}, \citenamefont {{McCauley}}, \citenamefont {{McFarland}},
  \citenamefont {{McGrew}}, \citenamefont {{Mezzetto}}, \citenamefont
  {{Minakata}}, \citenamefont {{Minamino}}, \citenamefont {{Mine}},
  \citenamefont {{Mineev}}, \citenamefont {{Miura}}, \citenamefont {{Monroe}},
  \citenamefont {{Mori}}, \citenamefont {{Moriyama}}, \citenamefont
  {{Mueller}}, \citenamefont {{Muheim}}, \citenamefont {{Nakahata}},
  \citenamefont {{Nakamura}}, \citenamefont {{Nakaya}}, \citenamefont
  {{Nakayama}}, \citenamefont {{Needham}}, \citenamefont {{Nicholls}},
  \citenamefont {{Nirkko}}, \citenamefont {{Nishimura}}, \citenamefont
  {{Noah}}, \citenamefont {{Nowak}}, \citenamefont {{Nunokawa}}, \citenamefont
  {{O'Keeffe}}, \citenamefont {{Okajima}}, \citenamefont {{Okumura}},
  \citenamefont {{Oser}}, \citenamefont {{O'Sullivan}}, \citenamefont
  {{Ovsiannikova}}, \citenamefont {{Owen}}, \citenamefont {{Oyama}},
  \citenamefont {{P{\'e}rez}}, \citenamefont {{Pac}}, \citenamefont
  {{Palladino}}, \citenamefont {{Palomino}}, \citenamefont {{Paolone}},
  \citenamefont {{Payne}}, \citenamefont {{Perevozchikov}}, \citenamefont
  {{Perkin}}, \citenamefont {{Pistillo}}, \citenamefont {{Playfer}},
  \citenamefont {{Posiadala-Zezula}}, \citenamefont {{Poutissou}},
  \citenamefont {{Quilain}}, \citenamefont {{Quinto}}, \citenamefont
  {{Radicioni}}, \citenamefont {{Ratoff}}, \citenamefont {{Ravonel}},
  \citenamefont {{Rayner}}, \citenamefont {{Redij}}, \citenamefont {{Retiere}},
  \citenamefont {{Riccio}}, \citenamefont {{Richard}}, \citenamefont
  {{Rondio}}, \citenamefont {{Rose}}, \citenamefont {{Ross-Lonergan}},
  \citenamefont {{Rott}}, \citenamefont {{Rountree}}, \citenamefont {{Rubbia}},
  \citenamefont {{Sacco}}, \citenamefont {{Sakuda}}, \citenamefont {{Sanchez}},
  \citenamefont {{Scantamburlo}}, \citenamefont {{Scholberg}}, \citenamefont
  {{Scott}}, \citenamefont {{Seiya}}, \citenamefont {{Sekiguchi}},
  \citenamefont {{Sekiya}}, \citenamefont {{Shaikhiev}}, \citenamefont
  {{Shimizu}}, \citenamefont {{Shiozawa}}, \citenamefont {{Short}},
  \citenamefont {{Sinnis}}, \citenamefont {{Smy}}, \citenamefont {{Sobczyk}},
  \citenamefont {{Sobel}}, \citenamefont {{Stewart}}, \citenamefont {{Stone}},
  \citenamefont {{Suda}}, \citenamefont {{Suzuki}}, \citenamefont {{Suzuki}},
  \citenamefont {{Svoboda}}, \citenamefont {{Tacik}}, \citenamefont {{Takeda}},
  \citenamefont {{Taketa}}, \citenamefont {{Takeuchi}}, \citenamefont
  {{Tanaka}}, \citenamefont {{Tanaka}}, \citenamefont {{Tanaka}}, \citenamefont
  {{Terri}}, \citenamefont {{Thompson}}, \citenamefont {{Thorpe}},
  \citenamefont {{Tobayama}}, \citenamefont {{Tolich}}, \citenamefont
  {{Tomura}}, \citenamefont {{Touramanis}}, \citenamefont {{Tsukamoto}},
  \citenamefont {{Tzanov}}, \citenamefont {{Uchida}}, \citenamefont {{Vagins}},
  \citenamefont {{Vasseur}}, \citenamefont {{Vogelaar}}, \citenamefont
  {{Walter}}, \citenamefont {{Wark}}, \citenamefont {{Wascko}}, \citenamefont
  {{Weber}}, \citenamefont {{Wendell}}, \citenamefont {{Wilkes}}, \citenamefont
  {{Wilking}}, \citenamefont {{Wilson}}, \citenamefont {{Xin}}, \citenamefont
  {{Yamamoto}}, \citenamefont {{Yanagisawa}}, \citenamefont {{Yano}},
  \citenamefont {{Yen}}, \citenamefont {{Yershov}}, \citenamefont
  {{Yokoyama}},\ and\ \citenamefont {{Zito}}}]{abe15a}%
  \BibitemOpen
  \bibfield  {author} {\bibinfo {author} {\bibfnamefont {K.}~\bibnamefont
  {{Abe}}}, \bibinfo {author} {\bibfnamefont {H.}~\bibnamefont {{Aihara}}},
  \bibinfo {author} {\bibfnamefont {C.}~\bibnamefont {{Andreopoulos}}},
  \bibinfo {author} {\bibfnamefont {I.}~\bibnamefont {{Anghel}}}, \bibinfo
  {author} {\bibfnamefont {A.}~\bibnamefont {{Ariga}}}, \bibinfo {author}
  {\bibfnamefont {T.}~\bibnamefont {{Ariga}}}, \bibinfo {author} {\bibfnamefont
  {R.}~\bibnamefont {{Asfandiyarov}}}, \bibinfo {author} {\bibfnamefont
  {M.}~\bibnamefont {{Askins}}}, \bibinfo {author} {\bibfnamefont {J.~J.}\
  \bibnamefont {{Back}}}, \bibinfo {author} {\bibfnamefont {P.}~\bibnamefont
  {{Ballett}}}, \bibinfo {author} {\bibfnamefont {M.}~\bibnamefont {{Barbi}}},
  \bibinfo {author} {\bibfnamefont {G.~J.}\ \bibnamefont {{Barker}}}, \bibinfo
  {author} {\bibfnamefont {G.}~\bibnamefont {{Barr}}}, \bibinfo {author}
  {\bibfnamefont {F.}~\bibnamefont {{Bay}}}, \bibinfo {author} {\bibfnamefont
  {P.}~\bibnamefont {{Beltrame}}}, \bibinfo {author} {\bibfnamefont
  {V.}~\bibnamefont {{Berardi}}}, \bibinfo {author} {\bibfnamefont
  {M.}~\bibnamefont {{Bergevin}}}, \bibinfo {author} {\bibfnamefont
  {S.}~\bibnamefont {{Berkman}}}, \bibinfo {author} {\bibfnamefont
  {T.}~\bibnamefont {{Berry}}}, \bibinfo {author} {\bibfnamefont
  {S.}~\bibnamefont {{Bhadra}}}, \bibinfo {author} {\bibfnamefont {F.~d.~M.}\
  \bibnamefont {{Blaszczyk}}}, \bibinfo {author} {\bibfnamefont
  {A.}~\bibnamefont {{Blondel}}}, \bibinfo {author} {\bibfnamefont
  {S.}~\bibnamefont {{Bolognesi}}}, \bibinfo {author} {\bibfnamefont {S.~B.}\
  \bibnamefont {{Boyd}}}, \bibinfo {author} {\bibfnamefont {A.}~\bibnamefont
  {{Bravar}}}, \bibinfo {author} {\bibfnamefont {C.}~\bibnamefont {{Bronner}}},
  \bibinfo {author} {\bibfnamefont {F.~S.}\ \bibnamefont {{Cafagna}}}, \bibinfo
  {author} {\bibfnamefont {G.}~\bibnamefont {{Carminati}}}, \bibinfo {author}
  {\bibfnamefont {S.~L.}\ \bibnamefont {{Cartwright}}}, \bibinfo {author}
  {\bibfnamefont {M.~G.}\ \bibnamefont {{Catanesi}}}, \bibinfo {author}
  {\bibfnamefont {K.}~\bibnamefont {{Choi}}}, \bibinfo {author} {\bibfnamefont
  {J.~H.}\ \bibnamefont {{Choi}}}, \bibinfo {author} {\bibfnamefont
  {G.}~\bibnamefont {{Collazuol}}}, \bibinfo {author} {\bibfnamefont
  {G.}~\bibnamefont {{Cowan}}}, \bibinfo {author} {\bibfnamefont
  {L.}~\bibnamefont {{Cremonesi}}}, \bibinfo {author} {\bibfnamefont
  {G.}~\bibnamefont {{Davies}}}, \bibinfo {author} {\bibfnamefont
  {G.}~\bibnamefont {{De Rosa}}}, \bibinfo {author} {\bibfnamefont
  {C.}~\bibnamefont {{Densham}}}, \bibinfo {author} {\bibfnamefont
  {J.}~\bibnamefont {{Detwiler}}}, \bibinfo {author} {\bibfnamefont
  {D.}~\bibnamefont {{Dewhurst}}}, \bibinfo {author} {\bibfnamefont
  {F.}~\bibnamefont {{Di Lodovico}}}, \bibinfo {author} {\bibfnamefont
  {S.}~\bibnamefont {{Di Luise}}}, \bibinfo {author} {\bibfnamefont
  {O.}~\bibnamefont {{Drapier}}}, \bibinfo {author} {\bibfnamefont
  {S.}~\bibnamefont {{Emery}}}, \bibinfo {author} {\bibfnamefont
  {A.}~\bibnamefont {{Ereditato}}}, \bibinfo {author} {\bibfnamefont
  {P.}~\bibnamefont {{Fern{\'a}ndez}}}, \bibinfo {author} {\bibfnamefont
  {T.}~\bibnamefont {{Feusels}}}, \bibinfo {author} {\bibfnamefont
  {A.}~\bibnamefont {{Finch}}}, \bibinfo {author} {\bibfnamefont
  {M.}~\bibnamefont {{Fitton}}}, \bibinfo {author} {\bibfnamefont
  {M.}~\bibnamefont {{Friend}}}, \bibinfo {author} {\bibfnamefont
  {Y.}~\bibnamefont {{Fujii}}}, \bibinfo {author} {\bibfnamefont
  {Y.}~\bibnamefont {{Fukuda}}}, \bibinfo {author} {\bibfnamefont
  {D.}~\bibnamefont {{Fukuda}}}, \bibinfo {author} {\bibfnamefont
  {V.}~\bibnamefont {{Galymov}}}, \bibinfo {author} {\bibfnamefont
  {K.}~\bibnamefont {{Ganezer}}}, \bibinfo {author} {\bibfnamefont
  {M.}~\bibnamefont {{Gonin}}}, \bibinfo {author} {\bibfnamefont
  {P.}~\bibnamefont {{Gumplinger}}}, \bibinfo {author} {\bibfnamefont {D.~R.}\
  \bibnamefont {{Hadley}}}, \bibinfo {author} {\bibfnamefont {L.}~\bibnamefont
  {{Haegel}}}, \bibinfo {author} {\bibfnamefont {A.}~\bibnamefont {{Haesler}}},
  \bibinfo {author} {\bibfnamefont {Y.}~\bibnamefont {{Haga}}}, \bibinfo
  {author} {\bibfnamefont {B.}~\bibnamefont {{Hartfiel}}}, \bibinfo {author}
  {\bibfnamefont {M.}~\bibnamefont {{Hartz}}}, \bibinfo {author} {\bibfnamefont
  {Y.}~\bibnamefont {{Hayato}}}, \bibinfo {author} {\bibfnamefont
  {M.}~\bibnamefont {{Hierholzer}}}, \bibinfo {author} {\bibfnamefont
  {J.}~\bibnamefont {{Hill}}}, \bibinfo {author} {\bibfnamefont
  {A.}~\bibnamefont {{Himmel}}}, \bibinfo {author} {\bibfnamefont
  {S.}~\bibnamefont {{Hirota}}}, \bibinfo {author} {\bibfnamefont
  {S.}~\bibnamefont {{Horiuchi}}}, \bibinfo {author} {\bibfnamefont
  {K.}~\bibnamefont {{Huang}}}, \bibinfo {author} {\bibfnamefont {A.~K.}\
  \bibnamefont {{Ichikawa}}}, \bibinfo {author} {\bibfnamefont
  {T.}~\bibnamefont {{Iijima}}}, \bibinfo {author} {\bibfnamefont
  {M.}~\bibnamefont {{Ikeda}}}, \bibinfo {author} {\bibfnamefont
  {J.}~\bibnamefont {{Imber}}}, \bibinfo {author} {\bibfnamefont
  {K.}~\bibnamefont {{Inoue}}}, \bibinfo {author} {\bibfnamefont
  {J.}~\bibnamefont {{Insler}}}, \bibinfo {author} {\bibfnamefont {R.~A.}\
  \bibnamefont {{Intonti}}}, \bibinfo {author} {\bibfnamefont {T.}~\bibnamefont
  {{Irvine}}}, \bibinfo {author} {\bibfnamefont {T.}~\bibnamefont {{Ishida}}},
  \bibinfo {author} {\bibfnamefont {H.}~\bibnamefont {{Ishino}}}, \bibinfo
  {author} {\bibfnamefont {M.}~\bibnamefont {{Ishitsuka}}}, \bibinfo {author}
  {\bibfnamefont {Y.}~\bibnamefont {{Itow}}}, \bibinfo {author} {\bibfnamefont
  {A.}~\bibnamefont {{Izmaylov}}}, \bibinfo {author} {\bibfnamefont
  {B.}~\bibnamefont {{Jamieson}}}, \bibinfo {author} {\bibfnamefont {H.~I.}\
  \bibnamefont {{Jang}}}, \bibinfo {author} {\bibfnamefont {M.}~\bibnamefont
  {{Jiang}}}, \bibinfo {author} {\bibfnamefont {K.~K.}\ \bibnamefont {{Joo}}},
  \bibinfo {author} {\bibfnamefont {C.~K.}\ \bibnamefont {{Jung}}}, \bibinfo
  {author} {\bibfnamefont {A.}~\bibnamefont {{Kaboth}}}, \bibinfo {author}
  {\bibfnamefont {T.}~\bibnamefont {{Kajita}}}, \bibinfo {author}
  {\bibfnamefont {J.}~\bibnamefont {{Kameda}}}, \bibinfo {author}
  {\bibfnamefont {Y.}~\bibnamefont {{Karadhzov}}}, \bibinfo {author}
  {\bibfnamefont {T.}~\bibnamefont {{Katori}}}, \bibinfo {author}
  {\bibfnamefont {E.}~\bibnamefont {{Kearns}}}, \bibinfo {author}
  {\bibfnamefont {M.}~\bibnamefont {{Khabibullin}}}, \bibinfo {author}
  {\bibfnamefont {A.}~\bibnamefont {{Khotjantsev}}}, \bibinfo {author}
  {\bibfnamefont {J.~Y.}\ \bibnamefont {{Kim}}}, \bibinfo {author}
  {\bibfnamefont {S.~B.}\ \bibnamefont {{Kim}}}, \bibinfo {author}
  {\bibfnamefont {Y.}~\bibnamefont {{Kishimoto}}}, \bibinfo {author}
  {\bibfnamefont {T.}~\bibnamefont {{Kobayashi}}}, \bibinfo {author}
  {\bibfnamefont {M.}~\bibnamefont {{Koga}}}, \bibinfo {author} {\bibfnamefont
  {A.}~\bibnamefont {{Konaka}}}, \bibinfo {author} {\bibfnamefont {L.~L.}\
  \bibnamefont {{Kormos}}}, \bibinfo {author} {\bibfnamefont {A.}~\bibnamefont
  {{Korzenev}}}, \bibinfo {author} {\bibfnamefont {Y.}~\bibnamefont
  {{Koshio}}}, \bibinfo {author} {\bibfnamefont {W.~R.}\ \bibnamefont
  {{Kropp}}}, \bibinfo {author} {\bibfnamefont {Y.}~\bibnamefont {{Kudenko}}},
  \bibinfo {author} {\bibfnamefont {T.}~\bibnamefont {{Kutter}}}, \bibinfo
  {author} {\bibfnamefont {M.}~\bibnamefont {{Kuze}}}, \bibinfo {author}
  {\bibfnamefont {L.}~\bibnamefont {{Labarga}}}, \bibinfo {author}
  {\bibfnamefont {J.}~\bibnamefont {{Lagoda}}}, \bibinfo {author}
  {\bibfnamefont {M.}~\bibnamefont {{Laveder}}}, \bibinfo {author}
  {\bibfnamefont {M.}~\bibnamefont {{Lawe}}}, \bibinfo {author} {\bibfnamefont
  {J.~G.}\ \bibnamefont {{Learned}}}, \bibinfo {author} {\bibfnamefont {I.~T.}\
  \bibnamefont {{Lim}}}, \bibinfo {author} {\bibfnamefont {T.}~\bibnamefont
  {{Lindner}}}, \bibinfo {author} {\bibfnamefont {A.}~\bibnamefont
  {{Longhin}}}, \bibinfo {author} {\bibfnamefont {L.}~\bibnamefont
  {{Ludovici}}}, \bibinfo {author} {\bibfnamefont {W.}~\bibnamefont {{Ma}}},
  \bibinfo {author} {\bibfnamefont {L.}~\bibnamefont {{Magaletti}}}, \bibinfo
  {author} {\bibfnamefont {K.}~\bibnamefont {{Mahn}}}, \bibinfo {author}
  {\bibfnamefont {M.}~\bibnamefont {{Malek}}}, \bibinfo {author} {\bibfnamefont
  {C.}~\bibnamefont {{Mariani}}}, \bibinfo {author} {\bibfnamefont
  {L.}~\bibnamefont {{Marti}}}, \bibinfo {author} {\bibfnamefont {J.~F.}\
  \bibnamefont {{Martin}}}, \bibinfo {author} {\bibfnamefont {C.}~\bibnamefont
  {{Martin}}}, \bibinfo {author} {\bibfnamefont {P.~P.~J.}\ \bibnamefont
  {{Martins}}}, \bibinfo {author} {\bibfnamefont {E.}~\bibnamefont
  {{Mazzucato}}}, \bibinfo {author} {\bibfnamefont {N.}~\bibnamefont
  {{McCauley}}}, \bibinfo {author} {\bibfnamefont {K.~S.}\ \bibnamefont
  {{McFarland}}}, \bibinfo {author} {\bibfnamefont {C.}~\bibnamefont
  {{McGrew}}}, \bibinfo {author} {\bibfnamefont {M.}~\bibnamefont
  {{Mezzetto}}}, \bibinfo {author} {\bibfnamefont {H.}~\bibnamefont
  {{Minakata}}}, \bibinfo {author} {\bibfnamefont {A.}~\bibnamefont
  {{Minamino}}}, \bibinfo {author} {\bibfnamefont {S.}~\bibnamefont {{Mine}}},
  \bibinfo {author} {\bibfnamefont {O.}~\bibnamefont {{Mineev}}}, \bibinfo
  {author} {\bibfnamefont {M.}~\bibnamefont {{Miura}}}, \bibinfo {author}
  {\bibfnamefont {J.}~\bibnamefont {{Monroe}}}, \bibinfo {author}
  {\bibfnamefont {T.}~\bibnamefont {{Mori}}}, \bibinfo {author} {\bibfnamefont
  {S.}~\bibnamefont {{Moriyama}}}, \bibinfo {author} {\bibfnamefont
  {T.}~\bibnamefont {{Mueller}}}, \bibinfo {author} {\bibfnamefont
  {F.}~\bibnamefont {{Muheim}}}, \bibinfo {author} {\bibfnamefont
  {M.}~\bibnamefont {{Nakahata}}}, \bibinfo {author} {\bibfnamefont
  {K.}~\bibnamefont {{Nakamura}}}, \bibinfo {author} {\bibfnamefont
  {T.}~\bibnamefont {{Nakaya}}}, \bibinfo {author} {\bibfnamefont
  {S.}~\bibnamefont {{Nakayama}}}, \bibinfo {author} {\bibfnamefont
  {M.}~\bibnamefont {{Needham}}}, \bibinfo {author} {\bibfnamefont
  {T.}~\bibnamefont {{Nicholls}}}, \bibinfo {author} {\bibfnamefont
  {M.}~\bibnamefont {{Nirkko}}}, \bibinfo {author} {\bibfnamefont
  {Y.}~\bibnamefont {{Nishimura}}}, \bibinfo {author} {\bibfnamefont
  {E.}~\bibnamefont {{Noah}}}, \bibinfo {author} {\bibfnamefont
  {J.}~\bibnamefont {{Nowak}}}, \bibinfo {author} {\bibfnamefont
  {H.}~\bibnamefont {{Nunokawa}}}, \bibinfo {author} {\bibfnamefont {H.~M.}\
  \bibnamefont {{O'Keeffe}}}, \bibinfo {author} {\bibfnamefont
  {Y.}~\bibnamefont {{Okajima}}}, \bibinfo {author} {\bibfnamefont
  {K.}~\bibnamefont {{Okumura}}}, \bibinfo {author} {\bibfnamefont {S.~M.}\
  \bibnamefont {{Oser}}}, \bibinfo {author} {\bibfnamefont {E.}~\bibnamefont
  {{O'Sullivan}}}, \bibinfo {author} {\bibfnamefont {T.}~\bibnamefont
  {{Ovsiannikova}}}, \bibinfo {author} {\bibfnamefont {R.~A.}\ \bibnamefont
  {{Owen}}}, \bibinfo {author} {\bibfnamefont {Y.}~\bibnamefont {{Oyama}}},
  \bibinfo {author} {\bibfnamefont {J.}~\bibnamefont {{P{\'e}rez}}}, \bibinfo
  {author} {\bibfnamefont {M.~Y.}\ \bibnamefont {{Pac}}}, \bibinfo {author}
  {\bibfnamefont {V.}~\bibnamefont {{Palladino}}}, \bibinfo {author}
  {\bibfnamefont {J.~L.}\ \bibnamefont {{Palomino}}}, \bibinfo {author}
  {\bibfnamefont {V.}~\bibnamefont {{Paolone}}}, \bibinfo {author}
  {\bibfnamefont {D.}~\bibnamefont {{Payne}}}, \bibinfo {author} {\bibfnamefont
  {O.}~\bibnamefont {{Perevozchikov}}}, \bibinfo {author} {\bibfnamefont
  {J.~D.}\ \bibnamefont {{Perkin}}}, \bibinfo {author} {\bibfnamefont
  {C.}~\bibnamefont {{Pistillo}}}, \bibinfo {author} {\bibfnamefont
  {S.}~\bibnamefont {{Playfer}}}, \bibinfo {author} {\bibfnamefont
  {M.}~\bibnamefont {{Posiadala-Zezula}}}, \bibinfo {author} {\bibfnamefont
  {J.~M.}\ \bibnamefont {{Poutissou}}}, \bibinfo {author} {\bibfnamefont
  {B.}~\bibnamefont {{Quilain}}}, \bibinfo {author} {\bibfnamefont
  {M.}~\bibnamefont {{Quinto}}}, \bibinfo {author} {\bibfnamefont
  {E.}~\bibnamefont {{Radicioni}}}, \bibinfo {author} {\bibfnamefont {P.~N.}\
  \bibnamefont {{Ratoff}}}, \bibinfo {author} {\bibfnamefont {M.}~\bibnamefont
  {{Ravonel}}}, \bibinfo {author} {\bibfnamefont {M.~A.}\ \bibnamefont
  {{Rayner}}}, \bibinfo {author} {\bibfnamefont {A.}~\bibnamefont {{Redij}}},
  \bibinfo {author} {\bibfnamefont {F.}~\bibnamefont {{Retiere}}}, \bibinfo
  {author} {\bibfnamefont {C.}~\bibnamefont {{Riccio}}}, \bibinfo {author}
  {\bibfnamefont {E.}~\bibnamefont {{Richard}}}, \bibinfo {author}
  {\bibfnamefont {E.}~\bibnamefont {{Rondio}}}, \bibinfo {author}
  {\bibfnamefont {H.~J.}\ \bibnamefont {{Rose}}}, \bibinfo {author}
  {\bibfnamefont {M.}~\bibnamefont {{Ross-Lonergan}}}, \bibinfo {author}
  {\bibfnamefont {C.}~\bibnamefont {{Rott}}}, \bibinfo {author} {\bibfnamefont
  {S.~D.}\ \bibnamefont {{Rountree}}}, \bibinfo {author} {\bibfnamefont
  {A.}~\bibnamefont {{Rubbia}}}, \bibinfo {author} {\bibfnamefont
  {R.}~\bibnamefont {{Sacco}}}, \bibinfo {author} {\bibfnamefont
  {M.}~\bibnamefont {{Sakuda}}}, \bibinfo {author} {\bibfnamefont {M.~C.}\
  \bibnamefont {{Sanchez}}}, \bibinfo {author} {\bibfnamefont {E.}~\bibnamefont
  {{Scantamburlo}}}, \bibinfo {author} {\bibfnamefont {K.}~\bibnamefont
  {{Scholberg}}}, \bibinfo {author} {\bibfnamefont {M.}~\bibnamefont
  {{Scott}}}, \bibinfo {author} {\bibfnamefont {Y.}~\bibnamefont {{Seiya}}},
  \bibinfo {author} {\bibfnamefont {T.}~\bibnamefont {{Sekiguchi}}}, \bibinfo
  {author} {\bibfnamefont {H.}~\bibnamefont {{Sekiya}}}, \bibinfo {author}
  {\bibfnamefont {A.}~\bibnamefont {{Shaikhiev}}}, \bibinfo {author}
  {\bibfnamefont {I.}~\bibnamefont {{Shimizu}}}, \bibinfo {author}
  {\bibfnamefont {M.}~\bibnamefont {{Shiozawa}}}, \bibinfo {author}
  {\bibfnamefont {S.}~\bibnamefont {{Short}}}, \bibinfo {author} {\bibfnamefont
  {G.}~\bibnamefont {{Sinnis}}}, \bibinfo {author} {\bibfnamefont {M.~B.}\
  \bibnamefont {{Smy}}}, \bibinfo {author} {\bibfnamefont {J.}~\bibnamefont
  {{Sobczyk}}}, \bibinfo {author} {\bibfnamefont {H.~W.}\ \bibnamefont
  {{Sobel}}}, \bibinfo {author} {\bibfnamefont {T.}~\bibnamefont {{Stewart}}},
  \bibinfo {author} {\bibfnamefont {J.~L.}\ \bibnamefont {{Stone}}}, \bibinfo
  {author} {\bibfnamefont {Y.}~\bibnamefont {{Suda}}}, \bibinfo {author}
  {\bibfnamefont {Y.}~\bibnamefont {{Suzuki}}}, \bibinfo {author}
  {\bibfnamefont {A.~T.}\ \bibnamefont {{Suzuki}}}, \bibinfo {author}
  {\bibfnamefont {R.}~\bibnamefont {{Svoboda}}}, \bibinfo {author}
  {\bibfnamefont {R.}~\bibnamefont {{Tacik}}}, \bibinfo {author} {\bibfnamefont
  {A.}~\bibnamefont {{Takeda}}}, \bibinfo {author} {\bibfnamefont
  {A.}~\bibnamefont {{Taketa}}}, \bibinfo {author} {\bibfnamefont
  {Y.}~\bibnamefont {{Takeuchi}}}, \bibinfo {author} {\bibfnamefont {H.~A.}\
  \bibnamefont {{Tanaka}}}, \bibinfo {author} {\bibfnamefont {H.~K.~M.}\
  \bibnamefont {{Tanaka}}}, \bibinfo {author} {\bibfnamefont {H.}~\bibnamefont
  {{Tanaka}}}, \bibinfo {author} {\bibfnamefont {R.}~\bibnamefont {{Terri}}},
  \bibinfo {author} {\bibfnamefont {L.~F.}\ \bibnamefont {{Thompson}}},
  \bibinfo {author} {\bibfnamefont {M.}~\bibnamefont {{Thorpe}}}, \bibinfo
  {author} {\bibfnamefont {S.}~\bibnamefont {{Tobayama}}}, \bibinfo {author}
  {\bibfnamefont {N.}~\bibnamefont {{Tolich}}}, \bibinfo {author}
  {\bibfnamefont {T.}~\bibnamefont {{Tomura}}}, \bibinfo {author}
  {\bibfnamefont {C.}~\bibnamefont {{Touramanis}}}, \bibinfo {author}
  {\bibfnamefont {T.}~\bibnamefont {{Tsukamoto}}}, \bibinfo {author}
  {\bibfnamefont {M.}~\bibnamefont {{Tzanov}}}, \bibinfo {author}
  {\bibfnamefont {Y.}~\bibnamefont {{Uchida}}}, \bibinfo {author}
  {\bibfnamefont {M.~R.}\ \bibnamefont {{Vagins}}}, \bibinfo {author}
  {\bibfnamefont {G.}~\bibnamefont {{Vasseur}}}, \bibinfo {author}
  {\bibfnamefont {R.~B.}\ \bibnamefont {{Vogelaar}}}, \bibinfo {author}
  {\bibfnamefont {C.~W.}\ \bibnamefont {{Walter}}}, \bibinfo {author}
  {\bibfnamefont {D.}~\bibnamefont {{Wark}}}, \bibinfo {author} {\bibfnamefont
  {M.~O.}\ \bibnamefont {{Wascko}}}, \bibinfo {author} {\bibfnamefont
  {A.}~\bibnamefont {{Weber}}}, \bibinfo {author} {\bibfnamefont
  {R.}~\bibnamefont {{Wendell}}}, \bibinfo {author} {\bibfnamefont {R.~J.}\
  \bibnamefont {{Wilkes}}}, \bibinfo {author} {\bibfnamefont {M.~J.}\
  \bibnamefont {{Wilking}}}, \bibinfo {author} {\bibfnamefont {J.~R.}\
  \bibnamefont {{Wilson}}}, \bibinfo {author} {\bibfnamefont {T.}~\bibnamefont
  {{Xin}}}, \bibinfo {author} {\bibfnamefont {K.}~\bibnamefont {{Yamamoto}}},
  \bibinfo {author} {\bibfnamefont {C.}~\bibnamefont {{Yanagisawa}}}, \bibinfo
  {author} {\bibfnamefont {T.}~\bibnamefont {{Yano}}}, \bibinfo {author}
  {\bibfnamefont {S.}~\bibnamefont {{Yen}}}, \bibinfo {author} {\bibfnamefont
  {N.}~\bibnamefont {{Yershov}}}, \bibinfo {author} {\bibfnamefont
  {M.}~\bibnamefont {{Yokoyama}}}, \ and\ \bibinfo {author} {\bibfnamefont
  {M.}~\bibnamefont {{Zito}}},\ }\href {\doibase 10.1093/ptep/ptv061}
  {\bibfield  {journal} {\bibinfo  {journal} {Progress of Theoretical and
  Experimental Physics}\ }\textbf {\bibinfo {volume} {2015}},\ \bibinfo {eid}
  {053C02} (\bibinfo {year} {2015})}\BibitemShut {NoStop}%
\bibitem [{\citenamefont {{Aker}}\ \emph {et~al.}(2019)\citenamefont {{Aker}},
  \citenamefont {{Altenm{\"u}ller}}, \citenamefont {{Arenz}}, \citenamefont
  {{Babutzka}}, \citenamefont {{Barrett}}, \citenamefont {{Bauer}},
  \citenamefont {{Beck}}, \citenamefont {{Beglarian}}, \citenamefont
  {{Behrens}}, \citenamefont {{Bergmann}},\ and\ \citenamefont
  {et~al.}}]{1909.06048}%
  \BibitemOpen
  \bibfield  {author} {\bibinfo {author} {\bibfnamefont {M.}~\bibnamefont
  {{Aker}}}, \bibinfo {author} {\bibfnamefont {K.}~\bibnamefont
  {{Altenm{\"u}ller}}}, \bibinfo {author} {\bibfnamefont {M.}~\bibnamefont
  {{Arenz}}}, \bibinfo {author} {\bibfnamefont {M.}~\bibnamefont {{Babutzka}}},
  \bibinfo {author} {\bibfnamefont {J.}~\bibnamefont {{Barrett}}}, \bibinfo
  {author} {\bibfnamefont {S.}~\bibnamefont {{Bauer}}}, \bibinfo {author}
  {\bibfnamefont {M.}~\bibnamefont {{Beck}}}, \bibinfo {author} {\bibfnamefont
  {A.}~\bibnamefont {{Beglarian}}}, \bibinfo {author} {\bibfnamefont
  {J.}~\bibnamefont {{Behrens}}}, \bibinfo {author} {\bibfnamefont
  {T.}~\bibnamefont {{Bergmann}}}, \ and\ \bibinfo {author} {\bibnamefont
  {et~al.}},\ }\href {\doibase 10.1103/PhysRevLett.123.221802} {\bibfield
  {journal} {\bibinfo  {journal} {\prl}\ }\textbf {\bibinfo {volume} {123}},\
  \bibinfo {eid} {221802} (\bibinfo {year} {2019})},\ \Eprint
  {http://arxiv.org/abs/1909.06048} {arXiv:1909.06048 [hep-ex]} \BibitemShut
  {NoStop}%
\bibitem [{\citenamefont {{Hannestad}}(2005)}]{astro-ph/0505551}%
  \BibitemOpen
  \bibfield  {author} {\bibinfo {author} {\bibfnamefont {S.}~\bibnamefont
  {{Hannestad}}},\ }\href {\doibase 10.1103/PhysRevLett.95.221301} {\bibfield
  {journal} {\bibinfo  {journal} {\prl}\ }\textbf {\bibinfo {volume} {95}},\
  \bibinfo {eid} {221301} (\bibinfo {year} {2005})},\ \Eprint
  {http://arxiv.org/abs/astro-ph/0505551} {arXiv:astro-ph/0505551 [astro-ph]}
  \BibitemShut {NoStop}%
\bibitem [{\citenamefont {{Weinberg}}(1989)}]{weinberg89a}%
  \BibitemOpen
  \bibfield  {author} {\bibinfo {author} {\bibfnamefont {S.}~\bibnamefont
  {{Weinberg}}},\ }\href@noop {} {\bibfield  {journal} {\bibinfo  {journal}
  {Reviews of Modern Physics}\ }\textbf {\bibinfo {volume} {61}},\ \bibinfo
  {pages} {1} (\bibinfo {year} {1989})}\BibitemShut {NoStop}%
\bibitem [{\citenamefont {{Brax}}\ and\ \citenamefont
  {{Valageas}}(2019)}]{brax19a}%
  \BibitemOpen
  \bibfield  {author} {\bibinfo {author} {\bibfnamefont {P.}~\bibnamefont
  {{Brax}}}\ and\ \bibinfo {author} {\bibfnamefont {P.}~\bibnamefont
  {{Valageas}}},\ }\href {\doibase 10.1103/PhysRevD.99.123506} {\bibfield
  {journal} {\bibinfo  {journal} {\prd}\ }\textbf {\bibinfo {volume} {99}},\
  \bibinfo {eid} {123506} (\bibinfo {year} {2019})},\ \Eprint
  {http://arxiv.org/abs/1903.04825} {arXiv:1903.04825 [astro-ph.CO]}
  \BibitemShut {NoStop}%
\bibitem [{\citenamefont {{Percival}}\ \emph {et~al.}(2010)\citenamefont
  {{Percival}}, \citenamefont {{Reid}}, \citenamefont {{Eisenstein}},
  \citenamefont {{Bahcall}}, \citenamefont {{Budavari}}, \citenamefont
  {{Frieman}}, \citenamefont {{Fukugita}}, \citenamefont {{Gunn}},
  \citenamefont {{Ivezi{\'c}}}, \citenamefont {{Knapp}}, \citenamefont
  {{Kron}}, \citenamefont {{Loveday}}, \citenamefont {{Lupton}}, \citenamefont
  {{McKay}}, \citenamefont {{Meiksin}}, \citenamefont {{Nichol}}, \citenamefont
  {{Pope}}, \citenamefont {{Schlegel}}, \citenamefont {{Schneider}},
  \citenamefont {{Spergel}}, \citenamefont {{Stoughton}}, \citenamefont
  {{Strauss}}, \citenamefont {{Szalay}}, \citenamefont {{Tegmark}},
  \citenamefont {{Vogeley}}, \citenamefont {{Weinberg}}, \citenamefont
  {{York}},\ and\ \citenamefont {{Zehavi}}}]{percival10a}%
  \BibitemOpen
  \bibfield  {author} {\bibinfo {author} {\bibfnamefont {W.~J.}\ \bibnamefont
  {{Percival}}}, \bibinfo {author} {\bibfnamefont {B.~A.}\ \bibnamefont
  {{Reid}}}, \bibinfo {author} {\bibfnamefont {D.~J.}\ \bibnamefont
  {{Eisenstein}}}, \bibinfo {author} {\bibfnamefont {N.~A.}\ \bibnamefont
  {{Bahcall}}}, \bibinfo {author} {\bibfnamefont {T.}~\bibnamefont
  {{Budavari}}}, \bibinfo {author} {\bibfnamefont {J.~A.}\ \bibnamefont
  {{Frieman}}}, \bibinfo {author} {\bibfnamefont {M.}~\bibnamefont
  {{Fukugita}}}, \bibinfo {author} {\bibfnamefont {J.~E.}\ \bibnamefont
  {{Gunn}}}, \bibinfo {author} {\bibfnamefont {{\v Z}.}~\bibnamefont
  {{Ivezi{\'c}}}}, \bibinfo {author} {\bibfnamefont {G.~R.}\ \bibnamefont
  {{Knapp}}}, \bibinfo {author} {\bibfnamefont {R.~G.}\ \bibnamefont {{Kron}}},
  \bibinfo {author} {\bibfnamefont {J.}~\bibnamefont {{Loveday}}}, \bibinfo
  {author} {\bibfnamefont {R.~H.}\ \bibnamefont {{Lupton}}}, \bibinfo {author}
  {\bibfnamefont {T.~A.}\ \bibnamefont {{McKay}}}, \bibinfo {author}
  {\bibfnamefont {A.}~\bibnamefont {{Meiksin}}}, \bibinfo {author}
  {\bibfnamefont {R.~C.}\ \bibnamefont {{Nichol}}}, \bibinfo {author}
  {\bibfnamefont {A.~C.}\ \bibnamefont {{Pope}}}, \bibinfo {author}
  {\bibfnamefont {D.~J.}\ \bibnamefont {{Schlegel}}}, \bibinfo {author}
  {\bibfnamefont {D.~P.}\ \bibnamefont {{Schneider}}}, \bibinfo {author}
  {\bibfnamefont {D.~N.}\ \bibnamefont {{Spergel}}}, \bibinfo {author}
  {\bibfnamefont {C.}~\bibnamefont {{Stoughton}}}, \bibinfo {author}
  {\bibfnamefont {M.~A.}\ \bibnamefont {{Strauss}}}, \bibinfo {author}
  {\bibfnamefont {A.~S.}\ \bibnamefont {{Szalay}}}, \bibinfo {author}
  {\bibfnamefont {M.}~\bibnamefont {{Tegmark}}}, \bibinfo {author}
  {\bibfnamefont {M.~S.}\ \bibnamefont {{Vogeley}}}, \bibinfo {author}
  {\bibfnamefont {D.~H.}\ \bibnamefont {{Weinberg}}}, \bibinfo {author}
  {\bibfnamefont {D.~G.}\ \bibnamefont {{York}}}, \ and\ \bibinfo {author}
  {\bibfnamefont {I.}~\bibnamefont {{Zehavi}}},\ }\href {\doibase
  10.1111/j.1365-2966.2009.15812.x} {\bibfield  {journal} {\bibinfo  {journal}
  {\mnras}\ }\textbf {\bibinfo {volume} {401}},\ \bibinfo {pages} {2148}
  (\bibinfo {year} {2010})},\ \Eprint {http://arxiv.org/abs/0907.1660}
  {arXiv:0907.1660} \BibitemShut {NoStop}%
\bibitem [{\citenamefont {{Palanque-Delabrouille}}\ \emph
  {et~al.}(2013)\citenamefont {{Palanque-Delabrouille}}, \citenamefont
  {{Y{\`e}che}}, \citenamefont {{Borde}}, \citenamefont {{Le Goff}},
  \citenamefont {{Rossi}}, \citenamefont {{Viel}}, \citenamefont {{Aubourg}},
  \citenamefont {{Bailey}}, \citenamefont {{Bautista}}, \citenamefont
  {{Blomqvist}}, \citenamefont {{Bolton}}, \citenamefont {{Bolton}},
  \citenamefont {{Busca}}, \citenamefont {{Carithers}}, \citenamefont
  {{Croft}}, \citenamefont {{Dawson}}, \citenamefont {{Delubac}}, \citenamefont
  {{Font-Ribera}}, \citenamefont {{Ho}}, \citenamefont {{Kirkby}},
  \citenamefont {{Lee}}, \citenamefont {{Margala}}, \citenamefont
  {{Miralda-Escud{\'e}}}, \citenamefont {{Muna}}, \citenamefont {{Myers}},
  \citenamefont {{Noterdaeme}}, \citenamefont {{P{\^a}ris}}, \citenamefont
  {{Petitjean}}, \citenamefont {{Pieri}}, \citenamefont {{Rich}}, \citenamefont
  {{Rollinde}}, \citenamefont {{Ross}}, \citenamefont {{Schlegel}},
  \citenamefont {{Schneider}}, \citenamefont {{Slosar}},\ and\ \citenamefont
  {{Weinberg}}}]{palanque-delabrouille13b}%
  \BibitemOpen
  \bibfield  {author} {\bibinfo {author} {\bibfnamefont {N.}~\bibnamefont
  {{Palanque-Delabrouille}}}, \bibinfo {author} {\bibfnamefont
  {C.}~\bibnamefont {{Y{\`e}che}}}, \bibinfo {author} {\bibfnamefont
  {A.}~\bibnamefont {{Borde}}}, \bibinfo {author} {\bibfnamefont {J.-M.}\
  \bibnamefont {{Le Goff}}}, \bibinfo {author} {\bibfnamefont {G.}~\bibnamefont
  {{Rossi}}}, \bibinfo {author} {\bibfnamefont {M.}~\bibnamefont {{Viel}}},
  \bibinfo {author} {\bibfnamefont {{\'E}.}~\bibnamefont {{Aubourg}}}, \bibinfo
  {author} {\bibfnamefont {S.}~\bibnamefont {{Bailey}}}, \bibinfo {author}
  {\bibfnamefont {J.}~\bibnamefont {{Bautista}}}, \bibinfo {author}
  {\bibfnamefont {M.}~\bibnamefont {{Blomqvist}}}, \bibinfo {author}
  {\bibfnamefont {A.}~\bibnamefont {{Bolton}}}, \bibinfo {author}
  {\bibfnamefont {J.~S.}\ \bibnamefont {{Bolton}}}, \bibinfo {author}
  {\bibfnamefont {N.~G.}\ \bibnamefont {{Busca}}}, \bibinfo {author}
  {\bibfnamefont {B.}~\bibnamefont {{Carithers}}}, \bibinfo {author}
  {\bibfnamefont {R.~A.~C.}\ \bibnamefont {{Croft}}}, \bibinfo {author}
  {\bibfnamefont {K.~S.}\ \bibnamefont {{Dawson}}}, \bibinfo {author}
  {\bibfnamefont {T.}~\bibnamefont {{Delubac}}}, \bibinfo {author}
  {\bibfnamefont {A.}~\bibnamefont {{Font-Ribera}}}, \bibinfo {author}
  {\bibfnamefont {S.}~\bibnamefont {{Ho}}}, \bibinfo {author} {\bibfnamefont
  {D.}~\bibnamefont {{Kirkby}}}, \bibinfo {author} {\bibfnamefont {K.-G.}\
  \bibnamefont {{Lee}}}, \bibinfo {author} {\bibfnamefont {D.}~\bibnamefont
  {{Margala}}}, \bibinfo {author} {\bibfnamefont {J.}~\bibnamefont
  {{Miralda-Escud{\'e}}}}, \bibinfo {author} {\bibfnamefont {D.}~\bibnamefont
  {{Muna}}}, \bibinfo {author} {\bibfnamefont {A.~D.}\ \bibnamefont {{Myers}}},
  \bibinfo {author} {\bibfnamefont {P.}~\bibnamefont {{Noterdaeme}}}, \bibinfo
  {author} {\bibfnamefont {I.}~\bibnamefont {{P{\^a}ris}}}, \bibinfo {author}
  {\bibfnamefont {P.}~\bibnamefont {{Petitjean}}}, \bibinfo {author}
  {\bibfnamefont {M.~M.}\ \bibnamefont {{Pieri}}}, \bibinfo {author}
  {\bibfnamefont {J.}~\bibnamefont {{Rich}}}, \bibinfo {author} {\bibfnamefont
  {E.}~\bibnamefont {{Rollinde}}}, \bibinfo {author} {\bibfnamefont {N.~P.}\
  \bibnamefont {{Ross}}}, \bibinfo {author} {\bibfnamefont {D.~J.}\
  \bibnamefont {{Schlegel}}}, \bibinfo {author} {\bibfnamefont {D.~P.}\
  \bibnamefont {{Schneider}}}, \bibinfo {author} {\bibfnamefont
  {A.}~\bibnamefont {{Slosar}}}, \ and\ \bibinfo {author} {\bibfnamefont
  {D.~H.}\ \bibnamefont {{Weinberg}}},\ }\href {\doibase
  10.1051/0004-6361/201322130} {\bibfield  {journal} {\bibinfo  {journal}
  {\aap}\ }\textbf {\bibinfo {volume} {559}},\ \bibinfo {eid} {A85} (\bibinfo
  {year} {2013})},\ \Eprint {http://arxiv.org/abs/1306.5896} {arXiv:1306.5896
  [astro-ph.CO]} \BibitemShut {NoStop}%
\bibitem [{\citenamefont {{Chabanier}}\ \emph {et~al.}(2019)\citenamefont
  {{Chabanier}}, \citenamefont {{Palanque-Delabrouille}}, \citenamefont
  {{Y{\`e}che}}, \citenamefont {{Le Goff}}, \citenamefont {{Armengaud}},
  \citenamefont {{Bautista}}, \citenamefont {{Blomqvist}}, \citenamefont
  {{Busca}}, \citenamefont {{Dawson}}, \citenamefont {{Etourneau}},
  \citenamefont {{Font-Ribera}}, \citenamefont {{Lee}}, \citenamefont {{du Mas
  des Bourboux}}, \citenamefont {{Pieri}}, \citenamefont {{Rich}},
  \citenamefont {{Rossi}}, \citenamefont {{Schneider}},\ and\ \citenamefont
  {{Slosar}}}]{chabanier19a}%
  \BibitemOpen
  \bibfield  {author} {\bibinfo {author} {\bibfnamefont {S.}~\bibnamefont
  {{Chabanier}}}, \bibinfo {author} {\bibfnamefont {N.}~\bibnamefont
  {{Palanque-Delabrouille}}}, \bibinfo {author} {\bibfnamefont
  {C.}~\bibnamefont {{Y{\`e}che}}}, \bibinfo {author} {\bibfnamefont {J.-M.}\
  \bibnamefont {{Le Goff}}}, \bibinfo {author} {\bibfnamefont {E.}~\bibnamefont
  {{Armengaud}}}, \bibinfo {author} {\bibfnamefont {J.}~\bibnamefont
  {{Bautista}}}, \bibinfo {author} {\bibfnamefont {M.}~\bibnamefont
  {{Blomqvist}}}, \bibinfo {author} {\bibfnamefont {N.}~\bibnamefont
  {{Busca}}}, \bibinfo {author} {\bibfnamefont {K.}~\bibnamefont {{Dawson}}},
  \bibinfo {author} {\bibfnamefont {T.}~\bibnamefont {{Etourneau}}}, \bibinfo
  {author} {\bibfnamefont {A.}~\bibnamefont {{Font-Ribera}}}, \bibinfo {author}
  {\bibfnamefont {Y.}~\bibnamefont {{Lee}}}, \bibinfo {author} {\bibfnamefont
  {H.}~\bibnamefont {{du Mas des Bourboux}}}, \bibinfo {author} {\bibfnamefont
  {M.}~\bibnamefont {{Pieri}}}, \bibinfo {author} {\bibfnamefont
  {J.}~\bibnamefont {{Rich}}}, \bibinfo {author} {\bibfnamefont
  {G.}~\bibnamefont {{Rossi}}}, \bibinfo {author} {\bibfnamefont
  {D.}~\bibnamefont {{Schneider}}}, \ and\ \bibinfo {author} {\bibfnamefont
  {A.}~\bibnamefont {{Slosar}}},\ }\href {\doibase
  10.1088/1475-7516/2019/07/017} {\bibfield  {journal} {\bibinfo  {journal}
  {\jcap}\ }\textbf {\bibinfo {volume} {2019}},\ \bibinfo {eid} {017} (\bibinfo
  {year} {2019})},\ \Eprint {http://arxiv.org/abs/1812.03554} {arXiv:1812.03554
  [astro-ph.CO]} \BibitemShut {NoStop}%
\bibitem [{\citenamefont {{Palanque-Delabrouille}}\ \emph
  {et~al.}(2020)\citenamefont {{Palanque-Delabrouille}}, \citenamefont
  {{Y{\`e}che}}, \citenamefont {{Sch{\"o}neberg}}, \citenamefont
  {{Lesgourgues}}, \citenamefont {{Walther}}, \citenamefont {{Chabanier}},\
  and\ \citenamefont {{Armengaud}}}]{palanque-delabrouille20a}%
  \BibitemOpen
  \bibfield  {author} {\bibinfo {author} {\bibfnamefont {N.}~\bibnamefont
  {{Palanque-Delabrouille}}}, \bibinfo {author} {\bibfnamefont
  {C.}~\bibnamefont {{Y{\`e}che}}}, \bibinfo {author} {\bibfnamefont
  {N.}~\bibnamefont {{Sch{\"o}neberg}}}, \bibinfo {author} {\bibfnamefont
  {J.}~\bibnamefont {{Lesgourgues}}}, \bibinfo {author} {\bibfnamefont
  {M.}~\bibnamefont {{Walther}}}, \bibinfo {author} {\bibfnamefont
  {S.}~\bibnamefont {{Chabanier}}}, \ and\ \bibinfo {author} {\bibfnamefont
  {E.}~\bibnamefont {{Armengaud}}},\ }\href {\doibase
  10.1088/1475-7516/2020/04/038} {\bibfield  {journal} {\bibinfo  {journal}
  {\jcap}\ }\textbf {\bibinfo {volume} {2020}},\ \bibinfo {eid} {038} (\bibinfo
  {year} {2020})},\ \Eprint {http://arxiv.org/abs/1911.09073} {arXiv:1911.09073
  [astro-ph.CO]} \BibitemShut {NoStop}%
\bibitem [{\citenamefont {{Castorina}}\ \emph {et~al.}(2019)\citenamefont
  {{Castorina}}, \citenamefont {{Hand}}, \citenamefont {{Seljak}},
  \citenamefont {{Beutler}}, \citenamefont {{Chuang}}, \citenamefont {{Zhao}},
  \citenamefont {{Gil-Mar{\'\i}n}}, \citenamefont {{Percival}}, \citenamefont
  {{Ross}}, \citenamefont {{Choi}}, \citenamefont {{Dawson}}, \citenamefont
  {{de la Macorra}}, \citenamefont {{Rossi}}, \citenamefont {{Ruggeri}},
  \citenamefont {{Schneider}},\ and\ \citenamefont {{Zhao}}}]{castorina19a}%
  \BibitemOpen
  \bibfield  {author} {\bibinfo {author} {\bibfnamefont {E.}~\bibnamefont
  {{Castorina}}}, \bibinfo {author} {\bibfnamefont {N.}~\bibnamefont {{Hand}}},
  \bibinfo {author} {\bibfnamefont {U.}~\bibnamefont {{Seljak}}}, \bibinfo
  {author} {\bibfnamefont {F.}~\bibnamefont {{Beutler}}}, \bibinfo {author}
  {\bibfnamefont {C.-H.}\ \bibnamefont {{Chuang}}}, \bibinfo {author}
  {\bibfnamefont {C.}~\bibnamefont {{Zhao}}}, \bibinfo {author} {\bibfnamefont
  {H.}~\bibnamefont {{Gil-Mar{\'\i}n}}}, \bibinfo {author} {\bibfnamefont
  {W.~J.}\ \bibnamefont {{Percival}}}, \bibinfo {author} {\bibfnamefont
  {A.~J.}\ \bibnamefont {{Ross}}}, \bibinfo {author} {\bibfnamefont {P.~D.}\
  \bibnamefont {{Choi}}}, \bibinfo {author} {\bibfnamefont {K.}~\bibnamefont
  {{Dawson}}}, \bibinfo {author} {\bibfnamefont {A.}~\bibnamefont {{de la
  Macorra}}}, \bibinfo {author} {\bibfnamefont {G.}~\bibnamefont {{Rossi}}},
  \bibinfo {author} {\bibfnamefont {R.}~\bibnamefont {{Ruggeri}}}, \bibinfo
  {author} {\bibfnamefont {D.}~\bibnamefont {{Schneider}}}, \ and\ \bibinfo
  {author} {\bibfnamefont {G.-B.}\ \bibnamefont {{Zhao}}},\ }\href {\doibase
  10.1088/1475-7516/2019/09/010} {\bibfield  {journal} {\bibinfo  {journal}
  {\jcap}\ }\textbf {\bibinfo {volume} {2019}},\ \bibinfo {eid} {010} (\bibinfo
  {year} {2019})},\ \Eprint {http://arxiv.org/abs/1904.08859} {arXiv:1904.08859
  [astro-ph.CO]} \BibitemShut {NoStop}%
\bibitem [{\citenamefont {{McDonald}}\ and\ \citenamefont
  {{Seljak}}(2009)}]{mcdonald09a}%
  \BibitemOpen
  \bibfield  {author} {\bibinfo {author} {\bibfnamefont {P.}~\bibnamefont
  {{McDonald}}}\ and\ \bibinfo {author} {\bibfnamefont {U.}~\bibnamefont
  {{Seljak}}},\ }\href {\doibase 10.1088/1475-7516/2009/10/007} {\bibfield
  {journal} {\bibinfo  {journal} {\jcap}\ }\textbf {\bibinfo {volume} {10}},\
  \bibinfo {eid} {007} (\bibinfo {year} {2009})},\ \Eprint
  {http://arxiv.org/abs/0810.0323} {arXiv:0810.0323} \BibitemShut {NoStop}%
\bibitem [{\citenamefont {{Seljak}}(2009)}]{seljak09a}%
  \BibitemOpen
  \bibfield  {author} {\bibinfo {author} {\bibfnamefont {U.}~\bibnamefont
  {{Seljak}}},\ }\href {\doibase 10.1103/PhysRevLett.102.021302} {\bibfield
  {journal} {\bibinfo  {journal} {Physical Review Letters}\ }\textbf {\bibinfo
  {volume} {102}},\ \bibinfo {eid} {021302} (\bibinfo {year} {2009})},\ \Eprint
  {http://arxiv.org/abs/0807.1770} {arXiv:0807.1770} \BibitemShut {NoStop}%
\bibitem [{\citenamefont {{Zhao}}\ \emph {et~al.}(2016)\citenamefont {{Zhao}},
  \citenamefont {{Wang}}, \citenamefont {{Ross}}, \citenamefont {{Shandera}},
  \citenamefont {{Percival}}, \citenamefont {{Dawson}}, \citenamefont
  {{Kneib}}, \citenamefont {{Myers}}, \citenamefont {{Brownstein}},
  \citenamefont {{Comparat}}, \citenamefont {{Delubac}}, \citenamefont {{Gao}},
  \citenamefont {{Hojjati}}, \citenamefont {{Koyama}}, \citenamefont
  {{McBride}}, \citenamefont {{Meza}}, \citenamefont {{Newman}}, \citenamefont
  {{Palanque-Delabrouille}}, \citenamefont {{Pogosian}}, \citenamefont
  {{Prada}}, \citenamefont {{Rossi}}, \citenamefont {{Schneider}},
  \citenamefont {{Seo}}, \citenamefont {{Tao}}, \citenamefont {{Wang}},
  \citenamefont {{Y{\`e}che}}, \citenamefont {{Zhang}}, \citenamefont
  {{Zhang}}, \citenamefont {{Zhou}}, \citenamefont {{Zhu}},\ and\ \citenamefont
  {{Zou}}}]{zhao16a}%
  \BibitemOpen
  \bibfield  {author} {\bibinfo {author} {\bibfnamefont {G.-B.}\ \bibnamefont
  {{Zhao}}}, \bibinfo {author} {\bibfnamefont {Y.}~\bibnamefont {{Wang}}},
  \bibinfo {author} {\bibfnamefont {A.~J.}\ \bibnamefont {{Ross}}}, \bibinfo
  {author} {\bibfnamefont {S.}~\bibnamefont {{Shandera}}}, \bibinfo {author}
  {\bibfnamefont {W.~J.}\ \bibnamefont {{Percival}}}, \bibinfo {author}
  {\bibfnamefont {K.~S.}\ \bibnamefont {{Dawson}}}, \bibinfo {author}
  {\bibfnamefont {J.-P.}\ \bibnamefont {{Kneib}}}, \bibinfo {author}
  {\bibfnamefont {A.~D.}\ \bibnamefont {{Myers}}}, \bibinfo {author}
  {\bibfnamefont {J.~R.}\ \bibnamefont {{Brownstein}}}, \bibinfo {author}
  {\bibfnamefont {J.}~\bibnamefont {{Comparat}}}, \bibinfo {author}
  {\bibfnamefont {T.}~\bibnamefont {{Delubac}}}, \bibinfo {author}
  {\bibfnamefont {P.}~\bibnamefont {{Gao}}}, \bibinfo {author} {\bibfnamefont
  {A.}~\bibnamefont {{Hojjati}}}, \bibinfo {author} {\bibfnamefont
  {K.}~\bibnamefont {{Koyama}}}, \bibinfo {author} {\bibfnamefont {C.~K.}\
  \bibnamefont {{McBride}}}, \bibinfo {author} {\bibfnamefont {A.}~\bibnamefont
  {{Meza}}}, \bibinfo {author} {\bibfnamefont {J.~A.}\ \bibnamefont
  {{Newman}}}, \bibinfo {author} {\bibfnamefont {N.}~\bibnamefont
  {{Palanque-Delabrouille}}}, \bibinfo {author} {\bibfnamefont
  {L.}~\bibnamefont {{Pogosian}}}, \bibinfo {author} {\bibfnamefont
  {F.}~\bibnamefont {{Prada}}}, \bibinfo {author} {\bibfnamefont
  {G.}~\bibnamefont {{Rossi}}}, \bibinfo {author} {\bibfnamefont {D.~P.}\
  \bibnamefont {{Schneider}}}, \bibinfo {author} {\bibfnamefont {H.-J.}\
  \bibnamefont {{Seo}}}, \bibinfo {author} {\bibfnamefont {C.}~\bibnamefont
  {{Tao}}}, \bibinfo {author} {\bibfnamefont {D.}~\bibnamefont {{Wang}}},
  \bibinfo {author} {\bibfnamefont {C.}~\bibnamefont {{Y{\`e}che}}}, \bibinfo
  {author} {\bibfnamefont {H.}~\bibnamefont {{Zhang}}}, \bibinfo {author}
  {\bibfnamefont {Y.}~\bibnamefont {{Zhang}}}, \bibinfo {author} {\bibfnamefont
  {X.}~\bibnamefont {{Zhou}}}, \bibinfo {author} {\bibfnamefont
  {F.}~\bibnamefont {{Zhu}}}, \ and\ \bibinfo {author} {\bibfnamefont
  {H.}~\bibnamefont {{Zou}}},\ }\href {\doibase 10.1093/mnras/stw135}
  {\bibfield  {journal} {\bibinfo  {journal} {\mnras}\ }\textbf {\bibinfo
  {volume} {457}},\ \bibinfo {pages} {2377} (\bibinfo {year} {2016})},\ \Eprint
  {http://arxiv.org/abs/1510.08216} {arXiv:1510.08216} \BibitemShut {NoStop}%
\bibitem [{\citenamefont {{Alam}}\ \emph
  {et~al.}(2020{\natexlab{b}})\citenamefont {{Alam}}, \citenamefont
  {{Peacock}}, \citenamefont {{Kraljic}}, \citenamefont {{Ross}},\ and\
  \citenamefont {{Comparat}}}]{alam19a}%
  \BibitemOpen
  \bibfield  {author} {\bibinfo {author} {\bibfnamefont {S.}~\bibnamefont
  {{Alam}}}, \bibinfo {author} {\bibfnamefont {J.~A.}\ \bibnamefont
  {{Peacock}}}, \bibinfo {author} {\bibfnamefont {K.}~\bibnamefont
  {{Kraljic}}}, \bibinfo {author} {\bibfnamefont {A.~J.}\ \bibnamefont
  {{Ross}}}, \ and\ \bibinfo {author} {\bibfnamefont {J.}~\bibnamefont
  {{Comparat}}},\ }\href {\doibase 10.1093/mnras/staa1956} {\bibfield
  {journal} {\bibinfo  {journal} {\mnras}\ } (\bibinfo {year}
  {2020}{\natexlab{b}}),\ 10.1093/mnras/staa1956},\ \Eprint
  {http://arxiv.org/abs/1910.05095} {arXiv:1910.05095 [astro-ph.CO]}
  \BibitemShut {NoStop}%
\bibitem [{\citenamefont {{Wang}}\ \emph {et~al.}(2020)\citenamefont {{Wang}},
  \citenamefont {{Zhao}}, \citenamefont {{Zhao}}, \citenamefont {{Philcox}},
  \citenamefont {{Alam}}, \citenamefont {{Tamone}}, \citenamefont {{de
  Mattia}}, \citenamefont {{Ross}}, \citenamefont {{Raichoor}}, \citenamefont
  {{Burtin}}, \citenamefont {{Paviot}}, \citenamefont {{de la Torre}},
  \citenamefont {{Percival}}, \citenamefont {{Dawson}}, \citenamefont
  {{Gil-Mar{\'\i}n}}, \citenamefont {{Bautista}}, \citenamefont {{Hou}},
  \citenamefont {{Koyama}}, \citenamefont {{Peacock}}, \citenamefont
  {{Ruhlmann-Kleider}}, \citenamefont {{Bourboux}}, \citenamefont {{Chuang}},
  \citenamefont {{Comparat}}, \citenamefont {{Escoffier}}, \citenamefont
  {{Kneib}}, \citenamefont {{Mueller}}, \citenamefont {{Newman}}, \citenamefont
  {{Rossi}}, \citenamefont {{Shafieloo}},\ and\ \citenamefont
  {{Schneider}}}]{wang20a}%
  \BibitemOpen
  \bibfield  {author} {\bibinfo {author} {\bibfnamefont {Y.}~\bibnamefont
  {{Wang}}}, \bibinfo {author} {\bibfnamefont {G.-B.}\ \bibnamefont {{Zhao}}},
  \bibinfo {author} {\bibfnamefont {C.}~\bibnamefont {{Zhao}}}, \bibinfo
  {author} {\bibfnamefont {O.~H.~E.}\ \bibnamefont {{Philcox}}}, \bibinfo
  {author} {\bibfnamefont {S.}~\bibnamefont {{Alam}}}, \bibinfo {author}
  {\bibfnamefont {A.}~\bibnamefont {{Tamone}}}, \bibinfo {author}
  {\bibfnamefont {A.}~\bibnamefont {{de Mattia}}}, \bibinfo {author}
  {\bibfnamefont {A.~J.}\ \bibnamefont {{Ross}}}, \bibinfo {author}
  {\bibfnamefont {A.}~\bibnamefont {{Raichoor}}}, \bibinfo {author}
  {\bibfnamefont {E.}~\bibnamefont {{Burtin}}}, \bibinfo {author}
  {\bibfnamefont {R.}~\bibnamefont {{Paviot}}}, \bibinfo {author}
  {\bibfnamefont {S.}~\bibnamefont {{de la Torre}}}, \bibinfo {author}
  {\bibfnamefont {W.~J.}\ \bibnamefont {{Percival}}}, \bibinfo {author}
  {\bibfnamefont {K.~S.}\ \bibnamefont {{Dawson}}}, \bibinfo {author}
  {\bibfnamefont {H.}~\bibnamefont {{Gil-Mar{\'\i}n}}}, \bibinfo {author}
  {\bibfnamefont {J.~E.}\ \bibnamefont {{Bautista}}}, \bibinfo {author}
  {\bibfnamefont {J.}~\bibnamefont {{Hou}}}, \bibinfo {author} {\bibfnamefont
  {K.}~\bibnamefont {{Koyama}}}, \bibinfo {author} {\bibfnamefont {J.~A.}\
  \bibnamefont {{Peacock}}}, \bibinfo {author} {\bibfnamefont {V.}~\bibnamefont
  {{Ruhlmann-Kleider}}}, \bibinfo {author} {\bibfnamefont {H.~d. M.~d.}\
  \bibnamefont {{Bourboux}}}, \bibinfo {author} {\bibfnamefont {C.-H.}\
  \bibnamefont {{Chuang}}}, \bibinfo {author} {\bibfnamefont {J.}~\bibnamefont
  {{Comparat}}}, \bibinfo {author} {\bibfnamefont {S.}~\bibnamefont
  {{Escoffier}}}, \bibinfo {author} {\bibfnamefont {J.-P.}\ \bibnamefont
  {{Kneib}}}, \bibinfo {author} {\bibfnamefont {E.-M.}\ \bibnamefont
  {{Mueller}}}, \bibinfo {author} {\bibfnamefont {J.~A.}\ \bibnamefont
  {{Newman}}}, \bibinfo {author} {\bibfnamefont {G.}~\bibnamefont {{Rossi}}},
  \bibinfo {author} {\bibfnamefont {A.}~\bibnamefont {{Shafieloo}}}, \ and\
  \bibinfo {author} {\bibfnamefont {D.~P.}\ \bibnamefont {{Schneider}}},\
  }\href {\doibase 10.1093/mnras/staa2593} {\bibfield  {journal} {\bibinfo
  {journal} {\mnras}\ }\textbf {\bibinfo {volume} {498}},\ \bibinfo {pages}
  {3470} (\bibinfo {year} {2020})},\ \Eprint {http://arxiv.org/abs/2007.09010}
  {arXiv:2007.09010 [astro-ph.CO]} \BibitemShut {NoStop}%
\bibitem [{\citenamefont {{du Mas des Bourboux}}\ \emph
  {et~al.}(2019)\citenamefont {{du Mas des Bourboux}}, \citenamefont
  {{Dawson}}, \citenamefont {{Busca}}, \citenamefont {{Blomqvist}},
  \citenamefont {{de Sainte Agathe}}, \citenamefont {{Balland}}, \citenamefont
  {{Bautista}}, \citenamefont {{Guy}}, \citenamefont {{Kamble}}, \citenamefont
  {{Myers}}, \citenamefont {{P{\'e}rez-R{\`a}fols}}, \citenamefont {{Pieri}},
  \citenamefont {{Rich}}, \citenamefont {{Schneider}},\ and\ \citenamefont
  {{Slosar}}}]{masdesbourboux19a}%
  \BibitemOpen
  \bibfield  {author} {\bibinfo {author} {\bibfnamefont {H.}~\bibnamefont {{du
  Mas des Bourboux}}}, \bibinfo {author} {\bibfnamefont {K.~S.}\ \bibnamefont
  {{Dawson}}}, \bibinfo {author} {\bibfnamefont {N.~G.}\ \bibnamefont
  {{Busca}}}, \bibinfo {author} {\bibfnamefont {M.}~\bibnamefont
  {{Blomqvist}}}, \bibinfo {author} {\bibfnamefont {V.}~\bibnamefont {{de
  Sainte Agathe}}}, \bibinfo {author} {\bibfnamefont {C.}~\bibnamefont
  {{Balland}}}, \bibinfo {author} {\bibfnamefont {J.~E.}\ \bibnamefont
  {{Bautista}}}, \bibinfo {author} {\bibfnamefont {J.}~\bibnamefont {{Guy}}},
  \bibinfo {author} {\bibfnamefont {V.}~\bibnamefont {{Kamble}}}, \bibinfo
  {author} {\bibfnamefont {A.~D.}\ \bibnamefont {{Myers}}}, \bibinfo {author}
  {\bibfnamefont {I.}~\bibnamefont {{P{\'e}rez-R{\`a}fols}}}, \bibinfo {author}
  {\bibfnamefont {M.~M.}\ \bibnamefont {{Pieri}}}, \bibinfo {author}
  {\bibfnamefont {J.}~\bibnamefont {{Rich}}}, \bibinfo {author} {\bibfnamefont
  {D.~P.}\ \bibnamefont {{Schneider}}}, \ and\ \bibinfo {author} {\bibfnamefont
  {A.}~\bibnamefont {{Slosar}}},\ }\href {\doibase 10.3847/1538-4357/ab1d49}
  {\bibfield  {journal} {\bibinfo  {journal} {\apj}\ }\textbf {\bibinfo
  {volume} {878}},\ \bibinfo {eid} {47} (\bibinfo {year} {2019})},\ \Eprint
  {http://arxiv.org/abs/1901.01950} {arXiv:1901.01950 [astro-ph.CO]}
  \BibitemShut {NoStop}%
\bibitem [{\citenamefont {{Blomqvist}}\ \emph {et~al.}(2019)\citenamefont
  {{Blomqvist}}, \citenamefont {{du Mas des Bourboux}}, \citenamefont
  {{Busca}}, \citenamefont {{de Sainte Agathe}}, \citenamefont {{Rich}},
  \citenamefont {{Balland}}, \citenamefont {{Bautista}}, \citenamefont
  {{Dawson}}, \citenamefont {{Font-Ribera}}, \citenamefont {{Guy}},
  \citenamefont {{Le Goff}}, \citenamefont {{Palanque-Delabrouille}},
  \citenamefont {{Percival}}, \citenamefont {{P{\'e}rez-R{\`a}fols}},
  \citenamefont {{Pieri}}, \citenamefont {{Schneider}}, \citenamefont
  {{Slosar}},\ and\ \citenamefont {{Y{\`e}che}}}]{blomqvist18a}%
  \BibitemOpen
  \bibfield  {author} {\bibinfo {author} {\bibfnamefont {M.}~\bibnamefont
  {{Blomqvist}}}, \bibinfo {author} {\bibfnamefont {H.}~\bibnamefont {{du Mas
  des Bourboux}}}, \bibinfo {author} {\bibfnamefont {N.~G.}\ \bibnamefont
  {{Busca}}}, \bibinfo {author} {\bibfnamefont {V.}~\bibnamefont {{de Sainte
  Agathe}}}, \bibinfo {author} {\bibfnamefont {J.}~\bibnamefont {{Rich}}},
  \bibinfo {author} {\bibfnamefont {C.}~\bibnamefont {{Balland}}}, \bibinfo
  {author} {\bibfnamefont {J.~E.}\ \bibnamefont {{Bautista}}}, \bibinfo
  {author} {\bibfnamefont {K.}~\bibnamefont {{Dawson}}}, \bibinfo {author}
  {\bibfnamefont {A.}~\bibnamefont {{Font-Ribera}}}, \bibinfo {author}
  {\bibfnamefont {J.}~\bibnamefont {{Guy}}}, \bibinfo {author} {\bibfnamefont
  {J.-M.}\ \bibnamefont {{Le Goff}}}, \bibinfo {author} {\bibfnamefont
  {N.}~\bibnamefont {{Palanque-Delabrouille}}}, \bibinfo {author}
  {\bibfnamefont {W.~J.}\ \bibnamefont {{Percival}}}, \bibinfo {author}
  {\bibfnamefont {I.}~\bibnamefont {{P{\'e}rez-R{\`a}fols}}}, \bibinfo {author}
  {\bibfnamefont {M.~M.}\ \bibnamefont {{Pieri}}}, \bibinfo {author}
  {\bibfnamefont {D.~P.}\ \bibnamefont {{Schneider}}}, \bibinfo {author}
  {\bibfnamefont {A.}~\bibnamefont {{Slosar}}}, \ and\ \bibinfo {author}
  {\bibfnamefont {C.}~\bibnamefont {{Y{\`e}che}}},\ }\href {\doibase
  10.1051/0004-6361/201935641} {\bibfield  {journal} {\bibinfo  {journal}
  {\aap}\ }\textbf {\bibinfo {volume} {629}},\ \bibinfo {eid} {A86} (\bibinfo
  {year} {2019})},\ \Eprint {http://arxiv.org/abs/1904.03430} {arXiv:1904.03430
  [astro-ph.CO]} \BibitemShut {NoStop}%
\bibitem [{\citenamefont {{Zarrouk}}\ \emph {et~al.}(2020)\citenamefont
  {{Zarrouk}}, \citenamefont {{Rezaie}}, \citenamefont {{Raichoor}},
  \citenamefont {{Ross}}, \citenamefont {{Alam}}, \citenamefont {{Blum}},
  \citenamefont {{Brookes}}, \citenamefont {{Chuang}}, \citenamefont {{Cole}},
  \citenamefont {{Dawson}}, \citenamefont {{Eisenstein}}, \citenamefont
  {{Kehoe}}, \citenamefont {{Landriau}}, \citenamefont {{Moustakas}},
  \citenamefont {{Myers}}, \citenamefont {{Norberg}}, \citenamefont
  {{Percival}}, \citenamefont {{Prada}}, \citenamefont {{Schubnell}},
  \citenamefont {{Seo}}, \citenamefont {{Tarl{\'e}}},\ and\ \citenamefont
  {{Zhao}}}]{zarrouk20a}%
  \BibitemOpen
  \bibfield  {author} {\bibinfo {author} {\bibfnamefont {P.}~\bibnamefont
  {{Zarrouk}}}, \bibinfo {author} {\bibfnamefont {M.}~\bibnamefont {{Rezaie}}},
  \bibinfo {author} {\bibfnamefont {A.}~\bibnamefont {{Raichoor}}}, \bibinfo
  {author} {\bibfnamefont {A.~J.}\ \bibnamefont {{Ross}}}, \bibinfo {author}
  {\bibfnamefont {S.}~\bibnamefont {{Alam}}}, \bibinfo {author} {\bibfnamefont
  {R.}~\bibnamefont {{Blum}}}, \bibinfo {author} {\bibfnamefont
  {D.}~\bibnamefont {{Brookes}}}, \bibinfo {author} {\bibfnamefont {C.-H.}\
  \bibnamefont {{Chuang}}}, \bibinfo {author} {\bibfnamefont {S.}~\bibnamefont
  {{Cole}}}, \bibinfo {author} {\bibfnamefont {K.~S.}\ \bibnamefont
  {{Dawson}}}, \bibinfo {author} {\bibfnamefont {D.~J.}\ \bibnamefont
  {{Eisenstein}}}, \bibinfo {author} {\bibfnamefont {R.}~\bibnamefont
  {{Kehoe}}}, \bibinfo {author} {\bibfnamefont {M.}~\bibnamefont {{Landriau}}},
  \bibinfo {author} {\bibfnamefont {J.}~\bibnamefont {{Moustakas}}}, \bibinfo
  {author} {\bibfnamefont {A.~D.}\ \bibnamefont {{Myers}}}, \bibinfo {author}
  {\bibfnamefont {P.}~\bibnamefont {{Norberg}}}, \bibinfo {author}
  {\bibfnamefont {W.~J.}\ \bibnamefont {{Percival}}}, \bibinfo {author}
  {\bibfnamefont {F.}~\bibnamefont {{Prada}}}, \bibinfo {author} {\bibfnamefont
  {M.}~\bibnamefont {{Schubnell}}}, \bibinfo {author} {\bibfnamefont {H.-J.}\
  \bibnamefont {{Seo}}}, \bibinfo {author} {\bibfnamefont {G.}~\bibnamefont
  {{Tarl{\'e}}}}, \ and\ \bibinfo {author} {\bibfnamefont {C.}~\bibnamefont
  {{Zhao}}},\ }\href@noop {} {\bibfield  {journal} {\bibinfo  {journal} {arXiv
  e-prints}\ ,\ \bibinfo {eid} {arXiv:2009.02308}} (\bibinfo {year} {2020})},\
  \Eprint {http://arxiv.org/abs/2009.02308} {arXiv:2009.02308 [astro-ph.CO]}
  \BibitemShut {NoStop}%
\bibitem [{\citenamefont {{Dey}}\ \emph {et~al.}(2019)\citenamefont {{Dey}},
  \citenamefont {{Schlegel}}, \citenamefont {{Lang}}, \citenamefont {{Blum}},
  \citenamefont {{Burleigh}}, \citenamefont {{Fan}}, \citenamefont {{Findlay}},
  \citenamefont {{Finkbeiner}}, \citenamefont {{Herrera}}, \citenamefont
  {{Juneau}}, \citenamefont {{Landriau}}, \citenamefont {{Levi}}, \citenamefont
  {{McGreer}}, \citenamefont {{Meisner}}, \citenamefont {{Myers}},
  \citenamefont {{Moustakas}}, \citenamefont {{Nugent}}, \citenamefont
  {{Patej}}, \citenamefont {{Schlafly}}, \citenamefont {{Walker}},
  \citenamefont {{Valdes}}, \citenamefont {{Weaver}}, \citenamefont
  {{Y{\`e}che}}, \citenamefont {{Zou}}, \citenamefont {{Zhou}}, \citenamefont
  {{Abareshi}}, \citenamefont {{Abbott}}, \citenamefont {{Abolfathi}},
  \citenamefont {{Aguilera}}, \citenamefont {{Alam}}, \citenamefont {{Allen}},
  \citenamefont {{Alvarez}}, \citenamefont {{Annis}}, \citenamefont
  {{Ansarinejad}}, \citenamefont {{Aubert}}, \citenamefont {{Beechert}},
  \citenamefont {{Bell}}, \citenamefont {{BenZvi}}, \citenamefont {{Beutler}},
  \citenamefont {{Bielby}}, \citenamefont {{Bolton}}, \citenamefont
  {{Brice{\~n}o}}, \citenamefont {{Buckley-Geer}}, \citenamefont {{Butler}},
  \citenamefont {{Calamida}}, \citenamefont {{Carlberg}}, \citenamefont
  {{Carter}}, \citenamefont {{Casas}}, \citenamefont {{Castander}},
  \citenamefont {{Choi}}, \citenamefont {{Comparat}}, \citenamefont
  {{Cukanovaite}}, \citenamefont {{Delubac}}, \citenamefont {{DeVries}},
  \citenamefont {{Dey}}, \citenamefont {{Dhungana}}, \citenamefont
  {{Dickinson}}, \citenamefont {{Ding}}, \citenamefont {{Donaldson}},
  \citenamefont {{Duan}}, \citenamefont {{Duckworth}}, \citenamefont
  {{Eftekharzadeh}}, \citenamefont {{Eisenstein}}, \citenamefont {{Etourneau}},
  \citenamefont {{Fagrelius}}, \citenamefont {{Farihi}}, \citenamefont
  {{Fitzpatrick}}, \citenamefont {{Font-Ribera}}, \citenamefont {{Fulmer}},
  \citenamefont {{G{\"a}nsicke}}, \citenamefont {{Gaztanaga}}, \citenamefont
  {{George}}, \citenamefont {{Gerdes}}, \citenamefont {{Gontcho}},
  \citenamefont {{Gorgoni}}, \citenamefont {{Green}}, \citenamefont {{Guy}},
  \citenamefont {{Harmer}}, \citenamefont {{Hernand ez}}, \citenamefont
  {{Honscheid}}, \citenamefont {{Huang}}, \citenamefont {{James}},
  \citenamefont {{Jannuzi}}, \citenamefont {{Jiang}}, \citenamefont {{Joyce}},
  \citenamefont {{Karcher}}, \citenamefont {{Karkar}}, \citenamefont {{Kehoe}},
  \citenamefont {{Kneib}}, \citenamefont {{Kueter-Young}}, \citenamefont
  {{Lan}}, \citenamefont {{Lauer}}, \citenamefont {{Le Guillou}}, \citenamefont
  {{Le Van Suu}}, \citenamefont {{Lee}}, \citenamefont {{Lesser}},
  \citenamefont {{Perreault Levasseur}}, \citenamefont {{Li}}, \citenamefont
  {{Mann}}, \citenamefont {{Marshall}}, \citenamefont
  {{Mart{\'\i}nez-V{\'a}zquez}}, \citenamefont {{Martini}}, \citenamefont {{du
  Mas des Bourboux}}, \citenamefont {{McManus}}, \citenamefont {{Meier}},
  \citenamefont {{M{\'e}nard}}, \citenamefont {{Metcalfe}}, \citenamefont
  {{Mu{\~n}oz-Guti{\'e}rrez}}, \citenamefont {{Najita}}, \citenamefont
  {{Napier}}, \citenamefont {{Narayan}}, \citenamefont {{Newman}},
  \citenamefont {{Nie}}, \citenamefont {{Nord}}, \citenamefont {{Norman}},
  \citenamefont {{Olsen}}, \citenamefont {{Paat}}, \citenamefont
  {{Palanque-Delabrouille}}, \citenamefont {{Peng}}, \citenamefont {{Poppett}},
  \citenamefont {{Poremba}}, \citenamefont {{Prakash}}, \citenamefont
  {{Rabinowitz}}, \citenamefont {{Raichoor}}, \citenamefont {{Rezaie}},
  \citenamefont {{Robertson}}, \citenamefont {{Roe}}, \citenamefont {{Ross}},
  \citenamefont {{Ross}}, \citenamefont {{Rudnick}}, \citenamefont
  {{Safonova}}, \citenamefont {{Saha}}, \citenamefont {{S{\'a}nchez}},
  \citenamefont {{Savary}}, \citenamefont {{Schweiker}}, \citenamefont
  {{Scott}}, \citenamefont {{Seo}}, \citenamefont {{Shan}}, \citenamefont
  {{Silva}}, \citenamefont {{Slepian}}, \citenamefont {{Soto}}, \citenamefont
  {{Sprayberry}}, \citenamefont {{Staten}}, \citenamefont {{Stillman}},
  \citenamefont {{Stupak}}, \citenamefont {{Summers}}, \citenamefont {{Sien
  Tie}}, \citenamefont {{Tirado}}, \citenamefont {{Vargas-Maga{\~n}a}},
  \citenamefont {{Vivas}}, \citenamefont {{Wechsler}}, \citenamefont
  {{Williams}}, \citenamefont {{Yang}}, \citenamefont {{Yang}}, \citenamefont
  {{Yapici}}, \citenamefont {{Zaritsky}}, \citenamefont {{Zenteno}},
  \citenamefont {{Zhang}}, \citenamefont {{Zhang}}, \citenamefont {{Zhou}},\
  and\ \citenamefont {{Zhou}}}]{dey19a}%
  \BibitemOpen
  \bibfield  {author} {\bibinfo {author} {\bibfnamefont {A.}~\bibnamefont
  {{Dey}}}, \bibinfo {author} {\bibfnamefont {D.~J.}\ \bibnamefont
  {{Schlegel}}}, \bibinfo {author} {\bibfnamefont {D.}~\bibnamefont {{Lang}}},
  \bibinfo {author} {\bibfnamefont {R.}~\bibnamefont {{Blum}}}, \bibinfo
  {author} {\bibfnamefont {K.}~\bibnamefont {{Burleigh}}}, \bibinfo {author}
  {\bibfnamefont {X.}~\bibnamefont {{Fan}}}, \bibinfo {author} {\bibfnamefont
  {J.~R.}\ \bibnamefont {{Findlay}}}, \bibinfo {author} {\bibfnamefont
  {D.}~\bibnamefont {{Finkbeiner}}}, \bibinfo {author} {\bibfnamefont
  {D.}~\bibnamefont {{Herrera}}}, \bibinfo {author} {\bibfnamefont
  {S.}~\bibnamefont {{Juneau}}}, \bibinfo {author} {\bibfnamefont
  {M.}~\bibnamefont {{Landriau}}}, \bibinfo {author} {\bibfnamefont
  {M.}~\bibnamefont {{Levi}}}, \bibinfo {author} {\bibfnamefont
  {I.}~\bibnamefont {{McGreer}}}, \bibinfo {author} {\bibfnamefont
  {A.}~\bibnamefont {{Meisner}}}, \bibinfo {author} {\bibfnamefont {A.~D.}\
  \bibnamefont {{Myers}}}, \bibinfo {author} {\bibfnamefont {J.}~\bibnamefont
  {{Moustakas}}}, \bibinfo {author} {\bibfnamefont {P.}~\bibnamefont
  {{Nugent}}}, \bibinfo {author} {\bibfnamefont {A.}~\bibnamefont {{Patej}}},
  \bibinfo {author} {\bibfnamefont {E.~F.}\ \bibnamefont {{Schlafly}}},
  \bibinfo {author} {\bibfnamefont {A.~R.}\ \bibnamefont {{Walker}}}, \bibinfo
  {author} {\bibfnamefont {F.}~\bibnamefont {{Valdes}}}, \bibinfo {author}
  {\bibfnamefont {B.~A.}\ \bibnamefont {{Weaver}}}, \bibinfo {author}
  {\bibfnamefont {C.}~\bibnamefont {{Y{\`e}che}}}, \bibinfo {author}
  {\bibfnamefont {H.}~\bibnamefont {{Zou}}}, \bibinfo {author} {\bibfnamefont
  {X.}~\bibnamefont {{Zhou}}}, \bibinfo {author} {\bibfnamefont
  {B.}~\bibnamefont {{Abareshi}}}, \bibinfo {author} {\bibfnamefont {T.~M.~C.}\
  \bibnamefont {{Abbott}}}, \bibinfo {author} {\bibfnamefont {B.}~\bibnamefont
  {{Abolfathi}}}, \bibinfo {author} {\bibfnamefont {C.}~\bibnamefont
  {{Aguilera}}}, \bibinfo {author} {\bibfnamefont {S.}~\bibnamefont {{Alam}}},
  \bibinfo {author} {\bibfnamefont {L.}~\bibnamefont {{Allen}}}, \bibinfo
  {author} {\bibfnamefont {A.}~\bibnamefont {{Alvarez}}}, \bibinfo {author}
  {\bibfnamefont {J.}~\bibnamefont {{Annis}}}, \bibinfo {author} {\bibfnamefont
  {B.}~\bibnamefont {{Ansarinejad}}}, \bibinfo {author} {\bibfnamefont
  {M.}~\bibnamefont {{Aubert}}}, \bibinfo {author} {\bibfnamefont
  {J.}~\bibnamefont {{Beechert}}}, \bibinfo {author} {\bibfnamefont {E.~F.}\
  \bibnamefont {{Bell}}}, \bibinfo {author} {\bibfnamefont {S.~Y.}\
  \bibnamefont {{BenZvi}}}, \bibinfo {author} {\bibfnamefont {F.}~\bibnamefont
  {{Beutler}}}, \bibinfo {author} {\bibfnamefont {R.~M.}\ \bibnamefont
  {{Bielby}}}, \bibinfo {author} {\bibfnamefont {A.~S.}\ \bibnamefont
  {{Bolton}}}, \bibinfo {author} {\bibfnamefont {C.}~\bibnamefont
  {{Brice{\~n}o}}}, \bibinfo {author} {\bibfnamefont {E.~J.}\ \bibnamefont
  {{Buckley-Geer}}}, \bibinfo {author} {\bibfnamefont {K.}~\bibnamefont
  {{Butler}}}, \bibinfo {author} {\bibfnamefont {A.}~\bibnamefont
  {{Calamida}}}, \bibinfo {author} {\bibfnamefont {R.~G.}\ \bibnamefont
  {{Carlberg}}}, \bibinfo {author} {\bibfnamefont {P.}~\bibnamefont
  {{Carter}}}, \bibinfo {author} {\bibfnamefont {R.}~\bibnamefont {{Casas}}},
  \bibinfo {author} {\bibfnamefont {F.~J.}\ \bibnamefont {{Castander}}},
  \bibinfo {author} {\bibfnamefont {Y.}~\bibnamefont {{Choi}}}, \bibinfo
  {author} {\bibfnamefont {J.}~\bibnamefont {{Comparat}}}, \bibinfo {author}
  {\bibfnamefont {E.}~\bibnamefont {{Cukanovaite}}}, \bibinfo {author}
  {\bibfnamefont {T.}~\bibnamefont {{Delubac}}}, \bibinfo {author}
  {\bibfnamefont {K.}~\bibnamefont {{DeVries}}}, \bibinfo {author}
  {\bibfnamefont {S.}~\bibnamefont {{Dey}}}, \bibinfo {author} {\bibfnamefont
  {G.}~\bibnamefont {{Dhungana}}}, \bibinfo {author} {\bibfnamefont
  {M.}~\bibnamefont {{Dickinson}}}, \bibinfo {author} {\bibfnamefont
  {Z.}~\bibnamefont {{Ding}}}, \bibinfo {author} {\bibfnamefont {J.~B.}\
  \bibnamefont {{Donaldson}}}, \bibinfo {author} {\bibfnamefont
  {Y.}~\bibnamefont {{Duan}}}, \bibinfo {author} {\bibfnamefont {C.~J.}\
  \bibnamefont {{Duckworth}}}, \bibinfo {author} {\bibfnamefont
  {S.}~\bibnamefont {{Eftekharzadeh}}}, \bibinfo {author} {\bibfnamefont
  {D.~J.}\ \bibnamefont {{Eisenstein}}}, \bibinfo {author} {\bibfnamefont
  {T.}~\bibnamefont {{Etourneau}}}, \bibinfo {author} {\bibfnamefont {P.~A.}\
  \bibnamefont {{Fagrelius}}}, \bibinfo {author} {\bibfnamefont
  {J.}~\bibnamefont {{Farihi}}}, \bibinfo {author} {\bibfnamefont
  {M.}~\bibnamefont {{Fitzpatrick}}}, \bibinfo {author} {\bibfnamefont
  {A.}~\bibnamefont {{Font-Ribera}}}, \bibinfo {author} {\bibfnamefont
  {L.}~\bibnamefont {{Fulmer}}}, \bibinfo {author} {\bibfnamefont {B.~T.}\
  \bibnamefont {{G{\"a}nsicke}}}, \bibinfo {author} {\bibfnamefont
  {E.}~\bibnamefont {{Gaztanaga}}}, \bibinfo {author} {\bibfnamefont
  {K.}~\bibnamefont {{George}}}, \bibinfo {author} {\bibfnamefont {D.~W.}\
  \bibnamefont {{Gerdes}}}, \bibinfo {author} {\bibfnamefont {S.~G.~A.}\
  \bibnamefont {{Gontcho}}}, \bibinfo {author} {\bibfnamefont {C.}~\bibnamefont
  {{Gorgoni}}}, \bibinfo {author} {\bibfnamefont {G.}~\bibnamefont {{Green}}},
  \bibinfo {author} {\bibfnamefont {J.}~\bibnamefont {{Guy}}}, \bibinfo
  {author} {\bibfnamefont {D.}~\bibnamefont {{Harmer}}}, \bibinfo {author}
  {\bibfnamefont {M.}~\bibnamefont {{Hernand ez}}}, \bibinfo {author}
  {\bibfnamefont {K.}~\bibnamefont {{Honscheid}}}, \bibinfo {author}
  {\bibfnamefont {L.~W.}\ \bibnamefont {{Huang}}}, \bibinfo {author}
  {\bibfnamefont {D.~J.}\ \bibnamefont {{James}}}, \bibinfo {author}
  {\bibfnamefont {B.~T.}\ \bibnamefont {{Jannuzi}}}, \bibinfo {author}
  {\bibfnamefont {L.}~\bibnamefont {{Jiang}}}, \bibinfo {author} {\bibfnamefont
  {R.}~\bibnamefont {{Joyce}}}, \bibinfo {author} {\bibfnamefont
  {A.}~\bibnamefont {{Karcher}}}, \bibinfo {author} {\bibfnamefont
  {S.}~\bibnamefont {{Karkar}}}, \bibinfo {author} {\bibfnamefont
  {R.}~\bibnamefont {{Kehoe}}}, \bibinfo {author} {\bibfnamefont {J.-P.}\
  \bibnamefont {{Kneib}}}, \bibinfo {author} {\bibfnamefont {A.}~\bibnamefont
  {{Kueter-Young}}}, \bibinfo {author} {\bibfnamefont {T.-W.}\ \bibnamefont
  {{Lan}}}, \bibinfo {author} {\bibfnamefont {T.~R.}\ \bibnamefont {{Lauer}}},
  \bibinfo {author} {\bibfnamefont {L.}~\bibnamefont {{Le Guillou}}}, \bibinfo
  {author} {\bibfnamefont {A.}~\bibnamefont {{Le Van Suu}}}, \bibinfo {author}
  {\bibfnamefont {J.~H.}\ \bibnamefont {{Lee}}}, \bibinfo {author}
  {\bibfnamefont {M.}~\bibnamefont {{Lesser}}}, \bibinfo {author}
  {\bibfnamefont {L.}~\bibnamefont {{Perreault Levasseur}}}, \bibinfo {author}
  {\bibfnamefont {T.~S.}\ \bibnamefont {{Li}}}, \bibinfo {author}
  {\bibfnamefont {J.~L.}\ \bibnamefont {{Mann}}}, \bibinfo {author}
  {\bibfnamefont {R.}~\bibnamefont {{Marshall}}}, \bibinfo {author}
  {\bibfnamefont {C.~E.}\ \bibnamefont {{Mart{\'\i}nez-V{\'a}zquez}}}, \bibinfo
  {author} {\bibfnamefont {P.}~\bibnamefont {{Martini}}}, \bibinfo {author}
  {\bibfnamefont {H.}~\bibnamefont {{du Mas des Bourboux}}}, \bibinfo {author}
  {\bibfnamefont {S.}~\bibnamefont {{McManus}}}, \bibinfo {author}
  {\bibfnamefont {T.~G.}\ \bibnamefont {{Meier}}}, \bibinfo {author}
  {\bibfnamefont {B.}~\bibnamefont {{M{\'e}nard}}}, \bibinfo {author}
  {\bibfnamefont {N.}~\bibnamefont {{Metcalfe}}}, \bibinfo {author}
  {\bibfnamefont {A.}~\bibnamefont {{Mu{\~n}oz-Guti{\'e}rrez}}}, \bibinfo
  {author} {\bibfnamefont {J.}~\bibnamefont {{Najita}}}, \bibinfo {author}
  {\bibfnamefont {K.}~\bibnamefont {{Napier}}}, \bibinfo {author}
  {\bibfnamefont {G.}~\bibnamefont {{Narayan}}}, \bibinfo {author}
  {\bibfnamefont {J.~A.}\ \bibnamefont {{Newman}}}, \bibinfo {author}
  {\bibfnamefont {J.}~\bibnamefont {{Nie}}}, \bibinfo {author} {\bibfnamefont
  {B.}~\bibnamefont {{Nord}}}, \bibinfo {author} {\bibfnamefont {D.~J.}\
  \bibnamefont {{Norman}}}, \bibinfo {author} {\bibfnamefont {K.~A.~G.}\
  \bibnamefont {{Olsen}}}, \bibinfo {author} {\bibfnamefont {A.}~\bibnamefont
  {{Paat}}}, \bibinfo {author} {\bibfnamefont {N.}~\bibnamefont
  {{Palanque-Delabrouille}}}, \bibinfo {author} {\bibfnamefont
  {X.}~\bibnamefont {{Peng}}}, \bibinfo {author} {\bibfnamefont {C.~L.}\
  \bibnamefont {{Poppett}}}, \bibinfo {author} {\bibfnamefont {M.~R.}\
  \bibnamefont {{Poremba}}}, \bibinfo {author} {\bibfnamefont {A.}~\bibnamefont
  {{Prakash}}}, \bibinfo {author} {\bibfnamefont {D.}~\bibnamefont
  {{Rabinowitz}}}, \bibinfo {author} {\bibfnamefont {A.}~\bibnamefont
  {{Raichoor}}}, \bibinfo {author} {\bibfnamefont {M.}~\bibnamefont
  {{Rezaie}}}, \bibinfo {author} {\bibfnamefont {A.~N.}\ \bibnamefont
  {{Robertson}}}, \bibinfo {author} {\bibfnamefont {N.~A.}\ \bibnamefont
  {{Roe}}}, \bibinfo {author} {\bibfnamefont {A.~J.}\ \bibnamefont {{Ross}}},
  \bibinfo {author} {\bibfnamefont {N.~P.}\ \bibnamefont {{Ross}}}, \bibinfo
  {author} {\bibfnamefont {G.}~\bibnamefont {{Rudnick}}}, \bibinfo {author}
  {\bibfnamefont {S.}~\bibnamefont {{Safonova}}}, \bibinfo {author}
  {\bibfnamefont {A.}~\bibnamefont {{Saha}}}, \bibinfo {author} {\bibfnamefont
  {F.~J.}\ \bibnamefont {{S{\'a}nchez}}}, \bibinfo {author} {\bibfnamefont
  {E.}~\bibnamefont {{Savary}}}, \bibinfo {author} {\bibfnamefont
  {H.}~\bibnamefont {{Schweiker}}}, \bibinfo {author} {\bibfnamefont
  {A.}~\bibnamefont {{Scott}}}, \bibinfo {author} {\bibfnamefont {H.-J.}\
  \bibnamefont {{Seo}}}, \bibinfo {author} {\bibfnamefont {H.}~\bibnamefont
  {{Shan}}}, \bibinfo {author} {\bibfnamefont {D.~R.}\ \bibnamefont {{Silva}}},
  \bibinfo {author} {\bibfnamefont {Z.}~\bibnamefont {{Slepian}}}, \bibinfo
  {author} {\bibfnamefont {C.}~\bibnamefont {{Soto}}}, \bibinfo {author}
  {\bibfnamefont {D.}~\bibnamefont {{Sprayberry}}}, \bibinfo {author}
  {\bibfnamefont {R.}~\bibnamefont {{Staten}}}, \bibinfo {author}
  {\bibfnamefont {C.~M.}\ \bibnamefont {{Stillman}}}, \bibinfo {author}
  {\bibfnamefont {R.~J.}\ \bibnamefont {{Stupak}}}, \bibinfo {author}
  {\bibfnamefont {D.~L.}\ \bibnamefont {{Summers}}}, \bibinfo {author}
  {\bibfnamefont {S.}~\bibnamefont {{Sien Tie}}}, \bibinfo {author}
  {\bibfnamefont {H.}~\bibnamefont {{Tirado}}}, \bibinfo {author}
  {\bibfnamefont {M.}~\bibnamefont {{Vargas-Maga{\~n}a}}}, \bibinfo {author}
  {\bibfnamefont {A.~K.}\ \bibnamefont {{Vivas}}}, \bibinfo {author}
  {\bibfnamefont {R.~H.}\ \bibnamefont {{Wechsler}}}, \bibinfo {author}
  {\bibfnamefont {D.}~\bibnamefont {{Williams}}}, \bibinfo {author}
  {\bibfnamefont {J.}~\bibnamefont {{Yang}}}, \bibinfo {author} {\bibfnamefont
  {Q.}~\bibnamefont {{Yang}}}, \bibinfo {author} {\bibfnamefont
  {T.}~\bibnamefont {{Yapici}}}, \bibinfo {author} {\bibfnamefont
  {D.}~\bibnamefont {{Zaritsky}}}, \bibinfo {author} {\bibfnamefont
  {A.}~\bibnamefont {{Zenteno}}}, \bibinfo {author} {\bibfnamefont
  {K.}~\bibnamefont {{Zhang}}}, \bibinfo {author} {\bibfnamefont
  {T.}~\bibnamefont {{Zhang}}}, \bibinfo {author} {\bibfnamefont
  {R.}~\bibnamefont {{Zhou}}}, \ and\ \bibinfo {author} {\bibfnamefont
  {Z.}~\bibnamefont {{Zhou}}},\ }\href {\doibase 10.3847/1538-3881/ab089d}
  {\bibfield  {journal} {\bibinfo  {journal} {\aj}\ }\textbf {\bibinfo {volume}
  {157}},\ \bibinfo {eid} {168} (\bibinfo {year} {2019})},\ \Eprint
  {http://arxiv.org/abs/1804.08657} {arXiv:1804.08657 [astro-ph.IM]}
  \BibitemShut {NoStop}%
\bibitem [{\citenamefont {Hawken}\ \emph {et~al.}(2020)\citenamefont {Hawken},
  \citenamefont {Aubert}, \citenamefont {Pisani}, \citenamefont {Cousinou},
  \citenamefont {Escoffier}, \citenamefont {Nadathur}, \citenamefont {Rossi},\
  and\ \citenamefont {Schneider}}]{hawken19a}%
  \BibitemOpen
  \bibfield  {author} {\bibinfo {author} {\bibfnamefont {A.~J.}\ \bibnamefont
  {Hawken}}, \bibinfo {author} {\bibfnamefont {M.}~\bibnamefont {Aubert}},
  \bibinfo {author} {\bibfnamefont {A.}~\bibnamefont {Pisani}}, \bibinfo
  {author} {\bibfnamefont {M.-C.}\ \bibnamefont {Cousinou}}, \bibinfo {author}
  {\bibfnamefont {S.}~\bibnamefont {Escoffier}}, \bibinfo {author}
  {\bibfnamefont {S.}~\bibnamefont {Nadathur}}, \bibinfo {author}
  {\bibfnamefont {G.}~\bibnamefont {Rossi}}, \ and\ \bibinfo {author}
  {\bibfnamefont {D.~P.}\ \bibnamefont {Schneider}},\ }\href {\doibase
  10.1088/1475-7516/2020/06/012} {\bibfield  {journal} {\bibinfo  {journal}
  {Journal of Cosmology and Astroparticle Physics}\ }\textbf {\bibinfo {volume}
  {2020}},\ \bibinfo {pages} {012} (\bibinfo {year} {2020})}\BibitemShut
  {NoStop}%
\bibitem [{\citenamefont {{Aubert}}\ \emph {et~al.}(2020)\citenamefont
  {{Aubert}}, \citenamefont {{Cousinou}}, \citenamefont {{Escoffier}},
  \citenamefont {{Hawken}}, \citenamefont {{Nadathur}}, \citenamefont {{Alam}},
  \citenamefont {{Bautista}}, \citenamefont {{Burtin}}, \citenamefont {{de
  Mattia}}, \citenamefont {{Gil-Mar{\'\i}n}}, \citenamefont {{Hou}},
  \citenamefont {{Jullo}}, \citenamefont {{Neveux}}, \citenamefont {{Rossi}},
  \citenamefont {{Smith}}, \citenamefont {{Tamone}},\ and\ \citenamefont
  {{Vargas Maga{\~n}a}}}]{aubert20}%
  \BibitemOpen
  \bibfield  {author} {\bibinfo {author} {\bibfnamefont {M.}~\bibnamefont
  {{Aubert}}}, \bibinfo {author} {\bibfnamefont {M.-C.}\ \bibnamefont
  {{Cousinou}}}, \bibinfo {author} {\bibfnamefont {S.}~\bibnamefont
  {{Escoffier}}}, \bibinfo {author} {\bibfnamefont {A.~J.}\ \bibnamefont
  {{Hawken}}}, \bibinfo {author} {\bibfnamefont {S.}~\bibnamefont
  {{Nadathur}}}, \bibinfo {author} {\bibfnamefont {S.}~\bibnamefont {{Alam}}},
  \bibinfo {author} {\bibfnamefont {J.}~\bibnamefont {{Bautista}}}, \bibinfo
  {author} {\bibfnamefont {E.}~\bibnamefont {{Burtin}}}, \bibinfo {author}
  {\bibfnamefont {A.}~\bibnamefont {{de Mattia}}}, \bibinfo {author}
  {\bibfnamefont {H.}~\bibnamefont {{Gil-Mar{\'\i}n}}}, \bibinfo {author}
  {\bibfnamefont {J.}~\bibnamefont {{Hou}}}, \bibinfo {author} {\bibfnamefont
  {E.}~\bibnamefont {{Jullo}}}, \bibinfo {author} {\bibfnamefont
  {R.}~\bibnamefont {{Neveux}}}, \bibinfo {author} {\bibfnamefont
  {G.}~\bibnamefont {{Rossi}}}, \bibinfo {author} {\bibfnamefont
  {A.}~\bibnamefont {{Smith}}}, \bibinfo {author} {\bibfnamefont
  {A.}~\bibnamefont {{Tamone}}}, \ and\ \bibinfo {author} {\bibfnamefont
  {M.}~\bibnamefont {{Vargas Maga{\~n}a}}},\ }\href@noop {} {\bibfield
  {journal} {\bibinfo  {journal} {arXiv e-prints}\ ,\ \bibinfo {eid}
  {arXiv:2007.09013}} (\bibinfo {year} {2020})},\ \Eprint
  {http://arxiv.org/abs/2007.09013} {arXiv:2007.09013 [astro-ph.CO]}
  \BibitemShut {NoStop}%
\bibitem [{\citenamefont {{Ravoux}}\ \emph {et~al.}(2020)\citenamefont
  {{Ravoux}}, \citenamefont {{Armengaud}}, \citenamefont {{Walther}},
  \citenamefont {{Etourneau}}, \citenamefont {{Pomar{\`e}de}}, \citenamefont
  {{Palanque-Delabrouille}}, \citenamefont {{Y{\`e}che}}, \citenamefont
  {{Bautista}}, \citenamefont {{du Mas des Bourboux}}, \citenamefont
  {{Chabanier}}, \citenamefont {{Dawson}}, \citenamefont {{Le Goff}},
  \citenamefont {{Lyke}}, \citenamefont {{Myers}}, \citenamefont {{Petitjean}},
  \citenamefont {{Pieri}}, \citenamefont {{Rich}}, \citenamefont {{Rossi}},\
  and\ \citenamefont {{Schneider}}}]{ravoux20a}%
  \BibitemOpen
  \bibfield  {author} {\bibinfo {author} {\bibfnamefont {C.}~\bibnamefont
  {{Ravoux}}}, \bibinfo {author} {\bibfnamefont {E.}~\bibnamefont
  {{Armengaud}}}, \bibinfo {author} {\bibfnamefont {M.}~\bibnamefont
  {{Walther}}}, \bibinfo {author} {\bibfnamefont {T.}~\bibnamefont
  {{Etourneau}}}, \bibinfo {author} {\bibfnamefont {D.}~\bibnamefont
  {{Pomar{\`e}de}}}, \bibinfo {author} {\bibfnamefont {N.}~\bibnamefont
  {{Palanque-Delabrouille}}}, \bibinfo {author} {\bibfnamefont
  {C.}~\bibnamefont {{Y{\`e}che}}}, \bibinfo {author} {\bibfnamefont
  {J.}~\bibnamefont {{Bautista}}}, \bibinfo {author} {\bibfnamefont
  {H.}~\bibnamefont {{du Mas des Bourboux}}}, \bibinfo {author} {\bibfnamefont
  {S.}~\bibnamefont {{Chabanier}}}, \bibinfo {author} {\bibfnamefont
  {K.}~\bibnamefont {{Dawson}}}, \bibinfo {author} {\bibfnamefont {J.~M.}\
  \bibnamefont {{Le Goff}}}, \bibinfo {author} {\bibfnamefont {B.}~\bibnamefont
  {{Lyke}}}, \bibinfo {author} {\bibfnamefont {A.~D.}\ \bibnamefont {{Myers}}},
  \bibinfo {author} {\bibfnamefont {P.}~\bibnamefont {{Petitjean}}}, \bibinfo
  {author} {\bibfnamefont {M.~M.}\ \bibnamefont {{Pieri}}}, \bibinfo {author}
  {\bibfnamefont {J.}~\bibnamefont {{Rich}}}, \bibinfo {author} {\bibfnamefont
  {G.}~\bibnamefont {{Rossi}}}, \ and\ \bibinfo {author} {\bibfnamefont
  {D.~P.}\ \bibnamefont {{Schneider}}},\ }\href {\doibase
  10.1088/1475-7516/2020/07/010} {\bibfield  {journal} {\bibinfo  {journal}
  {\jcap}\ }\textbf {\bibinfo {volume} {2020}},\ \bibinfo {eid} {010} (\bibinfo
  {year} {2020})},\ \Eprint {http://arxiv.org/abs/2004.01448} {arXiv:2004.01448
  [astro-ph.CO]} \BibitemShut {NoStop}%
\bibitem [{\citenamefont {{Kong}}\ \emph {et~al.}(2020)\citenamefont {{Kong}},
  \citenamefont {{Burleigh}}, \citenamefont {{Ross}}, \citenamefont
  {{Moustakas}}, \citenamefont {{Chuang}}, \citenamefont {{Comparat}},
  \citenamefont {{de Mattia}}, \citenamefont {{du Mas des Bourboux}},
  \citenamefont {{Honscheid}}, \citenamefont {{Lin}}, \citenamefont
  {{Raichoor}}, \citenamefont {{Rossi}},\ and\ \citenamefont
  {{Zhao}}}]{kong20a}%
  \BibitemOpen
  \bibfield  {author} {\bibinfo {author} {\bibfnamefont {H.}~\bibnamefont
  {{Kong}}}, \bibinfo {author} {\bibfnamefont {K.~J.}\ \bibnamefont
  {{Burleigh}}}, \bibinfo {author} {\bibfnamefont {A.}~\bibnamefont {{Ross}}},
  \bibinfo {author} {\bibfnamefont {J.}~\bibnamefont {{Moustakas}}}, \bibinfo
  {author} {\bibfnamefont {C.-H.}\ \bibnamefont {{Chuang}}}, \bibinfo {author}
  {\bibfnamefont {J.}~\bibnamefont {{Comparat}}}, \bibinfo {author}
  {\bibfnamefont {A.}~\bibnamefont {{de Mattia}}}, \bibinfo {author}
  {\bibfnamefont {H.}~\bibnamefont {{du Mas des Bourboux}}}, \bibinfo {author}
  {\bibfnamefont {K.}~\bibnamefont {{Honscheid}}}, \bibinfo {author}
  {\bibfnamefont {S.}~\bibnamefont {{Lin}}}, \bibinfo {author} {\bibfnamefont
  {A.}~\bibnamefont {{Raichoor}}}, \bibinfo {author} {\bibfnamefont
  {G.}~\bibnamefont {{Rossi}}}, \ and\ \bibinfo {author} {\bibfnamefont
  {C.}~\bibnamefont {{Zhao}}},\ }\href {\doibase 10.1093/mnras/staa2742}
  {\bibfield  {journal} {\bibinfo  {journal} {\mnras}\ }\textbf {\bibinfo
  {volume} {499}},\ \bibinfo {pages} {3943} (\bibinfo {year} {2020})},\ \Eprint
  {http://arxiv.org/abs/2007.08992} {arXiv:2007.08992 [astro-ph.CO]}
  \BibitemShut {NoStop}%
\bibitem [{\citenamefont {{Mohammad}}\ \emph {et~al.}(2020)\citenamefont
  {{Mohammad}}, \citenamefont {{Percival}}, \citenamefont {{Seo}},
  \citenamefont {{Chapman}}, \citenamefont {{Bianchi}}, \citenamefont {{Ross}},
  \citenamefont {{Zhao}}, \citenamefont {{Lang}}, \citenamefont {{Bautista}},
  \citenamefont {{Brinkmann}}, \citenamefont {{Brownstein}}, \citenamefont
  {{Burtin}}, \citenamefont {{Chuang}}, \citenamefont {{Dawson}}, \citenamefont
  {{de la Torre}}, \citenamefont {{de Mattia}}, \citenamefont
  {{Eftekharzadeh}}, \citenamefont {{Fromenteau}}, \citenamefont
  {{Gil-Mar{\'\i}n}}, \citenamefont {{Hou}}, \citenamefont {{Mueller}},
  \citenamefont {{Neveux}}, \citenamefont {{Paviot}}, \citenamefont
  {{Raichoor}}, \citenamefont {{Rossi}}, \citenamefont {{Schneider}},
  \citenamefont {{Tamone}}, \citenamefont {{Tinker}}, \citenamefont
  {{Tojeiro}}, \citenamefont {{Vargas Maga{\~n}a}},\ and\ \citenamefont
  {{Zhao}}}]{mohammad20a}%
  \BibitemOpen
  \bibfield  {author} {\bibinfo {author} {\bibfnamefont {F.~G.}\ \bibnamefont
  {{Mohammad}}}, \bibinfo {author} {\bibfnamefont {W.~J.}\ \bibnamefont
  {{Percival}}}, \bibinfo {author} {\bibfnamefont {H.-J.}\ \bibnamefont
  {{Seo}}}, \bibinfo {author} {\bibfnamefont {M.~J.}\ \bibnamefont
  {{Chapman}}}, \bibinfo {author} {\bibfnamefont {D.}~\bibnamefont
  {{Bianchi}}}, \bibinfo {author} {\bibfnamefont {A.~J.}\ \bibnamefont
  {{Ross}}}, \bibinfo {author} {\bibfnamefont {C.}~\bibnamefont {{Zhao}}},
  \bibinfo {author} {\bibfnamefont {D.}~\bibnamefont {{Lang}}}, \bibinfo
  {author} {\bibfnamefont {J.}~\bibnamefont {{Bautista}}}, \bibinfo {author}
  {\bibfnamefont {J.}~\bibnamefont {{Brinkmann}}}, \bibinfo {author}
  {\bibfnamefont {J.~R.}\ \bibnamefont {{Brownstein}}}, \bibinfo {author}
  {\bibfnamefont {E.}~\bibnamefont {{Burtin}}}, \bibinfo {author}
  {\bibfnamefont {C.-H.}\ \bibnamefont {{Chuang}}}, \bibinfo {author}
  {\bibfnamefont {K.~S.}\ \bibnamefont {{Dawson}}}, \bibinfo {author}
  {\bibfnamefont {S.}~\bibnamefont {{de la Torre}}}, \bibinfo {author}
  {\bibfnamefont {A.}~\bibnamefont {{de Mattia}}}, \bibinfo {author}
  {\bibfnamefont {S.}~\bibnamefont {{Eftekharzadeh}}}, \bibinfo {author}
  {\bibfnamefont {S.}~\bibnamefont {{Fromenteau}}}, \bibinfo {author}
  {\bibfnamefont {H.}~\bibnamefont {{Gil-Mar{\'\i}n}}}, \bibinfo {author}
  {\bibfnamefont {J.}~\bibnamefont {{Hou}}}, \bibinfo {author} {\bibfnamefont
  {E.-M.}\ \bibnamefont {{Mueller}}}, \bibinfo {author} {\bibfnamefont
  {R.}~\bibnamefont {{Neveux}}}, \bibinfo {author} {\bibfnamefont
  {R.}~\bibnamefont {{Paviot}}}, \bibinfo {author} {\bibfnamefont
  {A.}~\bibnamefont {{Raichoor}}}, \bibinfo {author} {\bibfnamefont
  {G.}~\bibnamefont {{Rossi}}}, \bibinfo {author} {\bibfnamefont {D.~P.}\
  \bibnamefont {{Schneider}}}, \bibinfo {author} {\bibfnamefont
  {A.}~\bibnamefont {{Tamone}}}, \bibinfo {author} {\bibfnamefont {J.~L.}\
  \bibnamefont {{Tinker}}}, \bibinfo {author} {\bibfnamefont {R.}~\bibnamefont
  {{Tojeiro}}}, \bibinfo {author} {\bibfnamefont {M.}~\bibnamefont {{Vargas
  Maga{\~n}a}}}, \ and\ \bibinfo {author} {\bibfnamefont {G.-B.}\ \bibnamefont
  {{Zhao}}},\ }\href {\doibase 10.1093/mnras/staa2344} {\bibfield  {journal}
  {\bibinfo  {journal} {\mnras}\ }\textbf {\bibinfo {volume} {498}},\ \bibinfo
  {pages} {128} (\bibinfo {year} {2020})},\ \Eprint
  {http://arxiv.org/abs/2007.09005} {arXiv:2007.09005 [astro-ph.CO]}
  \BibitemShut {NoStop}%
\bibitem [{\citenamefont {{DESI Collaboration}}\ \emph
  {et~al.}(2016{\natexlab{a}})\citenamefont {{DESI Collaboration}},
  \citenamefont {{Aghamousa}}, \citenamefont {{Aguilar}}, \citenamefont
  {{Ahlen}}, \citenamefont {{Alam}}, \citenamefont {{Allen}}, \citenamefont
  {{Allende Prieto}}, \citenamefont {{Annis}}, \citenamefont {{Bailey}},
  \citenamefont {{Balland}},\ and\ \citenamefont {et~al.}}]{desi16a}%
  \BibitemOpen
  \bibfield  {author} {\bibinfo {author} {\bibnamefont {{DESI Collaboration}}},
  \bibinfo {author} {\bibfnamefont {A.}~\bibnamefont {{Aghamousa}}}, \bibinfo
  {author} {\bibfnamefont {J.}~\bibnamefont {{Aguilar}}}, \bibinfo {author}
  {\bibfnamefont {S.}~\bibnamefont {{Ahlen}}}, \bibinfo {author} {\bibfnamefont
  {S.}~\bibnamefont {{Alam}}}, \bibinfo {author} {\bibfnamefont {L.~E.}\
  \bibnamefont {{Allen}}}, \bibinfo {author} {\bibfnamefont {C.}~\bibnamefont
  {{Allende Prieto}}}, \bibinfo {author} {\bibfnamefont {J.}~\bibnamefont
  {{Annis}}}, \bibinfo {author} {\bibfnamefont {S.}~\bibnamefont {{Bailey}}},
  \bibinfo {author} {\bibfnamefont {C.}~\bibnamefont {{Balland}}}, \ and\
  \bibinfo {author} {\bibnamefont {et~al.}},\ }\href@noop {} {\bibfield
  {journal} {\bibinfo  {journal} {ArXiv e-prints}\ } (\bibinfo {year}
  {2016}{\natexlab{a}})},\ \Eprint {http://arxiv.org/abs/1611.00036}
  {arXiv:1611.00036 [astro-ph.IM]} \BibitemShut {NoStop}%
\bibitem [{\citenamefont {{DESI Collaboration}}\ \emph
  {et~al.}(2016{\natexlab{b}})\citenamefont {{DESI Collaboration}},
  \citenamefont {{Aghamousa}}, \citenamefont {{Aguilar}}, \citenamefont
  {{Ahlen}}, \citenamefont {{Alam}}, \citenamefont {{Allen}}, \citenamefont
  {{Allende Prieto}}, \citenamefont {{Annis}}, \citenamefont {{Bailey}},
  \citenamefont {{Balland}},\ and\ \citenamefont {et~al.}}]{desi16b}%
  \BibitemOpen
  \bibfield  {author} {\bibinfo {author} {\bibnamefont {{DESI Collaboration}}},
  \bibinfo {author} {\bibfnamefont {A.}~\bibnamefont {{Aghamousa}}}, \bibinfo
  {author} {\bibfnamefont {J.}~\bibnamefont {{Aguilar}}}, \bibinfo {author}
  {\bibfnamefont {S.}~\bibnamefont {{Ahlen}}}, \bibinfo {author} {\bibfnamefont
  {S.}~\bibnamefont {{Alam}}}, \bibinfo {author} {\bibfnamefont {L.~E.}\
  \bibnamefont {{Allen}}}, \bibinfo {author} {\bibfnamefont {C.}~\bibnamefont
  {{Allende Prieto}}}, \bibinfo {author} {\bibfnamefont {J.}~\bibnamefont
  {{Annis}}}, \bibinfo {author} {\bibfnamefont {S.}~\bibnamefont {{Bailey}}},
  \bibinfo {author} {\bibfnamefont {C.}~\bibnamefont {{Balland}}}, \ and\
  \bibinfo {author} {\bibnamefont {et~al.}},\ }\href@noop {} {\bibfield
  {journal} {\bibinfo  {journal} {ArXiv e-prints}\ } (\bibinfo {year}
  {2016}{\natexlab{b}})},\ \Eprint {http://arxiv.org/abs/1611.00037}
  {arXiv:1611.00037 [astro-ph.IM]} \BibitemShut {NoStop}%
\bibitem [{\citenamefont {{Ahumada}}\ \emph {et~al.}(2020)\citenamefont
  {{Ahumada}}, \citenamefont {{Prieto}}, \citenamefont {{Almeida}},
  \citenamefont {{Anders}}, \citenamefont {{Anderson}}, \citenamefont
  {{Andrews}}, \citenamefont {{Anguiano}}, \citenamefont {{Arcodia}},
  \citenamefont {{Armengaud}}, \citenamefont {{Aubert}}, \citenamefont
  {{Avila}}, \citenamefont {{Avila-Reese}}, \citenamefont {{Badenes}},
  \citenamefont {{Balland }}, \citenamefont {{Barger}}, \citenamefont
  {{Barrera-Ballesteros}}, \citenamefont {{Basu}}, \citenamefont {{Bautista}},
  \citenamefont {{Beaton}}, \citenamefont {{Beers}}, \citenamefont
  {{Benavides}}, \citenamefont {{Bender}}, \citenamefont {{Bernardi}},
  \citenamefont {{Bershady}}, \citenamefont {{Beutler}}, \citenamefont
  {{Bidin}}, \citenamefont {{Bird}}, \citenamefont {{Bizyaev}}, \citenamefont
  {{Blanc}}, \citenamefont {{Blanton}}, \citenamefont {{Boquien}},
  \citenamefont {{Borissova}}, \citenamefont {{Bovy}}, \citenamefont {{Brand
  t}}, \citenamefont {{Brinkmann}}, \citenamefont {{Brownstein}}, \citenamefont
  {{Bundy}}, \citenamefont {{Bureau}}, \citenamefont {{Burgasser}},
  \citenamefont {{Burtin}}, \citenamefont {{Cano-D{\'\i}az}}, \citenamefont
  {{Capasso}}, \citenamefont {{Cappellari}}, \citenamefont {{Carrera}},
  \citenamefont {{Chabanier}}, \citenamefont {{Chaplin}}, \citenamefont
  {{Chapman}}, \citenamefont {{Cherinka}}, \citenamefont {{Chiappini}},
  \citenamefont {{Doohyun Choi}}, \citenamefont {{Chojnowski}}, \citenamefont
  {{Chung}}, \citenamefont {{Clerc}}, \citenamefont {{Coffey}}, \citenamefont
  {{Comerford}}, \citenamefont {{Comparat}}, \citenamefont {{da Costa}},
  \citenamefont {{Cousinou}}, \citenamefont {{Covey}}, \citenamefont {{Crane}},
  \citenamefont {{Cunha}}, \citenamefont {{Ilha}}, \citenamefont {{Dai}},
  \citenamefont {{Damsted}}, \citenamefont {{Darling}}, \citenamefont
  {{Davidson}}, \citenamefont {{Davies}}, \citenamefont {{Dawson}},
  \citenamefont {{De}}, \citenamefont {{de la Macorra}}, \citenamefont {{De
  Lee}}, \citenamefont {{Queiroz}}, \citenamefont {{Deconto Machado}},
  \citenamefont {{de la Torre}}, \citenamefont {{Dell'Agli}}, \citenamefont
  {{du Mas des Bourboux}}, \citenamefont {{Diamond-Stanic}}, \citenamefont
  {{Dillon}}, \citenamefont {{Donor}}, \citenamefont {{Drory}}, \citenamefont
  {{Thomas}}, \citenamefont {{Thomas}}, \citenamefont {{Tinker}}, \citenamefont
  {{Tojeiro}}, \citenamefont {{Toledo}}, \citenamefont {{Tremonti}},
  \citenamefont {{Troup}}, \citenamefont {{Tuttle}}, \citenamefont
  {{Unda-Sanzana}}, \citenamefont {{Valentini}}, \citenamefont
  {{Vargas-Gonz{\'a}lez}}, \citenamefont {{Vargas-Maga{\~n}a}}, \citenamefont
  {{V{\'a}zquez-Mata}}, \citenamefont {{Vivek}}, \citenamefont {{Wake}},
  \citenamefont {{Wang}}, \citenamefont {{Weaver}}, \citenamefont {{Weijmans}},
  \citenamefont {{Wild}}, \citenamefont {{Wilson}}, \citenamefont {{Wilson}},
  \citenamefont {{Wolthuis}}, \citenamefont {{Wood-Vasey}}, \citenamefont
  {{Yan}}, \citenamefont {{Yang}}, \citenamefont {{Y{\`e}che}}, \citenamefont
  {{Zamora}}, \citenamefont {{Zarrouk}}, \citenamefont {{Zasowski}},
  \citenamefont {{Zhang}}, \citenamefont {{Zhao}}, \citenamefont {{Zhao}},
  \citenamefont {{Zheng}}, \citenamefont {{Zheng}}, \citenamefont {{Zhu}},\
  and\ \citenamefont {{Zou}}}]{dr16}%
  \BibitemOpen
  \bibfield  {author} {\bibinfo {author} {\bibfnamefont {R.}~\bibnamefont
  {{Ahumada}}}, \bibinfo {author} {\bibfnamefont {C.~A.}\ \bibnamefont
  {{Prieto}}}, \bibinfo {author} {\bibfnamefont {A.}~\bibnamefont {{Almeida}}},
  \bibinfo {author} {\bibfnamefont {F.}~\bibnamefont {{Anders}}}, \bibinfo
  {author} {\bibfnamefont {S.~F.}\ \bibnamefont {{Anderson}}}, \bibinfo
  {author} {\bibfnamefont {B.~H.}\ \bibnamefont {{Andrews}}}, \bibinfo {author}
  {\bibfnamefont {B.}~\bibnamefont {{Anguiano}}}, \bibinfo {author}
  {\bibfnamefont {R.}~\bibnamefont {{Arcodia}}}, \bibinfo {author}
  {\bibfnamefont {E.}~\bibnamefont {{Armengaud}}}, \bibinfo {author}
  {\bibfnamefont {M.}~\bibnamefont {{Aubert}}}, \bibinfo {author}
  {\bibfnamefont {S.}~\bibnamefont {{Avila}}}, \bibinfo {author} {\bibfnamefont
  {V.}~\bibnamefont {{Avila-Reese}}}, \bibinfo {author} {\bibfnamefont
  {C.}~\bibnamefont {{Badenes}}}, \bibinfo {author} {\bibfnamefont
  {C.}~\bibnamefont {{Balland }}}, \bibinfo {author} {\bibfnamefont
  {K.}~\bibnamefont {{Barger}}}, \bibinfo {author} {\bibfnamefont {J.~K.}\
  \bibnamefont {{Barrera-Ballesteros}}}, \bibinfo {author} {\bibfnamefont
  {S.}~\bibnamefont {{Basu}}}, \bibinfo {author} {\bibfnamefont
  {J.}~\bibnamefont {{Bautista}}}, \bibinfo {author} {\bibfnamefont {R.~L.}\
  \bibnamefont {{Beaton}}}, \bibinfo {author} {\bibfnamefont {T.~C.}\
  \bibnamefont {{Beers}}}, \bibinfo {author} {\bibfnamefont {B.~I.~T.}\
  \bibnamefont {{Benavides}}}, \bibinfo {author} {\bibfnamefont {C.~F.}\
  \bibnamefont {{Bender}}}, \bibinfo {author} {\bibfnamefont {M.}~\bibnamefont
  {{Bernardi}}}, \bibinfo {author} {\bibfnamefont {M.}~\bibnamefont
  {{Bershady}}}, \bibinfo {author} {\bibfnamefont {F.}~\bibnamefont
  {{Beutler}}}, \bibinfo {author} {\bibfnamefont {C.~M.}\ \bibnamefont
  {{Bidin}}}, \bibinfo {author} {\bibfnamefont {J.}~\bibnamefont {{Bird}}},
  \bibinfo {author} {\bibfnamefont {D.}~\bibnamefont {{Bizyaev}}}, \bibinfo
  {author} {\bibfnamefont {G.~A.}\ \bibnamefont {{Blanc}}}, \bibinfo {author}
  {\bibfnamefont {M.~R.}\ \bibnamefont {{Blanton}}}, \bibinfo {author}
  {\bibfnamefont {M.}~\bibnamefont {{Boquien}}}, \bibinfo {author}
  {\bibfnamefont {J.}~\bibnamefont {{Borissova}}}, \bibinfo {author}
  {\bibfnamefont {J.}~\bibnamefont {{Bovy}}}, \bibinfo {author} {\bibfnamefont
  {W.~N.}\ \bibnamefont {{Brand t}}}, \bibinfo {author} {\bibfnamefont
  {J.}~\bibnamefont {{Brinkmann}}}, \bibinfo {author} {\bibfnamefont {J.~R.}\
  \bibnamefont {{Brownstein}}}, \bibinfo {author} {\bibfnamefont
  {K.}~\bibnamefont {{Bundy}}}, \bibinfo {author} {\bibfnamefont
  {M.}~\bibnamefont {{Bureau}}}, \bibinfo {author} {\bibfnamefont
  {A.}~\bibnamefont {{Burgasser}}}, \bibinfo {author} {\bibfnamefont
  {E.}~\bibnamefont {{Burtin}}}, \bibinfo {author} {\bibfnamefont
  {M.}~\bibnamefont {{Cano-D{\'\i}az}}}, \bibinfo {author} {\bibfnamefont
  {R.}~\bibnamefont {{Capasso}}}, \bibinfo {author} {\bibfnamefont
  {M.}~\bibnamefont {{Cappellari}}}, \bibinfo {author} {\bibfnamefont
  {R.}~\bibnamefont {{Carrera}}}, \bibinfo {author} {\bibfnamefont
  {S.}~\bibnamefont {{Chabanier}}}, \bibinfo {author} {\bibfnamefont
  {W.}~\bibnamefont {{Chaplin}}}, \bibinfo {author} {\bibfnamefont
  {M.}~\bibnamefont {{Chapman}}}, \bibinfo {author} {\bibfnamefont
  {B.}~\bibnamefont {{Cherinka}}}, \bibinfo {author} {\bibfnamefont
  {C.}~\bibnamefont {{Chiappini}}}, \bibinfo {author} {\bibfnamefont
  {P.}~\bibnamefont {{Doohyun Choi}}}, \bibinfo {author} {\bibfnamefont
  {S.~D.}\ \bibnamefont {{Chojnowski}}}, \bibinfo {author} {\bibfnamefont
  {H.}~\bibnamefont {{Chung}}}, \bibinfo {author} {\bibfnamefont
  {N.}~\bibnamefont {{Clerc}}}, \bibinfo {author} {\bibfnamefont
  {D.}~\bibnamefont {{Coffey}}}, \bibinfo {author} {\bibfnamefont {J.~M.}\
  \bibnamefont {{Comerford}}}, \bibinfo {author} {\bibfnamefont
  {J.}~\bibnamefont {{Comparat}}}, \bibinfo {author} {\bibfnamefont
  {L.}~\bibnamefont {{da Costa}}}, \bibinfo {author} {\bibfnamefont {M.-C.}\
  \bibnamefont {{Cousinou}}}, \bibinfo {author} {\bibfnamefont
  {K.}~\bibnamefont {{Covey}}}, \bibinfo {author} {\bibfnamefont {J.~D.}\
  \bibnamefont {{Crane}}}, \bibinfo {author} {\bibfnamefont {K.}~\bibnamefont
  {{Cunha}}}, \bibinfo {author} {\bibfnamefont {G.~d.~S.}\ \bibnamefont
  {{Ilha}}}, \bibinfo {author} {\bibfnamefont {Y.~S.}\ \bibnamefont {{Dai}}},
  \bibinfo {author} {\bibfnamefont {S.~B.}\ \bibnamefont {{Damsted}}}, \bibinfo
  {author} {\bibfnamefont {J.}~\bibnamefont {{Darling}}}, \bibinfo {author}
  {\bibfnamefont {J.}~\bibnamefont {{Davidson}}, \bibfnamefont {James~W.}},
  \bibinfo {author} {\bibfnamefont {R.}~\bibnamefont {{Davies}}}, \bibinfo
  {author} {\bibfnamefont {K.}~\bibnamefont {{Dawson}}}, \bibinfo {author}
  {\bibfnamefont {N.}~\bibnamefont {{De}}}, \bibinfo {author} {\bibfnamefont
  {A.}~\bibnamefont {{de la Macorra}}}, \bibinfo {author} {\bibfnamefont
  {N.}~\bibnamefont {{De Lee}}}, \bibinfo {author} {\bibfnamefont {A.~B.
  d.~A.}\ \bibnamefont {{Queiroz}}}, \bibinfo {author} {\bibfnamefont
  {A.}~\bibnamefont {{Deconto Machado}}}, \bibinfo {author} {\bibfnamefont
  {S.}~\bibnamefont {{de la Torre}}}, \bibinfo {author} {\bibfnamefont
  {F.}~\bibnamefont {{Dell'Agli}}}, \bibinfo {author} {\bibfnamefont
  {H.}~\bibnamefont {{du Mas des Bourboux}}}, \bibinfo {author} {\bibfnamefont
  {A.~M.}\ \bibnamefont {{Diamond-Stanic}}}, \bibinfo {author} {\bibfnamefont
  {S.}~\bibnamefont {{Dillon}}}, \bibinfo {author} {\bibfnamefont
  {J.}~\bibnamefont {{Donor}}}, \bibinfo {author} {\bibfnamefont
  {N.}~\bibnamefont {{Drory}}}, \bibinfo {author} {\bibfnamefont
  {D.}~\bibnamefont {{Thomas}}}, \bibinfo {author} {\bibfnamefont {Z.~C.}\
  \bibnamefont {{Thomas}}}, \bibinfo {author} {\bibfnamefont {J.}~\bibnamefont
  {{Tinker}}}, \bibinfo {author} {\bibfnamefont {R.}~\bibnamefont {{Tojeiro}}},
  \bibinfo {author} {\bibfnamefont {H.~H.}\ \bibnamefont {{Toledo}}}, \bibinfo
  {author} {\bibfnamefont {C.~A.}\ \bibnamefont {{Tremonti}}}, \bibinfo
  {author} {\bibfnamefont {N.~W.}\ \bibnamefont {{Troup}}}, \bibinfo {author}
  {\bibfnamefont {S.}~\bibnamefont {{Tuttle}}}, \bibinfo {author}
  {\bibfnamefont {E.}~\bibnamefont {{Unda-Sanzana}}}, \bibinfo {author}
  {\bibfnamefont {M.}~\bibnamefont {{Valentini}}}, \bibinfo {author}
  {\bibfnamefont {J.}~\bibnamefont {{Vargas-Gonz{\'a}lez}}}, \bibinfo {author}
  {\bibfnamefont {M.}~\bibnamefont {{Vargas-Maga{\~n}a}}}, \bibinfo {author}
  {\bibfnamefont {J.~A.}\ \bibnamefont {{V{\'a}zquez-Mata}}}, \bibinfo {author}
  {\bibfnamefont {M.}~\bibnamefont {{Vivek}}}, \bibinfo {author} {\bibfnamefont
  {D.}~\bibnamefont {{Wake}}}, \bibinfo {author} {\bibfnamefont
  {Y.}~\bibnamefont {{Wang}}}, \bibinfo {author} {\bibfnamefont {B.~A.}\
  \bibnamefont {{Weaver}}}, \bibinfo {author} {\bibfnamefont {A.-M.}\
  \bibnamefont {{Weijmans}}}, \bibinfo {author} {\bibfnamefont
  {V.}~\bibnamefont {{Wild}}}, \bibinfo {author} {\bibfnamefont {J.~C.}\
  \bibnamefont {{Wilson}}}, \bibinfo {author} {\bibfnamefont {R.~F.}\
  \bibnamefont {{Wilson}}}, \bibinfo {author} {\bibfnamefont {N.}~\bibnamefont
  {{Wolthuis}}}, \bibinfo {author} {\bibfnamefont {W.~M.}\ \bibnamefont
  {{Wood-Vasey}}}, \bibinfo {author} {\bibfnamefont {R.}~\bibnamefont {{Yan}}},
  \bibinfo {author} {\bibfnamefont {M.}~\bibnamefont {{Yang}}}, \bibinfo
  {author} {\bibfnamefont {C.}~\bibnamefont {{Y{\`e}che}}}, \bibinfo {author}
  {\bibfnamefont {O.}~\bibnamefont {{Zamora}}}, \bibinfo {author}
  {\bibfnamefont {P.}~\bibnamefont {{Zarrouk}}}, \bibinfo {author}
  {\bibfnamefont {G.}~\bibnamefont {{Zasowski}}}, \bibinfo {author}
  {\bibfnamefont {K.}~\bibnamefont {{Zhang}}}, \bibinfo {author} {\bibfnamefont
  {C.}~\bibnamefont {{Zhao}}}, \bibinfo {author} {\bibfnamefont
  {G.}~\bibnamefont {{Zhao}}}, \bibinfo {author} {\bibfnamefont
  {Z.}~\bibnamefont {{Zheng}}}, \bibinfo {author} {\bibfnamefont
  {Z.}~\bibnamefont {{Zheng}}}, \bibinfo {author} {\bibfnamefont
  {G.}~\bibnamefont {{Zhu}}}, \ and\ \bibinfo {author} {\bibfnamefont
  {H.}~\bibnamefont {{Zou}}},\ }\href {\doibase 10.3847/1538-4365/ab929e}
  {\bibfield  {journal} {\bibinfo  {journal} {\apjs}\ }\textbf {\bibinfo
  {volume} {249}},\ \bibinfo {eid} {3} (\bibinfo {year} {2020})},\ \Eprint
  {http://arxiv.org/abs/1912.02905} {arXiv:1912.02905 [astro-ph.GA]}
  \BibitemShut {NoStop}%
\bibitem [{\citenamefont {{Zhu}}\ \emph {et~al.}(2015)\citenamefont {{Zhu}},
  \citenamefont {{Comparat}}, \citenamefont {{Kneib}}, \citenamefont
  {{Delubac}}, \citenamefont {{Raichoor}}, \citenamefont {{Dawson}},
  \citenamefont {{Newman}}, \citenamefont {{Y{\`e}che}}, \citenamefont
  {{Zhou}},\ and\ \citenamefont {{Schneider}}}]{zhu15a}%
  \BibitemOpen
  \bibfield  {author} {\bibinfo {author} {\bibfnamefont {G.~B.}\ \bibnamefont
  {{Zhu}}}, \bibinfo {author} {\bibfnamefont {J.}~\bibnamefont {{Comparat}}},
  \bibinfo {author} {\bibfnamefont {J.-P.}\ \bibnamefont {{Kneib}}}, \bibinfo
  {author} {\bibfnamefont {T.}~\bibnamefont {{Delubac}}}, \bibinfo {author}
  {\bibfnamefont {A.}~\bibnamefont {{Raichoor}}}, \bibinfo {author}
  {\bibfnamefont {K.~S.}\ \bibnamefont {{Dawson}}}, \bibinfo {author}
  {\bibfnamefont {J.}~\bibnamefont {{Newman}}}, \bibinfo {author}
  {\bibfnamefont {C.}~\bibnamefont {{Y{\`e}che}}}, \bibinfo {author}
  {\bibfnamefont {X.}~\bibnamefont {{Zhou}}}, \ and\ \bibinfo {author}
  {\bibfnamefont {D.~P.}\ \bibnamefont {{Schneider}}},\ }\href {\doibase
  10.1088/0004-637X/815/1/48} {\bibfield  {journal} {\bibinfo  {journal}
  {\apj}\ }\textbf {\bibinfo {volume} {815}},\ \bibinfo {eid} {48} (\bibinfo
  {year} {2015})},\ \Eprint {http://arxiv.org/abs/1507.07979} {arXiv:1507.07979
  [astro-ph.GA]} \BibitemShut {NoStop}%
\bibitem [{\citenamefont {{Huang}}\ \emph {et~al.}(2019)\citenamefont
  {{Huang}}, \citenamefont {{Zou}}, \citenamefont {{Kong}}, \citenamefont
  {{Comparat}}, \citenamefont {{Lin}}, \citenamefont {{Gao}}, \citenamefont
  {{Liang}}, \citenamefont {{Delubac}}, \citenamefont {{Raichoor}},
  \citenamefont {{Kneib}}, \citenamefont {{Schneider}}, \citenamefont {{Zhou}},
  \citenamefont {{Yuan}},\ and\ \citenamefont {{Bershady}}}]{huang19a}%
  \BibitemOpen
  \bibfield  {author} {\bibinfo {author} {\bibfnamefont {C.}~\bibnamefont
  {{Huang}}}, \bibinfo {author} {\bibfnamefont {H.}~\bibnamefont {{Zou}}},
  \bibinfo {author} {\bibfnamefont {X.}~\bibnamefont {{Kong}}}, \bibinfo
  {author} {\bibfnamefont {J.}~\bibnamefont {{Comparat}}}, \bibinfo {author}
  {\bibfnamefont {Z.}~\bibnamefont {{Lin}}}, \bibinfo {author} {\bibfnamefont
  {Y.}~\bibnamefont {{Gao}}}, \bibinfo {author} {\bibfnamefont
  {Z.}~\bibnamefont {{Liang}}}, \bibinfo {author} {\bibfnamefont
  {T.}~\bibnamefont {{Delubac}}}, \bibinfo {author} {\bibfnamefont
  {A.}~\bibnamefont {{Raichoor}}}, \bibinfo {author} {\bibfnamefont {J.-P.}\
  \bibnamefont {{Kneib}}}, \bibinfo {author} {\bibfnamefont {D.~P.}\
  \bibnamefont {{Schneider}}}, \bibinfo {author} {\bibfnamefont
  {X.}~\bibnamefont {{Zhou}}}, \bibinfo {author} {\bibfnamefont
  {Q.}~\bibnamefont {{Yuan}}}, \ and\ \bibinfo {author} {\bibfnamefont {M.~A.}\
  \bibnamefont {{Bershady}}},\ }\href {\doibase 10.3847/1538-4357/ab4902}
  {\bibfield  {journal} {\bibinfo  {journal} {\apj}\ }\textbf {\bibinfo
  {volume} {886}},\ \bibinfo {eid} {31} (\bibinfo {year} {2019})},\ \Eprint
  {http://arxiv.org/abs/1910.08689} {arXiv:1910.08689 [astro-ph.GA]}
  \BibitemShut {NoStop}%
\bibitem [{\citenamefont {{Lan}}\ and\ \citenamefont {{Mo}}(2018)}]{lan18a}%
  \BibitemOpen
  \bibfield  {author} {\bibinfo {author} {\bibfnamefont {T.-W.}\ \bibnamefont
  {{Lan}}}\ and\ \bibinfo {author} {\bibfnamefont {H.}~\bibnamefont {{Mo}}},\
  }\href {\doibase 10.3847/1538-4357/aadc08} {\bibfield  {journal} {\bibinfo
  {journal} {\apj}\ }\textbf {\bibinfo {volume} {866}},\ \bibinfo {eid} {36}
  (\bibinfo {year} {2018})},\ \Eprint {http://arxiv.org/abs/1806.05786}
  {arXiv:1806.05786 [astro-ph.GA]} \BibitemShut {NoStop}%
\bibitem [{\citenamefont {{Talbot}}\ \emph {et~al.}(2020)\citenamefont
  {{Talbot}}, \citenamefont {{Brownstein}}, \citenamefont {{Dawson}},
  \citenamefont {{Kneib}},\ and\ \citenamefont {{Bautista}}}]{talbot20a}%
  \BibitemOpen
  \bibfield  {author} {\bibinfo {author} {\bibfnamefont {M.~S.}\ \bibnamefont
  {{Talbot}}}, \bibinfo {author} {\bibfnamefont {J.~R.}\ \bibnamefont
  {{Brownstein}}}, \bibinfo {author} {\bibfnamefont {K.~S.}\ \bibnamefont
  {{Dawson}}}, \bibinfo {author} {\bibfnamefont {J.-P.}\ \bibnamefont
  {{Kneib}}}, \ and\ \bibinfo {author} {\bibfnamefont {J.}~\bibnamefont
  {{Bautista}}},\ }\href@noop {} {\bibfield  {journal} {\bibinfo  {journal}
  {arXiv e-prints}\ ,\ \bibinfo {eid} {arXiv:2007.09006}} (\bibinfo {year}
  {2020})},\ \Eprint {http://arxiv.org/abs/2007.09006} {arXiv:2007.09006
  [astro-ph.GA]} \BibitemShut {NoStop}%
\bibitem [{\citenamefont {{Morganson}}\ \emph {et~al.}(2015)\citenamefont
  {{Morganson}}, \citenamefont {{Green}}, \citenamefont {{Anderson}},
  \citenamefont {{Ruan}}, \citenamefont {{Myers}}, \citenamefont {{Eracleous}},
  \citenamefont {{Kelly}}, \citenamefont {{Badenes}}, \citenamefont
  {{Ba{\~n}ados}}, \citenamefont {{Blanton}}, \citenamefont {{Bershady}},
  \citenamefont {{Borissova}}, \citenamefont {{Brandt}}, \citenamefont
  {{Burgett}}, \citenamefont {{Chambers}}, \citenamefont {{Draper}},
  \citenamefont {{Davenport}}, \citenamefont {{Flewelling}}, \citenamefont
  {{Garnavich}}, \citenamefont {{Hawley}}, \citenamefont {{Hodapp}},
  \citenamefont {{Isler}}, \citenamefont {{Kaiser}}, \citenamefont
  {{Kinemuchi}}, \citenamefont {{Kudritzki}}, \citenamefont {{Metcalfe}},
  \citenamefont {{Morgan}}, \citenamefont {{P{\^a}ris}}, \citenamefont
  {{Parvizi}}, \citenamefont {{Poleski}}, \citenamefont {{Price}},
  \citenamefont {{Salvato}}, \citenamefont {{Shanks}}, \citenamefont
  {{Schlafly}}, \citenamefont {{Schneider}}, \citenamefont {{Shen}},
  \citenamefont {{Stassun}}, \citenamefont {{Tonry}}, \citenamefont
  {{Walter}},\ and\ \citenamefont {{Waters}}}]{morganson15a}%
  \BibitemOpen
  \bibfield  {author} {\bibinfo {author} {\bibfnamefont {E.}~\bibnamefont
  {{Morganson}}}, \bibinfo {author} {\bibfnamefont {P.~J.}\ \bibnamefont
  {{Green}}}, \bibinfo {author} {\bibfnamefont {S.~F.}\ \bibnamefont
  {{Anderson}}}, \bibinfo {author} {\bibfnamefont {J.~J.}\ \bibnamefont
  {{Ruan}}}, \bibinfo {author} {\bibfnamefont {A.~D.}\ \bibnamefont {{Myers}}},
  \bibinfo {author} {\bibfnamefont {M.}~\bibnamefont {{Eracleous}}}, \bibinfo
  {author} {\bibfnamefont {B.}~\bibnamefont {{Kelly}}}, \bibinfo {author}
  {\bibfnamefont {C.}~\bibnamefont {{Badenes}}}, \bibinfo {author}
  {\bibfnamefont {E.}~\bibnamefont {{Ba{\~n}ados}}}, \bibinfo {author}
  {\bibfnamefont {M.~R.}\ \bibnamefont {{Blanton}}}, \bibinfo {author}
  {\bibfnamefont {M.~A.}\ \bibnamefont {{Bershady}}}, \bibinfo {author}
  {\bibfnamefont {J.}~\bibnamefont {{Borissova}}}, \bibinfo {author}
  {\bibfnamefont {W.~N.}\ \bibnamefont {{Brandt}}}, \bibinfo {author}
  {\bibfnamefont {W.~S.}\ \bibnamefont {{Burgett}}}, \bibinfo {author}
  {\bibfnamefont {K.}~\bibnamefont {{Chambers}}}, \bibinfo {author}
  {\bibfnamefont {P.~W.}\ \bibnamefont {{Draper}}}, \bibinfo {author}
  {\bibfnamefont {J.~R.~A.}\ \bibnamefont {{Davenport}}}, \bibinfo {author}
  {\bibfnamefont {H.}~\bibnamefont {{Flewelling}}}, \bibinfo {author}
  {\bibfnamefont {P.}~\bibnamefont {{Garnavich}}}, \bibinfo {author}
  {\bibfnamefont {S.~L.}\ \bibnamefont {{Hawley}}}, \bibinfo {author}
  {\bibfnamefont {K.~W.}\ \bibnamefont {{Hodapp}}}, \bibinfo {author}
  {\bibfnamefont {J.~C.}\ \bibnamefont {{Isler}}}, \bibinfo {author}
  {\bibfnamefont {N.}~\bibnamefont {{Kaiser}}}, \bibinfo {author}
  {\bibfnamefont {K.}~\bibnamefont {{Kinemuchi}}}, \bibinfo {author}
  {\bibfnamefont {R.~P.}\ \bibnamefont {{Kudritzki}}}, \bibinfo {author}
  {\bibfnamefont {N.}~\bibnamefont {{Metcalfe}}}, \bibinfo {author}
  {\bibfnamefont {J.~S.}\ \bibnamefont {{Morgan}}}, \bibinfo {author}
  {\bibfnamefont {I.}~\bibnamefont {{P{\^a}ris}}}, \bibinfo {author}
  {\bibfnamefont {M.}~\bibnamefont {{Parvizi}}}, \bibinfo {author}
  {\bibfnamefont {R.}~\bibnamefont {{Poleski}}}, \bibinfo {author}
  {\bibfnamefont {P.~A.}\ \bibnamefont {{Price}}}, \bibinfo {author}
  {\bibfnamefont {M.}~\bibnamefont {{Salvato}}}, \bibinfo {author}
  {\bibfnamefont {T.}~\bibnamefont {{Shanks}}}, \bibinfo {author}
  {\bibfnamefont {E.~F.}\ \bibnamefont {{Schlafly}}}, \bibinfo {author}
  {\bibfnamefont {D.~P.}\ \bibnamefont {{Schneider}}}, \bibinfo {author}
  {\bibfnamefont {Y.}~\bibnamefont {{Shen}}}, \bibinfo {author} {\bibfnamefont
  {K.}~\bibnamefont {{Stassun}}}, \bibinfo {author} {\bibfnamefont {J.~T.}\
  \bibnamefont {{Tonry}}}, \bibinfo {author} {\bibfnamefont {F.}~\bibnamefont
  {{Walter}}}, \ and\ \bibinfo {author} {\bibfnamefont {C.~Z.}\ \bibnamefont
  {{Waters}}},\ }\href {\doibase 10.1088/0004-637X/806/2/244} {\bibfield
  {journal} {\bibinfo  {journal} {\apj}\ }\textbf {\bibinfo {volume} {806}},\
  \bibinfo {eid} {244} (\bibinfo {year} {2015})},\ \Eprint
  {http://arxiv.org/abs/1505.00760} {arXiv:1505.00760 [astro-ph.HE]}
  \BibitemShut {NoStop}%
\bibitem [{\citenamefont {{MacLeod}}\ \emph {et~al.}(2018)\citenamefont
  {{MacLeod}}, \citenamefont {{Green}}, \citenamefont {{Anderson}},
  \citenamefont {{Eracleous}}, \citenamefont {{Ruan}}, \citenamefont
  {{Runnoe}}, \citenamefont {{Brand t}}, \citenamefont {{Badenes}},
  \citenamefont {{Greene}}, \citenamefont {{Morganson}}, \citenamefont
  {{Schmidt}}, \citenamefont {{Schwope}}, \citenamefont {{Shen}}, \citenamefont
  {{Amaro}}, \citenamefont {{Lebleu}}, \citenamefont {{Filiz Ak}},
  \citenamefont {{Grier}}, \citenamefont {{Hoover}}, \citenamefont {{McGraw}},
  \citenamefont {{Dawson}}, \citenamefont {{Hall}}, \citenamefont {{Hawley}},
  \citenamefont {{Mariappan}}, \citenamefont {{Myers}}, \citenamefont
  {{P{\^a}ris}}, \citenamefont {{Schneider}}, \citenamefont {{Stassun}},
  \citenamefont {{Bershady}}, \citenamefont {{Blanton}}, \citenamefont {{Seo}},
  \citenamefont {{Tinker}}, \citenamefont {{Fern{\'a}ndez-Trincado}},
  \citenamefont {{Chambers}}, \citenamefont {{Kaiser}}, \citenamefont
  {{Kudritzki}}, \citenamefont {{Magnier}}, \citenamefont {{Metcalfe}},\ and\
  \citenamefont {{Waters}}}]{macleod18a}%
  \BibitemOpen
  \bibfield  {author} {\bibinfo {author} {\bibfnamefont {C.~L.}\ \bibnamefont
  {{MacLeod}}}, \bibinfo {author} {\bibfnamefont {P.~J.}\ \bibnamefont
  {{Green}}}, \bibinfo {author} {\bibfnamefont {S.~F.}\ \bibnamefont
  {{Anderson}}}, \bibinfo {author} {\bibfnamefont {M.}~\bibnamefont
  {{Eracleous}}}, \bibinfo {author} {\bibfnamefont {J.~J.}\ \bibnamefont
  {{Ruan}}}, \bibinfo {author} {\bibfnamefont {J.}~\bibnamefont {{Runnoe}}},
  \bibinfo {author} {\bibfnamefont {W.~N.}\ \bibnamefont {{Brand t}}}, \bibinfo
  {author} {\bibfnamefont {C.}~\bibnamefont {{Badenes}}}, \bibinfo {author}
  {\bibfnamefont {J.}~\bibnamefont {{Greene}}}, \bibinfo {author}
  {\bibfnamefont {E.}~\bibnamefont {{Morganson}}}, \bibinfo {author}
  {\bibfnamefont {S.~J.}\ \bibnamefont {{Schmidt}}}, \bibinfo {author}
  {\bibfnamefont {A.}~\bibnamefont {{Schwope}}}, \bibinfo {author}
  {\bibfnamefont {Y.}~\bibnamefont {{Shen}}}, \bibinfo {author} {\bibfnamefont
  {R.}~\bibnamefont {{Amaro}}}, \bibinfo {author} {\bibfnamefont
  {A.}~\bibnamefont {{Lebleu}}}, \bibinfo {author} {\bibfnamefont
  {N.}~\bibnamefont {{Filiz Ak}}}, \bibinfo {author} {\bibfnamefont {C.~J.}\
  \bibnamefont {{Grier}}}, \bibinfo {author} {\bibfnamefont {D.}~\bibnamefont
  {{Hoover}}}, \bibinfo {author} {\bibfnamefont {S.~M.}\ \bibnamefont
  {{McGraw}}}, \bibinfo {author} {\bibfnamefont {K.}~\bibnamefont {{Dawson}}},
  \bibinfo {author} {\bibfnamefont {P.~B.}\ \bibnamefont {{Hall}}}, \bibinfo
  {author} {\bibfnamefont {S.~L.}\ \bibnamefont {{Hawley}}}, \bibinfo {author}
  {\bibfnamefont {V.}~\bibnamefont {{Mariappan}}}, \bibinfo {author}
  {\bibfnamefont {A.~D.}\ \bibnamefont {{Myers}}}, \bibinfo {author}
  {\bibfnamefont {I.}~\bibnamefont {{P{\^a}ris}}}, \bibinfo {author}
  {\bibfnamefont {D.~P.}\ \bibnamefont {{Schneider}}}, \bibinfo {author}
  {\bibfnamefont {K.~G.}\ \bibnamefont {{Stassun}}}, \bibinfo {author}
  {\bibfnamefont {M.~A.}\ \bibnamefont {{Bershady}}}, \bibinfo {author}
  {\bibfnamefont {M.~R.}\ \bibnamefont {{Blanton}}}, \bibinfo {author}
  {\bibfnamefont {H.-J.}\ \bibnamefont {{Seo}}}, \bibinfo {author}
  {\bibfnamefont {J.}~\bibnamefont {{Tinker}}}, \bibinfo {author}
  {\bibfnamefont {J.~G.}\ \bibnamefont {{Fern{\'a}ndez-Trincado}}}, \bibinfo
  {author} {\bibfnamefont {K.}~\bibnamefont {{Chambers}}}, \bibinfo {author}
  {\bibfnamefont {N.}~\bibnamefont {{Kaiser}}}, \bibinfo {author}
  {\bibfnamefont {R.~P.}\ \bibnamefont {{Kudritzki}}}, \bibinfo {author}
  {\bibfnamefont {E.}~\bibnamefont {{Magnier}}}, \bibinfo {author}
  {\bibfnamefont {N.}~\bibnamefont {{Metcalfe}}}, \ and\ \bibinfo {author}
  {\bibfnamefont {C.~Z.}\ \bibnamefont {{Waters}}},\ }\href {\doibase
  10.3847/1538-3881/aa99da} {\bibfield  {journal} {\bibinfo  {journal} {\aj}\
  }\textbf {\bibinfo {volume} {155}},\ \bibinfo {eid} {6} (\bibinfo {year}
  {2018})},\ \Eprint {http://arxiv.org/abs/1706.04240} {arXiv:1706.04240
  [astro-ph.GA]} \BibitemShut {NoStop}%
\bibitem [{\citenamefont {{Clerc}}\ \emph {et~al.}(2016)\citenamefont
  {{Clerc}}, \citenamefont {{Merloni}}, \citenamefont {{Zhang}}, \citenamefont
  {{Finoguenov}}, \citenamefont {{Dwelly}}, \citenamefont {{Nandra}},
  \citenamefont {{Collins}}, \citenamefont {{Dawson}}, \citenamefont {{Kneib}},
  \citenamefont {{Rozo}}, \citenamefont {{Rykoff}}, \citenamefont
  {{Sadibekova}}, \citenamefont {{Brownstein}}, \citenamefont {{Lin}},
  \citenamefont {{Ridl}}, \citenamefont {{Salvato}}, \citenamefont {{Schwope}},
  \citenamefont {{Steinmetz}}, \citenamefont {{Seo}},\ and\ \citenamefont
  {{Tinker}}}]{clerc16a}%
  \BibitemOpen
  \bibfield  {author} {\bibinfo {author} {\bibfnamefont {N.}~\bibnamefont
  {{Clerc}}}, \bibinfo {author} {\bibfnamefont {A.}~\bibnamefont {{Merloni}}},
  \bibinfo {author} {\bibfnamefont {Y.~Y.}\ \bibnamefont {{Zhang}}}, \bibinfo
  {author} {\bibfnamefont {A.}~\bibnamefont {{Finoguenov}}}, \bibinfo {author}
  {\bibfnamefont {T.}~\bibnamefont {{Dwelly}}}, \bibinfo {author}
  {\bibfnamefont {K.}~\bibnamefont {{Nandra}}}, \bibinfo {author}
  {\bibfnamefont {C.}~\bibnamefont {{Collins}}}, \bibinfo {author}
  {\bibfnamefont {K.}~\bibnamefont {{Dawson}}}, \bibinfo {author}
  {\bibfnamefont {J.~P.}\ \bibnamefont {{Kneib}}}, \bibinfo {author}
  {\bibfnamefont {E.}~\bibnamefont {{Rozo}}}, \bibinfo {author} {\bibfnamefont
  {E.}~\bibnamefont {{Rykoff}}}, \bibinfo {author} {\bibfnamefont
  {T.}~\bibnamefont {{Sadibekova}}}, \bibinfo {author} {\bibfnamefont
  {J.}~\bibnamefont {{Brownstein}}}, \bibinfo {author} {\bibfnamefont {Y.~T.}\
  \bibnamefont {{Lin}}}, \bibinfo {author} {\bibfnamefont {J.}~\bibnamefont
  {{Ridl}}}, \bibinfo {author} {\bibfnamefont {M.}~\bibnamefont {{Salvato}}},
  \bibinfo {author} {\bibfnamefont {A.}~\bibnamefont {{Schwope}}}, \bibinfo
  {author} {\bibfnamefont {M.}~\bibnamefont {{Steinmetz}}}, \bibinfo {author}
  {\bibfnamefont {H.~J.}\ \bibnamefont {{Seo}}}, \ and\ \bibinfo {author}
  {\bibfnamefont {J.}~\bibnamefont {{Tinker}}},\ }\href {\doibase
  10.1093/mnras/stw2214} {\bibfield  {journal} {\bibinfo  {journal} {\mnras}\
  }\textbf {\bibinfo {volume} {463}},\ \bibinfo {pages} {4490} (\bibinfo {year}
  {2016})},\ \Eprint {http://arxiv.org/abs/1608.08963} {arXiv:1608.08963
  [astro-ph.CO]} \BibitemShut {NoStop}%
\bibitem [{\citenamefont {{Dwelly}}\ \emph {et~al.}(2017)\citenamefont
  {{Dwelly}}, \citenamefont {{Salvato}}, \citenamefont {{Merloni}},
  \citenamefont {{Brusa}}, \citenamefont {{Buchner}}, \citenamefont
  {{Anderson}}, \citenamefont {{Boller}}, \citenamefont {{Brand t}},
  \citenamefont {{Budav{\'a}ri}}, \citenamefont {{Clerc}}, \citenamefont
  {{Coffey}}, \citenamefont {{Del Moro}}, \citenamefont {{Georgakakis}},
  \citenamefont {{Green}}, \citenamefont {{Jin}}, \citenamefont {{Menzel}},
  \citenamefont {{Myers}}, \citenamefont {{Nandra}}, \citenamefont {{Nichol}},
  \citenamefont {{Ridl}}, \citenamefont {{Schwope}},\ and\ \citenamefont
  {{Simm}}}]{dwelly17a}%
  \BibitemOpen
  \bibfield  {author} {\bibinfo {author} {\bibfnamefont {T.}~\bibnamefont
  {{Dwelly}}}, \bibinfo {author} {\bibfnamefont {M.}~\bibnamefont {{Salvato}}},
  \bibinfo {author} {\bibfnamefont {A.}~\bibnamefont {{Merloni}}}, \bibinfo
  {author} {\bibfnamefont {M.}~\bibnamefont {{Brusa}}}, \bibinfo {author}
  {\bibfnamefont {J.}~\bibnamefont {{Buchner}}}, \bibinfo {author}
  {\bibfnamefont {S.~F.}\ \bibnamefont {{Anderson}}}, \bibinfo {author}
  {\bibfnamefont {T.}~\bibnamefont {{Boller}}}, \bibinfo {author}
  {\bibfnamefont {W.~N.}\ \bibnamefont {{Brand t}}}, \bibinfo {author}
  {\bibfnamefont {T.}~\bibnamefont {{Budav{\'a}ri}}}, \bibinfo {author}
  {\bibfnamefont {N.}~\bibnamefont {{Clerc}}}, \bibinfo {author} {\bibfnamefont
  {D.}~\bibnamefont {{Coffey}}}, \bibinfo {author} {\bibfnamefont
  {A.}~\bibnamefont {{Del Moro}}}, \bibinfo {author} {\bibfnamefont
  {A.}~\bibnamefont {{Georgakakis}}}, \bibinfo {author} {\bibfnamefont {P.~J.}\
  \bibnamefont {{Green}}}, \bibinfo {author} {\bibfnamefont {C.}~\bibnamefont
  {{Jin}}}, \bibinfo {author} {\bibfnamefont {M.~L.}\ \bibnamefont {{Menzel}}},
  \bibinfo {author} {\bibfnamefont {A.~D.}\ \bibnamefont {{Myers}}}, \bibinfo
  {author} {\bibfnamefont {K.}~\bibnamefont {{Nandra}}}, \bibinfo {author}
  {\bibfnamefont {R.~C.}\ \bibnamefont {{Nichol}}}, \bibinfo {author}
  {\bibfnamefont {J.}~\bibnamefont {{Ridl}}}, \bibinfo {author} {\bibfnamefont
  {A.~D.}\ \bibnamefont {{Schwope}}}, \ and\ \bibinfo {author} {\bibfnamefont
  {T.}~\bibnamefont {{Simm}}},\ }\href {\doibase 10.1093/mnras/stx864}
  {\bibfield  {journal} {\bibinfo  {journal} {\mnras}\ }\textbf {\bibinfo
  {volume} {469}},\ \bibinfo {pages} {1065} (\bibinfo {year} {2017})},\ \Eprint
  {http://arxiv.org/abs/1704.01796} {arXiv:1704.01796 [astro-ph.GA]}
  \BibitemShut {NoStop}%
\bibitem [{\citenamefont {{Voges}}\ \emph {et~al.}(1999)\citenamefont
  {{Voges}}, \citenamefont {{Aschenbach}}, \citenamefont {{Boller}},
  \citenamefont {{Br{\"a}uninger}}, \citenamefont {{Briel}}, \citenamefont
  {{Burkert}}, \citenamefont {{Dennerl}}, \citenamefont {{Englhauser}},
  \citenamefont {{Gruber}}, \citenamefont {{Haberl}}, \citenamefont
  {{Hartner}}, \citenamefont {{Hasinger}}, \citenamefont {{K{\"u}rster}},
  \citenamefont {{Pfeffermann}}, \citenamefont {{Pietsch}}, \citenamefont
  {{Predehl}}, \citenamefont {{Rosso}}, \citenamefont {{Schmitt}},
  \citenamefont {{Tr{\"u}mper}},\ and\ \citenamefont
  {{Zimmermann}}}]{voges99a}%
  \BibitemOpen
  \bibfield  {author} {\bibinfo {author} {\bibfnamefont {W.}~\bibnamefont
  {{Voges}}}, \bibinfo {author} {\bibfnamefont {B.}~\bibnamefont
  {{Aschenbach}}}, \bibinfo {author} {\bibfnamefont {T.}~\bibnamefont
  {{Boller}}}, \bibinfo {author} {\bibfnamefont {H.}~\bibnamefont
  {{Br{\"a}uninger}}}, \bibinfo {author} {\bibfnamefont {U.}~\bibnamefont
  {{Briel}}}, \bibinfo {author} {\bibfnamefont {W.}~\bibnamefont {{Burkert}}},
  \bibinfo {author} {\bibfnamefont {K.}~\bibnamefont {{Dennerl}}}, \bibinfo
  {author} {\bibfnamefont {J.}~\bibnamefont {{Englhauser}}}, \bibinfo {author}
  {\bibfnamefont {R.}~\bibnamefont {{Gruber}}}, \bibinfo {author}
  {\bibfnamefont {F.}~\bibnamefont {{Haberl}}}, \bibinfo {author}
  {\bibfnamefont {G.}~\bibnamefont {{Hartner}}}, \bibinfo {author}
  {\bibfnamefont {G.}~\bibnamefont {{Hasinger}}}, \bibinfo {author}
  {\bibfnamefont {M.}~\bibnamefont {{K{\"u}rster}}}, \bibinfo {author}
  {\bibfnamefont {E.}~\bibnamefont {{Pfeffermann}}}, \bibinfo {author}
  {\bibfnamefont {W.}~\bibnamefont {{Pietsch}}}, \bibinfo {author}
  {\bibfnamefont {P.}~\bibnamefont {{Predehl}}}, \bibinfo {author}
  {\bibfnamefont {C.}~\bibnamefont {{Rosso}}}, \bibinfo {author} {\bibfnamefont
  {J.~H.~M.~M.}\ \bibnamefont {{Schmitt}}}, \bibinfo {author} {\bibfnamefont
  {J.}~\bibnamefont {{Tr{\"u}mper}}}, \ and\ \bibinfo {author} {\bibfnamefont
  {H.~U.}\ \bibnamefont {{Zimmermann}}},\ }\href@noop {} {\bibfield  {journal}
  {\bibinfo  {journal} {\aap}\ }\textbf {\bibinfo {volume} {349}},\ \bibinfo
  {pages} {389} (\bibinfo {year} {1999})},\ \Eprint
  {http://arxiv.org/abs/astro-ph/9909315} {arXiv:astro-ph/9909315 [astro-ph]}
  \BibitemShut {NoStop}%
\bibitem [{\citenamefont {{Voges}}\ \emph {et~al.}(2000)\citenamefont
  {{Voges}}, \citenamefont {{Aschenbach}}, \citenamefont {{Boller}},
  \citenamefont {{Brauninger}}, \citenamefont {{Briel}}, \citenamefont
  {{Burkert}}, \citenamefont {{Dennerl}}, \citenamefont {{Englhauser}},
  \citenamefont {{Gruber}}, \citenamefont {{Haberl}}, \citenamefont
  {{Hartner}}, \citenamefont {{Hasinger}}, \citenamefont {{Pfeffermann}},
  \citenamefont {{Pietsch}}, \citenamefont {{Predehl}}, \citenamefont
  {{Schmitt}}, \citenamefont {{Trumper}},\ and\ \citenamefont
  {{Zimmermann}}}]{voges00a}%
  \BibitemOpen
  \bibfield  {author} {\bibinfo {author} {\bibfnamefont {W.}~\bibnamefont
  {{Voges}}}, \bibinfo {author} {\bibfnamefont {B.}~\bibnamefont
  {{Aschenbach}}}, \bibinfo {author} {\bibfnamefont {T.}~\bibnamefont
  {{Boller}}}, \bibinfo {author} {\bibfnamefont {H.}~\bibnamefont
  {{Brauninger}}}, \bibinfo {author} {\bibfnamefont {U.}~\bibnamefont
  {{Briel}}}, \bibinfo {author} {\bibfnamefont {W.}~\bibnamefont {{Burkert}}},
  \bibinfo {author} {\bibfnamefont {K.}~\bibnamefont {{Dennerl}}}, \bibinfo
  {author} {\bibfnamefont {J.}~\bibnamefont {{Englhauser}}}, \bibinfo {author}
  {\bibfnamefont {R.}~\bibnamefont {{Gruber}}}, \bibinfo {author}
  {\bibfnamefont {F.}~\bibnamefont {{Haberl}}}, \bibinfo {author}
  {\bibfnamefont {G.}~\bibnamefont {{Hartner}}}, \bibinfo {author}
  {\bibfnamefont {G.}~\bibnamefont {{Hasinger}}}, \bibinfo {author}
  {\bibfnamefont {E.}~\bibnamefont {{Pfeffermann}}}, \bibinfo {author}
  {\bibfnamefont {W.}~\bibnamefont {{Pietsch}}}, \bibinfo {author}
  {\bibfnamefont {P.}~\bibnamefont {{Predehl}}}, \bibinfo {author}
  {\bibfnamefont {J.}~\bibnamefont {{Schmitt}}}, \bibinfo {author}
  {\bibfnamefont {J.}~\bibnamefont {{Trumper}}}, \ and\ \bibinfo {author}
  {\bibfnamefont {U.}~\bibnamefont {{Zimmermann}}},\ }\href@noop {} {\bibfield
  {journal} {\bibinfo  {journal} {VizieR Online Data Catalog}\ ,\ \bibinfo
  {eid} {IX/29}} (\bibinfo {year} {2000})}\BibitemShut {NoStop}%
\bibitem [{\citenamefont {{Jansen}}\ \emph {et~al.}(2001)\citenamefont
  {{Jansen}}, \citenamefont {{Lumb}}, \citenamefont {{Altieri}}, \citenamefont
  {{Clavel}}, \citenamefont {{Ehle}}, \citenamefont {{Erd}}, \citenamefont
  {{Gabriel}}, \citenamefont {{Guainazzi}}, \citenamefont {{Gondoin}},
  \citenamefont {{Much}}, \citenamefont {{Munoz}}, \citenamefont {{Santos}},
  \citenamefont {{Schartel}}, \citenamefont {{Texier}},\ and\ \citenamefont
  {{Vacanti}}}]{jansen01a}%
  \BibitemOpen
  \bibfield  {author} {\bibinfo {author} {\bibfnamefont {F.}~\bibnamefont
  {{Jansen}}}, \bibinfo {author} {\bibfnamefont {D.}~\bibnamefont {{Lumb}}},
  \bibinfo {author} {\bibfnamefont {B.}~\bibnamefont {{Altieri}}}, \bibinfo
  {author} {\bibfnamefont {J.}~\bibnamefont {{Clavel}}}, \bibinfo {author}
  {\bibfnamefont {M.}~\bibnamefont {{Ehle}}}, \bibinfo {author} {\bibfnamefont
  {C.}~\bibnamefont {{Erd}}}, \bibinfo {author} {\bibfnamefont
  {C.}~\bibnamefont {{Gabriel}}}, \bibinfo {author} {\bibfnamefont
  {M.}~\bibnamefont {{Guainazzi}}}, \bibinfo {author} {\bibfnamefont
  {P.}~\bibnamefont {{Gondoin}}}, \bibinfo {author} {\bibfnamefont
  {R.}~\bibnamefont {{Much}}}, \bibinfo {author} {\bibfnamefont
  {R.}~\bibnamefont {{Munoz}}}, \bibinfo {author} {\bibfnamefont
  {M.}~\bibnamefont {{Santos}}}, \bibinfo {author} {\bibfnamefont
  {N.}~\bibnamefont {{Schartel}}}, \bibinfo {author} {\bibfnamefont
  {D.}~\bibnamefont {{Texier}}}, \ and\ \bibinfo {author} {\bibfnamefont
  {G.}~\bibnamefont {{Vacanti}}},\ }\href {\doibase 10.1051/0004-6361:20000036}
  {\bibfield  {journal} {\bibinfo  {journal} {\aap}\ }\textbf {\bibinfo
  {volume} {365}},\ \bibinfo {pages} {L1} (\bibinfo {year} {2001})}\BibitemShut
  {NoStop}%
\bibitem [{\citenamefont {{Shen}}\ \emph {et~al.}(2015)\citenamefont {{Shen}},
  \citenamefont {{Brandt}}, \citenamefont {{Dawson}}, \citenamefont {{Hall}},
  \citenamefont {{McGreer}}, \citenamefont {{Anderson}}, \citenamefont
  {{Chen}}, \citenamefont {{Denney}}, \citenamefont {{Eftekharzadeh}},
  \citenamefont {{Fan}}, \citenamefont {{Gao}}, \citenamefont {{Green}},
  \citenamefont {{Greene}}, \citenamefont {{Ho}}, \citenamefont {{Horne}},
  \citenamefont {{Jiang}}, \citenamefont {{Kelly}}, \citenamefont
  {{Kinemuchi}}, \citenamefont {{Kochanek}}, \citenamefont {{P{\^a}ris}},
  \citenamefont {{Peters}}, \citenamefont {{Peterson}}, \citenamefont
  {{Petitjean}}, \citenamefont {{Ponder}}, \citenamefont {{Richards}},
  \citenamefont {{Schneider}}, \citenamefont {{Seth}}, \citenamefont {{Smith}},
  \citenamefont {{Strauss}}, \citenamefont {{Tao}}, \citenamefont {{Trump}},
  \citenamefont {{Wood-Vasey}}, \citenamefont {{Zu}}, \citenamefont
  {{Eisenstein}}, \citenamefont {{Pan}}, \citenamefont {{Bizyaev}},
  \citenamefont {{Malanushenko}}, \citenamefont {{Malanushenko}},\ and\
  \citenamefont {{Oravetz}}}]{shen15a}%
  \BibitemOpen
  \bibfield  {author} {\bibinfo {author} {\bibfnamefont {Y.}~\bibnamefont
  {{Shen}}}, \bibinfo {author} {\bibfnamefont {W.~N.}\ \bibnamefont
  {{Brandt}}}, \bibinfo {author} {\bibfnamefont {K.~S.}\ \bibnamefont
  {{Dawson}}}, \bibinfo {author} {\bibfnamefont {P.~B.}\ \bibnamefont
  {{Hall}}}, \bibinfo {author} {\bibfnamefont {I.~D.}\ \bibnamefont
  {{McGreer}}}, \bibinfo {author} {\bibfnamefont {S.~F.}\ \bibnamefont
  {{Anderson}}}, \bibinfo {author} {\bibfnamefont {Y.}~\bibnamefont {{Chen}}},
  \bibinfo {author} {\bibfnamefont {K.~D.}\ \bibnamefont {{Denney}}}, \bibinfo
  {author} {\bibfnamefont {S.}~\bibnamefont {{Eftekharzadeh}}}, \bibinfo
  {author} {\bibfnamefont {X.}~\bibnamefont {{Fan}}}, \bibinfo {author}
  {\bibfnamefont {Y.}~\bibnamefont {{Gao}}}, \bibinfo {author} {\bibfnamefont
  {P.~J.}\ \bibnamefont {{Green}}}, \bibinfo {author} {\bibfnamefont {J.~E.}\
  \bibnamefont {{Greene}}}, \bibinfo {author} {\bibfnamefont {L.~C.}\
  \bibnamefont {{Ho}}}, \bibinfo {author} {\bibfnamefont {K.}~\bibnamefont
  {{Horne}}}, \bibinfo {author} {\bibfnamefont {L.}~\bibnamefont {{Jiang}}},
  \bibinfo {author} {\bibfnamefont {B.~C.}\ \bibnamefont {{Kelly}}}, \bibinfo
  {author} {\bibfnamefont {K.}~\bibnamefont {{Kinemuchi}}}, \bibinfo {author}
  {\bibfnamefont {C.~S.}\ \bibnamefont {{Kochanek}}}, \bibinfo {author}
  {\bibfnamefont {I.}~\bibnamefont {{P{\^a}ris}}}, \bibinfo {author}
  {\bibfnamefont {C.~M.}\ \bibnamefont {{Peters}}}, \bibinfo {author}
  {\bibfnamefont {B.~M.}\ \bibnamefont {{Peterson}}}, \bibinfo {author}
  {\bibfnamefont {P.}~\bibnamefont {{Petitjean}}}, \bibinfo {author}
  {\bibfnamefont {K.}~\bibnamefont {{Ponder}}}, \bibinfo {author}
  {\bibfnamefont {G.~T.}\ \bibnamefont {{Richards}}}, \bibinfo {author}
  {\bibfnamefont {D.~P.}\ \bibnamefont {{Schneider}}}, \bibinfo {author}
  {\bibfnamefont {A.}~\bibnamefont {{Seth}}}, \bibinfo {author} {\bibfnamefont
  {R.~N.}\ \bibnamefont {{Smith}}}, \bibinfo {author} {\bibfnamefont {M.~A.}\
  \bibnamefont {{Strauss}}}, \bibinfo {author} {\bibfnamefont {C.}~\bibnamefont
  {{Tao}}}, \bibinfo {author} {\bibfnamefont {J.~R.}\ \bibnamefont {{Trump}}},
  \bibinfo {author} {\bibfnamefont {W.~M.}\ \bibnamefont {{Wood-Vasey}}},
  \bibinfo {author} {\bibfnamefont {Y.}~\bibnamefont {{Zu}}}, \bibinfo {author}
  {\bibfnamefont {D.~J.}\ \bibnamefont {{Eisenstein}}}, \bibinfo {author}
  {\bibfnamefont {K.}~\bibnamefont {{Pan}}}, \bibinfo {author} {\bibfnamefont
  {D.}~\bibnamefont {{Bizyaev}}}, \bibinfo {author} {\bibfnamefont
  {V.}~\bibnamefont {{Malanushenko}}}, \bibinfo {author} {\bibfnamefont
  {E.}~\bibnamefont {{Malanushenko}}}, \ and\ \bibinfo {author} {\bibfnamefont
  {D.}~\bibnamefont {{Oravetz}}},\ }\href {\doibase 10.1088/0067-0049/216/1/4}
  {\bibfield  {journal} {\bibinfo  {journal} {\apjs}\ }\textbf {\bibinfo
  {volume} {216}},\ \bibinfo {eid} {4} (\bibinfo {year} {2015})},\ \Eprint
  {http://arxiv.org/abs/1408.5970} {arXiv:1408.5970 [astro-ph.IM]} \BibitemShut
  {NoStop}%
\bibitem [{\citenamefont {{Grier}}\ \emph {et~al.}(2017)\citenamefont
  {{Grier}}, \citenamefont {{Trump}}, \citenamefont {{Shen}}, \citenamefont
  {{Horne}}, \citenamefont {{Kinemuchi}}, \citenamefont {{McGreer}},
  \citenamefont {{Starkey}}, \citenamefont {{Brand t}}, \citenamefont {{Hall}},
  \citenamefont {{Kochanek}}, \citenamefont {{Chen}}, \citenamefont {{Denney}},
  \citenamefont {{Greene}}, \citenamefont {{Ho}}, \citenamefont {{Homayouni}},
  \citenamefont {{I-Hsiu Li}}, \citenamefont {{Pei}}, \citenamefont
  {{Peterson}}, \citenamefont {{Petitjean}}, \citenamefont {{Schneider}},
  \citenamefont {{Sun}}, \citenamefont {{AlSayyad}}, \citenamefont {{Bizyaev}},
  \citenamefont {{Brinkmann}}, \citenamefont {{Brownstein}}, \citenamefont
  {{Bundy}}, \citenamefont {{Dawson}}, \citenamefont {{Eftekharzadeh}},
  \citenamefont {{Fernand ez-Trincado}}, \citenamefont {{Gao}}, \citenamefont
  {{Hutchinson}}, \citenamefont {{Jia}}, \citenamefont {{Jiang}}, \citenamefont
  {{Oravetz}}, \citenamefont {{Pan}}, \citenamefont {{Paris}}, \citenamefont
  {{Ponder}}, \citenamefont {{Peters}}, \citenamefont {{Rogerson}},
  \citenamefont {{Simmons}}, \citenamefont {{Smith}},\ and\ \citenamefont
  {{Wang}}}]{grier17a}%
  \BibitemOpen
  \bibfield  {author} {\bibinfo {author} {\bibfnamefont {C.~J.}\ \bibnamefont
  {{Grier}}}, \bibinfo {author} {\bibfnamefont {J.~R.}\ \bibnamefont
  {{Trump}}}, \bibinfo {author} {\bibfnamefont {Y.}~\bibnamefont {{Shen}}},
  \bibinfo {author} {\bibfnamefont {K.}~\bibnamefont {{Horne}}}, \bibinfo
  {author} {\bibfnamefont {K.}~\bibnamefont {{Kinemuchi}}}, \bibinfo {author}
  {\bibfnamefont {I.~D.}\ \bibnamefont {{McGreer}}}, \bibinfo {author}
  {\bibfnamefont {D.~A.}\ \bibnamefont {{Starkey}}}, \bibinfo {author}
  {\bibfnamefont {W.~N.}\ \bibnamefont {{Brand t}}}, \bibinfo {author}
  {\bibfnamefont {P.~B.}\ \bibnamefont {{Hall}}}, \bibinfo {author}
  {\bibfnamefont {C.~S.}\ \bibnamefont {{Kochanek}}}, \bibinfo {author}
  {\bibfnamefont {Y.}~\bibnamefont {{Chen}}}, \bibinfo {author} {\bibfnamefont
  {K.~D.}\ \bibnamefont {{Denney}}}, \bibinfo {author} {\bibfnamefont {J.~E.}\
  \bibnamefont {{Greene}}}, \bibinfo {author} {\bibfnamefont {L.~C.}\
  \bibnamefont {{Ho}}}, \bibinfo {author} {\bibfnamefont {Y.}~\bibnamefont
  {{Homayouni}}}, \bibinfo {author} {\bibfnamefont {J.}~\bibnamefont {{I-Hsiu
  Li}}}, \bibinfo {author} {\bibfnamefont {L.}~\bibnamefont {{Pei}}}, \bibinfo
  {author} {\bibfnamefont {B.~M.}\ \bibnamefont {{Peterson}}}, \bibinfo
  {author} {\bibfnamefont {P.}~\bibnamefont {{Petitjean}}}, \bibinfo {author}
  {\bibfnamefont {D.~P.}\ \bibnamefont {{Schneider}}}, \bibinfo {author}
  {\bibfnamefont {M.}~\bibnamefont {{Sun}}}, \bibinfo {author} {\bibfnamefont
  {Y.}~\bibnamefont {{AlSayyad}}}, \bibinfo {author} {\bibfnamefont
  {D.}~\bibnamefont {{Bizyaev}}}, \bibinfo {author} {\bibfnamefont
  {J.}~\bibnamefont {{Brinkmann}}}, \bibinfo {author} {\bibfnamefont {J.~R.}\
  \bibnamefont {{Brownstein}}}, \bibinfo {author} {\bibfnamefont
  {K.}~\bibnamefont {{Bundy}}}, \bibinfo {author} {\bibfnamefont {K.~S.}\
  \bibnamefont {{Dawson}}}, \bibinfo {author} {\bibfnamefont {S.}~\bibnamefont
  {{Eftekharzadeh}}}, \bibinfo {author} {\bibfnamefont {J.~G.}\ \bibnamefont
  {{Fernand ez-Trincado}}}, \bibinfo {author} {\bibfnamefont {Y.}~\bibnamefont
  {{Gao}}}, \bibinfo {author} {\bibfnamefont {T.~A.}\ \bibnamefont
  {{Hutchinson}}}, \bibinfo {author} {\bibfnamefont {S.}~\bibnamefont {{Jia}}},
  \bibinfo {author} {\bibfnamefont {L.}~\bibnamefont {{Jiang}}}, \bibinfo
  {author} {\bibfnamefont {D.}~\bibnamefont {{Oravetz}}}, \bibinfo {author}
  {\bibfnamefont {K.}~\bibnamefont {{Pan}}}, \bibinfo {author} {\bibfnamefont
  {I.}~\bibnamefont {{Paris}}}, \bibinfo {author} {\bibfnamefont {K.~A.}\
  \bibnamefont {{Ponder}}}, \bibinfo {author} {\bibfnamefont {C.}~\bibnamefont
  {{Peters}}}, \bibinfo {author} {\bibfnamefont {J.}~\bibnamefont
  {{Rogerson}}}, \bibinfo {author} {\bibfnamefont {A.}~\bibnamefont
  {{Simmons}}}, \bibinfo {author} {\bibfnamefont {R.}~\bibnamefont {{Smith}}},
  \ and\ \bibinfo {author} {\bibfnamefont {R.}~\bibnamefont {{Wang}}},\ }\href
  {\doibase 10.3847/1538-4357/aa98dc} {\bibfield  {journal} {\bibinfo
  {journal} {\apj}\ }\textbf {\bibinfo {volume} {851}},\ \bibinfo {eid} {21}
  (\bibinfo {year} {2017})},\ \Eprint {http://arxiv.org/abs/1711.03114}
  {arXiv:1711.03114 [astro-ph.GA]} \BibitemShut {NoStop}%
\bibitem [{\citenamefont {{Homayouni}}\ \emph {et~al.}(2020)\citenamefont
  {{Homayouni}}, \citenamefont {{Trump}}, \citenamefont {{Grier}},
  \citenamefont {{Horne}}, \citenamefont {{Shen}}, \citenamefont {{Brandt}},
  \citenamefont {{Dawson}}, \citenamefont {{Alvarez}}, \citenamefont {{Green}},
  \citenamefont {{Hall}}, \citenamefont {{Hern{\'a}ndez Santisteban}},
  \citenamefont {{Ho}}, \citenamefont {{Kinemuchi}}, \citenamefont
  {{Kochanek}}, \citenamefont {{Li}}, \citenamefont {{Peterson}}, \citenamefont
  {{Schneider}}, \citenamefont {{Starkey}}, \citenamefont {{Bizyaev}},
  \citenamefont {{Pan}}, \citenamefont {{Oravetz}},\ and\ \citenamefont
  {{Simmons}}}]{homayouni20a}%
  \BibitemOpen
  \bibfield  {author} {\bibinfo {author} {\bibfnamefont {Y.}~\bibnamefont
  {{Homayouni}}}, \bibinfo {author} {\bibfnamefont {J.~R.}\ \bibnamefont
  {{Trump}}}, \bibinfo {author} {\bibfnamefont {C.~J.}\ \bibnamefont
  {{Grier}}}, \bibinfo {author} {\bibfnamefont {K.}~\bibnamefont {{Horne}}},
  \bibinfo {author} {\bibfnamefont {Y.}~\bibnamefont {{Shen}}}, \bibinfo
  {author} {\bibfnamefont {W.~N.}\ \bibnamefont {{Brandt}}}, \bibinfo {author}
  {\bibfnamefont {K.~S.}\ \bibnamefont {{Dawson}}}, \bibinfo {author}
  {\bibfnamefont {G.~F.}\ \bibnamefont {{Alvarez}}}, \bibinfo {author}
  {\bibfnamefont {P.~J.}\ \bibnamefont {{Green}}}, \bibinfo {author}
  {\bibfnamefont {P.~B.}\ \bibnamefont {{Hall}}}, \bibinfo {author}
  {\bibfnamefont {J.~V.}\ \bibnamefont {{Hern{\'a}ndez Santisteban}}}, \bibinfo
  {author} {\bibfnamefont {L.~C.}\ \bibnamefont {{Ho}}}, \bibinfo {author}
  {\bibfnamefont {K.}~\bibnamefont {{Kinemuchi}}}, \bibinfo {author}
  {\bibfnamefont {C.~S.}\ \bibnamefont {{Kochanek}}}, \bibinfo {author}
  {\bibfnamefont {J.~I.~H.}\ \bibnamefont {{Li}}}, \bibinfo {author}
  {\bibfnamefont {B.~M.}\ \bibnamefont {{Peterson}}}, \bibinfo {author}
  {\bibfnamefont {D.~P.}\ \bibnamefont {{Schneider}}}, \bibinfo {author}
  {\bibfnamefont {D.~A.}\ \bibnamefont {{Starkey}}}, \bibinfo {author}
  {\bibfnamefont {D.}~\bibnamefont {{Bizyaev}}}, \bibinfo {author}
  {\bibfnamefont {K.}~\bibnamefont {{Pan}}}, \bibinfo {author} {\bibfnamefont
  {D.}~\bibnamefont {{Oravetz}}}, \ and\ \bibinfo {author} {\bibfnamefont
  {A.}~\bibnamefont {{Simmons}}},\ }\href {\doibase 10.3847/1538-4357/ababa9}
  {\bibfield  {journal} {\bibinfo  {journal} {\apj}\ }\textbf {\bibinfo
  {volume} {901}},\ \bibinfo {eid} {55} (\bibinfo {year} {2020})},\ \Eprint
  {http://arxiv.org/abs/2005.03663} {arXiv:2005.03663 [astro-ph.GA]}
  \BibitemShut {NoStop}%
\bibitem [{\citenamefont {{Grier}}\ \emph {et~al.}(2019)\citenamefont
  {{Grier}}, \citenamefont {{Shen}}, \citenamefont {{Horne}}, \citenamefont
  {{Brandt}}, \citenamefont {{Trump}}, \citenamefont {{Hall}}, \citenamefont
  {{Kinemuchi}}, \citenamefont {{Starkey}}, \citenamefont {{Schneider}},
  \citenamefont {{Ho}}, \citenamefont {{Homayouni}}, \citenamefont {{I-Hsiu
  Li}}, \citenamefont {{McGreer}}, \citenamefont {{Peterson}}, \citenamefont
  {{Bizyaev}}, \citenamefont {{Chen}}, \citenamefont {{Dawson}}, \citenamefont
  {{Eftekharzadeh}}, \citenamefont {{Guo}}, \citenamefont {{Jia}},
  \citenamefont {{Jiang}}, \citenamefont {{Kneib}}, \citenamefont {{Li}},
  \citenamefont {{Li}}, \citenamefont {{Nie}}, \citenamefont {{Oravetz}},
  \citenamefont {{Oravetz}}, \citenamefont {{Pan}}, \citenamefont
  {{Petitjean}}, \citenamefont {{Ponder}}, \citenamefont {{Rogerson}},
  \citenamefont {{Vivek}}, \citenamefont {{Zhang}},\ and\ \citenamefont
  {{Zou}}}]{grier19a}%
  \BibitemOpen
  \bibfield  {author} {\bibinfo {author} {\bibfnamefont {C.~J.}\ \bibnamefont
  {{Grier}}}, \bibinfo {author} {\bibfnamefont {Y.}~\bibnamefont {{Shen}}},
  \bibinfo {author} {\bibfnamefont {K.}~\bibnamefont {{Horne}}}, \bibinfo
  {author} {\bibfnamefont {W.~N.}\ \bibnamefont {{Brandt}}}, \bibinfo {author}
  {\bibfnamefont {J.~R.}\ \bibnamefont {{Trump}}}, \bibinfo {author}
  {\bibfnamefont {P.~B.}\ \bibnamefont {{Hall}}}, \bibinfo {author}
  {\bibfnamefont {K.}~\bibnamefont {{Kinemuchi}}}, \bibinfo {author}
  {\bibfnamefont {D.}~\bibnamefont {{Starkey}}}, \bibinfo {author}
  {\bibfnamefont {D.~P.}\ \bibnamefont {{Schneider}}}, \bibinfo {author}
  {\bibfnamefont {L.~C.}\ \bibnamefont {{Ho}}}, \bibinfo {author}
  {\bibfnamefont {Y.}~\bibnamefont {{Homayouni}}}, \bibinfo {author}
  {\bibfnamefont {J.}~\bibnamefont {{I-Hsiu Li}}}, \bibinfo {author}
  {\bibfnamefont {I.~D.}\ \bibnamefont {{McGreer}}}, \bibinfo {author}
  {\bibfnamefont {B.~M.}\ \bibnamefont {{Peterson}}}, \bibinfo {author}
  {\bibfnamefont {D.}~\bibnamefont {{Bizyaev}}}, \bibinfo {author}
  {\bibfnamefont {Y.}~\bibnamefont {{Chen}}}, \bibinfo {author} {\bibfnamefont
  {K.~S.}\ \bibnamefont {{Dawson}}}, \bibinfo {author} {\bibfnamefont
  {S.}~\bibnamefont {{Eftekharzadeh}}}, \bibinfo {author} {\bibfnamefont
  {Y.}~\bibnamefont {{Guo}}}, \bibinfo {author} {\bibfnamefont
  {S.}~\bibnamefont {{Jia}}}, \bibinfo {author} {\bibfnamefont
  {L.}~\bibnamefont {{Jiang}}}, \bibinfo {author} {\bibfnamefont {J.-P.}\
  \bibnamefont {{Kneib}}}, \bibinfo {author} {\bibfnamefont {F.}~\bibnamefont
  {{Li}}}, \bibinfo {author} {\bibfnamefont {Z.}~\bibnamefont {{Li}}}, \bibinfo
  {author} {\bibfnamefont {J.}~\bibnamefont {{Nie}}}, \bibinfo {author}
  {\bibfnamefont {A.}~\bibnamefont {{Oravetz}}}, \bibinfo {author}
  {\bibfnamefont {D.}~\bibnamefont {{Oravetz}}}, \bibinfo {author}
  {\bibfnamefont {K.}~\bibnamefont {{Pan}}}, \bibinfo {author} {\bibfnamefont
  {P.}~\bibnamefont {{Petitjean}}}, \bibinfo {author} {\bibfnamefont {K.~A.}\
  \bibnamefont {{Ponder}}}, \bibinfo {author} {\bibfnamefont {J.}~\bibnamefont
  {{Rogerson}}}, \bibinfo {author} {\bibfnamefont {M.}~\bibnamefont {{Vivek}}},
  \bibinfo {author} {\bibfnamefont {T.}~\bibnamefont {{Zhang}}}, \ and\
  \bibinfo {author} {\bibfnamefont {H.}~\bibnamefont {{Zou}}},\ }\href
  {\doibase 10.3847/1538-4357/ab4ea5} {\bibfield  {journal} {\bibinfo
  {journal} {\apj}\ }\textbf {\bibinfo {volume} {887}},\ \bibinfo {eid} {38}
  (\bibinfo {year} {2019})},\ \Eprint {http://arxiv.org/abs/1904.03199}
  {arXiv:1904.03199 [astro-ph.GA]} \BibitemShut {NoStop}%
\bibitem [{\citenamefont {{LaMassa}}\ \emph {et~al.}(2015)\citenamefont
  {{LaMassa}}, \citenamefont {{Cales}}, \citenamefont {{Moran}}, \citenamefont
  {{Myers}}, \citenamefont {{Richards}}, \citenamefont {{Eracleous}},
  \citenamefont {{Heckman}}, \citenamefont {{Gallo}},\ and\ \citenamefont
  {{Urry}}}]{lamassa15a}%
  \BibitemOpen
  \bibfield  {author} {\bibinfo {author} {\bibfnamefont {S.~M.}\ \bibnamefont
  {{LaMassa}}}, \bibinfo {author} {\bibfnamefont {S.}~\bibnamefont {{Cales}}},
  \bibinfo {author} {\bibfnamefont {E.~C.}\ \bibnamefont {{Moran}}}, \bibinfo
  {author} {\bibfnamefont {A.~D.}\ \bibnamefont {{Myers}}}, \bibinfo {author}
  {\bibfnamefont {G.~T.}\ \bibnamefont {{Richards}}}, \bibinfo {author}
  {\bibfnamefont {M.}~\bibnamefont {{Eracleous}}}, \bibinfo {author}
  {\bibfnamefont {T.~M.}\ \bibnamefont {{Heckman}}}, \bibinfo {author}
  {\bibfnamefont {L.}~\bibnamefont {{Gallo}}}, \ and\ \bibinfo {author}
  {\bibfnamefont {C.~M.}\ \bibnamefont {{Urry}}},\ }\href {\doibase
  10.1088/0004-637X/800/2/144} {\bibfield  {journal} {\bibinfo  {journal}
  {\apj}\ }\textbf {\bibinfo {volume} {800}},\ \bibinfo {eid} {144} (\bibinfo
  {year} {2015})},\ \Eprint {http://arxiv.org/abs/1412.2136} {arXiv:1412.2136
  [astro-ph.GA]} \BibitemShut {NoStop}%
\bibitem [{\citenamefont {{MacLeod}}\ \emph {et~al.}(2016)\citenamefont
  {{MacLeod}}, \citenamefont {{Ross}}, \citenamefont {{Lawrence}},
  \citenamefont {{Goad}}, \citenamefont {{Horne}}, \citenamefont {{Burgett}},
  \citenamefont {{Chambers}}, \citenamefont {{Flewelling}}, \citenamefont
  {{Hodapp}}, \citenamefont {{Kaiser}}, \citenamefont {{Magnier}},
  \citenamefont {{Wainscoat}},\ and\ \citenamefont {{Waters}}}]{macleod16a}%
  \BibitemOpen
  \bibfield  {author} {\bibinfo {author} {\bibfnamefont {C.~L.}\ \bibnamefont
  {{MacLeod}}}, \bibinfo {author} {\bibfnamefont {N.~P.}\ \bibnamefont
  {{Ross}}}, \bibinfo {author} {\bibfnamefont {A.}~\bibnamefont {{Lawrence}}},
  \bibinfo {author} {\bibfnamefont {M.}~\bibnamefont {{Goad}}}, \bibinfo
  {author} {\bibfnamefont {K.}~\bibnamefont {{Horne}}}, \bibinfo {author}
  {\bibfnamefont {W.}~\bibnamefont {{Burgett}}}, \bibinfo {author}
  {\bibfnamefont {K.~C.}\ \bibnamefont {{Chambers}}}, \bibinfo {author}
  {\bibfnamefont {H.}~\bibnamefont {{Flewelling}}}, \bibinfo {author}
  {\bibfnamefont {K.}~\bibnamefont {{Hodapp}}}, \bibinfo {author}
  {\bibfnamefont {N.}~\bibnamefont {{Kaiser}}}, \bibinfo {author}
  {\bibfnamefont {E.}~\bibnamefont {{Magnier}}}, \bibinfo {author}
  {\bibfnamefont {R.}~\bibnamefont {{Wainscoat}}}, \ and\ \bibinfo {author}
  {\bibfnamefont {C.}~\bibnamefont {{Waters}}},\ }\href {\doibase
  10.1093/mnras/stv2997} {\bibfield  {journal} {\bibinfo  {journal} {\mnras}\
  }\textbf {\bibinfo {volume} {457}},\ \bibinfo {pages} {389} (\bibinfo {year}
  {2016})},\ \Eprint {http://arxiv.org/abs/1509.08393} {arXiv:1509.08393
  [astro-ph.GA]} \BibitemShut {NoStop}%
\bibitem [{\citenamefont {{Dexter}}\ \emph {et~al.}(2019)\citenamefont
  {{Dexter}}, \citenamefont {{Xin}}, \citenamefont {{Shen}}, \citenamefont
  {{Grier}}, \citenamefont {{Liu}}, \citenamefont {{Gezari}}, \citenamefont
  {{McGreer}}, \citenamefont {{Brand t}}, \citenamefont {{Hall}}, \citenamefont
  {{Horne}}, \citenamefont {{Simm}}, \citenamefont {{Merloni}}, \citenamefont
  {{Green}}, \citenamefont {{Vivek}}, \citenamefont {{Trump}}, \citenamefont
  {{Homayouni}}, \citenamefont {{Peterson}}, \citenamefont {{Schneider}},
  \citenamefont {{Kinemuchi}}, \citenamefont {{Pan}},\ and\ \citenamefont
  {{Bizyaev}}}]{dexter19a}%
  \BibitemOpen
  \bibfield  {author} {\bibinfo {author} {\bibfnamefont {J.}~\bibnamefont
  {{Dexter}}}, \bibinfo {author} {\bibfnamefont {S.}~\bibnamefont {{Xin}}},
  \bibinfo {author} {\bibfnamefont {Y.}~\bibnamefont {{Shen}}}, \bibinfo
  {author} {\bibfnamefont {C.~J.}\ \bibnamefont {{Grier}}}, \bibinfo {author}
  {\bibfnamefont {T.}~\bibnamefont {{Liu}}}, \bibinfo {author} {\bibfnamefont
  {S.}~\bibnamefont {{Gezari}}}, \bibinfo {author} {\bibfnamefont {I.~D.}\
  \bibnamefont {{McGreer}}}, \bibinfo {author} {\bibfnamefont {W.~N.}\
  \bibnamefont {{Brand t}}}, \bibinfo {author} {\bibfnamefont {P.~B.}\
  \bibnamefont {{Hall}}}, \bibinfo {author} {\bibfnamefont {K.}~\bibnamefont
  {{Horne}}}, \bibinfo {author} {\bibfnamefont {T.}~\bibnamefont {{Simm}}},
  \bibinfo {author} {\bibfnamefont {A.}~\bibnamefont {{Merloni}}}, \bibinfo
  {author} {\bibfnamefont {P.~J.}\ \bibnamefont {{Green}}}, \bibinfo {author}
  {\bibfnamefont {M.}~\bibnamefont {{Vivek}}}, \bibinfo {author} {\bibfnamefont
  {J.~R.}\ \bibnamefont {{Trump}}}, \bibinfo {author} {\bibfnamefont
  {Y.}~\bibnamefont {{Homayouni}}}, \bibinfo {author} {\bibfnamefont {B.~M.}\
  \bibnamefont {{Peterson}}}, \bibinfo {author} {\bibfnamefont {D.~P.}\
  \bibnamefont {{Schneider}}}, \bibinfo {author} {\bibfnamefont
  {K.}~\bibnamefont {{Kinemuchi}}}, \bibinfo {author} {\bibfnamefont
  {K.}~\bibnamefont {{Pan}}}, \ and\ \bibinfo {author} {\bibfnamefont
  {D.}~\bibnamefont {{Bizyaev}}},\ }\href {\doibase 10.3847/1538-4357/ab4354}
  {\bibfield  {journal} {\bibinfo  {journal} {\apj}\ }\textbf {\bibinfo
  {volume} {885}},\ \bibinfo {eid} {44} (\bibinfo {year} {2019})},\ \Eprint
  {http://arxiv.org/abs/1906.10138} {arXiv:1906.10138 [astro-ph.GA]}
  \BibitemShut {NoStop}%
\bibitem [{\citenamefont {{Majewski}}\ \emph {et~al.}(2017)\citenamefont
  {{Majewski}}, \citenamefont {{Schiavon}}, \citenamefont {{Frinchaboy}},
  \citenamefont {{Allende Prieto}}, \citenamefont {{Barkhouser}}, \citenamefont
  {{Bizyaev}}, \citenamefont {{Blank}}, \citenamefont {{Brunner}},
  \citenamefont {{Burton}}, \citenamefont {{Carrera}}, \citenamefont
  {{Chojnowski}}, \citenamefont {{Cunha}}, \citenamefont {{Epstein}},
  \citenamefont {{Fitzgerald}}, \citenamefont {{Garc{\'{\i}}a P{\'e}rez}},
  \citenamefont {{Hearty}}, \citenamefont {{Henderson}}, \citenamefont
  {{Holtzman}}, \citenamefont {{Johnson}}, \citenamefont {{Lam}}, \citenamefont
  {{Lawler}}, \citenamefont {{Maseman}}, \citenamefont {{M{\'e}sz{\'a}ros}},
  \citenamefont {{Nelson}}, \citenamefont {{Nguyen}}, \citenamefont
  {{Nidever}}, \citenamefont {{Pinsonneault}}, \citenamefont {{Shetrone}},
  \citenamefont {{Smee}}, \citenamefont {{Smith}}, \citenamefont {{Stolberg}},
  \citenamefont {{Skrutskie}}, \citenamefont {{Walker}}, \citenamefont
  {{Wilson}}, \citenamefont {{Zasowski}}, \citenamefont {{Anders}},
  \citenamefont {{Basu}}, \citenamefont {{Beland}}, \citenamefont {{Blanton}},
  \citenamefont {{Bovy}}, \citenamefont {{Brownstein}}, \citenamefont
  {{Carlberg}}, \citenamefont {{Chaplin}}, \citenamefont {{Chiappini}},
  \citenamefont {{Eisenstein}}, \citenamefont {{Elsworth}}, \citenamefont
  {{Feuillet}}, \citenamefont {{Fleming}}, \citenamefont {{Galbraith-Frew}},
  \citenamefont {{Garc{\'{\i}}a}}, \citenamefont
  {{Garc{\'{\i}}a-Hern{\'a}ndez}}, \citenamefont {{Gillespie}}, \citenamefont
  {{Girardi}}, \citenamefont {{Gunn}}, \citenamefont {{Hasselquist}},
  \citenamefont {{Hayden}}, \citenamefont {{Hekker}}, \citenamefont {{Ivans}},
  \citenamefont {{Kinemuchi}}, \citenamefont {{Klaene}}, \citenamefont
  {{Mahadevan}}, \citenamefont {{Mathur}}, \citenamefont {{Mosser}},
  \citenamefont {{Muna}}, \citenamefont {{Munn}}, \citenamefont {{Nichol}},
  \citenamefont {{O'Connell}}, \citenamefont {{Parejko}}, \citenamefont
  {{Robin}}, \citenamefont {{Rocha-Pinto}}, \citenamefont {{Schultheis}},
  \citenamefont {{Serenelli}}, \citenamefont {{Shane}}, \citenamefont {{Silva
  Aguirre}}, \citenamefont {{Sobeck}}, \citenamefont {{Thompson}},
  \citenamefont {{Troup}}, \citenamefont {{Weinberg}},\ and\ \citenamefont
  {{Zamora}}}]{majewski17a}%
  \BibitemOpen
  \bibfield  {author} {\bibinfo {author} {\bibfnamefont {S.~R.}\ \bibnamefont
  {{Majewski}}}, \bibinfo {author} {\bibfnamefont {R.~P.}\ \bibnamefont
  {{Schiavon}}}, \bibinfo {author} {\bibfnamefont {P.~M.}\ \bibnamefont
  {{Frinchaboy}}}, \bibinfo {author} {\bibfnamefont {C.}~\bibnamefont {{Allende
  Prieto}}}, \bibinfo {author} {\bibfnamefont {R.}~\bibnamefont
  {{Barkhouser}}}, \bibinfo {author} {\bibfnamefont {D.}~\bibnamefont
  {{Bizyaev}}}, \bibinfo {author} {\bibfnamefont {B.}~\bibnamefont {{Blank}}},
  \bibinfo {author} {\bibfnamefont {S.}~\bibnamefont {{Brunner}}}, \bibinfo
  {author} {\bibfnamefont {A.}~\bibnamefont {{Burton}}}, \bibinfo {author}
  {\bibfnamefont {R.}~\bibnamefont {{Carrera}}}, \bibinfo {author}
  {\bibfnamefont {S.~D.}\ \bibnamefont {{Chojnowski}}}, \bibinfo {author}
  {\bibfnamefont {K.}~\bibnamefont {{Cunha}}}, \bibinfo {author} {\bibfnamefont
  {C.}~\bibnamefont {{Epstein}}}, \bibinfo {author} {\bibfnamefont
  {G.}~\bibnamefont {{Fitzgerald}}}, \bibinfo {author} {\bibfnamefont {A.~E.}\
  \bibnamefont {{Garc{\'{\i}}a P{\'e}rez}}}, \bibinfo {author} {\bibfnamefont
  {F.~R.}\ \bibnamefont {{Hearty}}}, \bibinfo {author} {\bibfnamefont
  {C.}~\bibnamefont {{Henderson}}}, \bibinfo {author} {\bibfnamefont {J.~A.}\
  \bibnamefont {{Holtzman}}}, \bibinfo {author} {\bibfnamefont {J.~A.}\
  \bibnamefont {{Johnson}}}, \bibinfo {author} {\bibfnamefont {C.~R.}\
  \bibnamefont {{Lam}}}, \bibinfo {author} {\bibfnamefont {J.~E.}\ \bibnamefont
  {{Lawler}}}, \bibinfo {author} {\bibfnamefont {P.}~\bibnamefont {{Maseman}}},
  \bibinfo {author} {\bibfnamefont {S.}~\bibnamefont {{M{\'e}sz{\'a}ros}}},
  \bibinfo {author} {\bibfnamefont {M.}~\bibnamefont {{Nelson}}}, \bibinfo
  {author} {\bibfnamefont {D.~C.}\ \bibnamefont {{Nguyen}}}, \bibinfo {author}
  {\bibfnamefont {D.~L.}\ \bibnamefont {{Nidever}}}, \bibinfo {author}
  {\bibfnamefont {M.}~\bibnamefont {{Pinsonneault}}}, \bibinfo {author}
  {\bibfnamefont {M.}~\bibnamefont {{Shetrone}}}, \bibinfo {author}
  {\bibfnamefont {S.}~\bibnamefont {{Smee}}}, \bibinfo {author} {\bibfnamefont
  {V.~V.}\ \bibnamefont {{Smith}}}, \bibinfo {author} {\bibfnamefont
  {T.}~\bibnamefont {{Stolberg}}}, \bibinfo {author} {\bibfnamefont {M.~F.}\
  \bibnamefont {{Skrutskie}}}, \bibinfo {author} {\bibfnamefont
  {E.}~\bibnamefont {{Walker}}}, \bibinfo {author} {\bibfnamefont {J.~C.}\
  \bibnamefont {{Wilson}}}, \bibinfo {author} {\bibfnamefont {G.}~\bibnamefont
  {{Zasowski}}}, \bibinfo {author} {\bibfnamefont {F.}~\bibnamefont
  {{Anders}}}, \bibinfo {author} {\bibfnamefont {S.}~\bibnamefont {{Basu}}},
  \bibinfo {author} {\bibfnamefont {S.}~\bibnamefont {{Beland}}}, \bibinfo
  {author} {\bibfnamefont {M.~R.}\ \bibnamefont {{Blanton}}}, \bibinfo {author}
  {\bibfnamefont {J.}~\bibnamefont {{Bovy}}}, \bibinfo {author} {\bibfnamefont
  {J.~R.}\ \bibnamefont {{Brownstein}}}, \bibinfo {author} {\bibfnamefont
  {J.}~\bibnamefont {{Carlberg}}}, \bibinfo {author} {\bibfnamefont
  {W.}~\bibnamefont {{Chaplin}}}, \bibinfo {author} {\bibfnamefont
  {C.}~\bibnamefont {{Chiappini}}}, \bibinfo {author} {\bibfnamefont {D.~J.}\
  \bibnamefont {{Eisenstein}}}, \bibinfo {author} {\bibfnamefont
  {Y.}~\bibnamefont {{Elsworth}}}, \bibinfo {author} {\bibfnamefont
  {D.}~\bibnamefont {{Feuillet}}}, \bibinfo {author} {\bibfnamefont {S.~W.}\
  \bibnamefont {{Fleming}}}, \bibinfo {author} {\bibfnamefont {J.}~\bibnamefont
  {{Galbraith-Frew}}}, \bibinfo {author} {\bibfnamefont {R.~A.}\ \bibnamefont
  {{Garc{\'{\i}}a}}}, \bibinfo {author} {\bibfnamefont {D.~A.}\ \bibnamefont
  {{Garc{\'{\i}}a-Hern{\'a}ndez}}}, \bibinfo {author} {\bibfnamefont {B.~A.}\
  \bibnamefont {{Gillespie}}}, \bibinfo {author} {\bibfnamefont
  {L.}~\bibnamefont {{Girardi}}}, \bibinfo {author} {\bibfnamefont {J.~E.}\
  \bibnamefont {{Gunn}}}, \bibinfo {author} {\bibfnamefont {S.}~\bibnamefont
  {{Hasselquist}}}, \bibinfo {author} {\bibfnamefont {M.~R.}\ \bibnamefont
  {{Hayden}}}, \bibinfo {author} {\bibfnamefont {S.}~\bibnamefont {{Hekker}}},
  \bibinfo {author} {\bibfnamefont {I.}~\bibnamefont {{Ivans}}}, \bibinfo
  {author} {\bibfnamefont {K.}~\bibnamefont {{Kinemuchi}}}, \bibinfo {author}
  {\bibfnamefont {M.}~\bibnamefont {{Klaene}}}, \bibinfo {author}
  {\bibfnamefont {S.}~\bibnamefont {{Mahadevan}}}, \bibinfo {author}
  {\bibfnamefont {S.}~\bibnamefont {{Mathur}}}, \bibinfo {author}
  {\bibfnamefont {B.}~\bibnamefont {{Mosser}}}, \bibinfo {author}
  {\bibfnamefont {D.}~\bibnamefont {{Muna}}}, \bibinfo {author} {\bibfnamefont
  {J.~A.}\ \bibnamefont {{Munn}}}, \bibinfo {author} {\bibfnamefont {R.~C.}\
  \bibnamefont {{Nichol}}}, \bibinfo {author} {\bibfnamefont {R.~W.}\
  \bibnamefont {{O'Connell}}}, \bibinfo {author} {\bibfnamefont {J.~K.}\
  \bibnamefont {{Parejko}}}, \bibinfo {author} {\bibfnamefont {A.~C.}\
  \bibnamefont {{Robin}}}, \bibinfo {author} {\bibfnamefont {H.}~\bibnamefont
  {{Rocha-Pinto}}}, \bibinfo {author} {\bibfnamefont {M.}~\bibnamefont
  {{Schultheis}}}, \bibinfo {author} {\bibfnamefont {A.~M.}\ \bibnamefont
  {{Serenelli}}}, \bibinfo {author} {\bibfnamefont {N.}~\bibnamefont
  {{Shane}}}, \bibinfo {author} {\bibfnamefont {V.}~\bibnamefont {{Silva
  Aguirre}}}, \bibinfo {author} {\bibfnamefont {J.~S.}\ \bibnamefont
  {{Sobeck}}}, \bibinfo {author} {\bibfnamefont {B.}~\bibnamefont
  {{Thompson}}}, \bibinfo {author} {\bibfnamefont {N.~W.}\ \bibnamefont
  {{Troup}}}, \bibinfo {author} {\bibfnamefont {D.~H.}\ \bibnamefont
  {{Weinberg}}}, \ and\ \bibinfo {author} {\bibfnamefont {O.}~\bibnamefont
  {{Zamora}}},\ }\href {\doibase 10.3847/1538-3881/aa784d} {\bibfield
  {journal} {\bibinfo  {journal} {\aj}\ }\textbf {\bibinfo {volume} {154}},\
  \bibinfo {eid} {94} (\bibinfo {year} {2017})},\ \Eprint
  {http://arxiv.org/abs/1509.05420} {arXiv:1509.05420 [astro-ph.IM]}
  \BibitemShut {NoStop}%
\bibitem [{\citenamefont {{Wilson}}\ \emph {et~al.}(2019)\citenamefont
  {{Wilson}}, \citenamefont {{Hearty}}, \citenamefont {{Skrutskie}},
  \citenamefont {{Majewski}}, \citenamefont {{Holtzman}}, \citenamefont
  {{Eisenstein}}, \citenamefont {{Gunn}}, \citenamefont {{Blank}},
  \citenamefont {{Henderson}}, \citenamefont {{Smee}}, \citenamefont
  {{Nelson}}, \citenamefont {{Nidever}}, \citenamefont {{Arns}}, \citenamefont
  {{Barkhouser}}, \citenamefont {{Barr}}, \citenamefont {{Beland}},
  \citenamefont {{Bershady}}, \citenamefont {{Blanton}}, \citenamefont
  {{Brunner}}, \citenamefont {{Burton}}, \citenamefont {{Carey}}, \citenamefont
  {{Carr}}, \citenamefont {{Colque}}, \citenamefont {{Crane}}, \citenamefont
  {{Damke}}, \citenamefont {{Davidson}}, \citenamefont {{Dean}}, \citenamefont
  {{Di Mille}}, \citenamefont {{Don}}, \citenamefont {{Ebelke}}, \citenamefont
  {{Evans}}, \citenamefont {{Fitzgerald}}, \citenamefont {{Gillespie}},
  \citenamefont {{Hall}}, \citenamefont {{Harding}}, \citenamefont {{Harding}},
  \citenamefont {{Hammond}}, \citenamefont {{Hancock}}, \citenamefont
  {{Harrison}}, \citenamefont {{Hope}}, \citenamefont {{Horne}}, \citenamefont
  {{Karakla}}, \citenamefont {{Lam}}, \citenamefont {{Leger}}, \citenamefont
  {{MacDonald}}, \citenamefont {{Maseman}}, \citenamefont {{Matsunari}},
  \citenamefont {{Melton}}, \citenamefont {{Mitcheltree}}, \citenamefont
  {{O'Brien}}, \citenamefont {{O'Connell}}, \citenamefont {{Patten}},
  \citenamefont {{Richardson}}, \citenamefont {{Rieke}}, \citenamefont
  {{Rieke}}, \citenamefont {{Roman-Lopes}}, \citenamefont {{Schiavon}},
  \citenamefont {{Sobeck}}, \citenamefont {{Stolberg}}, \citenamefont
  {{Stoll}}, \citenamefont {{Tembe}}, \citenamefont {{Trujillo}}, \citenamefont
  {{Uomoto}}, \citenamefont {{Vernieri}}, \citenamefont {{Walker}},
  \citenamefont {{Weinberg}}, \citenamefont {{Young}}, \citenamefont
  {{Anthony-Brumfield}}, \citenamefont {{Bizyaev}}, \citenamefont
  {{Breslauer}}, \citenamefont {{De Lee}}, \citenamefont {{Downey}},
  \citenamefont {{Halverson}}, \citenamefont {{Huehnerhoff}}, \citenamefont
  {{Klaene}}, \citenamefont {{Leon}}, \citenamefont {{Long}}, \citenamefont
  {{Mahadevan}}, \citenamefont {{Malanushenko}}, \citenamefont {{Nguyen}},
  \citenamefont {{Owen}}, \citenamefont {{S{\'a}nchez-Gallego}}, \citenamefont
  {{Sayres}}, \citenamefont {{Shane}}, \citenamefont {{Shectman}},
  \citenamefont {{Shetrone}}, \citenamefont {{Skinner}}, \citenamefont
  {{Stauffer}},\ and\ \citenamefont {{Zhao}}}]{wilson19a}%
  \BibitemOpen
  \bibfield  {author} {\bibinfo {author} {\bibfnamefont {J.~C.}\ \bibnamefont
  {{Wilson}}}, \bibinfo {author} {\bibfnamefont {F.~R.}\ \bibnamefont
  {{Hearty}}}, \bibinfo {author} {\bibfnamefont {M.~F.}\ \bibnamefont
  {{Skrutskie}}}, \bibinfo {author} {\bibfnamefont {S.~R.}\ \bibnamefont
  {{Majewski}}}, \bibinfo {author} {\bibfnamefont {J.~A.}\ \bibnamefont
  {{Holtzman}}}, \bibinfo {author} {\bibfnamefont {D.}~\bibnamefont
  {{Eisenstein}}}, \bibinfo {author} {\bibfnamefont {J.}~\bibnamefont
  {{Gunn}}}, \bibinfo {author} {\bibfnamefont {B.}~\bibnamefont {{Blank}}},
  \bibinfo {author} {\bibfnamefont {C.}~\bibnamefont {{Henderson}}}, \bibinfo
  {author} {\bibfnamefont {S.}~\bibnamefont {{Smee}}}, \bibinfo {author}
  {\bibfnamefont {M.}~\bibnamefont {{Nelson}}}, \bibinfo {author}
  {\bibfnamefont {D.}~\bibnamefont {{Nidever}}}, \bibinfo {author}
  {\bibfnamefont {J.}~\bibnamefont {{Arns}}}, \bibinfo {author} {\bibfnamefont
  {R.}~\bibnamefont {{Barkhouser}}}, \bibinfo {author} {\bibfnamefont
  {J.}~\bibnamefont {{Barr}}}, \bibinfo {author} {\bibfnamefont
  {S.}~\bibnamefont {{Beland}}}, \bibinfo {author} {\bibfnamefont {M.~A.}\
  \bibnamefont {{Bershady}}}, \bibinfo {author} {\bibfnamefont {M.~R.}\
  \bibnamefont {{Blanton}}}, \bibinfo {author} {\bibfnamefont {S.}~\bibnamefont
  {{Brunner}}}, \bibinfo {author} {\bibfnamefont {A.}~\bibnamefont {{Burton}}},
  \bibinfo {author} {\bibfnamefont {L.}~\bibnamefont {{Carey}}}, \bibinfo
  {author} {\bibfnamefont {M.}~\bibnamefont {{Carr}}}, \bibinfo {author}
  {\bibfnamefont {J.~P.}\ \bibnamefont {{Colque}}}, \bibinfo {author}
  {\bibfnamefont {J.}~\bibnamefont {{Crane}}}, \bibinfo {author} {\bibfnamefont
  {G.~J.}\ \bibnamefont {{Damke}}}, \bibinfo {author} {\bibfnamefont
  {J.}~\bibnamefont {{Davidson}}, \bibfnamefont {J.~W.}}, \bibinfo {author}
  {\bibfnamefont {J.}~\bibnamefont {{Dean}}}, \bibinfo {author} {\bibfnamefont
  {F.}~\bibnamefont {{Di Mille}}}, \bibinfo {author} {\bibfnamefont {K.~W.}\
  \bibnamefont {{Don}}}, \bibinfo {author} {\bibfnamefont {G.}~\bibnamefont
  {{Ebelke}}}, \bibinfo {author} {\bibfnamefont {M.}~\bibnamefont {{Evans}}},
  \bibinfo {author} {\bibfnamefont {G.}~\bibnamefont {{Fitzgerald}}}, \bibinfo
  {author} {\bibfnamefont {B.}~\bibnamefont {{Gillespie}}}, \bibinfo {author}
  {\bibfnamefont {M.}~\bibnamefont {{Hall}}}, \bibinfo {author} {\bibfnamefont
  {A.}~\bibnamefont {{Harding}}}, \bibinfo {author} {\bibfnamefont
  {P.}~\bibnamefont {{Harding}}}, \bibinfo {author} {\bibfnamefont
  {R.}~\bibnamefont {{Hammond}}}, \bibinfo {author} {\bibfnamefont
  {D.}~\bibnamefont {{Hancock}}}, \bibinfo {author} {\bibfnamefont
  {C.}~\bibnamefont {{Harrison}}}, \bibinfo {author} {\bibfnamefont
  {S.}~\bibnamefont {{Hope}}}, \bibinfo {author} {\bibfnamefont
  {T.}~\bibnamefont {{Horne}}}, \bibinfo {author} {\bibfnamefont
  {J.}~\bibnamefont {{Karakla}}}, \bibinfo {author} {\bibfnamefont
  {C.}~\bibnamefont {{Lam}}}, \bibinfo {author} {\bibfnamefont
  {F.}~\bibnamefont {{Leger}}}, \bibinfo {author} {\bibfnamefont
  {N.}~\bibnamefont {{MacDonald}}}, \bibinfo {author} {\bibfnamefont
  {P.}~\bibnamefont {{Maseman}}}, \bibinfo {author} {\bibfnamefont
  {J.}~\bibnamefont {{Matsunari}}}, \bibinfo {author} {\bibfnamefont
  {S.}~\bibnamefont {{Melton}}}, \bibinfo {author} {\bibfnamefont
  {T.}~\bibnamefont {{Mitcheltree}}}, \bibinfo {author} {\bibfnamefont
  {T.}~\bibnamefont {{O'Brien}}}, \bibinfo {author} {\bibfnamefont {R.~W.}\
  \bibnamefont {{O'Connell}}}, \bibinfo {author} {\bibfnamefont
  {A.}~\bibnamefont {{Patten}}}, \bibinfo {author} {\bibfnamefont
  {W.}~\bibnamefont {{Richardson}}}, \bibinfo {author} {\bibfnamefont
  {G.}~\bibnamefont {{Rieke}}}, \bibinfo {author} {\bibfnamefont
  {M.}~\bibnamefont {{Rieke}}}, \bibinfo {author} {\bibfnamefont
  {A.}~\bibnamefont {{Roman-Lopes}}}, \bibinfo {author} {\bibfnamefont {R.~P.}\
  \bibnamefont {{Schiavon}}}, \bibinfo {author} {\bibfnamefont {J.~S.}\
  \bibnamefont {{Sobeck}}}, \bibinfo {author} {\bibfnamefont {T.}~\bibnamefont
  {{Stolberg}}}, \bibinfo {author} {\bibfnamefont {R.}~\bibnamefont {{Stoll}}},
  \bibinfo {author} {\bibfnamefont {M.}~\bibnamefont {{Tembe}}}, \bibinfo
  {author} {\bibfnamefont {J.~D.}\ \bibnamefont {{Trujillo}}}, \bibinfo
  {author} {\bibfnamefont {A.}~\bibnamefont {{Uomoto}}}, \bibinfo {author}
  {\bibfnamefont {M.}~\bibnamefont {{Vernieri}}}, \bibinfo {author}
  {\bibfnamefont {E.}~\bibnamefont {{Walker}}}, \bibinfo {author}
  {\bibfnamefont {D.~H.}\ \bibnamefont {{Weinberg}}}, \bibinfo {author}
  {\bibfnamefont {E.}~\bibnamefont {{Young}}}, \bibinfo {author} {\bibfnamefont
  {B.}~\bibnamefont {{Anthony-Brumfield}}}, \bibinfo {author} {\bibfnamefont
  {D.}~\bibnamefont {{Bizyaev}}}, \bibinfo {author} {\bibfnamefont
  {B.}~\bibnamefont {{Breslauer}}}, \bibinfo {author} {\bibfnamefont
  {N.}~\bibnamefont {{De Lee}}}, \bibinfo {author} {\bibfnamefont
  {J.}~\bibnamefont {{Downey}}}, \bibinfo {author} {\bibfnamefont
  {S.}~\bibnamefont {{Halverson}}}, \bibinfo {author} {\bibfnamefont
  {J.}~\bibnamefont {{Huehnerhoff}}}, \bibinfo {author} {\bibfnamefont
  {M.}~\bibnamefont {{Klaene}}}, \bibinfo {author} {\bibfnamefont
  {E.}~\bibnamefont {{Leon}}}, \bibinfo {author} {\bibfnamefont
  {D.}~\bibnamefont {{Long}}}, \bibinfo {author} {\bibfnamefont
  {S.}~\bibnamefont {{Mahadevan}}}, \bibinfo {author} {\bibfnamefont
  {E.}~\bibnamefont {{Malanushenko}}}, \bibinfo {author} {\bibfnamefont
  {D.~C.}\ \bibnamefont {{Nguyen}}}, \bibinfo {author} {\bibfnamefont
  {R.}~\bibnamefont {{Owen}}}, \bibinfo {author} {\bibfnamefont {J.~R.}\
  \bibnamefont {{S{\'a}nchez-Gallego}}}, \bibinfo {author} {\bibfnamefont
  {C.}~\bibnamefont {{Sayres}}}, \bibinfo {author} {\bibfnamefont
  {N.}~\bibnamefont {{Shane}}}, \bibinfo {author} {\bibfnamefont {S.~A.}\
  \bibnamefont {{Shectman}}}, \bibinfo {author} {\bibfnamefont
  {M.}~\bibnamefont {{Shetrone}}}, \bibinfo {author} {\bibfnamefont
  {D.}~\bibnamefont {{Skinner}}}, \bibinfo {author} {\bibfnamefont
  {F.}~\bibnamefont {{Stauffer}}}, \ and\ \bibinfo {author} {\bibfnamefont
  {B.}~\bibnamefont {{Zhao}}},\ }\href {\doibase 10.1088/1538-3873/ab0075}
  {\bibfield  {journal} {\bibinfo  {journal} {Publications of the Astronomical
  Society of the Pacific}\ }\textbf {\bibinfo {volume} {131}},\ \bibinfo
  {pages} {055001} (\bibinfo {year} {2019})},\ \Eprint
  {http://arxiv.org/abs/1902.00928} {arXiv:1902.00928 [astro-ph.IM]}
  \BibitemShut {NoStop}%
\bibitem [{\citenamefont {{Merloni}}\ \emph {et~al.}(2012)\citenamefont
  {{Merloni}}, \citenamefont {{Predehl}}, \citenamefont {{Becker}},
  \citenamefont {{B{\"o}hringer}}, \citenamefont {{Boller}}, \citenamefont
  {{Brunner}}, \citenamefont {{Brusa}}, \citenamefont {{Dennerl}},
  \citenamefont {{Freyberg}}, \citenamefont {{Friedrich}}, \citenamefont
  {{Georgakakis}}, \citenamefont {{Haberl}}, \citenamefont {{Hasinger}},
  \citenamefont {{Meidinger}}, \citenamefont {{Mohr}}, \citenamefont
  {{Nandra}}, \citenamefont {{Rau}}, \citenamefont {{Reiprich}}, \citenamefont
  {{Robrade}}, \citenamefont {{Salvato}}, \citenamefont {{Santangelo}},
  \citenamefont {{Sasaki}}, \citenamefont {{Schwope}}, \citenamefont
  {{Wilms}},\ and\ \citenamefont {{German eROSITA Consortium}}}]{merloni12a}%
  \BibitemOpen
  \bibfield  {author} {\bibinfo {author} {\bibfnamefont {A.}~\bibnamefont
  {{Merloni}}}, \bibinfo {author} {\bibfnamefont {P.}~\bibnamefont
  {{Predehl}}}, \bibinfo {author} {\bibfnamefont {W.}~\bibnamefont {{Becker}}},
  \bibinfo {author} {\bibfnamefont {H.}~\bibnamefont {{B{\"o}hringer}}},
  \bibinfo {author} {\bibfnamefont {T.}~\bibnamefont {{Boller}}}, \bibinfo
  {author} {\bibfnamefont {H.}~\bibnamefont {{Brunner}}}, \bibinfo {author}
  {\bibfnamefont {M.}~\bibnamefont {{Brusa}}}, \bibinfo {author} {\bibfnamefont
  {K.}~\bibnamefont {{Dennerl}}}, \bibinfo {author} {\bibfnamefont
  {M.}~\bibnamefont {{Freyberg}}}, \bibinfo {author} {\bibfnamefont
  {P.}~\bibnamefont {{Friedrich}}}, \bibinfo {author} {\bibfnamefont
  {A.}~\bibnamefont {{Georgakakis}}}, \bibinfo {author} {\bibfnamefont
  {F.}~\bibnamefont {{Haberl}}}, \bibinfo {author} {\bibfnamefont
  {G.}~\bibnamefont {{Hasinger}}}, \bibinfo {author} {\bibfnamefont
  {N.}~\bibnamefont {{Meidinger}}}, \bibinfo {author} {\bibfnamefont
  {J.}~\bibnamefont {{Mohr}}}, \bibinfo {author} {\bibfnamefont
  {K.}~\bibnamefont {{Nandra}}}, \bibinfo {author} {\bibfnamefont
  {A.}~\bibnamefont {{Rau}}}, \bibinfo {author} {\bibfnamefont {T.~H.}\
  \bibnamefont {{Reiprich}}}, \bibinfo {author} {\bibfnamefont
  {J.}~\bibnamefont {{Robrade}}}, \bibinfo {author} {\bibfnamefont
  {M.}~\bibnamefont {{Salvato}}}, \bibinfo {author} {\bibfnamefont
  {A.}~\bibnamefont {{Santangelo}}}, \bibinfo {author} {\bibfnamefont
  {M.}~\bibnamefont {{Sasaki}}}, \bibinfo {author} {\bibfnamefont
  {A.}~\bibnamefont {{Schwope}}}, \bibinfo {author} {\bibfnamefont
  {J.}~\bibnamefont {{Wilms}}}, \ and\ \bibinfo {author} {\bibfnamefont
  {t.}~\bibnamefont {{German eROSITA Consortium}}},\ }\href@noop {} {\bibfield
  {journal} {\bibinfo  {journal} {ArXiv e-prints}\ ,\ \bibinfo {eid}
  {arXiv:1209.3114}} (\bibinfo {year} {2012})},\ \Eprint
  {http://arxiv.org/abs/1209.3114} {arXiv:1209.3114 [astro-ph.HE]} \BibitemShut
  {NoStop}%
\bibitem [{\citenamefont {{Stubbs}}\ \emph {et~al.}(2004)\citenamefont
  {{Stubbs}}, \citenamefont {{Sweeney}}, \citenamefont {{Tyson}},\ and\
  \citenamefont {{LSST}}}]{stubbs04a}%
  \BibitemOpen
  \bibfield  {author} {\bibinfo {author} {\bibfnamefont {C.~W.}\ \bibnamefont
  {{Stubbs}}}, \bibinfo {author} {\bibfnamefont {D.}~\bibnamefont {{Sweeney}}},
  \bibinfo {author} {\bibfnamefont {J.~A.}\ \bibnamefont {{Tyson}}}, \ and\
  \bibinfo {author} {\bibnamefont {{LSST}}},\ }in\ \href@noop {} {\emph
  {\bibinfo {booktitle} {Bulletin of the American Astronomical Society}}},\
  Vol.~\bibinfo {volume} {36}\ (\bibinfo {year} {2004})\ pp.\ \bibinfo {pages}
  {1527--+}\BibitemShut {NoStop}%
\end{thebibliography}%

\end{document}